\documentclass[12pt,a4paper,openany,twoside,fontset=fandol]{ctexbook}
\usepackage{graphicx}
\usepackage{color}
\usepackage{xcolor}
\definecolor{ceruleanblue}{rgb}{0.16, 0.32, 0.75}
\usepackage[CJKbookmarks=true,
            bookmarksnumbered=true,
            bookmarksopen=true,
            colorlinks=true,
            citecolor=ceruleanblue,
            linkcolor=ceruleanblue,
            anchorcolor=ceruleanblue,
            urlcolor=ceruleanblue
            ]{hyperref}
\usepackage{slashed}

\usepackage{subcaption}

\usepackage{indentfirst}

\usepackage[a4paper,left=2.6cm,right=2.6cm,top=3.15cm,headsep=0.4cm,headheight=0.75cm,,bottom=2.5cm,footskip=0.75cm,footnotesep=0.6cm]{geometry}

\usepackage{amsmath}
\usepackage{amssymb}

\usepackage{bm}

\usepackage[amsmath,thmmarks]{ntheorem}

\usepackage[below]{placeins}

\usepackage{type1cm}

\usepackage{titletoc}

\usepackage[perpage,symbol*,bottom]{footmisc}

\usepackage{fancyhdr}
\usepackage{fancyref}

\usepackage{cite}

\usepackage{caption} 

\usepackage{setspace}

\usepackage{newtxmath}
\usepackage{eso-pic}
\usepackage{utfsym}

\usepackage{url}
\usepackage{makeidx}

\usepackage[squaren]{SIunits}

\usepackage{multirow}

\usepackage{fancybox}

\usepackage{pifont}

\usepackage{pdfpages}

\xeCJKsetup{boldfont,EmboldenFactor=2}
\usepackage{CJKnumb}
\usepackage{fontspec}
\xeCJKsetwidth{[}{0.2em}

\DeclareMathAlphabet{\mathsfsl}{OT1}{cmss}{m}{sl}

\usepackage{cjkhl}
\usepackage{booktabs}
\usepackage{color}
\usepackage{multirow}
\usepackage[sort]{gbt7714}
\usepackage{setspace}  
\usepackage{listings}
\usepackage{bbm}
\usepackage{tikz}
\usetikzlibrary{arrows.meta}
\usetikzlibrary{positioning}
\usetikzlibrary{calc}
\usetikzlibrary{decorations}
\usetikzlibrary{decorations.markings}
\tikzset{global scale/.style={scale=#1, every node/.append style={scale=#1}}}
\tikzset{
myarrowc/.style={ultra thick, postaction={decorate, decoration={markings, mark=at position 0.6 with {\arrow{latex}}}}},
myarrowl/.style={thick, postaction={decorate, decoration={markings, mark=at position 0.6 with {\arrow{latex}}}}},
}
\usepackage{mathtools}
\usepackage{xspace}
\newcommand{\chapref}[1]{\ifhmode\unskip~\fi\hyperref[#1]{Chapter~\ref*{#1}}\xspace}
\newcommand{\mybf}[1]{{\bfseries\color{ceruleanblue}#1}}
\newcommand{\thesislistignorecite}[2][]{\unskip}
\allowdisplaybreaks

\begin{document}

\allowdisplaybreaks[4]

\renewcommand{\textfraction}{0.15}
\renewcommand{\topfraction}{0.85}
\renewcommand{\bottomfraction}{0.65}
\renewcommand{\floatpagefraction}{0.60}

\setlength{\abovedisplayskip}{6pt plus 2pt minus 2pt}
\setlength{\abovedisplayshortskip}{6pt plus 2pt minus 2pt}
\setlength{\belowdisplayskip}{6pt plus 2pt minus 2pt}
\setlength{\belowdisplayshortskip}{6pt plus 2pt minus 2pt}




\newcommand{\song}{\songti}                                  

\newcommand{\fs}{\fangsong}                                  

\newcommand{\kai}{\kaishu}                                   

\newcommand{\yh}{\heiti}                                    

\newcommand{\hei}{\heiti}                                   


\newcommand{\typethesis}{\zihao{-0}\hei}      
\newcommand{\typetitle}{\fontsize{26pt}{50pt}\hei\bfseries\selectfont}     
\newcommand{\typetimu}{\fontsize{22pt}{50pt}\selectfont}     
\newcommand{\typename}{\fontsize{16pt}{32pt}\fs\selectfont}         
\newcommand{\typexingming}{\fontsize{15pt}{32pt}\hei\selectfont}    
\newcommand{\typedate}{\zihao{3}}         

\newcommand{\typetitleen}{\fontsize{26pt}{48pt}\sffamily\bfseries\selectfont}     

\newcommand{\typeenglish}{\fontsize{16pt}{22pt}\selectfont}         
\newcommand{\typeenglishlarge}{\fontsize{16pt}{34pt}\selectfont}         
\newcommand{\typetitleab}{\fontsize{16pt}{32pt}\sffamily\selectfont}     
\newcommand{\typeab}{\fontsize{12pt}{12pt}\sffamily\selectfont}     

\newcommand{\typetext}{\fontsize{12pt}{20pt}\selectfont}       
\newcommand{\typechap}{\zihao{3}\linespread{1.64}\hei\bfseries}         
\newcommand{\typesec}{\zihao{4}\linespread{1.2}\hei}          
\newcommand{\typesubsecctex}{\fontsize{13pt}{13pt}\linespread{1.2}\hei} 
\newcommand{\typesubsubsecctex}{\fontsize{12pt}{12pt}\linespread{1.2}\hei} 

\newcommand{\typeyemei}{\fontsize{10.5pt}{20pt}\selectfont}      
\newcommand{\typeyemeien}{\fontsize{10.5pt}{20pt}\selectfont}      
\newcommand{\typeyejiao}{\fontsize{10.5pt}{20pt}\selectfont}          
\newcommand{\typefootnote}{\fontsize{9pt}{15.7pt}\selectfont}          
\newcommand{\typefootscript}{\fontsize{8pt}{8pt}\selectfont}          
\newcommand{\typetable}{\small\renewcommand\arraystretch{1.5}}          
\newcommand{\typebib}{\fontsize{10.5pt}{16pt}\selectfont}      

\newif\ifthesisinappendix
\let\thesisoldappendix\appendix
\renewcommand{\appendix}{\global\thesisinappendixtrue\thesisoldappendix}
\ctexset{
  contentsname   = {Contents},
  listfigurename = {List of Figures},
  listtablename  = {List of Tables},
  figurename     = {Figure},
  tablename      = {Table},
  abstractname   = {Abstract},
  indexname      = {Index},
  bibname        = {References},
  appendixname   = {Appendix},
  proofname      = {Proof},
  chapter = {
    name   = {Chapter~,{}},
    number = \arabic{chapter},
    tocline = {\protect\numberline{\ifthesisinappendix APPENDIX~\Alph{chapter}\else CHAPTER~\arabic{chapter}\fi\hspace{.3em}}#2}
  }
}
\CTEXsetup[format={\typechap\centering},nameformat={\typechap},aftername={\hspace*{1em}},titleformat={\typechap},beforeskip={-2pt},afterskip={24pt}]{chapter}
\CTEXsetup[format={\typesec},nameformat={\typesec},aftername={\hspace*{1em}},titleformat={\typesec},beforeskip={24pt},afterskip={6pt}]{section}
\CTEXsetup[format={\typesubsecctex},nameformat={\typesubsecctex},aftername={\hspace*{1em}},titleformat={\typesubsecctex},beforeskip={12pt},afterskip={6pt}]{subsection}
\CTEXsetup[format={\typesubsubsecctex},nameformat={\typesubsubsecctex},aftername={\hspace*{1em}},titleformat={\typesubsubsecctex},beforeskip={12pt},afterskip={6pt}]{subsubsection}

\theoremstyle{plain} \theorembodyfont{\song\rmfamily}
\theoremheaderfont{\hei\rmfamily} \theoremseparator{\ }
\newtheorem{definition}{\hei Definition}[chapter]
\newtheorem{proposition}[definition]{\hei Proposition}
\newtheorem{assumption}[definition]{\hei Assumption}
\newtheorem{lemma}[definition]{\hei Lemma}
\newtheorem{remark}[definition]{\hei Remark}
\newtheorem{theorem}[definition]{\hei Theorem}
\newtheorem{axiom}{\hei Axiom}
\newtheorem{corollary}[definition]{\hei Corollary}
\newtheorem{exercise}[definition]{}

\theoremheaderfont{\CJKfamily{hei}\rmfamily}\theorembodyfont{\rmfamily}
\theoremstyle{nonumberplain} \theoremseparator{:}
\theoremsymbol{$\blacksquare$}
\newtheorem{proof}{\hei Proof}

\theoremsymbol{$\square$}
\newtheorem{example}{\hei Example}

\renewcommand{\thefootnote}{}
\renewcommand{\footnoterule}{\noindent \rule{5cm}{0.6pt} \vspace*{6pt} }
\renewcommand{\footnotesize}{\typefootnote}
\renewcommand{\footnotesep}{11pt}

\newcommand\footnotecircle[1]{\footnote{\hangindent 1.5em \hspace*{-2.2em} \ding{\numexpr171+\value{footnote}}\hspace*{0.4em} #1}\textsuperscript{\typefootscript \ding{\numexpr171+\value{footnote}}}}


\setcounter{secnumdepth}{4}
\setcounter{tocdepth}{2}

\dottedcontents{chapter}[6.5em]{\hei\vspace{6pt}}{6.5em}{10pt}
\dottedcontents{section}[4.2em]{}{2.2em}{5pt}
\dottedcontents{subsection}[7em]{}{3em}{5pt}






\newcommand{\makeheadrule}{%
   \hrule width\headwidth height0.75pt}

\pagestyle{fancyplain}


\fancyhf{}

\makeatletter
\def\cleardoublepage{\clearpage\if@twoside \ifodd\c@page\else%
    \hbox{}%
    \thispagestyle{empty}
    \newpage%
    \if@twocolumn\hbox{}\newpage\fi\fi\fi}


\setlength{\parskip}{0pt}

\renewcommand{\baselinestretch}{1}

\let\orig@Itemize =\itemize
\let\orig@Enumerate =\enumerate
\let\orig@Description =\description

\def\Myspacing{\itemsep=2ex \topsep=-4ex \partopsep=-2ex \parskip=-1ex \parsep=2ex}

\def\newitemsep{
\renewenvironment{itemize}{\orig@Itemize\Myspacing}{\endlist}
\renewenvironment{enumerate}{\orig@Enumerate\Myspacing}{\endlist}
\renewenvironment{description}{\orig@Description\Myspacing}{\endlist}
}

\def\olditemsep{
\renewenvironment{itemize}{\orig@Itemize}{\endlist}
\renewenvironment{enumerate}{\orig@Enumerate}{\endlist}
\renewenvironment{description}{\orig@Description}{\endlist}
}

\newitemsep



\renewcommand{\@openbib@code}{\addtolength{\itemsep}{-7pt}}

\newcommand{\ucite}[1]{\textsuperscript{\mbox{\scriptsize \cite{#1}}}}

%
%

\setlength{\heavyrulewidth}{2.25pt}
\setlength{\lightrulewidth}{0.75pt}
\setlength{\abovetopsep}{0pt}
\setlength{\belowbottomsep}{-6pt}




\captionsetup[table]{labelsep=quad,format=hang,justification=centerlast,font={small,stretch=1.16},position=top,belowskip={15pt-\intextsep},aboveskip=6pt}

\captionsetup[figure]{labelsep=quad,format=hang,justification=centerlast,font={small,stretch=1.16},position=bottom,belowskip={10pt-\intextsep},aboveskip=6pt}

\captionsetup[subfigure]{labelsep=quad,format=hang,justification=centerlast,font={small,stretch=1.16},position=bottom,belowskip=0pt,aboveskip=6pt}


%

\newsavebox{\AphorismAuthor}
\newenvironment{Aphorism}[1]
{\vspace{0.5cm}\begin{sloppypar} \slshape
\sbox{\AphorismAuthor}{#1}
\begin{quote}\small\itshape }
{\\ \hspace*{\fill}------\hspace{0.2cm} \usebox{\AphorismAuthor}
\end{quote}
\end{sloppypar}\vspace{0.5cm}}

\newcommand{\comment}[1]{}

\newcommand{\longer}[2]{#1}

\newcommand{\ds}{\displaystyle}

\def\gs{1.0}

\newcounter{newlist} 
\newenvironment{denselist}[1][temp]{
\begin{list}{\textbf{#1} (\arabic{newlist})} 
    {
    \usecounter{newlist}
     \setlength{\labelwidth}{22pt} 
     \setlength{\labelsep}{0cm} 
     \setlength{\leftmargin}{0cm} 
     \setlength{\rightmargin}{0cm}
     \setlength{\parsep}{0ex} 
     \setlength{\itemsep}{0ex} 
     \setlength{\itemindent}{44pt} 
     \setlength{\listparindent}{44pt} 
    }}
{\end{list}}



\def\ctitlefirstline#1{\def\@ctitlefirstline{#1}}\def\@ctitlefirstline{}
\def\ctitlesecondline#1{\def\@ctitlesecondline{#1}}\def\@ctitlesecondline{}
\def\cdegree#1{\def\@cdegree{#1}}\def\@cdegree{}
\def\caffil#1{\def\@caffil{#1}}\def\@caffil{}
\def\csubject#1{\def\@csubject{#1}}\def\@csubject{}
\def\cauthor#1{\def\@cauthor{#1}}\def\@cauthor{}
\def\cnumber#1{\def\@cnumber{#1}}\def\@cnumber{}
\def\csubsub#1{\def\@csubsub{#1}}\def\@csubsub{}
\def\csupervisor#1{\def\@csupervisor{#1}}\def\@csupervisor{}
\def\cassosupervisor#1{\def\@cassosupervisor{~ & Associate Supervisor &:& #1\\}}\def\@cassosupervisor{}
\def\ccosupervisor#1{\def\@ccosupervisor{~ & Co-supervisor &:& #1\\}}\def\@ccosupervisor{}
\def\cdate#1{\def\@cdate{#1}}\def\@cdate{}
\long\def\cabstract#1{\long\def\@cabstract{#1}}\long\def\@cabstract{}
\def\ckeywords#1{\def\@ckeywords{#1}}\def\@ckeywords{}

\def\etitle#1{\def\@etitle{#1}}\def\@etitle{}
\def\edegree#1{\def\@edegree{#1}}\def\@edegree{}
\def\eaffil#1{\def\@eaffil{#1}}\def\@eaffil{}
\def\esubject#1{\def\@esubject{#1}}\def\@esubject{}
\def\edepartment#1{\def\@edepartment{#1}}\def\@edepartment{}
\def\eauthor#1{\def\@eauthor{#1}}\def\@eauthor{}
\def\exuewei#1{\def\@exuewei{#1}}\def\@exuewei{}
\def\esupervisor#1{\def\@esupervisor{#1}}\def\@esupervisor{}
\def\eassosupervisor#1{\def\@eassosupervisor{Associate Supervisor: & #1\\}}\def\@eassosupervisor{}
\def\ecosupervisor#1{\def\@ecosupervisor{~ & #1\\}}\def\@ecosupervisor{}
\def\edate#1{\def\@edate{#1}}\def\@edate{}
\long\def\eabstract#1{\long\def\@eabstract{#1}}\long\def\@eabstract{}
\def\ekeywords#1{\def\@ekeywords{#1}}\def\@ekeywords{}

\def\makecover{
    \begin{titlepage}
    \newpage
    \thispagestyle{empty}

    \begin{center}
      \vspace*{-1.2cm}
      {\typetitleen
      \begin{center} { \@etitle} \end{center}}
      \vspace*{1.8cm}
      {\typeenglish
      Dissertation Submitted to \\
      \textbf{Peking University} \\
      \vspace*{0.6cm}
      \textbf{In partial fulfillment of the requirements for the degree of} \\
      \@edegree \\
      \vspace*{2cm}
      By \\
      \textbf{\@eauthor}\textbf{\@exuewei}
      \\
      \textbf{(\@esubject)} \\}
      \vspace*{0.8cm}
      {\typeenglishlarge
       Dissertation Supervisor: \textbf{\@esupervisor} \\ \quad \\
      School of Physics \\
      Peking University \\
      \@edate, Beijing \\}
      \end{center}  

    \cleardoublepage
    \newpage
    \thispagestyle{empty}

    \begin{center} {
      \vspace*{3mm}
      \includegraphics[width=8.5cm]{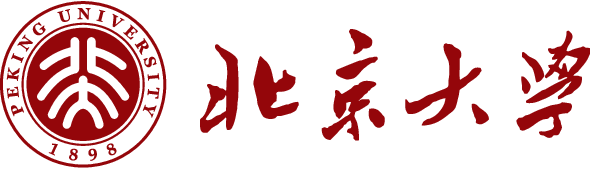} \\
      \vspace*{4mm}
      {\typethesis \@cdegree 研究生学位论文} \\
      \vspace*{8mm}
      {\typetimu \hspace*{-1cm}
      \begin{tabular}{lc}
      \raisebox{-2mm}{题目：} & {\typetitle \@ctitlefirstline} \\ \cline{2-2}
            & {\typetitle \@ctitlesecondline} \\ \cline{2-2}
      　 　 & {\hspace*{9cm}}　
      \end{tabular} } \\
      \vspace*{2.2cm}
      {\typexingming \hspace*{-2cm}
      \begin{tabular}{lc}
      姓\hphantom{姓名}名： \hspace*{1mm}& {\typename \@cauthor}\\ \cline{2-2}
      院\hphantom{姓名}系： & {\typename \@caffil}\\ \cline{2-2}
      专\hphantom{姓名}业： & {\typename \@csubject}\\ \cline{2-2}
      研究方向： & {\typename \@csubsub}\\ \cline{2-2}
      导师姓名： & {\typename \@csupervisor} \\ \cline{2-2}
      　 　 & {\hspace*{6.5cm}}　
      \end{tabular} } \\
     \typedate \usym{2611} 学术学位      \hspace{1.0cm}  \usym{2610} 专业学位\\
      \vspace*{1.2cm}
      \typedate \@cdate
      }
      \end{center}
      
    \cleardoublepage
     \newpage
    \thispagestyle{empty}
    \end{titlepage}
}

\def\makeabstract{
   \begin{titlepage}
   \newpage
    \thispagestyle{plain}

    \setcounter{page}{1}
    \markboth{\typeyemeien ABSTRACT}{\typeyemeien ABSTRACT}
   \begin{center}
    {
    \begin{center} 
    \end{center}}
    {\typeab \textbf{ABSTRACT}\footnotecircle{This research was supported by the National Natural Science Foundation of China Major Program "Key Scientific Problems in Lattice Quantum Chromodynamics Based on Domestic Supercomputing" (Grant Nos. 12293060, 12293061, and 12293063) and the National Natural Science Foundation of China Basic Research Program for Young Students (Doctoral Students) (Grant No. 124B2096).}\footnotecircle{This thesis was originally written in Chinese. The original Chinese version is available at Peking University Library, and this English translation will also be made publicly available on the arXiv preprint platform.}}
    \end{center}
    \vspace*{-2mm}
    \typetext
    \@eabstract

    \cleardoublepage
    \newpage

    \thispagestyle{plain}
    \chapter*{摘要 \markboth{摘要}{摘要}\footnotecircle{本研究得到国家自然科学基金重大项目“基于国产超算的格点量子色动力学关键科学问题研究”（编号：12293060、12293061、12293063）和国家自然科学基金青年学生基础研究项目（博士研究生）（编号：124B2096）的资助。}\footnotecircle{本论文中文版可于北京大学图书馆查阅，亦将译为英文版本提交至预印本平台 arXiv 公开发布。}}
    {\typetext \@cabstract }

    \cleardoublepage

     \end{titlepage}
}
\makeatother

\raggedbottom

\setlength{\multlinegap}{0mm}

\frontmatter
\fancyfoot[CO,CE]{\typeyejiao\thepage}
\sloppy
\pagenumbering{Roman}
\renewcommand{\headrulewidth}{0pt}
\renewcommand{\footrulewidth}{0pt}
%
%
%
%

\ctitlefirstline{多强子散射问题的} \cdegree{博士} \ctitlesecondline{格点量子色动力学研究}
\caffil{物理学院} \csubject{理论物理}
\cauthor{燕浩波} \cnumber{2101110113} \csubsub{粒子物理理论}
\csupervisor{刘川\ 教授}
\cdate{二〇二六年六月}

\etitle{Multi-Hadron Scattering from Lattice Quantum Chromodynamics}
\edegree{Doctor of Science}
\esubject{Theoretical Physics}
\eauthor{Haobo Yan}
\exuewei{, B. Sc.}
\esupervisor{Chuan Liu, Professor}
\edate{June, 2026}

\makecover

\includepdf[pages=-,fitpaper=true]{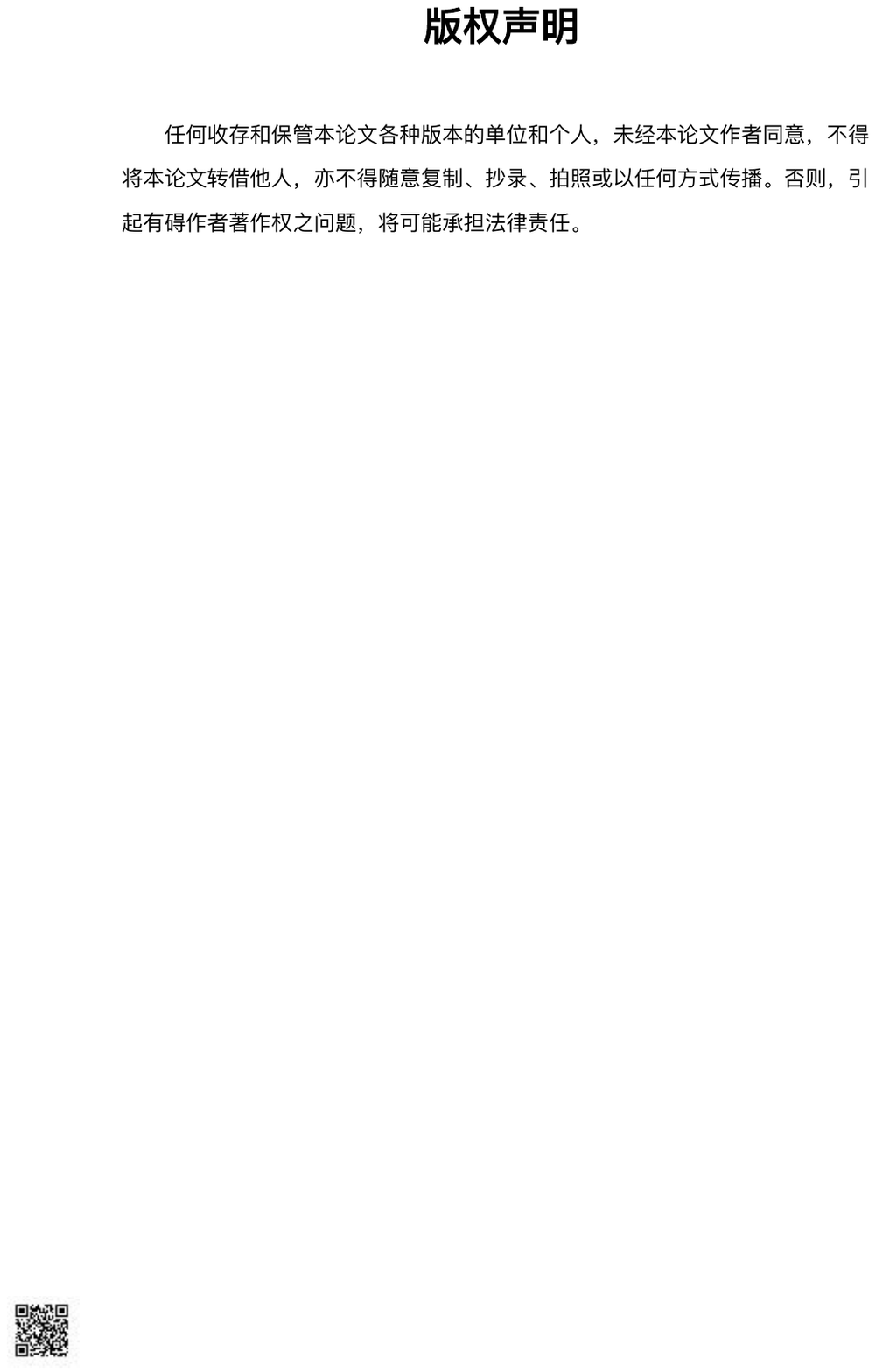}
\cleardoublepage
\renewcommand{\headrulewidth}{0.75pt}
\fancyhead[CE]{\typeyemei Doctoral Dissertation of Peking University}
\renewcommand{\chaptermark}[1]{\markboth{Chapter~\arabic{chapter}\enspace #1}{}}
\fancyhead[CO]{\typeyemei\leftmark}
\renewcommand{\headrule}{
{
    \makeheadrule}}
%
%
%
%

\newitemsep
\renewcommand{\labelenumi}{(\arabic{enumi})}
\cabstract{
\thispagestyle{plain}
在标准模型中，自然界的三种基本相互作用统一于量子场论之下。其中，强相互作用描述了质子、中子等强子内部结构以及夸克与胶子之间的动力学行为。在强子能标下，强相互作用耦合常数较大，导致传统微扰方法失效，从而使强子结构与相互作用的理论研究十分困难。格点量子色动力学（Lattice QCD）是目前唯一能够从第一性原理出发，系统研究强相互作用非微扰性质的理论工具。其计算误差是可控的，并能够随着计算资源与算法的进步不断系统改进。近年来，许多物理量的格点计算精度已可与实验结果相比，甚至在部分情形下超过实验精度，为实验与唯象模型提供了关键的理论输入。

强子性质的研究对于理解标准模型的动力学机制具有重要意义。在格点方法中，能谱的提取依赖于具有合适量子数的算符构造。本文开发了一个用于构造格点多体算符的 \texttt{Mathematica} 程序包 \texttt{OpTion}，简化了传统上繁琐的算符构造流程，并已在中国格点合作组中得到广泛应用。通过格点计算有限体积能谱，并结合量子化条件建立有限体积与无限体积之间的联系，研究衰变至两体末态的强子共振已成为可能。以粲物理为例，关于 $D_0^*(2300)$ 的极点结构长期存在争议，诸多理论研究认为该共振态应对应两个奇点。本文的相关工作通过对单道与多道 $D\pi$ 散射的格点研究，系统分析了极点随 $m_{\pi}$ 变化的轨迹，并将散射长度外推至物理点。多道分析结果支持了 $D_0^*(2300)$ 存在双极点结构的观点，为理解粲强子共振的内部结构提供了重要的第一性原理证据。

尽管两体问题的研究已取得显著进展，但其仅覆盖强子谱的一小部分，而绝大多数强子主要衰变至三个及以上粒子的末态。相比之下，三体问题的研究仍处于起步阶段，其困难主要来源于更加复杂的量子化条件、呈阶乘增长的计算复杂度，以及对两体散射输入的依赖。这些因素共同导致当前国际上利用格点 QCD 研究三体散射共振态的工作仍然十分有限，绝大多数三体格点研究集中于纯排斥相互作用体系。

本文考察了衰变至三个 $\pi$ 介子的 $\omega(782)$ 介子体系。该粒子是最轻的三体衰变强子，在光子-核子相互作用、重子电荷半径与同位旋标量电荷半径、缪子反常磁矩以及暗物质相关研究中具有重要作用。同时，作为同位旋标量，该体系的格点计算十分复杂。本文计算了相关的两体与三体能谱，发展了相应的三体量子化条件及有效场论描述，并通过求解积分方程提取共振极点位置，将其外推至物理点。最终得到的共振质量与衰变宽度与实验平均值一致。这一系列工作标志着人们在基于第一性原理研究相互作用三体问题方面取得了长足进展，并建立了配套的计算流程。

在此基础上，本文进一步尝试超越对已知共振态的事后验证。$\pi$ 介子作为自发手征对称性破缺产生的戈德斯通玻色子，在强相互作用理论中占据核心地位。然而，其首个径向激发态 $\pi(1300)$ 的存在性及性质至今仍存在争议。若该态确实存在，其质量约为基态 $\pi$ 介子的十倍，这一巨大的质量差为研究夸克禁闭机制提供了独特的窗口。此外，$\pi(1300)$ 也是少数具有显著三体相互作用效应的共振之一，可能包含混杂态成分，并在轻夸克质量的确定以及标准模型扩展中的 CP 破坏研究中具有潜在影响。本文发展了相应的三体量子化条件与有效场论方法，首次给出了该态在不同 $\pi$ 质量下共振极点位置的第一性原理预测。这为未来研究更困难的多体共振态奠定了重要的理论与计算基础。

此外，本文还研究了含粲介子的辐射衰变与半轻子衰变过程，发展了一种模型无关的跃迁因子提取方法，显著提高了跃迁矩阵元的计算精度，为高精度提取 CKM 矩阵元提供了新途径。



\vfill

关键词：格点量子色动力学，强子谱与强子结构，三体问题
}

%
%
%
%

%
%
\newitemsep
\renewcommand{\labelenumi}{(\arabic{enumi})}
\eabstract{
\thispagestyle{plain}

In the Standard Model, the three fundamental interactions of nature are unified within the framework of quantum field theory. Among them, the strong interaction governs the internal structure of hadrons such as protons and neutrons, as well as the dynamics between quarks and gluons. At hadronic energy scales, the large coupling constant renders perturbative methods inapplicable, making theoretical studies of hadron structure and interactions particularly challenging. Lattice Quantum Chromodynamics (Lattice QCD) is currently the only theoretical tool capable of systematically studying the non-perturbative properties of strong interactions from first principles. Its computational uncertainties are controllable and can be systematically improved with advances in computational resources and algorithms. In recent years, lattice calculations of many physical quantities have achieved precision comparable to, or even exceeding, experimental results, providing crucial theoretical input for experiments and phenomenological models.

The study of hadron properties is essential for understanding the dynamical mechanisms of the Standard Model. In lattice methods, spectral extraction relies on the construction of operators with appropriate quantum numbers. This thesis develops a \texttt{Mathematica} package \texttt{OpTion} for constructing lattice multi-hadron operators, which simplifies the traditionally tedious operator construction procedure and has been widely applied within the Chinese Lattice QCD Collaboration. By extracting finite-volume spectra from lattice calculations and establishing the connection between finite and infinite volumes through quantization conditions, it becomes possible to study hadron resonances decaying to two-body final states. Taking charm physics as an example, the pole structure of $D_0^*(2300)$ has long been controversial, with many theoretical studies suggesting that this resonance should correspond to two poles. The related work in this thesis systematically analyzes the pole trajectory as a function of $m_\pi$ through lattice studies of single-channel and coupled-channel $D\pi$ scattering, and extrapolates the scattering length to the physical point. The coupled-channel analysis supports the existence of a double-pole structure for $D_0^*(2300)$, providing important first-principles evidence for understanding the internal structure of charmed hadron resonances.

Although significant progress has been made in two-body problems, they cover only a small fraction of the hadron spectrum, while the majority of hadrons decay predominantly to final states with three or more particles. In contrast, research on three-body problems is still in its infancy, with difficulties arising from more complex quantization conditions, factorially growing computational complexity, and dependence on two-body scattering input. These factors collectively result in very limited lattice QCD studies of resonances in three-body scattering worldwide, with most three-body lattice studies focusing on systems with purely repulsive interactions.

This thesis examines the $\omega(782)$ meson system, which decays to three pions. This particle is the lightest hadron with a three-body decay, playing important roles in photon-nucleon interactions, baryon charge radii and isoscalar charge radii, the muon anomalous magnetic moment, and dark matter-related research. As an isoscalar system, its lattice calculation is highly complex. This thesis computes the relevant two-body and three-body spectra, develops the corresponding three-body quantization conditions and effective field theory descriptions, and extracts the resonance pole position by solving integral equations before extrapolating it to the physical point. The resulting resonance mass and decay width are consistent with experimental averages. This series of work marks significant progress in studying interacting three-body problems from first principles and establishes the corresponding computational procedures.

Building on this foundation, this thesis further attempts to go beyond post-diction of known resonances. As a Goldstone boson arising from spontaneous chiral symmetry breaking, the pion occupies a central position in strong interaction theory. However, the existence and properties of its first radial excitation, $\pi(1300)$, remain controversial. If this state indeed exists, its mass is approximately ten times that of the ground-state pion, providing a unique window for studying quark confinement mechanisms. Furthermore, $\pi(1300)$ is one of the few resonances with significant three-body interaction effects, may contain hybrid components, and has potential implications for the determination of light-quark masses and studies of CP violation in extensions of the Standard Model. This thesis develops the corresponding three-body quantization conditions and effective field theory methods and, for the first time, provides a first-principles prediction of the resonance pole position of this state at different pion masses. This lays important theoretical and computational foundations for future studies of more challenging multi-body resonances.

In addition, this thesis investigates radiative and semileptonic decays of charmed mesons, developing a model-independent method for extracting transition form factors that significantly improves the precision of transition matrix element calculations and provides a new avenue for the high-precision extraction of CKM matrix elements.

\vfill

}

\makeabstract

\renewcommand{\baselinestretch}{1}
\fontsize{12pt}{20pt}\selectfont

\tableofcontents

\cleardoublepage
\begingroup
\let\cite\thesislistignorecite
\listoftables
\endgroup
\thispagestyle{empty}

\begingroup
\let\cite\thesislistignorecite
\listoffigures
\endgroup
\thispagestyle{empty}


\cleardoublepage


\renewcommand{\labelenumi}{(\arabic{enumi})}
\newitemsep

\mainmatter
\renewcommand{\thechapter}{\arabic{chapter}}




\chapter{Introduction}
\label{chap:introduction}

{
\kaishu
\begin{center}
    道生一，一生二，二生三，三生万物。
\end{center}
\hfill ——《道德经》[春秋] 老子
}

\section{A Brief History of Particle Physics}
\mybf{Particle physics}, or \mybf{high-energy physics}, seeks to uncover the fundamental particles and interaction laws hidden beneath the diversity of natural phenomena. It is one of the deepest routes by which humankind has come to understand the structure of the universe. The development of high-energy physics forms one of the most magnificent chapters in the history of the exploration of nature. Over more than a century, physicists have repeatedly pushed the frontiers of energy and precision, while proposing new theories to explain the experimental phenomena that emerged. Each theoretical leap has been accompanied by major experimental breakthroughs, and each advance in experimental measurement has in turn driven further theoretical progress.

The story of modern particle physics began in 1897. In cathode-ray tube experiments, J. J. Thomson measured the charge-to-mass ratio of cathode rays by tuning electric and magnetic fields, and found it to be anomalously large. This led him to realize that the cathode ray was one of the constituents of the atom, thereby discovering the \mybf{electron}~\cite{Thomson:1897cm}\footnotecircle{The name ``electron'' was first proposed by Johnstone Stoney in 1891, originally meaning a unit of electricity, and was later used for this fundamental particle.}, and sounding the opening note of the particle-physics revolution\footnotecircle{When Thomson first announced this discovery at the Royal Society, more than half of the audience reportedly did not believe his conclusion.}. This discovery implied that the positive charge of the atom had to reside in other particles. After returning to Cambridge University, Thomson's student Ernest Rutherford discovered the atomic nucleus in 1911 through the gold-foil scattering experiment~\cite{Rutherford:1911zz}, revealing the internal structure of the atom. Between 1917 and 1925, Rutherford further confirmed the existence of the \mybf{proton} through a series of particle-bombardment experiments. In 1913, Bohr proposed the quantum model of the atom~\cite{Bohr:1913zba}, successfully explaining the spectrum of hydrogen. Studies of heavier nuclei then gradually showed that nuclei are composed of protons and neutrons. In 1932, Rutherford's student James Chadwick discovered the \mybf{neutron}~\cite{Chadwick:1932wcf}. By that point, a first unified picture of the structure of matter had emerged: all matter was composed of three elementary particles, protons, neutrons, and electrons\footnotecircle{At the energy scales of everyday life, this picture is already quite accurate. Many particles discovered later are not ordinarily present on Earth, and can only be produced by high-energy cosmic rays from distant astrophysical sources or in high-energy particle accelerators.}.

In 1928, Paul Dirac proposed the relativistic equation for the electron~\cite{Dirac:1928ej}, predicting the existence of antiparticles. In 1932, Carl Anderson discovered the \mybf{positron} in cosmic rays~\cite{Anderson:1932zza}, thereby confirming the existence of antimatter. This discovery marked the beginning of the era of quantum field theory. Quantum field theory combines quantum mechanics with special relativity and describes the creation and annihilation of elementary particles through the quantization of fields. Early quantum field theory, however, suffered from severe divergences in higher-order perturbative calculations. The Lamb shift, discovered in the fine structure of hydrogen by Willis Lamb and Robert Retherford in 1947~\cite{Lamb:1947zz}, showed that these divergences could not simply be ignored. Between 1947 and 1949, Sin-Itiro Tomonaga~\cite{Tomonaga:1946zz}, Julian Schwinger~\cite{Schwinger:1948yk}, and Richard Feynman~\cite{Feynman:1949hz} independently developed systematic renormalization methods, thereby establishing modern \mybf{quantum electrodynamics} (QED). QED describes the interaction between charged particles and photons, and successfully explains precision observables such as the anomalous magnetic moment of the electron and the Lamb shift, in agreement with experiment at extraordinarily high precision. It has therefore become one of the most successful and most precisely tested quantum field theories.

In 1930, while studying nuclear $\beta$ decay, Wolfgang Pauli proposed the concept of the \mybf{neutrino}~\cite{Pauli:1930pc}. In 1956, Cowan and Reines first observed neutrinos in a reactor experiment~\cite{Reines:1956rs}. In 1959, Raymond Davis used a related experiment to show lepton-number conservation, namely that neutrinos and antineutrinos are distinct particles\footnotecircle{This conclusion is not absolute. We still do not know whether the neutrino is its own antiparticle, that is, whether it is a Dirac or Majorana particle; this remains one of the important questions in contemporary particle physics.}. In 1934, Fermi proposed the celebrated four-fermion interaction theory to describe $\beta$ decay~\cite{Fermi:1934sk}, thereby establishing the earliest theoretical framework for the \mybf{weak interaction}\footnotecircle{At that time, the strong and weak interactions had not yet been distinguished. The very small coupling constant of Fermi's four-fermion theory was once viewed with suspicion.}.

In 1947 and 1949, George Rochester, Clifford Butler, and collaborators discovered the $\theta$ and $\tau$ particles in cosmic rays. Because their decay final states had different parities, they were initially thought to be different particles; nevertheless, their masses and lifetimes were identical. This became known as the $\theta$--$\tau$ puzzle. In 1956, Tsung-Dao Lee (李政道) and Chen-Ning Yang (杨振宁) proposed that parity might be violated in the weak interaction~\cite{Lee:1956qn}, a possibility confirmed in 1957 by the experiment of Chien-Shiung Wu (吴健雄) and collaborators~\cite{Wu:1957my}\footnotecircle{Both the theoretical proposal and experimental confirmation of parity violation involved decisive contributions by Chinese scientists.}. It was then understood that the $\theta$ and $\tau$ were in fact the same particle, the $K$ meson. The discovery of parity violation strongly stimulated the development of the theory of weak interactions. In 1958, Richard Feynman and Murray Gell-Mann~\cite{Gell-Mann:1958sai}, and independently Ennackal Sudarshan and Robert Marshak, proposed the $V-A$ structure of the weak interaction. In 1961, Sheldon Glashow proposed the \mybf{electroweak unification} framework based on an $SU(2)\times U(1)$ gauge symmetry~\cite{Glashow:1961ep}\footnotecircle{It is sometimes also called quantum flavordynamics (QFD).}.

At the same time, physicists realized that, under the $SU(2)\times U(1)$ electroweak symmetry, assigning masses directly to the \mybf{$W^\pm$ and $Z^0$ bosons} would break gauge invariance and render the theory nonrenormalizable. In 1964, Peter Higgs, François Englert, and their collaborators independently proposed the mechanism of spontaneous symmetry breaking (SSB)~\cite{Higgs:1964pj,Englert:1964et}. They showed that introducing a scalar field with a nonzero vacuum expectation value into a local gauge theory can give mass to gauge bosons while preserving the renormalizability of the theory, and predicted a new scalar particle associated with this mechanism, the \mybf{Higgs boson}. In 1967, Steven Weinberg incorporated this mechanism into electroweak unification~\cite{Weinberg:1967tq}, thereby giving masses to the $W^\pm$ and $Z^0$ bosons. In 1971, 't Hooft and Veltman proved that the electroweak theory with the Higgs mechanism is renormalizable~\cite{tHooft:1971akt}, establishing the theoretical foundation of the modern Standard Model. In 1983, the UA1 and UA2 experiments at CERN discovered the $W$ and $Z$ bosons in proton--antiproton collisions~\cite{Arnison:1983rp,UA2:1983tsx}, confirming the mechanism by which the electroweak gauge bosons acquire mass. In 2012, the ATLAS and CMS experiments at CERN announced the discovery of a new scalar particle with a mass of about $125\,\mathrm{GeV}$~\cite{ATLAS:2012:observation,CMS:2012qbp}. Its properties are highly consistent with those of the Standard Model Higgs boson, completing the experimental verification of the last elementary particle in the Standard Model.

On the other hand, physicists had long wondered why a strong short-range attraction exists between protons, overcoming electromagnetic repulsion and keeping nuclei stable. This interaction was called the \mybf{strong interaction}\footnotecircle{One may regard this as an example of physicists' limited talent for naming things.}. In 1935, Hideki Yukawa proposed the meson theory~\cite{Yukawa:1935}, in which the nuclear force arises from meson exchange between protons and neutrons. In 1947, C.~F.~Powell and collaborators discovered the $\pi$ meson in cosmic rays~\cite{Powell:1947}\footnotecircle{The name of the $\pi$ meson reflects its mass, which lies between the electron and proton masses.}\footnotecircle{We now know that the $\pi$ meson is a quark--antiquark bound state rather than an elementary particle. It is the Goldstone boson associated with spontaneous chiral symmetry breaking. In chiral perturbation theory, the $\pi$ meson is a low-energy effective field; in lattice quantum chromodynamics, it is the hadron with the cleanest signal and the most mature theoretical understanding; in high-energy experiments, it is also one of the most common hadronic decay products. The studies in this thesis are closely connected with the $\pi$ meson.}.

In 1952, the proton synchrotron at Brookhaven National Laboratory (BNL) was completed and brought into operation; it was the first modern high-energy particle accelerator. Experiments soon discovered a large number of new hadrons\footnotecircle{These are now collectively called hadrons, particles that participate in the strong interaction, including baryons and mesons.}, and particle physics entered the so-called ``particle zoo'' era\footnotecircle{The phrase ``particle zoo'' is said to have been first used by Julius Robert Oppenheimer at an international high-energy physics conference in 1956.}. The simple earlier picture of matter was broken, and physicists began to ask whether these particles were truly elementary. In his 1955 Nobel lecture, Lamb remarked that while the discovery of a new particle used to be rewarded with a Nobel Prize, now such a discovery perhaps ought to be punished by a ten-thousand-dollar fine.

In 1961, Gell-Mann and Yuval Ne'eman proposed the Eightfold Way based on $\mathrm{SU}(3)$ symmetry~\cite{GellMann:1961omu,Neeman:1961}, organizing hadrons into symmetry multiplets. In 1964, Gell-Mann and George Zweig independently proposed the \mybf{quark model}~\cite{Gell-Mann:1964ewy,Zweig:1964}, in which hadrons are composed of more elementary fermions, quarks. The original model contained three quark flavors: up, down, and strange. In 1970, Sheldon Glashow, John Iliopoulos, and Luciano Maiani proposed the GIM mechanism~\cite{Glashow:1970gm}, predicting the existence of the charm quark. The charm quark was discovered in 1974~\cite{E598:1974sol,SLAC-SP-017:1974ind}\footnotecircle{This event is known as the ``November Revolution''. The discovery of the $J/\psi$ greatly strengthened confidence in the quark model, and many charmed particles predicted by the quark model were subsequently observed experimentally. The spectrum of charmonium states can almost be placed in one-to-one correspondence with the energy levels of the hydrogen atom. These studies later also stimulated phenomenological theories such as nonrelativistic QCD (NRQCD), which remain active today.}, the bottom quark in 1977, and the top quark in 1995, leading to the six-flavor quark structure.

Nevertheless, all direct searches for free quarks failed. This phenomenon is known as \mybf{color confinement}. In 1973, Harald Fritzsch, Gell-Mann, and others proposed \mybf{quantum chromodynamics} (QCD), based on the $\mathrm{SU}(3)$ color symmetry and on the \mybf{non-Abelian gauge theory} introduced by Yang and Mills in 1954~\cite{Fritzsch:1973pi}. In the same year, David Gross, Frank Wilczek, and Hugh Politzer discovered the property of \mybf{asymptotic freedom}~\cite{Gross:1973ju,Politzer:1973fx}: the interaction becomes weak at high energies, while at low energies quarks are confined inside hadrons. Since then, extensive experimental and theoretical work has continued to confirm the predictions of QCD, establishing it as the fundamental theory of the strong interaction.

Around 1975, the gauge-symmetry-based theories describing the strong, electromagnetic, and weak interactions came to be known collectively as the \mybf{Standard Model}\footnotecircle{The origin of this term is disputed.}. All elementary particles predicted by the Standard Model have been discovered experimentally, and all experimental measurements to date are in excellent agreement with its predictions. It is therefore one of the most successful theoretical frameworks ever constructed for the microscopic world\footnotecircle{We emphasize that understanding elementary particles does not automatically imply understanding macroscopic systems, as captured by condensed-matter physicist Philip Anderson's famous statement, ``More is different.''}.

Dear reader, if you also find the history of high-energy physics poetic, allow the author to bring you closer to its more recent growth. Compared with the successive revolutionary discoveries of the last century, contemporary progress may appear calmer and more measured. Yet, driven by experiments of ever higher precision and theories of increasing depth, high-energy physics continues day by day to expand the boundary of human understanding. Each step may seem small, but each is solid, and each tells us a little more about the universe in which we live\footnotecircle{High-energy physics has already been associated with nearly thirty Nobel Prizes in Physics; it would be difficult even to list all of them here.}.

\section{The Standard Model of Particle Physics}
We have just introduced the language of modern particle physics and the most successful theory written in that language: quantum field theory and the Standard Model. They tell us that the fundamental building blocks of nature are $17$ quantum fields, namely $12$ matter fields, $4$ gauge fields, and the Higgs field. These quantum fields fill the spacetime of the universe like fluids, and particles are ripples in these fields. When one of the fields begins to oscillate and ripple, the corresponding particle is produced. These ripples then set in motion other fields with which they interact, and the universe performs a harmonious dance of interwoven, swaying fields\footnotecircle{This beautiful description is due to David Tong.}.

\begin{figure}[htbp]
    \centering
    \includegraphics[width=0.8\columnwidth]{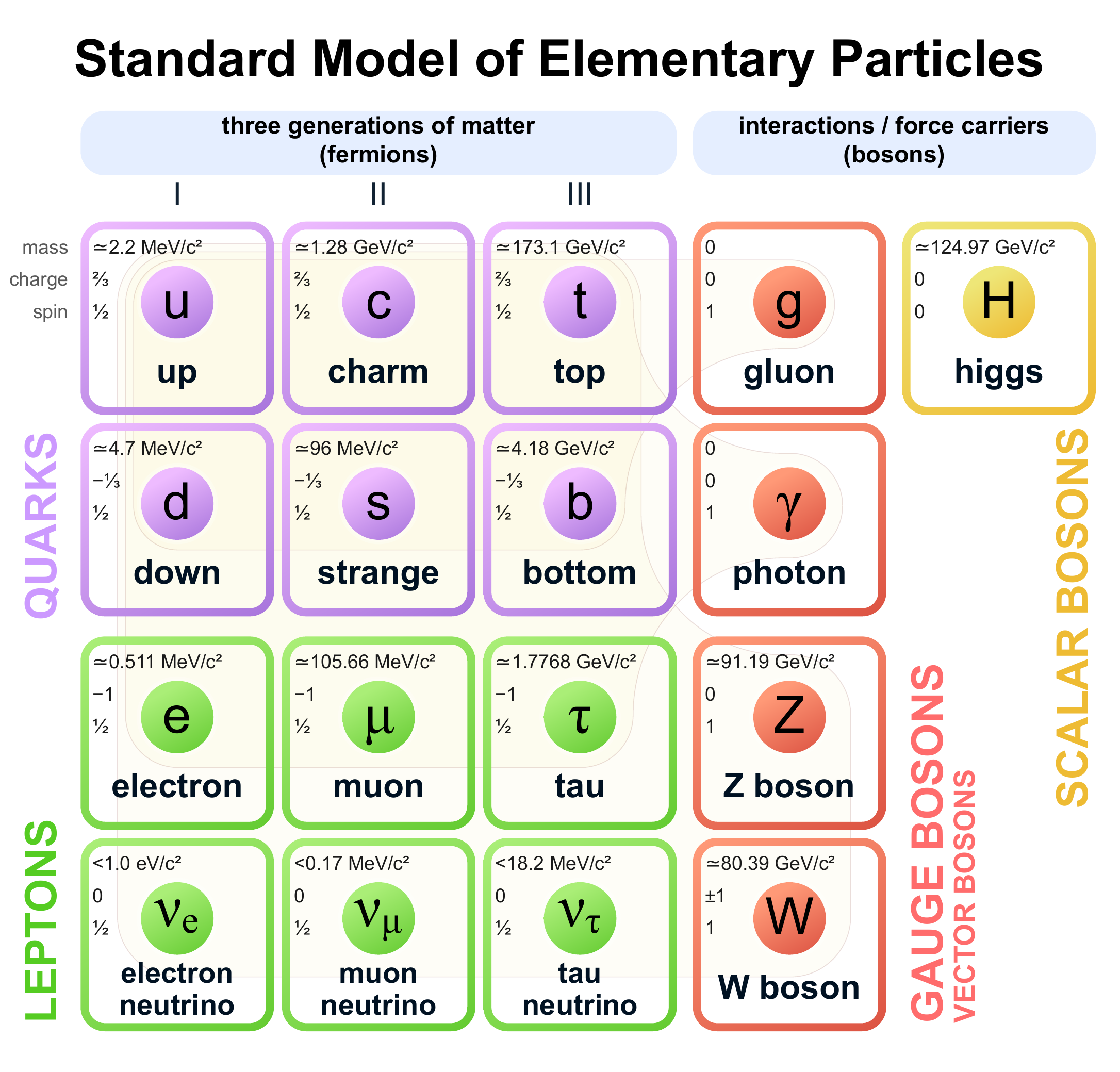}
    \caption{The Standard Model of particle physics. The masses of the up, down, and strange quarks are given in the $\overline{\text{MS}}$ scheme at the scale $2$ GeV; the charm and bottom quark masses are given in the $\overline{\text{MS}}$ scheme at their own $\overline{\text{MS}}$ masses; the top quark mass is extracted from event kinematics. Present determinations of the lighter quark masses rely mainly on lattice QCD calculations and QCD sum rules.}
\label{fig:SM}
\end{figure}

The elementary particles in the Standard Model are divided into fermions and bosons, as shown in Fig.~\ref{fig:SM}. The fermions come in three generations. The first-generation fermions do not decay and make up all ordinary matter\footnotecircle{Leaving aside dark matter and dark energy.}; each later generation is heavier than the previous one and decays into lighter particles. Fermions are further divided into quarks and leptons. Quarks carry color charge and participate in the strong interaction; because of color confinement, they cannot exist as free particles and can only form color-neutral hadrons. Leptons participate only in the electroweak interaction\footnotecircle{Leptons are not necessarily light; for example, the tau lepton is heavier than most quarks.}. In addition, all neutrinos are stable against decay. Bosons are divided into gauge bosons and scalar bosons. Gauge bosons mediate interactions: the photon for electromagnetism, the $W^\pm$ and $Z^0$ bosons for the weak interaction, and gluons for the strong interaction. The scalar boson is the Higgs boson, which gives mass to all elementary particles\footnotecircle{The Higgs mechanism, however, does not explain the masses of all composite particles. A famous example is the proton mass puzzle: the proton mass is about $938\,\mathrm{MeV}$, while the current masses of its constituent quarks are only a few $\mathrm{MeV}$. In fact, the overwhelming majority of the proton mass does not come from the quark masses themselves, but from \mybf{nonperturbative} QCD dynamics, including gluon contributions and the trace anomaly.}.

The Standard Model is based on the gauge group $\mathrm{SU}(3)_{\mathrm{QCD}}\times SU(2)_{\mathrm{weak}}\times U(1)_{\mathrm{hypercharge}}$, which is spontaneously broken at the scale $v=247\,\mathrm{GeV}$ to $\mathrm{SU}(3)_{\mathrm{QCD}}\times U(1)_{\mathrm{EM}}$. The Standard Model contains 19 parameters: three coupling constants, $g, g'$, and $g_s$; six quark masses and three charged-lepton masses; three mixing angles and one phase in the quark-sector CKM matrix; the Higgs mass $m_h$ and the vacuum expectation value $v$; and the QCD vacuum angle $\bar{\theta}$\footnotecircle{If neutrinos are assumed to have Dirac masses, one must also include three neutrino masses and three mixing angles plus one phase in the lepton-sector PMNS matrix, for a total of 26 parameters.}. Since the Standard Model is renormalizable, these parameters, though numerous, can be fixed by a finite set of experiments. In principle, infinitely many experimental observables can then be expressed as functions of these 19 parameters, making the Standard Model an overconstrained system. With sufficiently many and sufficiently precise measurements, it can therefore be tested rigorously.

\subsection{The Standard Model Lagrangian}
The Standard Model is remarkably compact; it can be written down in only a few lines, yet for decades almost all of its predictions have survived experimental tests\footnotecircle{What this means is that a sufficiently clever student, armed only with the Lagrangian and enough time, should in principle be able to compute quantities such as the electron anomalous magnetic moment, agreeing with experiment even at the $11$th digit after the decimal point.}. The Standard Model states that elementary particles correspond to particular representations of the Poincaré group and of the gauge group. We now briefly introduce its structure.
\begin{equation}
    \mathcal{L}_{\mathrm{SM}} = \mathcal{L}_{\mathrm{SU}(3)} + \mathcal{L}_{SU(2) \times U(1)} .
\end{equation}
The strong-interaction part of the Lagrangian is
\begin{equation}
    \mathcal{L}_{\mathrm{SU}(3)} = - \frac{1}{4} F^a_{\mu\nu} F^{a\mu\nu} + \sum_f \bar{q}^f i \slashed{D} q^f .
\label{eq:lagrangian_QCD}
\end{equation}
Here $q^f (f=1, \cdots, 6)$ denotes the quark field of flavor $f$, which transforms under $\mathrm{SU}(3)_C$ as the fundamental representation $\mathbf{3}$. The gluon field $A^a_\mu (a=1, \cdots, 8)$ is the gauge field of $\mathrm{SU}(3)$ in the eight-dimensional adjoint representation $\mathbf{8}$, and the field-strength tensor is defined by
\begin{equation}
F^a_{\mu\nu} = \partial_\mu A^a_\nu - \partial_\nu A^a_\mu - g_s f_{abc} A_\mu^b A_\nu^c .
\end{equation}
Here $g_s$ is the QCD coupling constant. The structure constants $f_{abc} (a,b,c=1,\cdots,8)$ carry color indices and are defined by
\begin{equation}
    [\frac{\lambda^a}{2},\frac{\lambda^b}{2}] = i f_{abc} \frac{\lambda^c}{2}.
\end{equation}
The covariant derivative is
\begin{equation}
D_{\mu} = \partial_\mu + i g_s A_\mu^a \frac{\lambda^a}{2},
\end{equation}
where $\frac{\lambda^a}{2}$ are the generators of the fundamental representation of $\mathrm{SU}(3)$. Notice that the Lagrangian written here contains no quark mass term. Such a term would break chiral symmetry and is therefore forbidden in the Standard Model\footnotecircle{More precisely, the quark mass term breaks the $\mathrm{SU}(3)_L \times \mathrm{SU}(3)_R$ chiral symmetry.}. The mass term is generated through the Yukawa interaction and spontaneous symmetry breaking of the Higgs field. When we discuss low-energy QCD in\chapref{chap:lattice_QCD}, we will write the mass term explicitly.

Moreover, because gauge fields contain infinitely many gauge-equivalent configurations, a gauge-fixing term must be introduced before perturbative calculations in order to remove this redundancy, and ghost fields are introduced through the Faddeev--Popov method. In\chapref{chap:lattice_QCD}, we will introduce the path-integral formulation of lattice QCD and explain why explicit gauge fixing and ghost fields are not needed in lattice QCD.

The electroweak part of the Lagrangian is
\begin{equation}
    \mathcal{L}_{SU(2) \times U(1)} = \mathcal{L}_{\rm gauge}+\mathcal{L}_\phi+\mathcal{L}_F+\mathcal{L}_{\rm Yukawa}.
\end{equation}
The gauge part has the same form as in QCD, but with a different gauge group:
\begin{equation}
    \mathcal{L}_{\rm gauge} = -\frac{1}{4} F^a_{\mu\nu}F^{a \mu\nu} - \frac{1}{4} B_{\mu\nu}B^{\mu\nu}.
\end{equation}
Here $W_\mu^a (i=1,2,3)$ and $B_\mu$ are the gauge fields associated with the adjoint representations of $SU(2)$ and $U(1)$, respectively:
\begin{equation}
    \begin{aligned}
        B_{\mu\nu} &= \partial_\mu B_\nu-\partial_\nu B_\mu, \\
        F_{\mu\nu} &= \partial_\mu W_\nu^i-\partial_\nu W_\mu^i-g\epsilon_{ijk}W_\mu^jW_\nu^k .
    \end{aligned}
\end{equation}
The constant $g(g')$ is the coupling constant of $SU(2)(U(1))$, respectively. The $B$ field carries weak hypercharge $Y=Q-T_3$, where $Q$ is the charge operator and $T_3$ is the third component of weak isospin. After spontaneous symmetry breaking, the $B$ and $W_3$ fields mix to form the photon and the $Z$ boson. The gluon field $A^a_\mu$ in the QCD sector is a singlet under $SU(2) \times U(1)$.

The scalar part is
\begin{equation}
    \mathcal{L}_\phi=(D^\mu\phi)^\dagger D_\mu\phi-V(\phi).
\end{equation}
Here
\begin{equation}
    \phi=\begin{pmatrix}\phi^+\\\phi^0\end{pmatrix}
\end{equation}
is a complex Higgs scalar, transforming as a doublet under $SU(2)$ and carrying weak hypercharge $Y=+1/2$. The covariant derivative is
\begin{equation}
    D_\mu = \partial_\mu + i g \frac{\tau^a}{2}W_\mu^a + i g' Y B_\mu .
\label{eq:SU2_covariant_derivative}
\end{equation}

The Higgs potential is
\begin{equation}
    V(\phi)=\mu^2\phi^\dagger\phi+\lambda(\phi^\dagger\phi)^2 .
\end{equation}
When $\mu^2<0$, spontaneous symmetry breaking occurs, and $\lambda$ describes the four-point self-interaction of the scalar field. Vacuum stability requires $\lambda>0$.

For the fermion part,
\begin{equation}
    \mathcal{L}_F = \bar{L}^i i \slashed{D} L^i + \bar{Q}^i i \slashed{D} Q^i + \bar{e}^i_R i \slashed{D} e^i_R + \bar{\nu}^i_R i \slashed{D} \nu^i_R + \bar{u}^i_R i \slashed{D} u^i_R + \bar{d}^i_R i \slashed{D} d^i_R .
\end{equation}
Although right-handed neutrinos have not yet been observed experimentally, they can be included theoretically, and we therefore include them here. The index $i=1,2,3$ labels the generation. Left-handed leptons and left-handed quarks are doublets under $SU(2)$, and transform under the Lorentz group as left-handed Weyl spinors, namely in the $(\frac{1}{2},0)$ representation:
\begin{equation}
    L^i = \left\{ \begin{pmatrix}\nu_{e L} \\ e_L\end{pmatrix}, \begin{pmatrix}\nu_{\mu L} \\ \mu_L\end{pmatrix}, \begin{pmatrix}\nu_{\tau L} \\ \tau_L\end{pmatrix} \right\}, \quad Q^i = \left\{ \begin{pmatrix}u_L \\ d_L\end{pmatrix}, \begin{pmatrix}c_L \\ s_L\end{pmatrix}, \begin{pmatrix}t_L \\ b_L\end{pmatrix} \right\}.
\end{equation}
Right-handed leptons and right-handed quarks are singlets under $SU(2)$. Under the Lorentz group, they transform as right-handed Weyl spinors, in the $(0,\frac{1}{2})$ representation:
\begin{equation}
    e^i_R = \{ e_R, \mu_R, \tau_R \}, \quad \nu^i_R = \{ \nu_{e R}, \nu_{\mu R}, \nu_{\tau R} \}, \quad u^i_R = \{ u_R, c_R, t_R \}, \quad d^i_R = \{ d_R, s_R, b_R \}.
\end{equation}
Here $L^i$ and $Q^i$ are $SU(2)$ doublets, whereas the right-handed fields $u^i_R,d^i_R,e^i_R,\nu^i_R$ are singlets. For $SU(2)$ doublets, the covariant derivative is given in Eq.~\ref{eq:SU2_covariant_derivative}; for $SU(2)$ singlets, the coupling to the $W$ field is absent. The different transformation properties of left- and right-handed fields are responsible for parity violation in the electroweak interaction.

The Yukawa coupling term is
\begin{equation}
    \mathcal{L}_{\rm Yukawa} = - Y_{ij}^d \bar{Q}^i H d_R^j - Y_{ij}^u \bar{Q}^i (i \tau_2 H^*) u_R^j + h.c.
\end{equation}
The matrices $Y_{ij}$ describe the Yukawa couplings between the Higgs doublet $\phi$ and quarks of different flavors; the lepton part can be written in an analogous form. After symmetry breaking, the Higgs field acquires a vacuum expectation value and gives masses to the quarks\footnotecircle{The electroweak transition experienced by the early universe is in fact a smooth crossover. When the temperature of the universe dropped to about $T \sim 160~\mathrm{GeV}$, the vacuum expectation value $v(T)$ of the Higgs field grew smoothly from zero to its zero-temperature value $v \approx 246~\mathrm{GeV}$, and electroweak symmetry was gradually broken spontaneously. Correspondingly, the masses of the W and Z bosons and of the fermions also increased smoothly with $v(T)$; before the onset of the transition, they were nearly massless.}. These mass terms are not diagonal in the flavor basis. Introducing two matrices $U_{u/d}$,
\begin{equation}
    Y_{u/d} Y_{u/d}^{\dagger} = U_{u/d} M_{u/d}^2 U_{u/d}^{\dagger},
\end{equation}
and transforming the quark basis, one finds that in the mass-diagonal basis the interactions of different quark flavors with the $W$ boson are mixed. This mixing is described by the Cabibbo--Kobayashi--Maskawa (CKM) matrix,
\begin{equation}
    V \equiv U_u^\dagger U_d =
    \begin{pmatrix}
    V_{ud} & V_{us} & V_{ub} \\
    V_{cd} & V_{cs} & V_{cb} \\
    V_{td} & V_{ts} & V_{tb}
    \end{pmatrix}.
\end{equation}
When neutrino oscillations are considered, the Pontecorvo--Maki--Nakagawa--Sakata (PMNS) mixing matrix plays a role analogous to that of the CKM matrix.

Although the Standard Model has achieved enormous success in describing the electromagnetic, weak, and strong interactions, gravity is not included in this framework. The most successful description of gravity is still Einstein's \mybf{general relativity}\footnotecircle{Among the three major theoretical breakthroughs of the early twentieth century, quantum mechanics, special relativity, and general relativity, two and a half were due to Einstein.}. In this view, gravity is not a force but a geometric effect of spacetime. Matter and energy determine the curvature of spacetime, and the curvature of spacetime governs the motion of matter. Its dynamics are given by the famous Einstein field equation,
\begin{equation}
    R_{\mu\nu} - \frac{1}{2} g_{\mu\nu} R = 8\pi G\, T_{\mu\nu}.
\end{equation}
Here $R_{\mu\nu}$ and $R$ are the Ricci tensor and scalar curvature, respectively, $g_{\mu\nu}$ is the metric tensor, $T_{\mu\nu}$ is the energy-momentum tensor, and $G$ is Newton's gravitational constant.

General relativity is a classical field theory and is in deep tension with quantum field theory. At high energy scales, quantum effects of gravity cannot be neglected. General relativity nevertheless remains the most successful theory of gravity, despite many attempts to modify it\footnotecircle{Examples include scalar--tensor theory~\cite{Yagi:2009zm}, Einstein--dilaton--Gauss--Bonnet gravity~\cite{Yagi:2011xp}, dynamical Chern--Simons gravity~\cite{Yagi:2012vf}, Einstein-Æther theory~\cite{Jacobson:2007veq}, Khronometric theory~\cite{Hansen:2014ewa}, varying-gravitational-constant theories~\cite{Chamberlain:2017fjl}, and noncommutative gravity~\cite{Chamseddine:2000si, Song:2025dbf}. More systematic reviews can be found in Ref.~\cite{Tahura:2018zuq}.}, and it has passed stringent tests from the Solar System, binary pulsars, stellar orbits around the Galactic center, and cosmological observations with extremely high precision; see, for example, Ref.~\cite{Will:2014kxa}. In attempts to reconcile quantum field theory with gravity, candidate theories such as superstring theory and loop quantum gravity have been proposed, and their possible Lorentz-symmetry violation at very high energy scales has been explored~\cite{Song:2025ksi,Song:2025myx}.

\subsection{Open Questions in the Standard Model}
Although the Standard Model has been extraordinarily successful experimentally, it is unlikely to be the final theory. Several unresolved problems remain, including:

\begin{itemize}
    \item \mybf{The hierarchy problem}. The mass of the Higgs scalar receives quadratically divergent quantum corrections proportional to the ultraviolet cutoff. Keeping the Higgs mass at the electroweak scale, $\sim 10^2~\mathrm{GeV}$, requires an extremely delicate cancellation between the bare mass parameter and quantum corrections. Equivalently, we do not understand why gravity is so weak compared with the other fundamental interactions. This severe fine-tuning is unnatural and may indicate new physics at higher energy scales, such as supersymmetry, a composite Higgs, or extra dimensions.
    \item \mybf{The dark-matter problem}. The Standard Model cannot explain the existence of the dark-matter component that accounts for about 85\% of the matter in the universe. Although abundant experimental evidence supports the existence of dark matter, including galaxy rotation curves, cosmic-microwave-background observations, and the Bullet Cluster, the Standard Model contains no suitable particle candidate.
    \item \mybf{The neutrino-mass problem}. The original Standard Model assumes massless neutrinos, but experiments have clearly shown that neutrinos have nonzero masses and can oscillate among different flavors. This requires new mechanisms in the Standard Model, such as right-handed neutrinos and the seesaw mechanism, or effective mass terms generated by dimension-five operators. Although these extensions can explain neutrino masses, they are not naturally contained in the original Standard Model framework.
    \item \mybf{The matter--antimatter asymmetry problem}. The universe contains far more matter than antimatter\footnotecircle{One may say that we are the one part in a billion that survived the annihilation of the much larger initial amounts of matter and antimatter in the early universe.}, but the amount of charge-parity (CP) violation in the Standard Model is far too small to explain this asymmetry. This points to possible new sources of CP violation or new physical processes, such as leptogenesis or baryogenesis.
    \item \mybf{The strong CP problem}. The strong interaction allows a CP-violating parameter \(\theta_{\rm QCD}\), whose value is theoretically allowed to lie in \([0,2\pi]\). Experimental measurements, such as the upper bound on the neutron electric dipole moment, imply \(|\theta_{\rm QCD}| \lesssim 10^{-10}\), far smaller than its natural expectation. A possible natural solution is the Peccei--Quinn symmetry and the associated axion mechanism.
    \item \mybf{The cosmological-constant problem}. The Standard Model cannot explain the mechanism behind the accelerated expansion of the universe, namely the dark-energy problem. Observations show that about 70\% of the energy density of the universe is dark energy, whose nature may be closely related to vacuum energy. Theoretically, the vacuum energy density of quantum fields should be near the Planck scale, of order \(\rho_{\rm vac} \sim M_{\rm Pl}^4 \sim 10^{76}\,\mathrm{GeV}^4\), whereas the observed dark-energy density is only \(\rho_{\Lambda} \sim 10^{-47}\,\mathrm{GeV}^4\), about 120 orders of magnitude smaller.
    \item \mybf{The flavor problem}. Along the vertical direction of the Standard Model particle table, we understand the dynamics and interactions quite well. Along the horizontal direction, however, the Standard Model does not explain why leptons and quarks come in three generations\footnotecircle{Rather than, say, thirty.}, nor why the masses differ so dramatically among generations.
\end{itemize}

\section{History and Current Status of Hadron Spectroscopy}
Among the four fundamental interactions in nature, the strong interaction is the strongest, and at the same time the one we understand least completely. Its fundamental theory is quantum chromodynamics (QCD), which describes the interactions among quarks and gluons. Because QCD is asymptotically free, observables at high energy scales can be computed through perturbative expansions. In the low-energy region, however, the coupling becomes large, quarks are confined inside hadrons, and perturbation theory breaks down. The nonperturbative nature of QCD in this regime makes \mybf{hadron spectroscopy} one of the last frontiers of the Standard Model, with many directions still driven by experiment.

In high-energy experiments, final states often appear as hadronic jets with large transverse momentum. This means that even if quark--quark scattering at high momentum transfer can be factorized and computed using perturbative QCD (pQCD), fragmentation functions are still needed to describe long-distance nonperturbative QCD effects. If the initial state is hadronic, nonperturbative parton distribution functions are also required. These functions must be determined through fits to experimental data or through model calculations. Moreover, even in processes without hadrons in the initial or final state, virtual quantum fluctuations can introduce hadronic contributions that have non-negligible effects on precision theoretical predictions.

The term ``spectroscopy'' in hadron spectroscopy originates from optical spectroscopy in atomic physics. In analogy with the way the emission and absorption spectra of atoms and molecules reveal the energy structure of microscopic particles, hadron spectroscopy studies the energy eigenstates of hadronic systems in order to reveal their internal structure. However, because $\mathrm{SU}(3)$ is a non-Abelian group, hadron phenomenology differs greatly from atomic phenomenology. The most immediate difference is that photons carry no color charge, whereas gluons do and also self-interact. Historically, the study of the hadron spectrum led to the quark model and then to QCD. Today, the main goal of hadron spectroscopy is to understand, by solving QCD directly or through QCD-based models, which hadrons exist, what their masses are, how they decay, and what internal structures they possess.

\subsection{Current Experimental Progress}
Revealing the spectral structure of QCD is one of the central goals of contemporary theoretical research, and it also motivates major experimental programs. Recent reviews can be found in Refs.~\cite{Hyodo:2020czb, Chen:2022asf, Mai:2022eur, Pelaez:2025wma}.

According to the quark model, hadrons are usually divided into ``conventional'' and ``exotic'' classes. Conventional hadrons are states that can be described by the simplest quark configurations: mesons made of a quark and an antiquark ($q\bar{q}$), and baryons made of three quarks ($qqq$). By contrast, \mybf{exotic hadrons} contain more complicated configurations, such as tetraquark states ($qq\bar{q}\bar{q}$), pentaquark states ($qqqq\bar{q}$), hybrid states with valence-gluon components (such as $q\bar{q}g$ and $qqqg$), and glueballs composed entirely of gluons (such as $gg$)\footnotecircle{Strictly speaking, glueballs exist only in pure gauge theory; in real QCD with dynamical quarks, glueballs inevitably mix with various quark-containing hadrons.}.

In the light-flavor sector, broadly accepted experimental confirmation has long been lacking; see Ref.~\cite{Meyer:2015eta} for a review. In recent years, as experimental precision has improved, some candidate states have attracted increasing attention. For example, the BESIII Collaboration has determined the spin-parity quantum numbers of the $X(2370)$; the result is consistent with the expected properties of a pseudoscalar glueball, making it a possible glueball candidate, although its nature remains to be confirmed~\cite{BESIII:2023wfi}.

By contrast, studies of heavy-flavor exotic hadrons have made more striking progress. Because of OZI suppression, when a $Q\bar{Q}$ pair is observed among the decay products of a resonance, it is often regarded as part of the intrinsic structure of that resonance. If the properties of the state cannot be accommodated within the conventional $Q\bar{Q}$ spectrum, this indicates an internal structure beyond the simple quark--antiquark system. In addition, heavy quarkonia usually have relatively narrow decay widths and can be identified more cleanly in experiment. In particular, charged states decaying into heavy quarkonia must contain both a $Q\bar{Q}$ pair and additional light-quark content, and are therefore strong exotic-hadron candidates.

Since the beginning of the twenty-first century, exotic-hadron studies have undergone breakthrough progress. In 2003, the Belle Collaboration first reported the discovery of the $X(3872)$, which is widely regarded as one of the first clear exotic-hadron candidates. Since then, many new candidates have been observed, launching a new wave of hadron-spectroscopy research\footnotecircle{In the 1960s, when the quark model was being established, no hadrons beyond mesons and baryons had yet been observed.}, including tetraquark states with charm and strangeness, doubly charmed and fully charmed tetraquark states, and charmed pentaquark states.

At present, exotic-hadron studies rely on several different types of experimental facilities. For $e^+e^-$ collision experiments, the main advantage is a relatively clean background environment, and exotic states can be produced through various mechanisms, including direct production, associated production with charmonium, two-photon processes, initial-state radiation (ISR), and decays of bottomonium and $B$ mesons. Important early results came mainly from BaBar, Belle, and CLEO. Representative experiments that are still running include Belle II and BESIII.

\begin{itemize}
    \item \mybf{The Belle II experiment}: Belle II operates at the SuperKEKB collider in Japan, mainly near the $\Upsilon(4S)$ resonance, and produces large numbers of $B\bar{B}$ pairs for precision flavor physics. Its asymmetric beam design enables time-dependent measurements of CP violation. Thanks to its broad effective energy coverage and diverse production mechanisms, Belle II can study exotic hadrons through associated charmonium production, through which most currently observed states were first discovered, as well as through two-photon processes, ISR, bottomonium decays, and $B$-meson decays. Compared with LHCb, Belle II has advantages in reconstructing final states containing photons, $\pi^0$, and $\eta^{(\prime)}$.
    \item \mybf{The BESIII experiment}: BESIII operates at BEPCII, the Beijing Electron--Positron Collider located about $10$ km from Peking University~\cite{Ketzer:2019wmd}, and covers center-of-mass energies from $1.84$ to $4.95~\mathrm{GeV}$. BESIII has collected the world's largest samples of charmonium data. Unlike $B$ factories, BESIII can tune the collision energy directly to the target resonance region, especially near $Y$ states, and can therefore produce exotic hadrons directly, with high detection efficiency and precise measurements of resonance line shapes. In addition, at the tau-charm threshold it can produce clean hadron-pair events, enabling precise measurements of hadron decay constants, form factors, and other basic physical quantities.
\end{itemize}

For $pp$ collisions, the Large Hadron Collider (LHC) is located at CERN on the border between Switzerland and France, and is the highest-energy particle accelerator ever built. Its design proton--proton center-of-mass energy is $\sqrt{s} = 14~\mathrm{TeV}$. The LHC is currently near the end of Run~3 and will then enter a high-luminosity upgrade lasting about five years. The upgraded machine, the High-Luminosity LHC (HL-LHC), aims to deliver an integrated luminosity of about $250~\mathrm{fb}^{-1}$ per year and to accumulate a total data set of $\mathcal{O}(3000$-$4000)~\mathrm{fb}^{-1}$ over roughly a decade of running.

Nine large experiments operate at the LHC. We briefly introduce ALICE, ATLAS, CMS, and LHCb:
\begin{itemize}
    \item \mybf{The LHCb experiment}: LHCb is designed specifically for heavy-flavor physics~\cite{LHCb:2008vvz}, focusing on CP violation and rare decays in bottom- and charm-quark systems.\footnotecircle{In 2025, the LHCb Collaboration observed CP violation in a baryon system for the first time in $\Lambda_b^-$ decays~\cite{LHCb:2025ray}, an important milestone in exploring the origin of the matter--antimatter asymmetry of the universe and searching for new physics. The Peking University LHCb group participated in the analysis.} The LHCb detector is a single-arm forward spectrometer covering the pseudorapidity range $2 < \eta < 5$, with unique advantages in the study of heavy-hadron production and decay. Because the LHC provides extremely high center-of-mass energies, heavy-quark production cross sections are very large. Compared with $e^+e^-$ collisions, however, the background in $pp$ collisions is much more complicated, and exotic hadrons are therefore often reconstructed through decays of $B$ mesons or heavy-flavor hadrons\footnotecircle{The Peking University LHCb group has participated in many data analyses, including leading contributions to the discovery of the fully charmed tetraquark state $X(6900)$, the strange-charm states $Z_{cs}(4000)$ and $Z_{cs}(4220)$, and the first charmed pentaquark states $P_c(4440)$ and $P_c(4457)$.}.
    \item \mybf{The CMS}, \mybf{ATLAS}, and \mybf{ALICE} \mybf{experiments}: CMS and ATLAS are general-purpose detectors, designed primarily to search for physics beyond the Standard Model. ALICE mainly studies the quark--gluon plasma (QGP) formed in heavy-ion collisions, recreating the extreme conditions roughly one microsecond after the Big Bang. Although their primary physics goals differ, these experiments also contribute to studies of exotic hadron states.
\end{itemize}
The remaining LHC experiments, LHCf, TOTEM, MoEDAL-MAPP, FASER, and SND@LHC, are devoted to cosmic-ray modeling, small-angle proton scattering, searches for magnetic monopoles, searches for extremely weakly interacting particles, and measurements of neutrinos produced in proton--proton scattering.

Several medium-energy experiments also play important roles in exotic-hadron studies. For example, the PANDA experiment at the Facility for Antiproton and Ion Research (FAIR) in Germany will use antiproton beams for precision spectroscopy~\cite{PANDA:2009yku}. The GlueX experiment at Jefferson Lab (JLab) in the United States focuses on searching for hybrid states with exotic quantum numbers through photoproduction~\cite{GlueX:2020idb}.

\begin{figure}[htbp]
    \centering
    \includegraphics[width=\columnwidth]{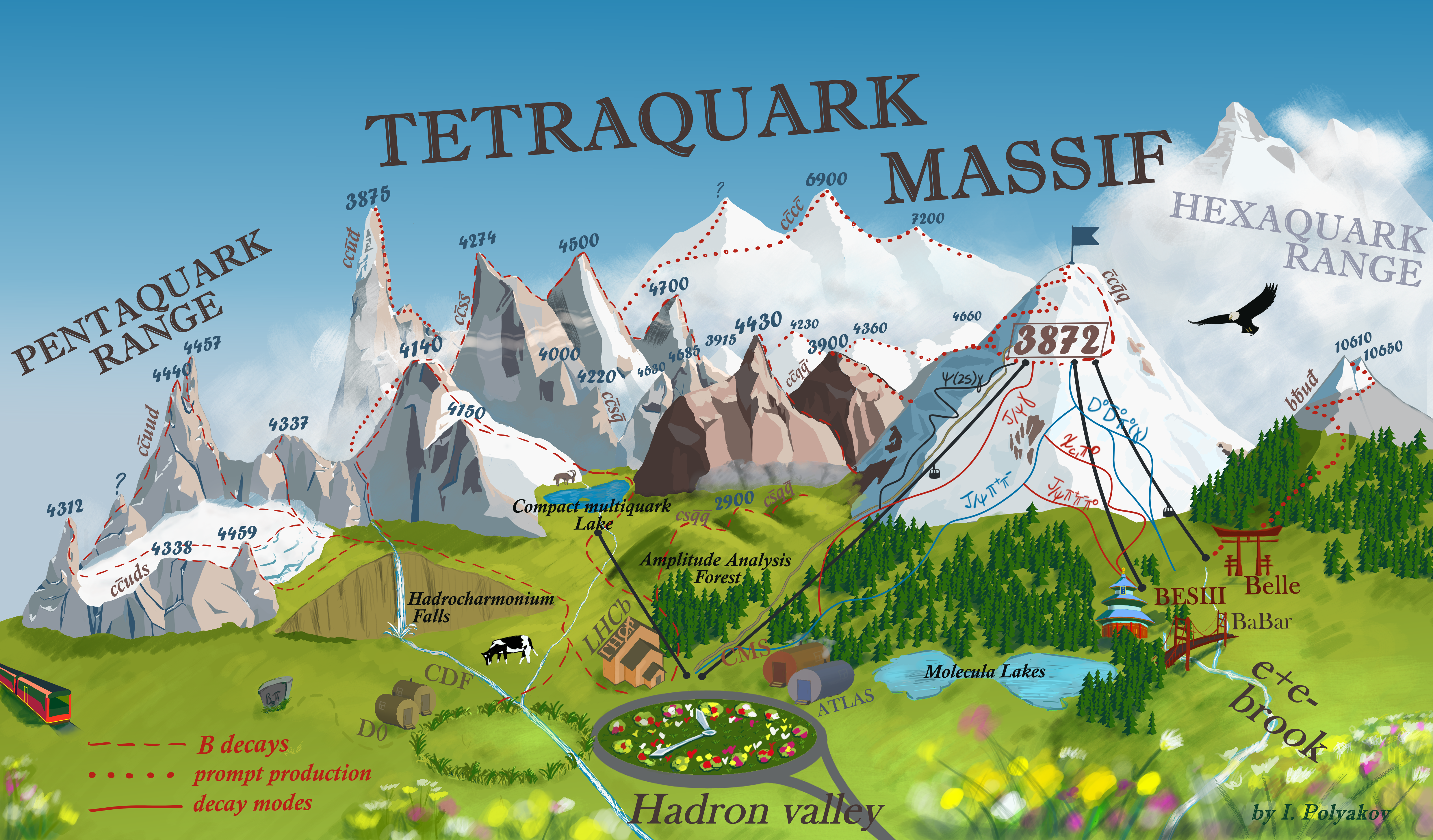}
    \caption{The exotic-hadron spectrum.}
\label{fig:zoo}
\end{figure}

Overall, exotic-hadron research is gradually entering a stage reminiscent of the ``particle zoo'' of the 1950s, and the LHCb Collaboration has even proposed naming conventions for these states~\cite{Gershon:2022xnn}. Owing to space limitations, this thesis will not discuss all candidate states one by one. An incomplete schematic overview is shown in Fig.~\ref{fig:zoo}; detailed reviews can be found in Ref.~\cite{Guo:2017jvc}.

In addition, several experimental projects are being planned or are about to begin operation, such as AMBER at the LHC~\cite{Friedrich:2024ylw}, SIS100 at GSI/FAIR~\cite{Messchendorp:2025men}, and Klong at JLab~\cite{KLF:2020gai}. Two future colliders proposed in China are\footnotecircle{The future development of these two machines remains to be clarified.}:
\begin{itemize}
    \item The \mybf{Electron-Ion Collider in China (EicC)} is a proposed high-luminosity electron--ion collider~\cite{Anderle:2021wcy}, designed to probe the internal structure of hadrons with high precision through deep-inelastic scattering of polarized electron beams off proton beams. Its main scientific goals include precision measurements of nucleon structure in the sea-quark region, three-dimensional hadron structure through TMDs and GPDs, exotic hadron states, and the origin of mass.
    \item The \mybf{Super Tau-Charm Facility (STCF)}: STCF is a proposed high-luminosity $e^+e^-$ collider~\cite{Achasov:2023gey, Ai:2025xop}, designed to operate in the energy range $2$--$7~\mathrm{GeV}$. Compared with BEPCII, it will substantially improve both luminosity and energy coverage, and is planned to include beam polarization. STCF will play an important role in precision charm-physics measurements, rare decays, and studies of CP violation.
    \item The \mybf{Circular Electron Positron Collider (CEPC)}: CEPC is a proposed high-energy circular $e^+e^-$ collider~\cite{CEPCStudyGroup:2025kmw}. Its initial operating energy is $\sqrt{s} \simeq 240~\mathrm{GeV}$, where it will serve as a Higgs factory, while also functioning as a $Z$ and $W$ factory and producing large samples of $B$ hadrons and $\tau$ leptons. Its tunnel could also accommodate a future high-energy proton--proton collider, the Super Proton--Proton Collider (SPPC), whose design center-of-mass energy is $125~\mathrm{TeV}$, nearly ten times that of the LHC.
\end{itemize}

\subsection{Theoretical Methods}
Because of the difficulties of QCD in the strong-coupling regime described above, many nonperturbative methods have been developed for hadron spectroscopy. These methods have different domains of applicability and different limitations, but they play important roles in classifying states, building physical intuition, and providing qualitative or semiquantitative estimates of relevant observables when more rigorous tools are unavailable. They include, but are not limited to, the following.
\subsubsection*{Quark models}
Quark models are very effective for classifying hadrons and describing the properties of states near or above open-flavor thresholds. Although the quark model was historically successful for ground-state hadrons, it has clear limitations for excited states. For example, in the scalar nonet, these mesons should be $P$-wave excitations of $q\bar{q}$ states, but this picture does not explain why the masses of $a_0(980)$ and $f_0(980)$ lie so close to the $K\bar{K}$ threshold, nor why the $f_0(500)$ is so light~\cite{Achasov:2008ut}. In the baryon sector, a typical example is the first radial excitation of the nucleon, the Roper resonance $N(1440)$, whose mass is significantly lower than expected in traditional quark models. In\chapref{chap:three_body_problems}, we will discuss the influence of its three-body decay channel and possible resolutions.
\subsubsection*{Phenomenological models}
Phenomenological methods provide intuitive interpretations of hadron structure and serve as references in identifying new states. Typical pictures include compact tetraquark states and hadronic molecules. The former are usually based on diquark--antidiquark configurations; see Ref.~\cite{Selem:2006nd} for a review. The latter regard exotic hadrons as weakly bound states of two hadrons; see Ref.~\cite{Guo:2017jvc} for a review. To distinguish different structures, one may use methods such as Weinberg's compositeness criterion~\cite{Weinberg:1965zz}. However, such models depend on specific assumptions, and no definitive picture exists for each exotic state.
\subsubsection*{QCD sum rules}
QCD sum rules; see Ref.~\cite{Albuquerque:2018jkn} for a review; use quark--hadron duality to connect hadron properties with the nonperturbative structure of the QCD vacuum. One constructs current operators with the desired quantum numbers and computes two-point \mybf{correlation functions}. The operator product expansion (OPE) expresses these correlation functions as a series of perturbative contributions and vacuum condensates. The truncation of condensates of different dimensions depends on the system under study; for exotic states, higher-dimensional terms are often required to ensure convergence. Combined with dispersion relations, the correlation functions can then be related to the hadron spectrum. This method can quickly estimate hadron properties, but its results depend on condensate parameters and truncation schemes, making systematic errors difficult to control.
\subsubsection*{Heavy-quark effective theory (HQET)}
HQET is an effective field theory for systems containing a single heavy particle and light degrees of freedom; textbook references include Ref.~\cite{Manohar:2000dt}. The heavy particle may be a heavy quark or, under suitable conditions, a composite system containing heavy quarks. Using the scale separation $m_h \gg \Lambda_{\rm QCD}$, one integrates out the high-energy degrees of freedom near the scale $m_h$ to obtain an effective theory containing only low-energy degrees of freedom. HQET provides a systematic expansion for heavy-flavor dynamics, but it does not directly include many-body structures or coupled-channel effects, and therefore has limited use in the spectroscopy of near-threshold exotic states.
\subsubsection*{Nonrelativistic QCD (NRQCD)}
NRQCD is an effective theory for systems composed of two or more nonrelativistic heavy quarks, such as heavy quarkonia~\cite{Caswell:1985ui,Brambilla:2014jmp}. Such systems contain multiple scales, $m_h \gg m_h v \gg m_h v^2$, corresponding respectively to the heavy-quark mass, the typical momentum, and the binding energy. NRQCD constructs the low-energy theory by integrating out the scale $m_h$, while retaining the kinetic term to describe nonrelativistic motion. Integrating out the scale $m_h v$ further leads to potential NRQCD (pNRQCD)~\cite{Brambilla:1999xf}, which reduces the problem to a potential-based description. These methods are well suited to heavy-quarkonium spectroscopy, although their validity can be limited in strongly coupled or near-threshold regions.

Against this background, it is especially important to have a method that starts from \mybf{first principles}, performs nonperturbative calculations, and allows theoretical errors to be improved systematically. \mybf{Lattice quantum chromodynamics} (Lattice QCD, LQCD) is precisely the theoretical tool developed to meet this need.

The central goal of this thesis is therefore to study, from first principles, the dynamical structure of near-threshold hadron resonances and to provide scattering and pole information that can be tested experimentally. To this end, the thesis focuses on three representative systems: the $D_0^*(2300)$ in two-body $D\pi$ scattering, and the $\omega(782)$ and $\pi(1300)$ in three-body systems. The first is connected with the double-pole puzzle of charmed scalar mesons, while the latter two represent, respectively, a known narrow resonance and a controversial broad resonance. They are ideal examples for testing and extending three-body formalism.

\section{History and Current Status of Lattice Field Theory}
The most systematic method for studying nonperturbative problems is lattice field theory\footnotecircle{For basic material on lattice field theory, see the Chinese textbook~\cite{liuchuan}.}. Perturbation theory is based on an asymptotic expansion in the coupling constant and is generally nonconvergent. It therefore does not constitute a well-defined quantum field theory by itself, nor can it handle intrinsically nonperturbative phenomena such as color confinement or spontaneous symmetry breaking\footnotecircle{Nonperturbative phenomena appear in many forms in different field theories, such as solitons and their scattering in scalar field theory~\cite{Yan:2021zwg,Yan:2020bdn}.}. A nonperturbative definition, by contrast, naturally contains perturbative results as limiting information.

Physically, we do not expect spacetime to be resolved with infinite precision, so introducing cutoffs is both natural and reasonable. Lattice field theory defines a field theory on a discrete spacetime lattice, with lattice spacing $a$ and volume $L^3$, thereby introducing an ultraviolet cutoff $\Lambda \sim 1/a$ and an infrared cutoff $\lambda \sim 2\pi/L$\footnotecircle{Strictly speaking, constructive quantum field theory (CQFT) has not yet constructed a nontrivial interacting field theory in $3+1$-dimensional Minkowski spacetime.}. In the continuum limit $a \to 0$ and infinite-volume limit $L \to \infty$, lattice field theory gives a nonperturbative definition of the field theory, and is currently the only nonperturbative definition available for quantum field theory. Lattice field theory is therefore both a \mybf{worldview}, as a nonperturbative definition of field theory, and a \mybf{methodology}, as an operational nonperturbative computational method~\cite{liuchuan}.

The main application of lattice field theory is lattice quantum chromodynamics\footnotecircle{The hep-lat archive was one of the earliest sections of the well-known arXiv preprint server.}, although its methods can also be extended to other quantum field theories. Both its statistical and systematic uncertainties can be estimated with considerable rigor, and can be improved systematically as algorithms and computational resources advance.

Lattice field theory originated with Wilson's pioneering work in 1974~\cite{Wilson:1974sk}\footnotecircle{Kenneth Wilson received the 1982 Nobel Prize in Physics for his pioneering work on the renormalization group in statistical physics and quantum field theory.}. Early studies focused mainly on strong-coupling expansions and pure gauge theory. In the 1980s, much work used the quenched approximation, neglecting fermion-loop effects, including some of the earliest hadron-spectrum calculations~\cite{Hamber:1981zn}\footnotecircle{Giorgio Parisi received the 2021 Nobel Prize in Physics for his pioneering contributions to the theory of complex systems.}.

In the twenty-first century, lattice QCD has made remarkable progress and can now perform calculations with dynamical quarks and at physical quark masses. Today, lattice QCD has become one of the central tools for studying many physical quantities in the Standard Model, with the precision of many results even surpassing experimental measurements.

We now briefly review several important directions in recent lattice-QCD research, with particular attention to its application to hadron spectroscopy.

\begin{figure}[htbp]
    \centering
    \includegraphics[width=0.6\columnwidth]{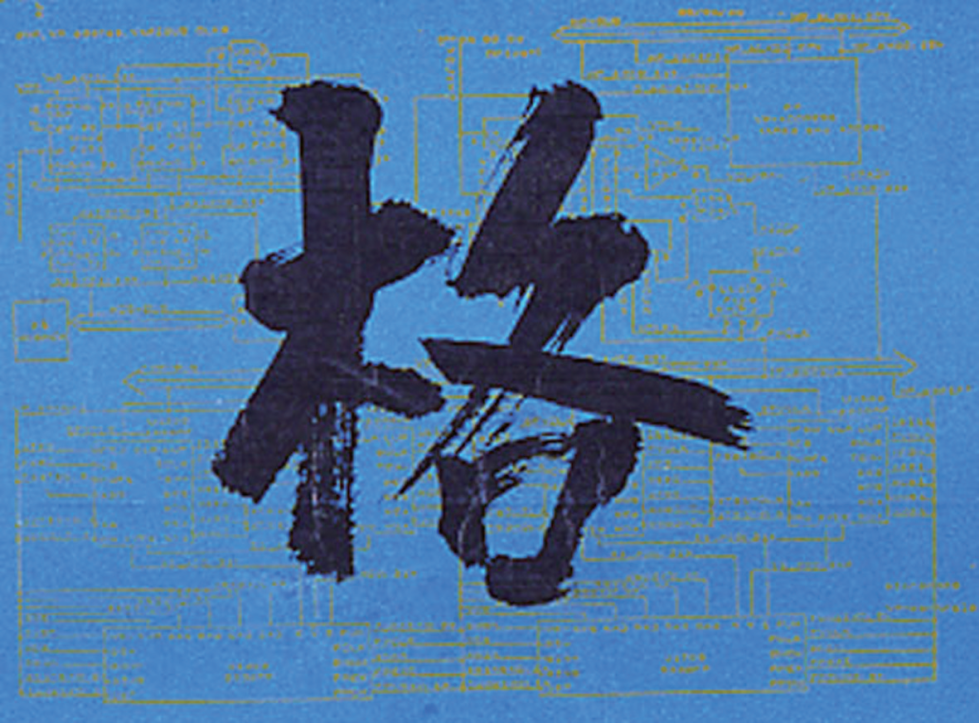}
    \caption{A poster prepared by Tsung-Dao Lee for a domestic academic meeting whose theme was ``lattice gauge theory on parallel computers''. The background shows a circuit diagram of the parallel computer used for lattice calculations at Columbia University, where he was working at the time. The large Chinese character ``格 (gé)'' both directly denotes the lattice and alludes to the classical phrase ``格物 (gé wù)'', the investigation of things.}
\end{figure}

\subsection{Lattice Hadron Spectroscopy}
Current experiments show that the proton is the only stable hadron in the universe. If only the strong interaction is considered\footnotecircle{Unless otherwise stated, the hadron masses discussed in this thesis neglect electromagnetic effects and the mass splittings caused by isospin breaking. Since QED is a long-range interaction, it cannot simply be placed directly on a finite lattice. Such corrections can be computed by several schemes, including $\mathrm{QED}_{\mathrm{TL}}$~\cite{Duncan:1996xy}, $\mathrm{QED}_{\mathrm{L}}$~\cite{BMW:2014pzb}, $\mathrm{QED}_{\mathrm{C}}$~\cite{Lucini:2015hfa}, $\mathrm{QED}_{\mathrm{M}}$~\cite{Endres:2015gda, Clark:2022wjy}, or the infinite-volume reconstruction (IVR) method~\cite{Feng:2018qpx}.}, nucleons and various ground-state pseudoscalar mesons are stable. In addition, $D_s^*$, $B^*$, and $B_s^*$ lie below their corresponding lowest strong-decay thresholds and are also stable. They are eigenstates of the QCD Hamiltonian. Most hadrons, however, are unstable under the strong interaction and appear as resonances or virtual states. They are not eigenstates of QCD, but rather poles in the analytic continuation of decay-particle scattering amplitudes onto unphysical Riemann sheets, and they thereby affect physical amplitudes indirectly. The invariant-mass spectra observed experimentally are influenced by these poles and appear as broad or narrow peaks.

With unphysical quark masses, some resonances may also appear as bound states\footnotecircle{We will discuss a typical example involving a charmed scalar meson in\chapref{chap:two_body_problems}.}. For single-particle states below strong-decay thresholds, masses, decay constants, and form factors can be extracted directly by analyzing two-point or three-point functions on the lattice. Some lattice calculations have poor signal-to-noise ratios; in\chapref{chap:lattice_QCD}, we will introduce an efficient method for dealing with this problem.

Most states have at least one strong decay channel. For such states, the \mybf{Lüscher method}~\cite{Luscher:1990ux} is the earliest and most widely used approach, relating finite-volume energy spectra of scattering final states to infinite-volume scattering amplitudes. The Lüscher method has been generalized to many more complicated situations, including coupled channels~\cite{He:2005ey}, systems with nonzero total momentum~\cite{Christ:2005gi, Kim:2005gf, Rummukainen:1995vs}, asymmetric volumes~\cite{Li:2003jn}, inelastic scattering channels~\cite{Lage:2009zv}, systems of nonidentical particles~\cite{Fu:2011xz, Leskovec:2012gb}, meson and baryon systems involving multiple partial waves~\cite{Gockeler:2012yj}, moving-frame multichannel systems~\cite{Briceno:2012yi}, multichannel baryon--meson systems~\cite{Li:2012bi}, and systems of arbitrary spin~\cite{Briceno:2014oea}.

Other approaches also exist, such as methods based on effective field theory and chiral unitary theory~\cite{Bernard:2007cm, Doring:2011vk, Doring:2012eu, Doring:2011nd, Bernard:2010fp, Guo:2008nc, Hanhart:2008mx}, as well as the HAL QCD method, which solves a nonrelativistic potential through the Bethe--Salpeter equation (BSE)\footnotecircle{HAL stands for ``Hadrons to Atomic nuclei''.}~\cite{Ishii:2006ec, Ishii:2012ssm, Aoki:2020bew}.

In\chapref{chap:lattice_QCD}, we will briefly introduce these methods, and in\chapref{chap:two_body_problems},\chapref{chap:three_body_problems}, and\chapref{chap:three_body_problems2}, we will apply them to concrete resonance problems.

As early as 1992, the $I=2$ $\pi\pi$ scattering length had been computed in the quenched approximation~\cite{Sharpe:1992pp}. Since 2010, energy-dependent scattering phase shifts away from threshold have been calculated, and in 2013 the $\rho$ resonance was successfully extracted~\cite{Dudek:2012xn}. Over the past decade, these methods have been widely applied to meson--meson, meson--baryon, baryon--baryon, and many other systems, with representative studies including Refs.~\cite{Dudek:2012xn, Dudek:2014qha, Wilson:2014cna, Briceno:2016mjc, Green:2021qol, Padmanath:2022cvl, Lyu:2023xro, Rodas:2023nec, BaryonScatteringBaSc:2023ori, BaryonScatteringBaSc:2023zvt, Wilson:2023anv, Wilson:2023hzu, Collins:2024sfi, Boyle:2024hvv, Whyte:2024ihh, Boyle:2024grr, Lang:2025pjq, Yan:2024yuq, Liu:2026gxr}. The most complicated system to date is the 2024 study of the charmonium channel by the Hadron Spectrum Collaboration (HadSpec), involving seven dynamically coupled channels.

In particular, two-nucleon systems are extremely challenging even in single-channel studies because of signal-to-noise problems. Before 2025, it was still unclear whether the HAL QCD method and the Lüscher method would yield consistent results. In 2025, the Baryon Scattering Collaboration (BaSc) examined both approaches on the same $\mathrm{SU}(3)$-symmetric ensemble at $M_{\pi} = 714\,\mathrm{MeV}$, confirming that at this pion mass the two methods agree and that both two-nucleon channels are virtual states rather than bound states~\cite{BaryonScattering:2025ziz}. Future calculations at more physical quark masses will be crucial.

Spectroscopic studies of many two-body systems have significantly improved our understanding of \mybf{hadron resonances}, but they still cover only a small subset of hadrons, since most hadrons decay into three or more hadronic final states. In the 2020s, it has become possible to compute scattering amplitudes for three-body systems, and this has gradually become an active research area. The finite-volume formalism for three-body problems is also developing rapidly~\cite{Polejaeva:2012ut, Briceno:2012rv,Briceno:2018aml,Briceno:2024ehy,Alexandru:2020xqf,Alotaibi:2025pxz,Meng:2017jgx,Mai:2018djl,Blanton:2019vdk,Culver:2019vvu,Hansen:2020otl,Fischer:2020jzp,Alexandru:2020xqf,Draper:2023boj,Blanton:2021llb,Dawid:2025doq,Dawid:2025zxc,Hansen:2025oag,Briceno:2025yuq,Feng:2026ixm,Severt:2022jtg}. Reviews of the formalism can be found in Refs.~\cite{Hansen:2019nir, Mai:2021lwb, Briceno:2017max}, with recent updates in Refs.~\cite{Green:2026jgv,Sharpe:2026mtt}.

One of the most widely studied three-body systems is the three-meson ($\pi\pi\pi$) system, which has been investigated in several isospin channels, including the maximal-isospin channel~\cite{Hansen:2020otl, Mai:2018djl, Blanton:2019vdk, Culver:2019vvu, Fischer:2020jzp, Beane:2007es, Horz:2019rrn, Mai:2019fba, Brett:2021wyd}, the isotensor channel~\cite{Briceno:2025yuq, Feng:2026ixm}, the isovector channel~\cite{Mai:2021nul,Yan:2025mdm}, and the isoscalar channel~\cite{Yan:2024gwp}. Other three-meson systems, such as $KK\pi$, $K\pi\pi$~\cite{Dawid:2025zxc, Blanton:2021llb, Draper:2023boj}, and $KKK$~\cite{Alexandru:2020xqf}, are also receiving increasing attention. Studies of high-isospin-density systems~\cite{Abbott:2023coj, Detmold:2008fn, Detmold:2008yn, Detmold:2011kw} and three-hadron systems containing charm quarks~\cite{Fu:2025joa} have likewise provided new insights into many-body systems.

A particularly important example is the doubly charmed tetraquark candidate $T_{cc}$ discovered by the LHCb Collaboration~\cite{LHCb:2021auc, LHCb:2021vvq}. Several pioneering lattice studies already exist~\cite{Collins:2024sfi, Chen:2022vpo, Lyu:2023xro, Padmanath:2022cvl, Prelovsek:2025vbr, Du:2023hlu, Meng:2023bmz}, mainly based on two-body $DD^*$ scattering analyses. A complete understanding of this state, however, also requires treating the complications associated with nearby left-hand cuts~\cite{Prelovsek:2025vbr, Du:2023hlu, Meng:2023bmz} and the presence of the three-body $DD\pi$ channel~\cite{Dawid:2024dgy, Hansen:2024ffk}. Another famous example is the Roper resonance $N(1440)$, which couples strongly to both the two-body $N\pi$ channel and the three-body $N\pi\pi$ channel. Its experimental and theoretical status is reviewed in Ref.~\cite{Burkert:2017djo}.

In summary, hadron spectroscopy has remained one of the most important applications of lattice field theory since the birth of the field. With rapid advances in algorithms, artificial-intelligence methods, and computing power, many long-standing spectroscopic challenges are gradually becoming solvable. We believe that these problems will see major breakthroughs in the near future.

\subsection{Progress in Other Areas of Lattice Field Theory}
Beyond spectroscopy, lattice methods have had a profound impact on many other areas of strong-interaction physics, including hadron structure, precision tests of the Standard Model, QCD at finite temperature and density, and nuclear lattice effective field theory. We cannot review all progress in these fields here, and instead select several representative directions.

\subsubsection*{Hadron structure}
The study of hadron states is the foundation for matrix-element calculations. In the Lüscher framework, the finite-volume correction to hadronic matrix elements is known as the Lellouch--Lüscher factor~\cite{Lellouch:2000pv}, and can be used to study weak decay matrix elements such as those in $K\to \pi\pi$~\cite{RBC:2023ynh}. The Lellouch--Lüscher factor has also been generalized to more complicated processes involving two-particle systems in the initial or final state; see Ref.~\cite{Briceno:2017max} for a review; and to cases involving three hadrons~\cite{Hansen:2021ofl, Briceno:2024txg}. After precise lattice determinations of basic hadron properties such as masses and decay constants through spectroscopy, one can study more detailed internal-structure information, including nucleon form factors~\cite{Capitani:2012gj,Bali:2014nma,Bhattacharya:2016zcn,Capitani:2017qpc,Chang:2018uxx,Harris:2019bih,Bali:2023sdi,Jang:2023zts,ChenChen:2025amm}, hadron radii and spin distributions; see Ref.~\cite{Syritsyn:2014saa} for a review; parton distribution functions (PDFs), generalized parton distributions (GPDs), and transverse-momentum-dependent distributions (TMDs). Methods for extracting PDFs include the heavy-quark OPE method~\cite{Aglietti:1998mz}, the quasi-PDF method~\cite{Ji:2013dva}, and the pseudo-PDF method~\cite{Radyushkin:2016hsy}. In particular, computing Mellin moments of PDFs was one of the earliest approaches, but because of operator mixing and poor signal-to-noise ratios, only the lowest few moments can usually be computed reliably. In recent years, the gradient-flow method~\cite{Shindler:2023xpd} has offered a possible way to alleviate these problems and may make higher-moment calculations feasible~\cite{Francis:2025pgf,Francis:2025rya}.

\subsubsection*{Flavor physics and precision tests of the Standard Model}
Lattice QCD calculations with precise and well-controlled uncertainties can serve as high-precision tests of the Standard Model. By computing specific hadronic matrix elements and combining them with experimental input, one can precisely determine quantities relevant to flavor physics, including light-quark masses, the strong coupling constant, decay constants, CKM matrix elements, the CP-violation parameter $\epsilon'/\epsilon$, and matrix elements of effective operators beyond the Standard Model; see Ref.~\cite{FlavourLatticeAveragingGroupFLAG:2024oxs} for a review. In particular, the muon anomalous magnetic moment $a_\mu$ has long been an important observable for testing the Standard Model. Both experiment and Standard Model theory have reached sub-part-per-million precision, and their comparison provides a stringent test of the Standard Model. The Fermilab E989 experiment reached a final precision of 127 ppb in 2025~\cite{Muong-2:2025xyk}. On the theory side, long-distance QCD contributions from hadronic vacuum polarization (HVP) and hadronic light-by-light scattering (HLbL) are the dominant sources of uncertainty. Lattice-QCD calculations have now reached a precision beyond that of data-driven methods based on dispersion relations; at the current ultra-high precision, no evidence for new physics remains, essentially resolving the earlier puzzle. Reviews of the theoretical side can be found in Ref.~\cite{Aliberti:2025beg}.

\subsubsection*{QCD at finite temperature and finite density}
Lattice methods can be used to study the properties of the strong interaction under extreme conditions, including the QCD equation of state, the physics of the crossover region at zero baryon chemical potential, and the QCD phase structure under external conditions such as strong magnetic fields, isospin chemical potential, rotation, acceleration, and quark spin polarization. These studies provide important first-principles theoretical input for understanding the evolution of the early universe and the hadronic physics of heavy-ion collisions. A recent review can be found in Ref.~\cite{Ding:2026gao}.

In addition, lattice field-theory methods can be applied to theories beyond the Standard Model, such as composite Higgs models and dark-matter models~\cite{DeGrand:2015zxa}, as well as to nuclear physics~\cite{Lahde:2019npb} and condensed-matter physics~\cite{Drut:2012md}. In recent years, algorithmic advances, together with applications of quantum computing~\cite{Zohar:2015hwa} and artificial-intelligence methods, such as generative models and normalizing flows~\cite{Albergo:2019eim,Kanwar:2020xzo}, to lattice generation and data analysis have continued to expand the boundaries of lattice methods. In short, lattice field theory is now a very active research field and plays an important role across many directions in theoretical physics.

This thesis summarizes part of the research work carried out by the author in lattice quantum chromodynamics during the author's doctoral studies. The chapters devoted to specific problems expand on published papers in greater detail and also include preliminary results from ongoing projects. The structure of the thesis is as follows:\chapref{chap:lattice_QCD} introduces the basic theory of lattice QCD;\chapref{chap:operators} presents methods for constructing lattice operators;\chapref{chap:two_body_problems} uses these methods to study two-body $D\pi$ scattering and extract the $D_0^*(2300)$ resonance;\chapref{chap:three_body_problems} addresses a three-body problem and extracts the $\omega(782)$ resonance;\chapref{chap:three_body_problems2} studies a more complicated three-body problem and extracts the $\pi(1300)$ resonance;\chapref{chap:decay} discusses precision measurements and hadron decays in lattice studies; and\chapref{chap:conclusion} summarizes the thesis and offers an outlook on future work.

\cleardoublepage
\chapter{Lattice Quantum Chromodynamics}
\label{chap:lattice_QCD}

{
\kaishu
\begin{center}
    天地虽大，其化均也；万物虽多，其治一也。
\end{center}
\hfill ——《天地》[战国] 庄子
}

\section{The Action of Lattice Quantum Chromodynamics}
Lattice field theory is a gauge theory quantized through the lattice path integral and formulated on discrete \mybf{Euclidean spacetime}\footnotecircle{For textbook references, see Refs.~\cite{Rothe:1992nt, DeGrand:2006zz, Creutz:1983njd, Gattringer:2010zz, liuchuan}.}. Quark fields are defined on lattice sites, while gluon fields are defined on the links connecting neighboring sites. As the lattice spacing approaches zero, the theory tends to continuum QCD. In practice, the continuum theory is obtained by repeating the calculation at several lattice spacings and extrapolating the results to the continuum limit.

In\chapref{chap:introduction}, we introduced the construction of the QCD Lagrangian, Eq.~\ref{eq:lagrangian_QCD}. In this section, we use the Lagrangian after symmetry breaking and discuss the gauge transformation properties of its fermion and gauge-field parts.

In continuum spacetime, the QCD action can be separated into a fermion part, $S_F[q, \bar{q}, A]$, which contains the quark fields and their interactions with gluons, and a gauge-field part, $S_G[A]$, which describes gluon propagation and self-interactions. The fermion part of the QCD action, $S_F[q,\bar{q},A]$, is a bilinear functional of the fields $q$ and $\bar{q}$:
\begin{equation}
    S_F[q,\bar{q},A] = \sum_{f=1}^{N_f} \int d^4x \, \bar{q}^{f}(x) \left[ \gamma_\mu (\partial_\mu + i g_s A_\mu^a(x) \frac{\lambda^a}{2}) + m^{f} \right] q^{f}(x).
\end{equation}
Here $m^{f}$ is the current quark mass of flavor $f$. The gauge action is
\begin{equation}
    S_G[A] = \frac{1}{4} \int d^4x \, F_{\mu\nu}^a(x) F_{\mu\nu}^a(x).
\end{equation}
Compared with Eq.~\ref{eq:lagrangian_QCD}, there is an additional minus sign, which originates from the Wick rotation from Minkowski space to Euclidean space.

Under a gauge transformation $\Omega(x)$, the quark and gauge fields transform as
\begin{equation}
\begin{aligned}
    q(x) &\to q'(x) = \Omega(x) q(x), \\
    \bar{q}(x) &\to \bar{q}'(x) = \bar{q}(x) \Omega(x)^\dagger, \\
    A_\mu(x) &\to A_\mu'(x) = \Omega(x) A_\mu(x) \Omega(x)^\dagger + i \partial_\mu \Omega(x) \Omega(x)^\dagger .
\end{aligned}
\end{equation}

\subsection{The Fermion Action}
Discrete Euclidean spacetime is defined on a four-dimensional lattice $\Lambda$:
\begin{equation}
    \Lambda = \{ n = (n_1, n_2, n_3, n_4) | n_1, n_2, n_3 = 0, 1, \ldots, N-1 ; n_4 = 0, 1, \ldots, N_T-1 \}.
\label{eq:lattice}
\end{equation}
The spacing between neighboring spacetime points is $a$. When a field is written as a function of $n$, it is evaluated at the spacetime point $x = an$. Sometimes one also uses anisotropic lattices, for which the temporal and spatial lattice spacings differ. Such lattices can improve the temporal resolution while keeping the spatial volume fixed. In addition, a smaller temporal lattice spacing provides denser sampling of the exponential decay of correlation functions, improving the extraction of spectra, especially excited states. Anisotropic lattices, however, require the tuning of an additional anisotropy parameter to ensure that gluons and fermions propagate with the same speed and that physical isotropy is maintained.

On discrete spacetime, the dynamical variables of QCD are link variables $U_\mu(n)_{ab} \in \mathrm{SU}(3)$, defined on the links between lattice sites, and Grassmann-valued fermion fields defined on the sites,
$q^{f}(n)_{\substack{\alpha \\ a}}$ and $\bar{q}^{f}(n)_{\substack{\alpha \\ a}}$. Their spacetime positions are denoted by $x$. The gauge field carries a Lorentz index $\mu = 1,2,3,4$ and color indices $a,b = 1,2,3$.

The quark field carries a Dirac index $\alpha = 1,2,3,4$, a color index $a = 1,2,3$, and a flavor index $f = 1,2,\ldots,6$. We denote the number of sea-quark flavors by $N_f$. Current studies often use degenerate up and down quarks, together with a strange quark or a charm quark, corresponding to $N_f = 2+1$ or $N_f = 2+1+1$. Sea-bottom effects are usually neglected, because the bottom-quark mass is much larger than the QCD scale~\cite{Athenodorou:2018wpk}.

If the link variables are assigned the gauge transformation law
\begin{equation}
    U_\mu(n) \to U'_\mu(n) = \Omega(n) U_\mu(n) \Omega(n+\hat{\mu})^\dagger,
\end{equation}
one can construct the following gauge-invariant, so-called naive fermion action:
\begin{equation}
    S_F[q,\bar{q},U] = a^4 \sum_{n\in\Lambda} \bar{q}(n) \left[ \gamma_\mu \frac{U_\mu(n)q(n+\hat{\mu}) - U_{-\mu}(n)q(n-\hat{\mu})}{2a} + m q(n) \right].
\label{eq:naive_fermion}
\end{equation}
For convenience, we have defined
\begin{equation}
    U_{-\mu}(n) \equiv U_\mu(n - \hat{\mu})^\dagger.
\end{equation}

It can be shown that, in the continuum limit, the fermion part of Eq.~\ref{eq:naive_fermion} indeed approaches the continuum fermion action. Note that the transformation behavior of the link variable is the same as that of the gauge transporter along the path $C_{xy}$ connecting two points $x$ and $y$,
\begin{equation}
    G(x,y) = \mathcal{P} \exp \left( i \int_{C_{xy}} A \cdot ds \right).
\end{equation}
The link variable $U_\mu(n)$ can therefore be viewed as a discrete gauge transporter connecting the points $n$ and $n+\hat{\mu}$. Gauge link variables differ from gauge fields in continuum spacetime: the former are group elements, while the latter are algebra elements. If the lattice is assumed to be embedded in continuum spacetime, they are naturally related by
\begin{equation}
    U_\mu(n) = \exp(i a A_\mu(n)).
\label{eq:relation_U_A}
\end{equation}

Assuming that the lattice spacing $a$ is small, one has
\begin{equation}
\begin{aligned}
    q(n \pm \hat{\mu}) &= q(n) + O(a), \\
    U_\mu(n) &= 1 + i a A_\mu(n) + O(a^2), \\
    A_\mu(n \pm \hat{\mu}) &= A_\mu(n) + O(a).
\end{aligned}
\end{equation}
Substituting these expressions into Eq.~\ref{eq:naive_fermion} gives the continuum fermion action.

The naive fermion action can be written as
\begin{equation}
    S_F[q, \bar{q}, U] = a^4 \sum_{f=1}^{N_f} \sum_{n, m \in \Lambda} \bar{q}^{f}(n) \, D^{f}(n | m) \, q^{f}(m),
\end{equation}
where the lattice Dirac operator is
\begin{equation}
    D^{f}(n | m)_{\substack{\alpha \beta \\ ab}} = m^f \delta_{\alpha \beta} \delta_{ab} \delta_{n, m} + \frac{1}{2 a} \sum_{\mu= \pm 1}^{ \pm 4} (\gamma_\mu)_{\alpha \beta} U_\mu(n)_{ab} \delta_{n+\hat{\mu}, m}.
\end{equation}
Here we have defined $\gamma_{-\mu}=-\gamma_\mu$.

Naive fermions have a serious problem: in the continuum limit they produce $16$ fermion species, or doublers, of which the additional $15$ are unphysical. This can be seen in the trivial gauge field $U_\mu(n) = 1$. The Fourier transform of the Dirac operator is
\begin{equation}
    D(p|q) = \frac{1}{L^3 N_T} \sum_{n,m\in \Lambda} e^{-ip\cdot n a} D(n|m) e^{iq\cdot m a} = \delta(p-q) D(p),
\end{equation}
where
\begin{equation}
    D(p) = m \mathbbm{1} + \frac{i}{a} \sum_{\mu=1}^{4} \gamma_\mu \sin(p_\mu a).
\end{equation}
Its inverse,
\begin{equation}
    D^{-1}(p) = \frac{m\mathbf{1}-i\sum_\mu \gamma_\mu \frac{\sin(p_\mu a)}{a}}{m^2 + \sum_\mu \frac{\sin^2(p_\mu a)}{a^2}},
\end{equation}
is the quark propagator in momentum space. If $m=0$, then in the continuum limit there is a pole at $p = (0,0,0,0)$, corresponding to the single fermion described by the continuum Dirac operator. However, $D^{-1}(p)$ also has $15$ unphysical poles located at the boundary of the Brillouin zone, namely
\begin{equation}
    p = (\pi/a,0,0,0), (0,\pi/a,0,0), \dots, (\pi/a,\pi/a,\pi/a,\pi/a).
\end{equation}
These unphysical poles are called \mybf{doublers}. One way to remove doublers in the continuum limit is to add an additional Wilson term\footnotecircle{In coordinate space, the Wilson term is proportional to the Laplacian.}, so that the momentum-space Dirac operator becomes
\begin{equation}
    D(p) = m\mathbf{1} + i \sum_{\mu=1}^4 \frac{\gamma_\mu \sin(p_\mu a)}{a} + \sum_{\mu=1}^4 \frac{1 - \cos(p_\mu a)}{a}.
\end{equation}
This term does not affect the physical fermion pole at $p_\mu=0$ in the continuum limit, but it acts as a mass term for doublers with $p_\mu = \pi/a$, giving them masses $m + \frac{2l}{a}$ that diverge in the continuum limit, where $l$ is the number of momentum components with $p_\mu=\pi/a$. Thus the doublers decouple from the theory.

The Wilson fermion action can finally be written as
\begin{equation}
    S_F[q, \bar{q}, U] = a^4 \sum_{f=1}^{N_f} \sum_{n, m \in \Lambda} \bar{q}^{f}(n) \, D^{f}(n | m) \, q^{f}(m),
\end{equation}
where the lattice Dirac operator is
\begin{equation}
    D^{f}(n | m)_{\substack{\alpha \beta \\ ab}} = \left(m^{f}+\frac{4}{a}\right) \delta_{\alpha \beta} \delta_{ab} \delta_{n, m} -\frac{1}{2 a} \sum_{\mu= \pm 1}^{ \pm 4} \left(\mathbbm{1}-\gamma_\mu\right)_{\alpha \beta} U_\mu(n)_{ab} \delta_{n+\hat{\mu}, m}.
\end{equation}

Starting from Wilson fermions, one can remove $O(a)$ discretization errors by introducing an additional \mybf{clover} term~\cite{Sheikholeslami:1985ij}; the coefficient $c_{\rm SW}$ of the clover term can be determined either at tree level or nonperturbatively. In addition, many other types of fermion actions are widely used in lattice QCD. Common choices include staggered fermions~\cite{Kogut:1974ag}, which reduce the number of doublers by a factor of $4$ by staggering the Dirac degrees of freedom and are computationally inexpensive; domain-wall fermions~\cite{Kaplan:1992bt, Shamir:1993zy} and overlap fermions~\cite{Neuberger:1997fp}, which realize approximate or exact chiral symmetry but are computationally more expensive; and twisted-mass fermions~\cite{Frezzotti:2000nk}, which achieve automatic $O(a)$ improvement at maximal twist but break isospin symmetry and parity. The Nielsen--Ninomiya theorem~\cite{Nielsen:1980rz} tells us that locality, translational invariance, chiral symmetry, and the absence of doublers cannot all be satisfied simultaneously on the lattice. Different fermion discretizations therefore make different compromises among these conditions. For different physical problems, one may choose different fermion discretizations.

\subsection{The Gauge-Field Action}
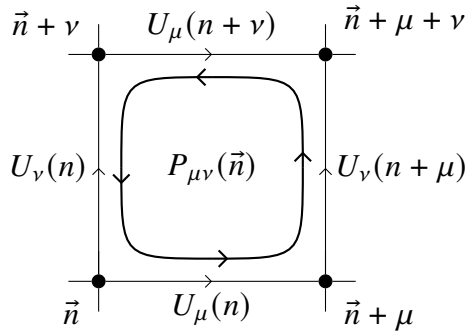
\begin{figure}[htbp]
    \centering
    \tikzset{
        gauge link/.style={
        },
        link arrow/.style={
            decoration={
                markings,
                mark=at position 0.1 with {\arrow{Straight Barb}},
            },
            postaction={decorate},
        },
        link arrow midway/.style={
            decoration={
                markings,
                mark=at position 0.5 with {\arrow{Straight Barb}},
            },
            postaction={decorate},
        },
        gauge link arrow/.style={
            gauge link,
            link arrow midway,
        },
        lattice point/.style={
            draw,
            fill,
            circle,
            minimum size=1.7mm,
            inner sep=0,
        },
    }
\begin{tikzpicture}
    \node[lattice point] (n) at (0, 0) {};
    \node[lattice point] (n-mu) at (30mm, 0) {};
    \node[lattice point] (n-nu) at (0, 30mm) {};
    \node[lattice point] (n-mu-nu) at (30mm, 30mm) {};

    \node[below left=0.5mm of n] {$\vec n$};
    \node[below right=0.5mm of n-mu] {$\vec n + \mu$};
    \node[above left=0.5mm of n-nu] {$\vec n + \nu$};
    \node[above right=0.5mm of n-mu-nu] {$\vec n + \mu + \nu$};

    \draw[gauge link] (n) -- ++(-4mm, 0);
    \draw[gauge link] (n) -- ++(0, 4mm);
    \draw[gauge link] (n) -- ++(0, -4mm);

    \draw[gauge link] (n-mu) -- ++(4mm, 0);
    \draw[gauge link] (n-mu) -- ++(0, 4mm);
    \draw[gauge link] (n-mu) -- ++(0, -4mm);

    \draw[gauge link] (n-nu) -- ++(-4mm, 0);
    \draw[gauge link] (n-nu) -- ++(0, 4mm);

    \draw[gauge link] (n-mu-nu) -- ++(4mm, 0);
    \draw[gauge link] (n-mu-nu) -- ++(0, 4mm);

    \draw[gauge link arrow] (n) -- (n-mu) node[below, midway] {$ U_\mu(n)$};
    \draw[gauge link arrow] (n) -- (n-nu) node[left, midway] {$ U_\nu(n)$};
    \draw[gauge link arrow] (n-mu) -- (n-mu-nu) node[right, midway] {$ U_\nu(n +
    \mu)$};
    \draw[gauge link arrow] (n-nu) -- (n-mu-nu) node[above, midway] {$ U_\mu(n +
    \nu)$};

    \newcommand\cloverlink{
        \draw[link arrow, thick] (1.5, 0.3) .. controls (2.7, 0.3) and (2.7, 0.3) ..
        (2.7, 1.5);
    }

    \cloverlink

    \begin{scope}[transform canvas={rotate around={90:(1.5, 1.5)}}]
        \cloverlink
    \end{scope}
    \begin{scope}[transform canvas={rotate around={180:(1.5, 1.5)}}]
        \cloverlink
    \end{scope}
    \begin{scope}[transform canvas={rotate around={270:(1.5, 1.5)}}]
        \cloverlink
    \end{scope}

    \node at (1.5, 1.5) {$ P_{\mu\nu}(\vec n)$};
\end{tikzpicture}
\caption{The plaquette in the gauge-field action.}
\label{fig:plaquette}
\end{figure}

For any closed lattice path $L$, the trace of the ordered product along the path,
\begin{equation}
    L[U] = \operatorname{tr} \left[ \prod_{(n,\mu)\in L} U_\mu(n) \right],
\end{equation}
is gauge invariant. The shortest nontrivial closed loop on the lattice is the so-called \mybf{plaquette}, shown in Fig.~\ref{fig:plaquette}:
\begin{equation}
    U_{\mu\nu}(n) = U_\mu(n) U_\nu(n+\hat{\mu}) U_{-\mu}(n+\hat{\mu}+\hat{\nu}) U_{-\nu}(n+\hat{\nu}) = U_\mu(n) U_\nu(n+\hat{\mu}) U_\mu(n+\hat{\nu})^\dagger U_\nu(n)^\dagger .
\end{equation}

This leads to the simplest gauge-field action, the Wilson gauge action:
\begin{equation}
    S_G[U] = \frac{2}{g^2} \sum_{n\in\Lambda} \sum_{\mu<\nu} \Re \, \operatorname{tr} [\mathbbm{1} - U_{\mu\nu}(n)].
\end{equation}
Using Eq.~\ref{eq:relation_U_A} and expanding the gauge field $A_\mu(n)$ to $O(a^2)$, one can show that this action indeed reproduces the continuum limit as $a\to0$.

In practical applications, additional operators are often added to reduce lattice artifacts and improve the action. For example, the Symanzik-improved action~\cite{Symanzik:1983dc,Luscher:1984xn} augments the Wilson plaquette gauge action with $1 \times 2$ or $2 \times 1$ rectangular loops, achieving $O(a^2)$ improvement. Tadpole improvement~\cite{Lepage:1992xa} further applies mean-field corrections to the coefficients in the Symanzik-improved action. The Chinese Lattice QCD Collaboration (CLQCD) uses this type of gauge action~\cite{CLQCD:2023sdb}, with the plaquette and rectangle coefficients determined at tree level. In addition, the Iwasaki action is a renormalization-group-improved action~\cite{Iwasaki:1983iya}, often used for ensembles with fermion actions designed to preserve chiral symmetry.

Once the action has been specified, vacuum expectation values of observables can be computed through the path integral,
\begin{equation}
    \langle O\rangle=\frac{1}{Z} \int \mathcal{D}[q, \bar{q}] \mathcal{D}[U] \, \mathrm{e}^{-S_F[q, \bar{q}, U]-S_G[U]} \, O[q, \bar{q}, U],
\label{eq:vev_exp}
\end{equation}
where the integration measures for the fermion and gauge fields are products of the measures over all field variables:
\begin{equation}
\begin{aligned}
    \mathcal{D}[q, \bar{q}] &=\prod_{n \in \Lambda} \prod_{f, \alpha, c}  \mathrm{d} q^{f}(n)_{\alpha c} \, \mathrm{d} \bar{q}^{f}(n)_{\alpha c}, \\
    \mathcal{D}[U] &=\prod_{n \in \Lambda} \prod_{\mu=1}^4 \mathrm{d} U_\mu(n).
\end{aligned}
\end{equation}
The partition function is
\begin{equation}
    Z=\int \mathcal{D}[q, \bar{q}] \mathcal{D}[U] \, \mathrm{e}^{-S_F[q, \bar{q}, U]-S_G[U]}.
\end{equation}
The measure $\mathrm{d} U$ for link variables is the Haar measure.

The mathematical structure of lattice field theory is very similar to that of statistical mechanics. Replacing the exponential $\exp(-S_E)$ by the Boltzmann factor $\exp(-\beta H)$, where $H$ is the Hamiltonian or energy functional of the statistical-mechanics system, Eq.~\ref{eq:vev_exp} gives the expectation value of an observable in that system.

The typical dimension of the integral above can reach hundreds of millions, making analytic evaluation impossible. Instead, one can use Monte Carlo simulations to generate $N$ gauge-field configurations $U_n$ with probability proportional to $\mathrm{e}^{-S[U]}$, and then measure the observable $O$ on a sufficiently large number of configurations:
\begin{equation}
    \langle O \rangle \approx \frac{1}{N} \sum_{n=1}^{N} O[U_n].
\label{eq:vev_MC}
\end{equation}
This gives a numerical estimate of $\langle O \rangle$. Statistical errors can be estimated rigorously using resampling methods, such as the bootstrap.

\section{Quark Propagators}
\subsection{Smearing Techniques}
The quark propagator is the inverse of the Dirac operator, $D^{-1}(y|x)_{\substack{\beta \alpha \\ b a}}$, and is the foundation of almost all lattice calculations, as we will see in concrete examples below. The operator $D$ is a high-dimensional sparse matrix of dimension $(L^3 \times N_T \times 4 \times 3)^2$. Computing the full inverse matrix directly is very expensive; this corresponds to an all-to-all propagator. In contrast, computing a single column of $D^{-1}$ is affordable, and many calculations require only such columns. These elements can be obtained by solving a linear system:
\begin{equation}
    D^{-1}(y|x_0)_{\substack{\beta \alpha_0 \\ b a_0}} = D^{-1}(y|x)_{\substack{\beta \alpha \\ b a}} \, S^{(m_0,\alpha_0,a_0)}_{0}(m)_{\substack{\alpha \\ a}}.
\label{eq:quark_propagator}
\end{equation}
Here we have introduced a point source,
\begin{equation}
    S^{(m_0,\alpha_0,a_0)}(m)_{\substack{\alpha \\ a}} = \delta_{m,m_0}\,\delta_{\alpha\alpha_0}\,\delta_{aa_0}.
\end{equation}
A single solve gives the point-source propagator from a specified spacetime point $x_0$, Dirac index $\alpha_0$, and color index $a_0$. What is usually called a point-source propagator is obtained by fixing the spacetime source point and using each of the $4 \times 3$ Dirac and color components as a source in turn. This is called a point-to-all propagator.

The source used in Eq.~\ref{eq:quark_propagator} is spatially localized. Thus the propagator is known to originate from $x_0$ and end at all sink points $y$. Hadrons, however, are not point particles; they contain quark and gluon degrees of freedom and have spatial structure. When constructing operators, one therefore often wants the operators to be spatially extended, so that they project more efficiently onto hadron states. This does not artificially change the physical states. With sufficiently long lattice time extent and sufficiently high precision, any operator with the same quantum numbers projects onto the same energy eigenstates. At finite precision, however, better-designed operators can project more strongly onto the desired states and thus improve the signal-to-noise ratio. If the smearing matrices at the source and sink are denoted by $S$ and $S^*$, respectively, then when the two ends of a propagator are connected to such smeared operators to create a given hadron, all spatial linear combinations contained in the smearing must be included:
\begin{equation}
    S^* D^{-1} S.
\label{eq:quark_propagator_smeared}
\end{equation}
If only one end is smeared, the corresponding factor $S$ or $S^*$ is removed. At the sink, since the propagator contains all spatial information, $S^* D^{-1}$ can be obtained relatively easily. At the source, however, if one still uses the point source in Eq.~\ref{eq:quark_propagator}, one must solve separately for all spatial linear combinations involved in the smearing, often amounting to some parametrized combination over all spatial sites. A convenient approach is to replace the source used in the propagator solve by $S$, thereby obtaining $D^{-1} S$ directly. This technique is called \mybf{smearing}\footnotecircle{The Chinese translation (涂摩) follows the textbook~\cite{liuchuan}.}. Common smearing methods include Gaussian smearing~\cite{Gusken:1989qx} and Jacobi smearing~\cite{UKQCD:1993gym}. Jacobi smearing filters out high-energy quark modes by iteratively applying the three-dimensional Laplacian. Gauge fields also have corresponding smearing techniques, used to reduce high-frequency quantum fluctuations. Common methods include APE smearing~\cite{APE:1987ehd}, hypercubic (HYP) smearing~\cite{Hasenfratz:2001hp}, stout smearing~\cite{Morningstar:2003gk}\footnotecircle{The author verified that the name originated from German dark beer.}, and the gradient flow~\cite{Luscher:2010iy}.

\subsection{The Distillation Algorithm}
This section briefly introduces one of the most advanced quark-field smearing methods currently used in spectroscopic calculations~\cite{HadronSpectrum:2009krc}. This method realizes all-to-all propagation within the low-energy subspace and is constructed from the Laplacian operator on a three-dimensional spatial lattice:
\begin{equation}
    \sum_{i=1}^3 U_i(\vec{x}, t) \delta_{\vec{x}+\hat{i}, \vec{y}} 
    + U_i^{\dagger}(\vec{x}-a\hat{i}, t) \delta_{\vec{x}-\hat{i}, \vec{y}} 
    - 6 \delta_{\vec{x}, \vec{y}} .
\end{equation}
By retaining only the lowest $N_v$ eigenmodes $V(t)$, one defines the \mybf{distillation} operator $\square(t) = V(t) V^{\dagger}(t)$. Acting with this operator on the quark field, $\square(t) \psi(\vec{y},t)$, is equivalent to smearing the field and suppressing high-momentum modes.

In this framework, contractions of correlation functions can be decomposed into momentum-projected eigenvector products and quark perambulators\footnotecircle{The term literally translates to “baby carriage.” The author consulted the researcher who coined the term, who confirmed that the name simply reflects its literal meaning. This is very much in the spirit of naming conventions in theoretical physics.}. As a smearing technique, the method replaces the propagator by
\begin{equation}
    S_{\substack{\beta \alpha\\ ba}}(y;x) 
    \to S^{\text{sm}}_{\substack{\beta \alpha\\ ba}}(y;x) 
    = (VV^{\dagger} S VV^{\dagger})_{\substack{\beta \alpha\\ ba}}(y;x) 
    = V_{bj}(t_y,\vec{y}) 
    \tau_{\substack{\beta \alpha\\ ji}}(t_y;t_x) 
    V^*_{ai}(t_x,\vec{x}),
\label{eq:semar_prop}
\end{equation}
where the perambulator is defined by
\begin{equation}
    \tau_{\alpha \beta}(t_1, t_2) 
    = V^{\dagger}(t_1) M^{-1}_{\alpha \beta}(t_1, t_2) V(t_2).
\end{equation}
Its form is precisely that of Eq.~\ref{eq:quark_propagator_smeared}. The key computational advantage is that the number of matrix inversions required for the perambulator scales only with the number of eigenmodes $N_v$, and $N_v$ is much smaller than the full spatial volume required for a true all-to-all propagator. When the number of eigenvectors equals the spatial volume, the distillation operator reduces to the identity, the quarks are no longer smeared, and the cost of solving all perambulators is the same as that of computing the full propagator. In practice, however, a true all-to-all propagator is not needed. The distillation algorithm both saves computational resources and improves the signal-to-noise ratio by retaining the physical information in the low-energy modes. In addition, perambulators do not depend on the structure or momentum of a hadron, and can therefore be reused in calculations of different correlation functions.

In contractions involving meson ends, one uses the eigenvector product $[V^{\dagger}V]$:
\begin{equation}
    [V^{\dagger}V]_{ij}(\vec{p},t) 
    = \sum_{\vec{x}} e^{-i \vec{p} \cdot \vec{x}} 
    V^{\dagger}_{ia}(\vec{x},t) V_{aj}(\vec{x},t).
\end{equation}
Here $i$ is an eigenmode index. For contractions containing baryons, one may construct
\begin{equation}
    [VVV]_{ijk}(\vec{p},t) 
    = \sum_{\vec{x}} e^{-i\vec{p} \cdot \vec{x}} 
    \epsilon_{abc} V_{ai}(x) V_{bj}(x) V_{ck}(x).
\end{equation}
The object $[VVV]_{ijk}$ has the same symmetry as $\epsilon^{ijk}$.

For contractions involving four-quark operators, one needs
\begin{equation}
    [VVV^{\dagger}V^{\dagger}]_{jklm}(\vec{p},t) 
    = \sum_{\vec{x}} e^{-i\vec{p} \cdot \vec{x}} 
    \epsilon_{abc} \epsilon_{ade} 
    V_{bj}(x) V_{ck}(x) V^*_{dl}(x) V^*_{em}(x).
\end{equation}
Its symmetry is the same as that of $\epsilon^{ijk} \epsilon^{ilm}$.

When the distillation algorithm is used in contractions of operators containing $n$ quark fields, the computational cost of the eigenvector products scales as $N_v^{n+1}$. For contractions containing multiquark operators, this cost can even exceed that of the ordinary propagator method. Using the idea of sparsening~\cite{Li:2020hbj}, one can construct $S^{\text{sm}}_{\substack{\beta \alpha\\ ba}}(y;x)$ on randomly selected $\vec{y}$ and $\vec{x}$ points~\cite{Stump:2025owq}.

In contractions involving nonlocal operators, such as operators containing link variables $U_l$, the link variables must be included in the eigenvector products. Below we list the eigenvector products for operators containing one and two covariant derivatives $\overleftrightarrow{\nabla}$:
\begin{equation}
\begin{aligned}
    [V^{\dagger}U_lV]_{ij}(\vec{p},t) = \frac{1}{4} \sum_{\vec{x}} e^{-i \vec{p} \cdot \vec{x}} [ +&V^{\dagger}_{ia}(\vec{x},t) U_l(\vec{x},t)_{ab} V_{bj}(\vec{x}+\hat{l},t) (+1 + e^{-i p_l}) \\
    +&V^{\dagger}_{ia}(\vec{x},t) U_l^{\dagger}(\vec{x}-\hat{l},t)_{ab} V_{bj}(\vec{x}-\hat{l},t) (-1 - e^{+i p_l})] .
\end{aligned}
\end{equation}

\begin{equation}
\begin{aligned}
    &[V^{\dagger}U_l U_m V]_{ij}(\vec{p},t) = \frac{1}{16} \sum_{\vec{x}} e^{-i \vec{p} \cdot \vec{x}} [ \\
    +&V^{\dagger}_{ia}(\vec{x},t) U_l(\vec{x},t)_{ab} U_m(\vec{x}+\hat{l},t)_{bc} V_{cj}(\vec{x}+\hat{l}+\hat{m},t) (+1 + e^{-i p_l} + e^{-i p_m} + e^{-i p_l} \cdot e^{-i p_m}) \\
    +&V^{\dagger}_{ia}(\vec{x},t) U_l^{\dagger}(\vec{x}-\hat{l},t)_{ab} U_m(\vec{x}-\hat{l},t)_{bc} V_{cj}(\vec{x}-\hat{l}+\hat{m},t) (-1 - e^{+i p_l} - e^{-i p_m} - e^{+i p_l} \cdot e^{-i p_m}) \\
    +&V^{\dagger}_{ia}(\vec{x},t) U_l(\vec{x},t)_{ab} U_m^{\dagger}(\vec{x}+\hat{l}-\hat{m},t)_{bc} V_{cj}(\vec{x}+\hat{l}-\hat{m},t) (-1 - e^{-i p_l} - e^{+i p_m} - e^{-i p_l} \cdot e^{+i p_m}) \\
    +&V^{\dagger}_{ia}(\vec{x},t) U_l^{\dagger}(\vec{x}-\hat{l},t)_{ab} U_m^{\dagger}(\vec{x}-\hat{l}-\hat{m},t)_{bc} V_{cj}(\vec{x}-\hat{l}-\hat{m},t) (+1 + e^{+i p_l} + e^{+i p_m} + e^{+i p_l} \cdot e^{+i p_m}) ] .
\end{aligned}
\end{equation}

These eigenvector products can be generalized to operators containing $N$ covariant derivatives $\overleftrightarrow{\nabla}$. The construction is as follows. Each term can be represented by a vector that starts at the center of an $N$-dimensional hypercube and points toward one of its vertices; the hypercube contains two lattice points in each dimension. This path can be understood as being built step by step along the various directions. We call such a vector a canonical path.

These canonical paths lie inside the unit hypercube $1^N$. Each path corresponds one to one to a term in the eigenvector product $[V^{\dagger} U U \cdots U V]$. The coefficient of each term can be obtained by translating these canonical paths into all $2^N$ sub-hypercubes; each translation introduces a phase factor that reflects the relative sign associated with the displacement direction.

\section{Hadron Correlation Functions}
\subsection{Wick Contractions}
As mentioned above, observables can be computed using Monte Carlo methods, Eq.~\ref{eq:vev_MC}. For hadron spectroscopy, the observable of interest is the two-point correlation function,
\begin{equation}
    C(\vec{p}_y;\vec{p}_x) = \langle \Omega | O_{\Omega}(\vec{p}_y) O_{\Gamma}^{\dagger}(\vec{p}_x) | \Omega \rangle,
\end{equation}
where $| \Omega \rangle$ is the QCD vacuum, and $O_{\Gamma}^{\dagger}(\vec{p}_x)$ and $O_{\Gamma}(\vec{p}_y)$ are creation and annihilation operators with the same quantum numbers. By measuring on sufficiently many configurations, one obtains a numerical estimate of $C(\vec{p}_y;\vec{p}_x)$. We will later explain how the time dependence of $C(\vec{p}_y;\vec{p}_x)$ is analyzed to extract the hadron spectrum. Taking a meson as an example, consider the annihilation operator $O_{\Gamma}(\vec{p}_x) = \sum_{\vec{x}} e^{-i\vec{p}_x \cdot \vec{x}} \bar{d}(x) \Gamma^i u(x)$, and define $O_{\Gamma}^{\dagger}(\vec{p}_x) = \eta_{\bar{O}_{\Gamma}} \sum_{\vec{x}} e^{-i\vec{p}_x \cdot \vec{x}} \bar{u}(x) \Gamma^i d(x)$. Here $\eta_{\bar{O}_{\Gamma}}$ is the sign arising when the Hermitian conjugate of the operator is taken; for the $\pi$ meson, for example, $\eta_{\bar{O}_{5}} = -1$. The \mybf{Wick contraction} of the two-point function is
\begin{equation}
    \begin{aligned}
        C(\vec{p}_y;\vec{p}_x) &= -\eta_{\bar{O}_{\Gamma}} \sum_{\vec{y}, \vec{x}} e^{-i\vec{p}_y \cdot \vec{y}} e^{-i\vec{p}_x \cdot \vec{x}} \operatorname{Tr}[S_c(y;x) \Gamma^i S_l(x;y) \Gamma^f] \\
        &\to - \eta_{\bar{O}_{\Gamma}} \sum_{\vec{y}, \vec{x}} e^{-i\vec{p}_y \cdot \vec{y}} e^{-i\vec{p}_x \cdot \vec{x}} \operatorname{Tr}[V(y) V^{\dagger} S_c V V^{\dagger}(x) \Gamma^i V(x) V^{\dagger} S_l V V^{\dagger}(y) \Gamma^f] \\
        &= - \eta_{\bar{O}_{\Gamma}} \operatorname{Tr}[\tau_c(t_y;t_x) [V^{\dagger}V](\vec{p}_x,t_x) \Gamma^i \tau_l(t_x;t_y) [V^{\dagger}V](\vec{p}_y,t_y) \Gamma^f].
    \end{aligned}
\label{eq:contraction_example}
\end{equation}
Because of momentum conservation, in principle one does not need to project momentum at both source and sink; the first line above can be evaluated using ordinary point-to-all propagators. In the distillation algorithm, because an effective all-to-all propagator is available, one can apply momentum projection at both source and sink to improve precision. The arrow in the equation denotes replacing the propagator by the smeared propagator, and the final line separates the eigenvector products from the perambulators\footnotecircle{For the $\pi$ meson, $- \eta_{\bar{O}_{\Gamma}}= +1$. The two-point function is positive definite and differs by a minus sign from the convention in the textbook~\cite{Gattringer:2010zz}.}. Momentum conservation imposes $\vec{p}_y = -\vec{p}_x$. This naming convention is useful because it keeps the data structures consistent when more hadrons and more complicated contractions are considered.

It can be shown that contractions of nonlocal operators containing covariant derivatives $\overleftrightarrow{\nabla}$ have the same structure as Eq.~\ref{eq:contraction_example}. The only difference is that $[V^{\dagger}V]$ is replaced by a sum over a series of $[V^{\dagger} U U \cdots U V]$ factors. In particular, the Hermitian-conjugation sign for such nonlocal operators is the same as for their local versions, namely $\eta_{\bar{O}_{\Gamma \nabla}} = \eta_{\bar{O}_{\Gamma}}$.

Multipoint functions can be contracted in an analogous way. For matrix-element extractions, the original distillation method smears all operators, but the current operator being measured usually cannot be smeared. Although this can be handled using generalized perambulators, the computational cost is high. A very promising recent approach is the blending propagator~\cite{Hu:2025vhd}, which combines low-energy-mode perambulators with high-energy-mode stochastic propagators. In some cases, this propagator yields more precise results at lower cost~\cite{Wang:2025nsd}.

\subsection{Correlation-Function Analysis}
The previous subsection introduced a correlation function built from a pair of creation and annihilation operators. For the physical system of interest, one can consider $N_{\text{op}}$ operators with the same quantum numbers. After computing the $(N_{\text{op}})^2$ correlation matrix $C_{ij}(t)$, one can extract information about the hadron spectrum. We first consider the spectral decomposition of the correlation function. For the two-point function $C_{ij}(t) = \langle \Omega | O_i(t) O_j^\dagger(0) | \Omega \rangle$, insert a complete set of Hamiltonian eigenstates, $\sum_n |n\rangle \langle n| = 1$, where $\hat{H}|n\rangle = E_n |n\rangle$. Assuming that the QCD vacuum $| \Omega \rangle$ has zero energy, one obtains
\begin{equation}
    \begin{aligned}
        C_{ij}(t) &= \frac{1}{Z_T} \operatorname{tr} \left[ e^{-\hat{H}T} e^{\hat{H}t} O_i(0) e^{-\hat{H}t} O_j^\dagger(0) \right] \\
        &= \frac{1}{Z_T} \sum_{nm} \langle m | e^{\hat{H}t} O_i(0) e^{-\hat{H}t} | n \rangle \langle n | O_j^\dagger(0) | m \rangle \\
        &= \frac{1}{Z_T} \sum_{nm} Z_i^{mn} Z_j^{*mn} e^{-E_n t} e^{-E_m (T-t)} .
    \end{aligned}
\end{equation}
Here the \mybf{overlap factor} is defined as $Z_i^{mn} \equiv \langle m | O_i | n \rangle$, and the correlation function can be written as
\begin{equation}
    C_{ij}(t) = \sum_n Z_i^n Z_j^{n*} e^{-E_n t}.
\end{equation}
Thus the correlation function is a sum of infinitely many exponentials. At sufficiently large time $t$, the lowest-energy ground state dominates, while contributions from higher excited states are exponentially suppressed. If one constructs the effective mass
\begin{equation}
    m_{\text{eff}}(t) = \ln \frac{C(t)}{C(t+1)},
\end{equation}
then for $T \gg t \gg 0$, $m_{\text{eff}}(t)$ approaches a constant \mybf{plateau}, namely the ground-state energy. If the temporal extent $T$ of the lattice is not large enough, the effective mass does not reach a plateau at large time but instead continues to rise or fall. This is because the contribution from particles propagating backward in time and wrapping around the temporal boundary is no longer negligible. This is the finite temporal extent effect, or \mybf{finite-temperature effect}\footnotecircle{Some literature calls it the rather awkward ``around-the-world effect''.}; terms that vanish as $T$ becomes large are called thermal-pollution terms. One can remove them by fitting the thermal-pollution contribution directly. This approach, however, does not combine easily with the method introduced below, so a more common approach is to use the weighting-and-shifting method~\cite{Dudek:2012gj}. We will encounter an example in\chapref{chap:two_body_problems}.

Although in principle the correlation function of a single operator can be used to extract excited-state energies, in practice the signal-to-noise problem causes information to deteriorate rapidly as time increases, making accurate excited-state extraction difficult. The \mybf{generalized eigenvalue problem} (GEVP)~\cite{Michael:1982gb,Luscher:1990ck,Blossier:2009kd,Fischer:2020bgv} is a method for extracting excited-state energies\footnotecircle{Other methods currently under development include the truncated Hankel correlator (THC) method~\cite{Ostmeyer:2025igc}, the Lanczos algorithm~\cite{Wagman:2024rid}, and Prony's method~\cite{Prony1795, Beane:2009kya}. These methods may outperform the traditional GEVP approach.}. It is defined by
\begin{equation}
    C(t) v_n(t,t_0) = \lambda_n(t,t_0) C(t_0) v_n(t,t_0),
\label{eq:gevp}
\end{equation}
where $t_0$ is a reference time. It can be shown that, when $t > t_0$ and $t$ is sufficiently large, and if only the target state and one effective contaminating state are retained, the eigenvalue $\lambda_n(t,t_0)$ behaves as
\begin{equation}
    \lambda_n(t, t_0) = (1-A_n) \operatorname{e}^{-E_n(t-t_0)} + A_n \operatorname{e}^{-E_n^{\prime}(t-t_0)}.
\end{equation}
Here $A$ is a parameter between $0$ and $1$. By fitting the eigenvalues $\lambda_n(t,t_0)$, one can extract the energy $E_n$ of each level. The eigenvalues obtained by solving the GEVP on different time slices must then be ordered. One may simply order them by magnitude, or order them according to the overlap of their eigenvectors with those at a reference time. Ref.~\cite{Fischer:2020bgv} proposed a more rigorous ordering scheme.

Because the linear space contains only $N_{\text{op}}$ basis operators, high-excited-state contamination comes from states outside the operator basis and typically satisfies $E_n^{\prime} \gtrsim E_{N_{\text{op}}+1}$. At large time, the eigenvectors $v_n$ become stable, encode the overlap information between operators and states, and satisfy the orthogonality relation
\begin{equation}
    v^{n\dagger} C(t_0) v^m = \delta_{nm}.
\end{equation}
They can be used to construct optimized operators for each state:
\begin{equation}
    \Omega_{n}^{\dagger}=\sum_i v_i^{n}(t_c) O_i^{\dagger}.
\end{equation}
It can be shown that $\Omega_{n}^{\dagger}$ creates only the state with energy $E_n$:
\begin{equation}
    \langle \Omega | \Omega_m(t) \Omega_n^{\dagger}(0) | \Omega \rangle = \delta_{nm} \lambda_n(t,t_0).
\end{equation}

\section{Finite-Volume Formalism}
The energy of a single particle in a finite volume differs from its infinite-volume value only by exponentially suppressed corrections. Lattice calculations can therefore extract the ground-state mass of a particle with given quantum numbers rather precisely. When the system contains two or more particles, however, finite-volume effects scale as powers of the volume and cannot be ignored. The \mybf{quantization condition} proposed by Lüscher in Ref.~\cite{Luscher:1990ux} relates finite-volume energy spectra to infinite-volume scattering amplitudes, making finite-volume effects not merely a correction to be handled systematically but also an important tool for studying scattering processes.

In this chapter, we first introduce one derivation of the Lüscher method in the simplified case of equal-mass, distinguishable, spinless particles. We then introduce another formal approach that derives the quantization condition from infinite-volume unitarity.

\subsection{The Lüscher Method}

This section briefly reviews the derivation in Ref.~\cite{Kim:2005gf}. Unlike the original derivation based on quantum mechanics, this approach starts from relativistic quantum field theory, expands finite-volume correlation functions in terms of the Bethe--Salpeter kernel, and determines the spectrum by analyzing the singularity structure.

For a function $f(\vec{k})$ that has no singularities on real $\vec{k}$ and decreases sufficiently rapidly as $|\vec{k}|\to\infty$, the difference between a finite-volume sum and an infinite-volume integral contains only exponentially suppressed terms. Neglecting such terms, the Poisson summation formula gives
\begin{equation}
\frac{1}{L^3}\sum_{\vec{k}} f(\vec{k}) = \int \frac{d^3k}{(2\pi)^3} f(\vec{k}).
\label{eq:poisson_f}
\end{equation}

Consider the sum
\begin{equation}
    S(\vec{q}\,)\equiv\frac{1}{L^3}\sum_{\vec{k}}\frac{f(\vec{k}\,)}{q^2-k^2}.
\end{equation}
The function $f$ can be expanded in spherical harmonics:
\begin{equation}
    f(\vec{k}\,)=\sum_{l=0}^{\infty}\sum_{m=-l}^l f_{lm}(k)\,k^l\,Y_{lm}(\theta,\phi).
\end{equation}
Thus
\begin{equation}
    S(\vec{q}\,)=\sum_{l,m} S_{lm}(q)
    =\frac{1}{L^3}\sum_{\vec{k}}\frac{f_{lm}(k)}{q^2-k^2}\,k^l\,Y_{lm}(\theta,\phi).
\end{equation}

To apply the Poisson summation formula and isolate the finite-volume correction, introduce a form-factor regularization,
\[
f_{lm}(k)\rightarrow f_{lm}(k)-f_{lm}(q)e^{\alpha(q^2-k^2)}\,,\quad \alpha\to0^+\,,
\]
which gives
\begin{equation}
    S_{lm}(q)=\delta_{l0}\, \mathcal{P}\int\frac{d^{3}k}{(2\pi)^3}
    \frac{f_{00}(k)}{q^2-k^2}\,Y_{00}
    + f_{lm}(q)\,{\cal Z}_{lm}(q),
\end{equation}
where
\begin{equation}
    {\cal Z}_{lm}(q)=\frac{1}{L^3}\sum_{\vec{k}}
    \frac{e^{\alpha(q^2-k^2)}}{q^2-k^2}\,k^l\,Y_{lm}
    -\delta_{l0}\,\mathcal{P}\int\frac{d^{3}k}{(2\pi)^3}
    \frac{e^{\alpha(q^2-k^2)}}{q^2-k^2}\,Y_{00}.
\end{equation}
Here $\mathcal{P}$ denotes the principal-value prescription.

As discussed above, lattice spectra are extracted from two-point correlation functions:
\begin{equation}
    C(t)=\langle \Omega | O(t) O^\dagger(0) | \Omega \rangle.
\end{equation}
Consider its energy representation,
\begin{equation}
    \tilde{C}(E)=\int dt\, e^{-iEt}\,C_{\vec{P}}(t).
\end{equation}
In infinite volume, the singularity of $\tilde{C}(E)$ is a branch cut extending from threshold to positive infinity. In finite volume, this cut decomposes into a series of discrete poles, corresponding to the finite-volume spectrum.

The correlation function $\tilde{C}(E)$ can be written as a series expansion in the Bethe--Salpeter kernel. In the approximation where $\mathcal{O}(e^{-mL})$ exponentially suppressed terms are neglected, $\tilde{K}_l(s)$ can be regarded as volume independent.

The finite-volume one-loop contribution is
\begin{equation}
\begin{aligned}
    I &\equiv\frac{1}{L^3}\sum_{\vec{k}}\int\frac{dk_0}{2\pi}
    \frac{f(k_0,\vec{k})}{(k^2-m^2+i\varepsilon)^2} \\
    &= -i\frac{1}{L^3}\sum_{\vec{k}}
    \frac{f(\omega_k,\vec{k})}{2\omega_k (E^2-2E\omega_k)}
    + \text{regular terms} \\
    &= -\frac{i}{2E}\,\mathcal{P}\int\frac{d^3k}{(2\pi)^3}
    \frac{f(\vec{k})}{q^{2}-k^{2}}\frac{E+2\omega_k}{4\omega_k}
    -\frac{i}{2E}\sum_{l,m} f_{lm}(q)\,c_{lm}(q^{2})
    + \text{regular terms}.
\end{aligned}
\end{equation}

The second equality uses the summation formula above. The $k_0$ integral shows that the singularity originates from the on-shell condition $E=2\omega_k$.

Using the Sokhotski--Plemelj theorem, the principal-value integral can be rewritten in the form of a Feynman integral with an $i\epsilon$ prescription. The first term above, together with the regular terms, then forms the infinite-volume one-loop integral, allowing the finite-volume singular contribution to be isolated as
\begin{equation}
    I_{\text{FV}} = \frac{q\,f_{00}(q)}{8\pi E} -\frac{i}{2E}\sum_{l,m}f_{lm}(q)\, c_{lm}(q^{2}).
\end{equation}

The correlation function can be expressed as a series of such loops. Retaining only the singular parts that produce power-law volume dependence, the series can be resummed as
\begin{equation}
    \tilde{C}^{\text{FV}}(E) = - A' F \frac{1}{1 + i \mathcal{M} F} A,
\end{equation}
where $\mathcal{M}$ is the scattering amplitude, $F$ is a purely kinematic quantity after removing the dynamical function $f$, and $A$ is related to the overlap of operators with states and is not important for the quantization condition.

The pole positions of the correlation function are therefore determined by the quantization condition, which separates dynamics from kinematics:
\begin{equation}
    \det[1 + i \mathcal{M} F] = 0.
\end{equation}

\subsection{The Finite-Volume Unitarity Method}
The infinite-volume unitarity (IVU) method~\cite{Mai:2017vot} imposes the unitarity constraints satisfied by scattering amplitudes in infinite volume, thereby establishing a formal framework for relativistic many-body scattering equations. The \mybf{finite-volume unitarity} (FVU) method~\cite{Mai:2017bge} applies these unitarity conditions to finite volume and derives the corresponding finite-volume quantization condition. We first briefly introduce the FVU approach for two-body systems; detailed discussions of three-body systems will be given in\chapref{chap:three_body_problems}and\chapref{chap:three_body_problems2}.

\subsubsection{Unitarity}

The scattering matrix $S$ describes the evolution of a system from an initial state to a final state. Its basic assumption is that, as $t\to \pm \infty$, the interacting particles are in free in- and out-states. The matrix elements of $S$ in the complete basis of scattering states,
$S_{fi} = \langle \text{out} \,|\, \text{in} \rangle$,
encode all scattering information. It is usually decomposed as
$S = \mathbbm{1} + i T$,
where the $T$ matrix contains the nontrivial scattering dynamics. The $T$-matrix element can be written as\footnotecircle{For systematic introductions to scattering theory, see the English textbook~\cite{Taylor:1972pty} and the Chinese textbook~\cite{zhenghanqing}.}
\begin{equation}
    \langle\{ \vec{q} \}|S|\{ \vec{p} \}\rangle = \langle\{ \vec{q} \}|\{ \vec{p} \}\rangle + i\langle\{ \vec{q} \}|T|\{ \vec{p} \}\rangle = \mathbbm{1} + i(2\pi)^4 \delta^{(4)}(\{ \vec{q}\}-\{\vec{p} \})\langle\{ \vec{q} \}|\mathcal{M}|\{ \vec{p} \}\rangle,
\end{equation}
where $\mathbbm{1}$ denotes the set of $\delta$ functions arising from four-momentum conservation, and $\mathcal{M}$ is the scattering amplitude. The \mybf{unitarity condition}, $S S^{\dagger} = \mathbbm{1}$, expresses probability conservation. Inserting a complete momentum basis,
$\int \prod_{i=1}^n \frac{d^3\vec{k}}{(2\pi)^3}\frac{1}{2E_{\vec{k}_i}} | \{ \vec{k} \} \rangle \langle \{ \vec{k} \} |$,
one can write the unitarity relation as
\begin{equation}
    \mathcal{M}^{fi} - \mathcal{M}^{if*} = i \int \prod_{i=1}^n \frac{d^3\vec{k}}{(2\pi)^3}\frac{1}{2E_{\vec{k}_i}} \mathcal{M}^{kf*} \mathcal{M}^{ki} (2\pi)^4 \delta^{(4)}(\{ \vec{p} \}-\{\vec{k} \}),
\end{equation}
where $i$, $f$, and $k$ denote the initial, final, and intermediate states, respectively.

Consider the scattering of two spinless particles with equal mass, $p_1 + p_2 \to q_1 + q_2$. Because of energy-momentum conservation, the scattering amplitude can be written as
$\mathcal{M}(\vec{q}_1, \vec{q}_2; \vec{p}_1, \vec{p}_2) = \mathcal{M}(\vec{q}, \vec{p}, s)$.
In the center-of-mass frame, it can be expanded in partial waves:
\begin{equation}
    \mathcal{M}(\vec{q},\vec{p},s) = 4\pi \sum_{l,m,l',m'}Y_{l' m'}(\hat{q})\mathcal{M}_{l'm',lm}(s)Y_{l m}^*(\hat{p}),
\end{equation}
where $Y_{l m}(\hat{p})$ are spherical harmonics and
$\mathcal{M}_{l'm',lm}(s) = \frac{1}{(4\pi)}\int d\Omega_{\hat{q}} d\Omega_{\hat{p}} \, Y_{l' m'}^*(\hat{q}) \mathcal{M}(\vec{q},\vec{p},s) Y_{l m}(\hat{p})$.
Rotational symmetry implies $l'=l$. If the incoming momentum direction is chosen as the $z$ axis, the orthogonality of spherical harmonics\footnotecircle{$\int d\Omega\, Y_{l m}(\Omega) Y_{l' m'}^*(\Omega) = \delta_{ll'} \delta_{mm'}$.} gives $m=m'=0$. The amplitude therefore depends only on $\cos\theta$, namely $\mathcal{M}(s,\cos\theta)$, and its partial-wave expansion is
\begin{equation}
    \mathcal{M}(s,\cos\theta) = 16 \pi \sum_l (2l+1)P_l(\cos\theta)\mathcal{M}_l(s),
\end{equation}
where $P_l$ are Legendre polynomials\footnotecircle{$\int_{-1}^1 P_l(x) P_{l'}(x) dx = \frac{2}{2l+1}\delta_{ll'}$.}, and
$\mathcal{M}_l(s) = \frac{1}{32\pi}\int_{-1}^1 d\cos\theta\, P_l(\cos\theta)\mathcal{M}(s,\cos\theta)$.

Combining the unitarity condition with this partial-wave expansion gives the partial-wave unitarity relation
\begin{equation}
    \mathcal{M}_l(s) - \mathcal{M}_l^{*}(s) = \frac{1}{32\pi}\int_{-1}^1 d\cos\theta\, P_l(\cos\theta)\left[\mathcal{M}(s,\cos\theta) - \mathcal{M}^{*}(s,\cos\theta)\right].
\end{equation}

If the energy lies below the first inelastic threshold, only a single two-body intermediate state exists. Using the unitarity condition further yields\footnotecircle{Here we use the product relation of Legendre functions, $\int d\cos\theta'\, P_{l'}(\cos\theta') P_{l^{\prime\prime}}(\cos(\theta-\theta')) = \frac{1}{2l'+1} \delta_{l' l^{\prime\prime}}$.}
\begin{equation}
    \begin{aligned}
        \mathcal{M}_l(s) - \mathcal{M}_l^{*}(s) &= \frac{1}{32\pi}\int_{-1}^1 d\cos\theta\, P_l(\cos\theta)\left[\mathcal{M}(s,\cos\theta) - \mathcal{M}^{*}(s,\cos\theta)\right] \\
        &= 8\pi \int_{-1}^1 d\cos\theta\, P_l(\cos\theta) \cdot i \int d \Pi_{\text{LIPS}} \\
        &\quad \cdot \sum_{l' l^{\prime\prime}} (2l'+1)(2l^{\prime\prime}+1) P_{l^{\prime\prime}}(\cos\theta^{\prime\prime}) P_{l'}(\cos\theta') \mathcal{M}_{l^{\prime\prime}}^{*}(s) \mathcal{M}_{l'}(s) \\
        &= 4i \frac{|\vec{k}|}{s} \theta(\sqrt{s} - 2m) \mathcal{M}_l(s) \mathcal{M}_l^{*}(s).
    \end{aligned}
\end{equation}
Here $\frac{\sqrt{s}}{2} = \sqrt{m^2 + |\vec{k}|^2}$, $m$ and $\vec{k}$ are the mass and scattering momentum, and $\theta(\sqrt{s} - 2m)$ is the step function. The two-body Lorentz-invariant phase space is $d \Pi_{\text{LIPS}} = \frac{1}{16 \pi^2} \frac{k}{\sqrt{s}} d\Omega \theta(\sqrt{s}-2m)$. The angle $\theta$ is the angle between the initial and final states, $\theta'$ is the angle between the initial and intermediate states, and $\theta^{\prime\prime}$ is the angle between the final and intermediate states. Defining the phase-space factor
\begin{equation}
    \rho(s) = \frac{2 k}{\sqrt{s}} \theta(\sqrt{s} - 2m),
\end{equation}
one obtains the partial-wave \mybf{optical theorem},
\begin{equation}
    \text{Im}\,\mathcal{M}_l^{-1}(s) = - \rho(s).
\end{equation}

This relation constrains only the imaginary part of the amplitude, while leaving the real part unrestricted. The amplitude can therefore be parametrized as
\begin{equation}
    \mathcal{M}_l(s) = \frac{1}{K_l^{-1}(s) - i \rho(s)} = \frac{\sqrt{s}}{2} \frac{1}{k \cot \delta_l(s) - i k}.
\end{equation}
Here the partial-wave $K$ matrix is defined by $K_l(s) = \frac{1}{\text{Re}\,\mathcal{M}_l^{-1}(s)}$, and $\delta_l(s)$ is the scattering phase shift. One may define the partial-wave $S_l$ matrix by
\begin{equation}
    S_l(s) = 1 + 2 i \rho(s) \mathcal{M}_l(s).
\end{equation}
It is straightforward to verify that it satisfies unitarity, $|S_l|^2=1$. The discussion above can be naturally generalized to scattering processes involving spin or multiple channels.

For a two-body scattering amplitude $\mathcal{M}$, the quadratic relation between momentum and energy introduces a square-root branch point at the threshold of each newly opened scattering channel. The scattering amplitude is therefore a multivalued function and can be analytically continued onto different \mybf{Riemann sheets}. In the simplest two-body scattering problem, the Riemann surface usually contains two sheets: the physical sheet and the unphysical sheet. The physical sheet corresponds to the physical region where the scattering process satisfies causality and unitarity; crossing the branch cut leads to the second Riemann sheet. Bound states correspond to below-threshold real-axis poles on the physical sheet, virtual states to below-threshold real-axis poles on the unphysical sheet, and resonances to complex poles on the second Riemann sheet. The mass $m$ and width $\Gamma$ of a resonance are determined by the pole position $\sqrt{s_0} = m - \frac{i}{2} \Gamma$. If an isolated pole exists above threshold and close to the real axis, the behavior of the $\mathcal{M}_l$ matrix near the pole is $\mathcal{M}_l \sim \frac{c^2}{s - s_0}$, where $s_0$ is the pole position.

For $N$ scattering channels, there are $2^N$ Riemann-sheet branches, connected to one another by crossing the branch cuts associated with the different thresholds. The distribution of poles on these different sheets carries model-independent information about the structure of a resonance.

\subsubsection{Quantization Condition}
We use \mybf{dispersion relations} to treat the analytic properties of $\mathcal{M}_l(s)$. Using a twice-subtracted dispersion relation,
\begin{equation}
    \frac{\mathcal{M}_l(s)-\mathcal{M}_l(s_0)-(s-s_0)\mathcal{M}_l'(s_0)}{(s-s_0)^2} = 32\pi \int\frac{d^3\vec{k}}{(2\pi)^3} \frac{1}{2E (s - 4E^2 + i\epsilon)},
\end{equation}
where $E = \sqrt{m^2 + \vec{k}^2}$.

One can define
\begin{equation}
    \Sigma^{\text{IV}}(s) = 32\pi \int\frac{d^3\vec{k}}{(2\pi)^3} \frac{1}{2E (s - 4E^2 + i\epsilon)}.
\end{equation}
Its imaginary part satisfies the unitarity requirement, but its real part is nonzero. For convenience, one can define a new $\tilde{K}$ matrix:
\begin{equation}
    \mathcal{M}_l^{-1}(s) = -\tilde{K}_l^{-1}(s) + \Sigma^{\text{IV}}(s).
\end{equation}

In a finite volume $L^3$, the momentum is quantized as
\begin{equation}
    \vec{k} = \frac{2\pi\vec{n}}{L}, \quad \vec{n}\in\mathbb{Z}^3.
\end{equation}

Accordingly, the integral $\int\frac{d^3\vec{k}}{(2\pi)^3}$ is replaced by
\begin{equation}
    \int\frac{d^3\vec{k}}{(2\pi)^3} \to \frac{1}{L^3}\sum_{\vec{n}}.
\end{equation}

This gives the finite-volume expression for $\Sigma^{\text{IV}}(s)$:
\begin{equation}
    \Sigma^{\text{FV}}(s) = 32\pi \frac{1}{L^3}\sum_{\vec{n}} \frac{1}{2E (s - 4E^2 + i\epsilon)}.
\end{equation}

The poles of the finite-volume scattering amplitude correspond to the discrete energy levels obtained from lattice calculations, leading to the quantization condition
\begin{equation}
    \tilde{K}_l^{-1}(s) = \Sigma^{\text{FV}}(s).
\end{equation}

\section{Ensemble Information}
The work in\chapref{chap:two_body_problems},\chapref{chap:three_body_problems}, and\chapref{chap:three_body_problems2} is based on $2+1$-flavor QCD ensembles generated by the Chinese Lattice QCD Collaboration. The gauge action is the tadpole-improved tree-level Symanzik gauge action (TITLS), and the fermion action is the tadpole-improved tree-level clover fermion action (TITLC). In the gauge action, the coupling parameter is written as $\hat{\beta}=10/(g_0^2u_0^4)$, where $u_0$ is the tadpole-improvement factor defined from the average thin-link plaquette. The fermion action uses once-stout-smeared link variables $V$, with smearing parameter $\rho=0.125$, and its clover coefficient is $c_{\rm sw}=1/v_0^3$, where $v_0$ is defined from the average plaquette of the stout-smeared links.

The ensemble information is listed in Table~\ref{tab:ensembles}. The volume in the table denotes the dimensionless lattice size $\tilde{L}^3\times\tilde{T}$ in lattice units, and the physical spatial length is $L=a\tilde{L}$. The lattice spacing $a$ is determined using the gradient-flow method~\cite{Luscher:2010iy}, with the Symanzik action and the scale parameter $w_0$~\cite{BMW:2012hcm}. The quantities $\tilde{m}_l^b$ and $\tilde{m}_s^b$ are the dimensionless bare mass parameters of the degenerate light quarks and the strange quark, respectively, namely $\tilde{m}_q^b=a m_q^b$. $M_\pi$ is the unitary pion mass on the corresponding ensemble, and $M_\pi L$ is the dimensionless combination that characterizes the size of finite-volume effects.

It should be noted that the negative bare mass parameters in the table do not imply negative physical quark masses. Wilson-type fermion actions explicitly break chiral symmetry, so the quark mass receives an additive renormalization. The critical bare mass $\tilde{m}_{\rm crit}$ at which the PCAC (partially conserved axial current) quark mass and the $\pi$ meson mass vanish is not zero. In this convention, the physically meaningful quantity is the offset from the critical mass, for example $\tilde{m}_q^{\rm PC}=k_m(\tilde{m}_q^b-\tilde{m}_{\rm crit})$, which becomes a positive physical quark mass after renormalization. Thus the negative numbers in the table are simply a consequence of the input convention for this clover action and of the additive mass shift.
\begin{table}
    \centering
    \caption{Ensemble information. $\tilde{L}^3\times\tilde{T}$ is the dimensionless volume in lattice units, $a$ is the lattice spacing, $\hat{\beta}=10/(g_0^2u_0^4)$, $\tilde{m}_{l,s}^b=a m_{l,s}^b$, $M_{\pi}$ is the unitary pion mass, $L=a\tilde{L}$, and $M_{\pi}L$ is the dimensionless finite-volume parameter.}
    \begin{tabular}{cccccccc}
        \toprule
        Ensemble & Volume & $a$ / fm & $\hat{\beta}$ & $\tilde{m}_l^b$ & $\tilde{m}_s^b$ & $M_{\pi}$ / MeV & $M_{\pi} L$ \\
        \midrule
        C24P34 & $24^3 \times 64$  & $0.10530(18)$ & $6.200$ & $-0.2770$ & $-0.2310$ & $340$ & $4.38$ \\
        C24P29 & $24^3 \times 72$  & $0.10530(18)$ & $6.200$ & $-0.2770$ & $-0.2400$ & $292$ & $3.75$ \\
        C32P29 & $32^3 \times 64$  & $0.10530(18)$ & $6.200$ & $-0.2770$ & $-0.2400$ & $292$ & $5.01$ \\
        C24P23 & $24^3 \times 64$  & $0.10530(18)$ & $6.200$ & $-0.2790$ & $-0.2400$ & $230$ & $2.93$ \\
        C32P23 & $32^3 \times 64$  & $0.10530(18)$ & $6.200$ & $-0.2790$ & $-0.2400$ & $228$ & $3.91$ \\
        C48P23 & $48^3 \times 96$  & $0.10530(18)$ & $6.200$ & $-0.2790$ & $-0.2400$ & $225$ & $5.79$ \\
        C48P14 & $48^3 \times 96$  & $0.10530(18)$ & $6.200$ & $-0.2825$ & $-0.2310$ & $133$ & $3.56$ \\
        C64P14 & $64^3 \times 128$ & $0.10530(18)$ & $6.200$ & $-0.2825$ & $-0.2310$ & $133$ & $4.63$ \\
        \midrule
        E28P35 & $28^3 \times 64$  & $0.08973(20)$ & $6.308$ & $-0.2490$ & $-0.2170$ & $351$ & $4.46$ \\
        E32P29 & $32^3 \times 64$  & $0.08973(20)$ & $6.308$ & $-0.2490$ & $-0.2170$ & $287$ & $4.19$ \\
        E32P22 & $32^3 \times 96$  & $0.08973(20)$ & $6.308$ & $-0.2490$ & $-0.2170$ & $215$ & $3.14$ \\
        \midrule
        F24P30 & $24^3 \times 96$  & $0.07746(18)$ & $6.410$ & $-0.2295$ & $-0.2050$ & $303$ & $2.86$ \\
        F32P30 & $32^3 \times 96$  & $0.07746(18)$ & $6.410$ & $-0.2295$ & $-0.2050$ & $303$ & $3.81$ \\
        F48P30 & $48^3 \times 96$  & $0.07746(18)$ & $6.410$ & $-0.2295$ & $-0.2050$ & $305$ & $5.72$ \\
        F32P21 & $32^3 \times 64$  & $0.07746(18)$ & $6.410$ & $-0.2320$ & $-0.2050$ & $210$ & $2.67$ \\
        F48P21 & $48^3 \times 96$  & $0.07746(18)$ & $6.410$ & $-0.2320$ & $-0.2050$ & $208$ & $3.91$ \\
        F64P14 & $64^3 \times 128$ & $0.07746(18)$ & $6.410$ & $-0.2320$ & $-0.2050$ & $136$ & $3.41$ \\
        \midrule
        G32P35 & $32^3 \times 96$  & $0.06887(12)$ & $6.498$ & $-0.2150$ & $-0.1926$ & $352$ & $3.94$ \\
        G36P29 & $36^3 \times 108$ & $0.06887(12)$ & $6.498$ & $-0.2150$ & $-0.1926$ & $297$ & $3.73$ \\
        \midrule
        H48P32 & $48^3 \times 144$ & $0.05187(26)$ & $6.72$  & $-0.1850$ & $-0.1700$ & $321$ & $4.06$ \\
        \midrule
        I64P31 & $64^3 \times 128$ & $0.03761(08)$ & $7.020$ & $-0.1569$ & $-0.1475$ & $312$ & $3.81$ \\
        I64P19 & $64^3 \times 128$ & $0.03761(08)$ & $7.020$ & $-0.1585$ & $-0.1475$ & $188$ & $2.29$ \\
        \bottomrule
    \end{tabular}
    \label{tab:ensembles}
\end{table}

The present ensembles cover several lattice spacings and volumes, making it possible to control discretization effects and finite-volume effects systematically. The lattice spacing ranges from approximately $a \sim 0.105 \,\mathrm{fm}$ to $0.038 \,\mathrm{fm}$. The light-quark masses range from the physical point, $M_{\pi}\approx 133 \,\mathrm{MeV}$, to a heavier region, $M_{\pi}\sim 350\ \mathrm{MeV}$, allowing the mass dependence to be studied and chiral extrapolations to be performed.

\cleardoublepage
\chapter{Construction of Lattice Operators}
\label{chap:operators}

{
\kaishu
\begin{center}
    春江潮水连海平，海上明月共潮生。
\end{center}
\hfill ——《春江花月夜》[唐] 张若虚
}

\section{Background}
\label{sec:operator_background}
\chapref{chap:lattice_QCD} introduced the basics of lattice field theory\footnotecircle{This chapter is based on the published paper~\cite{Yan:2025jlq}: H. Yan, C. Liu, L. Liu, and Y. Meng, Construction of general $N$-body lattice operators with arbitrary momenta, \textit{JHEP} 10, 210 (2025).}. Particles and their corresponding quantum fields are defined on a four-dimensional, discrete, hypercubic Euclidean spacetime lattice. Physical observables are estimated by Monte Carlo averaging of correlation functions in Euclidean spacetime. Usually, one of the four directions is chosen as the Euclidean time direction and given a longer extent so that excited states can decay sufficiently, while the remaining three directions are taken as spatial directions of equal length\footnotecircle{Some studies also use non-cubic lattices, for example Ref.~\cite{Mai:2019pqr}, to explore finite-volume spectra under different symmetries. Although the finite-volume spectrum changes when the lattice symmetry changes, the hadronic-interaction information encoded in it is invariant. This is rather like sketch training, where the same still life is observed from different angles while the object being drawn never changes.}. According to the spacetime symmetries of the ensembles used and the symmetries of the physical system of interest, one can construct a set of \mybf{interpolating operators} $\{O_i(t): i=1,\cdots N_{\mathrm{op}}\}$ with specified quantum numbers from the fundamental fields, namely quarks, antiquarks, and gauge fields. Most lattice calculations rely on the construction and analysis of multipoint correlation functions, whose general form can be written as
\begin{equation}
    C(t_1,\cdots,t_n) = \langle \Omega | O_1(t_1) O_2(t_2) \cdots O_n(t_n) | \Omega \rangle.
\label{eq:correlation_function_general}
\end{equation}
In hadron spectroscopy, the most common object is the two-point correlation matrix $C_{ij}(t)= \sum_{t^{\prime}} \langle O_i(t+t') O_j^{\dagger}(t^{\prime}) \rangle_T$, built from operators with the same quantum numbers. Here partial time-translation invariance of the ensemble is used to sum over $t^{\prime}$ on all time slices and increase statistics. By solving the generalized eigenvalue problem (GEVP)~\cite{Michael:1982gb,Luscher:1990ck,Blossier:2009kd,Fischer:2020bgv} for the correlation matrix, one can extract a set of finite-volume energy levels.\chapref{chap:lattice_QCD} showed that the overlap between an operator and an energy eigenstate is characterized by matrix elements such as $\langle \Omega | O_i | n\rangle$. We need to construct as many operators as possible to span the low-energy state space, and use the GEVP as a variational tool to extract the spectrum in an appropriate time window. As the number of operators increases, the energy spectrum in the region of interest becomes richer and more reliable. However, because of the lattice signal-to-noise problem, the signal may disappear before excited-state contamination has sufficiently decayed. Therefore, we need the operators to resemble the QCD energy eigenoperators as closely as possible; of course, we cannot achieve this exactly, since that would mean that we had already solved QCD. In the two- and three-body problems on which the author has worked, this can often be achieved by constructing operators that mimic noninteracting energy levels, and the operator basis should include all scattering channels as well as single-particle states. A typical example is the two-body spectrum of the isospin-$I = 1$ $\pi\pi$ system~\cite{Dudek:2012xn}: if two-hadron operators are omitted, the energy levels obtained from the GEVP reach plateaus only after times more than ten times the temporal lattice extent, which is clearly impractical. It is likewise observed numerically that single-hadron operators cannot be omitted either.

When the focus is on hadronic matrix elements rather than spectra, higher-point correlation functions, such as three- or four-point functions, are often required\footnotecircle{Peking University has accumulated substantial experience with higher-point correlation functions, for example in studies of the nucleon electric polarizability~\cite{Wang:2023omf} and $\pi$-meson electroproduction and weak production off the nucleon~\cite{Gao:2025loz}. In these works, to separate the contribution of the $N\pi$ excited state, two-body $N\pi$ operators must be introduced rather than simple single-nucleon operators.}. In short, regardless of the specific strong-interaction system being studied, constructing a set of hadronic operators ${O_i(t)}$ satisfying the required symmetries is the starting point of almost all lattice-QCD calculations.

This chapter focuses mainly on the \mybf{spatial symmetries} of cubic ensembles. The three-dimensional rotational symmetry of continuous space is broken on the lattice to cubic symmetry. Because of finite translation symmetry in the spatial directions, multihadron states can be classified by their discrete total three-momentum $\vec{P}$. For a given total momentum, only the spatial symmetry operations that leave $\vec{P}$ invariant survive in finite volume. These operations form subgroups of $O_h$, called \mybf{little groups}. Correspondingly, finite-volume spectra should be classified according to the irreducible representations, or irreps, of the little group, rather than according to total angular momentum in continuous space.

Center-of-mass three-momenta can be divided into the following classes:
\begin{equation}
    \vec{P} \in \Omega = \{[0,0,0],\ [0,0,n],\ [0,n,n],\ [n,n,n],\ [n,m,0],\ [n,n,m],\ [n,m,p] \}.
\label{eq:lattice_mom}
\end{equation}
The unit is $\frac{2\pi}{L}$, where $n,m,p$ are nonzero positive integers and are mutually distinct. Any three-momentum can be brought into one of the forms above by a rotation in the $O_h$ group. These classes represent all inequivalent momentum classes on the cubic lattice, and the corresponding little groups give all spatial symmetries that need to be considered. We can then construct operators that transform according to specified irreps. These $N$-hadron operators can be used in the numerical calculation of arbitrary lattice-QCD correlation functions of the kind described above.

For many years, several lattice groups have systematically studied the construction of lattice hadron operators relevant to this chapter; see Refs.~\cite{Basak:2005aq, Basak:2005ir, Basak:2007kj, Chen:2014afa, Prelovsek:2016iyo, CLQCD:2019npr, Detmold:2024ifm, Morningstar:2013bda, Wallace:2015pxa, Lyu:2022tsd}. Building on these works, this chapter organizes and synthesizes existing results, establishes a unified and self-consistent convention for operator construction, and extends it to general many-particle systems. The method applies to hadronic systems with arbitrary particle number, momentum, spin, and intrinsic quantum numbers, and treats different multihadron systems within a unified convention. We also discuss the construction of one-, two-, three-, and four-hadron operators. For two-hadron operators, we provide dictionaries up to relative momentum $1$ or $2$.

During the author's doctoral research, the author found that operator construction is often tedious and highly prone to errors. In particular, when a wider energy range is considered, the number of required operators can reach several tens, and a single operator is often itself a polynomial containing more than ten terms. The complexity of both the construction process and the resulting expressions can be intimidating to researchers entering lattice spectroscopy for the first time. One important goal of the work described in this chapter is therefore to develop and release a user-friendly \mybf{open-source package}, \texttt{OpTion}, short for Operator Construction\footnotecircle{The software is open-sourced on GitHub: \url{https://github.com/wittscien/OpTion}.}, in order to lower the entry barrier as much as possible. Today, \texttt{OpTion} has become one of the standard tools of the Chinese Lattice QCD Collaboration (CLQCD) and has been widely used in related international studies.

\section{Lattice Little Groups}
\label{sec:groups}
The symmetries of the point groups corresponding to the center-of-mass momentum set $\Omega$ decrease step by step in the order shown, eventually becoming the trivial group $G = \{E\}$, where all spatial symmetry operations are broken. In addition, for systems with half-integer total spin, one must use the corresponding double-cover groups. All symmetry groups and irreps appearing in this thesis are summarized in Table~\ref{tab:group}. Reference tables for one- and two-hadron operators are also listed\protect\footnotecircle{The WeChat group name of the Peking University lattice group is also ``lattice little group''.}.

\begin{table}[htbp]
\centering
\caption{Symmetry groups, double-cover groups, corresponding irreducible representations, and table links for one- and two-particle operators for center-of-mass momenta $\vec{P} \in \Omega$. Here $n \neq m \neq p$ are nonzero positive integers, and a superscript $D$ denotes a double-cover group.}
\addtolength{\tabcolsep}{-5pt}
\begin{tabular}{cccccccc}
\toprule
$\vec{P}$ & $[0,0,0]$ & $[0,0,n]$ & $[0,n,n]$ & $[n,n,n]$ & $[n,m,0]$ & $[n,n,m]$ & $[n,m,p]$ \\
\midrule
Point group & $O_h$ & $C_{4v}$ & $C_{2v}$ & $C_{3v}$ & $C_{2}$ & $C_{2}$ & $C_{1}$ \\
Irreps & Sec.~\ref{sec:oh} & Table~\ref{tab:c4v} & Table~\ref{tab:c2v} & Table~\ref{tab:c3v} & Table~\ref{tab:c2} & Table~\ref{tab:c2} & Table~\ref{tab:c1} \\
One-particle ops. & Table~\ref{tab:one-000} & Table~\ref{tab:one-00n} & Table~\ref{tab:one-0nn} & Table~\ref{tab:one-nnn} & Table~\ref{tab:one-nm0} & Table~\ref{tab:one-nnm} & Table~\ref{tab:one-nmp} \\
Meson-meson ops. & Table~\ref{tab:mm-000-A1+}-\ref{tab:mm-000-T2-} & Table~\ref{tab:mm-00n-A1}-\ref{tab:mm-00n-E} & Table~\ref{tab:mm-0nn-A1}-\ref{tab:mm-0nn-B2} & - & - & - & - \\
Baryon-baryon ops. & Table~\ref{tab:bb-000} & Table~\ref{tab:bb-00n} & Table~\ref{tab:bb-0nn} & Table~\ref{tab:bb-nnn} & Table~\ref{tab:bb-nm0} & Table~\ref{tab:bb-nnm} & - \\
\midrule
Double-cover point group & $O_h^D$ & $C_{4v}^D$ & $C_{2v}^D$ & $C_{3v}^D$ & $C_{2}^D$ & $C_{2}^D$ & $C_{1}^D$ \\
Irreps & Sec.~\ref{sec:oh} & Table~\ref{tab:c4vd} & Table~\ref{tab:c2vd} & Table~\ref{tab:c3vd} & Table~\ref{tab:c2d} & Table~\ref{tab:c2d} & Table~\ref{tab:c1d} \\
Meson-baryon ops. & Table~\ref{tab:mb-000-G1}-\ref{tab:mb-000-H-} & Table~\ref{tab:mb-00n} & Table~\ref{tab:mb-0nn} & - & - & - & - \\
\bottomrule
\end{tabular}
\addtolength{\tabcolsep}{5pt}
\label{tab:group}
\end{table}

\subsection{The Cubic Group}
\label{sec:oh}
In the rest frame, the lattice symmetry group is $O_h = O \otimes \{E, I\}$, where $O$ is the proper octahedral group. It contains $24$ group elements and has $5$ conjugacy classes:
$I$, $C_3$, $C_4$, $C_2^{\prime}$, and $C_2$. The group elements can be parametrized as~\cite{Bernard:2008ax}
\begin{equation}
    (R_i)_{\alpha \beta}
    = \exp\!\left(-i\,\vec{n}_i \cdot \vec{J}\,\omega_i\right)_{\alpha \beta}
    = \cos \omega_i \,\delta_{\alpha \beta}
    + \left(1-\cos \omega_i\right) n_{i\alpha} n_{i\beta}
    - \sin \omega_i \,\epsilon_{\alpha \beta \gamma} n_{i\gamma},
\end{equation}
where $(J_\gamma)_{\alpha \beta} = -i \epsilon_{\alpha \beta \gamma}$ are the generators of the group, $\vec{n}$ is the rotation axis, and $\omega$ is the rotation angle. This parametrization is also the fundamental representation $T_1$ of the group.

For systems containing half-integer spin, one must use the double cover $O^D$ of the cubic group. This group contains $48$ elements, divided into $8$ conjugacy classes: $I$, $C_6$, $C_3$, $C_8^{\prime}$, $C_8$, $C_4^{\prime}$, and $C_4$. The explicit form of each group element, its conjugacy class, and its construction are summarized in Table~\ref{tab:oh}. This table is equivalent to the results given in Ref.~\cite{Bernard:2008ax}. The ordering of the elements of the $O$ group is kept unchanged. For $O^D$, the first $24$ elements correspond one to one to the elements of $O$, while the remaining $24$ elements can be viewed as the original elements supplemented by an additional $\pm 2\pi$ rotation.

\begin{table}
\centering
\caption{Explicit forms and classes of the elements of the cubic group and its double cover. The first $24$ elements belong to the $O$ group, while the last $24$ are additional elements contained in $O^D$. The last column indicates how each element is generated by the rotations $C_{4x}$ and $C_{4y}$.}
\begin{tabular}{ccccc}
\toprule
Class & i & $\vec{n}$ & $\omega$ & \\
\midrule
$I(I/J)$ & $1 / 25$ & arbitrary & $0 /+2 \pi$ & - \\
\midrule
\multirow{8}{*}{$8 C_3\left(8 C_6 / 8 C_3\right)$} & $2 / 26$ & $\frac{1}{\sqrt{3}}(1,1,1)$ & $-\frac{2 \pi}{3} /+2 \pi$ & $C_{4 z} C_{4 y}$ \\
& $3 / 27$ & $\frac{1}{\sqrt{3}}(1,1,1)$ & $\frac{2 \pi}{3} /-2 \pi$ & $C_{4 y}^{-1} C_{4 z}^{-1}$ \\
& $4 / 28$ & $\frac{1}{\sqrt{3}}(-1,1,1)$ & $-\frac{2 \pi}{3} /+2 \pi$ & $C_{4 y} C_{4 z}$ \\
& $5 / 29$ & $\frac{1}{\sqrt{3}}(-1,1,1)$ & $\frac{2 \pi}{3} /-2 \pi$ & $C_{4 z}^{-1} C_{4 y}^{-1}$ \\
& $6 / 30$ & $\frac{1}{\sqrt{3}}(-1,-1,1)$ & $-\frac{2 \pi}{3} /+2 \pi$ & $C_{4 z} C_{4 y}^{-1}$ \\
& $7 / 31$ & $\frac{1}{\sqrt{3}}(-1,-1,1)$ & $\frac{2 \pi}{3} /-2 \pi$ & $C_{4 y} C_{4 z}^{-1}$ \\
& $8 / 32$ & $\frac{1}{\sqrt{3}}(1,-1,1)$ & $-\frac{2 \pi}{3} /+2 \pi$ & $C_{4 y}^{-1} C_{4 z}$ \\
& $9 / 33$ & $\frac{1}{\sqrt{3}}(1,-1,1)$ & $\frac{2 \pi}{3} /-2 \pi$ & $C_{4 z}^{-1} C_{4 y}$ \\
\midrule
\multirow{6}{*}{$6 C_4\left(6 C_8^{\prime} / 6 C_8\right)$} & $10 / 34$ & $(1,0,0)$ & $-\frac{\pi}{2} /+2 \pi$ & $C_{4 z} C_{4 y} C_{4 z}^{-1}$ \\
& $11 / 35$ & $(1,0,0)$ & $\frac{\pi}{2} /-2 \pi$ & $C_{4 z} C_{4 y}^{-1} C_{4 z}^{-1}$ \\
& $12 / 36$ & $(0,1,0)$ & $-\frac{\pi}{2} /+2 \pi$ & $C_{4 y}$ \\
& $13 / 37$ & $(0,1,0)$ & $\frac{\pi}{2} /-2 \pi$ & $C_{4 y}^{-1}$ \\
& $14 / 38$ & $(0,0,1)$ & $-\frac{\pi}{2} /+2 \pi$ & $C_{4 z}$ \\
& $15 / 39$ & $(0,0,1)$ & $\frac{\pi}{2} /-2 \pi$ & $C_{4 z}^{-1}$ \\
\midrule
\multirow{6}{*}{$6 C_2^{\prime}\left(12 C_4^{\prime}\right)$} & $16 / 40$ & $(0,1,1)$ & $-\pi /+2 \pi$ & $C_{4 y} C_{4 z} C_{4 y}$ \\
& $17 / 41$ & $(0,-1,1)$ & $-\pi /+2 \pi$ & $C_{4 y}^{-1} C_{4 z} C_{4 y}^{-1}$ \\
& $18 / 42$ & $(1,1,0)$ & $-\pi /+2 \pi$ & $C_{4 z} C_{4 y} C_{4 y}$ \\
& $19 / 43$ & $(1,-1,0)$ & $-\pi /+2 \pi$ & $C_{4 z}^{-1} C_{4 y}^{-1} C_{4 y}^{-1}$ \\
& $20 / 44$ & $(1,0,1)$ & $-\pi /+2 \pi$ & $C_{4 z} C_{4 z} C_{4 y}$ \\
& $21 / 45$ & $(-1,0,1)$ & $-\pi /+2 \pi$ & $C_{4 y} C_{4 z} C_{4 z}$ \\
\midrule
\multirow{3}{*}{$6 C_2\left(6 C_4\right)$} & $22 / 46$ & $(1,0,0)$ & $-\pi /+2 \pi$ & $C_{4 z} C_{4 z} C_{4 y} C_{4 y}$ \\
& $23 / 47$ & $(0,1,0)$ & $-\pi /+2 \pi$ & $C_{4 y} C_{4 y}$ \\
& $24 / 48$ & $(0,0,1)$ & $-\pi /+2 \pi$ & $C_{4 z} C_{4 z}$ \\
\bottomrule
\end{tabular}
\label{tab:oh}
\end{table}

The irreps of the group $O^{(D)}$ have been given in Ref.~\cite{Bernard:2008ax}. For completeness, we repeat them here and add several details relevant to practical operator construction.

$A_1$: $\Gamma_{A_1} = 1$

$A_2$: for classes $C_4$ and $C_2^{\prime}$, $\Gamma_{A_2}=-1$; for all other classes, $\Gamma_{A_2}=1$

$E$:
\begin{alignat*}{2}
& \Gamma_{E}=1 \quad && \text{for} \quad i=1,22,23,24, \\
& \Gamma_{E}=\sigma_3 \quad && \text{for} \quad i=14,15,18,19, \\
& \Gamma_{E}=-\cos \frac{\pi}{3} \mathbf{1}+i \sin \frac{\pi}{3} \sigma_2 && \text{for} \quad i=2,5,6,9, \\
& \Gamma_{E}=-\cos \frac{\pi}{3} \mathbf{1}-i \sin \frac{\pi}{3} \sigma_2 && \text{for} \quad i=3,4,7,8, \\
& \Gamma_{E}=-\cos \frac{\pi}{3} \sigma_3-\sin \frac{\pi}{3} \sigma_1 && \text{for} \quad i=10,11,16,17, \\
& \Gamma_{E}=-\cos \frac{\pi}{3} \sigma_3+\sin \frac{\pi}{3} \sigma_1 \quad && \text{for} \quad i=12,13,20,21
\end{alignat*}

$T_1$: $\Gamma_{T_1} = \exp (-i \vec{n}_i \cdot \vec{J} \omega_i)$

$T_2$: the same as $\Gamma_{T_1}$, but with an additional minus sign for classes $C_4$ and $C_2^{\prime}$

For the double-cover group $O^D$, the representations of the original cubic group $O$ remain, and the representations of the additional $O^D$ elements coincide with those of the group elements that differ by a $2\pi$ rotation. In addition, $O^D$ contains three irreps specific to the double-cover group:

$G_1$: $\Gamma_{G_1}\equiv Y=\exp (-\frac{i}{2} \vec{n}_i \cdot \vec{\sigma} \omega_i)$

$G_2$: the same as $\Gamma_{G_1}$, but with an additional minus sign for classes $C_8, C_8^{\prime}$ and $C_4^{\prime}$

$H$: $\Gamma_{H}=\exp (-i \vec{n}_i \cdot \vec{J}^{\frac{3}{2}} \omega_i)$, where $\vec{J}^{\frac{3}{2}}$ are the generators of the $J = 3 / 2$ representation of $SU(2)$:

\begin{equation}
S_x =\begin{pmatrix}
0 & \tfrac{\sqrt{3}}{2} & 0 & 0 \\
\tfrac{\sqrt{3}}{2} & 0 & 1 & 0 \\
0 & 1 & 0 & \tfrac{\sqrt{3}}{2} \\
0 & 0 & \tfrac{\sqrt{3}}{2} & 0
\end{pmatrix},\quad
S_y =\begin{pmatrix}
0 & -i \tfrac{\sqrt{3}}{2} & 0 & 0 \\
i \tfrac{\sqrt{3}}{2} & 0 & - i & 0 \\
0 &  i & 0 & -i \tfrac{\sqrt{3}}{2} \\
0 & 0 & i \tfrac{\sqrt{3}}{2} & 0
\end{pmatrix},\quad
S_z =\begin{pmatrix}
\tfrac{3}{2} & 0 & 0 & 0 \\
0 & \frac{1}{2} & 0 & 0 \\
0 & 0 & -\tfrac{1}{2} & 0 \\
0 & 0 & 0 & -\tfrac{3}{2}
\end{pmatrix}.
\end{equation}

\subsection{Little Groups}
Constructing operators in moving frames is important for several reasons. First, in lattice spectroscopy, the more energy levels one has, the more accurate the scattering analysis becomes. Generating new ensembles at different volumes to increase the number of levels is very resource intensive. By choosing different center-of-mass momenta on the same ensemble, however, one can use the same quark propagators to obtain more energy levels under different symmetries. Second, the QCD vacuum has zero momentum. In a moving frame, one can avoid the large uncertainties associated with vacuum subtraction in $S$-wave scattering channels that have vacuum quantum numbers. Third, moving-frame hadron states are often required in areas such as hadron structure and precision tests of the Standard Model. For example, the large-momentum effective theory approach to parton distribution functions (PDFs), reviewed in Ref.~\cite{ji:2020ect}, requires hadronic matrix elements at several nonzero center-of-mass momenta, which are then matched perturbatively to the corresponding effective theory.

When the total momentum is nonzero, the spatial symmetry of the lattice is further reduced, leaving only the little group that keeps the momentum invariant under group transformations. In particular, when the momentum class is $[n,m,0]$ or $[n,n,m]$, the little group is $C_2^{(D)}$ in both cases, although the specific symmetry operations differ slightly. It is especially important to note that parity is no longer a symmetry in moving frames, and therefore is no longer a good quantum number. This means that particles of different parity can mix in the same irrep, complicating finite-volume scattering analyses. This is nevertheless the unavoidable price of increasing the number of energy levels while keeping the computational cost acceptable. One exception is provided by the $A_1$ and $A_2$ irreps of these little groups, which can distinguish systems with zero helicity.

To simplify notation, this thesis adopts the $X$-matrix notation introduced in Ref.~\cite{Gockeler:2012yj}:
\begin{equation}
\begin{cases}
X_1=\mathbbm{1}, \\
X_2=-\frac{1}{2} \mathbbm{1}+i \frac{\sqrt{3}}{2} \sigma_2, \\
X_3=-\frac{1}{2} \mathbbm{1}-i \frac{\sqrt{3}}{2} \sigma_2, \\
X_4=-\frac{1}{2} \sigma_3-\frac{\sqrt{3}}{2} \sigma_1, \\
X_5=\sigma_3, \\
X_6=-\frac{1}{2} \sigma_3+\frac{\sqrt{3}}{2} \sigma_1, \\
X_7=i \frac{1}{\sqrt{2}}\left(\sigma_1+\sigma_2\right), \\
X_8=\frac{1}{\sqrt{2}}\left(\sigma_1-\sigma_2\right).
\end{cases}    
\end{equation}

In addition, we write $\omega = \operatorname{e}^{\frac{\pi i}{4}}$, and use $I$ to denote parity inversion. Tables~\ref{tab:c4v} through~\ref{tab:c1d} list all group elements in the various groups; the corresponding table numbers have already been encoded in Table~\ref{tab:oh}. For double-cover groups, the element $R_{25} \equiv J$ denotes a rotation by $2\pi$.

\begin{table}
\centering
\caption{Group elements and irreducible representations of the point group $C_{4v}$.}
\begin{tabular}{cccccc}
\toprule
$g$ & $E$ & $\left\{R_{14}, R_{15}\right\}$ & $\left\{I R_{18}, I R_{19}\right\}$ & $\left\{I R_{22}, I R_{23}\right\}$ & $R_{24}$ \\
\midrule
$A_1$ & $1$ & $1$ & $1$ & $1$ & $1$ \\
$A_2$ & $1$ & $1$ & $-1$ & $-1$ & $1$ \\
$B_1$ & $1$ & $-1$ & $-1$ & $1$ & $1$ \\
$B_2$ & $1$ & $-1$ & $1$ & $-1$ & $1$ \\
$E$ & $2$ & $0$ & $0$ & $0$ & $-2$ \\
\midrule
$E$ & $X_1$ & $\left\{-X_7, X_7\right\}$ & $\left\{X_5,-X_5\right\}$ & $\left\{X_8,-X_8\right\}$ & $-X_1$ \\
\bottomrule
\end{tabular}
\label{tab:c4v}
\end{table}

\begin{table}
\scriptsize
\centering
\caption{Group elements and irreducible representations of the point group $C_{4v}^D$.}
\begin{tabular}{cccccccc}
\toprule
$g$ & $E$ & $\left\{R_{48}, R_{24}\right\}$ & $\left\{R_{15}, R_{14}\right\}$ & $\left\{R_{38}, R_{39}\right\}$ & $\left\{I R_{46}, I R_{47}, I R_{22}, I R_{23}\right\}$ & $\left\{I R_{42}, I R_{43}, I R_{18}, I R_{19}\right\}$ & $R_{25}$ \\
\midrule
$A_1$ & $1$ & $1$ & $1$ & $1$ & $1$ & $1$ & $1$ \\
$A_2$ & $1$ & $1$ & $1$ & $1$ & $-1$ & $-1$ & $1$ \\
$B_1$ & $1$ & $1$ & $-1$ & $-1$ & $1$ & $-1$ & $1$ \\
$B_2$ & $1$ & $1$ & $-1$ & $-1$ & $-1$ & $1$ & $1$ \\
$E$ & $2$ & $-2$ & $0$ & $0$ & $0$ & $0$ & $2$ \\
$G_1$ & $2$ & $0$ & $\sqrt{2}$ & $-\sqrt{2}$ & $0$ & $0$ & $-2$ \\
$G_2$ & $2$ & $0$ & $-\sqrt{2}$ & $\sqrt{2}$ & $0$ & $0$ & $-2$ \\
\midrule
$E$ & $1$ & $\{-\mathbbm{1},-\mathbbm{1}\}$ & $\left\{i \sigma_3,-i \sigma_3\right\}$ & $\left\{-i \sigma_3, i \sigma_3\right\}$ & $\left\{\sigma_1,-\sigma_1, \sigma_1,-\sigma_1\right\}$ & $\left\{-\sigma_2, \sigma_2,-\sigma_2, \sigma_2\right\}$ & $2$ \\
$G_1$ & $Y_1$ & $\left\{Y_{48}, Y_{24}\right\}$ & $\left\{Y_{15}, Y_{14}\right\}$ & $\left\{Y_{38}, Y_{39}\right\}$ & $\left\{-Y_{46},-Y_{47},-Y_{22},-Y_{23}\right\}$ & $\left\{-Y_{42},-Y_{43},-Y_{18},-Y_{19}\right\}$ & $Y_{25}$ \\
$G_2$ & $Y_1$ & $\left\{Y_{48}, Y_{24}\right\}$ & $\left\{-Y_{15},-Y_{14}\right\}$ & $\left\{-Y_{38},-Y_{39}\right\}$ & $\left\{Y_{46}, Y_{47}, Y_{22}, Y_{23}\right\}$ & $\left\{-Y_{42},-Y_{43},-Y_{18},-Y_{19}\right\}$ & $Y_{25}$ \\
\bottomrule
\end{tabular}
\label{tab:c4vd}
\end{table}

\begin{table}
\centering
\caption{Group elements and irreducible representations of the point group $C_{2v}$.}
\begin{tabular}{ccccc}
\toprule
$g$ & $E$ & $R_{16}$ & $I R_{22}$ & $I R_{17}$ \\
\midrule
$A_1$ & $1$ & $1$ & $1$ & $1$ \\
$A_2$ & $1$ & $1$ & $-1$ & $-1$ \\
$B_1$ & $1$ & $-1$ & $1$ & $-1$ \\
$B_2$ & $1$ & $-1$ & $-1$ & $1$ \\
\bottomrule
\end{tabular}
\label{tab:c2v}
\end{table}

\begin{table}
\centering
\caption{Group elements and irreducible representations of the point group $C_{2v}^D$.}
\begin{tabular}{cccccc}
\toprule
$g$ & $E$ & $\left\{R_{40}, R_{16}\right\}$ & $\left\{I R_{41}, I R_{17}\right\}$ & $\left\{I R_{46}, I R_{22}\right\}$ & $R_{25}$ \\
\midrule
$A_1$ & $1$ & $1$ & $1$ & $1$ & $1$ \\
$A_2$ & $1$ & $1$ & $-1$ & $-1$ & $1$ \\
$B_1$ & $1$ & $-1$ & $-1$ & $1$ & $1$ \\
$B_2$ & $1$ & $-1$ & $1$ & $-1$ & $1$ \\
$G$ & $2$ & $0$ & $0$ & $0$ & $-2$ \\
\midrule
$G$ & $Y_1$ & $\left\{Y_{40}, Y_{16}\right\}$ & $\left\{-Y_{41},-Y_{17}\right\}$ & $\left\{-Y_{46},-Y_{22}\right\}$ & $Y_{25}$ \\
\bottomrule
\end{tabular}
\label{tab:c2vd}
\end{table}

\begin{table}
\centering
\caption{Group elements and irreducible representations of the point group $C_{3v}$.}
\begin{tabular}{cccc}
\toprule
$g$ & $E$ & $\left\{R_2, R_3\right\}$ & $\left\{I R_{17}, I R_{19}, I R_{21}\right\}$ \\
$A_1$ & $1$ & $1$ & $1$ \\
$A_2$ & $1$ & $1$ & $-1$ \\
$E$ & $2$ & $-1$ & $0$ \\
\midrule
$E$ & $X_1$ & $\left\{X_{2}, X_{3}\right\}$ & $\left\{-X_{4},-X_{5},-X_{6}\right\}$ \\
\bottomrule
\end{tabular}
\label{tab:c3v}
\end{table}

\begin{table}
\centering
\caption{Group elements and irreducible representations of the point group $C_{3v}^D$.}
\begin{tabular}{ccccccc}
\toprule
$g$ & $E$ & $\left\{R_{3}, R_{2}\right\}$ & $\left\{R_{26}, R_{27}\right\}$ & $\left\{I R_{41}, I R_{19}, I R_{21}\right\}$ & $\left\{I R_{43}, I R_{45}, I R_{17}\right\}$ & $R_{25}$ \\
\midrule
$A_1$ & $1$ & $1$ & $1$ & $1$ & $1$ & $1$ \\
$A_2$ & $1$ & $1$ & $1$ & $-1$ & $-1$ & $1$ \\
$K_1$ & $1$ & $-1$ & $1$ & $i$ & $-i$ & $-1$ \\
$K_2$ & $1$ & $-1$ & $1$ & $-i$ & $i$ & $-1$ \\
$E$ & $2$ & $-1$ & $-1$ & $0$ & $0$ & $2$ \\
$G$ & $2$ & $1$ & $-1$ & $0$ & $0$ & $-2$ \\
\midrule
$E$ & $X_1$ & $\left\{X_3, X_2\right\}$ & $\left\{X_2, X_3\right\}$ & $\left\{-X_4,-X_5,-X_6\right\}$ & $\left\{-X_5,-X_6,-X_4\right\}$ & $X_1$ \\
$G$ & $Y_1$ & $\left\{Y_{3}, Y_{2}\right\}$ & $\left\{Y_{26}, Y_{27}\right\}$ & $\left\{-Y_{41},-Y_{19},-Y_{21}\right\}$ & $\left\{-Y_{43},-Y_{45},-Y_{17}\right\}$ & $Y_{25}$ \\
\bottomrule
\end{tabular}
\label{tab:c3vd}
\end{table}

\begin{table}
\centering
\caption{Group elements and irreducible representations of the point group $C_{2}$.}
\begin{tabular}{ccc}
\toprule
$g (nm0)$ & $E$ & $I R_{24}$ \\
$g (nnm)$ & $E$ & $I R_{17}$ \\
\midrule
$A$ & $1$ & $1$ \\
$B$ & $1$ & $-1$ \\
\bottomrule
\end{tabular}
\label{tab:c2}
\end{table}

\begin{table}
\centering
\caption{Group elements and irreducible representations of the point group $C_{2}^D$.}
\begin{tabular}{ccccc}
\toprule
$g (nm0)$ & $E$ & $I R_{48}$ & $R_{25}$ & $I R_{24}$ \\
$g (nnm)$ & $E$ & $I R_{41}$ & $R_{25}$ & $I R_{17}$ \\
\midrule
$A_1$ & $1$ & $1$ & $1$ & $1$ \\ 
$A_2$ & $1$ & $-1$ & $1$ & $-1$ \\
$K_1$ & $1$ & $i$ & $-1$ & $-i$ \\
$K_2$ & $1$ & $-i$ & $-1$ & $i$ \\
\bottomrule
\end{tabular}
\label{tab:c2d}
\end{table}

\begin{table}
\centering
\caption{Group elements and irreducible representations of the point group $C_{1}$, the trivial group. In this case there are no remaining spatial rotational degrees of freedom. Of course, parity is not a symmetry transformation either.}
\begin{tabular}{cc}
\toprule
$g$ & $E$ \\
\midrule
$A$ & $1$ \\
\bottomrule
\end{tabular}
\label{tab:c1}
\end{table}

\begin{table}
\centering
\caption{Group elements and irreducible representations of the point group $C_{1}^D$.}
\begin{tabular}{ccc}
\toprule
$g$ & $E$ & $R_{25}$ \\
\midrule
$A$ & $1$ & $1$ \\ 
$K$ & $1$ & $-1$ \\
\bottomrule
\end{tabular}
\label{tab:c1d}
\end{table}

\subsection{Subduction of the Cubic Group}
When a representation of the continuous rotation group $SO(3)$ is broken to a representation of the cubic group $O$, a state carrying continuum spin $J$ decomposes into irreps $\Lambda$ of the cubic group. The first few decompositions are
\begin{equation}
    \begin{aligned}
        & J=0: A_1, \\
        & J=1: T_1, \\
        & J=2: T_2 \oplus E, \\
        & J=3: A_2 \oplus T_1 \oplus T_2, \\
        & J=4: A_1 \oplus E \oplus T_1 \oplus T_2, \\
        & \ldots.
    \end{aligned}
\end{equation}

Based on this decomposition, the subduction pattern can be summarized in Table~\ref{table:latticeirreps}. The table lists how states with $J \le 4$ and $|\lambda| \le 4$ decompose into irreps of different little groups~\cite{Dudek:2012gj}.

\begin{table}
    \centering
    \caption{Subduction patterns of representations of the continuous rotation group $SO(3)$ into representations of the cubic group $O$. Here $\tilde{\eta} \equiv P(-1)^J$.}
    \begin{tabular}{cc|cl}
        \toprule
        $\vec{P}$ & $LG(\vec{P})$ & $\Lambda^{(P)}$ & $J^{P}(|\lambda|^{(\tilde{\eta})})$  \\
        \midrule
        \multirow{5}{*}{$[0,0,0]$} & \multirow{5}{*}{$\text{O}^{\text{D}}_h$} 
        & $A_1^{\pm}$ & $0^{\pm},~ 4^{\pm},~ \ldots$ \\
        & & $T_1^{\pm}$ & $1^{\pm},~ 3^{\pm},~ 4^{\pm},~ \ldots$ \\
        & & $T_2^{\pm}$ & $2^{\pm},~ 3^{\pm},~ 4^{\pm},~ \ldots$ \\
        & & $E^{\pm}$   & $2^{\pm},~ 4^{\pm},~ \ldots$ \\
        & & $A_2^{\pm}$ & $3^{\pm},~ \ldots$ \\
        \midrule
        \multirow{5}{*}{$[0,0,n]$} & \multirow{5}{*}{$\text{Dic}_4$} 
        & $A_1$ & $0^+,~ 4,~ \ldots$ \\
        & & $A_2$ & $0^-,~ 4,~ \ldots$ \\
        & & $E_2$ & $1,~ 3,~ \ldots$ \\
        & & $B_1$ & $2,~ \ldots$ \\
        & & $B_2$ & $2,~ \ldots$ \\
        \hline
        \multirow{4}{*}{$[0,n,n]$} & \multirow{4}{*}{$\text{Dic}_2$} 
        & $A_1$ & $0^+,~ 2,~ 4,~ \ldots$ \\
        & & $A_2$ & $0^-,~ 2,~ 4,~ \ldots$ \\
        & & $B_1$ & $1,~ 3,~ \ldots$ \\
        & & $B_2$ & $1,~ 3,~ \ldots$ \\
        \hline
        \multirow{3}{*}{$[n,n,n]$} & \multirow{3}{*}{$\text{Dic}_3$} 
        & $A_1$ & $0^+,~ 3,~ \ldots$ \\
        & & $A_2$ & $0^-,~ 3,~ \ldots$ \\
        & & $E_2$ & $1,~ 2,~ 4,~ \ldots$ \\
        \hline
        $[n,m,0]$  & \multirow{2}{*}{$\text{C}_4$} & $A_1$    & $0^+,~ 1,~ 2,~ 3,~ 4,~ \ldots$ \\
        $[n,n,m]$ & & $A_2$ & $0^-,~ 1,~ 2,~ 3,~ 4,~ \ldots$ \\
        \bottomrule
    \end{tabular}
    \label{table:latticeirreps}
\end{table}

\section{Construction of Single-Hadron Operators}
\label{sec:one}
This section focuses mainly on the construction of meson-like single-hadron operators. For systematic constructions of single-baryon operators, see Refs.~\cite{Basak:2005ir, Basak:2005aq, Basak:2007kj}. Single-meson operators can be constructed from quark bilinears of the form $\bar{\psi}^{\prime}\,\Gamma\,\psi$, where the spin-parity quantum numbers of the operator are determined by the chosen gamma matrix $\Gamma$. For example, $\sum_{\vec{x}} e^{-i \vec{p} \cdot \vec{x}} \bar{d}(\vec{x}) \gamma_5 u(\vec{x})$ creates a $\pi^+$ particle with momentum $\vec{p}$, while $\sum_{f, \vec{x}} Q_f \bar{\psi}_f(\vec{x}) \gamma_t \psi_f(\vec{x})$ is the electromagnetic current operator, which can be used to measure hadron charges or compute QED corrections to hadronic matrix elements. However, such forms can only carry spin $0$ and $1$. To obtain higher spin or quantum numbers that cannot be obtained from simple gamma matrices alone, one can spatially separate the quark and antiquark and connect them by gauge fields. A common approach is to insert covariant derivatives between the quark fields~\cite{Dudek:2010wm, Thomas:2011rh}. This technique has been used in many studies, including Refs.~\cite{Wilson:2023hzu, Wilson:2023anv, Gayer:2021xzv, Yeo:2024chk, Whyte:2024ihh}. We will discuss the properties of such operators in detail in Sec.~\ref{sec:derivative}. Other related techniques can be found in Ref.~\cite{Morningstar:1999rf}.

Earlier literature usually first constructs operators with good quantum numbers $|jm\rangle$ and then projects them into the helicity basis. This procedure requires one to first choose a reference direction and then rotate operators with definite $|jm\rangle$ in the rest frame into operators with definite helicity $|j\lambda\rangle$ in a moving frame. It therefore involves two rotations. The helicity operators must then be subduced into the corresponding group irreps, which requires the Clebsch--Gordan coefficients (CG coefficients) to be computed separately for each lattice group and is rather cumbersome. This thesis adopts a more direct method: operators in continuous space are projected directly into irreps of the lattice group. The same method can also be naturally extended to the construction of $N$-hadron operators.

In the $SU(2)$ group, if an operator has a well-defined $|jm\rangle$, then under the action of a group element $g$ it transforms as
\begin{equation}
g A_{jm} g^{\dagger} = (-1)^{P} \sum_{m^{\prime}} A_{jm^{\prime}} \mathcal{D}_{m^{\prime}m}^{j}(g),
\label{eq:trans}
\end{equation}
where scalar and pseudoscalar mesons have angular momentum $J=0$, vector and axial-vector mesons have $J=1$, and nucleons have $J=\frac{1}{2}$. The factor $(-1)^P$ characterizes the intrinsic parity $P$ of the particle. For vector-type mesons, if the operators $A_x, A_y, A_z$ form a Cartesian vector, then one can construct an irreducible tensor by
\begin{equation}
\begin{aligned}
A_{1, \pm1} &= \mp \frac{i}{\sqrt{2}}(A_x \mp i A_y), \\
A_{1,0} &= i A_z,
\label{eq:ir_tensor}
\end{aligned}
\end{equation}
to obtain the rank-$1$ irreducible tensor $A_{1m}$, so that Eq.~\eqref{eq:trans} is satisfied for $j=1$. Higher-rank irreducible tensors can be constructed similarly. The gamma matrices $\Gamma_{1m}$ and the covariant derivatives $\overleftrightarrow{\nabla}_{1m}$ are both basic building blocks that can carry one unit of angular momentum and can be written in irreducible tensor form according to Eq.~\eqref{eq:ir_tensor}.

This procedure is formally equivalent to the method used in Refs.~\cite{Dudek:2010wm, Thomas:2011rh}, whose expression is
\begin{equation}
O_{J M}(\vec{P}) = \sum_{i} \operatorname{CGs}(m_i)
\times \sum_{\vec{x}} e^{-i \vec{P} \cdot \vec{x}}
\bar{\psi}^{\prime}(\vec{x}, t) \Gamma_{m_1}
\prod_{i>1} \overleftrightarrow{\nabla}_{m_i} \psi(\vec{x}, t).
\label{eq:OJM}
\end{equation}

We perform irrep projection directly from continuum-space operators, following Ref.~\cite{Prelovsek:2016iyo}, without first constructing helicity operators. The target operator with a given irrep $\Gamma$, row index $\mu$ (the row of an irrep plays the role analogous to the $z$ component of angular momentum), and center-of-mass momentum $\vec{P}$ is defined as
\begin{equation}
O_{\Gamma, \mu}(\vec{P}) =
\sum_{g \in G[\vec{P}]} T_{\mu, \mu}^{\Gamma}(g)\,
g O(\vec{P}) g^{\dagger},
\label{eq:projection}
\end{equation}
where $T_{\mu, \mu}^{\Gamma}(g)$ is the diagonal matrix element of the irrep $\Gamma$, and $g$ runs over all group elements of $G$. The correspondence between the group $G$ and the center-of-mass momentum is listed in Table~\ref{tab:group}. Here $O(\vec{P})$ is an operator of arbitrary form, and in this work it is chosen to be $O_{JM}(\vec{P})$ as defined in Eq.~\eqref{eq:OJM}. By scanning the number of covariant derivatives, $0 \leq n_D \leq N_{\nabla}$, the type of gamma matrix, and the total-spin range in continuous space, $|J^{\Gamma}-n_{\nabla}| \leq J \leq J^{\Gamma}+n_{\nabla}$, one obtains all operators for a given number of covariant derivatives. Here $J^{\Gamma}$ denotes the spin carried by the gamma matrix itself.

We use $S$, $P$, $V$, and $A$ to denote gamma matrices of scalar, pseudoscalar, vector, and axial-vector type, respectively. For example, $S$ can be either $\gamma_5$ or $\gamma_4\gamma_5$. For convenience, all operators in this work are projected to the first row of the corresponding irrep. We list only operators containing at most one covariant derivative. For each center-of-mass momentum class in $\Omega$, the constructed operators are summarized in Tables~\ref{tab:one-000}, \ref{tab:one-00n}, \ref{tab:one-0nn}, \ref{tab:one-nnn}, \ref{tab:one-nm0}, \ref{tab:one-nnm}, and \ref{tab:one-nmp} of Appendix~\ref{sec:append_list_one}.

In the following convention, the gamma matrices and covariant derivatives are always placed between $\bar{\psi}^{\prime}$ and $\psi$; for compactness, the left-right arrow notation on covariant derivatives is suppressed. The normalization of the listed operators is chosen so that the overall constant factor is $1$. For example, the operator denoted $P\nabla_x$ in the $T_1^+$ irrep of the $O_h$ group corresponds to
$\bar{\psi}^{\prime} \gamma_5 \overleftrightarrow{\nabla}_x \psi$
or
$\bar{\psi}^{\prime} \gamma_4 \gamma_5 \overleftrightarrow{\nabla}_x \psi$.

It should be noted that the $A_2^{\pm}$ irreps of $O_h$ have no operators with $N_{\nabla} \leq 1$.

As the symmetry is reduced step by step from $O_h$ to the trivial group $C_1=\{E\}$, the number of irreps also decreases. Operators that were originally distinguishable are gradually no longer protected by symmetry. The number of operators that can be constructed within the same irrep also increases. Specifically, when the momentum is $\vec{P}=[0,0,0]$, spatial inversion is a symmetry operation of the $O_h$ group, so parity $P$ is a conserved quantum number. Operators can then be classified by parity, and operators of different parity do not mix. For example, the scalar operator $S\sim \bar{\psi}^{\prime}\gamma_5\psi$ in $A_1^+$ does not mix with the vector operator $V_x\sim \bar{\psi}^{\prime}\gamma_x\psi$ in $T_1^-$. When the momentum is $\vec{P}=[0,0,n]$, however, a spatial axis has been selected and spatial inversion is no longer a symmetry operation. In this case, both $S$ and $V_z$ can create states in the $A_1$ irrep. Finally, for $\vec{P}=[n,m,p]$, no spatial symmetry remains, and all types of operators can mix. In that case, all states with different infinite-volume quantum numbers, including bound states, virtual states, resonances, and scattering states, appear in the unique irrep $A$. In addition, the symmetries of the $C_2$ and $C_1$ groups are too low, and operators obtained from Eq.~\eqref{eq:projection} are often rather lengthy. We therefore take appropriate linear combinations of these operators. This procedure loses no information, since the rank of the operator basis is unchanged.

As a consistency check, we verified the number of operators in each group. Quark bilinear operators constructed purely from gamma matrices of type $S$, $P$, $V$, and $A$ number $1,1,3,3$, respectively, since $A$ and $V$ each have three spatial directions. After introducing one covariant derivative, which has three possible directions, the number of operators becomes $(1+1+3+3)\times 3$. The total number of operators should therefore be $(1+1+3+3)\times 4=32$.

Since this thesis does not explicitly list operators with row index $\mu\neq 1$, the number of operators listed in each irrep must be multiplied by the dimension of that irrep when counting the total number of operators. For example, in the $C_{4v}$ group, the irreps $A_1$, $A_2$, $B_1$, and $B_2$ are all one-dimensional, while the irrep $E$ is two-dimensional. The total number of operators is therefore $6+6+2+2+8\times 2=32$, as expected. This shows that the projection method used here exhausts the complete set of operators allowed by the symmetry group.

\section{Single-Hadron Operators with Covariant Derivatives}
\label{sec:derivative}
In Sec.~\ref{sec:one}, we briefly introduced the construction of single-hadron operators containing covariant derivatives. In this section, we start from the simplest definitions, construct operators with covariant derivatives, and discuss their symmetry properties explicitly\footnotecircle{The material in this section is unpublished and will appear in the appendix of the coupled-channel $D\pi-D\eta-D_s\bar{K}$ scattering paper.}. The forward and backward covariant derivatives are usually defined as\footnotecircle{$\overrightarrow{\nabla}_i \psi(\vec{x})$ denotes $\sum_{\vec{y}} \overrightarrow{\nabla}_i(\vec{x}, \vec{y}) \psi(\vec{y})$.}
\begin{equation}
\begin{aligned}
    \overrightarrow{\nabla}_i^{\text{fwd}} \psi(\vec{x}) &= \frac{1}{a} \left[ U_i(\vec{x}) \psi(\vec{x}+a\hat{i}) - \psi(\vec{x}) \right], \\
    \overrightarrow{\nabla}_i^{\text{bwd}} \psi(\vec{x}) &= \frac{1}{a} \left[ \psi(\vec{x}) - U_i^{\dagger}(\vec{x}-a\hat{i}) \psi(\vec{x}-a\hat{i}) \right].
\end{aligned}
\end{equation}
We use the right arrow rather than vector notation to emphasize that the derivative acts on the quark field to its right. Setting the lattice spacing to $a=1$, we redefine the symmetric covariant derivative using the forward and backward derivatives:
\begin{equation}
    \overrightarrow{\nabla}_i \psi(\vec{x}) \equiv \frac{1}{2} (\overrightarrow{\nabla}_i^{\text{fwd}} + \overrightarrow{\nabla}_i^{\text{bwd}}) \psi(\vec{x}) = \frac{1}{2} \left[ U_i(\vec{x}) \psi(\vec{x}+\hat{i}) - U_i^{\dagger}(\vec{x}-\hat{i}) \psi(\vec{x}-\hat{i}) \right].
\end{equation}
Thus the explicit form of $\overrightarrow{\nabla}$ is
\begin{equation}
    \overrightarrow{\nabla}_i(\vec{x}, \vec{y}) = \frac{1}{2} \left[ U_i(\vec{x}, t) \delta_{\vec{x}+\hat{i}, \vec{y}}-U_i^{\dagger}(\vec{x}-\hat{i}, t) \delta_{\vec{x}-\hat{i}, \vec{y}} \right].
\end{equation}

In the discussion below, we will see that the right-acting covariant derivative alone cannot be projected to a definite $C$ parity. We therefore use the convention above and define the left-acting covariant derivative by integration by parts\footnotecircle{Sometimes one writes $\overrightarrow{\nabla}_i^{\text{fwd}}=\overrightarrow{\nabla}_i$ and $\overrightarrow{\nabla}_i^{\text{bwd}}=\overrightarrow{\nabla}_i^*$.}. On the lattice, integration by parts is equivalent to a translation:
\begin{equation}
\begin{aligned}
    \sum_{\vec{x}} \bar{\phi}(\vec{x}) \overrightarrow{\nabla}_i^{\text{fwd}} \psi(\vec{x}) &= \sum_{\vec{x}} \bar{\phi}(\vec{x}) \left[ U_i(\vec{x}) \psi(\vec{x}+\hat{i}) - \psi(\vec{x}) \right] \\
    &= \sum_{\vec{x}} \left[ \bar{\phi}(\vec{x}-\hat{i}) U_i(\vec{x}-\hat{i}) \psi(\vec{x}) - \bar{\phi}(\vec{x}) \psi(\vec{x}) \right], \\
    \sum_{\vec{x}} \bar{\phi}(\vec{x}) \overrightarrow{\nabla}_i^{\text{bwd}} \psi(\vec{x}) &= \sum_{\vec{x}} \bar{\phi}(\vec{x}) \left[ \psi(\vec{x}) - U_i^{\dagger}(\vec{x}-\hat{i}) \psi(\vec{x}-\hat{i}) \right] \\
    &= \sum_{\vec{x}} \left[ \bar{\phi}(\vec{x}) \psi(\vec{x}) - \bar{\phi}(\vec{x}+\hat{i}) U_i^{\dagger}(\vec{x}) \psi(\vec{x}) \right].
\end{aligned}
\end{equation}
We can therefore define\footnotecircle{$\bar{\phi}(\vec{x}) \overleftarrow{\nabla}_i$ denotes $\sum_{\vec{y}} \bar{\phi}(\vec{y}) \overleftarrow{\nabla}_i(\vec{y}, \vec{x})$.}
\begin{equation}
\begin{aligned}
    \bar{\phi}(\vec{x}) \overleftarrow{\nabla}_i^{\text{bwd}} &= \bar{\phi}(\vec{x}) - \bar{\phi}(\vec{x}-\hat{i}) U_i(\vec{x}-\hat{i}), \\
    \bar{\phi}(\vec{x}) \overleftarrow{\nabla}_i^{\text{fwd}} &= \bar{\phi}(\vec{x}+\hat{i})U_i^{\dagger}(\vec{x}) - \bar{\phi}(\vec{x}).
\end{aligned}
\end{equation}
Analogously to the right-acting derivative, we define the symmetric left-acting covariant derivative as
\begin{equation}
    \bar{\phi}(\vec{x}) \overleftarrow{\nabla}_i \equiv \frac{1}{2} \bar{\phi}(\vec{x}) (\overleftarrow{\nabla}_i^{\text{fwd}} + \overleftarrow{\nabla}_i^{\text{bwd}}) = \frac{1}{2} \left[ \bar{\phi}(\vec{x}+\hat{i})U_i^{\dagger}(\vec{x}) - \bar{\phi}(\vec{x}-\hat{i}) U_i(\vec{x}-\hat{i}) \right].
\end{equation}
Thus the explicit form of $\overleftarrow{\nabla}$ is
\begin{equation}
    \overleftarrow{\nabla}_i(\vec{y}, \vec{x}) = \frac{1}{2} \left[ U_i^{\dagger}(\vec{x}, t) \delta_{\vec{x}+\hat{i}, \vec{y}}-U_i(\vec{x}-\hat{i}, t) \delta_{\vec{x}-\hat{i}, \vec{y}} \right].
\end{equation}

Defining the left-right and forward-backward symmetric covariant derivative by $\overleftrightarrow{\nabla} = \frac{1}{2} [\overrightarrow{\nabla} - \overleftarrow{\nabla}]$, one obtains
\begin{equation}
    \overleftrightarrow{\nabla}_i(\vec{x}, \vec{y}) = \frac{1}{4} \left[ U_i(\vec{x}, t) \delta_{\vec{x}+\hat{i}, \vec{y}}-U_i^{\dagger}(\vec{x}-\hat{i}, t) \delta_{\vec{x}-\hat{i}, \vec{y}} - U_i^{\dagger}(\vec{y}, t) \delta_{\vec{x}, \vec{y}+\hat{i}} + U_i(\vec{y}-\hat{i}, t) \delta_{\vec{x}, \vec{y}-\hat{i}} \right].
\end{equation}
In a slightly informal notation, we may also write
\begin{equation}
\begin{aligned}
    \bar{\phi}(\vec{x}) \Gamma \overleftrightarrow{\nabla} \psi(\vec{x}) &= \left[ \bar{\phi}(\vec{x}) \Gamma U_i(\vec{x}) \psi(\vec{x}+\hat{i}) - \bar{\phi}(\vec{x}) \Gamma U_i^{\dagger}(\vec{x}-\hat{i}) \psi(\vec{x}-\hat{i}) \right. \\
    &\left. - \bar{\phi}(\vec{x}+\hat{i}) \Gamma U_i^{\dagger}(\vec{x})\psi(\vec{x}) + \bar{\phi}(\vec{x}-\hat{i}) \Gamma U_i(\vec{x}-\hat{i}) \psi(\vec{x}) \right].
\label{eq:covariant_derivative_operator}
\end{aligned}
\end{equation}    
This form of the covariant derivative is widely used not only in hadron spectroscopy, but also in studies involving quark renormalization with the gradient-flow method~\cite{Makino:2014taa} and extraction of parton distribution functions (PDFs) from the energy-momentum tensor~\cite{ExtendedTwistedMass:2024kjf}. We now discuss how to construct operators containing covariant derivatives and how their symmetries behave.

\subsection{Example: One Covariant Derivative}
An operator containing a single covariant derivative can be defined as
\begin{equation}
\begin{aligned}
    O_{\Gamma \nabla_i}(\vec{p}) &= \sum_{\vec{x}} \bar{\phi}(\vec{x}) \Gamma \overleftrightarrow{\nabla}_i \psi(\vec{x}) \operatorname{e}^{-i\vec{p}\cdot\vec{x}} \\
    &= \sum_{\vec{x}} \left[ \bar{\phi}(\vec{x}) \Gamma U_i(\vec{x}) \psi(\vec{x}+\hat{i}) - \bar{\phi}(\vec{x}) \Gamma U_i^{\dagger}(\vec{x}-\hat{i}) \psi(\vec{x}-\hat{i}) \right. \\
    &\left. - \bar{\phi}(\vec{x}+\hat{i}) \Gamma U_i^{\dagger}(\vec{x})\psi(\vec{x}) + \bar{\phi}(\vec{x}-\hat{i}) \Gamma U_i(\vec{x}-\hat{i}) \psi(\vec{x}) \right] \operatorname{e}^{-i\vec{p}\cdot\vec{x}} .
\end{aligned}
\end{equation}
Here $\vec{v}$ is the spatial index used in momentum projection, and it differs for right- and left-acting derivatives. With more covariant derivatives, the expression can become complicated. We will use a few simple tricks below to simplify the notation.

We next consider the conjugate transpose of the operator. For the first term,
\begin{equation}
    \sum_{\vec{x}} \bar{\phi}(\vec{x}) \Gamma U_i(\vec{x}) \psi(\vec{x}+\hat{i}) \operatorname{e}^{-i\vec{p}\cdot\vec{x}} \xrightarrow{\dagger} \mathrm{daggersign(\Gamma)} \sum_{\vec{x}} \bar{\psi}(\vec{x}+\hat{i}) \Gamma U_i^{\dagger}(\vec{x}) \phi(\vec{x}) \operatorname{e}^{+i\vec{p}\cdot\vec{x}}.
\end{equation}
Noting that this term carries a minus sign in the original operator as well, one obtains
\begin{equation}
    \sum_{\vec{x}} \bar{\phi}(\vec{x}) \Gamma \overleftrightarrow{\nabla}_i \psi(\vec{x}) \operatorname{e}^{-i\vec{p}\cdot\vec{x}} \xrightarrow{\dagger} (-1) \cdot \mathrm{daggersign(\Gamma)} \sum_{\vec{x}} \bar{\psi}(\vec{x}) \Gamma \overleftrightarrow{\nabla}_i \phi(\vec{x}) \operatorname{e}^{+i\vec{p}\cdot\vec{x}}.
\end{equation}
Here we define $\mathrm{daggersign(\Gamma)} = \gamma_4 \Gamma^{\dagger} \gamma_4$. It is important that the operator contains the structure $\overrightarrow{\nabla}-\overleftarrow{\nabla}$. If one keeps only a one-sided derivative, the relation is not simple and additional momentum phase factors appear.

A direct calculation shows that under $P$ parity transformation, $O_{\Gamma \nabla_i}(\vec{p}) \xrightarrow{P} (-1) \cdot \mathrm{Psign(\Gamma)} O_{\Gamma \nabla_i}(\vec{p})$, where $\mathrm{Psign(\Gamma)} = \gamma_4 \Gamma \gamma_4$. This conclusion holds even if only a one-sided derivative is used. Compared with an operator without a covariant derivative, the operator simply gains one additional minus sign, because $\overleftrightarrow{\nabla}$ is equivalent to a momentum operator in momentum space, and momentum changes sign under parity.

Under $C$ parity transformation, $O_{\Gamma \nabla_i}(\vec{p}) \xrightarrow{C} (-1) \cdot \mathrm{Csign(\Gamma)} O_{\Gamma \nabla_i}(\vec{p})$, where $\mathrm{Csign(\Gamma)} = C \Gamma C^{-1}$. This conclusion relies on the left-right symmetric derivative structure $\overleftrightarrow{\nabla}$. If a one-sided derivative is used, a $C$ transformation must be followed by a translation to restore the operator to its original form, introducing an additional momentum phase factor. In that case, the operator no longer has a well-defined $C$ parity. In the infinite-volume limit, this difference may formally vanish as the lattice momentum goes to zero, but it does not vanish in finite volume.

\subsection{Example: Two Covariant Derivatives}
In the example of the operator with one covariant derivative, we noticed that some terms are not symmetric around the summation point, making it difficult to determine the projected spatial position $\vec{v}$ unambiguously. In principle, one can check and analyze every term explicitly, but this is cumbersome. Here we directly use the form in Eq.~\ref{eq:covariant_derivative_operator}. This form is not rigorous, but after momentum projection it gives the correct and unique result, as can be verified by expanding all terms explicitly and adjusting $\vec{v}$ by hand. Expanding the operator with two covariant derivatives gives
\begin{align*}
    &O_{\Gamma \nabla_i \nabla_j}(\vec{p}) \\
    &= \sum_{\vec{x}} \bar{\phi}(\vec{x}) \Gamma \overleftrightarrow{\nabla}_i \overleftrightarrow{\nabla}_j \psi(\vec{x}) \operatorname{e}^{-i\vec{p}\cdot\vec{x}} \\
    &= \sum_{\vec{x}} \operatorname{e}^{-i\vec{p}\cdot\vec{x}} \Gamma_{\alpha \beta} \left[ \bar{\phi}_{\alpha}(\vec{x}) \overleftrightarrow{\nabla}_i U_j(\vec{x}) \psi_{\beta}(\vec{x}+\hat{j}) - \bar{\phi}_{\alpha}(\vec{x}) \overleftrightarrow{\nabla}_i U_j^{\dagger}(\vec{x}-\hat{j}) \psi_{\beta}(\vec{x}-\hat{j}) \right. \\
    &\qquad\qquad\qquad\quad \left. - \bar{\phi}_{\alpha}(\vec{x}+\hat{j}) \overleftrightarrow{\nabla}_i U_j^{\dagger}(\vec{x}) \psi_{\beta}(\vec{x}) + \bar{\phi}_{\alpha}(\vec{x}-\hat{j}) \overleftrightarrow{\nabla}_i U_j(\vec{x}-\hat{j}) \psi_{\beta}(\vec{x}) \right] \\
    &= \sum_{\vec{x}} \operatorname{e}^{-i\vec{p}\cdot\vec{x}} \Gamma_{\alpha \beta} \left[ \bar{\phi}_{\alpha}(\vec{x}) U_i(\vec{x}) U_j(\vec{x}+\hat{i}) \psi_{\beta}(\vec{x}+\hat{i}+\hat{j}) - \bar{\phi}_{\alpha}(\vec{x}) U_i^{\dagger}(\vec{x}-\hat{i}) U_j(\vec{x}-\hat{i}) \psi_{\beta}(\vec{x}-\hat{i}+\hat{j}) \right. \\
    &\quad - \bar{\phi}_{\alpha}(\vec{x}+\hat{i}) U_i^{\dagger}(\vec{x}) U_j(\vec{x}) \psi_{\beta}(\vec{x}+\hat{j}) + \bar{\phi}_{\alpha}(\vec{x}-\hat{i}) U_i(\vec{x}-\hat{i}) U_j(\vec{x}) \psi_{\beta}(\vec{x}+\hat{j}) \\
    &\quad - \bar{\phi}_{\alpha}(\vec{x}) U_i(\vec{x}) U_j^{\dagger}(\vec{x}+\hat{i}-\hat{j}) \psi_{\beta}(\vec{x}+\hat{i}-\hat{j}) + \bar{\phi}_{\alpha}(\vec{x}) U_i^{\dagger}(\vec{x}-\hat{i}) U_j^{\dagger}(\vec{x}-\hat{i}-\hat{j}) \psi_{\beta}(\vec{x}-\hat{i}-\hat{j}) \\
    &\quad + \bar{\phi}_{\alpha}(\vec{x}+\hat{i}) U_i^{\dagger}(\vec{x}) U_j^{\dagger}(\vec{x}-\hat{j}) \psi_{\beta}(\vec{x}-\hat{j}) - \bar{\phi}_{\alpha}(\vec{x}-\hat{i}) U_i(\vec{x}-\hat{i}) U_j^{\dagger}(\vec{x}-\hat{j}) \psi_{\beta}(\vec{x}-\hat{j}) \\
    &\quad - \bar{\phi}_{\alpha}(\vec{x}+\hat{j}) U_i(\vec{x}+\hat{j}) U_j^{\dagger}(\vec{x}+\hat{i}) \psi_{\beta}(\vec{x}+\hat{i}) + \bar{\phi}_{\alpha}(\vec{x}+\hat{j}) U_i^{\dagger}(\vec{x}-\hat{i}+\hat{j}) U_j^{\dagger}(\vec{x}-\hat{i}) \psi_{\beta}(\vec{x}-\hat{i}) \\
    &\quad + \bar{\phi}_{\alpha}(\vec{x}+\hat{i}+\hat{j}) U_i^{\dagger}(\vec{x}+\hat{j}) U_j^{\dagger}(\vec{x}) \psi_{\beta}(\vec{x}) - \bar{\phi}_{\alpha}(\vec{x}-\hat{i}+\hat{j}) U_i(\vec{x}-\hat{i}+\hat{j}) U_j^{\dagger}(\vec{x}) \psi_{\beta}(\vec{x}) \\
    &\quad + \bar{\phi}_{\alpha}(\vec{x}-\hat{j}) U_i(\vec{x}-\hat{j}) U_j(\vec{x}+\hat{i}-\hat{j}) \psi_{\beta}(\vec{x}+\hat{i}) - \bar{\phi}_{\alpha}(\vec{x}-\hat{j}) U_i^{\dagger}(\vec{x}-\hat{i}-\hat{j}) U_j(\vec{x}-\hat{i}-\hat{j}) \psi_{\beta}(\vec{x}-\hat{i}) \\
    &\quad \left. - \bar{\phi}_{\alpha}(\vec{x}+\hat{i}-\hat{j}) U_i^{\dagger}(\vec{x}-\hat{j}) U_j(\vec{x}-\hat{j}) \psi_{\beta}(\vec{x}) + \bar{\phi}_{\alpha}(\vec{x}-\hat{i}-\hat{j}) U_i(\vec{x}-\hat{i}-\hat{j}) U_j(\vec{x}-\hat{j}) \psi_{\beta}(\vec{x}) \right]. \\
\end{align*}

Under conjugate transpose,
\begin{equation}
    \sum_{\vec{x}} \bar{\phi}(\vec{x}) \Gamma \overleftrightarrow{\nabla}_i \overleftrightarrow{\nabla}_j \psi(\vec{x}) \operatorname{e}^{-i\vec{p}\cdot\vec{x}} \xrightarrow{\dagger} (+1) \cdot \mathrm{daggersign(\Gamma)} \sum_{\vec{x}} \bar{\psi}(\vec{x}) \Gamma \overleftrightarrow{\nabla}_j \overleftrightarrow{\nabla}_i \phi(\vec{x}) \operatorname{e}^{+i\vec{p}\cdot\vec{x}}.
\end{equation}

Two covariant derivatives are invariant under parity, so the $P$ parity of the operator is the same as that of $\sum_{\vec{x}} \bar{\phi}(\vec{x}) \Gamma \psi(\vec{x}) \operatorname{e}^{-i\vec{p}\cdot\vec{x}}$. A $C$ transformation exchanges the order of the two covariant derivatives, so $O_{\Gamma \nabla_i \nabla_j}(\vec{p}) \pm O_{\Gamma \nabla_j \nabla_i}(\vec{p})$ have $C$ parity $\pm 1$, respectively.

\section{Construction of $N$-Hadron Operators}
\label{sec:two}
In Sec.~\ref{sec:operator_background}, we emphasized that a broadly useful strategy for reliably extracting finite-volume spectra is to construct operators corresponding to all noninteracting energy levels. For example, in a two-body system, the free energy level can be written simply as $E = \sqrt{m_1^2 + p_1^2} + \sqrt{m_2^2 + p_2^2}$, where $p_{1,2} = \frac{2\pi}{L} n_{1,2}$ are lattice momenta. Below a given energy threshold, the number of free levels differs from volume to volume, so in principle one should choose a different number of two-body operators for different volumes. Larger volumes usually contain more low-lying free levels and thus require more momentum combinations to be contracted.

In practice, however, solving propagators on larger lattices requires substantially more computational resources. In the author's calculations, the additional cost caused by increasing the operator number is usually negligible compared with propagator solves and contractions. Based on this empirical observation, in the later chapters of this work we use operator bases of the same size for different volumes.

To construct many-body operators, we let $O(\vec{P})$ in Eq.~\eqref{eq:projection} be an $N$-body operator formed from a product of single-hadron operators in infinite volume:
\begin{equation}
    O(\vec{P})=\prod_{i=1}^{N} O_i(\vec{k}_i),
\end{equation}
where $O_i$ may denote a meson or baryon operator. Applying the same projection method to this expression gives
\begin{equation}
O_{\Gamma,\mu}(\vec{P})=\sum_{g\in G[\vec{P}]} T_{\mu,\mu}^{\Gamma}(g)\,
\prod_{i=1}^{N}\bigl(g\,O_i(\vec{k}_i)\,g^{\dagger}\bigr),
\end{equation}
where the transformation property $g O_i(\vec{k}_i) g^{\dagger}$ of the single-hadron operator is given by Eq.~\eqref{eq:trans}. Since constructing multiparticle states does not require the single-hadron operators to be projected to lattice-group irreps beforehand, $O_i(\vec{k}_i)$ can be chosen directly as the single-hadron operator $O_{JM}$ defined in continuous symmetry, as in Eq.~\eqref{eq:OJM}. Of course, the total momentum must satisfy the conservation condition $\sum_i \vec{k}_i=\vec{P}$, where $\vec{k}_i$ is the momentum of the $i$th hadron.

It is worth emphasizing that $\vec{k}_i$ appears only as a seed momentum in the construction, and its specific direction is unimportant. As long as this momentum belongs to one of the momentum combinations contained in the final operator, the projected operator $O_{\Gamma,\mu}(\vec{P})$ automatically contains the momentum combinations allowed in that irrep with the correct transformation properties, and these combinations appear as terms in the resulting operator. In principle, as long as $O(\vec{P})$ has a nonzero component in the irrep $\Gamma$, its angular-momentum or helicity quantum numbers are not additionally restricted.

We construct a complete operator basis up to maximum relative momentum $\max|\vec{k}_i|$ by the following procedure. For a given momentum shell, all $\vec{k}_i$ vectors are scanned systematically. To avoid redundant or linearly dependent operators, we construct an $(N_{\text{op}}+1)\times N_{\text{mon}}$ matrix, where $N_{\text{op}}$ is the number of operators already constructed and confirmed to be linearly independent, and $N_{\text{mon}}$ is the number of different monomials $O_i(\vec{k}_i)$. Let $M_{ij}$ be the coefficient of the $j$th monomial in the operator $O_{\Gamma,\mu}^{(i)}(\vec{P})$, where the first $N_{\text{op}}$ rows of the matrix correspond to the known linearly independent operators and the $(N_{\text{op}}+1)$th row corresponds to the new operator being tested. Let $M_{ij}^{\prime}$ denote the submatrix containing only the first $N_{\text{op}}$ rows. The new operator is accepted into the basis if and only if
$\operatorname{Rank}(M_{ij})>\operatorname{Rank}(M_{ij}^{\prime})$,
which guarantees that it is linearly independent of the existing operators.

In practical applications, the projected operators often need to be further linearly transformed into forms with clearer physical intuition. For example, in the rest frame $\vec{P}=0$, one can form linear combinations according to definite partial waves, so that the resulting operators mainly couple to the low partial waves of interest. This partial-wave projection technique has been systematically discussed in Ref.~\cite{Prelovsek:2016iyo} and is also implemented in the \texttt{OpTion} package~\cite{github}.

We use the symbols $P$, $S$, $V$, and $A$ to denote single-hadron operators constructed by the method described in Sec.~\ref{sec:one}. The operators may contain covariant derivatives of arbitrary order. These operators can be applied directly to studies of $S$-wave $D\pi$ scattering and the $D_0^*(2300)$ resonance~\cite{Yan:2023gvq, Yan:2024yuq}, and are also suitable for scattering processes such as $\pi\pi$ and $K\pi$. Because partial-wave mixing is minimal in the rest frame, the correspondence between operators and continuum partial waves is the cleanest there, and these operators provide the most direct and valuable input for determining scattering amplitudes. The operator lists for all point groups considered here, together with complete meson--meson, meson--baryon, and baryon--baryon operators, are systematically summarized in Appendices~\ref{sec:append_list_mm}, \ref{sec:append_list_bb}, and \ref{sec:append_list_mb}, providing a practical reference for future studies of hadron spectroscopy and matrix elements.

The construction of three-body and higher-body operators follows the same logic. Tables~\ref{tab:three-000} and \ref{tab:three-moving} in Appendix~\ref{sec:append_list_multi} give representative examples of three-pseudoscalar-meson operators in the rest frame and in moving frames, respectively. Operators with higher momentum shells or more particles can be constructed systematically using \texttt{OpTion}.

Table~\ref{tab:four-000} in Appendix~\ref{sec:append_list_multi} gives representative examples of four-body operators built from mutually distinguishable pseudoscalar mesons in the rest frame. These operators provide a foundation for future four-body spectroscopic calculations and can be applied directly once the corresponding finite-volume quantization conditions mature\footnotecircle{The author has no doubt that four-body quantization conditions will see breakthrough progress in the coming years and will be widely applied to a series of important spectroscopic problems. One example is the coupled effect between the $\pi\pi$ and $\pi\pi\pi\pi$ channels, and how much the position of the $\rho$-meson pole shifts as a result. In the author's view, this is precisely one of the key problems that hadron spectroscopy must confront as it moves toward precision measurements.}. For systems containing even more particles, we do not list all operators one by one, but instead point out a simple special case: an operator consisting of $N$ pseudoscalar mesons at rest, $\prod_i^N P_i(0)$, belongs to the $A_1^-$ irrep when $N$ is odd and to the $A_1^+$ irrep when it is even.

In addition to the operators discussed above, there is a special class of operators, namely local four-quark or six-quark operators used to study exotic states. In a series of studies of the $T_{cc}(3875)$, such operators have been found to be important for determining the energy~\cite{Stump:2025owq, Stump:2024lqx, Prelovsek:2025vbr, Vujmilovic:2024snz}.

\section{Projection of Intrinsic Quantum Numbers}
\label{sec:int}

In this section, we discuss how to use the intrinsic symmetries of hadron systems to further filter the spectrum, including isospin, $C$ parity, $G$ parity, and $P$ parity. For example, see Refs.~\cite{CLQCD:2019npr, Chen:2014afa} for a detailed discussion of the $C$ parity of the $Z_c(3900)$. For systems of identical particles such as $\pi\pi$, once spatial-group symmetry and isospin have been taken into account, Bose symmetry already gives the operators the correct $C$, $G$, and $P$ parities, and no additional projection is needed.

The ensembles used in this thesis have $N_{\text{f}} = 2 + 1$ quark flavors, with degenerate up and down quarks, and therefore possess isospin symmetry. We can further project operators to definite total isospin and its third component, $I$ and $I_z$. This symmetry can be used to project operators into a specified isospin channel. As an example, consider the $\pi$ meson, whose isospin triplet is
\begin{equation}
\begin{cases}
    | I=1, I_z=+1 \rangle &= -\bar{d} \gamma_5 u = - \pi^+, \\
    | I=1, I_z=0 \rangle &= \frac{1}{\sqrt{2}} (\bar{u} \gamma_5 u - \bar{d} \gamma_5 d) = \pi^0, \\
    | I=1, I_z=-1 \rangle &= \bar{u} \gamma_5 d = \pi^- .
\end{cases}
\end{equation}
Note the minus sign in front of $\pi^+$. It ensures that the antiquark doublet $(-\bar{d}, u)^T$ transforms under $SU(2)$ in the same way as the quark doublet $(u, d)^T$; see, for example, Ref.~\cite{Thomson:2013zua}.

The total isospin of a $\pi\pi$ system can be $0$, $1$, or $2$. If only a two-body problem is considered, one can choose to project to the maximal $I_z$, where the expressions are simplest. Applying CG coefficients, the operators in isospin space can be written as
\begin{equation}
\begin{cases}
    | \pi\pi \rangle_{I=2} &= \pi^+ \pi^+, \\
    | \pi\pi \rangle_{I=1} &= -\pi^+ \pi^0 + \pi^0 \pi^+, \\
    | \pi\pi \rangle_{I=0} &= -\frac{1}{\sqrt{3}} (\pi^+ \pi^- + \pi^- \pi^+ + \pi^0 \pi^0).
\end{cases}
\end{equation}

For three isospin-one particles,
\begin{equation}
1 \otimes 1 \otimes 1 = (0 \oplus 1 \oplus 2) \otimes 1 = 0 \oplus 1^3 \oplus 2^2 \oplus 3.
\end{equation}
The total-isospin states $I=1$ and $I=2$ have degeneracies of $3$ and $2$, respectively. The maximal ($I=3$) and minimal ($I=0$) isospin channels do not have this degeneracy complication. The explicit operator form depends on the order of isospin addition. Below, we first combine the isospins of the first two particles and then add the third, considering only operators with $I_z = I$. A straightforward calculation gives
\begin{equation}
\begin{cases}
    | \pi\pi\pi \rangle_{I=3} & = -\pi^+\pi^+\pi^+, \\
    | \pi\pi\pi \rangle_{I=2,J_{12}=1} & = \frac{1}{\sqrt{2}} (\pi^+\pi^0\pi^+ - \pi^0\pi^+\pi^+), \\
    | \pi\pi\pi \rangle_{I=2,J_{12}=2} & = \frac{1}{\sqrt{6}} (2 \pi^+\pi^+\pi^0 - \pi^+\pi^0\pi^+ - \pi^0\pi^+\pi^+), \\
    | \pi\pi\pi \rangle_{I=1,J_{12}=0} & = \frac{1}{\sqrt{3}} ( \pi^+\pi^-\pi^+ + \pi^0\pi^0\pi^+ + \pi^-\pi^+\pi^+ ), \\
    | \pi\pi\pi \rangle_{I=1,J_{12}=1} & = \frac{1}{2} ( - \pi^+\pi^0\pi^0 + \pi^0\pi^+\pi^0 - \pi^+\pi^-\pi^+ + \pi^-\pi^+\pi^+ ), \\
    | \pi\pi\pi \rangle_{I=1,J_{12}=2} & = \frac{1}{\sqrt{60}} ( 6 \pi^+\pi^+\pi^- + 3 \pi^+\pi^0\pi^0 + 3 \pi^0\pi^+\pi^0 + \pi^+\pi^-\pi^+ - 2 \pi^0\pi^0\pi^+ + \pi^-\pi^+\pi^+), \\
    | \pi\pi\pi \rangle_{I=0} & = \frac{1}{\sqrt{6}} (-\pi^+\pi^0\pi^- + \pi^0\pi^+\pi^- + \pi^+\pi^-\pi^0 - \pi^-\pi^+\pi^0 - \pi^0\pi^-\pi^+ + \pi^-\pi^0\pi^+).
\end{cases}
\label{eq:pipipi}
\end{equation}
Here $J_{12}$ denotes the isospin of the first two particles. These operators have been used in studies of $\pi\pi\pi$ scattering~\cite{Hansen:2020otl, Mai:2018djl, Blanton:2019vdk, Culver:2019vvu, Fischer:2020jzp, Mai:2021nul, Yan:2024gwp, Yan:2025mdm}.

Operators with higher isospin can be obtained analogously. For ensembles with different flavor symmetry groups, such as $\mathrm{SU}(3)_f$, one simply replaces the CG coefficients by those of the corresponding symmetry group.

Operator construction proceeds in two steps. First, the tables in Appendix~\ref{sec:append_list_mm} or \texttt{OpTion} are used to generate spatially projected operators. Then the appropriate intrinsic quantum numbers are assigned to each constituent particle. These operators can subsequently be projected to definite $I, I_z$, as described in this section.

Charge-conjugation parity, or $C$ parity, is a good quantum number in some scattering channels. Let the charge-conjugation transformation be denoted by $\mathcal{C}$, and its matrix representation by $C$. We first consider the definition and basic properties of Euclidean gamma matrices; see Ref.~\cite{Gattringer:2010zz}\footnotecircle{The $C$ parity of $\gamma_4\gamma_5$ given in the textbook~\cite{Gattringer:2010zz} is incorrect.}:
\begin{equation}
\{\gamma_\mu, \gamma_\nu \}=2 \delta_{\mu \nu} \mathbbm{1}, \,(\mu=1,2,3,4,5).
\end{equation}
Here $\gamma_5=\gamma_1 \gamma_2 \gamma_3 \gamma_4$, and $C=\mathrm{i} \gamma_2 \gamma_4$. The gamma matrices transform under $C$ as
\begin{equation}
\begin{aligned}
    C \mathbbm{1} C^{-1} &= \mathbbm{1}, \\
    C \gamma_5 C^{-1} &= \gamma_5, \\
    C \gamma_\mu C^{-1} &= -\gamma_\mu^T, \\
    C \gamma_i\gamma_4 C^{-1} &= -(\gamma_i\gamma_4)^T, \\
    C \gamma_{\mu}\gamma_5 C^{-1} &= (\gamma_{\mu}\gamma_5)^T, \\
    C \gamma_i\gamma_j C^{-1} &= -(\gamma_i\gamma_j)^T, \, (i \neq j), \\
    C \gamma_4\gamma_5 C^{-1} &= (\gamma_4\gamma_5)^T.
\end{aligned}
\end{equation}
Here $\mu \in [1,2,3,4]$ and $i \in [1,2,3]$.

Quark fields transform under charge conjugation as
\begin{equation}
\begin{aligned}
& \psi(\vec{x},t) \xrightarrow{\mathcal{C}} C^{-1} \bar{\psi}(\vec{x},t)^T, \\
& \bar{\psi}(\vec{x},t) \xrightarrow{\mathcal{C}} -\psi(\vec{x},t)^T C.
\end{aligned}
\end{equation}
Thus the quark bilinears transform under $C$ as
\begin{equation}
\begin{aligned}
    \psi_1 \mathbbm{1} \psi_2 &\xrightarrow{\mathcal{C}} + \psi_2 \mathbbm{1} \psi_1, \\
    \psi_1 \gamma_5 \psi_2 &\xrightarrow{\mathcal{C}} + \psi_2 \gamma_5 \psi_1, \\
    \psi_1 \gamma_\mu \psi_2 &\xrightarrow{\mathcal{C}} - \psi_2 \gamma_\mu \psi_1, \\
    \psi_1 \gamma_i\gamma_4 \psi_2 &\xrightarrow{\mathcal{C}} - \psi_2 \gamma_i\gamma_4 \psi_1, \\
    \psi_1 \gamma_{\mu}\gamma_5 \psi_2 &\xrightarrow{\mathcal{C}} + \psi_2 \gamma_{\mu}\gamma_5 \psi_1, \\
    \psi_1 \gamma_i\gamma_j \psi_2 &\xrightarrow{\mathcal{C}} - \psi_2 \gamma_i\gamma_j \psi_1, \\
    \psi_1 \gamma_4\gamma_5 \psi_2 &\xrightarrow{\mathcal{C}} + \psi_2 \gamma_4\gamma_5 \psi_1.
\end{aligned}
\end{equation}

Operators should satisfy the corresponding $\mathcal{C}$ transformation properties. For example, in the study of the $Z_c(3900)$ with $I^G(J^{PC}) = 1^+(1^{+-})$~\cite{CLQCD:2019npr}, linear combinations of operators are chosen to project to $C=-1$.

In systems with isospin symmetry, $C$ parity can be generalized to the $G$ parity of isospin multiplets, whose transformation property is analogous to $\mathcal{C}$:
\begin{equation}
    \mathcal{G} = \mathcal{C} \operatorname{e}^{i\pi I_y}.
\end{equation}
The corresponding quantum numbers satisfy $G = C (-1)^I$. For example, one can verify that the operators in Eq.~\eqref{eq:pipipi} have the correct $G$ parity under this transformation.

Unlike $C$ parity, $P$ parity depends not only on the intrinsic parities of the constituent particles, but also on their relative orbital angular momentum. Therefore, once an operator is projected to a specific irrep of the relevant little group, the resulting operator already has definite parity, consistent with the transformation property discussed in Eq.~\eqref{eq:trans}.

However, $P$ parity is a good quantum number only in the rest frame. In moving frames, as discussed in Refs.~\cite{Thomas:2011rh, Dudek:2012xn}, parity is generally no longer a symmetry of the finite-volume system because it does not commute with the little group. Energy levels of different parity may therefore mix, making the analysis more complicated.

\section{Summary}
\label{sec:con}

This chapter systematically reviewed the cubic group and its related subgroups in lattice QCD, together with their irreducible representations. We discussed projection techniques for operator construction and applied them to systems with arbitrary particle number, lattice momentum, and intrinsic quantum numbers. The method has been implemented in the open-source package \texttt{OpTion}, which automatically constructs operators with definite symmetry properties and is available at Ref.~\cite{github}. The appendices provide operators for one- and two-body systems in all irreps of the relevant lattice little groups, together with construction examples for three- and four-particle systems. The purpose of this chapter is to reduce repetitive symmetry-analysis work and to provide a convenient dictionary for future lattice studies.

This construction framework can be used to study increasingly complex systems and lays a foundation for lattice-QCD investigations of four-, five-, and even higher-particle interactions.

We also briefly discussed projections onto intrinsic quantum numbers such as isospin, so that the constructed operators carry good quantum numbers capable of distinguishing different physical channels. These operators can be used directly in spectroscopic and scattering analyses based on two-point and higher-point correlation functions.

The \texttt{OpTion} package has been widely used in our previous and ongoing works. In\chapref{chap:two_body_problems},\chapref{chap:three_body_problems}, and\chapref{chap:three_body_problems2}, we will use the methods developed in this chapter to construct operators for two- and three-body scattering analyses.

\cleardoublepage
\chapter{Two-Body Scattering: $D\pi$ Scattering and the $D_0^*(2300)$ Resonance}
\label{chap:two_body_problems}

{
\kaishu
\begin{center}
    晨野黍离春漠漠，水天星粲夜遥遥。
\end{center}
\hfill ——《荆渚逢禅友》[唐] 齐己
}

\section{Background}
The $D_0^*(2300)$ is the lightest \mybf{charmed} scalar resonance\footnotecircle{This chapter is based on the published works\\ \cite{Yan:2024yuq} H. Yan, C. Liu, L. Liu, Y. Meng and H. Xing, Pion mass dependence in $D\pi$ scattering and the $D_0^* (2300)$ resonance from lattice QCD, \textit{Phys. Rev. D} 111, 014503 (2025). \\ \cite{Yan:2023gvq} H. Yan, C. Liu, L. Liu, Y. Meng and H. Xing, Isospin-$1/2$ $D\pi$ scattering and the $D_0^*$ resonance, \textit{PoS LATTICE2023}, 055 (2024).}, with quantum numbers $(I)J^P = (\frac{1}{2})0^+$. Over the two decades since its discovery and confirmation by several experiments~\cite{Belle:2003nsh, FOCUS:2003gru, BaBar:2009pnd, LHCb:2022lzp, LHCb:2015klp}, joint efforts from hadron experiment and theory have clarified many aspects of its properties, while also revealing a number of puzzles. For example, the scalar resonance $D_0^*(2300)$ is very close in mass to its strange partner $D_{s0}^*(2317)$, whereas a naive quark model would usually predict the latter to be heavier. It has been argued that this feature originates from the strong coupling of the $D_{s0}^*(2317)$ to the $DK$ threshold~\cite{Chen:2016spr}. In addition, studies based on unitarized chiral perturbation theory (UChPT) have suggested that the relativistic Breit-Wigner (RBW) form used in experimental analyses is not suitable for the very broad $D_0^*(2300)$. By predicting or fitting the finite-volume levels obtained by the Hadron Spectrum Collaboration~\cite{Moir:2016srx}, these studies found a \mybf{double-pole} structure in this channel~\cite{Albaladejo:2016lbb, Asokan:2022usm}. More specifically, at the physical $\pi$ mass, the PDG $D_0^*(2300)$ splits into two resonances with comparable widths, denoted $D_0^*(2100)$ and $D_0^*(2450)$\footnotecircle{This naming convention is not yet widely used.}. On the lattice at $M_{\pi} = 391~\mathrm{MeV}$, the $D_0^*(2100)$ appears as a bound state, while the $D_0^*(2450)$ lies near the $D\eta$ threshold and appears as a resonance. As $M_{\pi}$ decreases, the $D_0^*(2100)$ becomes lighter and broader. This claim, however, must be tested at different quark masses, which is impossible in experiment. It is therefore essential to carry out a systematic first-principles scattering study of the $D\pi$ system using lattice QCD. We would like to know whether the double-pole structure can still be identified in a model-independent way at different values of $M_{\pi}$, and how this structure depends on $M_{\pi}$. This chapter aims to address these questions. The three-channel coupled system $D\pi-D\eta-D_s\bar{K}$ studied in the works mentioned above is, however, highly complex and computationally demanding. We therefore begin with the simplest single-channel $D\pi$ scattering problem and study how the single $D_0^*$ pole moves as $M_{\pi}$ changes. The final section of this chapter introduces our ongoing coupled-channel study.

In addition, the doubly charmed tetraquark candidate discovered by LHCb, $T_{cc}(3875) \to DD^* \to DD\pi$~\cite{LHCb:2021auc, LHCb:2021vvq}, has attracted broad attention in recent years. Since its discovery, several pioneering lattice studies of $DD^*$ scattering have identified states related to the $T_{cc}$ in scattering amplitudes at both physical and unphysical values of $M_{\pi}$~\cite{Padmanath:2022cvl, Chen:2022vpo, Lyu:2023xro, Collins:2024sfi, Stump:2025owq, Stump:2024lqx, Ortiz-Pacheco:2023ble}, including coupled-channel $DD^*-D^*D^*$ scattering~\cite{Whyte:2024ihh}. All of these lattice studies, however, were performed at unphysical $M_{\pi}$. Since the $D^*$ lies very close to the $D\pi$ threshold, an unphysical $M_{\pi}$ makes the $D^*$ a stable bound state, which is what makes a two-body $DD^*$ scattering study possible. At the physical point, however, the lowest decay channel of the $T_{cc}$ is the three-hadron channel $DD\pi$, and a two-body $DD^*$ study hides this complexity. A more rigorous approach is to study three-body scattering on the lattice, which requires the two-body $D\pi$ and $DD$ scattering amplitudes as input~\cite{Hansen:2024ffk, Dawid:2024dgy}. The systematic study of the $D\pi$ two-meson system presented here provides a key input for future three-body $DD\pi$ calculations.

The scattering parameters for $I=\frac{1}{2}$ $D\pi$ scattering are still mainly obtained indirectly. In Ref.~\cite{Flynn:2007ki}, the scattering length was extracted from lattice-determined form factors by jointly fitting the scattering length and the Omnes subtraction parameters. In Ref.~\cite{Liu:2012zya}, chiral perturbation theory, together with low-energy constants determined from related scattering channels, was used to obtain the $I=\frac{1}{2}$ $D\pi$ scattering length at the physical $M_{\pi}$. There are also studies based on phenomenological models, such as Refs.~\cite{Meng:2022ozq, Du:2017zvv, Guo:2009ct, Guo:2018tjx, Huang:2021fdt, Guo:2018kno, Korpa:2022voo, Lutz:2022enz}. At present, no first-principles calculation has directly determined this scattering length at the physical point.

The earliest lattice calculation of $I=\frac{1}{2}$ $D\pi$ scattering was performed on an $N_{\text{f}}=2$ ensemble with $M_{\pi} \approx 266~\mathrm{MeV}$~\cite{Mohler:2012na}, where a resonance pole was found. The HadSpec Collaboration subsequently carried out extensive work in the same scattering channel. First, they performed a coupled-channel analysis of $D\pi-D\eta-D_s\bar{K}$ on an $N_{\text{f}}=2+1$ anisotropic ensemble with $M_{\pi} \approx 391~\mathrm{MeV}$ and found the $D_0^*$ to be a bound state~\cite{Moir:2016srx}\footnotecircle{Although this lattice study considered three-channel scattering, it reported only one bound state. The location of the higher resonance was not stable under different parametrizations and was therefore not quoted.}. They then performed a single-channel study at $M_{\pi} \approx 239~\mathrm{MeV}$ and found a resonance~\cite{Gayer:2021xzv}. More recently, HadSpec also carried out a calculation at the $\mathrm{SU}(3)$ point~\cite{Yeo:2024chk}. For all values of $M_{\pi}$, however, the lattice-determined pole positions lie below the PDG average~\cite{ParticleDataGroup:2024cfk}. As $M_{\pi}$ is reduced toward the physical point, the pole positions obtained in these lattice studies do not show a monotonic trend. It is therefore especially important to examine the motion of the $D_0^*$ pole at different values of $M_{\pi}$, in particular at the physical point. Moreover, as emphasized in phenomenological studies, for broad resonances such as the $D_0^*(2300)$, the RBW form used in experimental measurements can lead to substantial differences between resonance parameters and pole positions. A more consistent comparison is to convert the lattice, phenomenological, and experimental results to pole positions.

This chapter uses ensembles generated by the Chinese Lattice QCD Collaboration with $M_{\pi} \approx 133, 208, 305$, and $317~\mathrm{MeV}$~\cite{CLQCD:2023sdb}; details are given in Table~\ref{tab:ensembles}. We study isospin-$I=\frac{1}{2}$ $D\pi$ scattering using Lüscher's method. We extract the $S$-wave and $P$-wave scattering phase shifts and analyze the trajectory of the single $D_0^*(2300)$ pole in the complex energy plane. We then extrapolate the scattering length to the physical $M_{\pi}$ and compare the resulting pole positions and scattering length with experimental measurements. The behavior of the $D_0^*(2300)$ pole shows an interesting pattern: at larger $M_{\pi}$, the $D_0^*$ first appears as a \mybf{bound state}; as $M_{\pi}$ decreases, it becomes lighter and turns into a \mybf{virtual state}; after $M_{\pi}$ is reduced further, the $D_0^*$ becomes a \mybf{resonance}, after which the pole follows a more complicated trajectory.

\section{Operator Lists}
\label{sec:supp:operators}
As discussed in\chapref{chap:operators}, in lattice calculations physical processes take place in a cubic spacetime volume. Continuous rotational symmetry is therefore reduced to the cubic group, and at nonzero total momentum it is further reduced to the corresponding little group. The finite-volume spectrum is accordingly organized into irreducible representations of these little groups. We construct operators using the method proposed in\chapref{chap:operators} and implemented in the \verb|OpTion| package~\cite{Yan:2025jlq}, considering all center-of-mass three-momenta
\begin{equation}
    \vec{P} = [0,0,0],\ [0,0,1],\ [0,1,1],\ [1,1,1],\ [0,0,2],
\end{equation}
in units of $\frac{2\pi}{L}$, corresponding respectively to the cubic group $O_h$ and its little groups $C_{4v}, C_{2v}, C_{3v}, C_{4v}$. For $D\pi$ scattering, we consider all irreps whose leading partial waves do not exceed the $P$ wave. Since parity is no longer a good quantum number in a moving frame, $S$ and $P$ waves can mix. The $A_1$ irreps of $C_{4v}, C_{2v}, C_{3v}$ contain both $S$ and $P$ waves; the $T_1^-$ irrep of $O_h$, the $E_2$ irreps of $C_{4v}, C_{3v}$, and the $B_1, B_2$ irreps of $C_{2v}$ contain only $P$ waves; the $A_1^+$ irrep of $O_h$ contains only the $S$ wave.

In the center-of-mass frame, the operators can be linearly transformed so that they couple predominantly to the low partial waves of interest. This construction follows the method of Ref.~\cite{Prelovsek:2016iyo}.

The strong interaction has isospin symmetry. When the up- and down-quark masses are not distinguished, the theory has an isospin $SU(2)$ symmetry. We further project these operators onto isospin $I=\frac{1}{2}$. There are many historical conventions for isospin states; here we use a convention based on charge conjugation~\cite{Thomson:2013zua}, which is more natural for the present purpose.

The up and down quarks can be written as the isospin doublet
\begin{equation}
q =
\begin{pmatrix}
    u \\
    d \\
\end{pmatrix}.
\end{equation}

Under a general $\mathrm{SU}(2)$ transformation $U$, one has $q \to q' = U q$, namely
\begin{equation}
    \begin{pmatrix}
        u \\
        d \\
    \end{pmatrix}
    \to
    \begin{pmatrix}
        u' \\
        d' \\
    \end{pmatrix}
    =
    U
    \begin{pmatrix}
        u \\
        d \\
    \end{pmatrix}
    =
    \begin{pmatrix}
        a & b \\
        -b^* & a^* \\
    \end{pmatrix}
    \begin{pmatrix}
        u \\
        d \\
    \end{pmatrix},
\label{eq:qtoqprime}
\end{equation}
where $a$ and $b$ are complex numbers satisfying $a a^{*} + b b^{*} = 1$. The charge-conjugation transformation is
\begin{equation}
    \psi' = \hat{C} \psi = i\gamma^{2} \gamma^{4} \bar{\psi}^T = i\gamma^{2} \psi^{*}.
\end{equation}
Thus, in flavor space, the wave function of a quark with the opposite charge, namely an antiquark, is the complex conjugate of the quark wave function. Taking the complex conjugate of Eq.~\ref{eq:qtoqprime} gives the transformation law of the antiquark wave function in flavor space:
\begin{equation}
    \begin{pmatrix}
        \bar{u}' \\
        \bar{d}' \\
    \end{pmatrix}
    = U^{*}
    \begin{pmatrix}
        \bar{u} \\
        \bar{d} \\
    \end{pmatrix}
    =
    \begin{pmatrix}
        a^{*} & b^{*} \\
        -b & a \\
    \end{pmatrix}
    \begin{pmatrix}
        \bar{u} \\
        \bar{d} \\
    \end{pmatrix}.
\label{eq:antiq_trans}
\end{equation}
For the $\mathrm{SU}(2)$ group, antiquarks can be arranged into a doublet that transforms in the same way as the quark doublet,
\begin{equation}
    \bar{q} \to \bar{q}' = U \bar{q},
\end{equation}
provided the antiquark doublet is defined as
\begin{equation}
    \bar{q} \equiv
    \begin{pmatrix}
        -\bar{d} \\
        \bar{u} \\
    \end{pmatrix}
    \equiv
    S
    \begin{pmatrix}
        \bar{u} \\
        \bar{d} \\
    \end{pmatrix}
    =
    \begin{pmatrix}
        0 & -1 \\
        1 & 0 \\
    \end{pmatrix}
    \begin{pmatrix}
        \bar{u} \\
        \bar{d} \\
    \end{pmatrix}.
\label{eq:S_def}
\end{equation}
Then
\begin{equation}
    \bar{q} \to \bar{q}' = S U^{*} S^{-1} \bar{q} = U \bar{q}.
\end{equation}

With this convention, antiquarks transform under the $\mathrm{SU}(2)$ flavor symmetry in exactly the same way as quarks. The ordering of $\bar{d}$ and $\bar{u}$ in the doublet, together with the minus sign in front of $\bar{d}$, ensures that quarks and antiquarks have identical transformation properties under $\mathrm{SU}(2)$ flavor rotations. Physical predictions are therefore invariant under simultaneous $u \leftrightarrow d$ and $\bar{u} \leftrightarrow \bar{d}$ transformations. It should be emphasized that, in general, quarks and antiquarks cannot be placed in the same representation. This property is special to $\mathrm{SU}(2)$, because the fundamental representation of $\mathrm{SU}(2)$ is equivalent to its conjugate representation. It does not, for example, generalize to $\mathrm{SU}(3)$ flavor symmetry. This is why one does not say that the spin of an antiquark is "anti-$\frac{1}{2}$", whereas color does have an anticolor representation.

Mesons are composed of a quark and an antiquark. In isospin space, the four possible states formed by the up and down quarks and their antiparticles can be written as the direct product of an $\mathrm{SU}(2)$ quark doublet and an $\mathrm{SU}(2)$ antiquark doublet. For mesons made of two light quarks, the decomposition $2 \otimes 2 = 3 \oplus 1$ gives an isospin triplet and an isospin singlet, which can then be identified with physical mesons. With our convention, the definitions used in this thesis are
\begin{equation}
\begin{gathered}
D:
\begin{cases}
    +\frac{1}{2}: -\bar{d} \gamma_5 c = -D^{+}, \\
    -\frac{1}{2}: \bar{u} \gamma_5 c = D^{0}, \\
\end{cases} \,
\bar{D}:
\begin{cases}
    +\frac{1}{2}: \bar{c} \gamma_5 u = \bar{D}^0 ,\\
    -\frac{1}{2}: \bar{c} \gamma_5 d = D^-, \\
\end{cases} \,
D_s:
\begin{cases}
    0: \bar{s} \gamma_5 c = D_s^{+}, \\
\end{cases} \\
K:
\begin{cases}
    +\frac{1}{2}: \bar{s} \gamma_5 u = K^+, \\
    -\frac{1}{2}: \bar{s} \gamma_5 d = K^0, \\
\end{cases} \,
\bar{K}:
\begin{cases}
    +\frac{1}{2}: -\bar{d} \gamma_5 s = -\bar{K}^{0}, \\
    -\frac{1}{2}: \bar{u} \gamma_5 s = K^-, \\
\end{cases} \\
\pi:
\begin{cases}
    +1: -\bar{d} \gamma_5 u = - \pi^+, \\
    0: \frac{1}{\sqrt{2}} (\bar{u} \gamma_5 u - \bar{d} \gamma_5 d) = \pi^0, \\
    -1: \bar{u} \gamma_5 d = \pi^-,
\end{cases} \,
\eta:
\begin{cases}
    0: \frac{1}{\sqrt{2}} (\bar{u} \gamma_5 u + \bar{d} \gamma_5 d) = \eta \equiv \frac{1}{\sqrt{2}} (\eta^u + \eta^d). \\
\end{cases}
\end{gathered}
\label{eq:multiplet}
\end{equation}
For later discussions in\chapref{chap:three_body_problems}and\chapref{chap:three_body_problems2}, we also define the isospin states of several other particles. In Eq.~\ref{eq:multiplet}, note that the conventions for the $K$ and $D$ mesons differ. These definitions are complete. For example, the single-particle $\omega$ state is obtained simply by replacing $\gamma_5$ in the $\eta$ operator with $\gamma_i$.

This chapter focuses on $I=\frac{1}{2}$ $D\pi$ scattering. Using the appropriate Clebsch-Gordan coefficients, one obtains
\begin{equation}
    O_{D^{(*)}\pi}^{I=\frac{1}{2}}
    = 2 D^0\pi^+ - \sqrt{2}\, D^{(*)+}\pi^0.
\label{eq:isospin}
\end{equation}
Single-hadron operators with different electric charges have the same flavor structure and differ only in isospin.

Each operator is either a quark bilinear or a linear combination of products of two bilinears, and can be written in the form
\begin{equation}
\begin{cases}
    O_{\text{one}}(\vec{P}) = \sum_{i} \eta_i\, D_{\mu_i}(\vec{P}), \\
    O_{\text{two}}(\vec{P}) = \sum_{i} \eta_i\, D_{\mu_i}(\vec{p}_i)\,
    \pi(\vec{P}-\vec{p}_i).
\end{cases}
\end{equation}
Thus, the one-meson and two-meson operators are uniquely specified by the parameters $\eta_i$, $\mu_i$, and $\vec{p}_i$. The Cartesian gamma-matrix labels $\mu_i \in \{ 0, 5, x, y, z, 5x, 5y, 5z \} $ correspond to $\gamma_0, \gamma_5, \gamma_x, \gamma_y, \gamma_z, \gamma_5\gamma_x, \gamma_5\gamma_y, \gamma_5\gamma_z$, while $\vec{p}_i$ is the relative momentum. These parameters are denoted collectively as
\begin{equation}
\eta_{\mu_i}^{\alpha_i},
\end{equation}
where $\alpha_i$ is in one-to-one correspondence with the momentum. A direction appearing in $\alpha_i$ denotes one unit of momentum in that direction; for example,
$\alpha = yz \rightarrow \vec{p} = [011]$,
$\alpha = -2x \rightarrow \vec{p} = [-200]$, and so on.
For the isospin-$I=\frac{1}{2}$ operators,
the case $\vec{P}=[000]$ is listed in Table~\ref{tab:operators000},
$\vec{P}=[001]$ in Table~\ref{tab:operators001},
$\vec{P}=[011]$ in Table~\ref{tab:operators011},
$\vec{P}=[111]$ in Table~\ref{tab:operators111},
and $\vec{P}=[002]$ in Table~\ref{tab:operators002}.

We also tested the effect of including operators with covariant derivatives in the operator basis. Although these operators can make the plateaus appear slightly earlier, they substantially worsen the signal-to-noise ratio. We therefore do not use them in this thesis.

\begin{table*}[htbp]
\centering
\caption{Operator basis with total momentum $\vec{P} = [000]$ used for $D\pi$ scattering. Each irrep contains both one-meson and two-meson operators. The notation $\eta_{\mu_i}^{\alpha_i}$ is defined in the text. Each operator is a one-meson operator or a linear combination of products of two meson bilinears, where $\eta$ denotes the coefficient of the corresponding operator, $\mu$ labels the Cartesian gamma matrix, and $\alpha$ denotes the momentum vector of a single meson or the relative momentum vector of a two-meson system.}
\begin{tabular}{ccc}
\toprule
Irrep & Type & Operator \\
\midrule
\multirow{4}{*}{$A_1^+$} & \multirow{1}{*}{one-meson} & $(+1)^{0}_{0}$ \\
\cmidrule(lr){2-3}
& \multirow{4}{*}{two-meson} & $(+1)^{0}_{5}$ \\
\cmidrule(lr){3-3}
& & $(+1)^{z}_{5}, (+1)^{y}_{5}, (+1)^{x}_{5}, (+1)^{-z}_{5}, (+1)^{-y}_{5}, (+1)^{-x}_{5}$ \\
\cmidrule(lr){3-3}
& & $(+1)^{yz}_{5}, (+1)^{xz}_{5}, (+1)^{xy}_{5}, (+1)^{-y,z}_{5}, (+1)^{-x,z}_{5}, (+1)^{-x,y}_{5}$ \\
& & $(+1)^{y,-z}_{5}, (+1)^{x,-z}_{5}, (+1)^{x,-y}_{5}, (+1)^{-y,-z}_{5}, (+1)^{-x,-z}_{5}, (+1)^{-x,-y}_{5}$ \\
\midrule
\multirow{4}{*}{$T_1^-$} & \multirow{1}{*}{one-meson} & $(+1)^{0}_{z}$ \\
\cmidrule(lr){2-3}
& \multirow{3}{*}{two-meson} & $(+1)^{z}_{5}, (-1)^{-z}_{5}$ \\
& & $(+1)^{xz}_{5}, (+1)^{-x,z}_{5}, (+1)^{yz}_{5}, (+1)^{-y,z}_{5}, (-1)^{x,-z}_{5}, (-1)^{-x,-z}_{5}, (-1)^{y,-z}_{5}, (-1)^{-y,-z}_{5}$ \\
\cmidrule(lr){3-3}
& & $(+1)^{y}_{x}, (-1)^{-y}_{x}, (+1)^{-x}_{y}, (-1)^{x}_{y}$ \\
\midrule
\multirow{4}{*}{$T_1^+$} & \multirow{1}{*}{one-meson} & $(+1)^{0}_{5z}$ \\
\cmidrule(lr){2-3}
& \multirow{3}{*}{two-meson} & $(+1)^{0}_{z}$ \\
\cmidrule(lr){3-3}
& & $(+1)^{z}_{z}, (+1)^{y}_{z}, (+1)^{x}_{z}, (+1)^{-z}_{z}, (+1)^{-y}_{z}, (+1)^{-x}_{z}$ \\
\cmidrule(lr){3-3}
& & $(-2)^{z}_{z}, (+1)^{y}_{z}, (+1)^{x}_{z}, (-2)^{-z}_{z}, (+1)^{-y}_{z}, (+1)^{-x}_{z}$ \\
\bottomrule
\end{tabular}
\label{tab:operators000}
\end{table*}

\begin{table}[htbp]
\centering
\caption{Operator basis with total momentum $\vec{P} = [001]$ used for $D\pi$ scattering. The notation is the same as in Table~\ref{tab:operators000}.}
\begin{tabular}{ccc}
\toprule
Irrep & Type & Operator \\
\midrule
\multirow{8}{*}{$A_1$} & \multirow{2}{*}{one-meson} & $(+1)^{z}_{0}$ \\
& & $(+1)^{z}_{z}$ \\
\cmidrule(lr){2-3}
& \multirow{6}{*}{two-meson} & $(+1)^{0}_{5}$ \\
& & $(+1)^{z}_{5}$ \\
& & $(+1)^{-x}_{5}, (+1)^{x}_{5}, (+1)^{-y}_{5}, (+1)^{y}_{5}$ \\
& & $(+1)^{xz}_{5}, (+1)^{-x,z}_{5}, (+1)^{yz}_{5}, (+1)^{-y,z}_{5}$ \\
& & $(+1)^{-z}_{5}$ \\
& & $(+1)^{2z}_{5}$ \\
\midrule
\multirow{6}{*}{$E_2$} & \multirow{2}{*}{one-meson} & $(+1)^{z}_{y}$ \\
& & $(+1)^{z}_{5x}$ \\
\cmidrule(lr){2-3}
& \multirow{4}{*}{two-meson} & $(+1)^{y}_{5}, (-1)^{-y}_{5}$ \\
& & $(+1)^{yz}_{5}, (-1)^{-y,z}_{5}$ \\
& & $(+1)^{0}_{x}$ \\
& & $(+1)^{z}_{x}$ \\
\bottomrule
\end{tabular}
\label{tab:operators001}
\end{table}

\begin{table}[htbp]
\centering
\caption{Operator basis with total momentum $\vec{P} = [011]$ used for $D\pi$ scattering. The notation is the same as in Table~\ref{tab:operators000}.}
\begin{tabular}{ccc}
\toprule
Irrep & Type & Operator \\
\midrule
\multirow{8}{*}{$A_1$} & \multirow{2}{*}{one-meson} & $(+1)^{yz}_{0}$ \\
& & $(+1)^{yz}_{y}, (+1)^{yz}_{z}$ \\
\cmidrule(lr){2-3}
& \multirow{6}{*}{two-meson} & $(+1)^{0}_{5}$ \\
& & $(+1)^{yz}_{5}$ \\
& & $(+1)^{y}_{5}, (+1)^{z}_{5}$ \\
& & $(+1)^{xy}_{5}, (+1)^{xz}_{5}, (+1)^{-x,y}_{5}, (+1)^{-x,z}_{5}$ \\
& & $(+1)^{xyz}_{5}, (+1)^{-x,yz}_{5}$ \\
& & $(+1)^{y}_{x}, (-1)^{z}_{x}$ \\
\midrule
\multirow{5}{*}{$B_1$} & \multirow{2}{*}{one-meson} & $(+1)^{yz}_{y}, (-1)^{yz}_{z}$ \\
& & $(+1)^{yz}_{5x}$ \\
\cmidrule(lr){2-3}
& \multirow{3}{*}{two-meson} & $(+1)^{y}_{5}, (-1)^{z}_{5}$ \\
& & $(+1)^{xy}_{5}, (-1)^{xz}_{5}, (+1)^{-x,y}_{5}, (-1)^{-x,z}_{5}$ \\
& & $(+1)^{y}_{x}, (+1)^{z}_{x}$ \\
\midrule
\multirow{9}{*}{$B_2$} & \multirow{2}{*}{one-meson} & $(+1)^{yz}_{x}$ \\
& & $(+1)^{yz}_{5y}, (-1)^{yz}_{5z}$ \\
\cmidrule(lr){2-3}
& \multirow{7}{*}{two-meson} & $(+1)^{xy}_{5}, (+1)^{xz}_{5}, (-1)^{-x,y}_{5}, (-1)^{-x,z}_{5}$ \\
& & $(+1)^{x}_{5}, (-1)^{-x}_{5}$ \\
& & $(+1)^{xyz}_{5}, (-1)^{-x,yz}_{5}$ \\
& & $(+1)^{y}_{y}, (-1)^{z}_{z}$ \\
& & $(+1)^{y}_{z}, (-1)^{z}_{y}$ \\
& & $(+1)^{0}_{y}, (-1)^{0}_{z}$ \\
& & $(+1)^{yz}_{y}, (-1)^{yz}_{z}$ \\
\bottomrule
\end{tabular}
\label{tab:operators011}
\end{table}

\begin{table}[htbp]
\centering
\caption{Operator basis with total momentum $\vec{P} = [111]$ used for $D\pi$ scattering. The notation is the same as in Table~\ref{tab:operators000}.}
\begin{tabular}{ccc}
\toprule
Irrep & Type & Operator \\
\midrule
\multirow{8}{*}{$A_1$} & \multirow{2}{*}{one-meson} & $(+1)^{xyz}_{0}$ \\
& & $(+1)^{xyz}_{x}, (+1)^{xyz}_{y}, (+1)^{xyz}_{z}$ \\
\cmidrule(lr){2-3}
& \multirow{6}{*}{two-meson} & $(+1)^{0}_{5}$ \\
& & $(+1)^{xyz}_{5}$ \\
& & $(+1)^{x}_{5}, (+1)^{y}_{5}, (+1)^{z}_{5}$ \\
& & $(+1)^{yz}_{5}, (+1)^{xz}_{5}, (+1)^{xy}_{5}$ \\
& & $(+1)^{x}_{y}, (-1)^{x}_{z}, (+1)^{y}_{z}, (-1)^{y}_{x}, (+1)^{z}_{x}, (-1)^{z}_{y}$ \\
& & $(+1)^{xy}_{x}, (-1)^{xy}_{y}, (+1)^{yz}_{y}, (-1)^{yz}_{z}, (+1)^{xz}_{z}, (-1)^{xz}_{x}$ \\
\midrule
\multirow{8}{*}{$E_2$} & \multirow{2}{*}{one-meson} & $(+1)^{xyz}_{x}, (-1)^{xyz}_{y}$ \\
& & $(+1)^{xyz}_{5x}, (+1)^{xyz}_{5y}, (-2)^{xyz}_{5z}$ \\
\cmidrule(lr){2-3}
& \multirow{6}{*}{two-meson} & $(+1)^{x}_{5}, (-1)^{y}_{5}$ \\
& & $(+1)^{yz}_{5}, (-1)^{xz}_{5}$ \\
& & $(+1)^{0}_{x}, (+1)^{0}_{y}, (-2)^{0}_{z}$ \\
& & $(+1)^{xyz}_{x}, (+1)^{xyz}_{y}, (-2)^{xyz}_{z}$ \\
& & $(+1)^{x}_{x}, (+1)^{y}_{y}, (-2)^{z}_{z}$ \\
& & $(+1)^{yz}_{x}, (+1)^{xz}_{y}, (-2)^{xy}_{z}$ \\
\bottomrule
\end{tabular}
\label{tab:operators111}
\end{table}

\begin{table}[htbp]
\centering
\caption{Operator basis with total momentum $\vec{P} = [002]$ used for $D\pi$ scattering. The notation is the same as in Table~\ref{tab:operators000}.}
\begin{tabular}{ccc}
\toprule
Irrep & Type & Operator \\
\midrule
\multirow{6}{*}{$A_1$} & \multirow{2}{*}{one-meson} & $(+1)^{2z}_{0}$ \\
& & $(+1)^{2z}_{z}$ \\
\cmidrule(lr){2-3}
& \multirow{4}{*}{two-meson} & $(+1)^{0}_{5}$ \\
& & $(+1)^{2z}_{5}$ \\
& & $(+1)^{z}_{5}$ \\
& & $(+1)^{xz}_{5}, (+1)^{-x,z}_{5}, (+1)^{yz}_{5}, (+1)^{-y,z}_{5}$ \\
\midrule
\multirow{6}{*}{$E_2$} & \multirow{2}{*}{one-meson} & $(+1)^{2z}_{y}$ \\
& & $(+1)^{2z}_{5x}$ \\
\cmidrule(lr){2-3}
& \multirow{4}{*}{two-meson} & $(+1)^{yz}_{5}, (-1)^{-y,z}_{5}$ \\
& & $(+1)^{0}_{x}$ \\
& & $(+1)^{z}_{x}$ \\
& & $(+1)^{yz}_{x}, (+1)^{-y,z}_{x}$ \\
\bottomrule
\end{tabular}
\label{tab:operators002}
\end{table}

\section{Two-Point Correlation Matrix and Finite-Volume Spectrum}
\label{sec:spectra}
The valence charm-quark action used in this chapter is the same as the light- and strange-quark action used to generate the gauge ensembles. In addition, to study discretization effects on the coarsest near-physical ensemble C48P14, with $M_{\pi} \approx 133~\mathrm{MeV}$, we repeat the analysis using two charm-quark actions: the Wilson-clover action used for the sea quarks, and the Fermilab action~\cite{El-Khadra:1996wdx}, which controls discretization effects of order $\mathcal{O}(am_c)^n$. The tuning of the Fermilab action follows the procedure of Ref.~\cite{Liu:2009jc}. The numerical calculations in this chapter are performed on six ensembles, corresponding to four pion masses, $M_{\pi} \approx 133, 208, 305$, and $317~\mathrm{MeV}$. The ensemble details are listed in Table~\ref{tab:ensembles}. Among these ensembles, F32P21/F48P21 and F32P30/F48P30 form two pairs with the same $M_{\pi}$ and lattice spacing but different volumes. They provide additional kinematic points in the finite-volume spectrum and thereby make the determination of scattering parameters more stable and precise.

The finite-volume spectrum is obtained from the matrix of lattice-operator correlation functions in Eq.~\ref{eq:correlation_function_general}. Quark propagators are computed using distillation smearing~\cite{HadronSpectrum:2009krc}. In this and the following chapters, we use $N_v=100$ eigenvectors for ensembles with volume $32^3$ and $N_v=200$ eigenvectors for ensembles with volume $48^3$. We have checked that the spectrum is insensitive to the choice of $N_v$, although the statistical uncertainties increase slightly when smaller values of $N_v$ are used.

We find that the effective masses associated with some eigenvalues do not exhibit plateaus at large times, but instead continue to rise or fall. Further analysis shows that this behavior originates from non-negligible \mybf{thermal pollution}, which can be understood by decomposing the correlation matrix. We attribute this effect to source and sink operators that are both two-body $D\pi$ operators with the same momentum structure:
\begin{equation}
\begin{aligned}
    &C_{[D\pi]_i,[D\pi]_j}(t) = \langle D_{\vec{k}}^{-}(t) \pi_{\vec{P} - \vec{k}}^{-}(t) D_{\vec{k}}^{+}(0) \pi_{\vec{P} - \vec{k}}^{+}(0) \rangle_T \\
    &\supset \frac{1}{Z_T} \left[ e^{-E_\pi(\vec{P} - \vec{k}) T} \langle \pi_{\vec{P} - \vec{k}}^{-} | D_{\vec{k}}^{-}(t) \pi_{\vec{P} - \vec{k}}^{-}(t) | D_{\vec{k}}^{+} \rangle \langle D_{\vec{k}}^{+} | D_{\vec{k}}^{+}(0) \pi_{\vec{P} - \vec{k}}^{+}(0) | \pi_{\vec{P} - \vec{k}}^{-} \rangle \right. \\
    &+ \left. e^{-E_D(\vec{k}) T} \langle D_{\vec{k}}^{-} | D_{\vec{k}}^{-}(t) \pi_{\vec{P} - \vec{k}}^{-}(t) | \pi_{\vec{P} - \vec{k}}^{+} \rangle \langle \pi_{\vec{P} - \vec{k}}^{+} | D_{\vec{k}}^{+}(0) \pi_{\vec{P} - \vec{k}}^{+}(0) | D_{\vec{k}}^{-} \rangle \right] \\
    &= \frac{1}{Z_T} |z_{\vec{k}}^D|^2 |z_{\vec{P} - \vec{k}}^{\pi}|^2  
    \left[ e^{-E_\pi(\vec{P} - \vec{k}) T -(E_D(\vec{k}) - E_\pi(\vec{P} - \vec{k})) t} 
    + e^{-E_D(\vec{k}) T -(E_\pi(\vec{P} - \vec{k}) - E_D(\vec{k})) t} \right] \\
    &\approx \frac{1}{Z_T} |z_{\vec{k}}^D|^2 |z_{\vec{P} - \vec{k}}^{\pi}|^2 
    e^{-E_\pi(\vec{P} - \vec{k}) T -(E_D(\vec{k}) - E_\pi(\vec{P} - \vec{k})) t},
\end{aligned}
\end{equation}
where $z_{\vec{k}}^X \equiv \langle X_{\vec{k}}^{+} | X_{\vec{k}}^{+} | \Omega \rangle$. In the last step, we have neglected the second thermal-state contribution, since it is subleading compared with the first term.

We extend the weighted-shift method proposed in Ref.~\cite{Dudek:2012gj} to the $D\pi$ system in order to remove the leading thermal pollution. By constructing the weighted-shifted correlation matrix, the leading thermal-pollution term can be eliminated:
\begin{equation}
    \tilde{C}_{ij}(t) = e^{(E_D(\vec{k}) - E_\pi(\vec{P} - \vec{k})) t} C_{ij}(t) - e^{(E_D(\vec{k}) - E_\pi(\vec{P} - \vec{k})) (t+1)} C_{ij}(t+1).
\end{equation}
It should be noted that most matrix elements do not contain the thermal pollution described above, but this treatment does not change their dependence on time $t$. To compensate for the overall shift introduced by the weighted operation, the extracted spectrum must be shifted upward by $(E_D(\vec{k}) - E_\pi(\vec{P} - \vec{k}))$. The weighted-shifted correlation matrix $\tilde{C}_{ij}(t)$ can then be used in the standard generalized eigenvalue problem analysis to obtain the correct energy levels. Correlated fits to these eigenvalues yield the spectra in the different irreps.

\subsection{Wick-Contraction Topologies}
We now give a concrete example of the Wick contractions. Consider the two-body $D\pi$ operator in the rest-frame $A_1^+$ irrep, namely $\langle O_{D\pi,A_1^+,p}^{[J=L=S=0]}(t^{\prime}) O_{D\pi,A_1^+,p}^{[J=L=S=0]\dagger}(t) \rangle$:
\begin{equation}
\begin{aligned}
    & \langle O_{D\pi,A_1^+,p}^{[J=L=S=0]}(t^{\prime}) O_{D\pi,A_1^+,p}^{[J=L=S=0]\dagger}(t) \rangle \\
    &= \left\langle \left\{\sum_{\beta} \left[ 2 D^0(\vec{p}_{\beta}) \pi^+(-\vec{p}_{\beta}) - \sqrt{2} D^+(\vec{p}_{\beta}) \pi^0(-\vec{p}_{\beta}) \right]\right\}(t^{\prime}) \right. \\
    &\left. \qquad \left\{\sum_{\alpha} \left[ 2 D^0(\vec{p}_{\alpha}) \pi^+(-\vec{p}_{\alpha}) - \sqrt{2} D^+(\vec{p}_{\alpha}) \pi^0(-\vec{p}_{\alpha}) \right]\right\}^{\dagger}(t) \right\rangle.
\label{eq:Dpi_contraction}
\end{aligned}
\end{equation}
Taking the first term as an example,
\begin{equation}
\begin{aligned}
    &\left\langle \sum_{\beta\alpha} \underbrace{\bar{u} \square e^{-i p_{\beta} \cdot x} \gamma_5 \square c \cdot \bar{d} \square e^{+i p_{\beta} \cdot x} \gamma_5 \square u}_{t^{\prime}} \cdot \underbrace{\bar{c} \square e^{+i p_{\alpha} \cdot x} \gamma_5 \square u \cdot \bar{u} \square e^{-i p_{\alpha} \cdot x} \gamma_5 \square d}_t \right\rangle \\
    &\equiv \sum_{\beta\alpha} \mathbbm{C}_{D\pi, [\gamma_5,\gamma_5,\gamma_5,\gamma_5], [-\beta, \beta, \alpha, -\alpha]}^{\bar{u}c\bar{d}u-\bar{c}u\bar{u}d} (t^{\prime},t) \\
    &\equiv \sum_{\beta\alpha} \mathbbm{E}_{D\pi, [\gamma_5,\gamma_5,\gamma_5,\gamma_5], [-\beta, \beta, \alpha, -\alpha]} + \mathbbm{F}_{D\pi, [\gamma_5,\gamma_5,\gamma_5,\gamma_5], [-\beta, \beta, \alpha, -\alpha]} (t^{\prime},t),
\label{eq:general}
\end{aligned}
\end{equation}
where the subscripts contain the ordering of momenta and gamma matrices. The ordering is first at the sink and then at the source, and within each end it is read from top to bottom. Similar contractions will appear more frequently in the three-body scattering studies below.

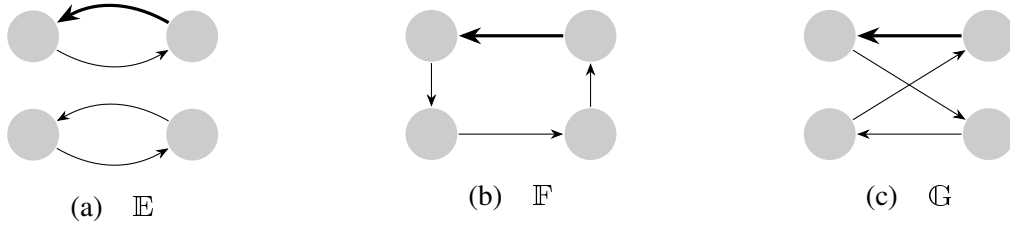
\begin{figure}[htbp]
\centering
\raisebox{-0.5\height}{\begin{subfigure}{0.3\linewidth}\centering\begin{tikzpicture}[global scale = 1.3]
    \node[circle, draw=black, fill=black!20, thick, draw=none, font = \scriptsize, anchor = east, minimum size=15pt, inner sep=0pt] (sink1) at (0, 1) {};
    \node[circle, draw=black, fill=black!20, thick, draw=none, font = \scriptsize, anchor = east, minimum size=15pt, inner sep=0pt] (sink2) at (0, 0) {};
    \node[circle, draw=black, fill=black!20, thick, draw=none, font = \scriptsize, anchor = east, minimum size=15pt, inner sep=0pt] (source1) at (1.618, 1) {};
    \node[circle, draw=black, fill=black!20, thick, draw=none, font = \scriptsize, anchor = east, minimum size=15pt, inner sep=0pt] (source2) at (1.618, 0) {};
    \draw[very thick, -Stealth, out=150, in=30] (source1) to (sink1);
    \draw[thin, -Stealth, out=330, in=210] (sink1) to (source1);
    \draw[thin, -Stealth, out=150, in=30] (source2) to (sink2);
    \draw[thin, -Stealth, out=330, in=210] (sink2) to (source2);
\end{tikzpicture}\caption{$\mathbbm{E}$}\end{subfigure}}
\quad
\raisebox{-0.5\height}{\begin{subfigure}{0.3\linewidth}\centering\begin{tikzpicture}[global scale = 1.3]
    \node[circle, draw=black, fill=black!20, thick, draw=none, font = \scriptsize, anchor = east, minimum size=15pt, inner sep=0pt] (sink1) at (0, 1) {};
    \node[circle, draw=black, fill=black!20, thick, draw=none, font = \scriptsize, anchor = east, minimum size=15pt, inner sep=0pt] (sink2) at (0, 0) {};
    \node[circle, draw=black, fill=black!20, thick, draw=none, font = \scriptsize, anchor = east, minimum size=15pt, inner sep=0pt] (source1) at (1.618, 1) {};
    \node[circle, draw=black, fill=black!20, thick, draw=none, font = \scriptsize, anchor = east, minimum size=15pt, inner sep=0pt] (source2) at (1.618, 0) {};
    \draw[very thick, -Stealth] (source1) to (sink1);
    \draw[thin, -Stealth] (sink1) to (sink2);
    \draw[thin, -Stealth] (sink2) to (source2);
    \draw[thin, -Stealth] (source2) to (source1);
\end{tikzpicture}\caption{$\mathbbm{F}$}\end{subfigure}}
\quad
\raisebox{-0.5\height}{\begin{subfigure}{0.3\linewidth}\centering\begin{tikzpicture}[global scale = 1.3]
    \node[circle, draw=black, fill=black!20, thick, draw=none, font = \scriptsize, anchor = east, minimum size=15pt, inner sep=0pt] (sink1) at (0, 1) {};
    \node[circle, draw=black, fill=black!20, thick, draw=none, font = \scriptsize, anchor = east, minimum size=15pt, inner sep=0pt] (sink2) at (0, 0) {};
    \node[circle, draw=black, fill=black!20, thick, draw=none, font = \scriptsize, anchor = east, minimum size=15pt, inner sep=0pt] (source1) at (1.618, 1) {};
    \node[circle, draw=black, fill=black!20, thick, draw=none, font = \scriptsize, anchor = east, minimum size=15pt, inner sep=0pt] (source2) at (1.618, 0) {};
    \draw[very thick, -Stealth] (source1) to (sink1);
    \draw[thin, -Stealth] (sink1) to (source2);
    \draw[thin, -Stealth] (source2) to (sink2);
    \draw[thin, -Stealth] (sink2) to (source1);
\end{tikzpicture}\caption{$\mathbbm{G}$}\end{subfigure}}
\caption{Three contraction diagrams for the isospin-$I=\frac{1}{2}$ $D\pi$ operator. Grey dots denote local single-particle operators. Thick and thin lines denote charm- and light-quark propagators, respectively.}
\label{fig:diagrams-Dpi}
\end{figure}

The first term contains two different Wick contractions, corresponding to the diagrams $\mathbbm{E}$ and $\mathbbm{F}$ shown in Fig.~\ref{fig:diagrams-Dpi}. The third contraction, $\mathbbm{G}$, arises from the other terms in the expansion. Since our ensembles use Wilson-clover fermions with degenerate light-quark masses, their propagators are numerically identical. The grey dots in the figure denote local single-particle operators with definite flavor, momentum, and gamma matrix, and the arrows indicate the direction of quark propagation.

For a general $\mathbbm{E}_{[\Omega \Xi \Lambda \Gamma], [\delta \gamma \beta \alpha]}(t^{\prime},t)$ diagram, the explicit contraction is
\begin{equation}
\begin{aligned}
    & \mathbbm{E} = \langle \mathrlap{\underbracket[1.3pt][7pt]{\phantom{\bar{u} \square e^{-i p_{\delta} \cdot x} \Omega \square c(t^{\prime}) \cdot \bar{d} \square e^{-i p_{\gamma} \cdot x} \Xi \square u(t^{\prime}) \cdot \bar{c} \square e^{-i p_{\beta} \cdot x} \Lambda \square u}}} \bar{u} \square e^{-i p_{\delta} \cdot x} \Omega \square \mathrlap{\underbracket{\phantom{c(t^{\prime}) \cdot \bar{d} \square e^{-i p_{\gamma} \cdot x} \Xi \square u(t^{\prime}) \cdot \bar{c}}}} c(t^{\prime}) \cdot \overbracket{\bar{d} \square e^{-i p_{\gamma} \cdot x} \Xi \square \overbracket{u(t^{\prime}) \cdot \bar{c} \square e^{-i p_{\beta} \cdot x} \Lambda \square u(t) \cdot \bar{u}} \square e^{-i p_{\alpha} \cdot x} \Gamma \square d}(t) \rangle \\
    &= + \langle {}^{D}\mathbbm{C}_{[\Omega \Lambda],[-\delta;\beta]}^{1B}(t^{\prime},t) \cdot {}^{\pi}\mathbbm{C}_{[\Xi \Gamma],[-\gamma;\alpha]}^{1B}(t^{\prime},t) \rangle,
\end{aligned}
\end{equation}
where $\mathbbm{C}^{1B}$ denotes a single-meson two-point function.

For the $\mathbbm{F}_{[\Omega \Xi \Lambda \Gamma], [\delta \gamma \beta \alpha]}(t^{\prime},t)$ diagram,
\begin{equation}
\begin{aligned}
    & \mathbbm{F} = \langle \mathrlap{\overbracket{\phantom{\bar{u} \square e^{-i p_{\delta} \cdot x} \Omega \square c(t^{\prime}) \cdot \bar{d} \square e^{-i p_{\gamma} \cdot x} \Xi \square u}}} \bar{u} \square e^{-i p_{\delta} \cdot x} \Omega \square \mathrlap{\underbracket{\phantom{c(t^{\prime}) \cdot \bar{d} \square e^{-i p_{\gamma} \cdot x} \Xi \square u(t^{\prime}) \cdot \bar{c}}}} c(t^{\prime}) \cdot \overbracket[1.3pt][7pt]{\bar{d} \square e^{-i p_{\gamma} \cdot x} \Xi \square u(t^{\prime}) \cdot \bar{c} \square e^{-i p_{\beta} \cdot x} \Lambda \square \underbracket{u(t) \cdot \bar{u}} \square e^{-i p_{\alpha} \cdot x} \Gamma \square d}(t) \rangle \\
    & \qquad \tau^u_{\substack{\mu\nu \\ ef}}(t^{\prime},t^{\prime}) \tilde{V}_{fg}(\vec{p}_{\delta},t^{\prime}) \Omega_{\nu\rho} \tau^c_{\substack{\rho\tau \\ gh}}(t^{\prime},t) \tilde{V}_{ha}(\vec{p}_{\beta},t) \Lambda_{\tau\alpha} \rangle \\
    &= - [(\Gamma \gamma_5)_{\beta\phi} (\gamma_5 \Xi)_{\theta\mu} \Omega_{\nu\rho} \Lambda_{\tau\alpha}] \\
    & \qquad \langle \tau^u_{\substack{\alpha\beta \\ ab}}(t,t) \tilde{V}_{bc}(\vec{p}_{\alpha},t) \tau^{d*}_{\substack{\theta\phi \\ dc}}(t^{\prime},t) \tilde{V}_{de}(\vec{p}_{\gamma},t^{\prime}) \tau^u_{\substack{\mu\nu \\ ef}}(t^{\prime},t^{\prime}) \tilde{V}_{fg}(\vec{p}_{\delta},t^{\prime}) \tau^c_{\substack{\rho\tau \\ gh}}(t^{\prime},t) \tilde{V}_{ha}(\vec{p}_{\beta},t) \rangle.
\end{aligned}
\end{equation}

For the $\mathbbm{G}_{[\Omega \Xi \Lambda \Gamma], [\delta \gamma \beta \alpha]}(t^{\prime},t)$ diagram,
\begin{equation}
\begin{aligned}
    & \mathbbm{G} = \langle \overbracket{\bar{u} \square e^{-i p_{\delta} \cdot x} \Omega \square \mathrlap{\underbracket{\phantom{c(t^{\prime}) \cdot \bar{d} \square e^{-i p_{\gamma} \cdot x} \Xi \square u(t^{\prime}) \cdot \bar{c}}}} c(t^{\prime}) \cdot \mathrlap{\overbracket{\phantom{\bar{d} \square e^{-i p_{\gamma} \cdot x} \Xi \square u(t^{\prime}) \cdot \bar{c} \square e^{-i p_{\beta} \cdot x} \Lambda \square d}}} \bar{d} \square e^{-i p_{\gamma} \cdot x} \Xi \square \underbracket[1.3pt][7pt]{u(t^{\prime}) \cdot \bar{c} \square e^{-i p_{\beta} \cdot x} \Lambda \square d(t) \cdot \bar{u}} \square e^{-i p_{\alpha} \cdot x} \Gamma \square u}(t) \rangle \\
    &= - [(\Lambda \gamma_5)_{\beta\phi} {(\gamma_5 \Xi)}_{\theta\mu} (\Gamma \gamma_5)_{\nu\lambda} {(\gamma_5 \Omega)}_{\chi\alpha}] \\
    & \qquad \langle \tau^c_{\substack{\alpha\beta \\ ab}}(t^{\prime},t) \tilde{V}_{bc}(\vec{p}_{\beta},t) \tau^{d*}_{\substack{\theta\phi \\ dc}}(t^{\prime},t) \tilde{V}_{de}(\vec{p}_{\gamma},t^{\prime}) \tau^u_{\substack{\mu\nu \\ ef}}(t^{\prime},t) \tilde{V}_{fg}(\vec{p}_{\alpha},t) \tau^{u*}_{\substack{\chi\lambda \\ hg}}(t^{\prime},t) \tilde{V}_{ha}(\vec{p}_{\delta},t^{\prime}) \rangle.
\end{aligned}
\end{equation}

The key steps in which $\gamma_5$ Hermiticity is used to reverse the propagator direction are indicated above in blue. If complete all-to-all propagators are available, this step can in fact be omitted.

It is important to emphasize that the quark-contraction diagrams describe only the connection topology of quark lines and do not explicitly draw gluon lines. Since QCD is a strongly coupled, nonperturbative theory, higher-order diagrams in a perturbative expansion are not necessarily smaller. In fact, the lattice calculation above equivalently includes Feynman-diagram contributions to all orders in perturbation theory, with any number of gluon lines implicitly connecting arbitrary quark lines. All such gluonic dynamics are fully included in the nonperturbative quark propagators, or equivalently in the background gauge-field configurations.

Expanding Eq.~\ref{eq:Dpi_contraction}, one obtains the full contraction expression:
\begin{align*}
    & \langle O_{D\pi,A_1^+,p}^{[J=L=S=0]}(t^{\prime}) O_{D\pi,A_1^+,p}^{[J=L=S=0]\dagger}(t) \rangle \\
    &= \left\langle \left\{\sum_{\beta} \left[ 2 D^0(\vec{p}_{\beta}) \pi^+(-\vec{p}_{\beta}) - \sqrt{2} D^+(\vec{p}_{\beta}) \pi^0(-\vec{p}_{\beta}) \right]\right\}(t^{\prime}) \right. \\
    &\left. \qquad \left\{\sum_{\alpha} \left[ 2 D^0(\vec{p}_{\alpha}) \pi^+(-\vec{p}_{\alpha}) - \sqrt{2} D^+(\vec{p}_{\alpha}) \pi^0(-\vec{p}_{\alpha}) \right]\right\}^{\dagger}(t) \right\rangle \\
    &= {\scriptstyle \left\langle \left\{\sum_{\beta} \left[ 2 \bar{u} \square e^{-i p_{\beta} \cdot x} \gamma_5 \square c \cdot \bar{d} \square e^{+i p_{\beta} \cdot x} \gamma_5 \square u - \bar{d} \square e^{-i p_{\beta} \cdot x} \gamma_5 \square c \cdot \bar{u} \square e^{+i p_{\beta} \cdot x} \gamma_5 \square u + \bar{d} \square e^{-i p_{\beta} \cdot x} \gamma_5 \square c \cdot \bar{d} \square e^{+i p_{\beta} \cdot x} \gamma_5 \square d \right]\right\}(t^{\prime}) \right.} \\
    & \quad {\scriptstyle \left. \left\{\sum_{\alpha} \left[ 2 \bar{u} \square e^{-i p_{\alpha} \cdot x} \gamma_5 \square c \cdot \bar{d} \square e^{+i p_{\alpha} \cdot x} \gamma_5 \square u - \bar{d} \square e^{-i p_{\alpha} \cdot x} \gamma_5 \square c \cdot \bar{u} \square e^{+i p_{\alpha} \cdot x} \gamma_5 \square u + \bar{d} \square e^{-i p_{\alpha} \cdot x} \gamma_5 \square c \cdot \bar{d} \square e^{+i p_{\alpha} \cdot x} \gamma_5 \square d \right]\right\}^{\dagger}(t) \right\rangle} \\
    &= \sum_{\beta\alpha} \left[ + \left(+ 4 \mathbbm{C}_{[\gamma_5,\gamma_5;\gamma_5,\gamma_5], [\beta,-\beta;-\alpha,\alpha]}^{\bar{u}c\bar{d}u-\bar{c}u\bar{u}d} - 2 \mathbbm{C}_{[\gamma_5,\gamma_5;\gamma_5,\gamma_5], [\beta,-\beta;-\alpha,\alpha]}^{\bar{u}c\bar{d}u-\bar{c}d\bar{u}u} + 2 \mathbbm{C}_{[\gamma_5,\gamma_5;\gamma_5,\gamma_5], [\beta,-\beta;-\alpha,\alpha]}^{\bar{u}c\bar{d}u-\bar{c}d\bar{d}d} \right) \right. \\
    & \qquad\quad + \left(- 2 \mathbbm{C}_{[\gamma_5,\gamma_5;\gamma_5,\gamma_5], [\beta,-\beta;-\alpha,\alpha]}^{\bar{d}c\bar{u}u-\bar{c}u\bar{u}d} + 1 \mathbbm{C}_{[\gamma_5,\gamma_5;\gamma_5,\gamma_5], [\beta,-\beta;-\alpha,\alpha]}^{\bar{d}c\bar{u}u-\bar{c}d\bar{u}u} - 1 \mathbbm{C}_{[\gamma_5,\gamma_5;\gamma_5,\gamma_5], [\beta,-\beta;-\alpha,\alpha]}^{\bar{d}c\bar{u}u-\bar{c}d\bar{d}d} \right) \\
    & \qquad\quad \left. + \left(+ 2 \mathbbm{C}_{[\gamma_5,\gamma_5;\gamma_5,\gamma_5], [\beta,-\beta;-\alpha,\alpha]}^{\bar{d}c\bar{d}d-\bar{c}u\bar{u}d} - 1 \mathbbm{C}_{[\gamma_5,\gamma_5;\gamma_5,\gamma_5], [\beta,-\beta;-\alpha,\alpha]}^{\bar{d}c\bar{d}d-\bar{c}d\bar{u}u} + 1 \mathbbm{C}_{[\gamma_5,\gamma_5;\gamma_5,\gamma_5], [\beta,-\beta;-\alpha,\alpha]}^{\bar{d}c\bar{d}d-\bar{c}d\bar{d}d} \right) \right] \\
    &= \sum_{\beta\alpha} \left( 6 \mathbbm{E} + 9 \mathbbm{F} - 3 \mathbbm{G} \right)_{[\gamma_5,\gamma_5;\gamma_5,\gamma_5], [\beta,-\beta;-\alpha,\alpha]}(t^{\prime},t).
\end{align*}

Applying the same contraction procedure to the other matrix elements of the correlation matrix provides all the information needed for the finite-volume spectrum of the isospin-$\frac{1}{2}$ $D\pi$ system. If the operators are chosen with $I_z = -\frac{1}{2}$, the contractions are identical by isospin symmetry.

\subsection{Finite-Volume Spectrum}
\begin{figure}[htbp]
\centering
\includegraphics[width=0.49\columnwidth]{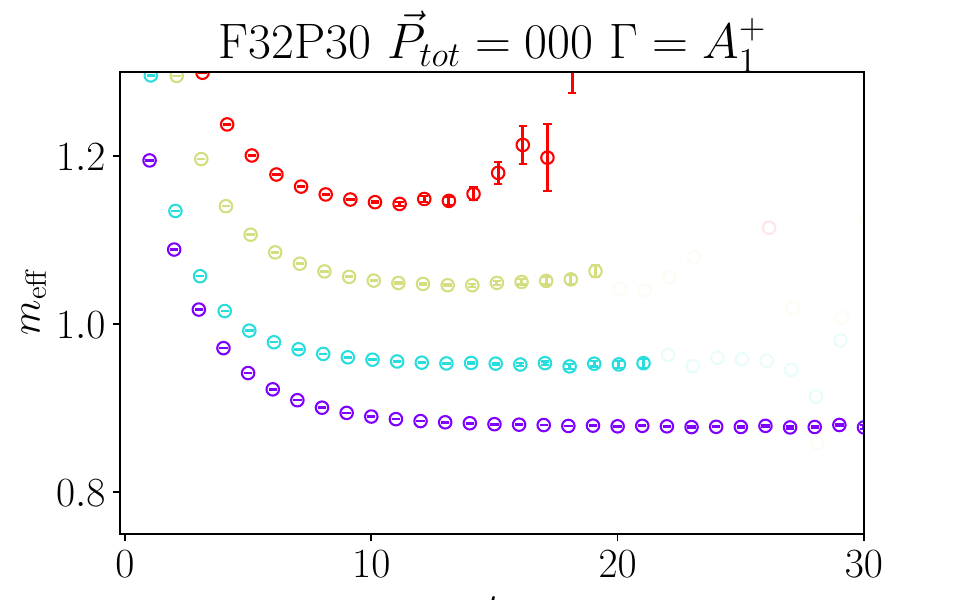}
\includegraphics[width=0.49\columnwidth]{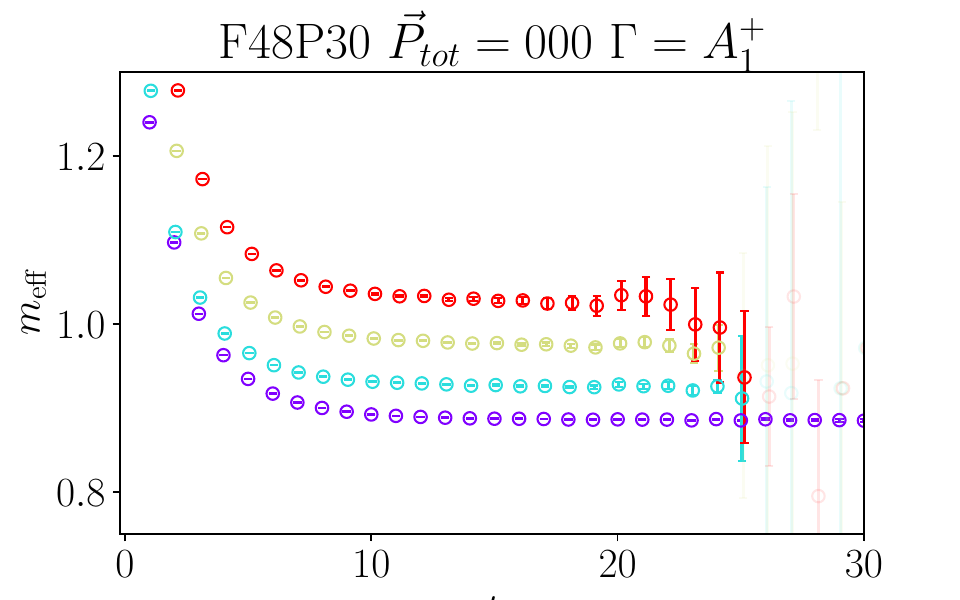}
\includegraphics[width=0.49\columnwidth]{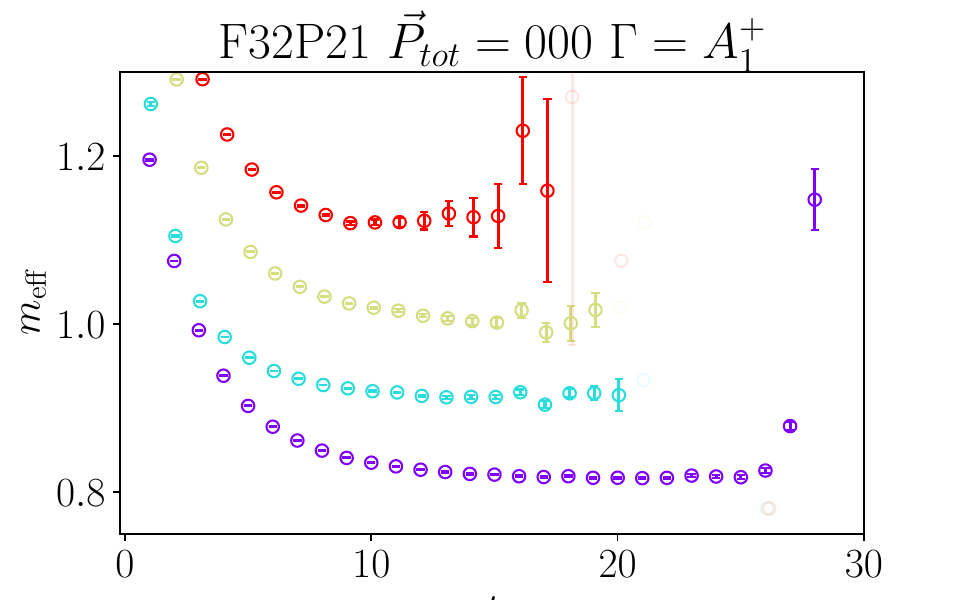}
\includegraphics[width=0.49\columnwidth]{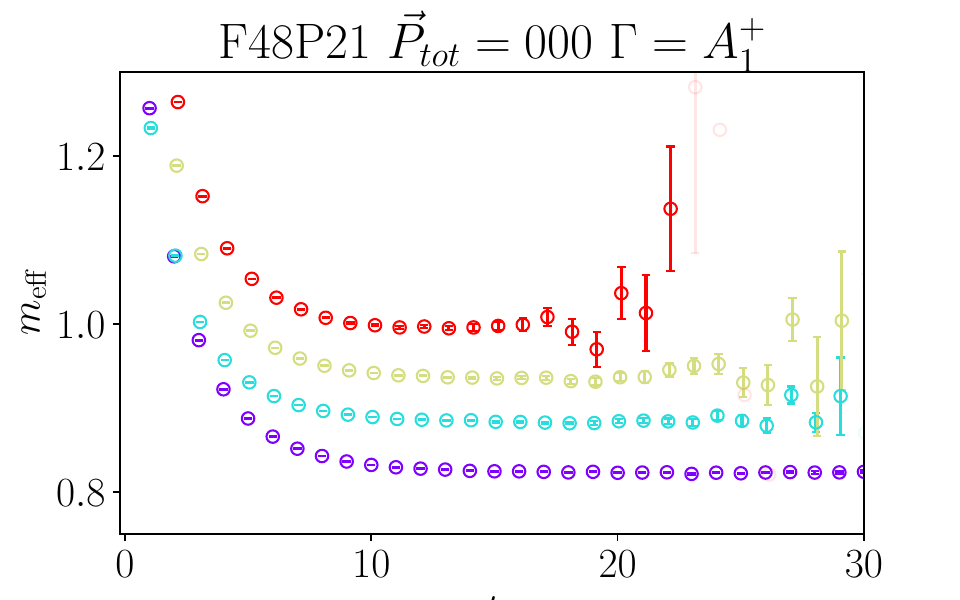}
\includegraphics[width=0.49\columnwidth]{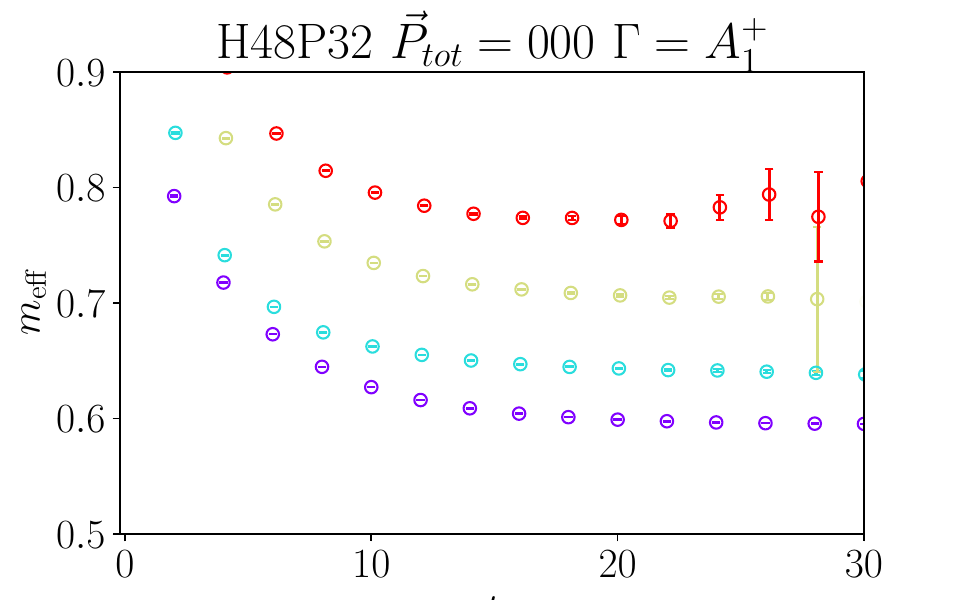}
\includegraphics[width=0.49\columnwidth]{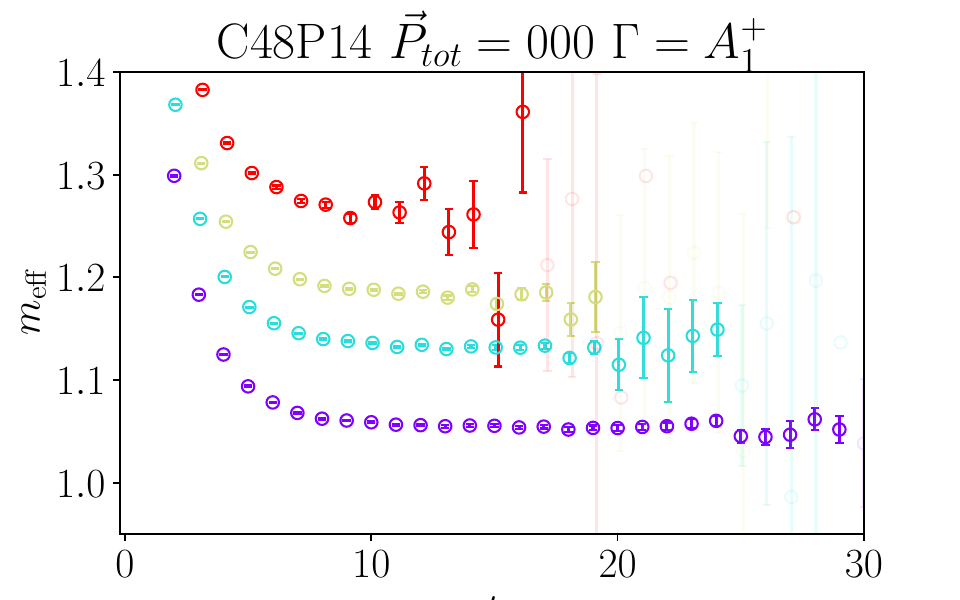}
\caption{Effective masses of the generalized eigenvalues $\lambda_n(t)$ of the two-point correlation functions for the $D\pi$ system in the center-of-mass-frame $A_1^+$ irrep on each ensemble. Different colors correspond to different energy levels. The vertical axis is in lattice units.}
\label{fig:Dpi-meff}
\end{figure}

The spectrum in each irrep is obtained by solving a generalized eigenvalue problem~\cite{Michael:1982gb,Luscher:1990ck,Blossier:2009kd,Fischer:2020bgv}. Figure~\ref{fig:Dpi-meff} shows effective-mass plots for the generalized eigenvalues $\lambda_n(t)$ of the two-point correlation functions in the $A_1^+$ irrep of the $D\pi$ system on each ensemble. Different colors correspond to different energy levels, and the vertical axis is in lattice units. For each level, the energy is extracted using a two-state fit. Fit stability is assessed by scanning different fit ranges. Effective masses for other ensembles and irreps are shown in Figs.~\ref{fig:Dpi-meff-A1+}--\ref{fig:Dpi-meff-002-E2} in Appendix~\ref{appendix:two_body_problems}; the detailed fit results and stability analyses for individual levels are shown in Figs.~\ref{fig:Dpi-fit-F32P30}--\ref{fig:Dpi-fit-C48P14}. For each level, we perform fits over different fit windows and with different initial guesses, and then examine the stability of the fitted result. Collecting all extracted energies gives the finite-volume spectrum of the $D\pi$ system. Statistical uncertainties are estimated using $2000$ bootstrap resamples.

\begin{figure}[htbp]
    \centering
    \includegraphics[width=\textwidth]{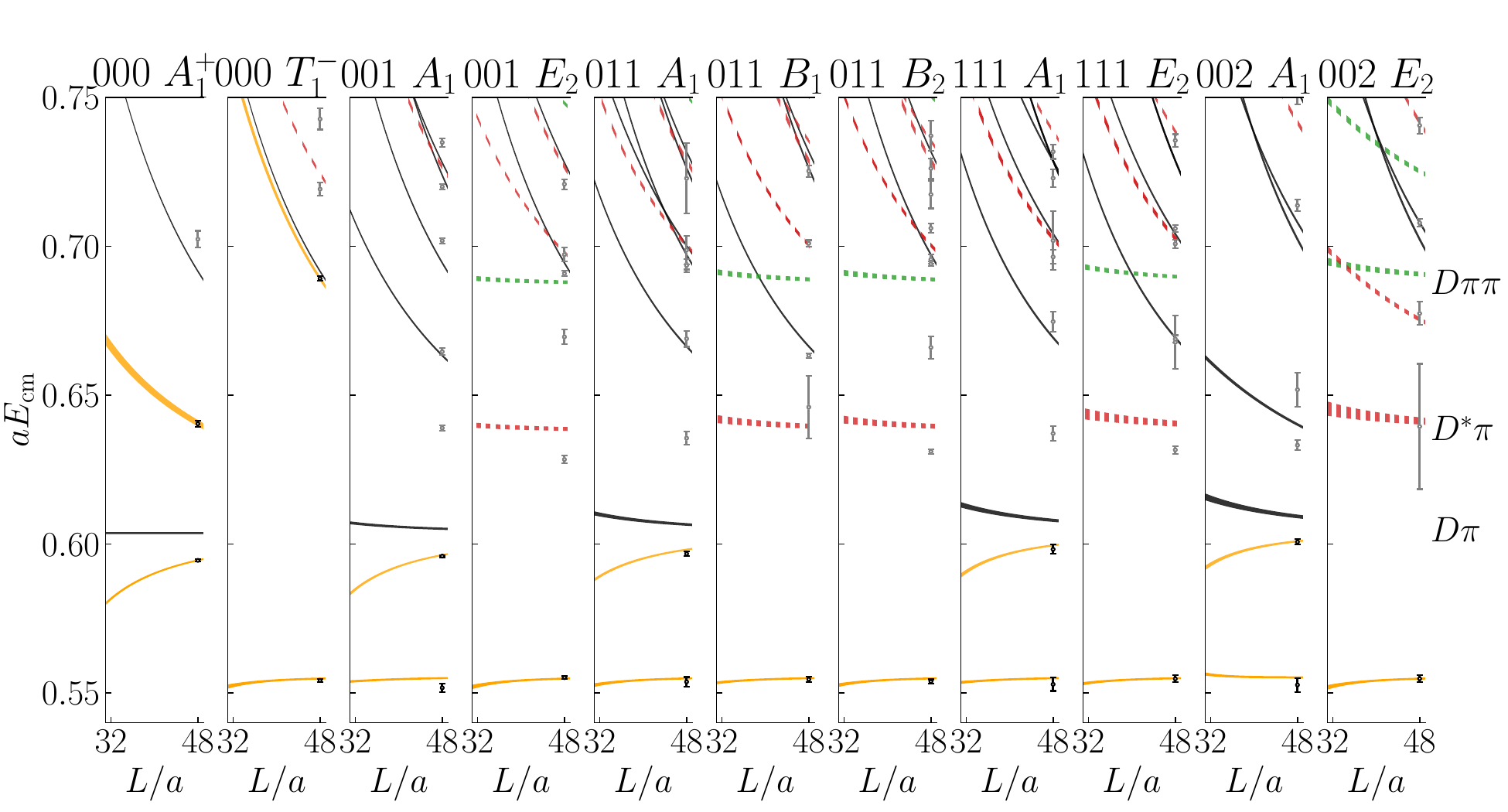}
    \caption{Finite-volume spectra at $M_{\pi} \approx 317~\mathrm{MeV}$ for all irreps whose leading partial waves are $S$ or $P$ waves. Black and grey points denote energy levels; black solid, red dashed, and green dashed bands denote the noninteracting $D\pi$, $D^*\pi$, and $D\pi\pi$ levels, respectively. Only the black points are used in the scattering analysis. Orange bands are solutions of Lüscher's equation.}
    \label{fig:spectra-Dpi-HH-1d2}
\end{figure}

\begin{figure}[htbp]
    \centering
    \includegraphics[width=\textwidth]{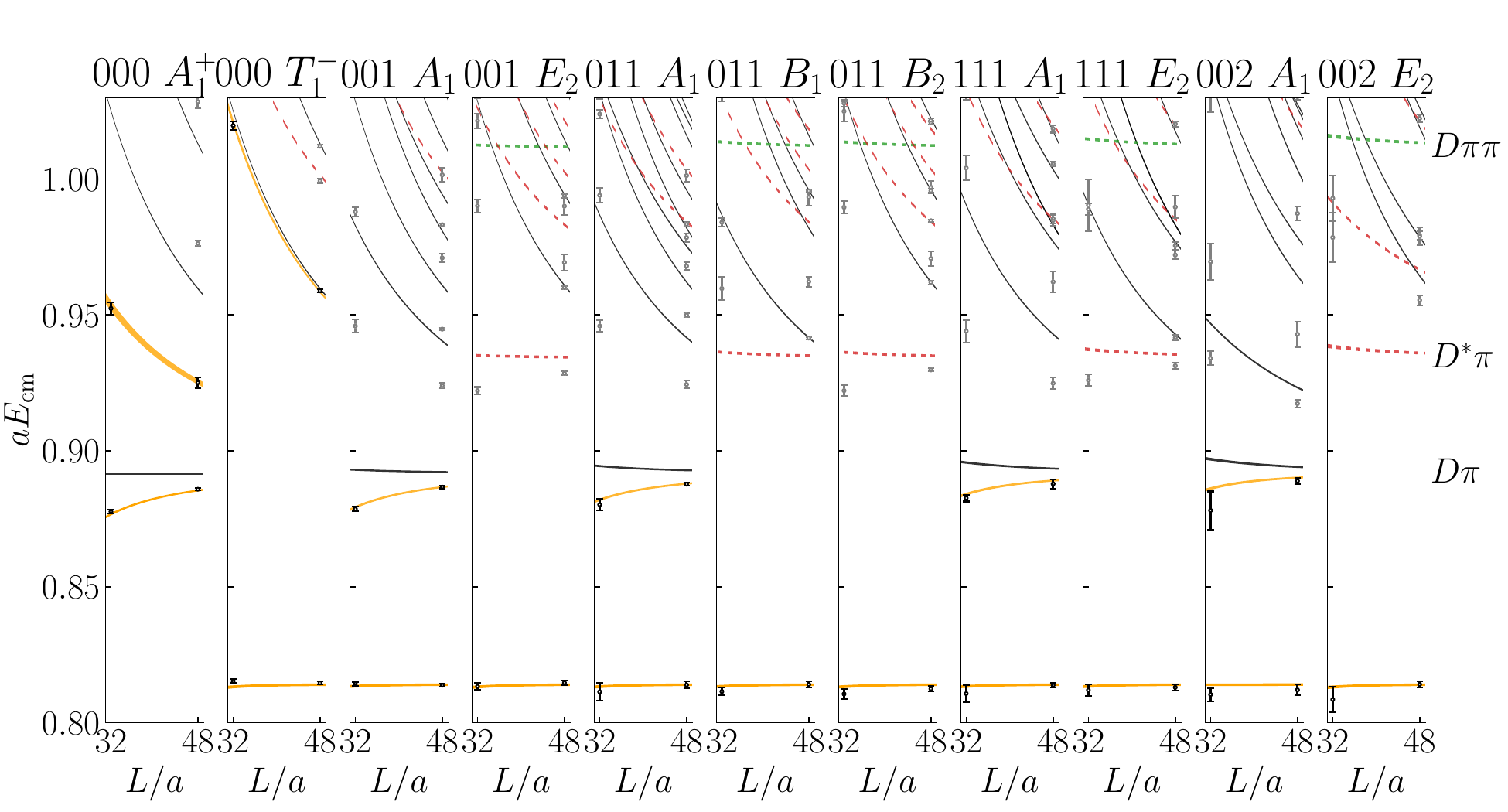}
    \caption{Finite-volume spectrum at $M_{\pi} \approx 305~\mathrm{MeV}$. The legend is the same as in Fig.~\ref{fig:spectra-Dpi-HH-1d2}.}
    \label{fig:spectra-Dpi-FH-1d2}
\end{figure}

\begin{figure}[htbp]
    \centering
    \includegraphics[width=\textwidth]{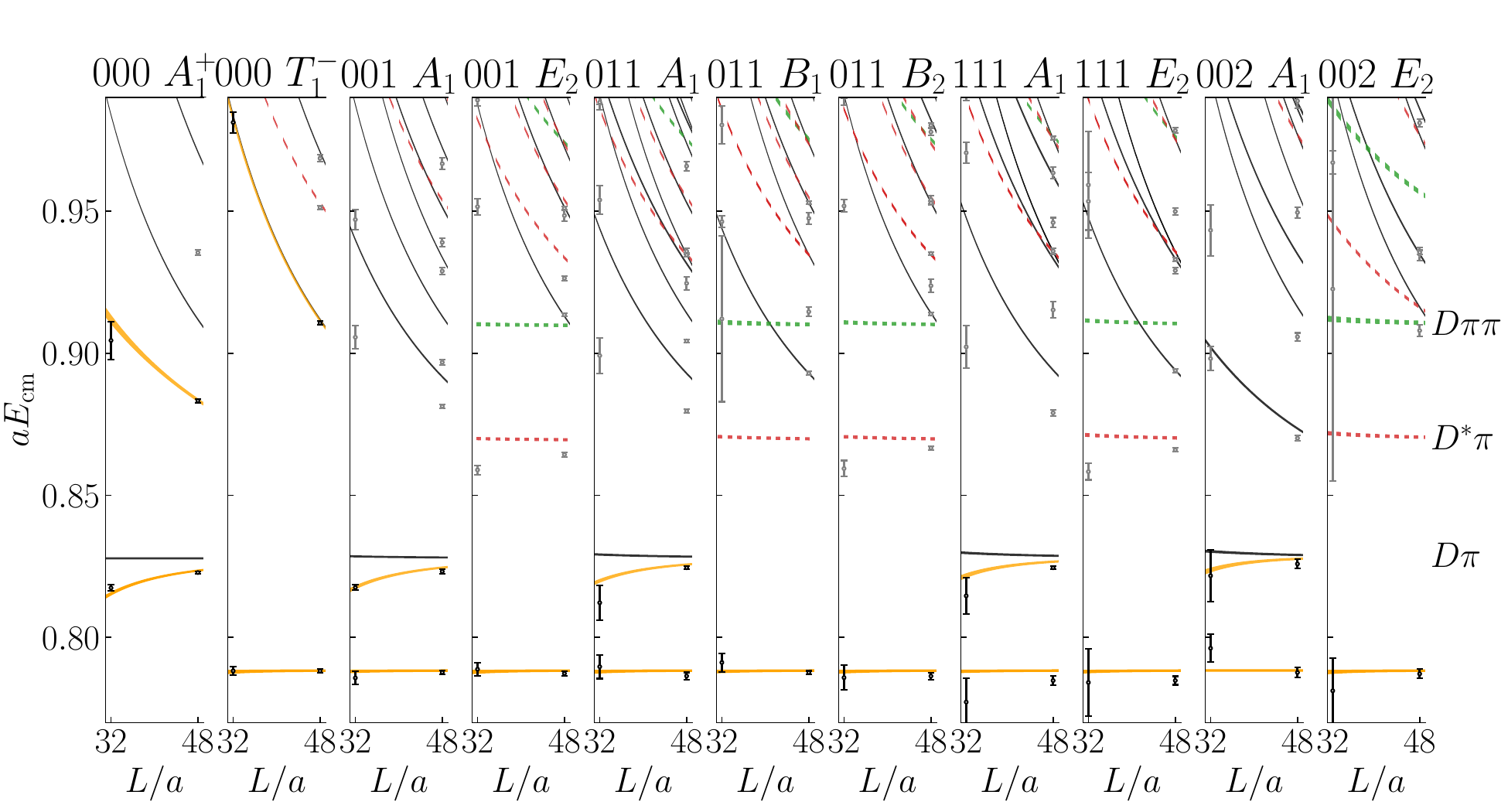}
    \caption{Finite-volume spectrum at $M_{\pi} \approx 208~\mathrm{MeV}$. The legend is the same as in Fig.~\ref{fig:spectra-Dpi-HH-1d2}.}
    \label{fig:spectra-Dpi-FL-1d2}
\end{figure}

\begin{figure}[htbp]
    \centering
    \includegraphics[width=\textwidth]{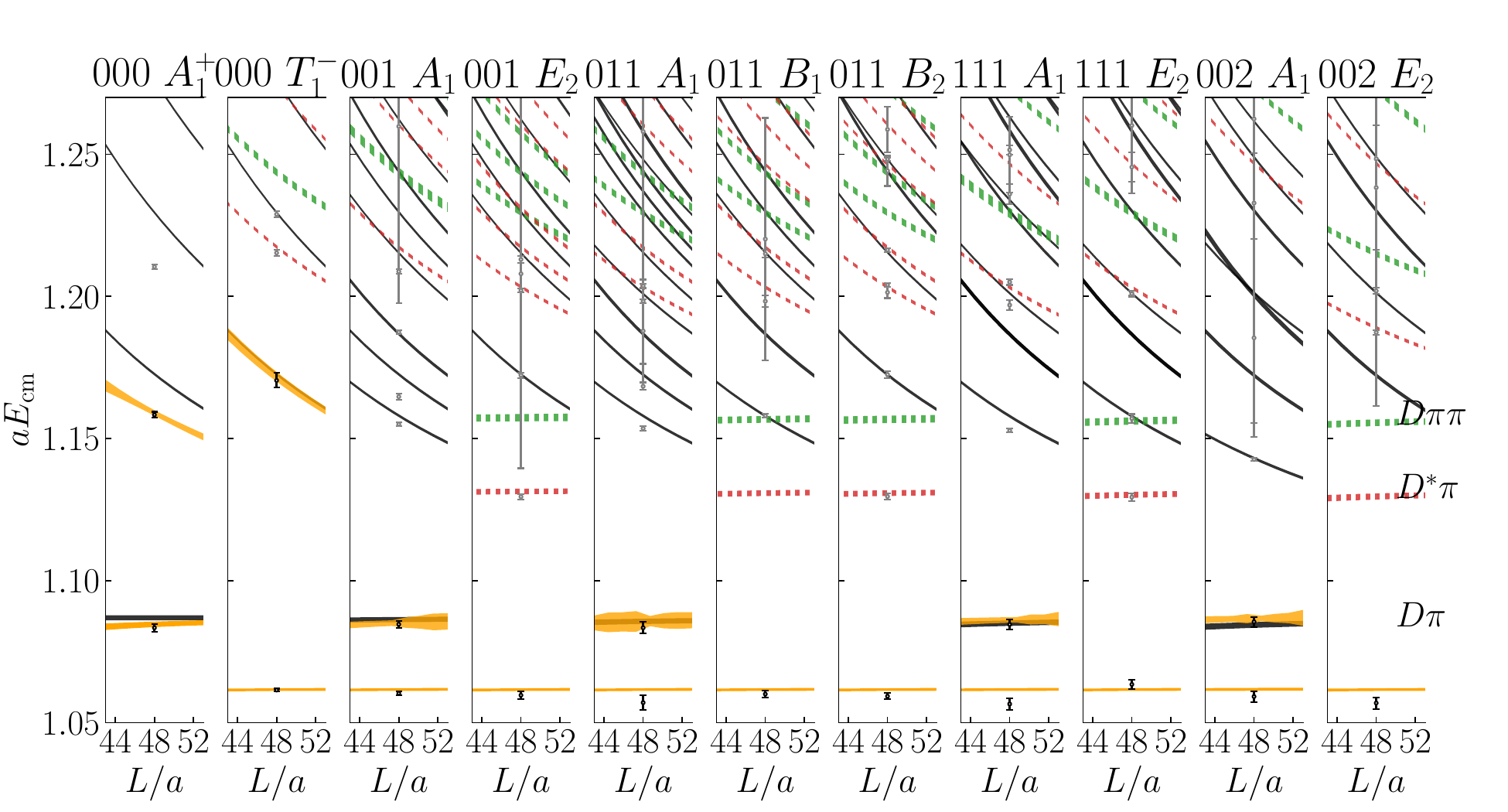}
    \caption{Finite-volume spectrum at $M_{\pi} \approx 133~\mathrm{MeV}$. The legend is the same as in Fig.~\ref{fig:spectra-Dpi-HH-1d2}.}
    \label{fig:spectra-Dpi-CP-1d2}
\end{figure}

The extracted finite-volume spectra are shown in Figs.~\ref{fig:spectra-Dpi-HH-1d2}--\ref{fig:spectra-Dpi-CP-1d2}. The black solid bands, red dashed bands, and green dashed bands denote the noninteracting $D\pi$, $D^*\pi$, and $D\pi\pi$ levels, respectively. Since the single-particle levels are determined very precisely, these noninteracting bands are narrow and appear nearly as lines. Black points denote energy levels obtained by fitting the eigenvalues of the correlation matrix. The orange bands are solutions of Lüscher's equation~(\ref{eq:luscher}), whose details are given below.

Near threshold, low partial waves dominate. In the single-channel scattering analysis in this chapter, we neglect $D$ waves and higher partial waves. In all irreps containing a $P$ wave, we find an energy level significantly below the $D\pi$ threshold, corresponding to the vector $D^*$ bound state. When $D$ waves and higher waves are neglected, the rest-frame $T_1^-$ irrep contains only the $D\pi$ $P$ wave. Apart from the $D^*$ bound state, all other levels are very close to the noninteracting levels, indicating that the $P$-wave interaction is weak. The remaining irreps are dominated by the $S$ wave, and their $D\pi$ scattering levels lie significantly below the corresponding free levels, indicating a strong attractive $S$-wave interaction. Figure~\ref{fig:spectra-Dpi-CP-1d2} shows the result on the $M_{\pi}=133~\mathrm{MeV}$ ensemble using the Fermilab charm-quark action. If the Wilson-clover action is used instead, the lowest few noninteracting levels increase as the volume $L$ is enlarged; this is a discretization effect.

\section{Scattering Analysis}
\label{sec:scattering}
Finite-volume spectra can be related to infinite-volume scattering phase shifts through Lüscher's quantization condition~\cite{Luscher:1986pf, Luscher:1990ux, Luscher:1990ck}, whose general form can be written as~\cite{Briceno:2014oea, Gockeler:2012yj}
\begin{equation}
\operatorname{det}[M(E, \vec{P} ; L)-\cot \delta(E)]=0,
\label{eq:luscher}
\end{equation}
where $M(E, \vec{P} ; L)$ contains the information on the finite-volume spectrum, and $\delta$ is the scattering phase shift.

\subsection{Dispersion Relation and Discretization Effects}
The scattering momentum $k$ is related to the scattering energy through the modified dispersion relation
\begin{equation}
    E(\vec{k})=\sqrt{m_D^2+Z_D\vec{k}^2} + \sqrt{M_{\pi}^2+Z_\pi\vec{k}^2},
\label{eq:dispersion}
\end{equation}
where $Z_X$ is the squared speed of light on the lattice. Because of discretization effects, $Z_X$ may deviate from the continuum value $1$. We compute single-particle energies with momenta satisfying $\frac{L}{2\pi} |\vec{P}| \leq 2$. We find that $Z_\pi$ is close to $1$ on all ensembles, whereas $Z_D$ and $Z_{D^*}$ are statistically smaller than $1$, indicating nonzero discretization effects. The extracted meson masses and the coefficients in the dispersion relation are listed in Table~\ref{tab:meson_masses} and Table~\ref{tab:disp}, respectively. The single-particle energies used in this work are the values measured in the rest frame rather than the masses obtained from fits to the dispersion relation; the difference is much smaller than the statistical uncertainty.

\begin{table*}[htbp]
\centering
\caption{Meson masses used in the single-channel $D\pi$ scattering analysis.}
\begin{tabular}{cccc}
\toprule
Ensemble & $M_{\pi}$ / MeV & $m_D$ / MeV & $m_{D^*}$ / MeV \\
\midrule
C48P14 & $133.1(1.6)$ & $1903.73(34)$ & $1988.15(72)$ \\
F32P21 & $206.8(2.1)$ & $1901.3(1.1)$ & $2007.2(2.2)$ \\
F48P21 & $208.12(70)$ & $1900.71(48)$ & $2006.3(1.2)$ \\
F32P30 & $305.81(71)$ & $1965.7(1.0)$ & $2076.6(1.5)$ \\
F48P30 & $304.98(50)$ & $1966.20(57)$ & $2074.1(1.1)$ \\
H48P32 & $317.00(68)$ & $1979.41(85)$ & $2108.9(2.0)$ \\
\bottomrule
\end{tabular}
\label{tab:meson_masses}
\end{table*}

\begin{table*}[htbp]
\centering
\caption{Coefficients in the dispersion relation~\ref{eq:dispersion}.}
\begin{tabular}{cccc}
\toprule
Ensemble & $Z_{\pi}$ & $Z_D$ & $Z_{D^*}$ \\
\midrule
C48P14 & $1.0103(33)$ & $0.861(14)$ & $0.865(11)$ \\
F32P21 & $1.0409(84)$ & $0.9283(48)$ & $0.905(16)$ \\
F48P21 & $1.0157(49)$ & $0.9302(31)$ & $0.9360(45)$ \\
F32P30 & $0.9977(56)$ & $0.9083(74)$ & $0.9027(93)$ \\
F48P30 & $1.0054(24)$ & $0.9262(41)$ & $0.9243(70)$ \\
H48P32 & $1.0115(41)$ & $0.9551(66)$ & $0.935(15)$ \\
\bottomrule
\end{tabular}
\label{tab:disp}
\end{table*}

The coarser the lattice spacing, the larger the deviation of $Z_X$ from $1$. We estimate the discretization uncertainty by comparing the modified dispersion relation with the continuum dispersion relation, namely by comparing the lattice-measured $Z_X$ with $1$. The resulting discretization uncertainty is small, and the uncertainty estimated in this way is combined in quadrature with the parametrization uncertainty as a systematic error.

Except for the coarse C48P14 ensemble, the scattering analysis on each ensemble is carried out using both the lattice-measured value of $Z_X$ and the continuum-limit choice $Z_X = 1$; the difference is taken as the systematic uncertainty associated with the lattice spacing. For C48P14, the corresponding uncertainty is obtained by comparing results from different charm-quark actions\footnotecircle{This procedure has also been adopted in many subsequent CLQCD spectroscopy studies.}. It should be noted that lattice-spacing effects are not exhausted by the speed of light, and that the difference in the speed of light has only a limited impact in the low-momentum region. This procedure may therefore underestimate discretization effects. However, CLQCD does not yet have enough ensembles with the small lattice spacings required for charm physics. Without calculations on ensembles at genuinely different lattice spacings, it is difficult to give a fully reliable estimate of the discretization uncertainty. We do not expect this possible underestimation to affect the final results. In the analysis, we also artificially enlarged the discretization uncertainty obtained from this procedure by several factors. After combining the statistical and discretization uncertainties in quadrature, the total uncertainty did not increase significantly, indicating that the current error budget is still dominated by statistics. As will be seen below, the results obtained at two different lattice spacings, $a=0.077~\mathrm{fm}$ and $a=0.052~\mathrm{fm}$, but at similar $\pi$ masses, are very close to one another. As for the coarsest lattice, $a=0.105~\mathrm{fm}$ with $M_{\pi}=133~\mathrm{MeV}$, its uncertainty is large enough that it carries little weight in the extrapolation to the physical point. Once more ensembles at different lattice spacings become available, discretization effects can be controlled more reliably.

\subsection{Fit Details}
All lattice studies using the Lüscher formalism must address two issues:
\begin{itemize}
    \item the determinant in Eq.~(\ref{eq:luscher}) contains all partial waves and is infinite-dimensional;
    \item each energy level supplies only one equation, so the problem is underdetermined.
\end{itemize}
The quantization condition contains scattering amplitudes in different partial waves and different scattering channels. Within the energy range of interest, each energy level provides only one constraint on several scattering amplitudes, so the problem is underdetermined. The only exception is single-channel scattering when only the lowest partial wave is retained; an example will be given below. There are many ways to address this problem. In this chapter, and also in the later studies of three-body systems, we parametrize the scattering matrix, so that the amplitudes at different energies can be described and related by a finite number of parameters\footnotecircle{Other approaches are discussed in Ref.~\cite{Dudek:2012gj}. My understanding of this issue owes much to that work.}. This inevitably introduces model dependence, and the parametrizations must therefore be studied systematically in order to estimate the associated uncertainty.

To address these two issues, we truncate the infinite-dimensional determinant in Eq.~(\ref{eq:luscher}) at the $P$ wave. The study in Ref.~\cite{Gayer:2021xzv} shows that the $D$-wave contribution is small in the near-threshold $D\pi$ scattering region. Partial waves with still higher angular momentum are suppressed by factors of $k^{2l}$, and can therefore be safely neglected.

For the irreps relevant to this chapter, when the dominant partial waves do not extend beyond the $P$ wave, Eq.~(\ref{eq:luscher}) reduces to~\cite{Gockeler:2012yj}
\begin{align}
O_h(A_1^+)
&\begin{cases}
    &l_{\mathrm{max}} = 3: \cot \delta_0 = \frac{1}{\pi^{\frac{3}{2}} q} \mathcal{Z}_{00}(1;q^2)
\end{cases}, \\
O_h(T_1^-)
&\begin{cases}
    &l_{\mathrm{max}} = 0: \cot \delta_0 = 0 \\
    &l_{\mathrm{max}} = 1: \cot \delta_1 = \frac{1}{\pi^{\frac{3}{2}} q} \mathcal{Z}_{00}
\end{cases}, \\
O_h(T_2^+)
&\begin{cases}
    &l_{\mathrm{max}} = 1: \cot \delta_2 = \frac{1}{7 \pi^{\frac{3}{2}} q^5} \left[ 7 q^4 \mathcal{Z}_{00}(1;q^2) - 4 \mathcal{Z}_{40}(1;q^2) \right]
\end{cases}, \\
C_{4v}(A_1)
&\begin{cases}
    &l_{\mathrm{max}} = 0: \cot \delta_0 = \frac{1}{\pi^{\frac{3}{2}} q \gamma} \mathcal{Z}_{00} \\
    &l_{\mathrm{max}} = 1: \cot \delta_0 = \frac{1}{\pi^{\frac{3}{2}} q \gamma} \left[ \mathcal{Z}_{00} + \frac{5 \mathcal{Z}_{10}^2}{5\pi^{\frac{3}{2}} q^3 \gamma \cot \delta_1 - 5 q^2 \mathcal{Z}_{00} - 2 \sqrt{5} \mathcal{Z}_{20}} \right]
\end{cases}, \\
C_{4v}(E_2)
&\begin{cases}
    &l_{\mathrm{max}} = 1: \cot \delta_1 = \frac{1}{5\pi^{\frac{3}{2}} q^3 \gamma} \left[ 5 q^2 \mathcal{Z}_{00} - \sqrt{5} \mathcal{Z}_{20} \right]
\end{cases}, \\
C_{4v}(B_1)
&\begin{cases}
    &l_{\mathrm{max}} = 2: \cot \delta_2 = \frac{1}{7\pi^{\frac{3}{2}} q^5 \gamma} \left[ 7 q^4 \mathcal{Z}_{00} - 2\sqrt{5} q^2 \mathcal{Z}_{20} + \mathcal{Z}_{40} + \sqrt{70} \mathcal{Z}_{44} \right]
\end{cases}, \\
C_{4v}(B_2)
&\begin{cases}
    &l_{\mathrm{max}} = 2: \cot \delta_2 = \frac{1}{7\pi^{\frac{3}{2}} q^5 \gamma} \left[ 7 q^4 \mathcal{Z}_{00} - 2\sqrt{5} q^2 \mathcal{Z}_{20} + \mathcal{Z}_{40} - \sqrt{70} \mathcal{Z}_{44} \right]
\end{cases}, \\
C_{2v}(A_1)
&\begin{cases}
    &l_{\mathrm{max}} = 0: \cot \delta_0 = \frac{1}{\pi^{\frac{3}{2}} q \gamma} \mathcal{Z}_{00} \\
    &l_{\mathrm{max}} = 1: \cot \delta_0 = \frac{1}{\pi^{\frac{3}{2}} q \gamma} \left[ \mathcal{Z}_{00} + \frac{20 (\Re\mathcal{Z}_{11})^2}{5\pi^{\frac{3}{2}} q^3 \gamma \cot \delta_1 - 5 q^2 \mathcal{Z}_{00} + \sqrt{5} \mathcal{Z}_{20} + i \sqrt{30} \mathcal{Z}_{22}} \right]
\end{cases}, \\
C_{2v}(B_1)
&\begin{cases}
    &l_{\mathrm{max}} = 1: \cot \delta_1 = \frac{1}{5\pi^{\frac{3}{2}} q^3 \gamma} \left[ 5 q^2 \mathcal{Z}_{00} + 2 \sqrt{5} \mathcal{Z}_{20} \right]
\end{cases}, \\
C_{2v}(B_2)
&\begin{cases}
    &l_{\mathrm{max}} = 1: \cot \delta_1 = \frac{1}{5\pi^{\frac{3}{2}} q^3 \gamma} \left[ 5 q^2 \mathcal{Z}_{00} - \sqrt{5} \mathcal{Z}_{20} + i\sqrt{30} \mathcal{Z}_{22} \right]
\end{cases}, \\
C_{2v}(A_2)
&\begin{cases}
    &l_{\mathrm{max}} = 2: \cot \delta_2 = \frac{1}{7\pi^{\frac{3}{2}} q^5 \gamma} \left[ 7 q^4 \mathcal{Z}_{00} + \sqrt{5} q^2 (\mathcal{Z}_{20} + i \sqrt{6} \mathcal{Z}_{22}) - 4 \mathcal{Z}_{40} + 2i \sqrt{10} \mathcal{Z}_{42} \right]
\end{cases}, \\
C_{3v}(A_1)
&\begin{cases}
    &l_{\mathrm{max}} = 0: \cot \delta_0 = \frac{1}{\pi^{\frac{3}{2}} q \gamma} \mathcal{Z}_{00} \\
    &l_{\mathrm{max}} = 1: \cot \delta_0 = \frac{1}{\pi^{\frac{3}{2}} q \gamma} \left[ \mathcal{Z}_{00} + \frac{15 \mathcal{Z}_{10}^2}{5\pi^{\frac{3}{2}} q^3 \gamma \cot \delta_1 - 5 q^2 \mathcal{Z}_{00} + 2i \sqrt{30} \mathcal{Z}_{22}} \right]
\end{cases}, \\
C_{3v}(E_2)
&\begin{cases}
    &l_{\mathrm{max}} = 1: \cot \delta_1 = \frac{1}{5\pi^{\frac{3}{2}} q^3 \gamma} \left[ 5 q^2 \mathcal{Z}_{00} + i \sqrt{30} \mathcal{Z}_{22} \right]
\end{cases}, \\
\end{align}
where $l_{\mathrm{max}}$ denotes the highest partial wave retained in the corresponding expression, $\gamma = \frac{E}{E_{\mathrm{c.m.}}}$ is the Lorentz factor, and $q$ is related to the scattering momentum $k$ by $q = \frac{k L}{2\pi}$. The function $ \mathcal{Z}_{lm}(1;q^2)$ is Lüscher's $\zeta$ function.

Even after the partial-wave truncation, the Lüscher equation remains underdetermined. We therefore use several different parametrizations and determine the model parameters by fitting the lattice spectra. The main results are based on the effective-range expansion (ERE),
\begin{equation}
    k^{2l+1} \cot \delta_{l}=\frac{1}{a_{l}}+\frac{1}{2} r_{l} k^2+P_2 k^4+\mathcal{O}(k^6),
\end{equation}
where for $l=0$, $a_0$ is the scattering length and $r_0$ is the effective range.

In addition to the ERE used in the main analysis, we also employ more general $K$-matrix parametrizations, which provide flexible forms for each partial wave. In this thesis, we mainly use $K$-matrix parametrizations with explicit poles. We do not use the RBW parametrization, since it is appropriate only for narrow resonances without nearby relevant thresholds or other resonances.

If an analytic form of the scattering amplitude is given, the lattice spectrum in any irrep below the inelastic threshold can be obtained by solving Lüscher's equation. Explicit expressions can be found in Ref.~\cite{Gockeler:2012yj}. Here, however, we need to solve the inverse problem: given lattice spectra in several irreps, how should the parameters of the scattering amplitude be determined? We define $\chi^2$ as~\cite{Dudek:2012gj}
\begin{equation}
\begin{aligned}
    \chi^2({\vec{a}}) = \sum_L \sum_{\substack{\vec{P} \Gamma n \ \vec{P}^{\prime} \Gamma^{\prime} n^{\prime}}} & [ E_{\text{CM}}(L;\vec{P} \Gamma n) - E_{\text{CM}}^{\text{det}}(L;\vec{P} \Gamma n;{\vec{a}}) ] \\
    & \operatorname{Cov}^{-1}(L;\vec{P} \Gamma n; \vec{P}^{\prime} \Gamma^{\prime} n^{\prime}) \\
    & [ E_{\text{CM}}(L;\vec{P}^{\prime} \Gamma^{\prime} n^{\prime}) - E_{\text{CM}}^{\text{det}}(L;\vec{P}^{\prime} \Gamma^{\prime} n^{\prime};{\vec{a}}) ],
\label{eq:def_chi2dof}
\end{aligned}
\end{equation}
where $\vec{a}$ denotes the set of parameters to be determined, and $\operatorname{Cov}(L;\vec{P} \Gamma n; \vec{P}^{\prime} \Gamma^{\prime} n^{\prime})$ is the covariance matrix among energy levels. A fit is obtained by varying the parameter set to minimize $\chi^2$. The fit result represents the constraint imposed by the lattice energy levels on the scattering amplitude. It should be noted that correlations between different irreps on the same ensemble are nonzero, whereas correlations between different ensembles vanish. Here $E_{\mathrm{c.m.}} = \sqrt{E^2 - |\vec{P}|^2}$ denotes the center-of-mass energy obtained from the GEVP fit, while $E_{\text{CM}}^{\text{det}}$ is the solution of Lüscher's equation for a given parameter set.

When the modified dispersion relation in Eq.~\ref{eq:dispersion} is used, the noninteracting energies and the pole positions of the $\zeta$ function differ slightly. To avoid missing any level, we divide the energy range into small intervals with step size $0.0005$ and search for solutions of Lüscher's equation.

\begin{figure}[htbp]
    \centering
    \includegraphics[width=0.49\columnwidth]{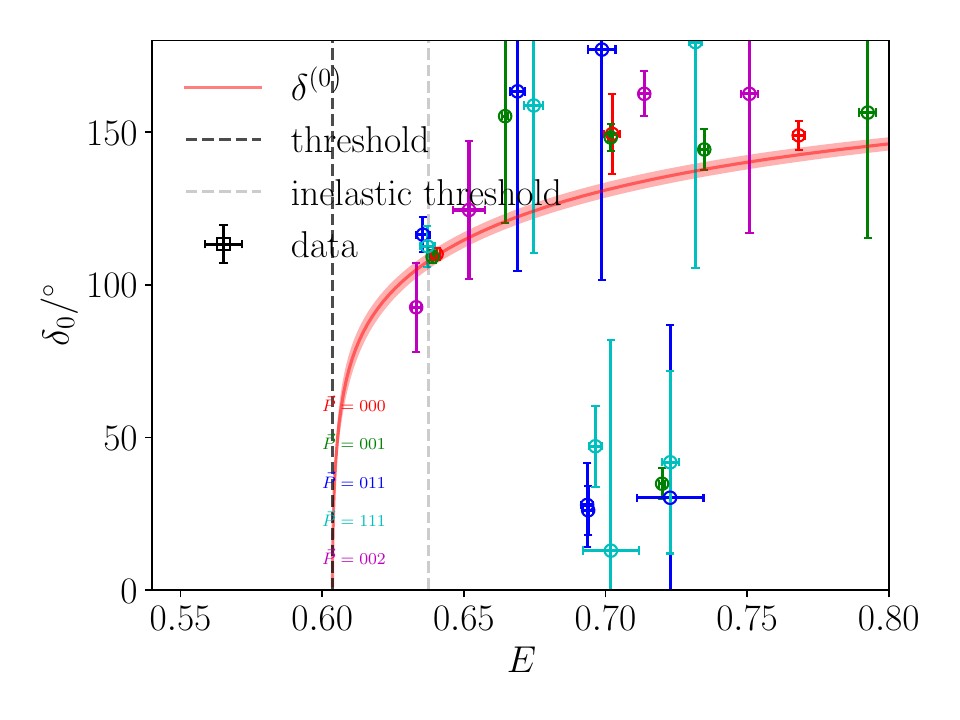}
    \hfill
    \includegraphics[width=0.49\columnwidth]{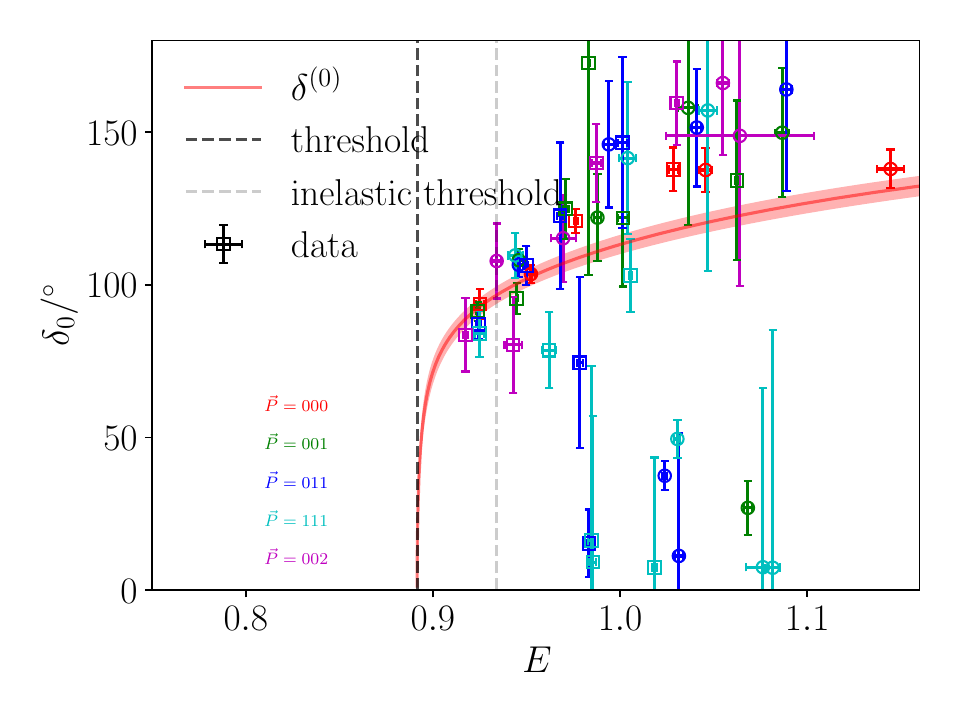}
    \hfill
    \includegraphics[width=0.49\columnwidth]{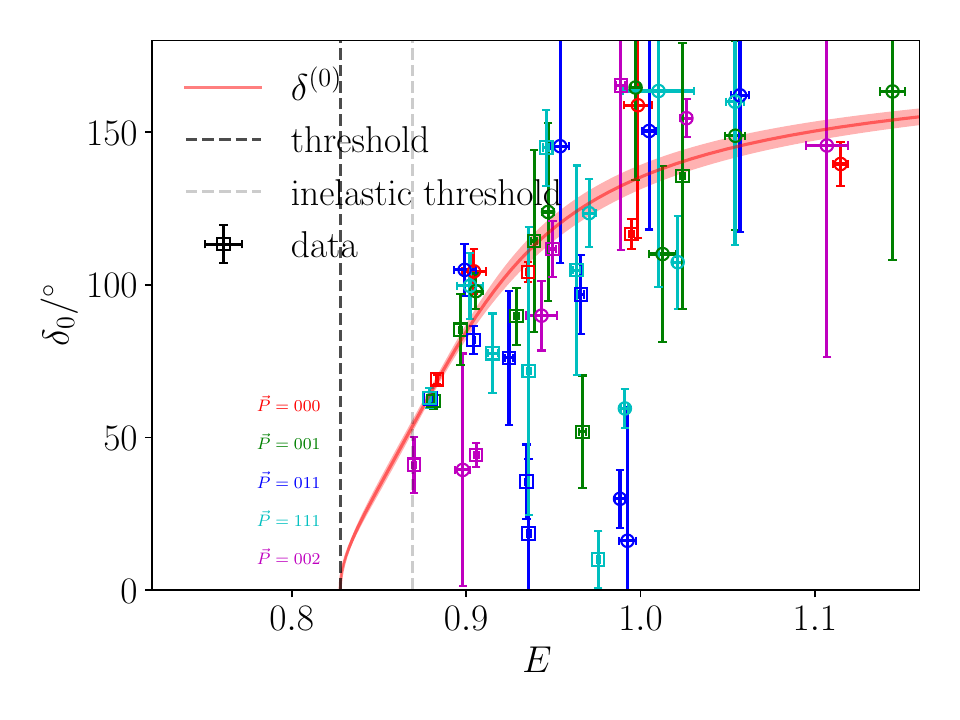}
    \hfill
    \includegraphics[width=0.49\columnwidth]{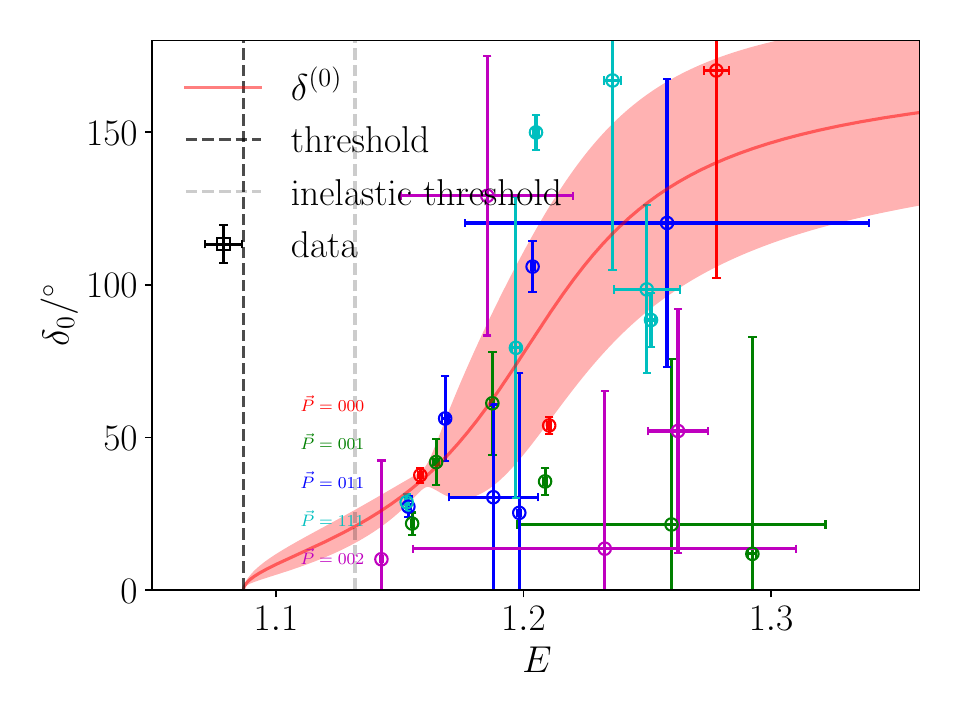}
\caption{The $S$-wave scattering phase shift $\delta_0$ obtained directly from the lattice energy levels. Other partial waves are neglected in this plot. The data points are obtained from the simplified Lüscher equation~\ref{eq:luscher}. The red band denotes the $S$-wave phase shift after removing the $P$-wave mixing in the analysis below. The red data points come from the center-of-mass $A_1^+$ irrep, while the other colors correspond to moving-frame $A_1$ irreps with different total momenta. The red data points are the most reliable, but the plot should still be viewed as qualitative. Because of the periodicity of the phase shift, the uncertainties shown for the data points are larger than the actual uncertainties.}
    \label{fig:phaseshift_plain}
\end{figure}

Before carrying out the fit, we can first assume that all partial waves above the $S$ wave are negligible and consider only single-channel $S$-wave scattering. In this case, Lüscher's equation reduces to a linear equation in one unknown. We can then use the rest-frame $A_1^+$ irrep and all moving-frame $A_1$ irreps to solve for the $S$-wave phase shift level by level. Figure~\ref{fig:phaseshift_plain} shows the $S$-wave phase shift for isospin-$I=\frac{1}{2}$ $D\pi$ scattering. The ERE fit including both $S$ and $P$ waves is also shown for comparison. The $S$-wave phase shift clearly exhibits resonant behavior. These phase shifts should be regarded only as qualitative guidance; moreover, the $2\pi$ ambiguity that arises when converting them into angles makes their displayed uncertainties larger than the actual uncertainties. It is therefore not appropriate to extract scattering parameters directly from this plot. The subsequent global fit gives more accurate results.

To extract the $S$- and $P$-wave phase shifts simultaneously, we perform global correlated fits to the energy levels in the irreps whose dominant partial waves extend up to the $P$ wave and whose energies lie below the lowest inelastic threshold. Each ensemble provides $17$ levels. For $M_{\pi} \approx 305$ and $208~\mathrm{MeV}$, there are two volumes, giving a total of $17 \times 2 = 34$ levels; for $133$ and $317~\mathrm{MeV}$, there are $17$ levels each. The parametrizations and fit results used for isospin-$I=\frac{1}{2}$ $D\pi$ scattering are summarized in Table~\ref{tab:para-Dpi-1d2}. We exclude parametrizations with excessively large $\chi^2/\text{d.o.f}$ or with poles near the left-hand cut.

\begin{table*}[htbp]
\centering
\caption{Parametrizations used in the analysis of $I=\frac{1}{2}$ $D\pi$ scattering.}
\addtolength{\tabcolsep}{2pt}
\begin{tabular}{c|ccc}
\toprule
Ensemble & $S$-wave parametrization & $P$-wave parametrization & $\chi^2/\text{d.o.f}$ \\
\midrule
\multirow{4}{*}{F32P30 / F48P30} & $k \cot \delta_0 = \frac{1}{a_0} + \frac{1}{2} r_0 k^2$ & $k^3 \cot \delta_1 = \frac{1}{a_1} + \frac{1}{2} r_1 k^2$ & $1.10$ \\
& $k \cot \delta_0 = \frac{1}{a_0} + \frac{1}{2} r_0 k^2 + P_{2,0} k^4$ & $k^3 \cot \delta_1 = \frac{1}{a_1} + \frac{1}{2} r_1 k^2$ & $1.14$ \\
& $k \cot \delta_0 = \frac{1}{a_0} + \frac{1}{2} r_0 k^2 + P_{2,0} k^4$ & $K_1 = \frac{g_1^2}{m_1^2 -s}$ & $1.55$ \\
& $K_0 = \frac{g_0^2}{m_0^2 -s}$ & $k^3 \cot \delta_1 = \frac{1}{a_1} + \frac{1}{2} r_1 k^2$ & $1.10$ \\
\midrule
\multirow{4}{*}{F32P21 / F48P21} & $k \cot \delta_0 = \frac{1}{a_0} + \frac{1}{2} r_0 k^2$ & $k^3 \cot \delta_1 = \frac{1}{a_1} + \frac{1}{2} r_1 k^2$ & $1.94$ \\
& $k \cot \delta_0 = \frac{1}{a_0} + \frac{1}{2} r_0 k^2 + P_{2,0} k^4$ & $k^3 \cot \delta_1 = \frac{1}{a_1} + \frac{1}{2} r_1 k^2$ & $1.76$ \\
& $k \cot \delta_0 = \frac{1}{a_0} + \frac{1}{2} r_0 k^2$ & $K_1 = \frac{g_1^2}{m_1^2 -s}$ & $2.01$ \\
& $k \cot \delta_0 = \frac{1}{a_0} + \frac{1}{2} r_0 k^2 + P_{2,0} k^4$ & $K_1 = \frac{g_1^2}{m_1^2 -s}$ & $1.82$ \\
\midrule
\multirow{2}{*}{H48P32} & $k \cot \delta_0 = \frac{1}{a_0} + \frac{1}{2} r_0 k^2$ & $k^3 \cot \delta_1 = \frac{1}{a_1} + \frac{1}{2} r_1 k^2$ & $1.17$ \\
& $k \cot \delta_0 = \frac{1}{a_0} + \frac{1}{2} r_0 k^2 + P_{2,0} k^4$ & $k^3 \cot \delta_1 = \frac{1}{a_1} + \frac{1}{2} r_1 k^2$ & $1.27$ \\
\midrule
\multirow{1}{*}{C48P14} & $k \cot \delta_0 = \frac{1}{a_0} + \frac{1}{2} r_0 k^2$ & $k^3 \cot \delta_1 = \frac{1}{a_1} + \frac{1}{2} r_1 k^2$ & $1.39$ \\
\bottomrule
\end{tabular}
\addtolength{\tabcolsep}{-2pt}
\label{tab:para-Dpi-1d2}
\end{table*}

\begin{figure}[htbp]
    \centering
    \includegraphics[width=0.49\columnwidth]{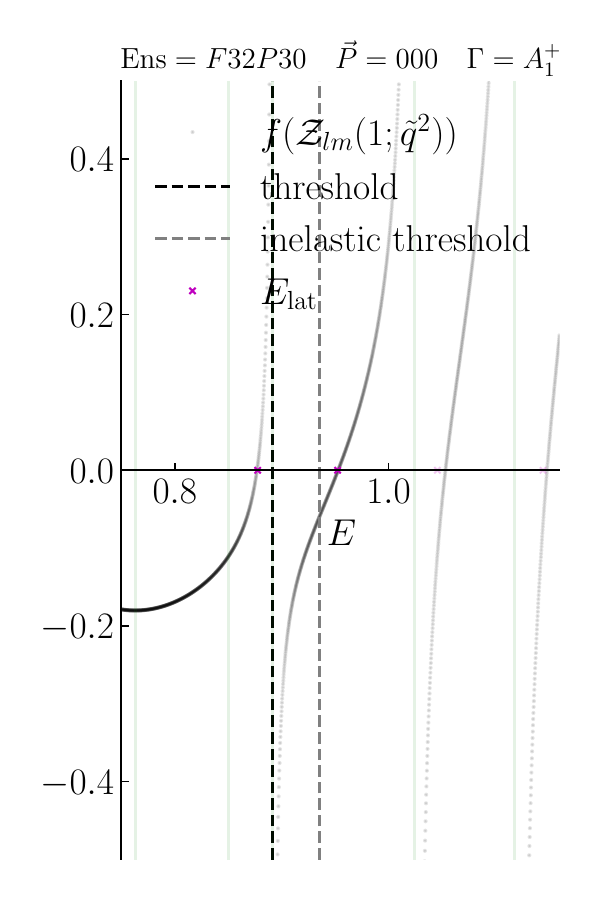}
    \hfill
    \includegraphics[width=0.49\columnwidth]{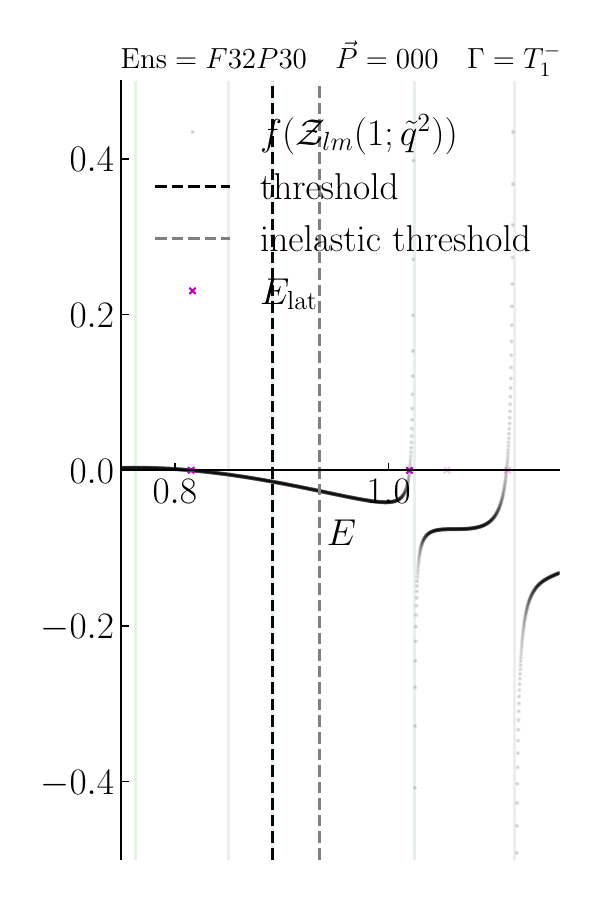}
    \caption{Lüscher equations for the $S$ and $P$ waves of $D\pi$ scattering on the F32P30 ensemble. The left and right panels correspond to the $A_1^+$ and $T_1^-$ irreps, respectively. The black points show the left-hand side of Eq.~\ref{eq:luscher}, and their intersections with the $x$ axis give the solutions. The vertical black and gray dashed lines denote the elastic and inelastic thresholds, respectively. The light-green vertical lines indicate noninteracting energy levels. The purple crosses denote the mean values of the lattice energy levels.}
    \label{fig:zeta_example1}
\end{figure}

\begin{figure}[htbp]
    \centering
    \includegraphics[width=0.49\columnwidth]{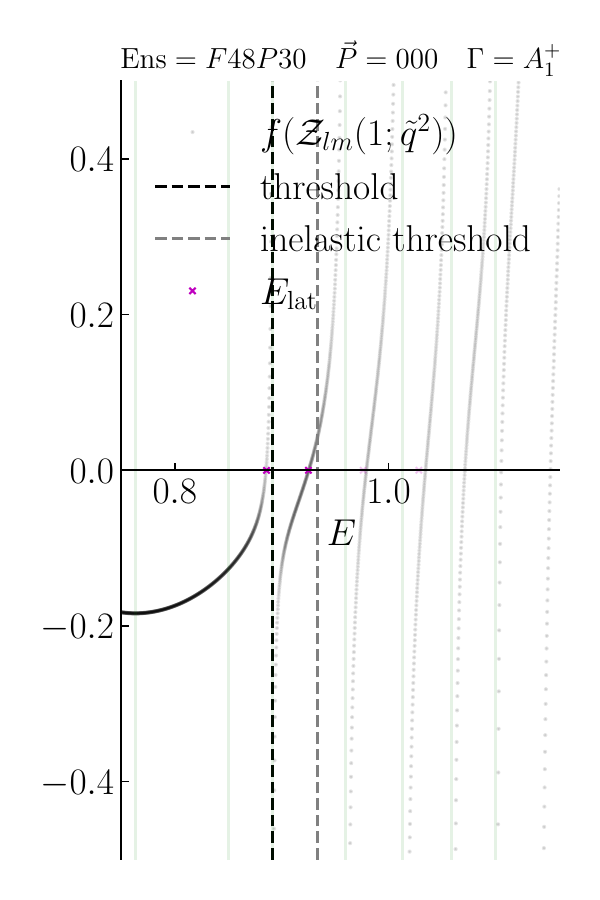}
    \hfill
    \includegraphics[width=0.49\columnwidth]{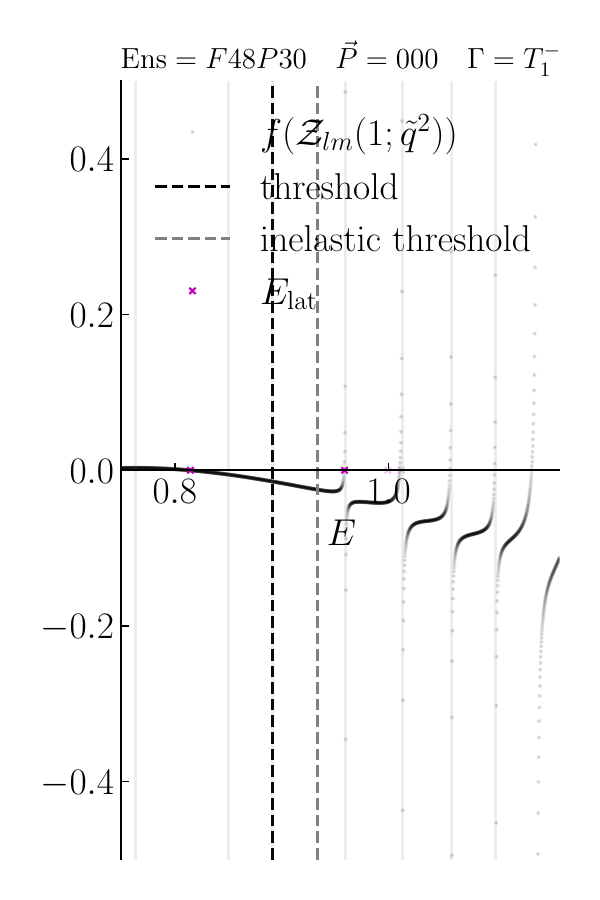}
    \caption{Example Lüscher equations for F48P30. The notation is the same as in Fig.~\ref{fig:zeta_example1}.}
    \label{fig:zeta_example2}
\end{figure}

Figure~\ref{fig:zeta_example1} and Fig.~\ref{fig:zeta_example2} show examples of the Lüscher equation. They display the equation after the fit has been obtained for the two volumes at $M_{\pi} = 305~\mathrm{MeV}$. The black points represent the value of the left-hand side of the equation, and their intersections with the $x$ axis give the solutions. The purple crosses denote the mean values of the lattice energy levels. The lighter crosses lie above the inelastic threshold and are therefore not used. The solutions of the equation lie very close to the lattice energy levels, showing that the fitted amplitude describes the lattice spectrum well.

Once the scattering matrix has been parametrized, the spectrum at any volume $L$ can be reconstructed using Lüscher's formula. The corresponding results are shown as orange bands in Figs.~\ref{fig:spectra-Dpi-HH-1d2}--\ref{fig:spectra-Dpi-CP-1d2}. We again see that when ensembles at different volumes are available, their energy levels constrain the scattering parameters more effectively. We also find that the $D^*$ corresponds to a volume-independent level, indicating that the $D^*$ is indeed a bound state.

\begin{figure}[htbp]
    \centering
    \includegraphics[width=0.8\columnwidth]{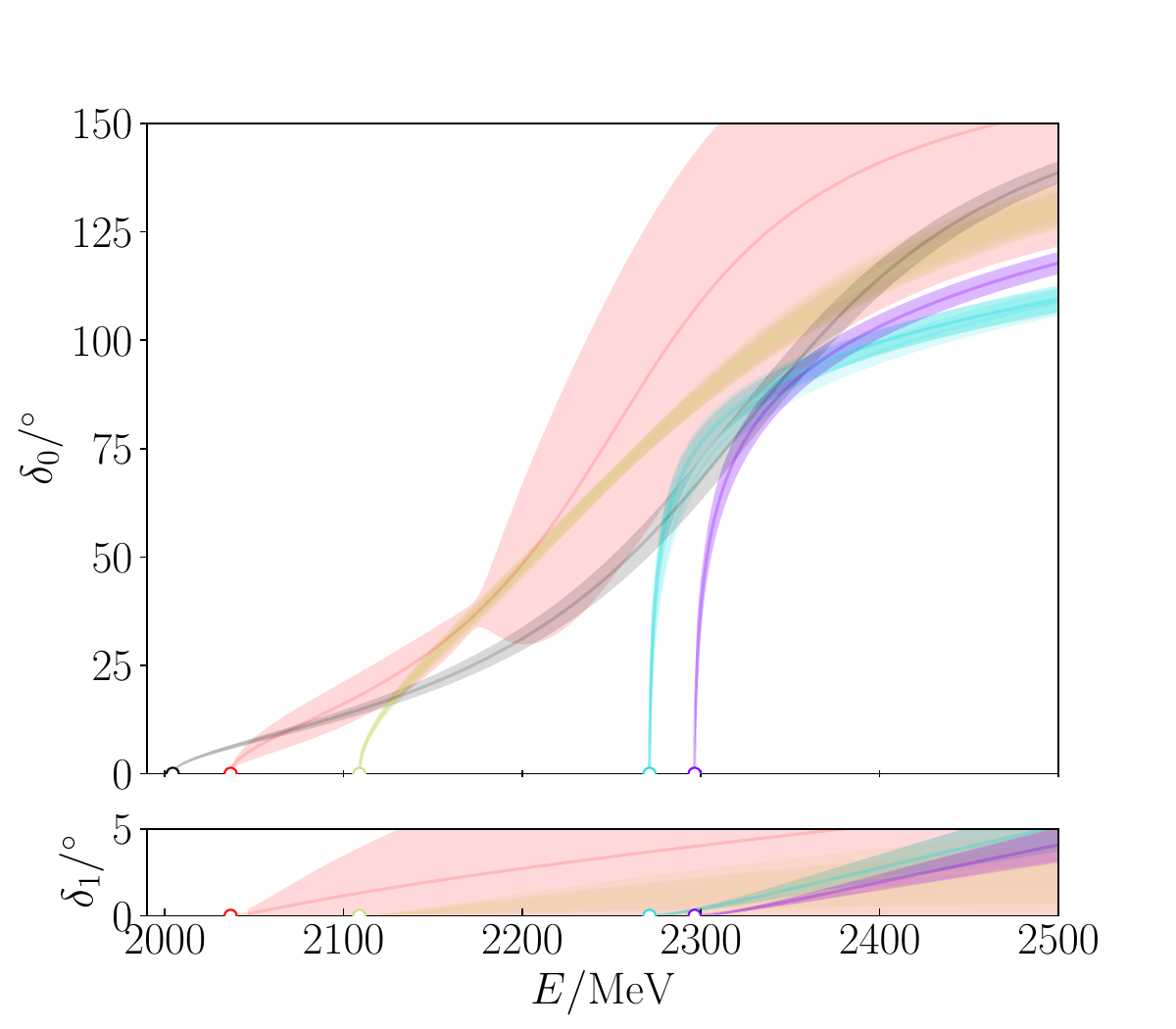}
    \caption{$S$- and $P$-wave scattering phase shifts $\delta_0$ (top) and $\delta_1$ (bottom). The red, yellow, cyan, and purple bands correspond to $M_{\pi} \approx 133, 208, 305$, and $317~\mathrm{MeV}$, respectively. The gray band is reconstructed from the PDG result. Bands of the same color show different parametrizations at the same value of $M_{\pi}$, and their widths represent statistical uncertainties.}
    \label{fig:phaseshift_SP_Dpi_1d2_normal}
\end{figure}

\begin{figure}[htbp]
    \centering
    \includegraphics[width=0.8\columnwidth]{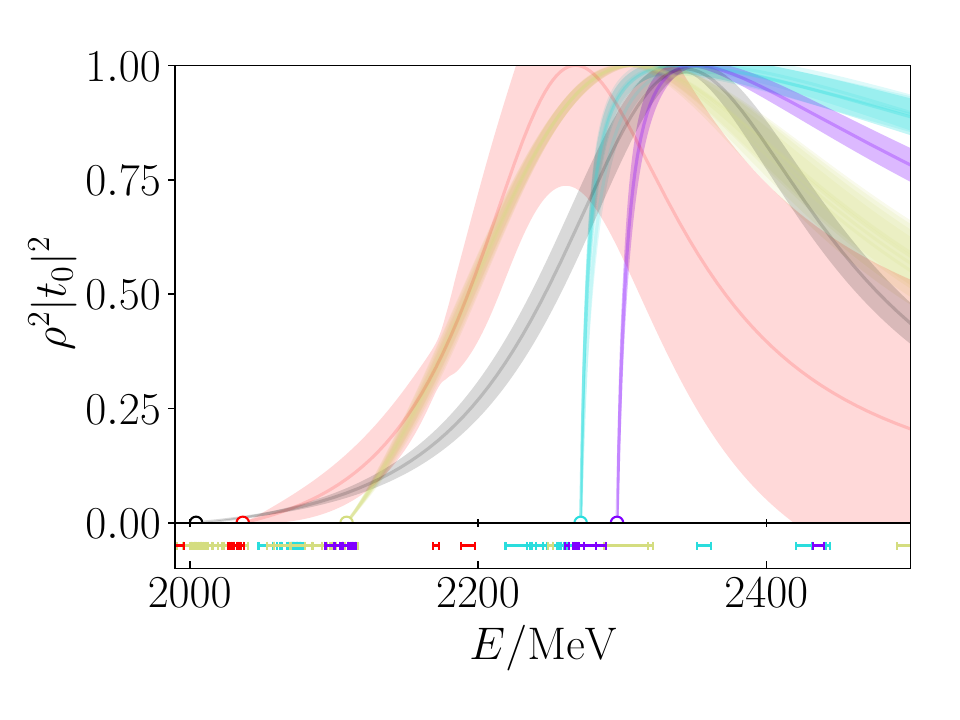}
    \caption{Normalized $S$-wave scattering cross section $\rho^2 |t_0|^2$. The error bars at the bottom denote the lattice energy levels used in this chapter. The color scheme is the same as in Fig.~\ref{fig:phaseshift_SP_Dpi_1d2_normal}.}
    \label{fig:rho2t2_S_Dpi_1d2_normal}
\end{figure}

The scattering phase shifts are shown in Fig.~\ref{fig:phaseshift_SP_Dpi_1d2_normal}. The red, yellow, cyan, and purple bands correspond to $M_{\pi} \approx 133, 208, 305$, and $317~\mathrm{MeV}$, respectively. Bands of the same color represent results from different parametrizations at a fixed value of $M_{\pi}$, and the width of each band denotes the statistical uncertainty. The differences between bands of the same color quantify the systematic uncertainty associated with the choice of parametrization; these differences are small. Overall, no significant model dependence is observed. The gray curve is reconstructed from the PDG average RBW parameters. For all values of $M_{\pi}$, the $S$-wave phase shift rises from $0$ to $180$ degrees, which is the characteristic behavior of a virtual state or a resonance. Near threshold, the $P$-wave phase shift $\delta_1$ is much smaller than the $S$-wave phase shift $\delta_0$. The normalized $S$-wave scattering cross section $\rho^2 |t_0|^2$ is shown in Fig.~\ref{fig:rho2t2_S_Dpi_1d2_normal}. The error bars at the bottom indicate the lattice energy levels used in this chapter. As $M_{\pi}$ decreases, the peak associated with the $S$-wave resonant behavior gradually moves to the left. This indicates that, when extrapolated toward the physical point, the mass of the $D_0^*(2300)$ resonance approaches the experimental $D\pi$ value.

The results obtained from the effective-range expansion are listed in Table~\ref{tab:ERE-Dpi-1d2}. On the $M_{\pi} \approx 133~\mathrm{MeV}$ ensemble, the result obtained with the Fermilab charm-quark action is denoted by $133^*$. The difference between the two charm-quark actions is significantly smaller than the statistical uncertainty. The total uncertainty is therefore still dominated by statistics.

\begin{table}[htbp]
\centering
\caption{Fit results for the $S$ and $P$ waves of $I=\frac{1}{2}$ $D\pi$ scattering obtained with the ERE parametrization. The entry $133^*$ denotes the result on the same ensemble as $133$, but using the Fermilab charm-quark action.}
\begin{tabular}{ccccc}
\toprule
$M_{\pi}/\text{MeV}$ & $a_0/\text{fm}$ & $r_0/\text{fm}$ & $a_1/\text{fm}^3$ & $r_1/\text{fm}^{-1}$ \\
\midrule
$317$ & $3.36(82)$ & $-0.734(60)$ & $0.0288(70)$ & $32(11)$ \\
$305$ & $3.9(1.2)$ & $-0.488(90)$ & $0.033(15)$ & $28(13)$ \\
$208$ & $0.663(60)$ & $-1.15(15)$ & $0.020(21)$ & $127(237)$ \\
$133$ & $0.33(16)$ & $-2.4(3.8)$ & $0.17(32)$ & $45(58)$ \\
$133^*$ & $0.26(14)$ & $-4.0(5.5)$ & $0.28(18)$ & $63(110)$ \\
\bottomrule
\end{tabular}
\label{tab:ERE-Dpi-1d2}
\end{table}

\section{Pole Positions}
\label{sec:pole}
To compare with previous lattice studies, we follow Ref.~\cite{Mohler:2012na} and consider the pole position after subtracting the spin-averaged mass of the $D$ and $D^*$ mesons:
\begin{equation}
    \sqrt{s_0^{\prime}} = \sqrt{s_0} - \frac{1}{4}(m_D + 3 m_{D^*}).
\end{equation}

\begin{figure}[htbp]
\centering
\includegraphics[width=\columnwidth]{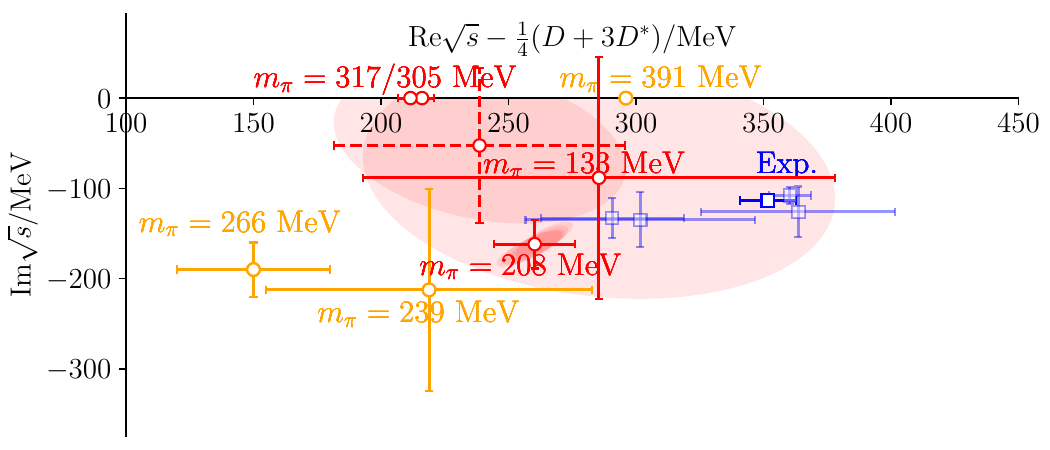}
\caption{Poles in the $S$ wave of $I=\frac{1}{2}$ $D\pi$ scattering. The results of this work are shown by covariance ellipses and red error bars. The dashed red error bar denotes the result obtained with the Fermilab action at $M_{\pi} \approx 133~\mathrm{MeV}$. The blue points are experimental results~\cite{Belle:2003nsh, BaBar:2009pnd, LHCb:2022lzp, LHCb:2015klp}, with the solid blue point denoting the PDG average~\cite{ParticleDataGroup:2024cfk}. The orange points are results from other lattice calculations~\cite{Mohler:2012na, Moir:2016srx, Gayer:2021xzv}.}
\label{fig:pole-Dpi}
\end{figure}

For the $S$ wave in $D\pi$ scattering, we find a virtual state in all parametrizations at $M_{\pi} \approx 305$ and $317~\mathrm{MeV}$, whereas at $M_{\pi} \approx 133$ and $208~\mathrm{MeV}$ we find a pair of resonance poles. These poles correspond to the experimentally observed scalar $D_0^*(2300)$ resonance. The pole positions on the Riemann sheets after subtraction of the spin-averaged mass are listed in Table~\ref{tab:pole} and shown in Fig.~\ref{fig:pole-Dpi}. The orange points show the subtracted pole positions from other lattice calculations at $M_{\pi} \approx 266$, $239$, and $391~\mathrm{MeV}$~\cite{Mohler:2012na, Moir:2016srx, Gayer:2021xzv}. The blue points show the pole positions corresponding to experimental descriptions of the $D_0^*$ using the RBW parametrization; the solid blue point is the PDG average~\cite{ParticleDataGroup:2024cfk}, while the translucent blue points are the individual experimental measurements entering that average~\cite{Belle:2003nsh, BaBar:2009pnd, LHCb:2022lzp, LHCb:2015klp}. Poles obtained from different parametrizations at the same $M_{\pi}$ are represented by overlapping covariance ellipses. The differences among these ellipses are very small, indicating that the model dependence of the pole position is small compared with the statistical uncertainty. The red error bars give a conservative estimate obtained by taking the envelope of the pole positions over all parametrizations and adding the discretization uncertainty in quadrature. The results at $M_{\pi} \approx 317~\mathrm{MeV}$ and $M_{\pi} \approx 305~\mathrm{MeV}$ are close to one another. Since these two calculations have similar $M_{\pi}$ values but different lattice spacings, this suggests that discretization effects are small at these lattice spacings. For the coarsest ensemble, C48P14 with $M_{\pi} \approx 133~\mathrm{MeV}$, we use both the conventional Wilson-clover action and the Fermilab prescription for the charm quark to estimate discretization effects. The dashed red error bar denotes the result obtained with the Fermilab action at $M_{\pi} \approx 133~\mathrm{MeV}$. Although we cannot completely exclude the possibility that $M_{\pi}$ effects and lattice-spacing effects partially cancel, within the current statistical precision we still expect discretization effects to be small.

\begin{table}[htbp]
\centering
\caption{Pole positions in the $S$ wave of $I=\frac{1}{2}$ $D\pi$ scattering. The uncertainties include statistical errors, parametrization systematics, and lattice-spacing uncertainties. The entry $133^*$ denotes the result on the same ensemble as $133$, but using the Fermilab charm-quark action.}
\addtolength{\tabcolsep}{6pt}
\begin{tabular}{ccc}
\toprule
$M_{\pi}/\text{MeV}$ & $\operatorname{Re} \sqrt{s} - \frac{1}{4} (D + 3 D^*) /\text{MeV}$ & $\operatorname{Im} \sqrt{s} /\text{MeV}$ \\
\midrule
$317$ & $211.6(4.9)$ & $0$ \\
$305$ & $216.1(4.8) $ & $0.00(11)$ \\
$208$ & $260(16)$ & $162(27)$ \\
$133$ & $285(93)$ & $88(134)$ \\
$133^*$ & $239(57)$ & $53(86)$ \\
\bottomrule
\end{tabular}
\addtolength{\tabcolsep}{-6pt}
\label{tab:pole}
\end{table}

Combining our results with previous lattice calculations~\cite{Mohler:2012na, Moir:2016srx, Gayer:2021xzv}, one observes an intriguing $M_{\pi}$ dependence of the $S$-wave pole associated with the $D_0^*$. At $M_{\pi} \approx 391~\mathrm{MeV}$, the pole lies on the real axis and corresponds to a bound state; as $M_{\pi}$ is lowered, it moves toward lower mass. In some region with $317~\mathrm{MeV}$ $\lesssim M_{\pi} \lesssim 391~\mathrm{MeV}$, the pole crosses the Riemann sheet and enters the unphysical sheet, becoming a virtual state. As $M_{\pi}$ is lowered further, for $266~\mathrm{MeV}$ $\lesssim M_{\pi} \lesssim 305~\mathrm{MeV}$ the pole moves into the complex plane and becomes a resonance, after which it turns around and moves to the right along a rather complicated trajectory. At the physical $M_{\pi}$, the pole position is compatible with experiment. Because the uncertainties remain large, however, no definitive conclusion can yet be drawn. It is worth noting that the trajectory of the $D_0^*$ pole is qualitatively similar to expectations from chiral perturbation theory~\cite{Guo:2009ct, Guo:2015dha}. For general discussions, see Refs.~\cite{Hanhart:2014ssa, Matuschek:2020gqe}.

For the $P$ wave, a bound state corresponding to the vector meson $D^*$ is found at all values of $M_{\pi}$. No clear pattern is observed in the motion of the pole as $M_{\pi}$ changes. It should be emphasized that, although the physical $D^*$ should be a resonance, on our physical ensemble C48P14 the hyperfine splitting $m_{D^*}-m_D$ becomes smaller than $M_{\pi}$ because of lattice-spacing effects, so that the $D^*$ still appears as a bound state. This indicates that the lattice spacing of this ensemble is not small enough for current charm-physics applications, and that the corresponding physical quantities are not well determined. Nevertheless, because this ensemble has large uncertainties, it provides only weak constraints in the scattering-length extrapolation described below and does not significantly affect the final result. Moreover, although the lattice artifact causes a visible qualitative deviation in the $P$ wave, it does not affect the reliability of the $S$-wave result. The $D^*$ lies close to threshold and is therefore more sensitive to discretization effects, whereas the mass of the $D_0^*$ is significantly higher and is comparatively insensitive to them.

We also fit the $S$- and $P$-wave spectra separately, using the $[000]\ A_1^+$ irrep for the $S$ wave and the $[000]\ T_1^-$ irrep for the $P$ wave. The results are consistent with those of the combined fit described above. This suggests that, at least for a single-channel scattering problem, the additional moving-frame levels do not make a major contribution to the scattering analysis, although this point has not often been emphasized in the literature.

\section{Chiral Extrapolation of $a_0^{-1}$}
\label{sec:interpolations}
In this section, we extrapolate the $S$-wave scattering length to the physical point, $M_{\pi}^{\text{phys}} = 135~\mathrm{MeV}$. Since the ensembles studied in this chapter include one ensemble slightly below the physical point, this procedure may also be viewed as an interpolation. We account for the model-dependent systematic uncertainty in the scattering length by taking the envelope of the distributions of $\lim_{k \to 0} k \cot \delta_0$ over all parametrizations. The discretization uncertainty is again estimated from the dispersion relation. We describe the chiral behavior with the polynomial form
\begin{equation}
a_0^{-1}(M_{\pi}) = c_0 + c_1 M_{\pi}^2 + c_2 M_{\pi}^4.
\label{eq:chiral}
\end{equation}
Similar extrapolations have also been used in Ref.~\cite{Lyu:2023xro}. To keep the action setup consistent, only the $a_0^{-1}$ results obtained with the Wilson-clover charm-quark action are included in the fit; the $133^*$ data are not used in this extrapolation.

\begin{figure}[htbp]
\centering
\includegraphics[width=\columnwidth]{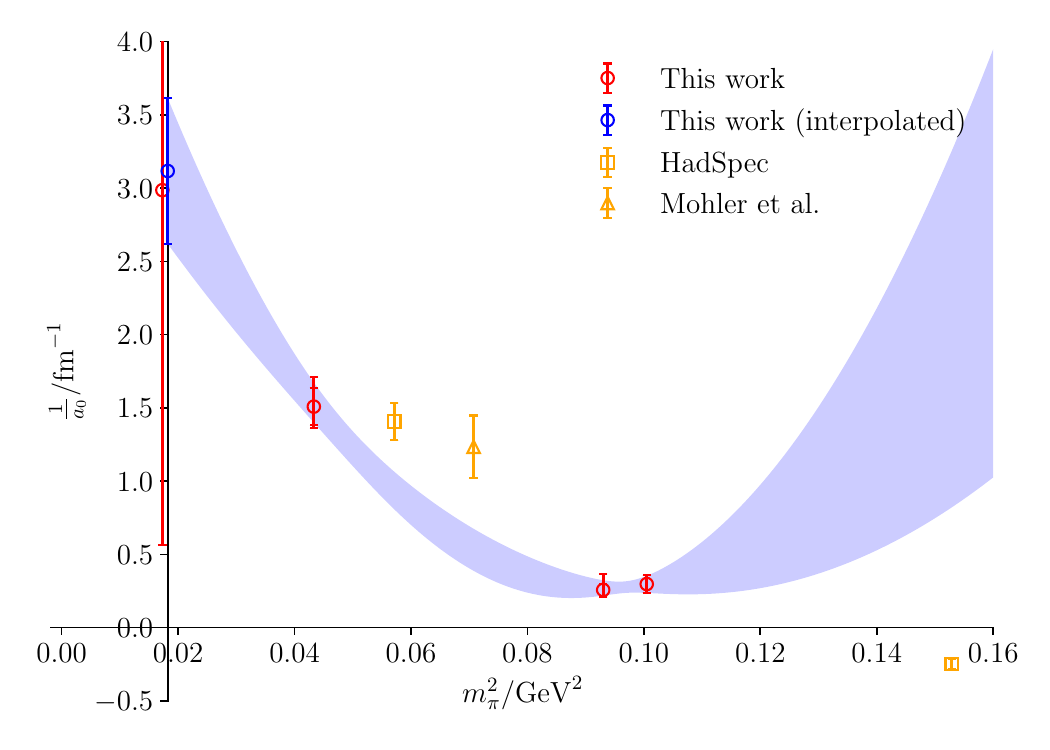}
\caption{Chiral interpolation of the inverse scattering length in $I=\frac{1}{2}$ $D\pi$ scattering. The fit uses the data from this work, shown as red points. The orange points are previous lattice results~\cite{Mohler:2012na, Moir:2016srx, Gayer:2021xzv}. The blue band is the reconstructed fit model, and the blue point is the prediction for $\frac{1}{a_0}$ at the physical $M_{\pi}$.}
\label{fig:interpolation-Dpi-1d2}
\end{figure}

The extrapolation is shown in Fig.~\ref{fig:interpolation-Dpi-1d2}, together with earlier lattice results from Refs.~\cite{Mohler:2012na, Moir:2016srx, Gayer:2021xzv} for comparison. The blue band denotes the extrapolation curve reconstructed from the fitted parameters, and the blue point gives the extrapolated value at the physical $M_{\pi}$.

The extrapolation is carried out as follows. First, for each value of $M_{\pi}$ and for each parametrization, we draw uniform samples of the inverse scattering length $a_0^{-1}$ from the corresponding distribution, without assuming that the distribution is Gaussian. We then combine the samples from all parametrizations to form the full distribution at that value of $M_{\pi}$. Next, each sampled set is extrapolated separately, yielding the distribution at the physical point. This procedure incorporates the uncertainties from all values of $M_{\pi}$ and from all parametrizations at the level of the fit. Since the different parametrizations have comparable $\chi^2/\text{d.o.f.}$ values, they are assigned equal weights in this work.

The fit gives
\begin{equation}
\begin{aligned}
    c_0 &= 4.66(95) \ \text{fm}^{-1}, \\
    c_1 &= -94(29) \ \text{MeV}^{-2} \cdot \text{fm}^{-1}, \\
    c_2 &= 502(203) \ \text{MeV}^{-4} \cdot \text{fm}^{-1}.
\end{aligned}
\end{equation}
The fit has $\chi^2/\text{d.o.f} = 1.02$. The resulting inverse scattering length at the physical point is
\begin{equation}
a_0^{-1}(M_{\pi}^{\text{phys}}) = 3.12(50) \ \text{fm}^{-1}.
\label{eq:a0}
\end{equation}

Adding an $a^2$ term to Eq.~(\ref{eq:chiral}) has little impact, indicating that the data points already contain a reasonable estimate of discretization effects. In addition, because the physical-point ensemble has a large statistical uncertainty, the chiral extrapolation/interpolation is in practice mainly constrained by the data at $208~\mathrm{MeV}$ and at $305/317~\mathrm{MeV}$.

\begin{figure}[htbp]
\centering
\begin{tikzpicture}
\pgftext{
\includegraphics[width=\columnwidth]{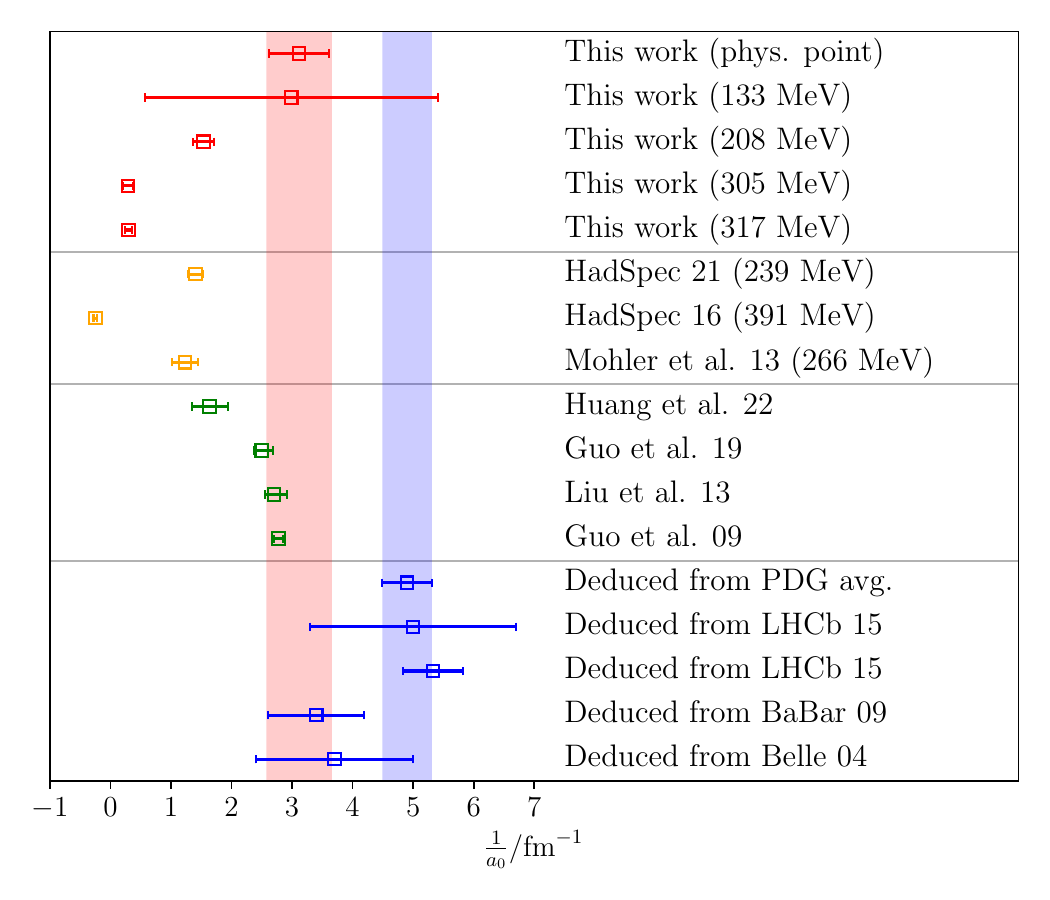}
}
\node[font = \small, anchor = west] at (6.5, 2.69) {\cite{Gayer:2021xzv}};
\node[font = \small, anchor = west] at (6.5, 2.69-1*0.67181818) {\cite{Moir:2016srx}};
\node[font = \small, anchor = west] at (6.5, 2.69-2*0.67181818) {\cite{Mohler:2012na}};
\node[font = \small, anchor = west] at (6.5, 2.69-3*0.67181818) {\cite{Huang:2021fdt}};
\node[font = \small, anchor = west] at (6.5, 2.69-4*0.67181818) {\cite{Guo:2018tjx}};
\node[font = \small, anchor = west] at (6.5, 2.69-5*0.67181818) {\cite{Liu:2012zya}};
\node[font = \small, anchor = west] at (6.5, 2.69-6*0.67181818) {\cite{Guo:2009ct}};
\node[font = \small, anchor = west] at (6.5, 2.69-7*0.67181818) {\cite{ParticleDataGroup:2024cfk}};
\node[font = \small, anchor = west] at (6.5, 2.69-8*0.67181818) {\cite{LHCb:2022lzp}};
\node[font = \small, anchor = west] at (6.5, 2.69-9*0.67181818) {\cite{LHCb:2015klp}};
\node[font = \small, anchor = west] at (6.5, 2.69-10*0.67181818) {\cite{BaBar:2009pnd}};
\node[font = \small, anchor = west] at (6.5, -4.7) {\cite{Belle:2003nsh}};
\end{tikzpicture}
\caption{Comparison of the $\frac{1}{a_0}$ result from this work (red), previous lattice results~\cite{Mohler:2012na,Moir:2016srx, Gayer:2021xzv} (orange), experimental results~\cite{ParticleDataGroup:2024cfk, Belle:2003nsh, BaBar:2009pnd, LHCb:2022lzp, LHCb:2015klp} (blue), and phenomenological models~\cite{Huang:2021fdt, Guo:2018tjx, Liu:2012zya, Guo:2009ct} (green). The experimental values are obtained by analytically continuing the RBW amplitudes constructed from the measured parameters.}
\label{fig:comparison-Dpi-1d2}
\end{figure}
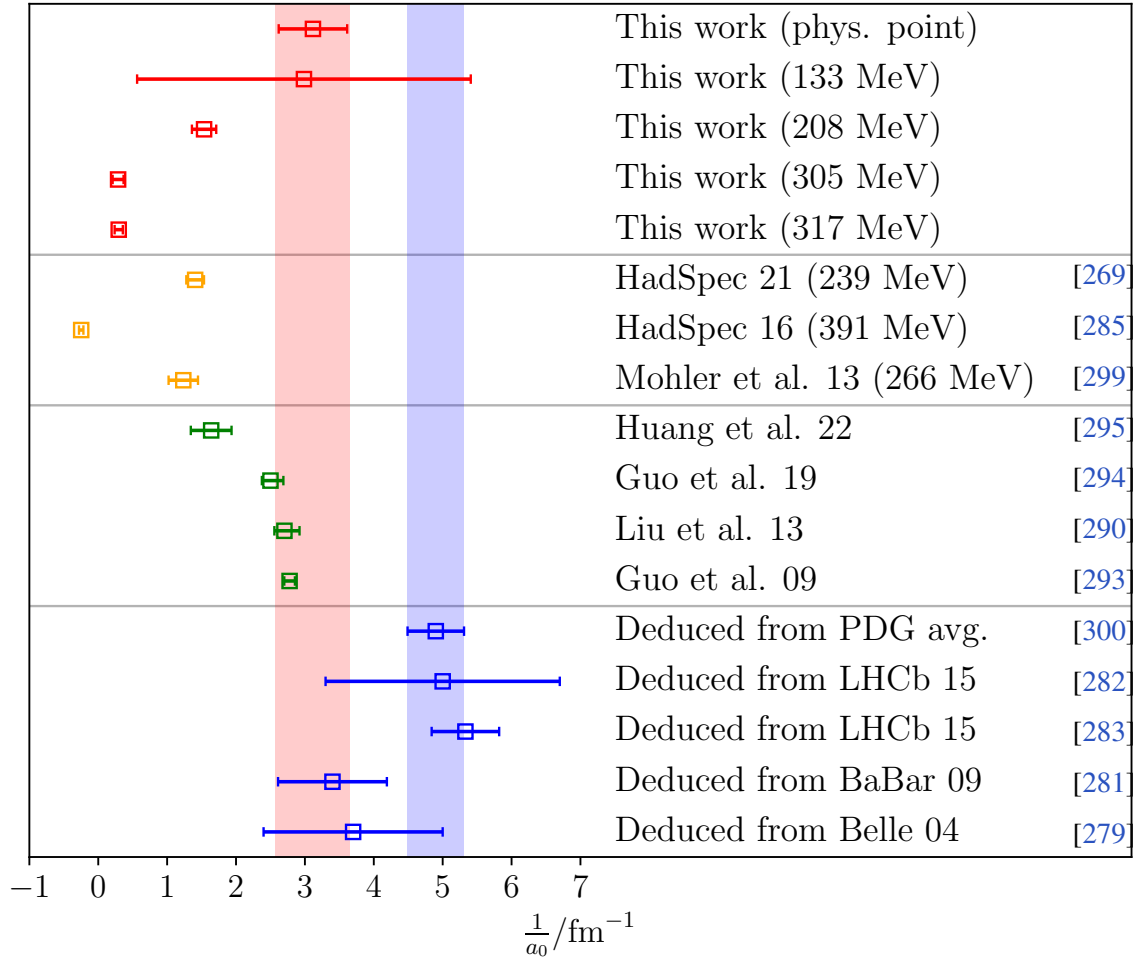

We note that the ALICE Collaboration recently reported an indirect measurement using femtoscopy and momentum correlation functions~\cite{ALICE:2024bhk}, and argued that the $D\pi$ interaction is dominated by electromagnetism. The quoted scattering length is $a_0 = 0.02(3)(1)$ fm, consistent with zero and roughly an order of magnitude smaller than our result. We do not yet understand the origin of this significant discrepancy. Apart from the ALICE result, no other experimental measurement directly determines the scattering length. In this chapter, scattering lengths inferred from experiment are obtained by analytically continuing the RBW amplitude and taking the threshold limit. The result in Ref.~\cite{Mohler:2012na} is obtained in a similar way. Figure~\ref{fig:comparison-Dpi-1d2} compares the scattering length from this work with previous lattice results~\cite{Mohler:2012na, Moir:2016srx, Gayer:2021xzv}, the value inferred from the PDG average~\cite{ParticleDataGroup:2024cfk}, experimental measurements~\cite{Belle:2003nsh, BaBar:2009pnd, LHCb:2022lzp, LHCb:2015klp}, and phenomenological models~\cite{Huang:2021fdt, Guo:2018tjx, Liu:2012zya, Guo:2009ct}. Since previous lattice calculations contain only one or two values of $M_{\pi}$, the pion-mass dependence is almost impossible to identify, limiting the understanding of their tension with experiment. With the red data points from this work included, the $M_{\pi}$ dependence becomes clear, and the discrepancy with the PDG value is also substantially reduced. In fact, within uncertainties, our result is consistent with the BABAR and Belle values, and also agrees with the phenomenological results of Refs.~\cite{Guo:2018tjx, Liu:2012zya, Guo:2009ct}. We emphasize again that the experimental values shown in the figure are not direct measurements in a strict sense. There are two reasons: first, these analyses should not use the RBW parametrization; second, the experiments do not directly measure the scattering length.

\section{Introduction to Three-Channel $D\pi-D\eta-D_s\bar{K}$ Scattering}
The lattice calculation of single-channel $D\pi$ scattering gives a picture of the near-threshold interaction and allows precise determinations of the scattering length and effective range. However, to resolve the two-pole puzzle of the $D_0^*(2300)$, namely whether this structure originates from a single pole or two poles on the complex Riemann sheets, one must study the higher-energy region away from the $D\pi$ threshold. As the energy increases, new inelastic channels open, primarily $D\eta$ and $D_s\bar{K}$. It is therefore necessary to extend the study to the three-coupled-channel scattering problem.

Although a lattice study of three-channel $D\pi$-$D\eta$-$D_s\bar{K}$ scattering already exists~\cite{Moir:2016srx}, only one stable pole structure was reported, due to the statistical precision and the limitations of the scattering-amplitude parametrization used in that work. Subsequent chiral-unitary studies based on the same lattice spectrum~\cite{Albaladejo:2016lbb, Asokan:2022usm} showed that, if approximate $\mathrm{SU}(3)$ flavor symmetry is imposed in the amplitude parametrization to reduce the number of free parameters, it may be possible to extract a second pole from the same lattice spectrum.

The goal of this section is to further clarify this two-pole puzzle. We carry out a systematic study on ensembles with several $\pi$ masses and consider amplitude parametrizations including both a general $K$ matrix and forms constrained by approximate symmetry. Our preliminary analysis supports the existence of a two-pole structure.

In studying three-channel $D\pi$--$D\eta$--$D_s\bar{K}$ scattering, we face several theoretical and computational challenges:
\begin{itemize}
    \item if moving-frame data are used, the $D^*\pi$ channel mixes with $D\pi$, $D\eta$, and $D_s\bar{K}$, turning the problem into a four-channel scattering system;
    \item the correlation-function matrix becomes substantially larger, leading to a higher computational cost;
    \item the signal quality of the $\eta$ meson is poor;
    \item the finite-volume quantization condition is more complicated for a multi-channel system;
    \item the scattering amplitude contains more free parameters;
    \item parametrization models based on $\mathrm{SU}(3)$ symmetry may no longer be adequate in the lighter-$M_{\pi}$ region.
\end{itemize}
We address these issues one by one.

\subsection{Discussion of Symmetry Reduction}
For the three-channel scattering problem, one of the main difficulties in lattice calculations is that, with only two volumes, the number of center-of-mass energy levels is often insufficient to constrain the scattering amplitude. If moving-frame data are introduced, however, the $D^*\pi$ channel mixes with $D\pi$, $D\eta$, and $D_s\bar{K}$, turning the system into a four-channel scattering problem. Since the $D^*\pi$ channel carries spin, this introduces more complicated angular-momentum and partial-wave mixing and substantially increases the complexity of the analysis.

Other coupled channels may also appear in the relevant energy region, such as $D^*\eta$, $D_s^*\bar{K}$, $D\pi\pi$, $D\omega$, and $D^*\pi\pi$. These channels are not included in the present analysis. To simplify the notation, we define
\begin{equation}
DX = \{D\pi,\, D\eta,\, D_s\bar{K}\}.
\end{equation}
The reduction procedure summarized in\chapref{chap:operators} determines the cubic-group irreps that need to be studied. We focus on irreps that contain only the $S$-wave contribution of the $DX$ system, since they provide the cleanest signals for constraining the $D_0^*(2300)$. In the present analysis strategy, whether in the center-of-mass frame or in moving frames, the energy levels extracted in each irrep either directly constrain the $DX$ scattering amplitude of interest or constrain it indirectly through the partial waves with which it mixes. We now discuss in detail the partial-wave content of the relevant irreps for center-of-mass momenta $|\vec{P}|^2 < 3$, in units of $\frac{2\pi}{L}$.

\begin{itemize}
    \item $\vec{P}=[000]$
    \begin{itemize}
        \item $A_1^+$: $DX$ $S$ wave
        \item $T_1^-$: $DX$ $P$ wave and $D^*\pi$ ${}^3P_1$ wave
        \item $E^+$: $DX$ $D$ wave and $D^*\pi$ ${}^3D_2$ wave
        \item $T_2^+$: $DX$ $D$ wave and $D^*\pi$ ${}^3D_2$, ${}^3D_3$ waves
        \item $T_1^+$: $D^*\pi$ ${}^3S_1$, ${}^3D_1$, ${}^3D_3$ waves
        \item $A_1^-$: $D^*\pi$ ${}^3P_0$ wave
        \item $E^-$: $D^*\pi$ ${}^3P_2$ wave
        \item $T_2^-$: $D^*\pi$ ${}^3P_2$ wave
        \item $A_2^+$: $D^*\pi$ ${}^3D_3$ wave
    \end{itemize}
    \item $\vec{P}=[001]$
    \begin{itemize}
        \item $A_1$: $DX$ $S$, $P$, and $D$ waves, and $D^*\pi$ ${}^3P_1$ and ${}^3D_2$ waves
        \item $E_2$: $DX$ $P$ and $D$ waves, and $D^*\pi$ ${}^3S_1$, ${}^3P_1$, ${}^3P_2$, and ${}^3D_1$ waves
        \item $B_1$: $DX$ $D$ wave, and $D^*\pi$ ${}^3P_2$ and ${}^3D_2$ waves
        \item $B_2$: $DX$ $D$ wave, and $D^*\pi$ ${}^3P_2$ and ${}^3D_2$ waves
        \item $E_2$: $D^*\pi$ ${}^3S_1$, ${}^3P_1$, ${}^3P_2$, ${}^3D_1$, ${}^3D_2$, and ${}^3D_3$ waves
        \item $A_2$: $D^*\pi$ ${}^3S_1$, ${}^3P_0$, and ${}^3P_2$ waves
    \end{itemize}
    \item $\vec{P}=[011]$
    \begin{itemize}
        \item $A_1$: $DX$ $S$, $P$, and $D$ waves, and $D^*\pi$ ${}^3P_1$, ${}^3P_2$, ${}^3D_2$, and ${}^3D_3$ waves
        \item $B_1$: $DX$ $P$ and $D$ waves, and $D^*\pi$ ${}^3S_1$, ${}^3P_1$, ${}^3P_2$, ${}^3D_1$, ${}^3D_2$, and ${}^3D_3$ waves
        \item $B_2$: $DX$ $P$ and $D$ waves, and $D^*\pi$ ${}^3S_1$, ${}^3P_1$, ${}^3P_2$, ${}^3D_1$, ${}^3D_2$, and ${}^3D_3$ waves
        \item $A_2$: $DX$ $D$ wave, and $D^*\pi$ ${}^3S_1$, ${}^3P_0$, ${}^3P_2$, ${}^3D_1$, ${}^3D_2$, and ${}^3D_3$ waves
    \end{itemize}
    \item $\vec{P}=[111]$
    \begin{itemize}
        \item $A_1$: $DX$ $S$, $P$, and $D$ waves, and $D^*\pi$ ${}^3P_1$, ${}^3D_2$, and ${}^3D_3$ waves
        \item $E_2$: $DX$ $P$ and $D$ waves, and $D^*\pi$ ${}^3S_1$, ${}^3P_1$, ${}^3P_2$, ${}^3D_1$, ${}^3D_2$, and ${}^3D_3$ waves
        \item $A_2$: $D^*\pi$ ${}^3S_1$, ${}^3P_0$, ${}^3P_2$, ${}^3D_1$, ${}^3D_2$, and ${}^3D_3$ waves
    \end{itemize}
\end{itemize}
Although the computational cost is substantially higher than in the single-channel problem, repeated calculations can be avoided by caching common subdiagrams, thereby reducing the overall cost considerably and accelerating the numerical computation.

\subsection{$K$-Matrix Structure}
For the systems of interest here, the $S$ wave corresponds to total quantum numbers $J^P = 0^+$, while the $P$ wave corresponds to $J^P = 1^-$.

For the $D^*\pi$ channel, the situation is different:
\begin{itemize}
\item the $S$ wave corresponds to $J^P = 1^+$;
\item the $P$ wave corresponds to $J^P = 0^-, 1^-, 2^-$.
\end{itemize}

For two pseudoscalar mesons in the $DX$ system, the relation between the partial wave and the total angular momentum is simple: the $S$ wave has total quantum numbers $J^P = 0^+$, the $P$ wave has $J^P = 1^-$, and so on. For $D^*\pi$, the partial waves are
\begin{equation}
\begin{cases}
    l = 0: 1^+ ({}^3 S_1) \\
    l = 1: 0^- ({}^3 P_0), 1^- ({}^3 P_1), 2^- ({}^3 P_2) \\
    l = 2: 1^+ ({}^3 D_1), 2^+ ({}^3 D_2), 3^+ ({}^3 D_3). \\
\end{cases}
\end{equation}
The multi-channel $K$ matrix $K_{L^{\prime} S^{\prime} a^{\prime} ; L S a} (\sqrt{s})$ is a function of the orbital angular momentum $L$, spin $S$, scattering channel $a$, and the total energy $\sqrt{s}$ of the incoming and outgoing two-body systems. Writing the quantum numbers as $[L, S, a]$ and considering total angular momentum $J<3$, the nonzero blocks of the $K$ matrix are
\begin{equation}
\begin{aligned}
&J^P=0^+\begin{cases}
[0, 0, 0]: D\pi \, (S) \\
[0, 0, 1]: D\eta \, (S) \\
[0, 0, 2]: D_s\bar{K} \, (S) \\
\end{cases}, \\
&J^P=0^-\begin{cases}
[1, 1, 3]: D^*\pi \, ({}^3 P_0) \\
\end{cases}, \\
&J^P=1^+\begin{cases}
[0, 1, 3]: D^*\pi \, ({}^3 S_1) \\
[2, 1, 3]: D^*\pi \, ({}^3 D_1) \\
\end{cases}, \\
&J^P=1^-\begin{cases}
[1, 0, 0]: D\pi \, (P) \\
{\color{gray}[1, 0, 1]: D\eta \, (P)} \\
{\color{gray}[1, 0, 2]: D_s\bar{K} \, (P)} \\
[1, 1, 3]: D^*\pi \, ({}^3 P_1) \\
\end{cases}, \\
&J^P=2^+\begin{cases}
[2, 0, 0]: D\pi \, (D) \\
{\color{gray}[2, 0, 1]: D\eta \, (D)} \\
{\color{gray}[2, 0, 2]: D_s\bar{K} \, (D)} \\
[2, 1, 3]: D^*\pi \, ({}^3 D_2) \\
\end{cases}, \\
&J^P=2^-\begin{cases}
[1, 1, 3]: D^*\pi \, ({}^3 P_2) \\
\end{cases}. \\
\end{aligned}
\end{equation}
All other submatrices vanish. The previous multi-channel scattering study~\cite{Moir:2016srx} used precisely the cleanest $A_1^+$ irrep corresponding to $0^+$. As will be seen in the next subsection, the other irreps contain many more energy levels. We expect the additional constraints from these levels to compensate for, and possibly outweigh, the added complexity from partial-wave mixing, thereby imposing stronger constraints on the scattering amplitude.

Although using moving-frame data introduces mixing with the $D^*\pi$ channel, it also provides a more complete picture of the charmed-meson spectrum. Experimentally, four low-lying positive-parity $D$-meson states have been observed~\cite{ParticleDataGroup:2024cfk}: the scalar state $D_0(2300)$, the two axial-vector states $D_1(2430)$ and $D_1(2420)$, and the tensor state $D_2(2460)$. The $D_0(2300)$ and $D_1(2430)$ are broad, whereas the $D_1(2420)$ and $D_2(2460)$ are relatively narrow. This pattern can be understood to some extent in heavy-quark effective theory (HQET). According to HQET, these states can be classified by the total angular momentum $j_\ell$ of the light degrees of freedom. In the heavy-quark limit $m_c\to\infty$, the heavy-quark spin decouples from the light degrees of freedom, and meson states are labeled by $j_\ell$ and parity $P$. For the $L=1$ excited states, the light degrees of freedom form two doublets, $j_\ell^{P}=1/2^{+}$ and $j_\ell^{P}=3/2^{+}$. The $j_\ell^{P}=1/2^{+}$ doublet corresponds to physical states with $J^{P}=0^{+}$ and $1^{+}$, namely $D_0(2300)$ and $D_1(2430)$; the $j_\ell^{P}=3/2^{+}$ doublet corresponds to $J^{P}=1^{+}$ and $2^{+}$, namely $D_1(2420)$ and $D_2(2460)$. Because of angular-momentum conservation, the $j_\ell^{P}=1/2^{+}$ states decay predominantly to $D^{(*)}\pi$ in an $S$ wave and are therefore broad. In contrast, the $j_\ell^{P}=3/2^{+}$ states can decay only through a $D$ wave and appear as narrow resonances because of the centrifugal barrier.

If only the $D^{(*)}\pi$ channels are considered, the $S$ wave of $D\pi$ generates the $D_0^*(2300)$. The ${}^3S_1$ and ${}^3D_1$ partial waves of $D^*\pi$ mix, but they couple predominantly to the $D_1(2430)$ and $D_1(2420)$, respectively, corresponding to one broad and one narrow resonance. At the same time, the $D$ wave of $D\pi$ mixes with the ${}^3D_2$ partial wave of $D^*\pi$ and forms the narrow tensor resonance $D_2^*(2460)$. Ref.~\cite{Lang:2022elg} also points out that the ${}^3S_1$ amplitude of $D^*\pi$ may have a zero near the $D^*\eta$ and $D_s^*\bar{K}$ thresholds and may contain a second pole, very much like the situation in the $D\pi$ system.

It is worth noting that existing studies of $D^*\pi$ scattering indicate~\cite{Lang:2022elg} that, in the energy region considered here, the $P$- and $D$-wave contributions from the $D\eta$ and $D_s\bar{K}$ channels can be neglected. These partial waves are marked in gray in the list above. In future scattering analyses, we will use this observation to reduce the number of free parameters in the scattering matrix. Unlike in the single-channel case, however, the $D$-wave contribution from $D^{(*)}\pi$ cannot be neglected in the multi-channel analysis. It is therefore necessary to construct nonlocal single-hadron operators that couple more efficiently to these narrow tensor resonances. Such operators are discussed in detail in\chapref{chap:lattice_QCD}.

After neglecting $F$ waves and higher partial waves, the $D^*\pi$ system still contains the ${}^3D_3$ partial wave. The experimentally observed $D_3^*(2750)$ lies above the energy region considered here, and its effect on the other partial waves is therefore expected to be small. More strictly, in the rest frame the $A_2^+$ irrep contains only this partial wave, so the energy levels in that irrep can be used to check whether it can indeed be neglected.

In summary, although the system contains rather complicated partial-wave mixing, a clear analysis strategy can still be formulated. Specifically, one can first use center-of-mass irreps that contain only a single $D^*\pi$ partial wave to determine the corresponding contribution and thereby fix part of the scattering amplitude. Once a stable description has been obtained, a global fit of the four-channel scattering amplitude can then be performed. The concrete steps are as follows:

\begin{itemize}
    \item use the center-of-mass $A_1^+$ irrep to determine the $S$-wave scattering of the $DX$ system;
    \item use the $A_1^-$ irrep to determine the ${}^3P_0$ partial wave of $D^*\pi$;
    \item use the $E^-$ and $T_2^-$ irreps to determine the ${}^3P_2$ partial wave of $D^*\pi$;
    \item use the $A_2^+$ irrep to determine the ${}^3D_3$ partial wave of $D^*\pi$;
    \item after fixing these partial-wave contributions, use the other center-of-mass irreps to determine the remaining partial waves;
    \item using this as the initial input, include the moving-frame levels and perform a global correlated fit to all ensembles;
    \item repeat this procedure for ensembles at different $\pi$ masses;
    \item finally perform a unified global fit to all ensembles, sharing the low-energy effective parameters.
\end{itemize}

\subsection{The $\eta$ and $\eta'$ Problem}
In the approximate $\mathrm{SU}(3)$ flavor-symmetric limit, the $\eta$ is one of the pseudoscalar octet states, with quantum numbers $0^+(0^{-+})$. When this symmetry is broken by the different light- and strange-quark masses, the $\eta$ mixes with the $\eta'$. The latter is the flavor singlet of $\mathrm{SU}(3)$, and the axial anomaly implies that there is no flavor-singlet Goldstone boson, explaining why the $\eta'$ is much heavier than the pseudoscalar octet. The Witten--Veneziano relation~\cite{Witten:1979vv, Veneziano:1979ec, Veneziano:1980xs} connects the $\eta'$ mass to the topological susceptibility $\chi_{\mathrm{top}}$, revealing a deep connection between the topological structure of the QCD vacuum and the hadron spectrum. To study $D\eta$ scattering, one must first determine the $\eta$-meson mass accurately. We therefore consider the following four operators:
\begin{equation}
\begin{cases}
    O_1 &= \frac{1}{2} (\bar{u} \gamma_5 u + \bar{d} \gamma_5 d) \\
    O_2 &= \bar{s} \gamma_5 s \\
    O_3 &= \frac{1}{2} (\bar{u} \gamma_4 \gamma_5 u + \bar{d} \gamma_4 \gamma_5 d) \\
    O_4 &= \bar{s} \gamma_4 \gamma_5 s.
\end{cases}
\end{equation}
These operators define a $4\times4$ correlation-function matrix.

Because the disconnected part of the $\eta$ correlation function has an extremely poor signal-to-noise ratio, extracting its mass is even more difficult than extracting typical two-body energy levels. To improve the signal, we adopt the strategy proposed in Ref.~\cite{Neff:2001zr}, which was later used in several ETMC studies~\cite{Jansen:2008wv,Michael:2013gka,Ottnad:2017bjt,Ottnad:2025zxq} and substantially improves the stability of the $\eta$ mass extraction. The basic idea is that the effective mass of the $\eta$ correlation function often reaches a plateau only shortly before the signal is lost; without increasing statistics, one can make the plateau appear earlier by approximately subtracting excited-state contributions.

Concretely, the connected parts of the diagonal correlators correspond to the $\pi$ meson or to its heavier pseudoscalar analogues, and therefore do not suffer from a signal-to-noise problem. We fit these connected parts; their excited-state decay rates are approximately the same as those of the excited states in the $\eta$ correlator. Keeping only the ground-state contribution and then recombining it with the disconnected part yields a reconstructed correlation-function matrix. Since the excited states are effectively suppressed, the reconstructed correlators typically form plateaus at earlier times, improving the precision of the mass extraction.

This procedure is equivalent to adding an $I=1$ $\pi(1300)$ contribution to the isospin-$I=0$ $\eta$ state\footnotecircle{The decay of the $\pi(1300)$ into the three-body final state $\pi\pi\pi$ will be discussed in detail in\chapref{chap:three_body_problems2}.}, thereby nearly canceling the coefficient of the excited-state term without significantly affecting the ground state. The author's interpretation is the following: although in the approximate $\mathrm{SU}(2)$ flavor-symmetric limit the $\eta$, as a flavor singlet, is much heavier than the triplet $\pi$, its excitation $\eta(1295)$ is almost degenerate with the $\pi(1300)$. The subtraction procedure therefore suppresses excited-state contamination very efficiently. Concrete numerical illustrations are given below.

In addition, on ensembles with insufficient sampling of topological sectors, the quark-disconnected contribution to the correlator mixes with the topological charge. At fixed topological charge, its large-time behavior in the $1/V$ expansion is
\begin{equation}
    C^{\mathrm{disc}}(t) \sim \frac{1}{V} \left(\chi_t - \frac{Q^2}{V} + \frac{c_4}{2 V \chi_t}\right) + \mathcal{O}(V^{-2}),
\end{equation}
where $Q$ is the topological charge, $\chi_t$ denotes the topological susceptibility, and $c_4$ is the kurtosis of the topological-charge distribution.

There are several ways to handle this issue, for example by including a constant term directly in the fit. After testing, we use a procedure analogous to finite-temperature analyses and take a temporal difference of the correlator to remove the constant term:
\begin{equation}
    C(t) \to \tilde{C}(t) = C(t) - C(t+1).
\end{equation}
Recent ETMC work also considers a second difference~\cite{Ottnad:2025zxq}:
\begin{equation}
    C(t) \to \tilde{\tilde{C}}(t)
    = \tilde{C}(t) - \tilde{C}(t+1)
    = C(t) - 2C(t+1) + C(t+2).
\end{equation}
The purpose is to further reduce temporal correlations. Our tests show that the second difference can indeed improve the final mass extraction on some ensembles.

After subtracting excited states and removing the topological constant term, the ground-state $\eta$ and the first excited state $\eta'$ can be extracted by solving the GEVP. Numerically, however, diagonalizing a matrix that simultaneously contains $\gamma_5$ and $\gamma_4\gamma_5$ operators often gives poor signal quality. Without excited-state subtraction, the better-performing $2\times2$ submatrix varies from ensemble to ensemble. After excited-state subtraction, we uniformly choose the $\gamma_5$ operators and use either the first or the second difference depending on the signal quality.

\begin{figure}[htbp]
\centering
\includegraphics[width=0.32\columnwidth]{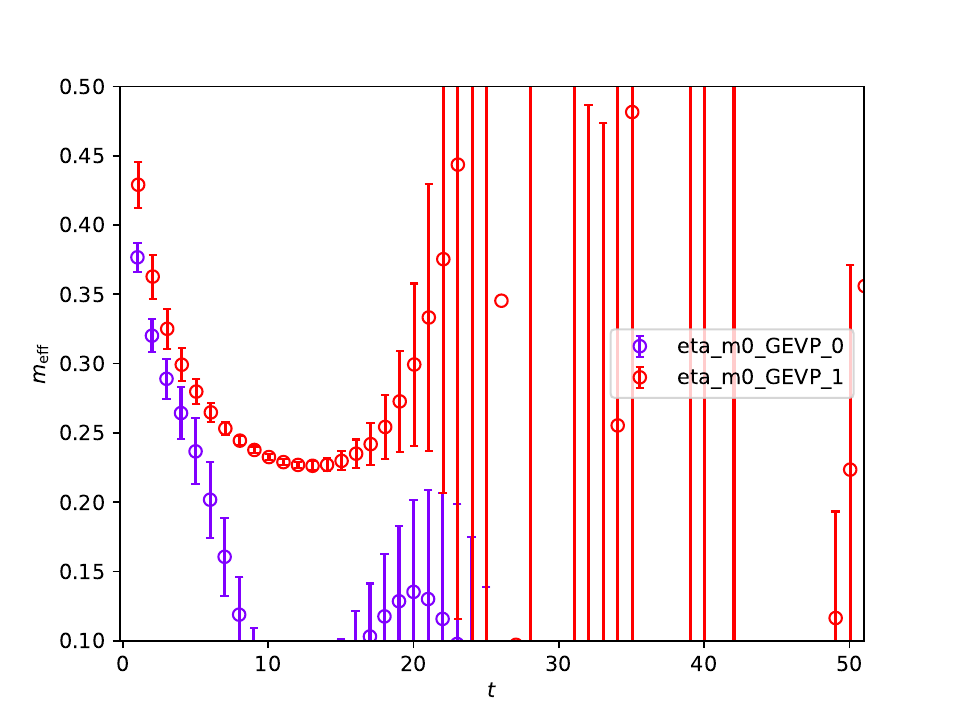}
\caption{Eigenvalue effective mass obtained from a GEVP constructed with two $\eta$ operators of the $\gamma_5$ type.}
\label{fig:eta}
\end{figure}

\begin{figure}[htbp]
\centering
\includegraphics[width=0.32\columnwidth]{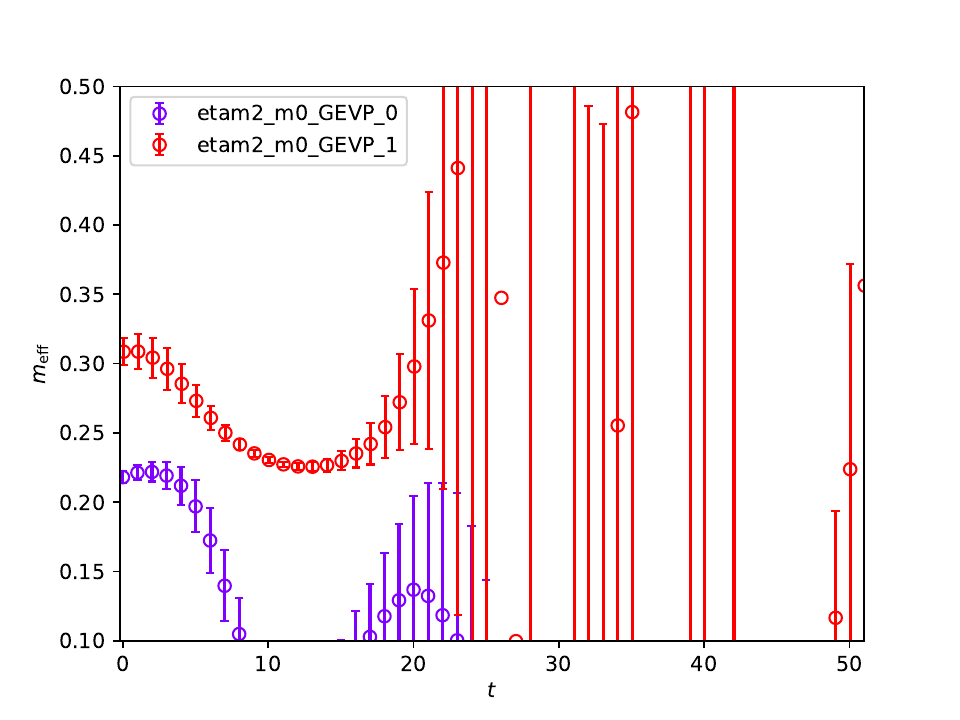}
\includegraphics[width=0.32\columnwidth]{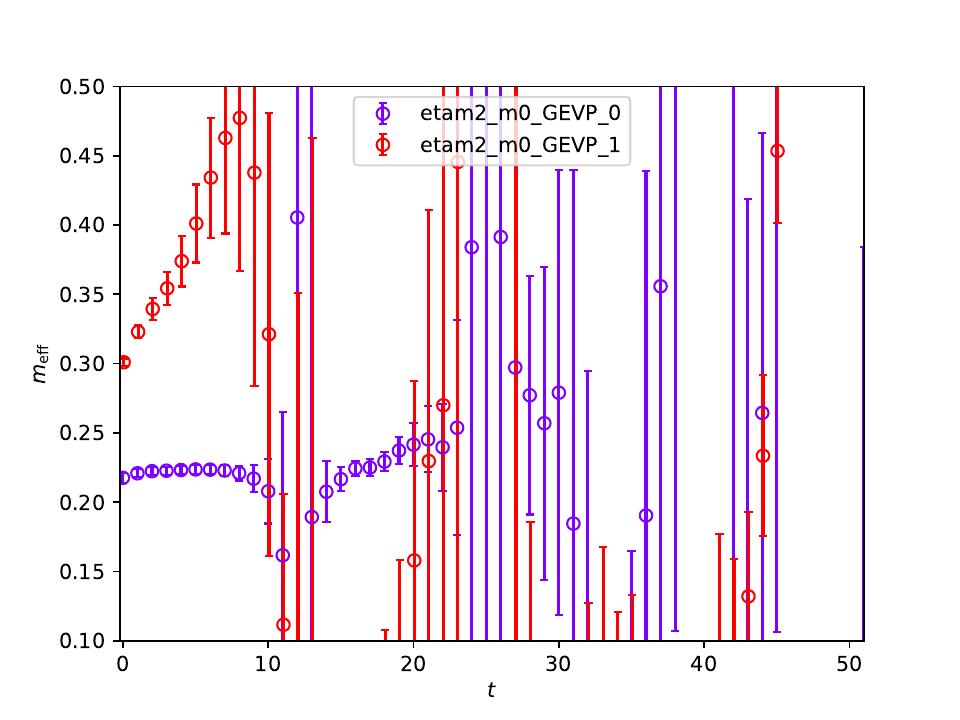}
\includegraphics[width=0.32\columnwidth]{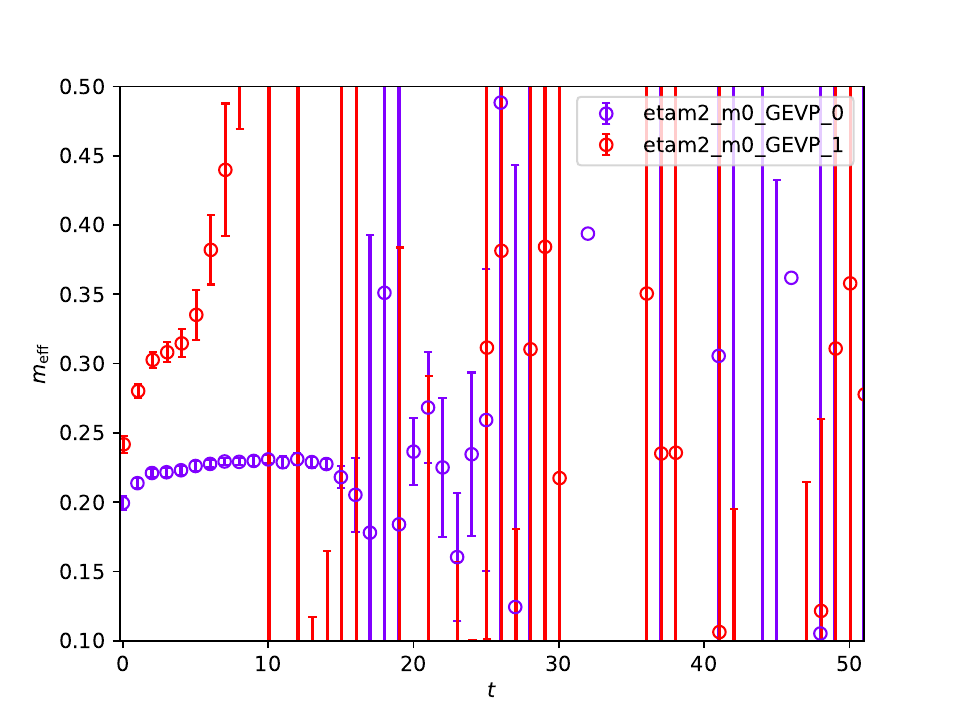}
\caption{Eigenvalue effective masses of the $\eta$ correlator after subtracting higher excited states and removing the constant term. From left to right, the panels correspond to no difference, the first difference, and the second difference.}
\label{fig:eta_good}
\end{figure}

Figure~\ref{fig:eta} shows the eigenvalue effective masses obtained from the GEVP using two $\gamma_5$ operators without any treatment. The ground-state effective mass keeps decreasing with time, indicating that the CLQCD ensembles still do not sample topological sectors sufficiently. Figure~\ref{fig:eta_good} shows the results after excited-state subtraction and removal of the constant term by temporal differences. From left to right, the panels correspond to no difference, the first difference, and the second difference. The plateau clearly appears earlier, allowing the $\eta$-meson mass to be extracted with higher precision.

In addition, the mixing between the operator and the topological charge density is weaker in moving frames. One can therefore measure the $\eta$ energy at larger momenta and use the dispersion relation to extrapolate to the $\eta$ rest mass, potentially achieving higher precision.

When constructing two-body $D\eta$ operators, we choose the $\eta$ single-hadron operator of either the $\gamma_5$ or the $\gamma_4\gamma_5$ type according to the signal quality. It should be noted that these $\eta$ operators may still contain components that couple to purely gluonic states, but this effect is relatively small. At the present level of precision, we do not consider this mixing effect in the two-body operators.

\subsection{Preliminary Spectral Results}
The single-channel scattering study shows that it is advantageous to use the Fermilab-type charm-quark action. Therefore, in the multi-channel scattering analysis presented here, this improved action is used on all ensembles. In this subsection, we briefly show the effective masses of the energy levels and the extracted spectra for several representative irreps.

For the $D\pi$-type operators, we include all allowed combinations with squared relative momentum less than $4$; for the various $D\eta$ and $D_s\bar{K}$ operators, we take the squared relative momentum to be less than $3$, while for the $D^*\pi$ system we take it to be less than $1$. This momentum cutoff is already sufficient to cover the energy region containing the $D_0^*(2300)$ and the other three positive-parity light-quark $L=1$ excitations.

\begin{figure}[htbp]
    \centering
    \includegraphics[width=0.49\columnwidth]{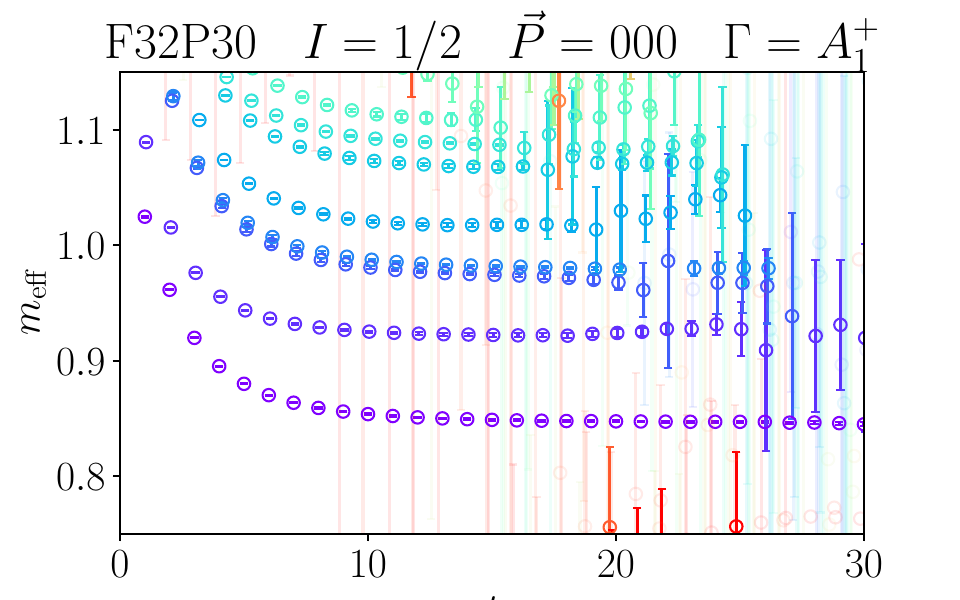}
    \includegraphics[width=0.49\columnwidth]{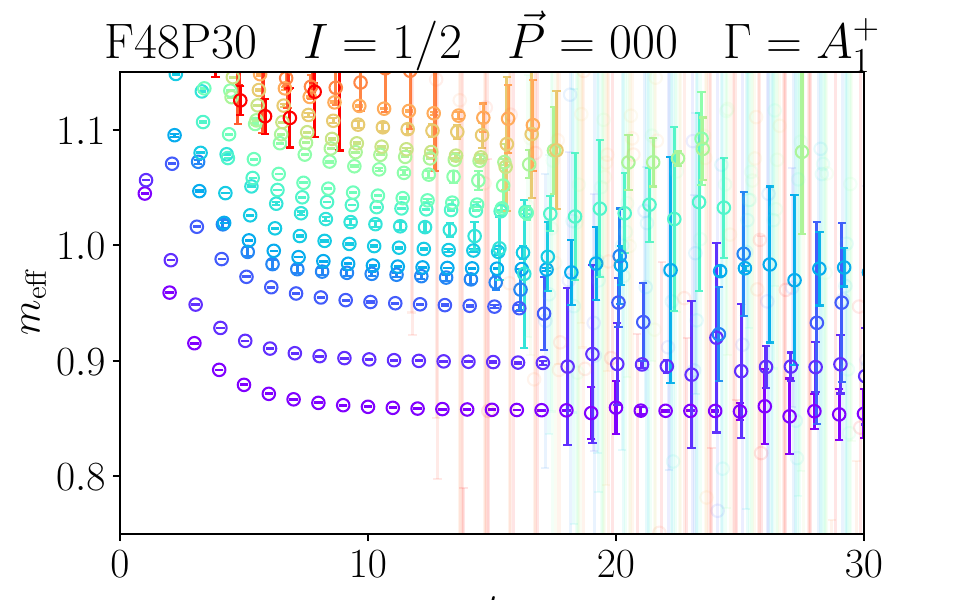}
    \includegraphics[width=0.49\columnwidth]{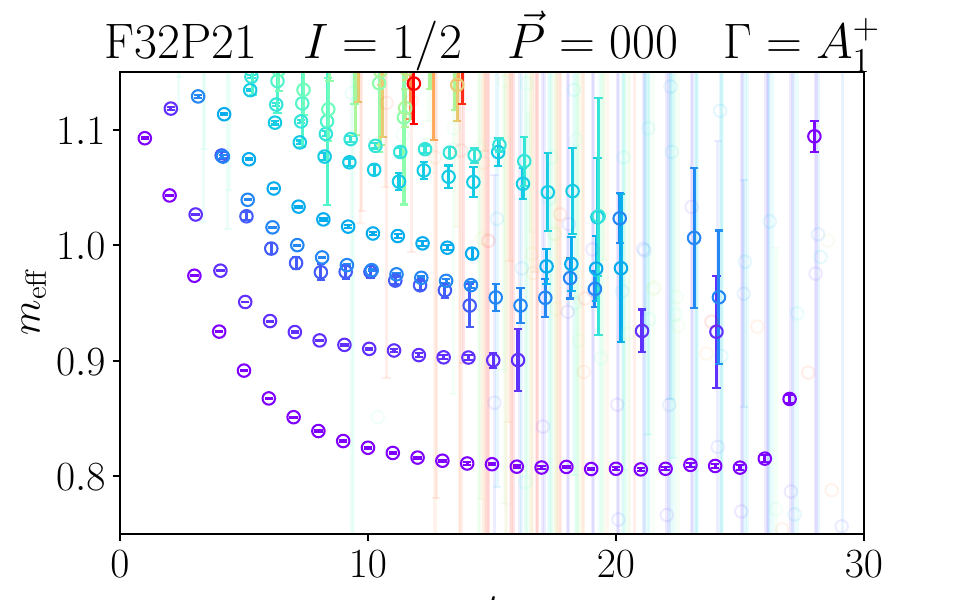}
    \includegraphics[width=0.49\columnwidth]{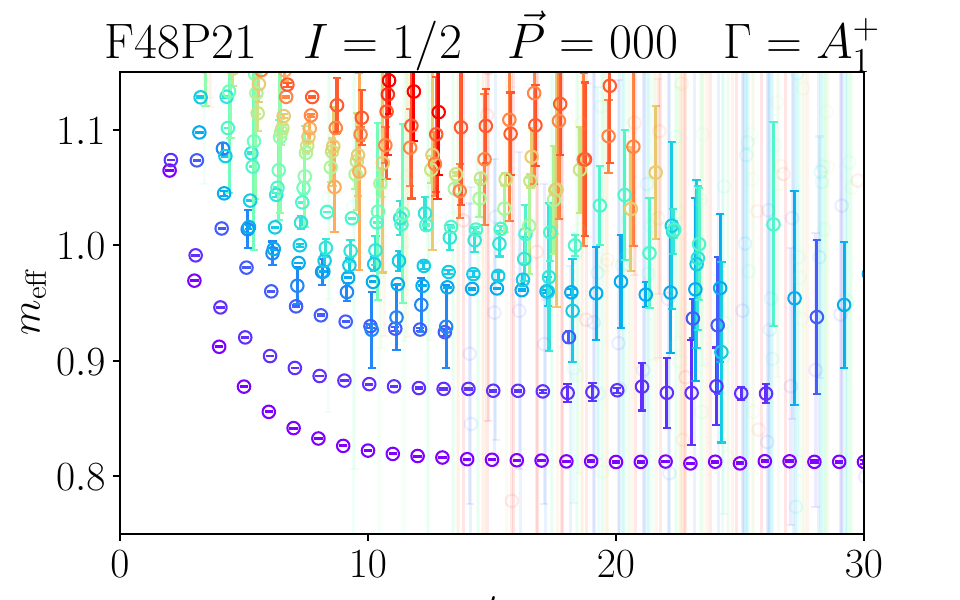}
    \caption{Effective-mass plots of the generalized eigenvalues $\lambda_n(t)$ for the two-point correlation functions of the four-channel scattering system in the $A_1^+$ irrep.}
    \label{fig:Dpi-meff-A1+_coupled}
\end{figure}

\begin{figure}[htbp]
    \centering
    \includegraphics[width=0.49\columnwidth]{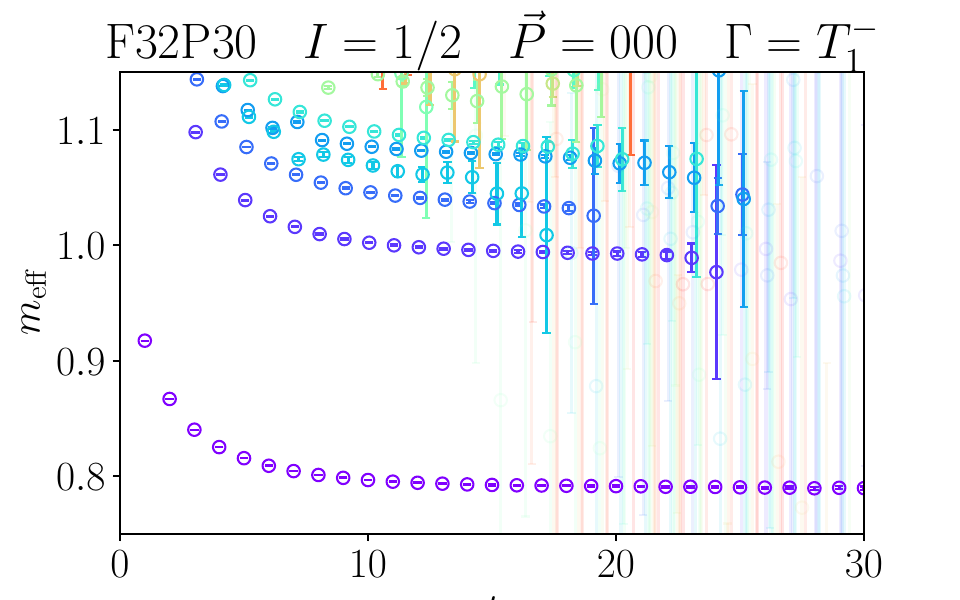}
    \includegraphics[width=0.49\columnwidth]{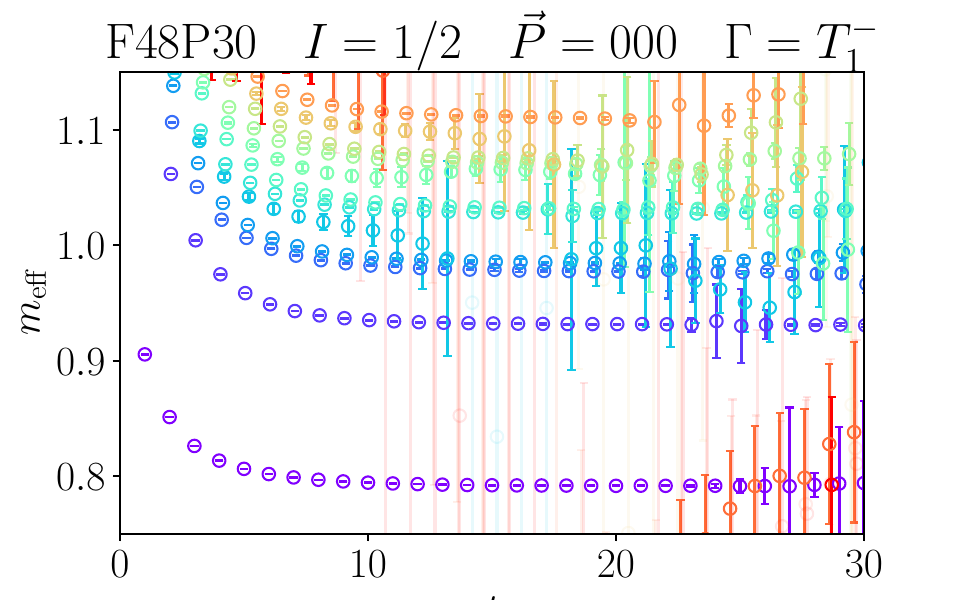}
    \includegraphics[width=0.49\columnwidth]{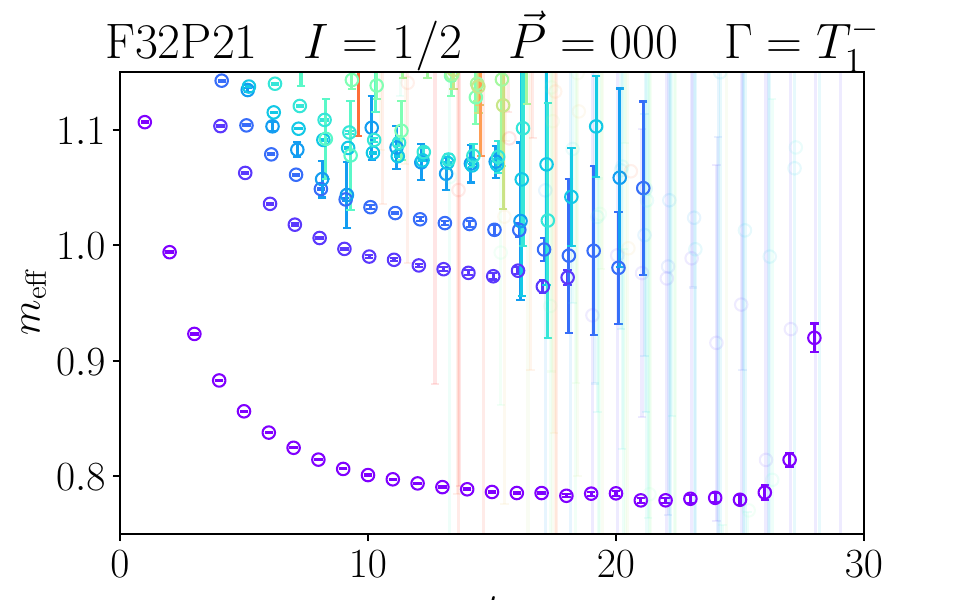}
    \includegraphics[width=0.49\columnwidth]{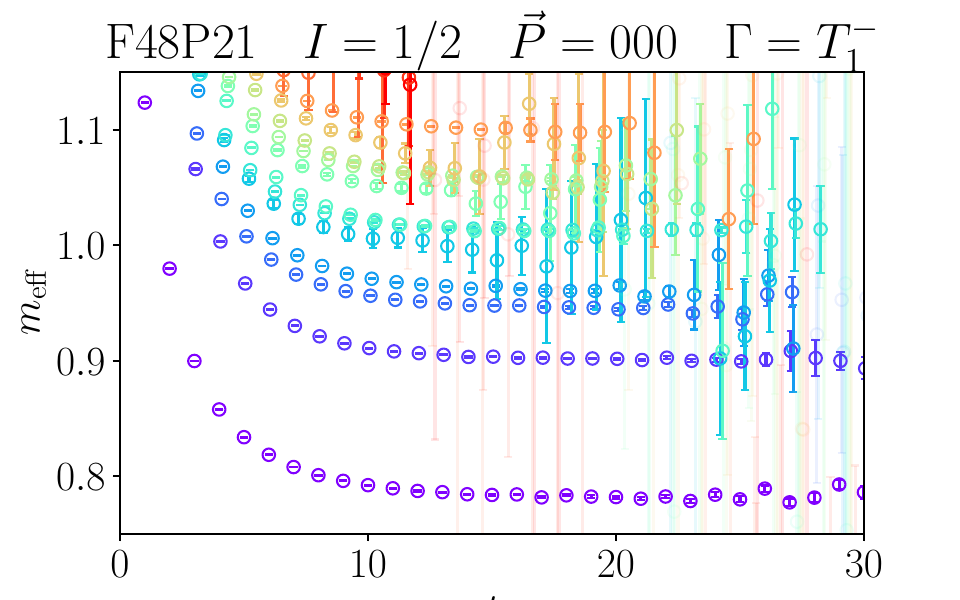}
    \caption{Effective-mass plots of the generalized eigenvalues $\lambda_n(t)$ for the two-point correlation functions of the four-channel scattering system in the $T_1^-$ irrep.}
    \label{fig:Dpi-meff-T1-_coupled}
\end{figure}

\begin{figure}[htbp]
    \centering
    \includegraphics[width=0.49\columnwidth]{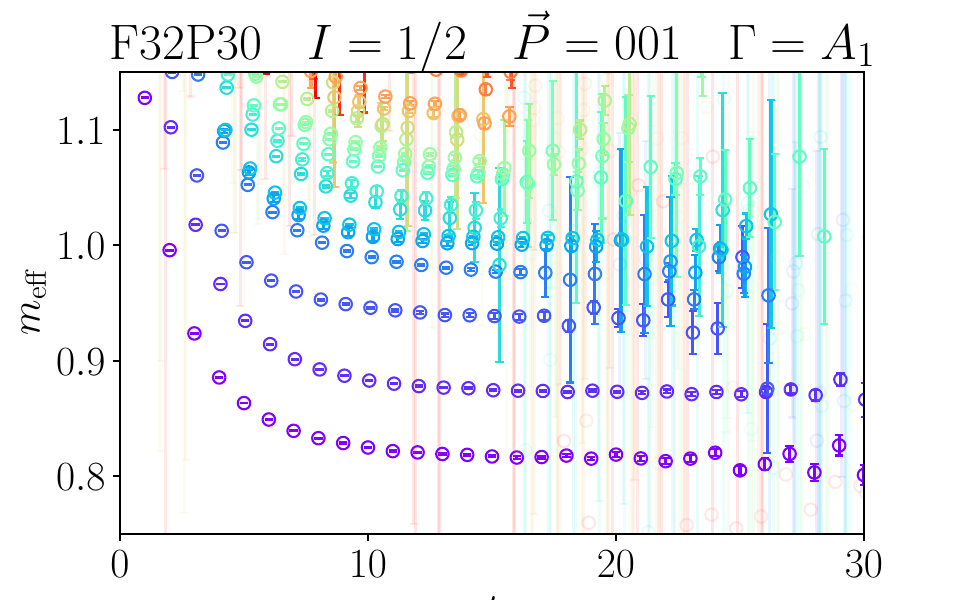}
    \includegraphics[width=0.49\columnwidth]{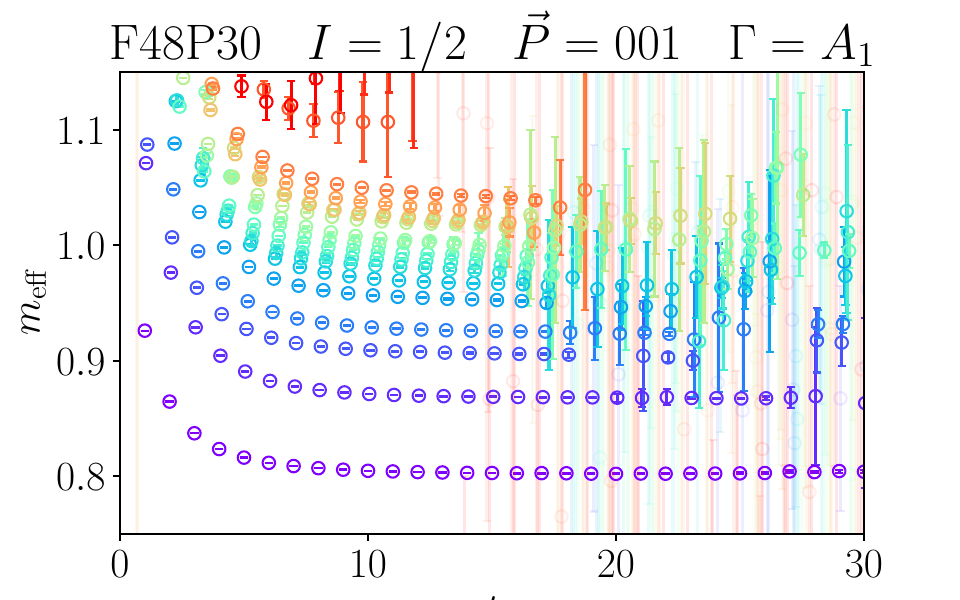}
    \includegraphics[width=0.49\columnwidth]{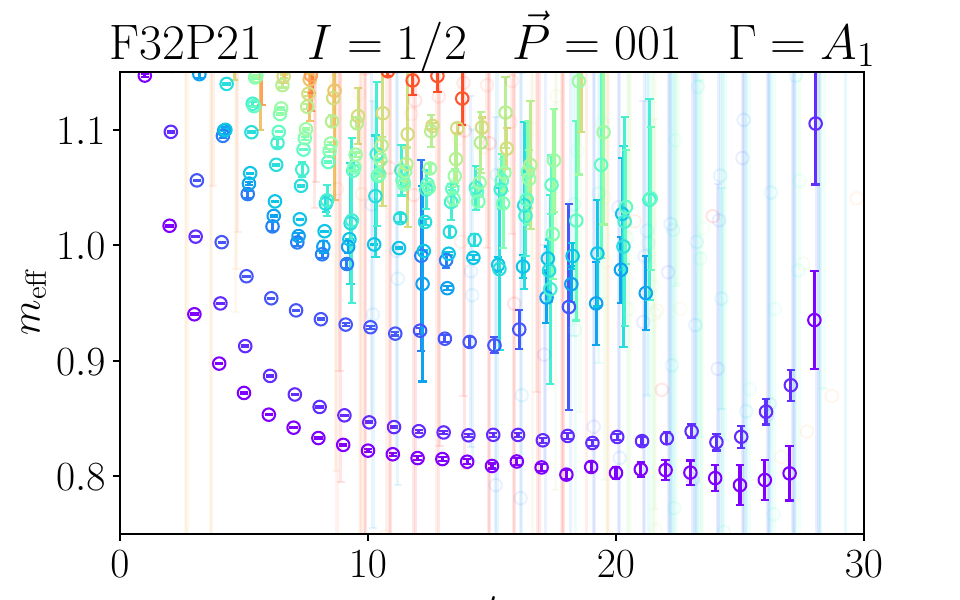}
    \includegraphics[width=0.49\columnwidth]{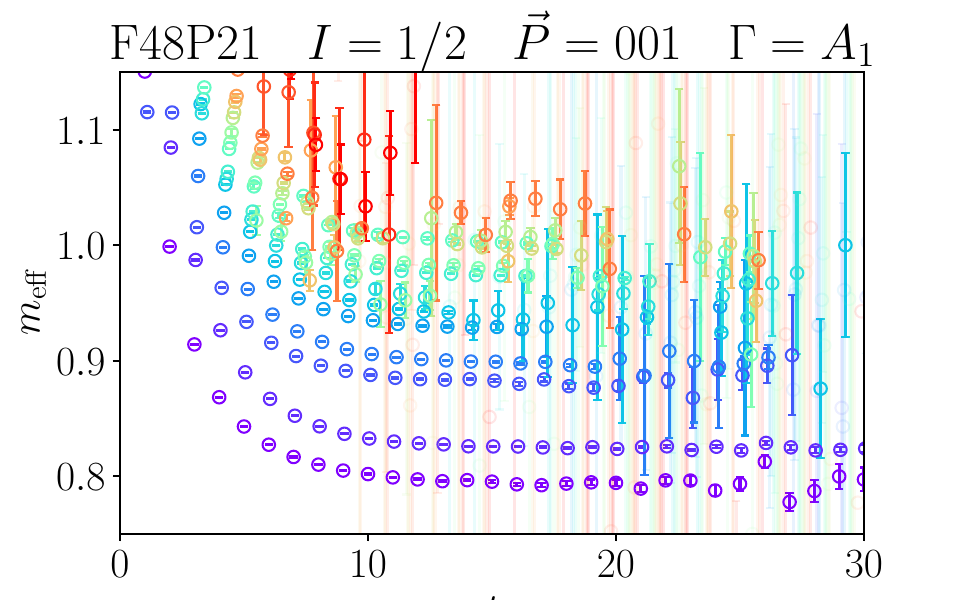}
    \caption{Effective-mass plots of the generalized eigenvalues $\lambda_n(t)$ for the two-point correlation functions of the four-channel scattering system in the $A_1$ irrep.}
    \label{fig:Dpi-meff-A1_coupled}
\end{figure}

\begin{figure}[htbp]
    \centering
    \includegraphics[width=0.49\columnwidth]{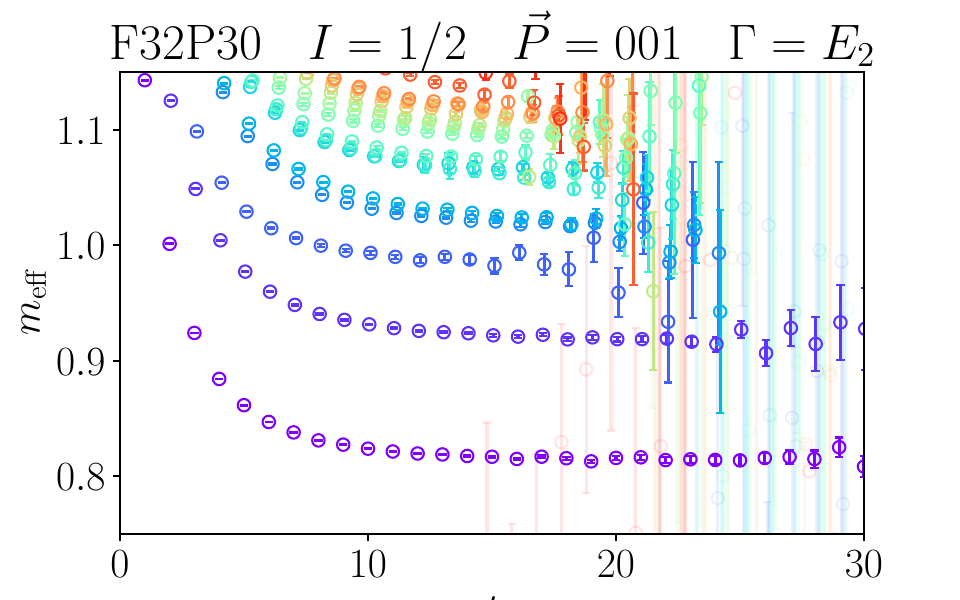}
    \includegraphics[width=0.49\columnwidth]{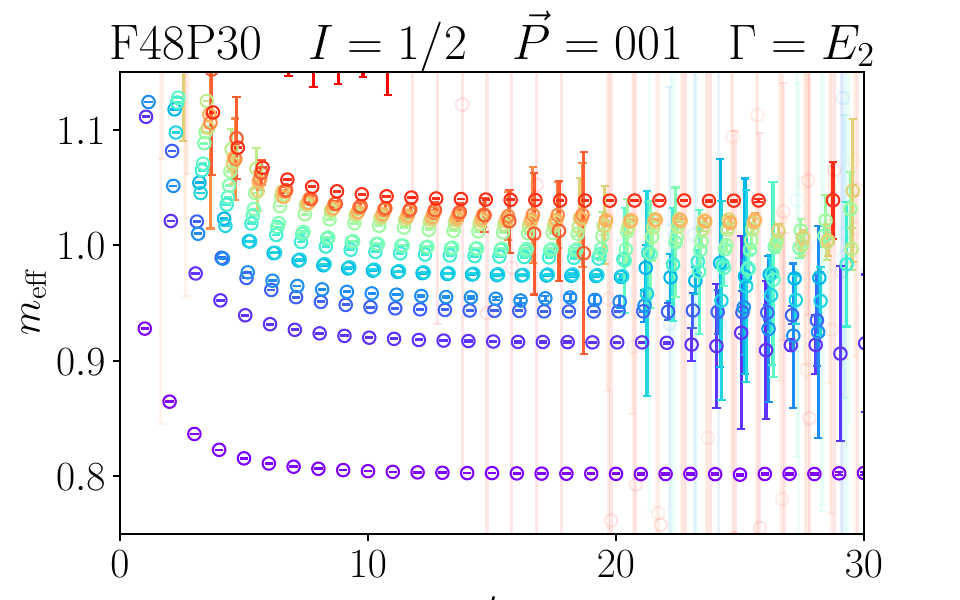}
    \includegraphics[width=0.49\columnwidth]{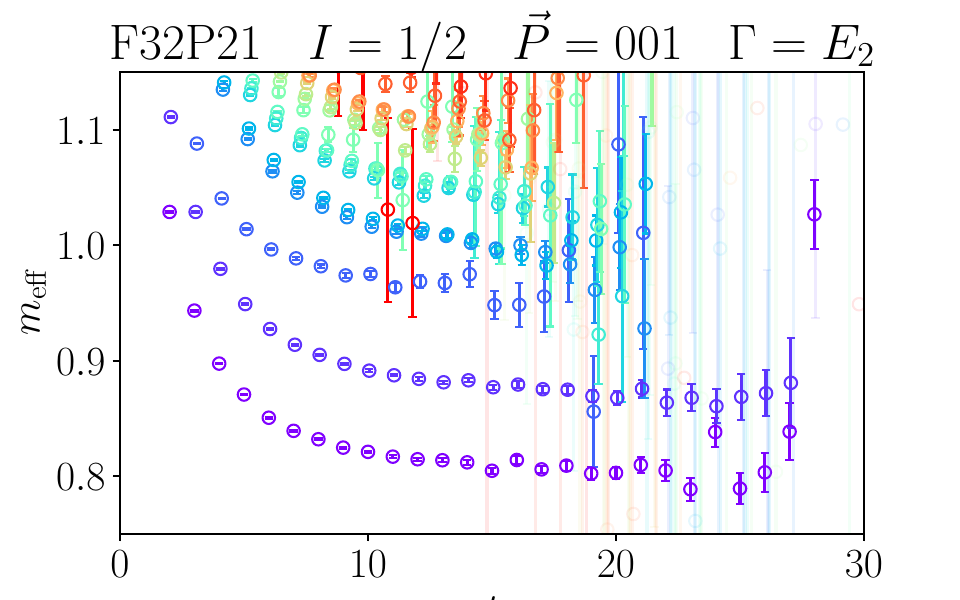}
    \includegraphics[width=0.49\columnwidth]{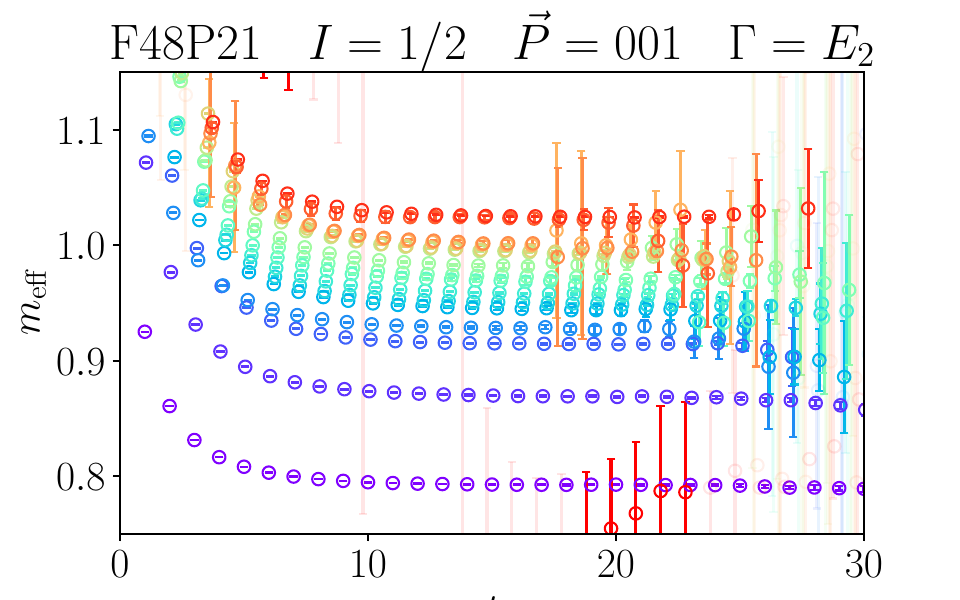}
    \caption{Effective-mass plots of the generalized eigenvalues $\lambda_n(t)$ for the two-point correlation functions of the four-channel scattering system in the $E_2$ irrep.}
    \label{fig:Dpi-meff-E2_coupled}
\end{figure}

Figures~\ref{fig:Dpi-meff-A1+_coupled}--\ref{fig:Dpi-meff-E2_coupled} show the GEVP energy levels obtained on the ensembles with lattice spacing approximately $0.07746\,\mathrm{fm}$, for the $A_1^+$ and $T_1^-$ irreps at $\vec{P}=[000]$ and for the $A_1$ and $E_2$ irreps at $\vec{P}=[001]$.

\begin{figure}[htbp]
\centering
\includegraphics[height=0.63\textheight,trim=0 0 2.1cm 0,clip]{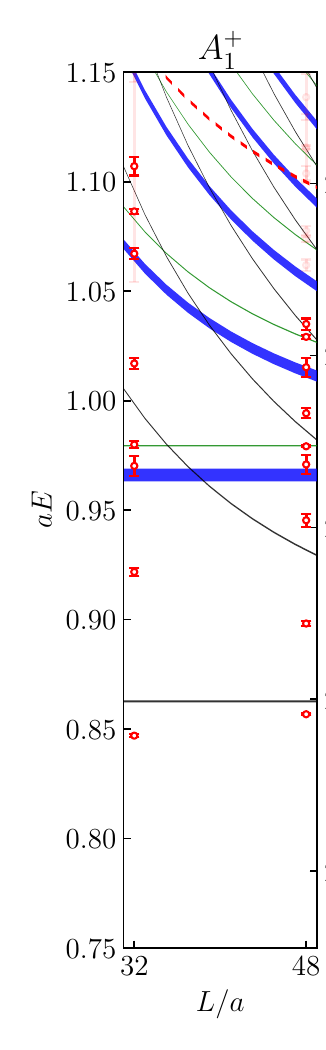}
\includegraphics[height=0.63\textheight,trim=2.0cm 0 2.1cm 0,clip]{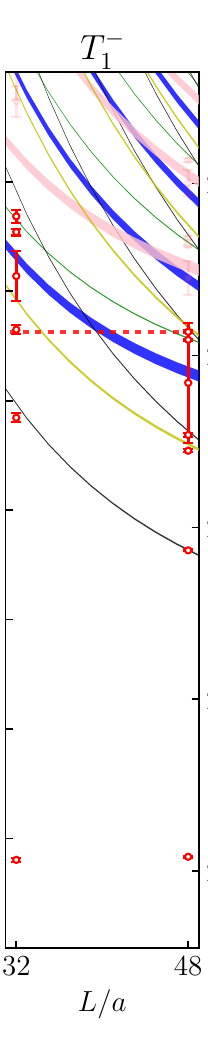}
\includegraphics[height=0.63\textheight,trim=2.0cm 0 2.1cm 0,clip]{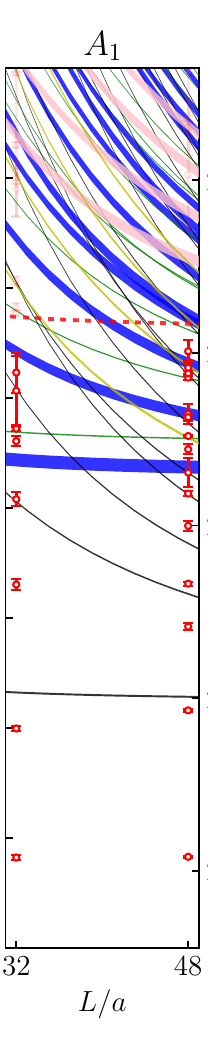}
\includegraphics[height=0.63\textheight,trim=2.0cm 0 0 0,clip]{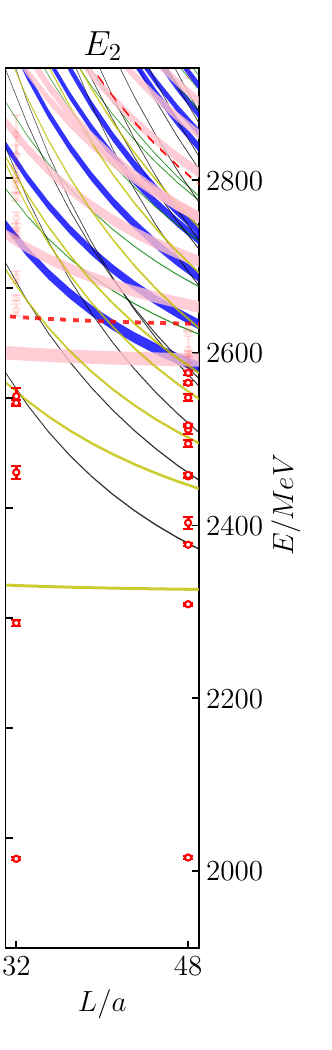}
\caption{Finite-volume spectra on the $M_{\pi} \approx 305 \,\mathrm{MeV}$ ensembles in the $A_1^+$ and $T_1^-$ irreps at $\vec{P} = [000]$, and in the $A_1$ and $E_2$ irreps at $\vec{P} = [001]$. The left and right vertical axes show the energies in lattice units and in physical units, respectively. Red points denote the levels obtained by diagonalizing operators for $D_0^*$, $D\pi$ with various back-to-back momenta, two types of $D\eta$, and $D_s\bar{K}$. Translucent red points denote levels lying above the four channels considered here and are not included in the subsequent analysis. The black, blue, green, and yellow bands denote the $D\pi$, $D\eta$, $D_s\bar{K}$, and $D^*\pi$ scattering channels, respectively. The three-body threshold is marked in red, and the pink band denotes all other inelastic channels.}
\label{fig:spectra_Dpi_coupled_FH}
\end{figure}

\begin{figure}[htbp]
\centering
\includegraphics[height=0.63\textheight,trim=0 0 2.1cm 0,clip]{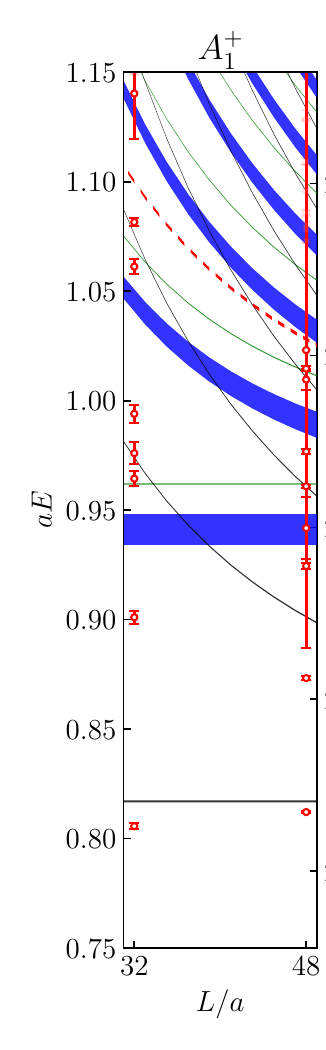}
\includegraphics[height=0.63\textheight,trim=2.0cm 0 2.1cm 0,clip]{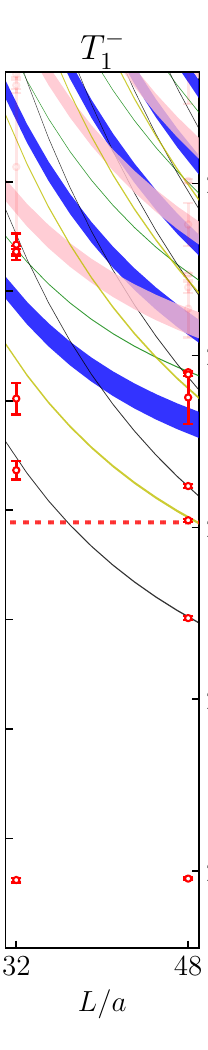}
\includegraphics[height=0.63\textheight,trim=2.0cm 0 2.1cm 0,clip]{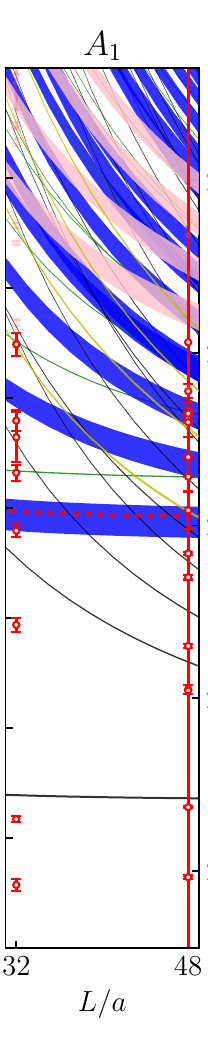}
\includegraphics[height=0.63\textheight,trim=2.0cm 0 0 0,clip]{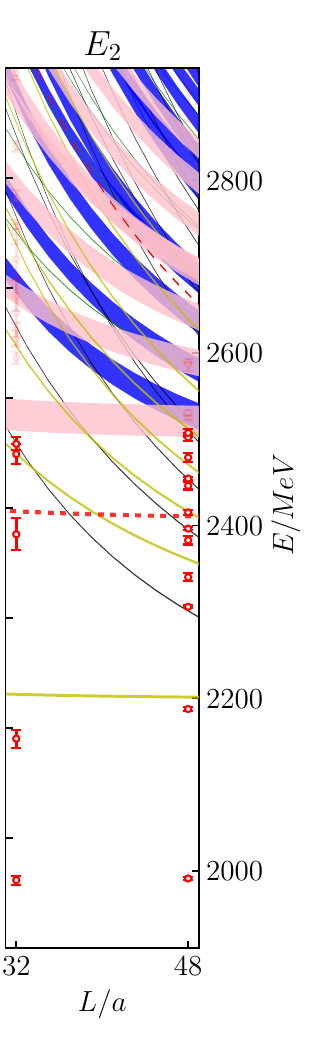}
\caption{Finite-volume spectra on the $M_{\pi} \approx 208 \,\mathrm{MeV}$ ensembles in the $A_1^+$ and $T_1^-$ irreps at $\vec{P} = [000]$, and in the $A_1$ and $E_2$ irreps at $\vec{P} = [001]$.}
\label{fig:spectra_Dpi_coupled_FL}
\end{figure}

It should be noted that the uncertainties of the levels near the noninteracting $D\eta$ energies are relatively large, which is difficult to avoid in the present analysis strategy. After fitting these levels, the resulting spectra are shown in Figs.~\ref{fig:spectra_Dpi_coupled_FH} and~\ref{fig:spectra_Dpi_coupled_FL}.

The leftmost $\vec{P}=[000]$ $A_1^+$ irrep contains the $S$-wave contributions from all $DX$ channels, and the corresponding energy levels exhibit clear volume dependence. The $\vec{P}=[000]$ $T_1^-$ irrep contains the $P$ wave of $DX$ and the ${}^3P_1$ partial wave of $D^*\pi$; since only two scattering channels are involved, this irrep is relatively easier to analyze. The two panels on the right correspond to the $A_1$ and $E_2$ irreps at $\vec{P}=[001]$. The $A_1$ irrep mixes all $DX$ partial waves starting from the $S$ wave, together with the ${}^3P_1$ and ${}^3D_2$ partial waves of $D^*\pi$; the $E_2$ irrep contains the higher $DX$ partial waves starting from the $P$ wave, together with the ${}^3S_1$, ${}^3P_1$, ${}^3P_2$, and ${}^3D_1$ partial waves of $D^*\pi$.

The noninteracting levels for the different scattering channels are also shown in the figure. Below the $D^*\pi$ threshold, all mesons in the scattering channels are pseudoscalars, and therefore, apart from isolated bound states, the interacting levels can usually be put into one-to-one correspondence with the corresponding noninteracting levels. For the higher $D^*\pi$ levels, the presence of degeneracies allows the interaction to split the levels. Below the first inelastic threshold, $D\eta$, the qualitative behavior of the spectrum is consistent with the previous single-channel study, including the position at which the $D^*$ bound state appears. This further indicates that the operator basis used in the single-channel scattering analysis above is complete in the low-energy region, especially near the $D\pi$ threshold.

In addition, the $D\eta$ and $D_s\bar{K}$ levels lie close to their noninteracting values. This suggests, on the one hand, that the corresponding operators couple only weakly to the other operators, and, on the other hand, that the interactions among these channels are relatively weak.

\subsection{Preliminary Finite-Volume Analysis}
In the preliminary finite-volume analysis, we first consider the most important rest-frame $A_1^+$ irrep. The scattering dynamics among different partial waves can be parametrized in a general $K$-matrix form as
\begin{equation}
K_{ij}^J = \sum_p 
\frac{\left(g_{p,i}^{(0)}+g_{p,i}^{(1)} s\right)
      \left(g_{p,j}^{(0)}+g_{p,j}^{(1)} s\right)}
     {m_p^2-s}
+\sum_{n}\gamma_{ij}^{(n)} s^n,
\label{eq:K_general}
\end{equation}
where $g_{p,i}^{(0)}$ denotes the coupling between the $i$th scattering channel and the $p$th pole, $m_p$ is a real parameter, and $\gamma_{ij}^{(n)}$ is a real symmetric constant matrix. Poles introduced explicitly in the $K$ matrix are usually used to describe resonance poles in the scattering amplitude; in some cases, however, poles of the scattering matrix may also be generated dynamically even without an explicit pole term. A related example in a three-body system will be given in\chapref{chap:three_body_problems2}.

Another parametrization is the $\mathrm{SU}(3)$-constrained form proposed in Ref.~\cite{Asokan:2022usm}, which imposes stronger constraints on the scattering amplitude. In the $\mathrm{SU}(3)$-symmetric limit, the coupled-channel system with $I=\frac{1}{2}$ formed by $D\pi$, $D\eta$, and $D_s\bar{K}$ decomposes into
\begin{equation}
\bar{\mathbf 3}\oplus \mathbf 6 \oplus \overline{\mathbf{15}},
\end{equation}
three $\mathrm{SU}(3)$ multiplets. The $\mathrm{SU}(3)$ basis and the isospin basis are related by
\begin{equation}
\begin{pmatrix}
| [\Bar{3}] \rangle \\ 
| [6] \rangle \\
| [\overline{15}] \rangle 
\end{pmatrix}
= U
\begin{pmatrix}
| D\pi \rangle \\ 
| D\eta \rangle \\
| D_{s}\Bar{K} \rangle 
\end{pmatrix}
,
\end{equation}    
where
\begin{equation}
U=
\begin{pmatrix}
    -3/4 & -1/4 & -\sqrt{3/8}
\\
\sqrt{3/8} & -\sqrt{3/8} & -1/2
\\
    1/4 & 3/4 & -\sqrt{3/8}
\\
\end{pmatrix}
.
\end{equation}
Using the rotation matrix $U$, one obtains the corresponding $\mathrm{SU}(3)$-symmetric coupling matrices $C_{\bar 3}$, $C_6$, and $C_{\overline{15}}$. Assuming that the $K$ matrix contains two poles, one can construct the form
\begin{equation}
K(s)=
\left(\frac{g_{\bar 3}^2}{m_{\bar 3}^2-s}+c_{\bar 3}\right)C_{\bar 3}
+
\left(\frac{g_{6}^2}{m_{6}^2-s}+c_{6}\right)C_{6}
+
c_{\overline{15}}\,C_{\overline{15}}
,
\end{equation}
where the two bare poles are associated with the $\bar{\mathbf 3}$ and $\mathbf 6$ representations, respectively, which are attractive in the $S$ wave at leading order in chiral dynamics~\cite{Albaladejo:2016lbb}. This parametrization contains slightly fewer free parameters than the general parametrization.

Unitarity requires only that the imaginary part of the scattering amplitude above threshold be equal to $-\delta_{ij} \rho_i(s)$. By specifying the real part, however, one can better preserve the analytic properties of the scattering amplitude. In this work we adopt two schemes: either setting the real part to zero, or using the Chew--Mandelstam prescription~\cite{Chew:1960iv}, which relates the real and imaginary parts through a dispersion relation. Following the review in Ref.~\cite{Wilson:2014cna}, the $\mathrm{M}$ matrix for partial wave $l$ is written as
\begin{equation}
    \mathrm{M}^{-1}_{ij}(s) = \frac{1}{(2 k_i)^l} \, K^{-1}_{ij}(s) \, \frac{1}{(2 k_j)^l} + I_{ij}(s),
\end{equation}
where the factor $(2 k_i)^{-l}$ ensures the correct behavior near the kinematic threshold. The $I$ matrix is diagonal:
\begin{equation}
I_{ij}(s) = \delta_{ij} I_i(s).
\end{equation}
If the two particles in channel $i$ have masses $m_1$ and $m_2$, respectively, the once-subtracted dispersion integral is
\begin{equation}
I(s) = I(s_{\mathrm{th}}) - \frac{s - s_{\mathrm{th}}}{\pi} \int_{s_{\mathrm{th}}}^{\infty} ds' \frac{\rho(s')}{(s' - s)(s' - s_{\mathrm{th}})},
\end{equation}
where the threshold is $s_{\mathrm{th}} = (m_1 + m_2)^2$, and
\begin{equation}
\rho(s) = \frac{2 k(s)}{\sqrt{s}} = \sqrt{1 - \frac{(m_1 + m_2)^2}{s}} \; \sqrt{1 - \frac{(m_1 - m_2)^2}{s}}.
\end{equation}

This form of the integral ensures that near $s + i\epsilon$, the real part is given by the principal-value integral, while the imaginary part satisfies the unitarity condition. Its analytic expression is
\begin{equation}
I(s) = I(s_{\mathrm{th}}) + \frac{\rho(s)}{\pi} \Bigg[
\xi(s) - \pi \log \xi(s) + \frac{\rho(s)}{\xi(s) - \rho(s)} \log \frac{m_2 - m_1}{m_2/m_1 + m_2/m_1} \Bigg]
,
\end{equation}
where $\xi(s) = 1 - \frac{(m_1 + m_2)^2}{s}$. In this work, $I(s_{\mathrm{th}})$ is chosen such that $\operatorname{Re} I(s=m_p^2) = 0$.

The present work is still in progress, and here we show only preliminary results for three-coupled-channel $D\pi$-$D\eta$-$D_s\bar{K}$ scattering using the general $K$-matrix parametrization. Specifically, we include one pole in Eq.~\eqref{eq:K_general} and constant terms in its residues, and introduce the constants $\gamma_{11}^{(0)}, \gamma_{12}^{(0)}, \gamma_{13}^{(0)}, \gamma_{22}^{(0)}, \gamma_{33}^{(0)}$, giving a total of $8$ free parameters.

The finite-volume quantization condition for a multi-channel system is rather complicated. Here we use the \verb|TwoHadronsInBox| package~\cite{Morningstar:2017spu} to implement the multi-channel Lüscher formula. In this implementation, the upper and lower bounds of the curves representing the finite-volume analysis are taken to be $\pm 1$. It should be noted that different references use different definitions of the $K$ matrix and the $\mathrm{M}$ matrix, and the corresponding conversions must be handled with care.

\begin{figure}[htbp]
    \centering
    \includegraphics[width=0.49\columnwidth]{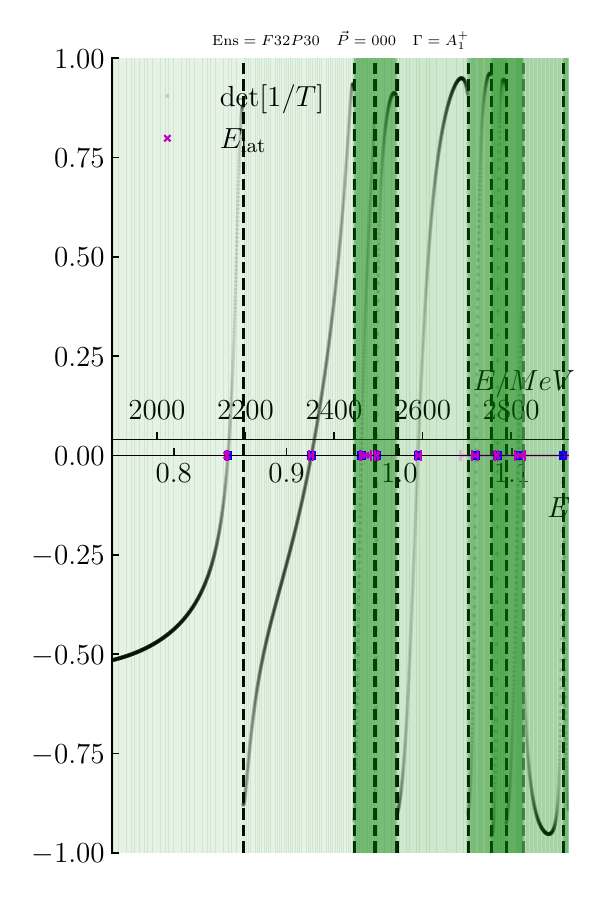}
    \includegraphics[width=0.49\columnwidth]{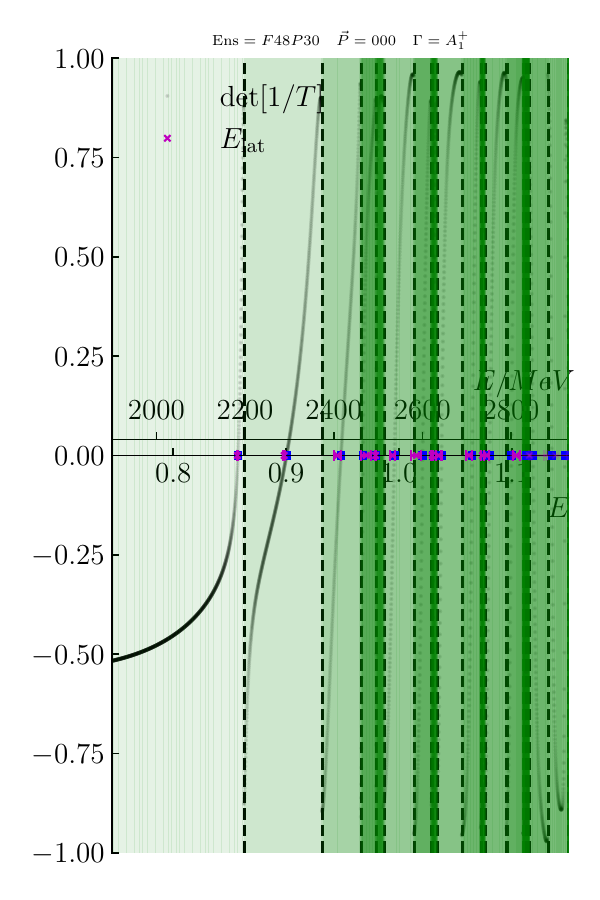}
    \caption{Finite-volume analysis of three-coupled-channel $D\pi-D\eta-D_s\bar{K}$ scattering in the rest-frame $A_1^+$ irrep on the F32P30 and F48P30 ensembles. The left and right panels correspond to F32P30 and F48P30, respectively. The data are described as in Fig.~\ref{fig:zeta_example1}. Here, thresholds and noninteracting levels are shown by black dashed lines, gray dashed lines indicate singularities introduced by the $K$-matrix parametrization, and green lines indicate the endpoints of the energy-search intervals. The size of each search interval is determined by two adjacent free levels: the closer the noninteracting levels are, the more rapidly Lüscher's $\zeta$ function varies between them, and the denser the search must be. The shade of the green regions represents the density of the search intervals.}
    \label{fig:zeta_coupled_FH}
\end{figure}

\begin{figure}[htbp]
    \centering
    \includegraphics[width=0.49\columnwidth]{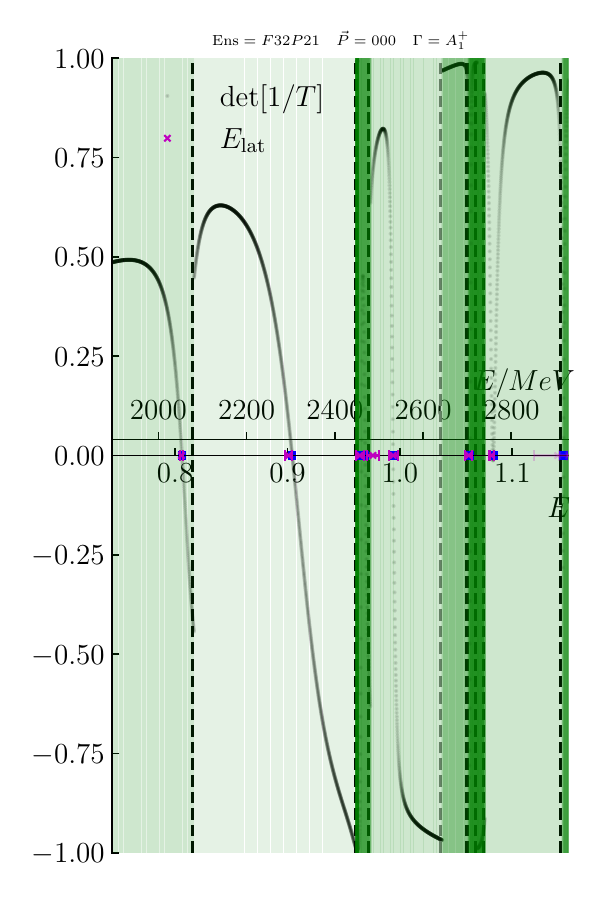}
    \includegraphics[width=0.49\columnwidth]{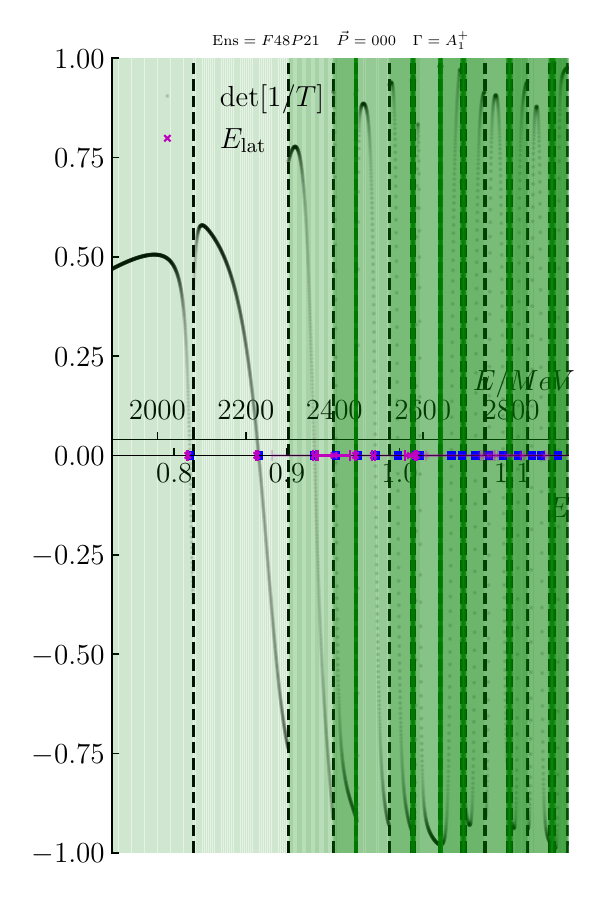}
    \caption{Finite-volume analysis of three-coupled-channel $D\pi-D\eta-D_s\bar{K}$ scattering in the rest-frame $A_1^+$ irrep on the F32P21 and F48P21 ensembles. The left and right panels correspond to F32P21 and F48P21, respectively.}
    \label{fig:zeta_coupled_FL}
\end{figure}

In this analysis we include only the $S$-wave contribution and perform a combined fit to the rest-frame $A_1^+$ irrep on the F32P30 and F48P30 ensembles. Figures~\ref{fig:zeta_coupled_FH} and~\ref{fig:zeta_coupled_FL} show the fit results at two $\pi$ masses. As in Fig.~\ref{fig:zeta_example1}, thresholds and noninteracting levels are shown by black dashed lines, while gray dashed lines indicate singularities introduced by the $K$-matrix parametrization. The green regions denote the energy intervals used in the level search; their shade indicates the search density. The size of each interval is determined by two adjacent free levels: the closer the free levels are, the more rapidly Lüscher's $\zeta$ function varies between them, and hence the denser the search must be.

The agreement between the purple curves and the blue data points shows that this $K$ matrix describes the $S$-wave interaction in $D\pi$-$D\eta$-$D_s\bar{K}$ reasonably well. At the two $\pi$ masses, $M_\pi = 305~\mathrm{MeV}$ and $208~\mathrm{MeV}$, the fitted $\chi^2/\text{d.o.f}$ values are $11.35/(21-9)=0.95$ and $8.10/(15-9)=1.35$, respectively.

A similar analysis procedure can be applied to the other ensembles. In the near future, we plan to use either the proposed $\mathrm{SU}(3)$-constrained $K$-matrix form~\cite{Asokan:2022usm} or scattering-amplitude parametrizations guided by chiral effective theory~\cite{Guo:2018tjx}, so as to relate ensembles at different values of $M_{\pi}$. If $\mathrm{SU}(3)$ symmetry is not sufficient to describe the data in the lighter-$M_{\pi}$ region, the general $K$-matrix parametrization is still expected to constrain the scattering amplitude adequately once more moving-frame energy levels are included.

\subsection{Pole Positions}
\begin{figure}[htbp]
    \centering
    \includegraphics[width=\columnwidth]{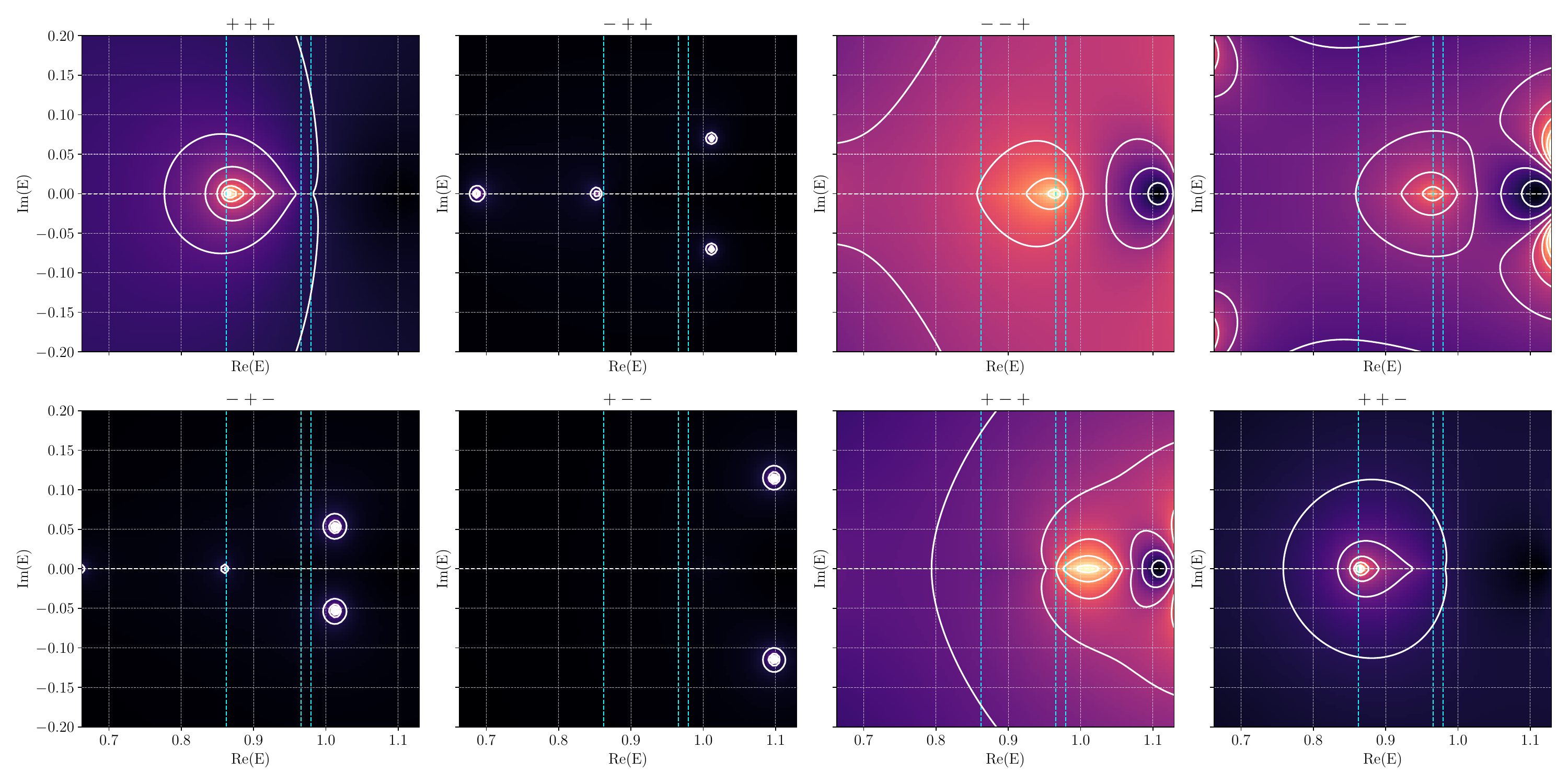}
    \caption{Analytically continued $|1 / \det[t^{-1}]|$ for three-coupled-channel $D\pi-D\eta-D_s\bar{K}$ scattering. The different panels correspond to different Riemann sheets. The first row shows all sheets directly connected to the real axis of the physical sheet, $-++$, $--+$, and $---$, which have the most direct impact on the physical scattering amplitude. The lower row shows other sheets farther away from the physical sheet.}
    \label{fig:image_FH}
\end{figure}

\begin{figure}[htbp]
    \centering
    \includegraphics[width=\columnwidth]{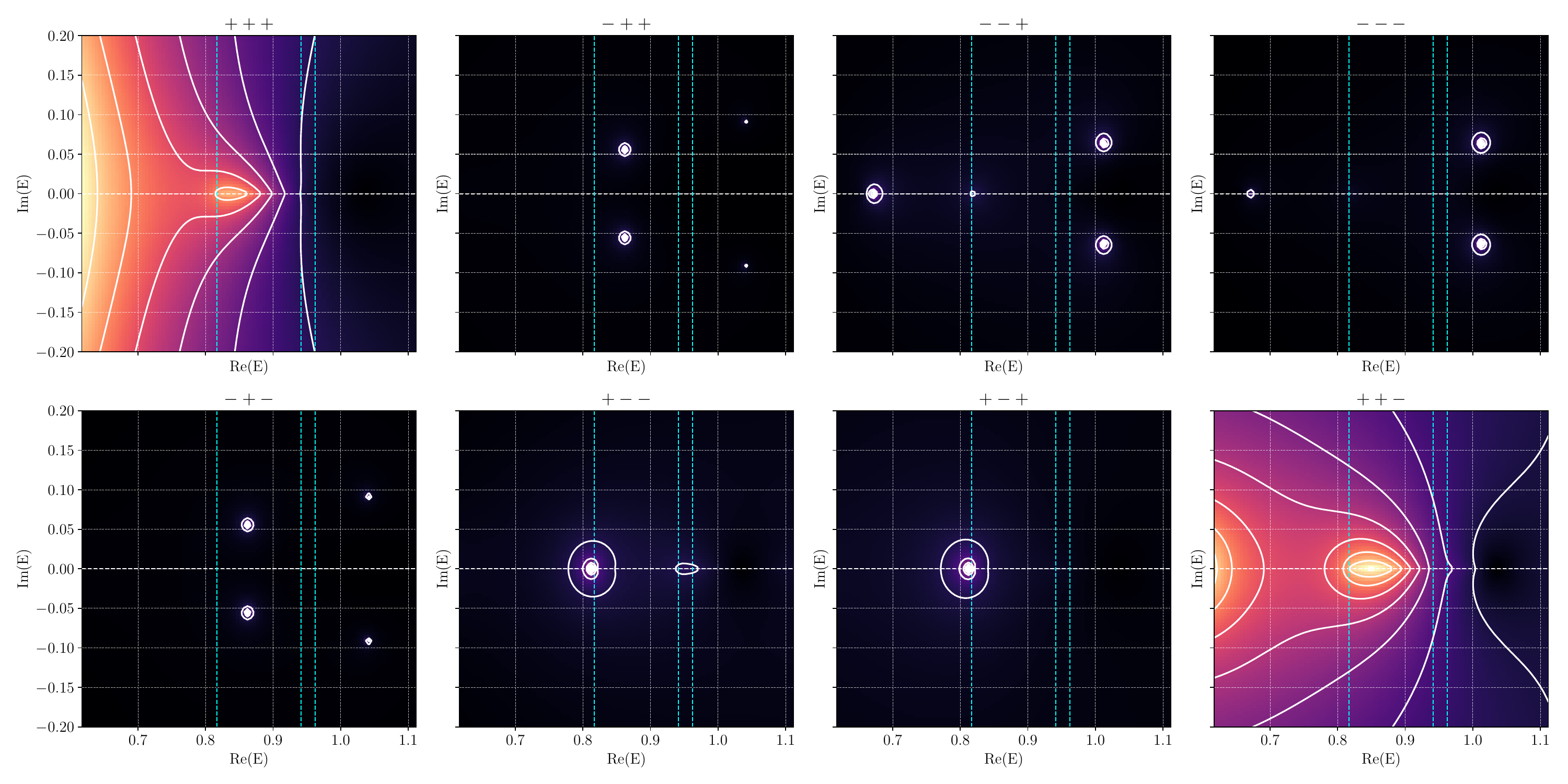}
    \caption{Analytically continued $|1 / \det[t^{-1}]|$ for three-coupled-channel $D\pi-D\eta-D_s\bar{K}$ scattering.}
    \label{fig:image_FL}
\end{figure}

In a three-channel scattering system, there are $2^3=8$ Riemann sheets, whose locations are specified by the signs of the imaginary parts of the three scattering momenta $k$. Figures~\ref{fig:image_FH} and~\ref{fig:image_FL} show the scattering amplitudes at two values of $M_{\pi}$, plotted as $|1 / \det[t^{-1}]|$. The upper-left panel of each large figure corresponds to the physical sheet $+++$; the first row shows all sheets directly connected to the real axis of the physical sheet, namely $-++$, $--+$, and $---$, which have the most direct impact on the physical scattering amplitude. The second row shows Riemann sheets farther away from the physical sheet, whose influence requires going around at least one branch point. Poles of the scattering amplitude appear as white points enclosed by white contours.

In Fig.~\ref{fig:image_FH}, for $M_{\pi} = 305~\mathrm{MeV}$, a virtual-state pole appears below threshold on the $-++$ sheet, consistent with the single-channel result. The deeper subthreshold pole is too far from the lattice energy levels to be well constrained, and its physical origin cannot be clearly identified. A pair of resonance poles appears near the third threshold, corresponding precisely to the second pole of interest. If this pole is present on ensembles at different values of $M_{\pi}$, the double-pole structure of the $D_0^*(2300)$ can be verified. Fortunately, at $M_{\pi} = 208~\mathrm{MeV}$, two pairs of poles remain on the $-++$ sheet. The pair near the $D_s\bar{K}$ threshold is the second pole being sought; the pole between the $D\pi$ and $D\eta$ thresholds is consistent with the single-channel result, and its motion with $M_{\pi}$ agrees with the expectation from chiral theory for interactions between pseudoscalar and charmed mesons.

On the other Riemann sheets, two pairs of poles appear on some sheets, while no clear pole is visible on others. It has been suggested in the literature~\cite{Albaladejo:2016lbb,Asokan:2022usm} that one can track the pole evolution on the $--+$ sheet, because at the physical value of $M_{\pi}$ the second pole lies between the $D\eta$ and $D_s\bar{K}$ thresholds. Since a complete analysis of model dependence is not yet available, we cannot at present provide a result for the second pole with controlled uncertainties; this will be reported separately in future work. It should be noted that no pole appears on the physical sheet at $M_{\pi} = 305~\mathrm{MeV}$, even though the corresponding curve may look close in the figure.

\section{Summary}
\label{sec:conclusions}
In this chapter, we used a set of $N_{\text{f}}=2+1$ Wilson-clover fermion ensembles at $M_{\pi} \approx 133, 208, 305$, and $317~\mathrm{MeV}$ to study the $S$- and $P$-wave phase shifts of single-channel $I=\frac{1}{2}$ $D\pi$ scattering using Lüscher's quantization condition. A variety of single-meson and two-meson operators were constructed, and a variational analysis of the correlation matrix yielded stable excited-state spectra. We observed a relatively strong attractive interaction in irreps containing the $S$ wave, while the $P$-wave interaction is weak. The scattering phase shifts were extracted by solving Lüscher's equation. Several parametrizations were used to analyze the model dependence, and we found that the systematic uncertainty from parametrization is much smaller than the statistical uncertainty.

The $D_0^*$ appears as a virtual state at $M_{\pi} \approx 305, 317~\mathrm{MeV}$, and as a resonance at $M_{\pi} \approx 133, 208~\mathrm{MeV}$. Combining our results with previous studies, we obtain the following picture for the evolution of the $D_0^*$ pole as $M_{\pi}$ changes: for $M_{\pi} \gtrsim 391~\mathrm{MeV}$, the $D_0^*$ pole is a bound state; it then evolves into a virtual state in the region $317~\mathrm{MeV}$ $\lesssim M_{\pi} \lesssim 391~\mathrm{MeV}$, and finally becomes a resonance in the region $266~\mathrm{MeV}$ $\lesssim M_{\pi} \lesssim 305~\mathrm{MeV}$. The pole first moves to the left, then turns around and moves to the right, displaying a complicated evolution. This trajectory of the $D_0^*$ pole with $M_{\pi}$ provides important clues for phenomenological models.

We extrapolated, or interpolated, the scattering length to the physical point and compared it with values inferred indirectly from experiment. Our result shows some tension with the PDG average, and an especially large discrepancy with the recent ALICE measurement~\cite{ALICE:2024bhk}. Although future calculations at finer lattice spacings and at more values of $M_{\pi}$ will help improve the precision of the lattice result, the author expects that using the same methodology would not substantially change the lattice determination. The discrepancy between the lattice result obtained here and the indirect ALICE measurement is likely to originate from systematic biases introduced either in the lattice method or in the experimental analysis. In addition, the near-threshold $D\pi$ scattering amplitude obtained in this work can provide useful input for future studies of the three-body $DD\pi$ system.

Physically, the systematic study of the $D_0^*$ pole evolution in this chapter reveals the central picture of chiral dynamics in charmed-meson scattering: as $M_\pi$ crosses critical values, the $D_0^*$ undergoes the full transition from a bound state to a virtual state and then to a resonance, reflecting the quantitative impact of the $\pi$ mass on interactions involving heavy-light mesons. This picture not only provides a model-independent test of the double-pole prediction from phenomenological models, but also offers methodological guidance for analogous chiral extrapolations in the three-body systems discussed in the next chapter: starting from calculations at unphysical $M_\pi$ and extrapolating to the physical point under effective-field-theory constraints is a key route to reliable predictions.

In single-channel scattering, only the motion of one pole can be observed. We therefore further studied three-coupled-channel $D\pi$-$D\eta$-$D_s\bar{K}$ scattering, considering frames with $|\vec{P}|^2<3$ and including all irreps of the corresponding little groups, while also exploring an analysis strategy for the extension to four-channel scattering. Using correlation matrices containing a large number of two-body and single-hadron operators, we extracted finite-volume spectra at $M_{\pi}\approx305~\mathrm{MeV}$ and $208~\mathrm{MeV}$. These spectra can be described well using a general $K$-matrix parametrization. In certain parametrizations, we observe the emergence of a double-pole structure, with the trajectory of the first pole consistent with that obtained from the single-channel scattering analysis. At present, however, the analysis uses only a small fraction of the available data: moving-frame data have not yet been included, and the coupling to the $D^*\pi$ channel has not yet been considered. The complete results, including a systematic analysis of four-channel scattering and the precise determination of the pole positions of the other three positive-parity excitations, will be reported in a separate future work.

To further investigate the mass ordering of the $D_0^*$ and $D_{s0}^*$, we are also carrying out studies of $I=0$ $DK-D_s\eta$, $I=1$ $DK-D_s\pi$, and $I=0/1$ $D\bar{K}$ scattering.

\cleardoublepage
\chapter{Three-Body Scattering: $\pi\pi\pi$ Scattering and the $\omega(782)$ Resonance}
\label{chap:three_body_problems}

{
\kaishu
\begin{center}
    俱怀逸兴壮思飞，欲上青天揽明月。
\end{center}
\hfill ——《宣州谢朓楼饯别校书叔云》[唐] 李白
}

\section{Background}
In celestial mechanics\footnotecircle{This chapter is based on the published paper~\cite{Yan:2025mdm}: H. Yan, M. Mai, M. Garofalo, U. G. Meißner, C. Liu, L. Liu and C. Urbach, $\omega$ Meson from Lattice QCD, \textit{Phys. Rev. Lett} (Editors' Suggestion) 133, 211906 (2024).}, the motion of three bodies under their mutual gravitational attraction is known as the \mybf{three-body problem}. The three-body problem is widely known for its chaotic behavior, its nonintegrability, and the science-fiction works bearing the same name~\cite{santi, santi2, santi3}. In the quantum domain, however, three-body systems are not only accessible at the level of their spectra, but can also provide deep physical insight into QCD.

In the hadron spectrum, many excited states have sizable strong-decay branching fractions into three-hadron final states. Over the past two decades, Lüscher's quantization condition and its various extensions~\cite{Luscher:1986pf, Luscher:1990ux, Luscher:1990ck} have enabled systematic studies of a wide range of two-body scattering and resonance problems, including the $\sigma$~\cite{Alford:2000mm, Prelovsek:2010kg, Fu:2012gf, Fu:2013ffa, Briceno:2016mjc, Doring:2016bdr, Liu:2016cba, Fu:2017apw, Guo:2018zss, Briceno:2017qmb} and $\rho$~\cite{Guo:2016zos, CP-PACS:2007wro, Bali:2015gji, Lang:2011mn, Feng:2010es, Fischer:2020yvw, Pelissier:2012pi, Erben:2019nmx, CS:2011vqf, Wilson:2015dqa, Bulava:2016mks, Fu:2016itp, Sun:2015enu, Alexandrou:2017mpi, Andersen:2018mau, Akahoshi:2021sxc, ExtendedTwistedMass:2019omo, Boyle:2024hvv, Boyle:2024grr}, which ultimately decay into two particles. Reliable results for many two-body resonances have thus been obtained from first principles and in a fully nonperturbative manner, providing important input for experimental amplitude analyses and phenomenology.

Lattice-QCD spectroscopy has now advanced to the point where one can attempt to study hadronic systems that decay strongly into three final-state particles. This is precisely the aim of this chapter and of\chapref{chap:three_body_problems2}: we use lattice QCD, presently the only viable first-principles framework for this purpose, to investigate the properties of hadrons with strong three-body decays. Since LQCD is formulated in Euclidean spacetime, unstable states can only be studied after mapping finite-volume spectra to scattering amplitudes. For earlier reviews and related work, see~\cite{Hansen:2019nir, Mai:2021lwb, Romero-Lopez:2022usb, Polejaeva:2012ut, Briceno:2012rv, Briceno:2018aml, Briceno:2024ehy, Alotaibi:2025pxz}.

The target of this chapter is the $\omega(782)$ meson, which decays strongly into three pions, $\omega \to 3\pi$, and has quantum numbers $I^G\left(J^{P C}\right)=0^{-}\left(1^{--}\right)$. It occupies an important place in the hadron spectrum. First, it is the lightest hadron that can decay into a three-particle final state. Second, in the vector-meson-dominance model (VMD) of photon-nucleon interactions, it dominates the isoscalar response~\cite{Sakurai:1960ju, Feynman:1973xc}. Combined with the topological-soliton picture of the nucleon, this helps explain the difference between the baryon charge radius and the isoscalar electric radius~\cite{Meissner:1986ka, Kaiser:2024vbc}. Third, in the one-boson-exchange picture of the nucleon-nucleon interaction, the $\omega$ generates the short-distance repulsion observed below $1$ fm; see, for example, Refs.~\cite{Erkelenz:1974uj, BJNN}. Fourth, when isospin is broken, it mixes with the $\rho(770)$ meson and produces visible effects in the meson vector form factor; see~\cite{Barkov:1985ac, OConnell:1995nse}. Finally, the $\omega-\rho$ mass splitting is also phenomenologically interesting, for instance through its relation to the muon anomalous magnetic moment~\cite{FermilabLattice:2021hzx, Hoferichter:2023sli, Stoffer:2023gba} and through its role in recent studies of dark matter and so-called mirror matter~\cite{Hippert:2021fch, Hippert:2022snq}. For these spectroscopic and broader phenomenological reasons, it is especially important to understand the $\omega(782)$ and to determine its mass and width precisely.

Precisely because the $\omega$ meson decays predominantly into a three-body final state, however, a first-principles determination of its pole position has so far been missing. The reason is that the formalism required to study three-body systems in finite-volume lattice calculations has only emerged over the past decade; see the recent reviews~\cite{Hansen:2019nir, Mai:2021lwb}. Two-body scattering studies provide the methodological foundation for three-body systems: the $\pi\pi$ scattering amplitude in the subsystem, in particular the $\rho$ resonance, enters as input to the three-body formalism, and the strategy of extrapolating from unphysical $M_\pi$ to the physical point carries over as well. This chapter describes how one of the three state-of-the-art formalisms~\cite{Mai:2017bge} can be used to study the $\omega(782)$ meson in lattice QCD and to determine the $\omega$ pole position, namely its mass and width, thereby filling the gap just described. Since, in the quantum numbers of the $\omega$, any two of the three final-state $\pi$ mesons can form a $\rho$ resonance~\cite{Gell-Mann:1962hpq}, we rely on a chiral Lagrangian with vector mesons to analyze the $\omega$ self-energy; see Ref.~\cite{Meissner:1987ge} for details. In particular, we use an effective-field-theory approach to extrapolate the $\omega$ to the physical $\pi$ mass.

\section{Isospin Structure}
In two-body $\pi\pi$ scattering, the computational and analytic difficulty of the various isospin channels is strongly anticorrelated with the value of the isospin. The $I=2$ channel is purely repulsive and weakly interacting, and it was among the first scattering channels used for proof-of-principle studies in lattice QCD. The $I=1$ channel is attractive and contains the relatively narrow $\rho$ resonance, which lies well away from threshold; as a result, its nonresonant background remains comparatively clean and the overall problem is still manageable. By contrast, the $I=0$ channel is much more complicated.

First, the interaction in this channel is very strong, giving the scalar $\sigma$ resonance a very large width and making both the extraction of the scattering phase and the amplitude analysis substantially more difficult. Second, the pole position lies close to the left-hand cut, which further complicates the analysis. Third, the two-point correlation functions in the $I=0$ channel contain many disconnected quark-contraction diagrams and require all-to-all propagators. This demands substantial computational resources, although it is not prohibitive when distillation is used, and the correlator signal is also significantly poorer. Fourth, this channel has the quantum numbers of the vacuum. Both single-hadron and two-hadron operators therefore couple to the vacuum, so vacuum subtraction must be performed when constructing the correlation functions; this procedure often further degrades the signal-to-noise ratio. Because of this combination of difficulties, $I=0$ $\pi\pi$ scattering was the last two-body channel to be solved systematically, nearly a decade after the $I=2$ channel. It is worth noting that some works have also combined information from all isospin channels, together with crossing symmetry and dispersion relations, to constrain the $I=0$ scattering amplitude more precisely in an indirect way~\cite{Rodas:2023gma, Mai:2019pqr}.

By comparison, lattice studies of three-body systems are still exploratory, and most existing calculations have focused on repulsive systems~\cite{Mai:2018djl, Blanton:2019vdk, Culver:2019vvu, Hansen:2020otl, Fischer:2020jzp, Alexandru:2020xqf, Draper:2023boj, Briceno:2025yuq, Feng:2026ixm}. Before the work presented in this chapter, the only exploratory lattice calculation involving a three-body resonance was the study of the axial-vector meson $a_1(1260)$ decaying into a three-$\pi$ final state~\cite{Mai:2021nul}.

Guided by the historical experience of two-body scattering, one might naturally expect that, among the three-body isospin channels $I=3,2,1,0$, the $I=0$ channel would again be the most difficult and therefore the last to be studied. Our exploratory work, however, suggests that in some cases the three-body problem may actually behave better numerically than the corresponding two-body problem. First, the $\omega(782)$ resonance is very narrow, which makes both the scattering analysis and the solution of the relevant integral equations numerically more stable. Second, with the aid of distillation, the disconnected diagrams appearing in the three-body system do not pose a fundamental computational obstacle. Third, the $\omega(782)$ is a vector meson and does not carry the quantum numbers of the vacuum, so no vacuum subtraction is required. In practice, the two-point functions of the $\omega(782)$ indeed have a better signal-to-noise ratio than those of the $\sigma$ meson. Based on these observations, we expect future studies of the three-body $I=1$ channel to be more challenging, and this is precisely the challenge addressed in\chapref{chap:three_body_problems2}\footnotecircle{In several talks, the author emphasized, based on the historical experience of two-body scattering, that the $I=0$ channel should be the most difficult one. In light of the present experience, this judgment was not appropriate.}.

Because of $G$-parity conservation, the $\pi\pi\pi$ system does not mix with the $\pi\pi$ system. The $\omega \to \pi^+\pi^-$ channel is forbidden by this symmetry, although it accounts for only about $2\%$ of the $\omega$ decay width. In nonidentical three-body systems, such as $T_{cc} \to DD\pi$, mixing with the $DD$ channel introduces additional complications. In addition, the $\omega(782)$ can mix with the $\phi(1020)$. A fully rigorous treatment should therefore consider the coupled scattering problem involving $\pi\pi\pi$ and $K\bar{K}$ simultaneously. As an exploratory study, this chapter temporarily neglects the complexity associated with strange-quark mixing. We will address $\omega-\phi$ mixing in future work.
It is important to note that the lattice calculation in this study strictly preserves isospin symmetry, so the isospin selection rules discussed above hold exactly in the calculation.

More explicitly, from the $CPT$ quantum numbers of the $\omega(782)$, we have
\begin{equation}
\begin{aligned}
    C(\pi^+\pi^-\pi^0) &= C(\pi^+\pi^-) C(\pi^0) = (-1)^{l_1} (+1), \\
    P(\pi^+\pi^-\pi^0) &= (-1)^{l_1} (-1)^{l_2} P(\pi)^3 = (-1)^{l_1+l_2} (-1), \\
    \vec{L}(\pi^+\pi^-\pi^0) &= \vec{l}_1 + \vec{l}_2 + \vec{0} \equiv \vec{1}.
\end{aligned}
\end{equation}
Here $l_1$ denotes the angular momentum between the $\pi^+\pi^-$ pair, while $l_2$ denotes the angular momentum between the $(\pi^+\pi^-)$ subsystem and the $\pi^0$. To satisfy the $C$ parity of the $\omega$, $l_1$ must be odd; the $P$-parity condition then implies that $l_2$ must also be odd. Thus, if partial waves of $F$ wave and higher are neglected, only $P$ waves occur between the $\pi$ mesons. If only the strong interaction is considered, all $\pi$ mesons are identical particles, and the same conclusion follows from Bose symmetry. In the $T_1^-$ irrep in the center-of-mass frame, the lowest noninteracting level satisfying this condition is $\pi[011] \pi[00-1] \pi[0-10]$, which lies far above threshold in our volumes. We therefore expect that, if the $\omega$ meson indeed exists and decays strongly to $\pi\pi\pi$, a significant energy-level shift should appear between this noninteracting level and the threshold. Equivalently, one may refer to this as the so-called ``extra level'' discussed in some of the literature\footnotecircle{An ``extra level'' usually refers to the appearance, in a lattice-QCD calculation at finite volume, of more levels than one would infer from the discretized free-particle momenta in the noninteracting theory. The author does not particularly like this terminology. First, the theory under study is the strongly interacting theory QCD, and its finite-volume energy levels are themselves direct manifestations of the interaction; in this sense there is no such thing as an ``extra'' level. Second, even in situations usually called ``extra-level'' cases, the interacting levels can often still be put in one-to-one correspondence with the corresponding free levels, with their energies shifted by the interaction. In this sense, too, there is nothing truly ``extra''.}.

By the same reasoning, the $\rho\pi$ system must also be in a $P$ wave in order to couple to the $\omega$. If moving frames are considered in the future, mixing between $S$ and $D$ waves will have to be included. Isospin symmetry also forbids mixing with channels such as $\sigma\pi$ and $\eta\pi$. For these reasons, the $\omega$ provides an excellent starting point for the study of three-body problems.

\section{Three-Body Operator Construction}
\label{sec:three_body_operators}
This chapter uses the gauge-field ensembles generated by the Chinese Lattice QCD Collaboration~\cite{CLQCD:2023sdb}. More specifically, we employ ensembles at a common lattice spacing and at two different $\pi$ masses, $M_\pi \approx 208$ and $305\,\mathrm{MeV}$. For each $\pi$ mass, two different spatial volumes are used in order to study finite-volume effects systematically, namely F32P30, F48P30, F32P21, and F48P21 in Table~\ref{tab:ensembles}.

As discussed in\chapref{chap:operators}, the finite cubic volume of the lattice reduces the continuous rotational symmetry to cubic symmetry in the rest frame. Unlike in\chapref{chap:two_body_problems}, this pioneering study of the $\omega(782)$ considers only the $T_1^-$ irrep in the rest frame. This choice is natural because both the $\rho$ and the $\omega$ are vector particles. To study a three-body problem, one must first understand its two-body subsystems, since information about the two-body interactions enters as input to the three-body quantization condition. The $\pi\pi$ subsystem couples to the $\rho$ meson through a $P$ wave, while the $\pi\pi\pi$ system couples to the $\omega$ meson through pairwise $P$ waves. The operators constructed here include one-meson, two-meson, and three-meson types. It is important to emphasize that all three types of operators are needed in order to couple well to all dynamical channels and to the $\omega$ meson, thereby obtaining a reliable and precise spectrum while avoiding contamination from higher-energy states. The spectrum is then obtained by solving the generalized eigenvalue problem with these operators~\cite{Michael:1982gb,Luscher:1990ck,Blossier:2009kd,Fischer:2020bgv}.

Three-body operators are more complicated than two-body operators. We now describe in detail how the different classes of operators are constructed in the relevant Hilbert spaces.
\subsection{Isospin Structure}
We first construct the isospin states of the two-body $\pi\pi$ subsystem. Using Clebsch-Gordan coefficients, one readily obtains
\begin{equation}
\begin{cases}
    | I=2, I_3=2 \rangle &= | 1; 1 \rangle, \\
    | I=2, I_3=1 \rangle &= \frac{1}{\sqrt{2}} (| 1; 0 \rangle + | 0; 1 \rangle), \\
    | I=2, I_3=0 \rangle &= \frac{1}{\sqrt{6}} (| 1; -1 \rangle + 2 | 0; 0 \rangle + | -1; 1 \rangle), \\
    | I=2, I_3=-1 \rangle &= \frac{1}{\sqrt{2}} (| 0; -1 \rangle + | -1; 0 \rangle), \\
    | I=2, I_3=-2 \rangle &= | -1; -1 \rangle, \\
    | I=1, I_3=1 \rangle &= \frac{1}{\sqrt{2}} (| 1; 0 \rangle - | 0; 1 \rangle), \\
    | I=1, I_3=0 \rangle &= \frac{1}{\sqrt{2}} (| 1; -1 \rangle - | -1; 1 \rangle), \\
    | I=1, I_3=-1 \rangle &= \frac{1}{\sqrt{2}} (| 0; -1 \rangle - | -1; 0 \rangle), \\
    | I=0, I_3=0 \rangle &= \frac{1}{\sqrt{3}} (| 1; -1 \rangle - | 0; 0 \rangle + | -1; 1 \rangle).
\end{cases}
\end{equation}

With the convention of Eq.~\ref{eq:multiplet}, the above isospin states can be expressed in terms of $\pi^+, \pi^-, \pi^u, \pi^d$ as
\begin{equation}
\begin{cases}
    | \pi\pi \rangle_{2,2} &= \pi^+ \pi^+, \\
    | \pi\pi \rangle_{1,1} &= \frac{1}{2} \left[ -\pi^+ \pi^u + \pi^+ \pi^d + \pi^u \pi^+ - \pi^d \pi^+ \right], \\
    | \pi\pi \rangle_{0,0} &= -\frac{1}{\sqrt{3}} \left[ \pi^+ \pi^- + \pi^- \pi^+ + \frac{1}{2} [\pi^u \pi^u - \pi^u \pi^d - \pi^d \pi^u + \pi^d \pi^d] \right],
\end{cases}
\end{equation}
or equivalently,
\begin{equation}
\begin{cases}
    | \pi\pi \rangle_{2,2} &= (\bar{d} \gamma_5 u) (\bar{d} \gamma_5 u), \\
    | \pi\pi \rangle_{1,1} &= \frac{1}{2} \left[-(\bar{d} \gamma_5 u) (\bar{u} \gamma_5 u) + (\bar{d} \gamma_5 u) (\bar{d} \gamma_5 d) + (\bar{u} \gamma_5 u) (\bar{d} \gamma_5 u) - (\bar{d} \gamma_5 d) (\bar{d} \gamma_5 u)\right], \\
    | \pi\pi \rangle_{0,0} &= -\frac{1}{\sqrt{3}} \left[(\bar{d} \gamma_5 u) (\bar{u} \gamma_5 d) + (\bar{u} \gamma_5 d) (\bar{d} \gamma_5 u) \right. \\
    &+ \left. \frac{1}{2} [(\bar{u} \gamma_5 u) (\bar{u} \gamma_5 u) - (\bar{u} \gamma_5 u) (\bar{d} \gamma_5 d) - (\bar{d} \gamma_5 d) (\bar{u} \gamma_5 u) + (\bar{d} \gamma_5 d) (\bar{d} \gamma_5 d)]\right].
\end{cases}
\end{equation}
One can verify that the $C$ parity of $| \pi\pi \rangle_{0,0}$ is $+1$.

For two-body $\pi\pi$ scattering, it is sufficient to choose any one $I_3$ component from the isospin multiplet above. For three-body $\pi\pi\pi$ systems, however, all $I_3$ components of the two-body isospin states are needed.

The same construction applies to $\rho\pi$, with $\pi$ in the above isospin states replaced by $\rho$, or equivalently with $\gamma_5$ replaced by $\gamma_i$. For example,
\begin{equation}
    | \rho\pi \rangle_{2,2} = \rho^+\pi^+ = (\bar{d} \gamma_i u) (\bar{d} \gamma_5 u).
\end{equation}

The isospin structure of $\sigma\pi$ is trivial:
\begin{equation}
\begin{cases}
    | \sigma\pi \rangle_{1,1} &= -\frac{1}{\sqrt{2}} \left[ \sigma^u \pi^+ + \sigma^d \pi^+ \right] = \frac{1}{\sqrt{2}} \left[ (\bar{u} \mathbbm{1} u)(\bar{d} \gamma_5 u) + (\bar{d} \mathbbm{1} d)(\bar{d} \gamma_5 u) \right]. \\
\end{cases}
\end{equation}

The isospin structure of three $\pi$ mesons is obtained in the same way. Combining the first two $\pi$ mesons with Clebsch-Gordan coefficients and then coupling the result to the third $\pi$ gives
\begin{equation}
    1 \otimes 1 \otimes 1 = (0 \oplus 1 \oplus 2) \otimes 1 = 0 \oplus 1^3 \oplus 2^2 \oplus 3.
\end{equation}
Here the $I=1$ and $I=2$ sectors have multiplicities three and two, respectively, corresponding to different coupling schemes among pairs of $\pi$ mesons, whereas the largest and smallest isospins, $I=3$ and $I=0$, have no multiplicity. More intuitively, we denote $(\pi\pi)_{I=2} \equiv G$, $(\pi\pi)_{I=1} \equiv \rho$, and $(\pi\pi)_{I=0} \equiv \sigma$. The symbol $G$ stands for ``isograviton'', referring to an object with isospin $I=2$~\cite{Feng:2026ixm}.
\begin{equation}
\begin{aligned}
    |I = 3 \rangle &=
    \begin{cases}
        | G \pi \rangle, \\
    \end{cases}, \\
    |I = 2 \rangle &=
    \begin{cases}
        | G \pi \rangle, \\
        | \rho \pi \rangle, \\
    \end{cases}, \\
    |I = 1 \rangle &=
    \begin{cases}
        | G \pi \rangle, \\
        | \rho \pi \rangle, \\
        | \sigma \pi \rangle, \\
    \end{cases}, \\
    |I = 0 \rangle &=
    \begin{cases}
        | \rho \pi \rangle. \\
    \end{cases}
\end{aligned}
\end{equation}

We can now write the explicit expressions for the isospin states one by one. For completeness, although this thesis studies only the $I=0$ and $I=1$ channels, all $I$ isospin states are listed below.

Starting from the maximal isospin,
\begin{equation}
    | \pi\pi\pi \rangle_{3,3} = | 1; 1; 1 \rangle = -\pi^+\pi^+\pi^+.
\end{equation}

For $I=2$, the two multiplicities correspond to different pairwise coupling schemes among the $\pi$ mesons.
\begin{equation}
\begin{aligned}
    | \pi\pi\pi \rangle_{2,2,I_{12}=1} &= | (1,1)_{\pi\pi}; (1,1)_{\pi} \rangle \\
    &= \frac{1}{2} \left[ \pi^+\pi^u\pi^+ - \pi^+\pi^d\pi^+ - \pi^u\pi^+\pi^+ + \pi^d\pi^+\pi^+ \right],
\end{aligned}
\end{equation}

\begin{equation}
\begin{aligned}
    | \pi\pi\pi \rangle_{2,2,I_{12}=2} &= \sqrt{\frac{2}{3}} | (2,2)_{\pi\pi}; (1,0)_{\pi} \rangle - \sqrt{\frac{1}{3}} | (2,1)_{\pi\pi}; (1,1)_{\pi} \rangle \\
    &= \frac{1}{\sqrt{12}} \left[ 2 \pi^+\pi^+\pi^u - 2 \pi^+\pi^+\pi^d - \pi^+\pi^u\pi^+ \right. \\
    & \qquad \left. + \pi^+\pi^d\pi^+ - \pi^u\pi^+\pi^+ + \pi^d\pi^+\pi^+ \right].
\end{aligned}
\end{equation}
Here $I_{12}$ denotes the isospin obtained by coupling the first two $\pi$ mesons.

The $I=1$ sector has three multiplicities. One of these channels will be studied in\chapref{chap:three_body_problems2}; here we first list the isospin states for all three multiplicities.
\begin{equation}
\begin{aligned}
    | \pi\pi\pi \rangle_{1,1,I_{12}=0} &= | (0,0)_{\pi\pi}; (1,1)_{\pi} \rangle \\
    &= \frac{1}{\sqrt{12}} \left[ 2 \pi^+\pi^-\pi^+ + \pi^u\pi^u\pi^+ - \pi^d\pi^u\pi^+ \right. \\
    & \qquad \left. - \pi^u\pi^d\pi^+ + \pi^d\pi^d\pi^+ + 2 \pi^-\pi^+\pi^+ \right],
\end{aligned}
\end{equation}

\begin{equation}
\begin{aligned}
    | \pi\pi\pi \rangle_{1,1,I_{12}=1} &= \frac{1}{\sqrt{2}} (| (1,1)_{\pi\pi}; (1,0)_{\pi} \rangle - | (1,0)_{\pi\pi}; (1,1)_{\pi} \rangle) \\
    &= \frac{1}{4} \left[ - \pi^+\pi^u\pi^u + \pi^+\pi^d\pi^u + \pi^+\pi^u\pi^d - \pi^+\pi^d\pi^d + \pi^u\pi^+\pi^u \right. \\
    & \qquad \left. - \pi^d\pi^+\pi^u - \pi^u\pi^+\pi^d + \pi^d\pi^+\pi^d - 2 \pi^+\pi^-\pi^+ + 2 \pi^-\pi^+\pi^+ \right],
\end{aligned}
\end{equation}

\begin{equation}
\begin{aligned}
    | \pi\pi\pi \rangle_{1,1,I_{12}=2} &= \sqrt{\frac{3}{5}} | (2,2)_{\pi\pi}; (1,-1)_{\pi} \rangle - \sqrt{\frac{3}{10}} | (2,1)_{\pi\pi}; (1,0)_{\pi} \rangle + \sqrt{\frac{1}{10}} | (2,0)_{\pi\pi}; (1,1)_{\pi} \rangle \\
    &= \frac{1}{\sqrt{240}} \left[ 12 \pi^+\pi^+\pi^- + 3 \pi^+\pi^u\pi^u - 3 \pi^+\pi^d\pi^u - 3 \pi^+\pi^u\pi^d + 3 \pi^+\pi^d\pi^d \right. \\
    & \qquad \left. + 3 \pi^u\pi^+\pi^u - 3 \pi^d\pi^+\pi^u - 3 \pi^u\pi^+\pi^d + 3 \pi^d\pi^+\pi^d + 2 \pi^+\pi^-\pi^+ \right. \\
    & \qquad \left. - 2 \pi^u\pi^u\pi^+ + 2 \pi^d\pi^u\pi^+ + 2 \pi^u\pi^d\pi^+ - 2 \pi^d\pi^d\pi^+ + 2 \pi^-\pi^+\pi^+ \right].
\end{aligned}
\end{equation}

The $I=0$ sector has no multiplicity,
\begin{equation}
\begin{aligned}
    | \pi\pi\pi \rangle_{0,0} &= \frac{1}{\sqrt{3}} (| (1,1)_{\pi\pi}; (1,-1)_{\pi} \rangle - | (1,0)_{\pi\pi}; (1,0)_{\pi} \rangle + | (1,-1)_{\pi\pi}; (1,1)_{\pi} \rangle) \\
    &= \frac{1}{\sqrt{12}} \left[ -\pi^+\pi^u\pi^- + \pi^+\pi^d\pi^- + \pi^u\pi^+\pi^- - \pi^d\pi^+\pi^- + \pi^+\pi^-\pi^u -\pi^+\pi^-\pi^d \right. \\
    & \qquad \left. -\pi^-\pi^+\pi^u + \pi^-\pi^+\pi^d - \pi^u\pi^-\pi^+ + \pi^d\pi^-\pi^+ + \pi^-\pi^u\pi^+ -\pi^-\pi^d\pi^+ \right].
\end{aligned}
\end{equation}

Whether the $\pi\pi\pi$ operators associated with different multiplicities are linearly independent depends on their momentum patterns. This point will be discussed further in\chapref{chap:three_body_problems2}.

We now consider all one-body, two-body, and three-body operators. The isospin structure determines the topology of the operator contractions, while the momentum combinations determine the linear combinations of these diagrams. The operator types for two-body $\pi\pi$ systems are summarized as follows:
\begin{align}
&I=2 \ \pi\pi
\begin{cases}
| \pi\pi \rangle_{2,2} = \pi^+\pi^+,
\end{cases}, \\
&I=1 \ \pi\pi
\begin{cases}
| \rho \rangle_{1,1} &= -\rho^+, \\
| \pi\pi \rangle_{1,1} &= \frac{1}{2} \left[ -\pi^+ \pi^u + \pi^+ \pi^d + \pi^u \pi^+ - \pi^d \pi^+ \right],
\end{cases}, \\
&I=0 \ \pi\pi
\begin{cases}
| \sigma \rangle_{0,0} &= \frac{1}{\sqrt{2}} (\sigma^u + \sigma^d), \\
| \pi\pi \rangle_{0,0} &= -\frac{1}{\sqrt{3}} \left[ \pi^+ \pi^- + \pi^- \pi^+ + \frac{1}{2} [\pi^u \pi^u - \pi^u \pi^d - \pi^d \pi^u + \pi^d \pi^d] \right].
\end{cases}
\label{eq:I_pipi}
\end{align}

The operator types appearing in the three-body channels are
\begin{align}
&I=3 \ \pi\pi\pi
\begin{cases}
| \pi\pi\pi \rangle_{3,3} = \pi^+\pi^+\pi^+,
\end{cases}, \\
&I=2 \ \pi\pi\pi
\begin{cases}
| \rho\pi \rangle_{2,2} &= \rho^+ \pi^+, \\
| \pi\pi\pi \rangle_{2,2,I_{12}=0} &= \frac{1}{2} \left[ \pi^+\pi^u\pi^+ - \pi^+\pi^d\pi^+ - \pi^u\pi^+\pi^+ + \pi^d\pi^+\pi^+ \right], \\
| \pi\pi\pi \rangle_{2,2,I_{12}=1} &= \frac{1}{\sqrt{12}} \left[ 2 \pi^+\pi^+\pi^u - 2 \pi^+\pi^+\pi^d - \pi^+\pi^u\pi^+ \right. \\
& \qquad \left. + \pi^+\pi^d\pi^+ - \pi^u\pi^+\pi^+ + \pi^d\pi^+\pi^+ \right],
\end{cases}, \\
&I=1 \ \pi\pi\pi
\begin{cases}
| a_1 \rangle_{1,1} &= -a_1^+, \\
| \rho\pi \rangle_{1,1} &= \frac{1}{2} \left[ -\rho^+ \pi^u + \rho^+ \pi^d + \rho^u \pi^+ - \rho^d \pi^+ \right], \\
| \sigma\pi \rangle_{1,1} &= -\frac{1}{\sqrt{2}} \left[ \sigma^u \pi^+ + \sigma^d \pi^+ \right], \\
| \pi\pi\pi \rangle_{1,1,I_{12}=0} &= \frac{1}{\sqrt{12}} \left[ 2 \pi^+\pi^-\pi^+ + \pi^u\pi^u\pi^+ - \pi^d\pi^u\pi^+ \right. \\
& \qquad \left. - \pi^u\pi^d\pi^+ + \pi^d\pi^d\pi^+ + 2 \pi^-\pi^+\pi^+ \right], \\
| \pi\pi\pi \rangle_{1,1,I_{12}=1} &= \frac{1}{4} \left[ - \pi^+\pi^u\pi^u + \pi^+\pi^d\pi^u + \pi^+\pi^u\pi^d - \pi^+\pi^d\pi^d \right. \\
& \qquad \left. + \pi^u\pi^+\pi^u - \pi^d\pi^+\pi^u - \pi^u\pi^+\pi^d + \pi^d\pi^+\pi^d \right. \\
& \qquad \left. - 2 \pi^+\pi^-\pi^+ + 2 \pi^-\pi^+\pi^+ \right], \\
| \pi\pi\pi \rangle_{1,1,I_{12}=2} &= \frac{1}{\sqrt{240}} \left[ 12 \pi^+\pi^+\pi^- + 3 \pi^+\pi^u\pi^u - 3 \pi^+\pi^d\pi^u \right. \\
& \quad \left. - 3 \pi^+\pi^u\pi^d + 3 \pi^+\pi^d\pi^d + 3 \pi^u\pi^+\pi^u - 3 \pi^d\pi^+\pi^u \right. \\
& \quad \left. - 3 \pi^u\pi^+\pi^d + 3 \pi^d\pi^+\pi^d + 2 \pi^+\pi^-\pi^+ - 2 \pi^u\pi^u\pi^+ \right. \\
& \quad \left. + 2 \pi^d\pi^u\pi^+ + 2 \pi^u\pi^d\pi^+ - 2 \pi^d\pi^d\pi^+ + 2 \pi^-\pi^+\pi^+ \right],
\end{cases}, \\
&I=0 \ \pi\pi\pi
\begin{cases}
| \omega \rangle_{0,0} &= \frac{1}{\sqrt{2}} (\omega^u + \omega^d), \\
| \rho\pi \rangle_{0,0} &= -\frac{1}{\sqrt{3}} \left[ \rho^+ \pi^- + \rho^- \pi^+ + \frac{1}{2} [\rho^u \pi^u - \rho^u \pi^d - \rho^d \pi^u + \rho^d \pi^d] \right], \\
| \pi\pi\pi \rangle_{0,0} &= \frac{1}{\sqrt{12}} \left[ -\pi^+\pi^u\pi^- + \pi^+\pi^d\pi^- + \pi^u\pi^+\pi^- - \pi^d\pi^+\pi^- \right. \\
& \qquad \left. + \pi^+\pi^-\pi^u -\pi^+\pi^-\pi^d -\pi^-\pi^+\pi^u + \pi^-\pi^+\pi^d \right. \\
& \qquad \left. - \pi^u\pi^-\pi^+ + \pi^d\pi^-\pi^+ + \pi^-\pi^u\pi^+ -\pi^-\pi^d\pi^+ \right].
\end{cases}
\label{eq:I_pipipi}
\end{align}

\subsection{Space-Group Projection}
Using the method described in\chapref{chap:operators} and the \verb|OpTion| package~\cite{Yan:2025jlq}, we construct the momentum combinations of the operators:
\begin{equation}
    \begin{cases}
        O_{\omega_1} &= \omega_z(0), \\
        O_{(\rho\pi)_1} &= - \rho_x(-e_y)\pi(e_y) + \rho_x(e_y)\pi(-e_y) + \rho_y(-e_x)\pi(e_x) - \rho_y(e_x)\pi(-e_x), \\
        O_{(\rho\pi)_2} &= + \rho_x(e_{-x,-y})\pi(e_{xy}) - \rho_x(e_{-x,y})\pi(e_{x,-y}) + \rho_x(e_{x,-y})\pi(e_{-x,y}) - \rho_x(e_{xy})\pi(e_{-x,-y}) \\
        & \quad - \rho_y(e_{-x,-y})\pi(e_{xy}) - \rho_y(e_{-x,y})\pi(e_{x,-y}) + \rho_y(e_{x,-y})\pi(e_{-x,y}) + \rho_y(e_{xy})\pi(e_{-x,-y}), \\
        O_{(\pi\pi\pi)_1} &= + \pi(e_{x,y})\pi(-e_y)\pi(-e_x) - \pi(e_{x,-y})\pi(e_y)\pi(-e_x) - \pi(e_{-x,y})\pi(-e_y)\pi(e_x) \\
        & \quad + \pi(e_{-x,-y})\pi(e_y)\pi(e_x) - \pi(e_{x,y})\pi(-e_x)\pi(-e_y) + \pi(e_{-x,y})\pi(e_x)\pi(-e_y) \\
        & \quad + \pi(e_{x,-y})\pi(-e_x)\pi(e_y) - \pi(e_{-x,-y})\pi(e_x)\pi(e_y). \\
        \end{cases}
\end{equation}
Here each hadron state denotes a state that has already been projected in isospin in the previous subsection.

As in\chapref{chap:two_body_problems}, one may define
\begin{equation}
\begin{cases}
    O_{\text{one}} = \sum_{i} \eta_i \omega_{\mu_i}(0), \\
    O_{\text{two}} = \sum_{i} \eta_i \rho_{\mu_i}(\vec{p}_i) \pi(-\vec{p}_i), \\
    O_{\text{three}} = \sum_{i} \eta_i \pi_{\mu_i}(\vec{p}_{i1}) \pi_{\mu_i}(\vec{p}_{i2}) \pi(-\vec{p}_{i1}-\vec{p}_{i2}).
\end{cases}
\label{eq:operator_table_label}
\end{equation}
These coefficients can be denoted as
\begin{equation}
\eta_{\mu_i}^{\alpha_{i1}(;\alpha_{i2})}.
\end{equation}
The operator coefficients are listed in Table~\ref{tab:operatorsomega}.

\begin{table*}[htbp]
    \centering
    \caption{Operators used to interpolate the $\pi\pi$ and $\pi\pi\pi$ systems. Each irrep contains one-meson, two-meson, and, where relevant, three-meson operators. The notation $\eta_{\mu_i}^{\alpha_{i1}(;\alpha_{i2})}$ is defined in the text. Unimportant overall normalization constants are omitted.}
    \addtolength{\tabcolsep}{6pt}
    \begin{tabular}{ccc}
    \toprule
    Scattering channel & Type & Operator \\
    \midrule
    \multirow{3}{*}{$\pi\pi$} & \multirow{1}{*}{one-meson} & $(+1)^{0}_{z}$ \\
    \cmidrule(lr){2-3}
    & \multirow{2}{*}{two-meson} & $(+1)^{0}_{5}$ \\
    \cmidrule(lr){3-3}
    & & $(+1)^{x}_{5}, (+1)^{-x}_{5}, (+1)^{y}_{5}, (+1)^{-y}_{5}, (+1)^{z}_{5}, (+1)^{-z}_{5}$ \\
    \midrule
    \multirow{5}{*}{$\pi\pi\pi$} & \multirow{1}{*}{one-meson} & $(+1)^{0}_{z}$ \\
    \cmidrule(lr){2-3}
    & \multirow{3}{*}{two-meson} & $(-1)^{-y}_{x}, (+1)^{y}_{x}, (+1)^{-x}_{y}, (-1)^{x}_{y}$ \\
    \cmidrule(lr){3-3}
    & & $(-1)^{xy}_{x}, (+1)^{xy}_{y}, (+1)^{x,-y}_{x}, (+1)^{x,-y}_{y}$ \\
    & & $(-1)^{-x,y}_{x}, (-1)^{-x,y}_{y}, (+1)^{-x,-y}_{x}, (-1)^{-x,-y}_{y}$ \\
    \cmidrule(lr){2-3}
    & \multirow{2}{*}{three-meson} & $(+1)^{xy;-y}_{5}, (-1)^{x,-y;y}_{5}, (-1)^{-x,y;-y}_{5}, (+1)^{-x,-y;y}_{5}$ \\
    & & $(-1)^{xy;-x}_{5}, (+1)^{-x,y;x}_{5}, (+1)^{x,-y;-x}_{5}, (-1)^{-x,-y;x}_{5}$ \\
    \bottomrule
    \end{tabular}
    \addtolength{\tabcolsep}{-6pt}
\label{tab:operatorsomega}
\end{table*}
We also constructed operators containing covariant derivatives. Since the spectrum showed no visible change, they are not listed here.

\section{Three-Body Correlation-Function Contractions}
The lattice energies $aE$ are extracted from the exponential decay of correlation functions in Euclidean time. Different types of operators constructed in the previous chapter give rise to different contraction diagrams. The number of quark-contraction diagrams grows factorially with the number of hadrons in the operator. For example, in $(I=0)$ $\pi\pi \to \pi\pi$, when both the source and the sink are two-body operators, there are $9$ diagrams, whereas in $(I=0)$ $\pi\pi\pi \to \pi\pi\pi$ there are $202$ diagrams. Most of these diagrams also contain disconnected quark-annihilation subdiagrams. We therefore continue to use the distillation method~\cite{HadronSpectrum:2009krc} to compute all quark propagators and construct the correlation-function matrix $C_{ij}$ from them.

We now discuss the topologies relevant for this chapter.
\subsection{Contraction Topologies}
\subsubsection{$I=1$ $\pi$ Single-Particle Operator}
\begin{figure}[htbp]
\centering
\raisebox{-0.5\height}{\begin{subfigure}{0.20\linewidth}\centering\begin{tikzpicture}[global scale = 1.3]
    \node[circle, draw=black, fill=black!20, thick, draw=none, font = \scriptsize, anchor = east, minimum size=15pt, inner sep=0pt] (sink1) at (0, 0) {$\pi$};
    \node[circle, draw=black, fill=black!20, thick, draw=none, font = \scriptsize, anchor = east, minimum size=15pt, inner sep=0pt] (source1) at (1.618, 0) {$\pi$};
    \draw[thin, -Stealth, out=150, in=30] (source1) to (sink1);
    \draw[thin, -Stealth, out=330, in=210] (sink1) to (source1);
\end{tikzpicture}\caption{$\mathbbm{C}$}\end{subfigure}}
\caption{Contraction diagram for the $I=1$ single-$\pi$ operator.}
\label{fig:diagrams-pi}
\end{figure}
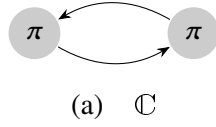

For completeness, the contraction diagram for the single $\pi$ meson operator is shown in Fig.~\ref{fig:diagrams-pi}. Each circular node represents a meson operator with definite momentum, and the arrows denote quark propagators. Since only light quarks appear in this chapter, it is unnecessary to label the flavor of the propagators explicitly.

\subsubsection{$I=2,1,0$ $\pi\pi / \rho\pi$ Two-Body Systems}
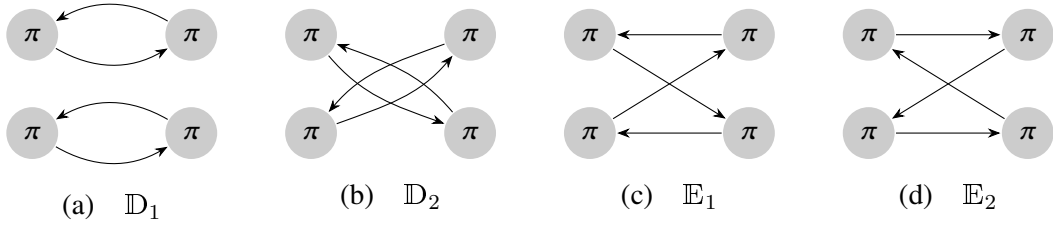
\begin{figure}[htbp]
\centering
\raisebox{-0.5\height}{\begin{subfigure}{0.20\linewidth}\centering\begin{tikzpicture}[global scale = 1.3]
    \node[circle, draw=black, fill=black!20, thick, draw=none, font = \scriptsize, anchor = east, minimum size=15pt, inner sep=0pt] (sink1) at (0, 1) {$\pi$};
    \node[circle, draw=black, fill=black!20, thick, draw=none, font = \scriptsize, anchor = east, minimum size=15pt, inner sep=0pt] (sink2) at (0, 0) {$\pi$};
    \node[circle, draw=black, fill=black!20, thick, draw=none, font = \scriptsize, anchor = east, minimum size=15pt, inner sep=0pt] (source1) at (1.618, 1) {$\pi$};
    \node[circle, draw=black, fill=black!20, thick, draw=none, font = \scriptsize, anchor = east, minimum size=15pt, inner sep=0pt] (source2) at (1.618, 0) {$\pi$};
    \draw[thin, -Stealth, out=150, in=30] (source1) to (sink1);
    \draw[thin, -Stealth, out=330, in=210] (sink1) to (source1);
    \draw[thin, -Stealth, out=150, in=30] (source2) to (sink2);
    \draw[thin, -Stealth, out=330, in=210] (sink2) to (source2);
\end{tikzpicture}\caption{$\mathbbm{D}_1$}\end{subfigure}}
\quad
\raisebox{-0.5\height}{\begin{subfigure}{0.20\linewidth}\centering\begin{tikzpicture}[global scale = 1.3]
    \node[circle, draw=black, fill=black!20, thick, draw=none, font = \scriptsize, anchor = east, minimum size=15pt, inner sep=0pt] (sink1) at (0, 1) {$\pi$};
    \node[circle, draw=black, fill=black!20, thick, draw=none, font = \scriptsize, anchor = east, minimum size=15pt, inner sep=0pt] (sink2) at (0, 0) {$\pi$};
    \node[circle, draw=black, fill=black!20, thick, draw=none, font = \scriptsize, anchor = east, minimum size=15pt, inner sep=0pt] (source1) at (1.618, 1) {$\pi$};
    \node[circle, draw=black, fill=black!20, thick, draw=none, font = \scriptsize, anchor = east, minimum size=15pt, inner sep=0pt] (source2) at (1.618, 0) {$\pi$};
    \draw[thin, -Stealth, out=200, in=50] (source1) to (sink2);
    \draw[thin, -Stealth, out=20, in=230] (sink2) to (source1);
    \draw[thin, -Stealth, out=130, in=340] (source2) to (sink1);
    \draw[thin, -Stealth, out=310, in=160] (sink1) to (source2);
\end{tikzpicture}\caption{$\mathbbm{D}_2$}\end{subfigure}}
\quad
\raisebox{-0.5\height}{\begin{subfigure}{0.20\linewidth}\centering\begin{tikzpicture}[global scale = 1.3]
    \node[circle, draw=black, fill=black!20, thick, draw=none, font = \scriptsize, anchor = east, minimum size=15pt, inner sep=0pt] (sink1) at (0, 1) {$\pi$};
    \node[circle, draw=black, fill=black!20, thick, draw=none, font = \scriptsize, anchor = east, minimum size=15pt, inner sep=0pt] (sink2) at (0, 0) {$\pi$};
    \node[circle, draw=black, fill=black!20, thick, draw=none, font = \scriptsize, anchor = east, minimum size=15pt, inner sep=0pt] (source1) at (1.618, 1) {$\pi$};
    \node[circle, draw=black, fill=black!20, thick, draw=none, font = \scriptsize, anchor = east, minimum size=15pt, inner sep=0pt] (source2) at (1.618, 0) {$\pi$};
    \draw[thin, -Stealth] (source1) to (sink1);
    \draw[thin, -Stealth] (sink1) to (source2);
    \draw[thin, -Stealth] (source2) to (sink2);
    \draw[thin, -Stealth] (sink2) to (source1);
\end{tikzpicture}\caption{$\mathbbm{E}_1$}\end{subfigure}}
\quad
\raisebox{-0.5\height}{\begin{subfigure}{0.20\linewidth}\centering\begin{tikzpicture}[global scale = 1.3]
    \node[circle, draw=black, fill=black!20, thick, draw=none, font = \scriptsize, anchor = east, minimum size=15pt, inner sep=0pt] (sink1) at (0, 1) {$\pi$};
    \node[circle, draw=black, fill=black!20, thick, draw=none, font = \scriptsize, anchor = east, minimum size=15pt, inner sep=0pt] (sink2) at (0, 0) {$\pi$};
    \node[circle, draw=black, fill=black!20, thick, draw=none, font = \scriptsize, anchor = east, minimum size=15pt, inner sep=0pt] (source1) at (1.618, 1) {$\pi$};
    \node[circle, draw=black, fill=black!20, thick, draw=none, font = \scriptsize, anchor = east, minimum size=15pt, inner sep=0pt] (source2) at (1.618, 0) {$\pi$};
    \draw[thin, -Stealth] (source1) to (sink2);
    \draw[thin, -Stealth] (sink2) to (source2);
    \draw[thin, -Stealth] (source2) to (sink1);
    \draw[thin, -Stealth] (sink1) to (source1);
\end{tikzpicture}\caption{$\mathbbm{E}_2$}\end{subfigure}}
\caption{Contraction diagrams for two-body operators in the $I=2$ $\pi\pi$ system.}
\label{fig:diagrams-pipi-I=2}
\end{figure}

For contractions between two-body operators in the $I=2$ $\pi\pi$ system, there are only the four diagrams shown in Fig.~\ref{fig:diagrams-pipi-I=2}. Here $\mathbbm{D}_1$ and $\mathbbm{D}_2$ have the same topology, as do $\mathbbm{E}_1$ and $\mathbbm{E}_2$\footnotecircle{The symbol $\mathbbm{D}$ indicates that the diagram is decomposable, while $\mathbbm{E}$ is used because the shape resembles the number $8$.}.

Since operators with covariant derivatives are not considered here, all nodes carry the same gamma matrix. With our convention, no additional minus sign is attached to the contraction diagrams; the signs arising from Grassmann algebra are absorbed into the coefficients of the individual diagrams, namely
\begin{equation}
    \langle \mathcal{O}_{\pi\pi,\Gamma,p}^{[I=2]}(t^{\prime}) \mathcal{O}_{\pi\pi,\Gamma,p}^{[I=2] \dagger}(t) \rangle = \sum_{\beta\alpha ji} \left( \mathbbm{D}_1 + \mathbbm{D}_2 - \mathbbm{E}_1 - \mathbbm{E}_2 \right)_{[\gamma_5,\gamma_5;\gamma_5,\gamma_5], [\beta,P-\beta;-\alpha,-(P-\alpha)]}(t^{\prime},t).
\end{equation}
The notation is the same as that used for the $D\pi$ system in\chapref{chap:two_body_problems}.

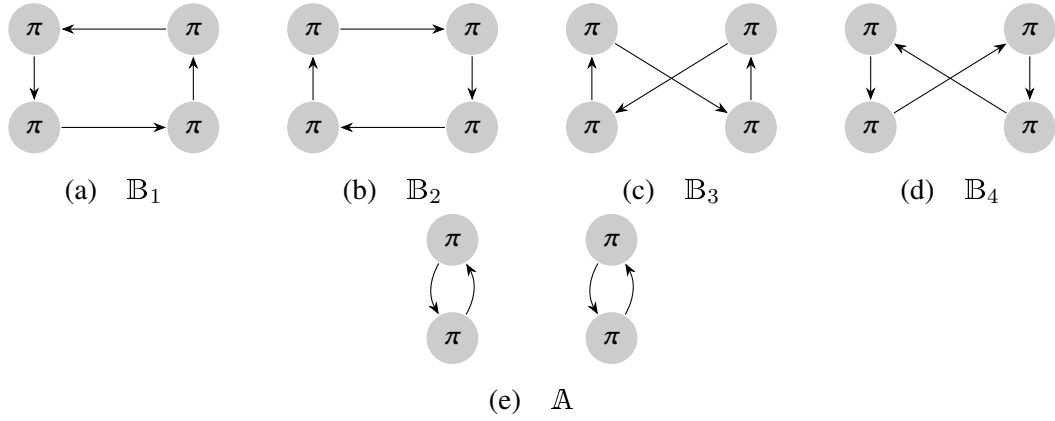
\begin{figure}[htbp]
\centering
\raisebox{-0.5\height}{\begin{subfigure}{0.20\linewidth}\centering\begin{tikzpicture}[global scale = 1.3]
    \node[circle, draw=black, fill=black!20, thick, draw=none, font = \scriptsize, anchor = east, minimum size=15pt, inner sep=0pt] (sink1) at (0, 1) {$\pi$};
    \node[circle, draw=black, fill=black!20, thick, draw=none, font = \scriptsize, anchor = east, minimum size=15pt, inner sep=0pt] (sink2) at (0, 0) {$\pi$};
    \node[circle, draw=black, fill=black!20, thick, draw=none, font = \scriptsize, anchor = east, minimum size=15pt, inner sep=0pt] (source1) at (1.618, 1) {$\pi$};
    \node[circle, draw=black, fill=black!20, thick, draw=none, font = \scriptsize, anchor = east, minimum size=15pt, inner sep=0pt] (source2) at (1.618, 0) {$\pi$};
    \draw[thin, -Stealth] (source1) to (sink1);
    \draw[thin, -Stealth] (sink1) to (sink2);
    \draw[thin, -Stealth] (sink2) to (source2);
    \draw[thin, -Stealth] (source2) to (source1);
\end{tikzpicture}\caption{$\mathbbm{B}_1$}\end{subfigure}}
\quad
\raisebox{-0.5\height}{\begin{subfigure}{0.20\linewidth}\centering\begin{tikzpicture}[global scale = 1.3]
    \node[circle, draw=black, fill=black!20, thick, draw=none, font = \scriptsize, anchor = east, minimum size=15pt, inner sep=0pt] (sink1) at (0, 1) {$\pi$};
    \node[circle, draw=black, fill=black!20, thick, draw=none, font = \scriptsize, anchor = east, minimum size=15pt, inner sep=0pt] (sink2) at (0, 0) {$\pi$};
    \node[circle, draw=black, fill=black!20, thick, draw=none, font = \scriptsize, anchor = east, minimum size=15pt, inner sep=0pt] (source1) at (1.618, 1) {$\pi$};
    \node[circle, draw=black, fill=black!20, thick, draw=none, font = \scriptsize, anchor = east, minimum size=15pt, inner sep=0pt] (source2) at (1.618, 0) {$\pi$};
    \draw[thin, -Stealth] (source1) to (source2);
    \draw[thin, -Stealth] (source2) to (sink2);
    \draw[thin, -Stealth] (sink2) to (sink1);
    \draw[thin, -Stealth] (sink1) to (source1);
\end{tikzpicture}\caption{$\mathbbm{B}_2$}\end{subfigure}}
\quad
\raisebox{-0.5\height}{\begin{subfigure}{0.20\linewidth}\centering\begin{tikzpicture}[global scale = 1.3]
    \node[circle, draw=black, fill=black!20, thick, draw=none, font = \scriptsize, anchor = east, minimum size=15pt, inner sep=0pt] (sink1) at (0, 1) {$\pi$};
    \node[circle, draw=black, fill=black!20, thick, draw=none, font = \scriptsize, anchor = east, minimum size=15pt, inner sep=0pt] (sink2) at (0, 0) {$\pi$};
    \node[circle, draw=black, fill=black!20, thick, draw=none, font = \scriptsize, anchor = east, minimum size=15pt, inner sep=0pt] (source1) at (1.618, 1) {$\pi$};
    \node[circle, draw=black, fill=black!20, thick, draw=none, font = \scriptsize, anchor = east, minimum size=15pt, inner sep=0pt] (source2) at (1.618, 0) {$\pi$};
    \draw[thin, -Stealth] (source1) to (sink2);
    \draw[thin, -Stealth] (sink2) to (sink1);
    \draw[thin, -Stealth] (sink1) to (source2);
    \draw[thin, -Stealth] (source2) to (source1);
\end{tikzpicture}\caption{$\mathbbm{B}_3$}\end{subfigure}}
\quad
\raisebox{-0.5\height}{\begin{subfigure}{0.20\linewidth}\centering\begin{tikzpicture}[global scale = 1.3]
    \node[circle, draw=black, fill=black!20, thick, draw=none, font = \scriptsize, anchor = east, minimum size=15pt, inner sep=0pt] (sink1) at (0, 1) {$\pi$};
    \node[circle, draw=black, fill=black!20, thick, draw=none, font = \scriptsize, anchor = east, minimum size=15pt, inner sep=0pt] (sink2) at (0, 0) {$\pi$};
    \node[circle, draw=black, fill=black!20, thick, draw=none, font = \scriptsize, anchor = east, minimum size=15pt, inner sep=0pt] (source1) at (1.618, 1) {$\pi$};
    \node[circle, draw=black, fill=black!20, thick, draw=none, font = \scriptsize, anchor = east, minimum size=15pt, inner sep=0pt] (source2) at (1.618, 0) {$\pi$};
    \draw[thin, -Stealth] (source1) to (source2);
    \draw[thin, -Stealth] (source2) to (sink1);
    \draw[thin, -Stealth] (sink1) to (sink2);
    \draw[thin, -Stealth] (sink2) to (source1);
\end{tikzpicture}\caption{$\mathbbm{B}_4$}\end{subfigure}}
\\
\raisebox{-0.5\height}{\begin{subfigure}{0.20\linewidth}\centering\begin{tikzpicture}[global scale = 1.3]
    \node[circle, draw=black, fill=black!20, thick, draw=none, font = \scriptsize, anchor = east, minimum size=15pt, inner sep=0pt] (sink1) at (0, 1) {$\pi$};
    \node[circle, draw=black, fill=black!20, thick, draw=none, font = \scriptsize, anchor = east, minimum size=15pt, inner sep=0pt] (sink2) at (0, 0) {$\pi$};
    \node[circle, draw=black, fill=black!20, thick, draw=none, font = \scriptsize, anchor = east, minimum size=15pt, inner sep=0pt] (source1) at (1.618, 1) {$\pi$};
    \node[circle, draw=black, fill=black!20, thick, draw=none, font = \scriptsize, anchor = east, minimum size=15pt, inner sep=0pt] (source2) at (1.618, 0) {$\pi$};
    \draw[thin, -Stealth, out=240, in=120] (source1) to (source2);
    \draw[thin, -Stealth, out=60, in=300] (source2) to (source1);
    \draw[thin, -Stealth, out=240, in=120] (sink1) to (sink2);
    \draw[thin, -Stealth, out=60, in=300] (sink2) to (sink1);
\end{tikzpicture}\caption{$\mathbbm{A}$}\end{subfigure}}
    \caption{Additional contraction topologies for two-body operators in the $I=1$ and $I=0$ $\pi\pi$ channels.}
\label{fig:diagrams-pipi-I=10}
\end{figure}

For the $I=1$ and $I=0$ channels, in addition to the diagrams in Fig.~\ref{fig:diagrams-pipi-I=2}, there are box diagrams $\mathbbm{B}$ and quark-annihilation diagrams $\mathbbm{A}$ at a single time slice, as shown in Fig.~\ref{fig:diagrams-pipi-I=10}.

We can therefore write the correlation functions for $I=1$ and $I=0$ $\pi\pi$ as
\begin{equation}
\begin{aligned}
    &\langle \mathcal{O}_{\pi\pi,\Gamma,p}^{[I=1]}(t^{\prime}) \mathcal{O}_{\pi\pi,\Gamma,p}^{[I=1] \dagger}(t) \rangle \\
    &= \sum_{\beta\alpha ji} \left( \mathbbm{D}_1 - \mathbbm{D}_2 - \mathbbm{B}_1 - \mathbbm{B}_2 + \mathbbm{B}_3 + \mathbbm{B}_4 \right)_{[\gamma_5,\gamma_5;\gamma_5,\gamma_5], [\beta,P-\beta;-\alpha,-(P-\alpha)]}(t^{\prime},t), \\
\end{aligned}
\end{equation}
and
\begin{equation}
\begin{aligned}
    &\langle \mathcal{O}_{\pi\pi,\Gamma,p}^{[I=0]}(t^{\prime}) \mathcal{O}_{\pi\pi,\Gamma,p}^{[I=0] \dagger}(t) \rangle \\
    &= \sum_{\beta\alpha ji} \left( \mathbbm{D}_1 + \mathbbm{D}_2 - \frac{3}{2} \mathbbm{B}_1 - \frac{3}{2}\mathbbm{B}_2 - \frac{3}{2}\mathbbm{B}_3 - \frac{3}{2}\mathbbm{B}_4 + \frac{1}{2} \mathbbm{E}_1 + \frac{1}{2} \mathbbm{E}_2 + 3\mathbbm{A} \right)_{[\gamma_5,\gamma_5;\gamma_5,\gamma_5], [\beta,P-\beta;-\alpha,-(P-\alpha)]}(t^{\prime},t).
\end{aligned}
\end{equation}

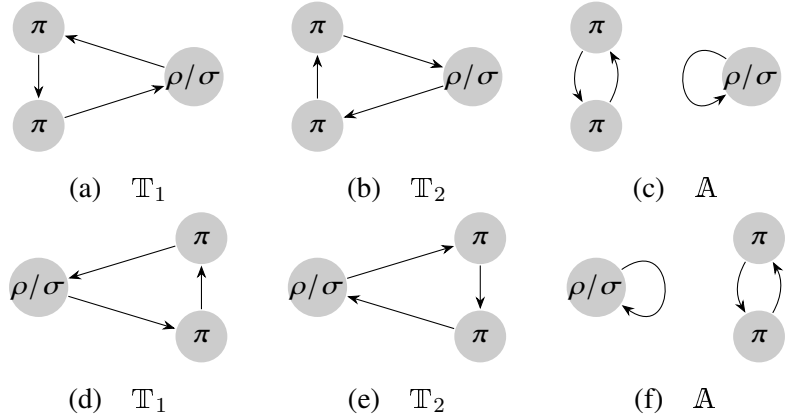
\begin{figure}[htbp]
\centering
\raisebox{-0.5\height}{\begin{subfigure}{0.20\linewidth}\centering\begin{tikzpicture}[global scale = 1.3]
    \node[circle, draw=black, fill=black!20, thick, draw=none, font = \scriptsize, anchor = east, minimum size=15pt, inner sep=0pt] (sink1) at (0, 1) {$\pi$};
    \node[circle, draw=black, fill=black!20, thick, draw=none, font = \scriptsize, anchor = east, minimum size=15pt, inner sep=0pt] (sink2) at (0, 0) {$\pi$};
    \node[circle, draw=black, fill=black!20, thick, draw=none, font = \scriptsize, anchor = east, minimum size=15pt, inner sep=0pt] (source1) at (1.618, 0.5) {$\rho / \sigma$};
    \draw[thin, -Stealth] (source1) to (sink1);
    \draw[thin, -Stealth] (sink1) to (sink2);
    \draw[thin, -Stealth] (sink2) to (source1);
\end{tikzpicture}\caption{$\mathbbm{T}_1$}\end{subfigure}}
\quad
\raisebox{-0.5\height}{\begin{subfigure}{0.20\linewidth}\centering\begin{tikzpicture}[global scale = 1.3]
    \node[circle, draw=black, fill=black!20, thick, draw=none, font = \scriptsize, anchor = east, minimum size=15pt, inner sep=0pt] (sink1) at (0, 1) {$\pi$};
    \node[circle, draw=black, fill=black!20, thick, draw=none, font = \scriptsize, anchor = east, minimum size=15pt, inner sep=0pt] (sink2) at (0, 0) {$\pi$};
    \node[circle, draw=black, fill=black!20, thick, draw=none, font = \scriptsize, anchor = east, minimum size=15pt, inner sep=0pt] (source1) at (1.618, 0.5) {$\rho / \sigma$};
    \draw[thin, -Stealth] (source1) to (sink2);
    \draw[thin, -Stealth] (sink2) to (sink1);
    \draw[thin, -Stealth] (sink1) to (source1);
\end{tikzpicture}\caption{$\mathbbm{T}_2$}\end{subfigure}}
\quad
\raisebox{-0.5\height}{\begin{subfigure}{0.20\linewidth}\centering\begin{tikzpicture}[global scale = 1.3]
    \node[circle, draw=black, fill=black!20, thick, draw=none, font = \scriptsize, anchor = east, minimum size=15pt, inner sep=0pt] (sink1) at (0, 1) {$\pi$};
    \node[circle, draw=black, fill=black!20, thick, draw=none, font = \scriptsize, anchor = east, minimum size=15pt, inner sep=0pt] (sink2) at (0, 0) {$\pi$};
    \node[circle, draw=black, fill=black!20, thick, draw=none, font = \scriptsize, anchor = east, minimum size=15pt, inner sep=0pt] (source1) at (1.618, 0.5) {$\rho / \sigma$};
    \draw[thin, -Stealth, out=240, in=120] (sink1) to (sink2);
    \draw[thin, -Stealth, out=60, in=300] (sink2) to (sink1);
    \draw[thin, -Stealth, out=145, in=215, looseness=5] (source1) to (source1);
\end{tikzpicture}\caption{$\mathbbm{A}$}\end{subfigure}}
\\
\raisebox{-0.5\height}{\begin{subfigure}{0.20\linewidth}\centering\begin{tikzpicture}[global scale = 1.3]
    \node[circle, draw=black, fill=black!20, thick, draw=none, font = \scriptsize, anchor = east, minimum size=15pt, inner sep=0pt] (sink1) at (0, 0.5) {$\rho / \sigma$};
    \node[circle, draw=black, fill=black!20, thick, draw=none, font = \scriptsize, anchor = east, minimum size=15pt, inner sep=0pt] (source1) at (1.618, 1) {$\pi$};
    \node[circle, draw=black, fill=black!20, thick, draw=none, font = \scriptsize, anchor = east, minimum size=15pt, inner sep=0pt] (source2) at (1.618, 0) {$\pi$};
    \draw[thin, -Stealth] (source1) to (sink1);
    \draw[thin, -Stealth] (sink1) to (source2);
    \draw[thin, -Stealth] (source2) to (source1);
\end{tikzpicture}\caption{$\mathbbm{T}_1$}\end{subfigure}}
\quad
\raisebox{-0.5\height}{\begin{subfigure}{0.20\linewidth}\centering\begin{tikzpicture}[global scale = 1.3]
    \node[circle, draw=black, fill=black!20, thick, draw=none, font = \scriptsize, anchor = east, minimum size=15pt, inner sep=0pt] (sink1) at (0, 0.5) {$\rho / \sigma$};
    \node[circle, draw=black, fill=black!20, thick, draw=none, font = \scriptsize, anchor = east, minimum size=15pt, inner sep=0pt] (source1) at (1.618, 1) {$\pi$};
    \node[circle, draw=black, fill=black!20, thick, draw=none, font = \scriptsize, anchor = east, minimum size=15pt, inner sep=0pt] (source2) at (1.618, 0) {$\pi$};
    \draw[thin, -Stealth] (source1) to (source2);
    \draw[thin, -Stealth] (source2) to (sink1);
    \draw[thin, -Stealth] (sink1) to (source1);
\end{tikzpicture}\caption{$\mathbbm{T}_2$}\end{subfigure}}
\quad
\raisebox{-0.5\height}{\begin{subfigure}{0.20\linewidth}\centering\begin{tikzpicture}[global scale = 1.3]
    \node[circle, draw=black, fill=black!20, thick, draw=none, font = \scriptsize, anchor = east, minimum size=15pt, inner sep=0pt] (sink1) at (0, 0.5) {$\rho / \sigma$};
    \node[circle, draw=black, fill=black!20, thick, draw=none, font = \scriptsize, anchor = east, minimum size=15pt, inner sep=0pt] (source1) at (1.618, 1) {$\pi$};
    \node[circle, draw=black, fill=black!20, thick, draw=none, font = \scriptsize, anchor = east, minimum size=15pt, inner sep=0pt] (source2) at (1.618, 0) {$\pi$};
    \draw[thin, -Stealth, out=240, in=120] (source1) to (source2);
    \draw[thin, -Stealth, out=60, in=300] (source2) to (source1);
    \draw[thin, -Stealth, out=35, in=325, looseness=5] (sink1) to (sink1);
\end{tikzpicture}\caption{$\mathbbm{A}$}\end{subfigure}}
\caption{Off-diagonal contractions between one-body and two-body operators in the $I=1$ and $I=0$ $\pi\pi$ channels.}
\label{fig:diagrams-pipi-off-I=10}
\end{figure}

The off-diagonal matrix elements are also needed; the corresponding diagrams are shown in Fig.~\ref{fig:diagrams-pipi-off-I=10}. The notation $\mathbbm{T}$ denotes triangle diagrams, with $\mathbbm{T}_1$ and $\mathbbm{T}_2$ representing the same topology, while $\mathbbm{A}$ denotes an annihilation diagram.

The correlation functions are
\begin{equation}
\begin{aligned}
    \langle \mathcal{O}_{\pi\pi,\Gamma,p}^{[I=1]}(t^{\prime}) \mathcal{O}_{\rho,\Gamma,p}^{[I=1] \dagger}(t) \rangle &= \sum_{\beta ji} \left( \mathbbm{T}_1 - \mathbbm{T}_2 \right)_{[\gamma_5,\gamma_5;\gamma_i], [\beta,P-\beta;-P]}(t^{\prime},t), \\
    \langle \mathcal{O}_{\rho,\Gamma,p}^{[I=1]}(t^{\prime}) \mathcal{O}_{\pi\pi,\Gamma,p}^{[I=1] \dagger}(t) \rangle &= \sum_{\alpha ji} \left( \mathbbm{T}_1 - \mathbbm{T}_2 \right)_{[\gamma_j;\gamma_5,\gamma_5], [P;-\alpha,-(P-\alpha)]}(t^{\prime},t), \\
\end{aligned}
\end{equation}
and
\begin{equation}
\begin{aligned}
    \langle \mathcal{O}_{\pi\pi,\Gamma,p}^{[I=0]}(t^{\prime}) \mathcal{O}_{\sigma,\Gamma,p}^{[I=0] \dagger}(t) \rangle &= \sum_{\beta ji} \left( \frac{\sqrt{6}}{2} \mathbbm{T}_1 + \frac{\sqrt{6}}{2} \mathbbm{T}_2 - \sqrt{6} \mathbbm{A} \right)_{[\gamma_5,\gamma_5;\gamma_i], [\beta,P-\beta;-P]}(t^{\prime},t), \\
    \langle \mathcal{O}_{\sigma,\Gamma,p}^{[I=0]}(t^{\prime}) \mathcal{O}_{\pi\pi,\Gamma,p}^{[I=0] \dagger}(t) \rangle &= \sum_{\alpha ji} \left( \frac{\sqrt{6}}{2} \mathbbm{T}_1 + \frac{\sqrt{6}}{2} \mathbbm{T}_2 - \sqrt{6} \mathbbm{A} \right)_{[\gamma_j;\gamma_5,\gamma_5], [P;-\alpha,-(P-\alpha)]}(t^{\prime},t). \\
\end{aligned}
\end{equation}

\subsubsection{Systems Related to $I=1$ $\sigma\pi$}
The $I=0$ $\sigma\pi$ system is not needed in this chapter, but for later use in\chapref{chap:three_body_problems2} we also give the correlation functions for the $I=1$ $\sigma\pi$ system. The contraction diagrams for $\sigma\pi$ and $\rho\pi$ are the same as those in the previous subsection; only the combinatorial coefficients differ. The $\sigma\pi-\sigma\pi$ correlation function is
\begin{equation}
    \langle \mathcal{O}_{\sigma\pi,\Gamma,p}^{[I=1]}(t^{\prime}) \mathcal{O}_{\sigma\pi,\Gamma,p}^{[I=1] \dagger}(t) \rangle \sim - \frac{1}{2} \mathbbm{B}_1 + 1 \mathbbm{D}_1 - \frac{1}{2} \mathbbm{E}_1 + \frac{1}{2} \mathbbm{B}_3 - \frac{1}{2} \mathbbm{E}_2 + \frac{1}{2} \mathbbm{B}_4 - \frac{1}{2} \mathbbm{B}_2.
\end{equation}
The tilde indicates that the summation symbols, gamma-matrix labels, and momentum-combination indices have been omitted.

The $\sigma\pi-\rho\pi$ and $\rho\pi-\sigma\pi$ correlation functions have the same form; here we display only the combinatorial coefficients:
\begin{equation}
\begin{aligned}
    &\langle \mathcal{O}_{\sigma\pi,\Gamma,p}^{[I=1]}(t^{\prime}) \mathcal{O}_{\rho\pi,\Gamma,p}^{[I=1] \dagger}(t) \rangle \sim \langle \mathcal{O}_{\rho\pi,\Gamma,p}^{[I=1]}(t^{\prime}) \mathcal{O}_{\sigma\pi,\Gamma,p}^{[I=1] \dagger}(t) \rangle \\
    &\sim \frac{1}{\sqrt{2}} \mathbbm{B}_1 - \frac{1}{\sqrt{2}} \mathbbm{D}_1 -\frac{1}{\sqrt{2}} \mathbbm{B}_3 + \frac{1}{\sqrt{2}} \mathbbm{D}_2 - \frac{1}{\sqrt{2}} \mathbbm{B}_4 + \frac{1}{\sqrt{2}} \mathbbm{B}_2.
\end{aligned}
\end{equation}
The first tilde indicates that their topologies are the same.

The correlation function between $\sigma\pi$ and the single-particle operator $\pi$ is
\begin{equation}
    \langle \mathcal{O}_{\sigma\pi,\Gamma,p}^{[I=1]}(t^{\prime}) \mathcal{O}_{\pi,\Gamma,p}^{[I=1] \dagger}(t) \rangle \sim \langle \mathcal{O}_{\pi,\Gamma,p}^{[I=1]}(t^{\prime}) \mathcal{O}_{\sigma\pi,\Gamma,p}^{[I=1] \dagger}(t) \rangle \sim - \frac{1}{\sqrt{2}} \mathbbm{T}_1 + \frac{1}{\sqrt{2}} \mathbbm{T}_2.
\end{equation}
The first tilde indicates that the two topologies are identical or related by symmetry, namely that both are the corresponding $\mathbbm{T}$ diagrams.

\subsubsection{Contractions Related to $I=0$ $\pi\pi\pi$ Operators}
\begin{figure}[htbp]
\centering
\raisebox{-0.5\height}{\begin{subfigure}{0.20\linewidth}\centering\begin{tikzpicture}[global scale = 1.3]
    \node[circle, draw=black, fill=black!20, thick, draw=none, font = \scriptsize, anchor = east, minimum size=15pt, inner sep=0pt] (sink1) at (0, 2) {$\pi$};
    \node[circle, draw=black, fill=black!20, thick, draw=none, font = \scriptsize, anchor = east, minimum size=15pt, inner sep=0pt] (sink2) at (0, 1) {$\pi$};
    \node[circle, draw=black, fill=black!20, thick, draw=none, font = \scriptsize, anchor = east, minimum size=15pt, inner sep=0pt] (sink3) at (0, 0) {$\pi$};
    \node[circle, draw=black, fill=black!20, thick, draw=none, font = \scriptsize, anchor = east, minimum size=15pt, inner sep=0pt] (source1) at (1.618, 2) {$\pi$};
    \node[circle, draw=black, fill=black!20, thick, draw=none, font = \scriptsize, anchor = east, minimum size=15pt, inner sep=0pt] (source2) at (1.618, 1) {$\pi$};
    \node[circle, draw=black, fill=black!20, thick, draw=none, font = \scriptsize, anchor = east, minimum size=15pt, inner sep=0pt] (source3) at (1.618, 0) {$\pi$};
    \draw[thin, -Stealth, out=150, in=30] (source1) to (sink1);
    \draw[thin, -Stealth, out=330, in=210] (sink1) to (source1);
    \draw[thin, -Stealth, out=150, in=30] (source2) to (sink2);
    \draw[thin, -Stealth, out=330, in=210] (sink2) to (source2);
    \draw[thin, -Stealth, out=150, in=30] (source3) to (sink3);
    \draw[thin, -Stealth, out=330, in=210] (sink3) to (source3);
\end{tikzpicture}\caption{$\mathbbm{TD}_1$}\end{subfigure}}
\quad
\raisebox{-0.5\height}{\begin{subfigure}{0.20\linewidth}\centering\begin{tikzpicture}[global scale = 1.3]
    \node[circle, draw=black, fill=black!20, thick, draw=none, font = \scriptsize, anchor = east, minimum size=15pt, inner sep=0pt] (sink1) at (0, 2) {$\pi$};
    \node[circle, draw=black, fill=black!20, thick, draw=none, font = \scriptsize, anchor = east, minimum size=15pt, inner sep=0pt] (sink2) at (0, 1) {$\pi$};
    \node[circle, draw=black, fill=black!20, thick, draw=none, font = \scriptsize, anchor = east, minimum size=15pt, inner sep=0pt] (sink3) at (0, 0) {$\pi$};
    \node[circle, draw=black, fill=black!20, thick, draw=none, font = \scriptsize, anchor = east, minimum size=15pt, inner sep=0pt] (source1) at (1.618, 2) {$\pi$};
    \node[circle, draw=black, fill=black!20, thick, draw=none, font = \scriptsize, anchor = east, minimum size=15pt, inner sep=0pt] (source2) at (1.618, 1) {$\pi$};
    \node[circle, draw=black, fill=black!20, thick, draw=none, font = \scriptsize, anchor = east, minimum size=15pt, inner sep=0pt] (source3) at (1.618, 0) {$\pi$};
    \draw[thin, -Stealth] (source1) to (sink2);
    \draw[thin, -Stealth] (sink2) to (source3);
    \draw[thin, -Stealth, out=120, in=240] (source3) to (source1);
    \draw[thin, -Stealth] (source2) to (sink1);
    \draw[thin, -Stealth, out=310, in=60] (sink1) to (sink3);
    \draw[thin, -Stealth] (sink3) to (source2);
\end{tikzpicture}\caption{$\mathbbm{TS}_1$}\end{subfigure}}
\quad
\raisebox{-0.5\height}{\begin{subfigure}{0.20\linewidth}\centering\begin{tikzpicture}[global scale = 1.3]
    \node[circle, draw=black, fill=black!20, thick, draw=none, font = \scriptsize, anchor = east, minimum size=15pt, inner sep=0pt] (sink1) at (0, 2) {$\pi$};
    \node[circle, draw=black, fill=black!20, thick, draw=none, font = \scriptsize, anchor = east, minimum size=15pt, inner sep=0pt] (sink2) at (0, 1) {$\pi$};
    \node[circle, draw=black, fill=black!20, thick, draw=none, font = \scriptsize, anchor = east, minimum size=15pt, inner sep=0pt] (sink3) at (0, 0) {$\pi$};
    \node[circle, draw=black, fill=black!20, thick, draw=none, font = \scriptsize, anchor = east, minimum size=15pt, inner sep=0pt] (source1) at (1.618, 2) {$\pi$};
    \node[circle, draw=black, fill=black!20, thick, draw=none, font = \scriptsize, anchor = east, minimum size=15pt, inner sep=0pt] (source2) at (1.618, 1) {$\pi$};
    \node[circle, draw=black, fill=black!20, thick, draw=none, font = \scriptsize, anchor = east, minimum size=15pt, inner sep=0pt] (source3) at (1.618, 0) {$\pi$};
    \draw[thin, -Stealth] (source1) to (sink1);
    \draw[thin, -Stealth] (sink1) to (sink2);
    \draw[thin, -Stealth] (sink2) to (source2);
    \draw[thin, -Stealth] (source2) to (source1);
    \draw[thin, -Stealth, out=150, in=30] (source3) to (sink3);
    \draw[thin, -Stealth, out=330, in=210] (sink3) to (source3);
\end{tikzpicture}\caption{$\mathbbm{TR}_1$}\end{subfigure}}
\quad
\raisebox{-0.5\height}{\begin{subfigure}{0.20\linewidth}\centering\begin{tikzpicture}[global scale = 1.3]
    \node[circle, draw=black, fill=black!20, thick, draw=none, font = \scriptsize, anchor = east, minimum size=15pt, inner sep=0pt] (sink1) at (0, 2) {$\pi$};
    \node[circle, draw=black, fill=black!20, thick, draw=none, font = \scriptsize, anchor = east, minimum size=15pt, inner sep=0pt] (sink2) at (0, 1) {$\pi$};
    \node[circle, draw=black, fill=black!20, thick, draw=none, font = \scriptsize, anchor = east, minimum size=15pt, inner sep=0pt] (sink3) at (0, 0) {$\pi$};
    \node[circle, draw=black, fill=black!20, thick, draw=none, font = \scriptsize, anchor = east, minimum size=15pt, inner sep=0pt] (source1) at (1.618, 2) {$\pi$};
    \node[circle, draw=black, fill=black!20, thick, draw=none, font = \scriptsize, anchor = east, minimum size=15pt, inner sep=0pt] (source2) at (1.618, 1) {$\pi$};
    \node[circle, draw=black, fill=black!20, thick, draw=none, font = \scriptsize, anchor = east, minimum size=15pt, inner sep=0pt] (source3) at (1.618, 0) {$\pi$};
    \draw[thin, -Stealth] (source1) to (source2);
    \draw[thin, -Stealth] (source2) to (source3);
    \draw[thin, -Stealth, out=140, in=220] (source3) to (source1);
    \draw[thin, -Stealth] (sink1) to (sink2);
    \draw[thin, -Stealth] (sink2) to (sink3);
    \draw[thin, -Stealth, out=40, in=320] (sink3) to (sink1);
\end{tikzpicture}\caption{$\mathbbm{TA}_1$}\end{subfigure}}
\\
\raisebox{-0.5\height}{\begin{subfigure}{0.20\linewidth}\centering\begin{tikzpicture}[global scale = 1.3]
    \node[circle, draw=black, fill=black!20, thick, draw=none, font = \scriptsize, anchor = east, minimum size=15pt, inner sep=0pt] (sink1) at (0, 2) {$\pi$};
    \node[circle, draw=black, fill=black!20, thick, draw=none, font = \scriptsize, anchor = east, minimum size=15pt, inner sep=0pt] (sink2) at (0, 1) {$\pi$};
    \node[circle, draw=black, fill=black!20, thick, draw=none, font = \scriptsize, anchor = east, minimum size=15pt, inner sep=0pt] (sink3) at (0, 0) {$\pi$};
    \node[circle, draw=black, fill=black!20, thick, draw=none, font = \scriptsize, anchor = east, minimum size=15pt, inner sep=0pt] (source1) at (1.618, 2) {$\pi$};
    \node[circle, draw=black, fill=black!20, thick, draw=none, font = \scriptsize, anchor = east, minimum size=15pt, inner sep=0pt] (source2) at (1.618, 1) {$\pi$};
    \node[circle, draw=black, fill=black!20, thick, draw=none, font = \scriptsize, anchor = east, minimum size=15pt, inner sep=0pt] (source3) at (1.618, 0) {$\pi$};
    \draw[thin, -Stealth] (source1) to (sink1);
    \draw[thin, -Stealth] (sink1) to (source2);
    \draw[thin, -Stealth] (source2) to (sink2);
    \draw[thin, -Stealth] (sink2) to (source3);
    \draw[thin, -Stealth] (source3) to (sink3);
    \draw[thin, -Stealth] (sink3) to (source1);
\end{tikzpicture}\caption{$\mathbbm{TZ}_1$}\end{subfigure}}
\quad
\raisebox{-0.5\height}{\begin{subfigure}{0.20\linewidth}\centering\begin{tikzpicture}[global scale = 1.3]
    \node[circle, draw=black, fill=black!20, thick, draw=none, font = \scriptsize, anchor = east, minimum size=15pt, inner sep=0pt] (sink1) at (0, 2) {$\pi$};
    \node[circle, draw=black, fill=black!20, thick, draw=none, font = \scriptsize, anchor = east, minimum size=15pt, inner sep=0pt] (sink2) at (0, 1) {$\pi$};
    \node[circle, draw=black, fill=black!20, thick, draw=none, font = \scriptsize, anchor = east, minimum size=15pt, inner sep=0pt] (sink3) at (0, 0) {$\pi$};
    \node[circle, draw=black, fill=black!20, thick, draw=none, font = \scriptsize, anchor = east, minimum size=15pt, inner sep=0pt] (source1) at (1.618, 2) {$\pi$};
    \node[circle, draw=black, fill=black!20, thick, draw=none, font = \scriptsize, anchor = east, minimum size=15pt, inner sep=0pt] (source2) at (1.618, 1) {$\pi$};
    \node[circle, draw=black, fill=black!20, thick, draw=none, font = \scriptsize, anchor = east, minimum size=15pt, inner sep=0pt] (source3) at (1.618, 0) {$\pi$};
    \draw[thin, -Stealth] (source1) to (sink1);
    \draw[thin, -Stealth] (sink1) to (sink2);
    \draw[thin, -Stealth] (sink2) to (sink3);
    \draw[thin, -Stealth] (sink3) to (source3);
    \draw[thin, -Stealth] (source3) to (source2);
    \draw[thin, -Stealth] (source2) to (source1);
\end{tikzpicture}\caption{$\mathbbm{TB}_1$}\end{subfigure}}
\quad
\raisebox{-0.5\height}{\begin{subfigure}{0.20\linewidth}\centering\begin{tikzpicture}[global scale = 1.3]
    \node[circle, draw=black, fill=black!20, thick, draw=none, font = \scriptsize, anchor = east, minimum size=15pt, inner sep=0pt] (sink1) at (0, 2) {$\pi$};
    \node[circle, draw=black, fill=black!20, thick, draw=none, font = \scriptsize, anchor = east, minimum size=15pt, inner sep=0pt] (sink2) at (0, 1) {$\pi$};
    \node[circle, draw=black, fill=black!20, thick, draw=none, font = \scriptsize, anchor = east, minimum size=15pt, inner sep=0pt] (sink3) at (0, 0) {$\pi$};
    \node[circle, draw=black, fill=black!20, thick, draw=none, font = \scriptsize, anchor = east, minimum size=15pt, inner sep=0pt] (source1) at (1.618, 2) {$\pi$};
    \node[circle, draw=black, fill=black!20, thick, draw=none, font = \scriptsize, anchor = east, minimum size=15pt, inner sep=0pt] (source2) at (1.618, 1) {$\pi$};
    \node[circle, draw=black, fill=black!20, thick, draw=none, font = \scriptsize, anchor = east, minimum size=15pt, inner sep=0pt] (source3) at (1.618, 0) {$\pi$};
    \draw[thin, -Stealth] (source1) to (sink1);
    \draw[thin, -Stealth] (sink1) to (source2);
    \draw[thin, -Stealth] (source2) to (source3);
    \draw[thin, -Stealth] (source3) to (sink3);
    \draw[thin, -Stealth] (sink3) to (sink2);
    \draw[thin, -Stealth] (sink2) to (source1);
\end{tikzpicture}\caption{$\mathbbm{TW}_1$}\end{subfigure}}
\caption{Several representative topologies for contractions of $I=0$ $\pi\pi\pi$ three-body operators.}
\label{fig:diagrams-pipipi-I=0}
\end{figure}
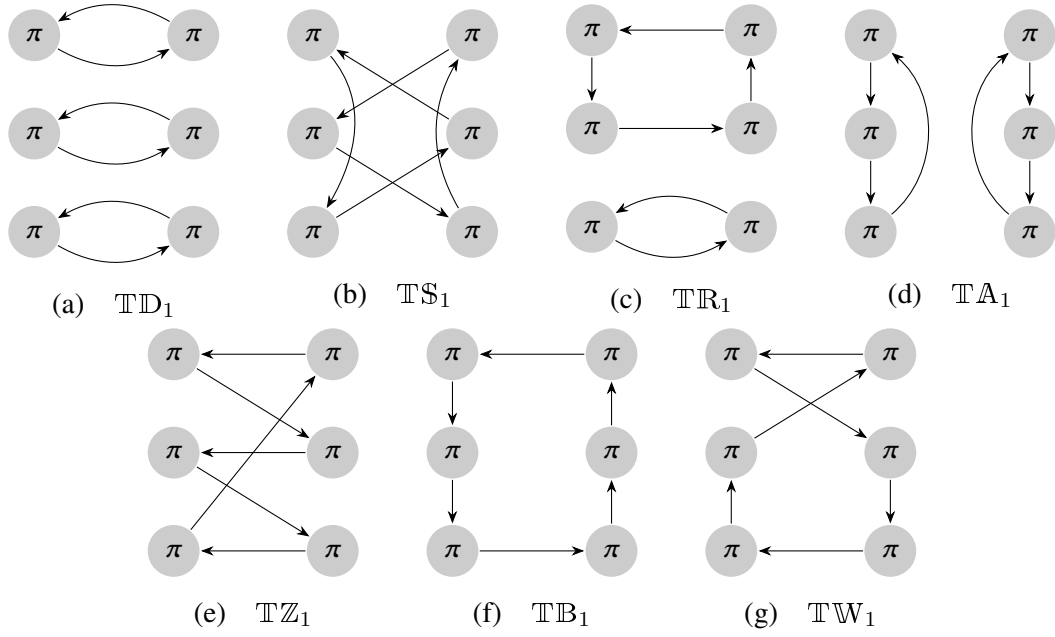

The number of contraction diagrams for three-body operators is much larger than in two-body and one-body systems. For example, the diagonal correlation function of $\pi\pi\pi$ operators contains $202$ diagrams~\footnotecircle{They are sometimes called six-point functions according to the number of hadrons.}. We wrote a \verb|Mathematica| code to generate all diagrams, their corresponding coefficients, and the contraction code. For reasons of space, we do not draw all diagrams here, but instead show their topologies in Fig.~\ref{fig:diagrams-pipipi-I=0}\footnotecircle{The naming convention is as follows: $\mathbbm{TD}$ denotes three-body decomposable, although many other topologies are also decomposable in practice; $\mathbbm{TZ}$ denotes a three-body zigzag topology; $\mathbbm{TS}$ denotes a three-body star topology; $\mathbbm{TB}$ denotes a box topology; $\mathbbm{TA}$ denotes an annihilation topology; $\mathbbm{TW}$ is named because the inverted diagram resembles a wine glass; and $\mathbbm{TR}$ is named because, if the two two-point correlated pairs are placed in the middle, the diagram resembles the Chinese character ``日 (rì)''. The same naming convention was also used in our published paper.}. The subscript $1$ denotes the first diagram of a given topology.

The topologies shown above exhaust all possible cases. The remaining quark-contraction diagrams may look more complicated because the quark lines are intertwined differently, but they can always be mapped back to these topologies by permuting the quark positions. All of these diagrams can be expressed as combinations of irreducible two-point, four-point, and six-point building blocks; in the actual computation, these building blocks are generated first and then assembled in the appropriate order. Moreover, exchange symmetry implies that the linear-combination coefficients associated with each topology are identical.

\begin{figure}[htbp]
\centering
\raisebox{-0.5\height}{\begin{subfigure}{0.20\linewidth}\centering\begin{tikzpicture}[global scale = 1.3]
    \node[circle, draw=black, fill=black!20, thick, draw=none, font = \scriptsize, anchor = east, minimum size=15pt, inner sep=0pt] (sink1) at (0, 2) {$\pi$};
    \node[circle, draw=black, fill=black!20, thick, draw=none, font = \scriptsize, anchor = east, minimum size=15pt, inner sep=0pt] (sink2) at (0, 1) {$\pi$};
    \node[circle, draw=black, fill=black!20, thick, draw=none, font = \scriptsize, anchor = east, minimum size=15pt, inner sep=0pt] (sink3) at (0, 0) {$\pi$};
    \node[circle, draw=black, fill=black!20, thick, draw=none, font = \scriptsize, anchor = east, minimum size=15pt, inner sep=0pt] (source1) at (1.618, 2) {$\pi$};
    \node[circle, draw=black, fill=black!20, thick, draw=none, font = \scriptsize, anchor = east, minimum size=15pt, inner sep=0pt] (source2) at (1.618, 1) {$\pi$};
    \node[circle, draw=black, fill=black!20, thick, draw=none, font = \scriptsize, anchor = east, minimum size=15pt, inner sep=0pt] (source3) at (1.618, 0) {$\pi$};
    \draw[thin, -Stealth] (source1) to (sink1);
    \draw[thin, -Stealth] (sink1) to (source2);
    \draw[thin, -Stealth] (source2) to (source1);
    \draw[thin, -Stealth] (source3) to (sink3);
    \draw[thin, -Stealth] (sink3) to (sink2);
    \draw[thin, -Stealth] (sink2) to (source3);
\end{tikzpicture}\end{subfigure}}
\quad
\raisebox{-0.5\height}{\begin{subfigure}{0.20\linewidth}\centering\begin{tikzpicture}[global scale = 1.3]
    \node[circle, draw=black, fill=black!20, thick, draw=none, font = \scriptsize, anchor = east, minimum size=15pt, inner sep=0pt] (sink1) at (0, 2) {$\pi$};
    \node[circle, draw=black, fill=black!20, thick, draw=none, font = \scriptsize, anchor = east, minimum size=15pt, inner sep=0pt] (sink2) at (0, 1) {$\pi$};
    \node[circle, draw=black, fill=black!20, thick, draw=none, font = \scriptsize, anchor = east, minimum size=15pt, inner sep=0pt] (sink3) at (0, 0) {$\pi$};
    \node[circle, draw=black, fill=black!20, thick, draw=none, font = \scriptsize, anchor = east, minimum size=15pt, inner sep=0pt] (source1) at (1.618, 2) {$\pi$};
    \node[circle, draw=black, fill=black!20, thick, draw=none, font = \scriptsize, anchor = east, minimum size=15pt, inner sep=0pt] (source2) at (1.618, 1) {$\pi$};
    \node[circle, draw=black, fill=black!20, thick, draw=none, font = \scriptsize, anchor = east, minimum size=15pt, inner sep=0pt] (source3) at (1.618, 0) {$\pi$};
    \draw[thin, -Stealth] (source1) to (sink1);
    \draw[thin, -Stealth] (sink1) to (sink2);
    \draw[thin, -Stealth] (sink2) to (source2);
    \draw[thin, -Stealth] (source2) to (source3);
    \draw[thin, -Stealth] (source3) to (sink3);
    \draw[thin, -Stealth] (sink3) to (source1);
\end{tikzpicture}\end{subfigure}}
\caption{Two examples of repeated topologies in the $I=0$ $\pi\pi\pi$ contractions.}
\label{fig:diagrams-pipipi-I=0-repeated}
\end{figure}
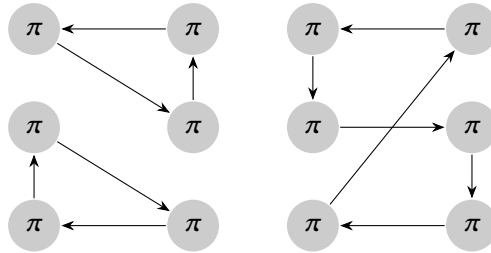

Using a \verb|Mathematica| topology-classification program for automatic identification, Fig.~\ref{fig:diagrams-pipipi-I=0-repeated} shows two examples that appear different at first sight but can be reduced, by permutations, to topologies already present in Fig.~\ref{fig:diagrams-pipipi-I=0}. The first diagram can be reduced to the topology $\mathbbm{TS}$ by exchanging the first and third $\pi$ at the source and the first and second $\pi$ at the sink; similarly, the second diagram can be reduced to the topology $\mathbbm{TW}$ by the corresponding cyclic permutation.

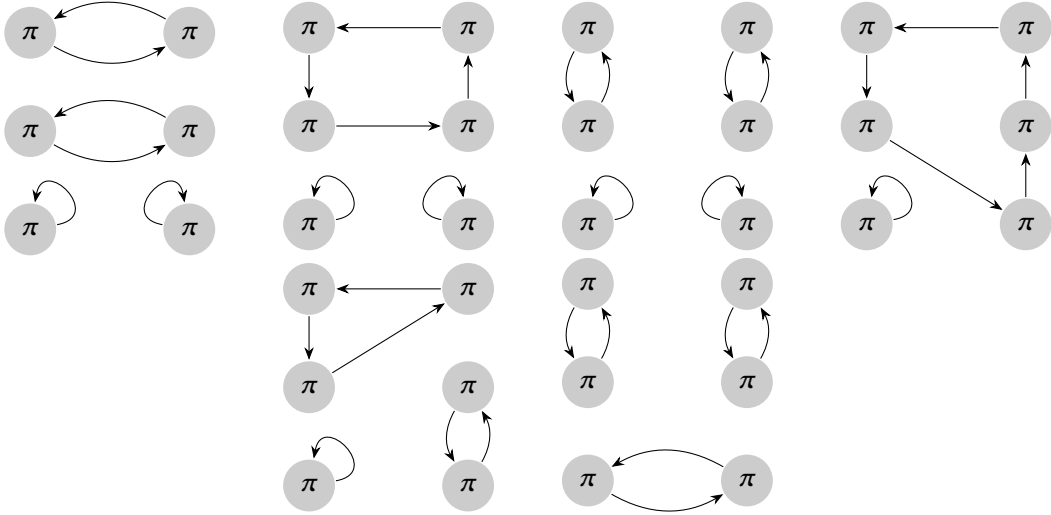
\begin{figure}[htbp]
\centering
\raisebox{-0.5\height}{\begin{subfigure}{0.20\linewidth}\centering\begin{tikzpicture}[global scale = 1.3]
    \node[circle, draw=black, fill=black!20, thick, draw=none, font = \scriptsize, anchor = east, minimum size=15pt, inner sep=0pt] (sink1) at (0, 2) {$\pi$};
    \node[circle, draw=black, fill=black!20, thick, draw=none, font = \scriptsize, anchor = east, minimum size=15pt, inner sep=0pt] (sink2) at (0, 1) {$\pi$};
    \node[circle, draw=black, fill=black!20, thick, draw=none, font = \scriptsize, anchor = east, minimum size=15pt, inner sep=0pt] (sink3) at (0, 0) {$\pi$};
    \node[circle, draw=black, fill=black!20, thick, draw=none, font = \scriptsize, anchor = east, minimum size=15pt, inner sep=0pt] (source1) at (1.618, 2) {$\pi$};
    \node[circle, draw=black, fill=black!20, thick, draw=none, font = \scriptsize, anchor = east, minimum size=15pt, inner sep=0pt] (source2) at (1.618, 1) {$\pi$};
    \node[circle, draw=black, fill=black!20, thick, draw=none, font = \scriptsize, anchor = east, minimum size=15pt, inner sep=0pt] (source3) at (1.618, 0) {$\pi$};
    \draw[thin, -Stealth, out=150, in=30] (source1) to (sink1);
    \draw[thin, -Stealth, out=330, in=210] (sink1) to (source1);
    \draw[thin, -Stealth, out=150, in=30] (source2) to (sink2);
    \draw[thin, -Stealth, out=330, in=210] (sink2) to (source2);
    \draw[thin, -Stealth, out=170, in=100, looseness=4] (source3) to (source3);
    \draw[thin, -Stealth, out=10, in=80, looseness=4] (sink3) to (sink3);
\end{tikzpicture}\end{subfigure}}
\quad
\raisebox{-0.5\height}{\begin{subfigure}{0.20\linewidth}\centering\begin{tikzpicture}[global scale = 1.3]
    \node[circle, draw=black, fill=black!20, thick, draw=none, font = \scriptsize, anchor = east, minimum size=15pt, inner sep=0pt] (sink1) at (0, 2) {$\pi$};
    \node[circle, draw=black, fill=black!20, thick, draw=none, font = \scriptsize, anchor = east, minimum size=15pt, inner sep=0pt] (sink2) at (0, 1) {$\pi$};
    \node[circle, draw=black, fill=black!20, thick, draw=none, font = \scriptsize, anchor = east, minimum size=15pt, inner sep=0pt] (sink3) at (0, 0) {$\pi$};
    \node[circle, draw=black, fill=black!20, thick, draw=none, font = \scriptsize, anchor = east, minimum size=15pt, inner sep=0pt] (source1) at (1.618, 2) {$\pi$};
    \node[circle, draw=black, fill=black!20, thick, draw=none, font = \scriptsize, anchor = east, minimum size=15pt, inner sep=0pt] (source2) at (1.618, 1) {$\pi$};
    \node[circle, draw=black, fill=black!20, thick, draw=none, font = \scriptsize, anchor = east, minimum size=15pt, inner sep=0pt] (source3) at (1.618, 0) {$\pi$};
    \draw[thin, -Stealth] (source1) to (sink1);
    \draw[thin, -Stealth] (sink1) to (sink2);
    \draw[thin, -Stealth] (sink2) to (source2);
    \draw[thin, -Stealth] (source2) to (source1);
    \draw[thin, -Stealth, out=170, in=100, looseness=4] (source3) to (source3);
    \draw[thin, -Stealth, out=10, in=80, looseness=4] (sink3) to (sink3);
\end{tikzpicture}\end{subfigure}}
\quad
\raisebox{-0.5\height}{\begin{subfigure}{0.20\linewidth}\centering\begin{tikzpicture}[global scale = 1.3]
    \node[circle, draw=black, fill=black!20, thick, draw=none, font = \scriptsize, anchor = east, minimum size=15pt, inner sep=0pt] (sink1) at (0, 2) {$\pi$};
    \node[circle, draw=black, fill=black!20, thick, draw=none, font = \scriptsize, anchor = east, minimum size=15pt, inner sep=0pt] (sink2) at (0, 1) {$\pi$};
    \node[circle, draw=black, fill=black!20, thick, draw=none, font = \scriptsize, anchor = east, minimum size=15pt, inner sep=0pt] (sink3) at (0, 0) {$\pi$};
    \node[circle, draw=black, fill=black!20, thick, draw=none, font = \scriptsize, anchor = east, minimum size=15pt, inner sep=0pt] (source1) at (1.618, 2) {$\pi$};
    \node[circle, draw=black, fill=black!20, thick, draw=none, font = \scriptsize, anchor = east, minimum size=15pt, inner sep=0pt] (source2) at (1.618, 1) {$\pi$};
    \node[circle, draw=black, fill=black!20, thick, draw=none, font = \scriptsize, anchor = east, minimum size=15pt, inner sep=0pt] (source3) at (1.618, 0) {$\pi$};
    \draw[thin, -Stealth, out=240, in=120] (source1) to (source2);
    \draw[thin, -Stealth, out=60, in=300] (source2) to (source1);
    \draw[thin, -Stealth, out=240, in=120] (sink1) to (sink2);
    \draw[thin, -Stealth, out=60, in=300] (sink2) to (sink1);
    \draw[thin, -Stealth, out=170, in=100, looseness=4] (source3) to (source3);
    \draw[thin, -Stealth, out=10, in=80, looseness=4] (sink3) to (sink3);
\end{tikzpicture}\end{subfigure}}
\quad
\raisebox{-0.5\height}{\begin{subfigure}{0.20\linewidth}\centering\begin{tikzpicture}[global scale = 1.3]
    \node[circle, draw=black, fill=black!20, thick, draw=none, font = \scriptsize, anchor = east, minimum size=15pt, inner sep=0pt] (sink1) at (0, 2) {$\pi$};
    \node[circle, draw=black, fill=black!20, thick, draw=none, font = \scriptsize, anchor = east, minimum size=15pt, inner sep=0pt] (sink2) at (0, 1) {$\pi$};
    \node[circle, draw=black, fill=black!20, thick, draw=none, font = \scriptsize, anchor = east, minimum size=15pt, inner sep=0pt] (sink3) at (0, 0) {$\pi$};
    \node[circle, draw=black, fill=black!20, thick, draw=none, font = \scriptsize, anchor = east, minimum size=15pt, inner sep=0pt] (source1) at (1.618, 2) {$\pi$};
    \node[circle, draw=black, fill=black!20, thick, draw=none, font = \scriptsize, anchor = east, minimum size=15pt, inner sep=0pt] (source2) at (1.618, 1) {$\pi$};
    \node[circle, draw=black, fill=black!20, thick, draw=none, font = \scriptsize, anchor = east, minimum size=15pt, inner sep=0pt] (source3) at (1.618, 0) {$\pi$};
    \draw[thin, -Stealth] (source1) to (sink1);
    \draw[thin, -Stealth] (sink1) to (sink2);
    \draw[thin, -Stealth] (sink2) to (source3);
    \draw[thin, -Stealth] (source3) to (source2);
    \draw[thin, -Stealth] (source2) to (source1);
    \draw[thin, -Stealth, out=10, in=80, looseness=4] (sink3) to (sink3);
\end{tikzpicture}\end{subfigure}}
\\
\raisebox{-0.5\height}{\begin{subfigure}{0.20\linewidth}\centering\begin{tikzpicture}[global scale = 1.3]
    \node[circle, draw=black, fill=black!20, thick, draw=none, font = \scriptsize, anchor = east, minimum size=15pt, inner sep=0pt] (sink1) at (0, 2) {$\pi$};
    \node[circle, draw=black, fill=black!20, thick, draw=none, font = \scriptsize, anchor = east, minimum size=15pt, inner sep=0pt] (sink2) at (0, 1) {$\pi$};
    \node[circle, draw=black, fill=black!20, thick, draw=none, font = \scriptsize, anchor = east, minimum size=15pt, inner sep=0pt] (sink3) at (0, 0) {$\pi$};
    \node[circle, draw=black, fill=black!20, thick, draw=none, font = \scriptsize, anchor = east, minimum size=15pt, inner sep=0pt] (source1) at (1.618, 2) {$\pi$};
    \node[circle, draw=black, fill=black!20, thick, draw=none, font = \scriptsize, anchor = east, minimum size=15pt, inner sep=0pt] (source2) at (1.618, 1) {$\pi$};
    \node[circle, draw=black, fill=black!20, thick, draw=none, font = \scriptsize, anchor = east, minimum size=15pt, inner sep=0pt] (source3) at (1.618, 0) {$\pi$};
    \draw[thin, -Stealth] (source1) to (sink1);
    \draw[thin, -Stealth] (sink1) to (sink2);
    \draw[thin, -Stealth] (sink2) to (source1);
    \draw[thin, -Stealth, out=240, in=120] (source2) to (source3);
    \draw[thin, -Stealth, out=60, in=300] (source3) to (source2);
    \draw[thin, -Stealth, out=10, in=80, looseness=4] (sink3) to (sink3);
\end{tikzpicture}\end{subfigure}}
\quad
\raisebox{-0.5\height}{\begin{subfigure}{0.20\linewidth}\centering\begin{tikzpicture}[global scale = 1.3]
    \node[circle, draw=black, fill=black!20, thick, draw=none, font = \scriptsize, anchor = east, minimum size=15pt, inner sep=0pt] (sink1) at (0, 2) {$\pi$};
    \node[circle, draw=black, fill=black!20, thick, draw=none, font = \scriptsize, anchor = east, minimum size=15pt, inner sep=0pt] (sink2) at (0, 1) {$\pi$};
    \node[circle, draw=black, fill=black!20, thick, draw=none, font = \scriptsize, anchor = east, minimum size=15pt, inner sep=0pt] (sink3) at (0, 0) {$\pi$};
    \node[circle, draw=black, fill=black!20, thick, draw=none, font = \scriptsize, anchor = east, minimum size=15pt, inner sep=0pt] (source1) at (1.618, 2) {$\pi$};
    \node[circle, draw=black, fill=black!20, thick, draw=none, font = \scriptsize, anchor = east, minimum size=15pt, inner sep=0pt] (source2) at (1.618, 1) {$\pi$};
    \node[circle, draw=black, fill=black!20, thick, draw=none, font = \scriptsize, anchor = east, minimum size=15pt, inner sep=0pt] (source3) at (1.618, 0) {$\pi$};
    \draw[thin, -Stealth, out=240, in=120] (source1) to (source2);
    \draw[thin, -Stealth, out=60, in=300] (source2) to (source1);
    \draw[thin, -Stealth, out=240, in=120] (sink1) to (sink2);
    \draw[thin, -Stealth, out=60, in=300] (sink2) to (sink1);
    \draw[thin, -Stealth, out=150, in=30] (source3) to (sink3);
    \draw[thin, -Stealth, out=330, in=210] (sink3) to (source3);
\end{tikzpicture}\end{subfigure}}
\caption{Examples of topologies appearing in individual terms of the $I=0$ $\pi\pi\pi$ contractions that cancel after isospin projection.}
\label{fig:diagrams-pipipi-bad}
\end{figure}

This illustrates how isospin symmetry protects the spectrum. If one inspects only some terms in the contraction of the $\pi\pi\pi$ operator, or if one selects a single component of the $\pi\pi\pi$ operator in isolation, a large number of topologies appear, including those shown in Fig.~\ref{fig:diagrams-pipipi-bad} (only a subset is displayed here), and these diagrams do contribute to the correlation function. After isospin projection, however, these contributions cancel exactly. If one were to use ensembles without isospin symmetry, operators with good isospin could not be constructed; the computational cost would then increase substantially, and many resonance states would mix with the target system.

It should be emphasized that some of these cancellations in the correlation functions do not, in a strict sense, originate from the $I=0$ isospin projection of the $\pi\pi\pi$ system. In particular, tadpole diagrams in which a single $\pi$ meson annihilates into itself arise from the $\bar u u$ or $\bar d d$ contractions of a $\pi^0$ at the same lattice site. Since the $\pi^0$ always appears as the linear combination $\bar u u - \bar d d$, these two types of contribution necessarily cancel. Thus this cancellation mechanism is already built into the isospin structure at the single-meson operator level, rather than being a feature unique to the three-body isospin projection.

We find that, when all $\pi$ mesons carry zero momentum and all diagrams with the same topology are assumed to contribute identically, the correlation function of the $\pi\pi\pi$ operator with itself vanishes exactly in the $I=0$ channel. This result implies that, in this limit, there is no $I=0$ state composed of three stationary $\pi$ mesons. This is not surprising: nonzero relative angular momentum among the three $\pi$ mesons is required in order to couple to the quantum numbers of the $\omega$.

This conclusion can also be understood, to some extent, from the structure of the operator itself. If $\pi^+$, $\pi^-$, and $\pi^0$ are treated as completely identical particles, the $I=0$ $\pi\pi\pi$ operator itself vanishes. By contrast, this phenomenon does not occur in the $I=0$ $\pi\pi$ system, because the corresponding operator contains an odd number ($3$) of terms. A similar structural cancellation can also be observed in the $I=1$ $\pi\pi$ channel, where the operator contains an even number ($2$) of terms; hence, at zero momentum, both the operator and its associated correlation function vanish exactly.

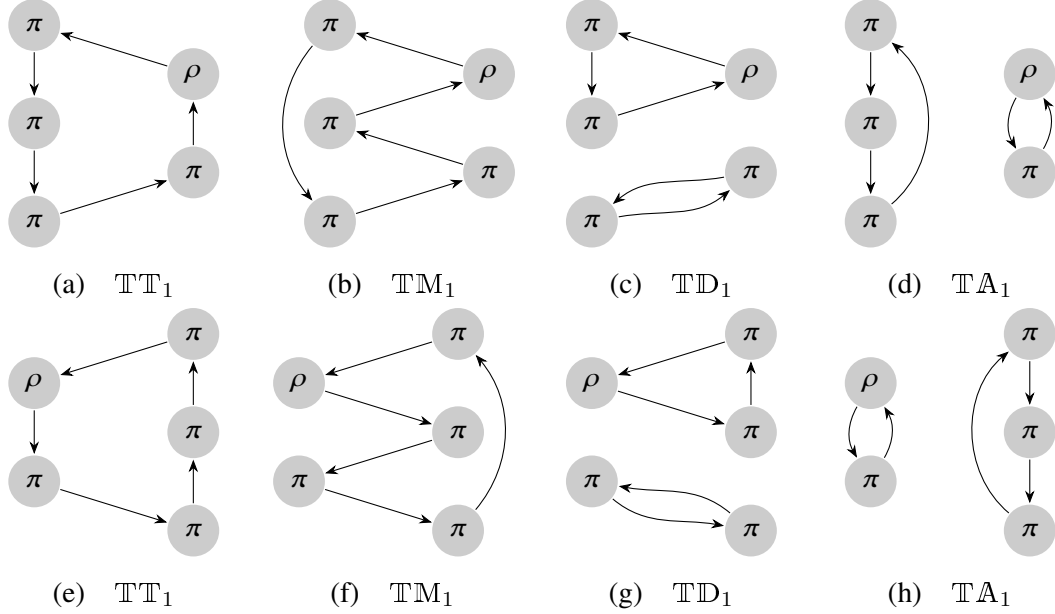
\begin{figure}[htbp]
\centering
\raisebox{-0.5\height}{\begin{subfigure}{0.20\linewidth}\centering\begin{tikzpicture}[global scale = 1.3]
    \node[circle, draw=black, fill=black!20, thick, draw=none, font = \scriptsize, anchor = east, minimum size=15pt, inner sep=0pt] (sink1) at (0, 2) {$\pi$};
    \node[circle, draw=black, fill=black!20, thick, draw=none, font = \scriptsize, anchor = east, minimum size=15pt, inner sep=0pt] (sink2) at (0, 1) {$\pi$};
    \node[circle, draw=black, fill=black!20, thick, draw=none, font = \scriptsize, anchor = east, minimum size=15pt, inner sep=0pt] (sink3) at (0, 0) {$\pi$};
    \node[circle, draw=black, fill=black!20, thick, draw=none, font = \scriptsize, anchor = east, minimum size=15pt, inner sep=0pt] (source1) at (1.618, 1.5) {$\rho$};
    \node[circle, draw=black, fill=black!20, thick, draw=none, font = \scriptsize, anchor = east, minimum size=15pt, inner sep=0pt] (source2) at (1.618, 0.5) {$\pi$};
    \draw[thin, -Stealth] (source1) to (sink1);
    \draw[thin, -Stealth] (sink1) to (sink2);
    \draw[thin, -Stealth] (sink2) to (sink3);
    \draw[thin, -Stealth] (sink3) to (source2);
    \draw[thin, -Stealth] (source2) to (source1);
\end{tikzpicture}\caption{$\mathbbm{TT}_1$}\end{subfigure}}
\quad
\raisebox{-0.5\height}{\begin{subfigure}{0.20\linewidth}\centering\begin{tikzpicture}[global scale = 1.3]
    \node[circle, draw=black, fill=black!20, thick, draw=none, font = \scriptsize, anchor = east, minimum size=15pt, inner sep=0pt] (sink1) at (0, 2) {$\pi$};
    \node[circle, draw=black, fill=black!20, thick, draw=none, font = \scriptsize, anchor = east, minimum size=15pt, inner sep=0pt] (sink2) at (0, 1) {$\pi$};
    \node[circle, draw=black, fill=black!20, thick, draw=none, font = \scriptsize, anchor = east, minimum size=15pt, inner sep=0pt] (sink3) at (0, 0) {$\pi$};
    \node[circle, draw=black, fill=black!20, thick, draw=none, font = \scriptsize, anchor = east, minimum size=15pt, inner sep=0pt] (source1) at (1.618, 1.5) {$\rho$};
    \node[circle, draw=black, fill=black!20, thick, draw=none, font = \scriptsize, anchor = east, minimum size=15pt, inner sep=0pt] (source2) at (1.618, 0.5) {$\pi$};
    \draw[thin, -Stealth] (source1) to (sink1);
    \draw[thin, -Stealth, out=230, in=130] (sink1) to (sink3);
    \draw[thin, -Stealth] (sink3) to (source2);
    \draw[thin, -Stealth] (source2) to (sink2);
    \draw[thin, -Stealth] (sink2) to (source1);
\end{tikzpicture}\caption{$\mathbbm{TM}_1$}\end{subfigure}}
\quad
\raisebox{-0.5\height}{\begin{subfigure}{0.20\linewidth}\centering\begin{tikzpicture}[global scale = 1.3]
    \node[circle, draw=black, fill=black!20, thick, draw=none, font = \scriptsize, anchor = east, minimum size=15pt, inner sep=0pt] (sink1) at (0, 2) {$\pi$};
    \node[circle, draw=black, fill=black!20, thick, draw=none, font = \scriptsize, anchor = east, minimum size=15pt, inner sep=0pt] (sink2) at (0, 1) {$\pi$};
    \node[circle, draw=black, fill=black!20, thick, draw=none, font = \scriptsize, anchor = east, minimum size=15pt, inner sep=0pt] (sink3) at (0, 0) {$\pi$};
    \node[circle, draw=black, fill=black!20, thick, draw=none, font = \scriptsize, anchor = east, minimum size=15pt, inner sep=0pt] (source1) at (1.618, 1.5) {$\rho$};
    \node[circle, draw=black, fill=black!20, thick, draw=none, font = \scriptsize, anchor = east, minimum size=15pt, inner sep=0pt] (source2) at (1.618, 0.5) {$\pi$};
    \draw[thin, -Stealth] (source1) to (sink1);
    \draw[thin, -Stealth] (sink1) to (sink2);
    \draw[thin, -Stealth] (sink2) to (source1);
    \draw[thin, -Stealth, out=190, in=40] (source2) to (sink3);
    \draw[thin, -Stealth, out=10, in=220] (sink3) to (source2);
\end{tikzpicture}\caption{$\mathbbm{TD}_1$}\end{subfigure}}
\quad
\raisebox{-0.5\height}{\begin{subfigure}{0.20\linewidth}\centering\begin{tikzpicture}[global scale = 1.3]
    \node[circle, draw=black, fill=black!20, thick, draw=none, font = \scriptsize, anchor = east, minimum size=15pt, inner sep=0pt] (sink1) at (0, 2) {$\pi$};
    \node[circle, draw=black, fill=black!20, thick, draw=none, font = \scriptsize, anchor = east, minimum size=15pt, inner sep=0pt] (sink2) at (0, 1) {$\pi$};
    \node[circle, draw=black, fill=black!20, thick, draw=none, font = \scriptsize, anchor = east, minimum size=15pt, inner sep=0pt] (sink3) at (0, 0) {$\pi$};
    \node[circle, draw=black, fill=black!20, thick, draw=none, font = \scriptsize, anchor = east, minimum size=15pt, inner sep=0pt] (source1) at (1.618, 1.5) {$\rho$};
    \node[circle, draw=black, fill=black!20, thick, draw=none, font = \scriptsize, anchor = east, minimum size=15pt, inner sep=0pt] (source2) at (1.618, 0.5) {$\pi$};
    \draw[thin, -Stealth] (sink1) to (sink2);
    \draw[thin, -Stealth] (sink2) to (sink3);
    \draw[thin, -Stealth, out=40, in=320] (sink3) to (sink1);
    \draw[thin, -Stealth, out=240, in=120] (source1) to (source2);
    \draw[thin, -Stealth, out=60, in=300] (source2) to (source1);
\end{tikzpicture}\caption{$\mathbbm{TA}_1$}\end{subfigure}}
\\
\raisebox{-0.5\height}{\begin{subfigure}{0.20\linewidth}\centering\begin{tikzpicture}[global scale = 1.3]
    \node[circle, draw=black, fill=black!20, thick, draw=none, font = \scriptsize, anchor = east, minimum size=15pt, inner sep=0pt] (sink1) at (0, 1.5) {$\rho$};
    \node[circle, draw=black, fill=black!20, thick, draw=none, font = \scriptsize, anchor = east, minimum size=15pt, inner sep=0pt] (sink2) at (0, 0.5) {$\pi$};
    \node[circle, draw=black, fill=black!20, thick, draw=none, font = \scriptsize, anchor = east, minimum size=15pt, inner sep=0pt] (source1) at (1.618, 2) {$\pi$};
    \node[circle, draw=black, fill=black!20, thick, draw=none, font = \scriptsize, anchor = east, minimum size=15pt, inner sep=0pt] (source2) at (1.618, 1) {$\pi$};
    \node[circle, draw=black, fill=black!20, thick, draw=none, font = \scriptsize, anchor = east, minimum size=15pt, inner sep=0pt] (source3) at (1.618, 0) {$\pi$};
    \draw[thin, -Stealth] (source1) to (sink1);
    \draw[thin, -Stealth] (sink1) to (sink2);
    \draw[thin, -Stealth] (sink2) to (source3);
    \draw[thin, -Stealth] (source3) to (source2);
    \draw[thin, -Stealth] (source2) to (source1);
\end{tikzpicture}\caption{$\mathbbm{TT}_1$}\end{subfigure}}
\quad
\raisebox{-0.5\height}{\begin{subfigure}{0.20\linewidth}\centering\begin{tikzpicture}[global scale = 1.3]
    \node[circle, draw=black, fill=black!20, thick, draw=none, font = \scriptsize, anchor = east, minimum size=15pt, inner sep=0pt] (sink1) at (0, 1.5) {$\rho$};
    \node[circle, draw=black, fill=black!20, thick, draw=none, font = \scriptsize, anchor = east, minimum size=15pt, inner sep=0pt] (sink2) at (0, 0.5) {$\pi$};
    \node[circle, draw=black, fill=black!20, thick, draw=none, font = \scriptsize, anchor = east, minimum size=15pt, inner sep=0pt] (source1) at (1.618, 2) {$\pi$};
    \node[circle, draw=black, fill=black!20, thick, draw=none, font = \scriptsize, anchor = east, minimum size=15pt, inner sep=0pt] (source2) at (1.618, 1) {$\pi$};
    \node[circle, draw=black, fill=black!20, thick, draw=none, font = \scriptsize, anchor = east, minimum size=15pt, inner sep=0pt] (source3) at (1.618, 0) {$\pi$};
    \draw[thin, -Stealth] (source1) to (sink1);
    \draw[thin, -Stealth] (sink1) to (source2);
    \draw[thin, -Stealth] (source2) to (sink2);
    \draw[thin, -Stealth] (sink2) to (source3);
    \draw[thin, -Stealth, out=50, in=310] (source3) to (source1);
\end{tikzpicture}\caption{$\mathbbm{TM}_1$}\end{subfigure}}
\quad
\raisebox{-0.5\height}{\begin{subfigure}{0.20\linewidth}\centering\begin{tikzpicture}[global scale = 1.3]
    \node[circle, draw=black, fill=black!20, thick, draw=none, font = \scriptsize, anchor = east, minimum size=15pt, inner sep=0pt] (sink1) at (0, 1.5) {$\rho$};
    \node[circle, draw=black, fill=black!20, thick, draw=none, font = \scriptsize, anchor = east, minimum size=15pt, inner sep=0pt] (sink2) at (0, 0.5) {$\pi$};
    \node[circle, draw=black, fill=black!20, thick, draw=none, font = \scriptsize, anchor = east, minimum size=15pt, inner sep=0pt] (source1) at (1.618, 2) {$\pi$};
    \node[circle, draw=black, fill=black!20, thick, draw=none, font = \scriptsize, anchor = east, minimum size=15pt, inner sep=0pt] (source2) at (1.618, 1) {$\pi$};
    \node[circle, draw=black, fill=black!20, thick, draw=none, font = \scriptsize, anchor = east, minimum size=15pt, inner sep=0pt] (source3) at (1.618, 0) {$\pi$};
    \draw[thin, -Stealth] (source1) to (sink1);
    \draw[thin, -Stealth] (sink1) to (source2);
    \draw[thin, -Stealth] (source2) to (source1);
    \draw[thin, -Stealth, out=140, in=350] (source3) to (sink2);
    \draw[thin, -Stealth, out=320, in=170] (sink2) to (source3);
\end{tikzpicture}\caption{$\mathbbm{TD}_1$}\end{subfigure}}
\quad
\raisebox{-0.5\height}{\begin{subfigure}{0.20\linewidth}\centering\begin{tikzpicture}[global scale = 1.3]
    \node[circle, draw=black, fill=black!20, thick, draw=none, font = \scriptsize, anchor = east, minimum size=15pt, inner sep=0pt] (sink1) at (0, 1.5) {$\rho$};
    \node[circle, draw=black, fill=black!20, thick, draw=none, font = \scriptsize, anchor = east, minimum size=15pt, inner sep=0pt] (sink2) at (0, 0.5) {$\pi$};
    \node[circle, draw=black, fill=black!20, thick, draw=none, font = \scriptsize, anchor = east, minimum size=15pt, inner sep=0pt] (source1) at (1.618, 2) {$\pi$};
    \node[circle, draw=black, fill=black!20, thick, draw=none, font = \scriptsize, anchor = east, minimum size=15pt, inner sep=0pt] (source2) at (1.618, 1) {$\pi$};
    \node[circle, draw=black, fill=black!20, thick, draw=none, font = \scriptsize, anchor = east, minimum size=15pt, inner sep=0pt] (source3) at (1.618, 0) {$\pi$};
    \draw[thin, -Stealth, out=240, in=120] (sink1) to (sink2);
    \draw[thin, -Stealth, out=60, in=300] (sink2) to (sink1);
    \draw[thin, -Stealth] (source1) to (source2);
    \draw[thin, -Stealth] (source2) to (source3);
    \draw[thin, -Stealth, out=140, in=220] (source3) to (source1);
\end{tikzpicture}\caption{$\mathbbm{TA}_1$}\end{subfigure}}
\caption{Topologies for the off-diagonal two-body--three-body contractions in the $I=0$ $\pi\pi\pi$ channel.}
\label{fig:diagrams-pipipi-two-off-I=0}
\end{figure}

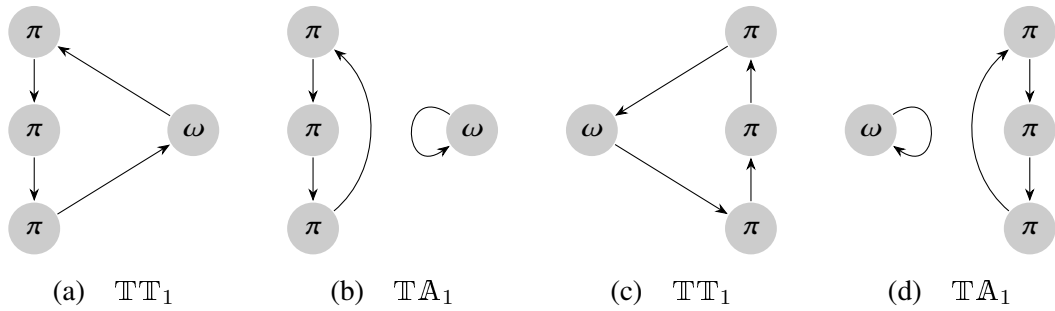
\begin{figure}[htbp]
\centering
\raisebox{-0.5\height}{\begin{subfigure}{0.20\linewidth}\centering\begin{tikzpicture}[global scale = 1.3]
    \node[circle, draw=black, fill=black!20, thick, draw=none, font = \scriptsize, anchor = east, minimum size=15pt, inner sep=0pt] (sink1) at (0, 2) {$\pi$};
    \node[circle, draw=black, fill=black!20, thick, draw=none, font = \scriptsize, anchor = east, minimum size=15pt, inner sep=0pt] (sink2) at (0, 1) {$\pi$};
    \node[circle, draw=black, fill=black!20, thick, draw=none, font = \scriptsize, anchor = east, minimum size=15pt, inner sep=0pt] (sink3) at (0, 0) {$\pi$};
    \node[circle, draw=black, fill=black!20, thick, draw=none, font = \scriptsize, anchor = east, minimum size=15pt, inner sep=0pt] (source1) at (1.618, 1) {$\omega$};
    \draw[thin, -Stealth] (source1) to (sink1);
    \draw[thin, -Stealth] (sink1) to (sink2);
    \draw[thin, -Stealth] (sink2) to (sink3);
    \draw[thin, -Stealth] (sink3) to (source1);
\end{tikzpicture}\caption{$\mathbbm{TT}_1$}\end{subfigure}}
\quad
\raisebox{-0.5\height}{\begin{subfigure}{0.20\linewidth}\centering\begin{tikzpicture}[global scale = 1.3]
    \node[circle, draw=black, fill=black!20, thick, draw=none, font = \scriptsize, anchor = east, minimum size=15pt, inner sep=0pt] (sink1) at (0, 2) {$\pi$};
    \node[circle, draw=black, fill=black!20, thick, draw=none, font = \scriptsize, anchor = east, minimum size=15pt, inner sep=0pt] (sink2) at (0, 1) {$\pi$};
    \node[circle, draw=black, fill=black!20, thick, draw=none, font = \scriptsize, anchor = east, minimum size=15pt, inner sep=0pt] (sink3) at (0, 0) {$\pi$};
    \node[circle, draw=black, fill=black!20, thick, draw=none, font = \scriptsize, anchor = east, minimum size=15pt, inner sep=0pt] (source1) at (1.618, 1) {$\omega$};
    \draw[thin, -Stealth] (sink1) to (sink2);
    \draw[thin, -Stealth] (sink2) to (sink3);
    \draw[thin, -Stealth, out=40, in=320] (sink3) to (sink1);
    \draw[thin, -Stealth, out=145, in=215, looseness=5] (source1) to (source1);
\end{tikzpicture}\caption{$\mathbbm{TA}_1$}\end{subfigure}}
\quad
\raisebox{-0.5\height}{\begin{subfigure}{0.20\linewidth}\centering\begin{tikzpicture}[global scale = 1.3]
    \node[circle, draw=black, fill=black!20, thick, draw=none, font = \scriptsize, anchor = east, minimum size=15pt, inner sep=0pt] (sink1) at (0, 1) {$\omega$};
    \node[circle, draw=black, fill=black!20, thick, draw=none, font = \scriptsize, anchor = east, minimum size=15pt, inner sep=0pt] (source1) at (1.618, 2) {$\pi$};
    \node[circle, draw=black, fill=black!20, thick, draw=none, font = \scriptsize, anchor = east, minimum size=15pt, inner sep=0pt] (source2) at (1.618, 1) {$\pi$};
    \node[circle, draw=black, fill=black!20, thick, draw=none, font = \scriptsize, anchor = east, minimum size=15pt, inner sep=0pt] (source3) at (1.618, 0) {$\pi$};
    \draw[thin, -Stealth] (source1) to (sink1);
    \draw[thin, -Stealth] (sink1) to (source3);
    \draw[thin, -Stealth] (source3) to (source2);
    \draw[thin, -Stealth] (source2) to (source1);
\end{tikzpicture}\caption{$\mathbbm{TT}_1$}\end{subfigure}}
\quad
\raisebox{-0.5\height}{\begin{subfigure}{0.20\linewidth}\centering\begin{tikzpicture}[global scale = 1.3]
    \node[circle, draw=black, fill=black!20, thick, draw=none, font = \scriptsize, anchor = east, minimum size=15pt, inner sep=0pt] (sink1) at (0, 1) {$\omega$};
    \node[circle, draw=black, fill=black!20, thick, draw=none, font = \scriptsize, anchor = east, minimum size=15pt, inner sep=0pt] (source1) at (1.618, 2) {$\pi$};
    \node[circle, draw=black, fill=black!20, thick, draw=none, font = \scriptsize, anchor = east, minimum size=15pt, inner sep=0pt] (source2) at (1.618, 1) {$\pi$};
    \node[circle, draw=black, fill=black!20, thick, draw=none, font = \scriptsize, anchor = east, minimum size=15pt, inner sep=0pt] (source3) at (1.618, 0) {$\pi$};
    \draw[thin, -Stealth] (source1) to (source2);
    \draw[thin, -Stealth] (source2) to (source3);
    \draw[thin, -Stealth, out=140, in=220] (source3) to (source1);
    \draw[thin, -Stealth, out=35, in=325, looseness=5] (sink1) to (sink1);
\end{tikzpicture}\caption{$\mathbbm{TA}_1$}\end{subfigure}}
\caption{Topologies for the off-diagonal one-body--three-body contractions in the $I=0$ $\pi\pi\pi$ channel.}
\label{fig:diagrams-pipipi-one-off-I=0}
\end{figure}

The topologies of the off-diagonal correlation functions are classified in Fig.~\ref{fig:diagrams-pipipi-two-off-I=0} and Fig.~\ref{fig:diagrams-pipipi-one-off-I=0}\footnotecircle{In contractions between three-body and one-body operators, $\mathbbm{TT}$ denotes a triangle topology. In contractions between three-body and two-body operators, $\mathbbm{TT}$ denotes a trapezoid topology, while the symbol $\mathbbm{TM}$ is motivated by its mouth-like shape. We used the same naming convention in our published paper.}.

\subsection{Definitions and Naming Conventions for Topological Subdiagrams}
In this subsection we define the building blocks of the various topological subdiagrams that enter our calculation. Since three-body contractions involve several layers of subdiagrams and several types of labels, we spell out the conventions used in naming these subdiagrams.

We first define a general topology with six nodes as follows:
\begin{equation}
\raisebox{-0.45\height}{\centering\begin{tikzpicture}[global scale = 1.5]
    \draw[fill=black!10, draw=none] (0, 0) rectangle (1.618, 2);
    \node[circle, fill, label = left:{\scriptsize $1, \Omega, \delta, l$}, inner sep=2pt] (sink1) at (0, 2) {};
    \node[circle, fill, label = left:{\scriptsize $2, \Xi, \gamma, k$}, inner sep=2pt] (sink2) at (0, 1) {};
    \node[circle, fill, label = left:{\scriptsize $3, \Phi, \nu, n$}, inner sep=2pt] (sink3) at (0, 0) {};
    \node[circle, fill, label = right:{\scriptsize $4, \Sigma, \mu, m$}, inner sep=2pt] (source1) at (1.618, 2) {};
    \node[circle, fill, label = right:{\scriptsize $5, \Lambda, \beta, j$}, inner sep=2pt] (source2) at (1.618, 1) {};
    \node[circle, fill, label = right:{\scriptsize $6, \Gamma, \alpha, i$}, inner sep=2pt] (source3) at (1.618, 0) {};
\end{tikzpicture}}. \\
\label{eq:general-6-node-diagram}
\end{equation}
Each node in the diagram carries four different labels, corresponding to the different types of indices used in our calculation. Specifically, the nodes on the left (hadron sinks) are labeled $1, 2, 3$, while the nodes on the right (hadron sources) are labeled $4, 5, 6$. In addition, each node is assigned an uppercase Greek letter, a lowercase Greek letter, and a Latin letter, denoting respectively the gamma-matrix type, the momentum magnitude, and, when the hadron operator is nonlocal, the Lorentz index of the inserted gauge field. Nonlocal operators are not used in any of the calculations in this thesis.

Using the node labels defined in Eq.~\ref{eq:general-6-node-diagram}, one can represent contractions with any number of sources and sinks satisfying $N \leq 3$ and assign a specific set of indices to each diagram. The ordering convention is
\begin{equation}
\begin{aligned}
    & [1, 2, 3, 4, 5, 6], \\
    & [\Omega, \Xi, \Phi, \Sigma, \Lambda, \Gamma], \\
    & [\delta, \gamma, \nu, \mu, \beta, \alpha], \\
    & [l, k, n, m, j, i].
\end{aligned}
\end{equation}

For the $\omega(782)$ problem, all required topologies can be assembled from the following subdiagram building blocks:
\begin{align}
\mathbbm{L}^2_{[\Omega], [\delta]}(t^{\prime})
\ &= \
\raisebox{-0.39\height}{\centering\begin{tikzpicture}[global scale = 1.3]
    \node[label = right:{\scriptsize $\Omega, \delta$}, inner sep=0pt] (sink1) at (0, 0) {};
    \draw[thin, -Stealth, out=145, in=215, looseness=60] (sink1) to (sink1);
\end{tikzpicture}}, \\
\mathbbm{L}^2_{[\Omega,\Xi], [\delta,\gamma]}(t^{\prime})
\ &= \
\raisebox{-0.43\height}{\centering\begin{tikzpicture}[global scale = 1.3]
    \node[label = right:{\scriptsize $\Omega, \delta$}, inner sep=0pt] (sink1) at (0, 1) {};
    \node[label = right:{\scriptsize $\Xi, \gamma$}, inner sep=0pt] (sink2) at (0, 0) {};
    \draw[thin, -Stealth, out=240, in=120] (sink1) to (sink2);
    \draw[thin, -Stealth, out=60, in=300] (sink2) to (sink1);
\end{tikzpicture}}, \\
\mathbbm{L}^3_{[\Omega,\Xi,\Phi], [\delta,\gamma,\nu]}(t^{\prime})
\ &= \
\raisebox{-0.46\height}{\centering\begin{tikzpicture}[global scale = 1.3]
    \node[label = right:{\scriptsize $\Omega, \delta$}, inner sep=0pt] (sink1) at (0, 2) {};
    \node[label = right:{\scriptsize $\Xi, \gamma$}, inner sep=0pt] (sink2) at (0, 1) {};
    \node[label = right:{\scriptsize $\Phi, \nu$}, inner sep=0pt] (sink3) at (0, 0) {};
    \draw[thin, -Stealth] (sink1) to (sink2);
    \draw[thin, -Stealth] (sink2) to (sink3);
    \draw[thin, -Stealth, out=40, in=320] (sink3) to (sink1);
\end{tikzpicture}}
\ = \
\raisebox{-0.43\height}{\centering\begin{tikzpicture}[global scale = 1.3]
    \node[label = right:{\scriptsize $\Omega, \delta$}, inner sep=0pt] (sink1) at (0.5, 0.8660254038) {};
    \node[label = left:{\scriptsize $\Xi, \gamma$}, inner sep=0pt] (sink2) at (0, 0) {};
    \node[label = right:{\scriptsize $\Phi, \nu$}, inner sep=0pt] (sink3) at (1, 0) {};
    \draw[thin, -Stealth] (sink1) to (sink2);
    \draw[thin, -Stealth] (sink2) to (sink3);
    \draw[thin, -Stealth] (sink3) to (sink1);
\end{tikzpicture}}, \\
1-1-\mathbbm{D}_{[\Omega;\Gamma], [\delta;\alpha]}(t^{\prime},t)
\ &= \
\raisebox{-0.32\height}{\centering\begin{tikzpicture}[global scale = 1.3]
    \node[label = left:{\scriptsize $\Omega, \delta, \bar{d} c$}, inner sep=0pt] (sink1) at (0, 0) {};
    \node[label = right:{\scriptsize $\Gamma, \alpha, \bar{c} d$}, inner sep=0pt] (source1) at (1.618, 0) {};
    \draw[thin, -Stealth, out=150, in=30] (source1) to (sink1);
    \draw[thin, -Stealth, out=330, in=210] (sink1) to (source1);
\end{tikzpicture}}, \\
2-1-\mathbbm{T}^1_{[\Omega,\Xi;\Gamma], [\delta,\gamma;\alpha]}(t^{\prime},t)
\ &= \
\raisebox{-0.44\height}{\centering\begin{tikzpicture}[global scale = 1.3]
    \node[label = left:{\scriptsize $\Omega, \delta$}, inner sep=0pt] (sink1) at (0, 1) {};
    \node[label = left:{\scriptsize $\Xi, \gamma$}, inner sep=0pt] (sink2) at (0, 0) {};
    \node[label = right:{\scriptsize $\Gamma, \alpha$}, inner sep=0pt] (source1) at (1.618, 0.5) {};
    \draw[thin, -Stealth] (source1) to (sink1);
    \draw[thin, -Stealth] (sink1) to (sink2);
    \draw[thin, -Stealth] (sink2) to (source1);
\end{tikzpicture}}, \\
1-2-\mathbbm{T}^1_{[\Omega;\Lambda,\Gamma], [\delta;\beta,\alpha]}(t^{\prime},t)
\ &= \
\raisebox{-0.44\height}{\centering\begin{tikzpicture}[global scale = 1.3]
    \node[label = left:{\scriptsize $\Omega, \delta$}, inner sep=0pt] (sink1) at (0, 0.5) {};
    \node[label = right:{\scriptsize $\Lambda, \beta$}, inner sep=0pt] (source1) at (1.618, 1) {};
    \node[label = right:{\scriptsize $\Gamma, \alpha$}, inner sep=0pt] (source2) at (1.618, 0) {};
    \draw[thin, -Stealth] (source1) to (sink1);
    \draw[thin, -Stealth] (sink1) to (source2);
    \draw[thin, -Stealth] (source2) to (source1);
\end{tikzpicture}}, \\
2-2-\mathbbm{E}^1_{[\Omega,\Xi;\Lambda,\Gamma], [\delta,\gamma;\beta,\alpha]}(t^{\prime},t)
\ &= \
\raisebox{-0.44\height}{\centering\begin{tikzpicture}[global scale = 1.3]
    \node[label = left:{\scriptsize $\Omega, \delta$}, inner sep=0pt] (sink1) at (0, 1) {};
    \node[label = left:{\scriptsize $\Xi, \gamma$}, inner sep=0pt] (sink2) at (0, 0) {};
    \node[label = right:{\scriptsize $\Lambda, \beta$}, inner sep=0pt] (source1) at (1.618, 1) {};
    \node[label = right:{\scriptsize $\Gamma, \alpha$}, inner sep=0pt] (source2) at (1.618, 0) {};
    \draw[thin, -Stealth] (source1) to (sink1);
    \draw[thin, -Stealth] (sink1) to (source2);
    \draw[thin, -Stealth] (source2) to (sink2);
    \draw[thin, -Stealth] (sink2) to (source1);
\end{tikzpicture}}, \\
2-2-\mathbbm{B}^1_{[\Omega,\Xi;\Lambda,\Gamma], [\delta,\gamma;\beta,\alpha]}(t^{\prime},t)
\ &= \
\raisebox{-0.44\height}{\centering\begin{tikzpicture}[global scale = 1.3]
    \node[label = left:{\scriptsize $\Omega, \delta$}, inner sep=0pt] (sink1) at (0, 1) {};
    \node[label = left:{\scriptsize $\Xi, \gamma$}, inner sep=0pt] (sink2) at (0, 0) {};
    \node[label = right:{\scriptsize $\Lambda, \beta$}, inner sep=0pt] (source1) at (1.618, 1) {};
    \node[label = right:{\scriptsize $\Gamma, \alpha$}, inner sep=0pt] (source2) at (1.618, 0) {};
    \draw[thin, -Stealth] (source1) to (sink1);
    \draw[thin, -Stealth] (sink1) to (sink2);
    \draw[thin, -Stealth] (sink2) to (source2);
    \draw[thin, -Stealth] (source2) to (source1);
\end{tikzpicture}}, \\
3-1-\mathbbm{T}^1_{[\Omega,\Xi,\Phi;\Gamma], [\delta,\gamma,\nu;\alpha]}(t^{\prime},t)
\ &= \
\raisebox{-0.44\height}{\centering\begin{tikzpicture}[global scale = 1.3]
    \node[label = left:{\scriptsize $\Omega, \delta$}, inner sep=0pt] (sink1) at (0, 2) {};
    \node[label = left:{\scriptsize $\Xi, \gamma$}, inner sep=0pt] (sink2) at (0, 1) {};
    \node[label = left:{\scriptsize $\Phi, \nu$}, inner sep=0pt] (sink3) at (0, 0) {};
    \node[label = right:{\scriptsize $\Sigma, \mu$}, inner sep=0pt] (source1) at (1.618, 1) {};
    \draw[thin, -Stealth] (source1) to (sink1);
    \draw[thin, -Stealth] (sink1) to (sink2);
    \draw[thin, -Stealth] (sink2) to (sink3);
    \draw[thin, -Stealth] (sink3) to (source1);
\end{tikzpicture}}, \\
1-3-\mathbbm{T}^1_{[\Omega;\Sigma,\Lambda,\Gamma], [\delta;\mu,\beta,\alpha]}(t^{\prime},t)
\ &= \
\raisebox{-0.44\height}{\centering\begin{tikzpicture}[global scale = 1.3]
    \node[label = left:{\scriptsize $\Omega, \delta$}, inner sep=0pt] (sink1) at (0, 1) {};
    \node[label = right:{\scriptsize $\Sigma, \mu$}, inner sep=0pt] (source1) at (1.618, 2) {};
    \node[label = right:{\scriptsize $\Lambda, \beta$}, inner sep=0pt] (source2) at (1.618, 1) {};
    \node[label = right:{\scriptsize $\Gamma, \alpha$}, inner sep=0pt] (source3) at (1.618, 0) {};
    \draw[thin, -Stealth] (source1) to (sink1);
    \draw[thin, -Stealth] (sink1) to (source3);
    \draw[thin, -Stealth] (source3) to (source2);
    \draw[thin, -Stealth] (source2) to (source1);
\end{tikzpicture}}, \\
3-2-\mathbbm{TT}^1_{[\Omega,\Xi,\Phi;\Lambda,\Gamma], [\delta,\gamma,\nu;\beta,\alpha]}(t^{\prime},t)
\ &= \
\raisebox{-0.44\height}{\centering\begin{tikzpicture}[global scale = 1.3]
    \node[label = left:{\scriptsize $\Omega, \delta$}, inner sep=0pt] (sink1) at (0, 2) {};
    \node[label = left:{\scriptsize $\Xi, \gamma$}, inner sep=0pt] (sink2) at (0, 1) {};
    \node[label = left:{\scriptsize $\Phi, \nu$}, inner sep=0pt] (sink3) at (0, 0) {};
    \node[label = right:{\scriptsize $\Lambda, \beta$}, inner sep=0pt] (source1) at (1.618, 1.5) {};
    \node[label = right:{\scriptsize $\Gamma, \alpha$}, inner sep=0pt] (source2) at (1.618, 0.5) {};
    \draw[thin, -Stealth] (source1) to (sink1);
    \draw[thin, -Stealth] (sink1) to (sink2);
    \draw[thin, -Stealth] (sink2) to (sink3);
    \draw[thin, -Stealth] (sink3) to (source2);
    \draw[thin, -Stealth] (source2) to (source1);
\end{tikzpicture}}, \\
3-2-\mathbbm{TM}^1_{[\Omega,\Xi,\Phi;\Lambda,\Gamma], [\delta,\gamma,\nu;\beta,\alpha]}(t^{\prime},t)
\ &= \
\raisebox{-0.44\height}{\centering\begin{tikzpicture}[global scale = 1.3]
    \node[label = left:{\scriptsize $\Omega, \delta$}, inner sep=0pt] (sink1) at (0, 2) {};
    \node[label = left:{\scriptsize $\Xi, \gamma$}, inner sep=0pt] (sink2) at (0, 1) {};
    \node[label = left:{\scriptsize $\Phi, \nu$}, inner sep=0pt] (sink3) at (0, 0) {};
    \node[label = right:{\scriptsize $\Lambda, \beta$}, inner sep=0pt] (source1) at (1.618, 1.5) {};
    \node[label = right:{\scriptsize $\Gamma, \alpha$}, inner sep=0pt] (source2) at (1.618, 0.5) {};
    \draw[thin, -Stealth] (source1) to (sink1);
    \draw[thin, -Stealth, out=230, in=130] (sink1) to (sink3);
    \draw[thin, -Stealth] (sink3) to (source2);
    \draw[thin, -Stealth] (source2) to (sink2);
    \draw[thin, -Stealth] (sink2) to (source1);
\end{tikzpicture}}, \\
2-3-\mathbbm{TT}^1_{[\Omega,\Xi;\Sigma,\Lambda,\Gamma], [\delta,\gamma;\mu,\beta,\alpha]}(t^{\prime},t)
\ &= \
\raisebox{-0.44\height}{\centering\begin{tikzpicture}[global scale = 1.3]
    \node[label = left:{\scriptsize $\Omega, \delta$}, inner sep=0pt] (sink1) at (0, 1.5) {};
    \node[label = left:{\scriptsize $\Xi, \gamma$}, inner sep=0pt] (sink2) at (0, 0.5) {};
    \node[label = right:{\scriptsize $\Sigma, \mu$}, inner sep=0pt] (source1) at (1.618, 2) {};
    \node[label = right:{\scriptsize $\Lambda, \beta$}, inner sep=0pt] (source2) at (1.618, 1) {};
    \node[label = right:{\scriptsize $\Gamma, \alpha$}, inner sep=0pt] (source3) at (1.618, 0) {};
    \draw[thin, -Stealth] (source1) to (sink1);
    \draw[thin, -Stealth] (sink1) to (sink2);
    \draw[thin, -Stealth] (sink2) to (source3);
    \draw[thin, -Stealth] (source3) to (source2);
    \draw[thin, -Stealth] (source2) to (source1);
\end{tikzpicture}}, \\
2-3-\mathbbm{TM}^1_{[\Omega,\Xi;\Sigma,\Lambda,\Gamma], [\delta,\gamma;\mu,\beta,\alpha]}(t^{\prime},t)
\ &= \
\raisebox{-0.44\height}{\centering\begin{tikzpicture}[global scale = 1.3]
    \node[label = left:{\scriptsize $\Omega, \delta$}, inner sep=0pt] (sink1) at (0, 1.5) {};
    \node[label = left:{\scriptsize $\Xi, \gamma$}, inner sep=0pt] (sink2) at (0, 0.5) {};
    \node[label = right:{\scriptsize $\Sigma, \mu$}, inner sep=0pt] (source1) at (1.618, 2) {};
    \node[label = right:{\scriptsize $\Lambda, \beta$}, inner sep=0pt] (source2) at (1.618, 1) {};
    \node[label = right:{\scriptsize $\Gamma, \alpha$}, inner sep=0pt] (source3) at (1.618, 0) {};
    \draw[thin, -Stealth] (source1) to (sink1);
    \draw[thin, -Stealth] (sink1) to (source2);
    \draw[thin, -Stealth] (source2) to (sink2);
    \draw[thin, -Stealth] (sink2) to (source3);
    \draw[thin, -Stealth, out=50, in=310] (source3) to (source1);
\end{tikzpicture}}, \\
3-3-\mathbbm{TZ}^1_{[\Omega,\Xi,\Phi;\Sigma,\Lambda,\Gamma], [\delta,\gamma,\nu;\mu,\beta,\alpha]}(t^{\prime},t)
\ &= \
\raisebox{-0.44\height}{\centering\begin{tikzpicture}[global scale = 1.3]
    \node[label = left:{\scriptsize $\Omega, \delta$}, inner sep=0pt] (sink1) at (0, 2) {};
    \node[label = left:{\scriptsize $\Xi, \gamma$}, inner sep=0pt] (sink2) at (0, 1) {};
    \node[label = left:{\scriptsize $\Phi, \nu$}, inner sep=0pt] (sink3) at (0, 0) {};
    \node[label = right:{\scriptsize $\Sigma, \mu$}, inner sep=0pt] (source1) at (1.618, 2) {};
    \node[label = right:{\scriptsize $\Lambda, \beta$}, inner sep=0pt] (source2) at (1.618, 1) {};
    \node[label = right:{\scriptsize $\Gamma, \alpha$}, inner sep=0pt] (source3) at (1.618, 0) {};
    \draw[thin, -Stealth] (source1) to (sink1);
    \draw[thin, -Stealth] (sink1) to (source2);
    \draw[thin, -Stealth] (source2) to (sink2);
    \draw[thin, -Stealth] (sink2) to (source3);
    \draw[thin, -Stealth] (source3) to (sink3);
    \draw[thin, -Stealth] (sink3) to (source1);
\end{tikzpicture}}, \\
3-3-\mathbbm{TB}^1_{[\Omega,\Xi,\Phi;\Sigma,\Lambda,\Gamma], [\delta,\gamma,\nu;\mu,\beta,\alpha]}(t^{\prime},t)
\ &= \
\raisebox{-0.44\height}{\centering\begin{tikzpicture}[global scale = 1.3]
    \node[label = left:{\scriptsize $\Omega, \delta$}, inner sep=0pt] (sink1) at (0, 2) {};
    \node[label = left:{\scriptsize $\Xi, \gamma$}, inner sep=0pt] (sink2) at (0, 1) {};
    \node[label = left:{\scriptsize $\Phi, \nu$}, inner sep=0pt] (sink3) at (0, 0) {};
    \node[label = right:{\scriptsize $\Sigma, \mu$}, inner sep=0pt] (source1) at (1.618, 2) {};
    \node[label = right:{\scriptsize $\Lambda, \beta$}, inner sep=0pt] (source2) at (1.618, 1) {};
    \node[label = right:{\scriptsize $\Gamma, \alpha$}, inner sep=0pt] (source3) at (1.618, 0) {};
    \draw[thin, -Stealth] (source1) to (sink1);
    \draw[thin, -Stealth] (sink1) to (sink2);
    \draw[thin, -Stealth] (sink2) to (sink3);
    \draw[thin, -Stealth] (sink3) to (source3);
    \draw[thin, -Stealth] (source3) to (source2);
    \draw[thin, -Stealth] (source2) to (source1);
\end{tikzpicture}}, \\
3-3-\mathbbm{TW}^1_{[\Omega,\Xi,\Phi;\Sigma,\Lambda,\Gamma], [\delta,\gamma,\nu;\mu,\beta,\alpha]}(t^{\prime},t)
\ &= \
\raisebox{-0.44\height}{\centering\begin{tikzpicture}[global scale = 1.3]
    \node[label = left:{\scriptsize $\Omega, \delta$}, inner sep=0pt] (sink1) at (0, 2) {};
    \node[label = left:{\scriptsize $\Xi, \gamma$}, inner sep=0pt] (sink2) at (0, 1) {};
    \node[label = left:{\scriptsize $\Phi, \nu$}, inner sep=0pt] (sink3) at (0, 0) {};
    \node[label = right:{\scriptsize $\Sigma, \mu$}, inner sep=0pt] (source1) at (1.618, 2) {};
    \node[label = right:{\scriptsize $\Lambda, \beta$}, inner sep=0pt] (source2) at (1.618, 1) {};
    \node[label = right:{\scriptsize $\Gamma, \alpha$}, inner sep=0pt] (source3) at (1.618, 0) {};
    \draw[thin, -Stealth] (source1) to (sink1);
    \draw[thin, -Stealth] (sink1) to (source2);
    \draw[thin, -Stealth] (source2) to (source3);
    \draw[thin, -Stealth] (source3) to (sink3);
    \draw[thin, -Stealth] (sink3) to (sink2);
    \draw[thin, -Stealth] (sink2) to (source1);
\end{tikzpicture}}, \\
3-3-\mathbbm{TWR}^1_{[\Omega,\Xi,\Phi;\Sigma,\Lambda,\Gamma], [\delta,\gamma,\nu;\mu,\beta,\alpha]}(t^{\prime},t)
\ &= \
\raisebox{-0.44\height}{\centering\begin{tikzpicture}[global scale = 1.3]
    \node[label = left:{\scriptsize $\Omega, \delta$}, inner sep=0pt] (sink1) at (0, 2) {};
    \node[label = left:{\scriptsize $\Xi, \gamma$}, inner sep=0pt] (sink2) at (0, 1) {};
    \node[label = left:{\scriptsize $\Phi, \nu$}, inner sep=0pt] (sink3) at (0, 0) {};
    \node[label = right:{\scriptsize $\Sigma, \mu$}, inner sep=0pt] (source1) at (1.618, 2) {};
    \node[label = right:{\scriptsize $\Lambda, \beta$}, inner sep=0pt] (source2) at (1.618, 1) {};
    \node[label = right:{\scriptsize $\Gamma, \alpha$}, inner sep=0pt] (source3) at (1.618, 0) {};
    \draw[thin, -Stealth] (source1) to (sink2);
    \draw[thin, -Stealth] (sink2) to (sink3);
    \draw[thin, -Stealth] (sink3) to (source3);
    \draw[thin, -Stealth] (source3) to (source2);
    \draw[thin, -Stealth] (source2) to (sink1);
    \draw[thin, -Stealth] (sink1) to (source1);
\end{tikzpicture}}.
\end{align}
Here, except for annihilation diagrams, the leading number $N$-$M$ assigned to each diagram indicates that the diagram contains $N$ hadron sinks and $M$ hadron sources; the letters $D$, $T$, $E$, $B$, $Z$, and $W$ denote different types of subdiagram building blocks; and the superscript $1$ indicates the first diagram of the topology associated with that building block.

In terms of the number of time slices, annihilation diagrams are one-point functions, whereas all other diagrams are two-point functions. The time indices in parentheses specify the time coordinates on which the building block depends.

For annihilation diagrams, the superscript denotes the number of points contained in the diagram; the letter $L$ indicates that the building block is an annihilation loop: $\mathbbm{L}^1$ denotes a one-point annihilation loop in space, $\mathbbm{L}^2$ denotes a two-point annihilation loop, and $\mathbbm{L}^3$ denotes a three-point annihilation loop with cyclic symmetry. The gamma-matrix type and momentum magnitude in the subscript of each diagram are specified by the labels in square brackets, with the ordering convention given in Eq.~\ref{eq:general-6-node-diagram}.

For a given topology, we choose a reference diagram denoted by $C({\vec{p}})$. We then identify the endpoints of the loop in the other diagrams. If the orientation of the loop affects the result, as in the topology $\mathbbm{TW}$, it must first be reversed. Denoting the permutation of this loop relative to the reference diagram by $\mathcal{P}$, these diagrams can be written as
\begin{equation}
    C^{\prime}(\{\vec{p}\}) = C^{(*)}(\mathcal{P}(\{\vec{p}\})).
\end{equation}

\subsubsection{Feynman Rules for Meson Contractions}
We summarize the Feynman rules for meson contractions in distillation as follows. These rules are identical to those for fermion propagators in perturbation theory, except that the vertex factor $i \lambda$ (where $\lambda$ is the coupling constant) is replaced by the product of a gamma matrix and an elemental.

Moreover, these diagrams have no ``external legs''; all diagrams are composed of closed loops.

\begin{enumerate}
    \item Propagator:
\begin{equation}
\raisebox{-0.35\height}{\centering\begin{tikzpicture}[global scale = 1.3]
    \node[thick, draw=none, font = \scriptsize, anchor = east] (sink1) at (0, 0) {$t^{\prime}$};
    \node[thick, draw=none, font = \scriptsize, anchor = west] (source1) at (1.618, 0) {$t$};
    \draw[thin, -Stealth] (source1) to (sink1);
\end{tikzpicture}}
\ = \ \tau(t^{\prime},t)
\end{equation}
    \item Vertex:
\begin{equation}
\raisebox{-0.18\height}{\centering\begin{tikzpicture}[global scale = 1.3]
    \node[circle, draw=black, fill=black!20, thick, draw=none, font = \scriptsize, minimum size=15pt, inner sep=0pt] (vertex) at (0.809, 0) {$\Gamma$};
    \node[above, font = \scriptsize] at (0.809, 0.15) {$\vec{p},t$};
    \node[thick, draw=none, font = \scriptsize, anchor = east] (sink1) at (0, 0) {};
    \node[thick, draw=none, font = \scriptsize, anchor = west] (source1) at (1.618, 0) {};
    \draw[thin] (source1) to (vertex);
    \draw[thin] (vertex) to (sink1);
\end{tikzpicture}}
\ = \ \tilde{V}(\vec{p},t) \Gamma
\end{equation}
    \item Trace
\begin{equation}
    \operatorname{Tr}[\cdot]
\end{equation}
\end{enumerate}

It should be noted that momentum conservation does not hold separately on each quark loop, but is enforced at the level of the full correlation function:
\begin{equation}
    \sum_{i} \vec{p}_i + \sum_{f} \vec{p}_f = 0.
\end{equation}

\section{Three-Body Correlation-Function Analysis}
The previous section discussed the topological structure of all correlation functions computed in this work. These contributions are then assembled into a correlation-function matrix\footnotecircle{To increase the statistics, we computed all time sources. After optimization, the topological diagrams in the two-body problem take essentially negligible time, whereas each topology of a three-body contraction costs roughly $10$ hours per ensemble on an NVIDIA A100 GPU.}. This section discusses the analysis of the correlation matrix and the resulting spectra.

\subsection{Effective Masses and Fits}
To determine the lowest energy levels, we solve the generalized eigenvalue problem on each ensemble,
\begin{equation}
C(t)\, v_n(t, t_0) = \lambda_n(t, t_0)\, C(t_0)\, v_n(t, t_0),
\end{equation}
where $\lambda_n(t, t_0)$ is the eigenvalue corresponding to the eigenvector $v_n(t, t_0)$. For the $\pi\pi$ channel, we use a $3 \times 3$ correlation matrix; for the $\pi\pi\pi$ channel, we use a $4 \times 4$ correlation matrix constructed from the operators in Table~\ref{tab:operatorsomega}. The choice of reference time $t_0$ has a negligible effect on the final spectrum.

The energy of the $n$th excited state is obtained by fitting $\lambda_n(t, t_0)$ to a two-state form:
\begin{equation}
\lambda_n(t, t_0) = (1-A_n)\,\mathrm{e}^{-E_n(t-t_0)} + A_n\,\mathrm{e}^{-E_n^{\prime}(t-t_0)},
\label{eq:2state_fit}
\end{equation}
where $E_n$ is the $n$th energy level.

\begin{figure}[htbp]
\centering
\includegraphics[width=0.49\columnwidth]{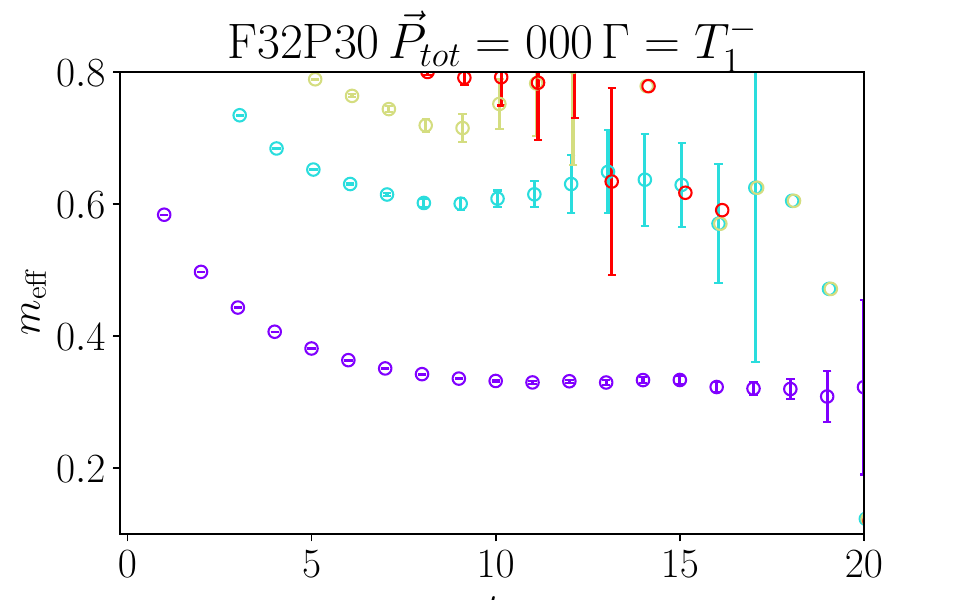}
\includegraphics[width=0.49\columnwidth]{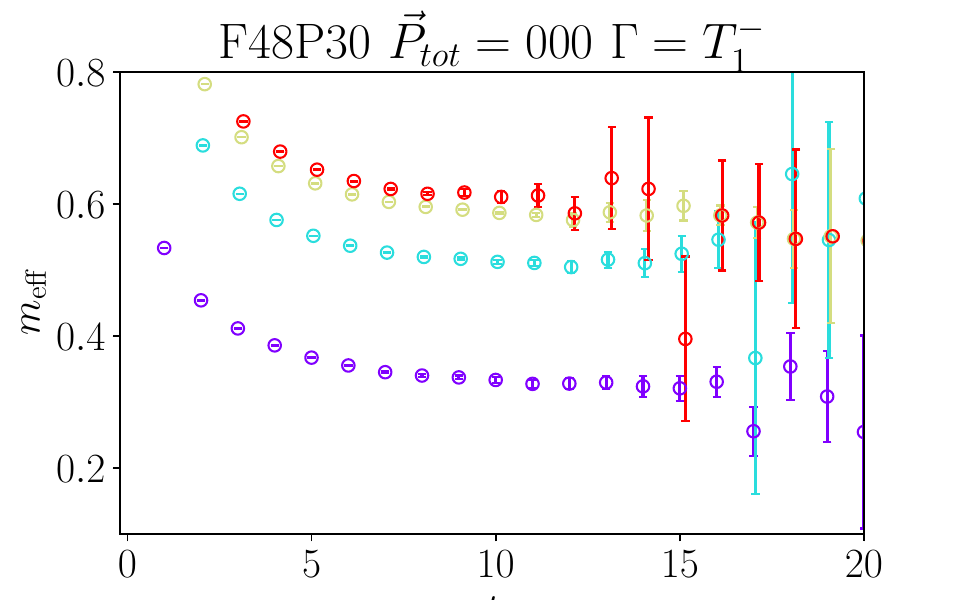}
\\
\includegraphics[width=0.49\columnwidth]{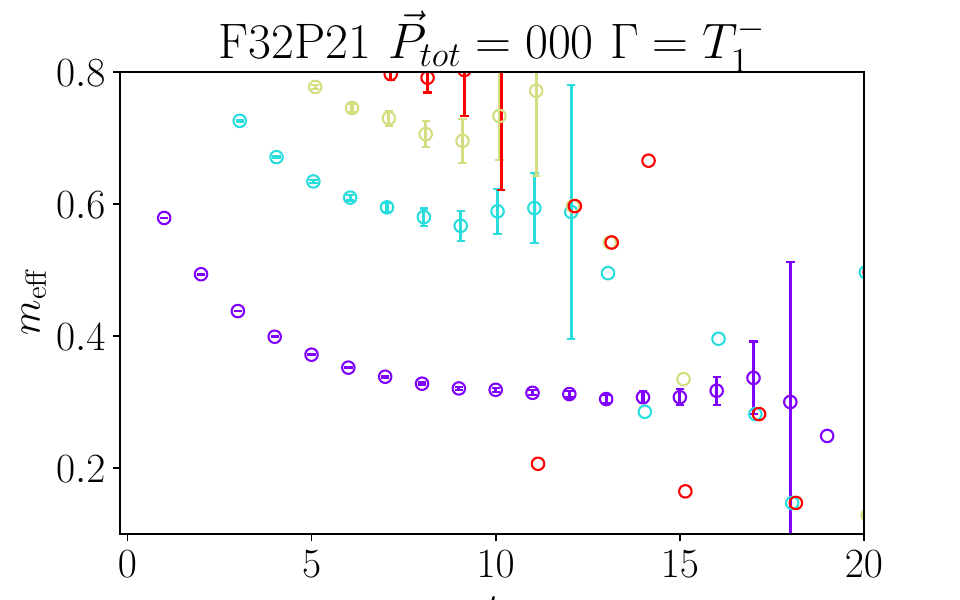}
\includegraphics[width=0.49\columnwidth]{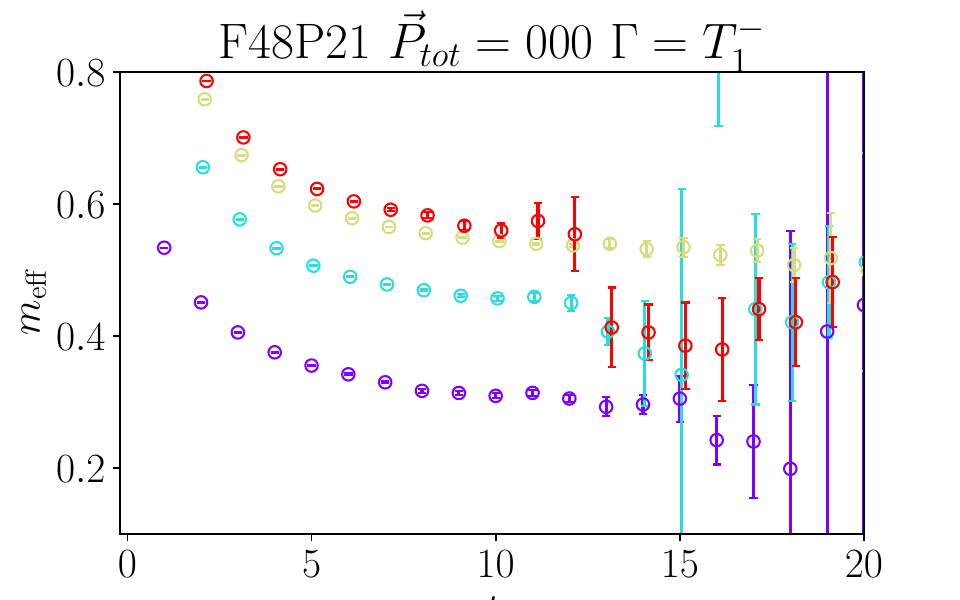}
\caption{Effective-mass plots for the eigenvalues $\lambda_n(t)$ in the $I=0$ $\pi\pi\pi$ channel. Different colors correspond to different energy levels. The vertical axis is in lattice units.}
\label{fig:pipipi-I=0-meff}
\end{figure}

Since, in\chapref{chap:three_body_problems2}, we will perform a more complete analysis of two-body $\pi\pi$ scattering in all isospin channels using the same ensembles, we avoid repetition here and show only the analysis results for the three-body correlation functions. The effective masses in the $I=0$ $\pi\pi\pi$ channel are shown in Fig.~\ref{fig:pipipi-I=0-meff}. Clear plateau behavior is observed at larger times $t$. The fit results for F32P30, F48P30, F32P21, and F48P21 are shown in Figs.~\ref{fig:pipipi-I=0-fit-F32P30}, \ref{fig:pipipi-I=0-fit-F48P30}, \ref{fig:pipipi-I=0-fit-F32P21}, and \ref{fig:pipipi-I=0-fit-F48P21} of Appendix~\ref{appendix:three_body_problems}. We studied the dependence of the levels on the fitting range and chose the starting time $t_{\mathrm{min}}$ such that the fits give reasonable $\chi^2_{\mathrm{dof}}$ values and remain stable under changes of the fit window. Statistical errors were estimated using $2000$ Bootstrap resamples.

\subsection{Spectral Decomposition}
We briefly discussed spectral decomposition in\chapref{chap:lattice_QCD} and used it in\chapref{chap:two_body_problems} to analyze the two-body problem. Here we examine the additional thermal-pollution effects that arise in the three-body problem:
\begin{equation}
\begin{aligned}
     \langle O(t)O^\dagger(0) \rangle &=
    \operatorname{Tr}[^{-T H} e^{tH}O(0)e^{-tH}O^\dagger(0)] \\
    &= \sum_{n,m} e^{-E_m (T-t)} e^{-E_n t} \langle m | O | n \rangle \langle n | O^\dagger | m \rangle \\
    &= |\langle 0 | O | \pi\pi\pi \rangle|^2 [e^{-E_{\pi\pi\pi}t}+e^{-E_{\pi\pi\pi}}(T-t)] + \text{excited $\pi\pi\pi$ states} \\
    &+ |\langle \pi | O | \pi\pi \rangle|^2 (e^{-E_{\pi}T-(E_{\pi\pi}-E_{\pi})t} + e^{-E_{\pi\pi}T-(E_{\pi}-E_{\pi\pi})t}) + \text{excited $\pi\pi$ states}.
\end{aligned}
\end{equation}
In the region $t \ll \frac{T}{2}$, it is sufficient to retain only the lightest $\pi\pi$ state,
\begin{equation}
\begin{aligned}
     \langle O(t) O^\dagger(0) \rangle &\approx |\langle 0 | O | \pi\pi\pi \rangle|^2 [e^{-E_{\pi\pi\pi}t}+e^{-E_{\pi\pi\pi}}(T-t)] \\
     &+ |\langle \pi | O | \pi\pi \rangle|^2 (e^{-E_{\pi}T-(E_{\pi\pi}-E_{\pi})t} + e^{-E_{\pi\pi}T-(E_{\pi}-E_{\pi\pi})t}).
\end{aligned}
\end{equation}
The thermal-pollution term here is induced by the energy level in the $I=1$ $\pi\pi \to \rho$ channel and can be removed using the method described in\chapref{chap:two_body_problems} and Ref.~\cite{Dudek:2012gj}. Since the three $\pi$ mesons carry momenta $[1,1,2]$, both $E_\pi$ and $E_{2\pi}$ are energies in moving frames. The leading contribution is then
\begin{equation}
    \operatorname{e}^{-E_{\pi}([-x]) T} \operatorname{e}^{-(E_{0(\pi\pi)_{I=1}}([xy,-y]) - E_{\pi}([-x])) t},
\end{equation}
where the two-body energy is $E_{0(\pi\pi)_{I=1}}([xy,-y]) = \sqrt{E_{0(\pi\pi)_{I=1}}^2 + (\frac{2\pi}{L})^2}$, and the one-body energy is $E_{\pi}([-x]) = \sqrt{M_{\pi}^2 + (\frac{2\pi}{L})^2}$.

In the $\pi\pi\pi$ channel, no significant thermal pollution is observed at the present level of statistical precision, and it is therefore neglected in the analysis. In the $\pi\pi$ channel, before solving the GEVP we apply a shifted-correlation-matrix procedure to remove thermal-pollution effects.

\subsection{Finite-Volume Spectra}
\begin{figure}[t]
    \centering
    \includegraphics[height=11.2cm,trim=0 0 1.0cm 0,clip]{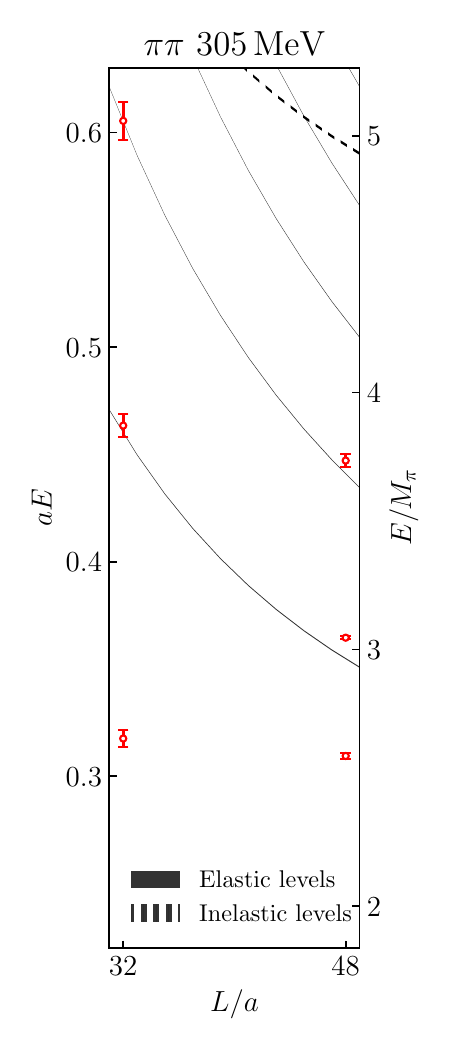}
    \includegraphics[height=11.2cm,trim=1.0cm 0 1.0cm 0,clip]{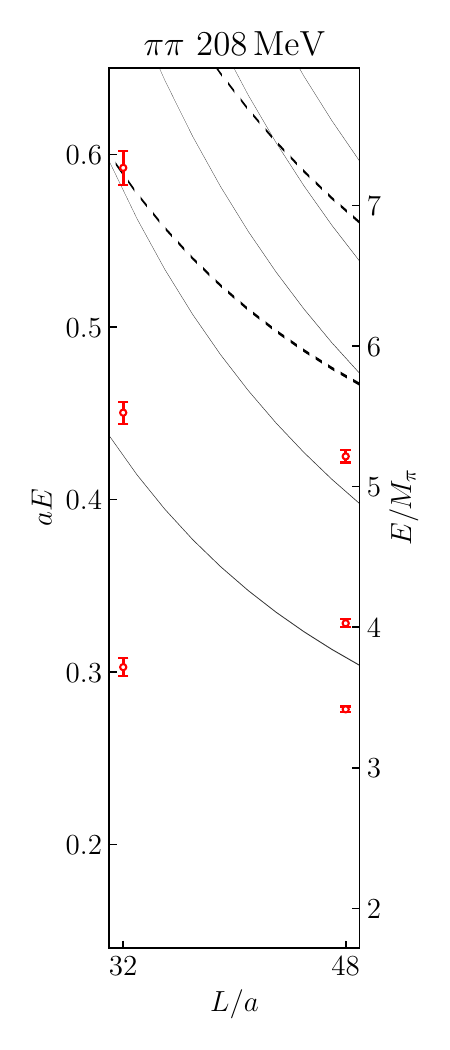}
    \includegraphics[height=11.2cm,trim=1.0cm 0 1.0cm 0,clip]{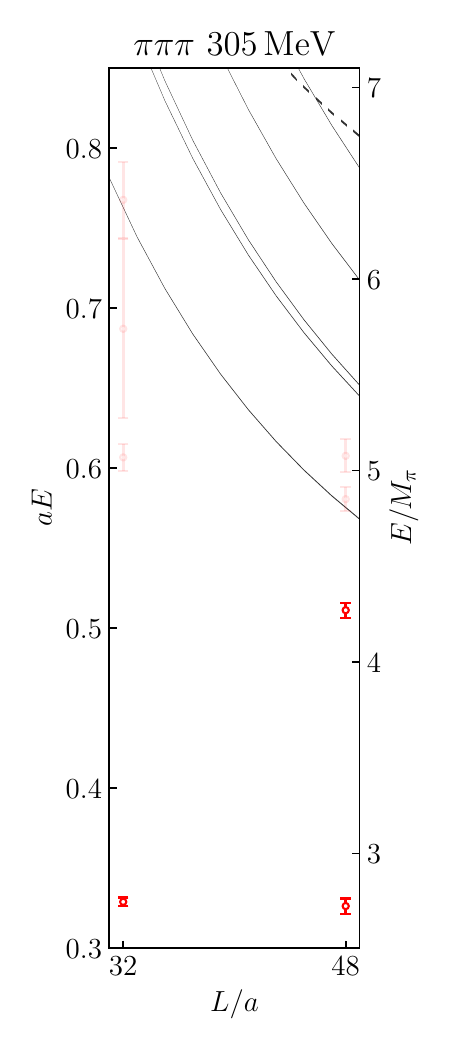}
    \includegraphics[height=11.2cm,trim=1.0cm 0 0 0,clip]{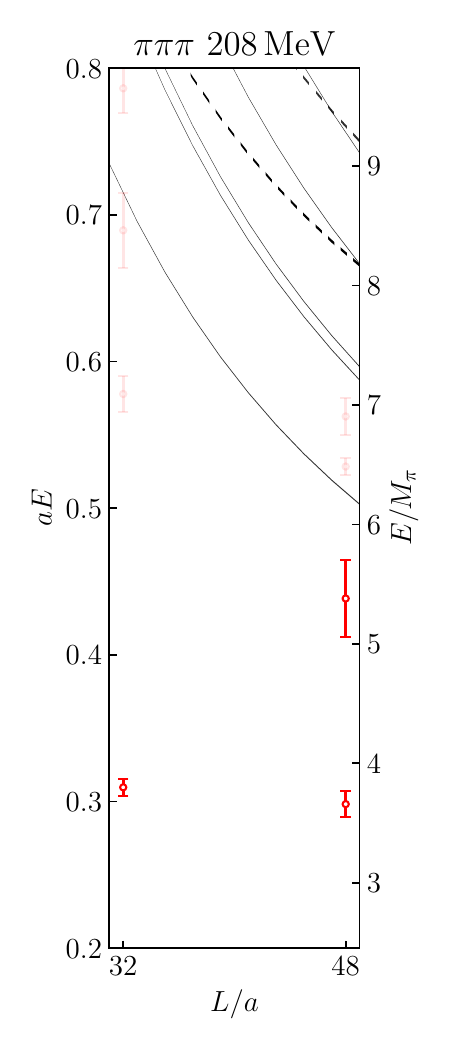}
    \caption{Finite-volume spectra of $\pi\pi \, (I=1)$ and $\pi\pi\pi \, (I=0)$ at $M_{\pi} = 305$ and $208$ MeV. Red points denote the interacting lattice energy levels $aE$ (in units of $M_{\pi}$); the lighter points are not included in the subsequent analysis. Black solid and dashed lines denote the noninteracting elastic and inelastic levels, respectively.}
    \label{fig:omega_spectra_plain}
\end{figure}

The final finite-volume spectra for $\pi\pi$ and $\pi\pi\pi$ are shown in Fig.~\ref{fig:omega_spectra_plain}. The spectrum in the $I=1$ $\pi\pi$ channel exhibits the characteristic pattern of a strongly attractive interaction, as expected from the well-known $\rho$ resonance in this channel. The ground state and first excited state in the $I=0$ $\pi\pi\pi$ channel both lie significantly below the lowest noninteracting level, also indicating strong attraction. In addition, based on the physical mass of the $\omega(782)$, this state can be roughly identified with the ground state. For $M_\pi \approx 208$ MeV, the ground state lies above threshold, implying nonzero phase space for decay and therefore a resonance. For $M_\pi \approx 305$ MeV, the ground-state energy is volume independent and lies below the $\pi\pi\pi$ threshold, suggesting that the $\omega$ is likely a bound state. This observation will be verified in Sec.~\ref{sec:omega_pole}.

\section{Three-Body Quantization Condition}
The finite-volume spectra considered in this chapter contain dynamical information about two- and three-body systems in infinite volume. These dynamical quantities must be extracted through an appropriate quantization condition. In this section we use the finite-volume unitarity (FVU) approach~\cite{Mai:2017bge}, which has been applied to a variety of three-body systems and validated in several studies~\cite{Mai:2018djl, Mai:2019fba, Alexandru:2020xqf, Brett:2021wyd, Mai:2021nul, Garofalo:2022pux, Feng:2024wyg}.

On the theoretical side, this approach has been shown to be equivalent to other established three-body formalisms~\cite{Jackura:2019bmu}. For example, Ref.~\cite{Garofalo:2022pux} showed that the FVU approach and the relativistic field theory (RFT) approach give mutually consistent results for resonances in scalar $\phi^4$ theory.

\subsection{Infinite-Volume Unitarity Method}
The dominant interaction channel of the $\omega$ system can be characterized by the $P$-wave $\pi\rho$ channel~\cite{Gell-Mann:1962hpq}. Therefore, in the FVU formalism, the three-body finite-volume spectrum in the center-of-mass frame is determined by the set of three-body energies $E_3=\sqrt{s}$ satisfying the quantization condition
\begin{equation}
    \det{[\tilde K^{-1}(s)-\Sigma^{FV}(s)]E_L-[\tilde B(s)+\tilde C(s)]}_{\vec{p}’\lambda’,\vec{p}\lambda}^\Gamma=0.
\label{eq:QC-tilde}
\end{equation}

We now describe in detail the matrices appearing in Eq.~\ref{eq:QC-tilde}. This quantization condition follows from the unitarity requirement for the three-body scattering amplitude in finite volume~\cite{Mai:2017vot, Mai:2017bge}. Its central idea is to decompose the three-body scattering process into interactions of two-body subsystems.

\subsubsection{Cluster Decomposition of the Three-Body Scattering Amplitude}
In infinite volume, the scattering amplitude must satisfy the $S$-matrix unitarity condition $\mathcal{S}\mathcal{S}^\dagger=\mathcal{S}^\dagger\mathcal{S}=1$. For the three-body scattering process $\pi(p_1)\pi(p_2)\pi(p_3)\to \pi(p'_1)\pi(p'_2)\pi(p'_3)$, the unitarity condition can be written as~\cite{Mai:2017vot}
\begin{equation}
\begin{aligned}
&\langle q_1,q_2,q_3|(\hat T-\hat T^\dagger)| p_1,p_2,p_3\rangle \\
&= i\int\prod_{\ell=1}^3\left[\frac{\mathrm{d}^4k_\ell}{(2\pi)^{4}}\,(2\pi)\delta^+(k_\ell^2-m^2)\right] \\
&\times(2\pi)^4\delta^4\left(P-\sum_{\ell=1}^3\,k_\ell\right)
\langle q_1,q_2,q_3|\hat T^\dagger|k_1,k_2,k_3\rangle\,
\langle k_1,k_2,k_3|\hat T| p_1,p_2,p_3\rangle,
\end{aligned}
\end{equation}
where $\hat T$ is the three-body scattering-amplitude operator, $P$ is the total four-momentum, and $s=P^2$ is the squared total energy.

\subsubsection{Isobar--Spectator Decomposition}
Using cluster decomposition, the amplitude can be rewritten in a helicity-isospin basis in terms of an isobar and a spectator particle. The three-body scattering amplitude is then separated into a connected part $\hat T_c$ and a disconnected part $\hat T_d$:
\begin{equation}
\begin{aligned}
&\langle q_1,q_2,q_3|\hat T_c(s)| p_1,p_2,p_3\rangle \\
&=\frac{1}{3!}\sum_{n=1}^3\sum_{m=1}^3
v(q_{\bar{n}},q_{\hat{n}}) \tilde \tau(\sigma(q_n))\,
\langle q_n|T(s)|p_m\rangle\,
\tilde \tau(\sigma(p_m))v(p_{\bar{m}},p_{\hat{m}}),
\end{aligned}
\end{equation}

\begin{equation}
\begin{aligned}
&\langle q_1,q_2,q_3|\hat T_d(s)| p_1,p_2,p_3\rangle \\
&=\frac{1}{3!}\sum_{n=1}^3\sum_{m=1}^3\,
2E_{q_n}(2\pi)^3\delta^3(\mathbf{q}_n-\mathbf{p}_m)\,
v(q_{\bar{n}},q_{\hat{n}}) \tilde \tau(\sigma(q_n)) v(p_{\bar{m}},p_{\hat{m}}).
\end{aligned}
\end{equation}
Here $v(p,p')$ is the isobar decay/formation vertex function, $\tilde \tau(\sigma)$ is the isobar propagator, and $T(s)$ is the isobar-spectator interaction amplitude. The indices are defined as above, and $\sigma(q):=(P-q)^2$ is the invariant mass squared of the isobar.

The full three-body scattering amplitude is the sum of the two contributions:
\begin{equation}
\begin{aligned}
&\langle q_1,q_2,q_3|\hat T(s)| p_1,p_2,p_3\rangle \\
&=\langle q_1,q_2,q_3|\hat T_c(s) |p_1,p_2,p_3\rangle+\langle q_1,q_2,q_3|\hat T_d(s)| p_1,p_2,p_3\rangle.
\end{aligned}
\end{equation}

\subsubsection{Bethe--Salpeter Equation and Integral Equation}
The fully connected isobar-spectator amplitude $T_{\lambda'\lambda}(s,\vec{p}',\vec{p})$ in the expression above is an unknown dynamical quantity and must be determined by matching to the unitarity condition. This idea was first introduced in Ref.~\cite{Aaron:1968aoz}, where the amplitude was parametrized in the form of a Bethe--Salpeter equation~\cite{Mai:2017vot}:
\begin{equation}
\begin{aligned}
    T_{\lambda'\lambda}(s,{\vec p}',\vec{p})=&
    B_{\lambda'\lambda}(s,{\vec p}',\vec{p})+
    C_{\lambda'\lambda}(s,{\vec p}',\vec{p}) \\
    &+\int\frac{\mathrm{d}^3l}{(2\pi)^3}
    \left(
    B_{\lambda'\lambda''}(s,{\vec p}',{\vec  l})+C_{\lambda'\lambda''}(s,{\vec p}',{\vec  l })
    \right)
    \frac{\tilde \tau(\sigma_l)}{2E_l} T_{\lambda''\lambda'}(s,{\vec l},{\vec{p}}).
\end{aligned}
\label{eq:T3-integral-equation}
\end{equation}
This equation iteratively sums over all possible intermediate states through an integral equation and thereby provides a complete description of the three-body dynamics. By expanding the BSE and comparing it term by term with the three-body unitarity condition, one can determine the functional forms of $B$, $C$, and $\tau$~\cite{Mai:2017vot, Mai:2017bge}.

\subsubsection{Basic Building Blocks}
The amplitude is built from the following three ingredients:

\subsubsection*{One-$\pi$-Exchange Diagram $B$}
Matching to the unitarity condition shows that $B$ must satisfy the following discontinuity relation:
\begin{equation}
    \langle q|B|p \rangle - \langle q|B^\dagger|p \rangle =iv(P-p-q,q)(2\pi)\delta^+((P-q-p)^2-m^2)v(P-p-q,p).
\end{equation}
Using a dispersion relation, one obtains the covariant form
\begin{equation}
    \tilde B_{\lambda'\lambda}(s,\vec{p}',\vec{p})
    =\tilde I_F\frac{
    v_{\lambda'}^*(p,P_3-p-p')
    v_{\lambda}(p',P_3-p-p')}
    {M_\pi^2-(P_3-p-p')^2-i\epsilon},
\label{btilde}
\end{equation}
where $\tilde I_F=-1$ corresponds to the isospin-$I=0$ channel.

\subsubsection*{Three-Body Force $C$}
In addition to the $B$ term, which can be decomposed into a sequence of two-body interactions, there is a contribution whose discontinuity is not fixed by unitarity. This term encodes genuinely three-body dynamics. For the quantum numbers studied in this work, only the $L=1$ orbital-angular-momentum component appears. Formally, it can be written as
\begin{equation}
\begin{aligned}
    \tilde C_{\lambda'\lambda}(s,\vec{p}^\prime,\vec{p})=&
    \frac{3}{4\pi}\sum^{1}_{M=-1}
    \mathcal{D}^{1*}_{M\lambda}(\phi_{-\vec{p'}},\theta_{-\vec{p'}},0) \times\tilde C_{\lambda'\lambda}(s,p',p)
    \mathcal{D}^{1}_{M\lambda}(\phi_{-\vec{p}},\theta_{-\vec{p}},0), 
\end{aligned}
\label{eq:Ctilde}
\end{equation}
where $\mathcal{D}^{1}_{M\lambda}$ is a Wigner matrix, and
\begin{equation}
\quad
\tilde C_{\lambda'\lambda}(s,p',p)=U_{\lambda'L'}
\left(
\left(p'\right)^{L'}\,
\cdot\tilde c_{L'L}\cdot
\left(p\right)^{L}
\right),
\end{equation}
with the Clebsch--Gordan coefficient combination
\begin{equation}
U_{L\lambda}=\begin{pmatrix}
    \frac{1}{\sqrt{3}}  & \frac{1}{\sqrt{3}}  &\frac{1}{\sqrt{3}}\\
    \frac{1}{\sqrt{2}}&0&-\frac{1}{\sqrt{2}}\\
    \frac{1}{\sqrt{6}}  &-\sqrt{\frac{2}{3}}   &\frac{1}{\sqrt{6}}.
\end{pmatrix}
\end{equation}

$\tilde c_{L'L}$ is a generic parametrization function to be fitted and can be expanded in a Laurent series. The simplest parametrization is
\begin{equation}
    \tilde c_{11}=\frac{c_0}{s-m_\omega^2}+c_1,
\end{equation}
which contains three fit parameters. This term can be interpreted either as the exchange of an infinitely heavy particle in the crossed channel, or as an irreducible genuine three-body force (interaction) that cannot be described by light-particle exchange. Through iteration in the scattering equation, such a contribution can generate a three-body resonance.

\subsubsection*{Two-Body $\tau$ Function}
Matching to the unitarity condition determines the isobar propagator through
\begin{equation}
\begin{aligned}
S(\sigma(k))-S^\dagger (\sigma(k))=&\,
i\,\frac{S(\sigma(k))S^\dagger (\sigma(k))}{2(2\pi)^2}
\int \mathrm{d}^4\bar K\,
\delta^+\left(\left(\frac{P-k}{2}+\bar K  \right)^2-m^2\right) \\
&\times\delta^+\left(\left(\frac{P-k}{2}-\bar K  \right)^2-m^2\right)
\left(v\left(\frac{P-k}{2}+\bar K,\frac{P-k}{2}-\bar K\right)\right)^2.
\end{aligned}
\end{equation}
This equation essentially describes unitarity in the two-body subsystem ($\pi\pi$ scattering), namely the unitarity of $T_{22}=vSv$. Using $S^{-1}(\sigma)= -D(\sigma)$, one obtains the inverse propagator of the $\rho$ resonance,
\begin{equation}
\begin{aligned}
    \tilde \tau(\sigma)
    =&\frac{1}{\tilde K^{-1}_{\rho}(\sigma)-\Sigma_\rho^\text{IV}(\sigma)}, \\
    \tilde K^{-1}_{\rho}(\sigma)=&-\frac{p^3\cot\delta^{11}}{12\pi\sqrt{\sigma}}+\Re\,\Sigma^\text{IV}_{\rho}(\sigma)
    \quad
    \Sigma^\text{IV}_{\rho}(\sigma)=\frac{1}{48\pi^2}\int dk \frac{\sigma^2}{E_k^5}\frac{k^4}{\sigma-4E_k^2},
\end{aligned}
\label{eq:Ktilde+SigmaIV_tau}
\end{equation}
where $p$ is the momentum in the $\pi\pi$ system, $\delta^{11}$ is the $\pi\pi$ phase shift in the $I=1$ channel, and $\Sigma_\rho^\text{IV}$ is the irreducible self-energy of the $\rho$ meson. This $\tau$ function is fully determined by the known $\pi\pi$ phase-shift data.

\subsection{Finite-Volume Unitarity Method}
In finite volume, the basic building blocks are given in the plane-wave-helicity (PWH) basis
\begin{equation}
    \{\pi(\vec{p})\rho_{\lambda}(-\vec{p}) \mid \vec{p}L/(2\pi)\in \mathbbm{Z}^3,\ \lambda\in\{-1,0,+1\}\},
\end{equation}
and each matrix element of the equation is projected onto the $T_1^-$ irrep of the $O_h$ group.

Two building blocks deserve special mention:
\subsubsection*{Finite-Volume Self-Energy $\Sigma^{\mathrm FV}$}
In finite volume, the self-energy of the $\rho$ meson (the $\pi\pi$ pair) must be computed in the Lorentz-boosted reference frame, because the lattice simulation breaks Lorentz invariance. More specifically, there is a Lorentz transformation between the three-body center-of-mass frame and the two-body center-of-mass frame of the $\pi\pi$ pair. Two energy regions must be distinguished:
\begin{itemize}
    \item \textbf{Matching region}: This is defined as $Reg = \big[\sqrt{s_{2}},\sqrt{s_{2}}+0.1(\sqrt{s_{\mathrm{phys}}}-\sqrt{s_{1}})\big]$, where $\sqrt{s_{1/2}} = E_p \pm |p|$ and $\sqrt{s_{phys}} = \sqrt{|\vec{p}|^2+4} + E_p$. In this region, $\Sigma^{\mathrm FV}$ is matched by interpolation to the infinite-volume expression $\Sigma^{\mathrm IV}$, so that errors induced by the breaking of Lorentz invariance are suppressed to exponential order $e^{-M_\pi L}$;
    \item \textbf{Exterior region}: Outside this region, one directly uses the finite-volume self-energy expression
    \begin{equation}
        \Sigma^{LP}_{\rho}(\sigma,\lambda',\lambda)=
        \frac{1}{2}
        \frac{J_P}{(M_\pi L)^3}
        \sum_{\vec{k}\in\mathcal{S}_{L}}
        \left(\frac{\sigma}{4E_{k^*}^2}\right)^2
        \frac{
        \epsilon_{\lambda'}^{*\mu}(\vec{0})~(-2k^*)_\mu
        \epsilon_{\lambda}^{\nu}(\vec{0})~(-2k^*)_\nu
        }{E_{k^*}(\sigma-4E_{k^*}^2)},
    \end{equation}
    where $k^*$ is the Lorentz-boosted momentum and $J_P$ is the Jacobian factor. The full expression for $\Sigma^{\mathrm FV}$ is
    \begin{equation}
        \left[\Sigma^{\mathrm FV}(s)\right]_{(\vec{p}',\lambda')(\vec{p},\lambda)}=
        \delta_{\vec{p}'\vec{p}}
        \begin{cases}
            \delta_{\lambda'\lambda}\Sigma^{\mathrm IV}_{\lambda}(\sigma_p), & \sqrt{s}\in Reg, \\
            \big[\Sigma^{LP}_{\rho}(\sigma_p)\big]_{\lambda'\lambda}, & \text{otherwise}.
        \end{cases}
    \end{equation}
\end{itemize}

\subsubsection*{Two-Body $K$-Matrix Parametrization}
$\tilde K^{-1}$ encodes the two-body dynamical information. Since the spectator particle can carry arbitrarily high momentum $p \in [0,\infty)$, the three-body equation depends explicitly on the behavior of the two-body $K$ matrix in the unphysical energy region $\sigma_p \in (-\infty, 4)$, i.e. the subthreshold region. We take the parametrization to be
\begin{equation}
    \left[\tilde K(s)^{-1}\right]_{(\vec{p}',\lambda')(\vec{p},\lambda)}=
    \delta_{\vec{p}'\vec{p}}
    \delta_{\lambda'\lambda}
    \big(1+e^{-(\sigma_{\vec{p}}(s)-\sigma_0)}\big)\big(a_0+a_1\sigma_{\vec{p}}(s)\big),
    \label{eq:Ktilde-inv-expanded}
\end{equation}
where $\sigma_{\vec{p}}(s) = s + M_\pi^2 - 2\sqrt{s}\,E_{\vec{p}}$ is the squared invariant mass of the two-body subsystem. The physical meaning of the cutoff parameter $\sigma_0$ is that, in the subthreshold region $\sigma_p < \sigma_0$, the contribution from the $K$ matrix is suppressed by the exponential factor $e^{-(\sigma_{\vec{p}}-\sigma_0)}$ to avoid unphysical singularities.

Combining all these building blocks, the quantization condition for the finite-volume spectrum can be written as the following matrix determinant equation:
\begin{equation}
    (\tilde K^{-1}(s)-\Sigma^{\mathrm FV}(s))E_L - (\tilde B(s)+\tilde C(s)) = 0.
\end{equation}
That is, the energy eigenvalues correspond to the zeros of the determinant of this matrix.

The matrices $\tilde B$, $\Sigma^{\mathrm FV}$, and $E_L$ represent all on-shell configurations of the three $\pi$ mesons and therefore contain all power-law volume dependence of the system; these matrices are completely fixed by unitarity. By contrast, if exponentially suppressed terms of order $e^{-M_\pi L}$ are neglected, $\tilde K^{-1}$ and $\tilde C$ are volume-independent quantities that encode two-body and three-body dynamics, respectively.

The F32P21 volume is small, with $M_{\pi} L$ around $2.6$, so exponentially small terms may not be negligible and may induce finite-volume uncertainties. We assess this systematic effect by removing the F32P21 energy levels from the EFT4 fit. The resulting shift relative to the original fit is comparable in size to the statistical uncertainty.

The two-body interaction can be obtained from the two-body levels whose energies satisfy
\begin{equation}
    \{E_2\in\mathbbm{R}\mid \tilde K^{-1}=\Sigma^{\mathrm FV},\ \vec{p}^{(\prime)}=\vec{0}\},
\end{equation}
which is equivalent to the usual Lüscher method~\cite{Luscher:1990ux} when exponentially suppressed terms are neglected. In general, $\tilde K^{-1}$ and $\tilde C$ are not known a priori. Below, guided by the physics relevant to the $\rho$ and $\omega$ systems, we discuss different choices for these quantities based on generic parametrizations and effective-field-theory frameworks.

\subsection{Irrep Projection}
Equation~\ref{eq:QC-tilde} contains all information about the cubic group, but it must still be projected onto a specific irrep before it can be applied to lattice data. We discuss two equivalent projection schemes below.

\subsubsection{Projector Method}
For any matrix $M\in\{Q, B, C, \dots\}$ appearing in Eq.~\ref{eq:QC-tilde}, one can show that it satisfies
\begin{equation}
M = U(R)\, M\, U(R)^\dagger, \quad R \in \{O_h\},
\end{equation}
where $U(R)$ is the unitary representation matrix corresponding to the rotation-group element $R$. Its structure is
\begin{equation}
    U_{\vec{p} \ell' m';\,\vec{q} \ell m}(R) 
    = \delta_{\vec{q}_R \vec{p}} \delta_{\ell'\ell}
    \mathcal{D}^{(\ell)}_{mm'}(R)
    =
    \begin{cases}
    \mathcal{D}^{(\ell)}_{mm'}(R), & R\vec{q} = \vec{p}, \\
    0, & \text{otherwise}.
    \end{cases}
\end{equation}

Since $\{U(R)\}_{R\in O_h}$ forms a representation of the group $O_h$, which is also the symmetry group of the matrix $M$, we can construct the projector for row $\gamma$ of the irrep $\Gamma$ using the standard method:
\begin{equation}
    P_{\Gamma\gamma} 
    = \frac{\dim(\Gamma)}{|G|}
    \sum_{R\in G} 
    S^{\Gamma *}_{\gamma\gamma}(R)\,
    U_{\vec{p}\vec{q}}(R),
    \label{eq:Projector}
\end{equation}
where $S^\Gamma_{\gamma\gamma}(R)$ is the matrix element representing the group element $R$ in the irrep $\Gamma$. One can verify that $P^\dagger = P$.

In the implementation, we take the orthonormalized set of eigenvectors $\{v_i\}$ of $P_{\Gamma\gamma}$ with eigenvalue $1$, and use them to construct the projected matrix in each irrep subspace:
\begin{equation}
    M^{\Gamma\gamma}_{i,j} 
    = v_i^\dagger \, M \, v_j.
\end{equation}

It should be noted that, in general, multiple linearly independent vectors $v_i$ exist for each irrep. Therefore, the left-hand side of the projected equation remains a matrix, and the spectrum is determined by solving for the zeros of its determinant.

\subsubsection{State-Vector Method}
Let the matrix inside the determinant in Eq.~\ref{eq:QC-tilde} be denoted by $T_{\mathrm{FVU}}$. The operators in Sec.~\ref{sec:three_body_operators} directly give the Cartesian-coordinate expression for the state vector $| \Gamma r \rangle$ corresponding to a given row of the irrep $\Gamma$. Using
\begin{equation}
\begin{cases}
    V_{+1} &= \frac{-V_x + i V_y}{\sqrt{2}}, \\
    V_{-1} &= \frac{V_x + i V_y}{\sqrt{2}}, \\
    V_{0} &= V_z, \\
\end{cases}
\end{equation}
to transform the state vector from the Cartesian basis $(x,y,z)$ of the canonical basis to the circular basis $(+1,0,-1)$, one can project directly:
\begin{equation}
    \langle \Gamma r | T | \Gamma r \rangle = \langle \Gamma r | \vec{p}^{\prime } \lambda^{\prime} \rangle \langle \vec{p}^{\prime } \lambda^{\prime} | T | \vec{p} \lambda \rangle \langle \vec{p} \lambda | \Gamma r \rangle.
\end{equation}

We verified that the two schemes give identical zero locations, i.e. identical energy solutions. Numerically, the projector method gives a flatter slope near the zero than the state-vector method, but it often produces excessively large numerical magnitudes, which can cause the root-finding algorithm to fail in some cases. The state-vector method involves matrix inversion and is therefore slightly slower. In the actual fits we use the state-vector method\footnotecircle{In internal discussions, we referred to it as the Mai/Yan vector.}.

\subsection{Parametrization Schemes}
\subsubsection{Laurent Expansion}
The two-body interaction can be generically parametrized as
\begin{equation}
[\tilde K^{-1}]_{\vec{p}'\lambda',\vec{p}\lambda}
= \delta_{\lambda'\lambda}\,\delta_{\vec{p}'\vec{p}}
\sum_{i=0}^{N} a_i \sigma_{p}^i,
\end{equation}
where the two-body invariant mass is defined by
\begin{equation}
\sigma_p := s + M_\pi^2 - 2\sqrt{s}\sqrt{\vec{p}^2+M_\pi^2},
\qquad (E_2=\sqrt{\sigma_{0}})
\end{equation}
$\vec{p}$ is the spectator momentum. We find that taking $N=1$ is sufficient to describe the lattice spectrum; mathematically, this choice is also equivalent to the conventional Breit--Wigner form.

Similarly, the three-body interaction term $\tilde C$ can be expanded generically in the orbital-angular-momentum (JLS) basis, corresponding to the relative $P$-wave $\pi\rho$ system:
\begin{equation}
\tilde c_{11}=\frac{c_0}{s-M_\omega^2}+c_1+\cdots.
\end{equation}
It is then mapped to the PWH basis~\cite{Chung:1971ri, Sadasivan:2020syi, Feng:2024wyg}. The order of the expansion depends on the amount and precision of the input data. In the present analysis, we find that the two-parameter fit $(c_0, M_\omega)$ already provides sufficient flexibility. We refer to this parametrization scheme as GEN (generic).

\subsubsection{Effective-Field-Theory Approach}
The GEN approach does not allow for a chiral extrapolation to the physical point; this limitation can be overcome using effective field theory (EFT). With EFT guidance, we can also perform a global fit to ensembles with different $\pi$ masses, using more data with fewer parameters. EFT has been widely used in two-body systems~\cite{Mai:2021lwb,Mai:2019pqr}, but it has not yet been systematically applied to lattice studies of three-hadron resonances. On the other hand, continuum results for $\omega\to\pi\pi\pi$ and $\rho\to\pi\pi$ have existed for decades; see the review in Ref.~\cite{Meissner:1987ge}. Using the results summarized there, we match at the level of the $2\to2$ and $3\to3$ scattering amplitudes. A somewhat lengthy but straightforward tree-level calculation gives
\begin{equation}
\begin{aligned}
    &\left[\tilde K^{-1}\right]_{\vec{p}'\lambda',\vec{p}\lambda}
    =\delta_{\lambda'\lambda}\,
    \delta_{\vec{p}'\vec{p}}\frac{\sigma_p-M_\rho^2}{2g^2}, \\
    &\tilde c_{11}
    =\frac{6s (M_\rho^2-\sigma_{q}+6g^2f_\pi^2)
    (M_\rho^2-\sigma_{p}+6g^2f_\pi^2)}
    {64g^2\pi^3 f_\pi^6(s-M_\omega^2)},
    \label{eq:c11-eft-matching}
\end{aligned}
\end{equation}
where the latter expression is given in the JLS basis and then projected to the PWH basis~\cite{Chung:1971ri, Sadasivan:2020syi, Feng:2024wyg}.

In the derivation above, we assume that both the two-$\pi$ and three-$\pi$ interactions are dominated by $s$-channel resonance exchange. This assumption is justified by the narrow widths of the $\rho$ and $\omega$ mesons. Following Ref.~\cite{Meissner:1987ge}, we take
\begin{equation}
    g_{\rho\pi\pi}=g_{\omega\rho\pi}=g.
\end{equation}
The matching relation in Eq.~\ref{eq:c11-eft-matching} shows that, once $(g,M_\rho,M_\omega)$ is specified, both the two-body and three-body interactions are determined.

Using the Kawarabayashi-Suzuki-Fayyazuddin-Riazuddin (KSFR) relation~\cite{Riazuddin:1966sw, Kawarabayashi:1966kd}
\begin{equation}
    M_\rho=\sqrt{2} g f_\pi,
\end{equation}
one can further reduce the number of free parameters. In practice, this provides a constraint on the chiral extrapolation through the $f_\pi(M_\pi)$ dependence from chiral perturbation theory~\cite{Gasser:1983yg}.

We define the EFT2 scheme by using the generalized KSFR relation while allowing for a shift independent of the $\pi$ meson mass,
\begin{equation}
    M_\omega = M_\rho+\delta = \sqrt{2} g f_\pi+\delta,
\end{equation}
so that, for a given $f_\pi$, the free parameters are $(g,\delta)$.

If the KSFR relation is abandoned completely, we refer to the scheme as EFT4:
\begin{equation}
\begin{aligned}
    M_\rho &= M_V+a\,M_\pi^2, \\
    M_\omega &= M_V+a\,M_\pi^2+\delta,
\end{aligned}
\end{equation}
see Refs.~\cite{Bruns:2004tj,Yu:2023xxf}. This scheme has four free parameters, $(g,M_V,a,\delta)$.

Clearly, the EFT schemes described above form only a subclass of a broader EFT parametrization with more degrees of freedom~\cite{Meissner:1986ka}. Ultimately, the validity of these schemes must be tested through comparison with lattice QCD results.

\section{Scattering-Parameter Fits}
\label{sec:omega_pole}
After fitting the lattice spectrum, the three-body parameters of the scattering amplitude can be determined, and the pole positions on the second Riemann sheet can be used to extract the universal parameters of the $\rho$ and $\omega$ mesons. The corresponding theoretical framework is provided by the infinite-volume unitarity (IVU) method~\cite{Mai:2017vot, Mai:2017bge}, which is the infinite-volume counterpart of the FVU quantization condition discussed above.

The part of the scattering amplitude responsible for generating resonance poles is given by the integral equation
\begin{equation}
\begin{aligned}
    T=\tilde B+\tilde C+\int\frac{d^3l}{(2\pi)^3}
    \frac{\tilde B+\tilde C}{2E_l(\tilde K^{-1}-\Sigma^{IV})}\,T.
    \label{eq:IVU}
\end{aligned}
\end{equation}
For compactness, the kinematic variables have been suppressed; see Eq.~\ref{eq:QC-tilde} and Refs.~\cite{Mai:2021nul,Sadasivan:2021emk, Feng:2024wyg} for details. Here, $\Sigma^{IV}$ denotes the usual $\rho$-meson self-energy integral.

The integral equation above can be solved in the JLS basis by deforming the complex integration contour~\cite{Hetherington:1965zza, Sadasivan:2020syi, Feng:2024wyg}, and the pole positions can then be extracted using the methods described in Refs.~\cite{Mai:2021nul, Sadasivan:2021emk, Garofalo:2022pux}.

For two-body resonance poles, the search is equivalent to solving, on the second Riemann sheet,
\begin{equation}
    \{E_2\in\mathbbm{C}\mid \tilde K^{-1}=\Sigma^{IV}\}.
\end{equation}

The three-body quantization condition is necessarily expressed as an infinite-dimensional determinant equation in the PWH basis~\cite{Mai:2021lwb}. In this work, for computational reasons, we truncate the spectator momentum at
\begin{equation}
    {\vec p}_{\mathrm max}=(2\pi)/L\,(0,1,1),
\end{equation}
and below.

A basic feature of three-body problems is that the spectator particle can carry arbitrarily large momentum, causing the squared invariant mass of the two-body subsystem to become negative. Since the exact form of $\tilde K^{-1}$ in the unphysical region is unknown, we introduce a simple shape factor to regulate it:
\begin{equation}
    \tilde K^{-1} \to (1+e^{-(\sigma-\sigma_0)/M_\pi^2} ) \tilde K^{-1},
\end{equation}
where $\sigma_0=2M_\pi^2$.

We tested other functional forms, different values of $\sigma_0$, and different choices of ${\vec p}_{\mathrm max}$. The results show that the truncation of the unphysical region has no significant effect on the extracted physical observables. Further details on truncation effects in three-body systems can be found in Refs.~\cite{Mai:2018djl, Sadasivan:2020syi, Feng:2024wyg}.

We consider both separate fits to ensembles at different $\pi$ meson masses and combined fits, using parametrizations of the two-body and three-body interactions denoted by GEN(305), GEN(208), EFT2(208/305), and EFT4(208/305). Statistical errors are estimated from $2000$ Bootstrap samples. The definition of $\chi^2$ is given in Eq.~\ref{eq:def_chi2dof} of\chapref{chap:two_body_problems}.

For the EFT4 fit, we also assess the systematic uncertainty from exponentially suppressed finite-volume effects by excluding the levels from the F32P21 ensemble, which has the smallest $M_\pi L$. The difference between fit parameters obtained with and without this ensemble is quoted as the second error in square brackets. This systematic uncertainty is comparable to, or slightly larger than, the statistical error, indicating that finite-volume effects do not significantly affect the fit results.

\begin{table}[h]
\centering
\caption{Parameters from separate and combined fits to the two-body and three-body spectra.}
\begin{tabular}{c|cccl}
\toprule
$M_{\pi}/\mathrm{MeV}$ & $\tilde{K}^{-1}$ & $c_{11}$ & $\chi^2_{\mathrm dof}$ & Parameters \\
\midrule
\multirow{2}{*}{305} & \multirow{2}{*}{$a_0 + a_1 \sigma$} & \multirow{2}{*}{$\frac{c_0}{s - M_\omega^2}$} & \multirow{2}{*}{$1.25$} & $a_0 = -9556(597) \,\mathrm{MeV}^2,\ a_1 = 0.01393(94)$ \\
& & & & $c_0 = 5.7(4.2) \times 10^6 \,\mathrm{MeV}^2,\ M_{\omega} = 841.5(7.3) \,\mathrm{MeV}$ \\
\midrule
\multirow{2}{*}{208} & \multirow{2}{*}{$a_0 + a_1 \sigma$} & \multirow{2}{*}{$\frac{c_0}{s - M_\omega^2}$} & \multirow{2}{*}{$1.57$} & $a_0 = -8068(718) \,\mathrm{MeV}^2,\ a_1 = 0.0130(14)$ \\
& & & & $c_0 = 2(17) \times 10^6 \,\mathrm{MeV}^2,\ M_{\omega} = 784(13) \,\mathrm{MeV}$ \\
\midrule
\multirow{3}{*}{305/208} & $\frac{\sigma - M_{\rho}^2}{2 g^2}$ & $c_{11}^{\mathrm{EFT}}$ & $3.16$ & $g = 5.432(29),\ \delta = 36.2(6.7) \ \mathrm{MeV}$ \\
\cmidrule(lr){2-5}
& \multirow{2}{*}{$\frac{\sigma - M_{\rho}^2(M_V, a)}{2 g^2}$} & \multirow{2}{*}{$c_{11}^{\mathrm{EFT}}$} & \multirow{2}{*}{$2.29$} & $g = 5.960(167)[5],\ \delta = 38.8(6.9)[7.3] \ \mathrm{MeV}$ \\
& & & & $M_V = 737(12)[2] \ \mathrm{MeV},\ a = 0.00096(14)[4] \ \mathrm{MeV}^{-1}$ \\
\bottomrule
\end{tabular}
\label{tab:23fit}
\end{table}

\begin{figure}[htbp]
\centering
\includegraphics[width=0.49\columnwidth]{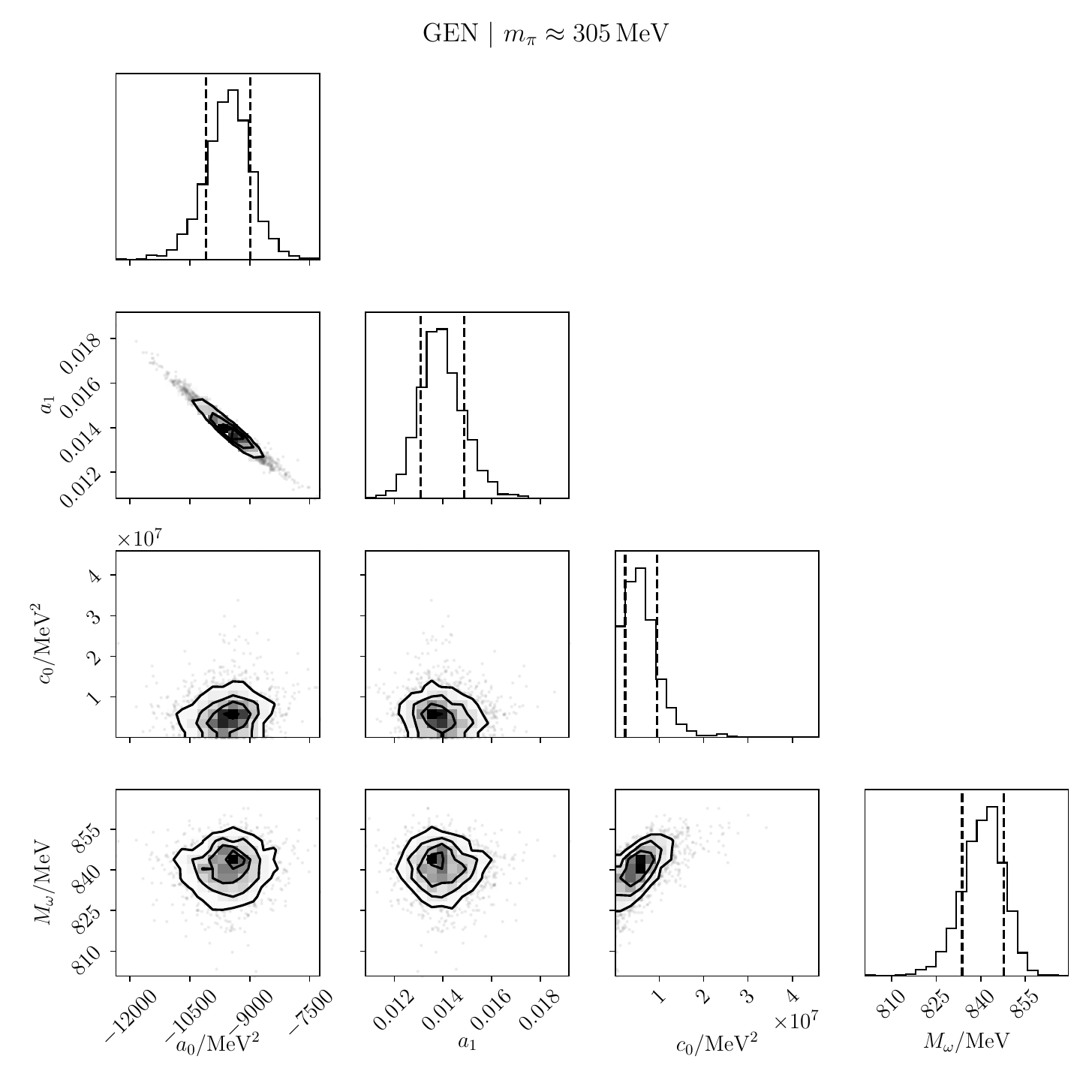}
\hfill
\includegraphics[width=0.49\columnwidth]{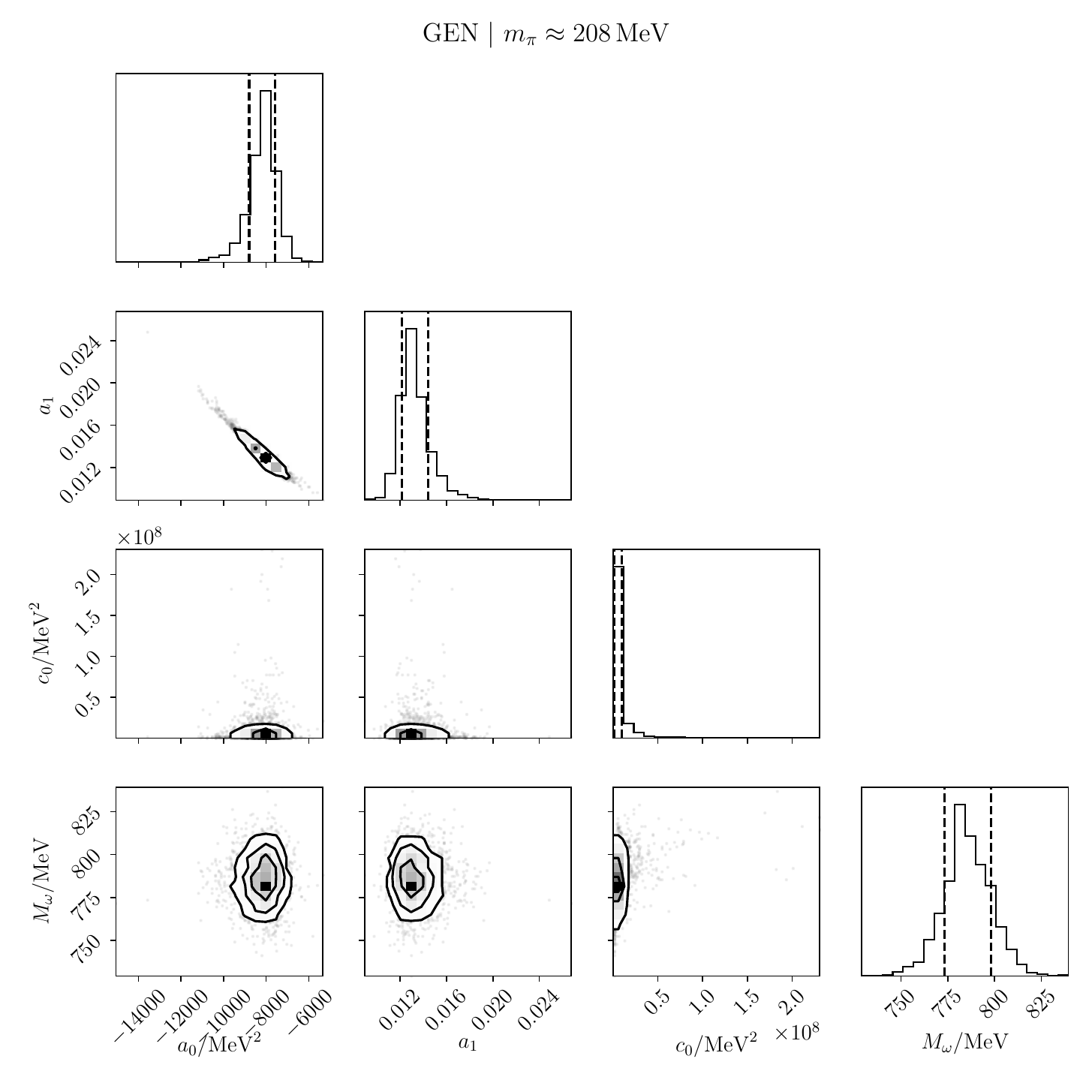}
\caption{Corner plots for the GEN-parametrization fit results at the heavier (left) and lighter (right) $\pi$ masses.}
\label{fig:GENcorner}
\end{figure}

\begin{figure}[htbp]
\centering
\includegraphics[width=0.30\columnwidth]{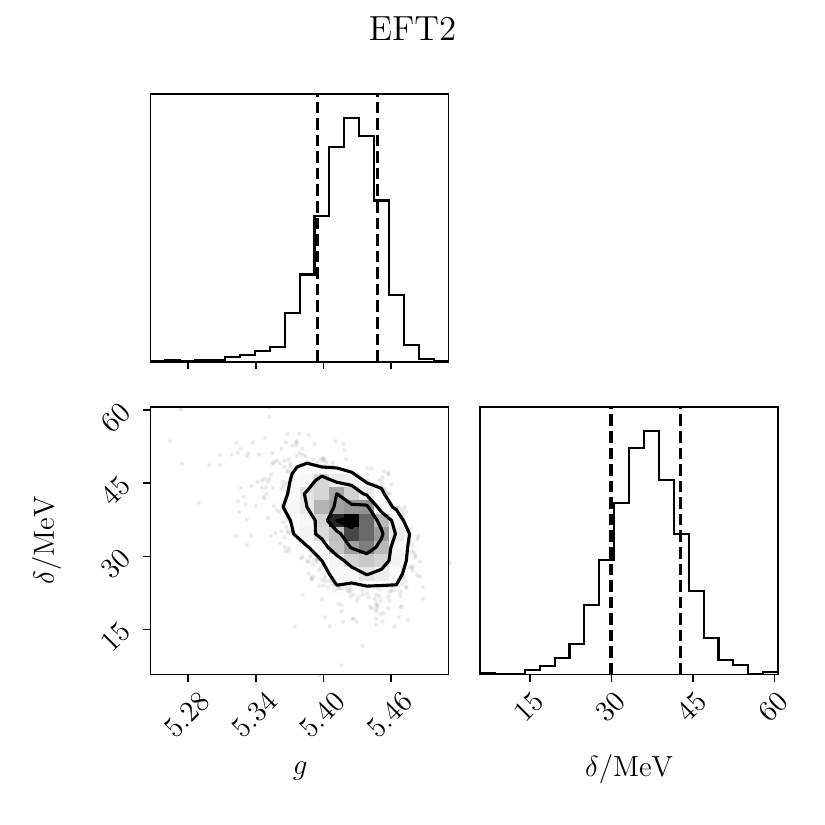}
\hfill
\includegraphics[width=0.6\columnwidth]{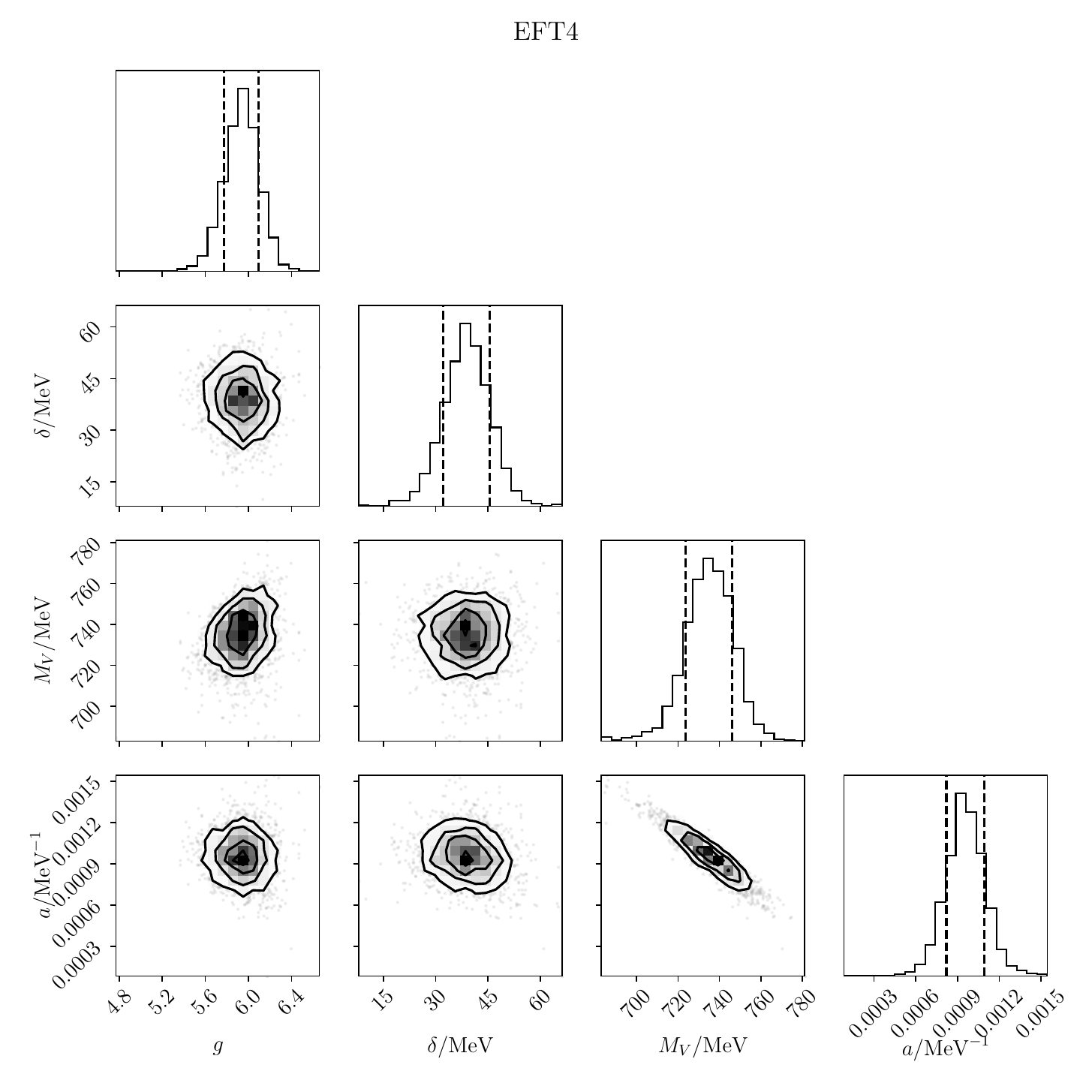}
\caption{Corner plots for the EFT2 (left) and EFT4 (right) parametrizations obtained from combined fits to the light- and heavy-$\pi$ ensembles.}
\label{fig:EFTcorner}
\end{figure}

The distributions and covariances of the fitted parameters in the GEN and EFT analyses are shown in the corner plots in Fig.~\ref{fig:GENcorner} and Fig.~\ref{fig:EFTcorner}, respectively.

\begin{figure}[t]
\centering
\includegraphics[height=11.2cm,trim=0 0 1.0cm 0,clip]{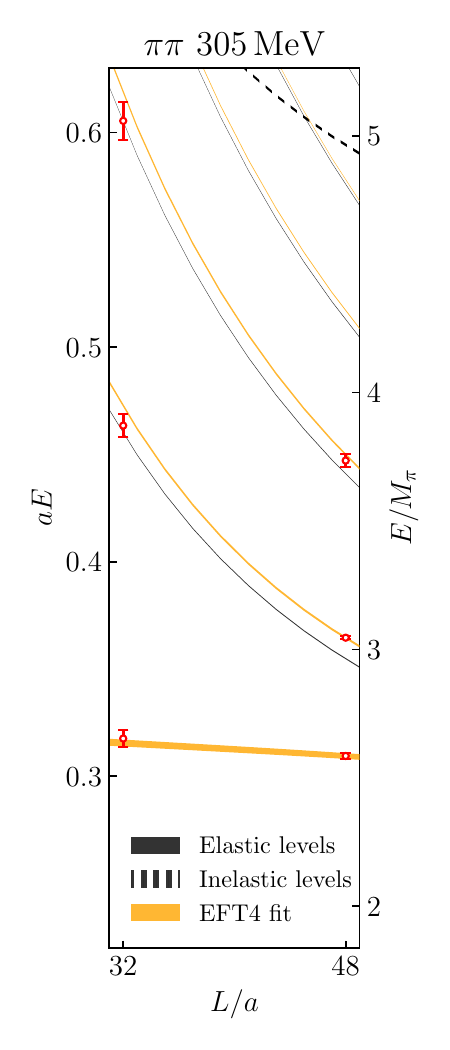}
\includegraphics[height=11.2cm,trim=1.0cm 0 1.0cm 0,clip]{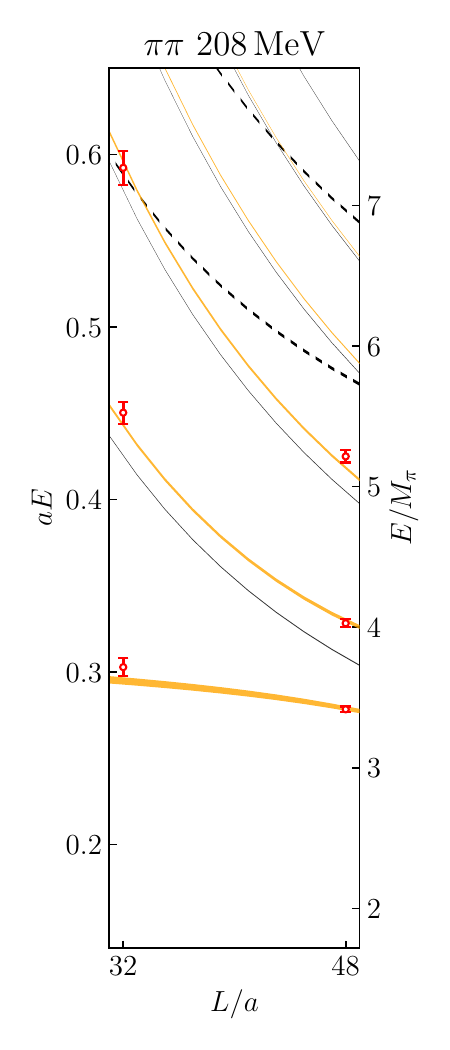}
\includegraphics[height=11.2cm,trim=1.0cm 0 1.0cm 0,clip]{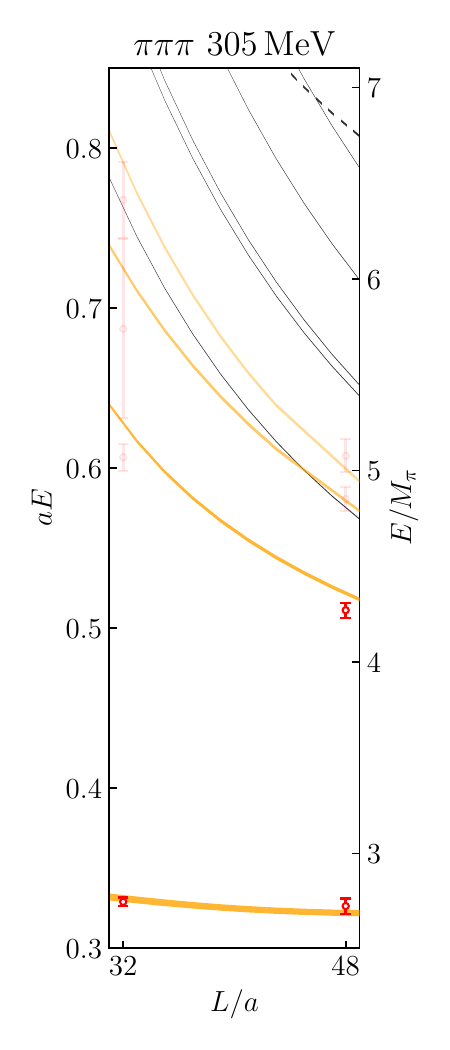}
\includegraphics[height=11.2cm,trim=1.0cm 0 0 0,clip]{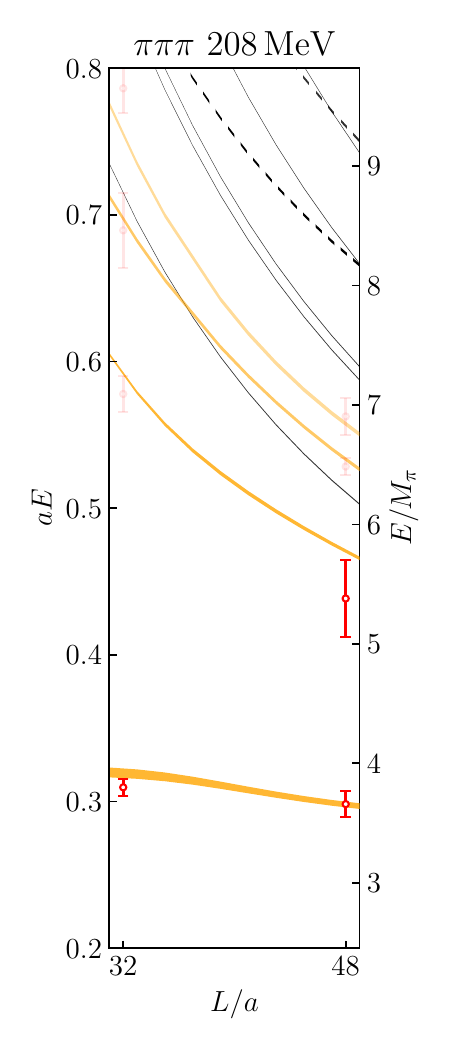}
\caption{
    Finite-volume spectra of $\pi\pi \, (I=1)$ and $\pi\pi\pi \, (I=0)$ at $M_{\pi} = 305$ and $208$ MeV. The orange bands denote the spectra reconstructed from the fit results and the quantization condition.
    }
\label{fig:omega_spectra_fit}
\end{figure}

Using the EFT4 parametrization and the fitted parameters, and inserting them into the quantization condition, one can obtain the finite-volume spectrum as a continuous function of the volume. The corresponding result is shown in Fig.~\ref{fig:omega_spectra_plain} and represented by the orange bands in Fig.~\ref{fig:omega_spectra_fit}. Notice that, for the three-body spectrum, several higher levels in addition to the lowest one or two levels used in the fit also satisfy the quantization condition. In this work, however, we do not use excessively high levels in the fits.

Several sources of systematic uncertainty have not yet been included here. First, the orange bands are based only on the EFT4 parametrization and do not include uncertainties from model dependence. Second, the analysis does not include coupling to the $K\bar{K}$ channel, although its threshold lies very close to the energy region studied here. A systematic investigation of this issue will be clarified in our forthcoming work.

We also tested several possible improvements, but found that none of them led to a significant enhancement:
\begin{itemize}
    \item simultaneously treating the low-energy constants (LECs) as free parameters in the fit;
    \item forming ratios of excited-state levels to the ground-state level, with the aim of reducing correlated errors between energy levels;
    \item directly fitting ratios of energy levels in the finite-volume fit, with the aim of reducing exponentially suppressed finite-volume effects;
    \item fitting energy levels relative to thresholds, namely replacing
    \begin{equation}
        E_{\mathrm{lat}} - E_{\mathrm{qc}}
        \;\longrightarrow\;
        (E_{\mathrm{lat}} - N M_{\pi}(L))
        - (E_{\mathrm{qc}} - N M_{\pi}(\infty))
        =
        (E_{\mathrm{lat}} - N (M_{\pi}(L) - M_{\pi}(\infty)))
    - E_{\mathrm{qc}},
    \end{equation}
    where the volume dependence of the $\pi$ meson mass can be parametrized as
    \begin{equation}
    \begin{cases}
        M_{\pi}(L) = M_{\pi}(\infty)
        \left(1 + c \operatorname{e}^{-M_{\pi}(\infty) L}\right), \\[6pt]
        M_{\pi}(L) = M_{\pi}(\infty)
        \left(1 + c \dfrac{\operatorname{e}^{-M_{\pi}(\infty) L}}
        {(M_{\pi} L)^{\frac{3}{2}}}\right).
    \end{cases}
    \end{equation}
\end{itemize}
All of the tests above changed the final results only within statistical uncertainties. We therefore do not adopt them.

\section{Pole Positions}
The two-body finite-volume spectrum contains three energy eigenvalues below the first inelastic threshold, because in the $T_1^-$ irrep the two $\pi$ mesons must carry at least one unit of momentum. For the three-body spectrum, we analyze only the ground state and the first excited state on the large-volume ensembles, as shown in Fig.~\ref{fig:omega_spectra_fit}. Although, at a qualitative level, the method appears capable of describing higher levels as well, a quantitative analysis would require a richer lattice-operator basis and more free parameters in the formalism. This is because the next excited state, $\omega(1420)$, lies close to the energy region under study. We also aim to avoid possible mixing with the $K\bar{K}$ scattering channel as much as possible.

For the data described above, the GEN approach gives the best fit results after cross-correlations are included:
\begin{equation}
\begin{aligned}
    \chi^2_{\mathrm d.o.f.}({\mathrm{GEN}},\, 305) &= 1.3, \\
    \chi^2_{\mathrm d.o.f.}({\mathrm{GEN}},\, 208) &= 1.6.
\end{aligned}
\end{equation}

The global EFT2 fit gives
\begin{equation}
    \chi^2_{\mathrm d.o.f.}({\mathrm{EFT2}},\, 305/208)=3.2.
\end{equation}
At the present level of precision, the data indicate that the EFT2 model does not contain enough parameters, or equivalently that the energy levels can constrain a more refined effective theory. In addition, the pole position obtained at the physical point deviates from the experimental value.

By contrast, EFT4 gives a better description of the finite-volume spectrum:
\begin{equation}
    \chi^2_{\mathrm d.o.f.}({\mathrm{EFT4}},\, 208/305)=2.3,
\end{equation}
with the corresponding parameters
\begin{equation}
\begin{aligned}
    g_{\mathrm{EFT4}} &= 5.96(17), \\
    \delta_{\mathrm{EFT4}} &= 38.8(6.9)\,\mathrm{MeV}, \\
    M_{V,\mathrm{EFT4}} &= 737(12)\,\mathrm{MeV}, \\
    a_{\mathrm{EFT4}} &= 0.96(14)\,\mathrm{GeV}^{-1}. \\
\end{aligned}
\end{equation}
These results are very close to the phenomenological values~\cite{Dax:2018rvs, Meissner:1986ka}.

We therefore take the EFT4 result as our main conclusion, while using the GEN and EFT2 results to estimate systematic uncertainties.

\begin{figure}
    \centering
    \includegraphics[width=\linewidth]{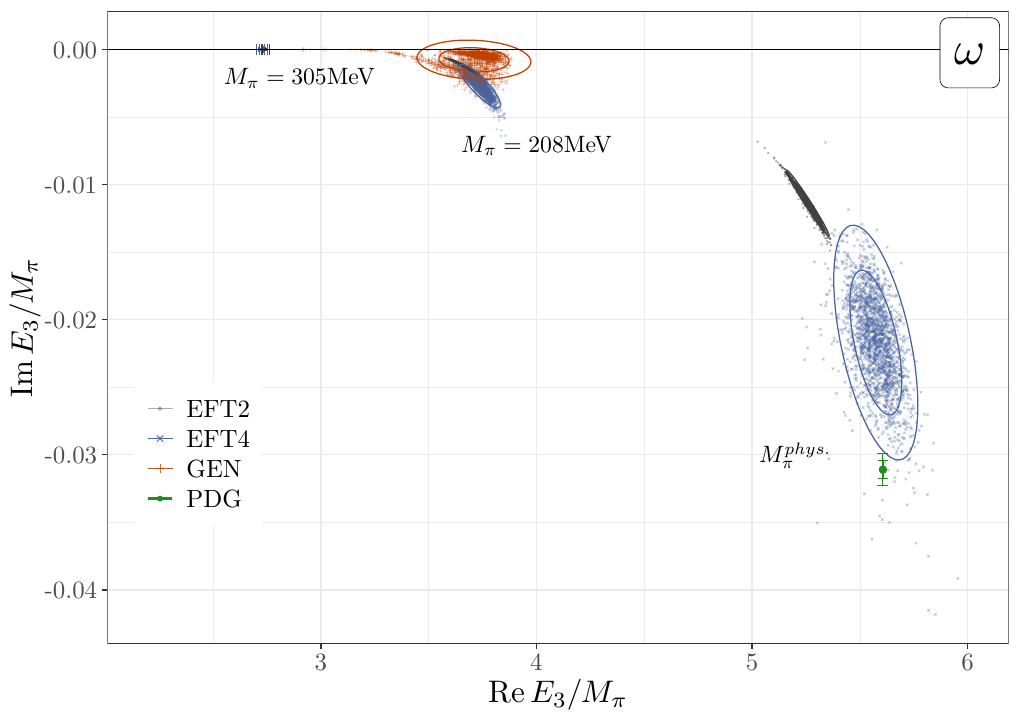}
    \caption{
        Pole positions of the $\omega$ meson obtained with the infinite-volume unitarity (IVU) method at different $\pi$ meson masses, using both generic (GEN) and effective-field-theory (EFT) forms for the two-body and three-body interactions. The scatter points denote the pole position for each Bootstrap sample, and the ellipses indicate the $1\sigma$ and $2\sigma$ confidence regions. The relevant parameters are determined by fitting the lattice data at $M_\pi=208$ and $M_\pi=305$ $\mathrm{MeV}$, allowing an extrapolation to the physical point. For comparison, the PDG result is shown as the green error bar~\cite{ParticleDataGroup:2024cfk, CMD-2:2003gqi, Achasov:2003ir}.
    }
    \label{fig:omega}
\end{figure}

\begin{figure}
    \centering
    \includegraphics[width=\linewidth]{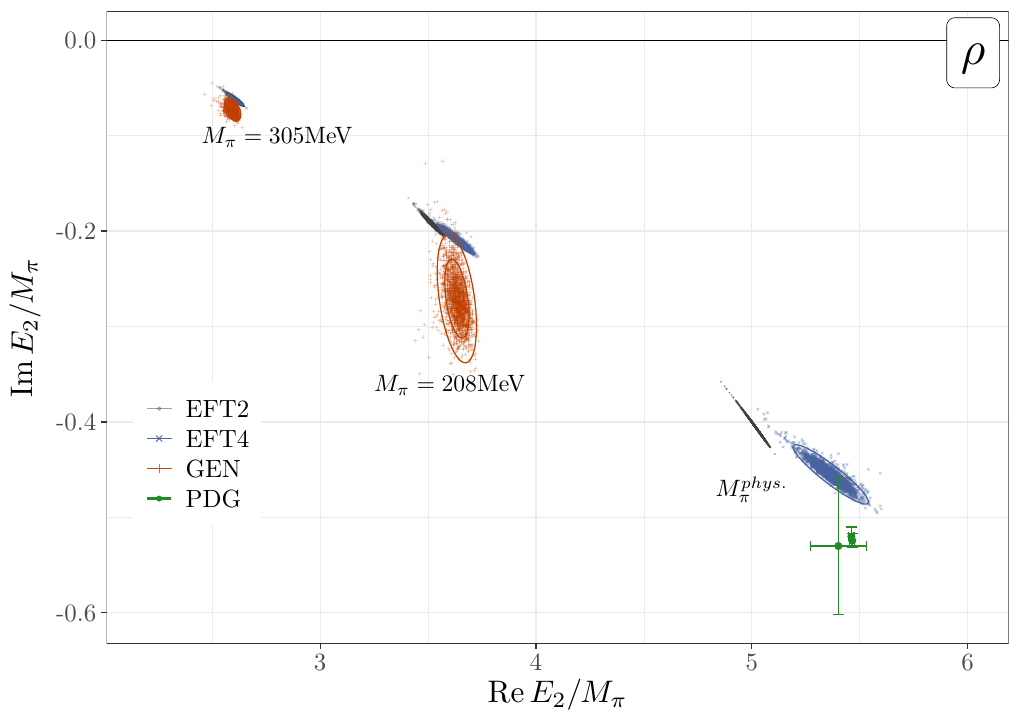}
    \caption{
        Pole positions of the $\rho$ meson. The description is the same as in Fig.~\ref{fig:omega}.
    }
    \label{fig:rho}
\end{figure}

The $\omega$ and $\rho$ meson pole positions obtained using the IVU formalism in Eq.~\ref{eq:IVU} together with the different methods described above are shown in Fig.~\ref{fig:omega} and Fig.~\ref{fig:rho}. For both mesons and at both $\pi$ meson masses, the different methods agree within $1\sigma$ in the real part of the pole position, and within $2\sigma$ in the imaginary part.

At the larger $\pi$ meson mass, the $\omega$ meson indeed appears as a bound state, with a binding energy of approximately $\sim 80\,\mathrm{MeV}$. It should be emphasized that the $\omega$ result from the GEN method should be treated with caution, because for each $\pi$ mass only two volumes are available in the $\omega$ channel, and only one data point lies in the relevant energy region. The EFT method avoids this problem by connecting data at different $\pi$ masses.

The EFT2 pole positions are constrained to a relatively small region and are described by only two parameters, $(g,\delta)$. EFT4 is more consistent with the lattice results both at the level of the finite-volume spectrum and at the level of the corresponding GEN pole positions. After extrapolation to the physical $\pi$ mass, the resulting $\rho$ and $\omega$ masses agree with the PDG values~\cite{ParticleDataGroup:2024cfk} within $<1\sigma$, and the widths are also consistent with experiment within $2\sigma$.

In physical units, the numerical results are
\begin{equation}
\begin{aligned}
    \sqrt{s_\rho}
    &= [748.9(10.0)-i\,63.5(1.8)] \, &&\mathrm{MeV}, \\
    \sqrt{s_\omega}
    &= [778.0(11.2)-i\,3.0(5)] \, &&\mathrm{MeV}. \\
\label{eq:main-result}
\end{aligned}
\end{equation}
The $\omega-\rho$ mass splitting is $29(15)\,\mathrm{MeV}$, consistent with the simple twisted-mass estimate of Ref.~\cite{McNeile:2009mx}. The small deviation from the experimental value may originate from the use of relatively small volumes in some cases (see Table~\ref{tab:ensembles}), which can introduce non-negligible exponential finite-volume effects as well as discretization errors.

\section{Conclusion and Outlook}
This chapter has presented the first lattice QCD calculation of the mass and width of the $\omega$ meson. One major challenge in this calculation is the precise extraction of the finite-volume spectrum. To resolve all eigenstates in the low-energy region, multihadron operators are essential~\cite{Wilson:2015dqa}. The three-particle operators used here require all relevant fermion contractions, which are computationally very demanding. Recent advances in algorithms, methodology, and computing power have made this task feasible.

Another challenge lies in the development of the formalism: one must not only map finite-volume results to infinite-volume scattering amplitudes, but also establish a reliable connection to the relevant effective field theory, thereby enabling the chiral extrapolation of three-body resonance parameters to the physical point, which had not previously been achieved. The final results show that this multistep theoretical framework gives $\rho$ and $\omega$ meson masses in good agreement with experiment~\cite{ParticleDataGroup:2024cfk}. The $\omega$ width is slightly smaller than the experimental value, but remains consistent within $2\sigma$. Possible sources of this difference include the omission of the coupled $K\bar{K}$ channel and systematic effects from neglecting five-body channels. For discussions of the current experimental values, see Refs.~\cite{Hoferichter:2023bjm, Hoferichter:2019mqg, Hoid:2020xjs, Colangelo:2022prz, Colangelo:2018mtw}.

This study marks a new milestone in lattice QCD hadron spectroscopy and paves the way toward understanding more complex hadronic systems. For lattice ensembles closer to the physical $\pi$ mass, the kinematic window for studying resonance properties becomes narrower because higher inelastic thresholds, such as $5\pi$, move closer. Therefore, using more stable and more readily accessible data at unphysical $\pi$ masses, combined with reliable effective-field-theory methods to extrapolate to the physical point, may offer a more advantageous strategy. Although this approach is already standard in two-body studies, it has not yet been established for three-body resonances. Future work includes assessing discretization errors and introducing larger volumes to further reduce systematic uncertainties. More challenging future applications include the Roper resonance and systems such as $T_{cc}$.

Finite-volume Lellouch--Lüscher form-factor corrections for three-body decay matrix elements have also been established~\cite{Muller:2020wjo, Hansen:2021ofl, Pang:2023jri}, although no lattice calculation has yet applied them directly. This direction is not part of the present work, but it will undoubtedly become one of the frontiers of lattice spectroscopy and precision calculations over the coming decade. Many exciting studies can be anticipated, such as $K \to \pi\pi\pi$ decays, which may provide new avenues for measuring CP violation. In addition, in mesonic systems there remain many important three-body decay problems, including three-body systems containing $K$ mesons and charmed three-body systems such as $DDK$, $D^*D^*D^*$, and the well-known $T_{cc}$. These systems are of substantial physical importance. Of course, one issue not included in the present work, but equally crucial from a physical perspective, is the coupled-channel effect of the $K\bar{K}$ channel. In realistic strong-interaction systems, the $\pi\pi\pi$ and $K\bar{K}$ channels are coupled, and this coupling directly introduces the dynamics of narrow resonances such as the $\phi(1020)$. Even if the $K\bar{K}$ channel is neglected, some mixing remains, though its effect is smaller. In particular, because the $\phi(1020)$ couples predominantly to the $K\bar{K}$ channel, its mixing with the $\omega$ is one of the central questions in light-vector-meson spectroscopy. The $\omega-\phi$ mixing not only reflects the breaking mechanism of $\mathrm{SU}(3)$ flavor symmetry, but is also closely related to OZI suppression, strange-quark components, and associated decay processes; it has therefore long been an important subject of phenomenological studies and experimental measurements. A systematic study of this scattering problem with coupled two-body and three-body channels, a first-principles understanding of the $\omega-\phi$ mixing mechanism, and a precise extraction of the mixing angle and associated coupling parameters will be a highly challenging and far-reaching direction for the future. We are steadily pursuing this program, and the reader may look forward to related results in the near future.

Although no lattice calculation of the $\omega(782)$ existed before the work presented in this chapter, its experimental value is accurately measured and its phenomenology is well understood. In the next chapter, we will take the techniques developed here one step further and use lattice QCD to address a particle that remains unclear both experimentally and phenomenologically: the $\pi(1300)$. This resonance has a much larger width and involves coupled two-body subsystems. It therefore provides a more challenging testing ground for our understanding of three-body resonances.

\cleardoublepage
\chapter{Three-Body Scattering: $\pi\pi\pi$ Scattering and the $\pi(1300)$ Resonance}
\label{chap:three_body_problems2}

{
\kaishu
\begin{center}
    欲穷千里目，更上一层楼。
\end{center}
\hfill ——《登鹳雀楼》[唐] 王之涣
}

\section{Background}
The previous chapter established a complete analysis workflow for the lattice three-body quantization condition through the study of $\omega(782)$. Building on that foundation, this chapter extends the method to the more challenging $\pi(1300)$, a broad resonance that remains unclear both experimentally and phenomenologically. The ultimate challenges in three-body hadron spectroscopy include, for example, the Roper resonance and three-nucleon systems. However, because these problems involve nucleons and more complicated quantization conditions, no lattice calculation has yet attempted them. Having addressed the important case of $\omega(782)$, we ask: what should be the next challenge?

The $\pi$ is the lightest hadron in nature\footnotecircle{This chapter is based on the published paper~\cite{Yan:2025jlq} H. Yan, M. Mai, M. Garofalo, Y. Feng, M. Döring, C. Liu, L. Liu, U. G. Meißner and C. Urbach, Emergence of the $\pi(1300)$ Resonance from Lattice QCD, arXiv:2510.09476}. Its mass is about one seventh of the proton mass, and it is understood as a Goldstone boson associated with spontaneous chiral symmetry breaking. It therefore plays a central role in understanding quark--gluon interactions. However, the properties of its first excited state, the $\pi(1300)$, remain far from settled. This state was first observed by Ananeva et al. in 1981 through diffractive dissociation on nuclei~\cite{Ananeva:1981sb}. It is generally believed to have a mass around $1300\pm100\,\mathrm{MeV}$ and a width between $200$--$600\,\mathrm{MeV}$~\cite{ParticleDataGroup:2024cfk}. Nevertheless, the very existence of the $\pi(1300)$ as a hadronic resonance remains debated. Although CLEO reported strong evidence~\cite{dArgent:2017gzv}, the $\pi(1300)$ peak disappears in the low-$t'$ region in the COMPASS analysis~\cite{COMPASS:2015gxz}, leading the PDG not to include the corresponding COMPASS result~\cite{ParticleDataGroup:2024cfk}.

If this state exists, its mass is roughly ten times that of the $\pi$ meson. This large energy gap provides a unique and important window into the mechanism of quark confinement and the internal structure of the lightest hadron. It is also approximately degenerate with other members of the first excited pseudoscalar octet, such as $\eta(1295)$ and $K(1460)$~\cite{Doring:2025phq}, and has a mass comparable to the nucleon excitation $N(1440)$, which is only about $1.5$ times heavier than the proton. Understanding this spectral pattern is essential for clarifying the formation mechanism of the hadron spectrum.

The $\pi(1300)$ also occupies a special place in three-body physics: its quantum numbers $J^{PC}=0^{-+}$ allow it to decay into a three-$\pi$ final state without a centrifugal barrier. This makes it one of the few resonances with significant three-body interactions, alongside the $N(1440)$, which involves the $\pi\pi N$ channel, and the lowest hadronic resonance that decays into three particles, the $\omega(782)$. In the decay of the $\pi(1300)$, all $\pi$ mesons are in a relative $S$ wave, distinguishing it from the $\omega$~\cite{Yan:2024gwp} and $a_1$~\cite{Mai:2021nul} resonances previously studied on the lattice. The $S$-wave system contains the well-known broad $f_0(500)$ resonance~\cite{Pelaez:2015qba} as a subchannel, and its strong $\pi\pi$ decay enhances three-body effects, making the $\pi(1300)$ a unique testing ground for few-body dynamics.

The $\pi(1300)$ is important in several contexts. First, although it is primarily regarded as a $q \bar{q}$ state, it may contain a sizable $q \bar{q} g$ hybrid component~\cite{Kalashnikova:1993xb, Donnachie:1999re}. Second, the \mybf{soft-$\pi$ theorem} predicts that the axial-current matrix element of the $\pi(1300)$ must vanish in the chiral limit, i.e. $f_{\pi(1300)} \to 0$ in that limit. Studying the $\pi(1300)$ provides a direct way to test this theorem. The uncertainty in $f_{\pi(1300)}$ is also one of the dominant uncertainties in determinations of the light-quark masses from QCD sum rules~\cite{Narison:2014vka, Boyle:2012jad}. Third, in $\tau \to 3\pi \nu_\tau$ decays, the large width of the $\pi(1300)$ may enhance CP-violating effects in extensions of the Standard Model~\cite{Choi:1994ch}.

\begin{figure}[t]
    \centering
    \includegraphics[width=1\linewidth]{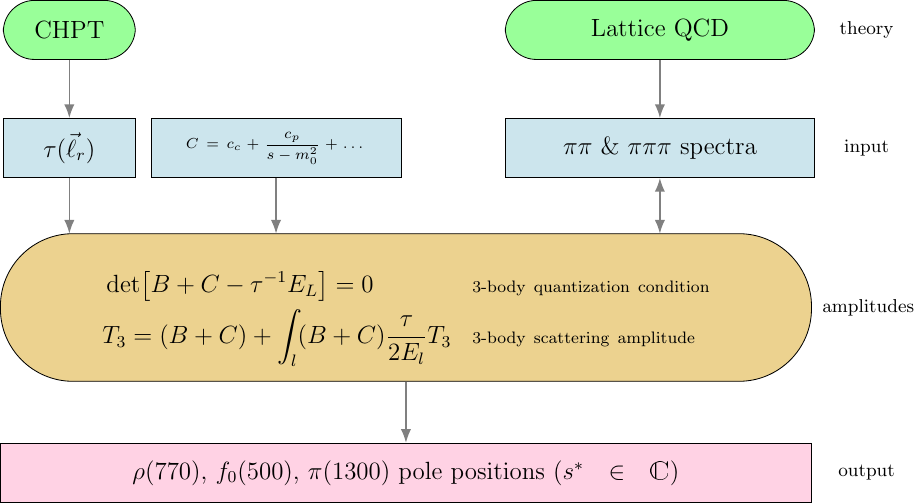}
    \caption{Workflow for extracting pole positions in two-body and three-body systems from lattice QCD. The finite-volume spectrum is obtained from lattice calculations, while information about short-range two-body/three-body interactions is supplemented by chiral perturbation theory.}
    \label{fig:Workflow}
\end{figure}

From the theoretical perspective, first-principles studies of the $\pi(1300)$ are still very limited, making such a calculation important\footnotecircle{Ref.~\cite{Gao:2021hvs} extracted finite-volume excited states of the $\pi$ from two-point functions constructed with point and Gaussian-smeared sources, interpreted one of them as the $\pi(1300)$, and attempted to compute the valence-quark parton distribution function of this state. Although rather exploratory, the state still approximately satisfied a single-particle dispersion relation after being boosted to a relatively large momentum of about $2.5\,\mathrm{GeV}$. This preliminary qualitative study suggests that the parton distributions of the $\pi$ meson and its radial excitation $\pi(1300)$ may differ significantly.}. In this chapter we use lattice QCD to investigate this state from first principles in a nonperturbative framework. The methodological workflow is shown in Fig.~\ref{fig:Workflow}.

In this chapter, we move beyond postdiction of an established resonance and carry out, for the first time, an exploratory study of hadron spectroscopy on the lattice. Previously, the $\pi(1300)$ had only been studied in frameworks where it was not treated as a resonance~\cite{Dudek:2010wm, McNeile:2006qy}; see Ref.~\cite{Kronfeld:2012uk} for a review. This work is the first to study the $\pi(1300)$ within a complete three-body finite-volume formalism, using calculations at two different $\pi$ masses. Combining this with an effective-field-theory approach, we give the first theoretical prediction of its pole position at the physical point. Other relevant theoretical studies can be found in Refs.~\cite{Steele:1997gh, Holl:2004fr, Wang:2007zzb, MartinezTorres:2011vh, Williams:2015cvx}.

We present the first lattice QCD determination of the resonance parameters of the $\pi(1300)$. To this end, we employ and further develop state-of-the-art methods both for constructing three-hadron operators and for mapping finite-volume spectra to infinite-volume scattering amplitudes, followed by analytic continuation to the complex energy plane. On the ensembles with the larger $\pi$ mass, we observe a clear resonance signal.

Assuming that the three-body interaction has negligible dependence on the $\pi$ mass, and simultaneously imposing constraints from chiral perturbation theory on all two-body channels, we extrapolate the result to the physical point. After model averaging, the extracted pole position is
\begin{equation}
    M_{\pi(1300)}=(1169\pm46)-i\,(62_{-62}^{+168})\,\text{MeV}.
\end{equation}
This result is consistent with the values obtained from phenomenological analyses.

\section{Lattice Spectrum}
This chapter continues to use the four CLQCD ensembles F32P21, F48P21, F32P30, and F48P30, with lattice spacing $a=0.07746(18)\,\mathrm{fm}$. They correspond to two $\pi$ meson masses, $M_\pi \approx 305\,\mathrm{MeV}$ (P30) and $M_\pi \approx 208\,\mathrm{MeV}$ (P21), and two volumes, $L=32$ (F32) and $L=48$ (F48). Details are given in Table~\ref{tab:ensembles}.

\subsection{Operator Construction}
We use the public package \verb|OpTion|~\cite{Yan:2025jlq} to construct all relevant single-particle and multiparticle operators in the two-$\pi$ ($I=0,1,2$) and three-$\pi$ ($I=1$) channels.

For example, in the three-$\pi$ channel, we work in the $A_1^-$ irrep with isospin $I=1$, and choose appropriate relative momenta to construct operators containing three $\pi$ mesons, $\rho\pi$, $\sigma\pi$, and a single $\pi$.

The operator construction should cover, as completely as possible, all relevant noninteracting levels below the inelastic threshold: in the $\pi\pi$ channel these include $\bar KK$ and $\eta\eta$, while in the $\pi\pi\pi$ channel they include $\bar KK\pi$, $\eta\eta\pi$, and $5\pi$. In the $\pi\pi$ channel, we construct two-body $\pi\pi$ operators for all isospins, and include a $\sigma$ operator in the $I=0$ channel and a $\rho$ operator in the $I=1$ channel. For the $A_1^-$ irrep of the $\pi\pi\pi$ channel, we construct a single-$\pi$ operator, two-body $\rho\pi$ (dominated by $P$ wave) and $\sigma\pi$ (dominated by $S$ wave) operators, and three-body $\pi\pi\pi$ operators with appropriate relative momenta.

The operators interpolating one-hadron, two-body, and three-body systems are built from linear combinations of quark bilinears. We use only local light-flavor operators of the form
\begin{equation}
O=\bar l' \Gamma l,
\end{equation}
where $l,l'\in\{u,d\}$, and $\Gamma$ determines the Dirac quantum numbers. We use the notation defined in\chapref{chap:two_body_problems}: the isospin projection is performed first, followed by the momentum-space projection.

\subsubsection{Isospin Space}
Part of this discussion has already appeared in the previous two chapters. For completeness, we nevertheless begin with the construction of one-body operators. A one-body $\sigma$ meson operator can appear in the $I=0$ $\pi\pi$ system and can be written as
\[
\frac{1}{\sqrt{2}}\left[\sigma^{u} + \sigma^{d}\right].
\]
For two-body operators, such as $\pi\pi$, $\sigma\pi$, and $\rho\pi$, the isospin decomposition is
\[
3 \otimes 3 = 1 \oplus 3 \oplus 5
\]
Accordingly, one can construct operators analogous to two-particle states:
\begin{equation}
\begin{cases}
O_{\rho\pi}^{I=2, I_z=2} &= \rho^+ \pi^+, \\
O_{\rho\pi}^{I=1, I_z=1} &= \frac{1}{2} \left[ -\rho^+ \pi^u + \rho^+ \pi^d + \rho^u \pi^+ - \rho^d \pi^+ \right], \\
O_{\rho\pi}^{I=0, I_z=0} &= -\frac{1}{\sqrt{3}} \left[ \rho^+ \pi^- + \rho^- \pi^+ + \frac{1}{2} [\rho^u \pi^u - \rho^u \pi^d - \rho^d \pi^u + \rho^d \pi^d] \right], \\
O_{\sigma\pi}^{I=1, I_z=1} &= -\frac{1}{\sqrt{2}} \left[ \sigma^u \pi^+ + \sigma^d \pi^+ \right], \\
O_{\sigma\pi}^{I=0, I_z=0} &= \frac{1}{2} \left[ \sigma^u \pi^u - \sigma^u \pi^d + \sigma^d \pi^u - \sigma^d \pi^d \right].
\end{cases}
\end{equation}
The $\pi\pi$ operators have the same flavor structure as the $\rho\pi$ operators listed above and are not repeated here.

For the three-$\pi$ system, the isospin decomposition is
\begin{equation}
3 \otimes 3 \otimes 3 = (1 \oplus 3 \oplus 5) \otimes 3 = 1 \oplus 3^3 \oplus 5^2 \oplus 7.
\end{equation}
$I=1$ is triply degenerate, implying that at most three independent flavor structures can be constructed. They can be distinguished by the intermediate isospin $I_{12}$ of the first two $\pi$ mesons. Their explicit forms are given in Eq.~\ref{eq:I_pipipi} of\chapref{chap:three_body_problems}. The number of physically independent operators depends on the momentum configuration. When all $\pi$ mesons are at rest,
($p_1^2 = p_2^2 = p_3^2 = 0$),
\[
O_{\pi\pi\pi}^{I=1, I_z=1, I_{12}=0}
=
O_{\pi\pi\pi}^{I=1, I_z=1, I_{12}=2},
\]
whereas
\[
O_{\pi\pi\pi}^{I=1, I_z=1, I_{12}=1}=0,
\]
so that only one independent operator remains.

For the momentum configuration $p_1^2 = p_2^2 = 1,\, p_3^2 = 0$, one has
\[
O_{\pi\pi\pi}^{I=1, I_z=1, I_{12}=0}
\neq
O_{\pi\pi\pi}^{I=1, I_z=1, I_{12}=2}
\neq 0,
\]
while
\[
O_{\pi\pi\pi}^{I=1, I_z=1, I_{12}=1}=0,
\]
so that two independent operators remain. In the previous chapter on $\omega(782)$, only one operator appeared near the physical energy region.

The charge-conjugation ($C$) parity and $G$ parity of the operators automatically satisfy the quantum numbers of the corresponding system.

\subsubsection{Spatial-Group Projection}
The operators are then projected onto irreps of the cubic group using \verb|OpTion|~\cite{Yan:2025jlq}. For the $I=1$ $\pi\pi\pi$ channel, the explicit operators obtained in the $A_1^-$ irrep are
\begin{equation}
\begin{cases}
O_{\pi_1} &= \pi(0), \\
O_{(\rho\pi)_1} &= \rho_x(-e_x)\pi(e_x) - \rho_x(e_x)\pi(-e_x) + \rho_y(-e_y)\pi(e_y) \\
& \quad - \rho_y(e_y)\pi(-e_y) + \rho_z(-e_z)\pi(e_z) - \rho_z(e_z)\pi(-e_z), \\
O_{(\sigma\pi)_1} &= \sigma(0)\pi(0), \\
O_{(\sigma\pi)_2} &= \sigma(-e_x)\pi(e_x) + \sigma(e_x)\pi(-e_x) + \sigma(-e_y)\pi(e_y) \\
& \quad + \sigma(e_y)\pi(-e_y) + \sigma(-e_z)\pi(e_z) + \sigma(e_z)\pi(-e_z), \\
O_{(\sigma\pi)_3} &= \sigma(e_{-x,-y})\pi(e_{x,y}) + \sigma(e_{-x,y})\pi(e_{x,-y})+ \sigma(e_{-x,-z})\pi(e_{x,z}) + \sigma(e_{-x,z})\pi(e_{x,-z}) \\
& \quad + \sigma(e_{x,-y})\pi(e_{-x,y}) + \sigma(e_{x,y})\pi(e_{-x,-y}) + \sigma(e_{x,-z})\pi(e_{-x,z}) + \sigma(e_{x,z})\pi(e_{-x,-z}) \\
& \quad + \sigma(e_{-y,-z})\pi(e_{y,z}) + \sigma(e_{-y,z})\pi(e_{y,-z}) + \sigma(e_{y,-z})\pi(e_{-y,z}) + \sigma(e_{y,z})\pi(e_{-y,-z}), \\
O_{(\pi\pi\pi)_1} &= \pi(0)\pi(0)\pi(0), \\
O_{(\pi\pi\pi)_2} &= \pi(e_x)\pi(-e_x)\pi(0) + \pi(-e_x)\pi(e_x)\pi(0) + \pi(e_y)\pi(-e_y)\pi(0) \\
& \quad + \pi(-e_y)\pi(e_y)\pi(0) + \pi(e_z)\pi(-e_z)\pi(0) + \pi(-e_z)\pi(e_z)\pi(0).
\end{cases}
\end{equation}

For the $I=0$ $\pi\pi$ channel, the operators in the $A_1^+$ irrep are
\begin{equation}
\begin{cases}
O_{\sigma_1} &= \sigma(0), \\
O_{(\pi\pi)_1} &= \pi(0)\pi(0), \\
O_{(\pi\pi)_2} &= \pi(-e_x)\pi(e_x) + \pi(e_x)\pi(-e_x) + \pi(-e_y)\pi(e_y) \\
& \quad + \pi(e_y)\pi(-e_y) + \pi(-e_z)\pi(e_z) + \pi(e_z)\pi(-e_z), \\
O_{(\pi\pi)_3} &= \pi(e_{-x,-y})\pi(e_{x,y}) + \pi(e_{-x,y})\pi(e_{x,-y})+ \pi(e_{-x,-z})\pi(e_{x,z}) + \pi(e_{-x,z})\pi(e_{x,-z}). \\
\end{cases}
\end{equation}
The operators in the $I=2$ $\pi\pi$ channel in the $A_1^+$ irrep have the same momentum structures as $O_{(\pi\pi){1,2,3}}$, but do not include the one-body operator $O_{\sigma_1}$.

For the $I=1$ $\pi\pi$ channel, the operators in the $T_1^-$ irrep are
\begin{equation}
\begin{cases}
O_{\rho_1} &= \rho_z(0), \\
O_{(\pi\pi)_1} &= \pi(e_x)\pi(-e_x) - \pi(-e_x)\pi(e_x), \\
O_{(\pi\pi)_2} &= \pi(e_{x,z})\pi(e_{-x,-z}) + \pi(e_{-x,z})\pi(e_{x,-z}) + \pi(e_{y,z})\pi(e_{-y,-z}) + \pi(e_{-y,z})\pi(e_{y,-z}) \\
& \quad - \pi(e_{x,-z})\pi(e_{-x,z}) - \pi(e_{-x,-z})\pi(e_{x,z}) - \pi(e_{y,-z})\pi(e_{-y,z}) - \pi(e_{-y,-z})\pi(e_{y,z}).
\end{cases}
\end{equation}

Using the notation defined in Eq.~\ref{eq:operator_table_label}, these operators can be written compactly and are listed in Table~\ref{tab:operators}.

\begin{table*}[htbp]
\centering
\caption{Operator list for the $\pi(1300) \to \pi\pi\pi$ study. For each channel, the operator basis, defined by the isospin $I$ and the cubic-group irrep, includes one-meson operators ($\pi$, $\rho$, $\sigma$), two-meson operators ($\pi\pi$, $\rho\pi$, $\sigma\pi$), and three-meson operators ($\pi\pi\pi$). The parameter $\eta_{\mu_i}^{\alpha_{i1}(;\alpha_{i2})}$ encodes the Dirac structure and momentum, as defined in the text; overall constants are omitted.}
\addtolength{\tabcolsep}{1pt}
\begin{tabular}{cccc}
\toprule
Scattering channel & Isospin & Type & Operator \\
\midrule
\multirow{13}{*}{$\pi\pi$} & \multirow{5}{*}{$I=0$} & \multirow{1}{*}{one-meson type} & $(+1)^{0}_{0}$ \\
\cmidrule(lr){3-4}
& & \multirow{4}{*}{two-meson type} & $(+1)^{0}_{0}$ \\
\cmidrule(lr){4-4}
& & & $(+1)^{x}_{5}, (+1)^{-x}_{5}, (+1)^{y}_{5}, (+1)^{-y}_{5}, (+1)^{z}_{5}, (+1)^{-z}_{5}$ \\
\cmidrule(lr){4-4}
& & & $(+1)^{yz}_{5}, (+1)^{xz}_{5}, (+1)^{xy}_{5}, (+1)^{-y,z}_{5}, (+1)^{-x,z}_{5}, (+1)^{-x,y}_{5},$ \\
& & & $ (+1)^{y,-z}_{5}, (+1)^{x,-z}_{5}, (+1)^{x,-y}_{5}, (+1)^{-y,-z}_{5}, (+1)^{-x,-z}_{5}, (+1)^{-x,-y}_{5}$ \\
\cmidrule(lr){2-4}
& \multirow{4}{*}{$I=1$} & \multirow{1}{*}{one-meson type} & $(+1)^{0}_{z}$ \\
\cmidrule(lr){3-4}
& & \multirow{3}{*}{two-meson type} & $(+1)^{z}_{5}, (-1)^{-z}_{5}$ \\
\cmidrule(lr){4-4}
& & & $(+1)^{xz}_{5}, (+1)^{-x,z}_{5}, (+1)^{yz}_{5}, (+1)^{-y,z}_{5},$ \\
& & & $ (-1)^{x,-z}_{5}, (-1)^{-x,-z}_{5}, (-1)^{y,-z}_{5}, (-1)^{-y,-z}_{5}$ \\
\cmidrule(lr){2-4}
& \multirow{4}{*}{$I=2$} & \multirow{4}{*}{two-meson type} & $(+1)^{0}_{0}$ \\
\cmidrule(lr){4-4}
& & & $(+1)^{x}_{5}, (+1)^{-x}_{5}, (+1)^{y}_{5}, (+1)^{-y}_{5}, (+1)^{z}_{5}, (+1)^{-z}_{5}$ \\
\cmidrule(lr){4-4}
& & & $(+1)^{yz}_{5}, (+1)^{xz}_{5}, (+1)^{xy}_{5}, (+1)^{-y,z}_{5}, (+1)^{-x,z}_{5}, (+1)^{-x,y}_{5},$ \\
& & & $ (+1)^{y,-z}_{5}, (+1)^{x,-z}_{5}, (+1)^{x,-y}_{5}, (+1)^{-y,-z}_{5}, (+1)^{-x,-z}_{5}, (+1)^{-x,-y}_{5}$ \\
\midrule
\multirow{9}{*}{$\pi\pi\pi$} & \multirow{9}{*}{$I=1$} & \multirow{1}{*}{one-meson type} & $(+1)^{0}_{5}$ \\
\cmidrule(lr){3-4}
& & \multirow{5}{*}{two-meson type} & $(+1)^{-x}_{x}, (-1)^{x}_{x}, (+1)^{-y}_{y}, (-1)^{y}_{y}, (+1)^{-z}_{z}, (-1)^{z}_{z}$ \\
\cmidrule(lr){4-4}
& & & $(+1)^{0}_{0}$ \\
\cmidrule(lr){4-4}
& & & $(+1)^{-x}_{0}, (+1)^{x}_{0}, (+1)^{-y}_{0}, (+1)^{y}_{0}, (+1)^{-z}_{0}, (+1)^{z}_{0}$ \\
\cmidrule(lr){4-4}
& & & $(+1)^{yz}_{0}, (+1)^{xz}_{0}, (+1)^{xy}_{0}, (+1)^{-y,z}_{0}, (+1)^{-x,z}_{0}, (+1)^{-x,y}_{0},$ \\
& & & $ (+1)^{y,-z}_{0}, (+1)^{x,-z}_{0}, (+1)^{x,-y}_{0}, (+1)^{-y,-z}_{0}, (+1)^{-x,-z}_{0}, (+1)^{-x,-y}_{0}$ \\
\cmidrule(lr){3-4}
& & \multirow{3}{*}{three-meson type} & $(+1)^{0;0}_{5}$ \\
\cmidrule(lr){4-4}
& & & $(+1)^{x;-x}_{5}, (+1)^{-x;x}_{5}, (+1)^{y;-y}_{5}, (+1)^{-y;y}_{5}, (+1)^{z;-z}_{5}, (+1)^{-z;z}_{5}$ \\
\cmidrule(lr){4-4}
& & & $(+1)^{x;-x}_{5}, (+1)^{-x;x}_{5}, (+1)^{y;-y}_{5}, (+1)^{-y;y}_{5}, (+1)^{z;-z}_{5}, (+1)^{-z;z}_{5}$ \\
\bottomrule
\end{tabular}
\addtolength{\tabcolsep}{-1pt}
\label{tab:operators}
\end{table*}

\subsection{Contraction Topologies}
\begin{figure}[htbp]
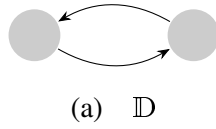

\centering
\raisebox{-0.5\height}{\begin{subfigure}{0.20\linewidth}\centering
\caption{$\mathbbm{D}$}\end{subfigure}}
\caption{Topology of the $\pi \to \pi$ contraction.}
\label{fig:topologies-1-1}
\end{figure}

\begin{figure}[htbp]
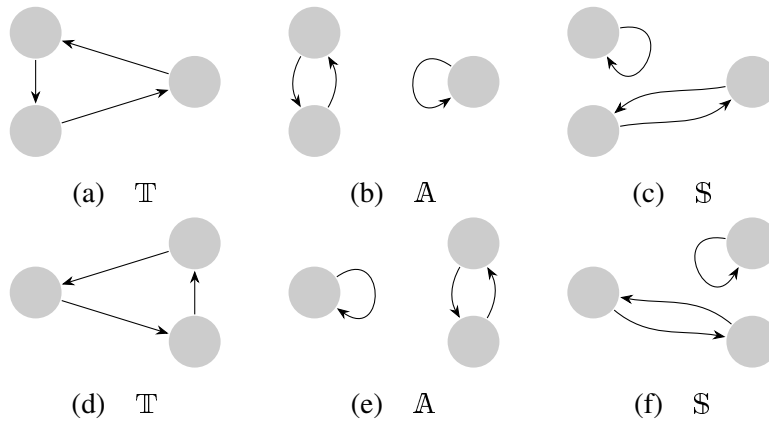

\centering
\raisebox{-0.5\height}{\begin{subfigure}{0.20\linewidth}\centering%
\caption{$\mathbbm{T}$}\end{subfigure}}
\quad
\raisebox{-0.5\height}{\begin{subfigure}{0.20\linewidth}\centering%
\caption{$\mathbbm{A}$}\end{subfigure}}
\quad
\raisebox{-0.5\height}{\begin{subfigure}{0.20\linewidth}\centering%
\caption{$\mathbbm{S}$}\end{subfigure}}
\\
\raisebox{-0.5\height}{\begin{subfigure}{0.20\linewidth}\centering%
\caption{$\mathbbm{T}$}\end{subfigure}}
\quad
\raisebox{-0.5\height}{\begin{subfigure}{0.20\linewidth}\centering%
\caption{$\mathbbm{A}$}\end{subfigure}}
\quad
\raisebox{-0.5\height}{\begin{subfigure}{0.20\linewidth}\centering%
\caption{$\mathbbm{S}$}\end{subfigure}}
\caption{Topologies of the $\pi \to \rho\pi, \sigma\pi$ contractions.}
\label{fig:topologies-1-2}
\end{figure}

\begin{figure}[htbp]
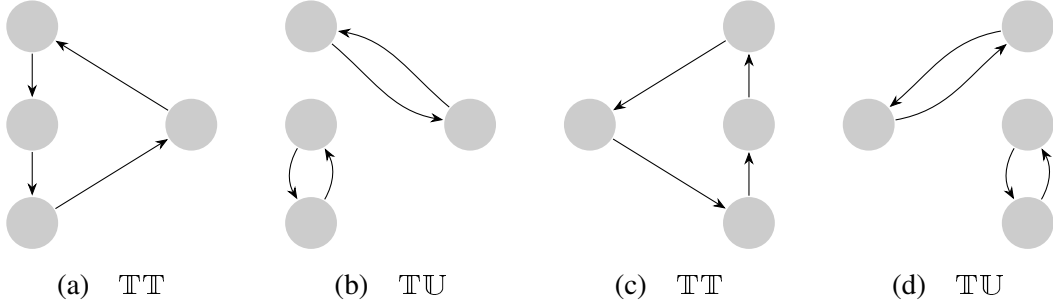

\centering
\raisebox{-0.5\height}{\begin{subfigure}{0.20\linewidth}\centering%
\caption{$\mathbbm{TT}$}\end{subfigure}}
\quad
\raisebox{-0.5\height}{\begin{subfigure}{0.20\linewidth}\centering%
\caption{$\mathbbm{TU}$}\end{subfigure}}
\quad
\raisebox{-0.5\height}{\begin{subfigure}{0.20\linewidth}\centering%
\caption{$\mathbbm{TT}$}\end{subfigure}}
\quad
\raisebox{-0.5\height}{\begin{subfigure}{0.20\linewidth}\centering%
\caption{$\mathbbm{TU}$}\end{subfigure}}
\caption{Topologies of the $\pi \to \pi\pi\pi$ contractions.}
\label{fig:topologies-1-3}
\end{figure}

\begin{figure}[htbp]
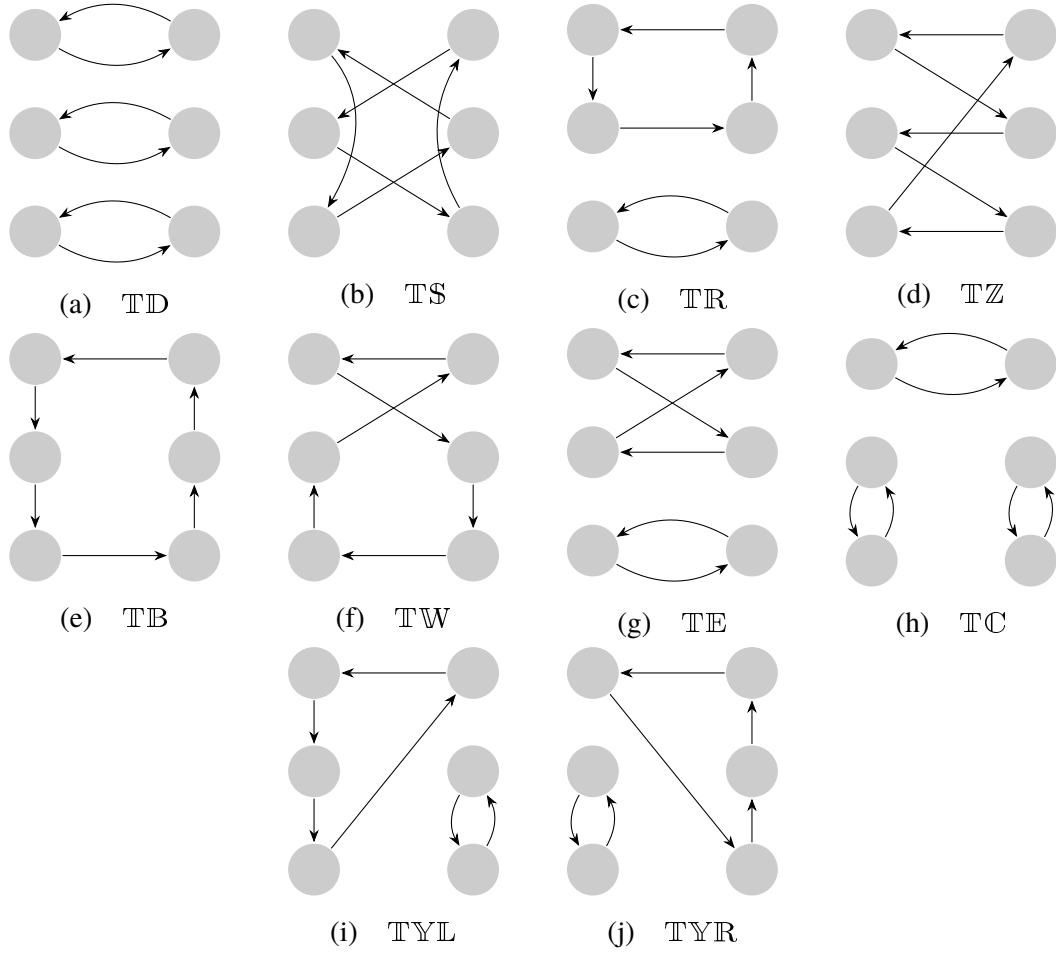

\centering
\raisebox{-0.5\height}{\begin{subfigure}{0.20\linewidth}\centering%
\caption{$\mathbbm{TD}$}\end{subfigure}}
\quad
\raisebox{-0.5\height}{\begin{subfigure}{0.20\linewidth}\centering%
\caption{$\mathbbm{TS}$}\end{subfigure}}
\quad
\raisebox{-0.5\height}{\begin{subfigure}{0.20\linewidth}\centering%
\caption{$\mathbbm{TR}$}\end{subfigure}}
\quad
\raisebox{-0.5\height}{\begin{subfigure}{0.20\linewidth}\centering%
\caption{$\mathbbm{TZ}$}\end{subfigure}}
\\
\raisebox{-0.5\height}{\begin{subfigure}{0.20\linewidth}\centering%
\caption{$\mathbbm{TB}$}\end{subfigure}}
\quad
\raisebox{-0.5\height}{\begin{subfigure}{0.20\linewidth}\centering%
\caption{$\mathbbm{TW}$}\end{subfigure}}
\quad
\raisebox{-0.5\height}{\begin{subfigure}{0.20\linewidth}\centering%
\caption{$\mathbbm{TE}$}\end{subfigure}}
\quad
\raisebox{-0.5\height}{\begin{subfigure}{0.20\linewidth}\centering%
\caption{$\mathbbm{TC}$}\end{subfigure}}
\\
\raisebox{-0.5\height}{\begin{subfigure}{0.20\linewidth}\centering%
\caption{$\mathbbm{TYL}$}\end{subfigure}}
\quad
\raisebox{-0.5\height}{\begin{subfigure}{0.20\linewidth}\centering%
\caption{$\mathbbm{TYR}$}\end{subfigure}}
\caption{Topologies of the $I=1$ $\pi\pi\pi \to \pi\pi\pi$ contractions. All diagrams with the same source and sink can be obtained from these basic topologies by permutations and recombinations.}
\label{fig:topologies-3-3}
\end{figure}

For the contraction calculation of the correlation matrix, we use the distillation method~\cite{HadronSpectrum:2009krc}. As in\chapref{chap:three_body_problems}, the number of contraction diagrams required in the three-$\pi$ channel grows factorially compared with the two-particle channel, making it impractical to write explicit expressions for all correlation functions one by one. We therefore systematically classify the topologies of all contraction diagrams. For the $I=1$ $\pi\pi\pi$ channel, the complete set of topologies is shown by operator type in Figs.~\ref{fig:topologies-1-1}, \ref{fig:topologies-1-2}, \ref{fig:topologies-2-2}, \ref{fig:topologies-1-3}, \ref{fig:topologies-2-3}, and \ref{fig:topologies-3-3}. All explicit contraction diagrams can be viewed as different permutations of these basic topologies.

\begin{figure}[htbp]
\centering
\raisebox{-0.5\height}{\begin{subfigure}{0.20\linewidth}\centering%
\caption{$\mathbbm{D}$}\end{subfigure}}
\quad
\raisebox{-0.5\height}{\begin{subfigure}{0.20\linewidth}\centering%
\caption{$\mathbbm{E}$}\end{subfigure}}
\quad
\raisebox{-0.5\height}{\begin{subfigure}{0.20\linewidth}\centering%
\caption{$\mathbbm{B}$}\end{subfigure}}
\\
\raisebox{-0.5\height}{\begin{subfigure}{0.20\linewidth}\centering%
\caption{$\mathbbm{A}$}\end{subfigure}}
\quad
\raisebox{-0.5\height}{\begin{subfigure}{0.20\linewidth}\centering%
\caption{$\mathbbm{ML}$}\end{subfigure}}
\quad
\raisebox{-0.5\height}{\begin{subfigure}{0.20\linewidth}\centering%
\caption{$\mathbbm{MR}$}\end{subfigure}}
\caption{Topologies of the $\rho\pi, \sigma\pi \to \rho\pi, \sigma\pi$ contractions.}
\label{fig:topologies-2-2}
\end{figure}

\begin{figure}[htbp]
\centering
\raisebox{-0.5\height}{\begin{subfigure}{0.20\linewidth}\centering%
\caption{$\mathbbm{TT}$}\end{subfigure}}
\quad
\raisebox{-0.5\height}{\begin{subfigure}{0.20\linewidth}\centering%
\caption{$\mathbbm{TM}$}\end{subfigure}}
\quad
\raisebox{-0.5\height}{\begin{subfigure}{0.20\linewidth}\centering%
\caption{$\mathbbm{TD}$}\end{subfigure}}
\\
\raisebox{-0.5\height}{\begin{subfigure}{0.20\linewidth}\centering%
\caption{$\mathbbm{TU}$}\end{subfigure}}
\quad
\raisebox{-0.5\height}{\begin{subfigure}{0.20\linewidth}\centering%
\caption{$\mathbbm{TO}$}\end{subfigure}}
\quad
\raisebox{-0.5\height}{\begin{subfigure}{0.20\linewidth}\centering%
\caption{$\mathbbm{TG}$}\end{subfigure}}
\\
\raisebox{-0.5\height}{\begin{subfigure}{0.20\linewidth}\centering%
\caption{$\mathbbm{TM}$}\end{subfigure}}
\quad
\raisebox{-0.5\height}{\begin{subfigure}{0.20\linewidth}\centering%
\caption{$\mathbbm{TD}$}\end{subfigure}}
\\
\raisebox{-0.5\height}{\begin{subfigure}{0.20\linewidth}\centering%
\caption{$\mathbbm{TU}$}\end{subfigure}}
\quad
\raisebox{-0.5\height}{\begin{subfigure}{0.20\linewidth}\centering%
\caption{$\mathbbm{TO}$}\end{subfigure}}
\quad
\raisebox{-0.5\height}{\begin{subfigure}{0.20\linewidth}\centering%
\caption{$\mathbbm{TG}$}\end{subfigure}}
\caption{Topologies of the $\rho\pi, \sigma\pi \to \pi\pi\pi$ contractions.}
\label{fig:topologies-2-3}
\end{figure}

\subsection{Finite-Volume Spectra}
In certain correlation functions, for example in the $I=0$ $\pi\pi$ channel, the operators carry the same quantum numbers as the vacuum, and a nonzero vacuum expectation value (VEV) contribution appears in the correlation function. This contribution is constant in time and is equivalent to mixing with a zero-mass level, thereby interfering with the extraction of the spectrum. To remove this vacuum contribution, we consider three procedures:
\begin{enumerate}
    \item remove the constant term by time-shifting the correlation function;
    \item compute the vacuum expectation value explicitly and subtract it;
    \item subtract the vacuum expectation value first and then apply the time shift.
\end{enumerate}

\begin{figure}[htbp]
\centering
\includegraphics[width=0.32\columnwidth]{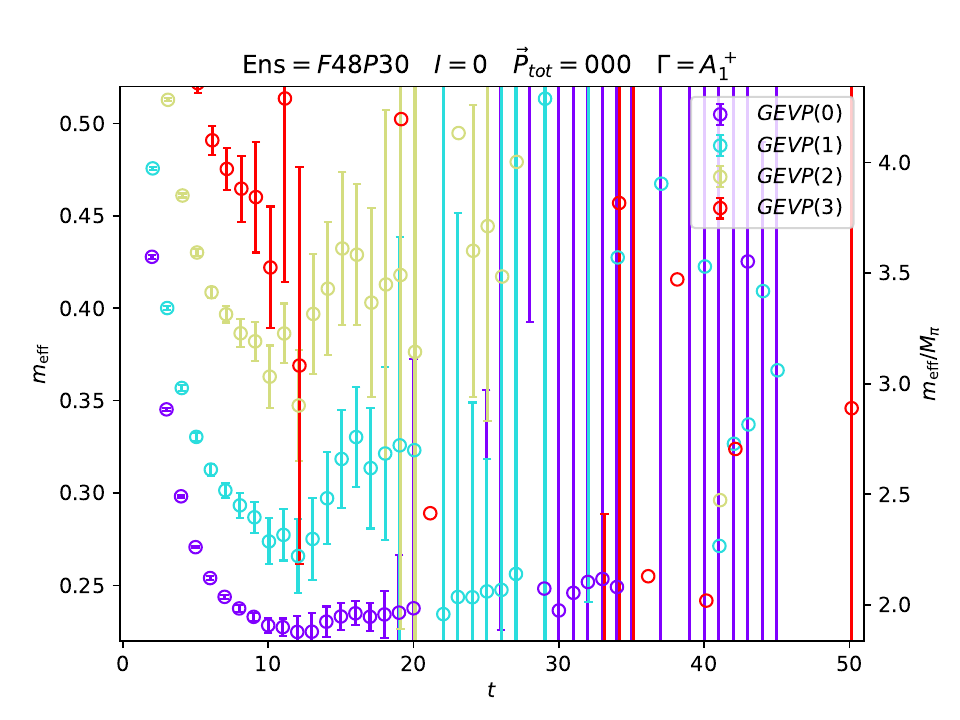}
\includegraphics[width=0.32\columnwidth]{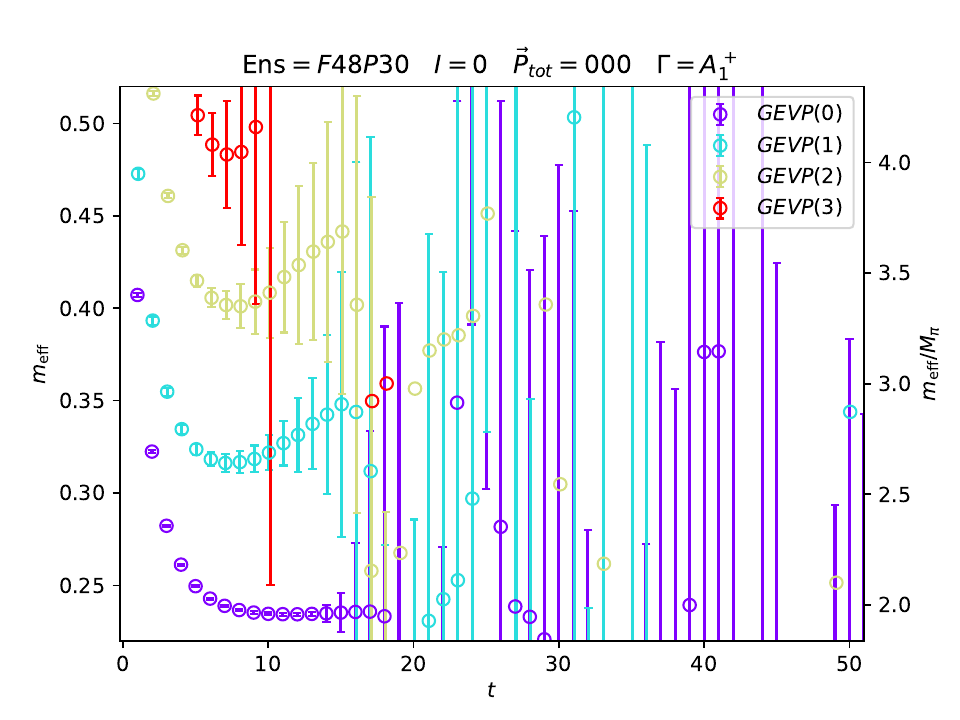}
\includegraphics[width=0.32\columnwidth]{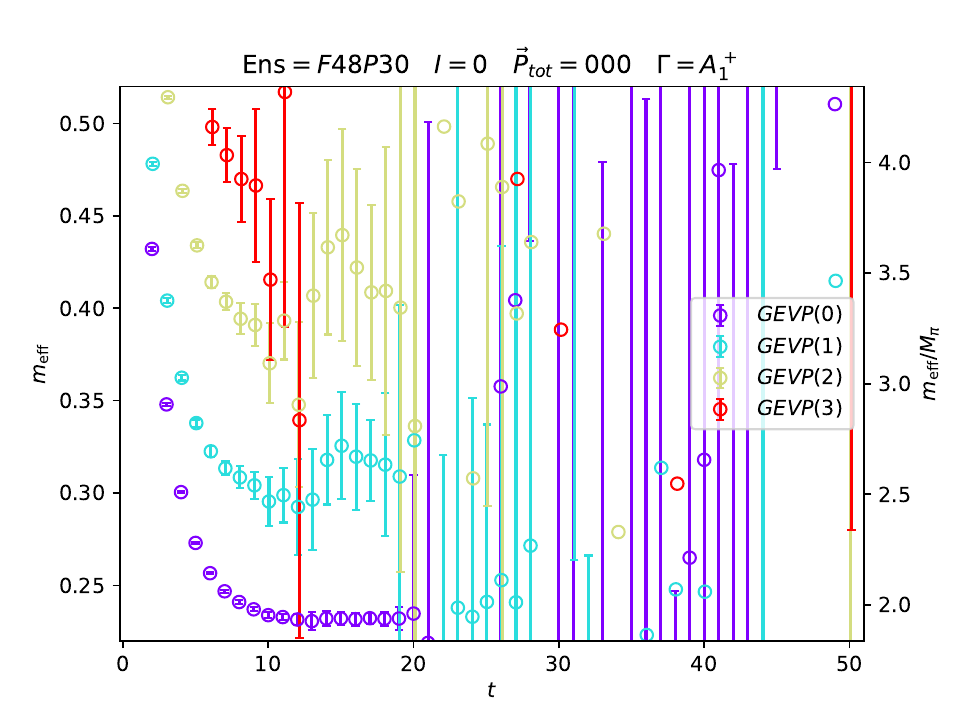}
\caption{Tests of vacuum-expectation-value subtraction in the $I=0$ $\pi\pi$ channel. From left to right, the panels correspond to time-shifting the correlation function, explicit subtraction, and subtraction followed by time-shifting.}
\label{fig:pipi-I-0-vev-tests}
\end{figure}

To compare the performance of these three methods, we carried out a systematic test on the F48P30 ensemble. The results are shown in Fig.~\ref{fig:pipi-I-0-vev-tests}. The figure shows that the different methods affect the behavior of the correlation function and the effective mass in different ways, providing guidance for choosing the most stable procedure in the subsequent analysis.

We find that direct subtraction of the vacuum expectation value gives the cleanest signal. In addition, compared with using a time-averaged vacuum expectation value, subtracting a time-dependent vacuum expectation value further improves the signal quality.

We observe a nonzero thermal-pollution contribution in the spectrum, which at leading order can be removed by applying a weighted shift to $C_{ij}$~\cite{Dudek:2012gj}:
\begin{equation}
\tilde{C}(t) = e^{-\mathcal{E} t} \big[ e^{\mathcal{E} t} C(t) - e^{\mathcal{E} (t+1)} C(t+1) \big].
\label{eq:thermal}
\end{equation}
In our calculation we take $\mathcal{E} = 0$. The resulting $\tilde{C}(t)$ is then processed with the standard GEVP procedure. This treatment is applied to the $I=0$ $\pi\pi$ channel on F32P30 and F32P21, and to the $I=0,2$ $\pi\pi$ channels on all ensembles. For the $I=0,1$ $\pi\pi$ and $I=1$ $\pi\pi\pi$ channels, the GEVP eigenvalues are ordered by their relative magnitudes; for the $I=2$ $\pi\pi$ channel, they are instead ordered according to their overlap with the reference time $t_0+1$. This strategy yields the clearest plateaus.

In the $I=1$ $\pi\pi\pi$ channel, since the ground-state single-$\pi$ energy is known precisely, it is sometimes useful to subtract a $\pi$ state before solving the GEVP. This corresponds to taking $\mathcal{E} = M_{\pi}$ in Eq.~\ref{eq:thermal}. We find that, on the smaller volumes (F32P30 and F32P21), this subtraction significantly improves the excited-state signal. For higher excited states, however, this trick is not effective because the signal decays very rapidly.

\begin{figure}[htbp]
\centering
\includegraphics[width=0.49\columnwidth]{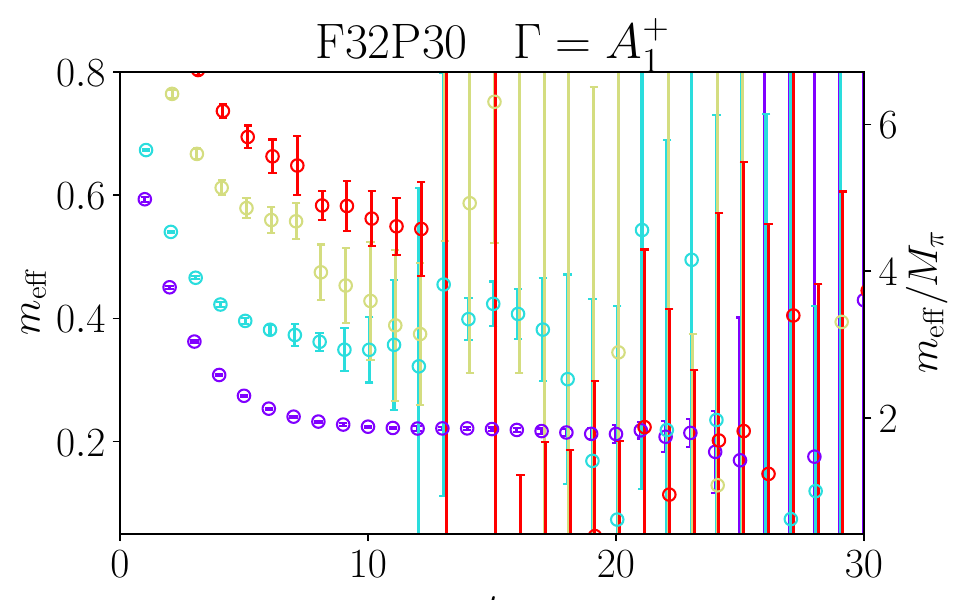}
\includegraphics[width=0.49\columnwidth]{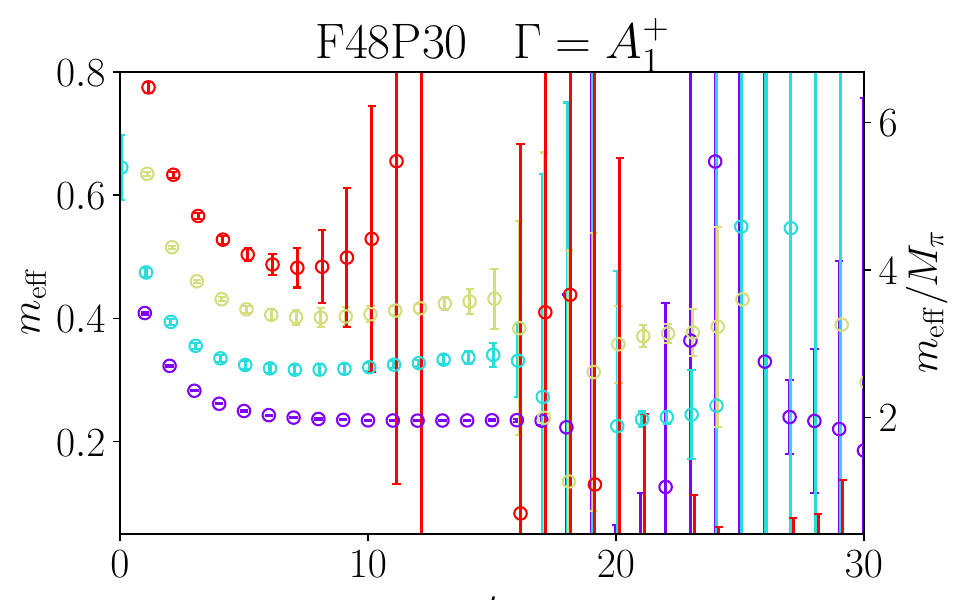}
\\
\includegraphics[width=0.49\columnwidth]{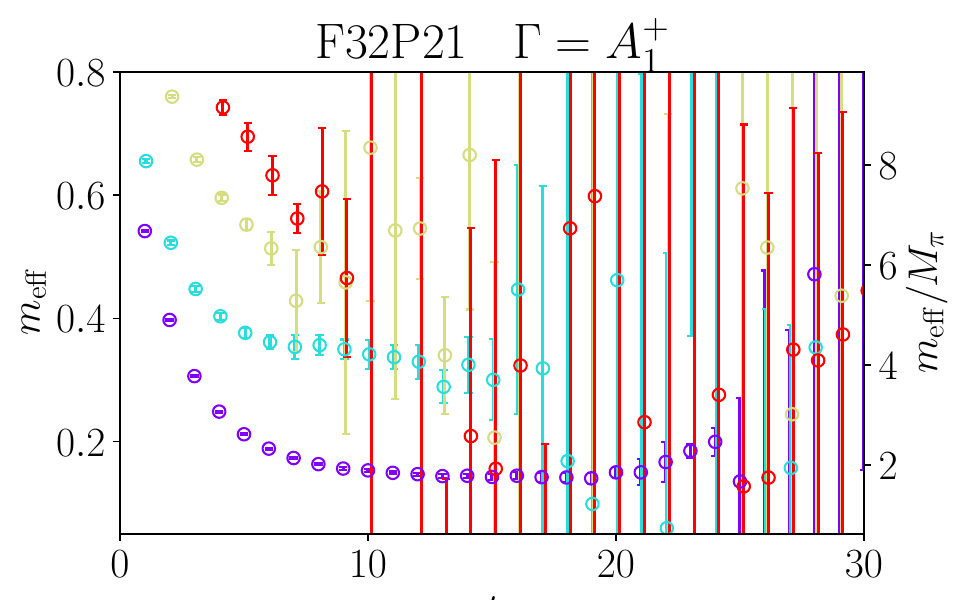}
\includegraphics[width=0.49\columnwidth]{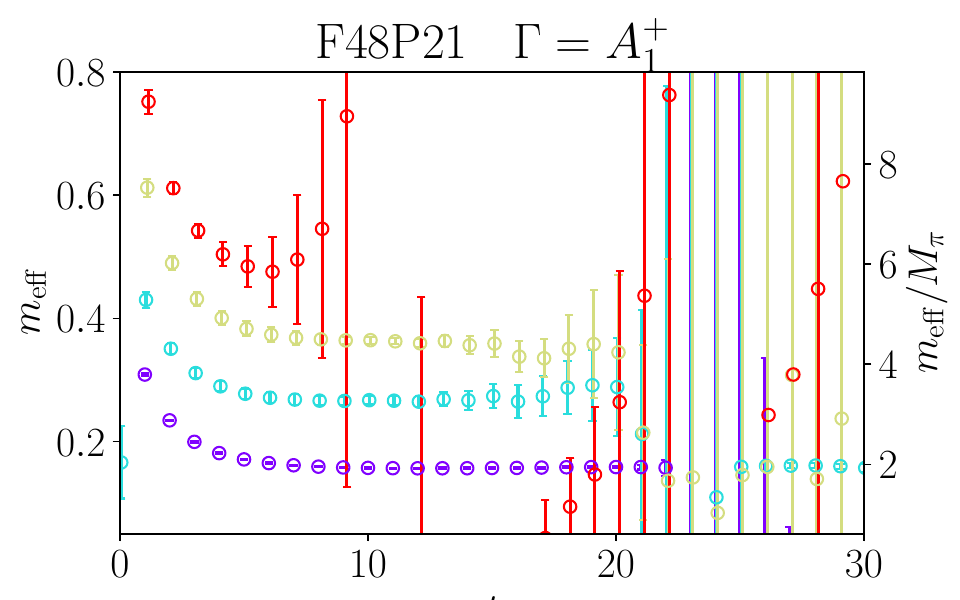}
\caption{Effective masses of the eigenvalues $\lambda_n(t)$ in the $I=0$ $\pi\pi$ channel. Different colors correspond to different energy levels. The left $y$ axis is in lattice units, while the right $y$ axis is in units of the $\pi$ meson mass.}
\label{fig:pipi-I=0-meff}
\end{figure}

\begin{figure}[htbp]
\centering
\includegraphics[width=0.49\columnwidth]{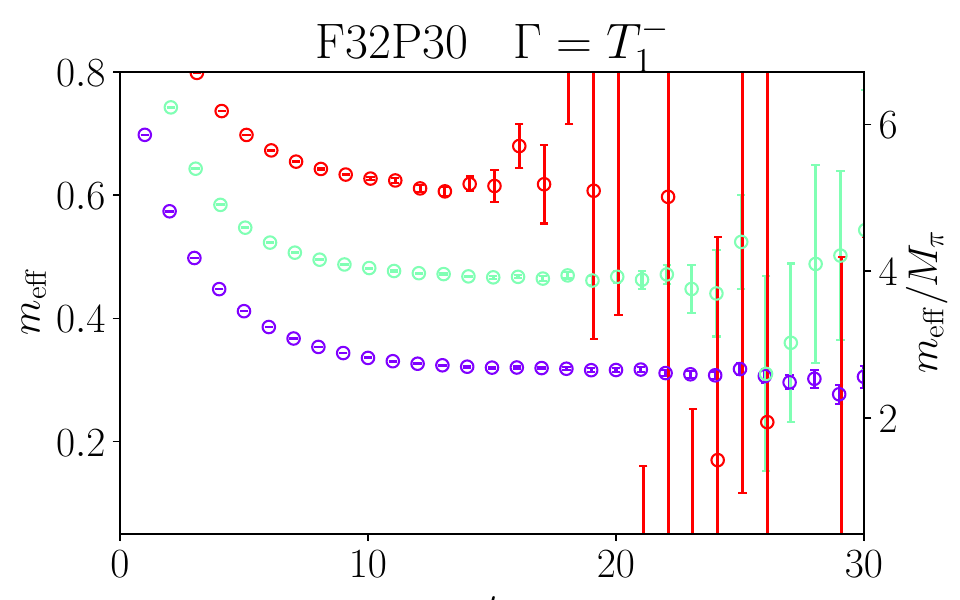}
\includegraphics[width=0.49\columnwidth]{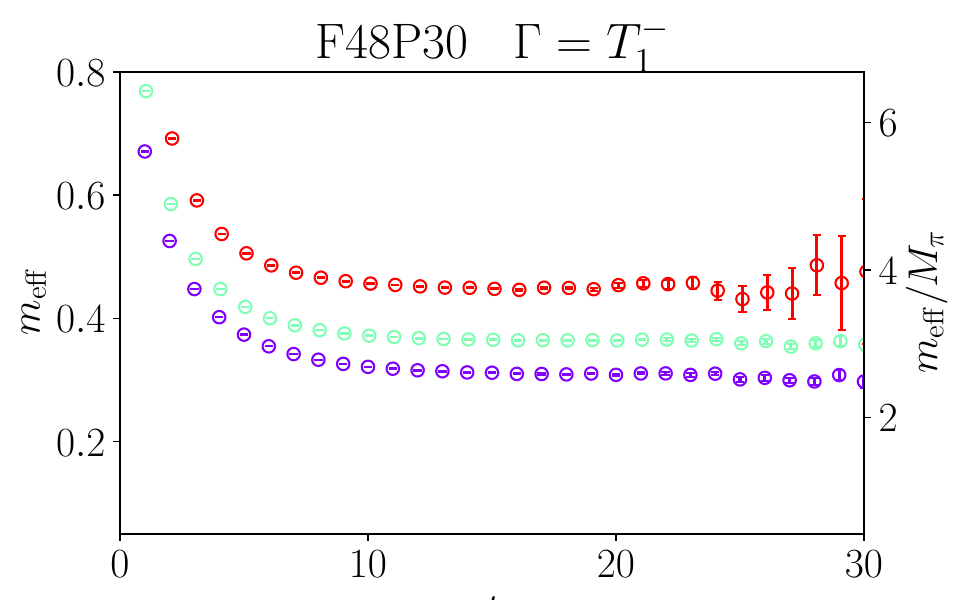}
\\
\includegraphics[width=0.49\columnwidth]{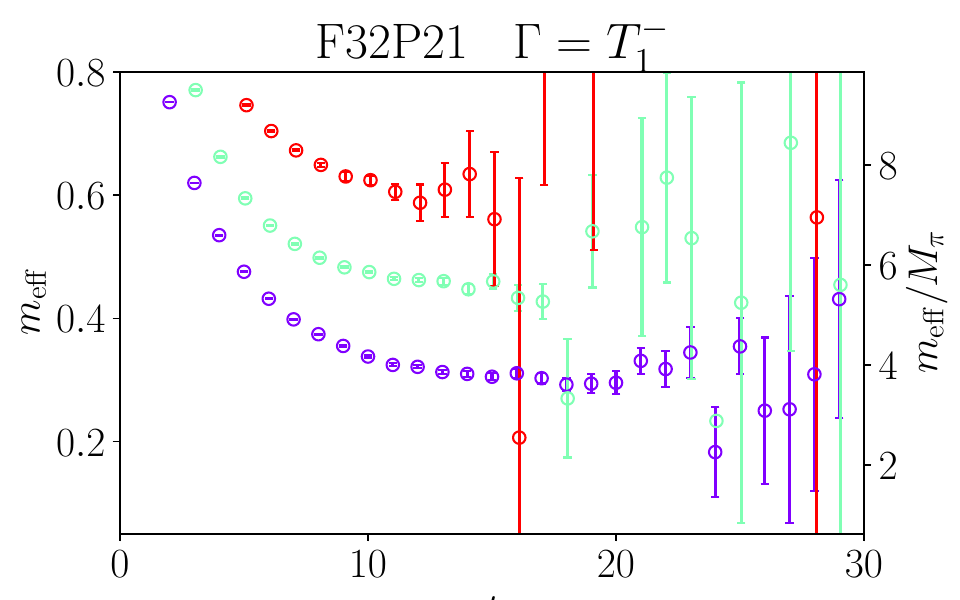}
\includegraphics[width=0.49\columnwidth]{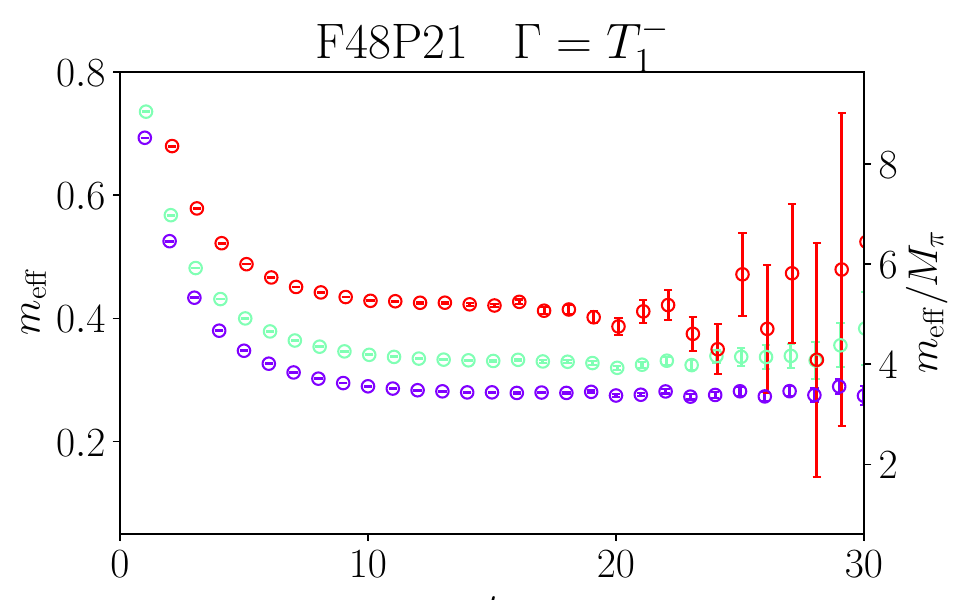}
\caption{Effective masses in the $I=1$ $\pi\pi$ channel.}
\label{fig:pipi-I=1-meff}
\end{figure}

\begin{figure}[htbp]
\centering
\includegraphics[width=0.49\columnwidth]{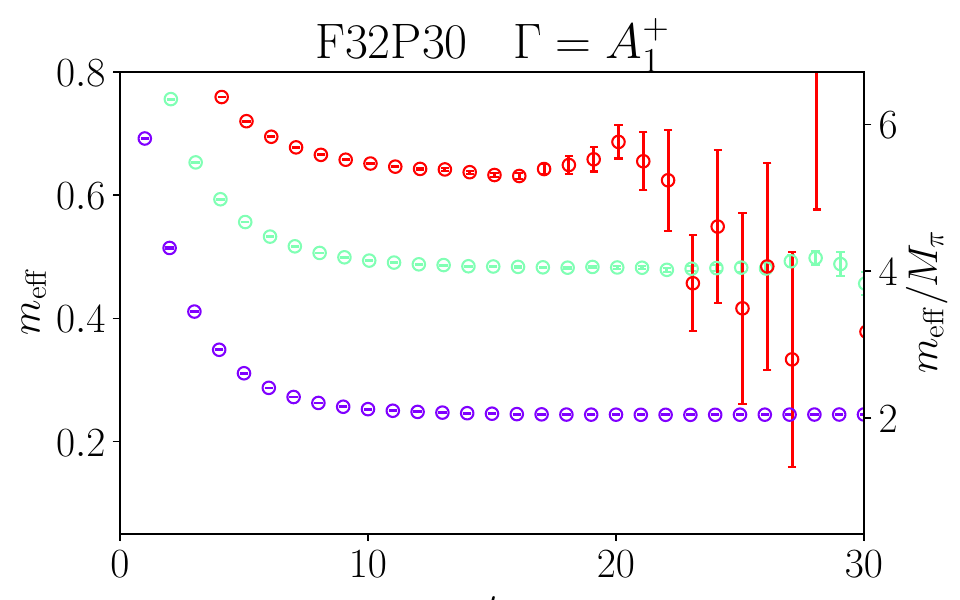}
\includegraphics[width=0.49\columnwidth]{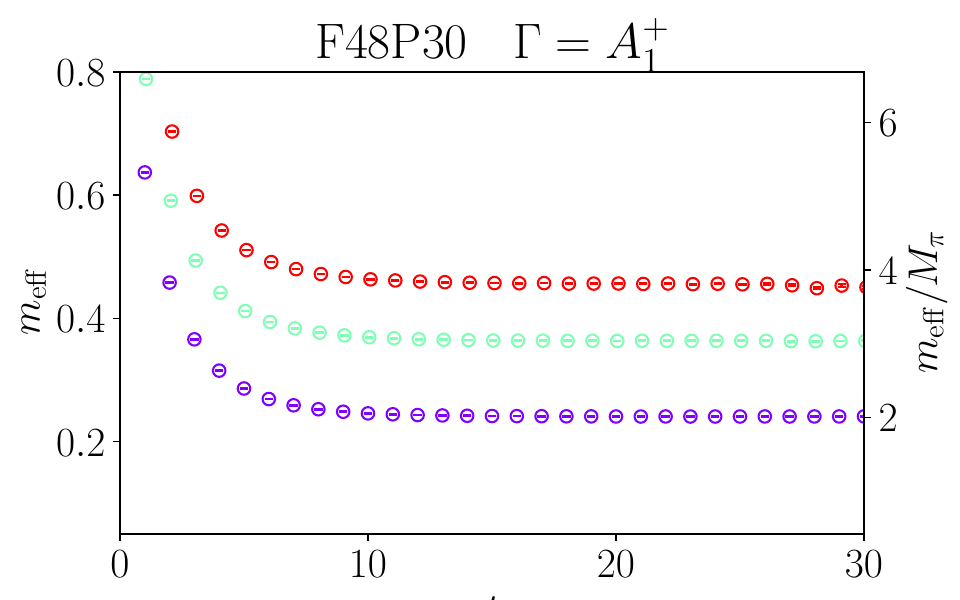}
\\
\includegraphics[width=0.49\columnwidth]{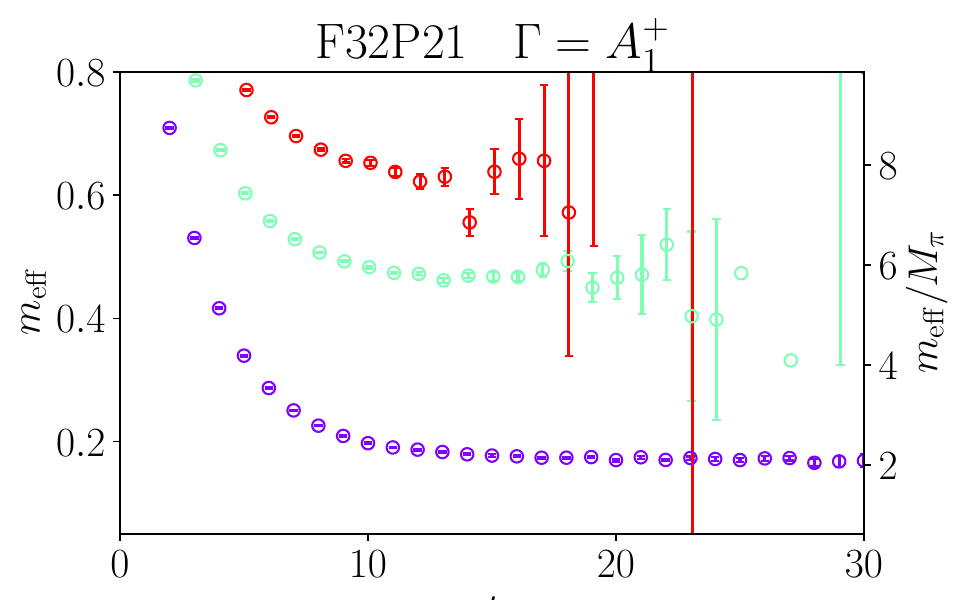}
\includegraphics[width=0.49\columnwidth]{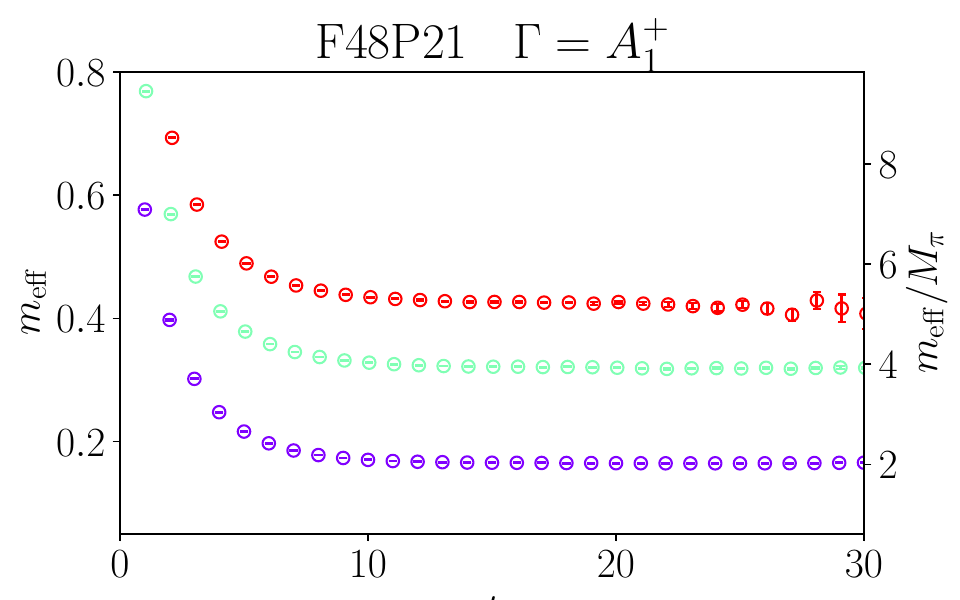}
\caption{Effective masses in the $I=2$ $\pi\pi$ channel.}
\label{fig:pipi-I=2-meff}
\end{figure}

\begin{figure}[htbp]
\centering
\includegraphics[width=0.49\columnwidth]{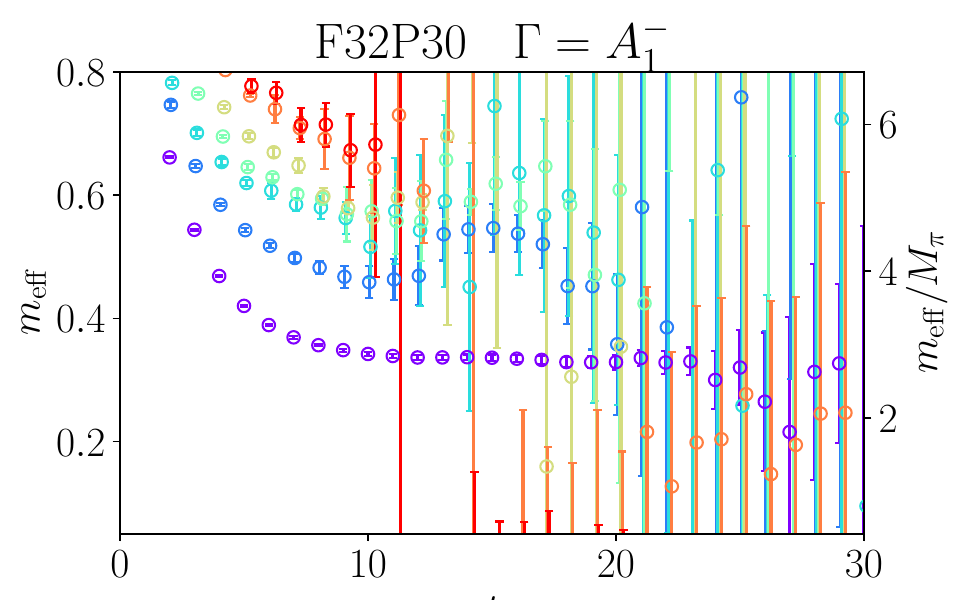}
\includegraphics[width=0.49\columnwidth]{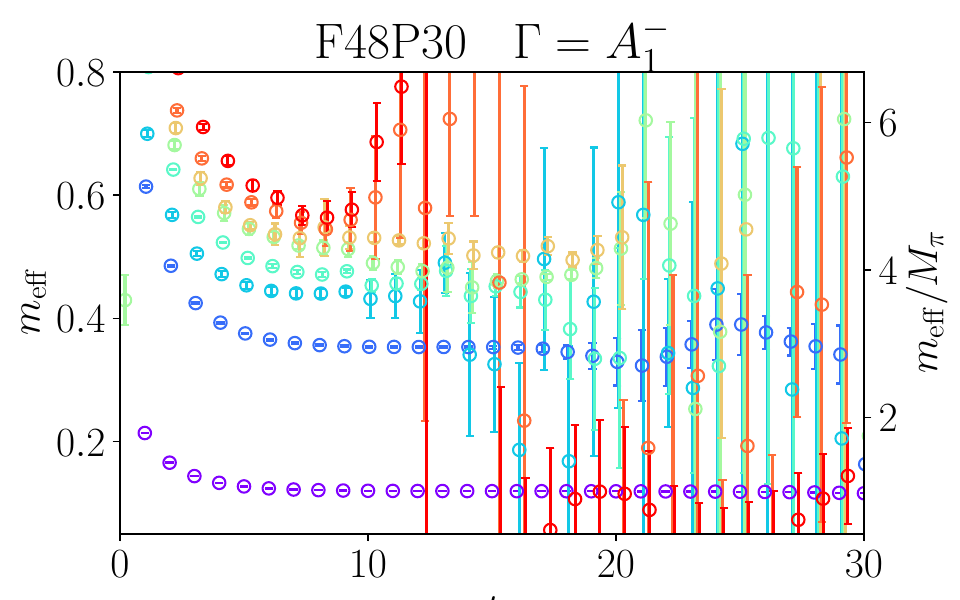}
\\
\includegraphics[width=0.49\columnwidth]{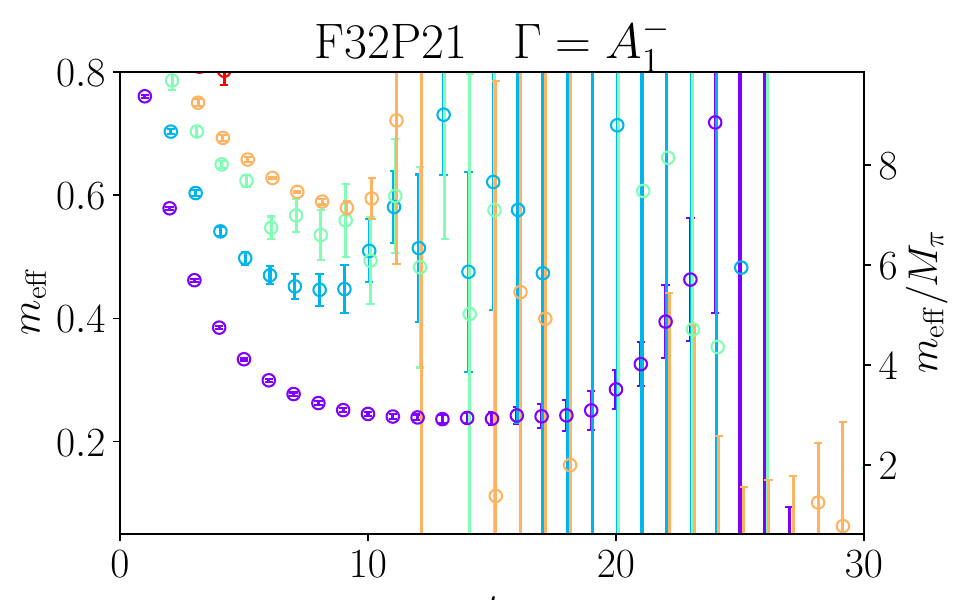}
\includegraphics[width=0.49\columnwidth]{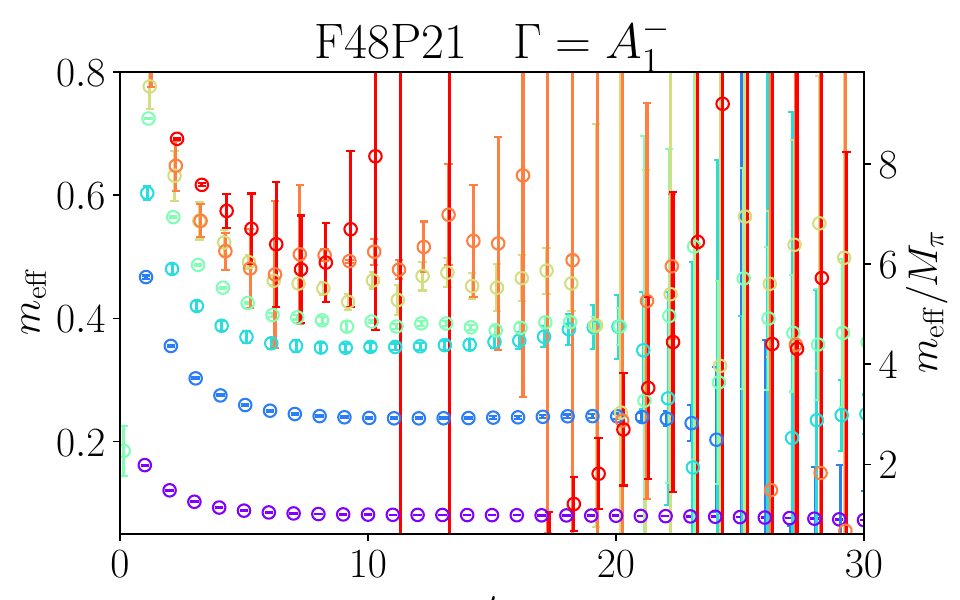}
\caption{Effective masses in the $I=1$ $\pi\pi\pi$ channel.}
\label{fig:pipipi-I=1-meff}
\end{figure}

The effective masses of the eigenvalues in the $\pi\pi$ and $\pi\pi\pi$ channels are shown in Figs.~\ref{fig:pipi-I=0-meff}, \ref{fig:pipi-I=1-meff}, \ref{fig:pipi-I=2-meff}, and \ref{fig:pipipi-I=1-meff}, respectively. To improve readability, data points with particularly large errors are made transparent.

\begin{figure}[t]
\centering
\includegraphics[width=\linewidth,trim=0.9cm 0.9cm 1.7cm 1.0cm,clip]{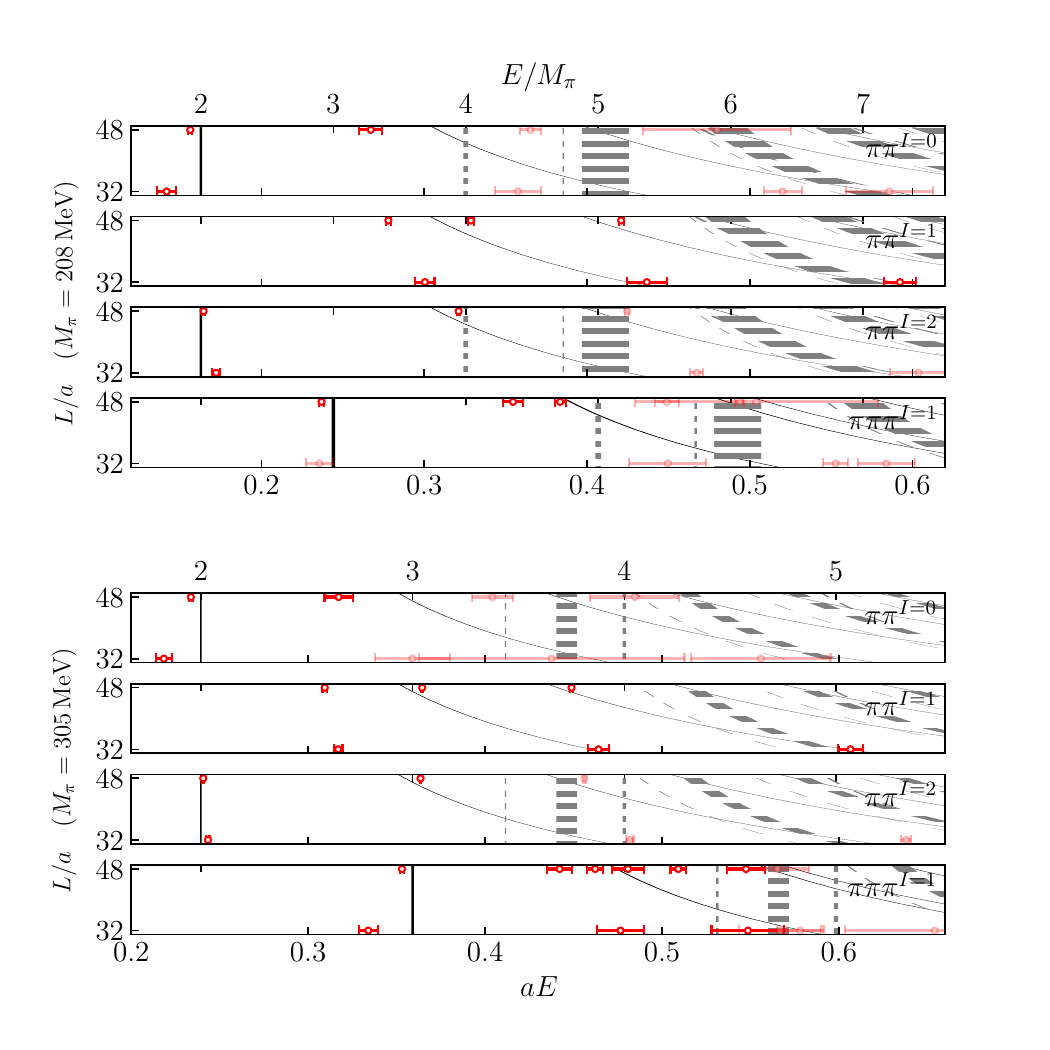}
\caption{
Finite-volume spectra at the heavier and lighter $\pi$ meson masses, including the $I=0,1,2$ $\pi\pi$ channels and the $I=1$ $\pi\pi\pi$ channel. Red points denote the interacting lattice energy levels; the lighter points are not included in the analysis. Solid bands denote noninteracting elastic levels, while dashed bands denote noninteracting inelastic levels.}
\label{fig:pi1300_spectra_non}
\end{figure}

The fit results and their dependence on the fit ranges are shown in Figs.~\ref{fig:pipi-I=0-fit-F32P30}--\ref{fig:pipipi-I=1-fit-F48P21} of Appendix~\ref{appendix:three_body_problems2}. Statistical errors are estimated using the Jackknife method. The extracted lattice energy levels are shown in Fig.~\ref{fig:pi1300_spectra_non}. The lower $x$ axis is in lattice units, while the upper $x$ axis is in units of $M_\pi$. Energy levels from different ensembles and channels are shown as red circles.

In the two-$\pi$ system, for $I=0,1$, the lowest $aE_n$ lies below the $2M_\pi$ threshold, indicating an attractive interaction. In the $I=2$ channel, by contrast, the value of $aE_n$ above this threshold is consistent with a weak repulsive interaction, as expected.

In the three-$\pi$ system, at both values of $M_\pi$, we observe a volume-independent ground-state level consistent with the $\pi$ mass; this level is not shown in the figure. Above the $3\pi$ threshold, the levels cannot be straightforwardly assigned to specific physical states. It is worth noting, however, that for the $\pi\pi\pi(I=1)$ system at $M_\pi \approx 305\,\mathrm{MeV}$, a clear clustering of levels appears near and above $4M_\pi$. These levels cannot be matched one-to-one with free levels, indicating a nontrivial and relatively strong attraction in this region, possibly associated with the presence of a resonance. This hypothesis will be tested below.

\section{From Lattice Spectra to Scattering Amplitudes}
The lattice spectrum provides first-principles information about two-body and three-body interactions. In practice, this information can be extracted through a quantization condition. To this end, we use the FVU three-body quantization condition~\cite{Mai:2017bge}; applications can be found in Refs.~\cite{Mai:2018djl, Mai:2019fba, Alexandru:2020xqf, Brett:2021wyd, Mai:2021nul, Garofalo:2022pux, Yan:2024gwp}.

Within this formalism, once the short-range two-body and three-body interaction terms $K$ and $C$ are specified, the quantization condition determines the finite-volume spectrum in a given irrep (here $A_1^-$) by solving for its roots. In other words, these interactions serve as inputs to the quantization condition, which then outputs finite-volume energy levels that can be fitted to the lattice levels obtained in the previous section. Through this procedure, we extract scattering amplitudes from lattice QCD. The general form is given in Eq.~\ref{eq:QC-tilde}.

For the $\pi(1300)$ channel, the linear space of the matrix is spanned by $\{ {\text{two-body isospin}} (I_2) \times {\text{helicity}} (\lambda) \times {\text{spectator momentum}} \}$. For the case $I^G(J^{PC})=1^-(0^{-+})$ in the center-of-mass frame, the following basis is obtained:

\begin{equation}
\{\sigma(-\bm{p})\pi(\bm{p}),
\rho(\lambda=-1,-\bm{p})\pi(\bm{p}),
\rho(\lambda=0,-\bm{p})\pi(\bm{p}),
\rho(\lambda=+1,-\bm{p})\pi(\bm{p}),
G(-\bm{p})\pi(\bm{p})\}.
\end{equation}
Here $\sigma$, $\rho$, and $G$ denote the two-$\pi$ subsystems with isospin $I_2=0,1,2$, respectively.

The matrices $\Sigma^{\mathrm{FV}}$, $B$, and $E_L$ depend only on momenta and masses and describe all possible on-shell kinematic configurations of the three $\pi$ mesons. The two-body and three-body interactions ($\tilde K^{-1}/C$), by contrast, contain the dynamical information specific to the physical channel~\cite{Mai:2021nul, Yan:2024gwp, Feng:2024wyg}. It should be noted that the zeros of $\tau^{-1}$ give the two-body energy eigenvalues in the corresponding isospin channel.

In a two-body system, each finite-volume energy eigenvalue determines one value of $\tilde K^{-1}$, provided higher partial waves are neglected and the level lies below the inelastic threshold. In a three-body system, however, one of the particles, the spectator, can carry away momentum. A more detailed understanding of the functional forms of $C$ and $K^{-1}$ is therefore required. Since the explicit form of the three-body force in this system is currently poorly known, we choose a generic parametrization in the orbital-angular-momentum basis for the transition $\alpha \overset{C}{\to} \beta$~\cite{Chung:1971ri}:
\begin{equation}
\begin{aligned}
    c^{\alpha\beta}=c_c^{\alpha\beta}+\frac{c_p^{\alpha\beta}}{s-m_0^2},
    \quad
    {\alpha=\beta}\in \{\sigma\pi,\rho\pi\}.
    \label{eq:C-term}
\end{aligned}
\end{equation}
Here $s$ denotes the squared total energy of the three-body system, and only the phenomenologically most important channels are included. By contrast, for the relevant two-body subsystems, we can make use of a large body of existing theoretical results and relate their dynamics directly to phase shifts:
\begin{equation}
\begin{aligned}
    &\tilde K^{-1}_{I_2}(\sigma)=K^{-1}_{I_2}(\sigma)+\Re\,\Sigma^{\mathrm{IV}}_{I_2}(\sigma), \\
    &K^{-1}_{I_2}(\sigma)=
    \begin{cases}
        -\frac{p(\sigma)\cot\delta^{I_20}(\sigma)}{16\pi\sqrt{\sigma}}
        \quad \text{for } I_2=0,2, \\
        -\frac{p^3(\sigma)\cot\delta^{11}(\sigma)}{12\pi\sqrt{\sigma}} \quad \text{for } I_2=1.
    \end{cases}
    \label{eq:Ktilde+SigmaIV}
\end{aligned}
\end{equation}
Here, $\sigma$, $p$, and $\Sigma^{\mathrm{IV}}$ denote the two-body invariant mass squared, the two-body center-of-mass momentum, and the two-body self-energy integral, respectively. Explicit formulae can be found in Refs.~\cite{Feng:2024wyg,Yan:2024gwp}. To parametrize the phase shift, we use the well-tested~\cite{Mai:2019pqr,Doring:2016bdr,Acharya:2015pya} modified inverse amplitude method (mIAM)~\cite{Hanhart:2008mx}, which combines $S$-matrix theory with chiral perturbation theory (ChPT). This amplitude remains well behaved far below the two-body threshold, for example near the Adler zero. At very low $\sigma$, however, the amplitude is no longer reliable. We therefore apply the cutoff
\begin{equation}
    K^{-1}_{I_2}(\sigma)\mapsto K^{-1}_{I_2}(\sigma_{\mathrm{MP}})\, e^{-(\sigma-\sigma_{\mathrm{MP}})},
\end{equation}
which is equivalent to smoothly turning off the interaction below the matching point $\sigma_{\mathrm{MP}}$. This matching point, together with the maximum spectator momentum $|l_{\mathrm{max}}|$ and the maximum self-energy momentum $|k_{\mathrm{max}}|$, constitutes the full set of cutoff parameters required in the three-body formalism.

\section{Brief Introduction to Chiral Perturbation Theory}
In the previous section, we used the \mybf{modified inverse amplitude method} (mIAM)~\cite{Hanhart:2008mx} to parametrize the phase shifts of the two-body subsystems. This section introduces the theoretical foundation of this method from the perspective of chiral perturbation theory (ChPT).

QCD is nonperturbative at low energies, so perturbation theory cannot be applied directly. In the limit where the light-quark masses are neglected, the QCD Lagrangian has an approximate $\mathrm{SU}(2)_L \times \mathrm{SU}(2)_R$ chiral symmetry. After this symmetry is spontaneously broken by the vacuum, it gives rise to nearly massless Goldstone bosons, namely the light mesons observed experimentally, such as $\pi$, $K$, and $\eta$.

\subsection{$\pi\pi$ Scattering}
In chiral perturbation theory, the light-meson fields are parametrized through a nonlinear $\sigma$ model. After external source fields are introduced, one can construct an effective Lagrangian consistent with chiral symmetry. The most commonly used form is written in terms of the $SU(2)$ matrix
\begin{equation}
U(x) = \exp\left(\frac{i}{f_\pi} \sum_{a=1}^3 \frac{\sigma_a}{2} \pi_a(x) \right),
\end{equation}
where $\sigma_a$ are the Pauli matrices and $\pi_a$ denotes the pseudoscalar-meson isotriplet field.

In the chiral expansion, where $p$ denotes a small momentum or meson mass, the first few terms of the effective Lagrangian are
\begin{equation}
    \mathcal{L}_2 = \frac{f_\pi^2}{4} \langle \partial_\mu U \partial^\mu U^\dagger + \chi U^\dagger + \chi^\dagger U \rangle,
\end{equation}
\begin{equation}
\begin{aligned}
    \mathcal{L}_4 &= l_1 \langle \partial_\mu U \partial^\mu U^\dagger \rangle^2 + l_2 \langle \partial_\mu U \partial_\nu U^\dagger \rangle \langle \partial^\mu U \partial^\nu U^\dagger \rangle \\
    &+ l_3 \langle \chi U^\dagger + U \chi^\dagger \rangle^2 + l_4 \langle \partial_\mu U \partial^\mu U^\dagger \rangle \langle \chi U^\dagger + U \chi^\dagger \rangle + \cdots,
\end{aligned}
\end{equation}
where $\chi = 2B_0 \mathcal{M}$ is the external source, $\mathcal{M}$ is the quark-mass matrix, $f_\pi$ is the $\pi$ decay constant, and $B_0$ is related to the quark condensate. $\mathcal{L}_2$ is the $O(p^2)$ contribution, while $\mathcal{L}_4$ is the $O(p^4)$ contribution. The low-energy constants (LECs) $l_i$ (or their renormalized counterparts $l_i^r$) encode short-distance physics and must be determined from experimental data or higher-order calculations. The power-counting rule organizes the perturbative hierarchy: the meson momentum satisfies $p \sim M_\pi \ll \Lambda_\chi \sim 1\,\mathrm{GeV}$, so higher-order terms are suppressed by $p^2/\Lambda_\chi^2$. Within this framework, the $\pi\pi$ scattering amplitude can be computed order by order.

At $O(p^2)$, chiral symmetry strongly constrains the form of the scattering amplitude~\cite{Weinberg:1966jm}. For the $S$-wave scattering lengths, one obtains the well-known Weinberg--Tomozawa formula:
\begin{equation}
a_0^0 = \frac{7M_\pi^2}{32\pi f_\pi^2} \approx 0.16 \quad
a_0^2 = -\frac{M_\pi^2}{16\pi f_\pi^2} \approx -0.045,
\end{equation}
where $a_0^I$ denotes the $S$-wave scattering length in the isospin-$I$ channel. The scattering length is positive in the $I=0$ channel, corresponding to attraction, and negative in the $I=2$ channel, corresponding to repulsion; this qualitative pattern agrees with experiment.

At $O(p^4)$, the pioneering work of Gasser and Leutwyler~\cite{Gasser:1983yg} gave the complete expression for the scattering amplitude. At this order, four low-energy constants $l_1^r, l_2^r, l_3^r, l_4^r$ must be introduced in a specified renormalization scheme. The $\pi\pi$ scattering lengths are
\begin{equation}
\begin{aligned}
a_0^0 &= \frac{7 M_\pi^2}{32 \pi f_\pi^2} 
\Bigg\{ 1 + \frac{M_\pi^2}{f_\pi^2} \Big[ 5 l_1^r + 2 l_2^r - \frac{3}{8} l_3^r + \frac{1}{8\pi^2} \ln\frac{M_\pi^2}{\mu^2} \Big] \Bigg\} + \mathcal{O}(M_\pi^4), \\
a_0^2 &= - \frac{M_\pi^2}{16 \pi f_\pi^2} 
\Bigg\{ 1 - \frac{2 M_\pi^2}{f_\pi^2} \Big[ l_1^r + l_2^r + \frac{1}{16\pi^2} \ln\frac{M_\pi^2}{\mu^2} \Big] \Bigg\} + \mathcal{O}(M_\pi^4),
\end{aligned}
\end{equation}
These expressions make clear how the low-energy constants bridge low-energy theorems and experimental data.

As a perturbative expansion, chiral perturbation theory is reliably convergent only at sufficiently low energies, typically $\sqrt{s} \lesssim 1\,\mathrm{GeV}$. At higher energies, especially near resonances such as the $\rho(770)$ and $\sigma(600)$, the perturbative expansion breaks down. More importantly, the chiral perturbation theory amplitudes do not satisfy exact unitarity of the $S$ matrix.

\subsection{Modified Inverse Amplitude Method}
The central idea of the inverse amplitude method (IAM) is to use the amplitude information obtained from the chiral expansion and reconstruct, through dispersion relations, a scattering amplitude that satisfies $S$-matrix unitarity~\cite{Truong:1988zp}.

Let $T_2^{Il}(s)$ be the $O(p^2)$ chiral partial-wave amplitude at leading order (LO), and let $T_4^{Il}(s)$ be the $O(p^4)$ amplitude at next-to-leading order (NLO), for isospin $I$ and angular momentum $l$.

At LO, the $\pi\pi$ $S$-wave amplitudes are
\begin{equation}
T_2^{I0}(s) =
\begin{cases}
\dfrac{2s - M_\pi^2}{32\pi f_\pi^2}, & I=0, \\
-\dfrac{s - 2M_\pi^2}{32\pi f_\pi^2}, & I=2,
\end{cases}
\end{equation}
where $M_\pi$ and $f_\pi$ are the mass and decay constant of the $\pi$ meson, respectively.

The NLO amplitude $T_4^{Il}(s)$ consists of three contributions: contact terms, which contain the low-energy constants, the $s$-channel loop diagrams, and the $t$- and $u$-channel loop diagrams. Their evaluation involves loop integrals and the renormalized low-energy constants $\{\bar{l}_i\}$~\cite{Gasser:1983yg}. The low-energy constants entering $T_4^{Il}(s)$ must be determined by fits to experimental or lattice QCD data.

The IAM gives the unitary scattering amplitude as
\begin{equation}
    T_{\text{IAM}}^{Il}(s) = \frac{(T_2^{Il}(s))^2}{T_2^{Il}(s) - T_4^{Il}(s)}.
\end{equation}
The $K$-matrix form used in FVU is
\begin{equation}
    (K^{\text{IAM}})^{-1} = \frac{1}{16\pi} \left( \frac{1}{T_2^{Il}(s)} - \frac{T_4^{Il}(s)}{(T_2^{Il}(s))^2} \right).
\end{equation}
This expression can be derived rigorously from dispersion relations~\cite{Truong:1988zp}. In the low-energy limit, $T_4^{Il}(s) \sim \mathcal{O}(p^4)$, so $T_{\text{IAM}}^{Il}(s)=T_2^{Il}(s)+T_4^{Il}(s)+\mathcal{O}(p^6)$ automatically satisfies the chiral low-energy theorem. As the energy increases, zeros of the denominator $T_2^{Il}(s)-T_4^{Il}(s)$ generate poles in the complex energy plane, corresponding to physical resonances such as the $\rho(770)$ resonance~\cite{Dobado:1989qm,Dobado:1989xz}.

Through dispersion relations, the IAM reconstructs a unitary scattering amplitude while preserving the low-energy behavior of ChPT, and therefore can describe the formation of resonances. In the subthreshold region, however, the chiral perturbation theory amplitude contains an Adler zero, and the standard IAM generates a spurious pole there. The mIAM introduces an improvement term into the amplitude, restoring the correct Adler-zero structure and removing this unphysical pole while retaining the desirable features of the IAM.

To eliminate the unphysical singularity, the modified inverse amplitude method (mIAM) introduces an additional term $A_m^{Il}(s)$ in the denominator~\cite{GomezNicola:2007qj}:
\begin{equation}
T_{mIAM}^{Il}(s) = \frac{(T_2^{Il}(s))^2}{T_2^{Il}(s) - T_4^{Il}(s) + A_m^{Il}(s)},
\end{equation}
where
\begin{equation}
A_m^{Il}(s) = T_4^{Il}(s_2) - \frac{(s_2 - s_A)(s - s_2)}{s - s_A} \left[ (T_2^{Il})'(s_2) - (T_4^{Il})'(s_2) \right].
\end{equation}
$s_A$ is the root of the equation $T_2^{Il}(s) - T_4^{Il}(s) = 0$, namely the Adler zero, and $s_2$ is the root of the equation $T_2^{Il}(s) = 0$. In the $\rho$ channel ($I=1$, $l=1$), $A_m^{Il} = 0$, so the mIAM is equivalent to the original IAM.

The amplitude produced by the mIAM is an analytic and unitary scattering amplitude. In the low-energy region it reduces to the ChPT amplitude at the corresponding order, while also satisfying crossing symmetry up to NLO~\cite{Hanhart:2008mx}. This method has been successfully applied in several lattice QCD calculations, including $\pi\pi$ scattering~\cite{Doring:2016bdr}, $D\pi$ scattering~\cite{Mai:2019pqr}, and related systems. It provides a reliable bridge between first-principles lattice QCD calculations and hadron phenomenology.

In the present analysis, we employ the mIAM4 scheme, whose inverse $K$ matrix, $\tilde K^{-1}$, is given in Eq.~\ref{eq:Ktilde+SigmaIV}. This parametrization remains consistent with the ChPT low-energy theorems while retaining enough flexibility to describe strong-interaction dynamics such as the $\rho$ resonance.

\section{Model-Averaging Scheme}
The next section discusses various model fits to the lattice spectra; before that, we briefly introduce the model-averaging scheme. To average the analysis results obtained from different models, we adopt the method proposed in Ref.~\cite{EuropeanTwistedMass:2014osg}. Specifically, suppose there are $N$ calculations, with mean values $x_k$ and uncertainties $\sigma_{x,k}$ ($k=1,\dots,N$). The weighted average $x$ and its total uncertainty $\sigma_x$ can then be written as
\begin{equation}
\begin{aligned}
    x &= \sum_{k=1}^N \omega_k x_k, \\
    \sigma_x^2 &= \sigma_{x,\mathrm{stat}}^2 + \sigma_{x,\mathrm{syst}}^2, \\
    \sigma_{x,\mathrm{stat}}^2 &= \sum_{k=1}^N \omega_k \sigma_{x,k}^2, \\
    \sigma_{x,\mathrm{syst}}^2 &= \sum_{k=1}^N \omega_k (x_k - x)^2.
    \label{eq:averaging}
\end{aligned}
\end{equation}
where $\omega_k$ is the weight assigned to the $k$-th result. The weights are determined using the Akaike information criterion (AIC)~\cite{Akaike}:
\begin{equation}
    \omega_k = A \operatorname{e}^{- \frac{1}{2} (\chi_k^2 + 2 N_\mathrm{parms} - N_\mathrm{data})},
\label{eq:AIC}
\end{equation}
where $\chi_k^2$ is the $\chi^2$ value of the $k$-th fit, $N_\mathrm{parms}$ is the number of free parameters, $N_\mathrm{data}$ is the number of data points, and $A$ is a normalization constant chosen such that $\sum_{k=1}^N \omega_k = 1$.

The above formula can be generalized to the multivariate case. For example, if one needs the covariance $\sigma_{xy}$ between $x$ and another variable $y$, it can be defined as
\begin{equation}
\begin{aligned}
    \sigma_{xy}^2 &= \sigma_{xy,\mathrm{stat}}^2 + \sigma_{xy,\mathrm{syst}}^2, \\
    \sigma_{xy,\mathrm{stat}}^2 &= \sum_{k=1}^N \omega_k \, \sigma_{xy,k}^2, \\
    \sigma_{xy,\mathrm{syst}}^2 &= \sum_{k=1}^N \omega_k \, (x_k - x)(y_k - y).
    \label{eq:averaging1}
\end{aligned}
\end{equation}

\section{Finite-Volume Analysis}
When applying the three-body quantization formalism to finite-volume spectra, one must make concrete choices for the parametrization and for the cutoff conditions, including the three-body force $C$. To define the model space, we specify the physical inputs and cutoff parameters and then systematically explore their effects through different fits to the LQCD data. In particular, we consider the following classes of scenarios:
\begin{itemize}
    \item [(a)] all possible combinations of the terms in Eq.~\ref{eq:c};
    \item [(b)] all two-body energy eigenvalues, supplemented by either the three-body energy eigenvalues at light $M_{\pi}$, the three-body energy eigenvalues at heavy $M_{\pi}$, or all three-body spectra;
    \item [(c)] $3\times 3\times 3$ different combinations of the cutoff parameters $\sigma_{\mathrm{MP}}, |l_{\mathrm{max}}|, |k_{\mathrm{max}}|$;
    \item [(d)] different choices for the chiral perturbation theory low-energy constants $l_i^r$ fixed to the FLAG values~\cite{FlavourLatticeAveragingGroupFLAG:2024oxs}: mIAM4 ($l_1^r,l_2^r,l_3^r,l_4^r$), mIAM3 ($l_1^r,l_2^r,l_3^r,l_4^{r,\mathrm{FLAG}}$), and mIAM2 ($l_1^r,l_2^r,l_3^{r,\mathrm{FLAG}},l_4^{r,\mathrm{FLAG}}$), in analogy with Ref.~\cite{Mai:2019pqr}.
\end{itemize}

To assess the associated systematic uncertainties, we tested approximately $2000$ different scenarios. The best-fit parameters obtained from simultaneous fits to the two- and three-body finite-volume spectra are listed in Table~\ref{tab:fit-results-all}. The fits include either only the heavy, only the light, or both $M_{\pi}$ ensembles. The two-body part is described using the modified inverse amplitude method~\cite{Hanhart:2008mx}, and the three-body force is taken in the general form
\begin{align}
    c^{\alpha\beta} = c_c^{\alpha\beta} + \frac{c_p^{\alpha\beta}}{s - m_0^2}
    \quad {\alpha=\beta} \in \{\sigma\pi, \rho\pi\},
\label{eq:c}
\end{align}
corresponding to the relevant two-body-pair--spectator channels. The fit results in Table~\ref{tab:fit-results-all} are obtained at the fixed cutoff values
\[
MP/M_\pi^2 = 1, \quad |k_{\mathrm{max}}| = \sqrt{3\cdot5^2}\frac{2\pi}{aL}, \quad |l_{\mathrm{max}}| = \sqrt{3}\frac{2\pi}{aL}.
\]
The systematic uncertainty associated with the cutoff choice is small and is subleading compared with the statistical and other systematic uncertainties discussed in the main text.

The above coefficients simultaneously specify the angular-momentum and helicity bases. The projected three-body force is defined as
\begin{align}
    \left[C_{\mathrm{HB}}\right]_{(\bm{p}^\prime,j)(\bm{p},i)} &= 
    \frac{1}{4\pi} 
    \mathfrak{D}^{0*}_{0,-\lambda(j)}(\phi_{-\bm{p'}},\theta_{-\bm{p'}},0)
    C_{ji}(s,p',p)
    \mathfrak{D}^{0}_{0,-\lambda(i)}(\phi_{-\bm{p}},\theta_{-\bm{p}},0),
    \\
    &\text{where}\quad
    C_{ji}(s,p',p) = U_{jL'} [C_{\mathrm{JLS}}(s,p',p)]_{L'L} U_{Li}, 
    \quad
    U_{Lj} = \begin{pmatrix}
        -1 & 0 & 0 \\
        0 & 1 & 0 \\
        0 & 0 & 1 \\
    \end{pmatrix}_{Lj},
    \\
    &\text{and}\quad
    C_{\mathrm{JLS}} =
    \begin{pmatrix}
       (p')^1 c_{\pi\rho,\pi\rho}(p)^1 & 0 & 0 \\
       0 & (p')^0 c_{\pi\sigma,\pi\sigma}(p)^0 & 0 \\
       0 & 0 & 0 \\
    \end{pmatrix}.
\end{align}

\begin{table}[b]
\centering
\tiny
\caption{Fit results obtained with cutoff parameters $\{1M_\pi^2, \sqrt{3} \frac{2\pi}{aL}, \sqrt{3\cdot5^2} \frac{2\pi}{aL}\}$. The top, middle, and bottom blocks correspond to the heavy, light, and combined $M_{\pi}$ ensembles, respectively. The fits include both two- and three-body energy levels and their correlations. The best fits used for pole-position extraction are $\{5,23,194,203\}$, and the overall best fit ($203$) is shown in bold.}
\addtolength{\tabcolsep}{-3pt}
\label{tab:fit-results-all}
\begin{tabular}{l|llll|rrrrr|rrrrr|cc}
\toprule
\# & \multicolumn{4}{c|}{$\{l_1^r,l_2^r,l_3^r,l_4^r\}\times1000$} & $c^\sigma_c/{M_\pi^2}$ & $c^\sigma_p$ & $c^\rho_c/{M_\pi^2}$ & $c^\rho_p$ & $m_0/{M_\pi}$ & $c^\sigma_c/{M_\pi^2}$ & $c^\sigma_p$ & $c^\rho_c/{M_\pi^2}$ & $c^\rho_p$ & $m_0/{M_\pi}$ & 
$\chi^2/{\rm d.o.f.}$ & $N_{\rm data}$ \\
\midrule
$2$ & $-4.56$ & $4.33$ & $11.69$ & $4.49$ & -- & -- & -- & -- & -- & -- & $10.71$ & -- & -- & $4.07$ & $1.22$ & $33$ \\
$5$ & $-4.57$ & $4.32$ & $11.68$ & -- & -- & -- & -- & -- & -- & -- & $10.71$ & -- & -- & $4.07$ & $1.18$ & $33$ \\
$8$ & $-4.78$ & $3.88$ & -- & -- & -- & -- & -- & -- & -- & -- & $2.23$ & -- & -- & $4.02$ & $2.32$ & $33$ \\
$11$ & $-5.05$ & $4.24$ & $10.78$ & $-15.24$ & -- & -- & -- & -- & -- & -- & -- & -- & $26.36$ & $4.14$ & $1.44$ & $33$ \\
$14$ & $-4.56$ & $4.35$ & $11.37$ & -- & -- & -- & -- & -- & -- & -- & -- & -- & $22.14$ & $4.11$ & $1.28$ & $33$ \\
$17$ & $-4.77$ & $3.92$ & -- & -- & -- & -- & -- & -- & -- & -- & -- & -- & $30.76$ & $4.21$ & $2.4$ & $33$ \\
$20$ & $-4.62$ & $4.21$ & $8.52$ & $4.58$ & -- & -- & -- & -- & -- & $19.57$ & $117.21$ & -- & -- & $4.4$ & $1.16$ & $33$ \\
$23$ & $-4.62$ & $4.21$ & $8.51$ & -- & -- & -- & -- & -- & -- & $19.6$ & $117.77$ & -- & -- & $4.4$ & $1.12$ & $33$ \\
$29$ & $-5.37$ & $4.17$ & $8.18$ & $-22.34$ & -- & -- & -- & -- & -- & -- & -- & $9.19$ & $38.14$ & $4.21$ & $1.72$ & $33$ \\
$32$ & $-4.57$ & $4.32$ & $11.34$ & -- & -- & -- & -- & -- & -- & -- & -- & $-0.01$ & $9.78$ & $3.98$ & $1.31$ & $33$ \\
$35$ & $-4.78$ & $3.9$ & -- & -- & -- & -- & -- & -- & -- & -- & -- & $4.7$ & $16.16$ & $4.$ & $2.46$ & $33$ \\
$38$ & $-4.84$ & $4.4$ & $10.85$ & $-12.99$ & -- & -- & -- & -- & -- & $0.$ & -- & -- & -- & -- & $2.26$ & $33$ \\
$41$ & $-4.31$ & $4.89$ & $14.03$ & -- & -- & -- & -- & -- & -- & $-7.75$ & -- & -- & -- & -- & $2.58$ & $33$ \\
$44$ & $-4.47$ & $4.57$ & -- & -- & -- & -- & -- & -- & -- & $15.27$ & -- & -- & -- & -- & $3.63$ & $33$ \\
$56$ & $-4.31$ & $4.88$ & $10.76$ & $2.8$ & -- & -- & -- & -- & -- & $0.32$ & -- & $-0.83$ & -- & -- & $2.81$ & $33$ \\
$59$ & $-4.55$ & $4.36$ & $12.47$ & -- & -- & -- & -- & -- & -- & $-4.1$ & -- & $-8.05$ & -- & -- & $1.42$ & $33$ \\
$62$ & $-4.66$ & $4.13$ & -- & -- & -- & -- & -- & -- & -- & $1.1$ & -- & $-8.52$ & -- & -- & $2.39$ & $33$ \\
\midrule
$65$ & $-4.65$ & $4.15$ & $6.64$ & $4.3$ & -- & $14.67$ & -- & -- & $4.95$ & -- & -- & -- & -- & -- & $1.03$ & $27$ \\
$68$ & $-4.65$ & $4.15$ & $6.62$ & -- & -- & $14.79$ & -- & -- & $4.95$ & -- & -- & -- & -- & -- & $0.99$ & $27$ \\
$71$ & $-4.76$ & $3.92$ & -- & -- & -- & $12.9$ & -- & -- & $4.92$ & -- & -- & -- & -- & -- & $1.14$ & $27$ \\
$74$ & $-4.63$ & $4.17$ & $6.59$ & $4.84$ & -- & -- & -- & $61720.$ & $70.78$ & -- & -- & -- & -- & -- & $1.06$ & $27$ \\
$77$ & $-4.64$ & $4.17$ & $6.55$ & -- & -- & -- & -- & $59071.6$ & $69.$ & -- & -- & -- & -- & -- & $1.02$ & $27$ \\
$80$ & $-4.76$ & $3.94$ & -- & -- & -- & -- & -- & $60050.6$ & $69.47$ & -- & -- & -- & -- & -- & $1.16$ & $27$ \\
$83$ & $-4.65$ & $4.13$ & $6.17$ & $4.77$ & $23.02$ & $761.8$ & -- & -- & $6.96$ & -- & -- & -- & -- & -- & $1.06$ & $27$ \\
$86$ & $-4.66$ & $4.13$ & $6.15$ & -- & $21.42$ & $668.27$ & -- & -- & $6.82$ & -- & -- & -- & -- & -- & $1.01$ & $27$ \\
$89$ & $-4.77$ & $3.92$ & -- & -- & $93.54$ & $9302.87$ & -- & -- & $10.75$ & -- & -- & -- & -- & -- & $1.13$ & $27$ \\
$92$ & $-4.63$ & $4.17$ & $6.59$ & $4.84$ & -- & -- & $-12.35$ & $763.47$ & $142.71$ & -- & -- & -- & -- & -- & $1.12$ & $27$ \\
$95$ & $-4.64$ & $4.17$ & $6.55$ & -- & -- & -- & $-12.47$ & $657.45$ & $874.9$ & -- & -- & -- & -- & -- & $1.07$ & $27$ \\
$98$ & $-4.76$ & $3.94$ & -- & -- & -- & -- & $-12.51$ & $30.47$ & $364.77$ & -- & -- & -- & -- & -- & $1.22$ & $27$ \\
$101$ & $-4.63$ & $4.19$ & $6.9$ & $4.3$ & $-2.92$ & -- & -- & -- & -- & -- & -- & -- & -- & -- & $1.03$ & $27$ \\
$104$ & $-4.63$ & $4.19$ & $6.89$ & -- & $-2.94$ & -- & -- & -- & -- & -- & -- & -- & -- & -- & $0.99$ & $27$ \\
$107$ & $-4.75$ & $3.95$ & -- & -- & $-2.64$ & -- & -- & -- & -- & -- & -- & -- & -- & -- & $1.15$ & $27$ \\
$119$ & $-4.64$ & $4.16$ & $6.76$ & $4.32$ & $-1.86$ & -- & $-10.92$ & -- & -- & -- & -- & -- & -- & -- & $1.05$ & $27$ \\
$122$ & $-4.65$ & $4.16$ & $6.75$ & -- & $-1.89$ & -- & $-10.93$ & -- & -- & -- & -- & -- & -- & -- & $1.$ & $27$ \\
$125$ & $-4.76$ & $3.93$ & -- & -- & $-1.26$ & -- & $-11.53$ & -- & -- & -- & -- & -- & -- & -- & $1.16$ & $27$ \\
\midrule
$191$ & $-4.47$ & $4.72$ & $12.31$ & $-2.02$ & -- & $103.16$ & -- & -- & $6.68$ & -- & $35.74$ & -- & -- & $4.54$ & $2.36$ & $36$ \\
$194$ & $-4.56$ & $4.34$ & $11.79$ & -- & -- & $55.34$ & -- & -- & $5.98$ & -- & $10.62$ & -- & -- & $4.07$ & $1.11$ & $36$ \\
$197$ & $-4.77$ & $3.89$ & -- & -- & -- & $41.22$ & -- & -- & $5.81$ & -- & $0.87$ & -- & -- & $3.95$ & $2.25$ & $36$ \\
$200$ & $-4.62$ & $4.21$ & $8.51$ & $4.61$ & $15.34$ & $416.83$ & -- & -- & $6.47$ & $19.45$ & $115.28$ & -- & -- & $4.4$ & $1.12$ & $36$ \\
$\bf 203$ & $\bf -4.62$ & $\bf 4.21$ & $\bf 8.5$ & -- & $\bf 15.24$ & $\bf 415.65$ & -- & -- & $\bf 6.47$ & $\bf 19.46$ & $\bf 115.5$ & -- & -- & $\bf 4.4$ & $\bf 1.08$ & $\bf 36$ \\
$206$ & $-4.76$ & $3.93$ & -- & -- & $129.73$ & $18028.$ & -- & -- & $12.45$ & $272.48$ & $15731.$ & -- & -- & $8.47$ & $1.31$ & $36$ \\
$209$ & $-4.47$ & $4.49$ & $11.48$ & $5.91$ & -- & -- & -- & $104.33$ & $6.24$ & -- & -- & -- & $34.24$ & $4.24$ & $1.35$ & $36$ \\
$212$ & $-4.53$ & $4.41$ & $11.43$ & -- & -- & -- & -- & $123.52$ & $5.97$ & -- & -- & -- & $17.19$ & $4.06$ & $1.26$ & $36$ \\
$215$ & $-4.74$ & $3.97$ & -- & -- & -- & -- & -- & $125.48$ & $5.68$ & -- & -- & -- & $0.02$ & $3.86$ & $2.92$ & $36$ \\
$218$ & $-4.63$ & $4.39$ & $10.56$ & $-2.85$ & -- & -- & $3.55$ & $317.27$ & $6.13$ & -- & -- & $9.28$ & $16.38$ & $4.17$ & $1.59$ & $36$ \\
$221$ & $-4.48$ & $4.5$ & $11.09$ & -- & -- & -- & $6.3$ & $427.06$ & $6.75$ & -- & -- & $69.13$ & $401.66$ & $4.59$ & $1.37$ & $36$ \\
$224$ & $-4.71$ & $4.02$ & -- & -- & -- & -- & $2.07$ & $915.38$ & $5.89$ & -- & -- & $11.12$ & $31.23$ & $4.01$ & $2.66$ & $36$ \\
\bottomrule
\end{tabular}
\addtolength{\tabcolsep}{3pt}
\end{table}

\begin{figure}[b]
    \centering
    \includegraphics[width=\linewidth]{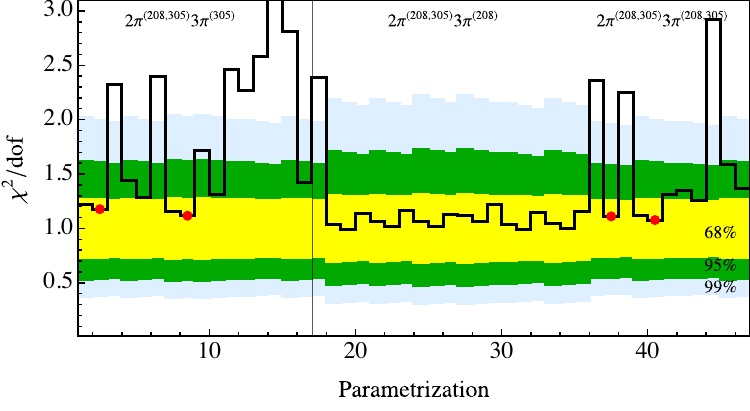}
    \caption{Fit quality of the 47 central fits. The lattice QCD inputs used in the fits are labeled as $2\pi^{(M_\pi)}$ and $3\pi^{(M_\pi)}$. The shaded bands indicate the $68\%$, $95\%$, and $99\%$ confidence intervals around the expectation value $\mathbb{E}[\chi^2]$ of the $\chi^2$ distribution, and the points denote the best fits selected for determining the pole positions shown in Fig.~\ref{fig:poles}.}
    \label{fig:fit-quality-all-fits}
\end{figure}

After fixing the cutoff choice, to which the results are insensitive, to $\{1M_\pi^2,\sqrt{3}\, \frac{2\pi}{aL},\sqrt{3\cdot5^2} \frac{2\pi}{aL}\}$, we are left with $47$ primary fits. Their fit quality is shown in Fig.~\ref{fig:fit-quality-all-fits}. The overall best fit selected in our analysis has $\chi^2/{\mathrm{dof}}=1.08$.

We observe that, when only the three-body inputs at the light $\pi$ mass are included, namely the F32P21 and F48P21 ensembles, almost all types of $C$ parametrizations give excellent fit quality within the $68\%$ confidence interval of the $\chi^2$ distribution. This is most likely because there are very few energy levels below the relevant threshold, as shown in Fig.~\ref{fig:pi1300_spectra_non}. In addition, we find that the mIAM3 models always have slightly better $\chi^2/{\mathrm{dof}}$ values, so in the subsequent analysis we use only the mIAM3 fits to avoid double counting similar models.

Using the selected best parametrization and cutoff choice, we redraw the finite-volume spectra in Fig.~\ref{fig:pi1300_spectra_non}. We note that, in the two-body system, the orange-band energy levels obtained from FVU are much more precise than the lattice energy levels. This is because the three-body data provide stronger constraints on the two-body scattering amplitude; without them, this part would have had substantially larger uncertainties, as has also been observed recently in Ref.~\cite{Dawid:2025zxc}\footnotecircle{The strong constraints imposed by mIAM on the amplitude provide another reason why the uncertainties of the two-body parameters and pole positions are small.}.
\begin{figure}[t]
\centering
\includegraphics[width=\linewidth,trim=0.9cm 0.9cm 1.7cm 1.0cm,clip]{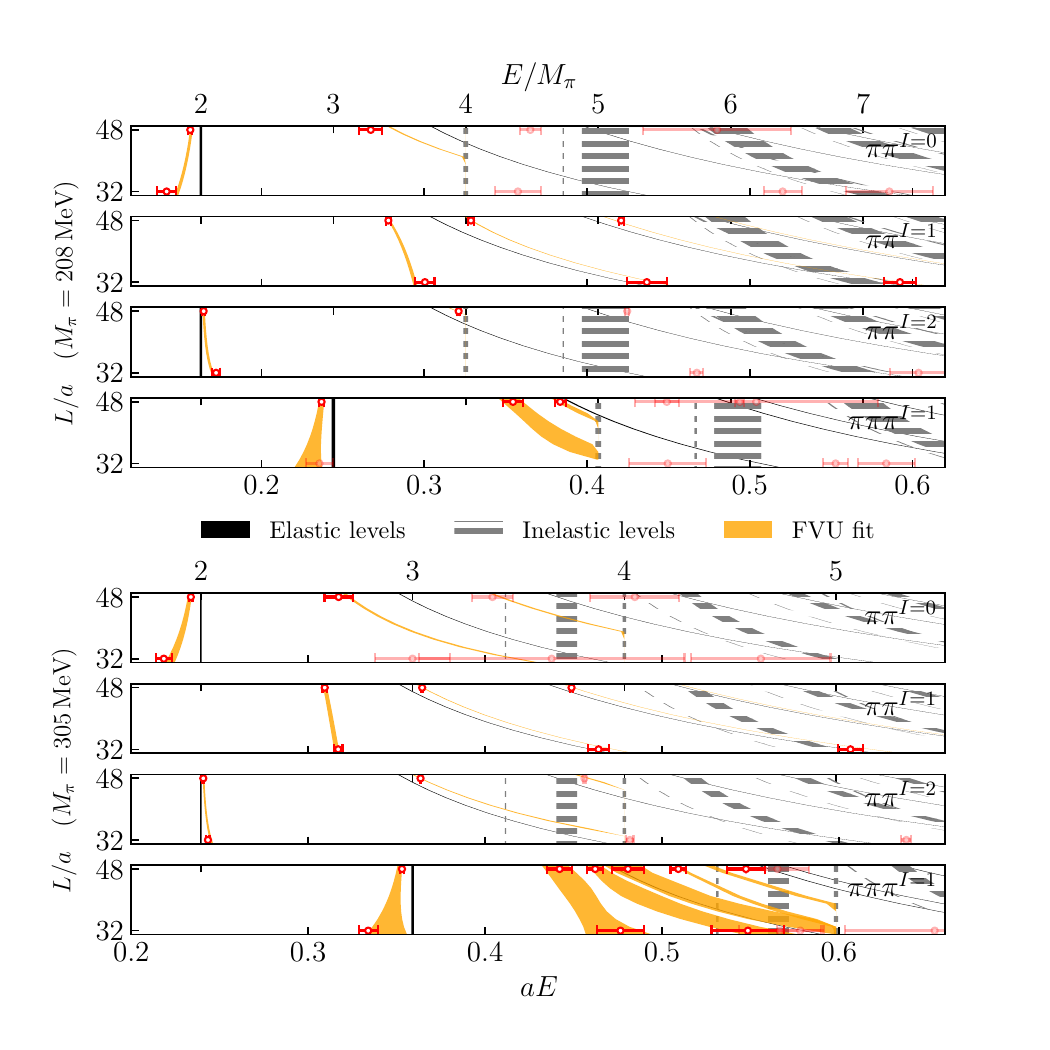}
\caption{
Same as Fig.~\ref{fig:pi1300_spectra_non}, with the orange bands added to show the finite-volume energy levels obtained from the optimal parametrization.}
\label{fig:pi1300_spectra}
\end{figure}

The parameters obtained from fits to the finite-volume spectra can be used to extract hadron-resonance pole positions through the infinite-volume unitarity (IVU) method~\cite{Sadasivan:2021emk, Garofalo:2022pux, Yan:2024gwp, Feng:2024wyg}; the overall workflow is shown in Fig.~\ref{fig:Workflow}. This requires solving integral equations, which we do efficiently using the complex-contour method. Further technical details and comparisons of methods can be found in Refs.~\cite{Feng:2024wyg, Sakthivasan:2024uwd, Doring:2025sgb}. We note that the spectator-momentum cutoff limits the maximum energy up to which the amplitude satisfies unitarity, namely $\sqrt{s_{\mathrm{max}}}=\sqrt{\sigma+l_{\mathrm{max}}^2}+\sqrt{M_\pi^2+l_{\mathrm{max}}^2}\big|_{\sigma=4M_\pi^2}$. On the other hand, for fixed $s$, a larger cutoff may lead to $\sigma\ll\sigma_{\mathrm{MP}}$, where the two-body input is no longer reliable. The cutoff set chosen here maximizes the domain of applicability of the method.

For fits using only the data at the light $\pi$ mass, we find that only a small number of fits generate a $\pi(1300)$ pole. Together with the sparse finite-volume spectrum in Fig.~\ref{fig:pi1300_spectra_non}, this suggests that the $\pi(1300)$ has only weak dependence on the meson mass. Its pole is expected near $M_{\pi(1300)}\approx 1300/208 \gtrsim 6 M_\pi$, which lies beyond the energy range currently accessible to lattice and finite-volume methods. By contrast, at the heavy $\pi$ mass, $M_{\pi(1300)}\approx 1300/305\approx 4.2 M_\pi$ lies below $5M_{\pi}$, and all fits within the $68\%$, $95\%$, and $99\%$ confidence intervals around $\chi^2_{\mathrm{obs}}$ do produce a $\pi(1300)$ pole. Interestingly, these fits include cases in which no explicit pole term is present in the three-body-force parametrization, indicating that the $\pi(1300)$ pole is dynamically generated.

\begin{figure}
    \centering
    \includegraphics[width=\linewidth]{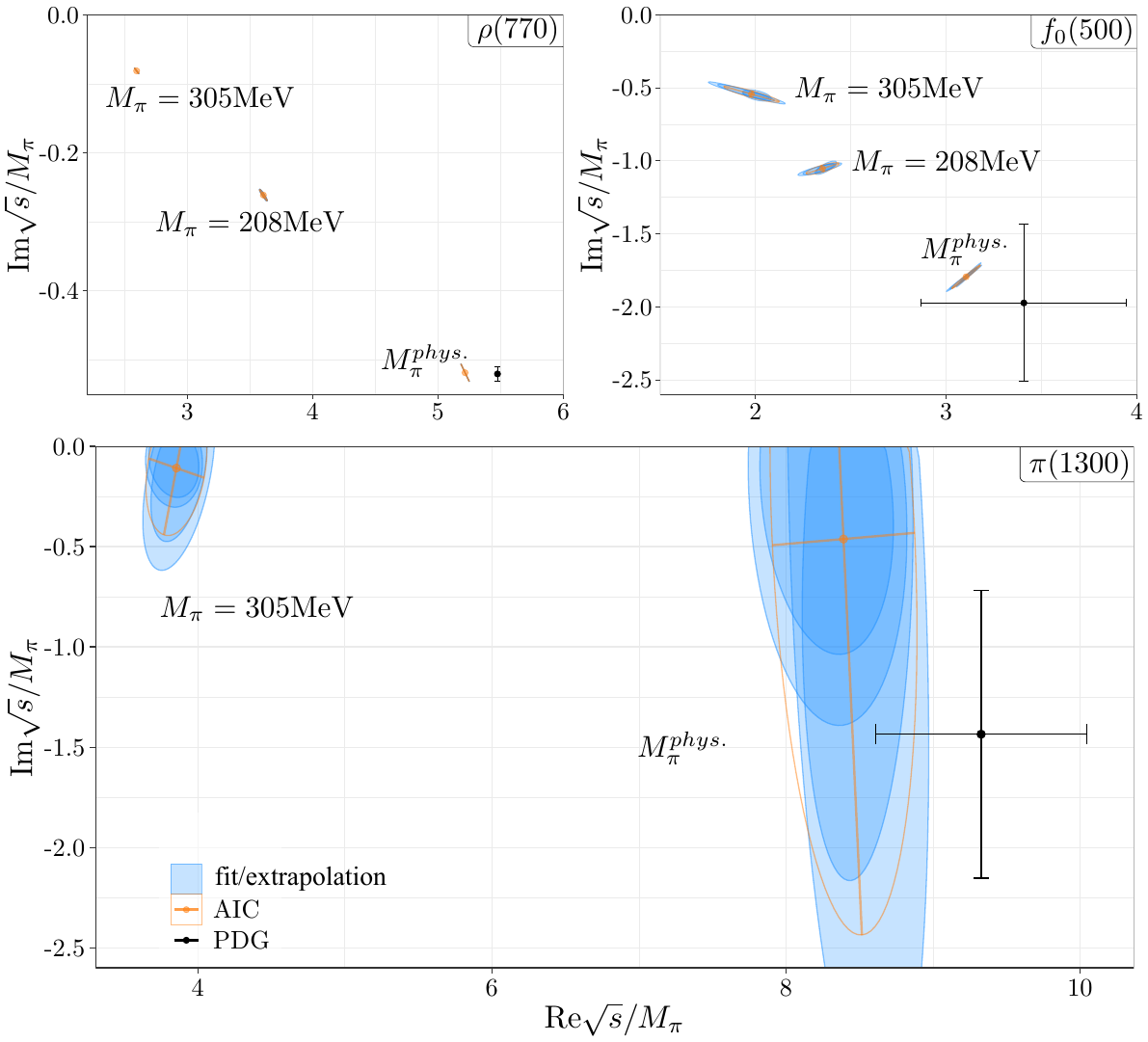}
    \caption{Pole positions of the two- and three-hadron systems on the second Riemann sheet, corresponding to the $\rho(770)$, $f_0(500)$, and $\pi(1300)$ states. The shaded regions show the results of individual fits within the $68\%$ confidence interval around the observed $\chi^2_{\mathrm{obs}}$. The orange crosses denote the AIC-based model-averaged results, while the black crosses show the PDG values for comparison.}
    \label{fig:poles}
\end{figure}

Considering the fits that lie within the $68\%$ confidence interval around $\mathbb{E}[\chi^2]$ of the $\chi^2$ distribution and include the heavy-$M_{\pi}$ spectra, shown as red points in Fig.~\ref{fig:fit-quality-all-fits}, we extract the pole positions shown in Fig.~\ref{fig:poles}. These include the two-body resonance poles in the $I=1$ vector $\rho$-meson subsystem and the $I=0$ scalar $\sigma$-meson subsystem. The extrapolation of the two-body sector to the physical point is performed by setting $M_\pi$ to its physical value, as required by chiral perturbation theory (ChPT). In the three-body case, the simplest procedure is to directly use the parameters obtained from the fits to the heavy-$M_\pi$ data, thereby assuming that all terms entering Eq.~\ref{eq:C-term} have negligible $M_\pi$ dependence in both physical and lattice units. The results from the individual fits, together with the model-averaged result based on the Akaike information criterion (AIC)~\cite{Akaike}, are shown in Fig.~\ref{fig:poles}. The main result of this chapter is the model-averaged pole position of the $\pi(1300)$:
\begin{equation}
\begin{aligned}
    M_{\pi(1300)} &= (1169\pm46)-i(62\pm169)\,\mathrm{MeV}, \\
    M_{\rho(770)} &= (727\pm3)-i(72\pm1)\,\mathrm{MeV}, \\
    M_{f_{0}(500)} &= (433\pm 7)-i(250\pm 7)\,\mathrm{MeV}.
\end{aligned}
\end{equation}
The two-body resonance results obtained here are generally in good agreement with the PDG values, except for $\Re M_{\rho(770)}$. Since $\Re M_{\rho(770)}$ is consistent with previous lattice studies on the same ensembles~\cite{Mai:2019pqr, Yan:2024gwp}, we attribute this discrepancy to finite-lattice-spacing effects; this issue has been discussed in detail in Ref.~\cite{Wang:2025hew} using the same lattice ensembles.

The new result of this chapter is a first-principles QCD prediction for the pole position of the radial excitation of the $\pi$ meson, which agrees with the PDG value within the $1\sigma$ uncertainty. The imaginary parts of all extracted pole positions are negative; the error ellipses appear to extend partly above the real axis only because a simple Gaussian representation is used in the plot.

\section{Conclusions and Outlook}
By constructing a complete set of one-, two-, and three-hadron interpolating operators, we used lattice QCD to compute all $\pi\pi$ two-body spectra and the three-body finite-volume spectra in the isospin $I=1$ and $A_1^-$ irreps relevant to the $\pi(1300)$ channel. Combining these results with the state-of-the-art three-body quantization condition, we determined the resonance parameters of the $\pi(1300)$ from QCD. At the lighter $M_\pi$, the position of the $\pi(1300)$ lies above the effective range of FVU, $5M_{\pi}$, and five-hadron interpolating operators were not included in the lattice calculation, so the information is insufficient for a reliable extraction of the resonance parameters. At the heavier $M_\pi$, however, all fits yield scattering amplitudes containing a resonance structure, thereby confirming the presence of the $\pi(1300)$ in QCD. On this basis, and under the assumption that the three-body force has no quark-mass dependence, we extrapolated the resonance parameters to the physical point and obtained results consistent with the PDG values. Physically, the $\pi(1300)$ is the first radial excitation of the $\pi$ meson, and its mass is roughly ten times that of the ground state. This large energy gap reflects the characteristic scale of the QCD confining potential and provides benchmark information for a quantitative understanding of quark confinement. In addition, this chapter validates the applicability of the three-body finite-volume formalism to broad resonances: even when the subchannel contains the well-known broad $f_0(500)$ resonance, the method still yields a reliable pole extraction. This lays the methodological foundation for future studies of more complicated three-body decay systems, such as the Roper resonance.

We tested several sources of systematic uncertainty, including different parametrizations of the two- and three-body inputs and the cutoff dependence, and found that they have little impact on the main conclusions. In future work, however, it will be important to study the quark-mass dependence of the $\pi(1300)$ more systematically, for example by carrying out calculations at heavier $M_\pi$. In addition, providing line shapes and Dalitz plots will facilitate comparison with phenomenological studies of multihadron final states.

The work presented in this chapter is a first step toward studying radial excitations of light mesons. In hadron spectroscopy, radial excitations provide a more sensitive and systematic probe of nonperturbative QCD dynamics. Compared with ground-state hadrons, they correspond to more highly excited radial wave functions with nodal structure, and are therefore more sensitive to the long-range behavior of the confining potential, the details of dynamical models, and the internal spatial structure of hadrons. They thus offer an important window for testing quark models and effective field theories.

In the light pseudoscalar-meson sector, the $\pi(1300)$, $K(1460)$, $\eta(1295)$, and $\eta(1475)$ form the most natural set of radial-excitation candidates. They correspond to the $2S$ excitation structure of an $\mathrm{SU}(3)$ flavor nonet and couple strongly to multibody decay channels such as $\pi\pi\pi$, $K\bar{K}\pi$, and $\eta\pi\pi$. The ground states and radial excitations of the light pseudoscalar mesons exhibit qualitatively different dynamical mechanisms for mass generation. The ground-state masses are mainly governed by spontaneous and explicit chiral-symmetry breaking, depend strongly on the light-quark masses, and display a pronounced pattern of $\mathrm{SU}(3)$ flavor breaking. By contrast, the masses of the radial excitations are dominated by the excitation-energy scale set by confinement dynamics, with the excitation energy of the color flux tube providing the leading contribution; light-quark mass differences enter only as subleading corrections. Thus, although the ground-state pseudoscalar mesons show significant $\mathrm{SU}(3)$ breaking, the radial excitations are more nearly governed by a common confining-level structure, leading to their approximate mass degeneracy.

Nevertheless, this spectral structure remains subject to long-standing experimental and theoretical uncertainties, so the completeness of the nonet has not yet been reliably established; the associated questions are also highly challenging. Experimentally, radial excitations usually have large decay widths and involve complicated coupled multibody channels, making their resonance signals difficult to isolate cleanly with finite statistics. Their extraction also depends strongly on the amplitude-analysis model, leading to sizable experimental uncertainties. Theoretically, such states are highly sensitive to dynamical details and model parameters, and they often couple strongly to the continuum and to multibody scattering states, which increases the complexity and model dependence of the spectroscopic description.

Furthermore, extending this program to vector mesons, such as the $\rho(1450)$, $\omega(1420)$, and $\phi(1680)$, will help build a unified picture of hadron spectroscopy across flavor sectors and provide systematic comparisons and constraints for possible exotic hadron states.

We are steadily pursuing this broader program.

\cleardoublepage
\chapter{Precision Lattice Calculations and Hadron Decays}
\label{chap:decay}

{
\kaishu
\begin{center}
    试玉要烧三日满，辨材须待七年期。
\end{center}
\hfill ——《放言五首·其三》[唐] 白居易
}

\section{Background}
The preceding chapters studied the properties of hadron resonances from the perspective of scattering. This chapter turns to decay processes, where precision lattice-QCD determinations of decay matrix elements provide a complementary route to understanding the strong interaction.

\mybf{Flavor physics} at high precision is one of the major frontiers of contemporary particle physics\footnotecircle{This chapter is based on the published works~\cite{Meng:2024axn} Y. Meng, C. Liu, T. Wang and H. Yan, Lattice study of $J/\psi\to \gamma \eta_c$ using a method without momentum extrapolation, \textit{Phys. Rev. D} 111, 014508 (2025). \\ \cite{Meng:2024gpd} Y. Meng, J. L. Dang, C. Liu, Z. Liu, T. Shen, H. Yan and K. L. Zhang, Lattice QCD calculation of the $D_s^*$ radiative decay with $(2+1)$-flavor Wilson-Clover ensembles, \textit{Phys. Rev. D} 109, 074511 (2024). \\ \cite{Meng:2024nyo} Y. Meng, J. L. Dang, C. Liu, X. Y. Tuo, H. Yan, Y. B. Yang and K. L. Zhang, First lattice QCD calculation of $J/\psi$ semileptonic decay containing $D$ and $D_s$ particles, \textit{Phys. Rev. D} 110, 074510 (2024).}. It is central not only to precision tests of the Standard Model, but also to the search for possible new physics. Such tests broadly proceed along two paths. One is to compare theoretical predictions directly with experimental measurements, looking for significant deviations in quantities such as particle masses, leptonic decay constants, and decay rates of specific processes. The other is to study flavor-changing weak decays, extract CKM matrix elements with high precision, and test the unitarity of CKM rows and columns~\cite{FlavourLatticeAveragingGroupFLAG:2024oxs}. In both approaches, high-precision theory and high-precision experiment are indispensable. Historically, however, the two have not always advanced in step; whichever side first achieves a breakthrough often provides an essential benchmark for the other.

Charmonium is a bound-state system composed of a charm quark and its antiquark. Since its discovery half a century ago, it has remained a central subject of both theoretical and experimental studies~\cite{E598:1974sol,SLAC-SP-017:1974ind}. On the experimental side, BESIII, currently the world's largest charmonium facility, has accumulated an enormous sample of $J/\psi$ events~\cite{BESIII:2021cxx}, and the future STCF is expected to further improve experimental precision~\cite{Achasov:2023gey,Ai:2025xop}. On the theoretical side, charmonium lies at an intermediate energy scale where both perturbative and nonperturbative methods are useful, making it an ideal system for testing different theoretical frameworks and deepening our understanding of the strong interaction.

The dominant decay modes of the $J/\psi$ are strong and electromagnetic decays, and their properties have been studied extensively over the past several decades. Among them, radiative transitions are among the simplest processes in the charmonium system. Before the work described in this chapter was published, however, the relevant experimental measurements were still limited~\cite{Gaiser:1985ix,CLEO:2008pln,Anashin:2014wva}, and the branching fraction carried a substantial uncertainty~\cite{CLEO:2008qfy,BESIII:2012lxx}. The most recent Particle Data Group (PDG) average updates the branching fraction for this process to $1.41(14)\%$~\cite{PDG24}, a significant improvement over the previous value $1.7(4)\%$~\cite{PDG22}. Recently, using the process $J/\psi \to \gamma p\bar p$ and a data sample of approximately $1.01\times10^{10}$ $J/\psi$ events, the BESIII Collaboration studied $J/\psi \to \gamma \eta_c$ while systematically accounting for interference between the $\eta_c$ resonance and nonresonant backgrounds~\cite{BESIII:2025vdn}. This substantially improved the precision of the product branching fraction $\mathcal{B}(J/\psi\to\gamma\eta_c)\mathcal{B}(\eta_c\to p\bar p)$. With external input, that analysis gives an updated value of $\mathcal{B}(J/\psi\to\gamma\eta_c)$, consistent with the lattice calculation presented in this chapter.

For weak decays of the $J/\psi$, early estimates placed the branching fractions at the level of $10^{-8}$~\cite{Sanchis-Lozano:1993vyw}, long below experimental reach. In recent years, these rare decays have attracted increasing attention from experiment~\cite{BES:2006mls,BESIII:2014pps,BESIII:2021mnd,BESIII:2023fqz} and from a variety of phenomenological approaches, including the Bauer-Stech-Wirbel (BSW) model~\cite{Dhir:2009rb}, QCD sum rules (QCDSR)~\cite{Wang:2007ys}, the Bethe-Salpeter equation~\cite{Wang:2016dkd}, and the covariant light-front quark model (CLFQM)~\cite{Sun:2023uyn}.

This chapter employs a new method that directly computes the off-shell transition form factor at zero spatial momentum and thereby extracts the on-shell transition form factor. In conventional lattice calculations, one usually computes the off-shell transition form factor at nonzero photon virtuality and then extrapolates in momentum to obtain the on-shell result, or introduces twisted boundary conditions to tune the momentum directly to the on-shell point. The former inevitably introduces model-dependent extrapolation uncertainty, while the latter requires additional dedicated propagators and therefore limits reusability in other calculations.

On-shell form factors obtained through continuous momentum extrapolation can often be determined with higher precision. In recent years, this strategy has been widely applied to many physical processes~\cite{Feng:2019geu,Feng:2020zdc,Tuo:2021ewr,Meng:2021ecs,Zou:2021mgf,Fu:2022fgh,Tuo:2022hft,Christ:2022rho,Meng:2023bjc}, leading to significant progress. We further consider the radiative decay of the vector meson $D_s^*$, where the theoretically calculated partial width can be combined with the experimental branching fraction to determine the total width. An existing lattice calculation~\cite{Donald:2013sra} gives an uncertainty of about $40\%$ for the total width, much larger than the experimental uncertainty of roughly $10\%$. It is therefore important to reduce the theoretical uncertainty in order to extract $f_{D_s^*}|V_{cs}|$ more precisely.

\section{Lattice Study of Charmonium Radiative Transitions}
\subsection{Method for the On-Shell Transition Form Factor}
This section presents in detail a new method for determining the on-shell transition form factor $V(0)$ directly from lattice data. The central idea is to construct scalar combinations of the hadronic function under suitable momentum projections, thereby eliminating the need for momentum extrapolation.

The Euclidean hadronic function in infinite volume is defined as
\begin{equation}
    H_{\mu\nu}(\vec{x},t)= \langle 0|\phi_{\eta_c}(\vec{x},t)J_{\nu}(0)|J/\psi_{\mu}(p')\rangle,
\end{equation}
where $t>0$, $|J/\psi_{\mu}(p') \rangle$ denotes a $J/\psi$ state with four-momentum $p'=(im_{J/\psi},\vec{0})$, and $\phi_{\eta_c}$ is the interpolating operator for $\eta_c$ that creates the $\eta_c$ meson. $J_{\nu}$ is the electromagnetic vector current, $J_{\nu}=\sum_qe_q\,\bar{q}\gamma_\nu q$, with $e_q=2/3,-1/3,-1/3,2/3$ for the $u$, $d$, $s$, and $c$ quarks, respectively.

At large $t$, the hadronic function is dominated by the single-$\eta_c$ state, while contributions from excited states can be neglected. Thus,
\begin{equation}
    \begin{aligned}
    H_{\mu \nu}(\vec{x}, t)&=\langle 0|\phi_{\eta_{c}}(\vec{x}, t) J_{\nu}(0)| J / \psi_{\mu}\left(p^{\prime}\right)\rangle \\
    &\approx \int \frac{d^{3} \vec{p}}{(2 \pi)^{3} 2E}\langle 0|\phi_{\eta_{c}}(\vec{x}, t)|\eta_c\rangle\langle \eta_c| J_{\nu}(0)| J / \psi_{\mu}\left(p^{\prime}\right)\rangle \\
    &=\int \frac{d^{3} \vec{p}}{(2 \pi)^{3} 2E}\langle 0|\phi_{\eta_{c}}(0)\mathrm{e}^{-iP\cdot x}|\eta_c\rangle\langle \eta_c| J_{\nu}(0)| J / \psi_{\mu}\left(p^{\prime}\right)\rangle \\
    &=\int \frac{d^{3} \vec{p}}{(2 \pi)^{3} 2E}\langle 0|\phi_{\eta_{c}}(0)\mathrm{e}^{-i(Et-\vec{p} \cdot \vec{x})}|\eta_c\rangle\langle \eta_c| J_{\nu}(0)| J / \psi_{\mu}\left(p^{\prime}\right)\rangle \\
    &\xrightarrow{\text{Wick rotation}}\int \frac{d^{3} \vec{p}}{(2 \pi)^{3} 2E}\langle 0|\phi_{\eta_{c}}(0)\mathrm{e}^{-i(E(-it)-\vec{p} \cdot \vec{x})}|\eta_c\rangle\langle \eta_c| J_{\nu}(0)| J / \psi_{\mu}\left(p^{\prime}\right)\rangle \\
    &=\frac{2 e_{c}}{m_{\eta_{c}}+m_{J / \psi}} \int \frac{d^{3} \vec{p}}{(2 \pi)^{3}} \frac{Z}{E} \epsilon_{\mu \nu \alpha \beta} p_{\alpha} p_{\beta}^{\prime}V\left(q^{2}\right) \mathrm{e}^{-E t+i \vec{p} \cdot \vec{x}},
    \end{aligned}
\end{equation}
where the overlap factor $Z$ and the transition form factor $V(q^2)$ are defined by
\begin{equation}
    \langle 0|\phi_{\eta_c}(0)|\eta_c(\vec{p})\rangle = Z,
\end{equation}
\begin{equation}
    \langle \eta_c(\vec{p})|J_{\nu}(0)|J/\psi_{\mu}(p')\rangle=\frac{4V(q^2)}{m_{\eta_c}+m_{J/\psi}}e_c\epsilon_{\mu\nu\alpha \beta}p_{\alpha}p'_{\beta}.
\end{equation}
Here $q^2=(m_{J/\psi}-E)^2- |\vec{p}|^2$ is the photon four-momentum squared. The form factor $V(q^2)$ is a function of $q^2$; the point $q^2=0$ corresponds to a real-photon radiative transition and defines the desired on-shell transition form factor $V(0)$.

Because of lattice discretization effects, the continuum dispersion relation $E^2=m_{\eta_c}^2+|\vec{p}|^2$ is no longer exact, and Lorentz symmetry is also broken. As a result, the overlap factor $Z$ acquires a momentum dependence. To describe these effects, we introduce the correction parameters $\xi$~\footnotecircle{Here $\xi$ is the same quantity as $Z_X$ in\chapref{chap:two_body_problems}.} and $\eta$:
\begin{equation}
    \begin{aligned}
    E^2 &= m_{\eta_c}^2+\xi\cdot |\vec{p}|^2, \\
    Z^2 &= Z_0^2+\eta\cdot |\vec{p}|^2.
    \end{aligned}
    \label{eq:xieta}
\end{equation}
Here $\xi$ describes the correction to the dispersion relation, with $\xi=1$ corresponding to the continuum limit, while $\eta$ describes the momentum dependence of the overlap factor. Both parameters are determined from lattice data.

We now describe how to extract the on-shell transition form factor $V(0)$ from the hadronic function. The strategy is to construct a \mybf{scalar function} whose specific momentum projection removes the model dependence associated with extrapolation.

The spatial Fourier transform of $H_{\mu\nu}(\vec{x},t)$ is
\begin{equation}
    \mathcal{H}(\vec{p},t)=\int d^{3} \vec{x} \,  \mathrm{e}^{-i\vec{p}\cdot\vec{x}}H_{\mu\nu}(\vec{x},t)=\frac{2 e_{c}}{m_{\eta_{c}}+m_{J / \psi}}\frac{Z}{E} \epsilon_{\mu \nu \alpha \beta} p_{\alpha} p_{\beta}^{\prime}V\left(q^{2}\right) \mathrm{e}^{-E t}.
\end{equation}
This motivates the scalar function
\begin{equation}
    \mathcal{I}_{0}(t,\vec{p}) =\epsilon_{\mu \nu \alpha^{\prime} \beta^{\prime}} p_{\alpha^{\prime}} p_{\beta^{\prime}}^{\prime} \frac{1}{m_{J / \psi}|\vec{p}|^{2}}\mathcal{H}(\vec{p},t),
\end{equation}
which gives
\begin{equation}
\mathcal{I}_0(t,\vec{p})
=-\frac{4e_cZm_{J/\psi}}{m_{\eta_c}+m_{J/\psi}}V(q^2)\frac{e^{-Et}}{E}
=\frac{1}{ |\vec{p}|^2}\int d^3\vec{x}e^{-i\vec{p}\cdot\vec{x}}\epsilon_{\mu\nu\alpha0}\frac{\partial H_{\mu\nu}(x) }{\partial x_{\alpha}}.
\end{equation}

Averaging over the spatial direction of $\vec{p}$ yields
\begin{equation}
\mathcal{I}_0(t,|\vec{p}|)=\int d^3\vec{x}\frac{j_1(|\vec{p}||\vec{x}|)}{|\vec{p}||\vec{x}|}\epsilon_{\mu\nu \alpha 0}x_{\alpha}H_{\mu\nu}(\vec{x},t),
\label{eq:I0}
\end{equation}
where $j_1(x)$ is the spherical Bessel function of the first kind.

Expanding $V(q^2)$ around $q^2=0$ gives
\begin{equation}
V(q^2)=\sum\limits_{n=0}^{\infty}c_n\left(\frac{q^2}{m_{J/\psi}^2}\right)^n \doteq c_0+c_1\cdot \frac{q^2}{m_{J/\psi}^2}+\mathcal{O}(q^4/m_{J/\psi}^4),
\end{equation}
where $c_0 \equiv V(0)$ is precisely the desired on-shell transition form factor, and $c_1$ describes the slope of $V(q^2)$ at $q^2=0$. We keep only the linear term because the value of $q^2$ at the zero-momentum point, $(\delta m)^2 \equiv (m_{J/\psi}-m_{\eta_c})^2$, is very small:
\begin{equation}
\frac{(\delta m)^2}{m_{J/\psi}^2} \sim 0.14\%.
\end{equation}
Thus contributions from higher-order terms $c_{n\geq 2}$ are suppressed by powers of $(0.14\%)^n$ and can be neglected.

By differentiating at $|\vec{p}|^2=0$, one can obtain information about both $c_0$ and $c_1$. We define
\begin{equation}
\mathcal{I}_1(t,0)\equiv -\frac{\partial \mathcal{I}_0(t,|\vec{p}|)}{\partial |\vec{p}|^2}\Big{|}_{|\vec{p}|^2=0}
=\frac{1}{30}\int d^3\vec{x}|\vec{x}|^2\epsilon_{\mu\nu\alpha 0}x_{\alpha}H_{\mu\nu}(\vec{x},t).
\label{eq:I1}
\end{equation}
Expressing $\mathcal{I}_0$ and $\mathcal{I}_1$ in terms of $c_0$ and $c_1$, one obtains
\begin{equation}
\begin{aligned}
\mathcal{I}_1(t,0)
=&\frac{-4e_c\xi Z_0 m_{J/\psi}}{m_{\eta_c}+m_{J/\psi}}\frac{e^{-m_{\eta_c}t}}{ m_{\eta_c}}
\Bigg{[} \frac{c_0}{2m_{\eta_c}^2}\Big{(}1+ m_{\eta_c}t -\frac{\eta m_{\eta_c}^2}{Z_0^2\xi}\Big{)}\\
&+\frac{c_1}{m^2_{J/\psi}}\Big{(}
\frac{1}{\xi}+\frac{\delta m}{m_{\eta_c}}
+\frac{(\delta m)^2}{2m_{\eta_c}^2}(1+m_{\eta_c}t-\frac{\eta m_{\eta_c}^2}{Z_0^2\xi})\Big{)} \Bigg{]},
\end{aligned}
\end{equation}
and
\begin{equation}
\mathcal{I}_0(t,0)=\frac{-4e_c Z_0 m_{J/\psi}}{m_{\eta_c}+m_{J/\psi}}\frac{e^{-m_{\eta_c}t}}{ m_{\eta_c}} \left( c_0 +c_1\frac{(\delta m)^2 }{m_{J/\psi}^2}\right),
\end{equation}
where we have set $E=m_{\eta_c}$ at $|\vec{p}|=0$. Combining the two equations, one can solve for $c_0$ and $c_1$:
\begin{equation}
c_1=\Bigg{[}\tilde{\mathcal{I}}_1(t)-\frac{\xi\tilde{\mathcal{I}}_0(t)}{2m_{\eta_c}^2}\Big{(}1 + m_{\eta_c}t-\frac{\eta m_{\eta_c}^2}{Z_0^2\xi }\Big{)}\Bigg{]}\frac{m_{J/\psi}^2 m_{\eta_c}}{m_{\eta_c}+\xi\delta m},
\end{equation}
and
\begin{equation}
c_0=\tilde{\mathcal{I}}_0(t)-c_1 \times \frac{(\delta m)^2}{m_{J/\psi}^2},
\end{equation}
where we have introduced the shorthand notation
\begin{equation}
\tilde{\mathcal{I}}_n(t)\equiv -\frac{(m_{\eta_c}+m_{J/\psi})}{4e_c Z_{0}m_{J/\psi}}m_{\eta_c}e^{m_{\eta_c}t}\mathcal{I}_n(t,0).
\end{equation}
In this way, the on-shell transition form factor $V(0)=c_0$ can be extracted directly from the hadronic function $H_{\mu\nu}(\vec{x},t)$, without any momentum extrapolation.

\subsection{Calculation of the Hadronic Function}
We use three two-flavor twisted-mass gauge ensembles generated by the ETMC Collaboration~\cite{ETM:2009ptp,Becirevic:2012dc}, with lattice spacings $a \simeq 0.0667,0.085,0.098$ fm, denoted a67, a85, and a98, respectively. The uncertainties in the lattice spacings are taken from Ref.~\cite{Blossier:2010cr}, where the lattice spacing was determined by matching the lattice $\pi$ decay constant to its physical value, giving the Sommer parameter $r_0=0.440(12)$ fm. The ensemble parameters are listed in Table~\ref{table:lat_ens}. The valence charm-quark mass is tuned by matching the lattice determination of the $J/\psi$ mass to its physical value~\cite{Meng:2021ecs}.

\begin{table}[htbp]
\centering
\caption{Parameters of the gauge ensembles used in this chapter. From left to right we list the ensemble name, lattice spacing $a$, spatial and temporal lattice sizes $L$ and $T$, number of correlation-function measurements $N_{\textrm{conf}}\times T$ on each ensemble, where $N_{\textrm{conf}}$ is the number of configurations used, the $\pi$ mass, the range of source-current time separations $t$, and the spatial lattice size $L$ in physical units.}
\begin{tabular}{ccccccc}
\toprule
Ensemble & $a$/fm & $L^3\times T$ & $N_{\textrm{conf}}\times T$ & $M_{\pi}$/MeV & $t$  & $L$/fm \\
\midrule
a67 & 0.0667(20) & $32^3 \times 64$& $197\times 64$ & 300 & 5$-$15 & 2.13 \\
a85 & 0.085(2) & $24^3 \times 48$ & $200\times 48$ & 315 & 3$-$12 & 2.04\\
a98 & 0.098(3) & $24^3 \times 48$ & $236\times 48$ & 365 & 2$-$11  & 2.35\\
\bottomrule
\end{tabular}
\label{table:lat_ens}
\end{table}

In this chapter, we compute the three-point correlation function $C^3_{\mu\nu}(\vec{x},t) \equiv\langle \phi_{\eta_c}(\vec{x},t)J_{\nu}(0)\phi_{J/\psi,\mu}^{\dagger}(-t)\rangle$, using a $Z_4$ stochastic wall-source $J/\psi$ operator $\phi_{J/\psi,\mu}=\bar{c}\gamma_{\mu}c$ and a point-source $\eta_c$ operator $\phi_{\eta_c}=\bar{c}\gamma_5c$. Only connected diagrams are included. To compute the connected correlation functions, we place the wall-source propagator at the initial $J/\psi$, the point-source propagator at the $\eta_c$, and take the current as the sink. All propagators are averaged over all time slices to improve statistics. The stochastic propagator reduces the mass uncertainty by nearly a factor of two. We also apply APE smearing~\cite{APE:1987ehd} and Gaussian smearing~\cite{Gusken:1989qx} to the $J/\psi$ field to suppress excited-state effects efficiently. Hadron masses are extracted from the two-point correlation function $C^2(t)=\langle \mathcal{O}_{h}(t) \mathcal{O}_{h}^{\dagger}(0)\rangle$ using a two-state fit,
\begin{equation}
C^2(t)= \sum_{i=0,1}\frac{(Z_i^{h})^2}{2E_i^{h}} \left(\textrm{e}^{-E_i^{h}t}+\textrm{e}^{-E_i^{h}(T-t)}\right),
\end{equation}
where $E_0^{h}$ is the ground-state energy of the meson $h=m_{\eta_c},m_{J/\psi}$, and $E_1^{h}$ is the first excited-state energy. The quantities $Z_i^{h}=\langle i|\mathcal{O}_{h}^\dagger|0\rangle$ with $i=0,1$ are the overlap amplitudes of the ground state and first excited state, respectively. Using $Z_0^{J/\psi}$ and $m_{J/\psi}$ as inputs, the hadronic function $H_{\mu\nu}(\vec{x},t)$ is then determined as
\begin{equation}
H_{\mu\nu}(\vec{x},t)=C^3_{\mu\nu}(\vec{x},t)/[(Z_0^{J/\psi}/2m_{J/\psi})e^{-m_{J/\psi}t}].
\end{equation}

In the calculation we use the local vector current $J_\nu(x)=Z_V e_c \bar{c}\gamma_{\nu}c$, where the additional renormalization factor $Z_V$ converts the local vector current to the conserved current up to corrections no larger than $O(a^2)$. For $a=0.098,0.085$, and $0.0667$ fm, we use $Z_V = 0.6047(19), 0.6257(21)$, and $0.6516(15)$, respectively.

By fitting two-point functions at different momenta, one can extract the corresponding energies and overlap factors, $E(\vec{p})$ and $Z$, which are then used to determine $\xi$ and $\eta$ in Eq.~\ref{eq:xieta}. We consider five center-of-mass momenta $|\vec{P}|^2=0,1,2,3,4$ in units of $(\frac{2\pi}{L})^2$. This parametrization describes the linear momentum dependence of $E$ and $Z$ well.

\begin{figure}[htbp]
\centering
\includegraphics[width=0.32\textwidth]{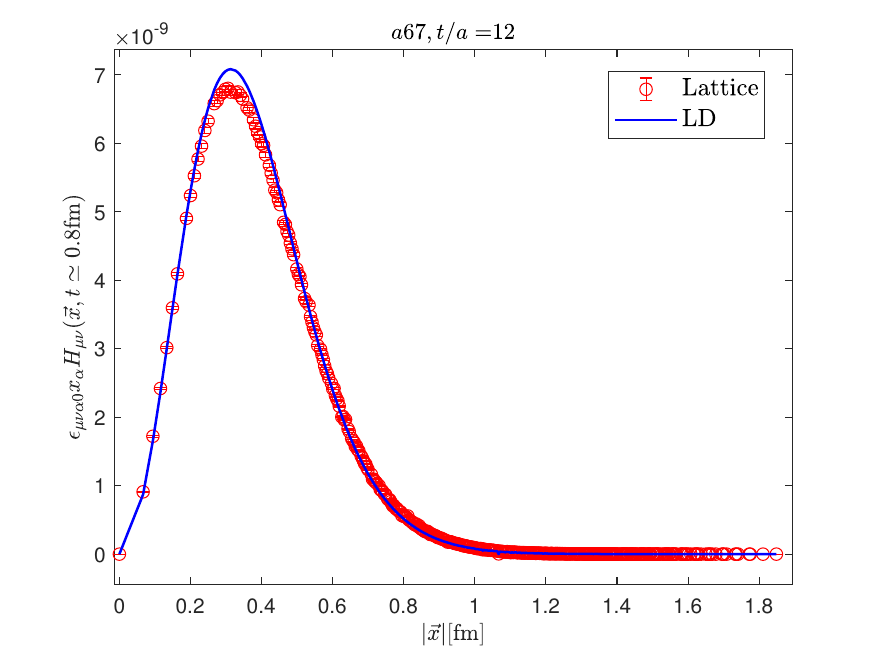}
\includegraphics[width=0.32\textwidth]{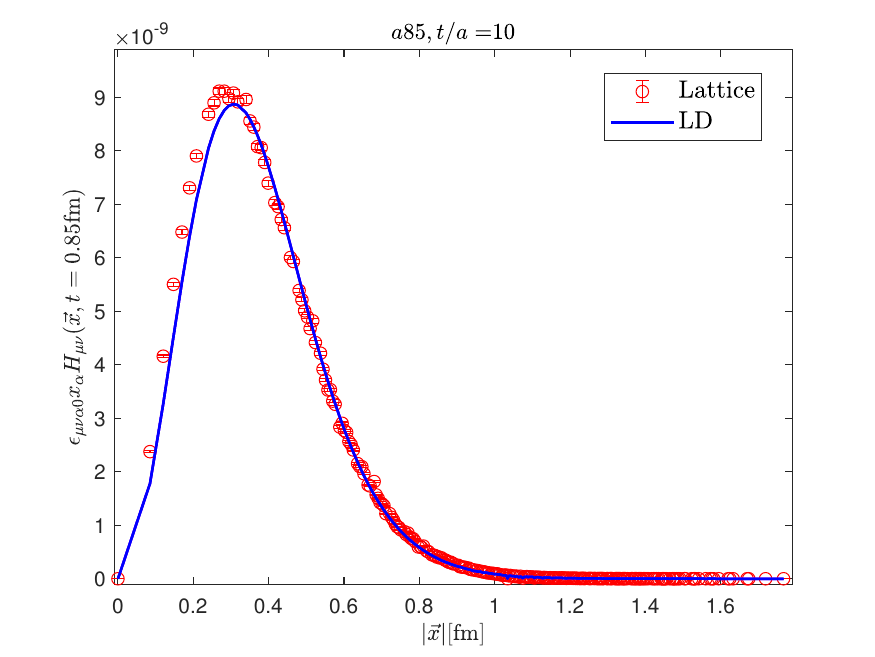}
\includegraphics[width=0.32\textwidth]{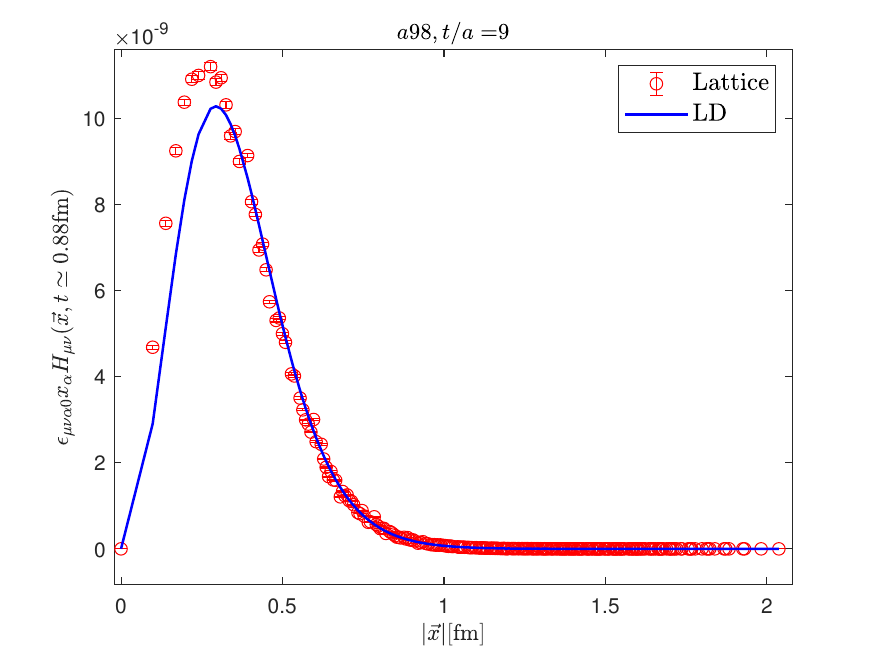}
\caption{Finite-volume matching. The figure shows the lattice data for $\epsilon_{\mu\nu\alpha 0}x_{\alpha}H_{\mu\nu}(\vec{x},t)$ together with the result from the long-distance model.}
\label{fig:LD}
\end{figure}

Because $H_{\mu\nu}(\vec{x},t)$ is dominated by the $\eta_c$ state, it is exponentially suppressed at large $|\vec{x}|$. However, for quantities containing rapidly growing factors such as $|\vec{x}|^2$, for instance $|\vec{x}|^2\epsilon_{\mu\nu\alpha 0}x_{\alpha}H_{\mu\nu}(\vec{x},t)$, finite-volume effects may give non-negligible contributions. To estimate these effects, we construct a long-distance hadronic function $H_{\mu\nu}^{LD}(\vec{x},t)$,
\begin{equation}
    H_{\mu\nu}^{LD}(\vec{x},t)\doteq\frac{-2e_cm_{J/\psi}}{m_{\eta_c}+m_{J/\psi}}\frac{1}{L^3}\sum\limits_{\vec{p}}Z(\vec{p})V(q^2) \epsilon_{\mu\nu\alpha 0}p_{\alpha}\sin(\vec{p}\cdot \vec{x})\frac{e^{-Et}}{E},
\end{equation}
where the transition form factor is taken as $V(q^2)=d_0/(1-d_1q^2)$. The input parameters $d_0$ and $d_1$ can be determined by matching the lattice hadronic function $H_{\mu\nu}(\vec{x},t)$ to the long-distance form $H_{\mu\nu}^{LD}(\vec{x},t)$ at sufficiently large $t$. This matching is shown in Fig.~\ref{fig:LD}. The spatial integrals in Eqs.~\ref{eq:I0} and~\ref{eq:I1} can then be separated into short- and long-distance parts:
\begin{equation}
\begin{aligned}
    \mathcal{I}_0(t,|\vec{p}|)&=\int_{V} d^3\vec{x}\frac{j_1(|\vec{p}||\vec{x}|)}{|\vec{p}||\vec{x}|}\epsilon_{\mu\nu \alpha 0}x_{\alpha}H_{\mu\nu}(\vec{x},t)+\int_{>V} d^3\vec{x}\frac{j_1(|\vec{p}||\vec{x}|)}{|\vec{p}||\vec{x}|}\epsilon_{\mu\nu \alpha 0}x_{\alpha}H_{\mu\nu}^{LD}(\vec{x},t), \\
    \mathcal{I}_1(t,0)&=\frac{1}{30}\int_V d^3\vec{x}|\vec{x}|^2\epsilon_{\mu\nu\alpha 0}x_{\alpha}H_{\mu\nu}(\vec{x},t)+\frac{1}{30}\int_{>V} d^3\vec{x}|\vec{x}|^2\epsilon_{\mu\nu\alpha 0}x_{\alpha}H_{\mu\nu}^{LD}(\vec{x},t).
\end{aligned}
\end{equation}

\subsection{Extraction of the Transition Form Factor}
\begin{figure}[htbp]
\centering
\includegraphics[width=0.49\textwidth]{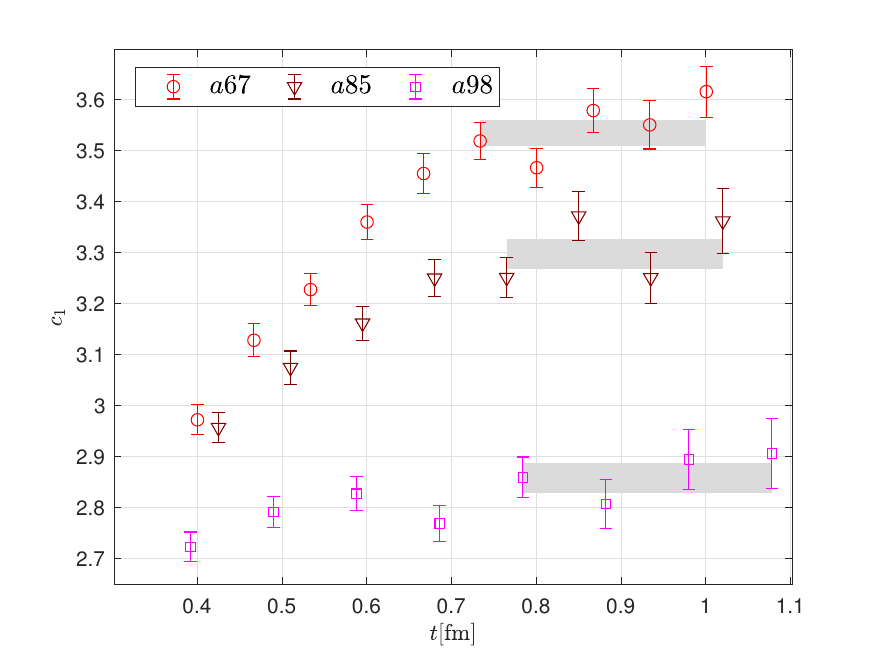}
\includegraphics[width=0.49\textwidth]{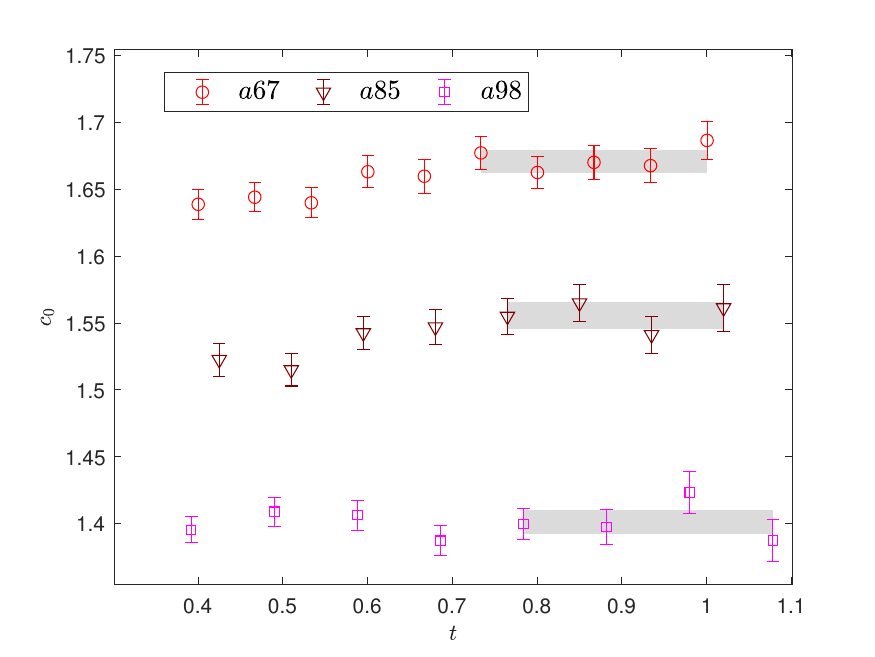}
\caption{Lattice results for $c_1$ and $c_0$ as functions of $t$.}
\label{fig:c01_FVC}
\end{figure}

The lattice results for $c_0$ and $c_1$ as functions of the time separation $t$ are shown in Fig.~\ref{fig:c01_FVC}. We perform constant fits to their large-$t$ plateaus. The resulting values are listed in Table~\ref{tab:c01_FVC}.

For comparison, we also analyze the data using the conventional method, in which $V(q^2)$ is given by
\begin{equation}
V(q^2)=-\frac{(m_{\eta_c}+m_{J/\psi})}{4e_cZm_{J/\psi}}Ee^{Et}
\times\frac{i}{|\vec{p}|^2}\epsilon_{\mu\nu\alpha0}p_{\alpha}\int d^3\vec{x}e^{-i\vec{p}\cdot \vec{x}}H_{\mu\nu}(\vec{x},t),
\end{equation}
where $\vec{p}=\frac{2\pi}{L}\vec{n}$ with $\vec{n}=(0,0,1),(0,1,1),(1,1,1)$ and $(0,0,2)$.

\begin{figure}[htbp]
\centering
\includegraphics[width=\textwidth]{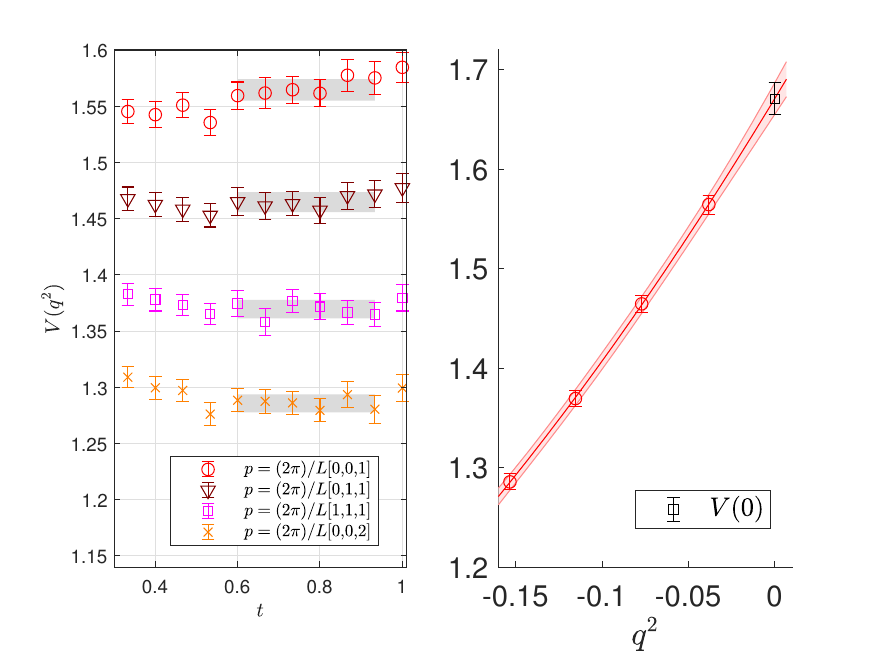}
\caption{Conventional analysis on the a67 ensemble. The left panel shows $V(q^2)$ as a function of $t$ for different momenta, and the right panel shows the continuous extrapolation in $q^2$.}
\label{fig:Vq2_old}
\end{figure}

The $t$ dependence of $V(q^2)$ is shown in the left panel of Fig.~\ref{fig:Vq2_old}, and its $q^2$ dependence is shown in the right panel. We use the a67 ensemble, which has the smallest lattice spacing, as an example. A continuous extrapolation in $q^2$ is performed using the polynomial form $V(q^2)=c_0+c_1q^2/m_{J/\psi}^2+c_2q^4/m_{J/\psi}^4$. The coefficients $c_i$ are then obtained from correlated fits to the lattice data, and the results are summarized in Table~\ref{tab:c01_FVC}. They are consistent with the new-method results, but the uncertainty in $c_0$ is larger by a factor of $1.7$--$1.8$, and the uncertainty in $c_1$ is larger by about an order of magnitude.

\begin{table}[htbp]
\centering
\caption{Numerical results for $c_0$ and $c_1$ on all ensembles. The coefficient $c_2$ from the conventional method is also shown. In the conventional method, a polynomial parametrization of $V(q^2)$ is used. The sign change of $c_2$ on a98 indicates possible model dependence for this ensemble.}
\begin{tabular}{ccc|ccc}
\toprule
 & \multicolumn{2}{c}{New method}& \multicolumn{3}{c}{Conventional method} \\
\cmidrule{2-6}
Ensemble & $c_0$ & $c_1$ & $c_0$ & $c_1$ & $c_2$ \\
\midrule
$a67$ & 1.670(09) &3.53(3) & 1.671(16) & 3.16(33) & 2.8(1.8) \\
$a85$ & 1.556(10) &3.30(3) & 1.547(17) & 2.78(37) & 2.1(1.9) \\
$a98$ & 1.401(09) &2.86(3) & 1.366(16) & 1.44(45) & $-$5.5(3.1) \\
\bottomrule
\end{tabular}
\label{tab:c01_FVC}
\end{table}

\begin{figure}[htbp]
\centering
\begin{subfigure}{0.48\textwidth}
\centering
\includegraphics[width=\textwidth]{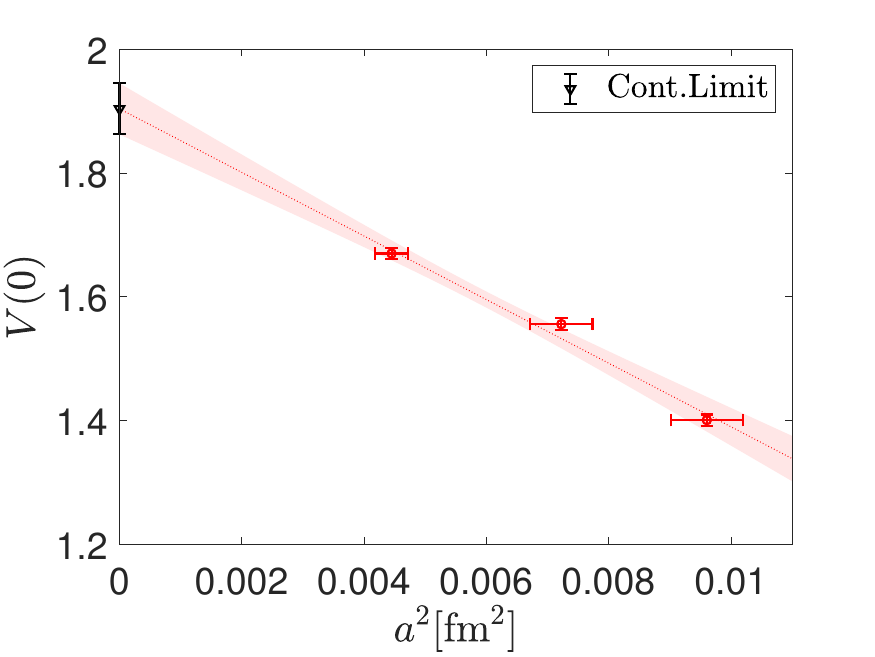}
\end{subfigure}
\caption{Lattice results for $V(0)$ as a function of the lattice spacing. The lattice-spacing uncertainty is included in the fit and shown as horizontal error bars. The red circles from left to right correspond to the a67, a85, and a98 ensembles. The black triangle denotes the continuum-limit result at $a^2\rightarrow 0$.}
\label{fig:cont_limit}
\end{figure}

For twisted-mass ensembles, lattice-spacing artifacts are automatically $\mathcal{O}(a)$ improved. Figure~\ref{fig:cont_limit} shows the linear $a^2$ extrapolation of $V(0)$ using the three lattice spacings. The results indicate that no residual $\mathcal{O}(a)$ effects remain on these ensembles. The final on-shell form factor is
\begin{equation}
V(0)=1.90(4),
\end{equation}
where the uncertainty includes the lattice-spacing uncertainty.

Using the on-shell transition form factor $V(0)$ as input, one obtains the partial width for $J/\psi\rightarrow \gamma\eta_c$,
\begin{equation}
\Gamma(J/\psi\rightarrow \gamma\eta_c)=\alpha e_c^2\frac{16}{3}\frac{|\vec{k}|^3} {(m_{J/\psi}+m_{\eta_c})^2}|V(0)|^2 = 2.30(10)~\textrm{keV},
\end{equation}
where $\alpha=1/137.036$ and $|\vec{k}|=(m_{J/\psi}^2-m_{\eta_c}^2)/(2m_{J/\psi})$. The masses $m_{\eta_c}$ and $m_{J/\psi}$ are the experimental values quoted by the PDG. The corresponding branching fraction is
\begin{equation}
\operatorname{Br}(J/\psi\rightarrow \gamma\eta_c)=2.49(11)_{\textrm{lat}}(5)_{\textrm{exp}}\%,
\end{equation}
where the first uncertainty is the statistical uncertainty from the lattice calculation, and the second comes from the uncertainty in the total $J/\psi$ width, $92.6(1.7)$ keV.

On the same ensembles, Ref.~\cite{Meng:2021ecs} computed the two-photon decay width of the $\eta_c$, obtaining $\Gamma(\eta_c\rightarrow 2\gamma)=6.67(16)(6)$ keV. Combining this with $\Gamma(J/\psi \rightarrow \gamma \eta_c)=2.30(10)$ keV and using the total $\eta_c$ width $\Gamma_{\eta_c}^{\textrm{total}}=30.5(0.5)$ MeV~\cite{PDG24}, we obtain the product branching fraction $\operatorname{Br}(J/\psi\rightarrow \gamma \eta_c)\times \operatorname{Br}(\eta_c\rightarrow 2\gamma)=5.43(42)\times 10^{-6}$. The recent BESIII result, $\operatorname{Br}(J/\psi\rightarrow \gamma \eta_c)\times \operatorname{Br}(\eta_c\rightarrow 2\gamma)=5.23(26)(30)\times 10^{-6}$~\cite{BESIII:2024rex}, agrees very well with the lattice calculation in this chapter.

\subsection{Summary}
This section introduced a model-independent method for studying radiative transitions and applied it to the charmonium transition $J/\psi \rightarrow \gamma\eta_c$. The method allows the on-shell transition form factor to be determined directly from lattice hadronic functions. With the new method, the uncertainty in the on-shell transition form factor is reduced by between $40\%$ and an order of magnitude. In the controlled continuum limit, the resulting on-shell transition form factor for $J/\psi\rightarrow \gamma\eta_c$ is $V(0)=1.90(4)$, in good agreement with both previous lattice studies and the latest experimental results.

The lattice result is consistent with earlier lattice calculations~\cite{Gui:2019dtm,Becirevic:2012dc,Donald:2012ga,Delaney:2023fsc,Colquhoun:2023zbc}. Some systematic effects remain, including the omission of disconnected diagrams, the omission of strange- and charm-quark annihilation diagrams, the unphysical $\pi$ masses, and finite-volume effects. A systematic study by HPQCD found these effects to be small~\cite{Colquhoun:2023zbc}.

\section{Lattice Calculation of the $D_s^*$ Radiative Decay}
The scalar-function method introduced in the previous section can also be applied to other radiative decay processes. In this section, we apply it to the radiative decay of the $D_s^*$.

\subsection{Scalar-Function Method}
We study the $D_s^*$ radiative decay using the scalar-function method described above. The Euclidean hadronic function for this process in infinite volume is defined as
\begin{equation}
H_{\mu\nu}(\vec{x},t)= \langle 0|\mathcal{O}_{D_s}(\vec{x},t)J_{\nu}^{\textrm{em}}(0)|D_{s,\mu}^*(p')\rangle, \quad t>0,
\label{eq:H-def_Ds}
\end{equation}
where $J_\nu^{\textrm{em}}=\sum_qe_q\,\bar{q}\gamma_\nu q$, with $e_q=2/3,-1/3,-1/3,2/3$ for the $u,d,s,c$ quarks, respectively. The state $|D_{s,\mu}^*(p')\rangle$ denotes a $D_s^*$ with momentum $p'=(im_{D^*_s},\vec{0})$, and $\mathcal{O}_{D_s}$ is the $D_s$ operator. At large time separation $t$, the hadronic function is dominated by the single-$D_s$ state. We define the following parametrization and the effective transition form factor $V_{\textrm{eff}}(q^2)$:
\begin{equation}
\label{eq:F_param_Ds}
\begin{aligned}
\langle 0|\mathcal{O}_{D_s}(0)|D_s(\vec{p})\rangle &= Z_{D_s}, \\
\langle D_s(p)|J_{\nu}^{\textrm{em}}(0)|D_{s,\mu}^*(p')\rangle&=\frac{2V_{\textrm{eff}}(q^2)}{m_{D_s}+m_{D_s^*}}\epsilon_{\mu\nu\alpha \beta}p_{\alpha}p'_{\beta},
\end{aligned}
\end{equation}
where the four-momentum transfer is $q^2=(m_{D_s^*}-E_{D_s})^2-|\vec{p}|^2$.

The conventional method for extracting the form factor $V_{\textrm{eff}}(q^2)$ uses the above equations at a set of nonzero lattice momenta $\vec{p}=2 \pi \vec{n}/L,\vec{n}\neq 0$. The on-shell transition form factor $V_{\textrm{eff}}(0)$ is then obtained by extrapolating these discrete $q^2$ data in momentum. Notably, this approach omits the momentum point closest to the on-shell condition $q^2=0$, namely $\vec{p}=0$. When both the $D_s$ and $D_s^*$ are at rest, $\tilde{H}_{\mu\nu}(\vec{p},t) = 0$, so the $\vec{p}=0$ data cannot be included through this quantity. Including this point would improve both the precision and, more importantly, the reliability of the momentum extrapolation. To achieve this, we construct the scalar function
\begin{equation}
\mathcal{I}(t,|\vec{p}|)=\frac{1}{m_{D_s^*}|\vec{p}|^2}\epsilon_{\mu\nu\alpha \beta}p_{\alpha}p'_{\beta}\tilde{H}_{\mu\nu}(\vec{p},t),
\end{equation}
which is related to the transition form factor $V_{\textrm{eff}}(q^2)$ by
\begin{equation}
\mathcal{I}(t,|\vec{p}|)=\frac{-2Z_{D_s}m_{D_s^*}}{m_{D_s}+m_{D_s^*}}V_{\textrm{eff}}(q^2)\frac{e^{-E_{D_s}t}}{E_{D_s}}.
\end{equation}
In the scalar-function method, one may set $|\vec{p}|=0$, since $j_1(x)/x$ approaches a finite value as $x \rightarrow 0$. The on-shell transition form factor $V_{\textrm{eff}}(0)$ can then be determined through a generic polynomial extrapolation:
\begin{equation}
\label{eq:mom_extra_Ds}
V_{\textrm{eff}}(q^2)=d_0+d_1 \cdot \frac{q^2}{m_{D_s^*}^2}+d_2\cdot \frac{q^4}{m_{D_s^*}^4}+\mathcal{O}(q^6/m_{D_s^*}^6).
\end{equation}
Here the coefficients $d_i$ are introduced, with $V_{\textrm{eff}}(0)\equiv d_0$. Since at $\vec{p}=0$ one has $q^2=(m_{D_s^*}-m_{D_s})^2\equiv (\delta m)^2$, and since $(\delta m)^2/m_{D_s^*}^2 \sim 0.46\%$, the difference between $V_{\textrm{eff}}(0)$ and $V_{\textrm{eff}}((\delta m)^2)$ is tiny. One therefore expects the extrapolation including the $\vec{p}=0$ point to be substantially more precise.

\subsection{Decay Width}
The amplitude for $D_s^{+,*}\rightarrow \gamma D_s^+$ can be written as
\begin{equation}
\label{eq:amplitude_Ds}
i\mathcal{M}(\lambda',\lambda)=ie\epsilon^{\mu}_{D_s^{*}}(p',\lambda')\epsilon^{\nu}_{\gamma}(q,\lambda)\langle D_s(p)|J_{\nu}^{\textrm{em}}(0)|D_{s,\mu}^*(p')\rangle,
\end{equation}
where $\epsilon^{\mu}_{D_s^{*}}(p',\lambda')$ is the polarization vector of the $D_s^*$, $\epsilon^{\nu}_{\gamma}(q,\lambda)$ is the photon polarization, and $q=p'-p$ is the photon four-momentum. These polarizations satisfy
\begin{equation}
\begin{aligned}
\sum\limits_{\lambda}\epsilon^{\mu}_{D_s^*}(p',\lambda)\epsilon^{\nu}_{D_s^*}(p',\lambda)&=-g_{\mu\nu}+\frac{p'_{\mu}p'_{\nu}}{m_{D_s^*}^2}, \\
\sum\limits_{\lambda}\epsilon^{\mu}_{\gamma}(p,\lambda)\epsilon^{\nu}_{\gamma}(p,\lambda)&=-g_{\mu\nu}.
\end{aligned}
\end{equation}
Combining this with the parametrization in Eq.~\ref{eq:F_param_Ds}, one obtains the decay width
\begin{equation}
\begin{aligned}
\Gamma(D_s^{*}\rightarrow \gamma D_s)
=&\frac{1}{2m_{D_s^*}}\int \frac{d^3\vec{q}}{(2\pi)^32|\vec{q}|}\int \frac{d^3\vec{p}}{(2\pi)^32E_{D_s}} \\
&\times (2\pi)^4\delta^{(4)}(p'-p-q) \times \frac{1}{3}\sum\limits_{\lambda'}\sum\limits_{\lambda}|\mathcal{M}(\lambda',\lambda)|^2 \\
=&\frac{4}{3}\frac{\alpha (\delta m)^3}{(m_{D_s}+m_{D_s^*})^2}|V_{\textrm{eff}}(0)|^2,
\end{aligned}
\end{equation}
where $\alpha\equiv e^2/4\pi$. The factor $1/3$ in the third line denotes the average over the three polarizations of the $D_s^*$ in its rest frame, while the photon polarizations have been summed.

Compared with the radiative-decay amplitude $D_s^*\rightarrow D_s\gamma$ in Eq.~\ref{eq:amplitude_Ds}, the amplitude for the Dalitz decay $D_s^*\rightarrow D_se^+e^-$ contains the additional photon propagator $-ig_{\nu\nu'}/q^2$ and the lepton current $\bar{u}_e\gamma_{\nu'}u_e$. A direct calculation analogous to the one above gives the Dalitz decay width normalized to the corresponding radiative decay width~\cite{LANDSBERG1985301}:
\begin{equation}
\label{eq:R_ee}
\begin{aligned}
R_{ee}&=\frac{\Gamma(D_s^*\rightarrow D_se^+e^-)}{\Gamma(D_s^*\rightarrow D_s\gamma)} \\
&=\frac{\alpha}{3\pi}\int \frac{dq^2}{q^2}\Big{|}\frac{V_{\textrm{eff}}(q^2)}{V_{\textrm{eff}}(0)}\Big{|}^2\left(1-\frac{4m_e^2}{q^2}\right)^{\frac{1}{2}}\left(1+\frac{2m_e^2}{q^2}\right) \\
&\times \left[\left(1+\frac{q^2}{m_{D_s^*}^2-m_{D_s}^2}\right)^2-\frac{4m_{D_s^*}^2q^2}{(m_{D_s^*}^2-m_{D_s}^2)^2}\right]^{\frac{3}{2}}.
\end{aligned}
\end{equation}

\subsection{Lattice Numerical Results for the Radiative Decay}
We use the C24P29, C32P30, and C48P32 ensembles from CLQCD~\cite{CLQCD:2023sdb}. The $D_s$ and $D_s^*$ mesons are found to satisfy the lattice dispersion relation well.

\begin{figure}[htbp]
\centering
\includegraphics[width=0.6\textwidth]{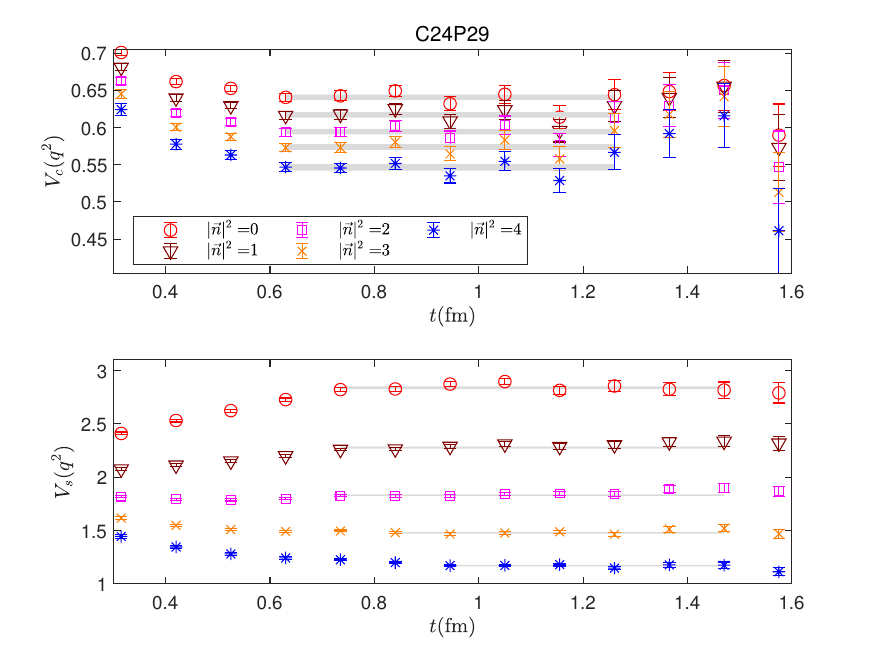}
\includegraphics[width=0.6\textwidth]{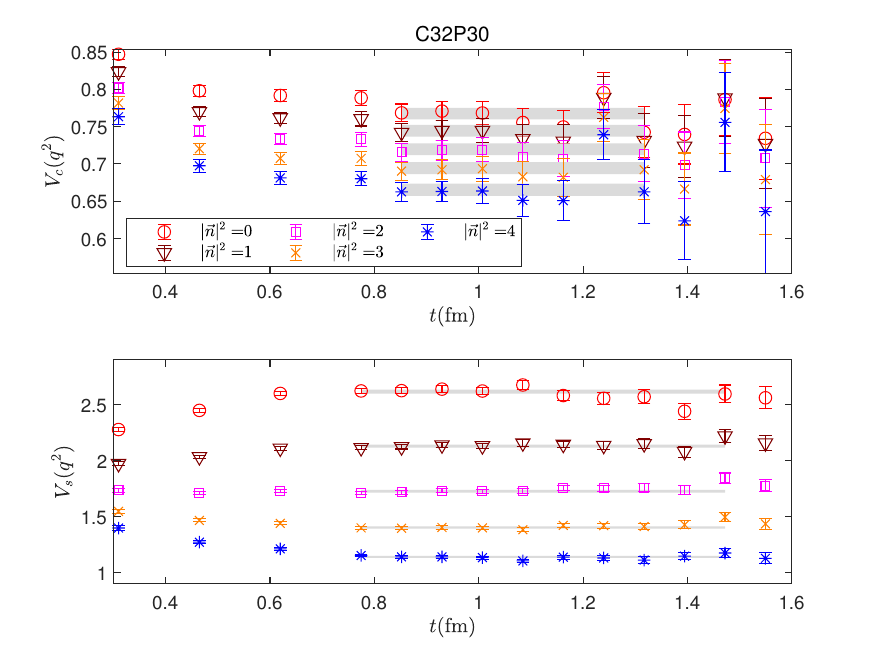}
\includegraphics[width=0.6\textwidth]{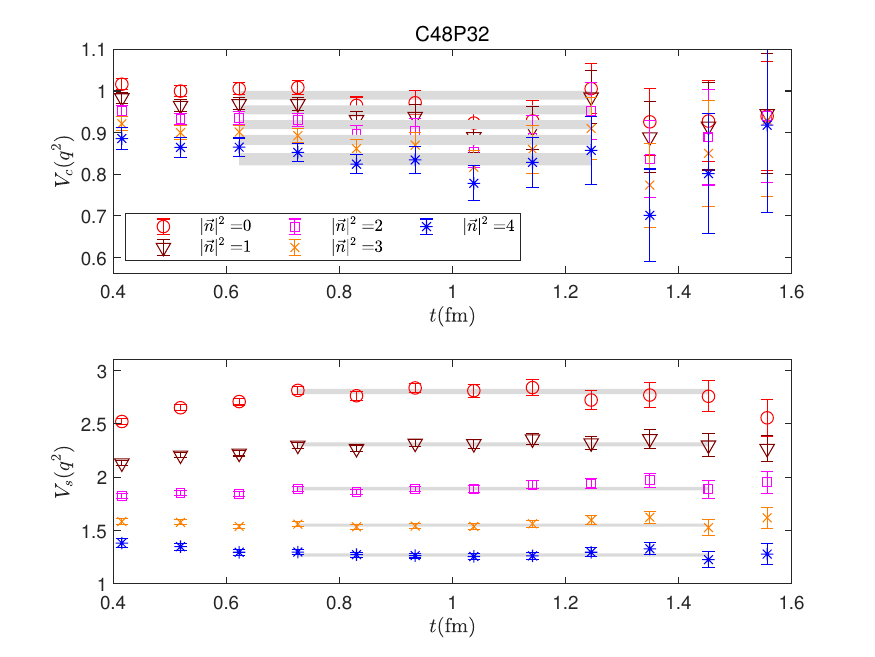}
\caption{\label{fig:Vcs_Ds} Transition form factors $V_{c/s}(q^2)$ as functions of the time separation $t$ for momenta $\vec{p}=2\pi\vec{n}/L,|\vec{n}|^2=0,1,2,3,4$. The grey bands show the fit results.}
\end{figure}

The effective transition form factor can be written as $V_{\textrm{eff}}(q^2)=\frac{1}{3}V_s(q^2)-\frac{2}{3}V_c(q^2)$, where $V_c(q^2)$ comes from photon radiation by the charm quark and $V_s(q^2)$ from photon radiation by the strange quark. Their $t$ dependence at the momenta $\vec{p}=2\pi\vec{n}/L,|\vec{n}|^2=0,1,2,3,4$ is shown in Fig.~\ref{fig:Vcs_Ds}. Significant excited-state effects are visible at small time separations, while clear plateaus appear at large time separations. Excited-state effects are therefore well controlled in the calculation. The grey bands indicate constant fits.

\begin{figure}[htbp]
\centering
\includegraphics[width=0.48\textwidth]{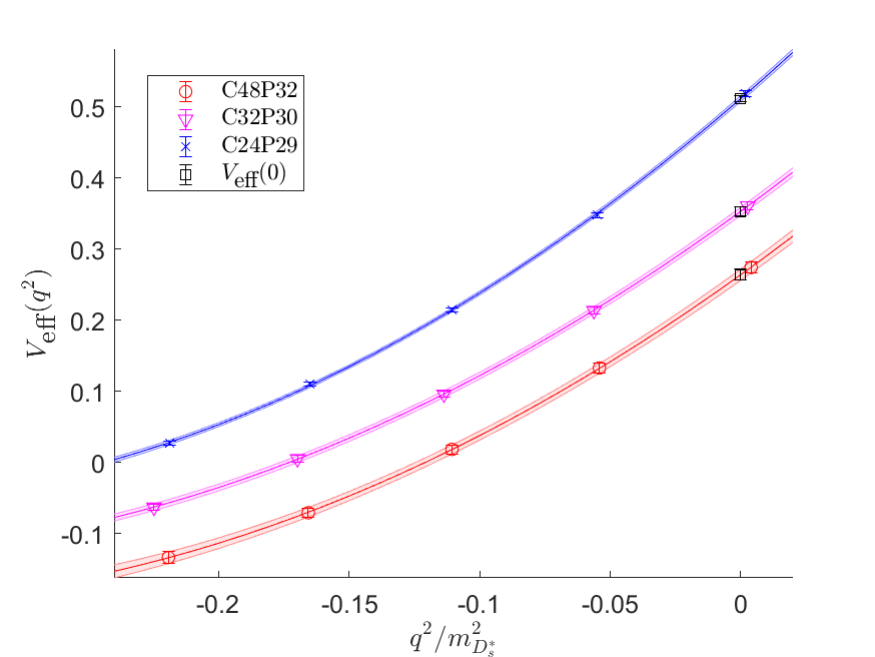}
\caption{\label{fig:Vcs_cont_Ds} Extrapolation of the transition form factor $V_{\textrm{eff}}(q^2)$. The black square denotes the on-shell transition form factor $V_{\textrm{eff}}(0)$.}
\end{figure}

The on-shell form factor $V_{\textrm{eff}}(0)$ is extracted by extrapolating in momentum to $q^2\rightarrow 0$. The polynomial form in Eq.~\ref{eq:mom_extra_Ds} describes the lattice data well, as shown in Fig.~\ref{fig:Vcs_cont_Ds}. Since the initial and final particles, $D_s^{*}$ and $D_s$, have similar masses, the on-shell transition form factor is very close to the zero-momentum transition form factor. Consequently, the statistical uncertainty of the on-shell transition form factor is almost entirely dominated by the uncertainty of the zero-momentum off-shell form factor.

\begin{figure}[htbp]
\centering
\includegraphics[width=0.49\textwidth]{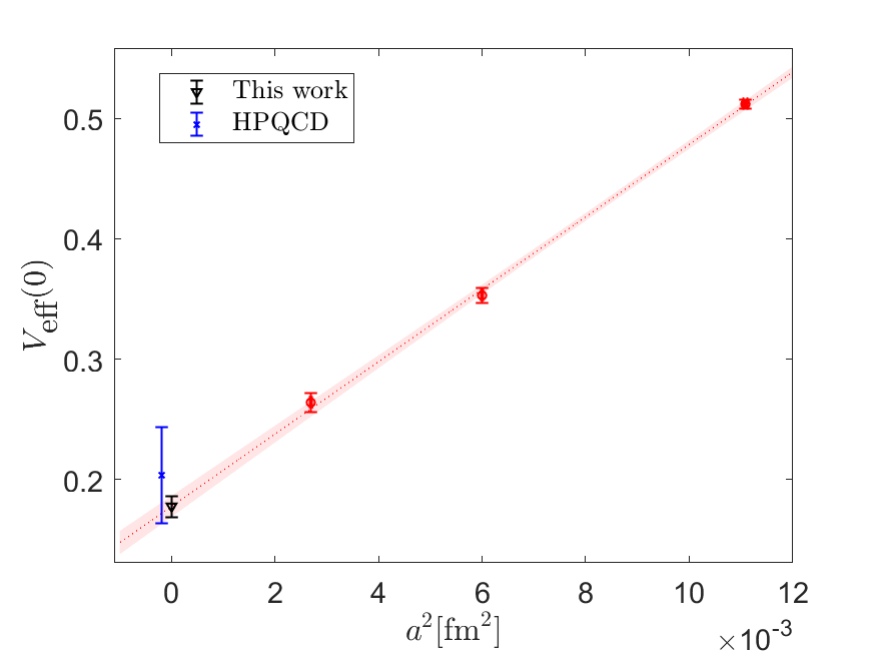}
\includegraphics[width=0.49\textwidth]{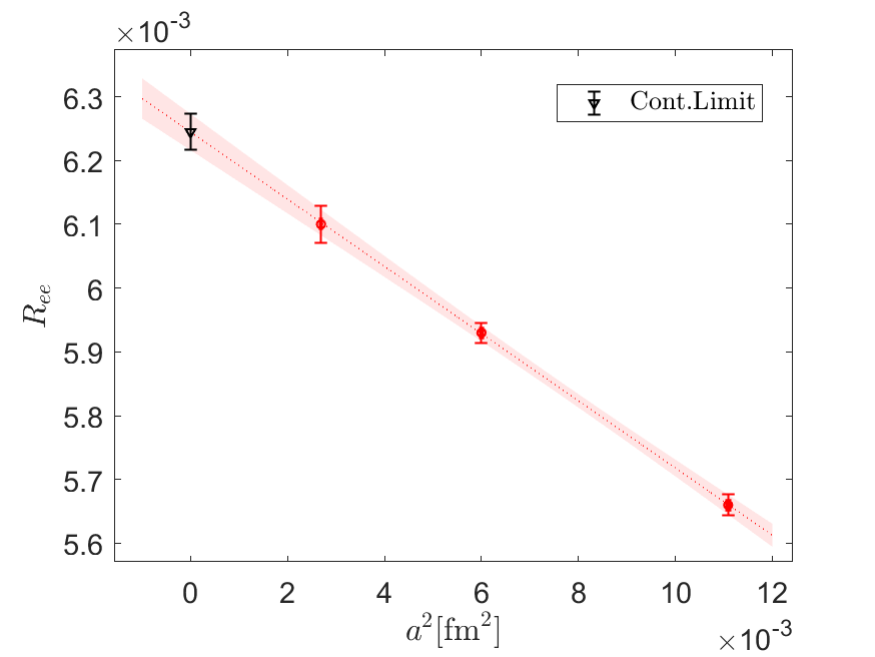}
\caption{\label{fig:decay_width} Continuum extrapolations of $V_{\textrm{eff}}(0)$ and $R_{ee}$. The black triangles denote the continuum-limit results at $a^2\rightarrow 0$.}
\end{figure}

The linear continuum extrapolation of the effective transition form factor in $a^2$ is shown in the left panel of Fig.~\ref{fig:decay_width}. The continuum-limit result is
\begin{equation}
    V_{\textrm{eff}}(0)=0.178(9),
\end{equation}
whose statistical uncertainty is more than four times smaller than that of the previous result~\cite{Donald:2013sra}. Taking the physical transition form factor $V_{\textrm{eff}}(0)=0.178(9)$ as input and using the physical masses of the $D_s^*$ and $D_s$, the radiative decay width for $D_s^* \rightarrow \gamma D_s$ is
\begin{equation}
    \Gamma(D_s^*\rightarrow \gamma D_s) = 0.0549(54) ~\textrm{keV}.
\label{eq:decay_width_value}
\end{equation}

Combining this with the branching fraction $\operatorname{Br}(D_s^*\rightarrow D_s\gamma)=93.5(7)\%$, we obtain the total decay width
\begin{equation}
\Gamma^{\textrm{total}}_{D_s^*}=0.0587(54)\ \textrm{keV}.
\end{equation}
Recently, BESIII reported the first experimental study of the purely leptonic decay $D_s^{+,*}\rightarrow e^+\nu_{e}$ and gave the branching fraction $(2.1^{+1.2}_{-0.9_{\textrm{stat.}}}\pm 0.2_{\textrm{syst.}})\times 10^{-5}$~\cite{BESIII:2023zjq}. Combining this branching fraction with the total decay width obtained from our lattice calculation gives
\begin{equation}
f_{D_s^*}|V_{cs}|=(190.5^{+55.1}_{-41.7_{\textrm{stat.}}}\pm 12.6_{\textrm{syst.}})\ \textrm{MeV},
\end{equation}
where stat. denotes only the experimental statistical uncertainty, while syst. includes both the experimental systematic uncertainty and the lattice statistical uncertainty.

In the Dalitz decay of the $D_s^*$, the virtual photon internally converts into a lepton pair $l^+l^-$. Since the $\mu$-lepton mass is larger than the mass splitting between the $D_s^*$ and the $D_s$, the only allowed channel is $e^+e^-$. Using the transition form factor $V_{\textrm{eff}}(q^2)$ already obtained in the calculation of $D_s^*\rightarrow D_s\gamma$, we compute the ratio $R_{ee}$ defined in Eq.~(\ref{eq:R_ee}). The result is shown in the right panel of Fig.~\ref{fig:decay_width}, and the final value is
\begin{equation}
    R_{ee}=\frac{\Gamma(D_s^*\to D_se^+e^-)}{\Gamma(D_s^*\to D_s\gamma)}=0.624(3)\%.
\end{equation}
This result is consistent with the PDG value $0.67(16)\%$. Because the uncertainties in $V_{\textrm{eff}}(q^2)$ and $V_{\textrm{eff}}(0)$ almost completely cancel in the ratio, the final precision reaches the per-mille level.

The PDG lists three main decay modes of the $D_s^*$: $D_s^*\rightarrow D_s\gamma$, $D_s^*\rightarrow D_s \pi^0$, and $D_s^*\rightarrow D_se^+e^-$. The branching fractions of these channels are in fact determined by only two relative ratios,
\begin{equation}
R_{ee}=\Gamma(D_s^*\rightarrow D_se^+e^-)/\Gamma(D_s^*\rightarrow D_s\gamma),
\end{equation}
and
\begin{equation}
R_{D_s\pi^0}=\Gamma(D_s^*\rightarrow D_s\pi^0)/\Gamma(D_s^*\rightarrow D_s\gamma).
\end{equation}
Assuming that the three branching fractions sum to $1$, one obtains
\begin{equation}
\operatorname{Br}(D_s^*\rightarrow D_s\gamma)=\frac{1}{1+R_{ee}+R_{D_s\pi^0}},
\end{equation}
\begin{equation}
\operatorname{Br}(D_s^*\rightarrow D_s \pi^0)=\frac{R_{D_s\pi^0}}{1+R_{ee}+R_{D_s\pi^0}},
\end{equation}
\begin{equation}
\operatorname{Br}(D_s^*\rightarrow D_se^+e^-)=\frac{R_{ee}}{1+R_{ee}+R_{D_s\pi^0}}.
\end{equation}
The quantity $R_{ee}$, which can be determined with very high precision, provides an ideal platform for testing the Standard Model.

\subsection{Summary}
This section presented the lattice calculation of the $D_s^*$ radiative decay, including both the transition $D_s^*\rightarrow D_s\gamma$ and the Dalitz decay $D_s^*\rightarrow D_s e^+e^-$. Using $(2+1)$-flavor Wilson-clover gauge ensembles at three different lattice spacings, we obtain
\begin{equation}
\Gamma(D_s^*\rightarrow D_s \gamma)=0.0549(54)\ \mathrm{keV},
\end{equation}
whose statistical uncertainty is substantially reduced compared with previous lattice calculations. We also study the Dalitz decay of the $D_s^*$ for the first time and obtain the ratio
\begin{equation}
R_{ee}=0.624(3)\%,
\end{equation}
with sub-percent precision. Several improvements are responsible for this precision. Most importantly, the scalar-function method is used to compute the effective transition form factor for $D_s^*\rightarrow D_s\gamma$, allowing the matrix element to be evaluated directly when both the initial and final states carry zero momentum. Because the mass splitting between the $D_s^*$ and the $D_s$ is small, the on-shell transition form factor is very close to the zero-momentum transition form factor. After momentum extrapolation, the precision of the on-shell transition form factor is therefore almost entirely controlled by that of the zero-momentum off-shell form factor.

A precise determination of the total $D_s^*$ decay width is important for extracting the CKM matrix element $V_{cs}$. Combining the result with the recent experimental measurement of $D_s^{*,+}\rightarrow e^+\nu_e$ gives
\begin{equation}
f_{D_s^*}|V_{cs}|=(190.5^{+55.1}_{-41.7_{\textrm{stat.}}}\pm 12.6_{\textrm{syst.}})\ \textrm{MeV}.
\end{equation}
Compared with the systematic uncertainty of $42.7$ obtained when using the earlier lattice result $\Gamma_{D_s^*}^{\textrm{total}}=0.070(28)$ keV as input, the systematic uncertainty here is substantially reduced. Future measurements of purely leptonic $D_s^*$ decays at next-generation facilities, such as the Super Tau Charm Facility~\cite{Achasov:2023gey}, are expected to further improve both statistical and systematic uncertainties.

\section{Lattice Study of $J/\psi$ Semileptonic Decays}
The previous two sections considered radiative decays governed by the electromagnetic interaction. We now turn to weak processes and study semileptonic decays of the $J/\psi$.

\subsection{Decay Width}
At leading order in the Standard Model, the transition amplitude for $J/\psi\rightarrow D/D_s l\nu_l$ can be written as
\begin{equation}
i\mathcal{M}=-i\frac{G_F}{\sqrt{2}}V_{cs(d)}H_{\mu}(p,p')g_{\mu\nu}\bar{u}_{l}\gamma_{\nu}(1-\gamma_5)u_{\nu_l},
\label{eq:amp}
\end{equation}
where $l=e,\mu$. The four-momenta of the $J/\psi$ and the $D/D_s$ meson are denoted by $p'$ and $p$, respectively. The quantity $H_{\mu}(p,p')$ encodes the nonperturbative hadronic interaction between the initial and final states,
\begin{equation}
    H_{\mu}(p,p')\equiv \langle D/D_s(p)| J_{\mu}^{W}|J/\psi(\epsilon,p')\rangle,
\end{equation}
and can be decomposed in terms of four invariant form factors as~\cite{Korner:1989qb}
\begin{equation}
    \begin{aligned}
    H_{\mu}(p,p')&=\epsilon_{\alpha}(p')H_{\mu\alpha}(p,p'), \\
    H_{\mu\alpha}(p,p')&=F_1(q^2)g_{\mu\alpha}+\frac{F_2(q^2)}{M m}p'_{\mu}p_{\alpha}+\frac{F_3(q^2)}{m^2}p_{\mu}p_{\alpha} \\
    &- \frac{iF_{0}(q^2)}{M m}\epsilon_{\mu\alpha\rho\sigma}p'_{\rho}p_{\sigma},
    \end{aligned}
\label{eq:FF}
\end{equation}
where $M$ is the $J/\psi$ mass, $m$ is the $D/D_s$ mass, and $\epsilon_{\alpha}(p')$ denotes the polarization vector of the $J/\psi$.

In the $J/\psi$ rest frame, Eqs.~(\ref{eq:amp}) and~(\ref{eq:FF}) give the decay width for $J/\psi \rightarrow D/D_s l \nu_l$:
\begin{equation}
    \begin{aligned}
        \Gamma&=\frac{1}{2 M}\int\frac{d^3\vec{p}}{(2\pi)^3 2E}\int\frac{d^3\vec{q_l}}{(2\pi)^32E_l}\int\frac{d^3\vec{q}_{\nu}}{(2\pi)^32E_{\nu_l}}  \\
        &\times(2\pi)^4\delta^4(p'-p-q_l-q_{\nu})\frac{1}{3}|\mathcal{M}|^2   \\
        &=\frac{G_F^2V_{cs(d)}^2}{12M^2}\frac{1}{32\pi^3}\int_{m_l^2}^{(M-m)^2} dq^2 \times \big{[} c_0(E_l^+-E_l^-)  \\
        &+\frac{c_1}{2}((E_l^+)^2-(E_l^-)^2)+\frac{c_2}{3}((E_l^+)^3-(E_l^-)^3)\big{]},
\end{aligned}
\label{eq:width}
\end{equation}
where $(q^2,E_l)$ are the two independent Dalitz variables. Here $q^2$ is the momentum transfer squared, $E$ is the energy of the $D/D_s$ meson, $m_l$ is the lepton mass, $q_l$ and $E_l$ are the four-momentum and energy of the charged lepton, and $q_{\nu_l}$ and $E_{\nu_l}$ are the four-momentum and energy of the neutrino. The variables $E$ and $q^2$ are related by
\begin{equation}
E=\frac{M^2+m^2-q^2}{2M}.
\end{equation}
The coefficients $c_i\,(i=0,1,2)$ are functions of $M,m,m_l,E$ and of the form factors $F_i\,(i=0,1,2,3)$, and therefore depend on $q^2$. The explicit expressions for these $c_i$ are
\begin{equation}
    \begin{aligned}
        c_0&= \frac{1}{m^4}\Big{[}-8E^3F_0^2m^2M+4E^3F_3m_l^2(2F_2m+F_3M)+4E^3M(F_2m+F_3M)^2 \\
        &+ 4E^2F_0^2m^2(m^2+3M^2+m_l^2) +2E^2m_l^2(m^2(F_2^2+F_3^2+4F_1F_3) - 2F_3F_2mM \\
        &-2F_3^2M^2)-2E^2(F_2m+F_3M)(-4F_1m^2M+(F_2m+F_3M)(m^2+M^2)) \\
        &-2E^2F_3^2m_l^4 - 4Em^2m_l^2(2F_0^2M +F_2m(F_1+2F_3)+F_3M(3F_1+F_3)) \\
        &-8EF_0^2m^2M^3-4EF_1^2m^4M -4EF_1F_2m^5-12EF_1F_2m^3M^2-4EF_1F_3m^4M \\
        &-12EF_1F_3m^2M^3-4EF_2^2m^4M-8EF_2F_3m^3M^2 -4EF_3^2m^2M^3 \\
        &+2F_0^2m^2((M^2+m_l^2)^2-m^4) + 2m^2(m^2+M^2)(F_1^2m^2+(2F_1M+F_2m+ \\
        &F_3M)(F_2m+F_3M)) -2m^2m_l^2(m^2((F_1+F_3)^2+F_2^2)-2F_2mM(F_1+F_3) \\
        &-2F_3M^2(2F_1+F_3)) +2F_3m^2m_l^4(2F_1+F_3)\Big{]} \\
        &+\frac{8}{m}F_0F_1\left((E-M)(2ME-M^2-m^2-m_l^2)\right),  \\
        c_1&=\frac{1}{m^4}\Big{[} -8E^3(F_2m+F_3M)^2 -(F_2m+F_3M)(F_2mM+F_3(M^2+m_l^2))\big{)} \\ 
        &+8Em^2\big{(}F_0^2(m^2+3M^2+m_l^2)-F_1^2m^2+4F_1M(F_2m+F_3M)+F_1F_3m_l^2 \\
        &+(F_2m+F_3M)^2 \big{)} +8m^2M \big{(}-F_0^2(m^2+M^2)+F_1^2m^2-2F_1M(F_2m+F_3M) \\
        &-(F_2m+F_3M)^2\big{)} - 8m^2m_l^2(M(F_0^2+F_3(2F_1+F_3))+F_2m(F_1+F_3)) \Big{]} \\
        &+ \frac{16}{m}F_0F_1(2ME-M^2-m^2), \\
        c_2&=\frac{1}{m^4}\Big{[}-8E^2(F_2m+F_3M)^2-16Em^2(F_0^2M+F_1F_2m+F_1F_3M)  \\
        &+8m^2\big{(}F_0^2(m^2+M^2)-F_1^2m^2+2F_1M(F_2m+F_3M)+(F_2m+F_3M)^2 \big{)}  \Big{]}.  \\
    \end{aligned}
\label{eq:c_coef}
\end{equation}

The phase-space boundaries in the $(q^2,E_l)$ plane satisfy~\cite{Korner:1989qb}
\begin{equation}
    E_l^{\pm}=\frac{1}{2M}\Big{[} q^2+m_l^2-\frac{1}{2q^2}\big{(}(q^2-M^2+m^2)(q^2+m_l^2) \mp 2M|\vec{p}|(q^2-m_l^2)\big{)}\Big{]},
\end{equation}
where $\vec{p}$ is the three-momentum of the $D/D_s$ meson.

\subsection{Method for Extracting the Form Factors}
This section describes the extraction of the form factors $F_i(i=0,1,2,3)$ using the scalar-function method. The Euclidean hadronic function in infinite volume is defined as
\begin{equation}
    H_{\mu\nu}(\vec{x},t)=\langle 0| \phi_h(\vec{x},t)J_{\mu}^{W}(0)| J/\psi_{\nu}(\epsilon,p') \rangle (t>0),
\end{equation}
where $\phi_h$ is the interpolating operator for the $D/D_s$ meson, with $h=D$ or $h=D_s$. For $h=D_s$, the weak current is
$J_\mu^{W}=\bar{s}\gamma_{\mu}(1-\gamma_5)c$,
whereas for $h=D$ it is
$J_\mu^{W}=\bar{l}\gamma_{\mu}(1-\gamma_5)c$,
where $l,s,c$ denote the light, strange, and charm quarks, respectively. The state $|J/\psi_{\nu}(\epsilon,p')\rangle$ denotes a $J/\psi$ with polarization direction $\nu$ and momentum $p'=(iM,\vec{0})$.

As in the previous two sections, this hadronic function is dominated at large $t$ by the single-particle $D$ or $D_s$ state:
\begin{equation}
    \begin{aligned}
        &H_{\mu\nu}(x)\doteq H_{\mu\nu}^{h}(x)
=\int \frac{d^3\vec{p}}{(2\pi)^3}\frac{1}{2E_{h}}e^{-E_{h}t+i\vec{p}\cdot \vec{x}}  \\
&\times \langle 0|\phi_h(0)|\phi_h(\vec{p})\rangle
\langle \phi_h(\vec{p})|J_{\mu}^W(0)|J/\psi_{\nu}(p')\rangle,
\end{aligned}
\end{equation}
where
\begin{equation}
    \begin{aligned}
\langle 0|\phi_h(0)|\phi_h(\vec{p})\rangle &=Z_{h}, \\
\langle\phi_h(\vec{p})|J_{\mu}^V(0)|J/\psi_{\nu}(\epsilon,p')\rangle
&=\frac{F_0(q^2)}{Mm}\epsilon_{\mu\nu\rho\sigma}p'_{\rho}p_{\sigma}, \\
\langle\phi_h(\vec{p})|J_{\mu}^A(0)|J/\psi_{\nu}(\epsilon,p')\rangle
&=-F_1(q^2)\delta_{\mu\nu}
-\frac{F_2(q^2)}{M m}p'_{\mu}p_{\nu} \\
&-\frac{F_3(q^2)}{m^2}p_{\mu}p_{\nu}.
\label{eq:F_param}
\end{aligned}
\end{equation}
The weak current is decomposed as
$J_{\mu}^{W}=J_{\mu}^{V}-J_{\mu}^{A}$. The four-momentum of the final state $\phi_h$ is denoted
$p=(iE_h,\vec{p})$. The momentum transfer squared is
\begin{equation}
q^2=(M-E_h)^2-|\vec{p}|^2.
\end{equation}

Taking the spatial Fourier transform of
$H_{\mu\nu}(\vec{x},t)\equiv V_{\mu\nu}(\vec{x},t)-A_{\mu\nu}(\vec{x},t)$
gives
\begin{equation}
\begin{aligned}
\tilde{V}_{\mu\nu}(\vec{p},t)&\doteq \tilde{V}_{\mu\nu}^{h}(\vec{p},t)
=\frac{F_0(q^2)}{Mm} \frac{Z_{h}}{2E_{h}}
e^{-E_{h}t}\epsilon_{\mu\nu\rho\sigma}p'_{\rho}p_{\sigma} \\
\tilde{A}_{\mu\nu}(\vec{p},t)&\doteq \tilde{A}_{\mu\nu}^{h}(\vec{p},t)
=\frac{Z_{h}}{2E_{h}}e^{-E_{h}t}
\big{(}-F_1(q^2)\delta_{\mu\nu} \\
&-\frac{F_2(q^2)}{M m}p'_{\mu}p_{\nu}
-\frac{F_3(q^2)}{m^2}p_{\mu}p_{\nu}\big{)}.
\end{aligned}
\end{equation}

The form factor $F_0$ can be obtained from the scalar function
\begin{equation}
\mathcal{I}_0\equiv 
\frac{1}{M|\vec{p}|^2}
\epsilon_{\mu\nu\rho \sigma}p'_{\rho}p_{\sigma}
\tilde{V}_{\mu\nu}(\vec{p},t),
\end{equation}
which yields
\begin{equation}
F_0(q^2)=\frac{m E_h }{Z_h} e^{E_ht}
\int d^3\vec{x}\frac{j_1(|\vec{p}||\vec{x}|)}{|\vec{p}||\vec{x}|}
\epsilon_{\mu\nu \alpha 0}x_{\alpha}
V_{\mu\nu}(\vec{x},t).
\label{eq:V_eff}
\end{equation}

We further construct the scalar functions
\begin{equation}
    \begin{aligned}
        \mathcal{I}_1&\equiv \delta_{\mu\nu}\tilde{A}_{\mu\nu}(\vec{p},t), \\
\mathcal{I}_2&\equiv \frac{E}{M}
\frac{ p'_{\mu}p_{\nu}}{|\vec{p}|^2}
\tilde{A}_{\mu\nu}(\vec{p},t),\\
\mathcal{I}_3&\equiv 
\frac{p_{\mu}p_{\nu}}{|\vec{p}|^2}
\tilde{A}_{\mu\nu}(\vec{p},t),
    \end{aligned}
\end{equation}
and obtain
\begin{equation}
    \begin{aligned}
        F_1(q^2)&=\frac{2E_h e^{E_ht}}{3m^2Z_h}
\left[E_h^2I_{2}-E_h|\vec{p}|^2(I_{3}+I_{4})-m^2I_1-|\vec{p}|^2I_5 \right] \\
F_2(q^2)&=\frac{2E_h e^{E_ht}}{mZ_h}
\left[E_hI_{2}-E_h^2I_{4}-E_hI_5-|\vec{p}|^2I_{3}\right] \\
F_3(q^2)&=\frac{2E_h e^{E_ht}}{3m^2Z_h}
\big{[}E_h^2I_{2}+3m_{h}^2(E_hI_{4}+I_5)-m^2I_1 \\
&-|\vec{p}|^2(E_hI_{3}+E_hI_{4}+I_5)\big{]},
    \end{aligned}
\end{equation}
where $I_i(i=1,2,3,4,5)$ are scalar functions constructed directly from the correlation function $A_{\mu\nu}(\vec{x},t)$. Their explicit forms are
\begin{equation}
    \begin{aligned}
I_1&=\int d^3\vec{x}j_0(|\vec{p}||\vec{x}|)\delta_{\mu\nu}A_{\mu\nu}(\vec{x},t),  \\
I_{2}&=\int d^3\vec{x}j_0(|\vec{p}||\vec{x}|)A_{00}(\vec{x},t),  \\
I_{3}&=\int d^3\vec{x}\frac{j_1(|\vec{p}||\vec{x}|)}{|\vec{p}||\vec{x}|}x_{i}A_{0i}(\vec{x},t),  \\
I_{4}&=\int d^3\vec{x}\frac{j_1(|\vec{p}||\vec{x}|)}{|\vec{p}||\vec{x}|}x_{i}A_{i0}(\vec{x},t),  \\
I_{5}&=\int d^3\vec{x}\Big{\{}\frac{j_1(|\vec{p}||\vec{x}|)}{|\vec{p}||\vec{x}|}\delta_{ij}-|\vec{p}|^2\frac{j_2(|\vec{p}||\vec{x}|)}{(|\vec{p}||\vec{x}|)^2}x_ix_j}{ \Big{\}}A_{ij}(\vec{x},t). \\
    \end{aligned}
    \label{eq:I_scalar}
\end{equation}

\begin{figure}[htbp]
\centering
\includegraphics[width=0.49\textwidth]{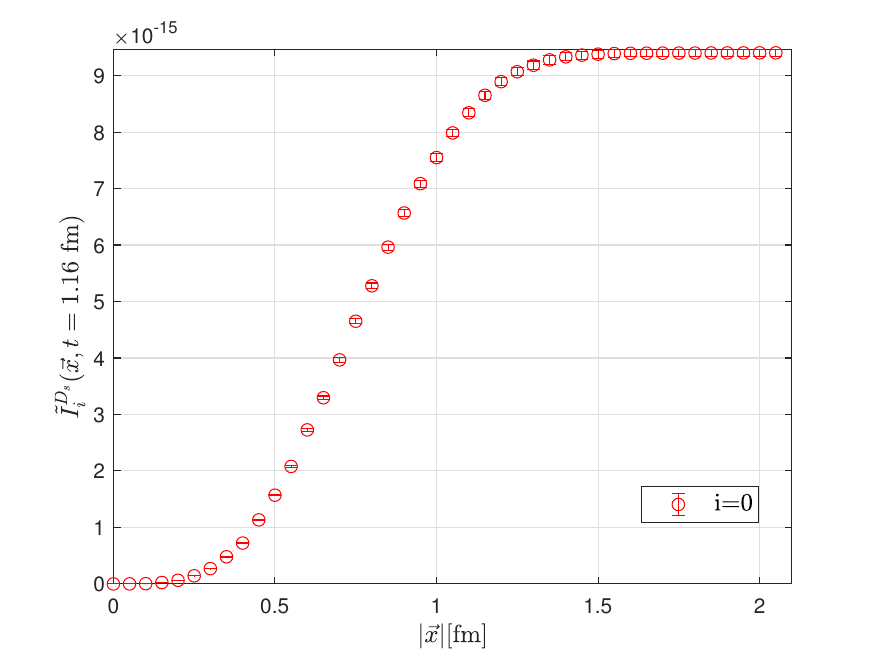}
\includegraphics[width=0.49\textwidth]{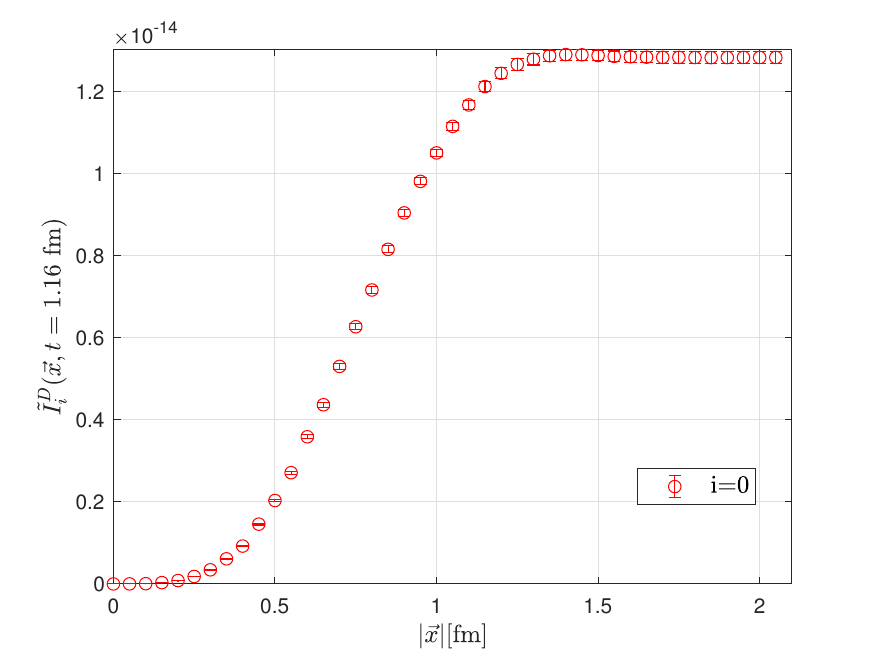}
\includegraphics[width=0.49\textwidth]{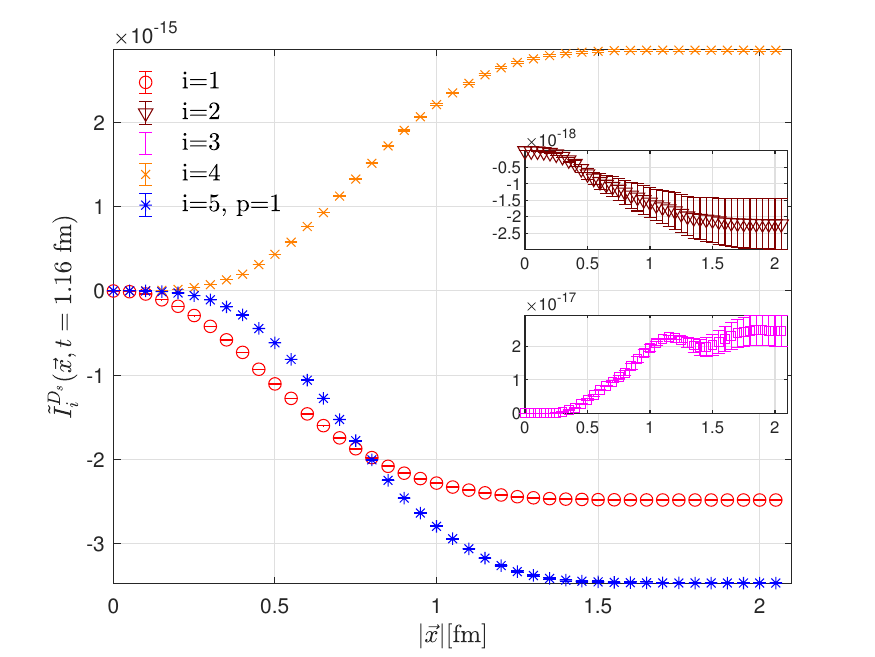}
\includegraphics[width=0.49\textwidth]{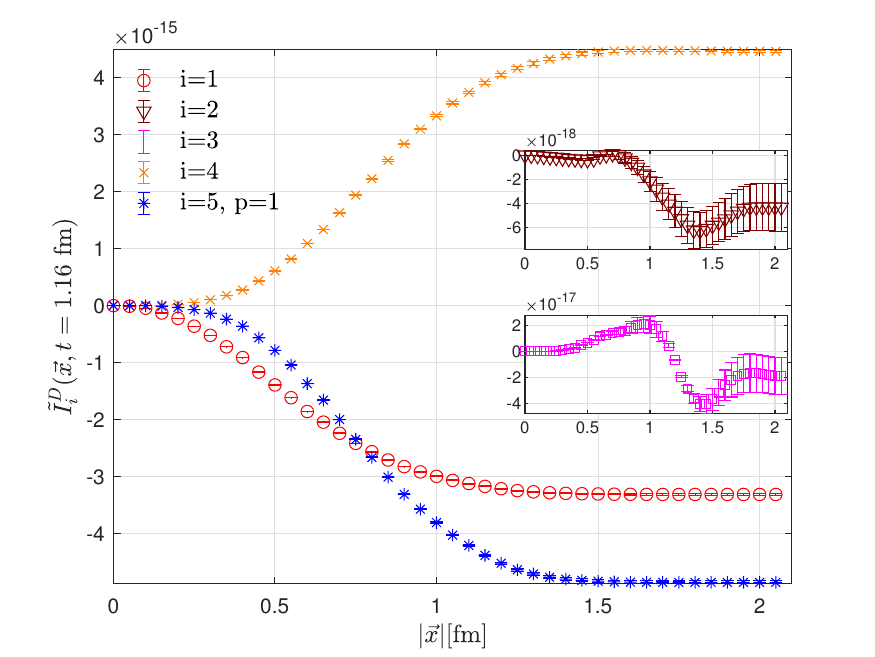}
\caption{\label{diag:I_V} Dependence of $\tilde{I}_i^{D_s}(\vec{x},t)$ (left) and $\tilde{I}_i^{D}(\vec{x},t)$ (right) on the spatial cutoff $|\vec{x}|$ on the F32P30 ensemble, with $t=1.16~\mathrm{fm}$ and $i=0,1,2,3,4,5$. Here $p=1$ denotes the projected momentum $|\vec{p}|=2\pi/L$.}
\end{figure}

Replacing the infinite-volume hadronic functions $V_{\mu\nu}(\vec{x},t)$ and $A_{\mu\nu}(\vec{x},t)$ by their finite-volume lattice counterparts $V_{\mu\nu}^{L}(\vec{x},t)$ and $A_{\mu\nu}^{L}(\vec{x},t)$ introduces exponentially suppressed finite-volume effects, since these effects decay exponentially at large $|\vec{x}|$. They can be checked by introducing a spatial integration cutoff $R=|\vec{x}|$ and verifying convergence at large $R$. We illustrate this using the smallest volume, $L=32$. The scalar structures include the following forms:
\begin{equation}
    \begin{aligned}
        \tilde{I}_0 &= \epsilon_{\mu\nu\alpha 0}\,x_{\alpha}V_{\mu\nu}(\vec{x},t), \\
        \tilde{I}_1 &= \delta_{ij}A_{ij}(\vec{x},t), \\
        \tilde{I}_2 &= A_{00}(\vec{x},t), \\
        \tilde{I}_3 &= x_iA_{0i}(\vec{x},t), \\
        \tilde{I}_4 &= x_iA_{i0}(\vec{x},t), \\
        \tilde{I}_5 &= 
        \frac{j_2(|\vec{p}||\vec{x}|)}{(|\vec{p}||\vec{x}|)^2}
        x_ix_jA_{ij}(\vec{x},t).
    \end{aligned}
\end{equation}
The spatial integrals of these structures as functions of the cutoff $|\vec{x}|$ are shown in Fig.~\ref{diag:I_V}. Clear plateaus are observed for $|\vec{x}| \gtrsim 1.5~\mathrm{fm}$, indicating that contributions from the hadronic functions beyond this distance are negligible. Finite-volume effects are therefore well controlled in our calculation.

In the present case the intermediate state is heavier than the initial state. When the intermediate state is lighter, however, finite-volume effects grow exponentially. In that situation, the infinite-volume reconstruction (IVR) method can be used to address the problem~\cite{Feng:2018qpx, Tuo:2024bhm}.

\subsection{Calculation of the Hadronic Function}
\begin{figure}[htbp]
\centering
\includegraphics[width=0.49\textwidth]{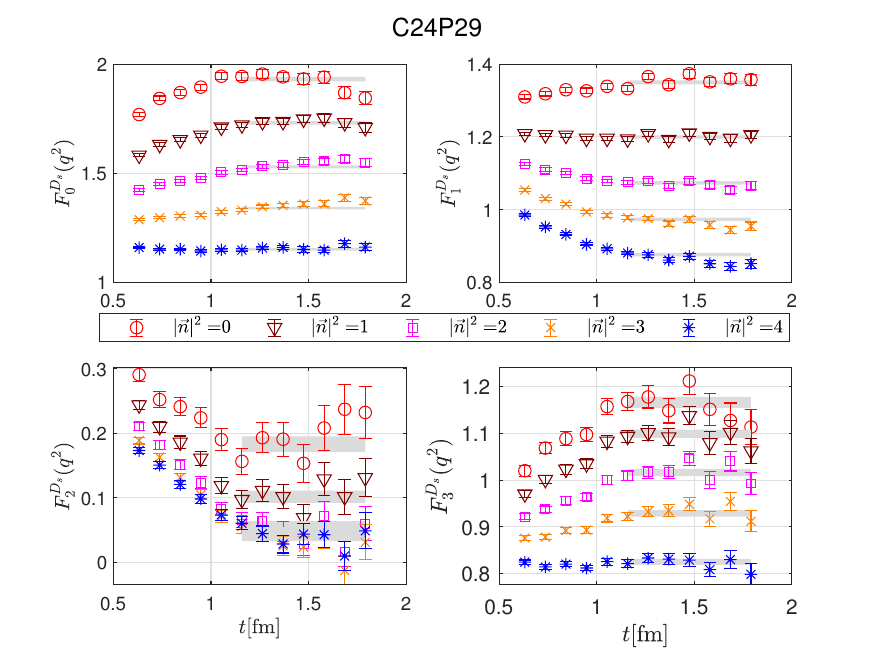}
\includegraphics[width=0.49\textwidth]{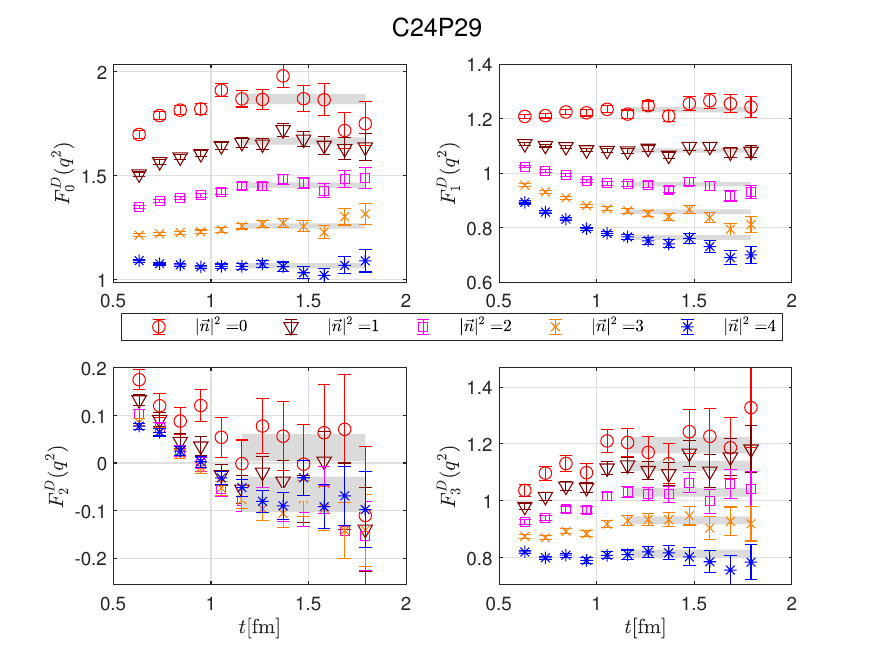}
\includegraphics[width=0.49\textwidth]{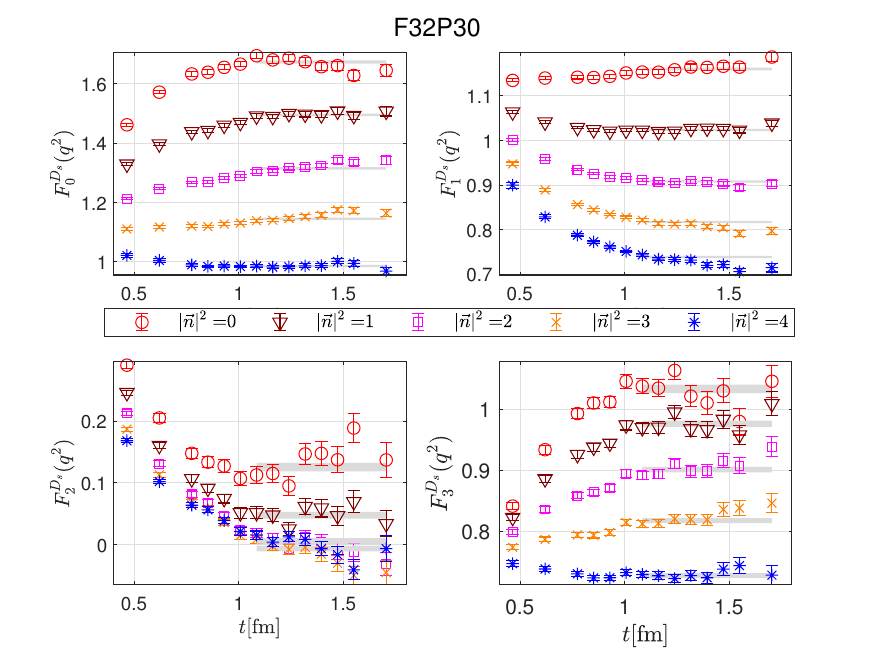}
\includegraphics[width=0.49\textwidth]{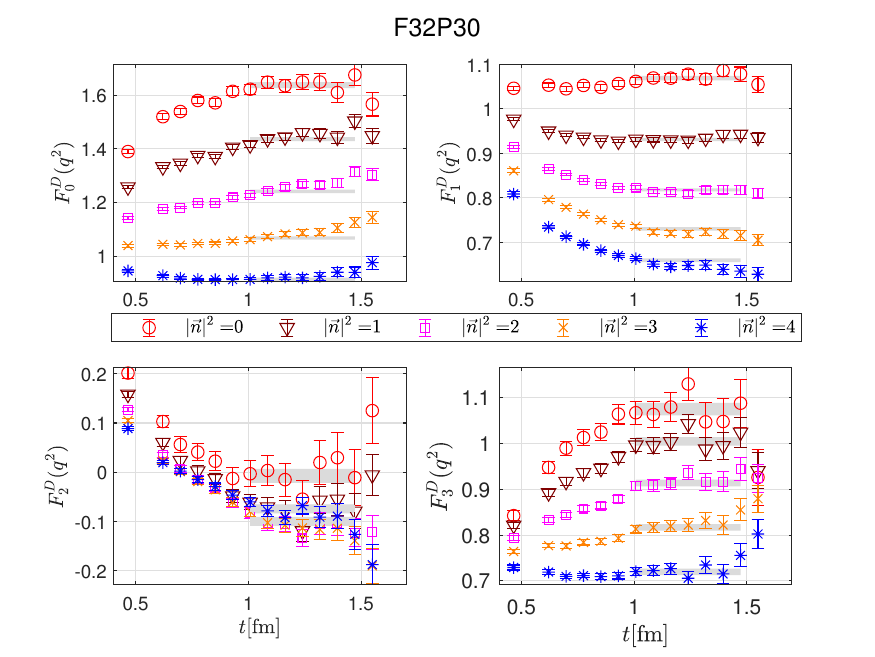}
\includegraphics[width=0.49\textwidth]{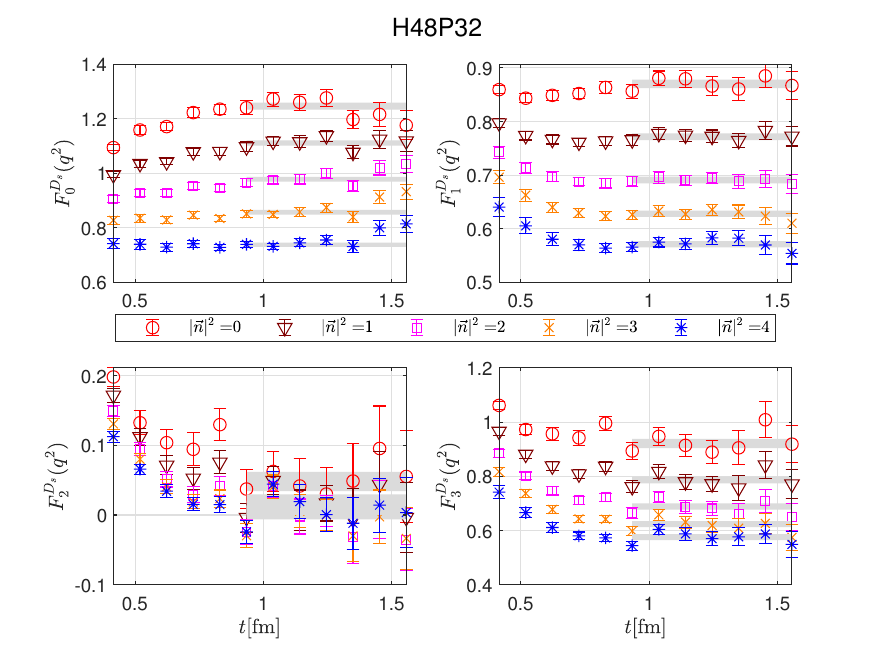}
\includegraphics[width=0.49\textwidth]{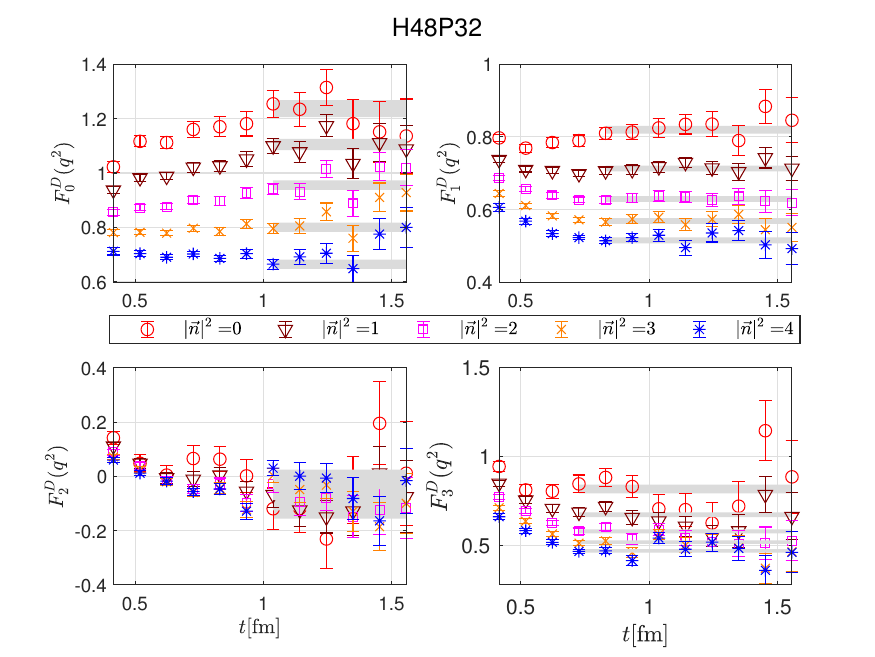}
\caption{\label{diag:F}
Form factors $F_i^{D_s}(q^2)$ (left) and $F_i^{D}(q^2)$ (right), with $i=0,1,2,3$, on the C24P29, F32P30, and H48P32 ensembles at momenta $\vec{p}=2\pi\vec{n}/L\,(|\vec{n}|^2=0,1,2,3,4)$. Here $F_i^{D_s}(q^2)$ denotes the form factors for $J/\psi\rightarrow D_s l\nu_l$, while $F_i^{D}(q^2)$ denotes those for $J/\psi\rightarrow D l\nu_l$.
}
\end{figure}

This section uses the CLQCD C24P29, F32P30, and H48P32 ensembles. The lattice results for the form factors $F_i^{D/D_s}(q^2)\,(i=0,1,2,3)$ as functions of the time separation $t$ at momenta $\vec{p}=2\pi\vec{n}/L,\;|\vec{n}|^2=0,1,2,3,4$ are shown in Fig.~\ref{diag:F}. Clear plateaus are observed for both $F_i^{D}$ and $F_i^{D_s}$ at large time separations, allowing constant fits, shown by the grey bands.

\begin{figure}[htbp]
\centering
\includegraphics[width=0.8\textwidth]{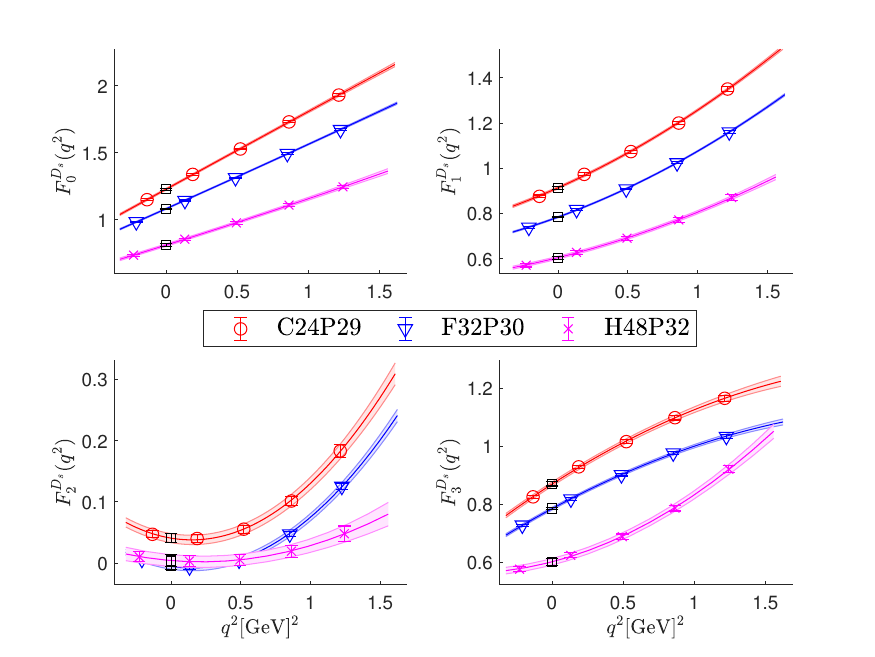}
\caption{\label{diag:F_cont}
Polynomial fits to the form factors $F_i^{D_s}(q^2)$ with $i=0,1,2,3$ on the C24P29, F32P30, and H48P32 ensembles at $\vec{p}=2\pi\vec{n}/L\,(|\vec{n}|^2=0,1,2,3,4)$. Here $F_i^{D_s}(q^2)$ denotes the form factors for $J/\psi\rightarrow D_s l\nu_l$. The black squares show the results at $q^2=0$.
}
\end{figure}

\begin{figure}[htbp]
\centering
\includegraphics[width=0.8\textwidth]{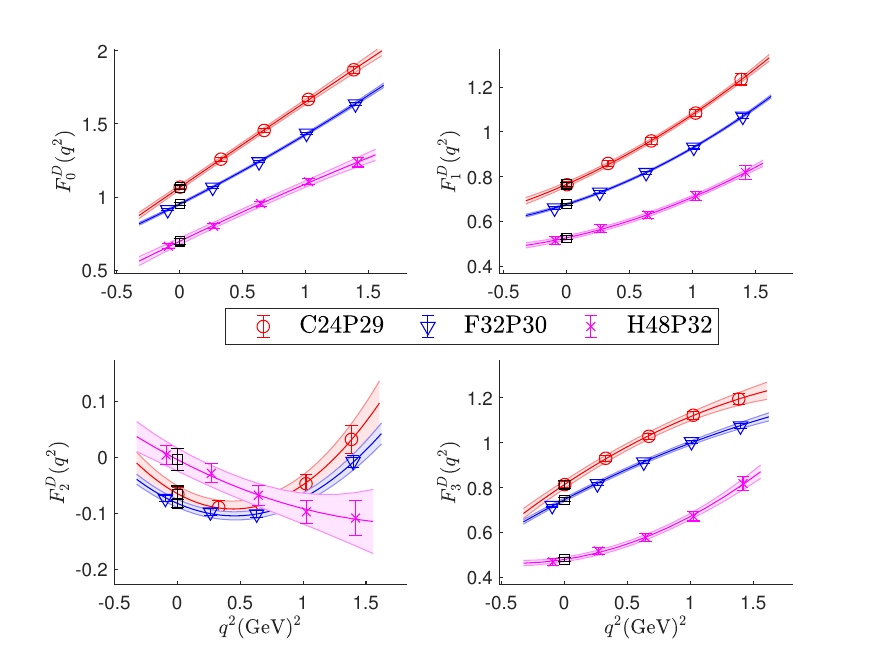}
\caption{\label{diag:F_cont_D}
Polynomial fits to the form factors $F_i^{D}(q^2)$ with $i=0,1,2,3$ on the C24P29, F32P30, and H48P32 ensembles at $\vec{p}=2\pi\vec{n}/L\,(|\vec{n}|^2=0,1,2,3,4)$. Here $F_i^{D}(q^2)$ denotes the form factors for $J/\psi\rightarrow D l\nu_l$. The black squares show the results at $q^2=0$.
}
\end{figure}

After considering the five momentum modes for the form factors $F_i^{D/D_s}(q^2)$, we perform the momentum extrapolation using the polynomial form
\begin{equation}\label{eq:mom_extra}
F_i^{h}(q^2)=d_i^{h,(0)}+d_i^{h,(1)}\, q^2+d_i^{h,(2)}\, q^4,
\end{equation}
where the coefficients $d_i^{h,(j)}$ with $j=0,1,2$ are introduced. We perform correlated fits using this form, as shown in Figs.~\ref{diag:F_cont} and~\ref{diag:F_cont_D}. The polynomial form in Eq.~(\ref{eq:mom_extra}) describes the lattice data well.

\subsection{Decay Widths and Differential Decay Widths}
Using the numerical results for the coefficients $d_i^{D/D_s,(j)}$ together with Eq.~(\ref{eq:width}), we obtain the differential decay rate $d\Gamma/(dq^2 V_{cs(d)}^2)$, with continuous $q^2$ covering the full phase space. Integrating over $q^2$ throughout the full phase-space region and using $V_{cs}=0.975(6)$ and $V_{cd}=0.221(4)$~\cite{ParticleDataGroup:2024cfk}, we obtain the total branching fractions $\operatorname{Br}(J/\psi\rightarrow D/D_s l\nu_l)$ for the corresponding decay channels.

\begin{figure}[htbp]
\centering
\includegraphics[width=0.49\textwidth]{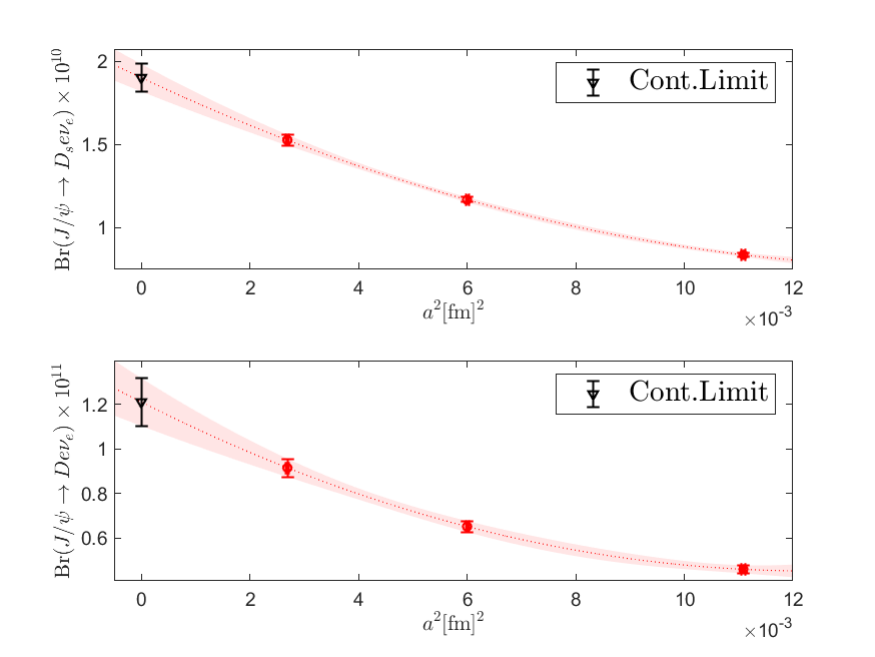}
\includegraphics[width=0.49\textwidth]{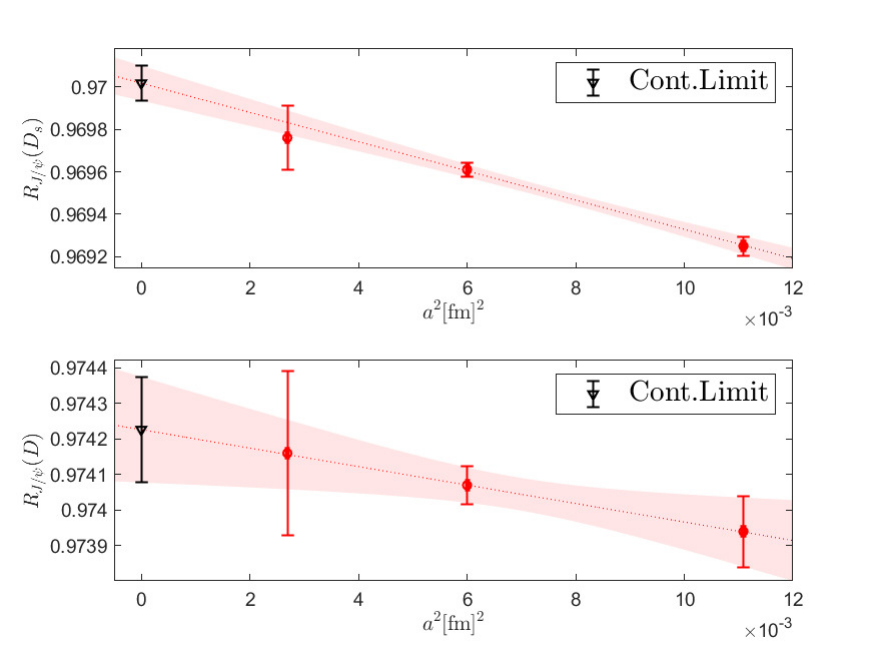}
\caption{\label{diag:Gamma_cont}
Lattice results for $\operatorname{Br}(J/\psi\rightarrow D/D_s e\nu_e)$ (left) and $R_{J/\psi}(D/D_s)$ (right) as functions of the lattice spacing. In the left panel, the continuum extrapolation includes an $a^4$ term. In the right panel, only linear $a^2$ behavior is included.
}
\end{figure}

The lattice results for the branching fractions of $J/\psi \rightarrow D/D_s e\nu_e$ at different lattice spacings are shown in Fig.~\ref{diag:Gamma_cont}, together with the ratios between the $\mu$ and $e$ modes,
\begin{equation}
    R_{J/\psi}(D/D_s)=\frac{\operatorname{Br}(J/\psi \rightarrow D/D_s \mu\nu_{\mu})}{\operatorname{Br}(J/\psi \rightarrow D/D_s e\nu_e)}.
\end{equation}

Figure~\ref{diag:Gamma_cont} shows that the branching fractions are well described by continuum extrapolations including an $a^4$ term, while the ratios require only an $a^2$ term. This is because discretization effects largely cancel in the ratios. For charmonium systems, the next-to-leading $a^4$ term can become visible on relatively coarse lattices, for example at lattice spacings larger than $0.1\,\mathrm{fm}$. Since this work uses only three lattice spacings, an extrapolation including an $a^4$ term cannot be fully controlled. We therefore use a linear $a^2$ extrapolation to ensure a controlled continuum limit.

\begin{equation}
    \begin{aligned}
        \operatorname{Br}(J/\psi \rightarrow D_s e\nu_e) &=1.90(6)_{\textrm{stat}}(5)_{V_{cs}}\times 10^{-10}, \\
        \operatorname{Br}(J/\psi \rightarrow D e\nu_e) &=1.21(6)_{\textrm{stat}}(9)_{V_{cd}}\times 10^{-11}, \\
        R_{J/\psi}(D_s) &=0.97002(8)_{\textrm{stat}}, \\
        R_{J/\psi}(D) &=0.97423(15)_{\textrm{stat}},
    \end{aligned}
    \label{eq:Br_value}
\end{equation}
where the first uncertainty is statistical, and the second arises from the uncertainty in the CKM matrix element $V_{cs(d)}$.

It should be noted that the light-quark masses in this chapter are not physical. This may affect the process $J/\psi\rightarrow Dl\nu_l$, where the light quark appears as a valence quark. For $J/\psi\rightarrow D_sl\nu_l$, by contrast, the unphysical light quark appears only in the sea, and its effect is expected to be small~\cite{FlavourLatticeAveragingGroupFLAG:2024oxs}.

\begin{figure}[htbp]
\centering
\includegraphics[width=0.8\textwidth]{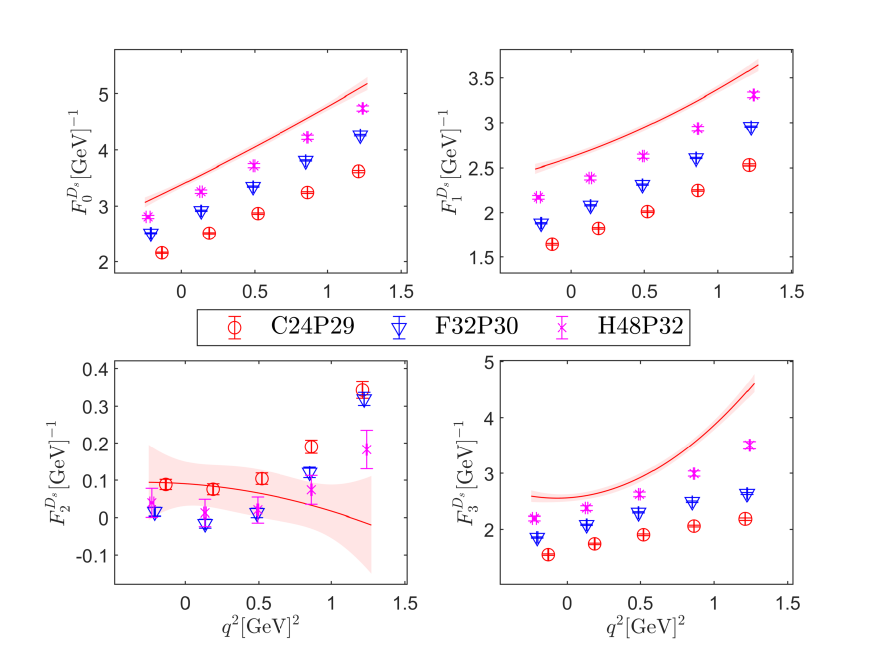}
\caption{\label{diag:F_cont_q}
$q^2$-expansion result for the $J/\psi\rightarrow D_s$ form factors. The red shaded region denotes the final result in the continuum limit $a\rightarrow 0$.
}
\end{figure}

\begin{figure}[htbp]
\centering
\includegraphics[width=0.8\textwidth]{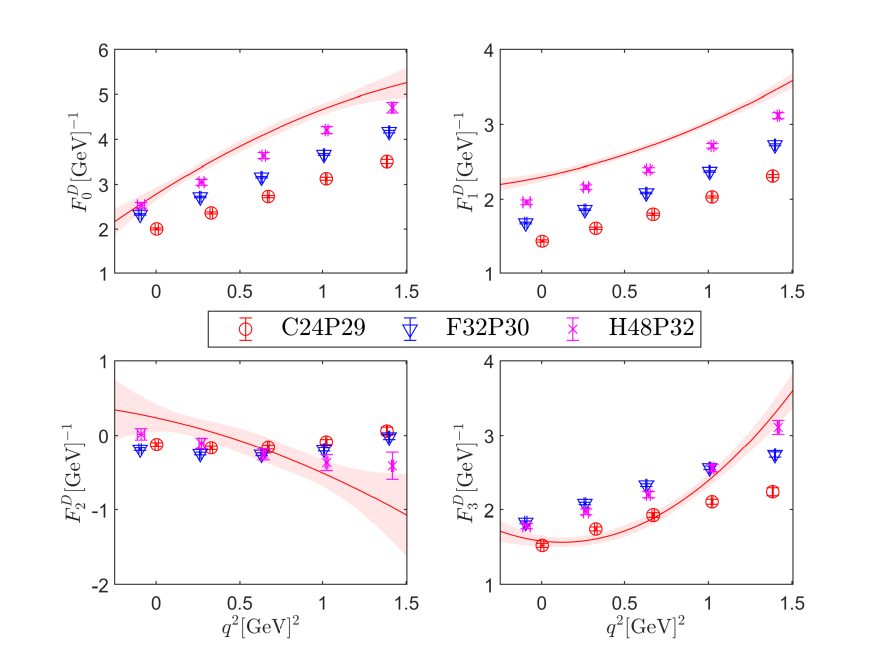}
\caption{\label{diag:F_cont_q_D}
$q^2$-expansion result for the $J/\psi\rightarrow D$ form factors. The red shaded region denotes the final result in the continuum limit $a\rightarrow 0$.
}
\end{figure}

For comparison with future experimental and phenomenological studies, we provide the differential distribution in $q^2$, $d\Gamma/dq^2$, of the decay width. To this end, we use a general $q^2$-expansion parametrization of the form factors,
\begin{equation}
    F(a^2,q^2)=\sum_{n=0}^{n_{\textrm{max}}}(c_n+d_n a^2+f_n a^4)q^{2n},
\end{equation}
where the $d_n$ and $f_n$ terms describe discretization effects.

\begin{figure}[htbp]
    \centering
    \includegraphics[width=0.8\textwidth]{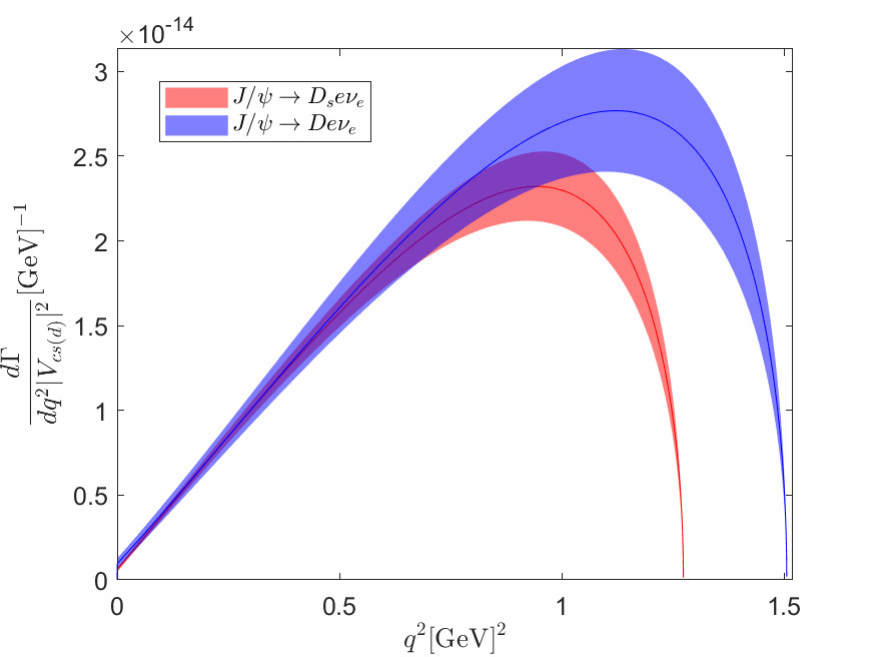}
    \caption{\label{diag:width_cont_q}
    Predictions for the differential decay widths of $J/\psi\rightarrow D_s e \nu_e$ and $J/\psi\rightarrow D e \nu_e$, divided by $V_{cs}$ and $V_{cd}$, respectively.
    }
\end{figure}

Using the PDG masses
$m_{J/\psi}=3.09690(1)\,\mathrm{GeV}$,
$m_{D_s}=1.96834(7)\,\mathrm{GeV}$,
and $m_{D}=1.86966(5)\,\mathrm{GeV}$,
we obtain the differential decay width $d\Gamma/dq^2$, shown in Fig.~\ref{diag:width_cont_q}.

Integrating over $q^2$ gives the branching fractions
\begin{equation}
\operatorname{Br}(J/\psi \rightarrow D_s e\nu_e)=1.99(14)\times 10^{-10},
\end{equation}
and
\begin{equation}
\operatorname{Br}(J/\psi \rightarrow D e\nu_e)=1.44(17)\times 10^{-11}.
\end{equation}
The uncertainties include both statistical uncertainties and the uncertainty in the CKM matrix element $V_{cs(d)}$. These results are broadly consistent with those in Eq.~(\ref{eq:Br_value}), but have larger statistical uncertainties. We therefore quote the results in Eq.~(\ref{eq:Br_value}) as the final values in this thesis.

Experimental searches for $J/\psi$ semileptonic decays have continued for many years, but no signal has yet been observed. The current most stringent upper limit comes from the BESIII study of $J/\psi \rightarrow D e\nu_e$, which used the full sample of $1.01\times 10^{10}$ $J/\psi$ events. This experimental limit is still about three orders of magnitude above the theoretical prediction for the $J/\psi \rightarrow D e\nu_e$ decay rate. Since no measurement of $J/\psi \rightarrow D_s l\nu_l$ using the full $J/\psi$ sample is currently available, previous experiments suggest that the upper limit for this channel could reach roughly the $10^{-8}$ level if the full $J/\psi$ data sample were used. The experimental results for $J/\psi$ semileptonic decays are summarized in Table~\ref{tab:BESIII} and compared with the lattice results of this thesis.

\begin{table}[htbp]
    \centering
    \caption{Upper limits on the branching fractions for $J/\psi \rightarrow D/D_s\, l\nu_l$ from BESIII, together with the branching fractions predicted by lattice QCD.}
    \begin{tabular}{cccc}
        \toprule
        Decay channel & Upper limit/branching fraction & Number of $J/\psi$ events &  Reference \\
        \midrule
        $J/\psi\rightarrow D_s e\nu_e$ & $4.9\times 10^{-5}$ & $5.8\times 10^{7}$ & \cite{BES:2006mls} \\
        $J/\psi\rightarrow D_s e\nu_e$ & $1.3\times 10^{-6}$ & $2.3\times 10^{8}$ & \cite{BESIII:2014pps} \\
        $J/\psi\rightarrow D e\nu_e$   & $7.1\times 10^{-8}$ &$1.01\times 10^{10}$ & \cite{BESIII:2021mnd} \\
        $J/\psi\rightarrow D \mu\nu_{\mu}$ & $5.6\times 10^{-7}$  & $1.01\times 10^{10}$ & \cite{BESIII:2023fqz} \\
        \midrule
        $J/\psi\rightarrow D_s e\nu_e$ & $1.90(8)\times 10^{-10}$ &  & This work \\
        $J/\psi\rightarrow D_s \mu\nu_{\mu}$ & $1.84(8)\times 10^{-10}$ &  & This work \\
        $J/\psi\rightarrow D e\nu_e$ & $1.21(11)\times 10^{-11}$ &  & This work \\
        $J/\psi\rightarrow D \mu\nu_{\mu}$ & $1.18(11)\times 10^{-11}$ &  & This work \\
        \bottomrule
    \end{tabular}
    \label{tab:BESIII}
\end{table}

Future facilities such as the Super Tau Charm Facility~\cite{Achasov:2023gey} are expected to improve experimental sensitivity to the $10^{-10}$ level. Once reliable experimental branching-fraction measurements become available, combining them with lattice-QCD form factors will allow the CKM matrix element $V_{cs(d)}$ to be extracted, providing a new high-precision test of the Standard Model.

\subsection{Summary}
This section presented the first lattice-QCD calculation of $J/\psi$ semileptonic decays, studying the weak decays $J/\psi\rightarrow D_s l\nu_l$ and $J/\psi \rightarrow D l\nu_l$. After a simple continuum extrapolation using three lattice spacings, we obtain the lattice predictions
\begin{equation}
    \begin{aligned}
        \operatorname{Br}(J/\psi\rightarrow D_s e\nu_e)
        =1.90(6)(5)_{V_{cs}}\times 10^{-10}, \\
        \operatorname{Br}(J/\psi\rightarrow D e\nu_e)
        =1.21(6)(9)_{V_{cd}}\times 10^{-11},
    \end{aligned}
\end{equation}
where the first uncertainty is statistical, and the second arises from the CKM matrix element $V_{cs(d)}$.

After obtaining the form factors for $J/\psi\rightarrow D$ and $J/\psi\rightarrow D_s$, we compute the branching-fraction ratios between the $\mu$ and $e$ modes. The results are
\begin{equation}
R_{J/\psi}(D_s)=0.97002(8) \\
R_{J/\psi}(D)=0.97423(15),
\end{equation}
where only statistical uncertainties are included.

In the future, combining these results with precise experimental measurements will enable stringent tests of lepton flavor universality and the extraction of the CKM matrix element $V_{cs(d)}$, thereby providing another high-precision probe of the Standard Model. The lattice method used in this thesis can also be extended directly to other $P\rightarrow V$ semileptonic decays, such as $D\rightarrow K^*$~\cite{BESIII:2024qnx},
$B\rightarrow K^*$~\cite{LHCb:2023gpo}, and $B\rightarrow D^*$~\cite{HFLAV:2022esi}.

\cleardoublepage
\chapter{Summary and Outlook}
\label{chap:conclusion}

{
\kaishu
\begin{center}
    长风破浪会有时，直挂云帆济沧海。
\end{center}
\hfill ——《行路难》[唐] 李白
}

The development of modern physics has always been driven by the attempt to connect macroscopic phenomena with the microscopic world. High-energy physics, as the discipline devoted to the most elementary constituents of matter and their interactions, uses both experimental and theoretical methods to explore fundamental particles and the forces among them. The strong interaction is one of the four fundamental interactions in nature and governs the dynamics of quarks and gluons inside nucleons. At low energies, the strong interaction exhibits the highly nontrivial phenomenon of color confinement and cannot be treated perturbatively. Lattice quantum chromodynamics provides a first-principles, nonperturbative framework in which the strong interaction can be studied quantitatively without relying on model assumptions. In this thesis, lattice methods are used to study hadron scattering and spectroscopy, with the aim of testing the Standard Model description and prediction of resonance phenomena. The work presented here was carried out at a historical turning point for lattice QCD, where two-body scattering methods have become relatively mature, while three-body formalism has only recently become practically applicable. At this critical stage, this thesis completes some of the first systematic studies of three-body resonances, providing both methodological benchmarks and physical reference points for future developments in the field.

To address scattering problems, this thesis discusses lattice symmetries and techniques for constructing lattice operators, and extends them to systems with arbitrary particle number, lattice momentum, and intrinsic quantum numbers. These methods have been implemented in the open-source package \texttt{OpTion}. The resulting operators can be used directly in spectroscopic and scattering analyses.

For two-body scattering, the thesis focuses primarily on single-channel $D\pi$ scattering. Using a set of $N_{\text{f}}=2+1$ Wilson-clover ensembles at four different values of $M_{\pi}$, the $D\pi$ phase shift and scattering length are extracted through Lüscher's quantization condition together with parametrizations such as the effective-range expansion, and the pole structure of the $D_0^*(2300)$ resonance is analyzed. In combination with earlier studies, we find that as the $\pi$ mass is lowered from $391\,\mathrm{MeV}$ to the physical point, the $D_0^*(2300)$ evolves continuously from a bound state into a virtual state and then into a resonance. In addition, the scattering length is extrapolated to the physical point, yielding the near-threshold $D\pi$ scattering amplitude and providing useful input for future studies of the $DD\pi$ three-body system. This thesis also presents a preliminary study of the coupled three-channel $D\pi$-$D\eta$-$D_s\bar{K}$ scattering system and explores an analysis strategy for four-channel scattering, where a double-pole structure is observed.

Many experimentally observed resonances have multi-particle decay channels, making it essential to study their resonance properties with lattice methods. With the operator-construction framework developed in this thesis, one can construct one-, two-, and three-particle operators. The first three-body system studied here is the isoscalar vector $\pi\pi\pi$ channel, which contains the $\omega(782)$ resonance. One computational challenge in three-body problems is the precise extraction of finite-volume spectra; in this work, the spectra are obtained through a carefully designed set of correlation-function contractions. Another challenge is the formulation of the three-body quantization condition. By solving infinite-volume integral equations and performing the analytic continuation, we determine the resonance pole associated with the $\omega(782)$ and extract its resonance parameters. When the three-body interaction is described either by a general parametrization or by chiral effective field theory, the two descriptions give mutually consistent results. In the effective-field-theory parametrization, the pole can be extrapolated to the physical point, where the resulting mass agrees well with experiment. The final pole position of the $\omega$ meson is $\sqrt{s_{\omega(782)}} = [778.0(11.2)-i\,3.0(5)]\,\mathrm{MeV}$, which represents the first complete lattice-QCD determination of the mass and width of the $\omega(782)$ resonance.

Building on this foundation, the thesis then studies the radial excitation of the $\pi$, the $\pi(1300)$ meson, whose experimental status remains unsettled and which represents an important step toward predictive calculations of resonances. The mass of the $\pi(1300)$ is roughly ten times the $\pi$ mass, making it a direct probe of the soft-$\pi$ theorem and an important system for understanding chiral dynamics. The decay pattern of the $\pi(1300)$ is analogous to that of the Roper resonance, providing a mesonic testing ground for questions relevant to the Roper; at the same time, its possible hybrid component gives it a distinctive role in studies of physics beyond the Standard Model and CP violation. The $\pi(1300)$ can decay into the three-body $\pi\pi\pi$ final state, including the $\rho\pi$, $\sigma\pi$, and $G\pi$ subchannels. In this thesis, calculations are performed at heavier $\pi$ masses so that the $\pi(1300)$ lies within the range of validity of the three-body quantization condition, and its existence is verified. Assuming a simple quark-mass dependence for the three-body force, the resonance parameters are extrapolated to the physical point, where they agree with the PDG values. The resulting pole position is $\sqrt{s_{\pi(1300)}} = (1169\pm46)-i(62\pm169)\,\mathrm{MeV}$, which provides the first first-principles verification of the existence of the $\pi(1300)$.

This thesis also studies radiative and semileptonic decays of charmed mesons. A model-independent method is proposed for analyzing radiative transitions and is applied to the processes $J/\psi \rightarrow \gamma\eta_c$, $D_s^*\rightarrow D_s\gamma$, $D_s^*\rightarrow D_s e^+e^-$, $J/\psi\rightarrow D_s l\nu_l$, and $J/\psi \rightarrow D l\nu_l$. This method determines on-shell transition form factors directly from lattice hadronic correlation functions and substantially reduces the uncertainties of transition matrix elements, opening a path toward high-precision determinations of CKM matrix elements.

For two-body scattering involving charmed hadrons, especially systems containing both charm and strangeness, our present understanding remains limited and further work is needed. Baryon-involving scattering still suffers from substantial uncertainties due to the deterioration of the signal-to-noise ratio, and its dynamics differ significantly from those of purely light-quark systems. These systems constitute an important frontier of two-body scattering. Progress on these questions in the coming years will have broad implications for the structure of nuclear forces, nuclear-reaction dynamics, and neutron-star physics.

The methods developed and the results obtained in the $D_0^*$, $\omega(782)$, and $\pi(1300)$ systems can be extended directly to more complicated hadronic systems. Future progress on three-body problems may proceed in several directions. The first is the extension of three-body quantization conditions: at present they are mainly applicable to single-channel scattering, and they should be generalized to coupled-channel and other more general situations. The second is the systematic study of three-body spectroscopy. The present work begins with the $\pi\pi\pi$ system, but future studies can address systems containing strange mesons, charmed mesons, and even baryons. The third direction is the study of electroweak processes that decay into three-body final states, together with the calculation of the corresponding matrix elements. Finally, it is equally important to extend two- and three-body methods to systems with still more particles, while also exploring new paradigms for treating scattering problems. The present era offers exceptional opportunities for lattice studies: we are fortunate to be working at a time when three-body physics is rapidly opening up, and the coming years will bring many results that will greatly enrich our understanding of the strong interaction, the Standard Model, and the evolution of the universe.

\cleardoublepage


\bibliographystyle{gbt7714-numerical}
\setlength{\bibsep}{3pt} 
\phantomsection
\addcontentsline{toc}{chapter}{References}

{\typebib
\bibliography{body/references}
}

\cleardoublepage
\begin{appendix}
\renewcommand{\chaptermark}[1]{\markboth{Appendix~\Alph{chapter}\enspace #1}{}}

\chapter{Operator Lists}
\label{appendix:operators}

\section{\texttt{OpTion} Tutorial}
Once the hadron types, target quantum numbers, derivative order, and maximum relative momentum are specified, the corresponding operator set can be generated with a single simple command.

This section briefly introduces the basic usage of \texttt{OpTion} and gives several concrete examples. The package is publicly available online~\cite{github}. We focus on the projection method discussed in the main text. For installation instructions and a detailed tutorial, see the \textit{Manual} folder in the repository.

The function \verb|OneHadronOperatorAll[Ptot, Rep, RepRow, MaxND]| searches for all one-particle operators whose covariant-derivative order is lower than \verb|MaxND| and returns the corresponding operator list. For example:

\begin{lstlisting}[language=Mathematica, breaklines=true, basicstyle=\ttfamily\small, frame=single, numbers=left, numberstyle=\tiny]
Ptot = {0, 0, 0};
Rep = "T1-";
RepRow = 3;
MaxND = 1;
Print/@OneHadronOperatorAll[Ptot, Rep, RepRow, MaxND];
\end{lstlisting}
Output:\\
\noindent $\small{\mathsf{V_z}}$ \\
\noindent $\small{\mathsf{D_z}}$ \\
\noindent $\small{\mathsf{A_y D_x - A_x D_y}}$ \\

These outputs correspond to one row of Table~\ref{tab:one-000}.

The function \verb|TwoHadronOperatorAll[Ptot, Rep, RepRow, MaxMom, Par1, Par2]| searches for all two-particle operators in which the individual particle momenta are smaller than \verb|MaxMom| and returns the corresponding operator list. For example:

\begin{lstlisting}[language=Mathematica, breaklines=true, basicstyle=\ttfamily\small, frame=single, numbers=left, numberstyle=\tiny]
Par1 = "P";
Par2 = "V";
Ptot = {0, 0, 0};
Rep = "T1-";
RepRow = 3;
MaxMom = 1;
Print/@TwoHadronOperatorAll[Ptot, Rep, RepRow, MaxMom, Par1, Par2];
\end{lstlisting}
Output:\\
\noindent $\small{\mathsf{-P_1\left[e_y\right] V_{2 x}\left[-e_y\right]+P_1\left[-e_y\right] V_{2 x}\left[e_y\right]+P_1\left[e_x\right] V_{2 y}\left[-e_x\right]-P_1\left[-e_x\right] V_{2 y}\left[e_x\right]}}$ \\

To construct operators with an arbitrary number of particles, the particle types can be specified through the argument \verb|ParTuple|. The following example shows a three-particle $\pi\pi\pi$ operator in the rest frame projected onto $T_1^+$:

\begin{lstlisting}[language=Mathematica, breaklines=true, basicstyle=\ttfamily\small, frame=single, numbers=left, numberstyle=\tiny]
Rep = "T1+";
ParTuple = {"P", "P", "P"};
Print/@NHadronOperatorAll[Ptot, Rep, RepRow, MaxMom, ParTuple];
\end{lstlisting}
Output:\\
\noindent $\small{\mathsf{P_1\left[e_z\right] P_2[0] P_3\left[-e_z\right]-P_1\left[-e_z\right] P_2[0] P_3\left[e_z\right]}}$ \\
\noindent $\small{\mathsf{P_1\left[e_z\right] P_2\left[-e_z\right] P_3[0]-P_1\left[-e_z\right] P_2\left[e_z\right] P_3[0]}}$ \\
\noindent $\small{\mathsf{P_1[0] P_2\left[e_z\right] P_3\left[-e_z\right]-P_1[0] P_2\left[-e_z\right] P_3\left[e_z\right]}}$ \\

These operators can be used to study three-meson interactions at maximal isospin.

\section{One-Meson Operator Lists}
\label{sec:append_list_one}
\begin{table}[htbp]
\centering
\caption{One-meson operators projected onto the irreps of the cubic group $O_h$. The symbols $S$, $P$, $V$, and $A$ denote scalar, pseudoscalar, vector, and axial-vector gamma-matrix structures, respectively. To simplify the notation, arrows on covariant derivatives and overall numerical factors are omitted.}
\addtolength{\tabcolsep}{6pt}
\begin{tabular}{cc}
\toprule
Irrep & Operator \\
\midrule
\multirow{2}{*}{$A_1^+$} & $S$ \\
& $V_x\nabla_x + V_y\nabla_y + V_z\nabla_z$ \\
\midrule
\multirow{2}{*}{$A_1^-$} & $P$ \\
& $A_x\nabla_x + A_y\nabla_y + A_z\nabla_z$ \\
\midrule
\multirow{1}{*}{$A_2^+$} & - \\
\midrule
\multirow{1}{*}{$A_2^-$} & - \\
\midrule
\multirow{1}{*}{$E^+$} & $V_x\nabla_x + V_y\nabla_y -2 V_z\nabla_z$ \\
\midrule
\multirow{1}{*}{$E^-$} & $A_x\nabla_x + A_y\nabla_y -2 A_z\nabla_z$ \\
\midrule
\multirow{3}{*}{$T_1^+$} & $A_x$ \\
& $P\nabla_x$ \\
& $-V_z\nabla_y + V_y\nabla_z$ \\
\midrule
\multirow{3}{*}{$T_1^-$} & $V_x$ \\
& $SD_x$ \\
& $A_z\nabla_y - A_y\nabla_z$ \\
\midrule
\multirow{1}{*}{$T_2^+$} & $V_z\nabla_y + V_y\nabla_z$ \\
\midrule
\multirow{1}{*}{$T_2^-$} & $A_z\nabla_y + A_y\nabla_z$ \\
\bottomrule
\end{tabular}
\addtolength{\tabcolsep}{-6pt}
\label{tab:one-000}
\end{table}

\begin{table}[htbp]
\centering
\caption{One-meson operators for point group $C_{4v}$ at total momentum $\vec{P} = [0,0,n]$. The notation follows Table~\ref{tab:one-000}.}
\addtolength{\tabcolsep}{6pt}
\begin{tabular}{cc}
\toprule
Irrep & Operator \\
\midrule
\multirow{6}{*}{$A_1$} & $S$ \\
& $V_z$ \\
& $S\nabla_z$ \\
& $V_x\nabla_x + V_y\nabla_y + V_z\nabla_z$ \\
& $V_x\nabla_x + V_y\nabla_y -2 V_z\nabla_z$ \\
& $A_y\nabla_x - A_x\nabla_y$ \\
\midrule
\multirow{6}{*}{$A_2$} & $P$ \\
& $A_z$ \\
& $P\nabla_z$ \\
& $A_x\nabla_x + A_y\nabla_y + A_z\nabla_z$ \\
& $A_x\nabla_x + A_y\nabla_y -2 A_z\nabla_z$ \\
& $V_y\nabla_x - V_x\nabla_y$ \\
\midrule
\multirow{2}{*}{$B_1$} & $V_x\nabla_x - V_y\nabla_y$ \\
& $A_y\nabla_x + A_x\nabla_y$ \\
\midrule
\multirow{2}{*}{$B_2$} & $A_x\nabla_x - A_y\nabla_y$ \\
& $V_y\nabla_x + V_x\nabla_y$ \\
\midrule
\multirow{8}{*}{$E$} & $V_y$ \\
& $A_x$ \\
& $S\nabla_y$ \\
& $P\nabla_x$ \\
& $-V_z\nabla_y + V_y\nabla_z$ \\
& $V_z\nabla_y + V_y\nabla_z$ \\
& $-A_z\nabla_x + A_x\nabla_z$ \\
& $A_z\nabla_x + A_x\nabla_z$ \\
\bottomrule
\end{tabular}
\addtolength{\tabcolsep}{-6pt}
\label{tab:one-00n}
\end{table}

\begin{table}[htbp]
\centering
\caption{One-meson operators for point group $C_{2v}$ at total momentum $\vec{P} = [0,n,n]$. The notation follows Table~\ref{tab:one-000}.}
\addtolength{\tabcolsep}{6pt}
\begin{tabular}{cc}
\toprule
Irrep & Operator \\
\midrule
\multirow{8}{*}{$A_1$} & $S$ \\
& $V_y+V_z$ \\
& $S (\nabla_y + \nabla_z)$ \\
& $V_x\nabla_x + V_y\nabla_y + V_z\nabla_z$ \\
& $2 V_x\nabla_x - V_y\nabla_y - V_z\nabla_z$ \\
& $V_z\nabla_y + V_y\nabla_z$ \\
& $A_y\nabla_x - A_z\nabla_x - A_x\nabla_y + A_x\nabla_z$ \\
& $A_y\nabla_x - A_z\nabla_x + A_x\nabla_y - A_x\nabla_z$ \\
\midrule
\multirow{8}{*}{$A_2$} & $P$ \\
& $A_y+A_z$ \\
& $P (\nabla_y + \nabla_z)$ \\
& $A_x\nabla_x + A_y\nabla_y + A_z\nabla_z$ \\
& $2 A_x\nabla_x - A_y\nabla_y - A_z\nabla_z$ \\
& $A_z\nabla_y + A_y\nabla_z$ \\
& $V_y\nabla_x - V_z\nabla_x - V_x\nabla_y + V_x\nabla_z$ \\
& $V_y\nabla_x - V_z\nabla_x + V_x\nabla_y - V_x\nabla_z$ \\
\midrule
\multirow{8}{*}{$B_1$} & $V_y - V_z$ \\
& $A_x$ \\
& $P\nabla_x$ \\
& $S (\nabla_y - \nabla_z)$ \\
& $V_z\nabla_y - V_y\nabla_z$ \\
& $V_y\nabla_y - V_z\nabla_z$ \\
& $A_y\nabla_x + A_z\nabla_x - A_x\nabla_y - A_x\nabla_z$ \\
& $A_y\nabla_x + A_z\nabla_x + A_x\nabla_y + A_x\nabla_z$ \\
\midrule
\multirow{8}{*}{$B_2$} & $A_y - A_z$ \\
& $V_x$ \\
& $S\nabla_x$ \\
& $P (\nabla_y - \nabla_z)$ \\
& $A_z\nabla_y - A_y\nabla_z$ \\
& $A_y\nabla_y - A_z\nabla_z$ \\
& $V_y\nabla_x + V_z\nabla_x - V_x\nabla_y - V_x\nabla_z$ \\
& $V_y\nabla_x + V_z\nabla_x + V_x\nabla_y + V_x\nabla_z$ \\
\bottomrule
\end{tabular}
\addtolength{\tabcolsep}{-6pt}
\label{tab:one-0nn}
\end{table}

\begin{table}[htbp]
\centering
\caption{One-meson operators for point group $C_{3v}$ at total momentum $\vec{P} = [n,n,n]$. The notation follows Table~\ref{tab:one-000}.}
\addtolength{\tabcolsep}{6pt}
\begin{tabular}{cc}
\toprule
Irrep & Operator \\
\midrule
\multirow{6}{*}{$A_1$} & $S$ \\
& $V_x + V_y + V_z$ \\
& $S (\nabla_x + \nabla_y + \nabla_z)$ \\
& $V_x\nabla_x + V_y\nabla_y + V_z\nabla_z$ \\
& $V_y\nabla_x + V_z\nabla_x + V_x\nabla_y + V_z\nabla_y + V_x\nabla_z + V_y\nabla_z$ \\
& $A_y\nabla_x - A_z\nabla_x - A_x\nabla_y + A_z\nabla_y + A_x\nabla_z - A_y\nabla_z$ \\
\midrule
\multirow{6}{*}{$A_2$} & $P$ \\
& $A_x + A_y + A_z$ \\
& $P (\nabla_x + \nabla_y + \nabla_z)$ \\
& $A_x\nabla_x + A_y\nabla_y + A_z\nabla_z$ \\
& $A_y\nabla_x + A_z\nabla_x + A_x\nabla_y + A_z\nabla_y + A_x\nabla_z + A_y\nabla_z$ \\
& $V_y\nabla_x - V_z\nabla_x - V_x\nabla_y + V_z\nabla_y + V_x\nabla_z - V_y\nabla_z$ \\
\midrule
\multirow{10}{*}{$E$} & $V_x - V_y$ \\
& $A_x + A_y -2 A_z$ \\
& $S (\nabla_x - \nabla_y)$ \\
& $P (\nabla_x + \nabla_y - 2 \nabla_z)$ \\
& $-2 V_y\nabla_x - V_z\nabla_x +2 V_x\nabla_y + V_z\nabla_y + V_x\nabla_z - V_y\nabla_z$ \\
& $V_x\nabla_x - V_y\nabla_y$ \\
& $V_z\nabla_x - V_z\nabla_y + V_x\nabla_z - V_y\nabla_z$ \\
& $A_z\nabla_x + A_z\nabla_y - A_x\nabla_z - A_y\nabla_z$ \\
& $2 A_y\nabla_x - A_z\nabla_x +2 A_x\nabla_y - A_z\nabla_y - A_x\nabla_z - A_y\nabla_z$ \\
& $A_x\nabla_x + A_y\nabla_y -2 A_z\nabla_z$ \\
\bottomrule
\end{tabular}
\addtolength{\tabcolsep}{-6pt}
\label{tab:one-nnn}
\end{table}

\begin{table}[htbp]
\centering
\caption{One-meson operators for point group $C_{2}$ at total momentum $\vec{P} = [n,m,0]$. The notation follows Table~\ref{tab:one-000}.}
\addtolength{\tabcolsep}{6pt}
\begin{tabular}{cc}
\toprule
Irrep & Operator \\
\midrule
\multirow{16}{*}{$A$} & $S$ \\
& $V_x$ \\
& $V_y$ \\
& $A_z$ \\
& $S\nabla_x$ \\
& $S\nabla_y$ \\
& $P\nabla_z$ \\
& $V_x\nabla_x + V_y\nabla_y + V_z\nabla_z$ \\
& $V_y\nabla_x$ \\
& $V_x\nabla_x - V_y\nabla_y$ \\
& $V_x\nabla_y$ \\
& $V_z\nabla_z$ \\
& $A_x\nabla_z$ \\
& $A_z\nabla_x$ \\
& $A_y\nabla_z$ \\
& $A_z\nabla_y$ \\
\midrule
\multirow{16}{*}{$B$} & $P$ \\
& $A_x$ \\
& $A_y$ \\
& $V_z$ \\
& $P\nabla_x$ \\
& $P\nabla_y$ \\
& $S\nabla_z$ \\
& $A_x\nabla_x + A_y\nabla_y + A_z\nabla_z$ \\
& $A_y\nabla_x$ \\
& $A_x\nabla_x - A_y\nabla_y$ \\
& $A_x\nabla_y$ \\
& $A_z\nabla_z$ \\
& $V_x\nabla_z$ \\
& $V_z\nabla_x$ \\
& $V_y\nabla_z$ \\
& $V_z\nabla_y$ \\
\bottomrule
\end{tabular}
\addtolength{\tabcolsep}{-6pt}
\label{tab:one-nm0}
\end{table}

\begin{table}[htbp]
\centering
\caption{One-meson operators for point group $C_{2}$ at total momentum $\vec{P} = [n,n,m]$. The notation follows Table~\ref{tab:one-000}.}
\addtolength{\tabcolsep}{6pt}
\begin{tabular}{cc}
\toprule
Irrep & Operator \\
\midrule
\multirow{16}{*}{$A$} & $S$ \\
& $V_x + V_y$ \\
& $V_z$ \\
& $A_x - A_y$ \\
& $S(D_x + D_y)$ \\
& $SD_z$ \\
& $P(D_x - D_y)$ \\
& $V_x\nabla_x + V_y\nabla_y$ \\
& $V_z\nabla_z$ \\
& $V_x\nabla_z + V_y\nabla_z$ \\
& $V_z\nabla_x + V_z\nabla_y$ \\
& $V_y\nabla_x + V_x\nabla_y$ \\
& $A_z\nabla_x - A_z\nabla_y$ \\
& $A_x\nabla_z - A_y\nabla_z$ \\
& $A_y\nabla_x - A_x\nabla_y$ \\
& $A_x\nabla_x - A_y\nabla_y$ \\
\midrule
\multirow{16}{*}{$B$} & $P$ \\
& $A_x + A_y$ \\
& $A_z$ \\
& $V_x - V_y$ \\
& $P(D_x + D_y)$ \\
& $PD_z$ \\
& $S(D_x - D_y)$ \\
& $A_x\nabla_x + A_y\nabla_y$ \\
& $A_z\nabla_z$ \\
& $A_x\nabla_z + A_y\nabla_z$ \\
& $A_z\nabla_x + A_z\nabla_y$ \\
& $A_y\nabla_x + A_x\nabla_y$ \\
& $V_z\nabla_x - V_z\nabla_y$ \\
& $V_x\nabla_z - V_y\nabla_z$ \\
& $V_y\nabla_x - V_x\nabla_y$ \\
& $V_x\nabla_x - V_y\nabla_y$ \\
\bottomrule
\end{tabular}
\addtolength{\tabcolsep}{-6pt}
\label{tab:one-nnm}
\end{table}

\begin{table}[htbp]
\centering
\caption{One-meson operators for point group $C_{1}$ at total momentum $\vec{P} = [n,m,p]$. The notation follows Table~\ref{tab:one-000}.}
\addtolength{\tabcolsep}{6pt}
\begin{tabular}{cc}
\toprule
Irrep & Operator \\
\midrule
\multirow{4}{*}{$A$} & $S, P, V_x, V_y, V_z, A_x, A_y, A_z$ \\
& $S\nabla_x, S\nabla_y, S\nabla_z, P\nabla_x, P\nabla_y, P\nabla_z$ \\
& $V_x\nabla_x, V_y\nabla_y, V_z\nabla_z, V_x\nabla_y, V_y\nabla_x, V_x\nabla_z, V_z\nabla_x, V_y\nabla_z, V_z\nabla_y$ \\
& $A_x\nabla_x, A_y\nabla_y, A_z\nabla_z, A_x\nabla_y, A_y\nabla_x, A_x\nabla_z, A_z\nabla_x, A_y\nabla_z, A_z\nabla_y$ \\
\bottomrule
\end{tabular}
\addtolength{\tabcolsep}{-6pt}
\label{tab:one-nmp}
\end{table}
    
\section{Meson-Meson Operator Lists}
\label{sec:append_list_mm}
Table~\ref{tab:mm-000-A1+} gives the complete set of meson-meson operators in the rest frame belonging to the $A_1^+$ irrep of the cubic group $O_h$, constructed with relative momenta up to one unit in $2\pi/L$. The table assumes that the hadrons are nonidentical.

Meson-meson operators containing all $P,S,V,A$ structures and satisfying $|\vec{P}|^2 \leq 2$ are listed in Tables~\ref{tab:mm-000-A1+}, \ref{tab:mm-000-A1-}, \ref{tab:mm-000-E+}, \ref{tab:mm-000-E-}, \ref{tab:mm-000-T1+}, \ref{tab:mm-000-T1-}, \ref{tab:mm-000-T2+}, and \ref{tab:mm-000-T2-}.

\begin{table}[htbp]
\centering
\caption{Meson-meson operator list for the $A_1^+$ irrep of $O_h$. The gamma-matrix structures $S$, $P$, $V$, and $A$ denote scalar, pseudoscalar, vector, and axial-vector one-body operators, respectively. For compactness, left-right arrows on covariant derivatives and overall normalization factors are omitted.}
\addtolength{\tabcolsep}{6pt}
\begin{tabular}{c}
\toprule
$P_1(0) P_2(0)$ \\
\midrule
$P_1(e_{x}) P_2(-e_{x}) + P_1(-e_{x}) P_2(e_{x}) + P_1(e_{y}) P_2(-e_{y}) + P_1(-e_{y}) P_2(e_{y})$ \\
$+ P_1(e_{z}) P_2(-e_{z}) + P_1(-e_{z}) P_2(e_{z})$ \\
\midrule
$P_2(e_{x}) A_{1,x}(-e_{x}) - P_2(-e_{x}) A_{1,x}(e_{x}) + P_2(e_{y}) A_{1,y}(-e_{y}) - P_2(-e_{y}) A_{1,y}(e_{y})$ \\
$+ P_2(e_{z}) A_{1,z}(-e_{z}) - P_2(-e_{z}) A_{1,z}(e_{z})$ \\
\midrule
$S_1(0) S_2(0)$ \\
\midrule
$S_1(e_{x}) S_2(-e_{x}) + S_1(-e_{x}) S_2(e_{x}) + S_1(e_{y}) S_2(-e_{y}) + S_1(-e_{y}) S_2(e_{y})$ \\
$+ S_1(e_{z}) S_2(-e_{z}) + S_1(-e_{z}) S_2(e_{z})$ \\
\bottomrule
\end{tabular}
\addtolength{\tabcolsep}{-6pt}
\label{tab:mm-000-A1+}
\end{table}
    
\begin{table}[htbp]
\centering
\caption{Same as Table~\ref{tab:mm-000-A1+}, but for the $A_1^-$ irrep of $O_h$.}
\addtolength{\tabcolsep}{6pt}
\begin{tabular}{c}
\toprule
$P_1[0] S_2[0]$ \\
\midrule
$P_1[e_x] S_2[-e_x] + P_1[-e_x] S_2[e_x] + P_1[e_y] S_2[-e_y] + P_1[-e_y] S_2[e_y]$ \\
$+ P_1[e_z] S_2[-e_z] + P_1[-e_z] S_2[e_z]$ \\
\midrule
$P_1[e_x] V_{2,x}[-e_x] - P_1[-e_x] V_{2,x}[e_x] + P_1[e_y] V_{2,y}[-e_y] - P_1[-e_y] V_{2,y}[e_y]$ \\
$+ P_1[e_z] V_{2,z}[-e_z] - P_1[-e_z] V_{2,z}[e_z]$ \\
\midrule
$S_2[e_x] A_{1,x}[-e_x] - S_2[-e_x] A_{1,x}[e_x] + S_2[e_y] A_{1,y}[-e_y] - S_2[-e_y] A_{1,y}[e_y]$ \\
$+ S_2[e_z] A_{1,z}[-e_z] - S_2[-e_z] A_{1,z}[e_z]$ \\
\midrule
$V_{1,z}[e_y] V_{2,x}[-e_y] - V_{1,z}[-e_y] V_{2,x}[e_y] - V_{1,y}[e_z] V_{2,x}[-e_z] + V_{1,y}[-e_z] V_{2,x}[e_z]$ \\
\bottomrule
\end{tabular}
\addtolength{\tabcolsep}{-6pt}
\label{tab:mm-000-A1-}
\end{table}

\begin{table}[htbp]
\centering
\small
\caption{Same as Table~\ref{tab:mm-000-A1+}, but for the $E^+$ irrep of $O_h$.}
\addtolength{\tabcolsep}{6pt}
\begin{tabular}{c}
\toprule
$P_1[e_x] P_2[-e_x] + P_1[-e_x] P_2[e_x] + P_1[e_y] P_2[-e_y] + P_1[-e_y] P_2[e_y]$ \\ 
$-2 P_1[e_z] P_2[-e_z] -2 P_1[-e_z] P_2[e_z]$ \\ 
\midrule
$P_2[e_x] A_{1,x}[-e_x] - P_2[-e_x] A_{1,x}[e_x] + P_2[e_y] A_{1,y}[-e_y] - P_2[-e_y] A_{1,y}[e_y]$ \\ 
$-2 P_2[e_z] A_{1,z}[-e_z] + 2 P_2[-e_z] A_{1,z}[e_z]$ \\ 
\midrule
$S_1[e_x] S_2[-e_x] + S_1[-e_x] S_2[e_x] + S_1[e_y] S_2[-e_y] + S_1[-e_y] S_2[e_y]$ \\ 
$-2 S_1[e_z] S_2[-e_z] -2 S_1[-e_z] S_2[e_z]$ \\ 
\midrule
$S_1[e_x] V_{2,x}[-e_x] - S_1[-e_x] V_{2,x}[e_x] + S_1[e_y] V_{2,y}[-e_y] - S_1[-e_y] V_{2,y}[e_y]$ \\ 
$-2 S_1[e_z] V_{2,z}[-e_z] + 2 S_1[-e_z] V_{2,z}[e_z]$ \\
\bottomrule
\end{tabular}
\addtolength{\tabcolsep}{-6pt}
\label{tab:mm-000-E+}
\end{table}

\begin{table}[htbp]
\centering
\caption{Same as Table~\ref{tab:mm-000-A1+}, but for the $E^-$ irrep of $O_h$.}
\addtolength{\tabcolsep}{6pt}
\begin{tabular}{c}
\toprule
$P_1(e_x) S_2(-e_x) P_1(-e_x) S_2(e_x) P_1(e_y) S_2(-e_y) P_1(-e_y) S_2(e_y)$ \\
$-2 P_1(e_z) S_2(-e_z) -2 P_1(-e_z) S_2(e_z)$ \\
\midrule
$P_1(e_x) V_{2,x}(-e_x) -P_1(-e_x) V_{2,x}(e_x) P_1(e_y) V_{2,y}(-e_y) -P_1(-e_y) V_{2,y}(e_y)$ \\
$-2 P_1(e_z) V_{2,z}(-e_z) +2 P_1(-e_z) V_{2,z}(e_z)$ \\
\midrule
$S_2(e_x) A_{1,x}(-e_x) -S_2(-e_x) A_{1,x}(e_x) S_2(e_y) A_{1,y}(-e_y) -S_2(-e_y) A_{1,y}(e_y)$ \\
$-2 S_2(e_z) A_{1,z}(-e_z) +2 S_2(-e_z) A_{1,z}(e_z)$ \\
\midrule
$-V_{1,z}(e_y) V_{2,x}(-e_y) +V_{1,z}(-e_y) V_{2,x}(e_y) +V_{1,y}(e_z) V_{2,x}(-e_z) -V_{1,y}(-e_z) V_{2,x}(e_z)$ \\
$+V_{1,z}(e_x) V_{2,y}(-e_x) -V_{1,z}(-e_x) V_{2,y}(e_x) -V_{1,x}(e_z) V_{2,y}(-e_z) +V_{1,x}(-e_z) V_{2,y}(e_z)$ \\
\bottomrule
\end{tabular}
\addtolength{\tabcolsep}{-6pt}
\label{tab:mm-000-E-}
\end{table}

\begin{table}[htbp]
\centering
\caption{Same as Table~\ref{tab:mm-000-A1+}, but for the $T_1^+$ irrep of $O_h$.}
\addtolength{\tabcolsep}{6pt}
\begin{tabular}{c}
\toprule
$P_1(e_x) S_2(-e_x) -P_1(-e_x) S_2(e_x)$ \\
\midrule
$P_1(0) V_{2,x}(0)$ \\
\midrule
$P_1(e_x) V_{2,x}(-e_x) +P_1(-e_x) V_{2,x}(e_x)$ \\
\midrule
$P_1(e_y) V_{2,x}(-e_y) +P_1(-e_y) V_{2,x}(e_y) +P_1(e_z) V_{2,x}(-e_z) +P_1(-e_z) V_{2,x}(e_z)$ \\
\midrule
$P_2(e_z) A_{1,y}(-e_z) -P_2(-e_z) A_{1,y}(e_z) -P_2(e_y) A_{1,z}(-e_y) +P_2(-e_y) A_{1,z}(e_y)$ \\
\midrule
$S_1(e_z) V_{2,y}(-e_z) -S_1(-e_z) V_{2,y}(e_z) -S_1(e_y) V_{2,z}(-e_y) +S_1(-e_y) V_{2,z}(e_y)$ \\
\midrule
$S_2(0) A_{1,x}(0)$ \\
\midrule
$S_2(e_x) A_{1,x}(-e_x) +S_2(-e_x) A_{1,x}(e_x)$ \\
\bottomrule
\end{tabular}
\addtolength{\tabcolsep}{-6pt}
\label{tab:mm-000-T1+}
\end{table}

\begin{table}[htbp]
\centering
\caption{Same as Table~\ref{tab:mm-000-A1+}, but for the $T_1^-$ irrep of $O_h$.}
\addtolength{\tabcolsep}{6pt}
\begin{tabular}{c}
\toprule
$P_1(e_x) P_2(-e_x) -P_1(-e_x) P_2(e_x)$ \\
\midrule
$P_1(e_z) V_{2,y}(-e_z) -P_1(-e_z) V_{2,y}(e_z) -P_1(e_y) V_{2,z}(-e_y) +P_1(-e_y) V_{2,z}(e_y)$ \\
\midrule
$P_2(0) A_{1,x}(0)$ \\
\midrule
$P_2(e_x) A_{1,x}(-e_x) +P_2(-e_x) A_{1,x}(e_x)$ \\
\midrule
$P_2(e_y) A_{1,x}(-e_y) +P_2(-e_y) A_{1,x}(e_y) +P_2(e_z) A_{1,x}(-e_z) +P_2(-e_z) A_{1,x}(e_z)$ \\
\midrule
$S_1(e_x) S_2(-e_x) -S_1(-e_x) S_2(e_x)$ \\
\midrule
$S_1(0) V_{2,x}(0)$ \\
\midrule
$S_1(e_x) V_{2,x}(-e_x) +S_1(-e_x) V_{2,x}(e_x)$ \\
\bottomrule
\end{tabular}
\addtolength{\tabcolsep}{-6pt}
\label{tab:mm-000-T1-}
\end{table}

\begin{table}[htbp]
\centering
\caption{Same as Table~\ref{tab:mm-000-A1+}, but for the $T_2^+$ irrep of $O_h$.}
\addtolength{\tabcolsep}{6pt}
\begin{tabular}{c}
\toprule
$P_1(e_y) V_{2,x}(-e_y) + P_1(-e_y) V_{2,x}(e_y) - P_1(e_z) V_{2,x}(-e_z) - P_1(-e_z) V_{2,x}(e_z)$ \\
\midrule
$P_2(e_z) A_{1,y}(-e_z) - P_2(-e_z) A_{1,y}(e_z) + P_2(e_y) A_{1,z}(-e_y) - P_2(-e_y) A_{1,z}(e_y)$ \\
\midrule
$S_1(e_z) V_{2,y}(-e_z) - S_1(-e_z) V_{2,y}(e_z) + S_1(e_y) V_{2,z}(-e_y) - S_1(-e_y) V_{2,z}(e_y)$ \\
\midrule
$S_2(e_y) A_{1,x}(-e_y) + S_2(-e_y) A_{1,x}(e_y) - S_2(e_z) A_{1,x}(-e_z) - S_2(-e_z) A_{1,x}(e_z)$ \\
\midrule
$V_{1,z}(0) V_{2,y}(0) + V_{1,y}(0) V_{2,z}(0)$ \\
\midrule
$V_{1,z}(e_x) V_{2,y}(-e_x) + V_{1,z}(-e_x) V_{2,y}(e_x) + V_{1,y}(e_x) V_{2,z}(-e_x) + V_{1,y}(-e_x) V_{2,z}(e_x)$ \\
\midrule
$V_{1,z}(e_z) V_{2,y}(-e_z) + V_{1,z}(-e_z) V_{2,y}(e_z) + V_{1,y}(e_y) V_{2,z}(-e_y) + V_{1,y}(-e_y) V_{2,z}(e_y)$ \\
\midrule
$V_{1,z}(e_y) V_{2,y}(-e_y) + V_{1,z}(-e_y) V_{2,y}(e_y) + V_{1,y}(e_z) V_{2,z}(-e_z) + V_{1,y}(-e_z) V_{2,z}(e_z)$ \\
\bottomrule
\end{tabular}
\addtolength{\tabcolsep}{-6pt}
\label{tab:mm-000-T2+}
\end{table}

\begin{table}[htbp]
\centering
\caption{Same as Table~\ref{tab:mm-000-A1+}, but for the $T_2^-$ irrep of $O_h$.}
\addtolength{\tabcolsep}{6pt}
\begin{tabular}{c}
\toprule
$P_1(e_z) V_{2,y}(-e_z) -P_1(-e_z) V_{2,y}(e_z) +P_1(e_y) V_{2,z}(-e_y) -P_1(-e_y) V_{2,z}(e_y)$ \\
\midrule
$P_2(e_y) A_{1,x}(-e_y) +P_2(-e_y) A_{1,x}(e_y) -P_2(e_z) A_{1,x}(-e_z) -P_2(-e_z) A_{1,x}(e_z)$ \\
\midrule
$S_1(e_y) V_{2,x}(-e_y) +S_1(-e_y) V_{2,x}(e_y) -S_1(e_z) V_{2,x}(-e_z) -S_1(-e_z) V_{2,x}(e_z)$ \\
\midrule
$S_2(e_z) A_{1,y}(-e_z) -S_2(-e_z) A_{1,y}(e_z) +S_2(e_y) A_{1,z}(-e_y) -S_2(-e_y) A_{1,z}(e_y)$ \\
\midrule
$V_{1,y}(e_x) V_{2,y}(-e_x) -V_{1,y}(-e_x) V_{2,y}(e_x) -V_{1,z}(e_x) V_{2,z}(-e_x) +V_{1,z}(-e_x) V_{2,z}(e_x)$ \\
\midrule
$V_{1,y}(e_y) V_{2,x}(-e_y) -V_{1,y}(-e_y) V_{2,x}(e_y) -V_{1,z}(e_z) V_{2,x}(-e_z) +V_{1,z}(-e_z) V_{2,x}(e_z)$ \\
$+V_{1,x}(e_y) V_{2,y}(-e_y) -V_{1,x}(-e_y) V_{2,y}(e_y) -V_{1,x}(e_z) V_{2,z}(-e_z) +V_{1,x}(-e_z) V_{2,z}(e_z)$ \\
\midrule
$V_{1,y}(e_y) V_{2,x}(-e_y) -V_{1,y}(-e_y) V_{2,x}(e_y) -V_{1,z}(e_z) V_{2,x}(-e_z) +V_{1,z}(-e_z) V_{2,x}(e_z)$ \\
\bottomrule
\end{tabular}
\addtolength{\tabcolsep}{-6pt}
\label{tab:mm-000-T2-}
\end{table}

\begin{table}[htbp]
\centering
\caption{Same as Table~\ref{tab:mm-000-A1+}, but for the $A_1$ irrep of the little group associated with $\vec{P}=[0,0,n]$.}
\addtolength{\tabcolsep}{6pt}
\begin{tabular}{c}
\toprule
$P_1(0) P_2(e_z)$ \\
\midrule
$P_1(e_z) P_2(0)$ \\
\midrule
$P_2(e_z) A_{1,z}(0)$ \\
\midrule
$P_2(0) A_{1,z}(e_z)$ \\
\midrule
$S_1(0) S_2(e_z)$ \\
\midrule
$S_1(e_z) S_2(0)$ \\
\midrule
$S_1(0) V_{2,z}(e_z)$ \\
\midrule
$S_1(e_z) V_{2,z}(0)$ \\
\midrule
$V_{1,x}(0) V_{2,x}(e_z) + V_{1,y}(0) V_{2,y}(e_z)$ \\
\midrule
$V_{1,z}(0) V_{2,z}(e_z)$ \\
\midrule
$V_{1,x}(e_z) V_{2,x}(0) + V_{1,y}(e_z) V_{2,y}(0)$ \\
\midrule
$V_{1,z}(e_z) V_{2,z}(0)$ \\
\midrule
$-A_{1,y}(0) V_{2,x}(e_z) + A_{1,x}(0) V_{2,y}(e_z)$ \\
\midrule
$-A_{1,y}(e_z) V_{2,x}(0) + A_{1,x}(e_z) V_{2,y}(0)$ \\
\midrule
$A_{1,x}(0) A_{2,x}(e_z) + A_{1,y}(0) A_{2,y}(e_z)$ \\
\midrule
$A_{1,z}(0) A_{2,z}(e_z)$ \\
\midrule
$A_{1,x}(e_z) A_{2,x}(0) + A_{1,y}(e_z) A_{2,y}(0)$ \\
\midrule
$A_{1,z}(e_z) A_{2,z}(0)$ \\
\bottomrule
\end{tabular}
\addtolength{\tabcolsep}{-6pt}
\label{tab:mm-00n-A1}
\end{table}

\begin{table}[htbp]
\centering
\caption{Same as Table~\ref{tab:mm-000-A1+}, but for the $A_2$ irrep of the little group associated with $\vec{P}=[0,0,n]$.}
\addtolength{\tabcolsep}{6pt}
\begin{tabular}{c}
\toprule
$P_1(0) S_2(e_z)$ \\
\midrule
$P_1(e_z) S_2(0)$ \\
\midrule
$P_1(0) V_{2,z}(e_z) $ \\
\midrule
$P_1(e_z) V_{2,z}(0)$ \\
\midrule
$S_2(e_z) A_{1,z}(0)$ \\
\midrule
$S_2(0) A_{1,z}(e_z)$ \\
\midrule
$-V_{1,y}(0) V_{2,x}(e_z) + V_{1,x}(0) V_{2,y}(e_z)$ \\
\midrule
$-V_{1,y}(e_z) V_{2,x}(0) + V_{1,x}(e_z) V_{2,y}(0)$ \\
\midrule
$A_{1,x}(0) V_{2,x}(e_z) + A_{1,y}(0) V_{2,y}(e_z)$ \\
\midrule
$A_{1,z}(0) V_{2,z}(e_z)$ \\
\midrule
$A_{1,x}(e_z) V_{2,x}(0) + A_{1,y}(e_z) V_{2,y}(0)$ \\
\midrule
$A_{1,z}(e_z) V_{2,z}(0)$ \\
\midrule
$-A_{1,y}(0) A_{2,x}(e_z) + A_{1,x}(0) A_{2,y}(e_z)$ \\
\midrule
$-A_{1,y}(e_z) A_{2,x}(0) + A_{1,x}(e_z) A_{2,y}(0)$ \\
\bottomrule
\end{tabular}
\addtolength{\tabcolsep}{-6pt}
\label{tab:mm-00n-A2}
\end{table}

\begin{table}[htbp]
\centering
\caption{Same as Table~\ref{tab:mm-000-A1+}, but for the $B_1$ irrep of the little group associated with $\vec{P}=[0,0,n]$.}
\addtolength{\tabcolsep}{6pt}
\begin{tabular}{c}
\toprule
$V_{1,x}(0) V_{2,x}(e_z) -V_{1,y}(0) V_{2,y}(e_z)$ \\
\midrule
$V_{1,x}(e_z) V_{2,x}(0) -V_{1,y}(e_z) V_{2,y}(0)$ \\
\midrule
$A_{1,y}(0) V_{2,x}(e_z) +A_{1,x}(0) V_{2,y}(e_z)$ \\
\midrule
$A_{1,y}(e_z) V_{2,x}(0) +A_{1,x}(e_z) V_{2,y}(0)$ \\
\midrule
$A_{1,x}(0) A_{2,x}(e_z) -A_{1,y}(0) A_{2,y}(e_z)$ \\
\midrule
$A_{1,x}(e_z) A_{2,x}(0) -A_{1,y}(e_z) A_{2,y}(0)$ \\
\bottomrule
\end{tabular}
\addtolength{\tabcolsep}{-6pt}
\label{tab:mm-00n-B1}
\end{table}

\begin{table}[htbp]
\centering
\caption{Same as Table~\ref{tab:mm-000-A1+}, but for the $B_2$ irrep of the little group associated with $\vec{P}=[0,0,n]$.}
\addtolength{\tabcolsep}{6pt}
\begin{tabular}{c}
\toprule
$V_{1,y}(0) V_{2,x}(e_z) +V_{1,x}(0) V_{2,y}(e_z)$ \\
\midrule
$V_{1,y}(e_z) V_{2,x}(0) +V_{1,x}(e_z) V_{2,y}(0)$ \\
\midrule
$A_{1,x}(0) V_{2,x}(e_z) -A_{1,y}(0) V_{2,y}(e_z)$ \\
\midrule
$A_{1,x}(e_z) V_{2,x}(0) -A_{1,y}(e_z) V_{2,y}(0)$ \\
\midrule
$A_{1,y}(0) A_{2,x}(e_z) +A_{1,x}(0) A_{2,y}(e_z)$ \\
\midrule
$A_{1,y}(e_z) A_{2,x}(0) +A_{1,x}(e_z) A_{2,y}(0)$ \\
\bottomrule
\end{tabular}
\addtolength{\tabcolsep}{-6pt}
\label{tab:mm-00n-B2}
\end{table}

\begin{table}[htbp]
\centering
\caption{Same as Table~\ref{tab:mm-000-A1+}, but for the $E$ irrep of the little group associated with $\vec{P}=[0,0,n]$.}
\addtolength{\tabcolsep}{6pt}
\begin{tabular}{c}
\toprule
$P_1(0) V_{2,x}(e_z)$ \\
\midrule
$P_1(e_z) V_{2,x}(0)$ \\
\midrule
$P_2(e_z) A_{1,y}(0)$ \\
\midrule
$P_2(0) A_{1,y}(e_z)$ \\
\midrule
$S_1(0) V_{2,y}(e_z)$ \\
\midrule
$S_1(e_z) V_{2,y}(0)$ \\
\midrule
$S_2(e_z) A_{1,x}(0)$ \\
\midrule
$S_2(0) A_{1,x}(e_z)$ \\
\midrule
$V_{1,y}(0) V_{2,z}(e_z)$ \\
\midrule
$V_{1,z}(0) V_{2,y}(e_z)$ \\
\midrule
$V_{1,y}(e_z) V_{2,z}(0)$ \\
\midrule
$V_{1,z}(e_z) V_{2,y}(0)$ \\
\midrule
$A_{1,x}(0) V_{2,z}(e_z)$ \\
\midrule
$A_{1,z}(0) V_{2,x}(e_z)$ \\
\midrule
$A_{1,x}(e_z) V_{2,z}(0)$ \\
\midrule
$A_{1,z}(e_z) V_{2,x}(0)$ \\
\midrule
$A_{1,y}(0) A_{2,z}(e_z)$ \\
\midrule
$A_{1,z}(0) A_{2,y}(e_z)$ \\
\midrule
$A_{1,y}(e_z) A_{2,z}(0)$ \\
\midrule
$A_{1,z}(e_z) A_{2,y}(0)$ \\
\midrule
$A_{1,x}(e_z) V_{2,x}(0) -A_{1,y}(e_z) V_{2,y}(0)$ \\
\midrule
$A_{1,y}(0) A_{2,x}(e_z) +A_{1,x}(0) A_{2,y}(e_z)$ \\
\midrule
$A_{1,y}(e_z) A_{2,x}(0) +A_{1,x}(e_z) A_{2,y}(0)$ \\
\bottomrule
\end{tabular}
\addtolength{\tabcolsep}{-6pt}
\label{tab:mm-00n-E}
\end{table}

\begin{table}[htbp]
\centering
\caption{Same as Table~\ref{tab:mm-000-A1+}, but for the $A_1$ irrep of the little group associated with $\vec{P}=[0,n,n]$.}
\addtolength{\tabcolsep}{6pt}
\begin{tabular}{c}
\toprule
$P_1(e_z) P_2(e_y) +P_1(e_y) P_2(e_z)$ \\
\midrule
$P_1(e_z) V_{2,x}(e_y) -P_1(e_y) V_{2,x}(e_z)$ \\
\midrule
$P_2(e_y) A_{1,y}(e_z) +P_2(e_z) A_{1,z}(e_y)$ \\
\midrule
$P_2(e_z) A_{1,y}(e_y) +P_2(e_y) A_{1,z}(e_z)$ \\
\midrule
$S_1(e_z) S_2(e_y) +S_1(e_y) S_2(e_z)$ \\
\midrule
$S_1(e_z) V_{2,y}(e_y) +S_1(e_y) V_{2,z}(e_z)$ \\
\midrule
$S_1(e_y) V_{2,y}(e_z) +S_1(e_z) V_{2,z}(e_y)$ \\
\midrule
$-S_2(e_z) A_{1,x}(e_y) +S_2(e_y) A_{1,x}(e_z)$ \\
\midrule
$V_{1,x}(e_z) V_{2,x}(e_y) +V_{1,x}(e_y) V_{2,x}(e_z) -V_{1,y}(e_z) V_{2,y}(e_y) -V_{1,z}(e_y) V_{2,z}(e_z)$ \\
\midrule
$V_{1,z}(e_y) V_{2,y}(e_z) +V_{1,y}(e_z) V_{2,z}(e_y)$ \\
\midrule
$-V_{1,x}(e_z) V_{2,x}(e_y) -V_{1,x}(e_y) V_{2,x}(e_z) -V_{1,y}(e_z) V_{2,y}(e_y) -V_{1,z}(e_y) V_{2,z}(e_z)$ \\
\midrule
$V_{1,z}(e_z) V_{2,y}(e_y) +V_{1,y}(e_y) V_{2,z}(e_z)$ \\
\midrule
$V_{1,y}(e_y) V_{2,y}(e_z) +V_{1,z}(e_z) V_{2,z}(e_y)$ \\
\midrule
$A_{1,y}(e_z) V_{2,x}(e_y) -A_{1,z}(e_y) V_{2,x}(e_z) +A_{1,x}(e_z) V_{2,y}(e_y) -A_{1,x}(e_y) V_{2,z}(e_z)$ \\
\midrule
$A_{1,x}(e_y) V_{2,y}(e_z) -A_{1,x}(e_z) V_{2,z}(e_y)$ \\
\midrule
$A_{1,y}(e_z) V_{2,x}(e_y) -A_{1,z}(e_y) V_{2,x}(e_z) -A_{1,x}(e_z) V_{2,y}(e_y) +A_{1,x}(e_y) V_{2,z}(e_z)$ \\
\midrule
$A_{1,z}(e_z) V_{2,x}(e_y) -A_{1,y}(e_y) V_{2,x}(e_z)$ \\
\midrule
$A_{1,x}(e_z) A_{2,x}(e_y) +A_{1,x}(e_y) A_{2,x}(e_z) -A_{1,y}(e_z) A_{2,y}(e_y) -A_{1,z}(e_y) A_{2,z}(e_z)$ \\
\midrule
$A_{1,z}(e_y) A_{2,y}(e_z) +A_{1,y}(e_z) A_{2,z}(e_y)$ \\
\midrule
$-A_{1,x}(e_z) A_{2,x}(e_y) -A_{1,x}(e_y) A_{2,x}(e_z) -A_{1,y}(e_z) A_{2,y}(e_y) -A_{1,z}(e_y) A_{2,z}(e_z)$ \\
\midrule
$A_{1,z}(e_z) A_{2,y}(e_y) +A_{1,y}(e_y) A_{2,z}(e_z)$ \\
\midrule
$A_{1,y}(e_y) A_{2,y}(e_z) +A_{1,z}(e_z) A_{2,z}(e_y)$ \\
\bottomrule
\end{tabular}
\addtolength{\tabcolsep}{-6pt}
\label{tab:mm-0nn-A1}
\end{table}

\begin{table}[htbp]
\centering
\caption{Same as Table~\ref{tab:mm-000-A1+}, but for the $A_2$ irrep of the little group associated with $\vec{P}=[0,n,n]$.}
\addtolength{\tabcolsep}{6pt}
\begin{tabular}{c}
\toprule
$P_1(e_z) S_2(e_y) +P_1(e_y) S_2(e_z)$ \\
\midrule
$P_1(e_z) V_{2,y}(e_y) +P_1(e_y) V_{2,z}(e_z)$ \\
\midrule
$P_1(e_y) V_{2,y}(e_z) +P_1(e_z) V_{2,z}(e_y)$ \\
\midrule
$-P_2(e_z) A_{1,x}(e_y) +P_2(e_y) A_{1,x}(e_z)$ \\
\midrule
$S_1(e_z) V_{2,x}(e_y) -S_1(e_y) V_{2,x}(e_z)$ \\
\midrule
$S_2(e_y) A_{1,y}(e_z) +S_2(e_z) A_{1,z}(e_y)$ \\
\midrule
$S_2(e_z) A_{1,y}(e_y) +S_2(e_y) A_{1,z}(e_z)$ \\
\midrule
$V_{1,y}(e_z) V_{2,x}(e_y) -V_{1,z}(e_y) V_{2,x}(e_z)$ \\
$+V_{1,x}(e_z) V_{2,y}(e_y) -V_{1,x}(e_y) V_{2,z}(e_z)$ \\
\midrule
$V_{1,x}(e_y) V_{2,y}(e_z) -V_{1,x}(e_z) V_{2,z}(e_y)$ \\
\midrule
$V_{1,y}(e_z) V_{2,x}(e_y) -V_{1,z}(e_y) V_{2,x}(e_z)$ \\
$-V_{1,x}(e_z) V_{2,y}(e_y) +V_{1,x}(e_y) V_{2,z}(e_z)$ \\
\midrule
$V_{1,z}(e_z) V_{2,x}(e_y) -V_{1,y}(e_y) V_{2,x}(e_z)$ \\
\midrule
$A_{1,x}(e_z) V_{2,x}(e_y) +A_{1,x}(e_y) V_{2,x}(e_z)$ \\
$-A_{1,y}(e_z) V_{2,y}(e_y) -A_{1,z}(e_y) V_{2,z}(e_z)$ \\
\midrule
$A_{1,z}(e_y) V_{2,y}(e_z) +A_{1,y}(e_z) V_{2,z}(e_y)$ \\
\midrule
$-A_{1,x}(e_z) V_{2,x}(e_y) -A_{1,x}(e_y) V_{2,x}(e_z)$ \\
$-A_{1,y}(e_z) V_{2,y}(e_y) -A_{1,z}(e_y) V_{2,z}(e_z)$ \\
\midrule
$A_{1,z}(e_z) V_{2,y}(e_y) +A_{1,y}(e_y) V_{2,z}(e_z)$ \\
\midrule
$A_{1,y}(e_y) V_{2,y}(e_z) +A_{1,z}(e_z) V_{2,z}(e_y)$ \\
\midrule
$A_{1,y}(e_z) A_{2,x}(e_y) -A_{1,z}(e_y) A_{2,x}(e_z)$ \\
$+A_{1,x}(e_z) A_{2,y}(e_y) -A_{1,x}(e_y) A_{2,z}(e_z)$ \\
\midrule
$A_{1,x}(e_y) A_{2,y}(e_z) -A_{1,x}(e_z) A_{2,z}(e_y)$ \\
\midrule
$A_{1,y}(e_z) A_{2,x}(e_y) -A_{1,z}(e_y) A_{2,x}(e_z)$ \\
$-A_{1,x}(e_z) A_{2,y}(e_y) +A_{1,x}(e_y) A_{2,z}(e_z)$ \\
\midrule
$A_{1,z}(e_z) A_{2,x}(e_y) -A_{1,y}(e_y) A_{2,x}(e_z)$ \\
\bottomrule
\end{tabular}
\addtolength{\tabcolsep}{-6pt}
\label{tab:mm-0nn-A2}
\end{table}

\begin{table}[htbp]
\centering
\caption{Same as Table~\ref{tab:mm-000-A1+}, but for the $B_1$ irrep of the little group associated with $\vec{P}=[0,n,n]$.}
\addtolength{\tabcolsep}{6pt}
\begin{tabular}{c}
\toprule
$P_1(e_z) P_2(e_y) -P_1(e_y) P_2(e_z)$ \\
\midrule
$P_1(e_z) V_{2,x}(e_y) +P_1(e_y) V_{2,x}(e_z)$ \\
\midrule
$P_2(e_y) A_{1,y}(e_z) -P_2(e_z) A_{1,z}(e_y)$ \\
\midrule
$P_2(e_z) A_{1,y}(e_y) -P_2(e_y) A_{1,z}(e_z)$ \\
\midrule
$S_1(e_z) S_2(e_y) -S_1(e_y) S_2(e_z)$ \\
\midrule
$S_1(e_z) V_{2,y}(e_y) -S_1(e_y) V_{2,z}(e_z)$ \\
\midrule
$S_1(e_y) V_{2,y}(e_z) -S_1(e_z) V_{2,z}(e_y)$ \\
\midrule
$S_2(e_z) A_{1,x}(e_y) +S_2(e_y) A_{1,x}(e_z)$ \\
\midrule
$V_{1,x}(e_z) V_{2,x}(e_y) -V_{1,x}(e_y) V_{2,x}(e_z)$ \\
$-V_{1,y}(e_z) V_{2,y}(e_y) +V_{1,z}(e_y) V_{2,z}(e_z)$ \\
\midrule
$V_{1,z}(e_y) V_{2,y}(e_z) -V_{1,y}(e_z) V_{2,z}(e_y)$ \\
\midrule
$-V_{1,x}(e_z) V_{2,x}(e_y) +V_{1,x}(e_y) V_{2,x}(e_z)$ \\
$-V_{1,y}(e_z) V_{2,y}(e_y) +V_{1,z}(e_y) V_{2,z}(e_z)$ \\
\midrule
$V_{1,z}(e_z) V_{2,y}(e_y) -V_{1,y}(e_y) V_{2,z}(e_z)$ \\
\midrule
$V_{1,y}(e_y) V_{2,y}(e_z) -V_{1,z}(e_z) V_{2,z}(e_y)$ \\
\midrule
$A_{1,y}(e_z) V_{2,x}(e_y) +A_{1,z}(e_y) V_{2,x}(e_z)$ \\
$+A_{1,x}(e_z) V_{2,y}(e_y) +A_{1,x}(e_y) V_{2,z}(e_z)$ \\
\midrule
$A_{1,x}(e_y) V_{2,y}(e_z) +A_{1,x}(e_z) V_{2,z}(e_y)$ \\
\midrule
$A_{1,y}(e_z) V_{2,x}(e_y) +A_{1,z}(e_y) V_{2,x}(e_z)$ \\
$-A_{1,x}(e_z) V_{2,y}(e_y) -A_{1,x}(e_y) V_{2,z}(e_z)$ \\
\midrule
$A_{1,z}(e_z) V_{2,x}(e_y) +A_{1,y}(e_y) V_{2,x}(e_z)$ \\
\midrule
$A_{1,x}(e_z) A_{2,x}(e_y) -A_{1,x}(e_y) A_{2,x}(e_z)$ \\
$-A_{1,y}(e_z) A_{2,y}(e_y) +A_{1,z}(e_y) A_{2,z}(e_z)$ \\
\midrule
$A_{1,z}(e_y) A_{2,y}(e_z) -A_{1,y}(e_z) A_{2,z}(e_y)$ \\
\midrule
$-A_{1,x}(e_z) A_{2,x}(e_y) +A_{1,x}(e_y) A_{2,x}(e_z)$ \\
$-A_{1,y}(e_z) A_{2,y}(e_y) +A_{1,z}(e_y) A_{2,z}(e_z)$ \\
\midrule
$A_{1,z}(e_z) A_{2,y}(e_y) -A_{1,y}(e_y) A_{2,z}(e_z)$ \\
\midrule
$A_{1,y}(e_y) A_{2,y}(e_z) -A_{1,z}(e_z) A_{2,z}(e_y)$ \\
\bottomrule
\end{tabular}
\addtolength{\tabcolsep}{-6pt}
\label{tab:mm-0nn-B1}
\end{table}

\begin{table}[htbp]
\centering
\caption{Same as Table~\ref{tab:mm-000-A1+}, but for the $B_2$ irrep of the little group associated with $\vec{P}=[0,n,n]$.}
\addtolength{\tabcolsep}{6pt}
\begin{tabular}{c}
\toprule
$P_1(e_z) S_2(e_y) -P_1(e_y) S_2(e_z)$ \\
\midrule
$P_1(e_z) V_{2,y}(e_y) -P_1(e_y) V_{2,z}(e_z)$ \\
\midrule
$P_1(e_y) V_{2,y}(e_z) -P_1(e_z) V_{2,z}(e_y)$ \\
\midrule
$P_2(e_z) A_{1,x}(e_y) +P_2(e_y) A_{1,x}(e_z)$ \\
\midrule
$S_1(e_z) V_{2,x}(e_y) +S_1(e_y) V_{2,x}(e_z)$ \\
\midrule
$S_2(e_y) A_{1,y}(e_z) -S_2(e_z) A_{1,z}(e_y)$ \\
\midrule
$S_2(e_z) A_{1,y}(e_y) -S_2(e_y) A_{1,z}(e_z)$ \\
\midrule
$V_{1,y}(e_z) V_{2,x}(e_y) +V_{1,z}(e_y) V_{2,x}(e_z)$ \\
$+V_{1,x}(e_z) V_{2,y}(e_y) +V_{1,x}(e_y) V_{2,z}(e_z)$ \\
\midrule
$V_{1,x}(e_y) V_{2,y}(e_z) +V_{1,x}(e_z) V_{2,z}(e_y)$ \\
\midrule
$V_{1,y}(e_z) V_{2,x}(e_y) +V_{1,z}(e_y) V_{2,x}(e_z)$ \\
$-V_{1,x}(e_z) V_{2,y}(e_y) -V_{1,x}(e_y) V_{2,z}(e_z)$ \\
\midrule
$V_{1,z}(e_z) V_{2,x}(e_y) +V_{1,y}(e_y) V_{2,x}(e_z)$ \\
\midrule
$A_{1,x}(e_z) V_{2,x}(e_y) -A_{1,x}(e_y) V_{2,x}(e_z)$ \\
$-A_{1,y}(e_z) V_{2,y}(e_y) +A_{1,z}(e_y) V_{2,z}(e_z)$ \\
\midrule
$A_{1,z}(e_y) V_{2,y}(e_z) -A_{1,y}(e_z) V_{2,z}(e_y)$ \\
\midrule
$-A_{1,x}(e_z) V_{2,x}(e_y) +A_{1,x}(e_y) V_{2,x}(e_z)$ \\
$-A_{1,y}(e_z) V_{2,y}(e_y) +A_{1,z}(e_y) V_{2,z}(e_z)$ \\
\midrule
$A_{1,z}(e_z) V_{2,y}(e_y) -A_{1,y}(e_y) V_{2,z}(e_z)$ \\
\midrule
$A_{1,y}(e_y) V_{2,y}(e_z) -A_{1,z}(e_z) V_{2,z}(e_y)$ \\
\midrule
$A_{1,y}(e_z) A_{2,x}(e_y) +A_{1,z}(e_y) A_{2,x}(e_z)$ \\
$+A_{1,x}(e_z) A_{2,y}(e_y) +A_{1,x}(e_y) A_{2,z}(e_z)$ \\
\midrule
$A_{1,x}(e_y) A_{2,y}(e_z) +A_{1,x}(e_z) A_{2,z}(e_y)$ \\
\midrule
$A_{1,y}(e_z) A_{2,x}(e_y) +A_{1,z}(e_y) A_{2,x}(e_z)$ \\
$-A_{1,x}(e_z) A_{2,y}(e_y) -A_{1,x}(e_y) A_{2,z}(e_z)$ \\
\midrule
$A_{1,z}(e_z) A_{2,x}(e_y) +A_{1,y}(e_y) A_{2,x}(e_z)$ \\
\bottomrule
\end{tabular}
\addtolength{\tabcolsep}{-6pt}
\label{tab:mm-0nn-B2}
\end{table}

As the total momentum increases, the lattice symmetry is gradually reduced. In particular, for momentum classes satisfying $|\vec{P}|^2 \geq 3$, the corresponding little groups become so small that the number of distinct operator structures grows rapidly. Since $|\vec{P}|^2 \leq 2$ is usually sufficient for most spectroscopic calculations, explicit operators beyond this range are not listed here. The reader may instead use the \texttt{OpTion} package, which can generate operator bases for arbitrary momenta and symmetry classes as needed.

For the total-momentum class $\vec{P} = [n,m,p]$, the little group degenerates to the trivial group $C_1$. In this case, because no symmetry protection remains, all operators whose particle momenta sum to $\vec{P}$ may contribute to the spectrum.

\section{Baryon-Baryon Operator Lists}
\label{sec:append_list_bb}
This section gives operator lists for distinguishable baryon-baryon systems, collected in Tables~\ref{tab:bb-000}, \ref{tab:bb-00n}, \ref{tab:bb-0nn}, \ref{tab:bb-nnn}, \ref{tab:bb-nm0}, and \ref{tab:bb-nnm}. These tables cover all relevant irreps except those of the trivial group.

\begin{table}[htbp]
\centering
\caption{Baryon-baryon operator list for the group $O_h$.}
\addtolength{\tabcolsep}{-5pt}
\begin{tabular}{cc}
\toprule
\multirow{4}{*}{$A_1^+$} & $N_{1,\frac{1}{2}}(0) N_{2,-\frac{1}{2}}(0) -N_{1,-\frac{1}{2}}(0) N_{2,\frac{1}{2}}(0)$ \\
\cmidrule(lr){2-2}
& $N_{1,\frac{1}{2}}(e_x) N_{2,-\frac{1}{2}}(-e_x) +N_{1,\frac{1}{2}}(-e_x) N_{2,-\frac{1}{2}}(e_x) +N_{1,\frac{1}{2}}(e_y) N_{2,-\frac{1}{2}}(-e_y) +N_{1,\frac{1}{2}}(-e_y) N_{2,-\frac{1}{2}}(e_y)$ \\
& $+N_{1,\frac{1}{2}}(e_z) N_{2,-\frac{1}{2}}(-e_z) +N_{1,\frac{1}{2}}(-e_z) N_{2,-\frac{1}{2}}(e_z) -N_{1,-\frac{1}{2}}(e_x) N_{2,\frac{1}{2}}(-e_x) -N_{1,-\frac{1}{2}}(-e_x) N_{2,\frac{1}{2}}(e_x)$ \\
& $-N_{1,-\frac{1}{2}}(e_y) N_{2,\frac{1}{2}}(-e_y) -N_{1,-\frac{1}{2}}(-e_y) N_{2,\frac{1}{2}}(e_y) -N_{1,-\frac{1}{2}}(e_z) N_{2,\frac{1}{2}}(-e_z) -N_{1,-\frac{1}{2}}(-e_z) N_{2,\frac{1}{2}}(e_z)$ \\
\midrule
\multirow{3}{*}{$A_1^-$} & $i N_{1,-\frac{1}{2}}(e_x) N_{2,-\frac{1}{2}}(-e_x) -i N_{1,-\frac{1}{2}}(-e_x) N_{2,-\frac{1}{2}}(e_x) +N_{1,-\frac{1}{2}}(e_y) N_{2,-\frac{1}{2}}(-e_y) -N_{1,-\frac{1}{2}}(-e_y) N_{2,-\frac{1}{2}}(e_y)$ \\
& $+i N_{1,\frac{1}{2}}(e_z) N_{2,-\frac{1}{2}}(-e_z) -i N_{1,\frac{1}{2}}(-e_z) N_{2,-\frac{1}{2}}(e_z) -i N_{1,\frac{1}{2}}(e_x) N_{2,\frac{1}{2}}(-e_x) +i N_{1,\frac{1}{2}}(-e_x) N_{2,\frac{1}{2}}(e_x)$ \\
& $+N_{1,\frac{1}{2}}(e_y) N_{2,\frac{1}{2}}(-e_y) -N_{1,\frac{1}{2}}(-e_y) N_{2,\frac{1}{2}}(e_y) +i N_{1,-\frac{1}{2}}(e_z) N_{2,\frac{1}{2}}(-e_z) -i N_{1,-\frac{1}{2}}(-e_z) N_{2,\frac{1}{2}}(e_z)$ \\
\midrule
\multirow{3}{*}{$E^+$} & $-N_{1,\frac{1}{2}}(e_x) N_{2,-\frac{1}{2}}(-e_x) -N_{1,\frac{1}{2}}(-e_x) N_{2,-\frac{1}{2}}(e_x) -N_{1,\frac{1}{2}}(e_y) N_{2,-\frac{1}{2}}(-e_y) -N_{1,\frac{1}{2}}(-e_y) N_{2,-\frac{1}{2}}(e_y)$ \\
\bottomrule
\end{tabular}
\addtolength{\tabcolsep}{5pt}
\label{tab:bb-000}
\end{table}

\begin{table}[htbp]
\centering
\caption{Baryon-baryon operator list for the group associated with $\vec{P} = [0,0,n]$.}
\addtolength{\tabcolsep}{6pt}
\begin{tabular}{cc}
\toprule
\multirow{2}{*}{$A_1$} & $N_{1,\frac{1}{2}}(0) N_{2,-\frac{1}{2}}(e_z) -N_{1,-\frac{1}{2}}(0) N_{2,\frac{1}{2}}(e_z)$ \\
\cmidrule(lr){2-2}
& $N_{1,\frac{1}{2}}(e_z) N_{2,-\frac{1}{2}}(0) -N_{1,-\frac{1}{2}}(e_z) N_{2,\frac{1}{2}}(0)$ \\
\midrule
\multirow{2}{*}{$A_2$} & $N_{1,\frac{1}{2}}(0) N_{2,-\frac{1}{2}}(e_z) +N_{1,-\frac{1}{2}}(0) N_{2,\frac{1}{2}}(e_z)$ \\
\cmidrule(lr){2-2}
& $N_{1,\frac{1}{2}}(e_z) N_{2,-\frac{1}{2}}(0) +N_{1,-\frac{1}{2}}(e_z) N_{2,\frac{1}{2}}(0)$ \\
\midrule
\multirow{2}{*}{$E$} & $N_{1,-\frac{1}{2}}(0) N_{2,-\frac{1}{2}}(e_z) -N_{1,\frac{1}{2}}(0) N_{2,\frac{1}{2}}(e_z)$ \\
\cmidrule(lr){2-2}
& $N_{1,-\frac{1}{2}}(e_z) N_{2,-\frac{1}{2}}(0) -N_{1,\frac{1}{2}}(e_z) N_{2,\frac{1}{2}}(0)$ \\
\bottomrule
\end{tabular}
\addtolength{\tabcolsep}{-6pt}
\label{tab:bb-00n}
\end{table}

\begin{table}[htbp]
\centering
\caption{Baryon-baryon operator list for the group associated with $\vec{P} = [0,n,n]$.}
\addtolength{\tabcolsep}{-5pt}
\begin{tabular}{cc}
\toprule
\multirow{2}{*}{$A_1$} & $N_{1,-\frac{1}{2}}(e_z) N_{2,-\frac{1}{2}}(e_y) -N_{1,-\frac{1}{2}}(e_y) N_{2,-\frac{1}{2}}(e_z) -N_{1,\frac{1}{2}}(e_z) N_{2,\frac{1}{2}}(e_y) +N_{1,\frac{1}{2}}(e_y) N_{2,\frac{1}{2}}(e_z)$ \\
\cmidrule(lr){2-2}
& $-N_{1,\frac{1}{2}}(e_z) N_{2,-\frac{1}{2}}(e_y) -N_{1,\frac{1}{2}}(e_y) N_{2,-\frac{1}{2}}(e_z) +N_{1,-\frac{1}{2}}(e_z) N_{2,\frac{1}{2}}(e_y) +N_{1,-\frac{1}{2}}(e_y) N_{2,\frac{1}{2}}(e_z)$ \\
\midrule
\multirow{2}{*}{$A_2$} & $N_{1,-\frac{1}{2}}(e_z) N_{2,-\frac{1}{2}}(e_y) +i N_{1,\frac{1}{2}}(e_y) N_{2,-\frac{1}{2}}(e_z) +N_{1,\frac{1}{2}}(e_z) N_{2,\frac{1}{2}}(e_y) +i N_{1,-\frac{1}{2}}(e_y) N_{2,\frac{1}{2}}(e_z)$ \\
\cmidrule(lr){2-2}
& $N_{1,\frac{1}{2}}(e_z) N_{2,-\frac{1}{2}}(e_y) -i N_{1,-\frac{1}{2}}(e_y) N_{2,-\frac{1}{2}}(e_z) +N_{1,-\frac{1}{2}}(e_z) N_{2,\frac{1}{2}}(e_y) -i N_{1,\frac{1}{2}}(e_y) N_{2,\frac{1}{2}}(e_z)$ \\
\midrule
\multirow{2}{*}{$B_1$} & $N_{1,-\frac{1}{2}}(e_z) N_{2,-\frac{1}{2}}(e_y) +N_{1,-\frac{1}{2}}(e_y) N_{2,-\frac{1}{2}}(e_z) -N_{1,\frac{1}{2}}(e_z) N_{2,\frac{1}{2}}(e_y) -N_{1,\frac{1}{2}}(e_y) N_{2,\frac{1}{2}}(e_z)$ \\
\cmidrule(lr){2-2}
& $-N_{1,\frac{1}{2}}(e_z) N_{2,-\frac{1}{2}}(e_y) +N_{1,\frac{1}{2}}(e_y) N_{2,-\frac{1}{2}}(e_z) +N_{1,-\frac{1}{2}}(e_z) N_{2,\frac{1}{2}}(e_y) -N_{1,-\frac{1}{2}}(e_y) N_{2,\frac{1}{2}}(e_z)$ \\
\midrule
\multirow{2}{*}{$B_2$} & $N_{1,-\frac{1}{2}}(e_z) N_{2,-\frac{1}{2}}(e_y) -i N_{1,\frac{1}{2}}(e_y) N_{2,-\frac{1}{2}}(e_z) +N_{1,\frac{1}{2}}(e_z) N_{2,\frac{1}{2}}(e_y) -i N_{1,-\frac{1}{2}}(e_y) N_{2,\frac{1}{2}}(e_z)$ \\
\cmidrule(lr){2-2}
& $N_{1,\frac{1}{2}}(e_z) N_{2,-\frac{1}{2}}(e_y) +i N_{1,-\frac{1}{2}}(e_y) N_{2,-\frac{1}{2}}(e_z) +N_{1,-\frac{1}{2}}(e_z) N_{2,\frac{1}{2}}(e_y) +i N_{1,\frac{1}{2}}(e_y) N_{2,\frac{1}{2}}(e_z)$ \\
\bottomrule
\end{tabular}
\addtolength{\tabcolsep}{5pt}
\label{tab:bb-0nn}
\end{table}

\begin{table}[htbp]
\centering
\scriptsize
\caption{Baryon-baryon operator list for the group associated with $\vec{P} = [n,n,n]$.}
\addtolength{\tabcolsep}{-5pt}
\begin{tabular}{cc}
\toprule
\multirow{13}{*}{$A_1$} & $N_{1,\frac{1}{2}}(0) N_{2,-\frac{1}{2}}(e_{x,y,z}) -N_{1,-\frac{1}{2}}(0) N_{2,\frac{1}{2}}(e_{x,y,z})$ \\
\cmidrule(lr){2-2}
& $(1+i) N_{1,-\frac{1}{2}}(e_z) N_{2,-\frac{1}{2}}(e_{x,y}) -N_{1,-\frac{1}{2}}(e_y) N_{2,-\frac{1}{2}}(e_{x,z}) +N_{1,\frac{1}{2}}(e_y) N_{2,-\frac{1}{2}}(e_{x,z}) -i N_{1,-\frac{1}{2}}(e_x) N_{2,-\frac{1}{2}}(e_{y,z})$ \\
& $-N_{1,\frac{1}{2}}(e_x) N_{2,-\frac{1}{2}}(e_{y,z}) -(1-i) N_{1,\frac{1}{2}}(e_z) N_{2,\frac{1}{2}}(e_{x,y}) +N_{1,-\frac{1}{2}}(e_y) N_{2,\frac{1}{2}}(e_{x,z}) +N_{1,\frac{1}{2}}(e_y) N_{2,\frac{1}{2}}(e_{x,z})$ \\
& $-N_{1,-\frac{1}{2}}(e_x) N_{2,\frac{1}{2}}(e_{y,z}) -i N_{1,\frac{1}{2}}(e_x) N_{2,\frac{1}{2}}(e_{y,z})$ \\
\cmidrule(lr){2-2}
& $-N_{1,\frac{1}{2}}(e_z) N_{2,-\frac{1}{2}}(e_{x,y}) -N_{1,\frac{1}{2}}(e_y) N_{2,-\frac{1}{2}}(e_{x,z}) -N_{1,\frac{1}{2}}(e_x) N_{2,-\frac{1}{2}}(e_{y,z}) +N_{1,-\frac{1}{2}}(e_z) N_{2,\frac{1}{2}}(e_{x,y})$ \\
& $+N_{1,-\frac{1}{2}}(e_y) N_{2,\frac{1}{2}}(e_{x,z}) +N_{1,-\frac{1}{2}}(e_x) N_{2,\frac{1}{2}}(e_{y,z})$ \\
\cmidrule(lr){2-2}
& $-i N_{1,-\frac{1}{2}}(e_{y,z}) N_{2,-\frac{1}{2}}(e_x) -N_{1,\frac{1}{2}}(e_{y,z}) N_{2,-\frac{1}{2}}(e_x) -N_{1,-\frac{1}{2}}(e_{x,z}) N_{2,-\frac{1}{2}}(e_y) +N_{1,\frac{1}{2}}(e_{x,z}) N_{2,-\frac{1}{2}}(e_y)$ \\
& $+(1+i) N_{1,-\frac{1}{2}}(e_{x,y}) N_{2,-\frac{1}{2}}(e_z) -N_{1,-\frac{1}{2}}(e_{y,z}) N_{2,\frac{1}{2}}(e_x) -i N_{1,\frac{1}{2}}(e_{y,z}) N_{2,\frac{1}{2}}(e_x) +N_{1,-\frac{1}{2}}(e_{x,z}) N_{2,\frac{1}{2}}(e_y)$ \\
\bottomrule
\end{tabular}
\addtolength{\tabcolsep}{5pt}
\label{tab:bb-nnn}
\end{table}

\begin{table}[htbp]
\centering
\caption{Baryon-baryon operator list for the group associated with $\vec{P} = [n,m,0]$.}
\addtolength{\tabcolsep}{6pt}
\begin{tabular}{cc}
\toprule
\multirow{6}{*}{$A$} & $N_{1,-\frac{1}{2}}(e_x) N_{2,\frac{1}{2}}(e_{x,y}) N_{1,\frac{1}{2}}(e_x) N_{2,-\frac{1}{2}}(e_{x,y})$ \\
\cmidrule(lr){2-2}
& $N_{1,-\frac{1}{2}}(e_{x,y}) N_{2,\frac{1}{2}}(e_x) N_{1,\frac{1}{2}}(e_{x,y}) N_{2,-\frac{1}{2}}(e_x)$ \\
\cmidrule(lr){2-2}
& $N_{1,-\frac{1}{2}}(e_y) N_{2,\frac{1}{2}}(e_{2x}) N_{1,\frac{1}{2}}(e_y) N_{2,-\frac{1}{2}}(e_{2x})$ \\
\cmidrule(lr){2-2}
& $-N_{1,-\frac{1}{2}}(e_{x,z}) N_{2,-\frac{1}{2}}(e_{x,y,-z}) +N_{1,-\frac{1}{2}}(e_{x,-z}) N_{2,-\frac{1}{2}}(e_{x,y,z})$ \\
\cmidrule(lr){2-2}
& $N_{1,-\frac{1}{2}}(e_{x,z}) N_{2,\frac{1}{2}}(e_{x,y,-z}) +N_{1,-\frac{1}{2}}(e_{x,-z}) N_{2,\frac{1}{2}}(e_{x,y,z})$ \\
\cmidrule(lr){2-2}
& $N_{1,\frac{1}{2}}(e_{x,z}) N_{2,-\frac{1}{2}}(e_{x,y,-z}) +N_{1,\frac{1}{2}}(e_{x,-z}) N_{2,-\frac{1}{2}}(e_{x,y,z})$ \\
\midrule
\multirow{6}{*}{$B$} & $N_{1,-\frac{1}{2}}(e_x) N_{2,-\frac{1}{2}}(e_{x,y}) N_{1,\frac{1}{2}}(e_x) N_{2,\frac{1}{2}}(e_{x,y})$ \\
\cmidrule(lr){2-2}
& $N_{1,-\frac{1}{2}}(e_{x,y}) N_{2,-\frac{1}{2}}(e_x) N_{1,\frac{1}{2}}(e_{x,y}) N_{2,\frac{1}{2}}(e_x)$ \\
\cmidrule(lr){2-2}
& $N_{1,-\frac{1}{2}}(e_y) N_{2,-\frac{1}{2}}(e_{2x}) N_{1,\frac{1}{2}}(e_y) N_{2,\frac{1}{2}}(e_{2x})$ \\
\cmidrule(lr){2-2}
& $N_{1,-\frac{1}{2}}(e_{x,z}) N_{2,-\frac{1}{2}}(e_{x,y,-z}) +N_{1,-\frac{1}{2}}(e_{x,-z}) N_{2,-\frac{1}{2}}(e_{x,y,z})$ \\
\cmidrule(lr){2-2}
& $-N_{1,-\frac{1}{2}}(e_{x,z}) N_{2,\frac{1}{2}}(e_{x,y,-z}) +N_{1,-\frac{1}{2}}(e_{x,-z}) N_{2,\frac{1}{2}}(e_{x,y,z})$ \\
\cmidrule(lr){2-2}
& $-N_{1,\frac{1}{2}}(e_{x,z}) N_{2,-\frac{1}{2}}(e_{x,y,-z}) +N_{1,\frac{1}{2}}(e_{x,-z}) N_{2,-\frac{1}{2}}(e_{x,y,z})$ \\
\bottomrule
\end{tabular}
\addtolength{\tabcolsep}{-6pt}
\label{tab:bb-nm0}
\end{table}

\begin{table}[htbp]
\centering
\caption{Baryon-baryon operator list for the group associated with $\vec{P} = [n,n,m]$.}
\addtolength{\tabcolsep}{6pt}
\begin{tabular}{cc}
\toprule
\multirow{8}{*}{$A$} & $N_{1,-\frac{1}{2}}(e_z) N_{2,-\frac{1}{2}}(e_{x,y,z}) +i N_{1,\frac{1}{2}}(e_z) N_{2,\frac{1}{2}}(e_{x,y,z})$ \\
\cmidrule(lr){2-2}
& $-N_{1,\frac{1}{2}}(e_z) N_{2,-\frac{1}{2}}(e_{x,y,z}) +N_{1,-\frac{1}{2}}(e_z) N_{2,\frac{1}{2}}(e_{x,y,z})$ \\
\cmidrule(lr){2-2}
& $N_{1,-\frac{1}{2}}(e_{y,z}) N_{2,-\frac{1}{2}}(e_{x,z}) +i N_{1,\frac{1}{2}}(e_{x,z}) N_{2,\frac{1}{2}}(e_{y,z})$ \\
\cmidrule(lr){2-2}
& $-N_{1,\frac{1}{2}}(e_{x,z}) N_{2,-\frac{1}{2}}(e_{y,z}) +N_{1,-\frac{1}{2}}(e_{y,z}) N_{2,\frac{1}{2}}(e_{x,z})$ \\
\cmidrule(lr){2-2}
& $N_{1,\frac{1}{2}}(e_{y,z}) N_{2,-\frac{1}{2}}(e_{x,z}) -N_{1,-\frac{1}{2}}(e_{x,z}) N_{2,\frac{1}{2}}(e_{y,z})$ \\
\cmidrule(lr){2-2}
& $-i N_{1,-\frac{1}{2}}(e_{x,z}) N_{2,-\frac{1}{2}}(e_{y,z}) +N_{1,\frac{1}{2}}(e_{y,z}) N_{2,\frac{1}{2}}(e_{x,z})$ \\
\cmidrule(lr){2-2}
& $N_{1,-\frac{1}{2}}(e_{x,y,z}) N_{2,-\frac{1}{2}}(e_z) +i N_{1,\frac{1}{2}}(e_{x,y,z}) N_{2,\frac{1}{2}}(e_z)$ \\
\cmidrule(lr){2-2}
& $-N_{1,\frac{1}{2}}(e_{x,y,z}) N_{2,-\frac{1}{2}}(e_z) +N_{1,-\frac{1}{2}}(e_{x,y,z}) N_{2,\frac{1}{2}}(e_z)$ \\
\bottomrule
\end{tabular}
\addtolength{\tabcolsep}{-6pt}
\label{tab:bb-nnm}
\end{table}

\section{Meson-Baryon Operator Lists}
\label{sec:append_list_mb}
This section gives meson-baryon operators, collected in Tables~\ref{tab:mb-000-G1}, \ref{tab:mb-000-G2}, \ref{tab:mb-000-H+}, \ref{tab:mb-000-H-}, \ref{tab:mb-00n}, and \ref{tab:mb-0nn}. Since these systems have half-integer total spin, the operators belong to irreps of the double-cover groups of the corresponding lattice symmetry groups.

\begin{table}[htbp]
\centering
\scriptsize
\caption{Meson-baryon operator list for the $G_1^{\pm}$ irreps of $O_h^D$.}
\addtolength{\tabcolsep}{-5pt}
\begin{tabular}{cc}
\toprule
\multirow{22}{*}{$G_1^+$} & $P_{1}(e_x) N_{2,-\frac{1}{2}}(-e_x) -P_{1}(-e_x) N_{2,-\frac{1}{2}}(e_x) -i P_{1}(e_y) N_{2,-\frac{1}{2}}(-e_y) +i P_{1}(-e_y) N_{2,-\frac{1}{2}}(e_y)$ \\
& $+P_{1}(e_z) N_{2,\frac{1}{2}}(-e_z) -P_{1}(-e_z) N_{2,\frac{1}{2}}(e_z)$ \\
\cmidrule(lr){2-2}
& $S_{1}(0) N_{2,\frac{1}{2}}(0)$ \\
\cmidrule(lr){2-2}
& $S_{1}(e_x) N_{2,\frac{1}{2}}(-e_x) +S_{1}(-e_x) N_{2,\frac{1}{2}}(e_x) +S_{1}(e_y) N_{2,\frac{1}{2}}(-e_y) +S_{1}(-e_y) N_{2,\frac{1}{2}}(e_y)$ \\
& $+S_{1}(e_z) N_{2,\frac{1}{2}}(-e_z) +S_{1}(-e_z) N_{2,\frac{1}{2}}(e_z)$ \\
\cmidrule(lr){2-2}
& $2 N_{2,\frac{1}{2}}(e_x) V_{1,x}(-e_x) -2 N_{2,\frac{1}{2}}(-e_x) V_{1,x}(e_x) -i N_{2,\frac{1}{2}}(e_y) V_{1,x}(-e_y) +i N_{2,\frac{1}{2}}(-e_y) V_{1,x}(e_y)$ \\
& $+N_{2,-\frac{1}{2}}(e_z) V_{1,x}(-e_z) -N_{2,-\frac{1}{2}}(-e_z) V_{1,x}(e_z) +i N_{2,\frac{1}{2}}(e_x) V_{1,y}(-e_x) -i N_{2,\frac{1}{2}}(-e_x) V_{1,y}(e_x)$ \\
& $+2 N_{2,\frac{1}{2}}(e_y) V_{1,y}(-e_y) -2 N_{2,\frac{1}{2}}(-e_y) V_{1,y}(e_y) -i N_{2,-\frac{1}{2}}(e_z) V_{1,y}(-e_z) +i N_{2,-\frac{1}{2}}(-e_z) V_{1,y}(e_z)$ \\
\bottomrule
\end{tabular}
\addtolength{\tabcolsep}{5pt}
\label{tab:mb-000-G1}
\end{table}

\begin{table}[htbp]
\centering
\scriptsize
\caption{Meson-baryon operator list for the $G_2^{\pm}$ irreps of $O_h^D$.}
\addtolength{\tabcolsep}{-5pt}
\begin{tabular}{cc}
\toprule
\multirow{8}{*}{$G_2^+$} & $i N_{2,\frac{1}{2}}(e_y) V_{1,x}(-e_y) -i N_{2,\frac{1}{2}}(-e_y) V_{1,x}(e_y) +N_{2,-\frac{1}{2}}(e_z) V_{1,x}(-e_z) -N_{2,-\frac{1}{2}}(-e_z) V_{1,x}(e_z)$ \\
& $+i N_{2,\frac{1}{2}}(e_x) V_{1,y}(-e_x) -i N_{2,\frac{1}{2}}(-e_x) V_{1,y}(e_x) +i N_{2,-\frac{1}{2}}(e_z) V_{1,y}(-e_z) -i N_{2,-\frac{1}{2}}(-e_z) V_{1,y}(e_z)$ \\
& $+N_{2,-\frac{1}{2}}(e_x) V_{1,z}(-e_x) -N_{2,-\frac{1}{2}}(-e_x) V_{1,z}(e_x) +i N_{2,-\frac{1}{2}}(e_y) V_{1,z}(-e_y) -i N_{2,-\frac{1}{2}}(-e_y) V_{1,z}(e_y)$ \\
\cmidrule(lr){2-2}
& $i A_{1,y}(e_x) N_{2,-\frac{1}{2}}(-e_x) +i A_{1,y}(-e_x) N_{2,-\frac{1}{2}}(e_x) +A_{1,x}(e_y) N_{2,-\frac{1}{2}}(-e_y) +A_{1,x}(-e_y) N_{2,-\frac{1}{2}}(e_y)$ \\
& $-A_{1,x}(e_z) N_{2,-\frac{1}{2}}(-e_z) -i A_{1,y}(e_z) N_{2,-\frac{1}{2}}(-e_z) -A_{1,x}(-e_z) N_{2,-\frac{1}{2}}(e_z) -i A_{1,y}(-e_z) N_{2,-\frac{1}{2}}(e_z)$ \\
& $+A_{1,z}(e_x) N_{2,\frac{1}{2}}(-e_x) +A_{1,z}(-e_x) N_{2,\frac{1}{2}}(e_x) -A_{1,z}(e_y) N_{2,\frac{1}{2}}(-e_y) -A_{1,z}(-e_y) N_{2,\frac{1}{2}}(e_y)$ \\
\midrule
\multirow{8}{*}{$G_2^-$} & $N_{2,-\frac{1}{2}}(e_y) V_{1,x}(-e_y) + N_{2,-\frac{1}{2}}(-e_y) V_{1,x}(e_y) - N_{2,-\frac{1}{2}}(e_z) V_{1,x}(-e_z) - N_{2,-\frac{1}{2}}(-e_z) V_{1,x}(e_z)$ \\
& $+ i N_{2,-\frac{1}{2}}(e_x) V_{1,y}(-e_x) + i N_{2,-\frac{1}{2}}(-e_x) V_{1,y}(e_x) - i N_{2,-\frac{1}{2}}(e_z) V_{1,y}(-e_z) - i N_{2,-\frac{1}{2}}(-e_z) V_{1,y}(e_z)$ \\
\bottomrule
\end{tabular}
\addtolength{\tabcolsep}{5pt}
\label{tab:mb-000-G2}
\end{table}

\begin{table}[htbp]
\centering
\small
\caption{Meson-baryon operator list for the $H^+$ irrep of $O_h^D$.}
\addtolength{\tabcolsep}{-5pt}
\begin{tabular}{cc}
\toprule
\multirow{25}{*}{$H^+$} & $P_{1}[e_x] N_{2,\frac{1}{2}}[-e_x] - P_{1}[-e_x] N_{2,\frac{1}{2}}[e_x] - i P_{1}[e_y] N_{2,\frac{1}{2}}[-e_y] + i P_{1}[-e_y] N_{2,\frac{1}{2}}[e_y]$ \\
\cmidrule(lr){2-2}
& $S_{1}[e_x] N_{2,-\frac{1}{2}}[-e_x] + S_{1}[-e_x] N_{2,-\frac{1}{2}}[e_x] - S_{1}[e_y] N_{2,-\frac{1}{2}}[-e_y] - S_{1}[-e_y] N_{2,-\frac{1}{2}}[e_y]$ \\
\cmidrule(lr){2-2}
& $3 N_{2,-\frac{1}{2}}[e_x] V_{1,x}[-e_x] - 3 N_{2,-\frac{1}{2}}[-e_x] V_{1,x}[e_x] + 2 i N_{2,-\frac{1}{2}}[e_y] V_{1,x}[-e_y] - 2 i N_{2,-\frac{1}{2}}[-e_y] V_{1,x}[e_y]$ \\
& $- N_{2,\frac{1}{2}}[e_z] V_{1,x}[-e_z] + N_{2,\frac{1}{2}}[-e_z] V_{1,x}[e_z] + 2 i N_{2,-\frac{1}{2}}[e_x] V_{1,y}[-e_x] - 2 i N_{2,-\frac{1}{2}}[-e_x] V_{1,y}[e_x]$ \\
& $- 3 N_{2,-\frac{1}{2}}[e_y] V_{1,y}[-e_y] + 3 N_{2,-\frac{1}{2}}[-e_y] V_{1,y}[e_y] + i N_{2,\frac{1}{2}}[e_z] V_{1,y}[-e_z] - i N_{2,\frac{1}{2}}[-e_z] V_{1,y}[e_z]$ \\
& $- N_{2,\frac{1}{2}}[e_x] V_{1,z}[-e_x] + N_{2,\frac{1}{2}}[-e_x] V_{1,z}[e_x] + i N_{2,\frac{1}{2}}[e_y] V_{1,z}[-e_y] - i N_{2,\frac{1}{2}}[-e_y] V_{1,z}[e_y]$ \\
\cmidrule(lr){2-2}
& $i N_{2,-\frac{1}{2}}[e_y] V_{1,x}[-e_y] - i N_{2,-\frac{1}{2}}[-e_y] V_{1,x}[e_y] + N_{2,\frac{1}{2}}[e_z] V_{1,x}[-e_z] - N_{2,\frac{1}{2}}[-e_z] V_{1,x}[e_z]$ \\
& $+ i N_{2,-\frac{1}{2}}[e_x] V_{1,y}[-e_x] - i N_{2,-\frac{1}{2}}[-e_x] V_{1,y}[e_x] - i N_{2,\frac{1}{2}}[e_z] V_{1,y}[-e_z] + i N_{2,\frac{1}{2}}[-e_z] V_{1,y}[e_z]$ \\
\bottomrule
\end{tabular}
\addtolength{\tabcolsep}{5pt}
\label{tab:mb-000-H+}
\end{table}

\begin{table}[htbp]
\centering
\small
\caption{Meson-baryon operator list for the $H^-$ irrep of $O_h^D$.}
\addtolength{\tabcolsep}{-5pt}
\begin{tabular}{cc}
\toprule
\multirow{25}{*}{$H^-$} & $P_1[e_x] N_{2,-\frac{1}{2}}[-e_x] + P_1[-e_x] N_{2,-\frac{1}{2}}[e_x] - P_1[e_y] N_{2,-\frac{1}{2}}[-e_y] - P_1[-e_y] N_{2,-\frac{1}{2}}[e_y]$ \\
\cmidrule(lr){2-2}
& $S_1[e_x] N_{2,\frac{1}{2}}[-e_x] - S_1[-e_x] N_{2,\frac{1}{2}}[e_x] - i S_1[e_y] N_{2,\frac{1}{2}}[-e_y] + i S_1[-e_y] N_{2,\frac{1}{2}}[e_y]$ \\
\cmidrule(lr){2-2}
& $i N_{2,\frac{1}{2}}[0] V_{1,x}[0] + N_{2,\frac{1}{2}}[0] V_{1,y}[0]$ \\
\cmidrule(lr){2-2}
& $3 N_{2,\frac{1}{2}}[e_x] V_{1,x}[-e_x] + 3 N_{2,\frac{1}{2}}[-e_x] V_{1,x}[e_x] - 2 N_{2,\frac{1}{2}}[e_y] V_{1,x}[-e_y] - 2 N_{2,\frac{1}{2}}[-e_y] V_{1,x}[e_y]$ \\
& $- N_{2,\frac{1}{2}}[e_z] V_{1,x}[-e_z] - N_{2,\frac{1}{2}}[-e_z] V_{1,x}[e_z] + 2 i N_{2,\frac{1}{2}}[e_x] V_{1,y}[-e_x] + 2 i N_{2,\frac{1}{2}}[-e_x] V_{1,y}[e_x]$ \\
& $- 3 i N_{2,\frac{1}{2}}[e_y] V_{1,y}[-e_y] - 3 i N_{2,\frac{1}{2}}[-e_y] V_{1,y}[e_y] + i N_{2,\frac{1}{2}}[e_z] V_{1,y}[-e_z] + i N_{2,\frac{1}{2}}[-e_z] V_{1,y}[e_z]$ \\
& $- N_{2,-\frac{1}{2}}[e_x] V_{1,z}[-e_x] - N_{2,-\frac{1}{2}}[-e_x] V_{1,z}[e_x] + N_{2,-\frac{1}{2}}[e_y] V_{1,z}[-e_y] + N_{2,-\frac{1}{2}}[-e_y] V_{1,z}[e_y]$ \\
\cmidrule(lr){2-2}
& $N_{2,\frac{1}{2}}[e_y] V_{1,x}[-e_y] + N_{2,\frac{1}{2}}[-e_y] V_{1,x}[e_y] - N_{2,\frac{1}{2}}[e_z] V_{1,x}[-e_z] - N_{2,\frac{1}{2}}[-e_z] V_{1,x}[e_z]$ \\
\bottomrule
\end{tabular}
\addtolength{\tabcolsep}{5pt}
\label{tab:mb-000-H-}
\end{table}

\begin{table}[htbp]
\centering
\caption{Meson-baryon operator list for the group associated with $\vec{P} = [0,0,n]$.}
\addtolength{\tabcolsep}{6pt}
\begin{tabular}{cc}
\toprule
\multirow{12}{*}{$G_1$} & $P_{1}(0) N_{2,\frac{1}{2}}(e_z)$ \\
\cmidrule(lr){2-2}
& $P_{1}(e_z) N_{2,\frac{1}{2}}(0)$ \\
\cmidrule(lr){2-2}
& $S_{1}(0) N_{2,\frac{1}{2}}(e_z)$ \\
\cmidrule(lr){2-2}
& $S_{1}(e_z) N_{2,\frac{1}{2}}(0)$ \\
\cmidrule(lr){2-2}
& $N_{2,\frac{1}{2}}(e_z) V_{1,z}(0)$ \\
\cmidrule(lr){2-2}
& $N_{2,-\frac{1}{2}}(e_z) V_{1,x}(0) -i N_{2,-\frac{1}{2}}(e_z) V_{1,y}(0)$ \\
\cmidrule(lr){2-2}
& $N_{2,\frac{1}{2}}(0) V_{1,z}(e_z)$ \\
\cmidrule(lr){2-2}
& $N_{2,-\frac{1}{2}}(0) V_{1,x}(e_z) -i N_{2,-\frac{1}{2}}(0) V_{1,y}(e_z)$ \\
\cmidrule(lr){2-2}
& $A_{1,z}(0) N_{2,\frac{1}{2}}(e_z)$ \\
\cmidrule(lr){2-2}
& $A_{1,x}(0) N_{2,-\frac{1}{2}}(e_z) -i A_{1,y}(0) N_{2,-\frac{1}{2}}(e_z)$ \\
\cmidrule(lr){2-2}
& $A_{1,z}(e_z) N_{2,\frac{1}{2}}(0)$ \\
\cmidrule(lr){2-2}
& $A_{1,x}(e_z) N_{2,-\frac{1}{2}}(0) -i A_{1,y}(e_z) N_{2,-\frac{1}{2}}(0)$ \\
\midrule
\multirow{4}{*}{$G_2$} & $N_{2,-\frac{1}{2}}(e_z) V_{1,x}(0) +i N_{2,-\frac{1}{2}}(e_z) V_{1,y}(0)$ \\
\cmidrule(lr){2-2}
& $N_{2,-\frac{1}{2}}(0) V_{1,x}(e_z) +i N_{2,-\frac{1}{2}}(0) V_{1,y}(e_z)$ \\
\cmidrule(lr){2-2}
& $A_{1,x}(0) N_{2,-\frac{1}{2}}(e_z) +i A_{1,y}(0) N_{2,-\frac{1}{2}}(e_z)$ \\
\cmidrule(lr){2-2}
& $A_{1,x}(e_z) N_{2,-\frac{1}{2}}(0) +i A_{1,y}(e_z) N_{2,-\frac{1}{2}}(0)$ \\
\bottomrule
\end{tabular}
\addtolength{\tabcolsep}{-6pt}
\label{tab:mb-00n}
\end{table}

\begin{table}[htbp]
\centering
\caption{Meson-baryon operator list for the group associated with $\vec{P} = [0,n,n]$.}
\addtolength{\tabcolsep}{6pt}
\begin{tabular}{cc}
\toprule
\multirow{16}{*}{$G$} & $P_{1}(e_z) N_{2,-\frac{1}{2}}(e_y) +i P_{1}(e_y) N_{2,\frac{1}{2}}(e_z)$ \\
\cmidrule(lr){2-2}
& $-i P_{1}(e_y) N_{2,-\frac{1}{2}}(e_z) +P_{1}(e_z) N_{2,\frac{1}{2}}(e_y)$ \\
\cmidrule(lr){2-2}
& $S_{1}(e_z) N_{2,-\frac{1}{2}}(e_y) -S_{1}(e_y) N_{2,-\frac{1}{2}}(e_z)$ \\
\cmidrule(lr){2-2}
& $S_{1}(e_z) N_{2,\frac{1}{2}}(e_y) +S_{1}(e_y) N_{2,\frac{1}{2}}(e_z)$ \\
\cmidrule(lr){2-2}
& $-i N_{2,\frac{1}{2}}(e_z) V_{1,x}(e_y) +N_{2,-\frac{1}{2}}(e_y) V_{1,x}(e_z) +i N_{2,-\frac{1}{2}}(e_y) V_{1,y}(e_z) -i N_{2,-\frac{1}{2}}(e_z) V_{1,z}(e_y)$ \\
\cmidrule(lr){2-2}
& $i N_{2,-\frac{1}{2}}(e_z) V_{1,x}(e_y) +N_{2,\frac{1}{2}}(e_y) V_{1,x}(e_z) +i N_{2,\frac{1}{2}}(e_y) V_{1,y}(e_z) +i N_{2,\frac{1}{2}}(e_z) V_{1,z}(e_y)$ \\
\cmidrule(lr){2-2}
& $N_{2,-\frac{1}{2}}(e_z) V_{1,y}(e_y) -N_{2,-\frac{1}{2}}(e_y) V_{1,z}(e_z)$ \\
\cmidrule(lr){2-2}
& $N_{2,\frac{1}{2}}(e_z) V_{1,y}(e_y) +N_{2,\frac{1}{2}}(e_y) V_{1,z}(e_z)$ \\
\bottomrule
\end{tabular}
\addtolength{\tabcolsep}{-6pt}
\label{tab:mb-0nn}
\end{table}

\section{Examples of Multihadron Operators}
\label{sec:append_list_multi}

\begin{table}[htbp]
\centering
\caption{Representative examples of three-pseudoscalar-meson operators for the group $O_h$. The notation follows Table~\ref{tab:mm-000-A1+}.}
\addtolength{\tabcolsep}{-4pt}
\begin{tabular}{cc}
\toprule
Irrep & Operator \\
\midrule
\multirow{16}{*}{$A_1^+$} & {\small $- P_{1}(e_{y,z})P_{2}(e_{x,-z})P_{3}(e_{-x,-y}) + P_{1}(e_{y,-z})P_{2}(e_{x,z})P_{3}(e_{-x,-y}) + P_{1}(e_{x,z})P_{2}(e_{y,-z})P_{3}(e_{-x,-y})$} \\
& {\small $- P_{1}(e_{x,-z})P_{2}(e_{y,z})P_{3}(e_{-x,-y}) + P_{1}(e_{-y,z})P_{2}(e_{x,-z})P_{3}(e_{-x,y}) - P_{1}(e_{-y,-z})P_{2}(e_{x,z})P_{3}(e_{-x,y})$} \\
& {\small $- P_{1}(e_{x,z})P_{2}(e_{-y,-z})P_{3}(e_{-x,y}) + P_{1}(e_{x,-z})P_{2}(e_{-y,z})P_{3}(e_{-x,y}) + P_{1}(e_{y,z})P_{2}(e_{x,-y})P_{3}(e_{-x,-z})$} \\
& {\small $- P_{1}(e_{-y,z})P_{2}(e_{x,y})P_{3}(e_{-x,-z}) - P_{1}(e_{x,y})P_{2}(e_{-y,z})P_{3}(e_{-x,-z}) + P_{1}(e_{x,-y})P_{2}(e_{y,z})P_{3}(e_{-x,-z})$} \\
& {\small $- P_{1}(e_{y,-z})P_{2}(e_{x,-y})P_{3}(e_{-x,z}) + P_{1}(e_{-y,-z})P_{2}(e_{x,y})P_{3}(e_{-x,z}) + P_{1}(e_{x,y})P_{2}(e_{-y,-z})P_{3}(e_{-x,z})$} \\
& {\small $- P_{1}(e_{x,-y})P_{2}(e_{y,-z})P_{3}(e_{-x,z}) + P_{1}(e_{y,z})P_{2}(e_{-x,-z})P_{3}(e_{x,-y}) - P_{1}(e_{y,-z})P_{2}(e_{-x,z})P_{3}(e_{x,-y})$} \\
& {\small $- P_{1}(e_{-x,z})P_{2}(e_{y,-z})P_{3}(e_{x,-y}) + P_{1}(e_{-x,-z})P_{2}(e_{y,z})P_{3}(e_{x,-y}) - P_{1}(e_{-y,z})P_{2}(e_{-x,-z})P_{3}(e_{x,y})$} \\
\bottomrule
\end{tabular}
\addtolength{\tabcolsep}{4pt}
\label{tab:three-000}
\end{table}

\begin{table}[htbp]
\centering
\caption{Representative examples of three-pseudoscalar-meson operators for the point groups $C_{4v}$, $C_{2v}$, and $C_{3v}$. The notation follows Table~\ref{tab:mm-000-A1+}.}
\addtolength{\tabcolsep}{6pt}
\begin{tabular}{ccc}
\toprule
Point group & Irrep & Operator \\
\midrule
\multirow{10}{*}{$C_{4v}$} & \multirow{4}{*}{$A_1$} & $P_{1}(e_{x,y}) P_{2}(-e_{y}) P_{3}(e_{-x,z}) - P_{1}(e_{x,-y}) P_{2}(e_{y}) P_{3}(e_{-x,z})$ \\
& & $- P_{1}(e_{-x,y}) P_{2}(-e_{y}) P_{3}(e_{x,z}) + P_{1}(e_{-x,-y}) P_{2}(e_{y}) P_{3}(e_{x,z})$ \\
& & $- P_{1}(e_{x,y}) P_{2}(-e_{x}) P_{3}(e_{-y,z}) + P_{1}(e_{-x,y}) P_{2}(e_{x}) P_{3}(e_{-y,z})$ \\
& & $+ P_{1}(e_{x,-y}) P_{2}(-e_{x}) P_{3}(e_{y,z}) - P_{1}(e_{-x,-y}) P_{2}(e_{x}) P_{3}(e_{y,z})$ \\
\cmidrule(lr){2-3}
& \multirow{1}{*}{$A_2$} & $P_{1}(0) P_{2}(0) P_{3}(e_{z})$ \\
\cmidrule(lr){2-3}
& \multirow{2}{*}{$B_1$} & $P_{1}(e_{x,y}) P_{2}(0) P_{3}(e_{-x,-y,z}) - P_{1}(e_{x,-y}) P_{2}(0) P_{3}(e_{-x,y,z})$ \\
& & $- P_{1}(e_{-x,y}) P_{2}(0) P_{3}(e_{x,-y,z}) + P_{1}(e_{-x,-y}) P_{2}(0) P_{3}(e_{x,y,z})$ \\
\bottomrule
\end{tabular}
\addtolength{\tabcolsep}{-6pt}
\label{tab:three-moving}
\end{table}

Table~\ref{tab:four-000} gives representative examples of four-body operators in the rest frame built from mutually distinguishable pseudoscalar mesons. These operators provide a basis for future four-body spectroscopic calculations and can be applied directly once the relevant finite-volume quantization conditions become available\footnotecircle{The author has no doubt that four-body quantization conditions will see major progress in the next few years and will be widely applied to a range of important spectroscopic problems. For example, by how much do the coupling effects between the $\pi\pi$ and $\pi\pi\pi\pi$ channels shift the pole position of the $\rho$ meson? The author believes that this is precisely one of the central questions that hadron spectroscopy must confront as it moves toward precision measurements.}. For systems containing still more particles, we do not list all operators individually. Instead, we point out a simple special case: if the operator formed by $N$ pseudoscalar mesons at rest is $\prod_i^N P_i(0)$, then it belongs to the $A_1^-$ irrep for odd $N$ and to the $A_1^+$ irrep for even $N$.

\begin{table}[htbp]
\centering
\caption{Representative examples of four-pseudoscalar-meson operators for the group $O_h$. Only the $A_1^+$ and $A_1^-$ irreps are listed. The notation follows Table~\ref{tab:mm-000-A1+}.}
\addtolength{\tabcolsep}{6pt}
\begin{tabular}{cc}
\toprule
Irrep & Operator \\
\midrule
\multirow{3}{*}{$A_1^+$} & $P_{1}(0) P_{2}(0) P_{3}(0) P_{4}(0)$ \\
\cmidrule(lr){2-2}
& $P_{1}(e_{x}) P_{2}(0) P_{3}(0) P_{4}(-e_{x}) + P_{1}(-e_{x}) P_{2}(0) P_{3}(0) P_{4}(e_{x})$ \\
& $+ P_{1}(e_{y}) P_{2}(0) P_{3}(0) P_{4}(-e_{y}) + P_{1}(-e_{y}) P_{2}(0) P_{3}(0) P_{4}(e_{y})$ \\
& $+ P_{1}(e_{z}) P_{2}(0) P_{3}(0) P_{4}(-e_{z}) + P_{1}(-e_{z}) P_{2}(0) P_{3}(0) P_{4}(e_{z})$ \\
\midrule
\multirow{3}{*}{$E^+$} & $P_{1}(e_{x}) P_{2}(0) P_{3}(0) P_{4}(-e_{x}) + P_{1}(-e_{x}) P_{2}(0) P_{3}(0) P_{4}(e_{x})$ \\
& $+ P_{1}(e_{y}) P_{2}(0) P_{3}(0) P_{4}(-e_{y}) + P_{1}(-e_{y}) P_{2}(0) P_{3}(0) P_{4}(e_{y})$ \\
& $- 2 P_{1}(e_{z}) P_{2}(0) P_{3}(0) P_{4}(-e_{z}) - 2 P_{1}(-e_{z}) P_{2}(0) P_{3}(0) P_{4}(e_{z})$ \\
\midrule
\multirow{4}{*}{$T_1^+$} & $P_{1}(e_{y}) P_{2}(e_{z}) P_{3}(-e_{z}) P_{4}(-e_{y}) - P_{1}(e_{y}) P_{2}(-e_{z}) P_{3}(e_{z}) P_{4}(-e_{y})$ \\
& $- P_{1}(-e_{y}) P_{2}(e_{z}) P_{3}(-e_{z}) P_{4}(e_{y}) + P_{1}(-e_{y}) P_{2}(-e_{z}) P_{3}(e_{z}) P_{4}(e_{y})$ \\
& $- P_{1}(e_{z}) P_{2}(e_{y}) P_{3}(-e_{y}) P_{4}(-e_{z}) + P_{1}(e_{z}) P_{2}(-e_{y}) P_{3}(e_{y}) P_{4}(-e_{z})$ \\
& $+ P_{1}(-e_{z}) P_{2}(e_{y}) P_{3}(-e_{y}) P_{4}(e_{z}) - P_{1}(-e_{z}) P_{2}(-e_{y}) P_{3}(e_{y}) P_{4}(e_{z})$ \\
\midrule
\multirow{1}{*}{$T_1^-$} & $P_{1}(e_{x}) P_{2}(0) P_{3}(0) P_{4}(-e_{x}) - P_{1}(-e_{x}) P_{2}(0) P_{3}(0) P_{4}(e_{x})$ \\
\midrule
\multirow{4}{*}{$T_2^+$} & $P_{1}(e_{y}) P_{2}(e_{z}) P_{3}(-e_{z}) P_{4}(-e_{y}) - P_{1}(e_{y}) P_{2}(-e_{z}) P_{3}(e_{z}) P_{4}(-e_{y})$ \\
& $- P_{1}(-e_{y}) P_{2}(e_{z}) P_{3}(-e_{z}) P_{4}(-e_{y}) + P_{1}(-e_{y}) P_{2}(-e_{z}) P_{3}(e_{z}) P_{4}(e_{y})$ \\
& $+ P_{1}(e_{z}) P_{2}(e_{y}) P_{3}(-e_{y}) P_{4}(-e_{z}) - P_{1}(e_{z}) P_{2}(-e_{y}) P_{3}(e_{y}) P_{4}(-e_{z})$ \\
& $- P_{1}(-e_{z}) P_{2}(e_{y}) P_{3}(-e_{y}) P_{4}(e_{z}) + P_{1}(-e_{z}) P_{2}(-e_{y}) P_{3}(e_{y}) P_{4}(e_{z})$ \\
\midrule
\multirow{4}{*}{$T_2^-$} & $P_{1}(e_{x}) P_{2}(e_{y}) P_{3}(-e_{y}) P_{4}(-e_{x}) + P_{1}(e_{x}) P_{2}(-e_{y}) P_{3}(e_{y}) P_{4}(-e_{x})$ \\
& $- P_{1}(e_{x}) P_{2}(e_{z}) P_{3}(-e_{z}) P_{4}(-e_{x}) - P_{1}(e_{x}) P_{2}(-e_{z}) P_{3}(e_{z}) P_{4}(-e_{x})$ \\
& $- P_{1}(-e_{x}) P_{2}(e_{y}) P_{3}(-e_{y}) P_{4}(e_{x}) - P_{1}(-e_{x}) P_{2}(-e_{y}) P_{3}(e_{y}) P_{4}(e_{x})$ \\
& $+ P_{1}(-e_{x}) P_{2}(e_{z}) P_{3}(-e_{z}) P_{4}(e_{x}) + P_{1}(-e_{x}) P_{2}(-e_{z}) P_{3}(e_{z}) P_{4}(e_{x})$ \\
\bottomrule
\end{tabular}
\addtolength{\tabcolsep}{-6pt}
\label{tab:four-000}
\end{table}

\cleardoublepage
\chapter{Effective-Mass Plots and Energy-Level Fits for $D\pi \to D_0^*(2300)$ Scattering}
\label{appendix:two_body_problems}

\begin{figure}[htbp]
\centering
\includegraphics[width=0.49\columnwidth]{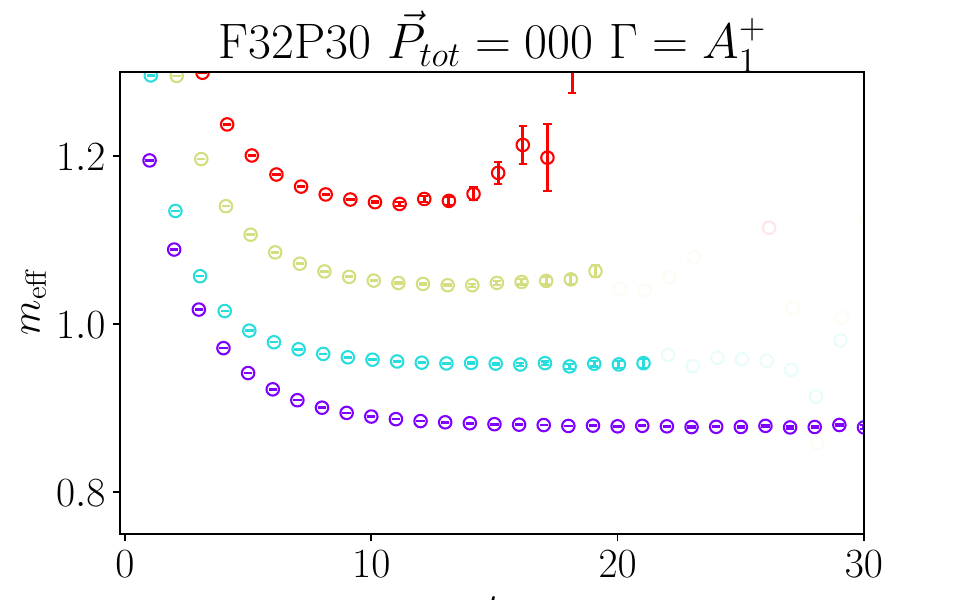}
\includegraphics[width=0.49\columnwidth]{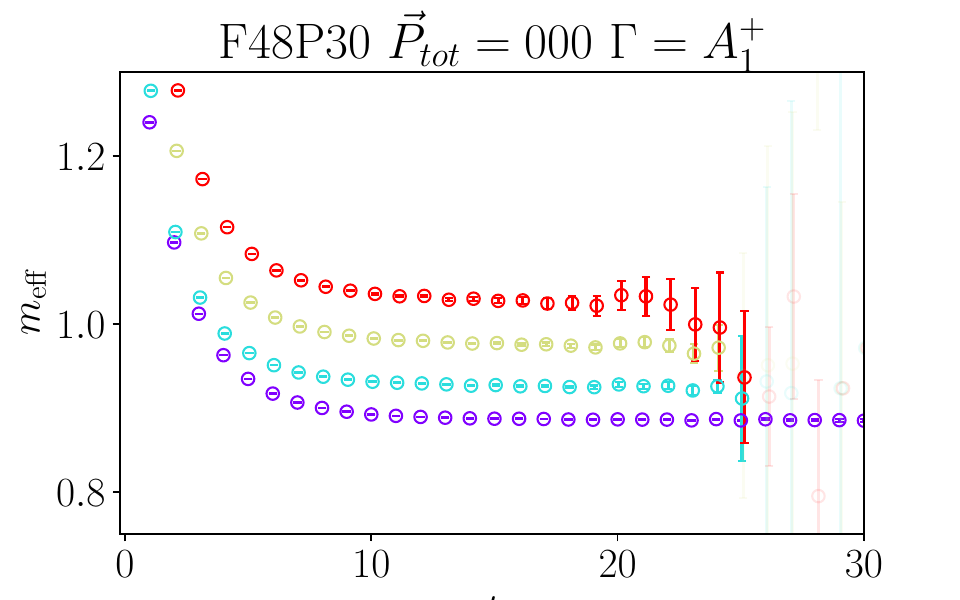}
\includegraphics[width=0.49\columnwidth]{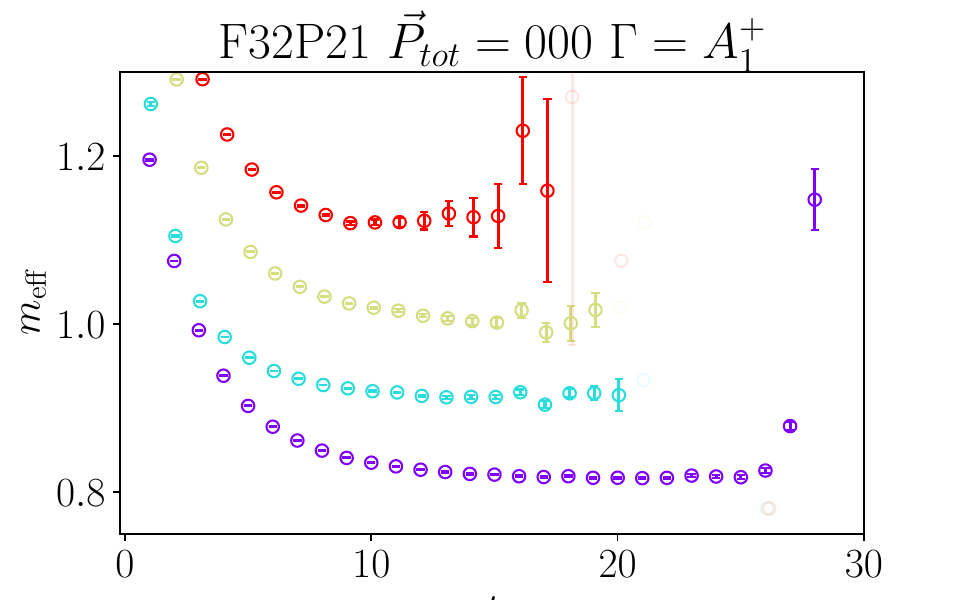}
\includegraphics[width=0.49\columnwidth]{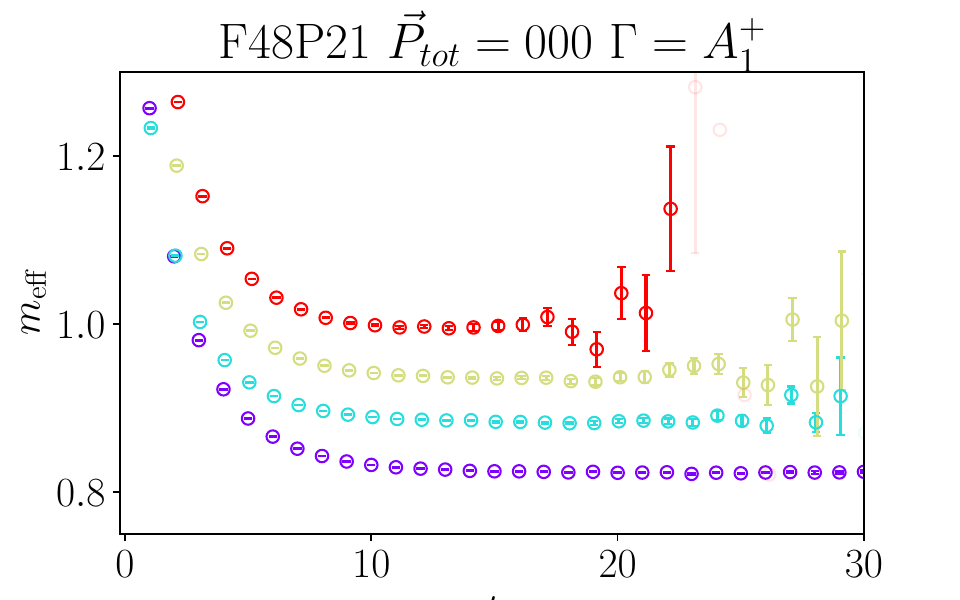}
\includegraphics[width=0.49\columnwidth]{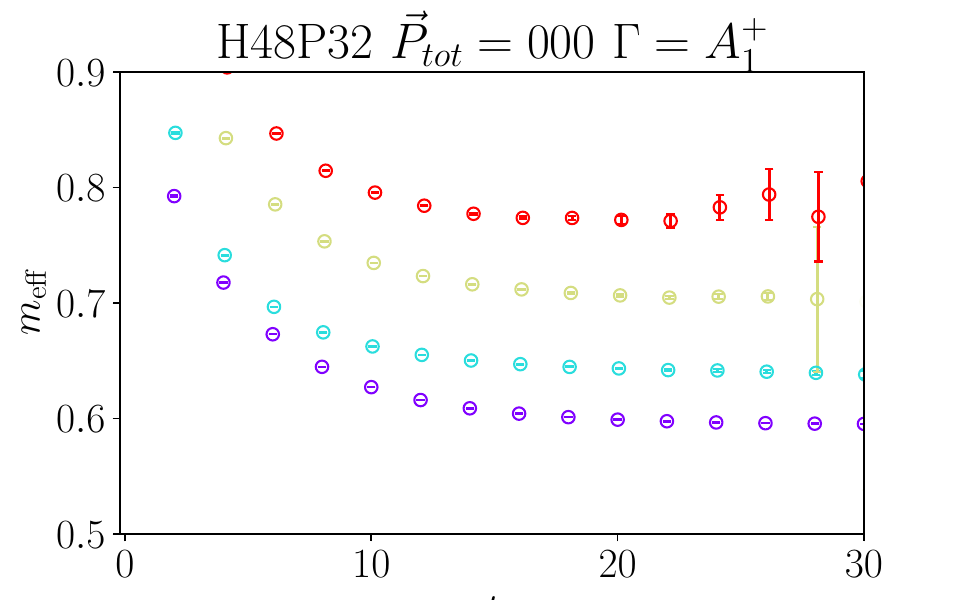}
\includegraphics[width=0.49\columnwidth]{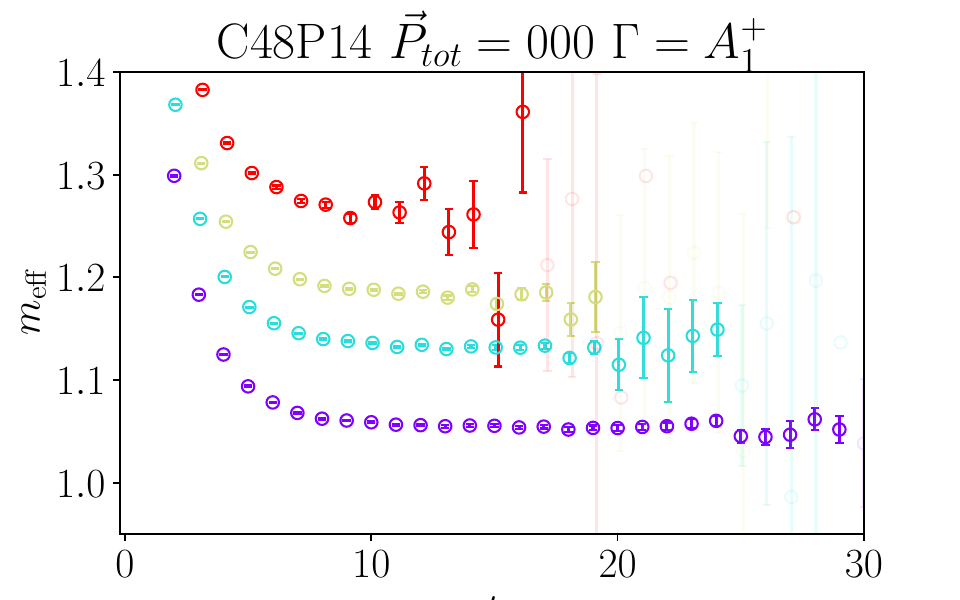}
\caption{Effective-mass plots of the generalized eigenvalues $\lambda_n(t)$ for the two-point correlation functions of the $D\pi$ system in the center-of-mass-frame $A_1^+$ irrep. Different colors correspond to different energy levels. The vertical axis is in lattice units.}
\label{fig:Dpi-meff-A1+}
\end{figure}

\begin{figure}[htbp]
\centering
\includegraphics[width=0.49\columnwidth]{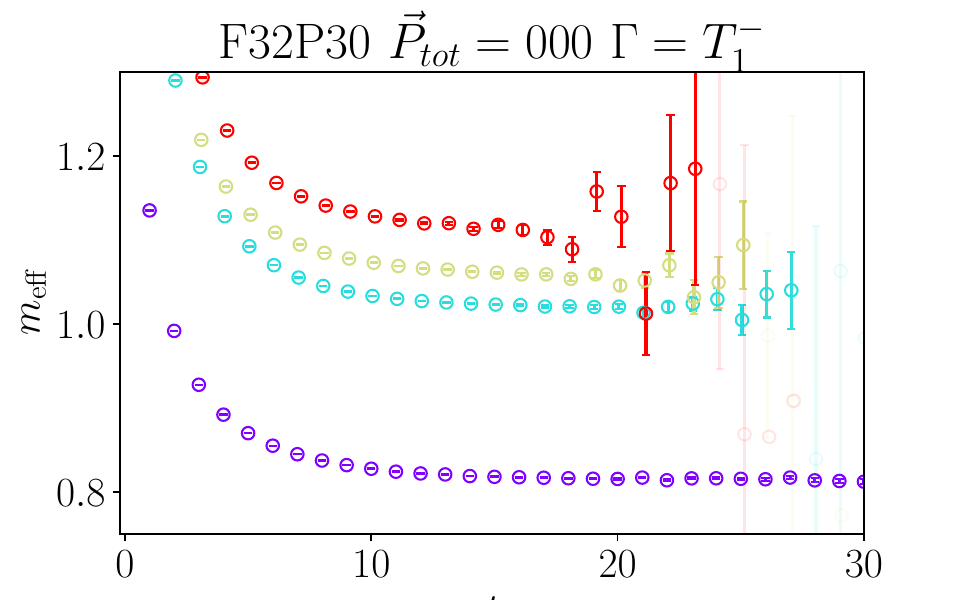}
\includegraphics[width=0.49\columnwidth]{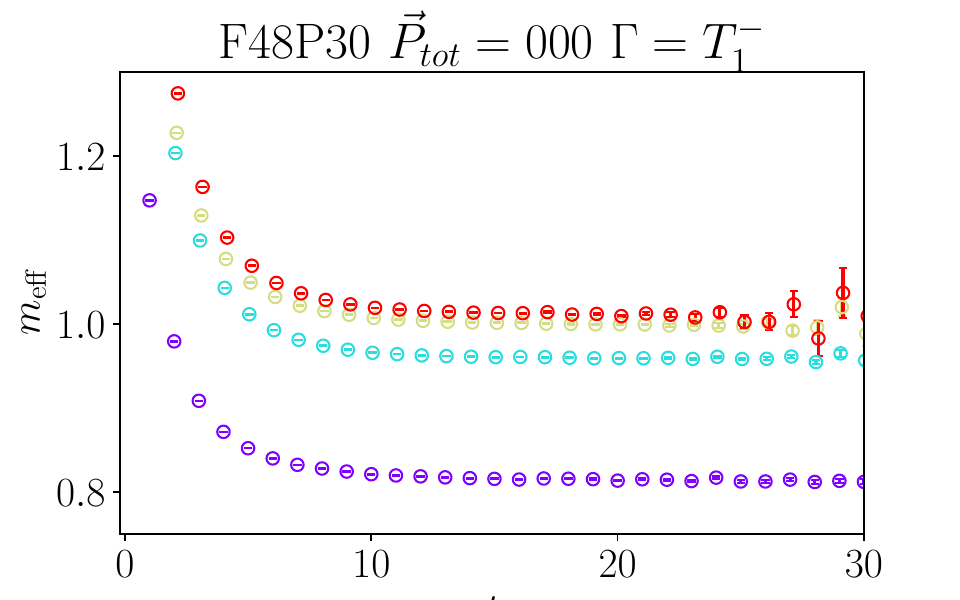}
\includegraphics[width=0.49\columnwidth]{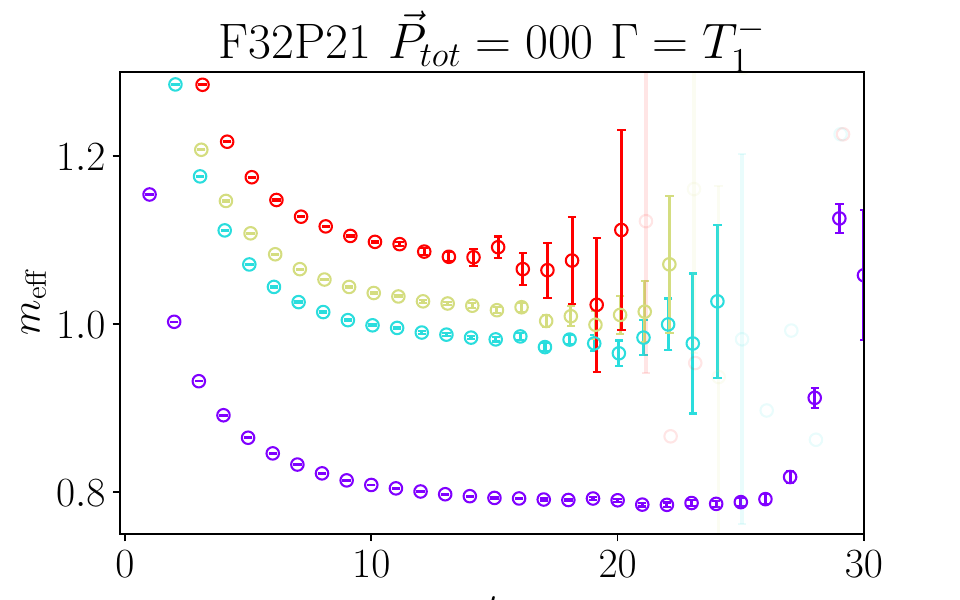}
\includegraphics[width=0.49\columnwidth]{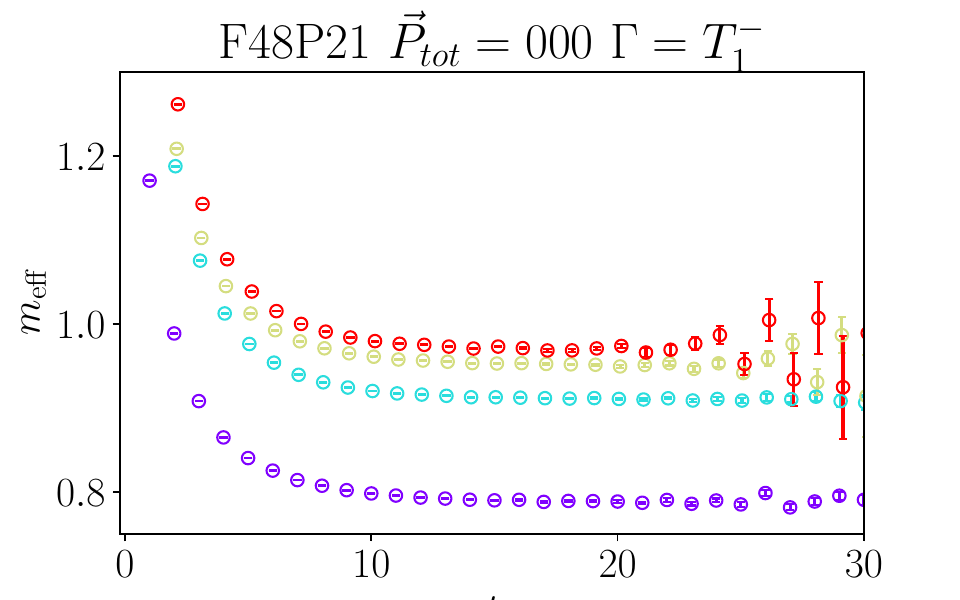}
\includegraphics[width=0.49\columnwidth]{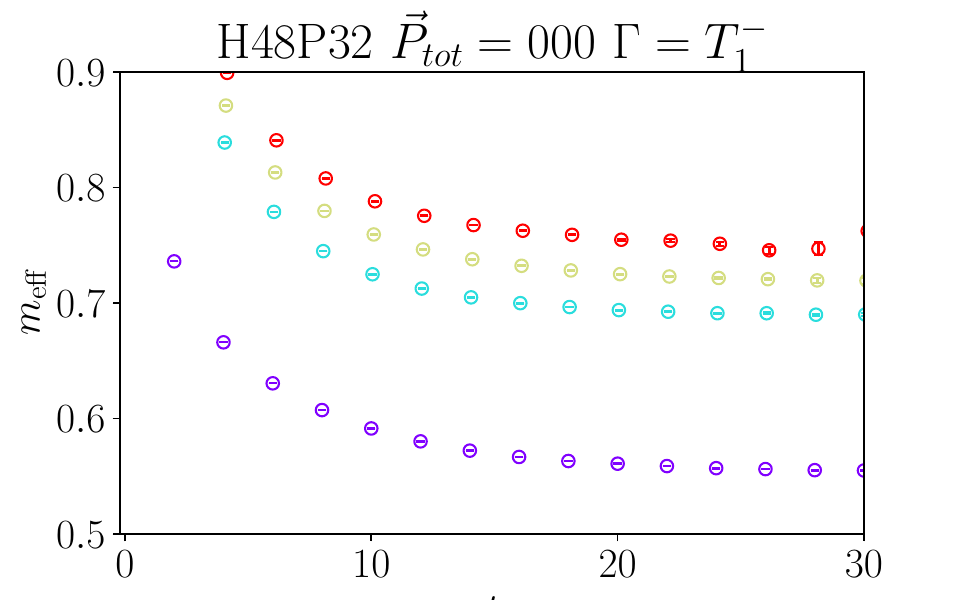}
\includegraphics[width=0.49\columnwidth]{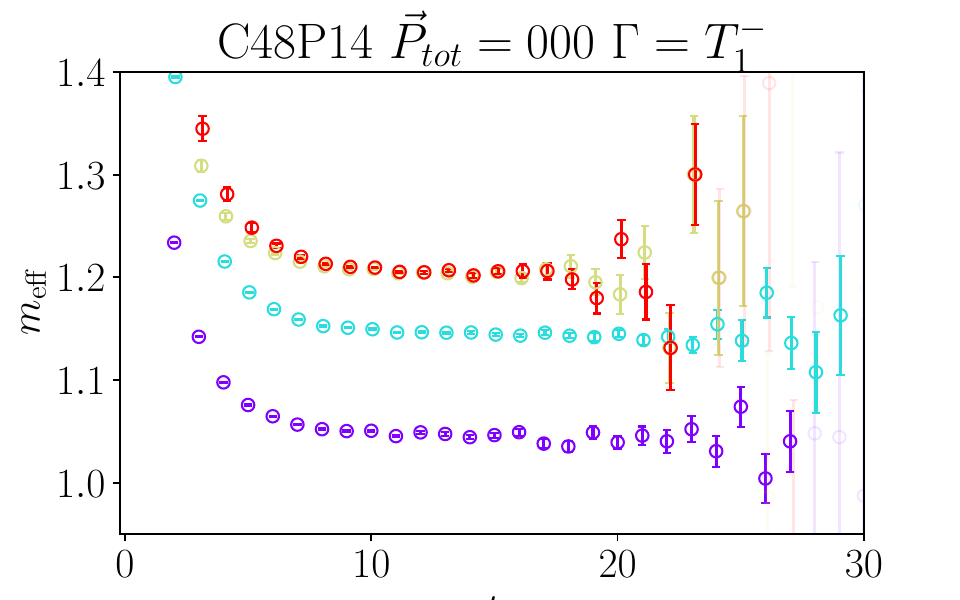}
\caption{Effective-mass plots of the generalized eigenvalues $\lambda_n(t)$ for the two-point correlation functions of the $D\pi$ system in the center-of-mass-frame $T_1^-$ irrep.}
\label{fig:Dpi-meff-T1-}
\end{figure}

\begin{figure}[htbp]
\centering
\includegraphics[width=0.49\columnwidth]{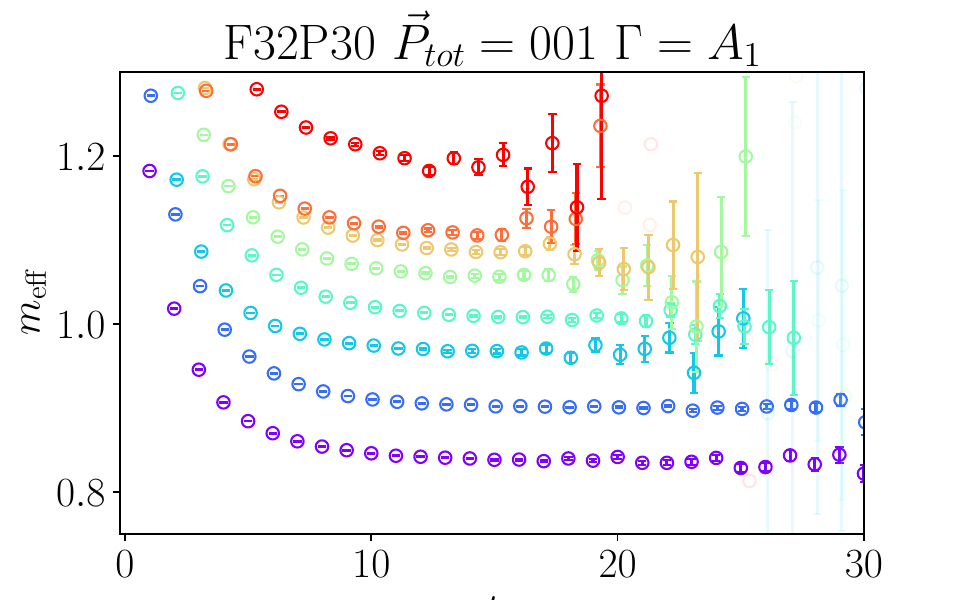}
\includegraphics[width=0.49\columnwidth]{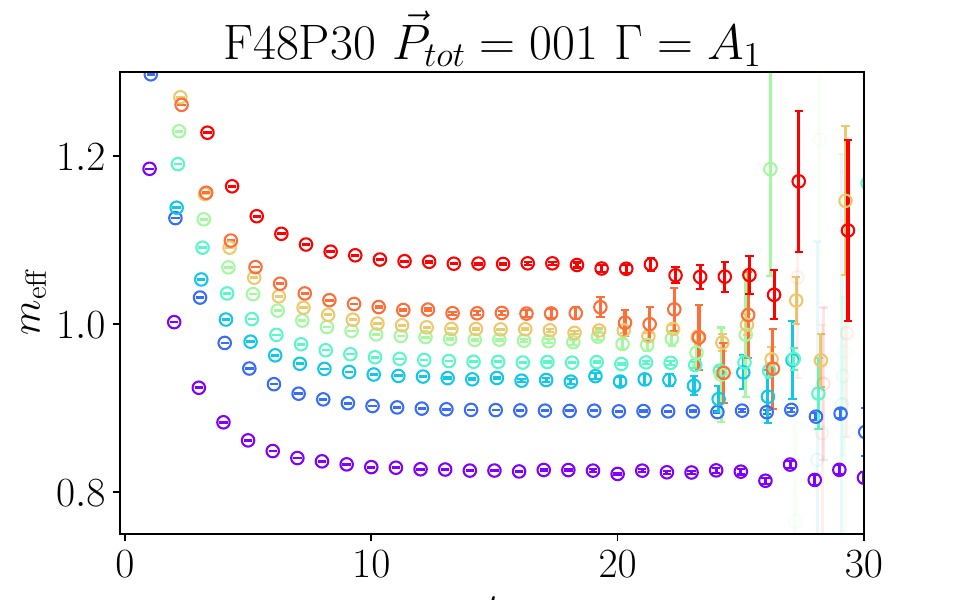}
\includegraphics[width=0.49\columnwidth]{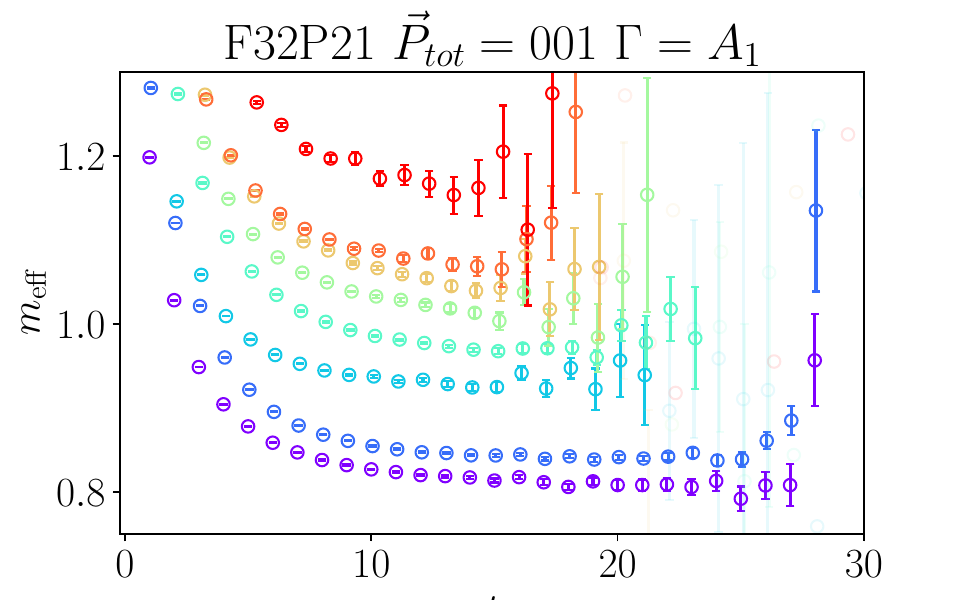}
\includegraphics[width=0.49\columnwidth]{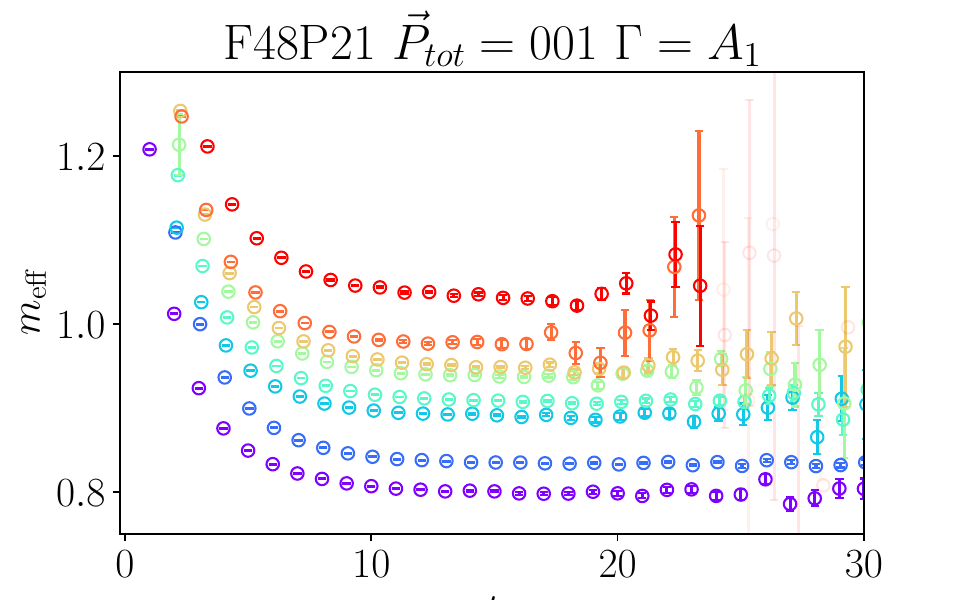}
\includegraphics[width=0.49\columnwidth]{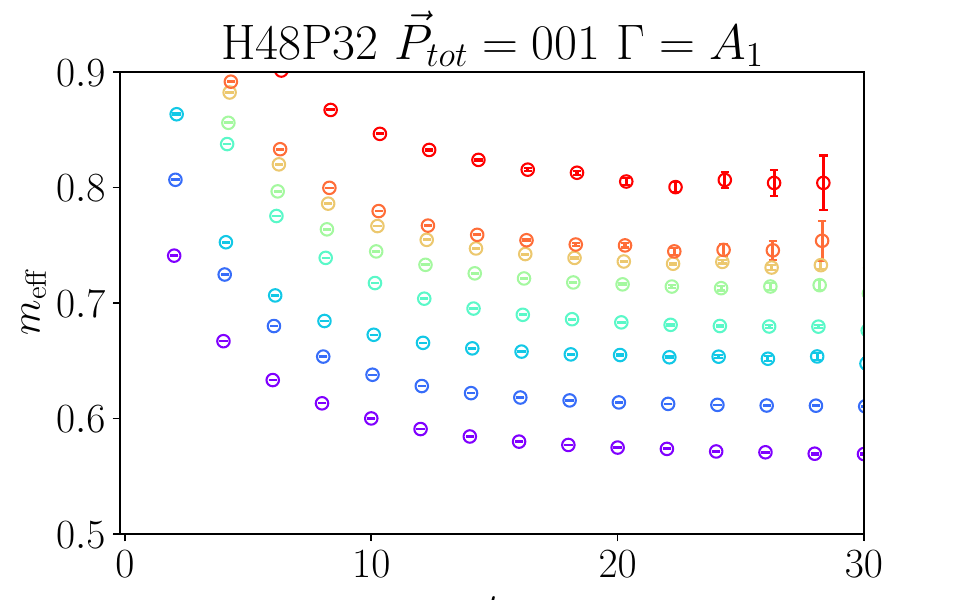}
\includegraphics[width=0.49\columnwidth]{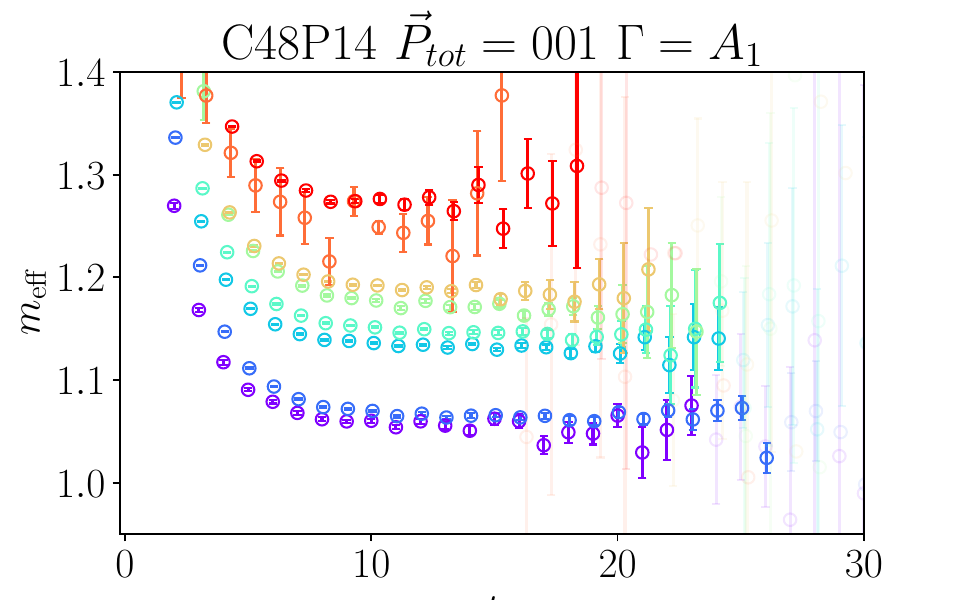}
\caption{Effective-mass plots of the generalized eigenvalues $\lambda_n(t)$ for the two-point correlation functions of the $D\pi$ system in the $A_1$ irrep at total momentum $\vec{P} = [0,0,1] \frac{2\pi}{L}$. Different colors correspond to different energy levels. The vertical axis is in lattice units.}
\label{fig:Dpi-meff-001-A1}
\end{figure}

\begin{figure}[htbp]
\centering
\includegraphics[width=0.49\columnwidth]{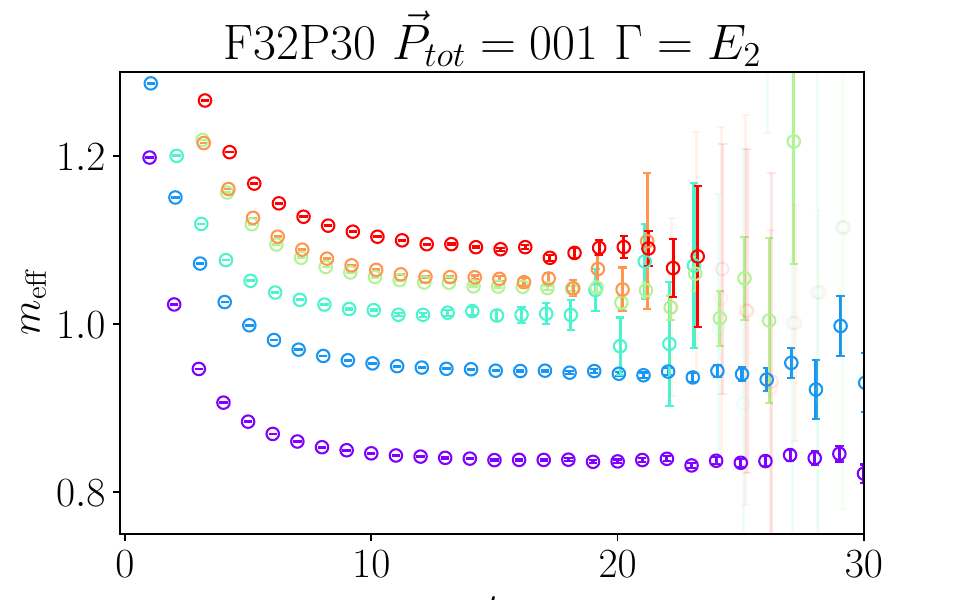}
\includegraphics[width=0.49\columnwidth]{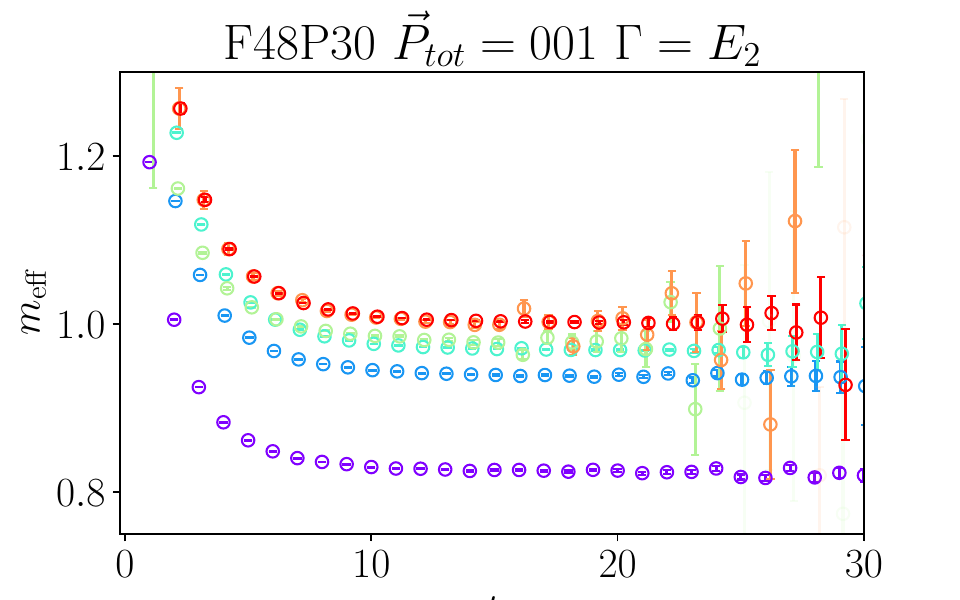}
\includegraphics[width=0.49\columnwidth]{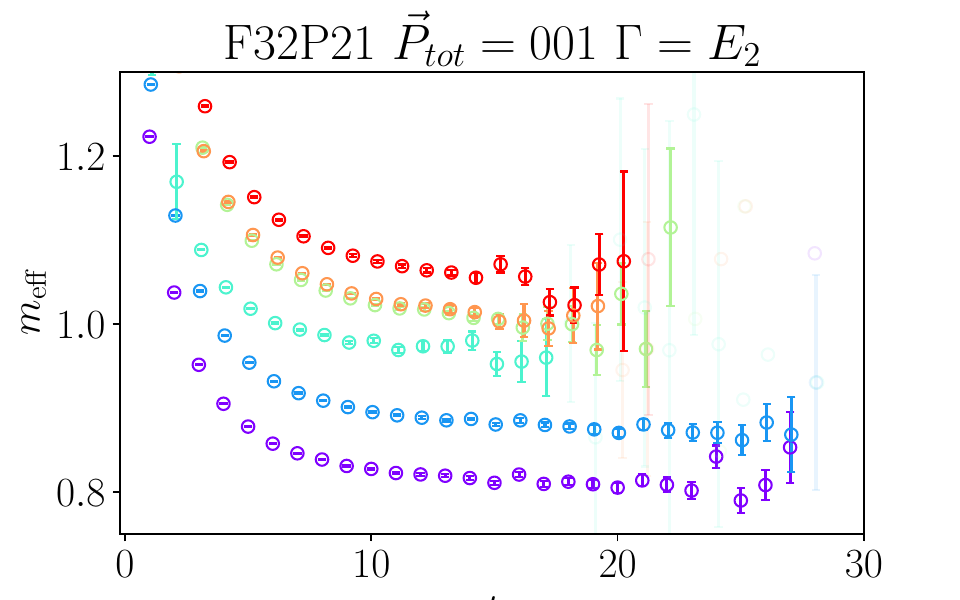}
\includegraphics[width=0.49\columnwidth]{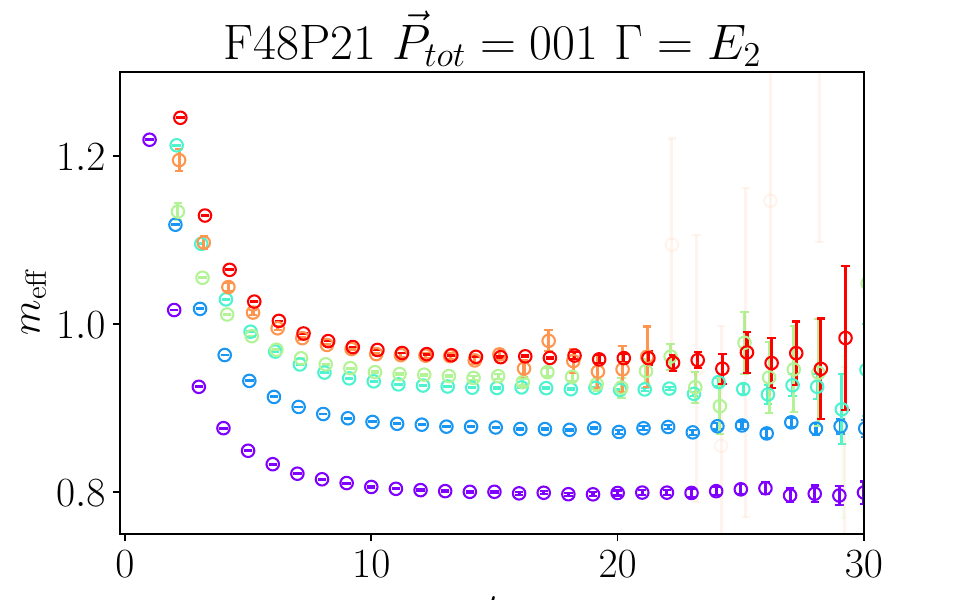}
\includegraphics[width=0.49\columnwidth]{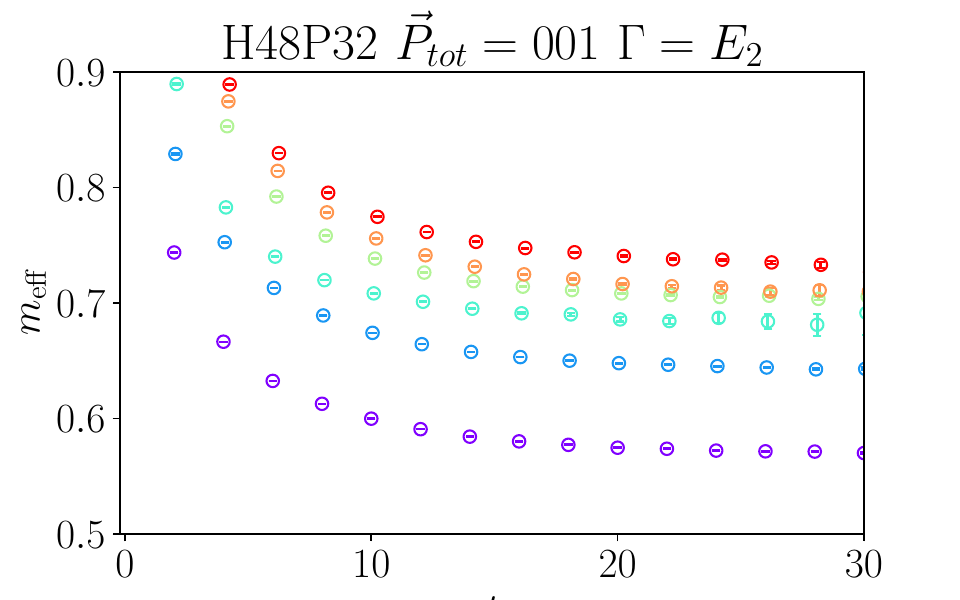}
\includegraphics[width=0.49\columnwidth]{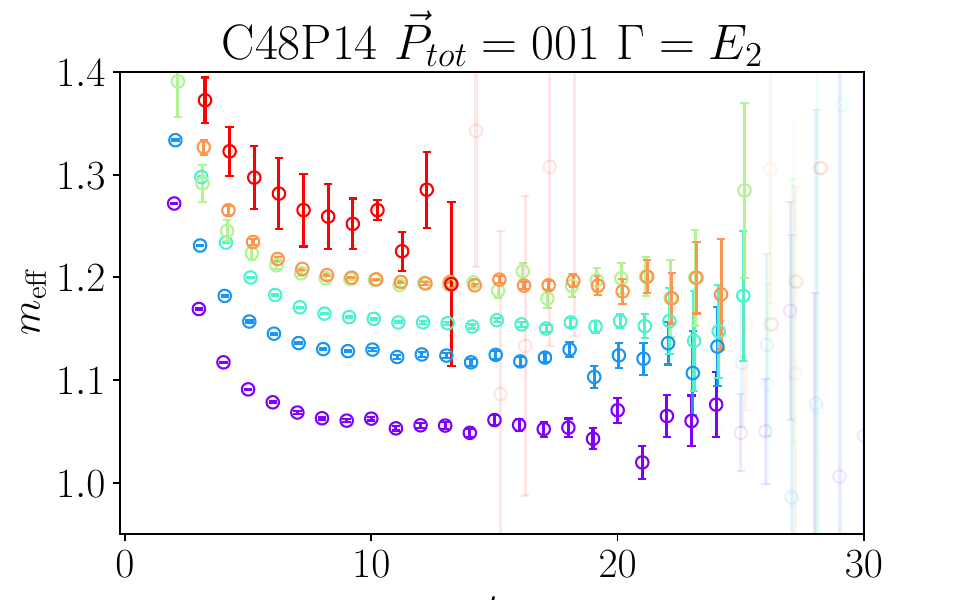}
\caption{Effective-mass plots of the generalized eigenvalues $\lambda_n(t)$ for the two-point correlation functions of the $D\pi$ system in the $E_2$ irrep at total momentum $\vec{P} = [0,0,1] \frac{2\pi}{L}$. Different colors correspond to different energy levels. The vertical axis is in lattice units.}
\label{fig:Dpi-meff-001-E2}
\end{figure}

\begin{figure}[htbp]
\centering
\includegraphics[width=0.49\columnwidth]{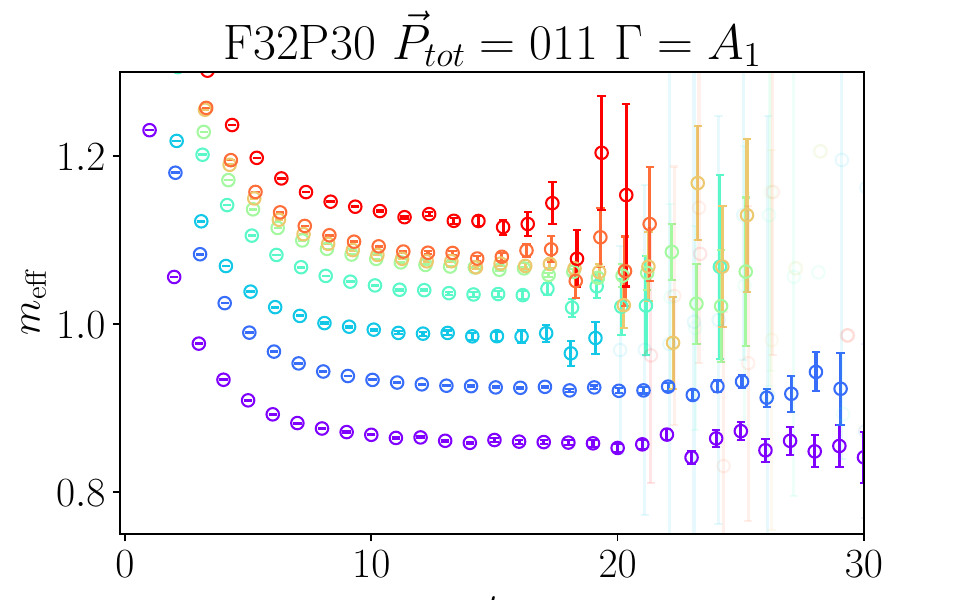}
\includegraphics[width=0.49\columnwidth]{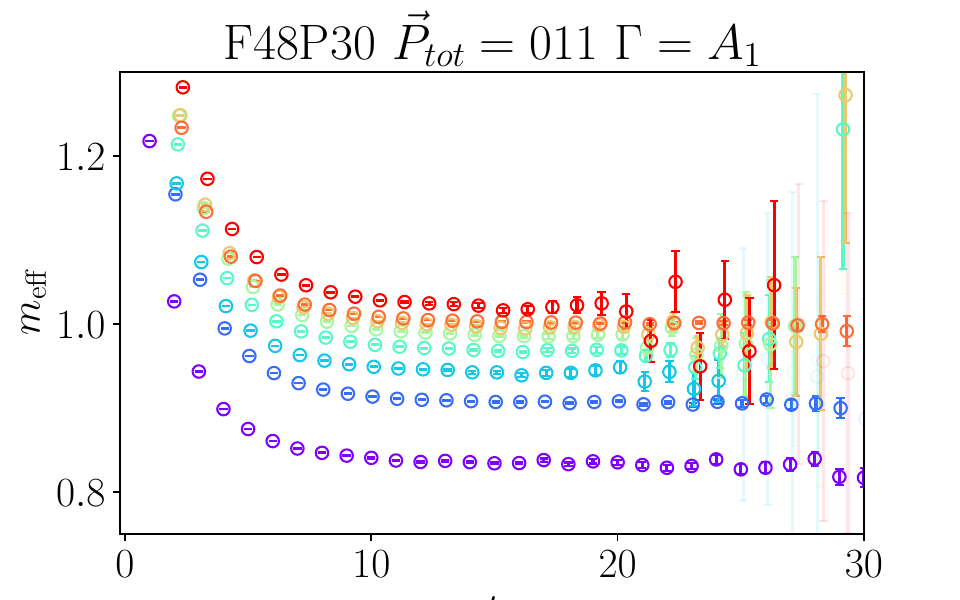}
\includegraphics[width=0.49\columnwidth]{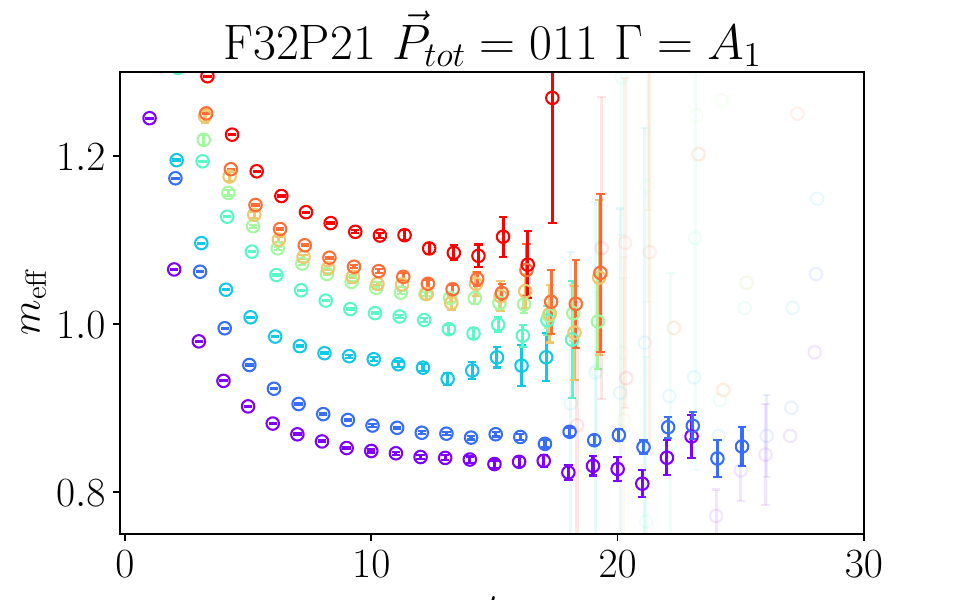}
\includegraphics[width=0.49\columnwidth]{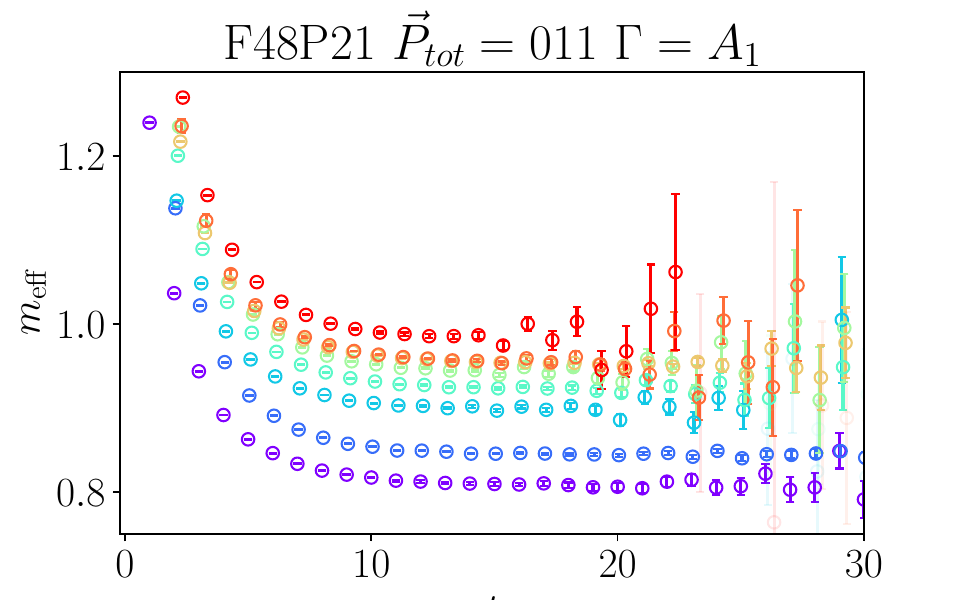}
\includegraphics[width=0.49\columnwidth]{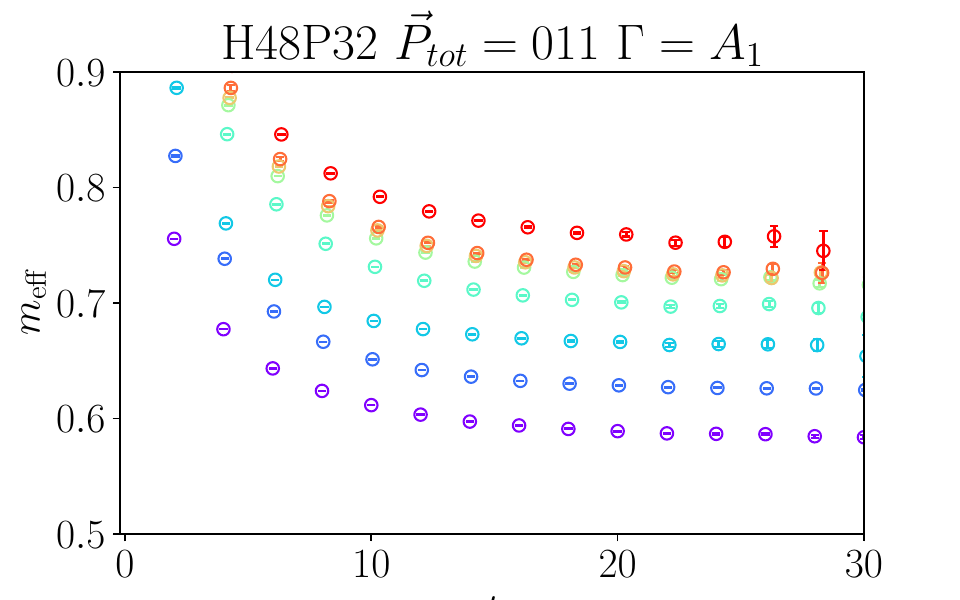}
\includegraphics[width=0.49\columnwidth]{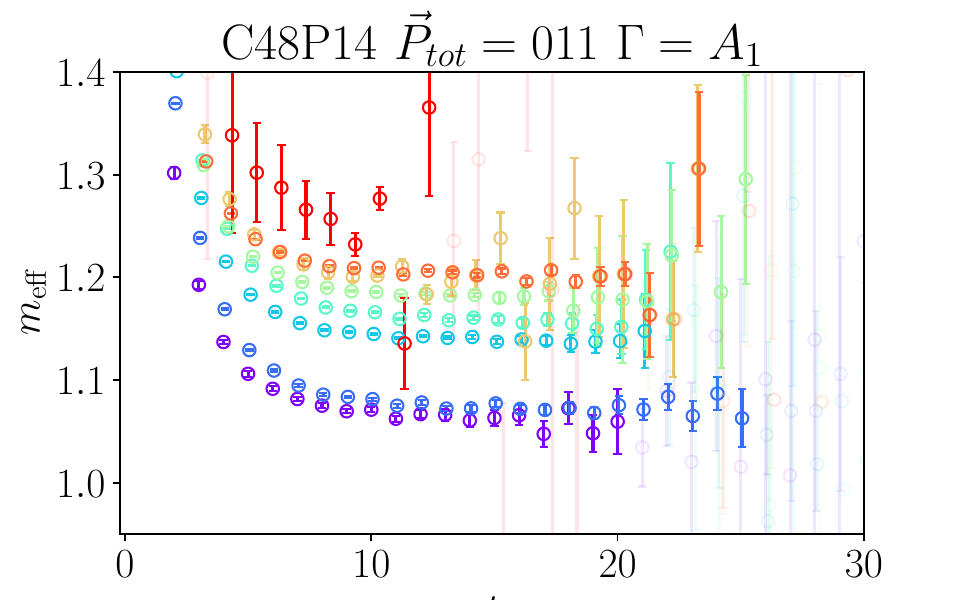}
\caption{Effective-mass plots of the generalized eigenvalues $\lambda_n(t)$ for the two-point correlation functions of the $D\pi$ system in the $A_1$ irrep at total momentum $\vec{P} = [0,1,1] \frac{2\pi}{L}$. Different colors correspond to different energy levels. The vertical axis is in lattice units.}
\label{fig:Dpi-meff-011-A1}
\end{figure}

\begin{figure}[htbp]
\centering
\includegraphics[width=0.49\columnwidth]{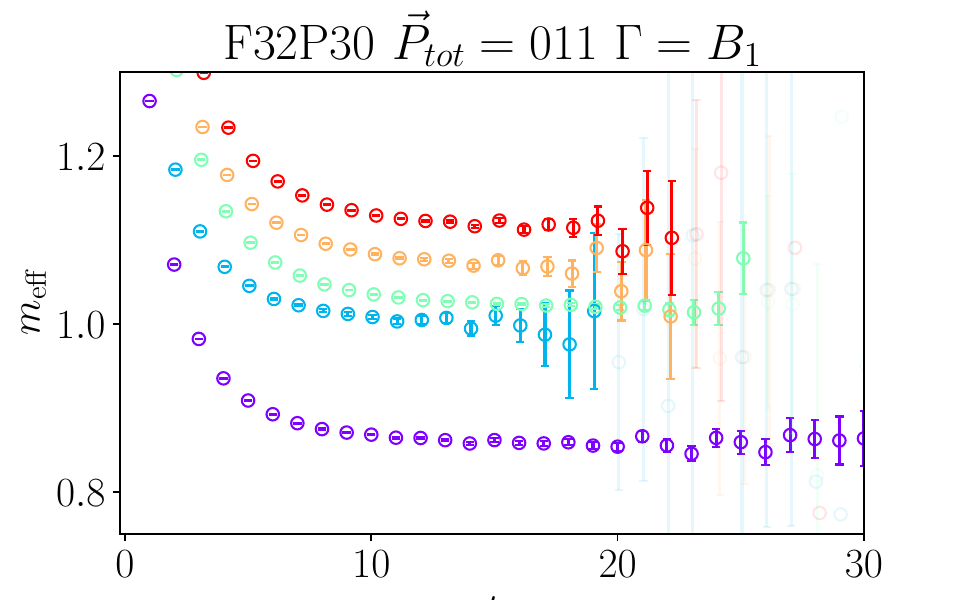}
\includegraphics[width=0.49\columnwidth]{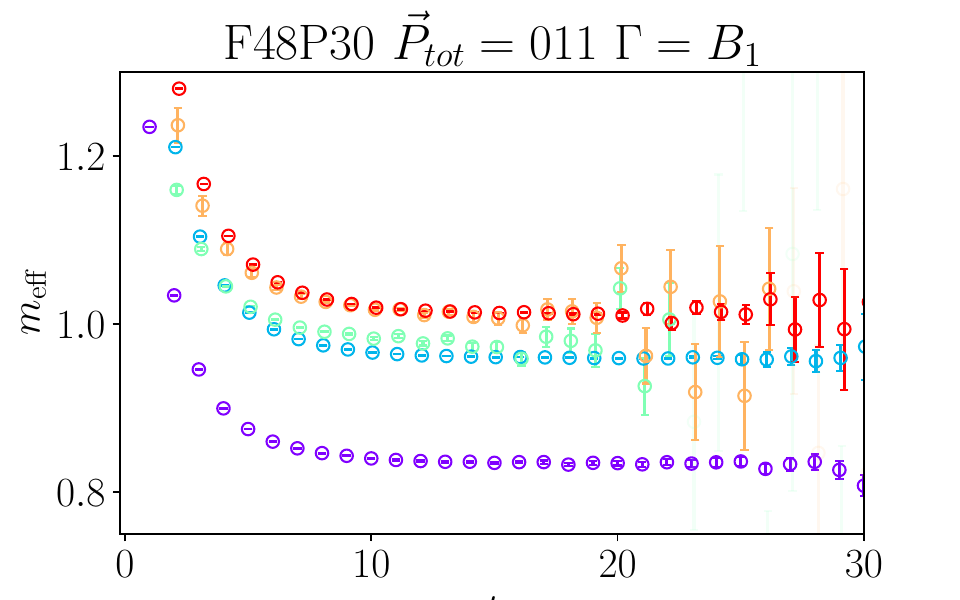}
\includegraphics[width=0.49\columnwidth]{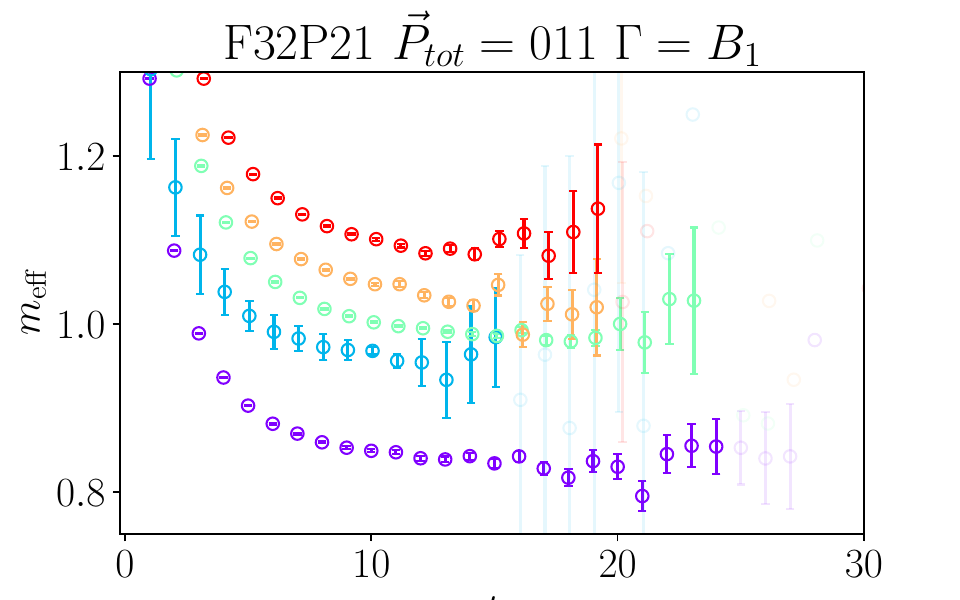}
\includegraphics[width=0.49\columnwidth]{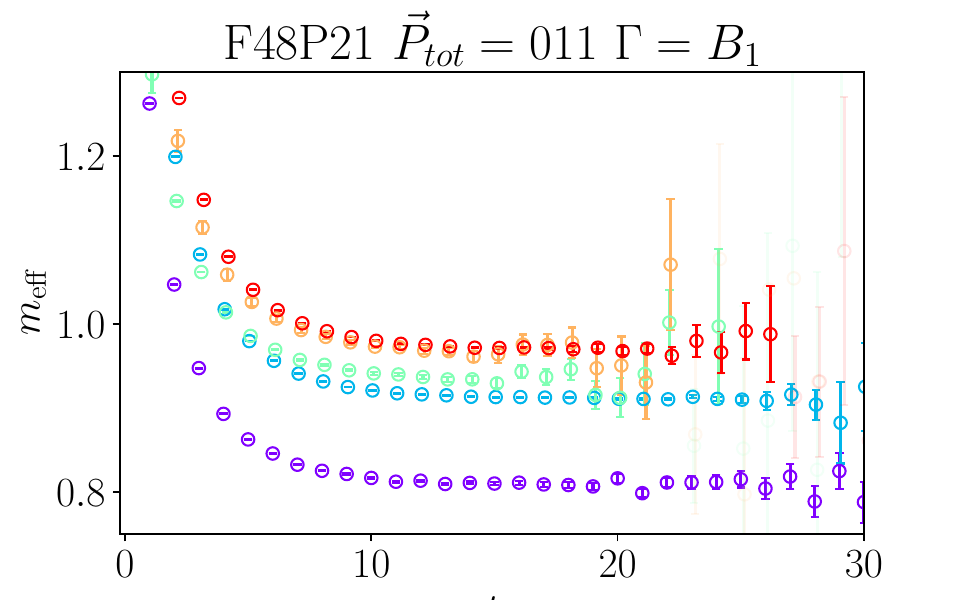}
\includegraphics[width=0.49\columnwidth]{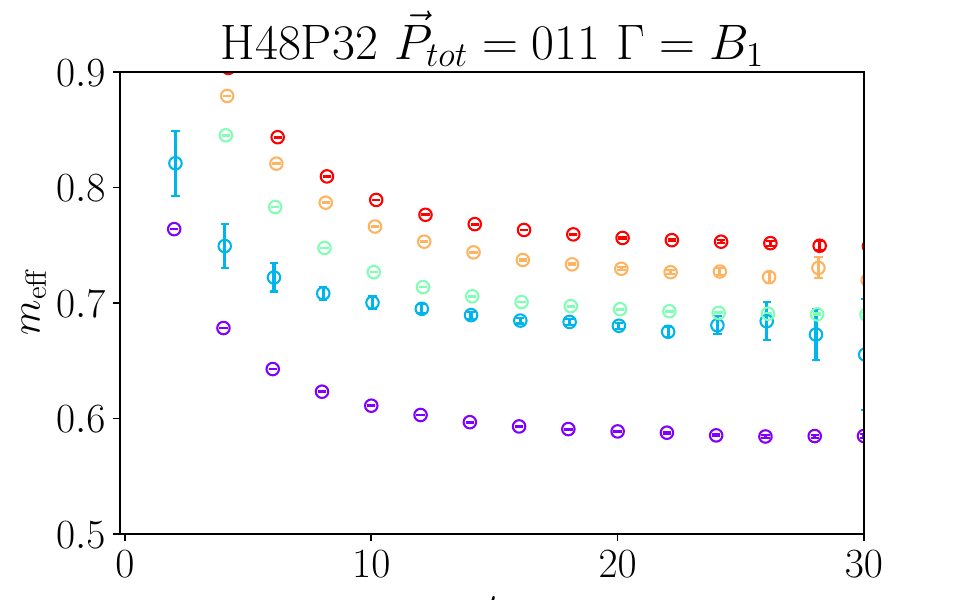}
\includegraphics[width=0.49\columnwidth]{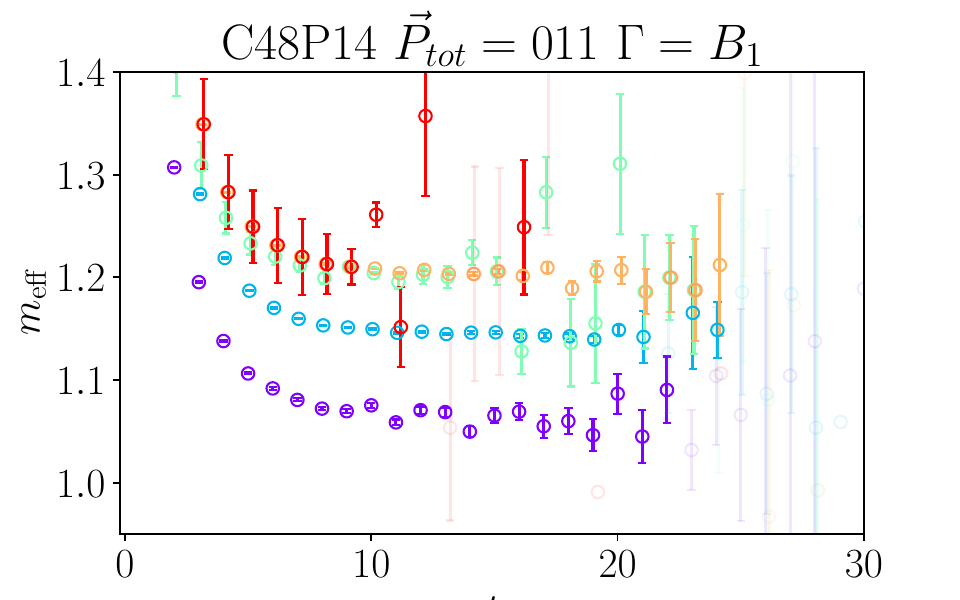}
\caption{Effective-mass plots of the generalized eigenvalues $\lambda_n(t)$ for the two-point correlation functions of the $D\pi$ system in the $B_1$ irrep at total momentum $\vec{P} = [0,1,1] \frac{2\pi}{L}$. Different colors correspond to different energy levels. The vertical axis is in lattice units.}
\label{fig:Dpi-meff-011-B1}
\end{figure}

\begin{figure}[htbp]
\centering
\includegraphics[width=0.49\columnwidth]{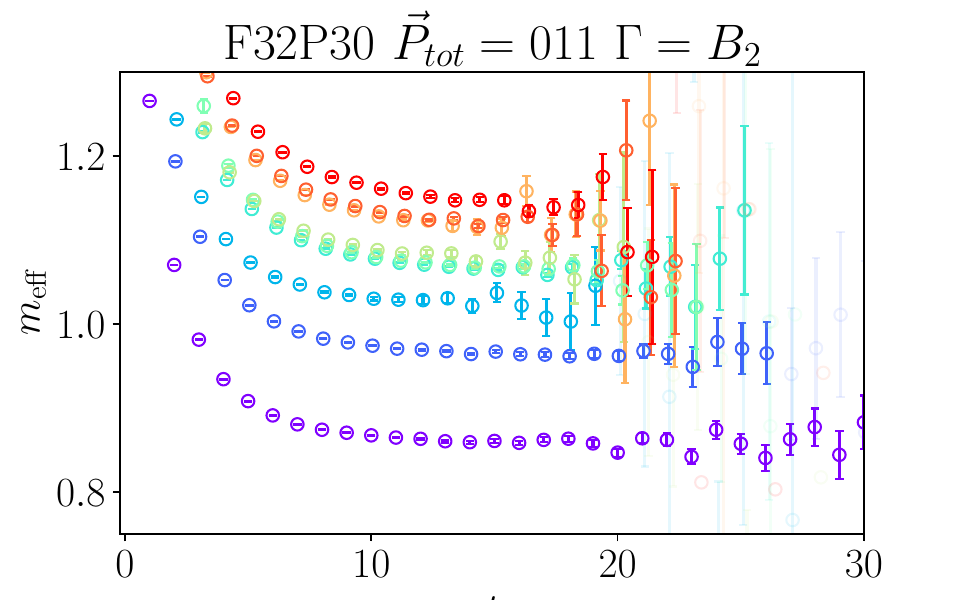}
\includegraphics[width=0.49\columnwidth]{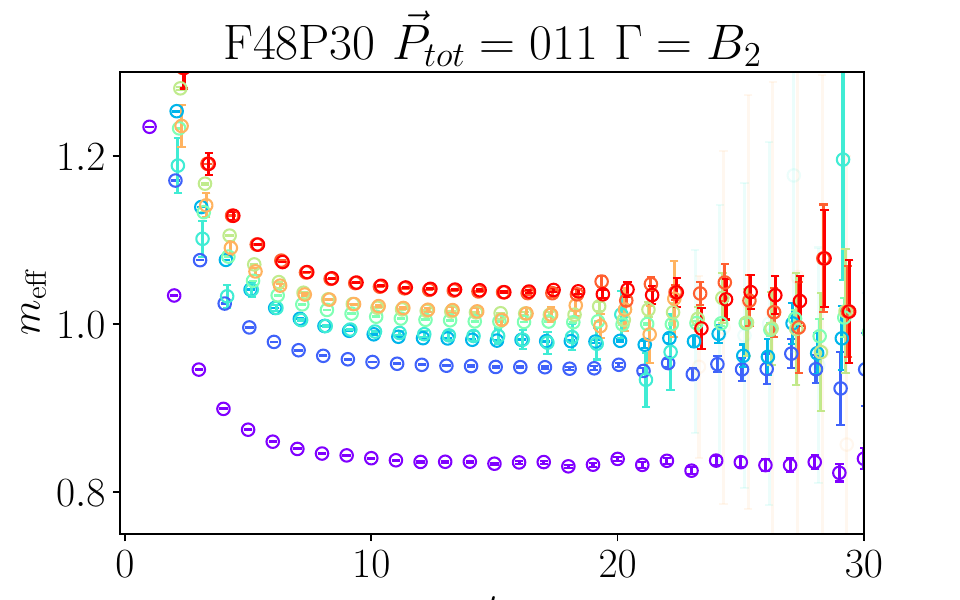}
\includegraphics[width=0.49\columnwidth]{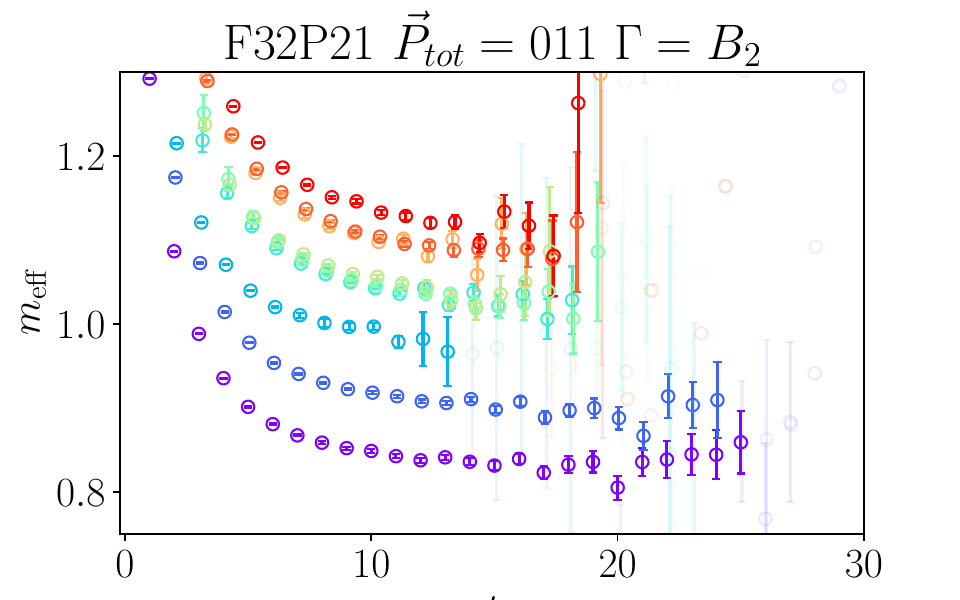}
\includegraphics[width=0.49\columnwidth]{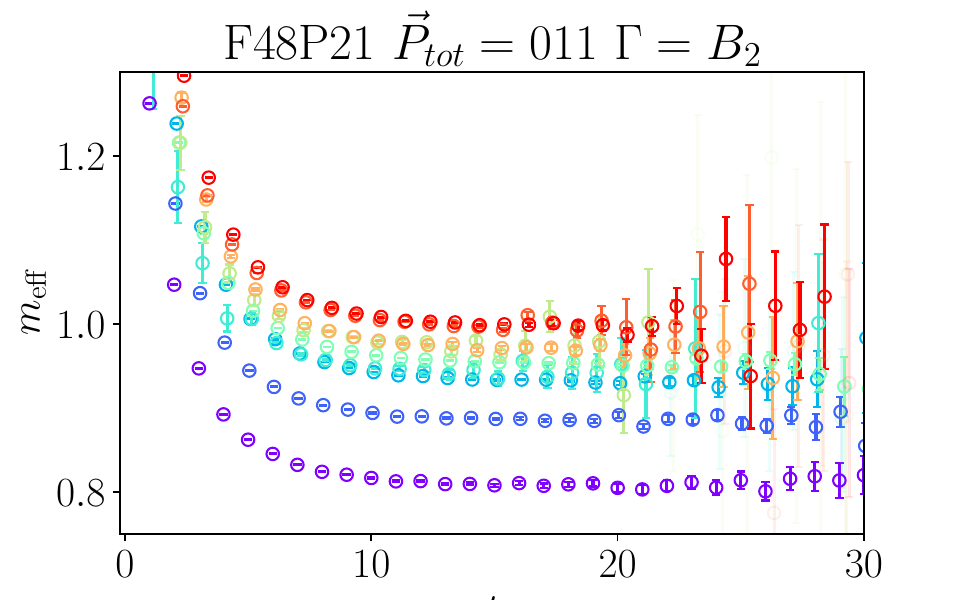}
\includegraphics[width=0.49\columnwidth]{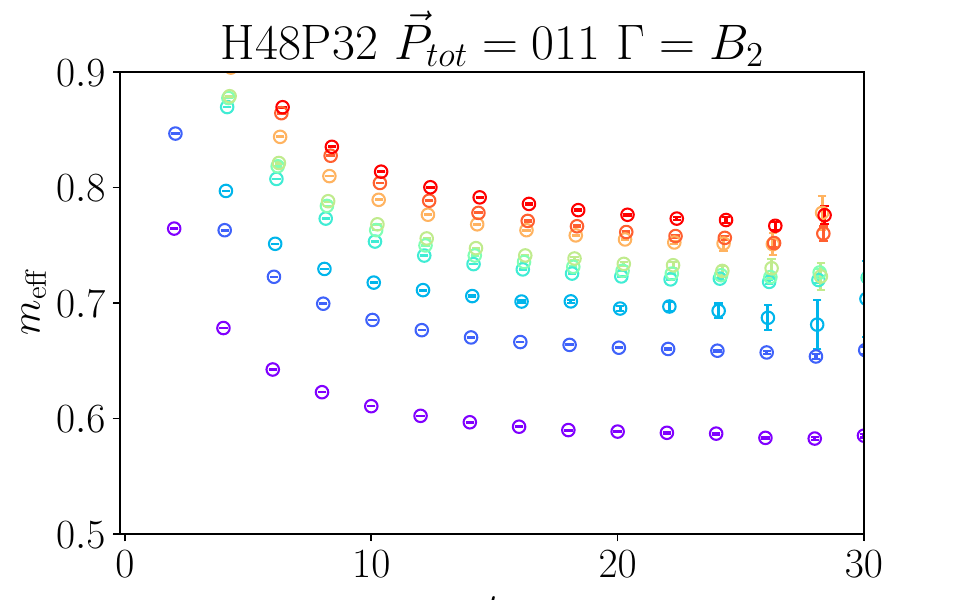}
\includegraphics[width=0.49\columnwidth]{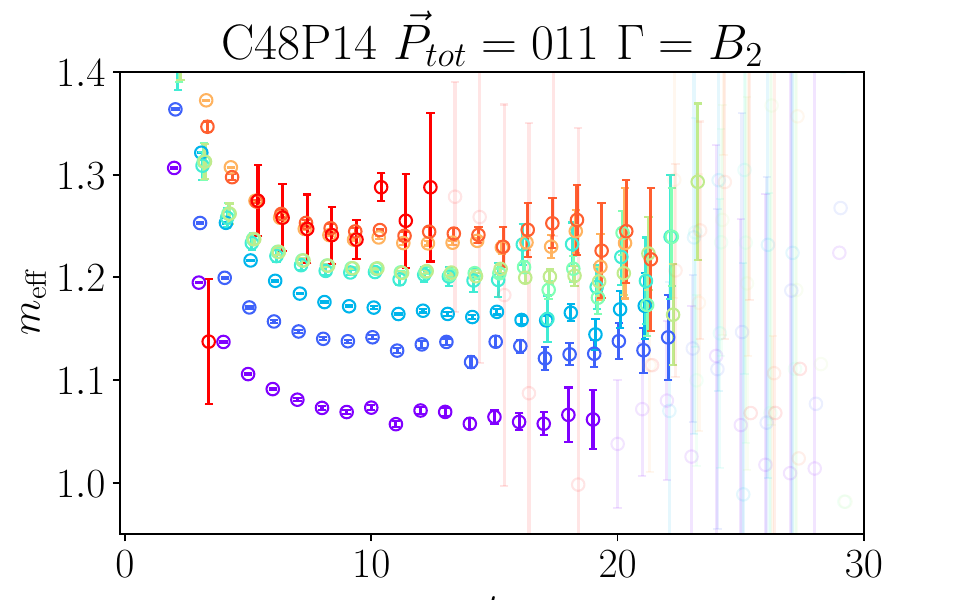}
\caption{Effective-mass plots of the generalized eigenvalues $\lambda_n(t)$ for the two-point correlation functions of the $D\pi$ system in the $B_2$ irrep at total momentum $\vec{P} = [0,1,1] \frac{2\pi}{L}$. Different colors correspond to different energy levels. The vertical axis is in lattice units.}
\label{fig:Dpi-meff-011-B2}
\end{figure}

\begin{figure}[htbp]
\centering
\includegraphics[width=0.49\columnwidth]{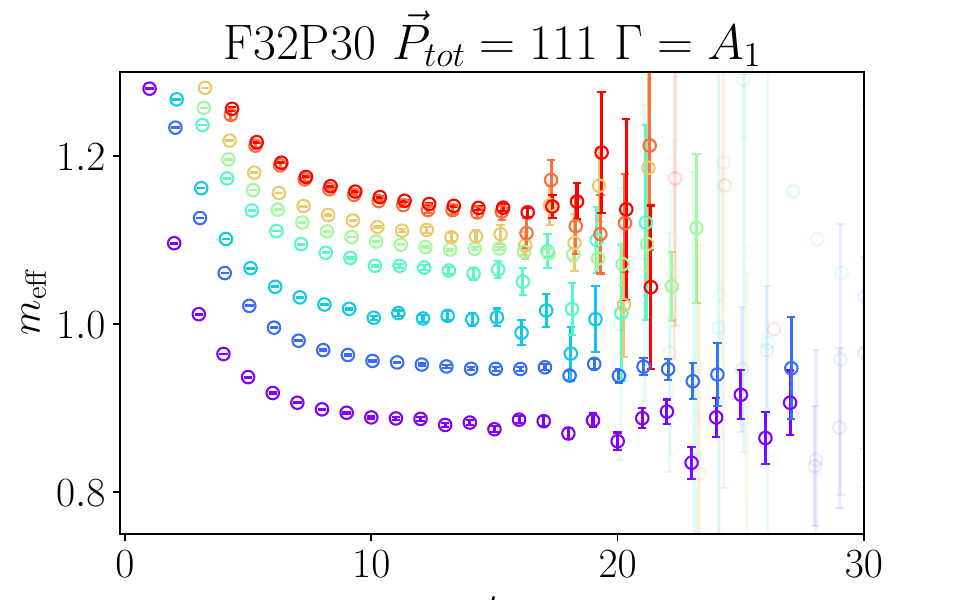}
\includegraphics[width=0.49\columnwidth]{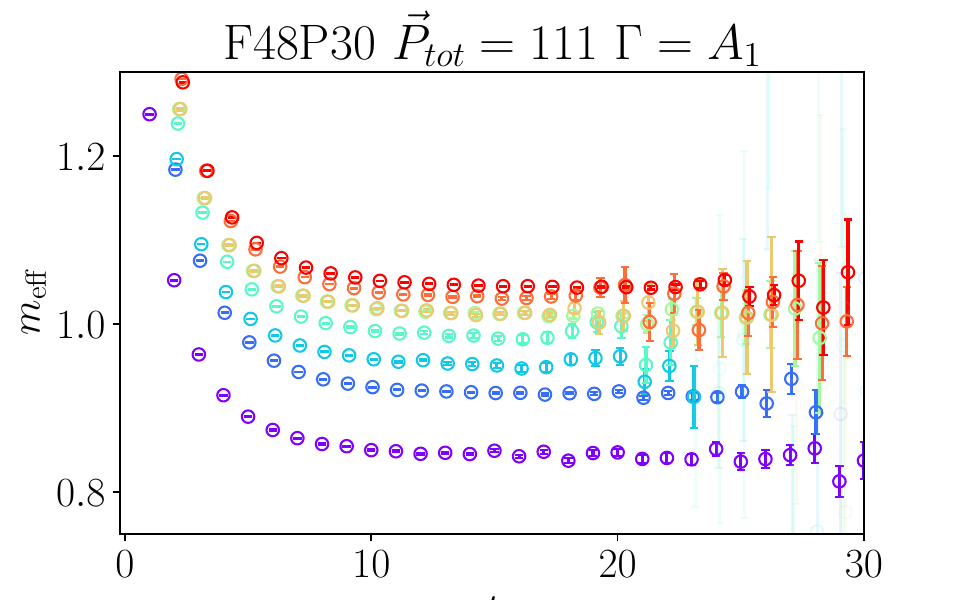}
\includegraphics[width=0.49\columnwidth]{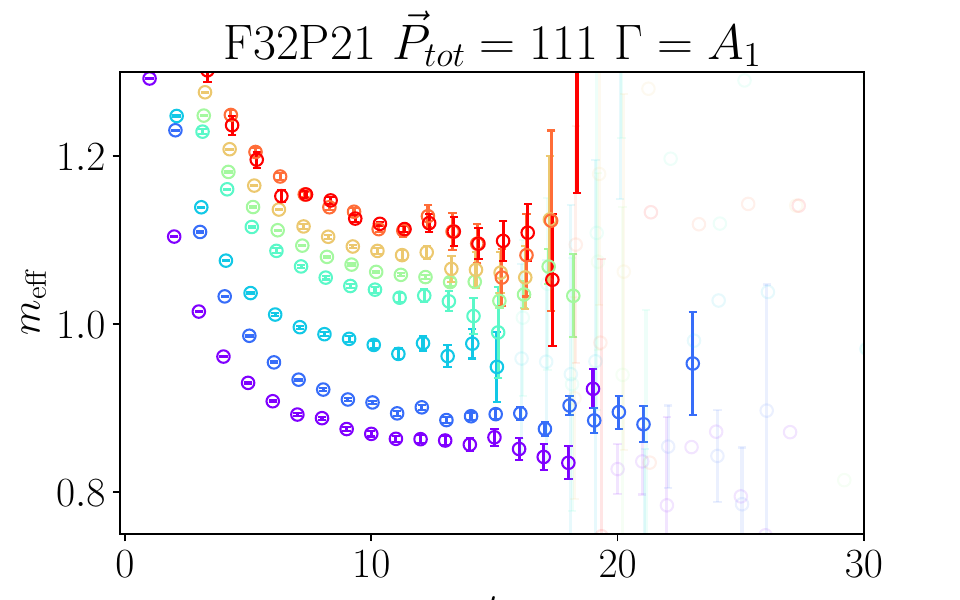}
\includegraphics[width=0.49\columnwidth]{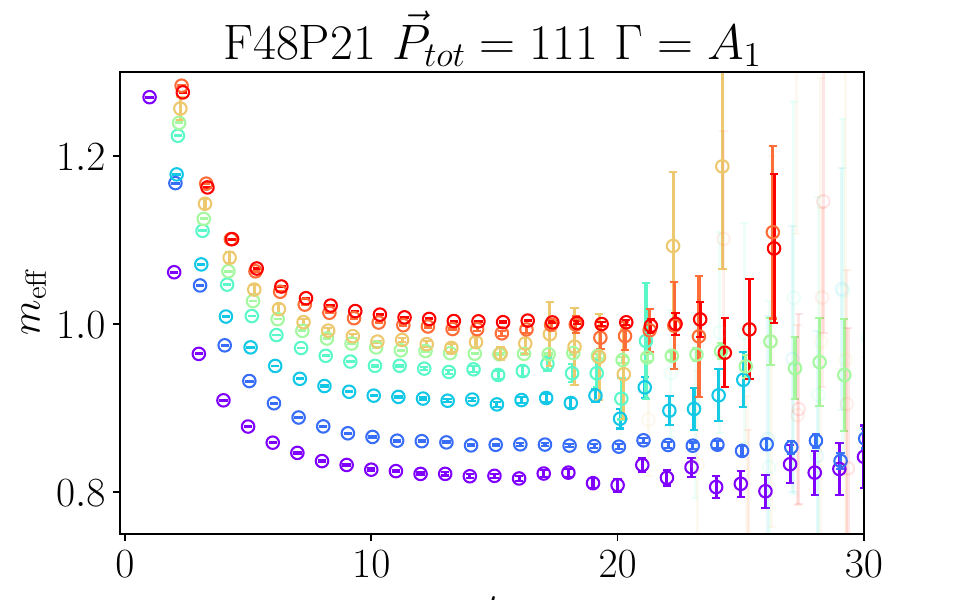}
\includegraphics[width=0.49\columnwidth]{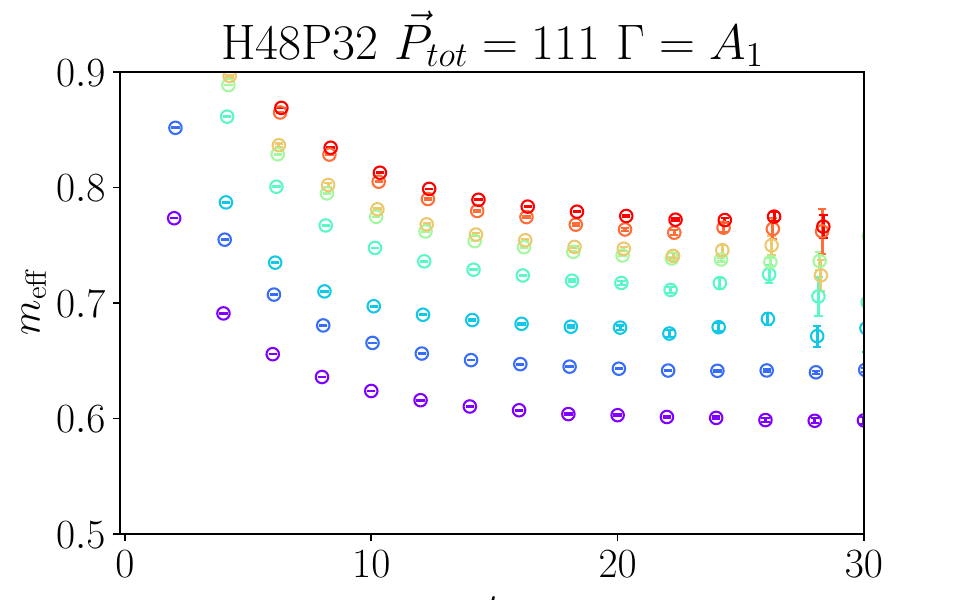}
\includegraphics[width=0.49\columnwidth]{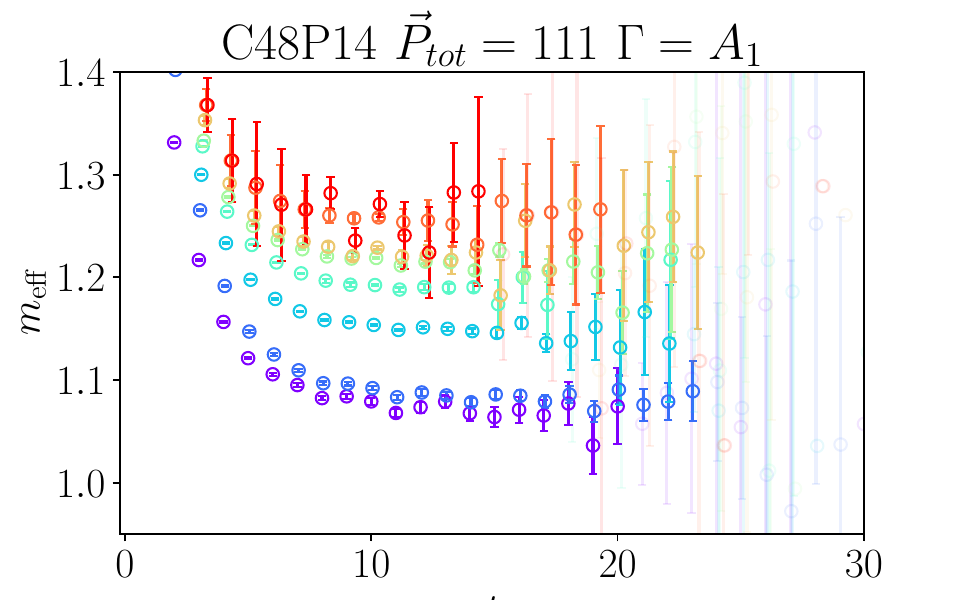}
\caption{Effective-mass plots of the generalized eigenvalues $\lambda_n(t)$ for the two-point correlation functions of the $D\pi$ system in the $A_1$ irrep at total momentum $\vec{P} = [1,1,1] \frac{2\pi}{L}$. Different colors correspond to different energy levels. The vertical axis is in lattice units.}
\label{fig:Dpi-meff-111-A1}
\end{figure}

\begin{figure}[htbp]
\centering
\includegraphics[width=0.49\columnwidth]{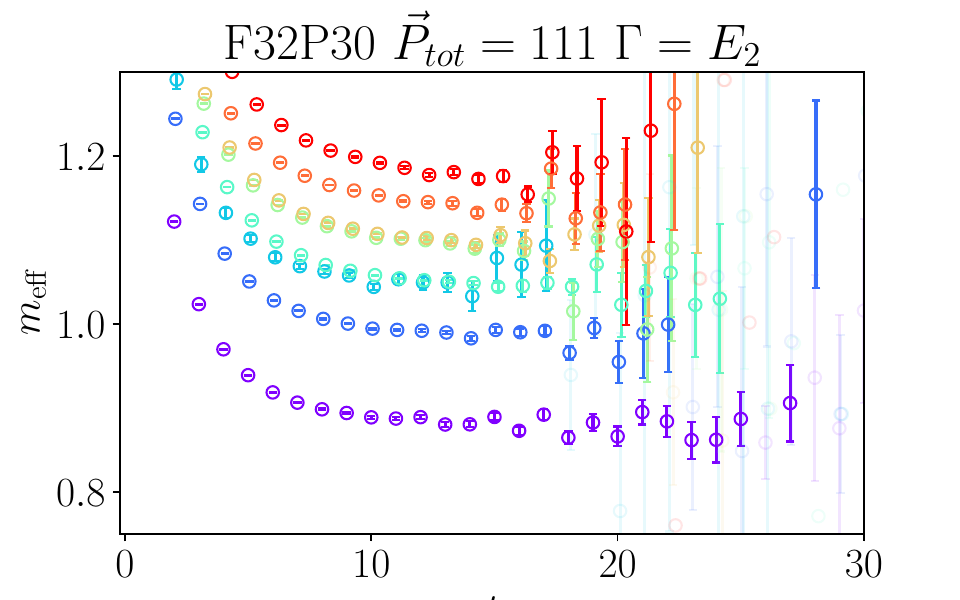}
\includegraphics[width=0.49\columnwidth]{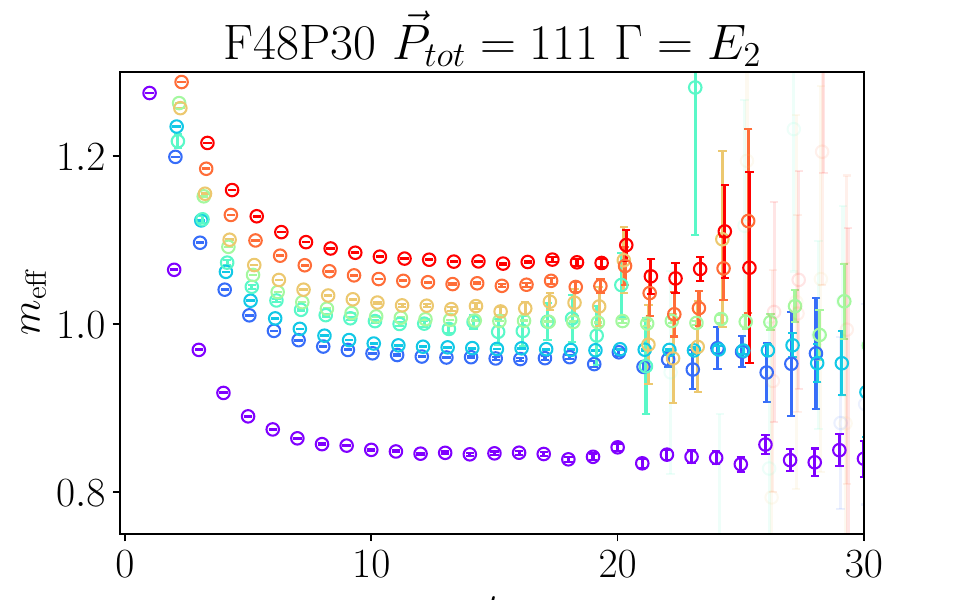}
\includegraphics[width=0.49\columnwidth]{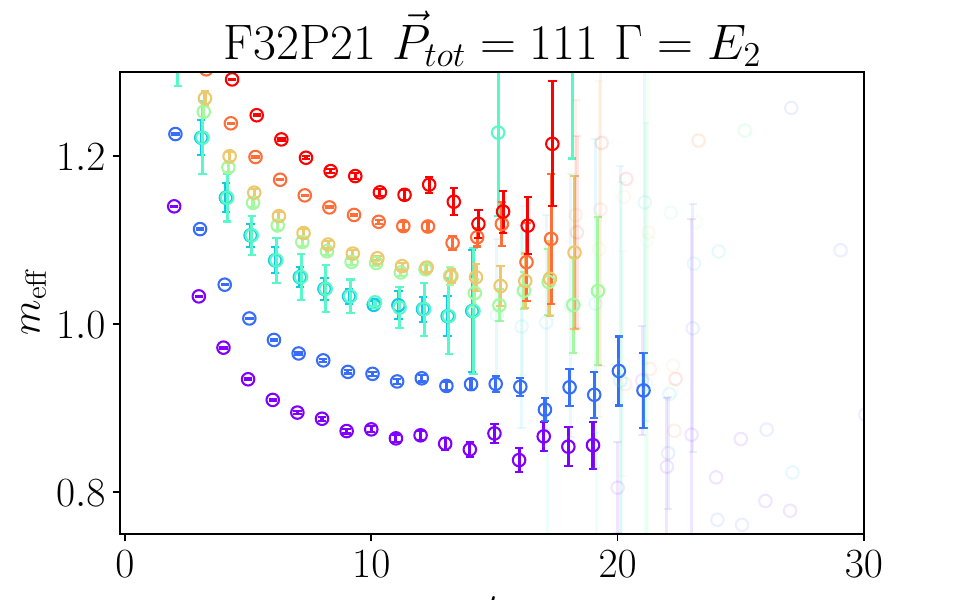}
\includegraphics[width=0.49\columnwidth]{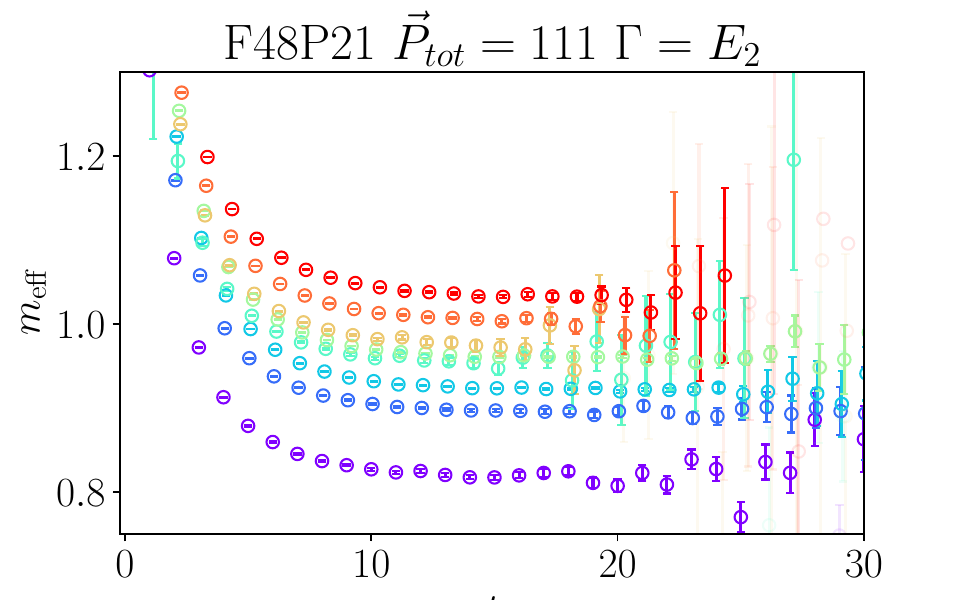}
\includegraphics[width=0.49\columnwidth]{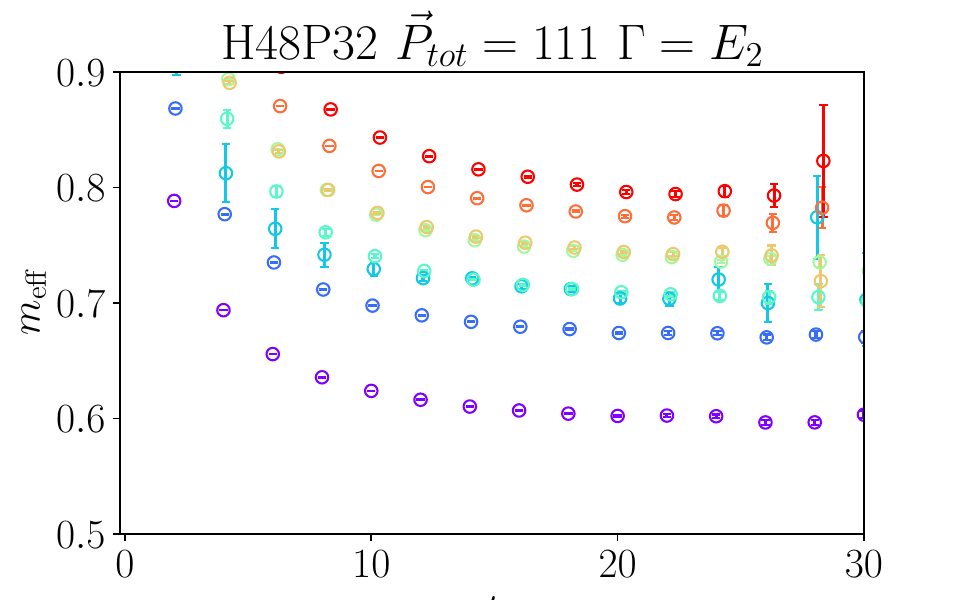}
\includegraphics[width=0.49\columnwidth]{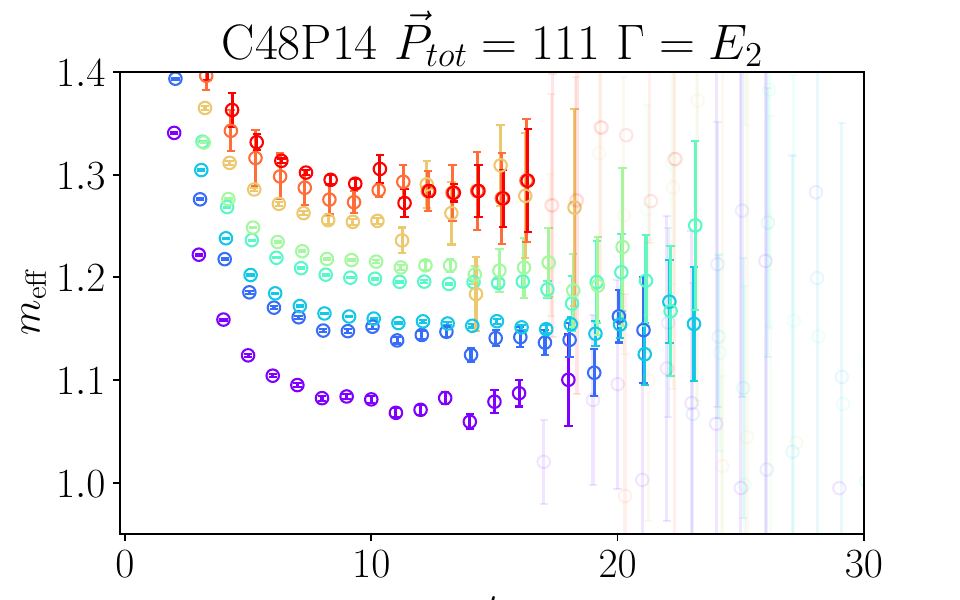}
\caption{Effective-mass plots of the generalized eigenvalues $\lambda_n(t)$ for the two-point correlation functions of the $D\pi$ system in the $E_2$ irrep at total momentum $\vec{P} = [1,1,1] \frac{2\pi}{L}$. Different colors correspond to different energy levels. The vertical axis is in lattice units.}
\label{fig:Dpi-meff-111-E2}
\end{figure}

\begin{figure}[htbp]
\centering
\includegraphics[width=0.49\columnwidth]{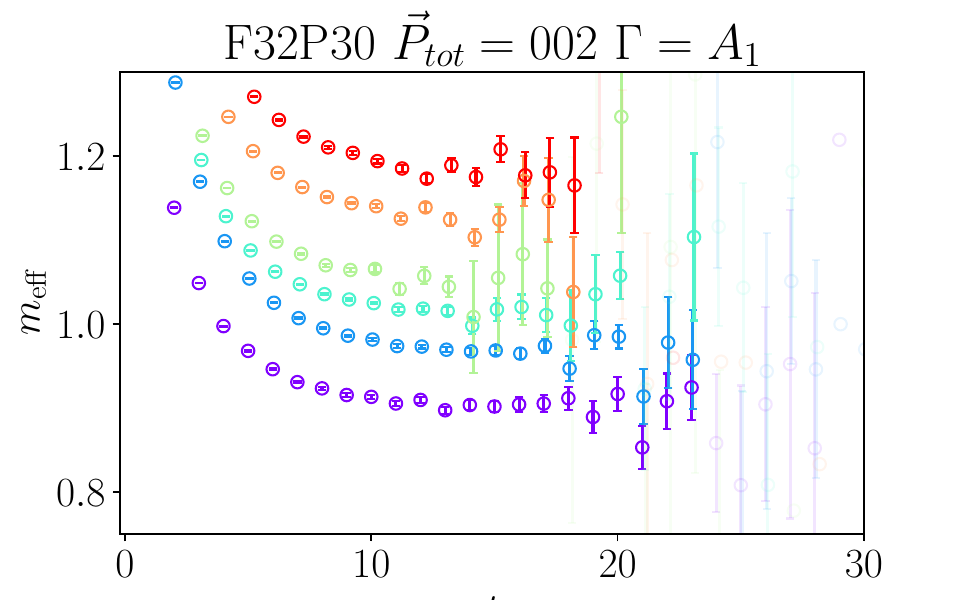}
\includegraphics[width=0.49\columnwidth]{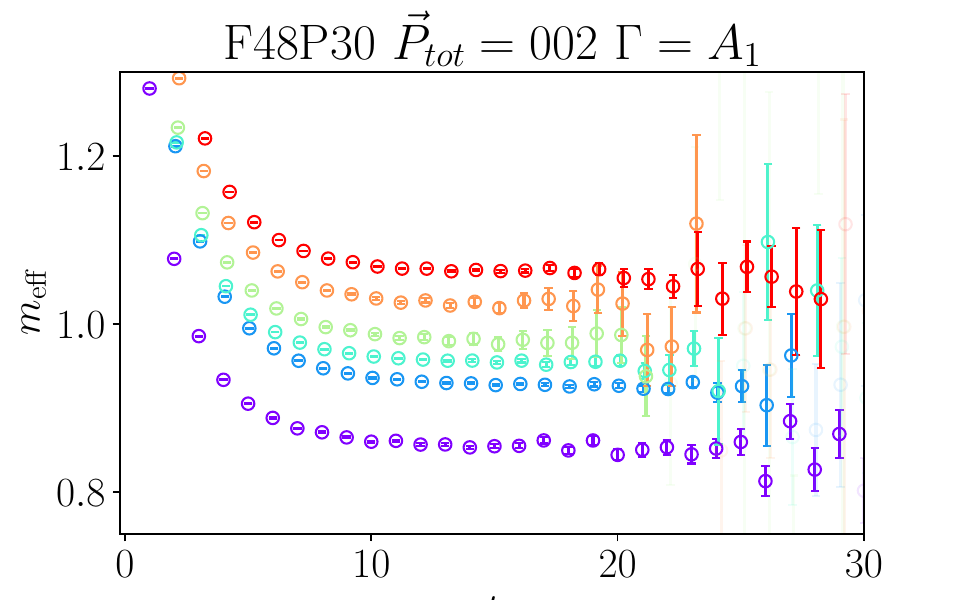}
\includegraphics[width=0.49\columnwidth]{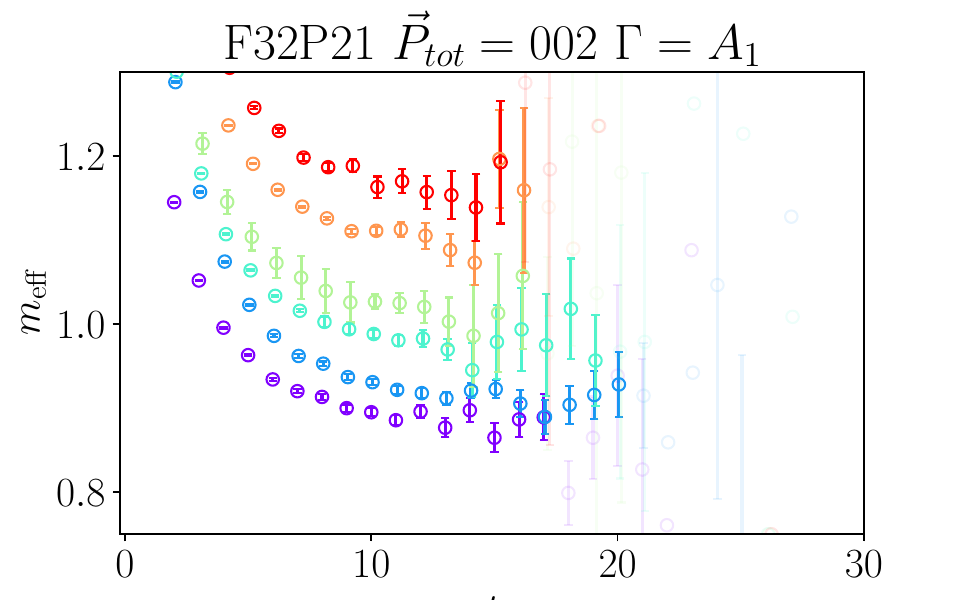}
\includegraphics[width=0.49\columnwidth]{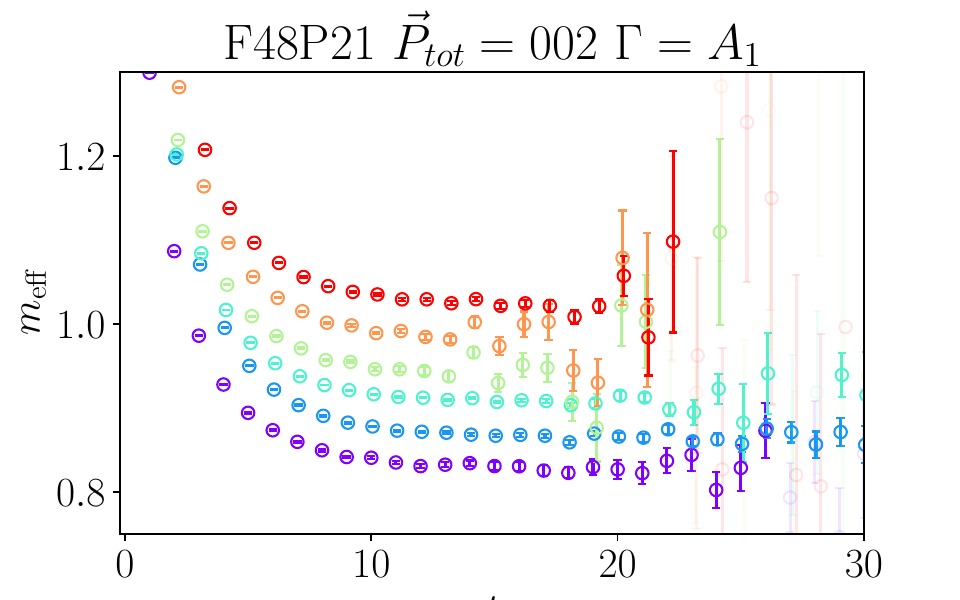}
\includegraphics[width=0.49\columnwidth]{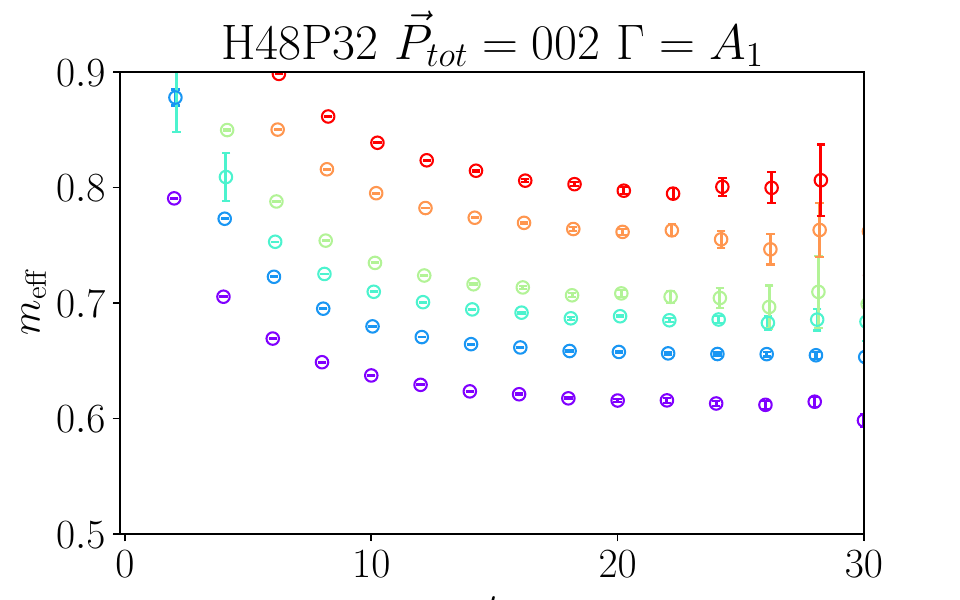}
\includegraphics[width=0.49\columnwidth]{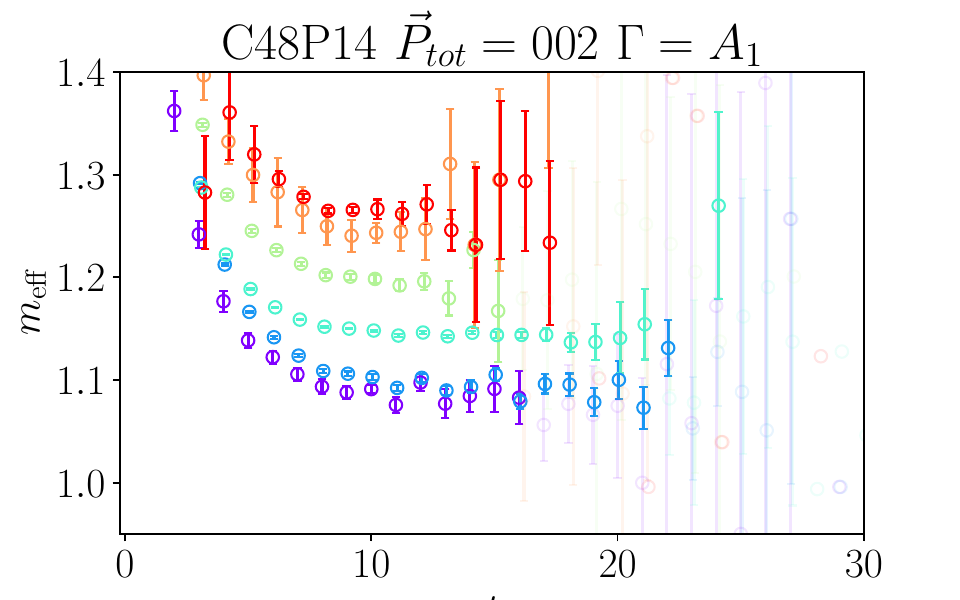}
\caption{Effective-mass plots of the generalized eigenvalues $\lambda_n(t)$ for the two-point correlation functions of the $D\pi$ system in the $A_1$ irrep at total momentum $\vec{P} = [0,0,2] \frac{2\pi}{L}$. Different colors correspond to different energy levels. The vertical axis is in lattice units.}
\label{fig:Dpi-meff-002-A1}
\end{figure}

\begin{figure}[htbp]
\centering
\includegraphics[width=0.49\columnwidth]{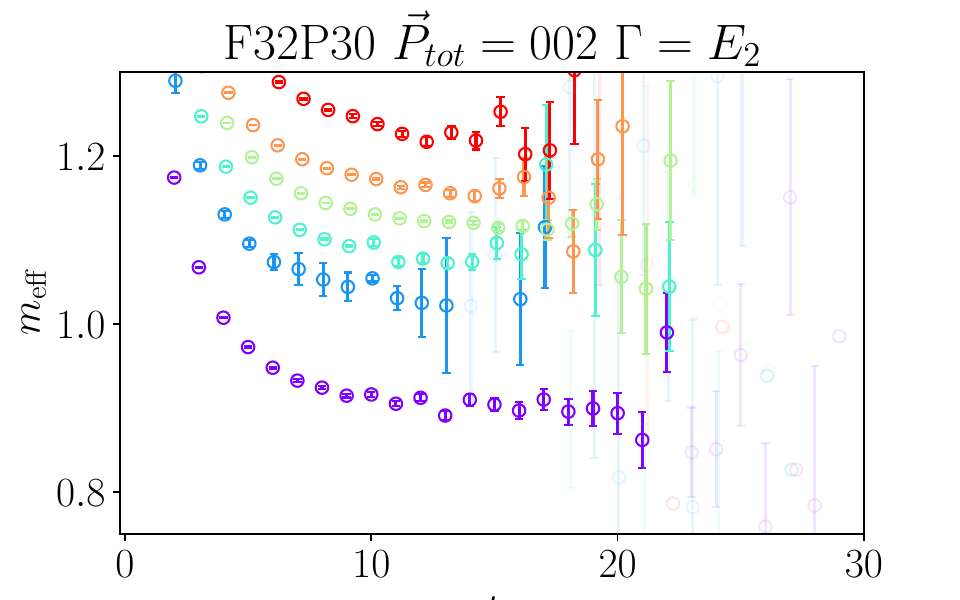}
\includegraphics[width=0.49\columnwidth]{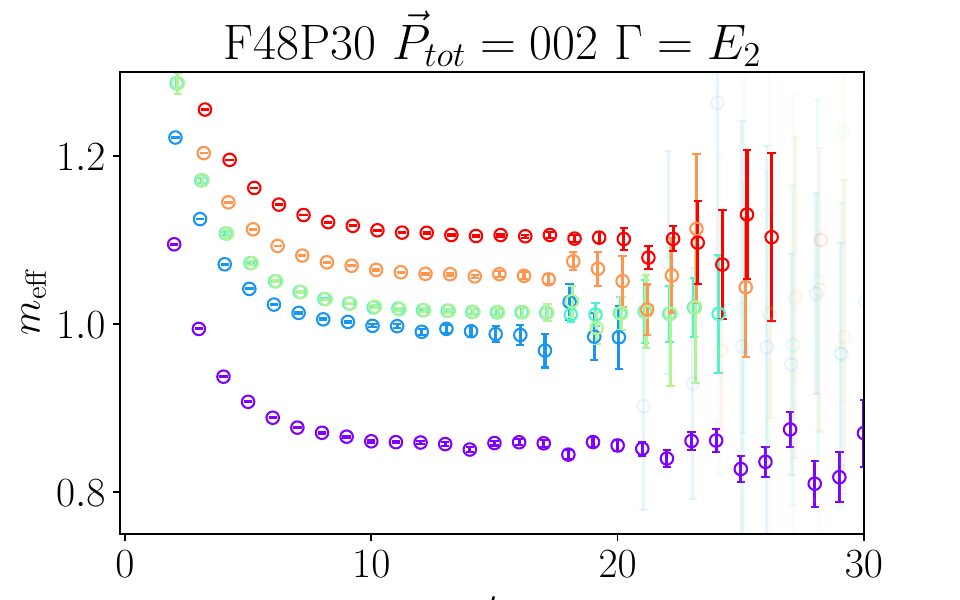}
\includegraphics[width=0.49\columnwidth]{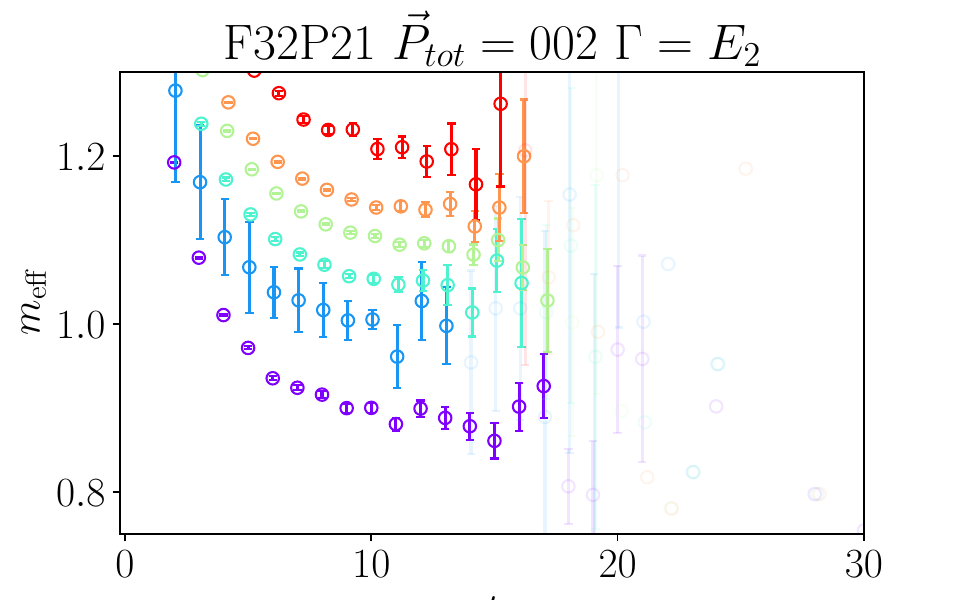}
\includegraphics[width=0.49\columnwidth]{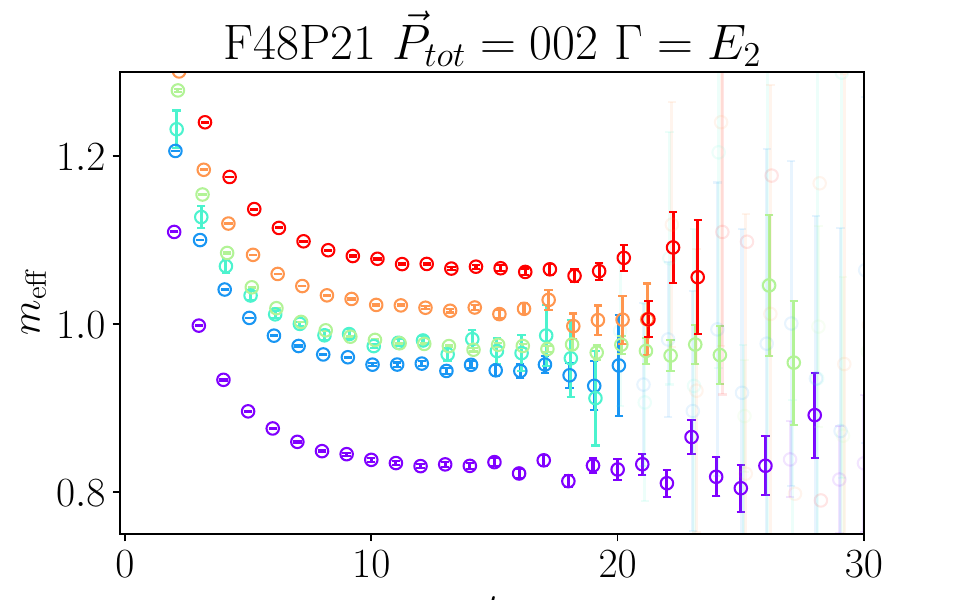}
\includegraphics[width=0.49\columnwidth]{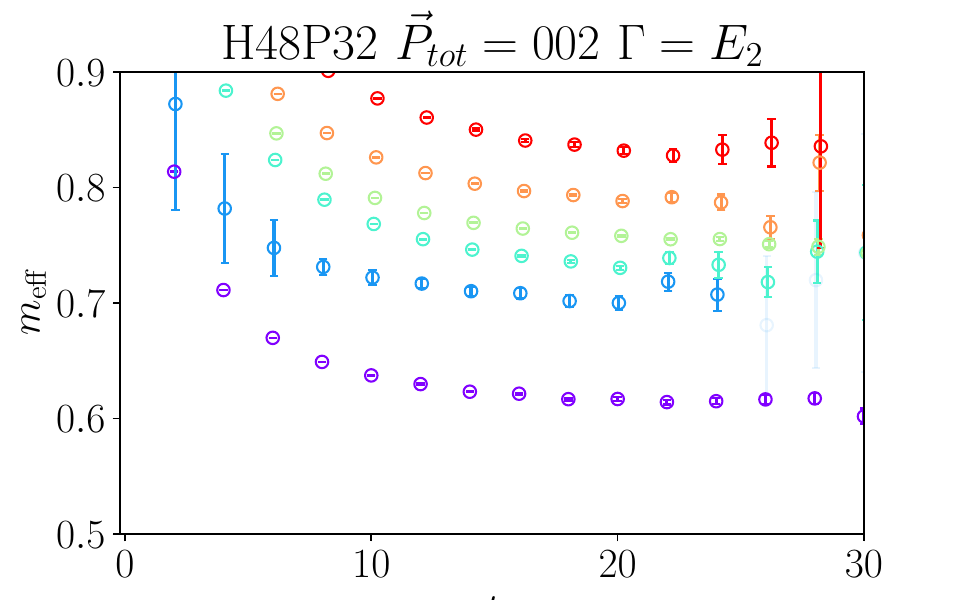}
\includegraphics[width=0.49\columnwidth]{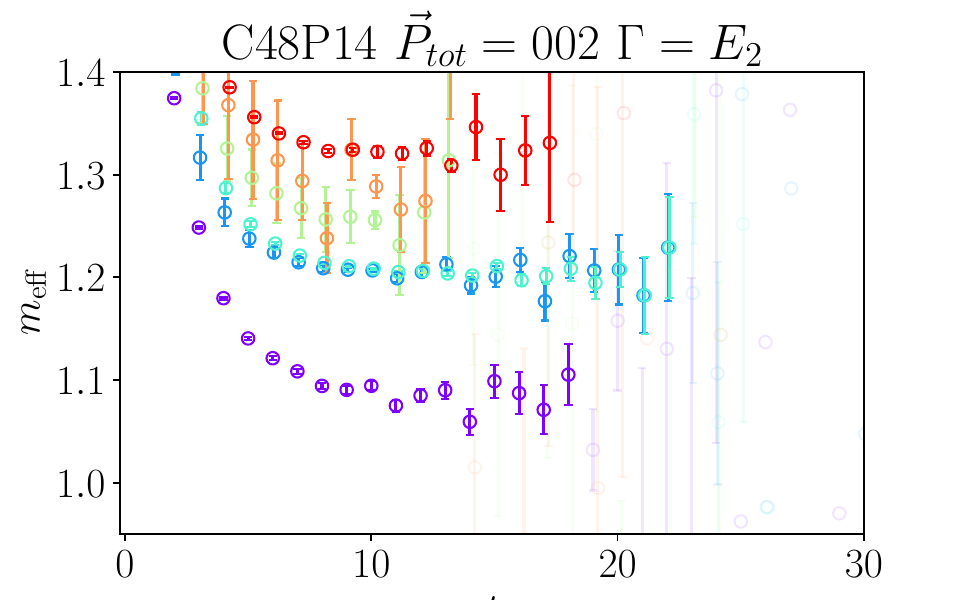}
\caption{Effective-mass plots of the generalized eigenvalues $\lambda_n(t)$ for the two-point correlation functions of the $D\pi$ system in the $E_2$ irrep at total momentum $\vec{P} = [0,0,2] \frac{2\pi}{L}$. Different colors correspond to different energy levels. The vertical axis is in lattice units.}
\label{fig:Dpi-meff-002-E2}
\end{figure}

Figures~\ref{fig:Dpi-meff-A1+}--\ref{fig:Dpi-meff-002-E2} show the GEVP and effective-mass plots in the relevant irreducible representations for all ensembles used in this work.

The fit results and stability analyses for individual energy levels are shown in Figures~\ref{fig:Dpi-fit-F32P30}--\ref{fig:Dpi-fit-C48P14}. For each level, we perform fits over different fit ranges and with different initial guesses, and examine the stability of the fitted results. Finally, we collect the energies of all levels to obtain the finite-volume spectrum of the $D\pi$ system.

To save space, only the lowest two energy levels in each irreducible representation are shown.

\begin{figure}[htbp]
\centering
\includegraphics[width=0.245\columnwidth]{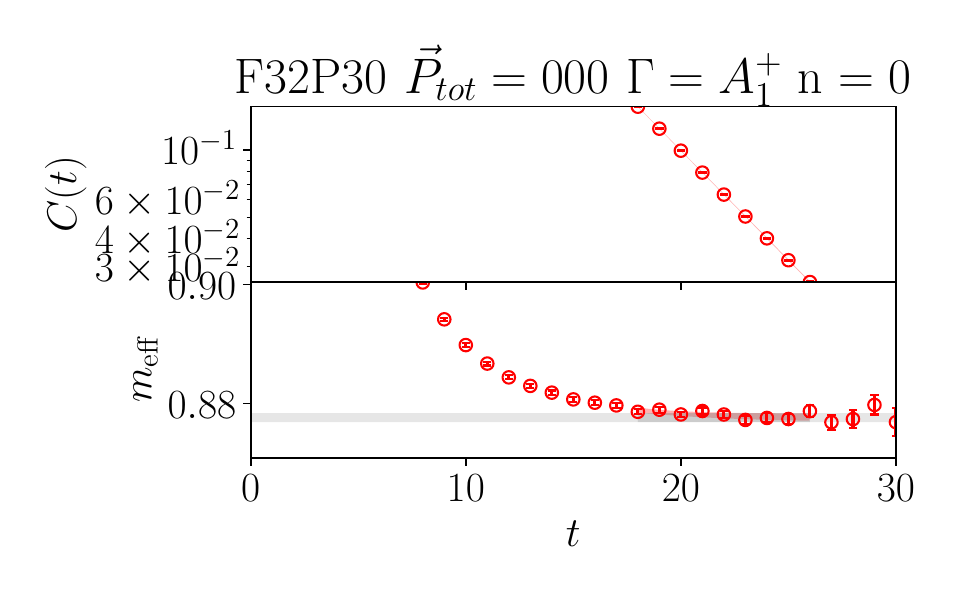}
\includegraphics[width=0.245\columnwidth]{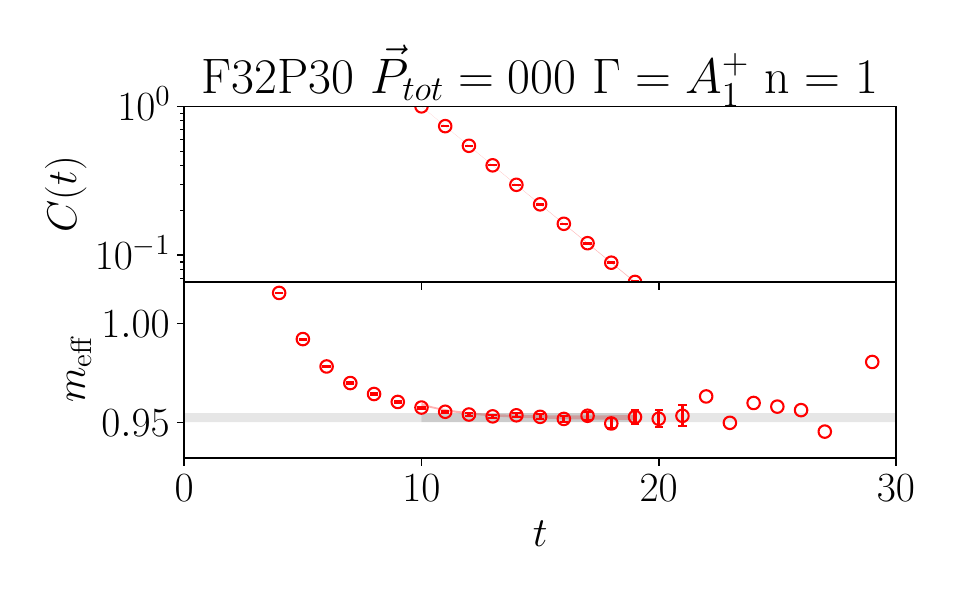}
\includegraphics[width=0.245\columnwidth]{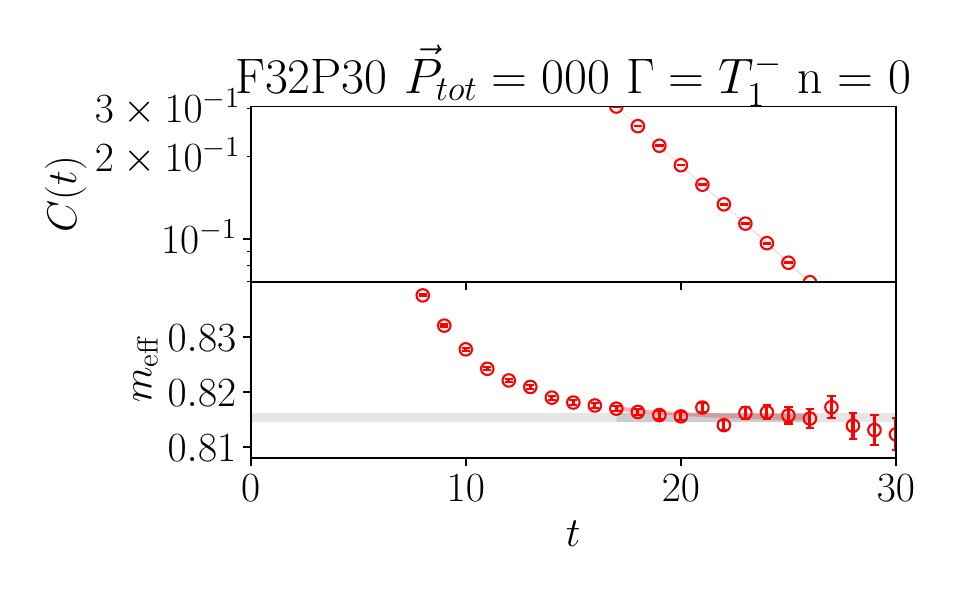}
\includegraphics[width=0.245\columnwidth]{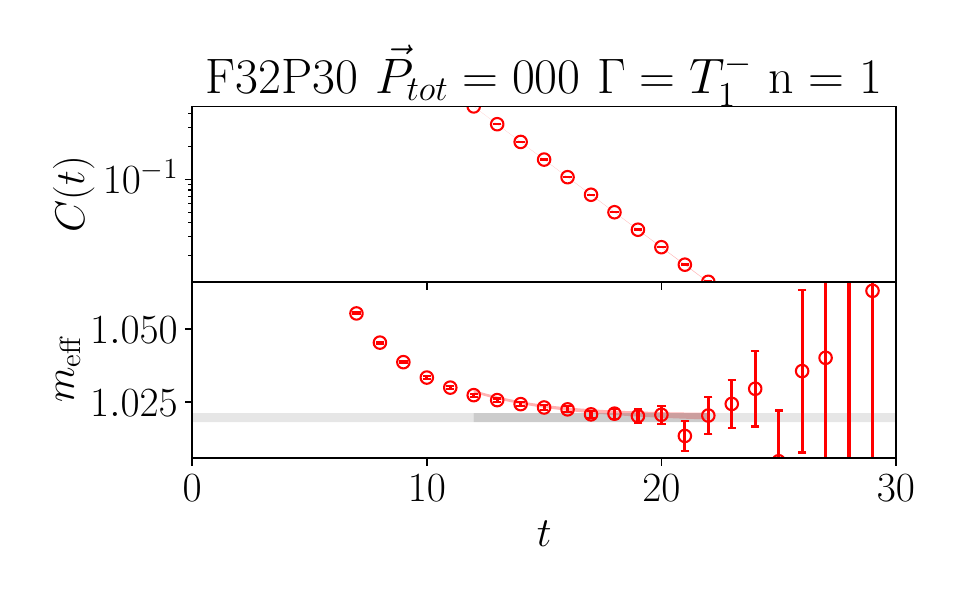}
\\
\includegraphics[width=0.245\columnwidth]{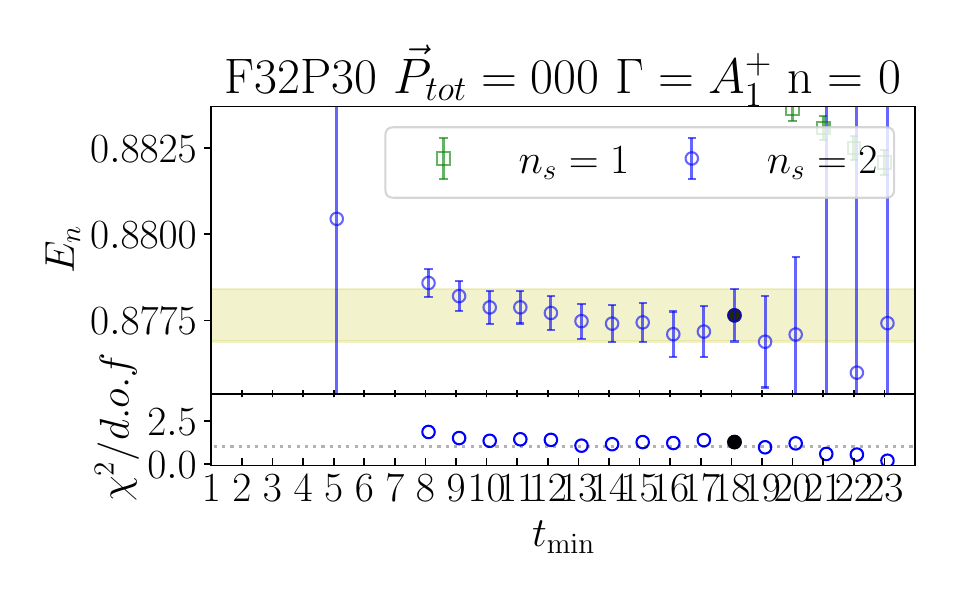}
\includegraphics[width=0.245\columnwidth]{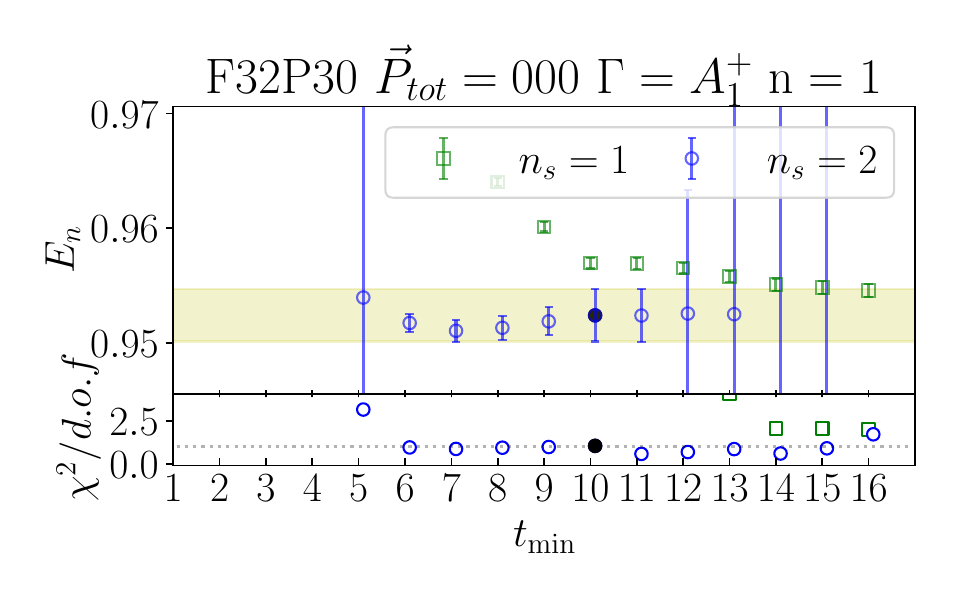}
\includegraphics[width=0.245\columnwidth]{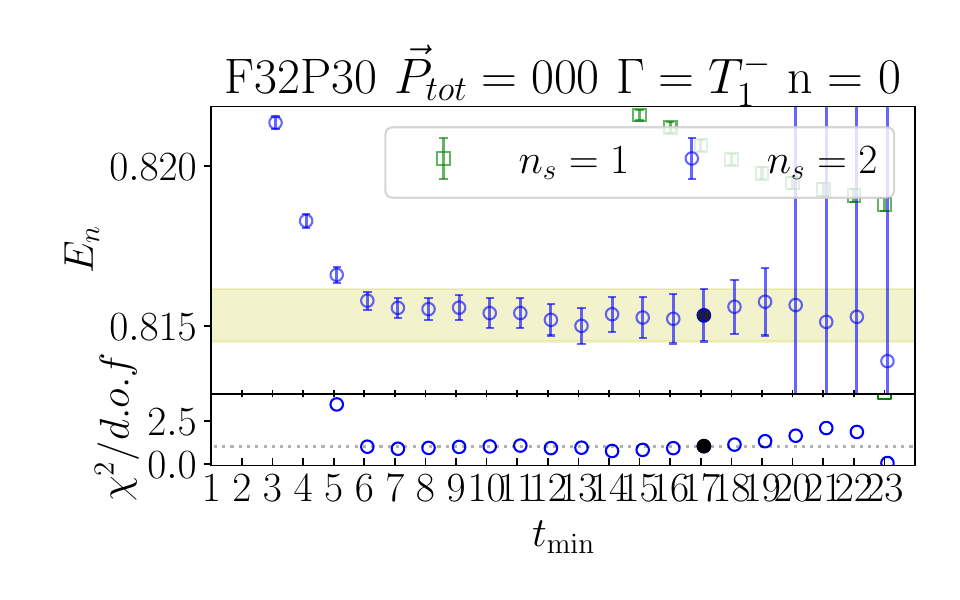}
\includegraphics[width=0.245\columnwidth]{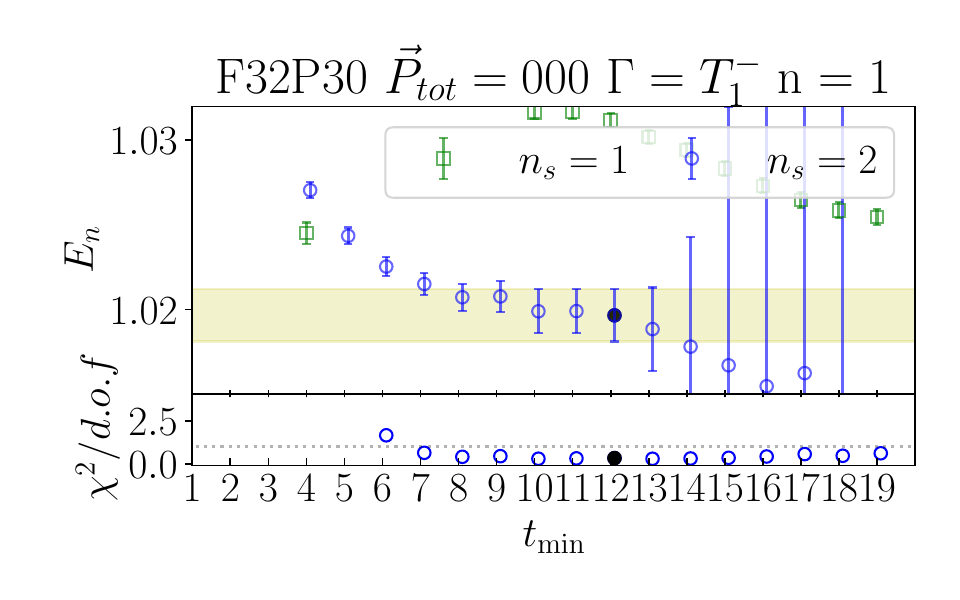}
\\
\includegraphics[width=0.245\columnwidth]{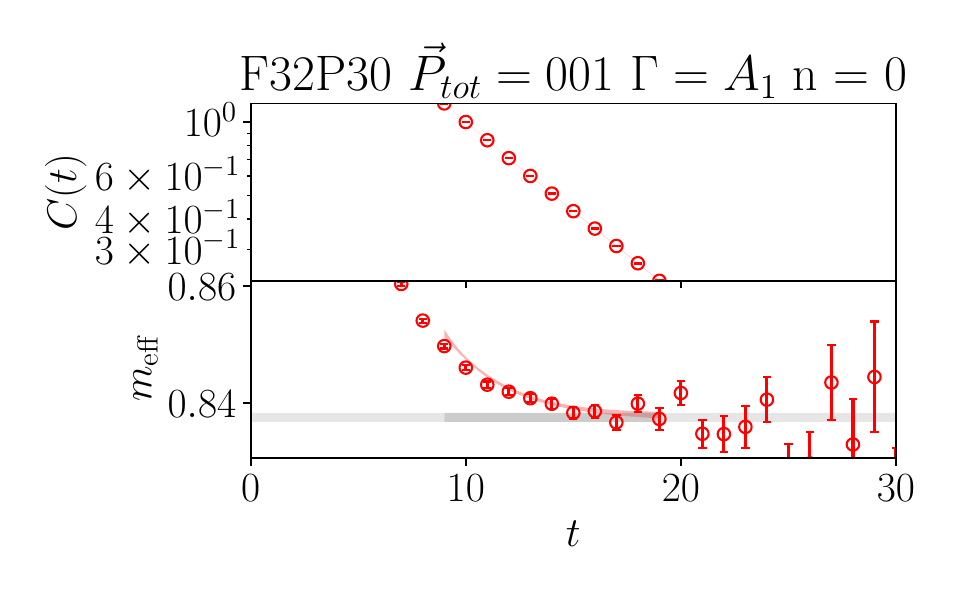}
\includegraphics[width=0.245\columnwidth]{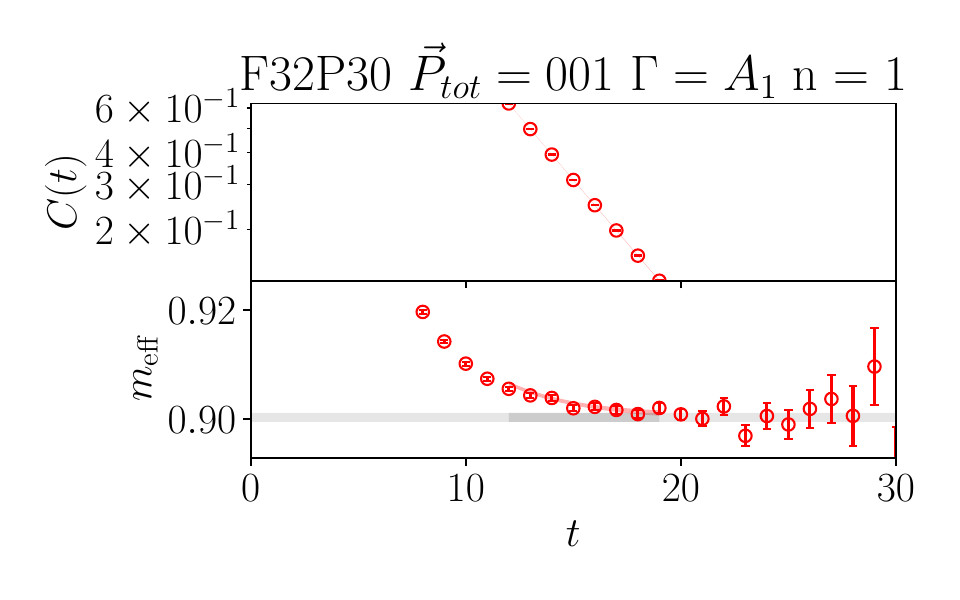}
\includegraphics[width=0.245\columnwidth]{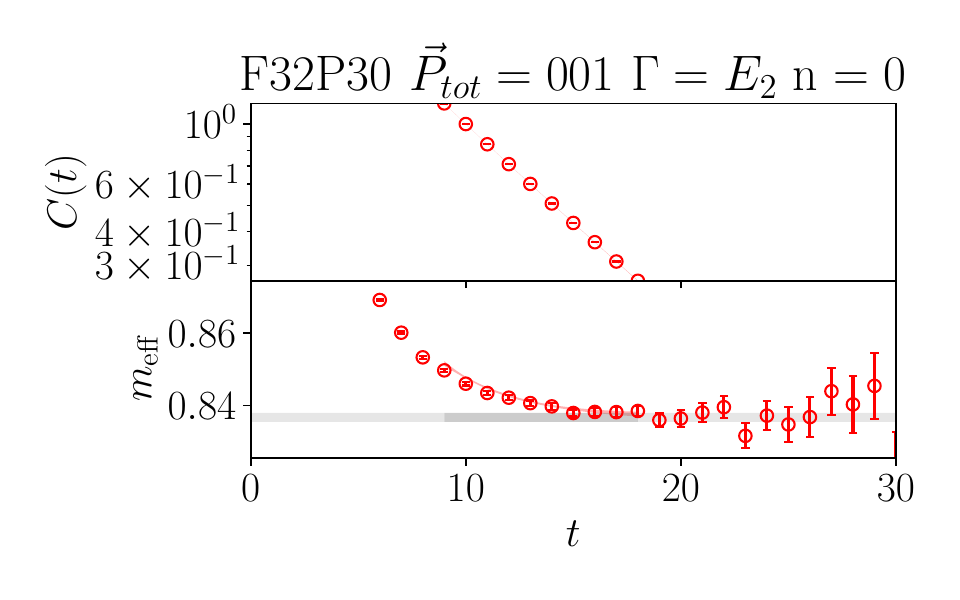}
\includegraphics[width=0.245\columnwidth]{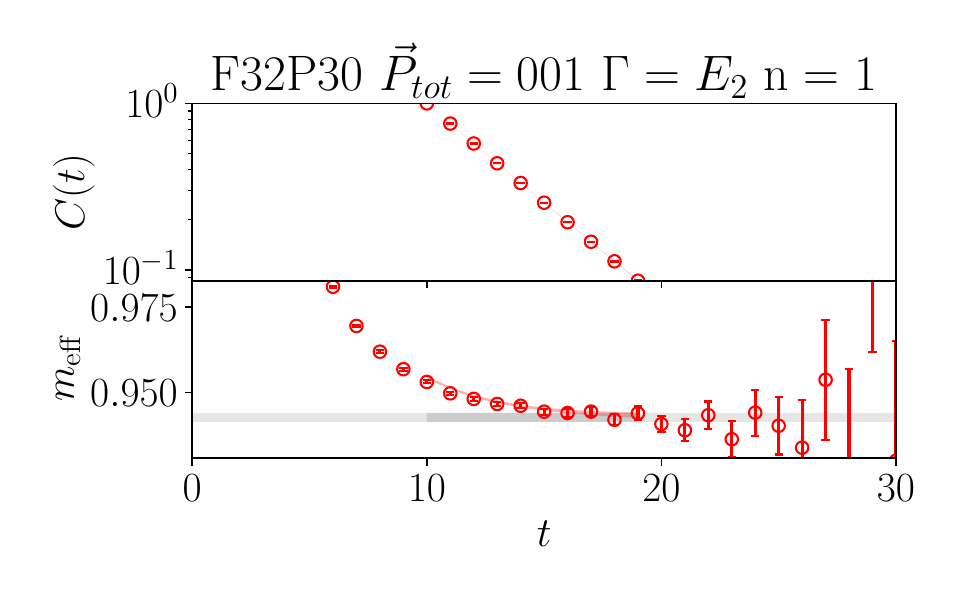}
\\
\includegraphics[width=0.245\columnwidth]{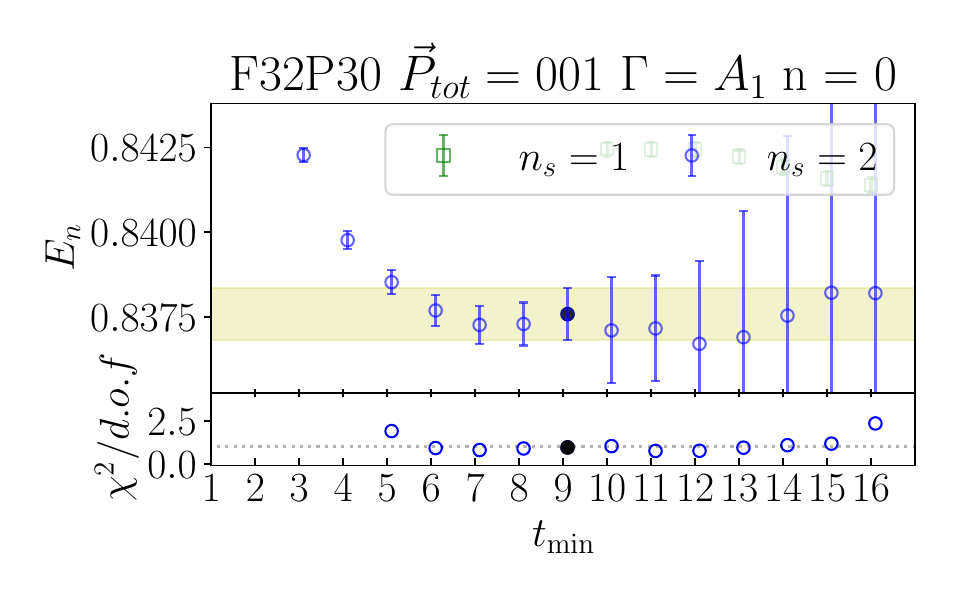}
\includegraphics[width=0.245\columnwidth]{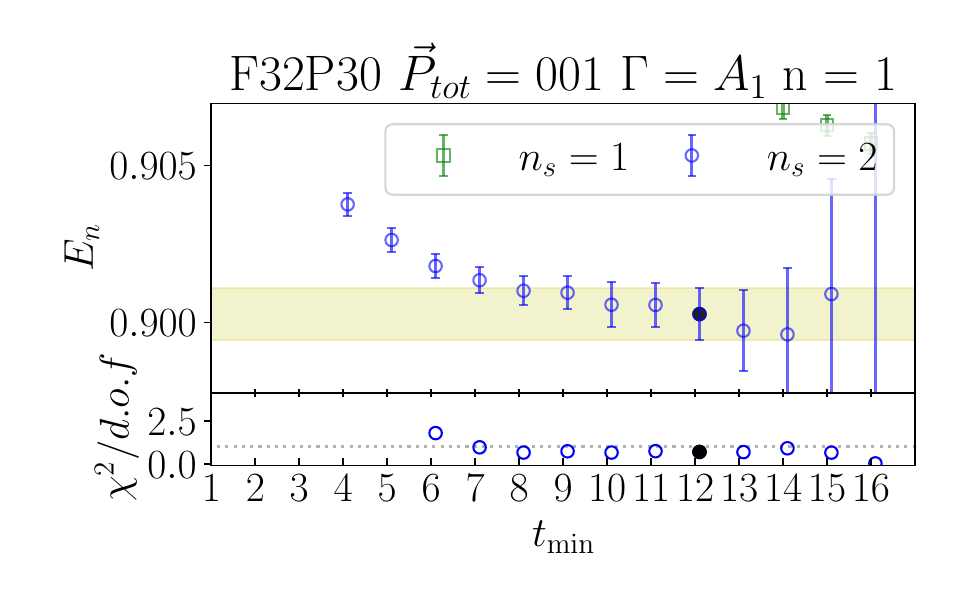}
\includegraphics[width=0.245\columnwidth]{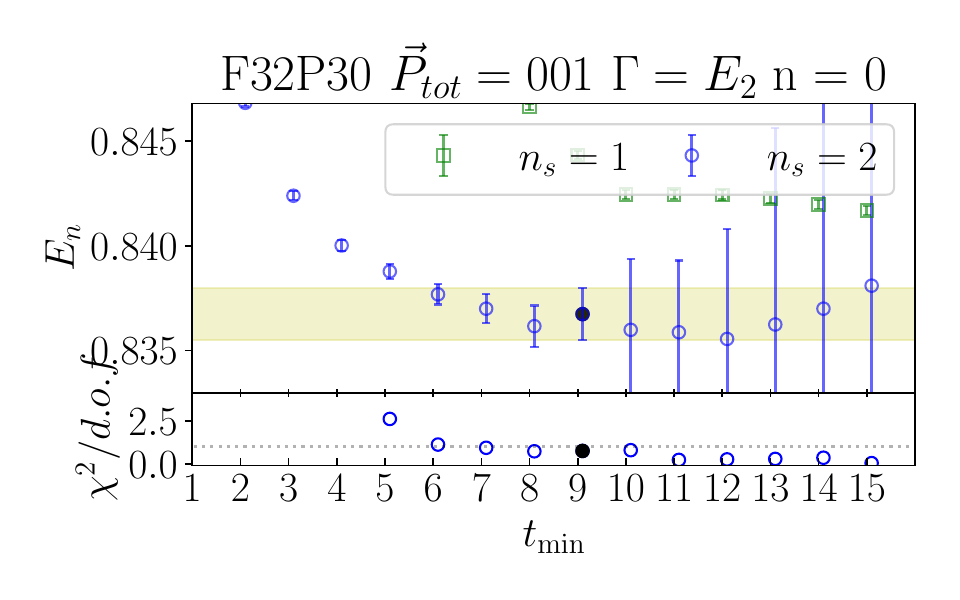}
\includegraphics[width=0.245\columnwidth]{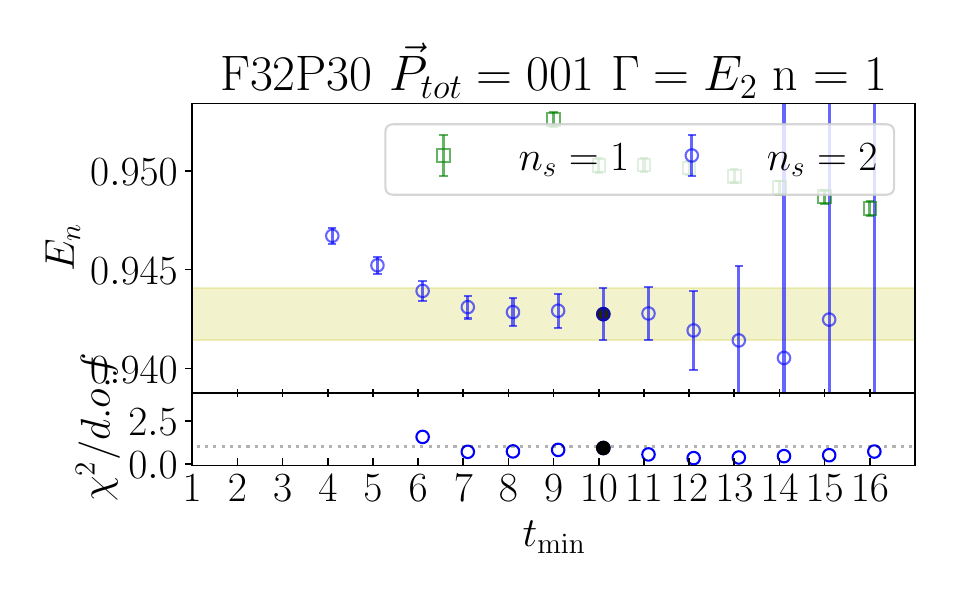}
\\
\includegraphics[width=0.245\columnwidth]{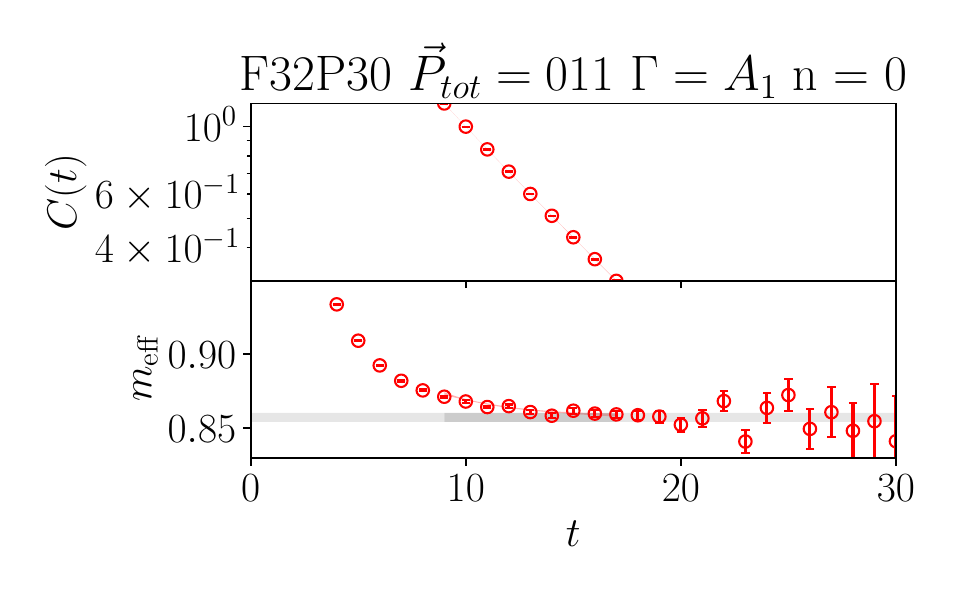}
\includegraphics[width=0.245\columnwidth]{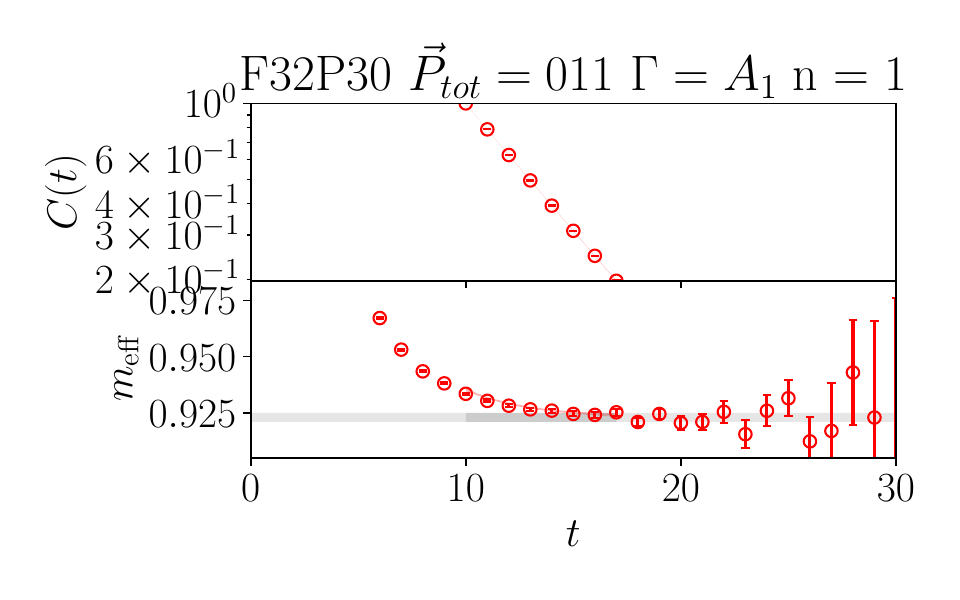}
\includegraphics[width=0.245\columnwidth]{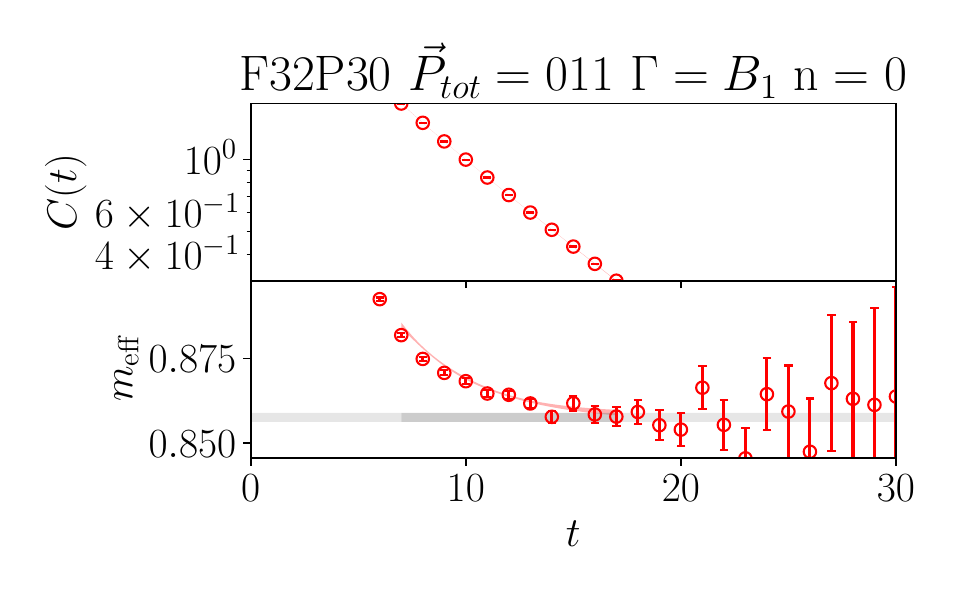}
\includegraphics[width=0.245\columnwidth]{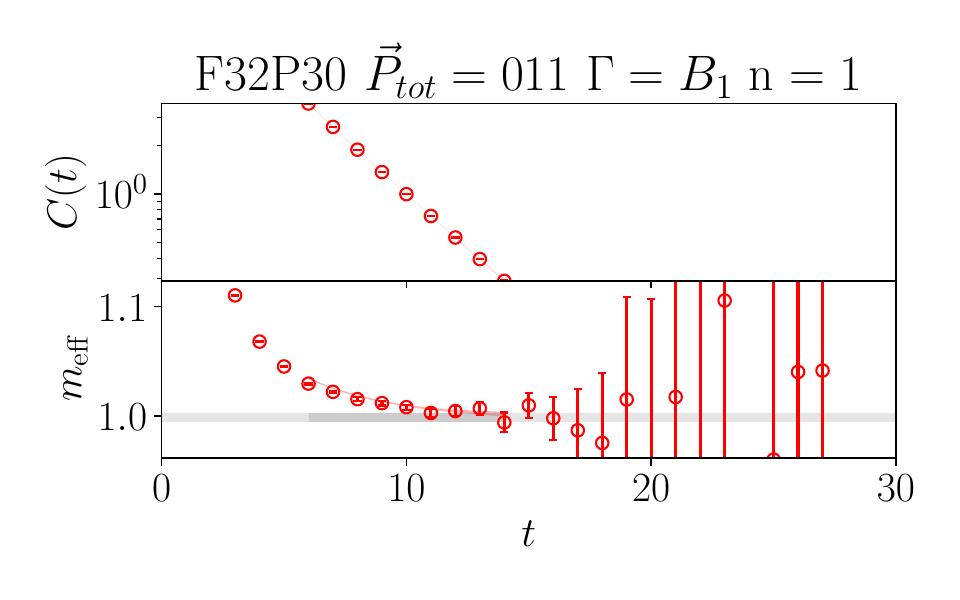}
\\
\includegraphics[width=0.245\columnwidth]{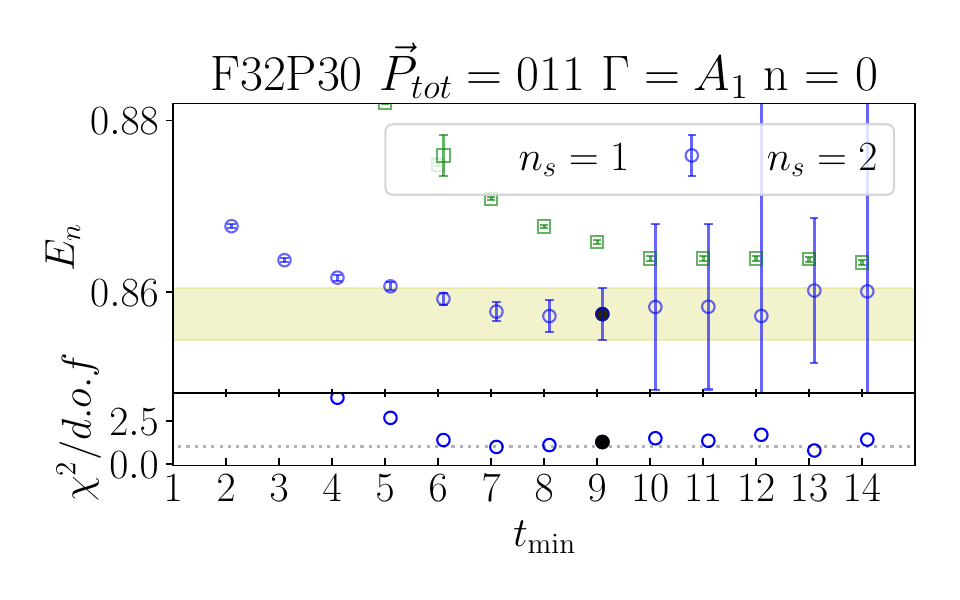}
\includegraphics[width=0.245\columnwidth]{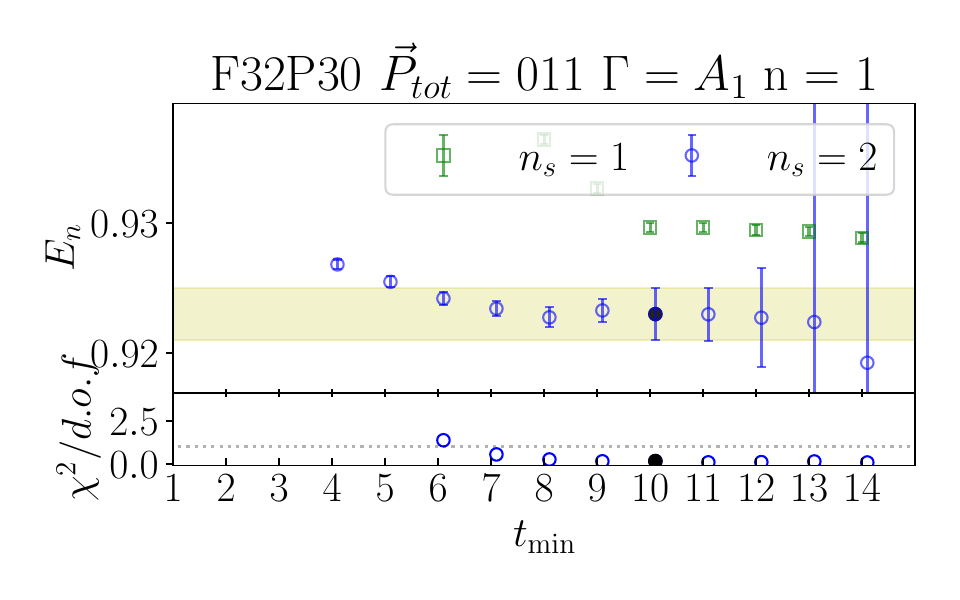}
\includegraphics[width=0.245\columnwidth]{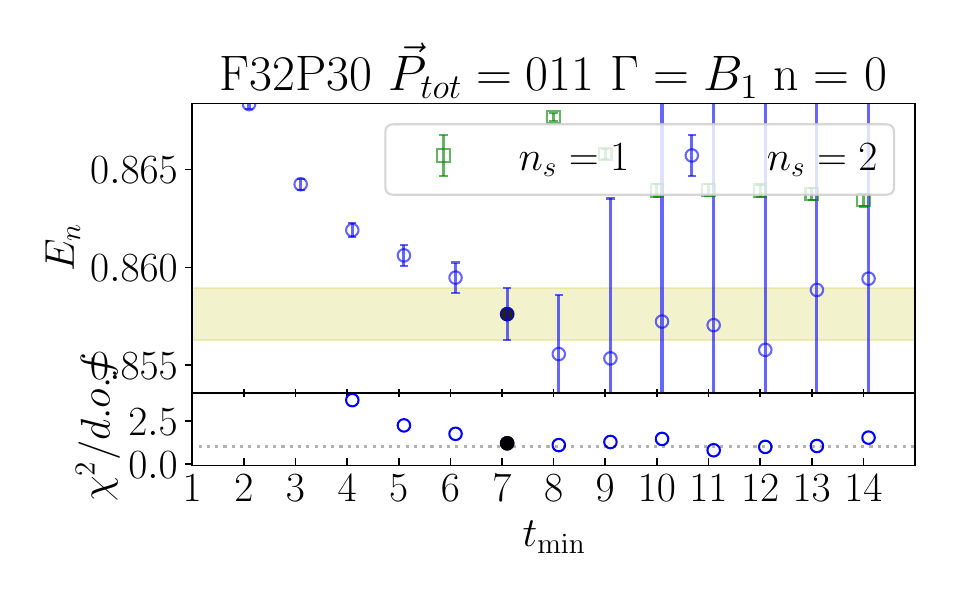}
\includegraphics[width=0.245\columnwidth]{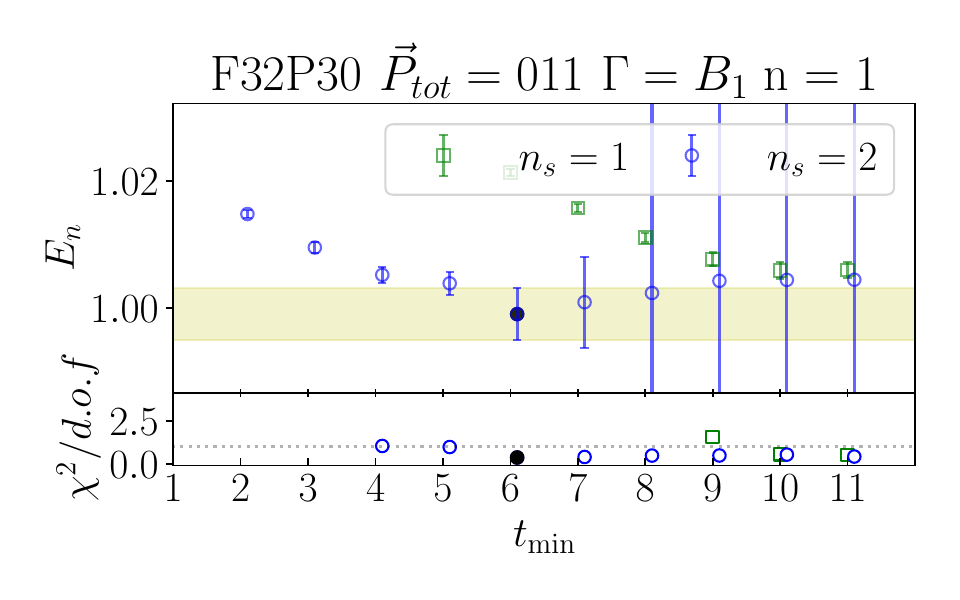}
\\
\includegraphics[width=0.245\columnwidth]{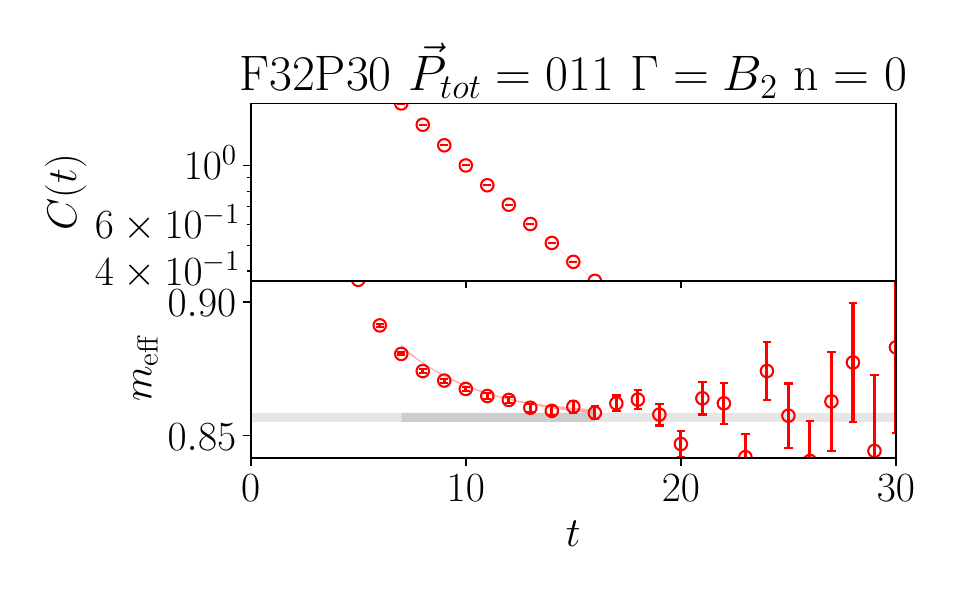}
\includegraphics[width=0.245\columnwidth]{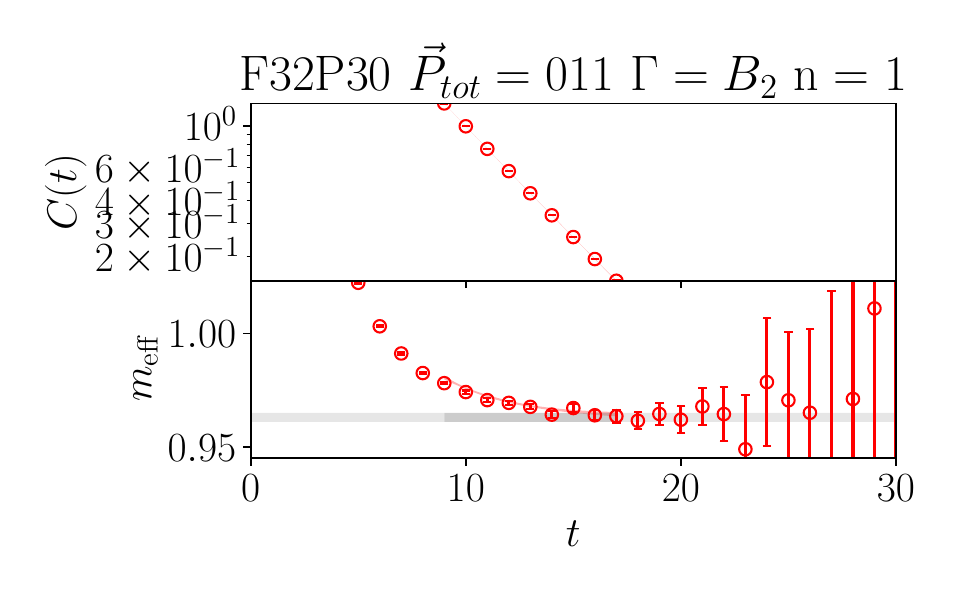}
\includegraphics[width=0.245\columnwidth]{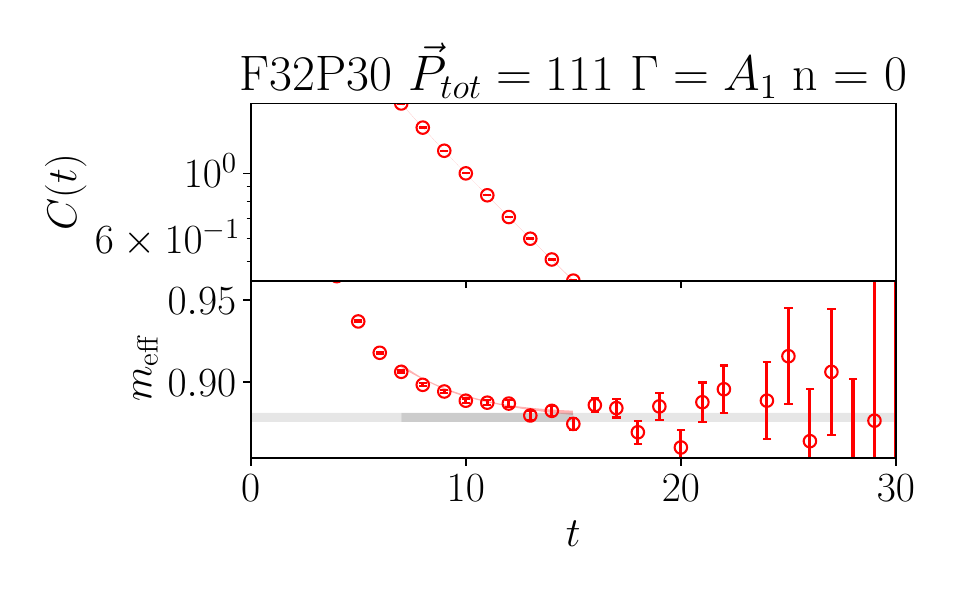}
\includegraphics[width=0.245\columnwidth]{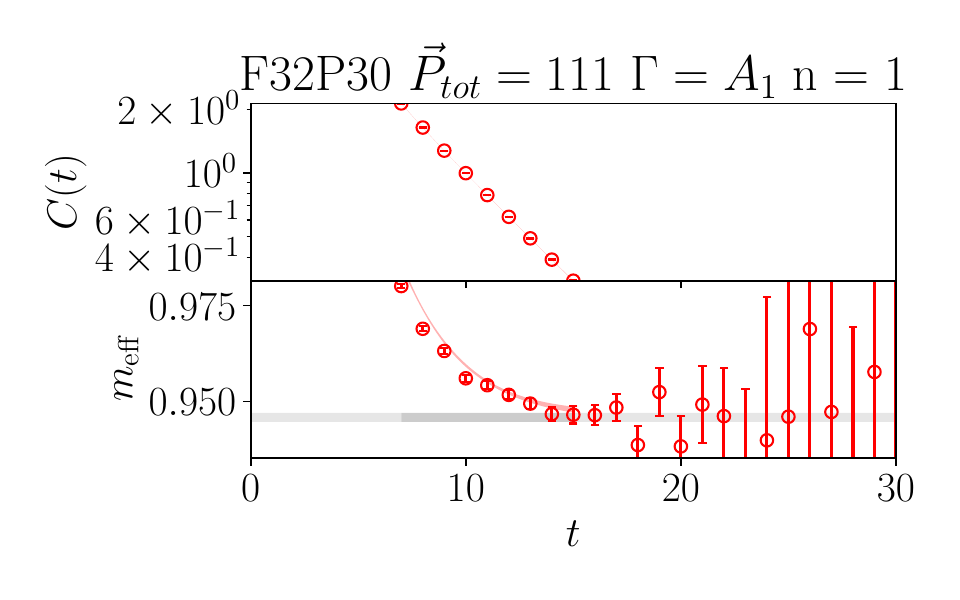}
\\
\includegraphics[width=0.245\columnwidth]{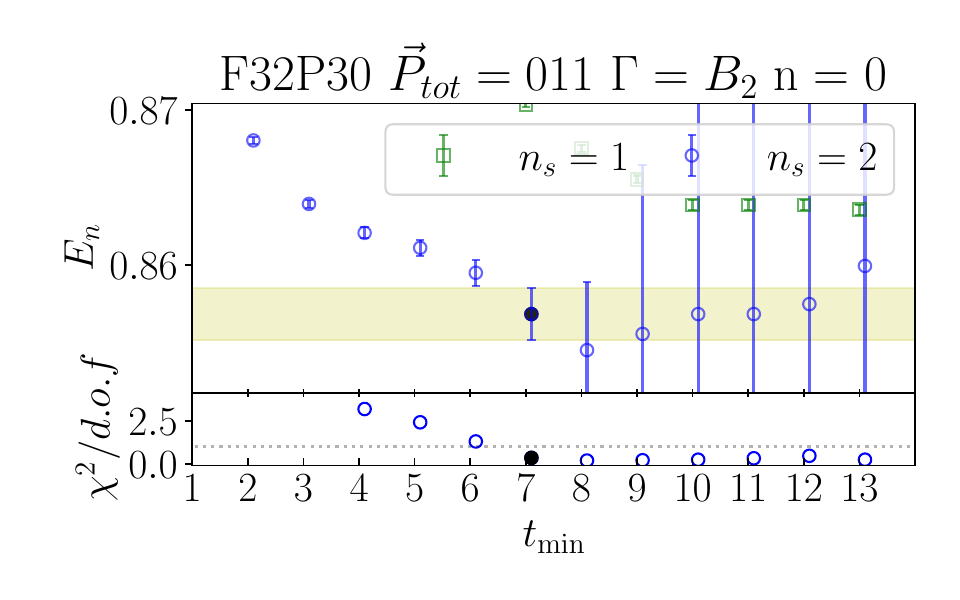}
\includegraphics[width=0.245\columnwidth]{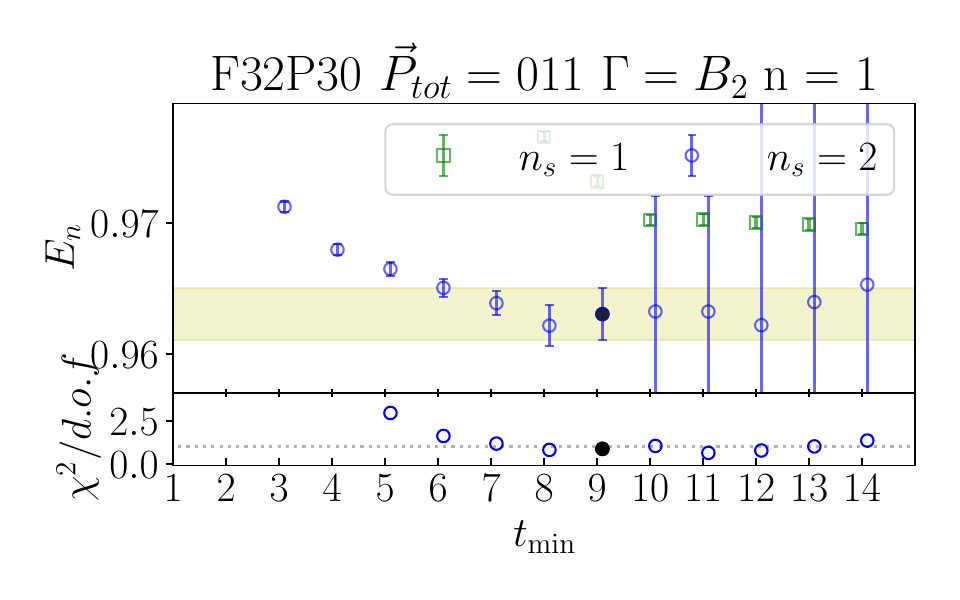}
\includegraphics[width=0.245\columnwidth]{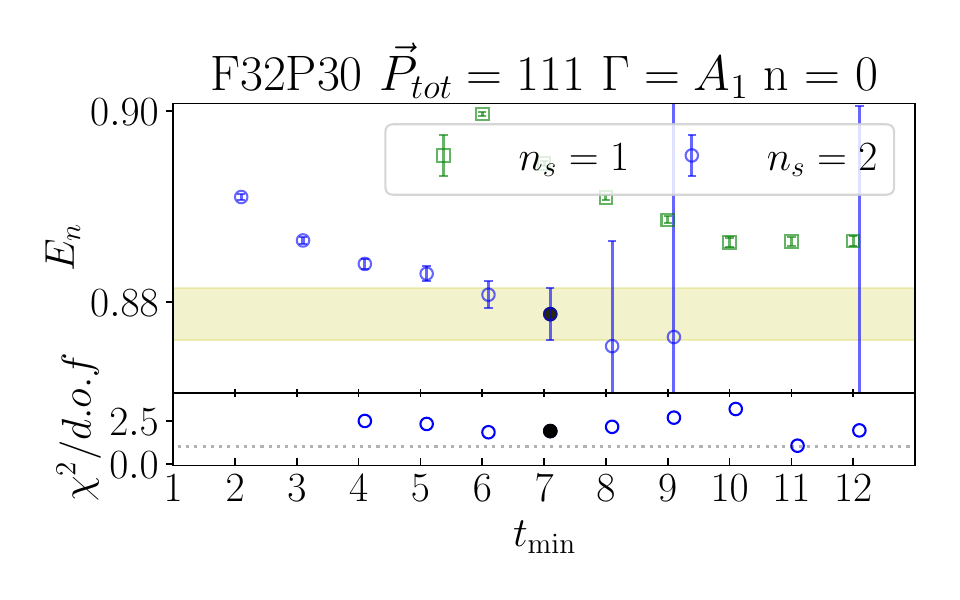}
\includegraphics[width=0.245\columnwidth]{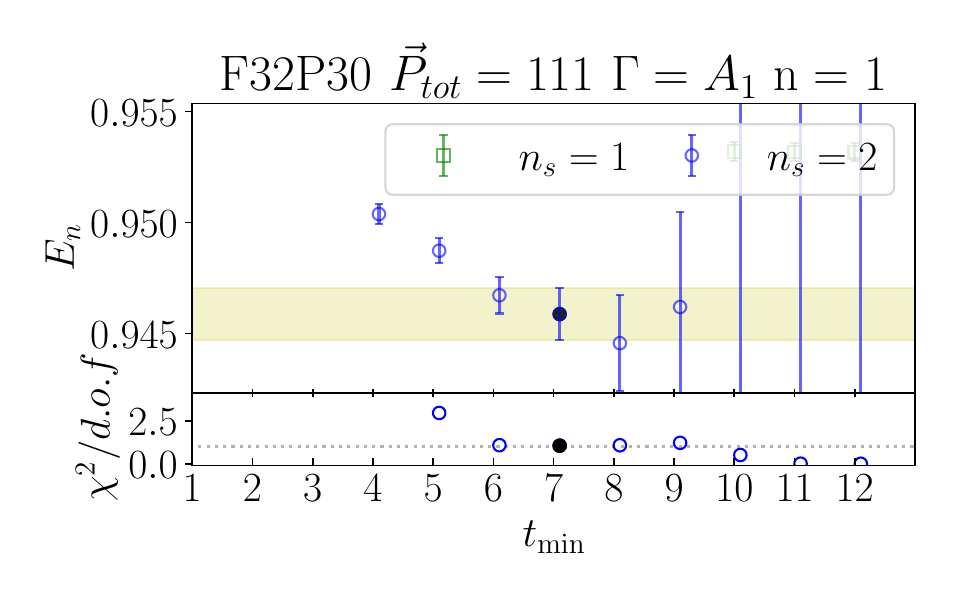}
\caption{Energy-level fit results for the $I=\frac{1}{2}$ $D\pi$ channel on the F32P30 ensemble. For each energy level, the fit is shown in upper and lower panels. In the upper panel, the red band denotes the correlator or the reconstructed effective mass, and the gray band denotes the extracted energy. The lower panel shows the stability of the fit with respect to the fit starting time, where the green and blue data points correspond to one-state and two-state fits, respectively. The black error bar marks the chosen fit starting time. The lower panel also gives the $\chi^2/\mathrm{d.o.f.}$ value.}
\label{fig:Dpi-fit-F32P30}
\end{figure}

\begin{figure}[htbp]
\centering
\includegraphics[width=0.245\columnwidth]{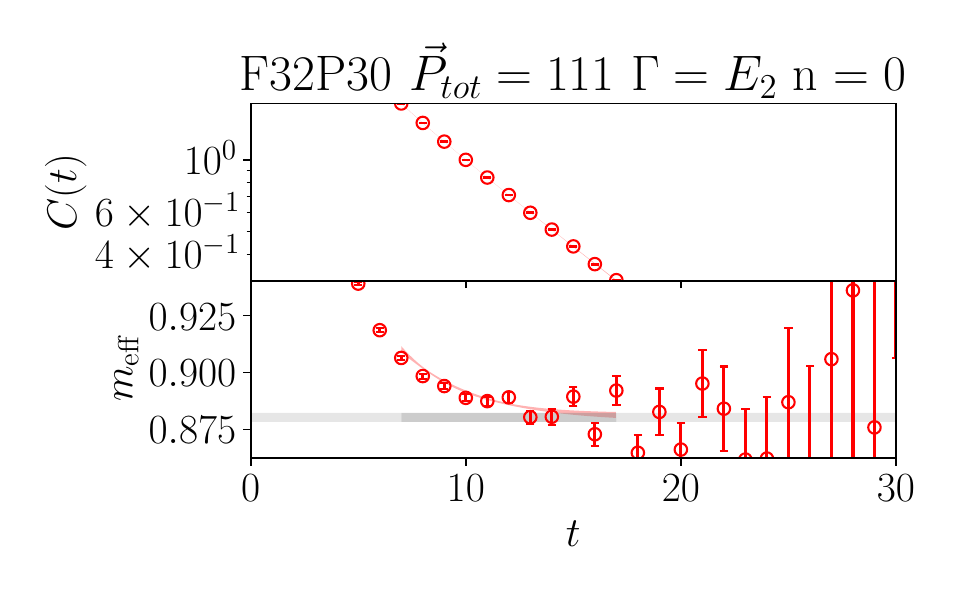}
\includegraphics[width=0.245\columnwidth]{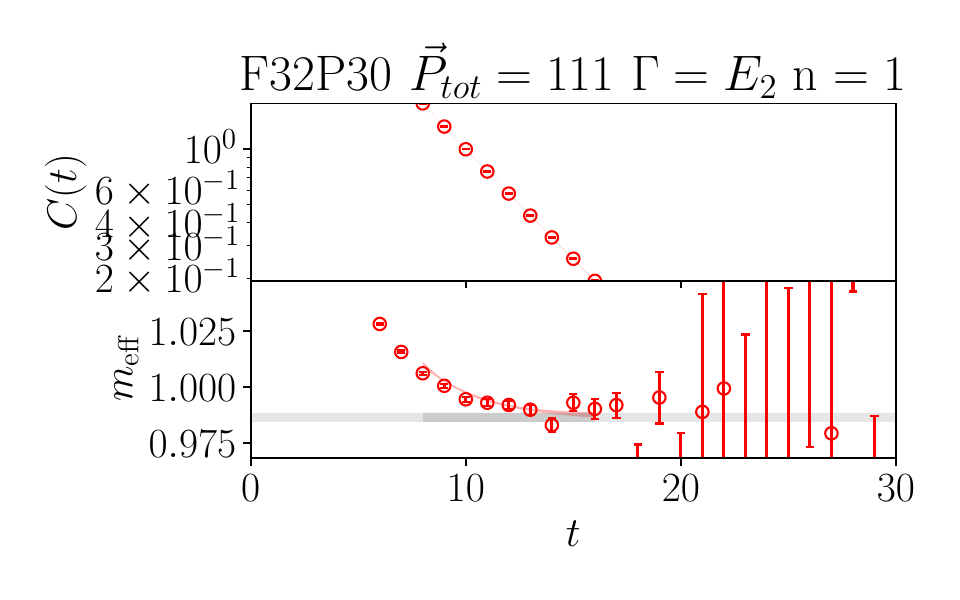}
\includegraphics[width=0.245\columnwidth]{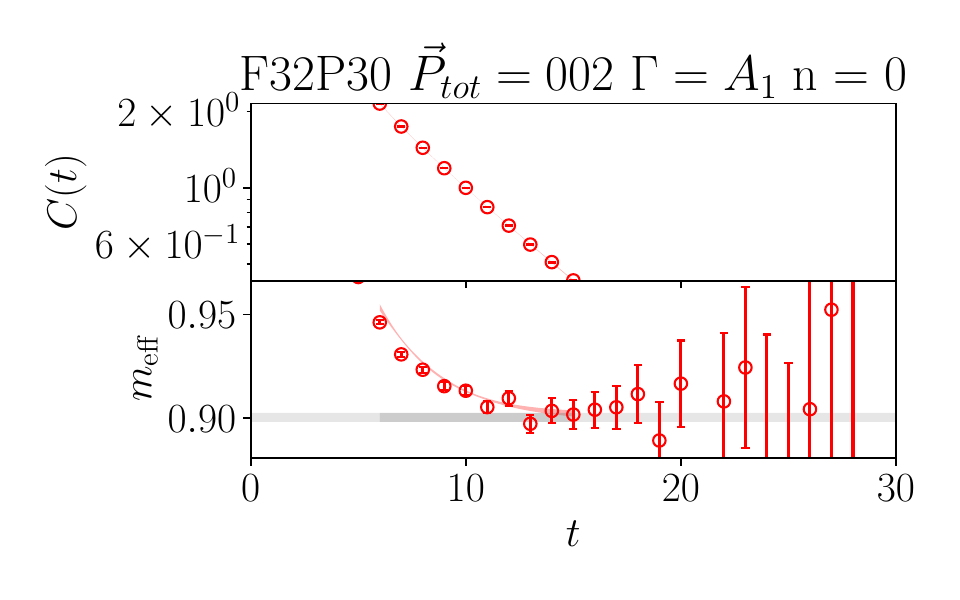}
\includegraphics[width=0.245\columnwidth]{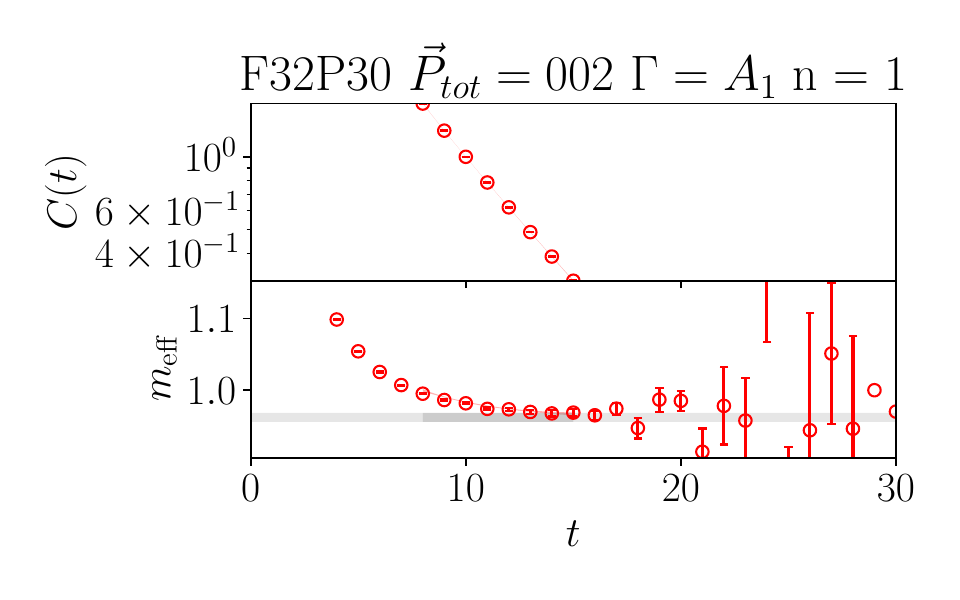}
\\
\includegraphics[width=0.245\columnwidth]{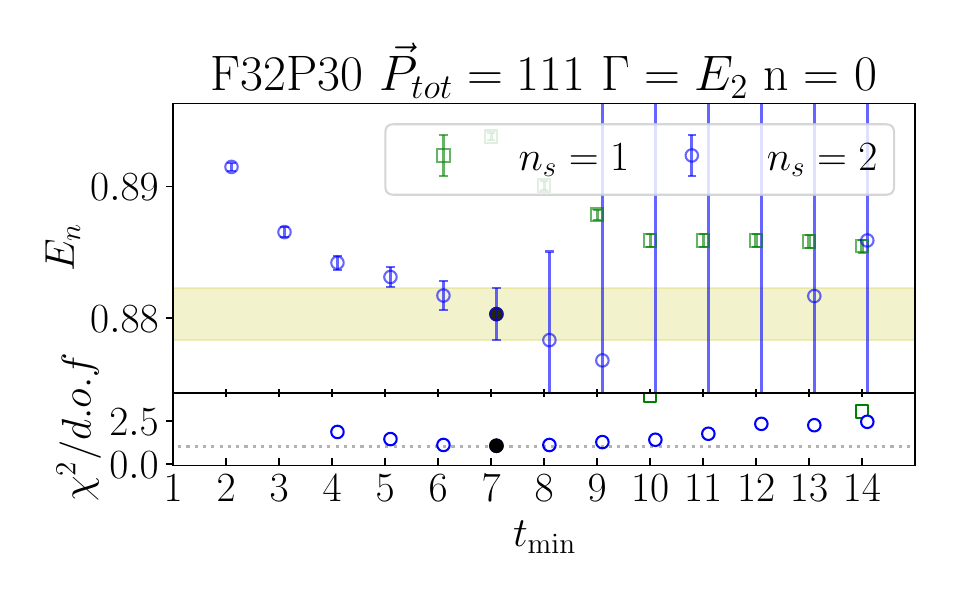}
\includegraphics[width=0.245\columnwidth]{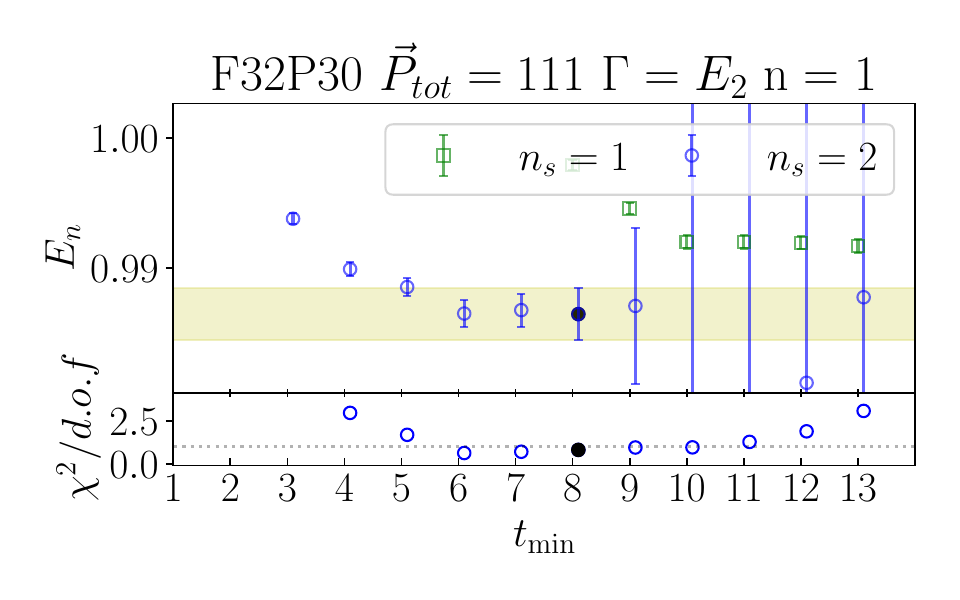}
\includegraphics[width=0.245\columnwidth]{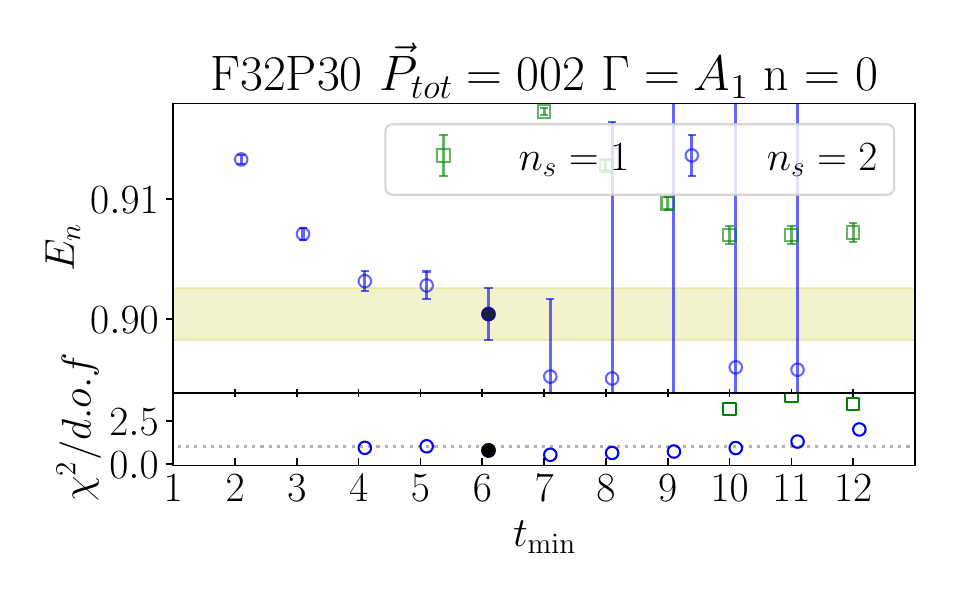}
\includegraphics[width=0.245\columnwidth]{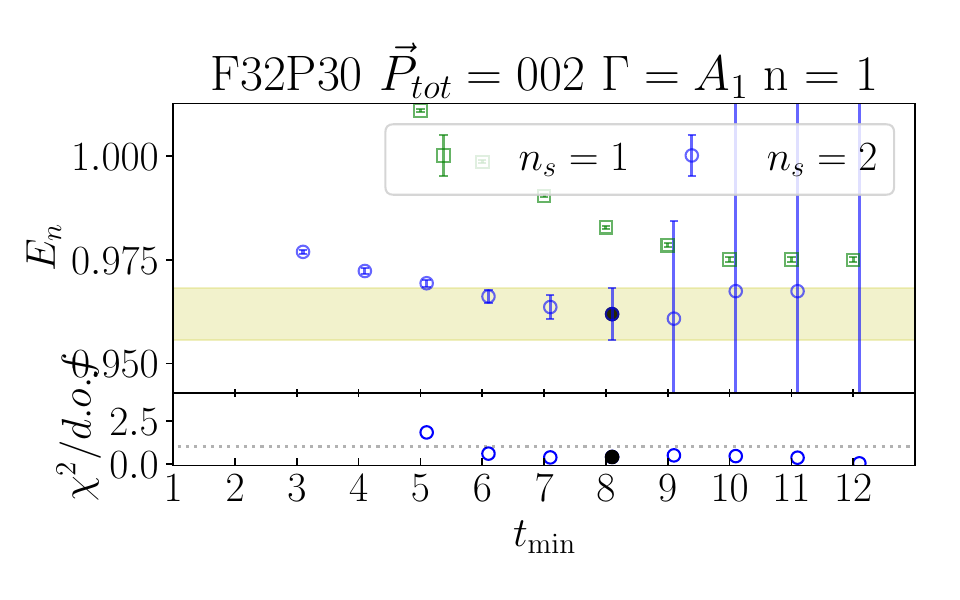}
\\
\includegraphics[width=0.245\columnwidth]{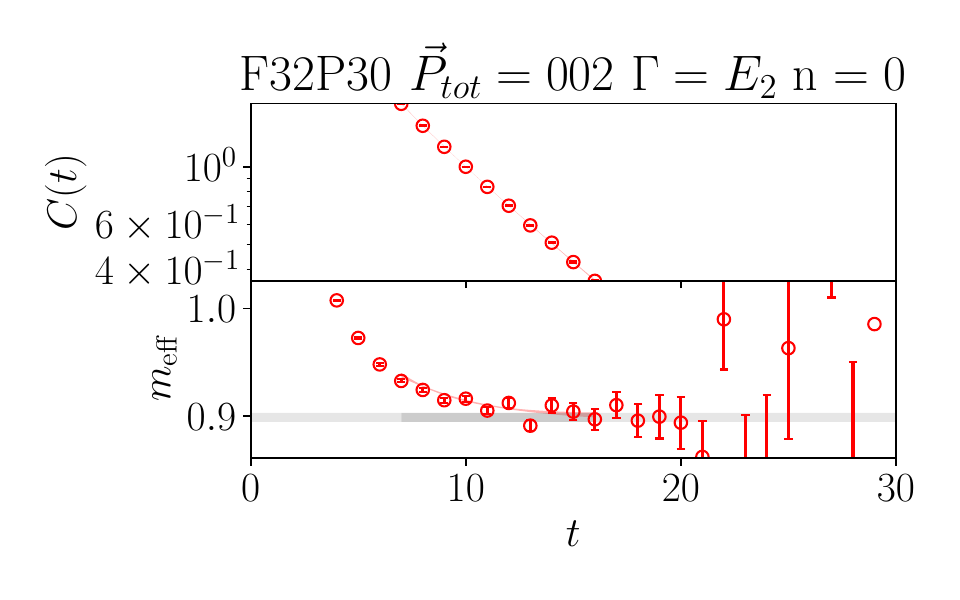}
\includegraphics[width=0.245\columnwidth]{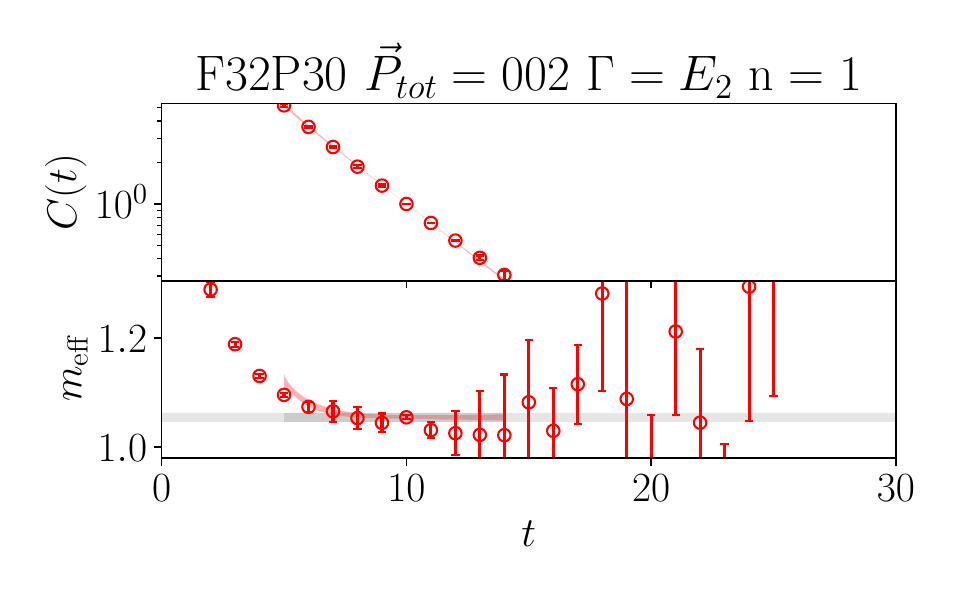}
\\
\includegraphics[width=0.245\columnwidth]{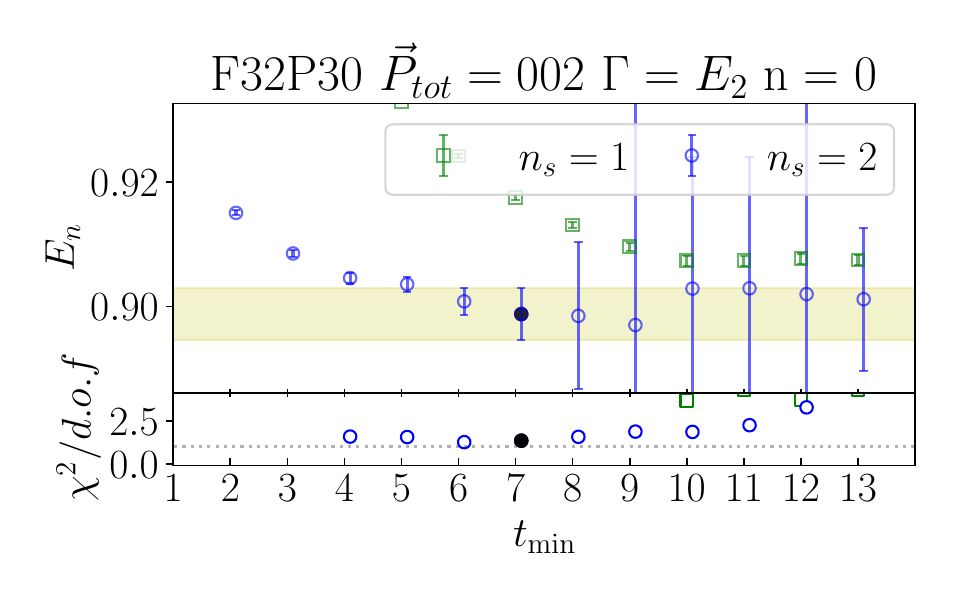}
\includegraphics[width=0.245\columnwidth]{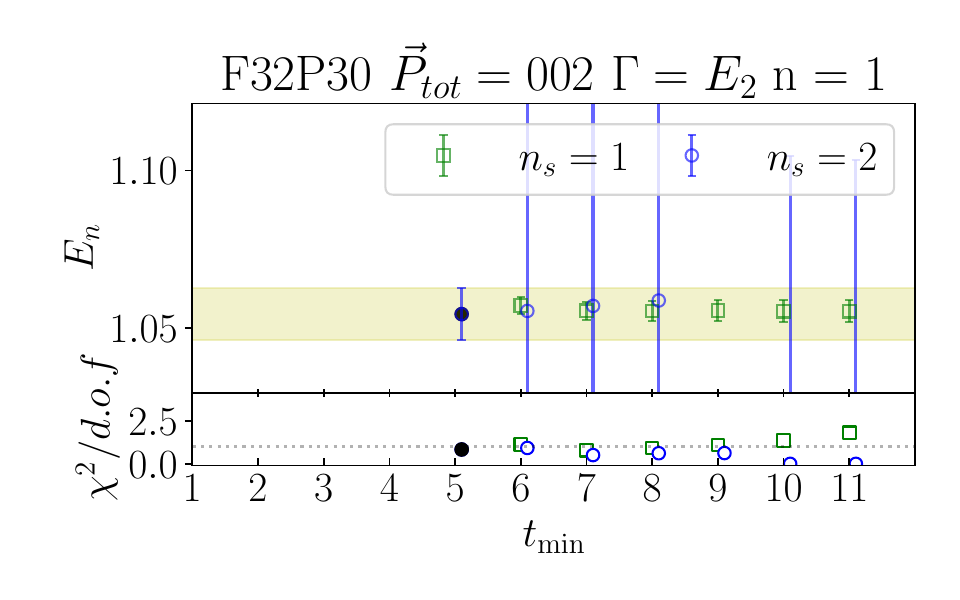}
\caption{Continued from Figure~\ref{fig:Dpi-fit-F32P30}. Energy-level fit results for the $I=\frac{1}{2}$ $D\pi$ channel on the F32P30 ensemble.}
\label{fig:Dpi-fit-F32P302}
\end{figure}

\begin{figure}[htbp]
\centering
\includegraphics[width=0.245\columnwidth]{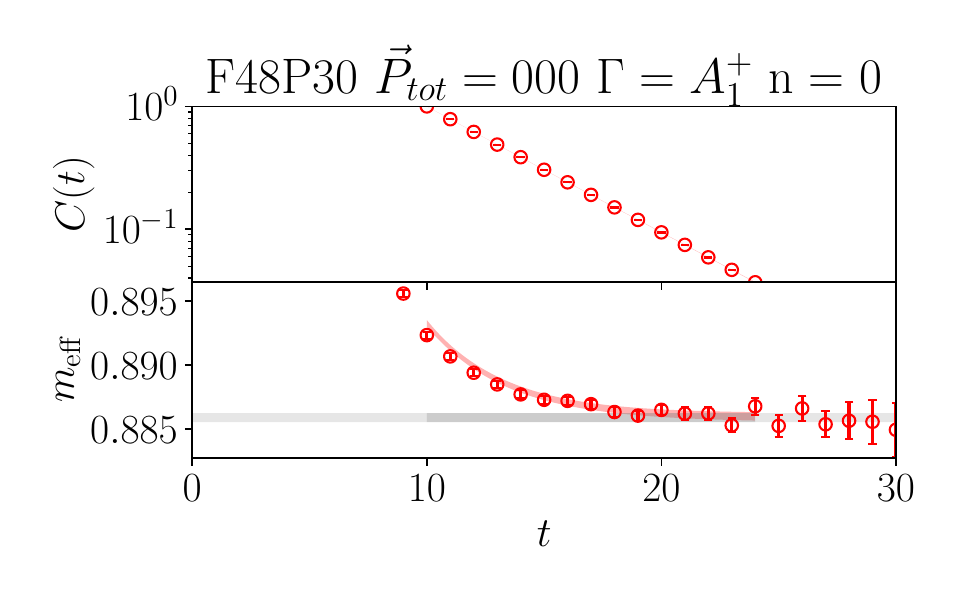}
\includegraphics[width=0.245\columnwidth]{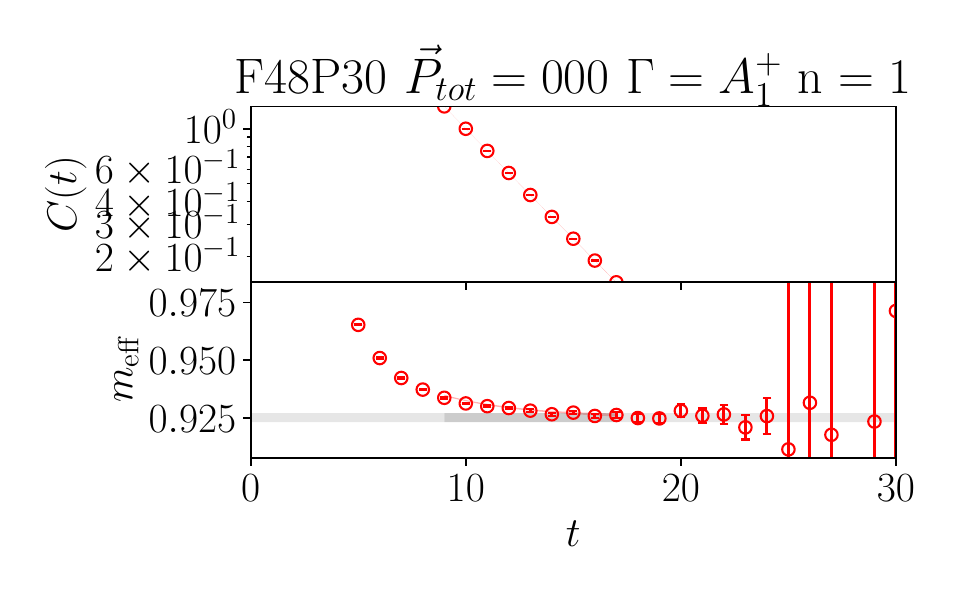}
\includegraphics[width=0.245\columnwidth]{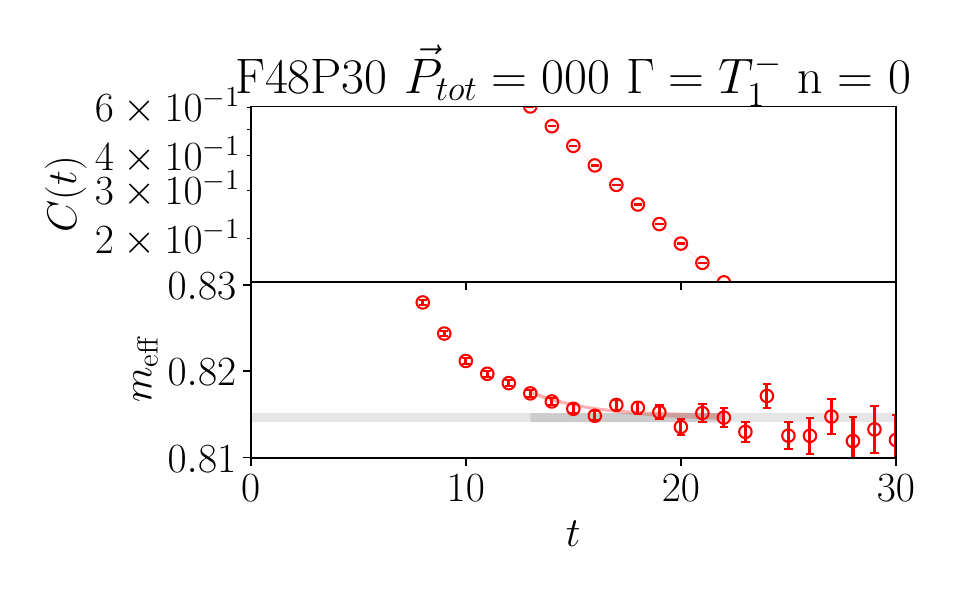}
\includegraphics[width=0.245\columnwidth]{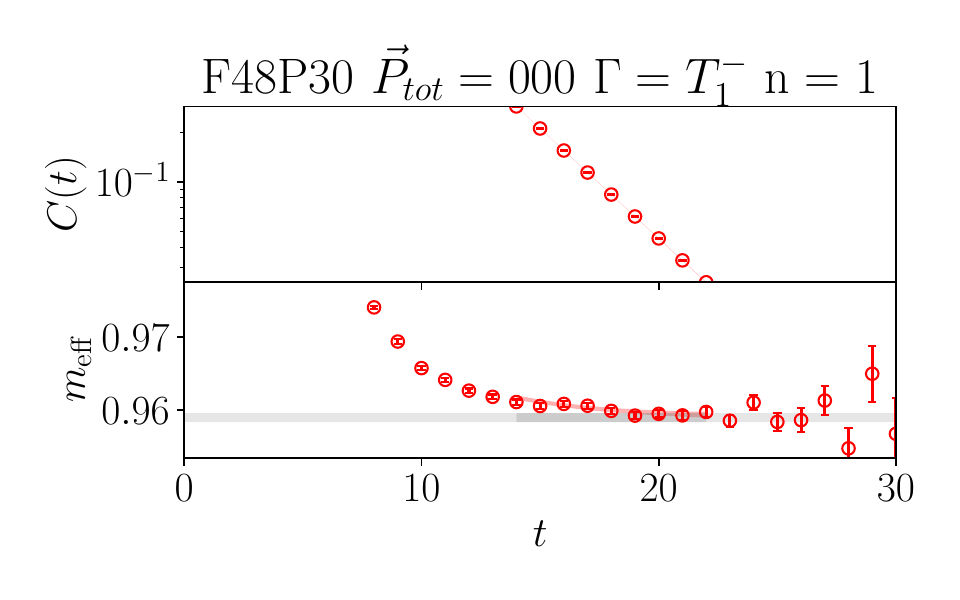}
\\
\includegraphics[width=0.245\columnwidth]{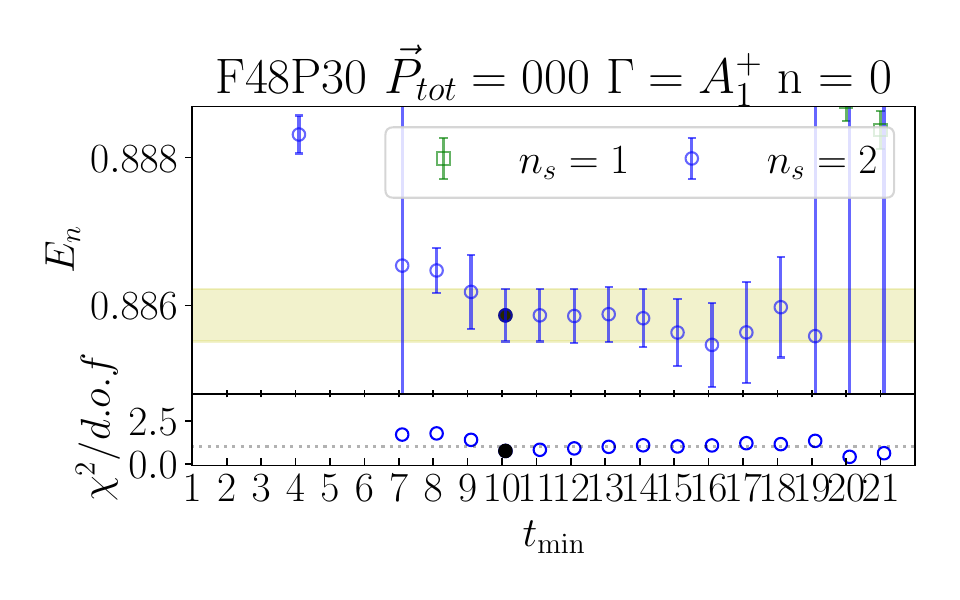}
\includegraphics[width=0.245\columnwidth]{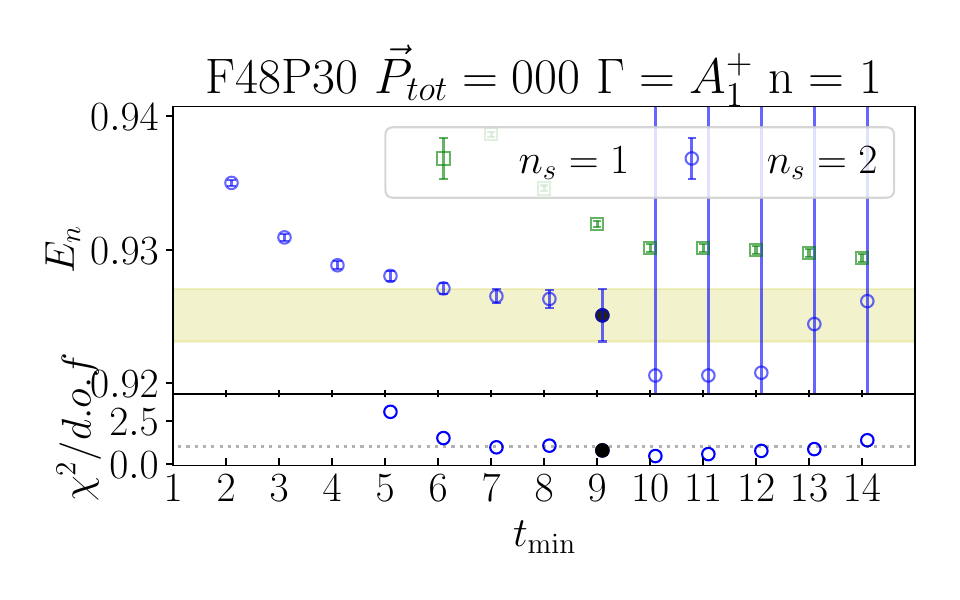}
\includegraphics[width=0.245\columnwidth]{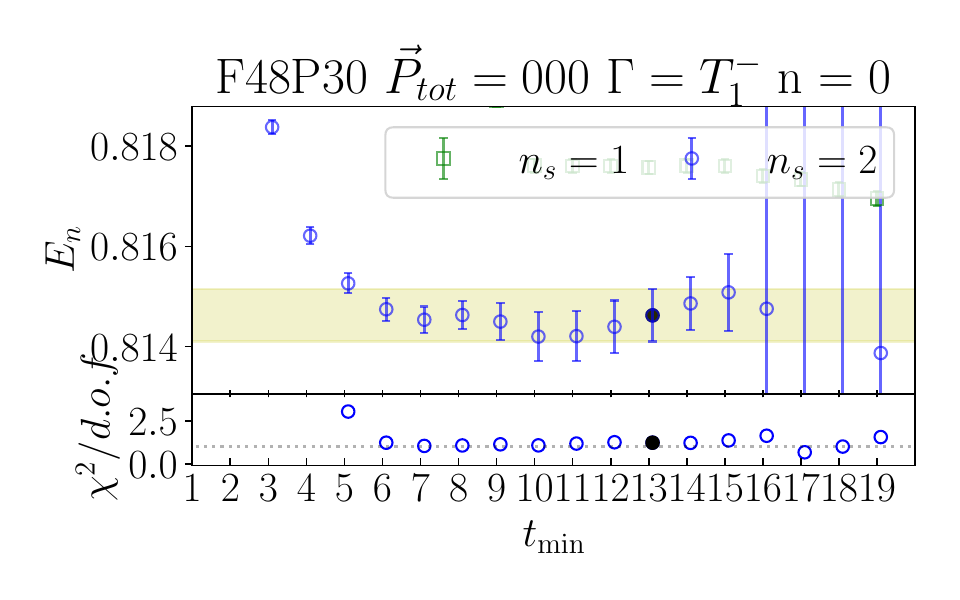}
\includegraphics[width=0.245\columnwidth]{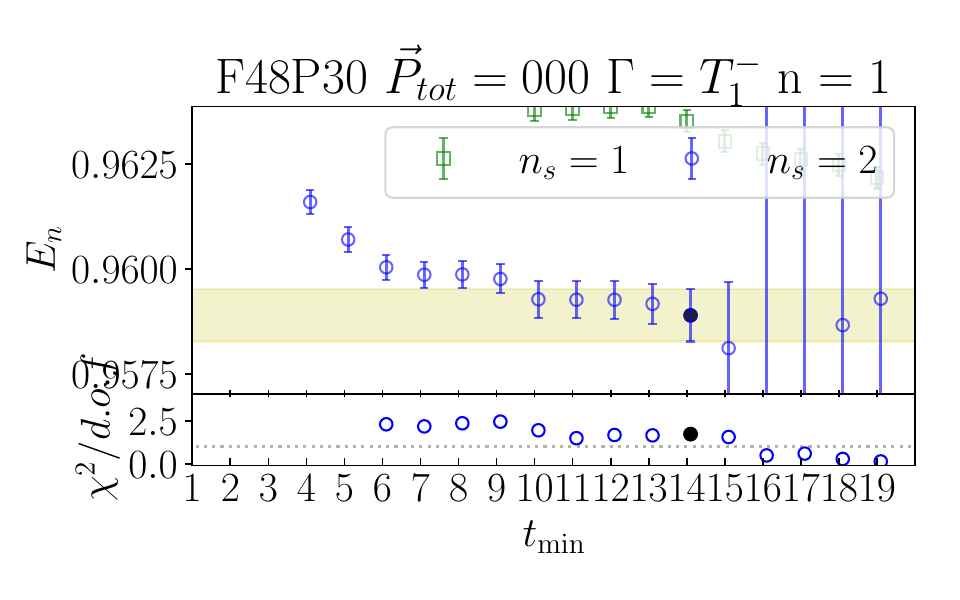}
\\
\includegraphics[width=0.245\columnwidth]{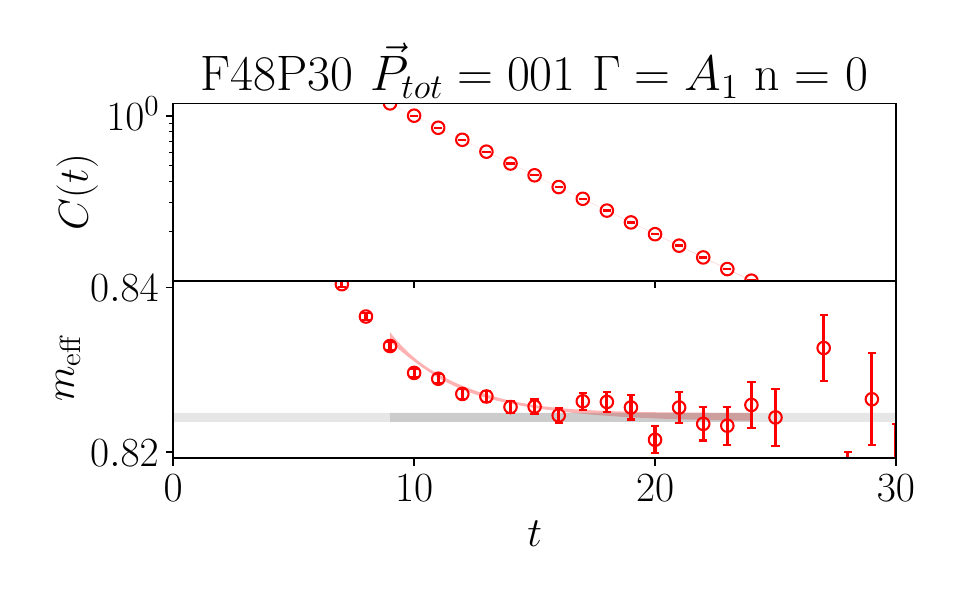}
\includegraphics[width=0.245\columnwidth]{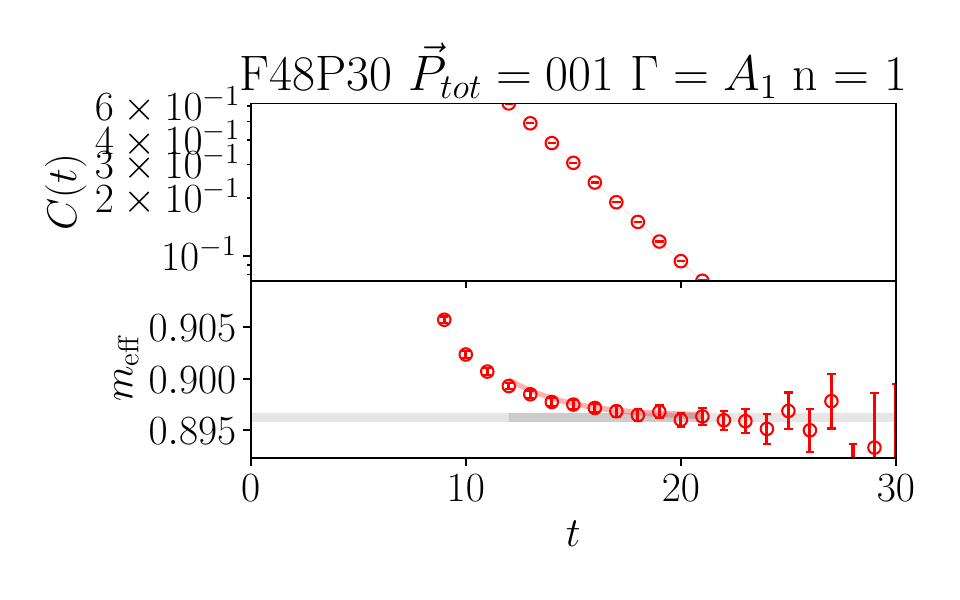}
\includegraphics[width=0.245\columnwidth]{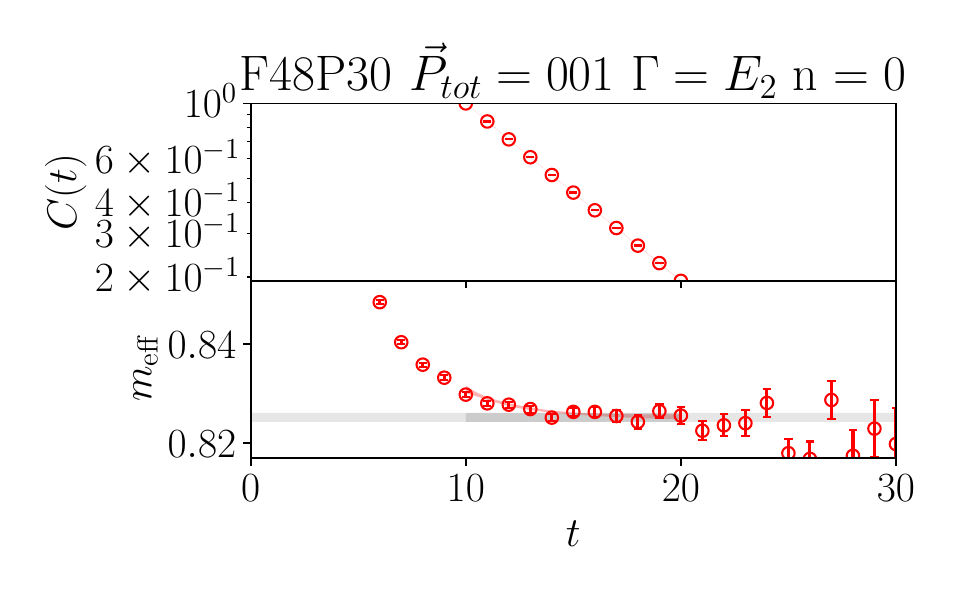}
\includegraphics[width=0.245\columnwidth]{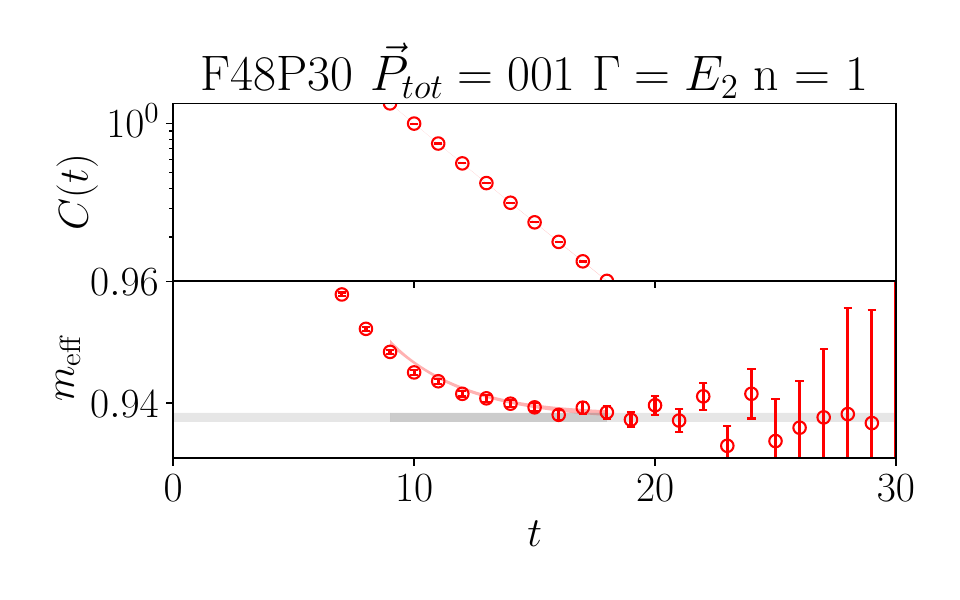}
\\
\includegraphics[width=0.245\columnwidth]{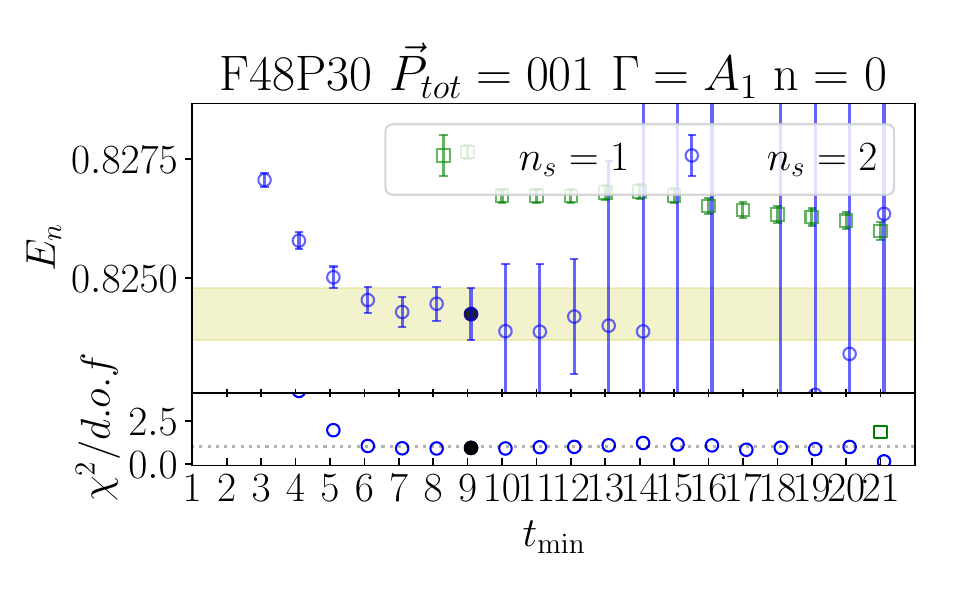}
\includegraphics[width=0.245\columnwidth]{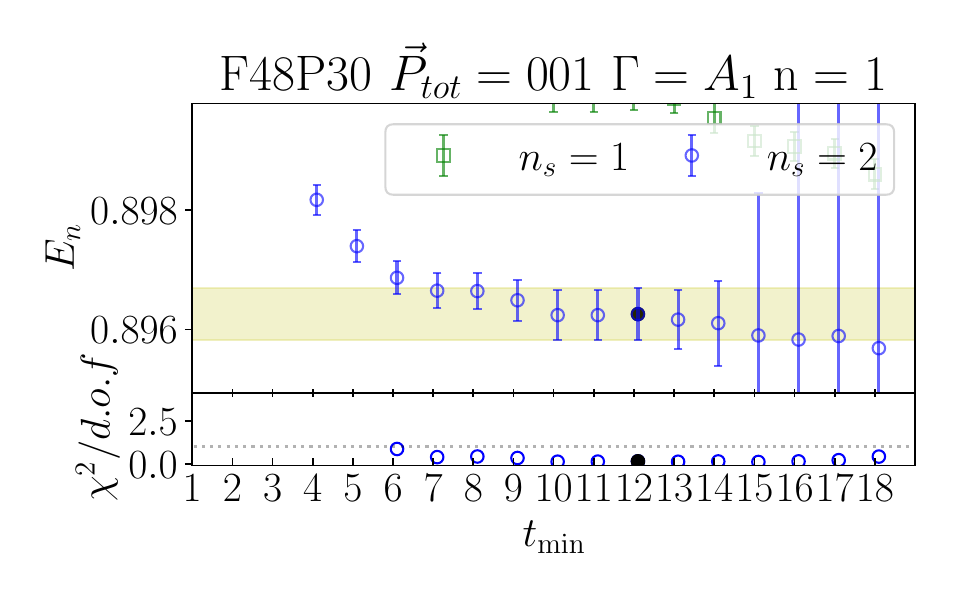}
\includegraphics[width=0.245\columnwidth]{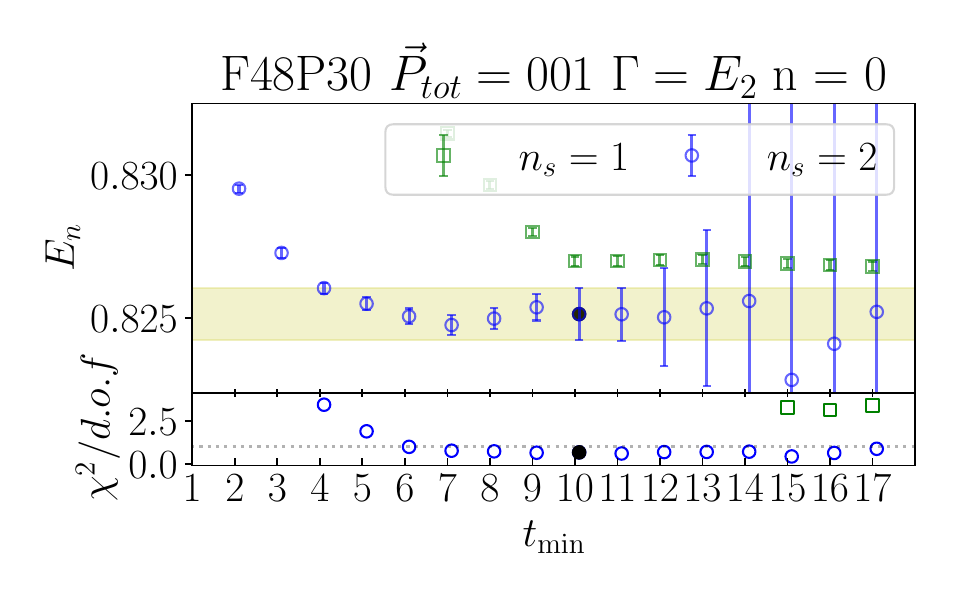}
\includegraphics[width=0.245\columnwidth]{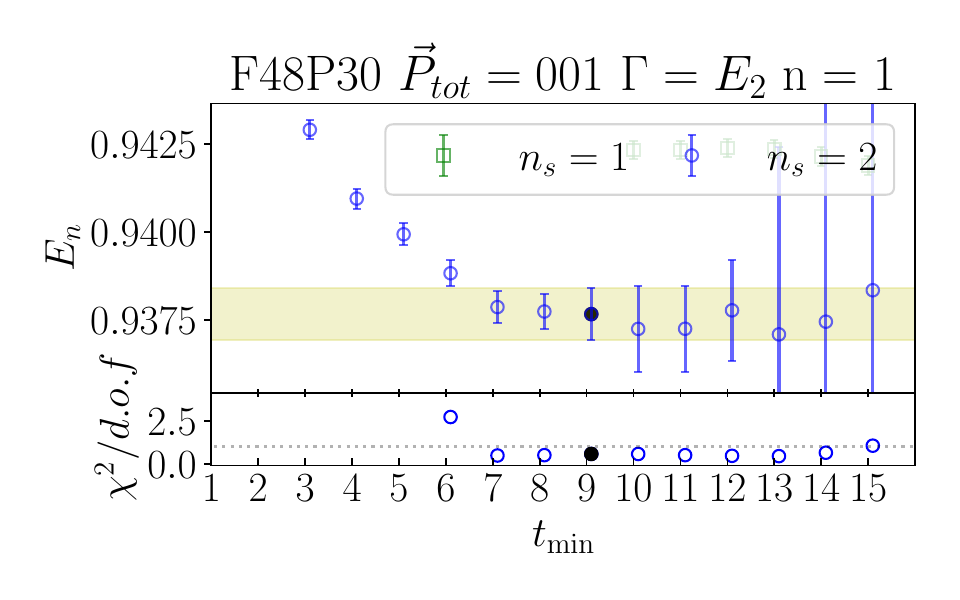}
\\
\includegraphics[width=0.245\columnwidth]{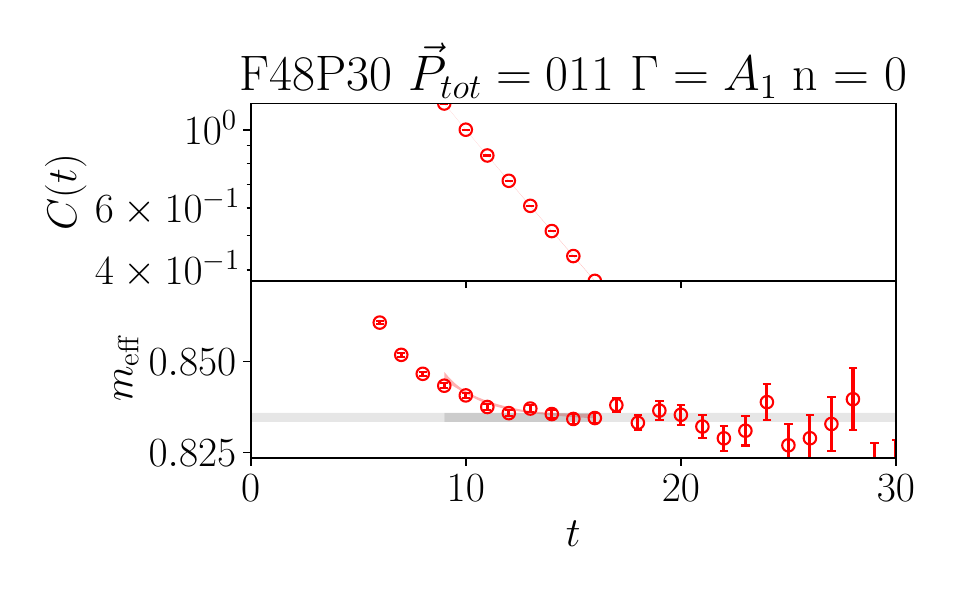}
\includegraphics[width=0.245\columnwidth]{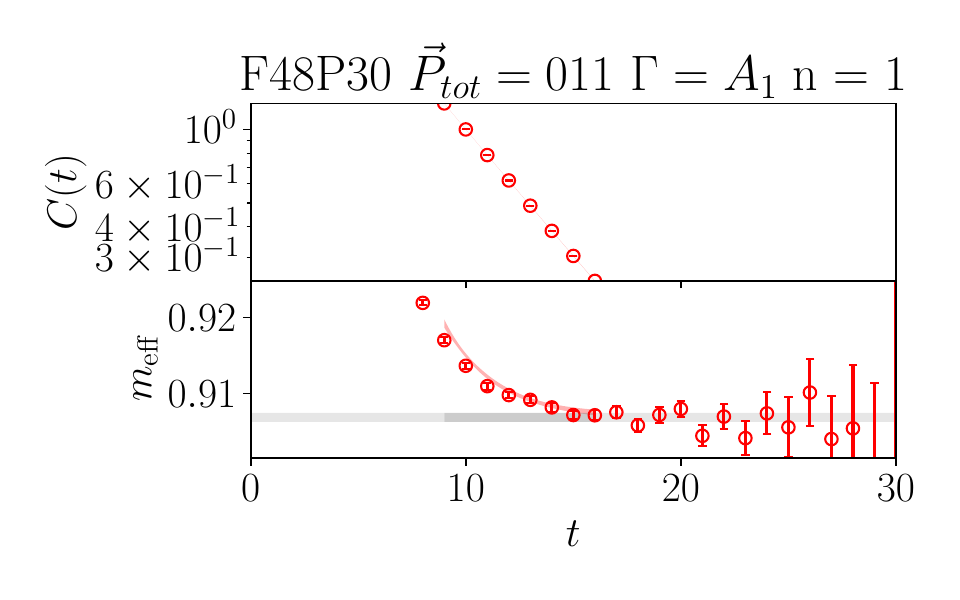}
\includegraphics[width=0.245\columnwidth]{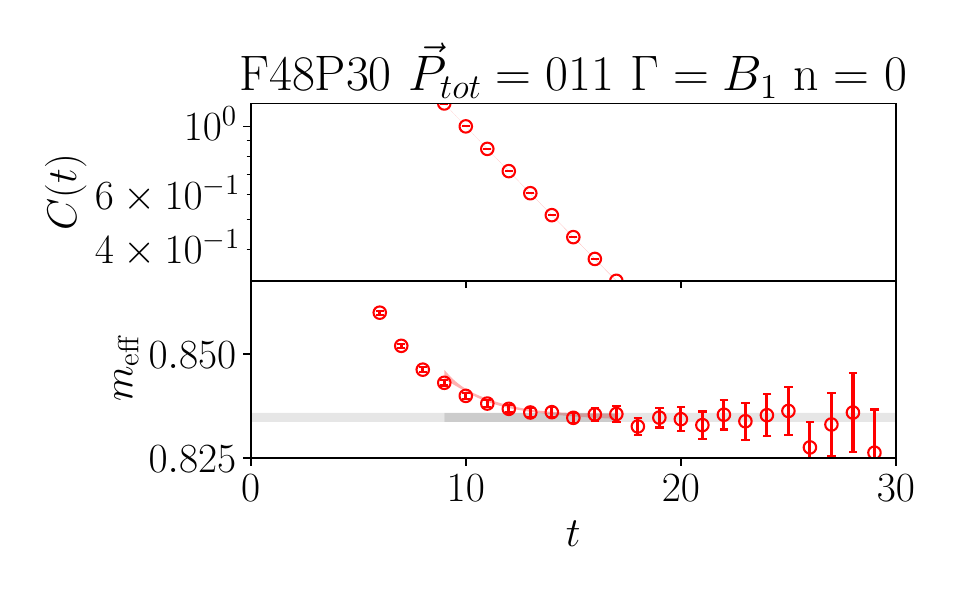}
\includegraphics[width=0.245\columnwidth]{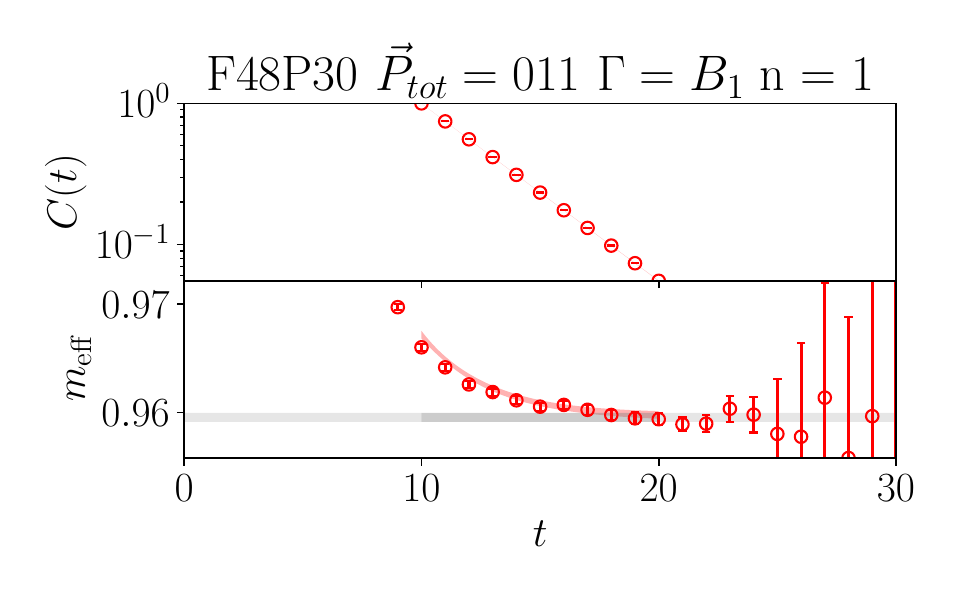}
\\
\includegraphics[width=0.245\columnwidth]{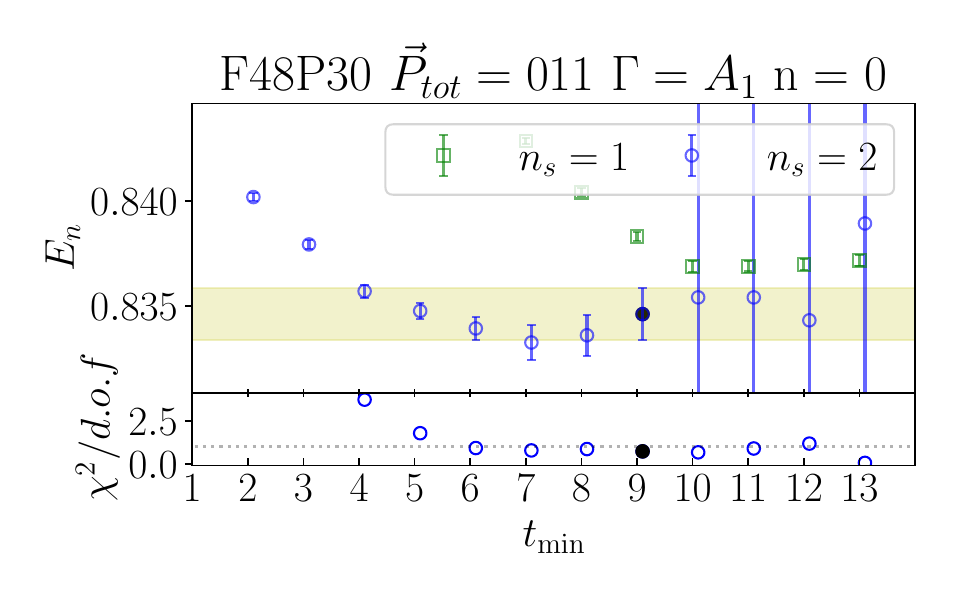}
\includegraphics[width=0.245\columnwidth]{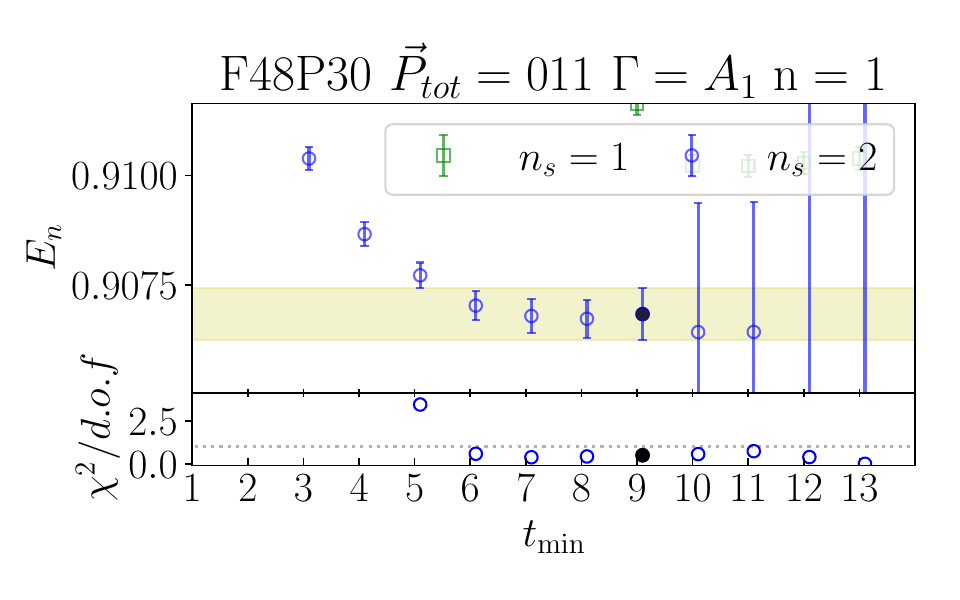}
\includegraphics[width=0.245\columnwidth]{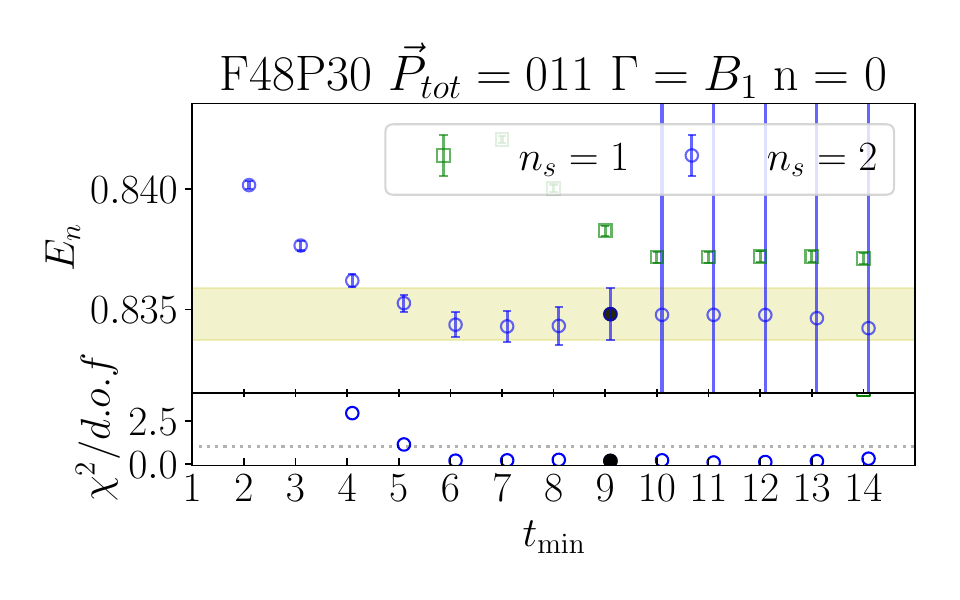}
\includegraphics[width=0.245\columnwidth]{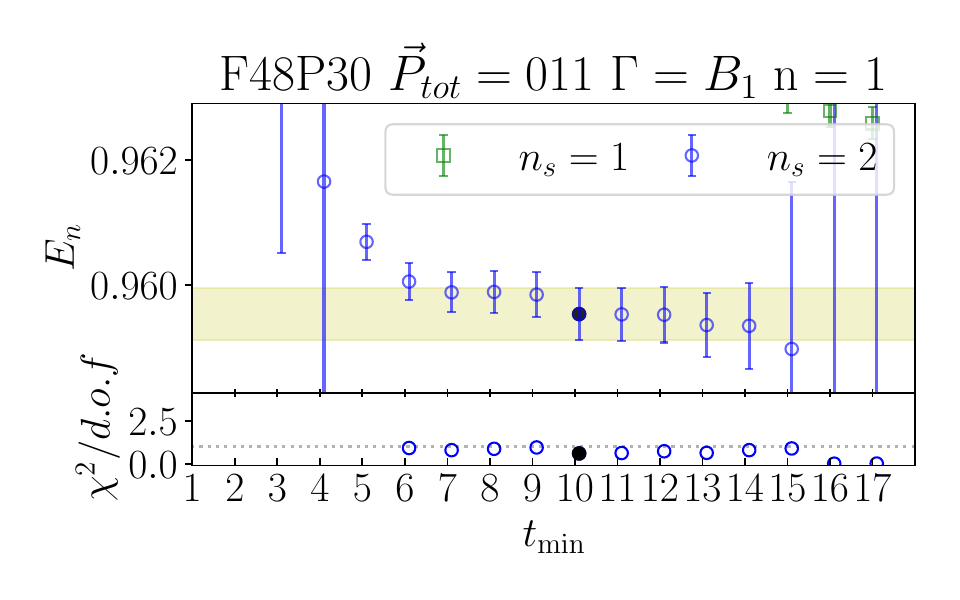}
\\
\includegraphics[width=0.245\columnwidth]{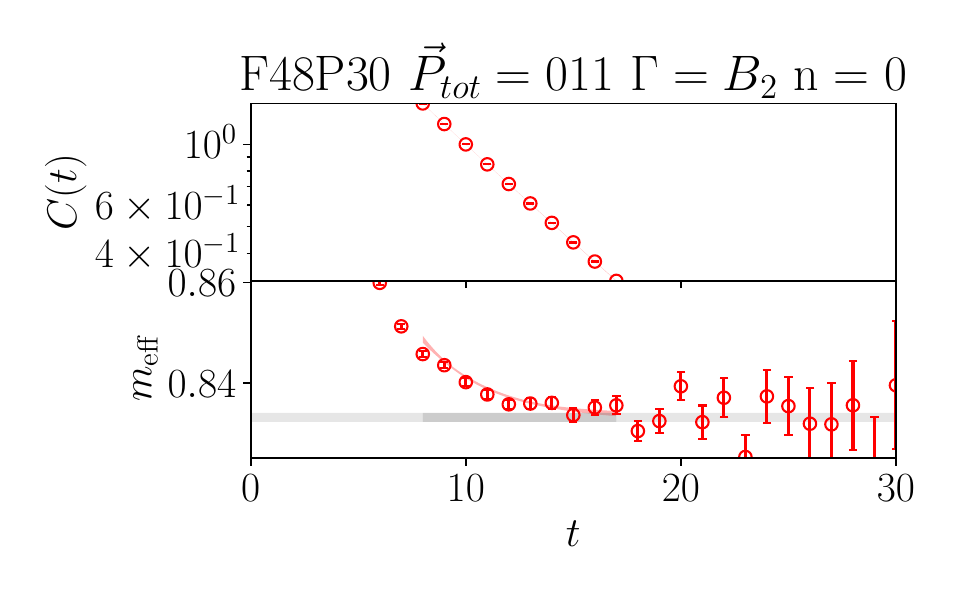}
\includegraphics[width=0.245\columnwidth]{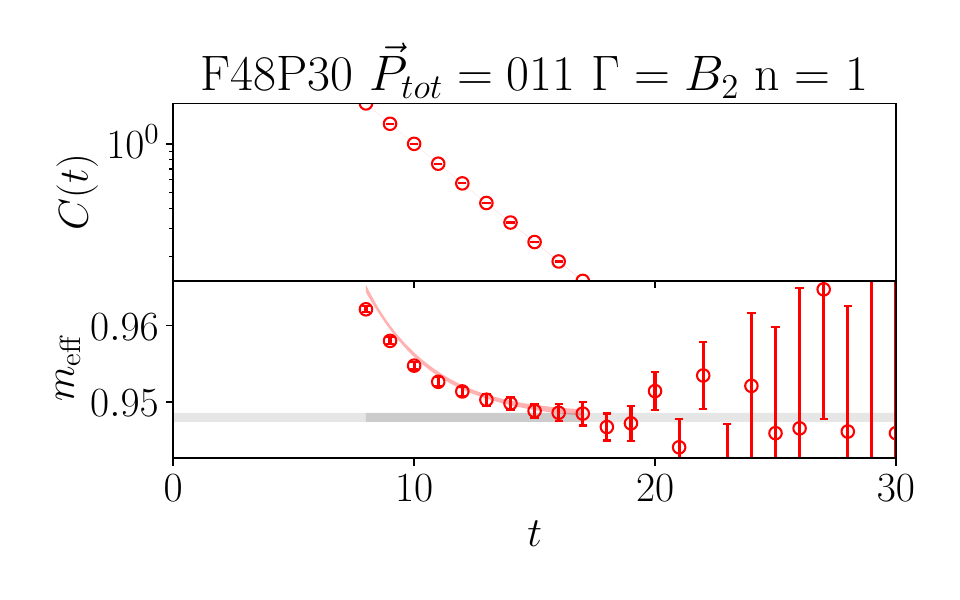}
\includegraphics[width=0.245\columnwidth]{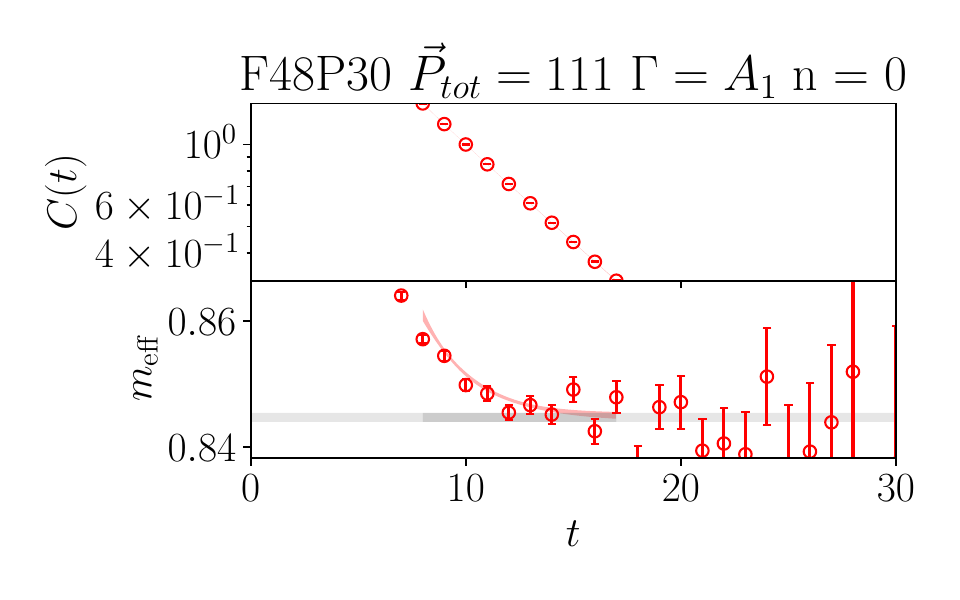}
\includegraphics[width=0.245\columnwidth]{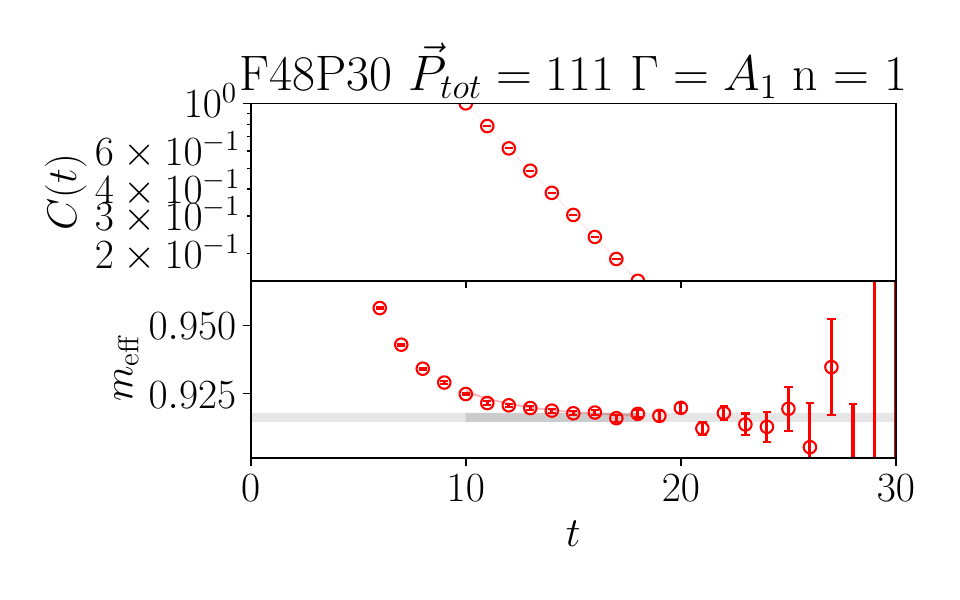}
\\
\includegraphics[width=0.245\columnwidth]{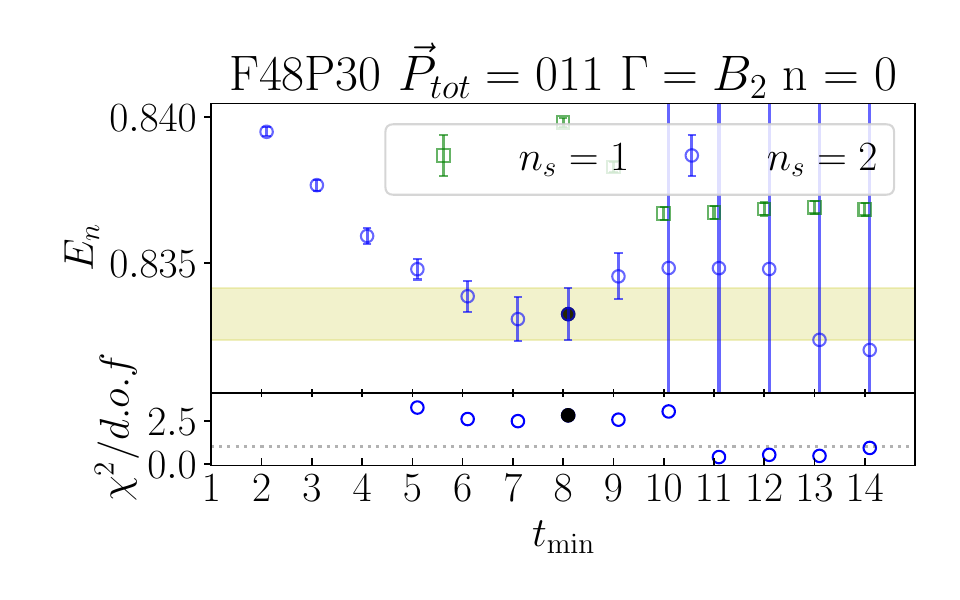}
\includegraphics[width=0.245\columnwidth]{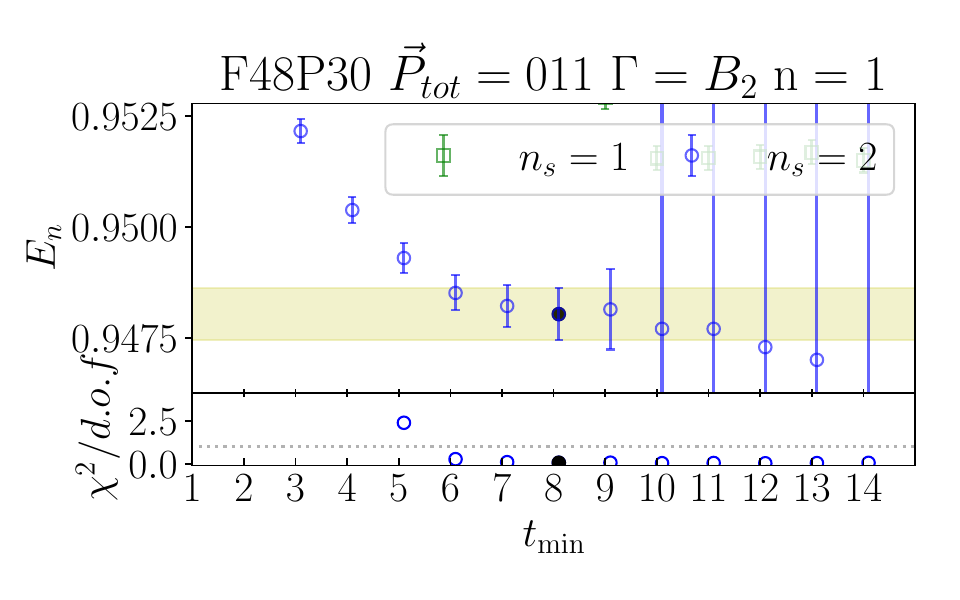}
\includegraphics[width=0.245\columnwidth]{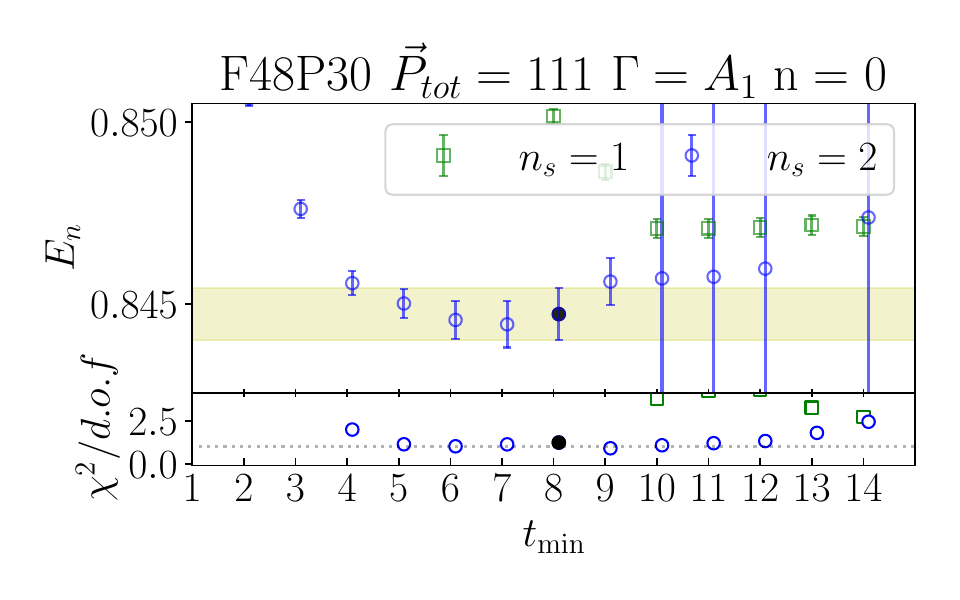}
\includegraphics[width=0.245\columnwidth]{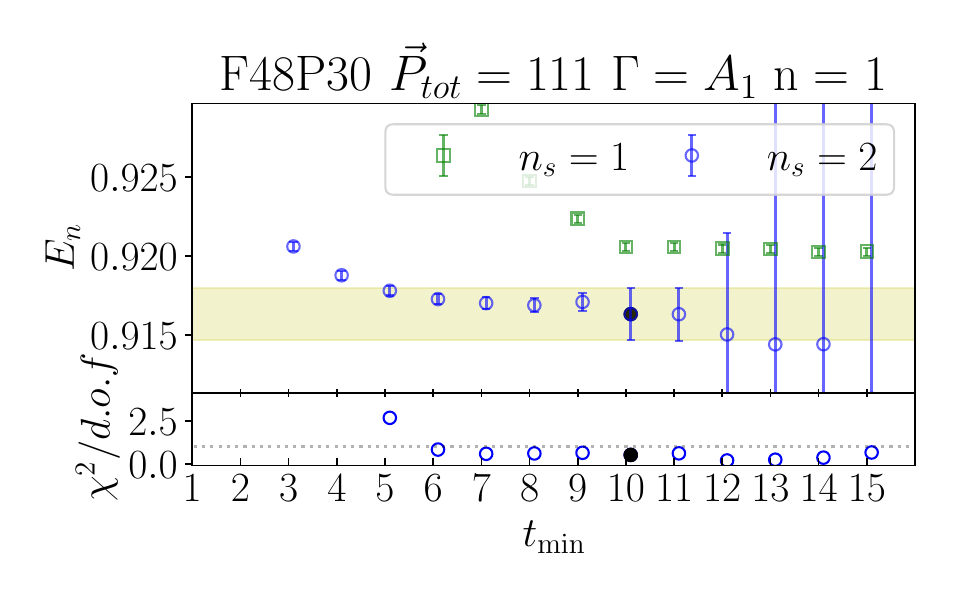}
\caption{Energy-level fit results for the $I=\frac{1}{2}$ $D\pi$ channel on the F48P30 ensemble. The description follows Figure~\ref{fig:Dpi-fit-F32P30}.}
\label{fig:Dpi-fit-F48P30}
\end{figure}

\begin{figure}[htbp]
\centering
\includegraphics[width=0.245\columnwidth]{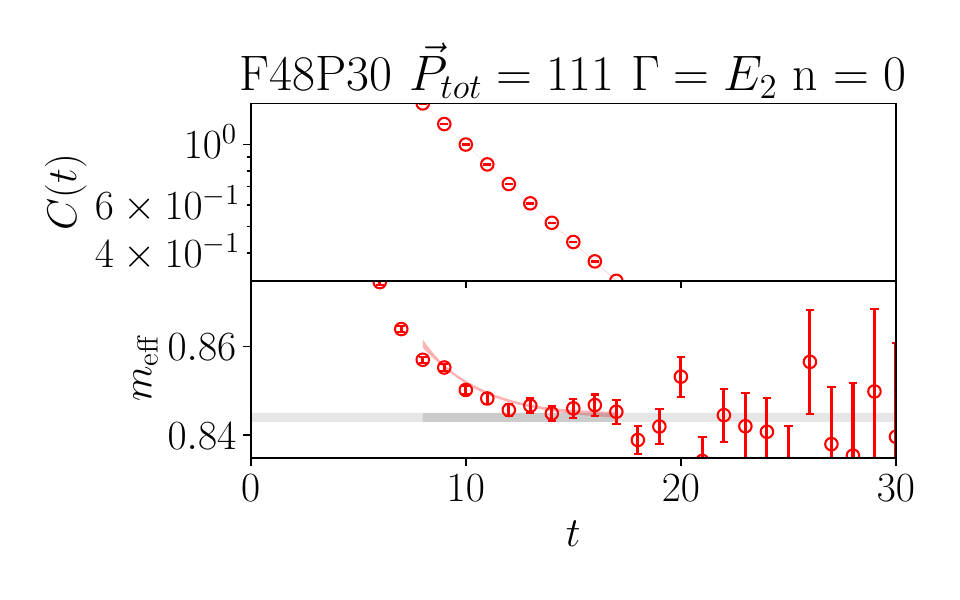}
\includegraphics[width=0.245\columnwidth]{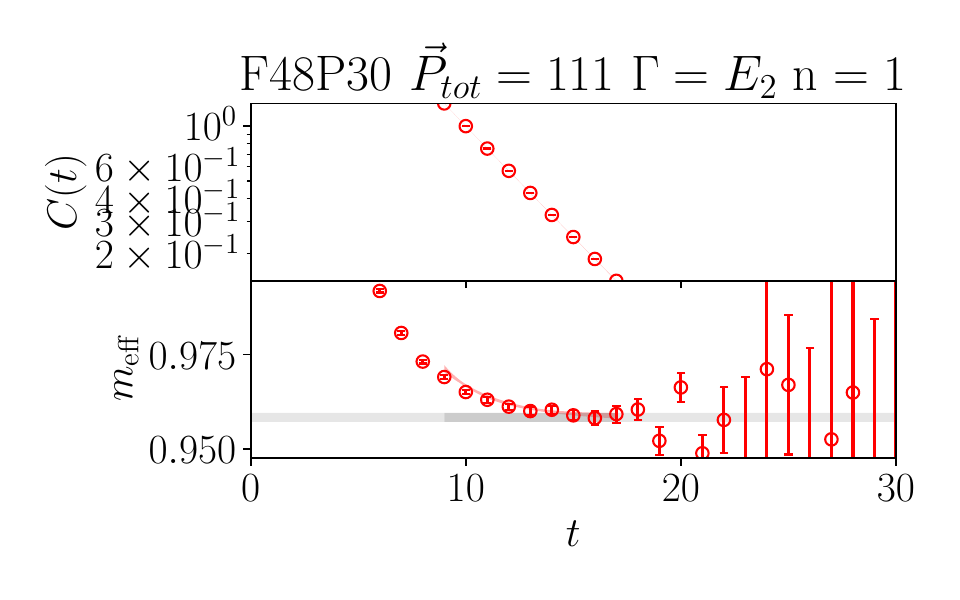}
\includegraphics[width=0.245\columnwidth]{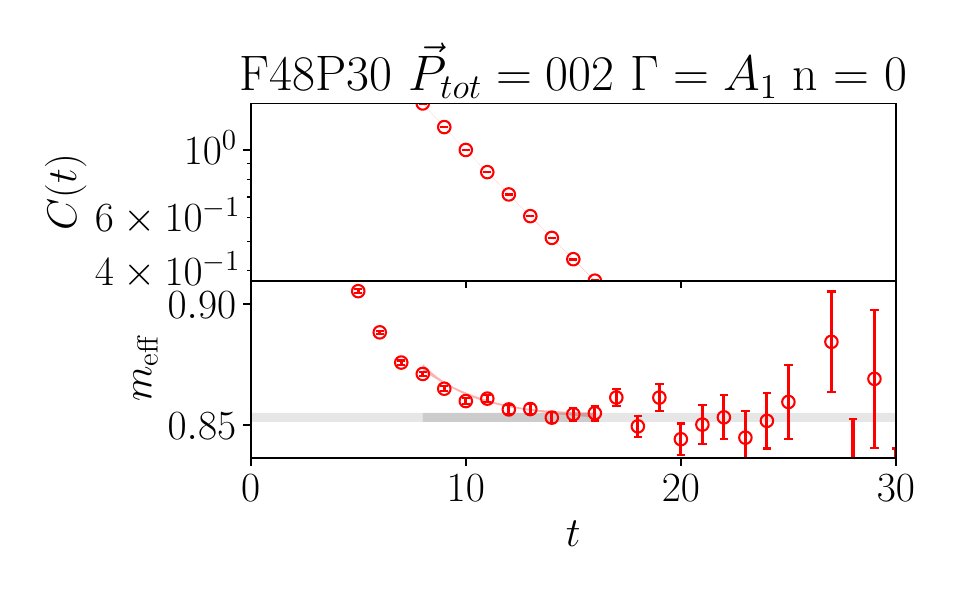}
\includegraphics[width=0.245\columnwidth]{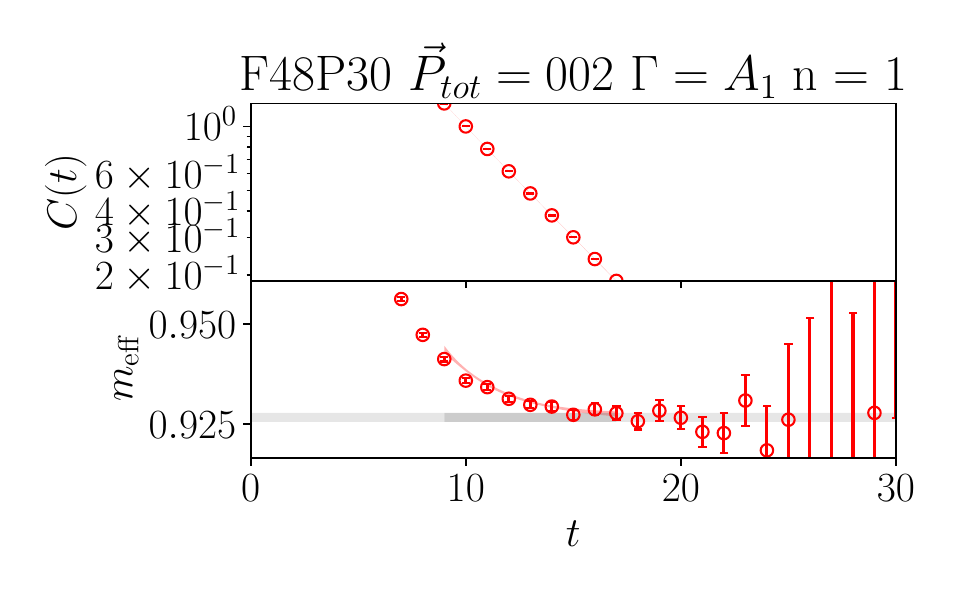}
\\
\includegraphics[width=0.245\columnwidth]{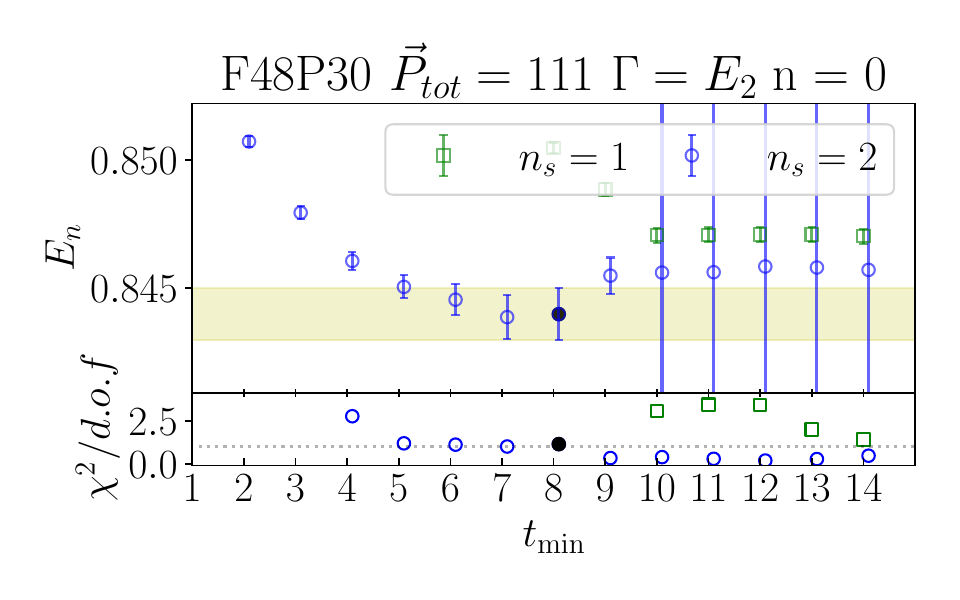}
\includegraphics[width=0.245\columnwidth]{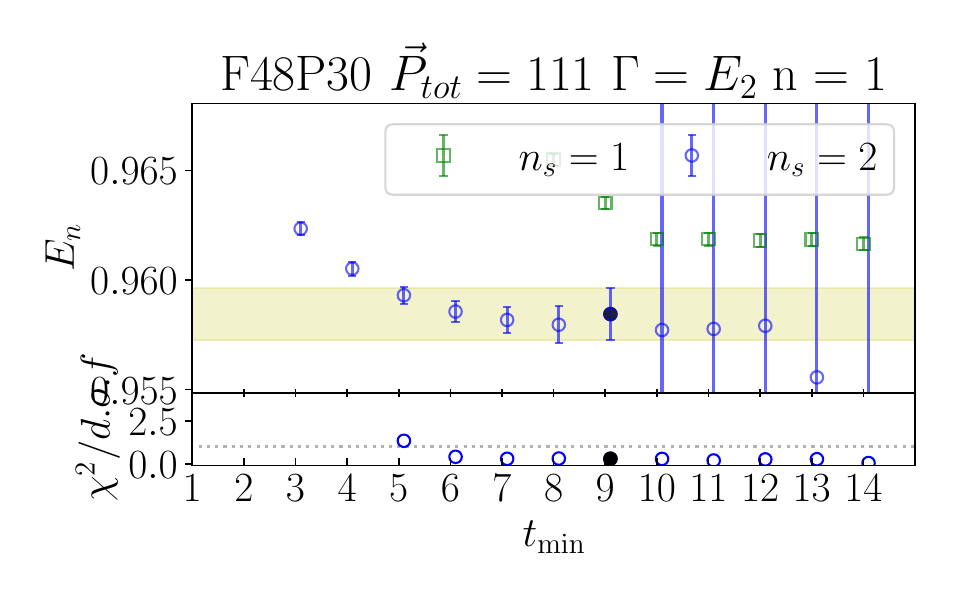}
\includegraphics[width=0.245\columnwidth]{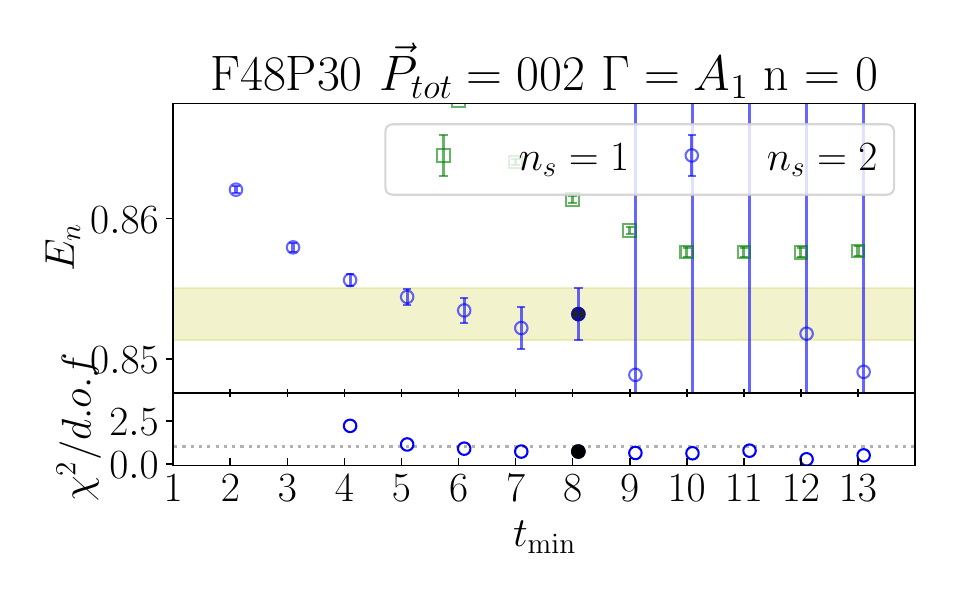}
\includegraphics[width=0.245\columnwidth]{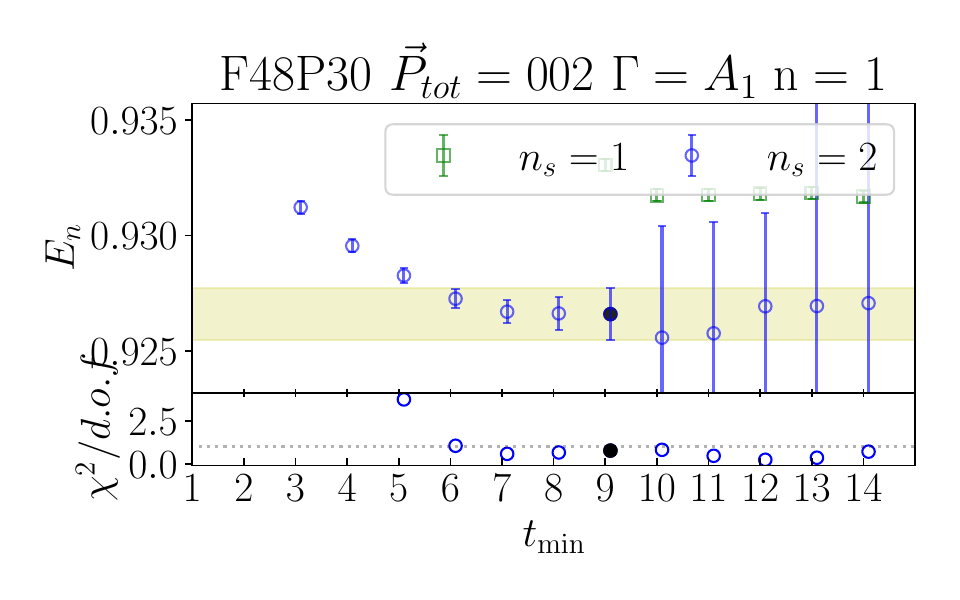}
\\
\includegraphics[width=0.245\columnwidth]{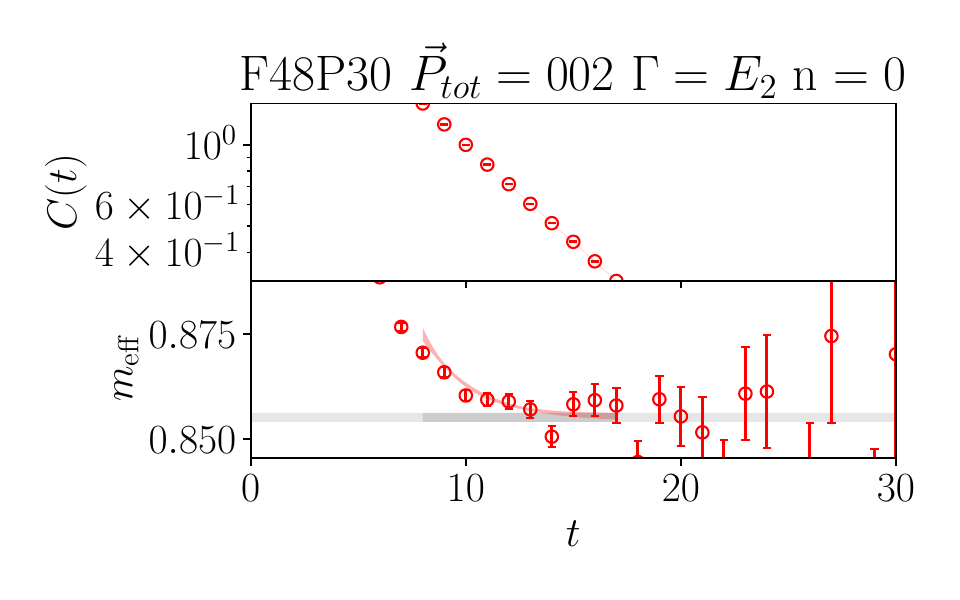}
\includegraphics[width=0.245\columnwidth]{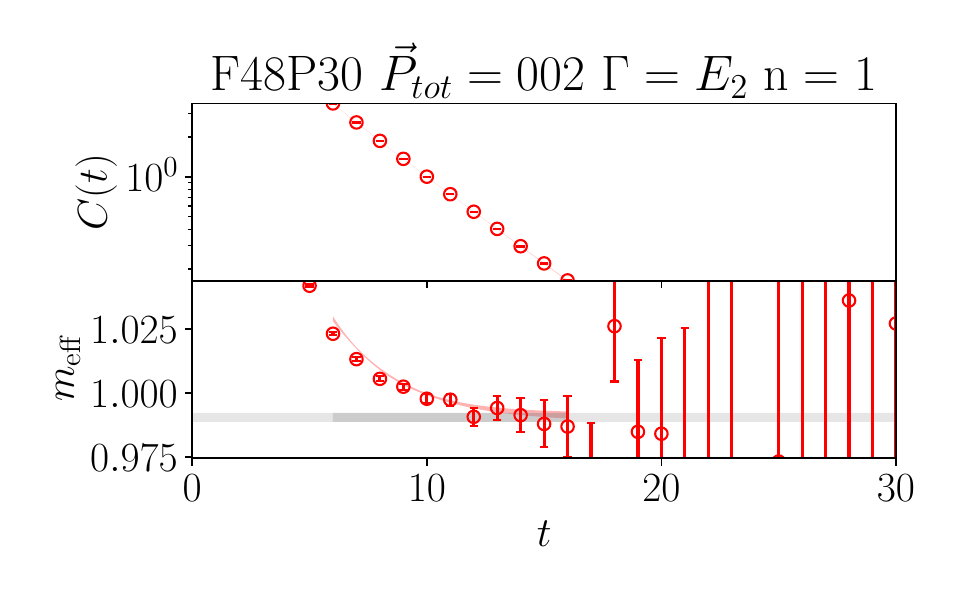}
\\
\includegraphics[width=0.245\columnwidth]{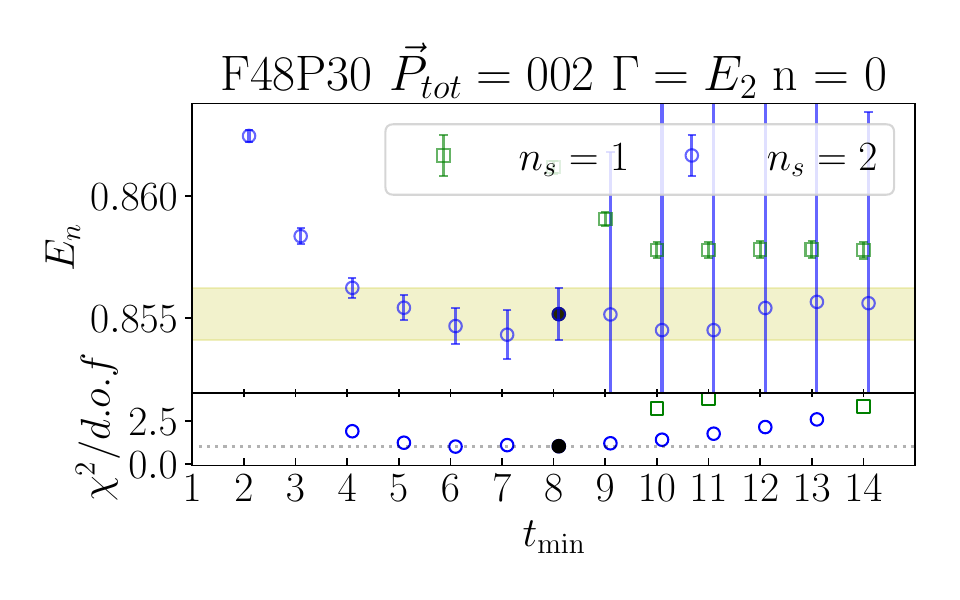}
\includegraphics[width=0.245\columnwidth]{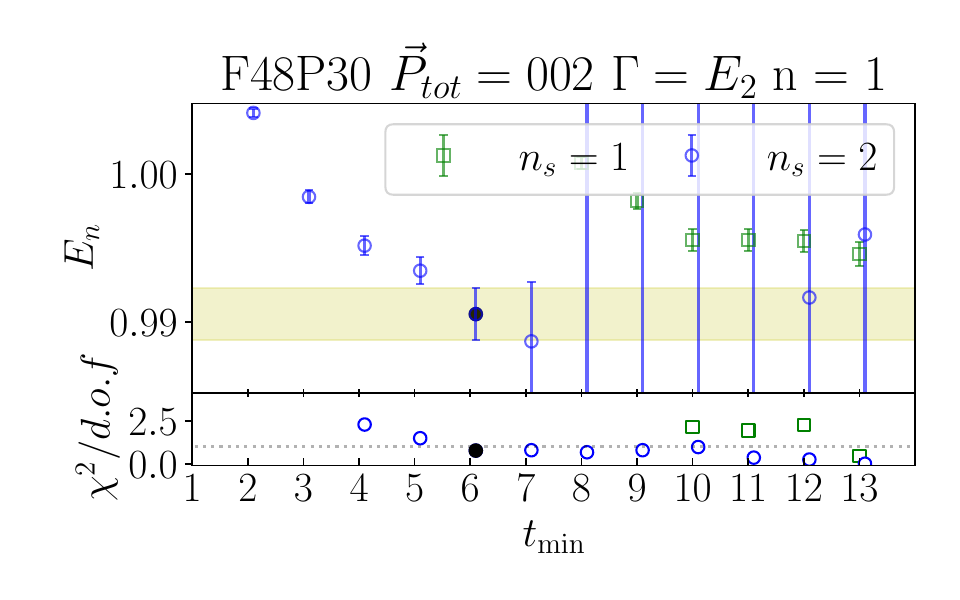}
\caption{Continued from Figure~\ref{fig:Dpi-fit-F48P30}. Energy-level fit results for the $I=\frac{1}{2}$ $D\pi$ channel on the F48P30 ensemble.}
\label{fig:Dpi-fit-F48P302}
\end{figure}

\begin{figure}[htbp]
\centering
\includegraphics[width=0.245\columnwidth]{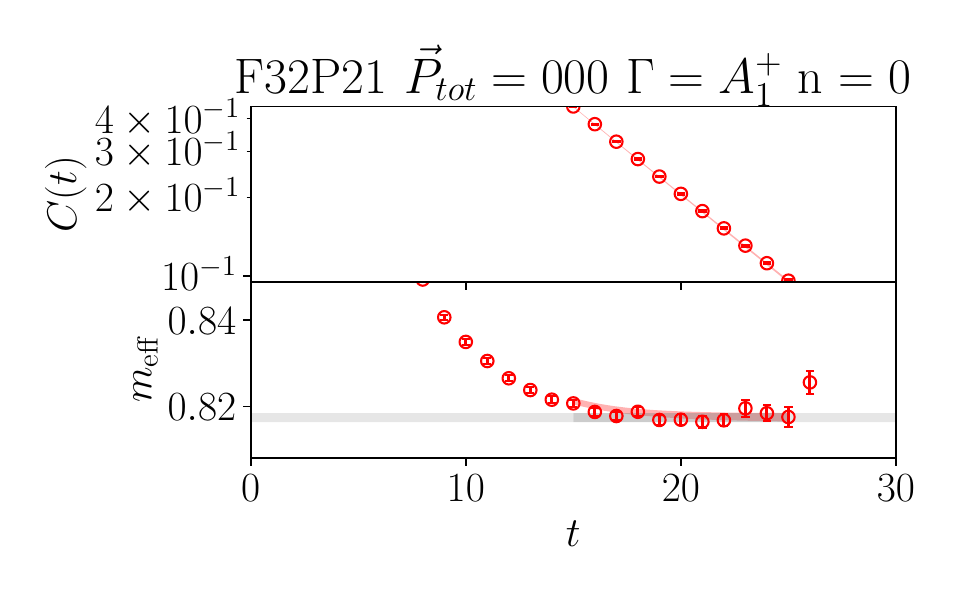}
\includegraphics[width=0.245\columnwidth]{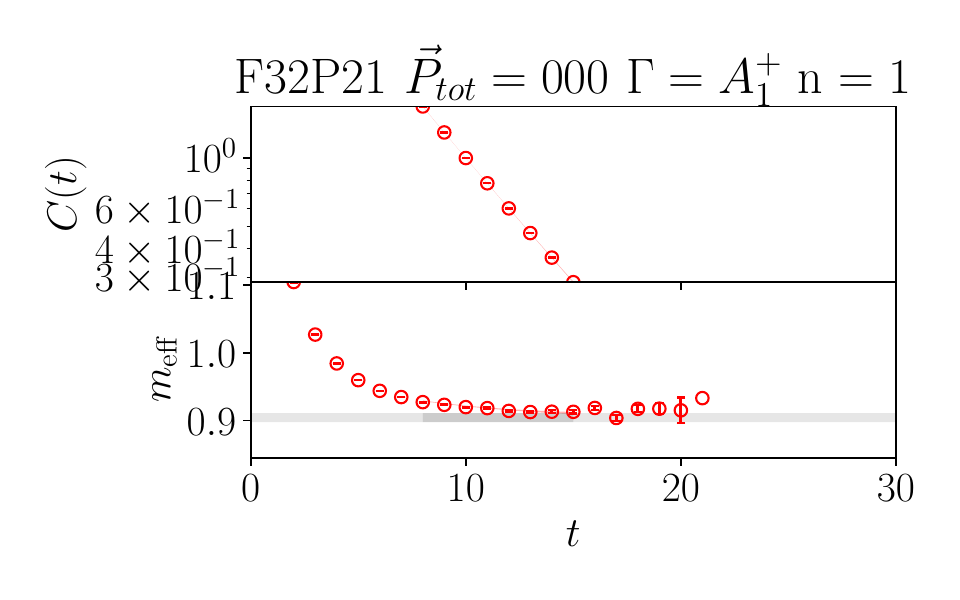}
\includegraphics[width=0.245\columnwidth]{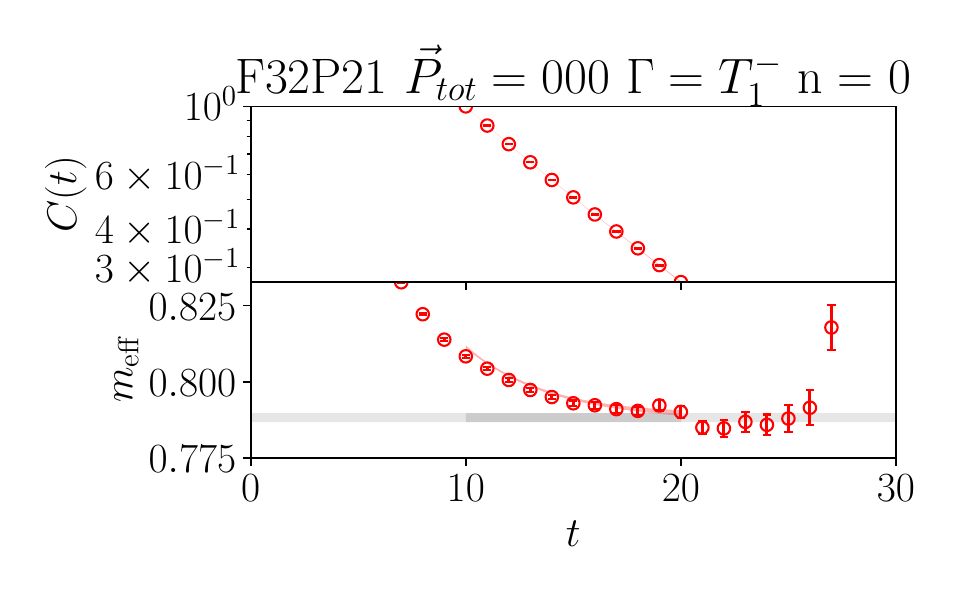}
\includegraphics[width=0.245\columnwidth]{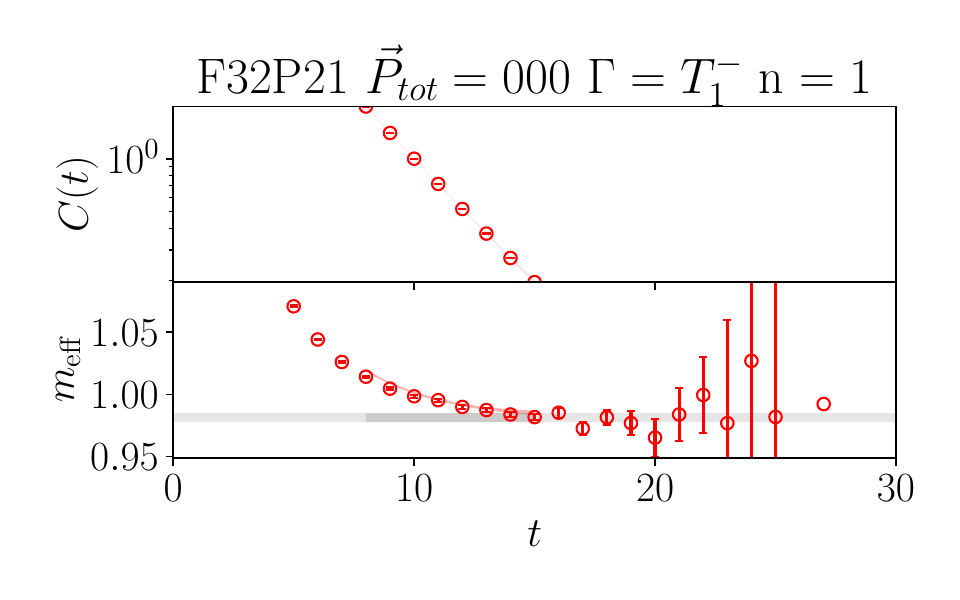}
\\
\includegraphics[width=0.245\columnwidth]{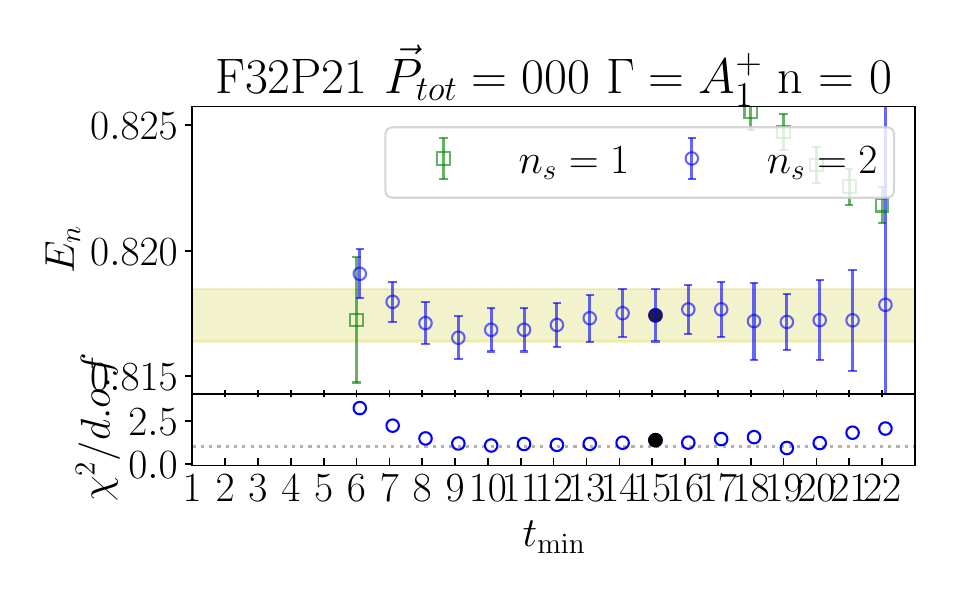}
\includegraphics[width=0.245\columnwidth]{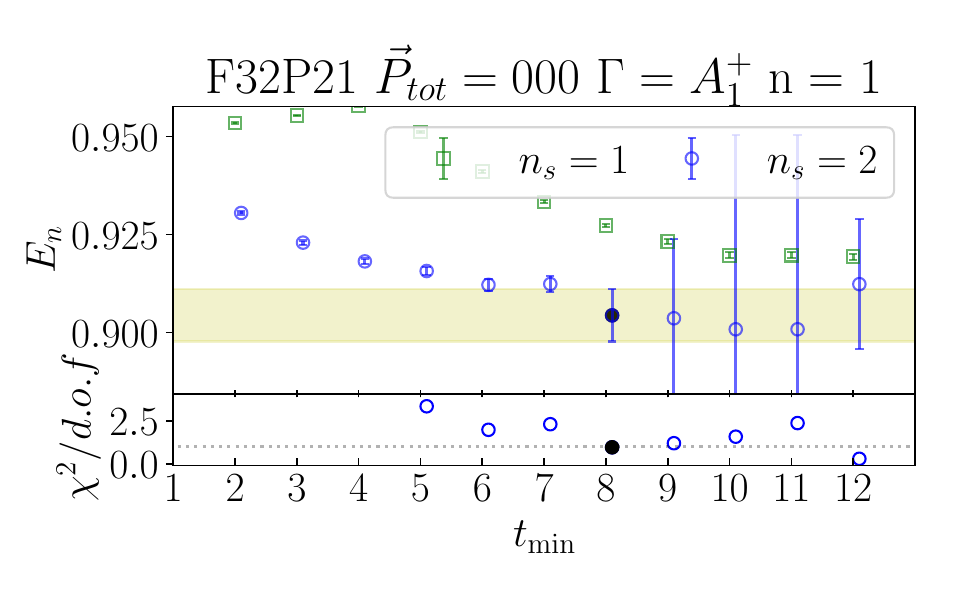}
\includegraphics[width=0.245\columnwidth]{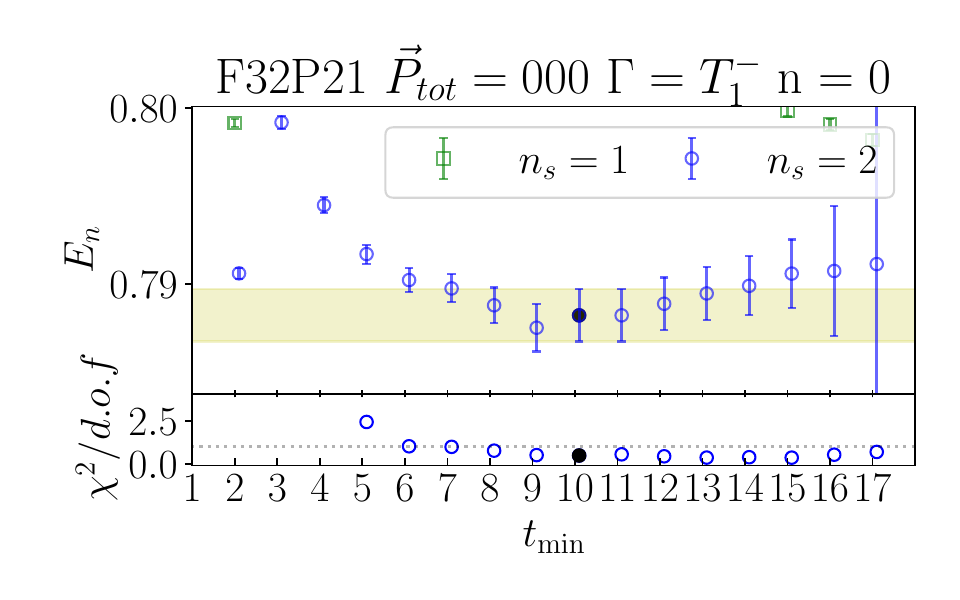}
\includegraphics[width=0.245\columnwidth]{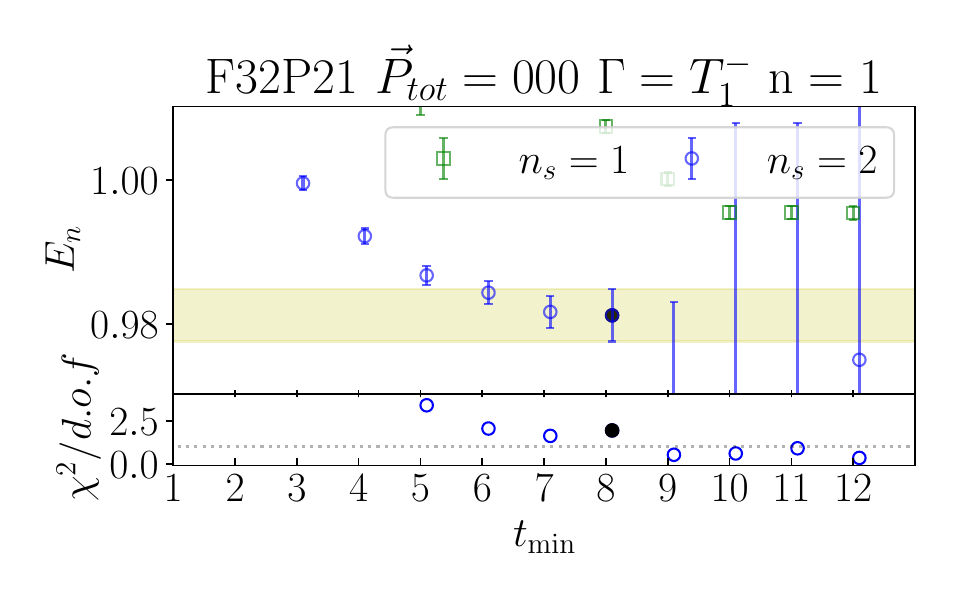}
\\
\includegraphics[width=0.245\columnwidth]{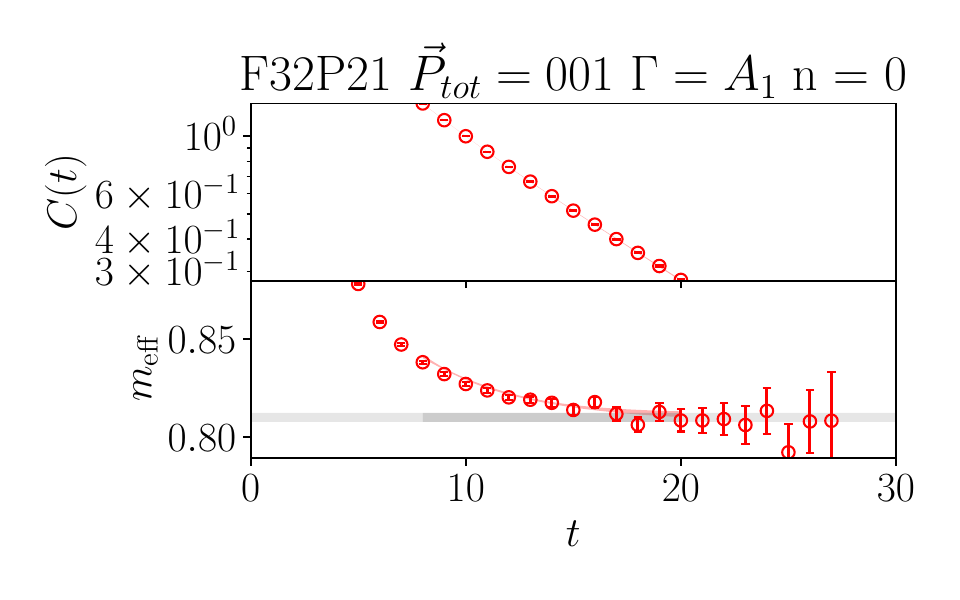}
\includegraphics[width=0.245\columnwidth]{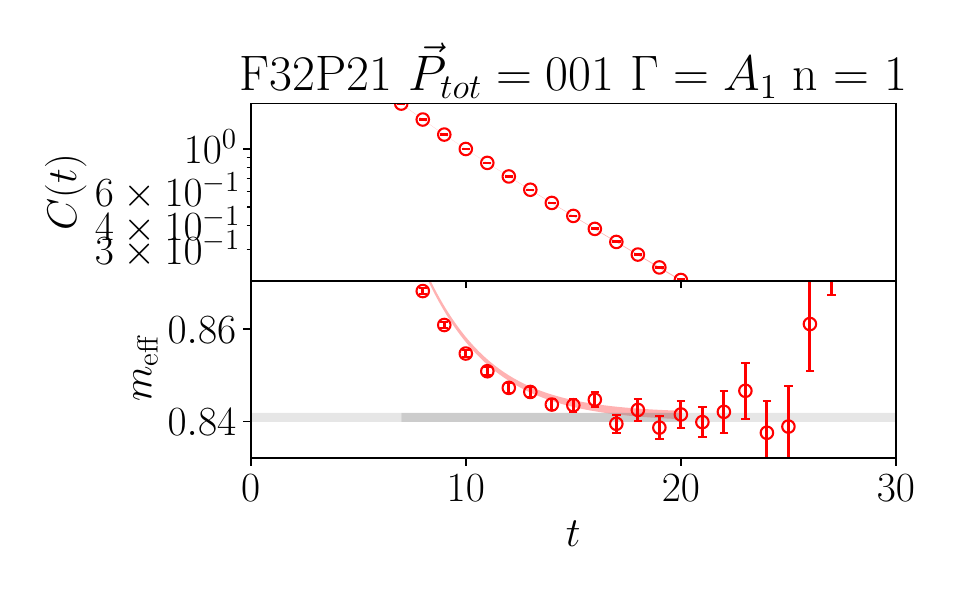}
\includegraphics[width=0.245\columnwidth]{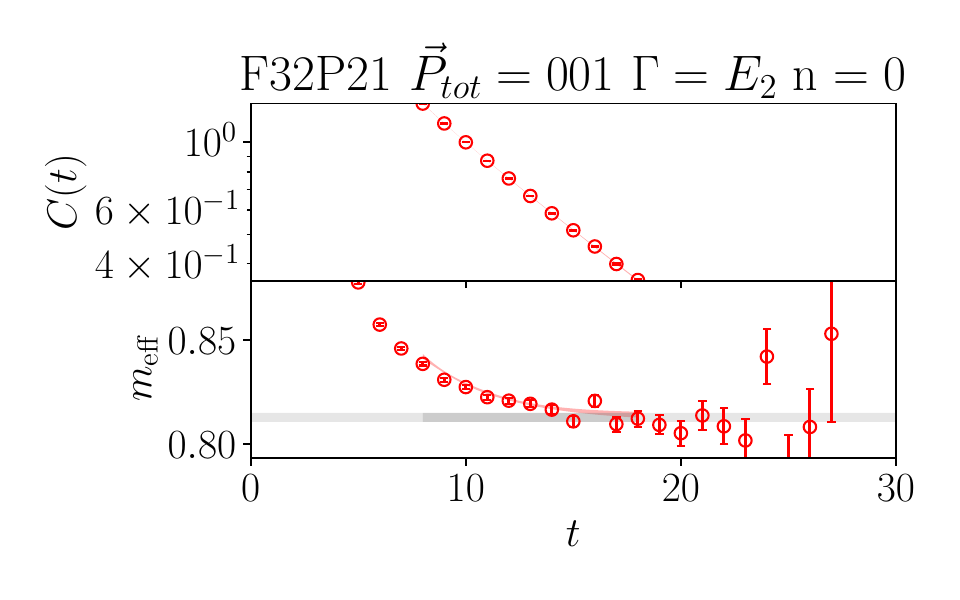}
\includegraphics[width=0.245\columnwidth]{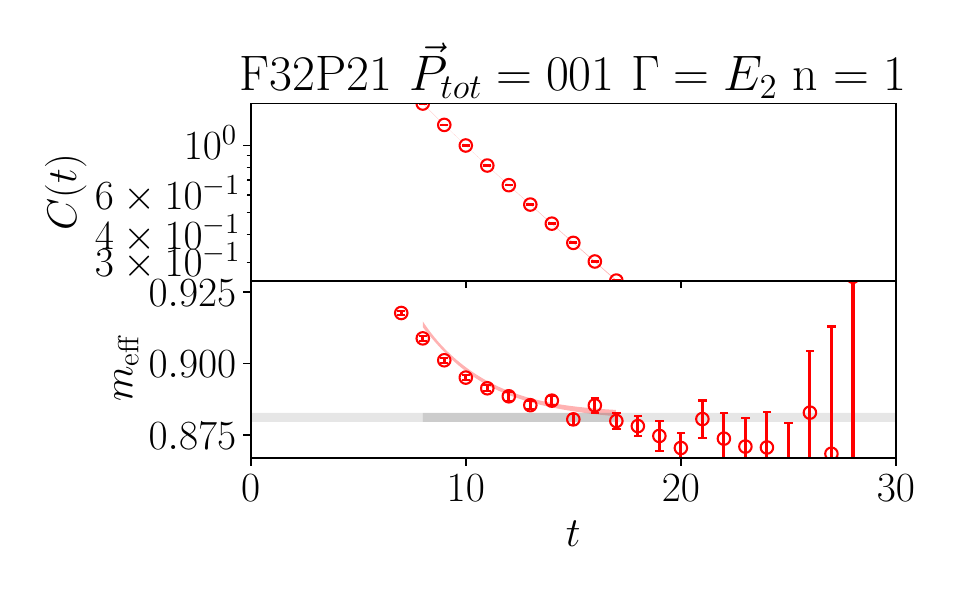}
\\
\includegraphics[width=0.245\columnwidth]{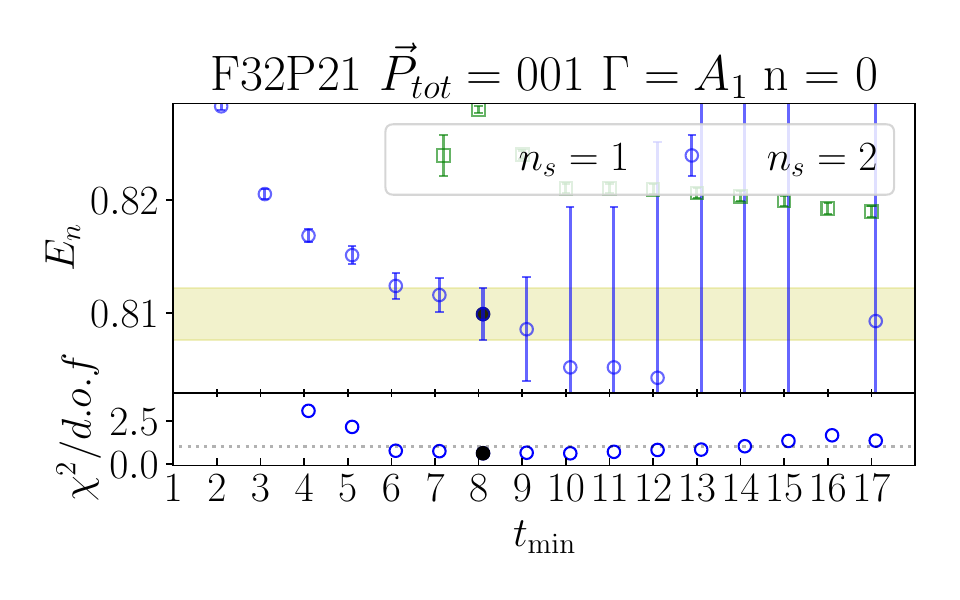}
\includegraphics[width=0.245\columnwidth]{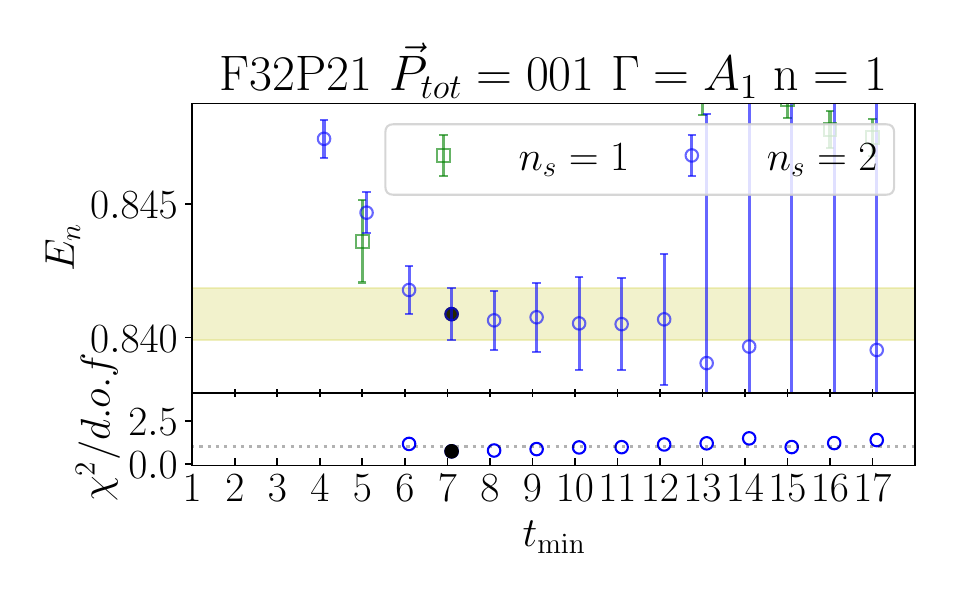}
\includegraphics[width=0.245\columnwidth]{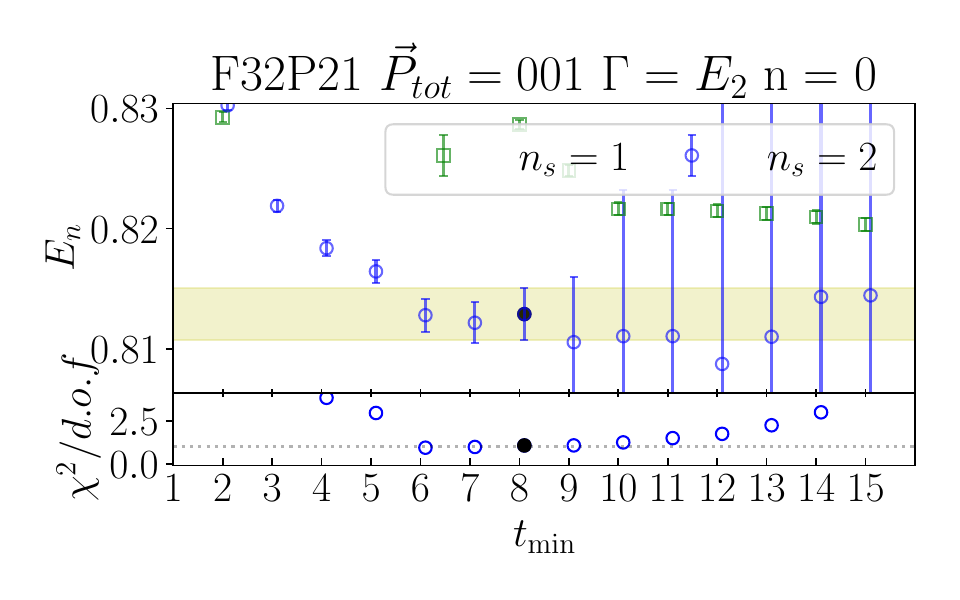}
\includegraphics[width=0.245\columnwidth]{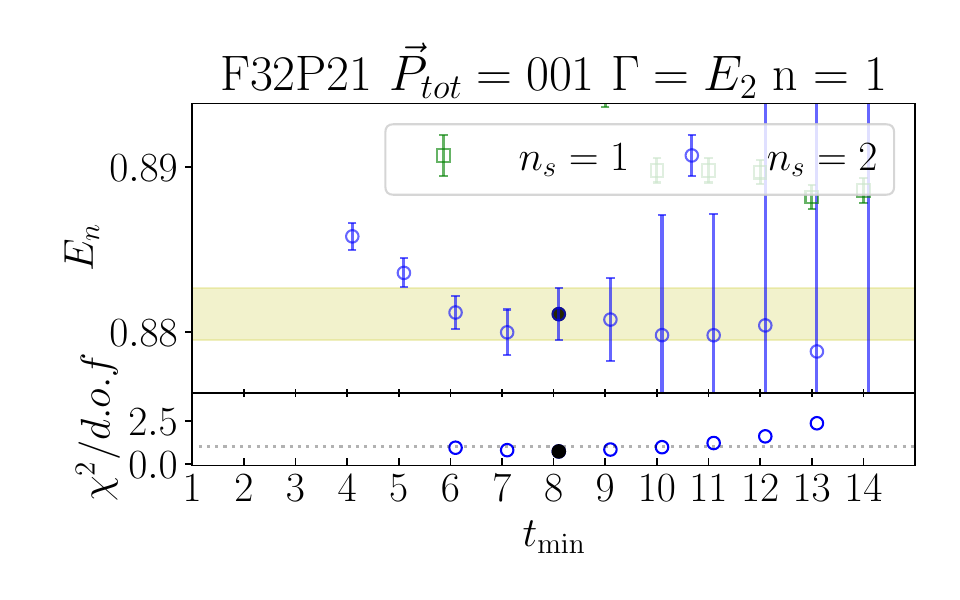}
\\
\includegraphics[width=0.245\columnwidth]{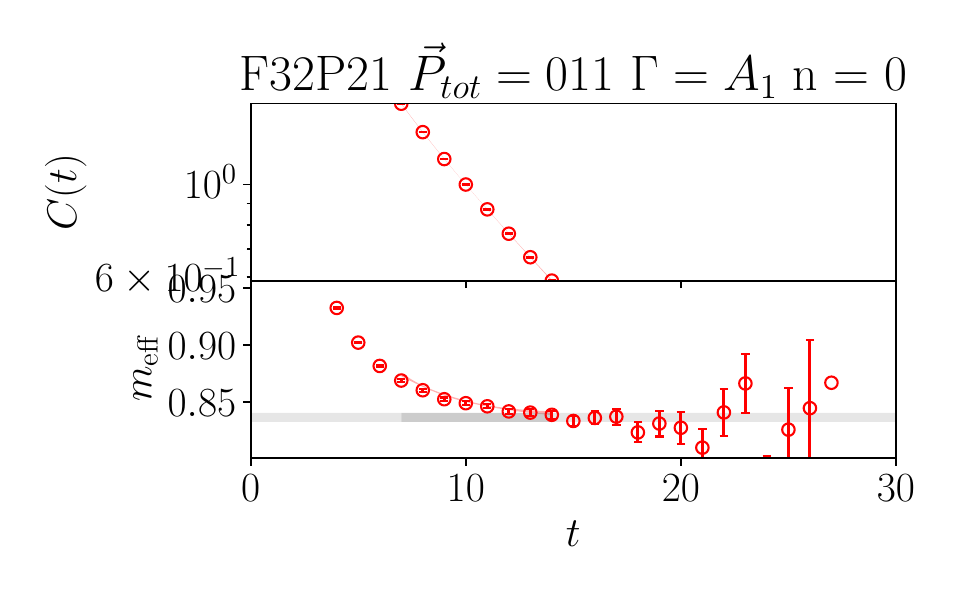}
\includegraphics[width=0.245\columnwidth]{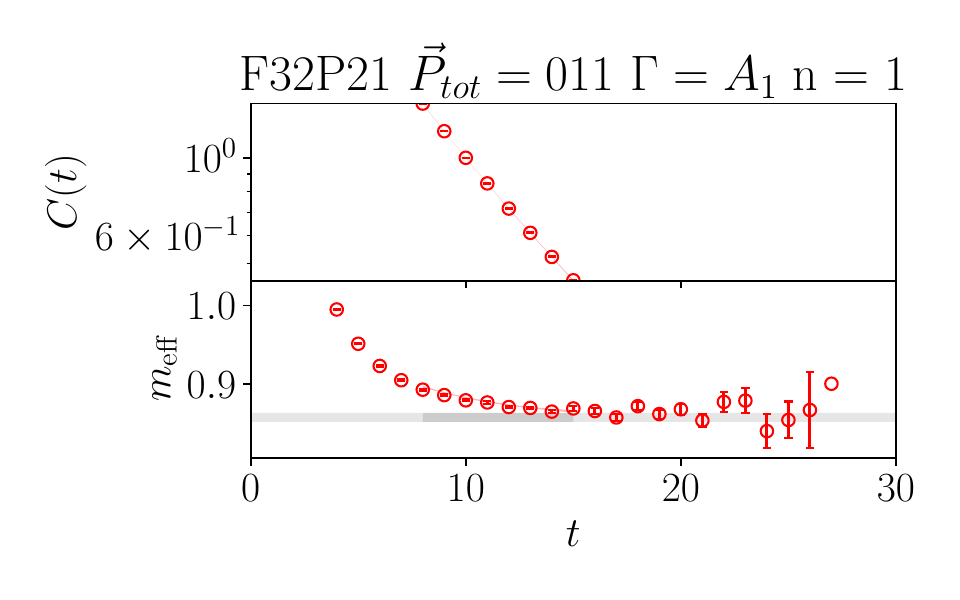}
\includegraphics[width=0.245\columnwidth]{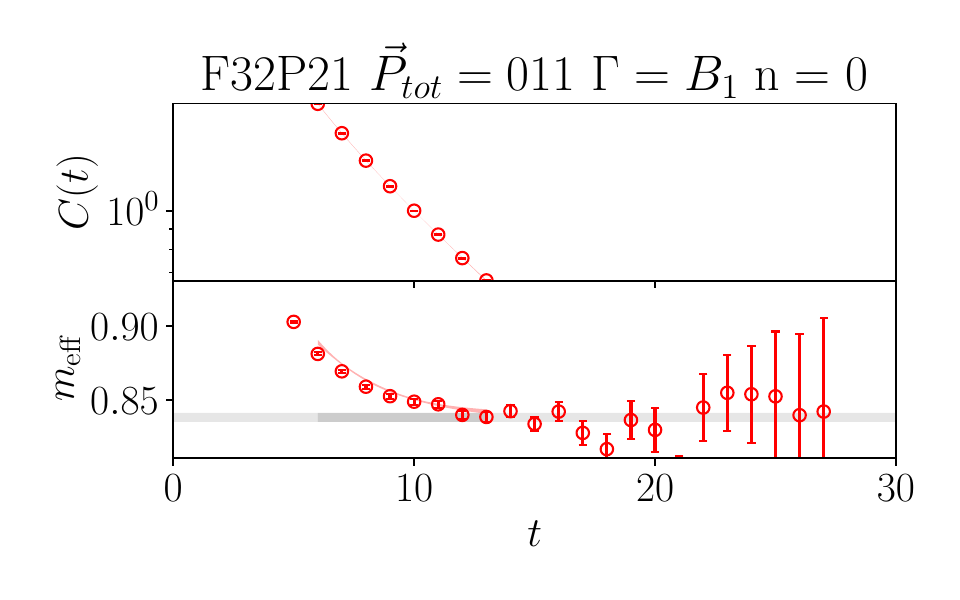}
\includegraphics[width=0.245\columnwidth]{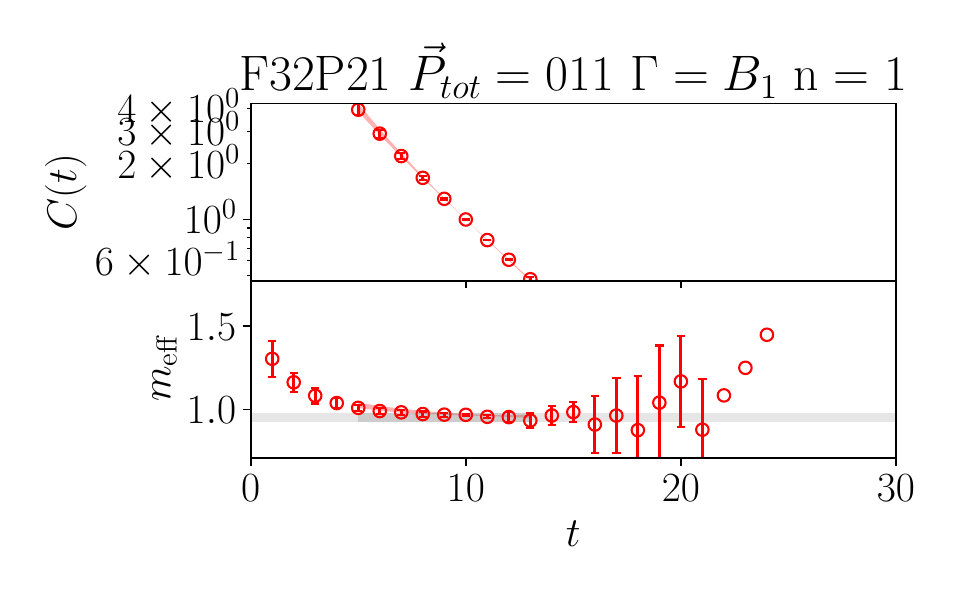}
\\
\includegraphics[width=0.245\columnwidth]{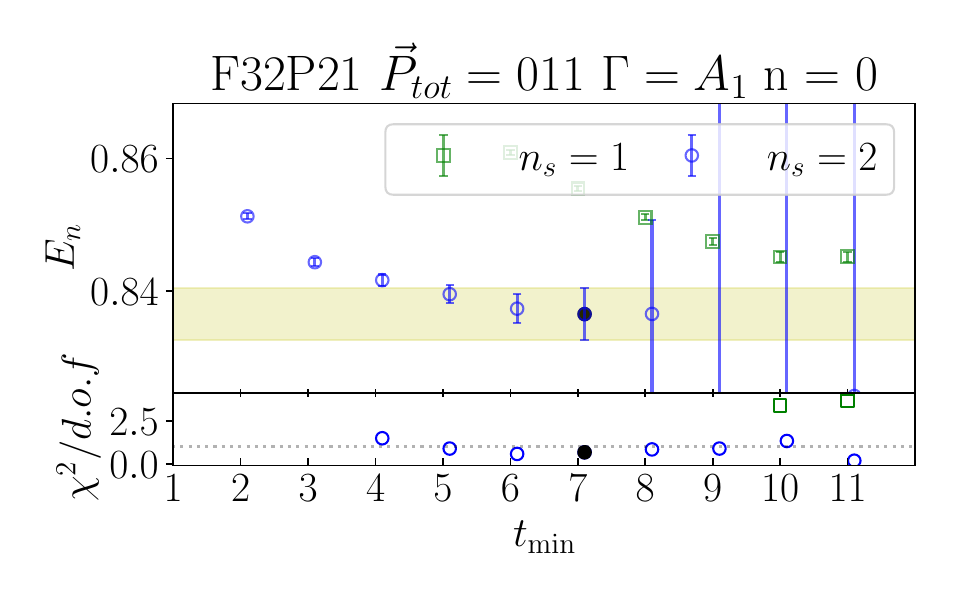}
\includegraphics[width=0.245\columnwidth]{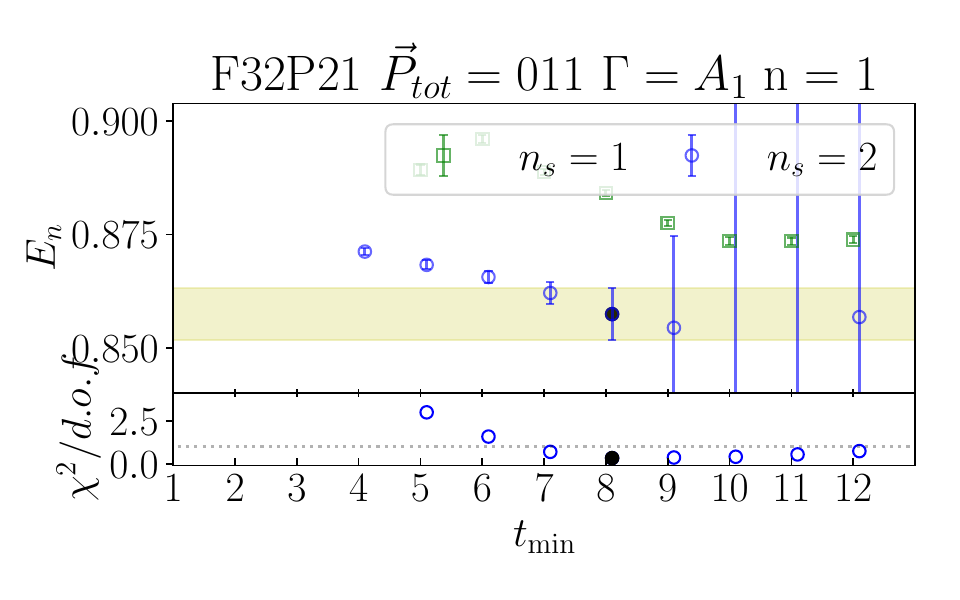}
\includegraphics[width=0.245\columnwidth]{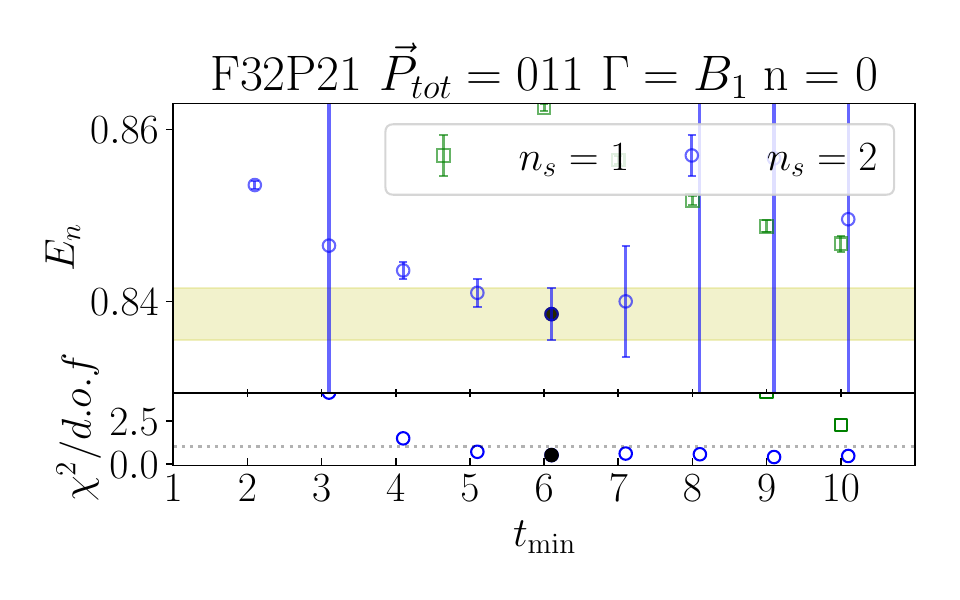}
\includegraphics[width=0.245\columnwidth]{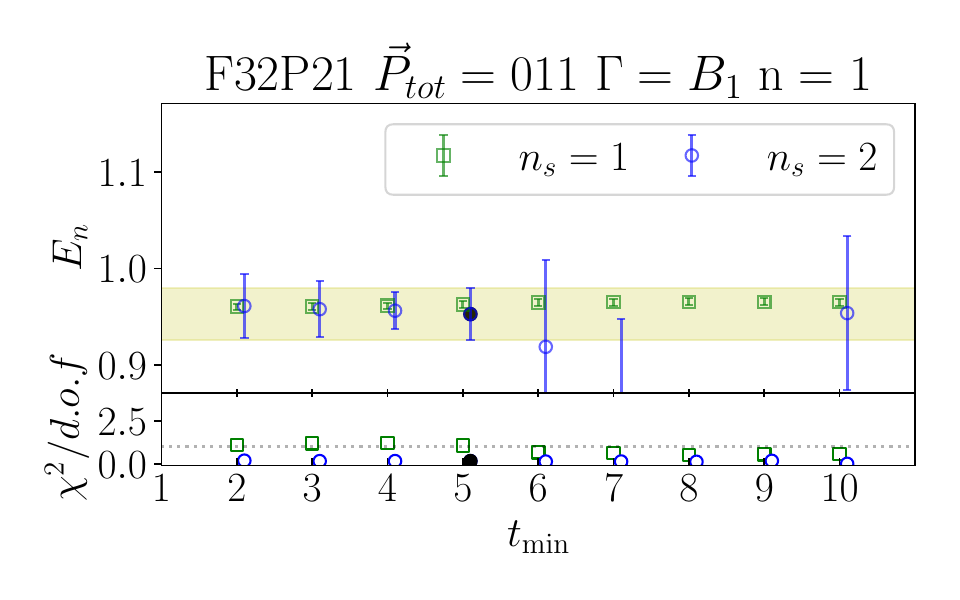}
\\
\includegraphics[width=0.245\columnwidth]{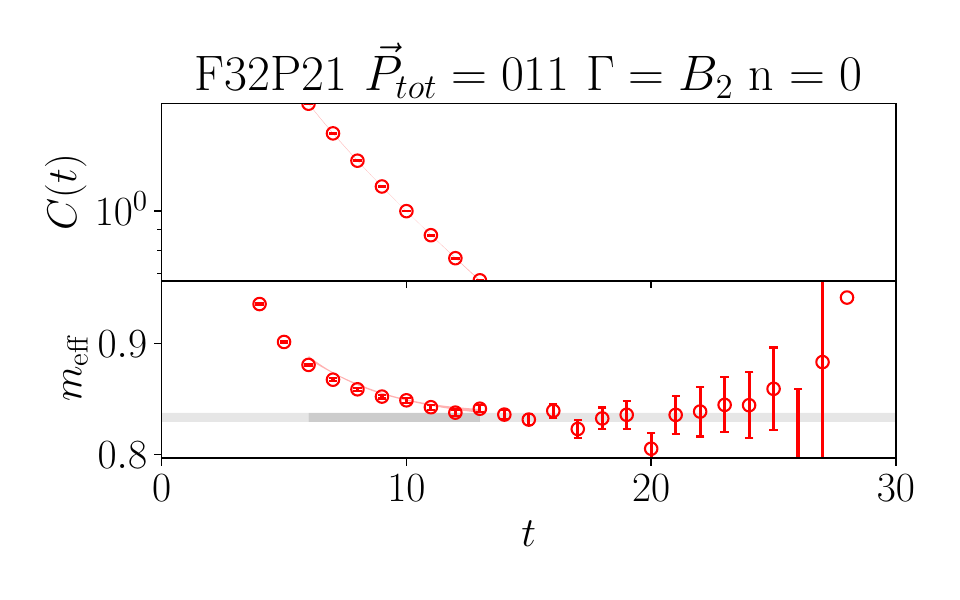}
\includegraphics[width=0.245\columnwidth]{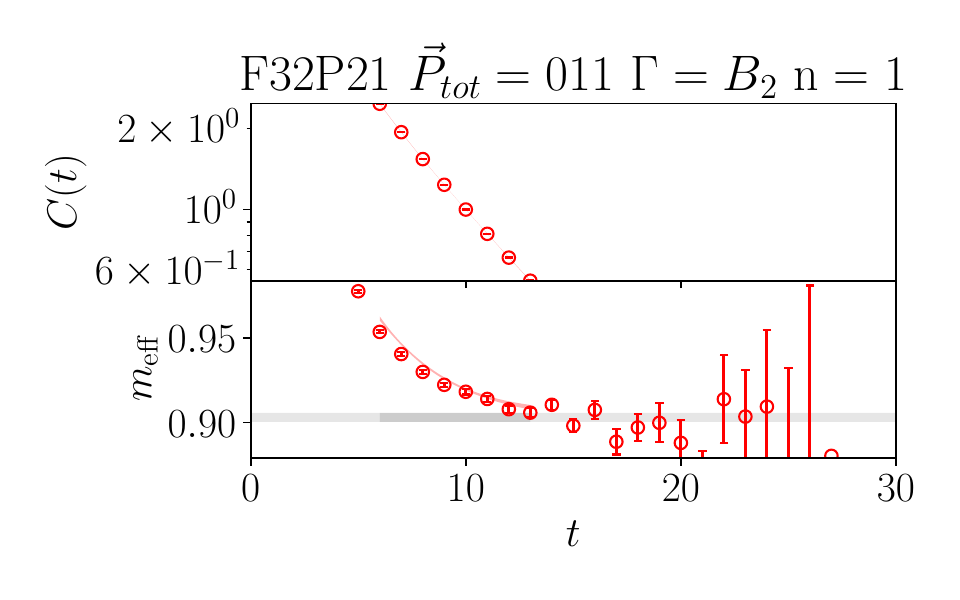}
\includegraphics[width=0.245\columnwidth]{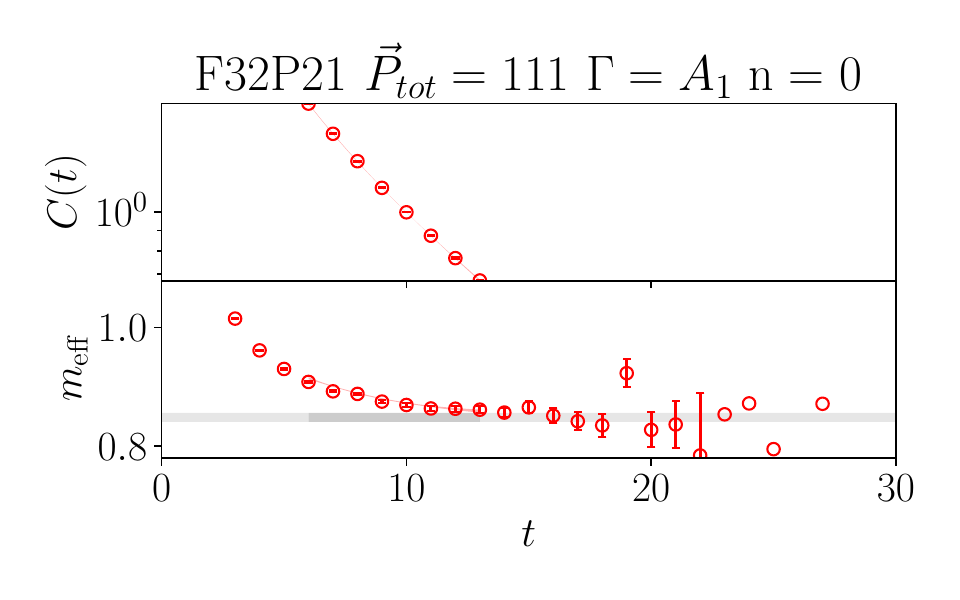}
\includegraphics[width=0.245\columnwidth]{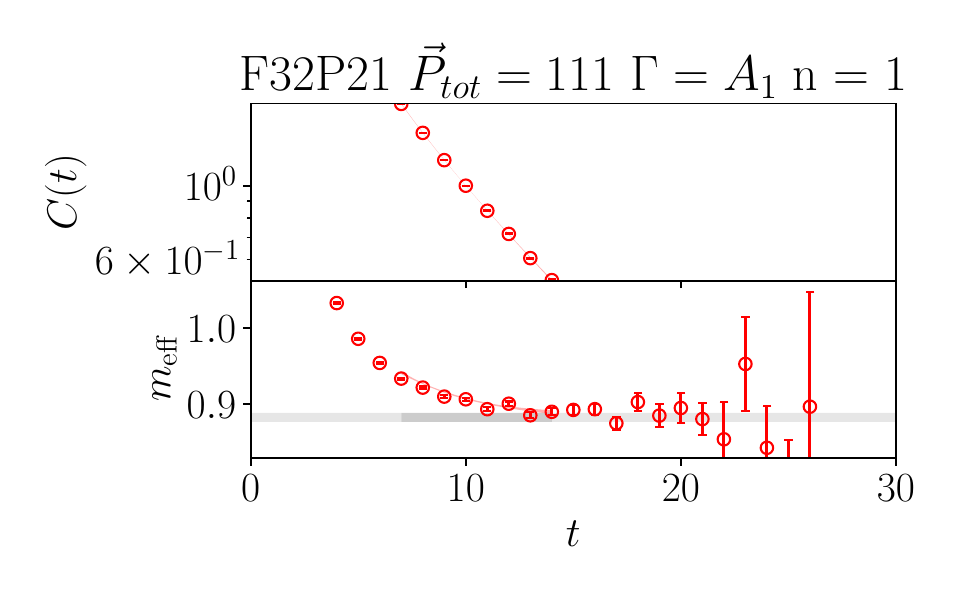}
\\
\includegraphics[width=0.245\columnwidth]{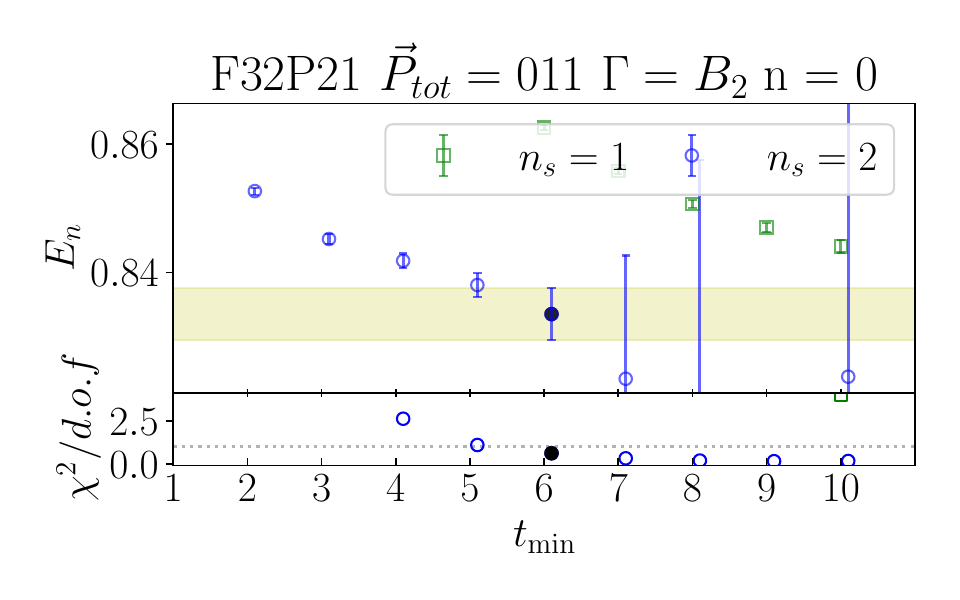}
\includegraphics[width=0.245\columnwidth]{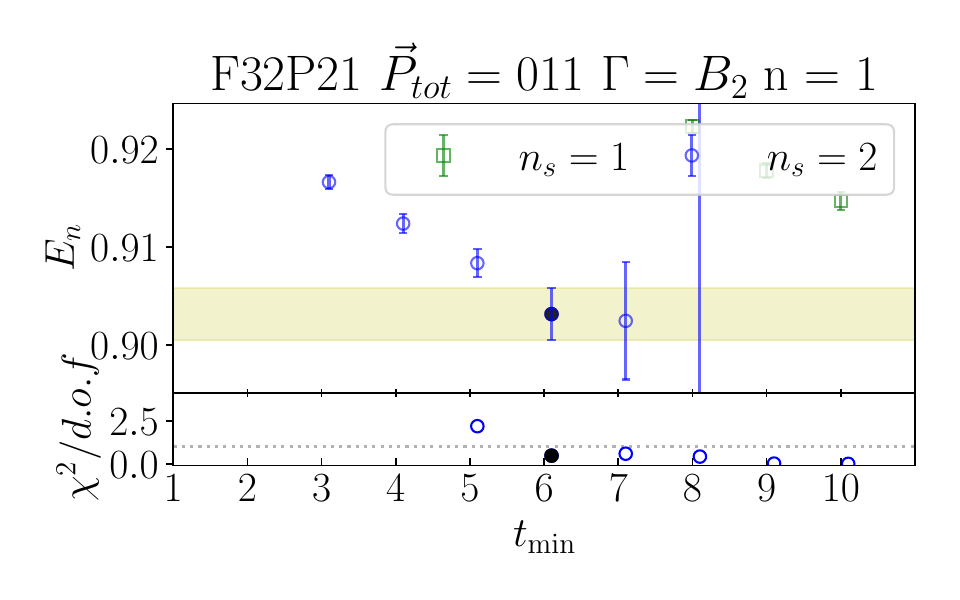}
\includegraphics[width=0.245\columnwidth]{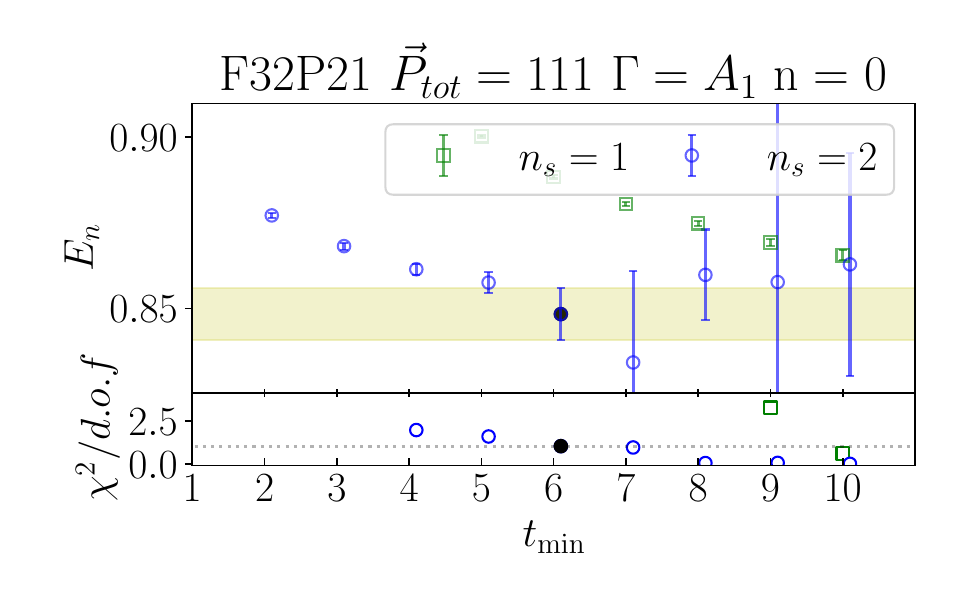}
\includegraphics[width=0.245\columnwidth]{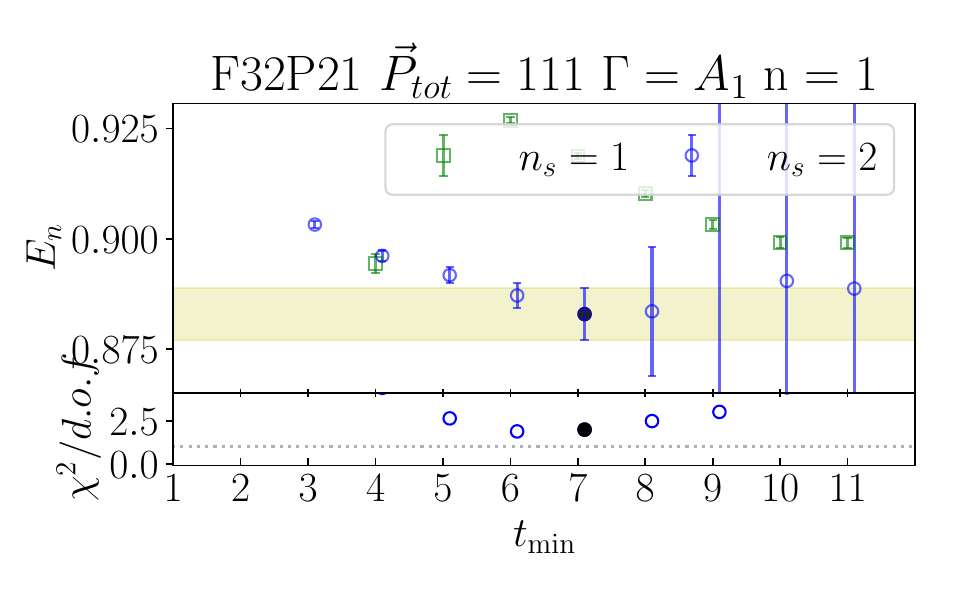}
\caption{Energy-level fit results for the $I=\frac{1}{2}$ $D\pi$ channel on the F32P21 ensemble. The description follows Figure~\ref{fig:Dpi-fit-F32P30}.}
\label{fig:Dpi-fit-F32P21}
\end{figure}

\begin{figure}[htbp]
\centering
\includegraphics[width=0.245\columnwidth]{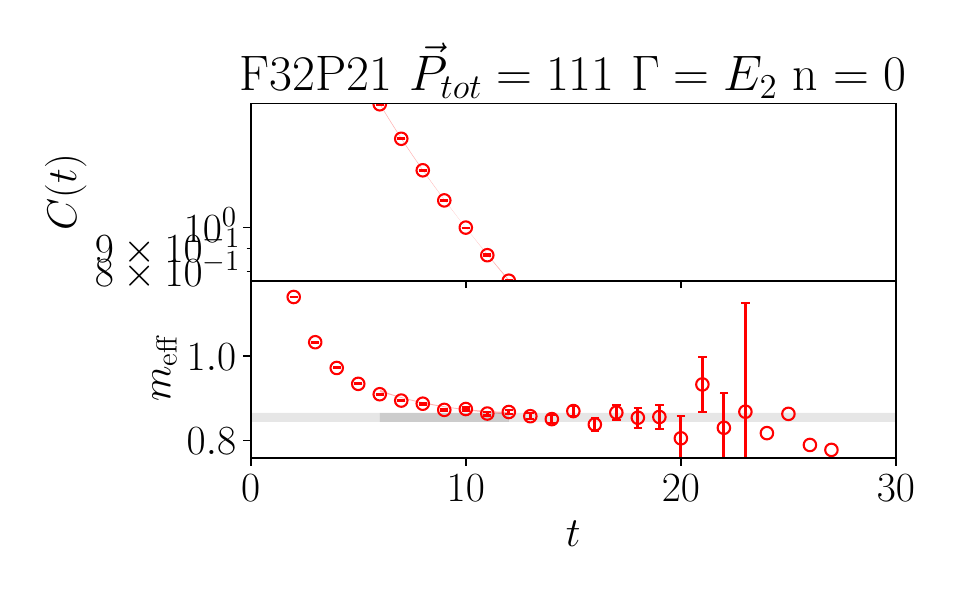}
\includegraphics[width=0.245\columnwidth]{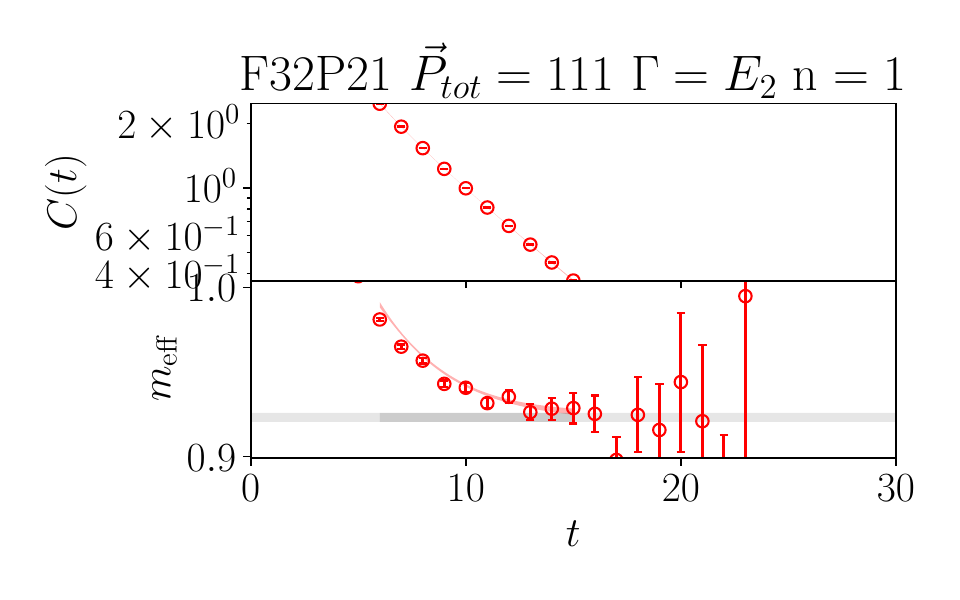}
\includegraphics[width=0.245\columnwidth]{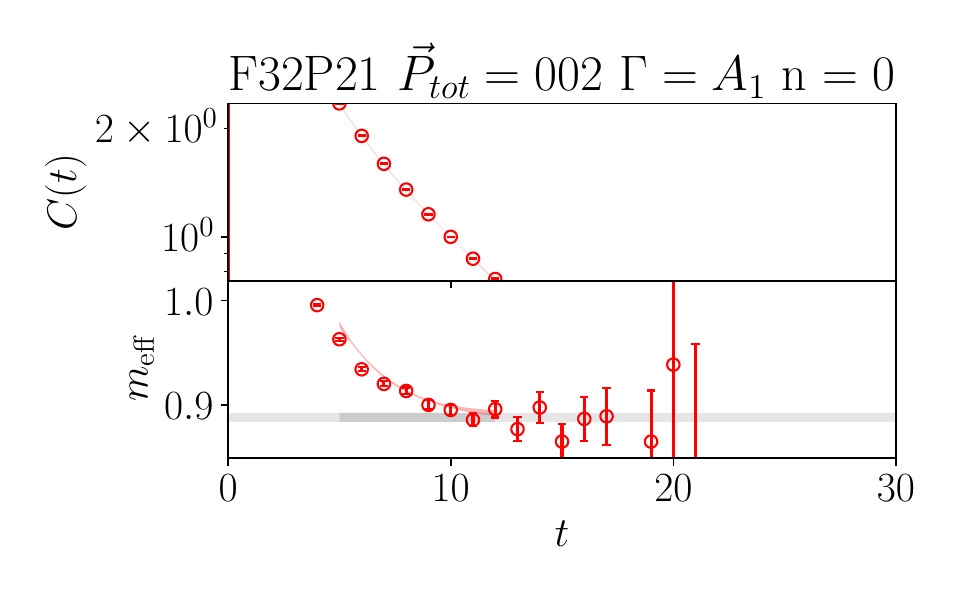}
\includegraphics[width=0.245\columnwidth]{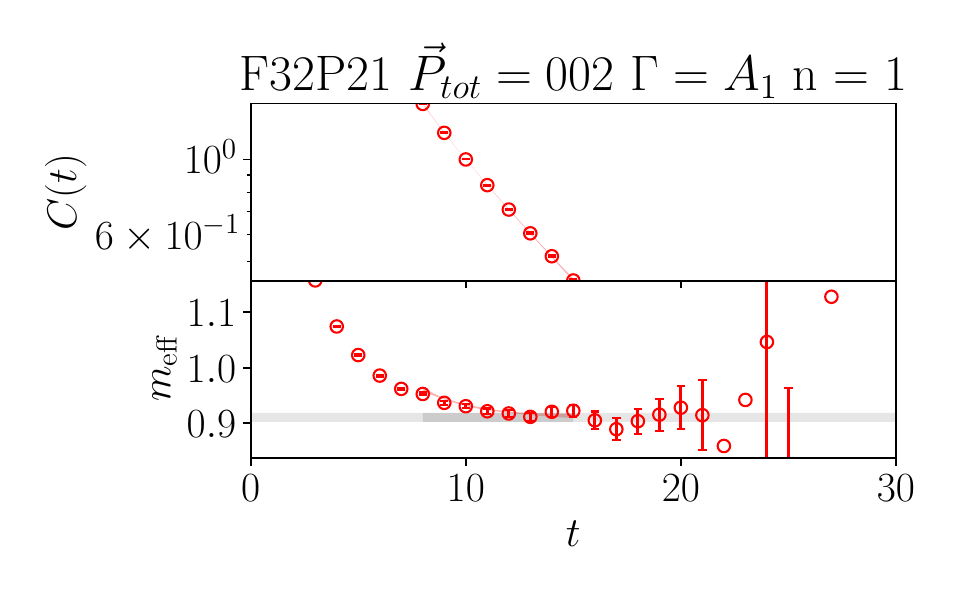}
\\
\includegraphics[width=0.245\columnwidth]{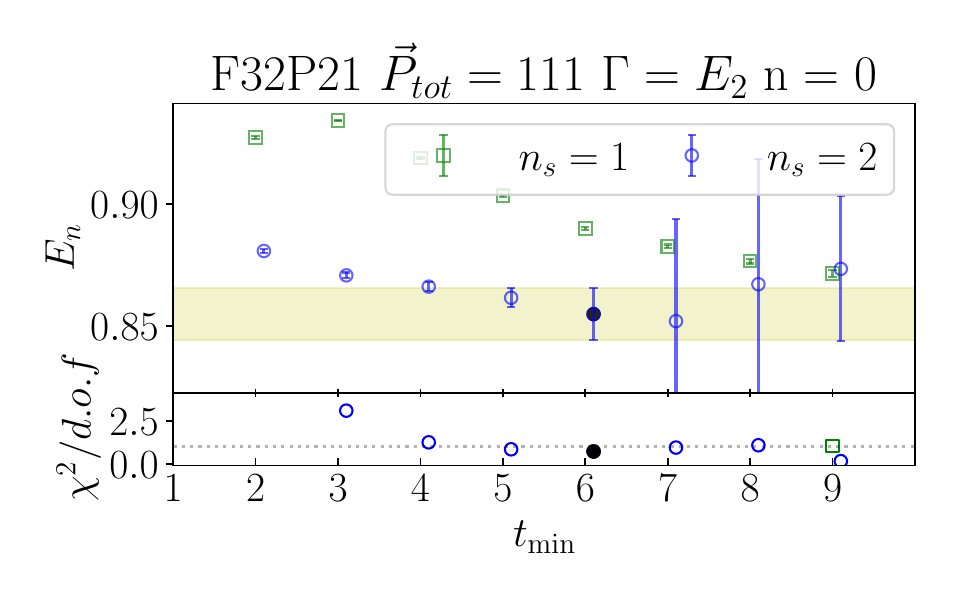}
\includegraphics[width=0.245\columnwidth]{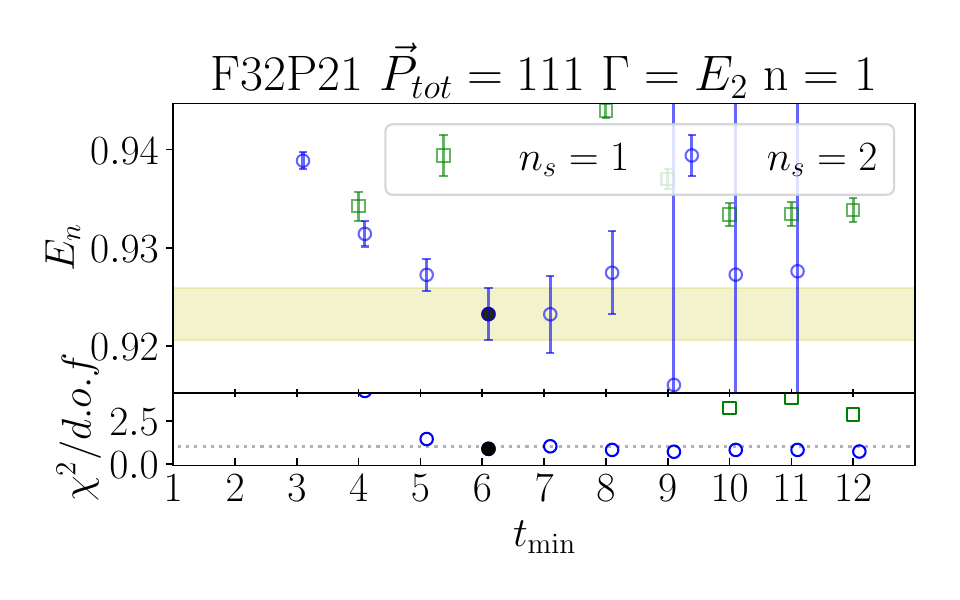}
\includegraphics[width=0.245\columnwidth]{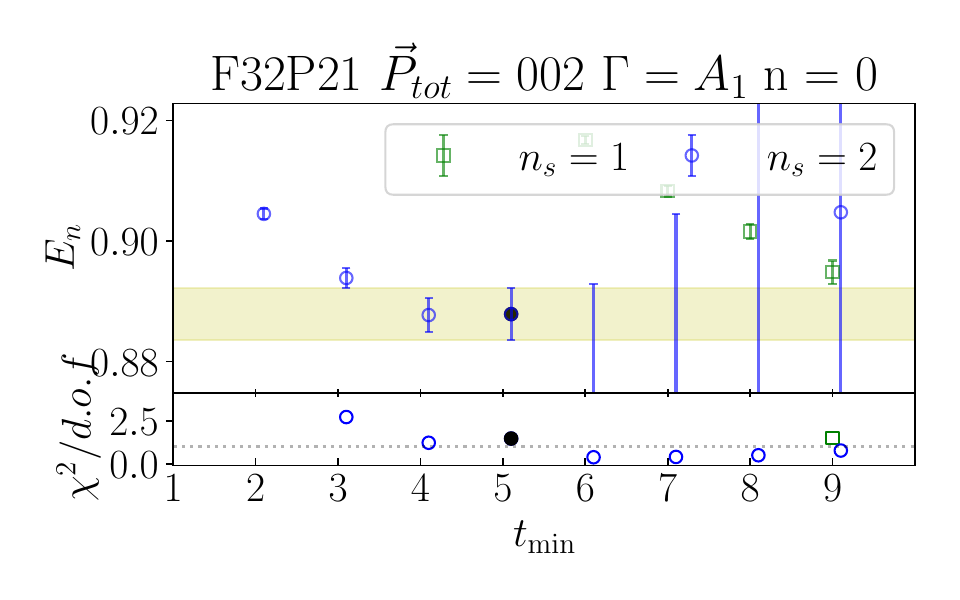}
\includegraphics[width=0.245\columnwidth]{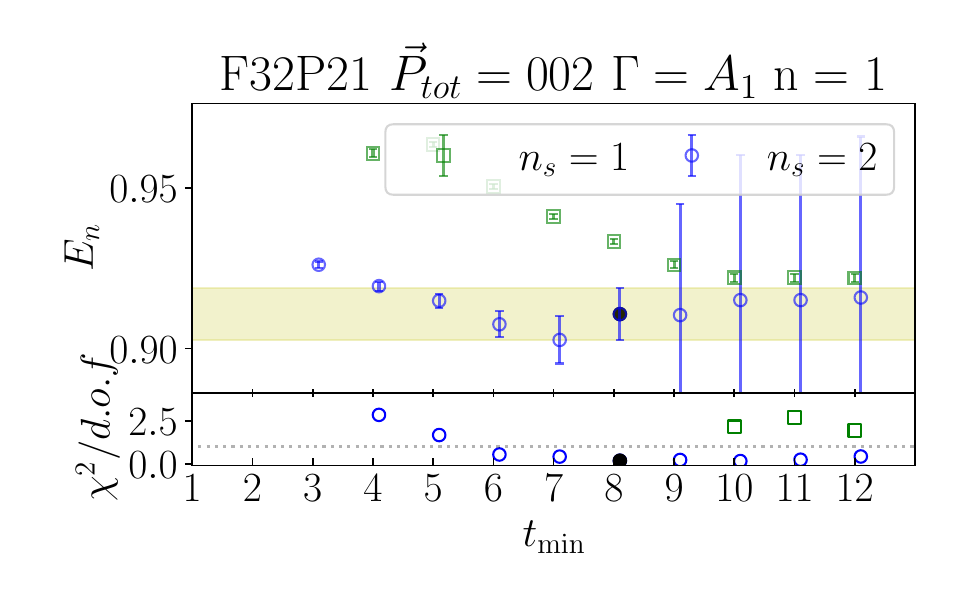}
\\
\includegraphics[width=0.245\columnwidth]{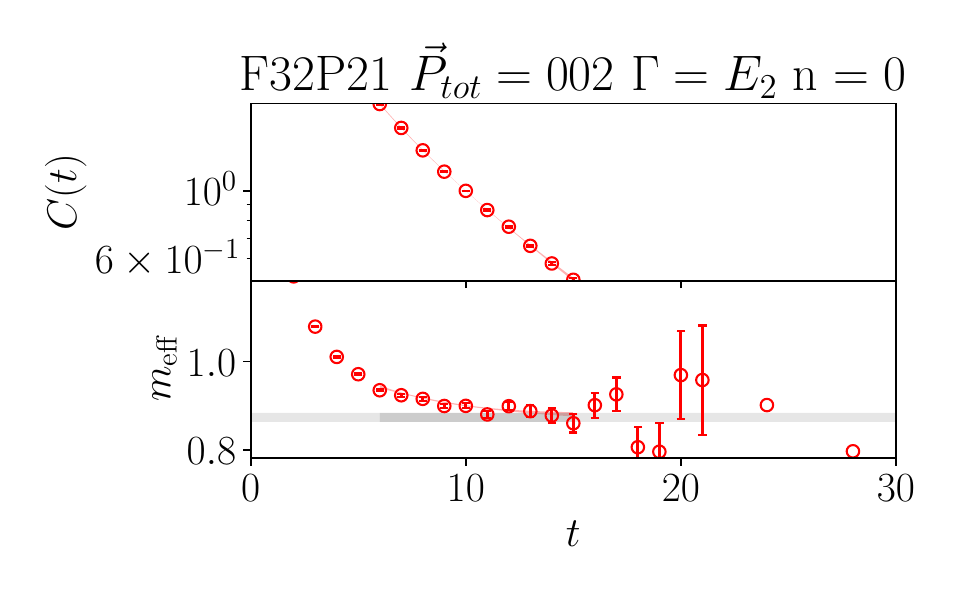}
\includegraphics[width=0.245\columnwidth]{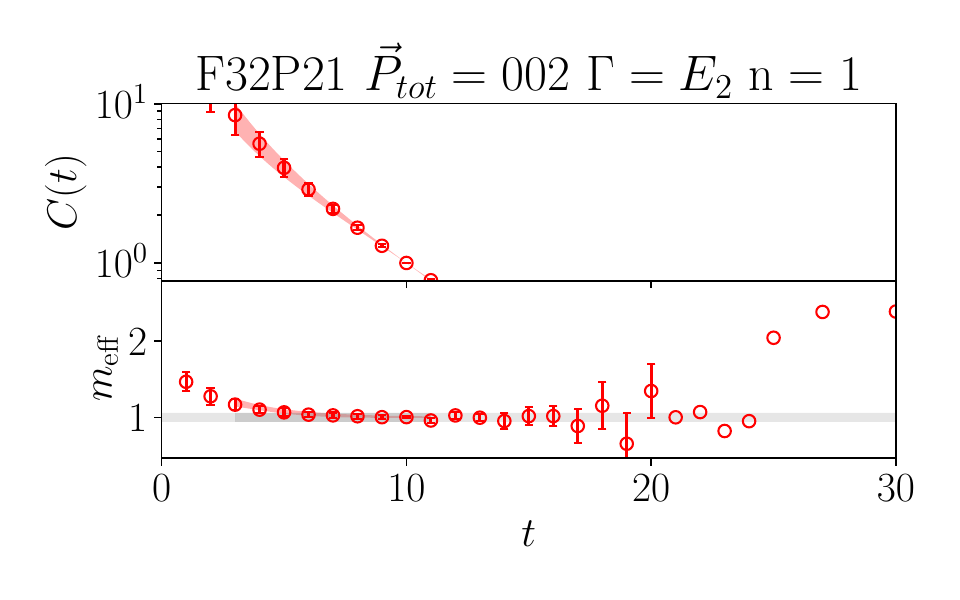}
\\
\includegraphics[width=0.245\columnwidth]{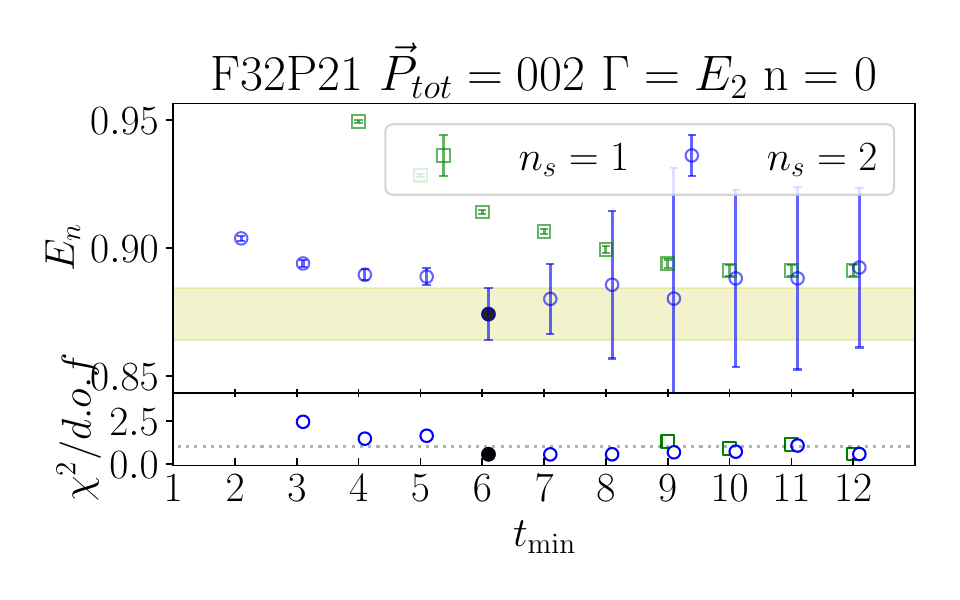}
\includegraphics[width=0.245\columnwidth]{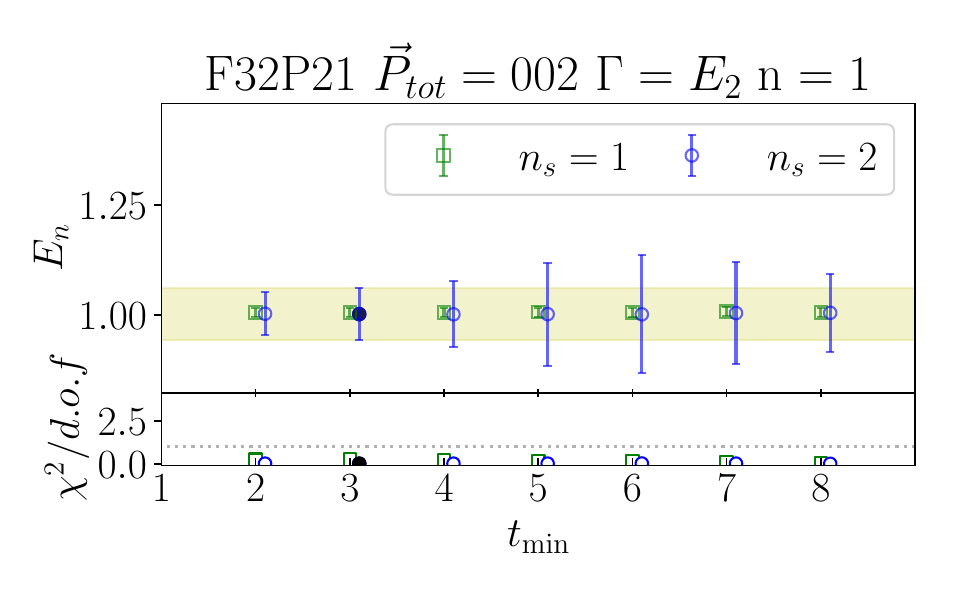}
\caption{Continued from Figure~\ref{fig:Dpi-fit-F32P21}. Energy-level fit results for the $I=\frac{1}{2}$ $D\pi$ channel on the F32P21 ensemble.}
\label{fig:Dpi-fit-F32P212}
\end{figure}

\begin{figure}[htbp]
\centering
\includegraphics[width=0.245\columnwidth]{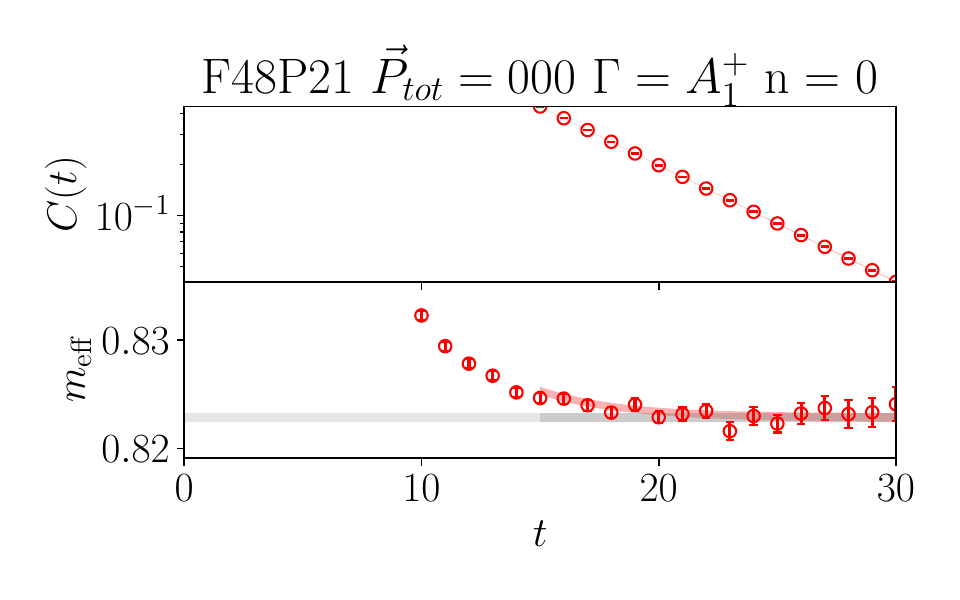}
\includegraphics[width=0.245\columnwidth]{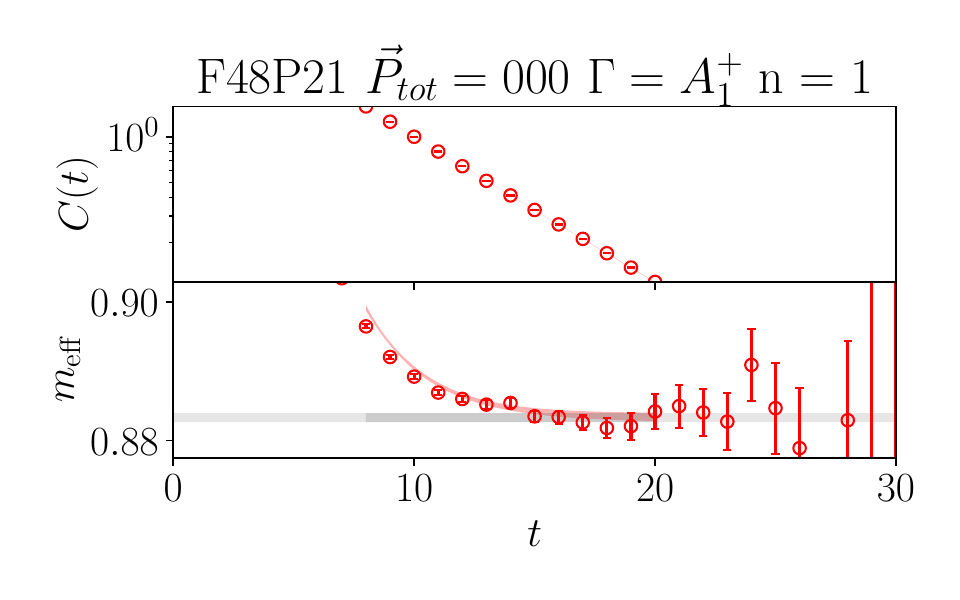}
\includegraphics[width=0.245\columnwidth]{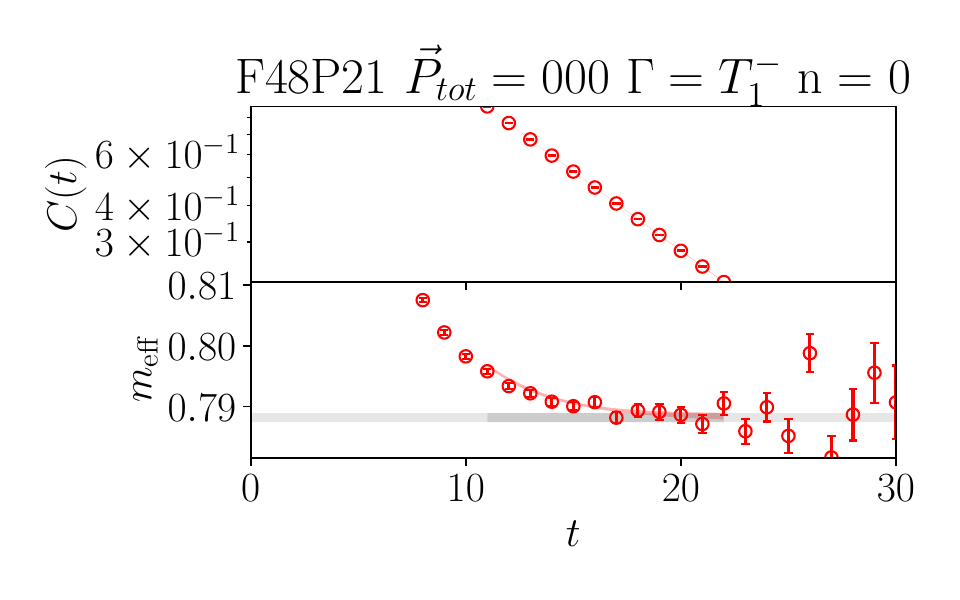}
\includegraphics[width=0.245\columnwidth]{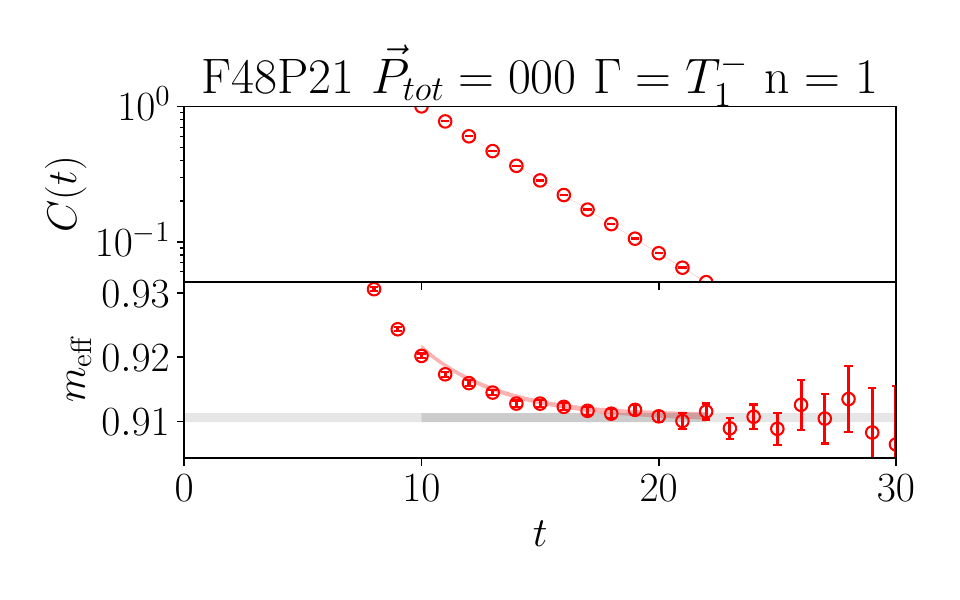}
\\
\includegraphics[width=0.245\columnwidth]{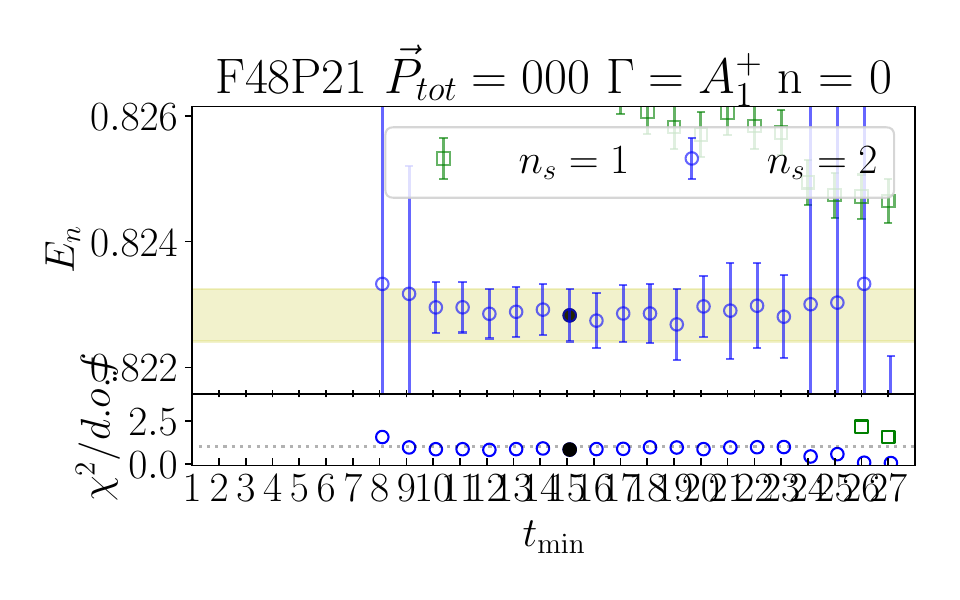}
\includegraphics[width=0.245\columnwidth]{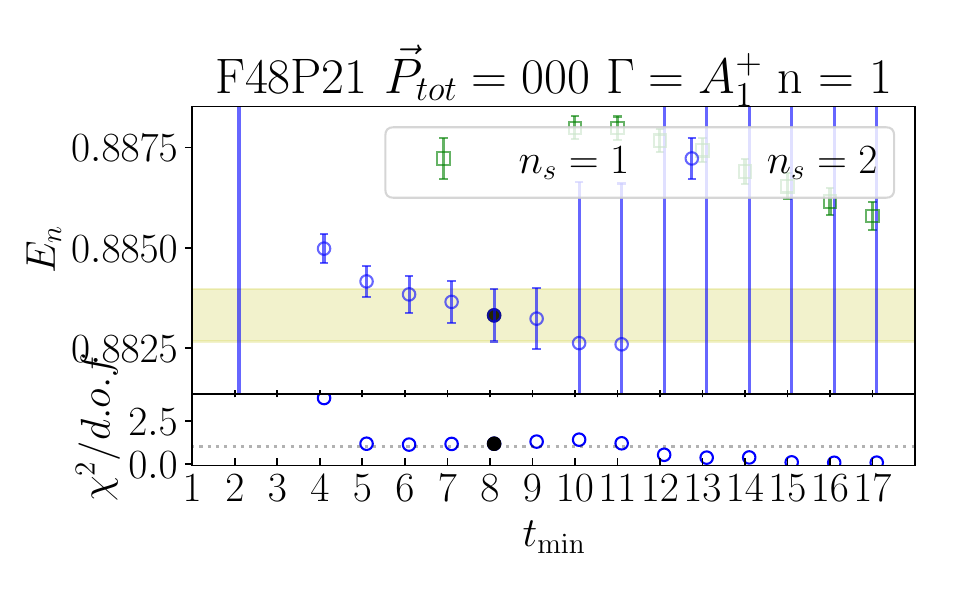}
\includegraphics[width=0.245\columnwidth]{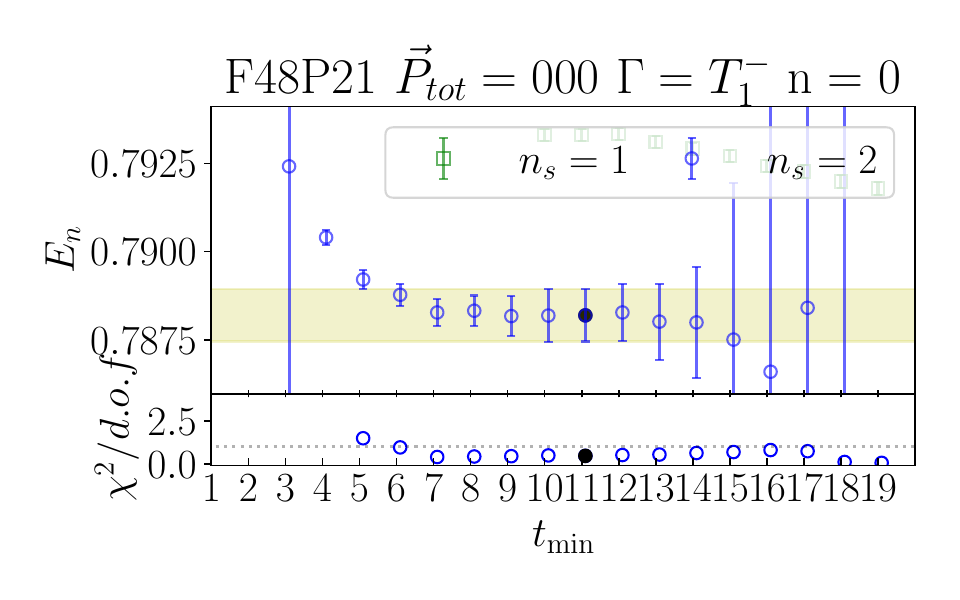}
\includegraphics[width=0.245\columnwidth]{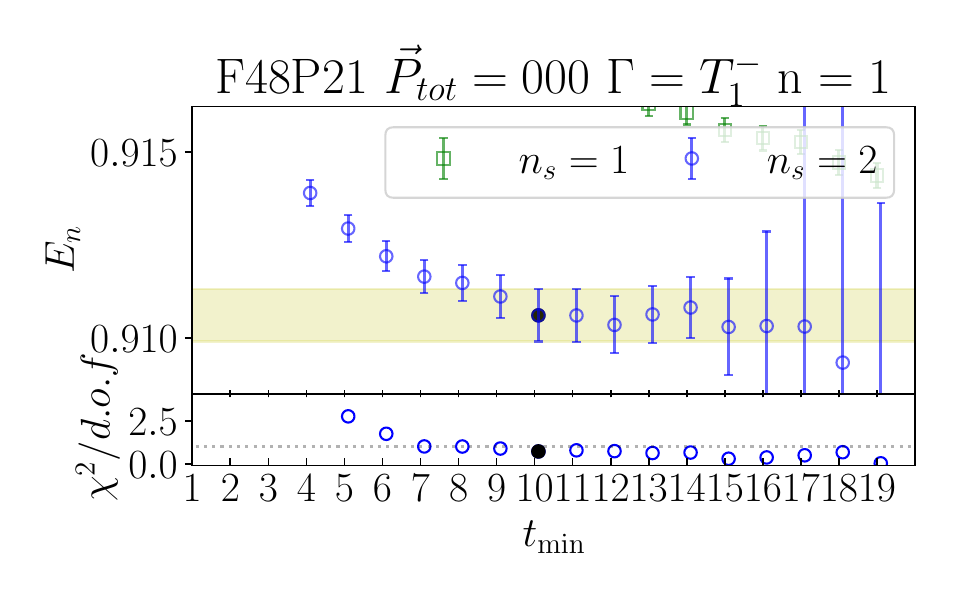}
\\
\includegraphics[width=0.245\columnwidth]{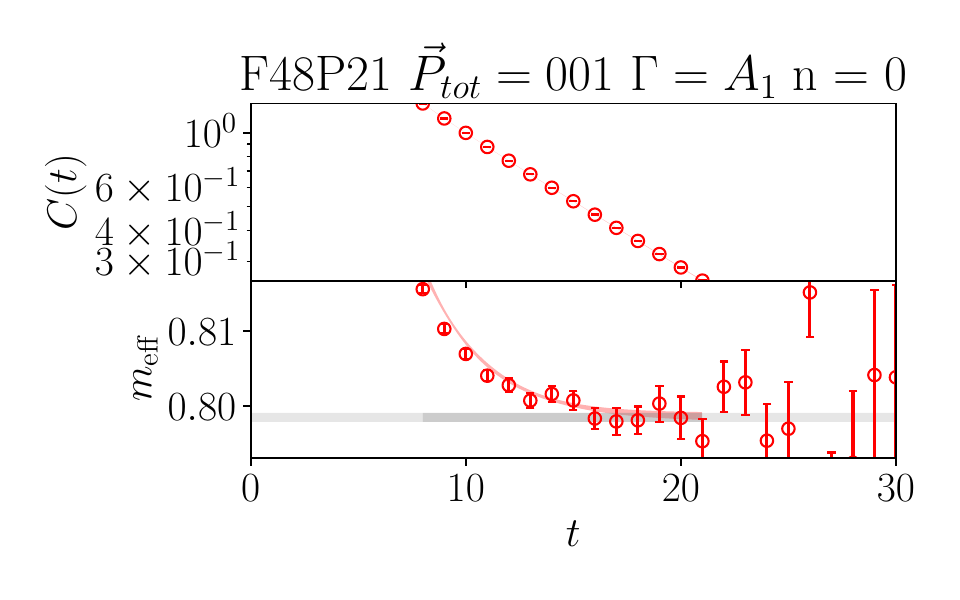}
\includegraphics[width=0.245\columnwidth]{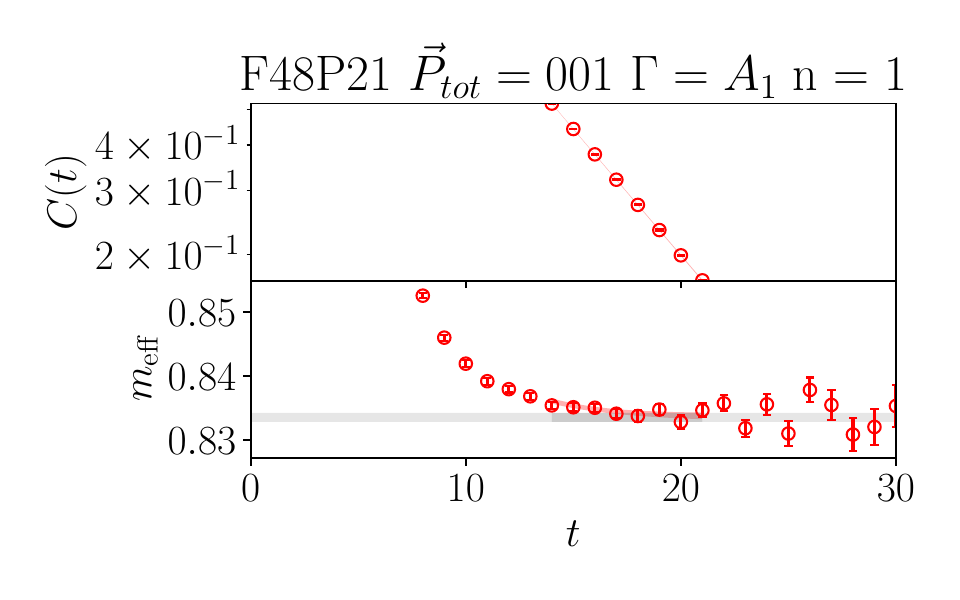}
\includegraphics[width=0.245\columnwidth]{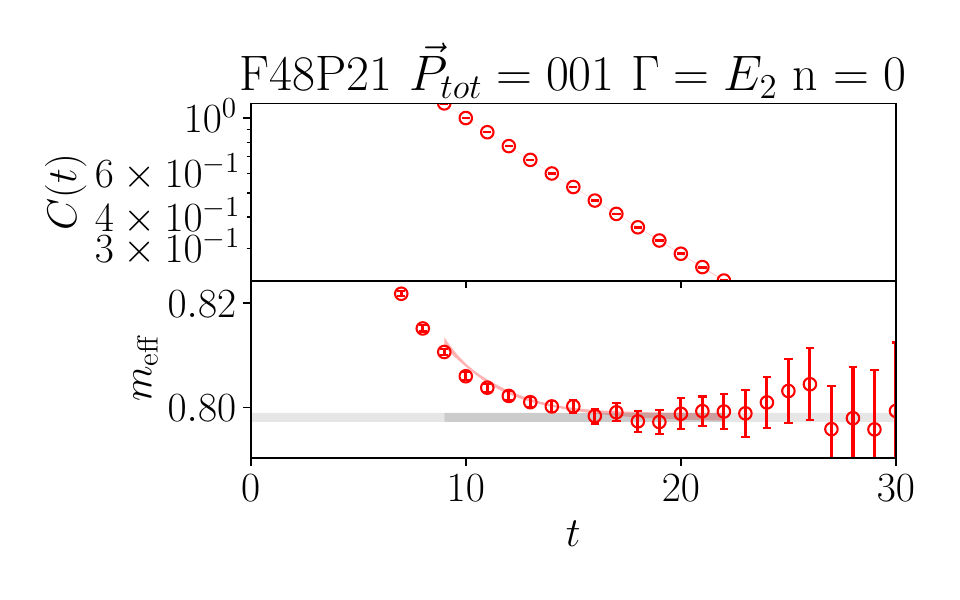}
\includegraphics[width=0.245\columnwidth]{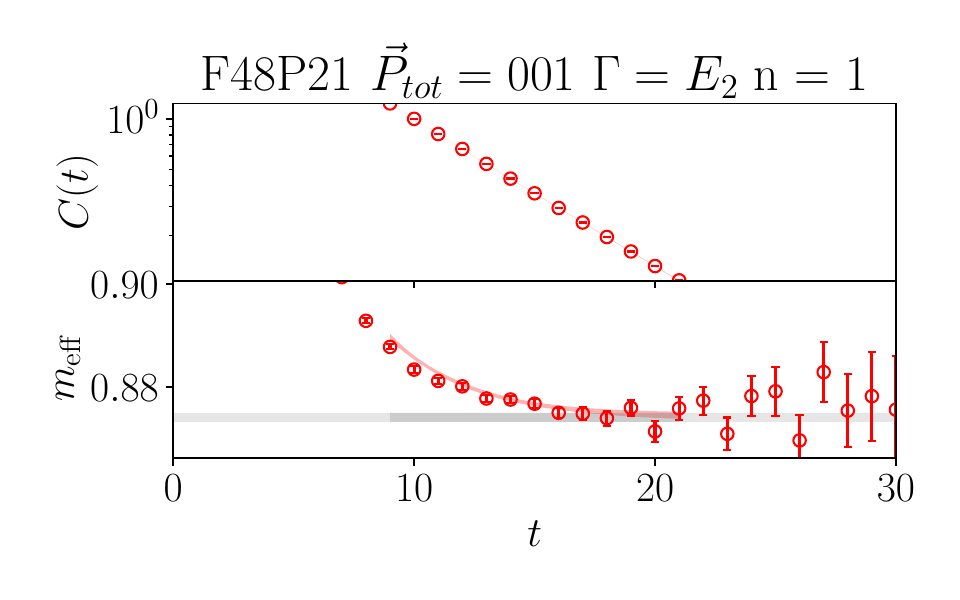}
\\
\includegraphics[width=0.245\columnwidth]{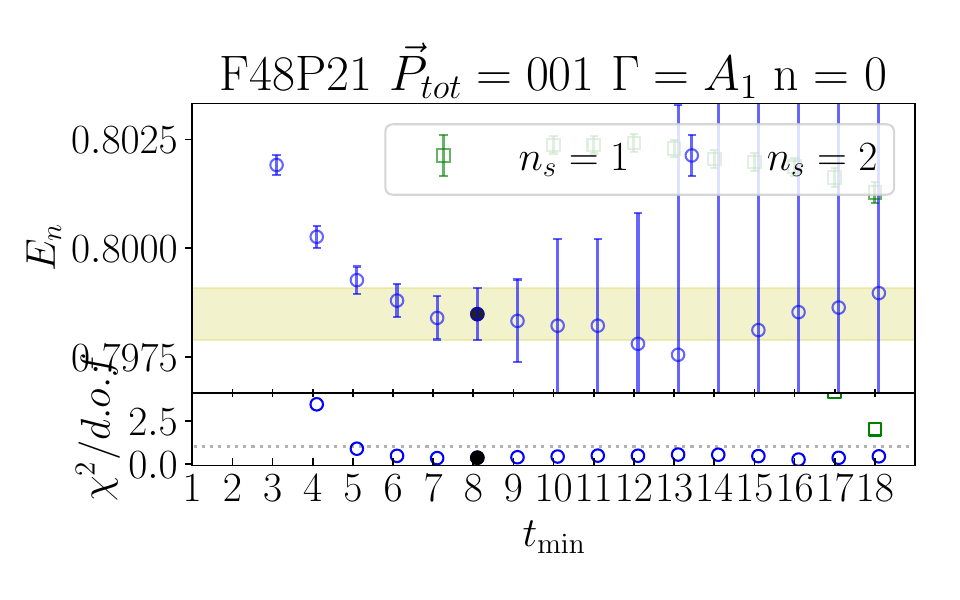}
\includegraphics[width=0.245\columnwidth]{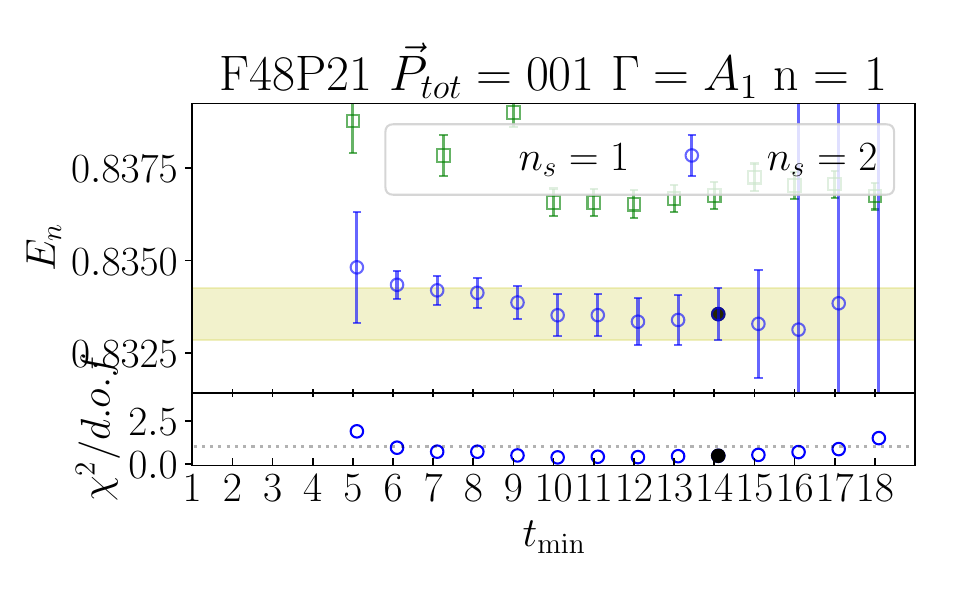}
\includegraphics[width=0.245\columnwidth]{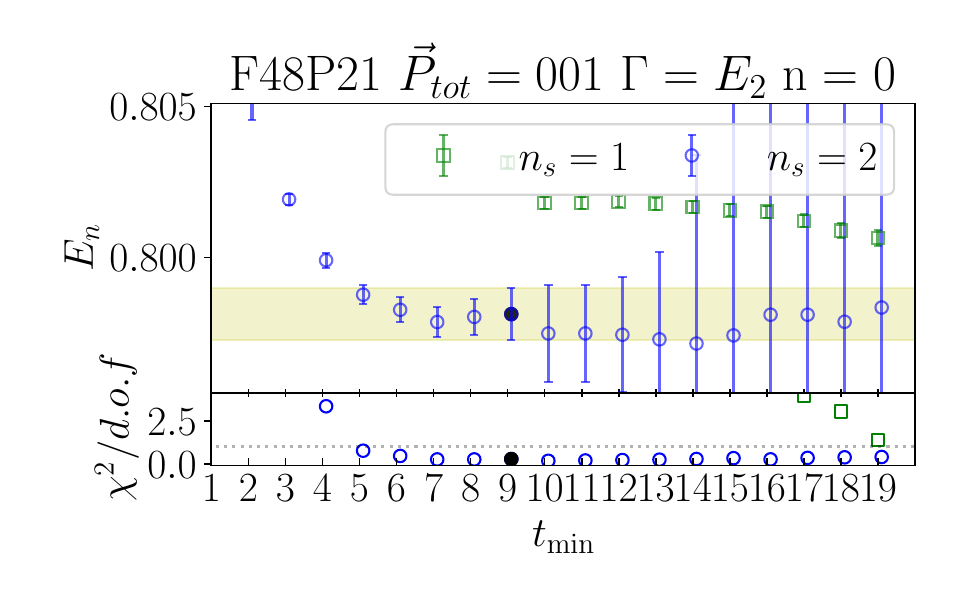}
\includegraphics[width=0.245\columnwidth]{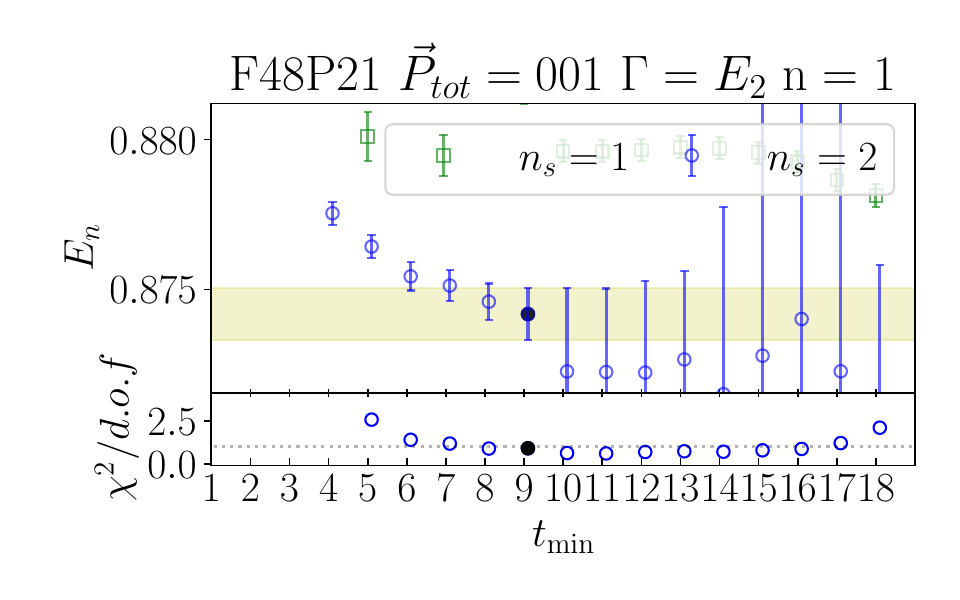}
\\
\includegraphics[width=0.245\columnwidth]{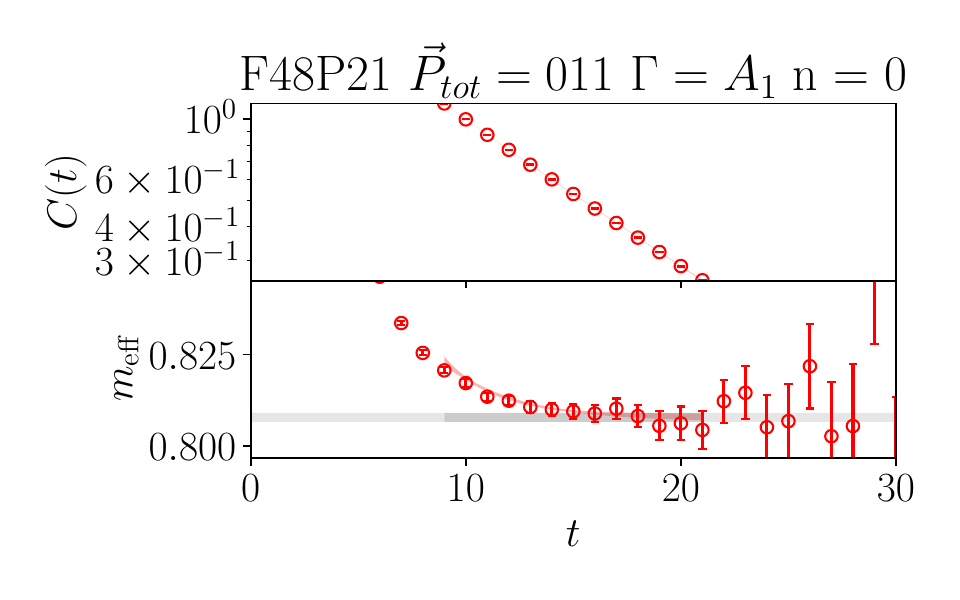}
\includegraphics[width=0.245\columnwidth]{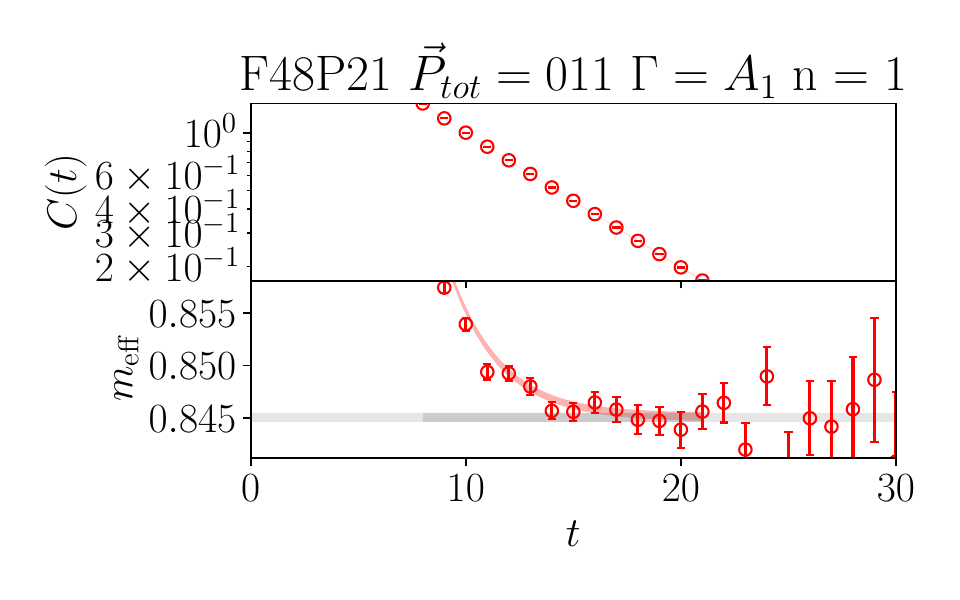}
\includegraphics[width=0.245\columnwidth]{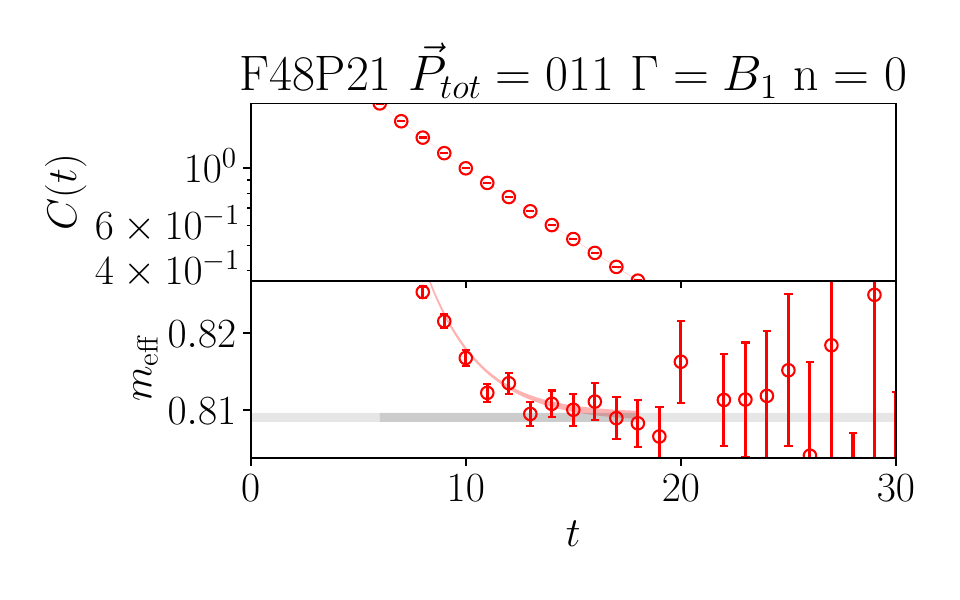}
\includegraphics[width=0.245\columnwidth]{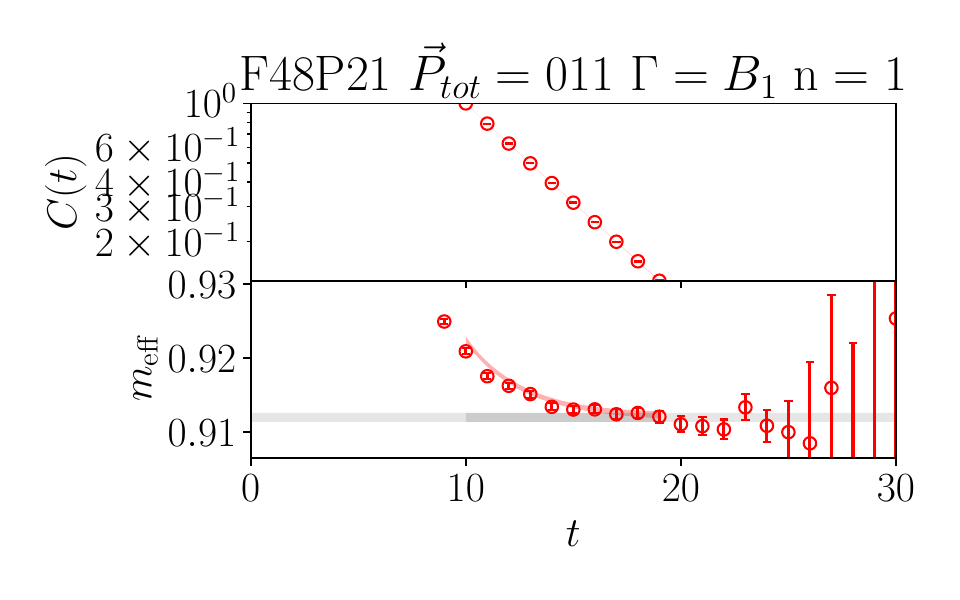}
\\
\includegraphics[width=0.245\columnwidth]{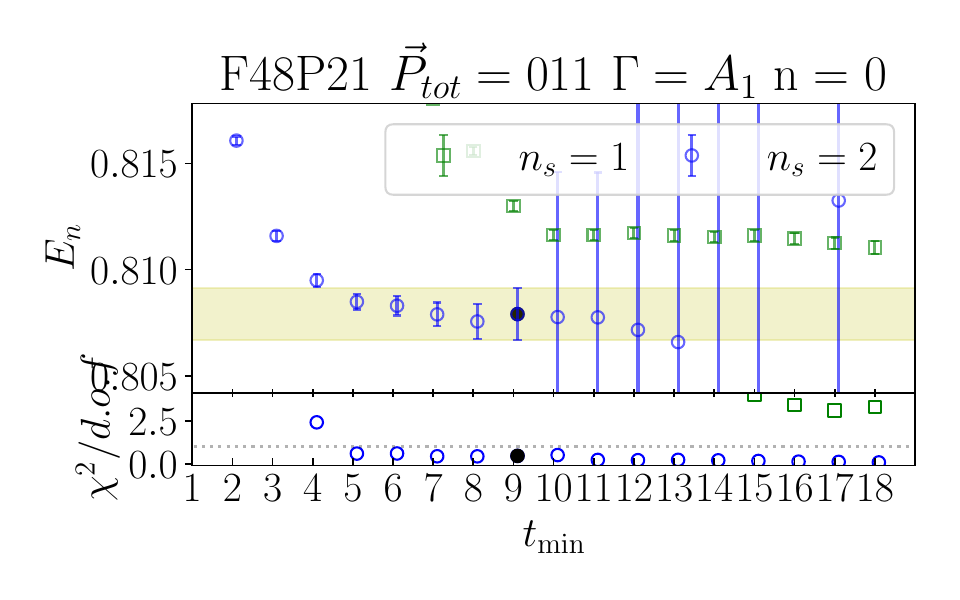}
\includegraphics[width=0.245\columnwidth]{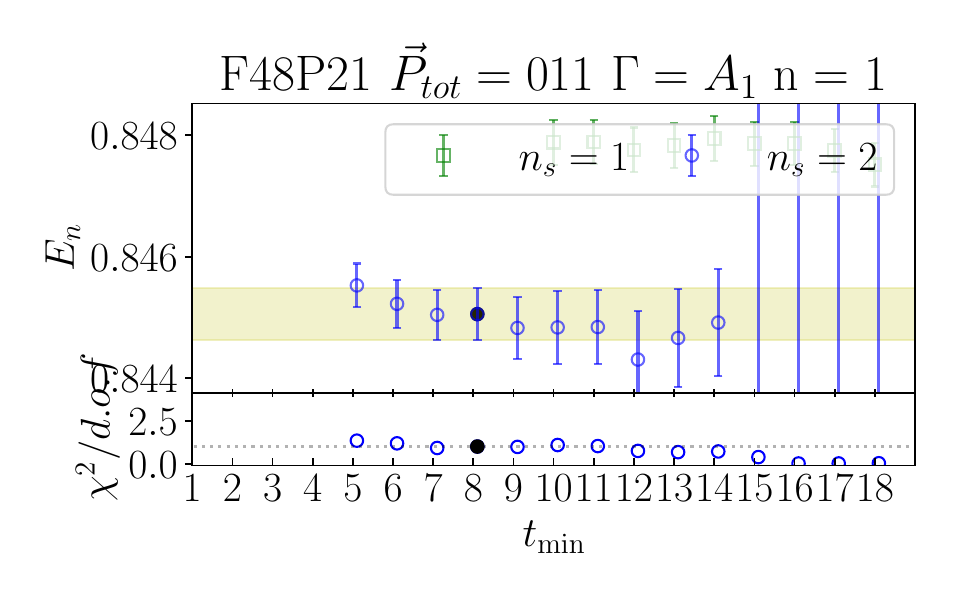}
\includegraphics[width=0.245\columnwidth]{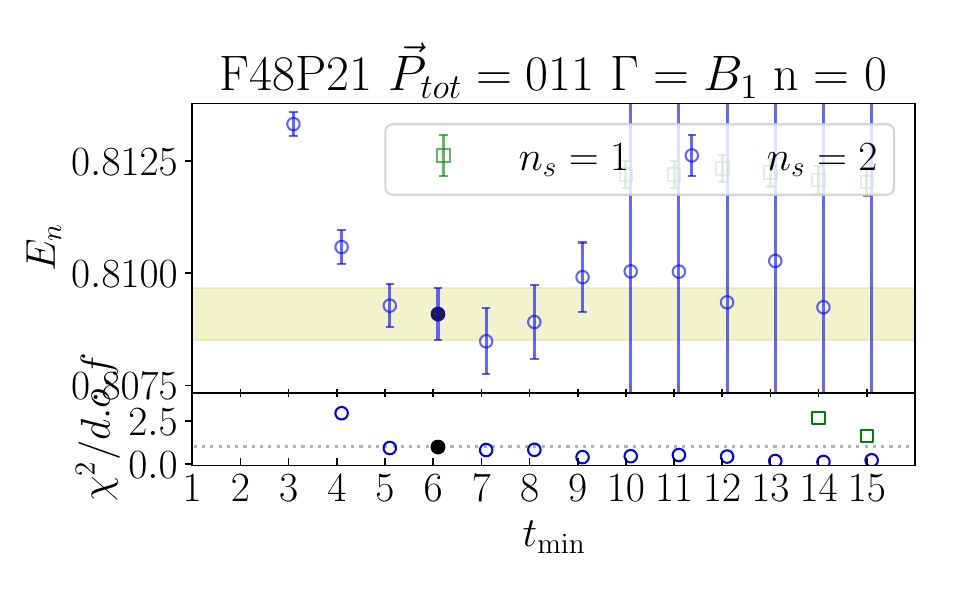}
\includegraphics[width=0.245\columnwidth]{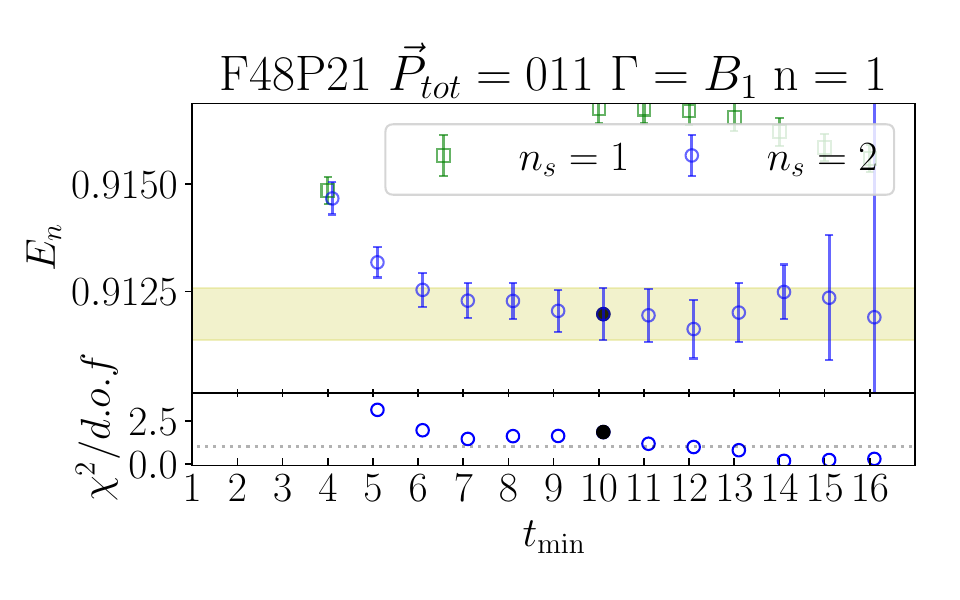}
\\
\includegraphics[width=0.245\columnwidth]{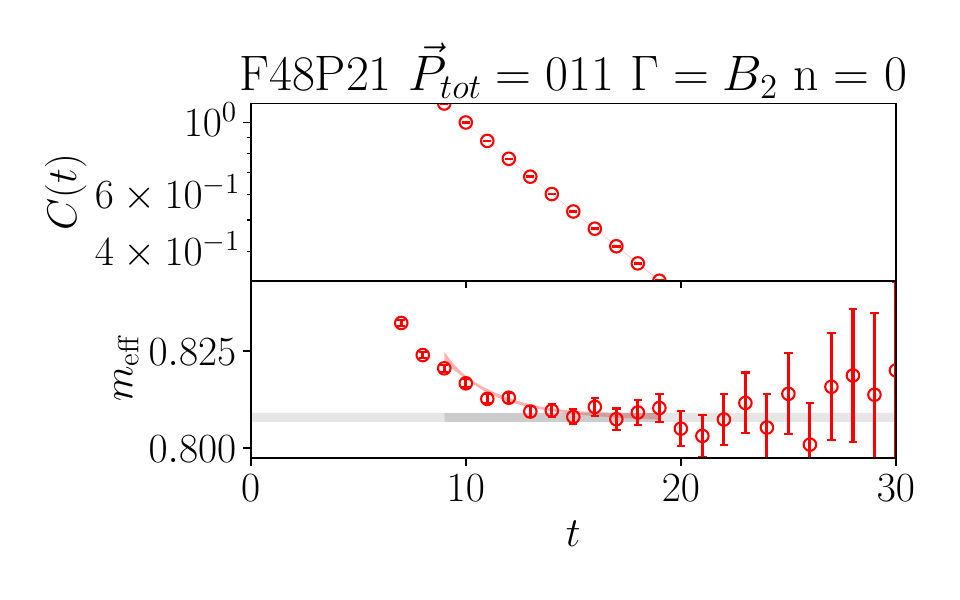}
\includegraphics[width=0.245\columnwidth]{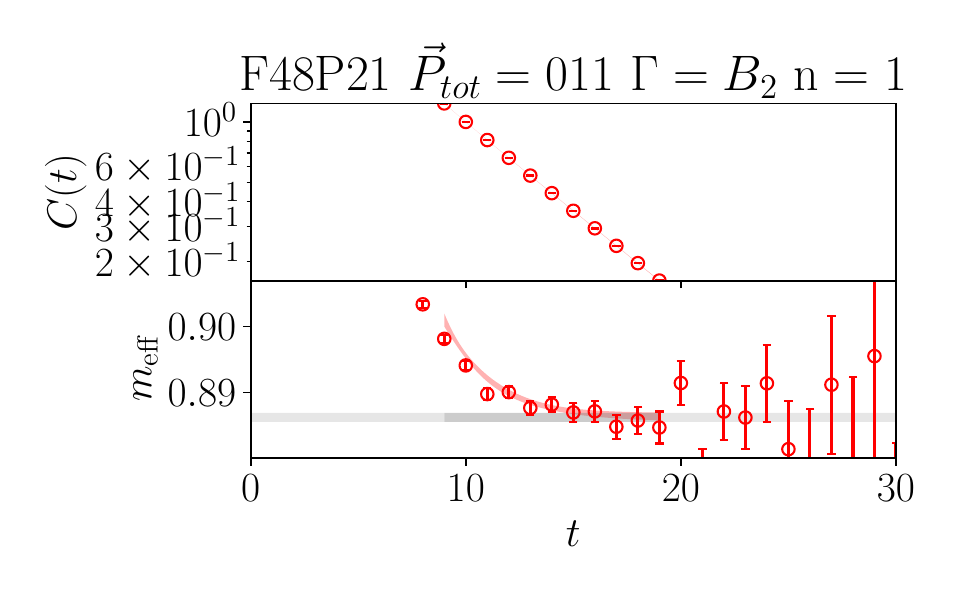}
\includegraphics[width=0.245\columnwidth]{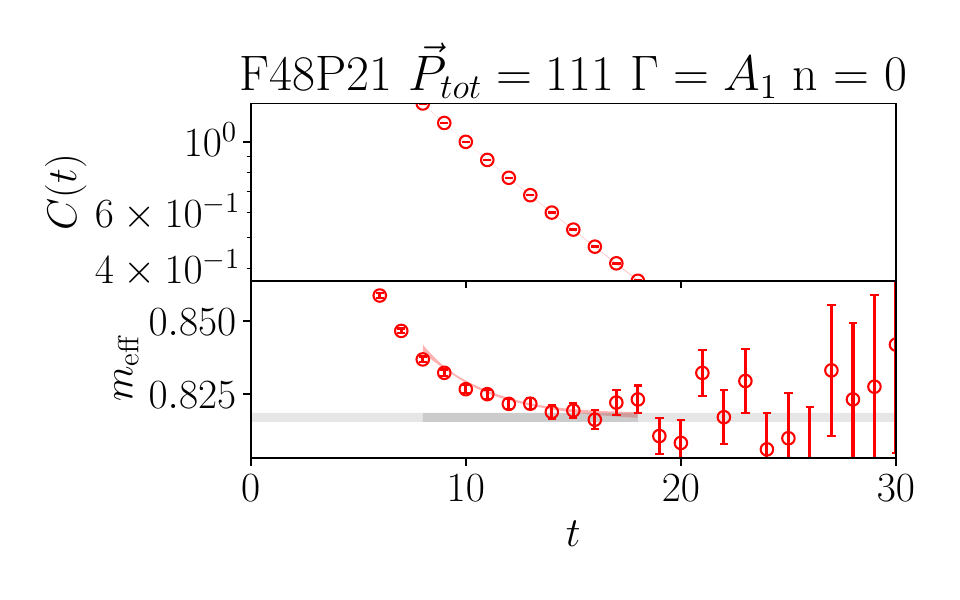}
\includegraphics[width=0.245\columnwidth]{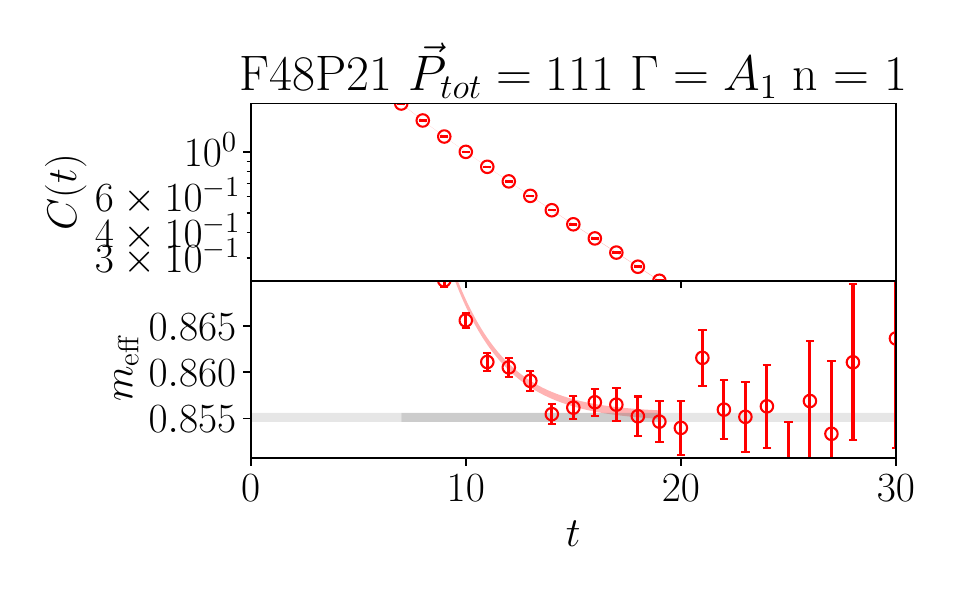}
\\
\includegraphics[width=0.245\columnwidth]{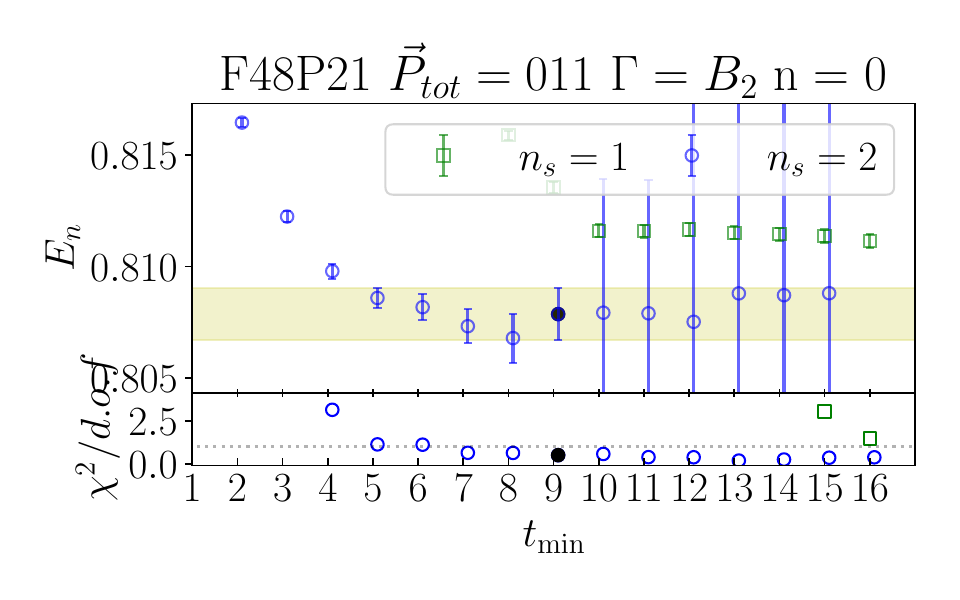}
\includegraphics[width=0.245\columnwidth]{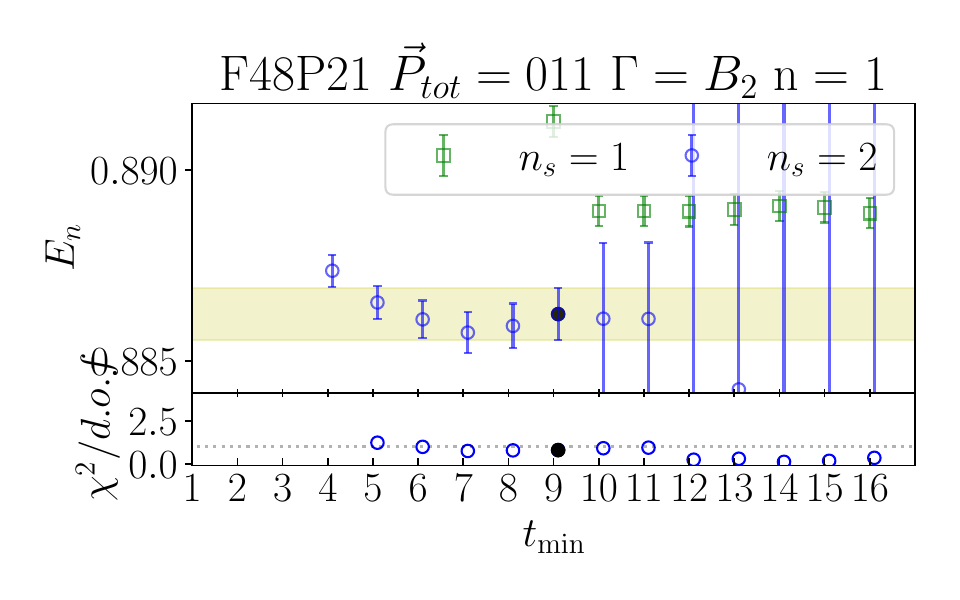}
\includegraphics[width=0.245\columnwidth]{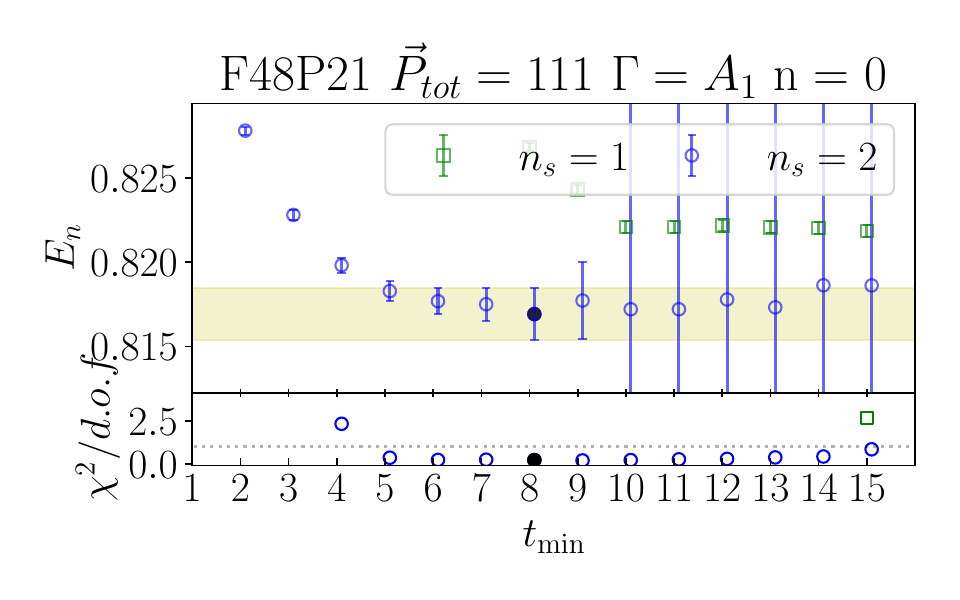}
\includegraphics[width=0.245\columnwidth]{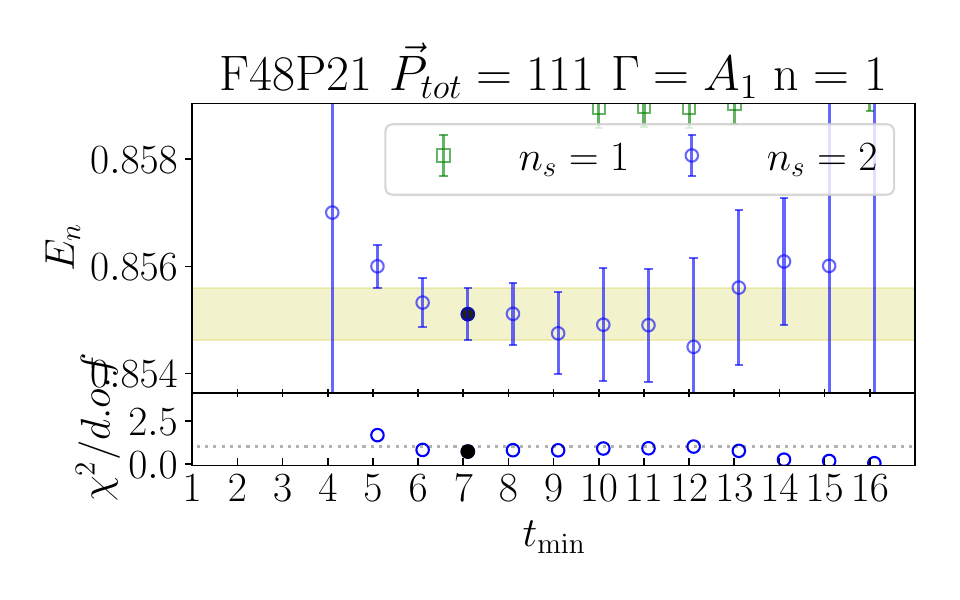}
\caption{Energy-level fit results for the $I=\frac{1}{2}$ $D\pi$ channel on the F48P21 ensemble. The description follows Figure~\ref{fig:Dpi-fit-F32P30}.}
\label{fig:Dpi-fit-F48P21}
\end{figure}

\begin{figure}[htbp]
\centering
\includegraphics[width=0.245\columnwidth]{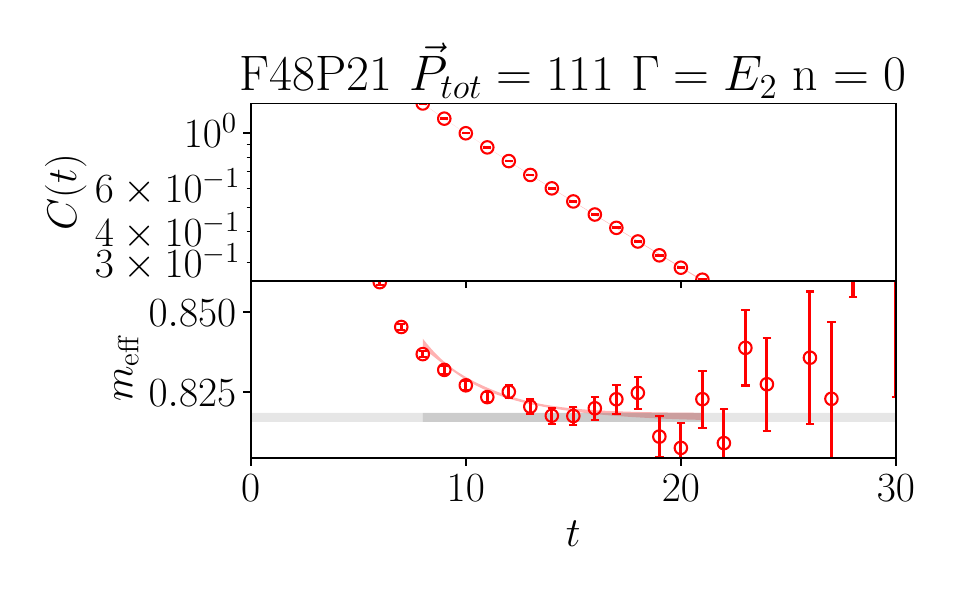}
\includegraphics[width=0.245\columnwidth]{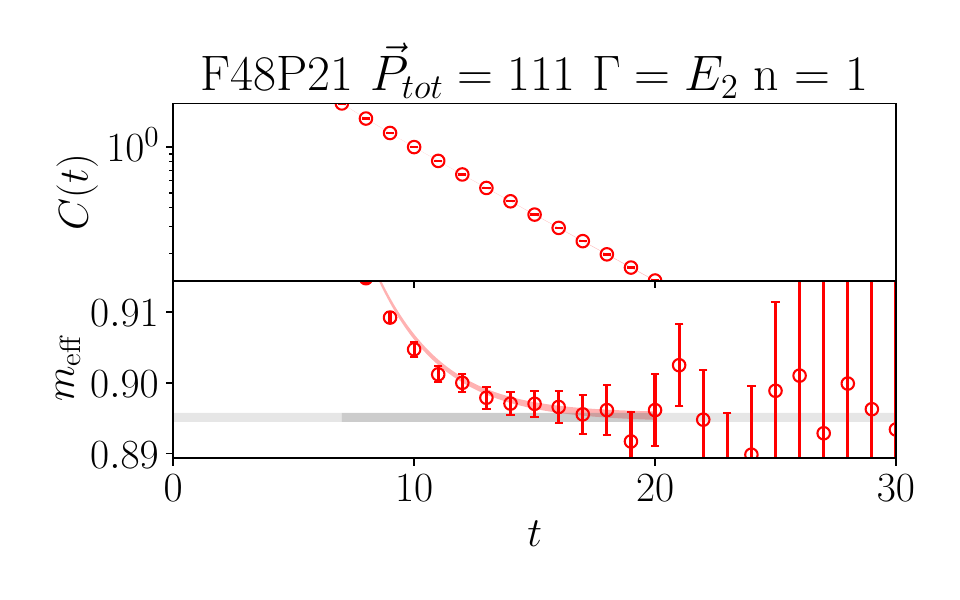}
\includegraphics[width=0.245\columnwidth]{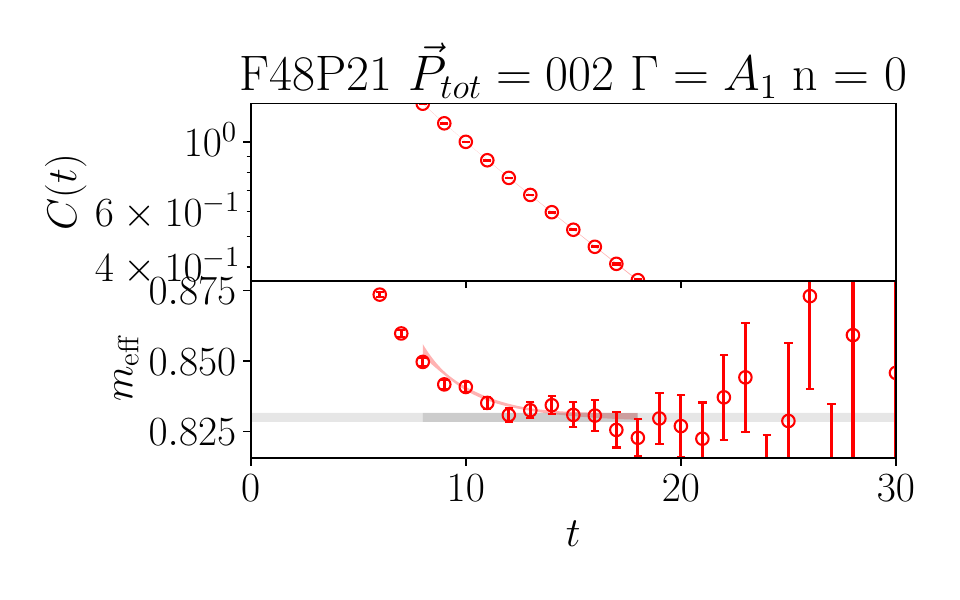}
\includegraphics[width=0.245\columnwidth]{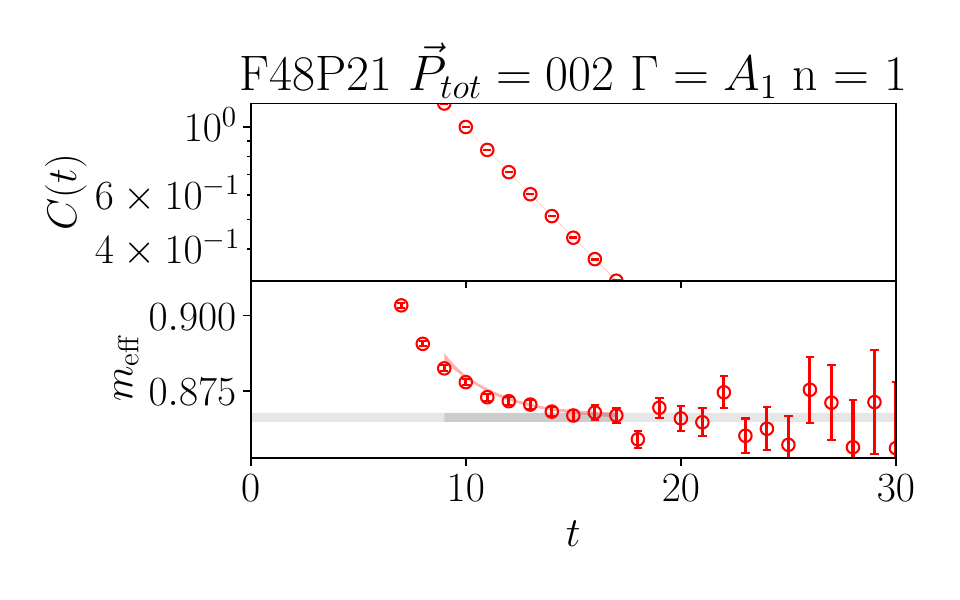}
\\
\includegraphics[width=0.245\columnwidth]{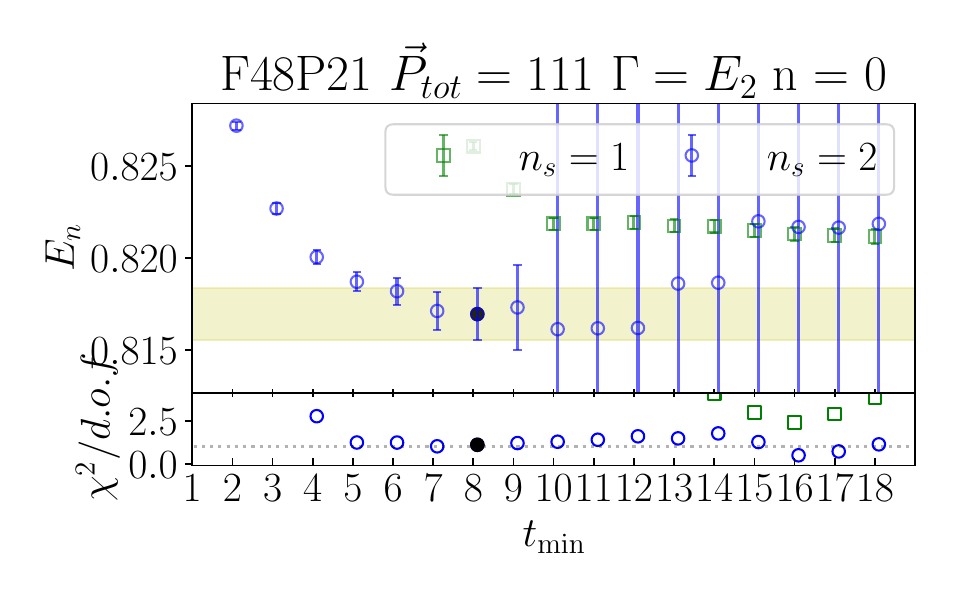}
\includegraphics[width=0.245\columnwidth]{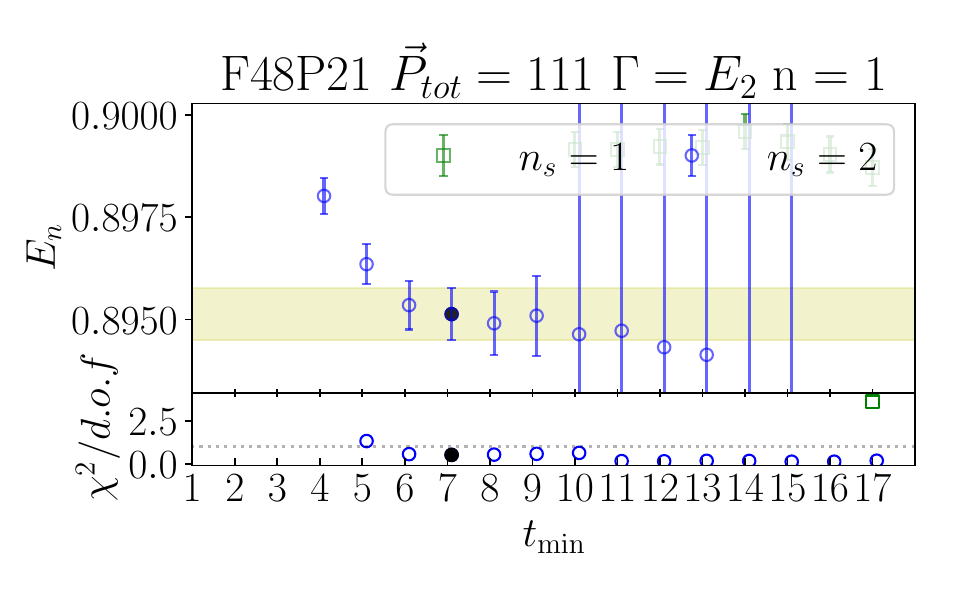}
\includegraphics[width=0.245\columnwidth]{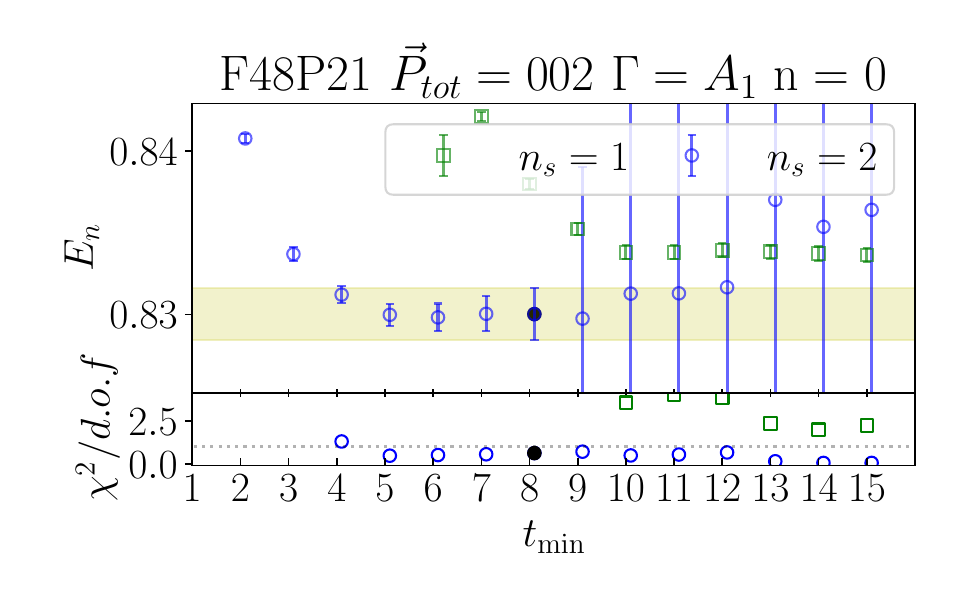}
\includegraphics[width=0.245\columnwidth]{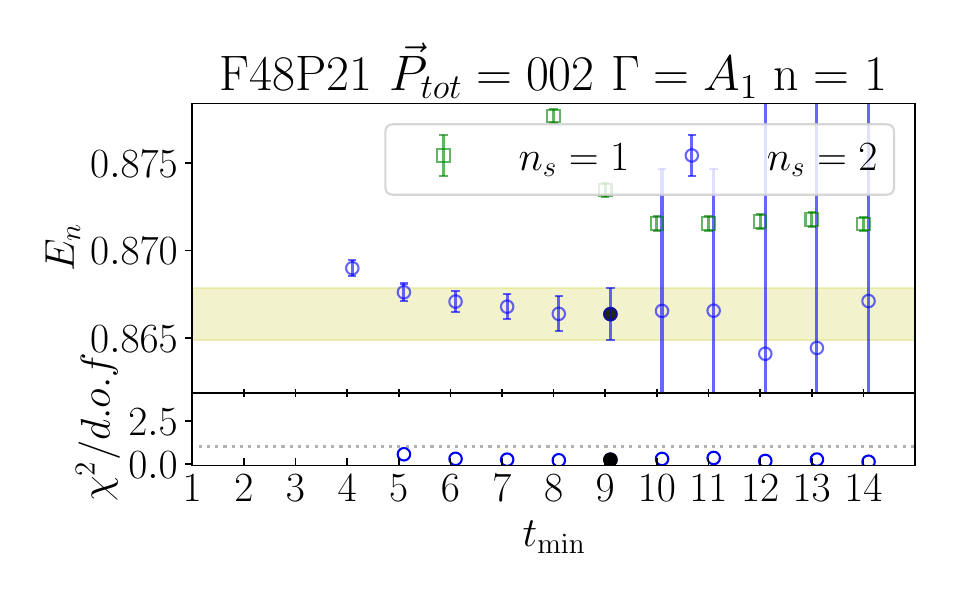}
\\
\includegraphics[width=0.245\columnwidth]{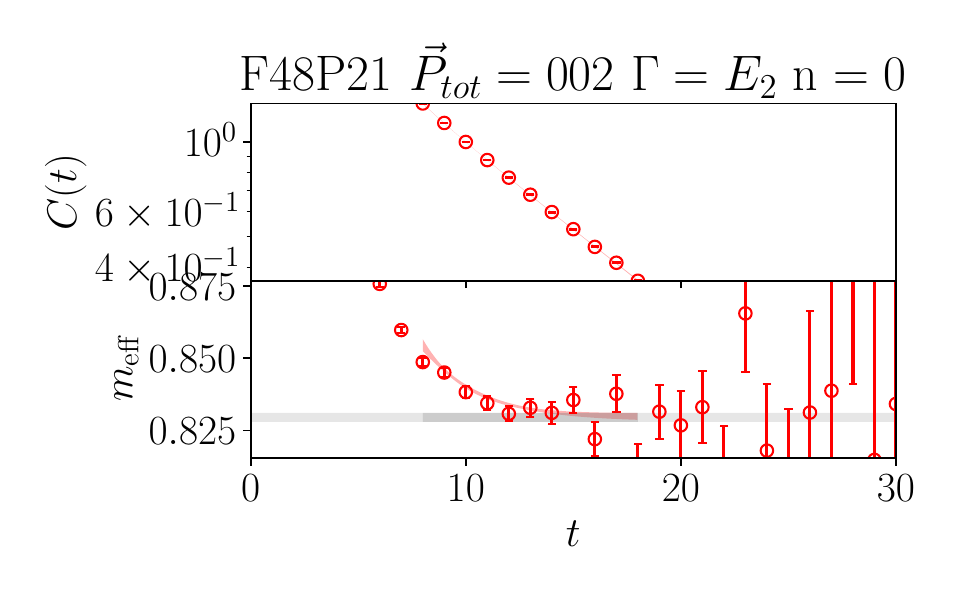}
\includegraphics[width=0.245\columnwidth]{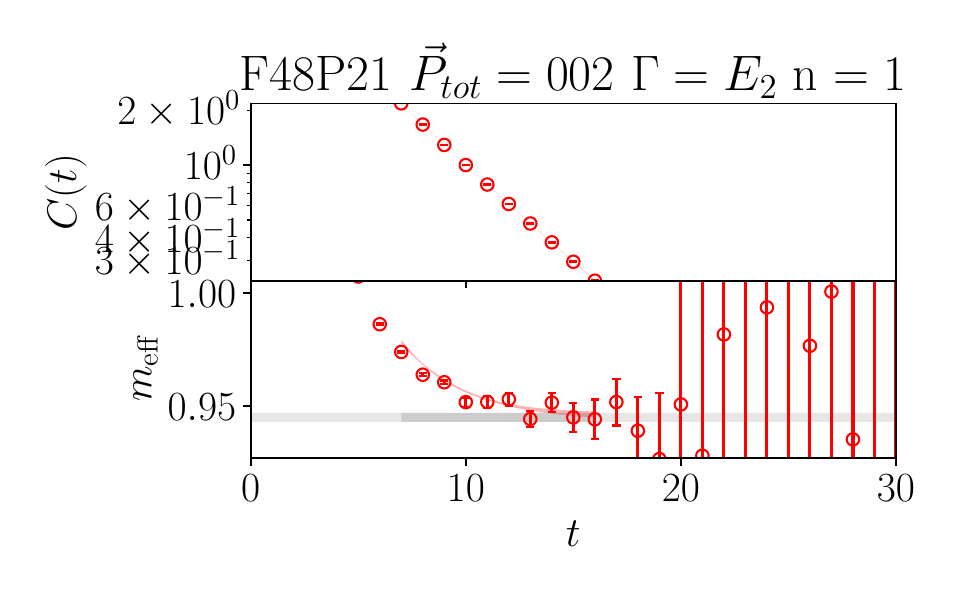}
\\
\includegraphics[width=0.245\columnwidth]{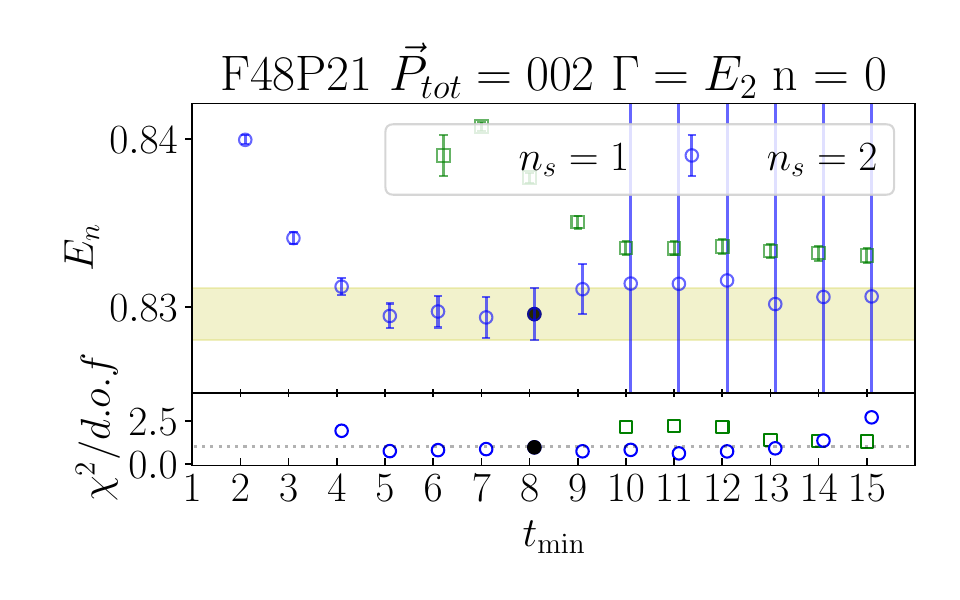}
\includegraphics[width=0.245\columnwidth]{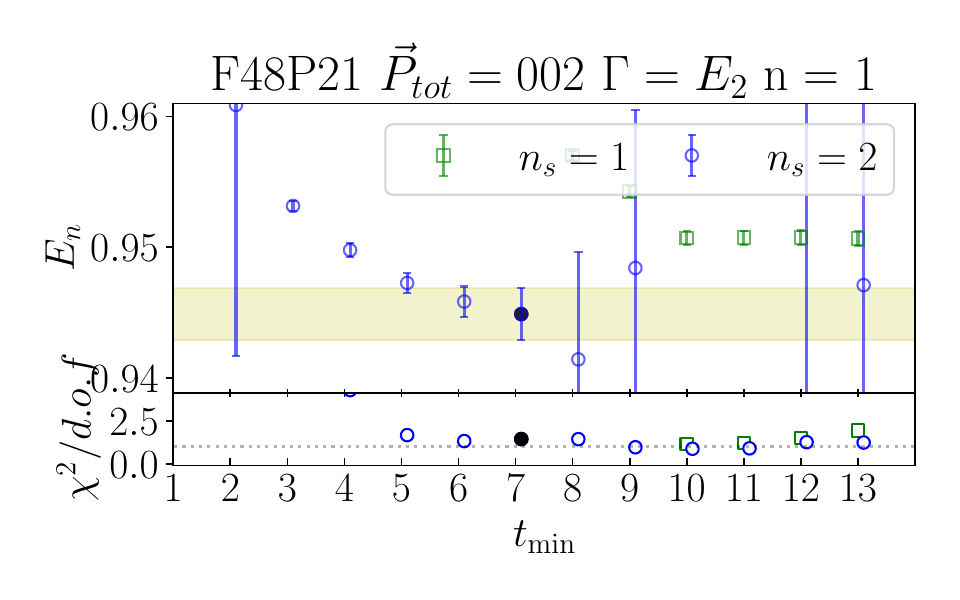}
\caption{Continued from Figure~\ref{fig:Dpi-fit-F48P21}. Energy-level fit results for the $I=\frac{1}{2}$ $D\pi$ channel on the F48P21 ensemble.}
\label{fig:Dpi-fit-F48P212}
\end{figure}

\begin{figure}[htbp]
\centering
\includegraphics[width=0.245\columnwidth]{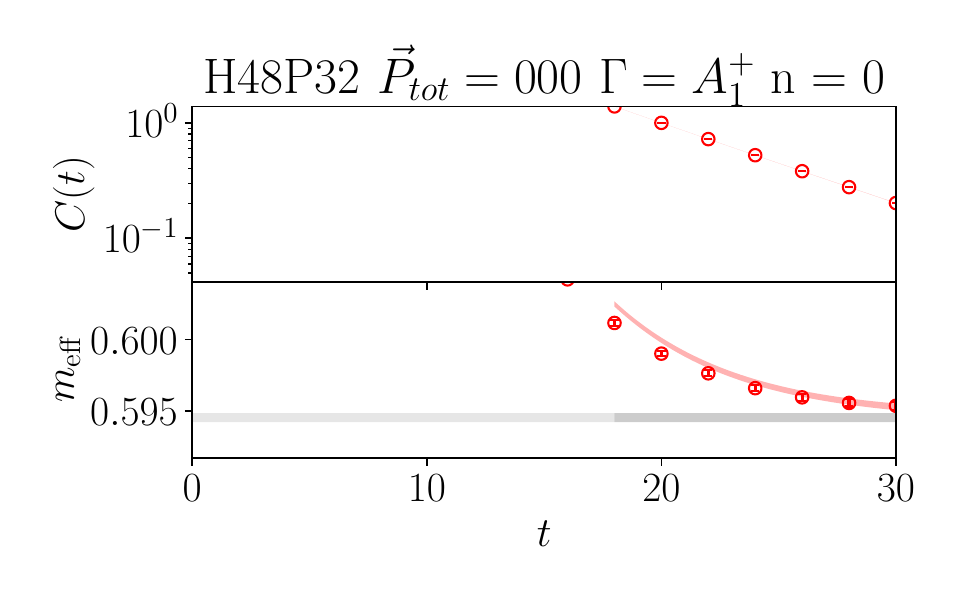}
\includegraphics[width=0.245\columnwidth]{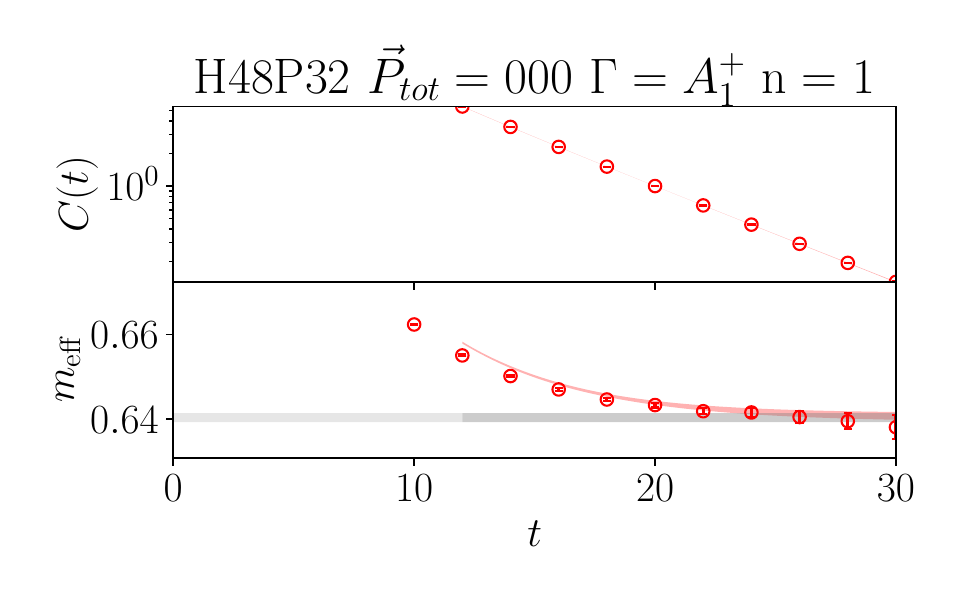}
\includegraphics[width=0.245\columnwidth]{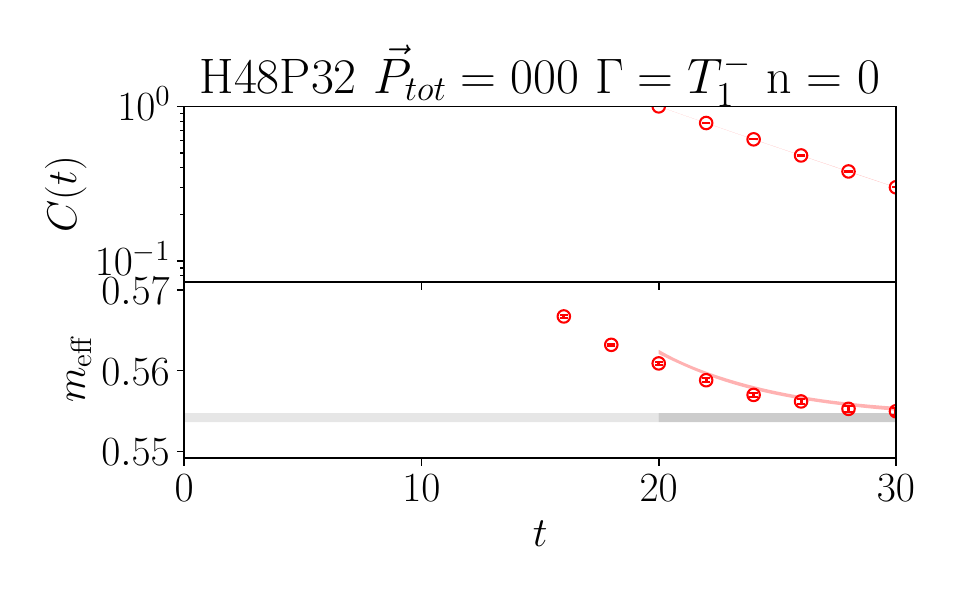}
\includegraphics[width=0.245\columnwidth]{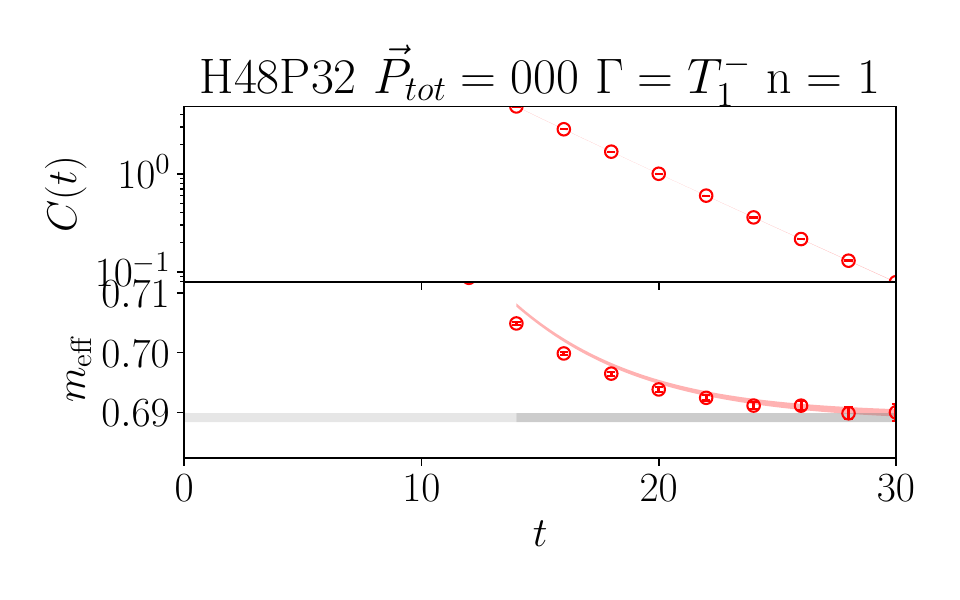}
\\
\includegraphics[width=0.245\columnwidth]{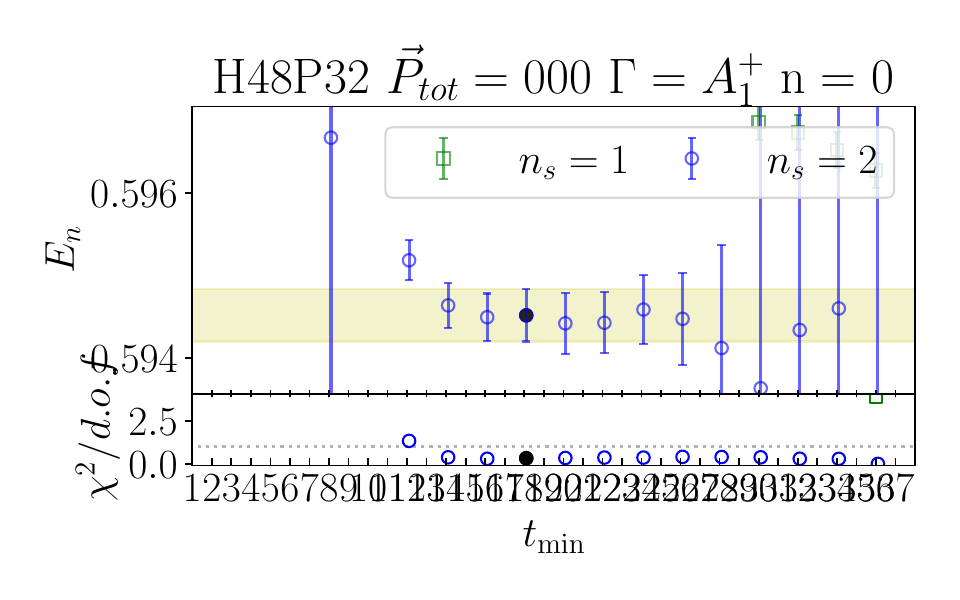}
\includegraphics[width=0.245\columnwidth]{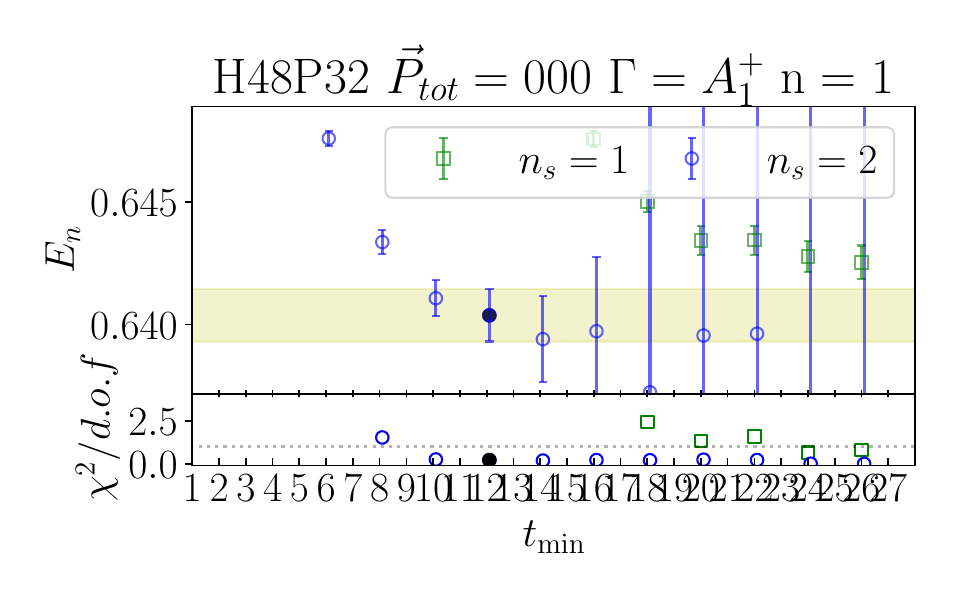}
\includegraphics[width=0.245\columnwidth]{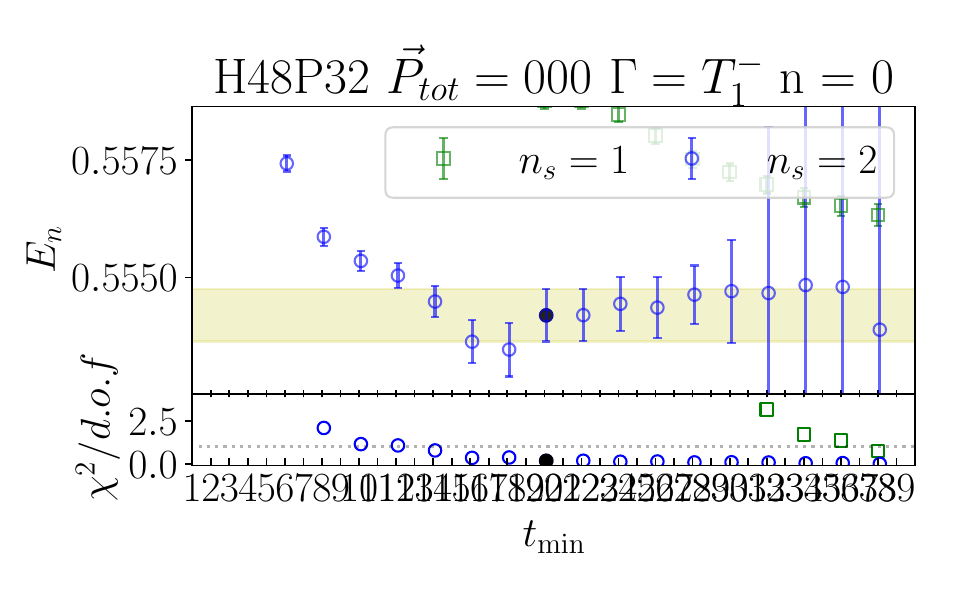}
\includegraphics[width=0.245\columnwidth]{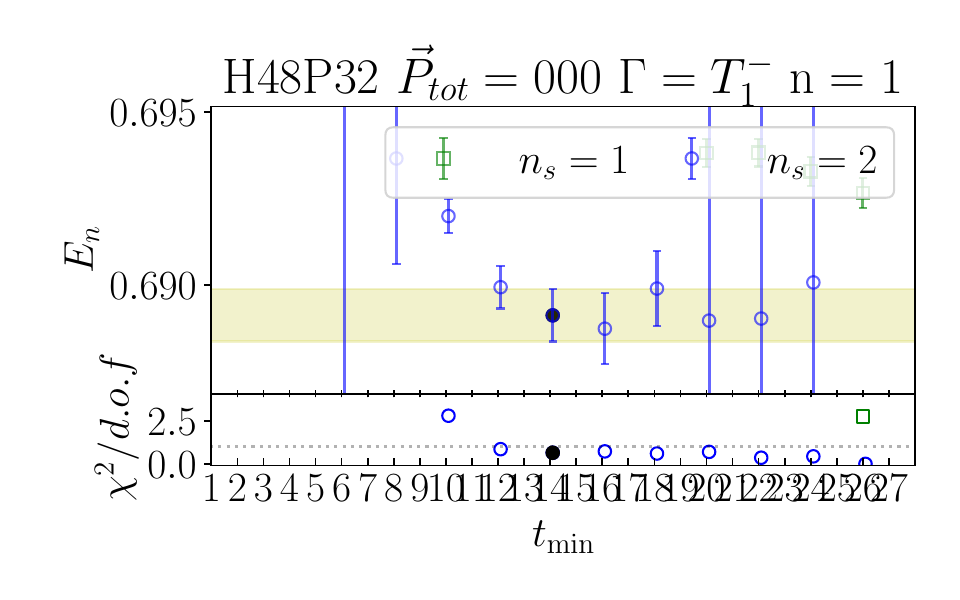}
\\
\includegraphics[width=0.245\columnwidth]{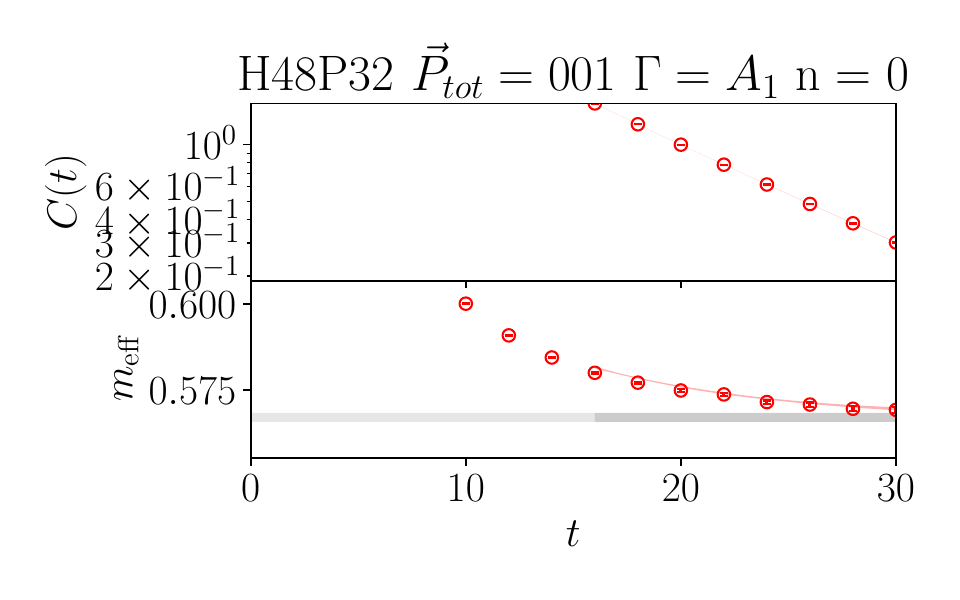}
\includegraphics[width=0.245\columnwidth]{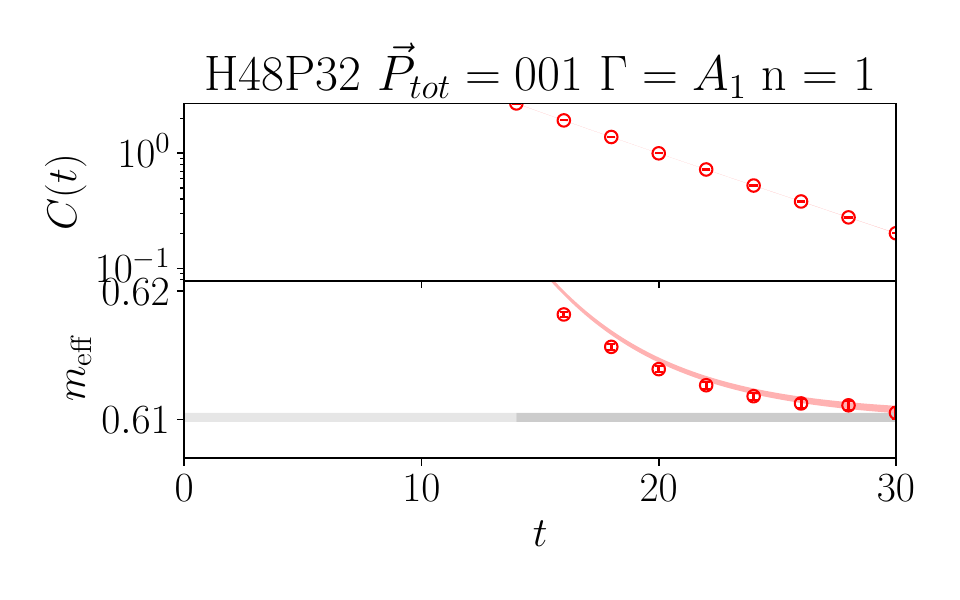}
\includegraphics[width=0.245\columnwidth]{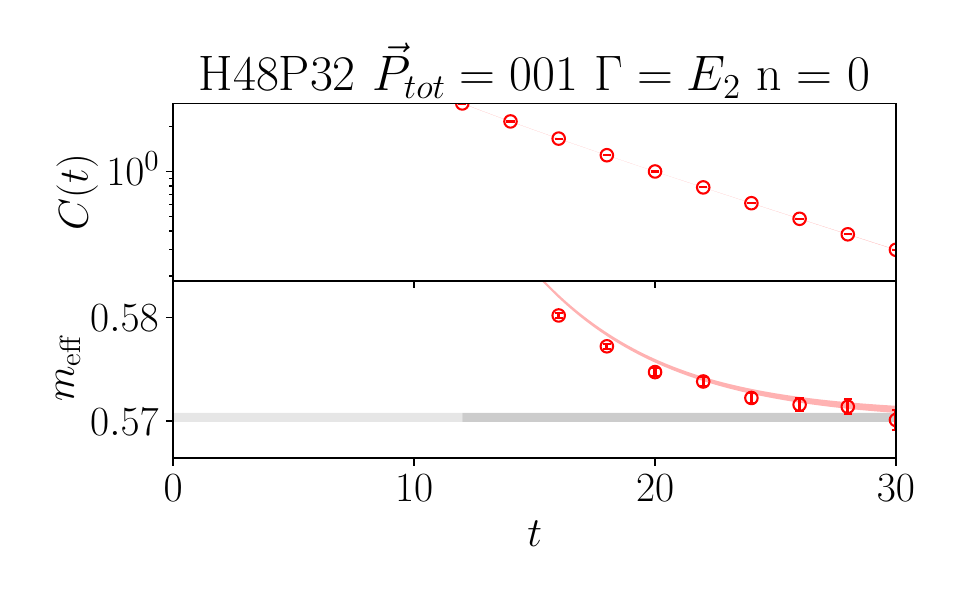}
\includegraphics[width=0.245\columnwidth]{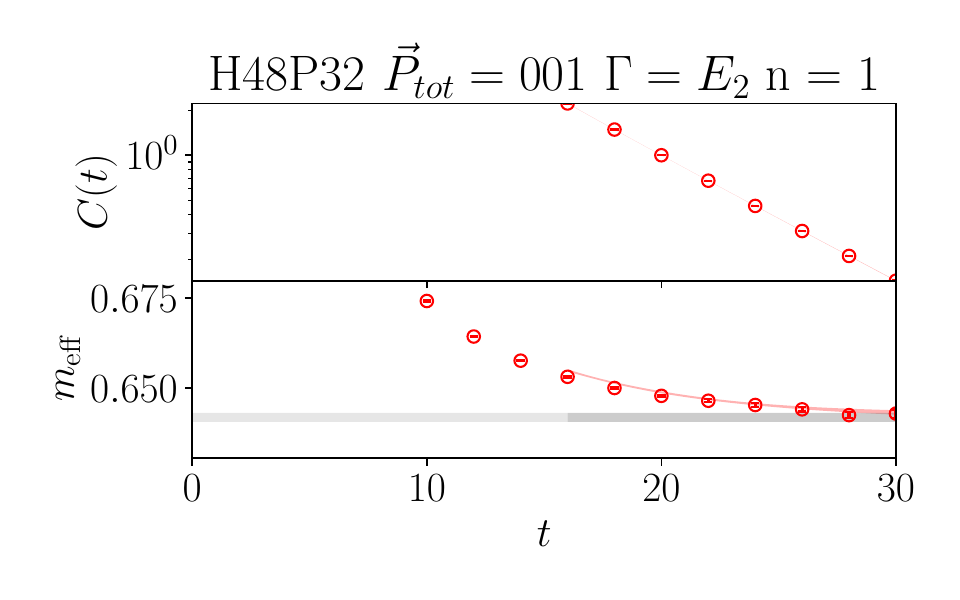}
\\
\includegraphics[width=0.245\columnwidth]{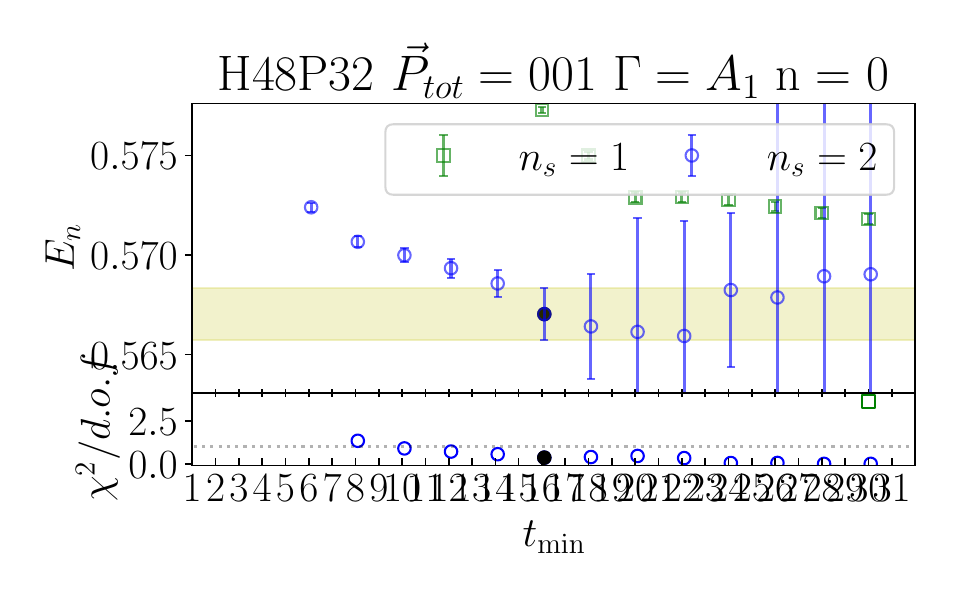}
\includegraphics[width=0.245\columnwidth]{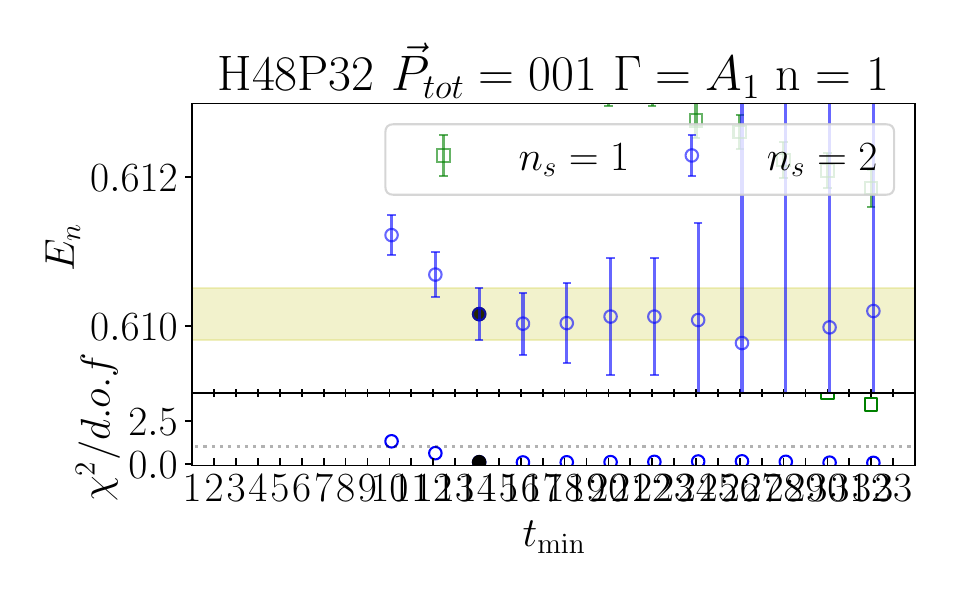}
\includegraphics[width=0.245\columnwidth]{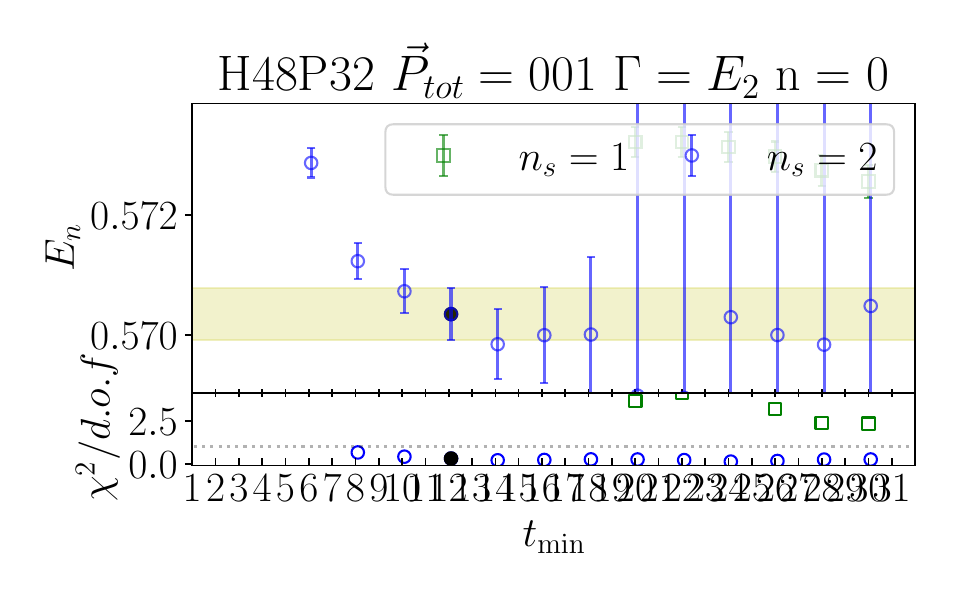}
\includegraphics[width=0.245\columnwidth]{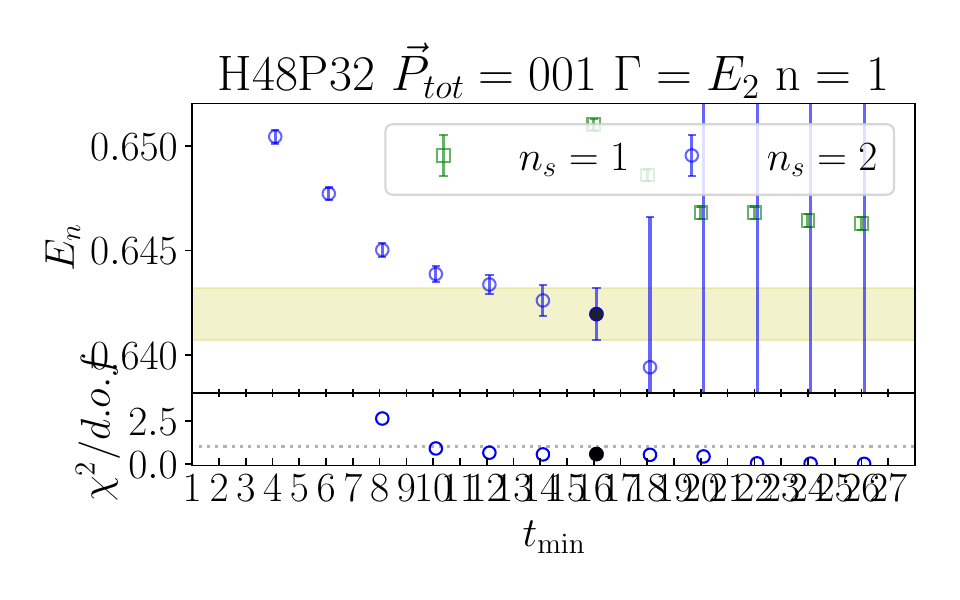}
\\
\includegraphics[width=0.245\columnwidth]{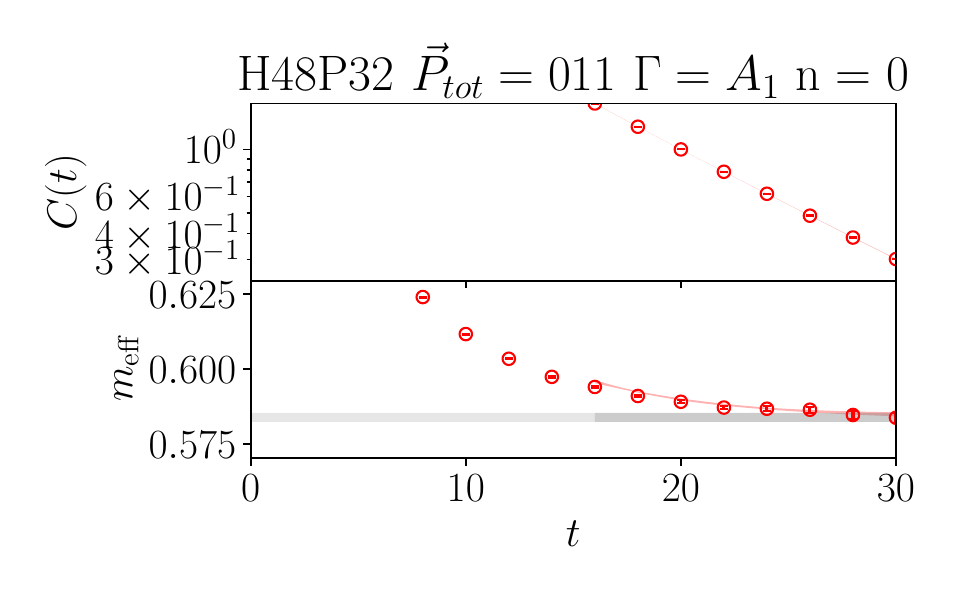}
\includegraphics[width=0.245\columnwidth]{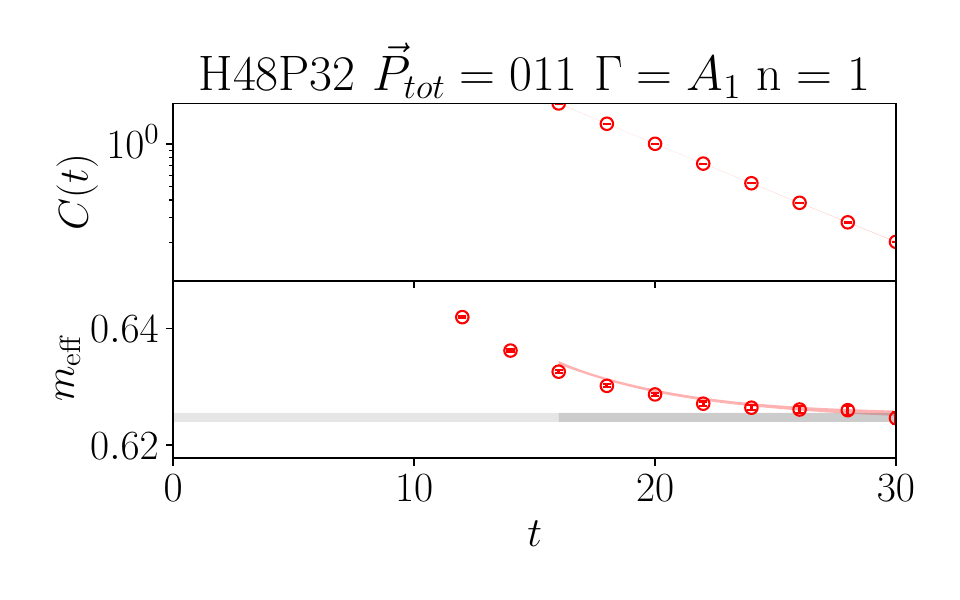}
\includegraphics[width=0.245\columnwidth]{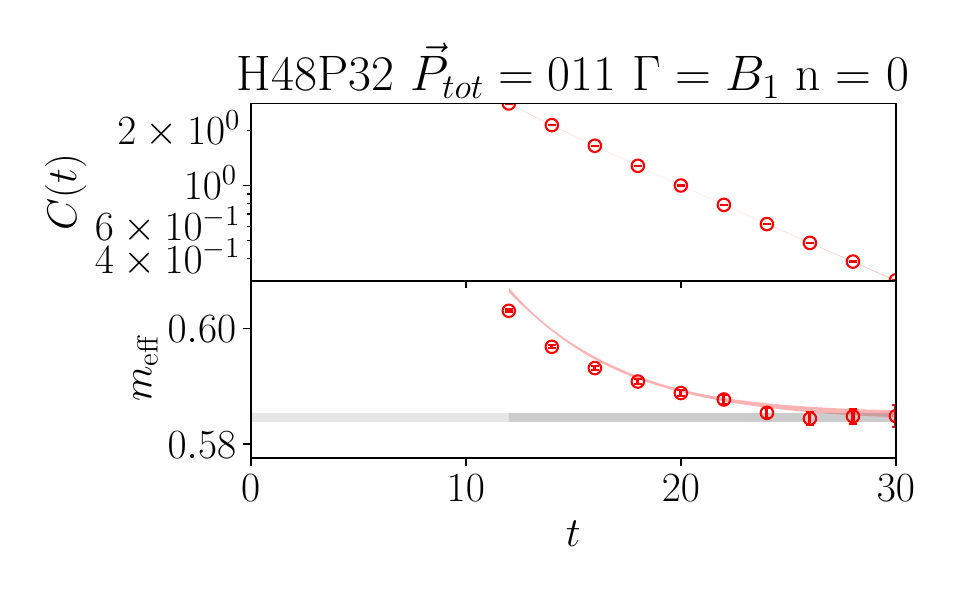}
\includegraphics[width=0.245\columnwidth]{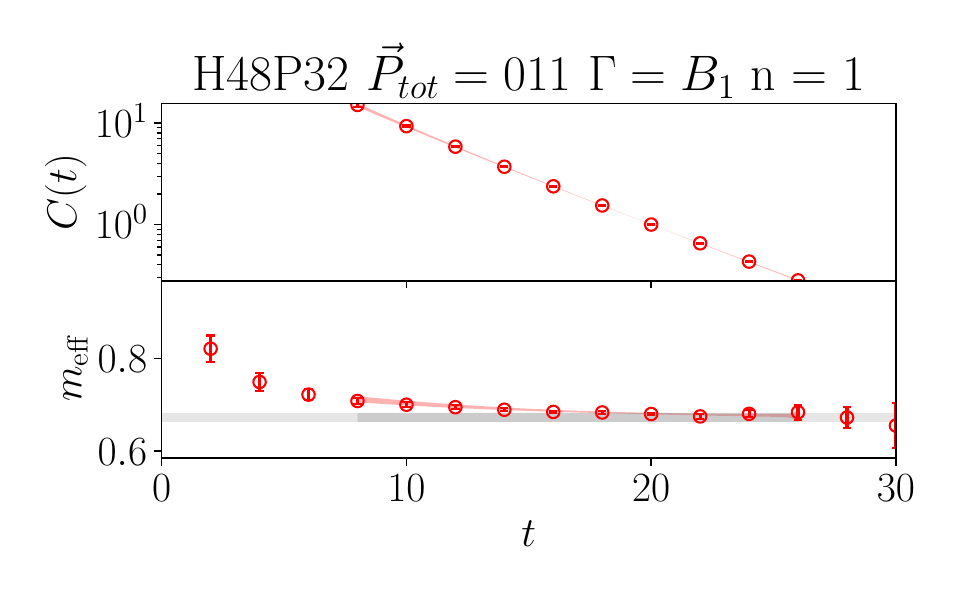}
\\
\includegraphics[width=0.245\columnwidth]{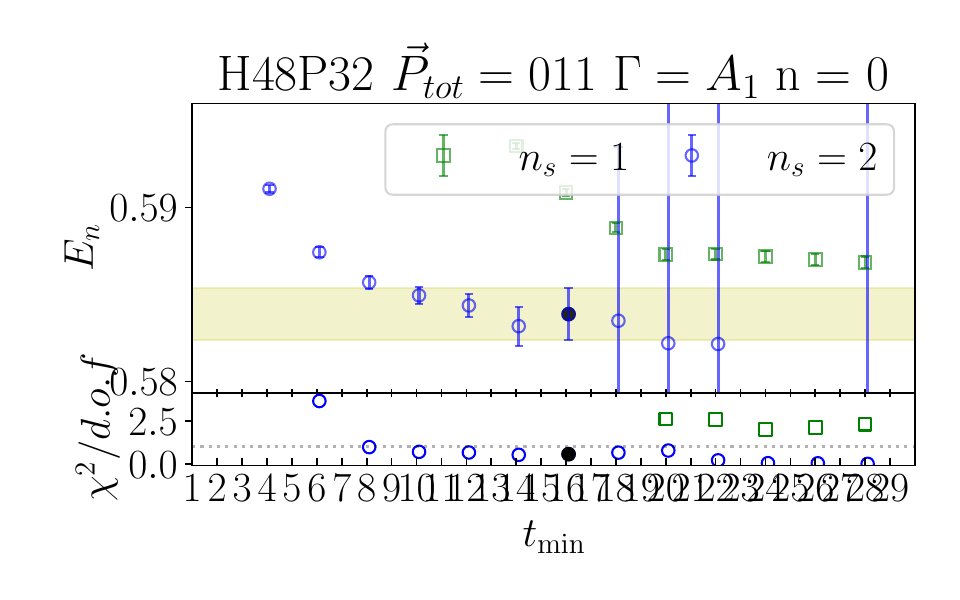}
\includegraphics[width=0.245\columnwidth]{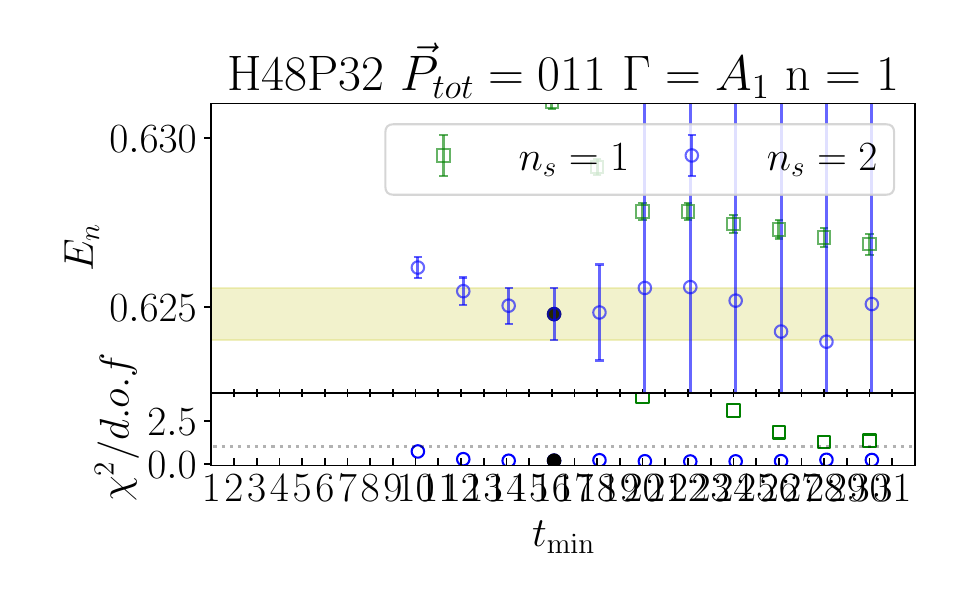}
\includegraphics[width=0.245\columnwidth]{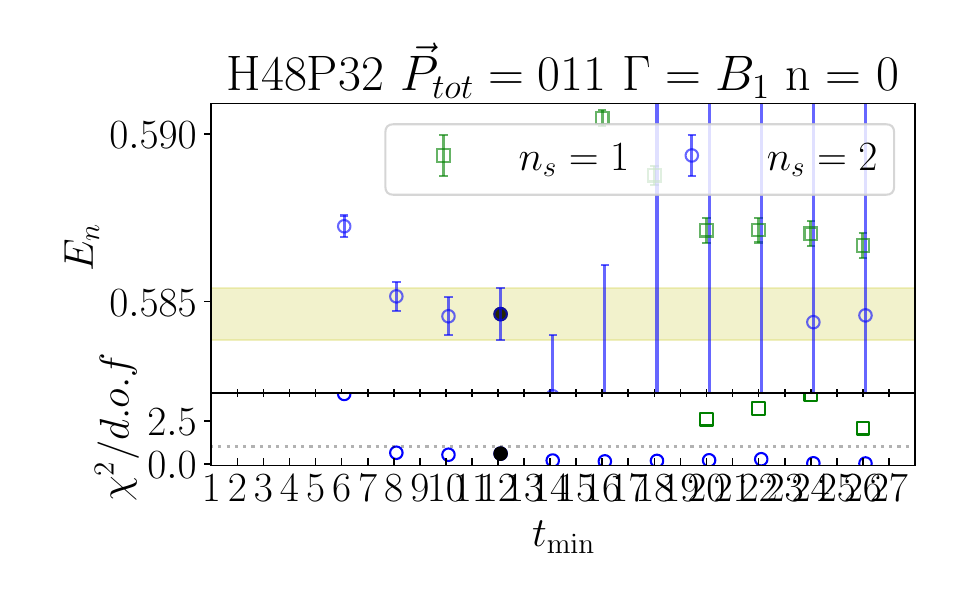}
\includegraphics[width=0.245\columnwidth]{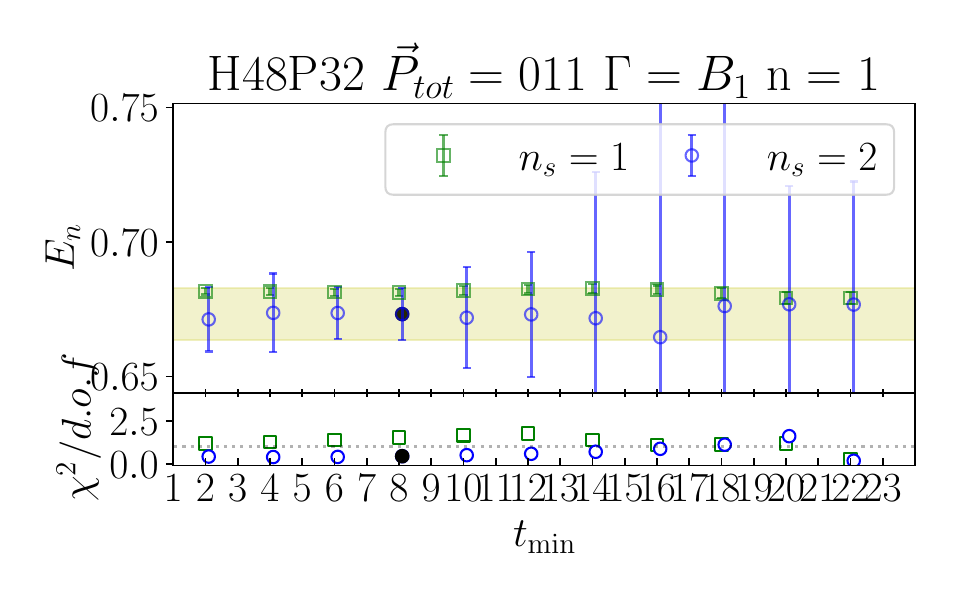}
\\
\includegraphics[width=0.245\columnwidth]{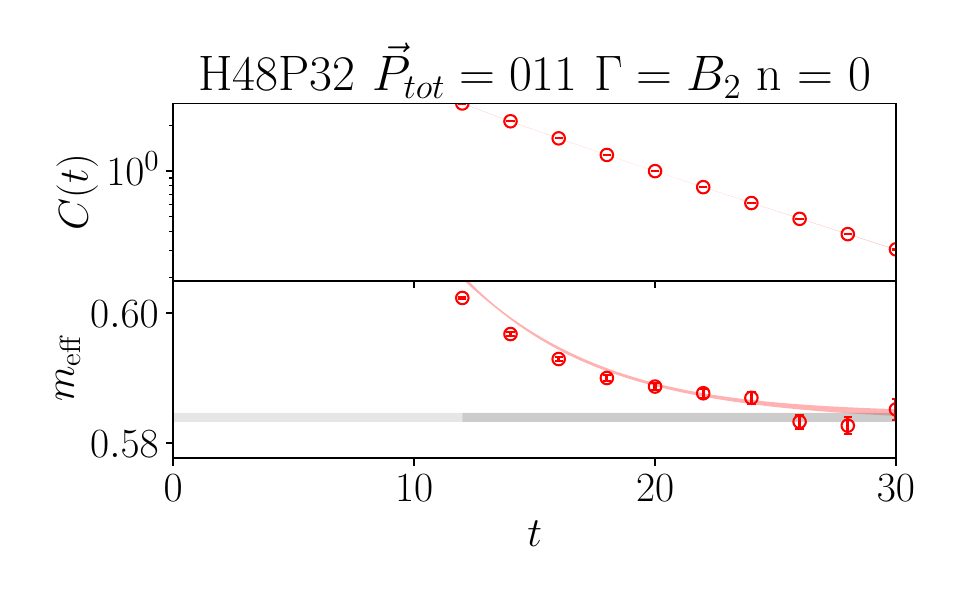}
\includegraphics[width=0.245\columnwidth]{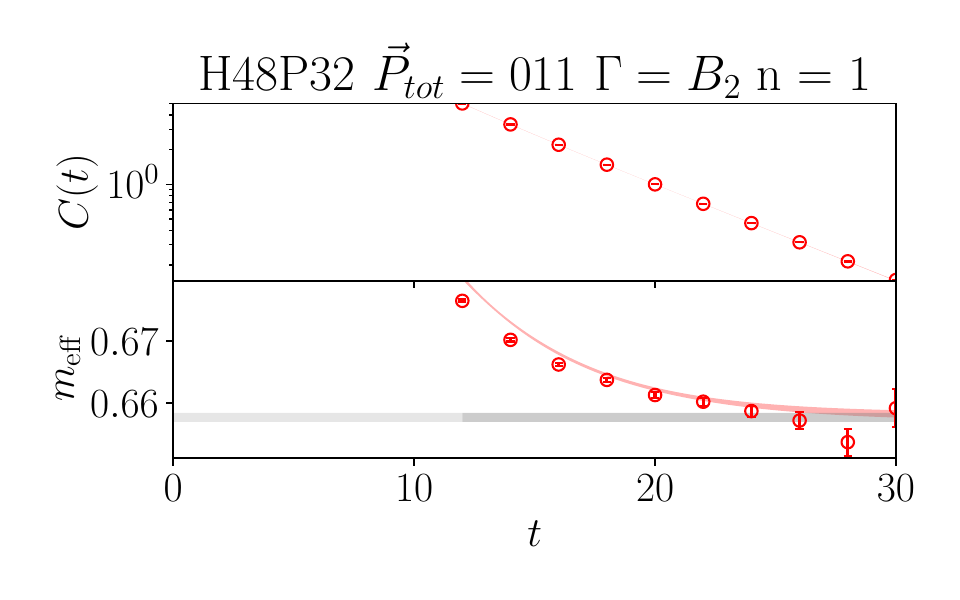}
\includegraphics[width=0.245\columnwidth]{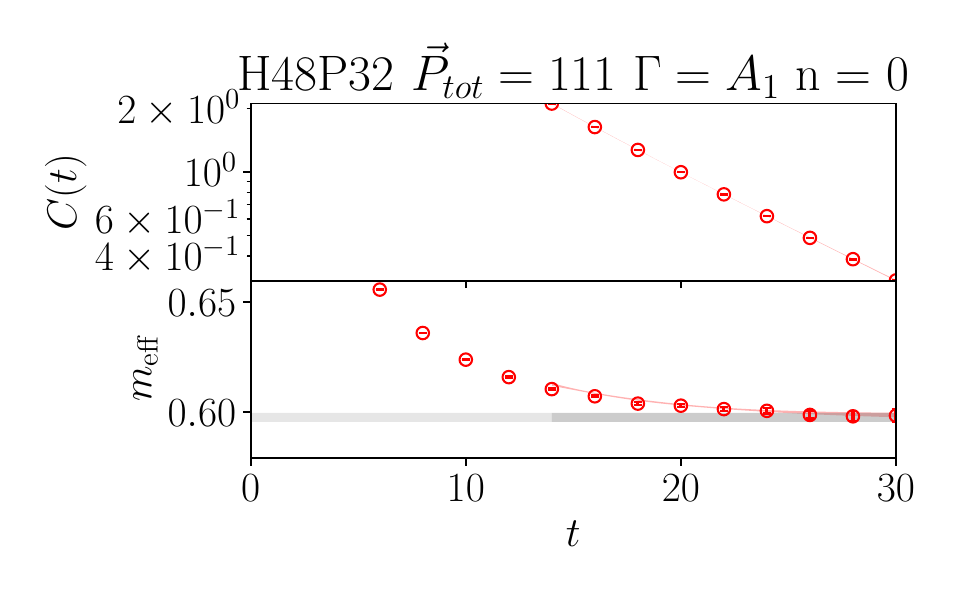}
\includegraphics[width=0.245\columnwidth]{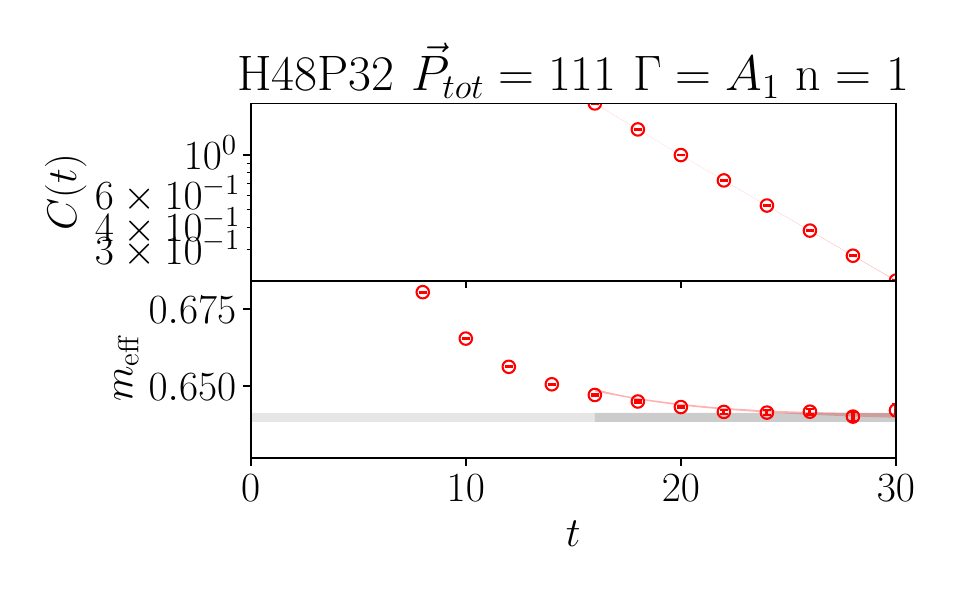}
\\
\includegraphics[width=0.245\columnwidth]{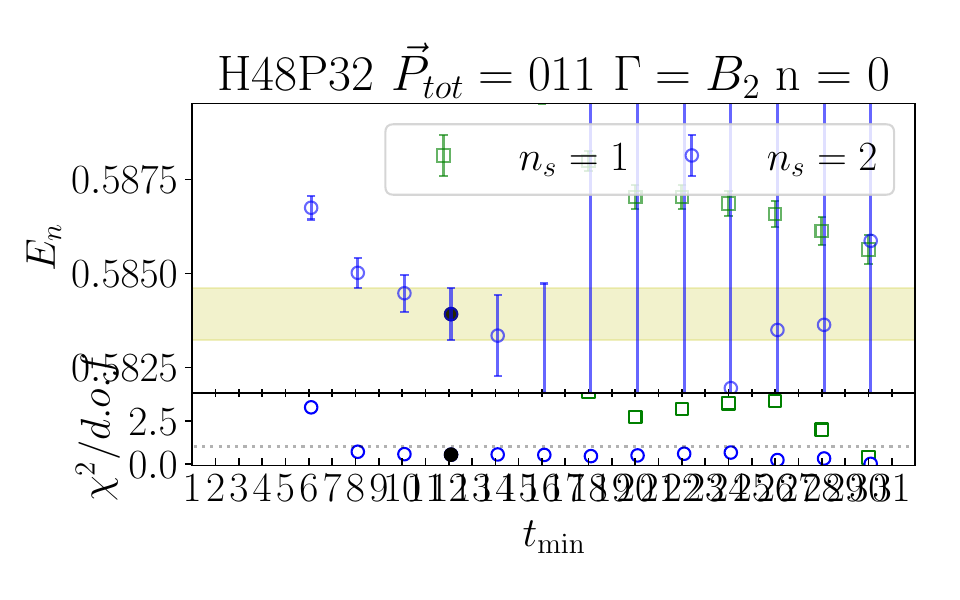}
\includegraphics[width=0.245\columnwidth]{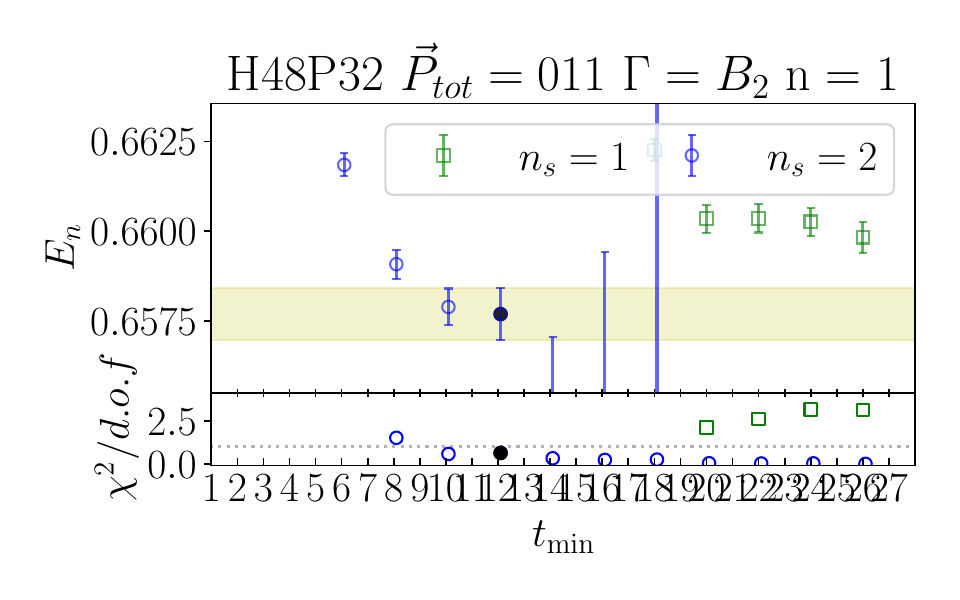}
\includegraphics[width=0.245\columnwidth]{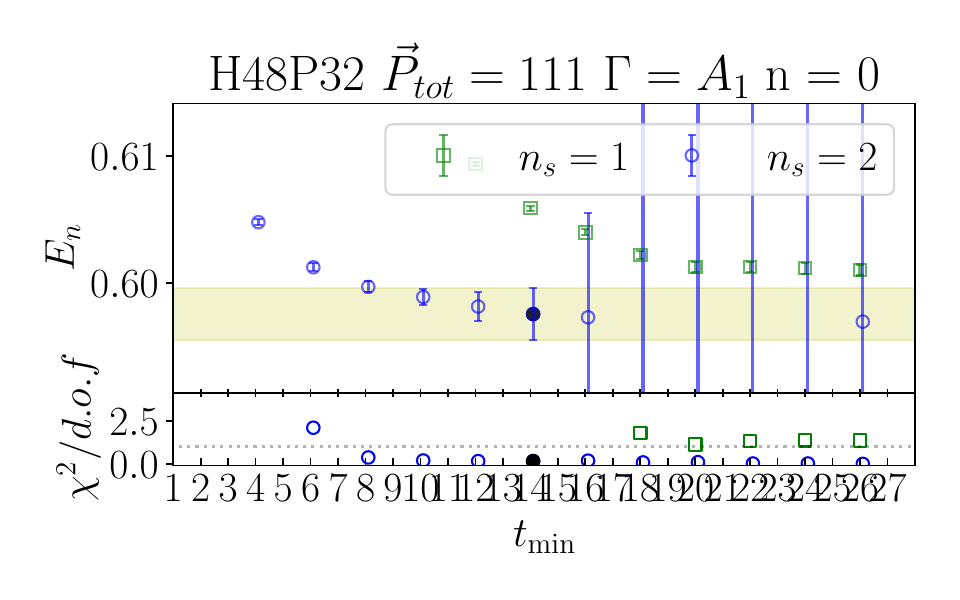}
\includegraphics[width=0.245\columnwidth]{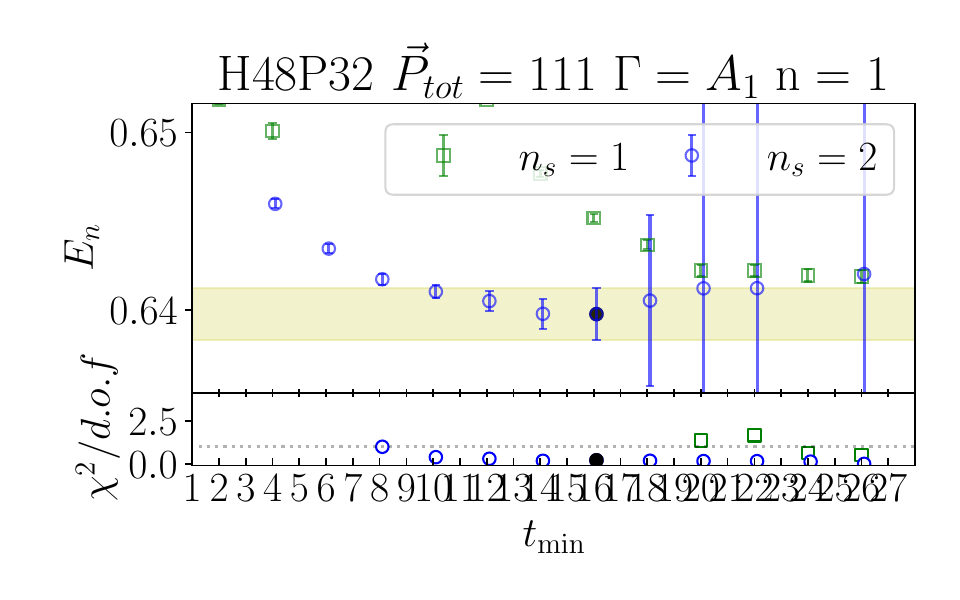}
\caption{Energy-level fit results for the $I=\frac{1}{2}$ $D\pi$ channel on the H48P32 ensemble. The description follows Figure~\ref{fig:Dpi-fit-F32P30}.}
\label{fig:Dpi-fit-H48P32}
\end{figure}

\begin{figure}[htbp]
\centering
\includegraphics[width=0.245\columnwidth]{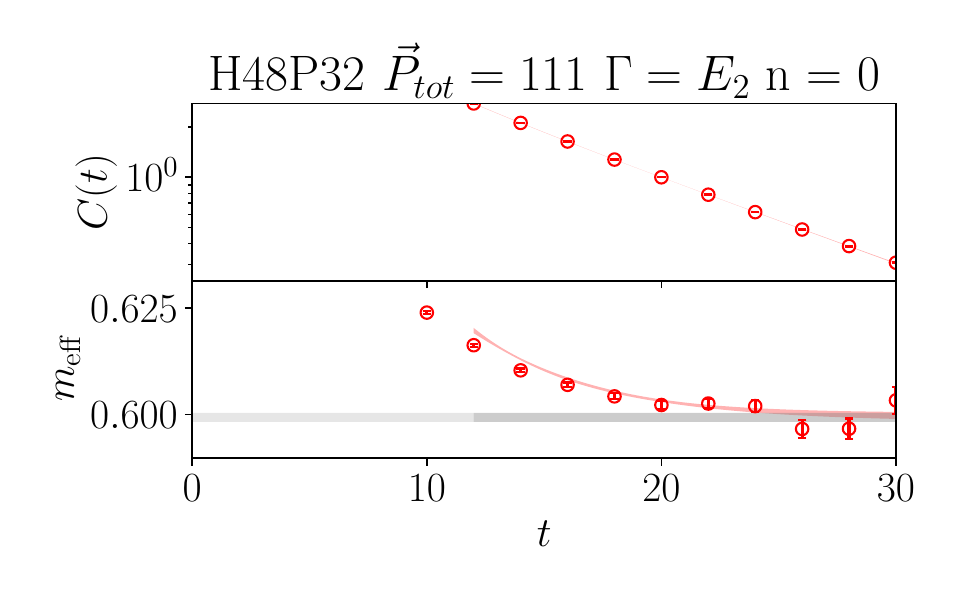}
\includegraphics[width=0.245\columnwidth]{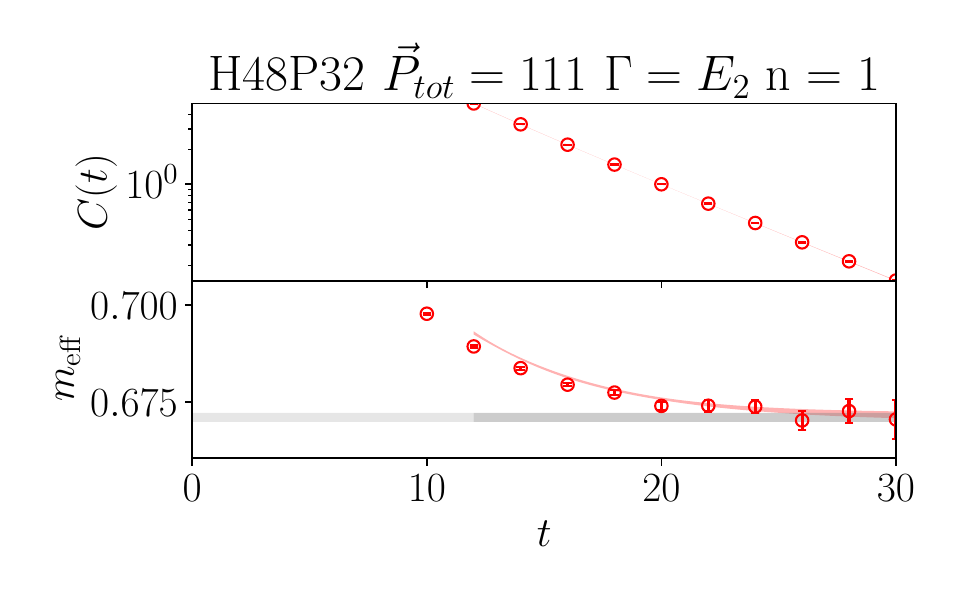}
\includegraphics[width=0.245\columnwidth]{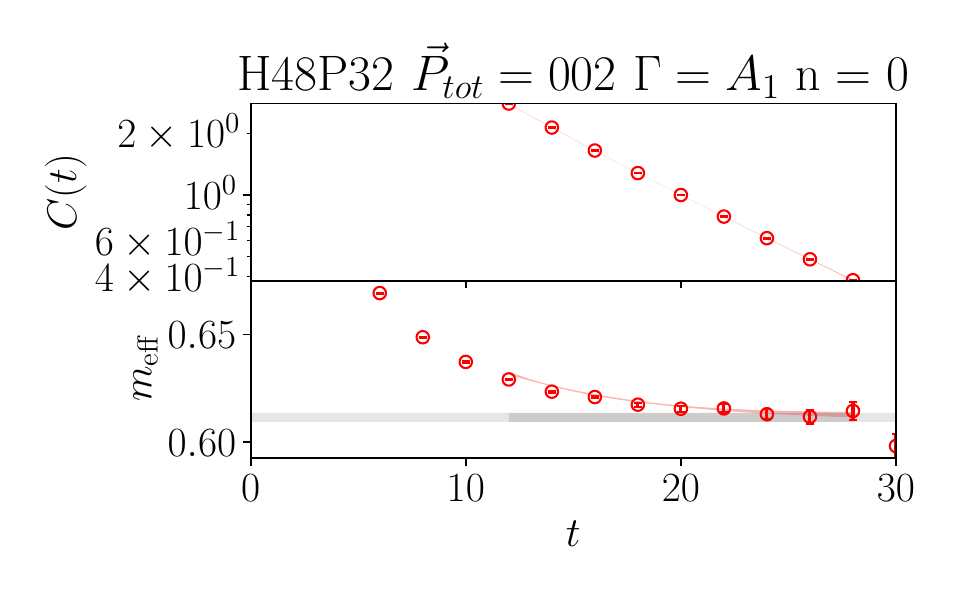}
\includegraphics[width=0.245\columnwidth]{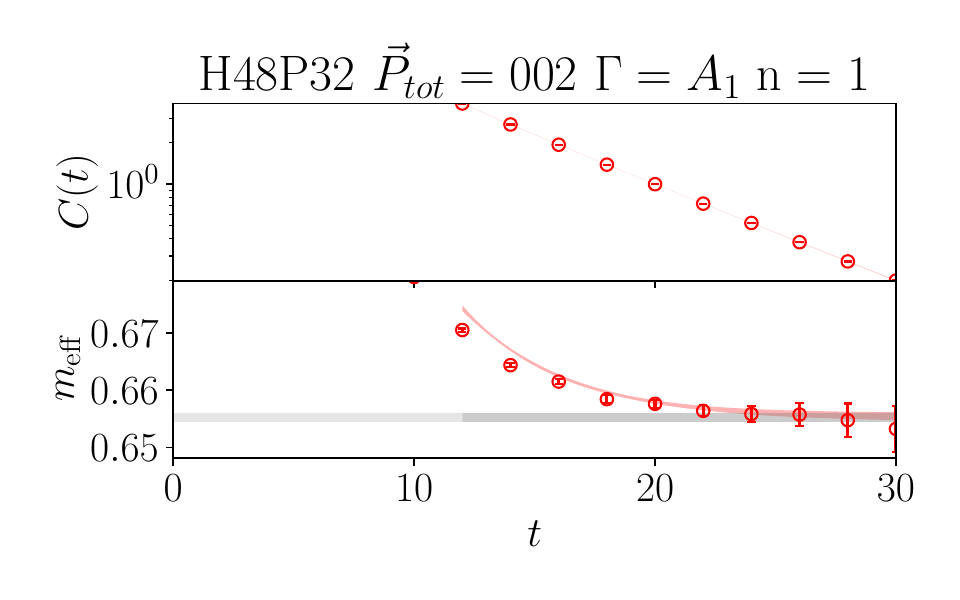}
\\
\includegraphics[width=0.245\columnwidth]{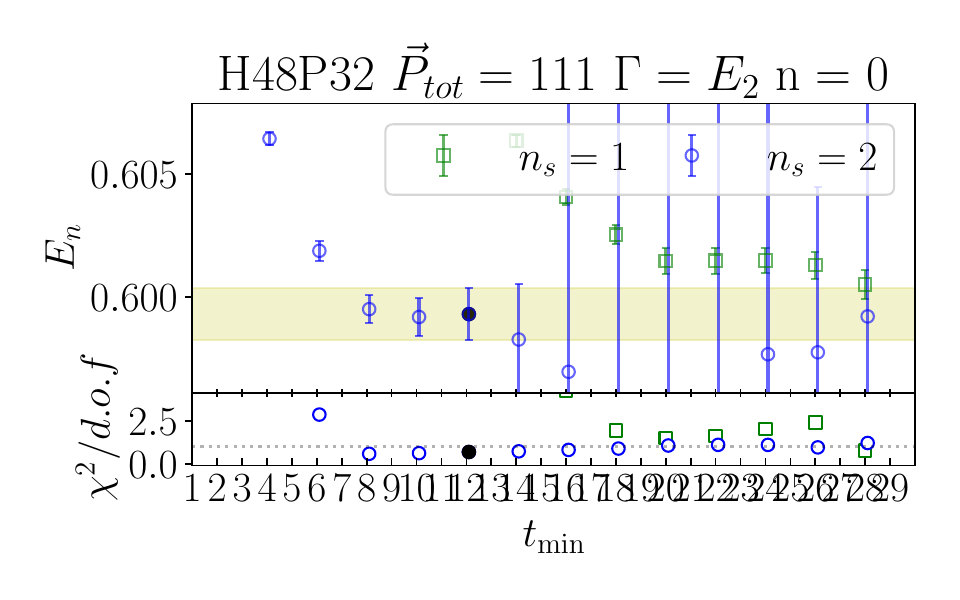}
\includegraphics[width=0.245\columnwidth]{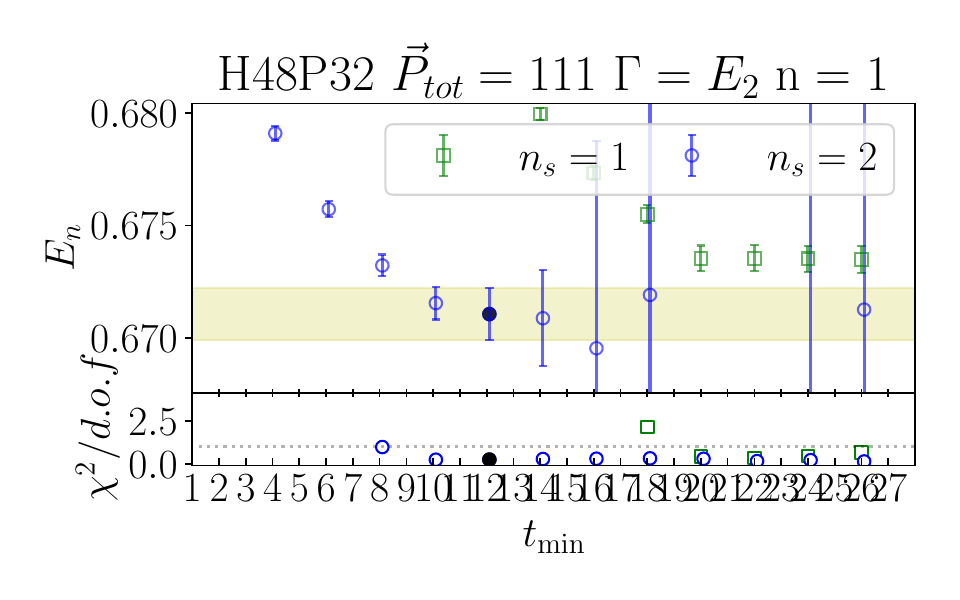}
\includegraphics[width=0.245\columnwidth]{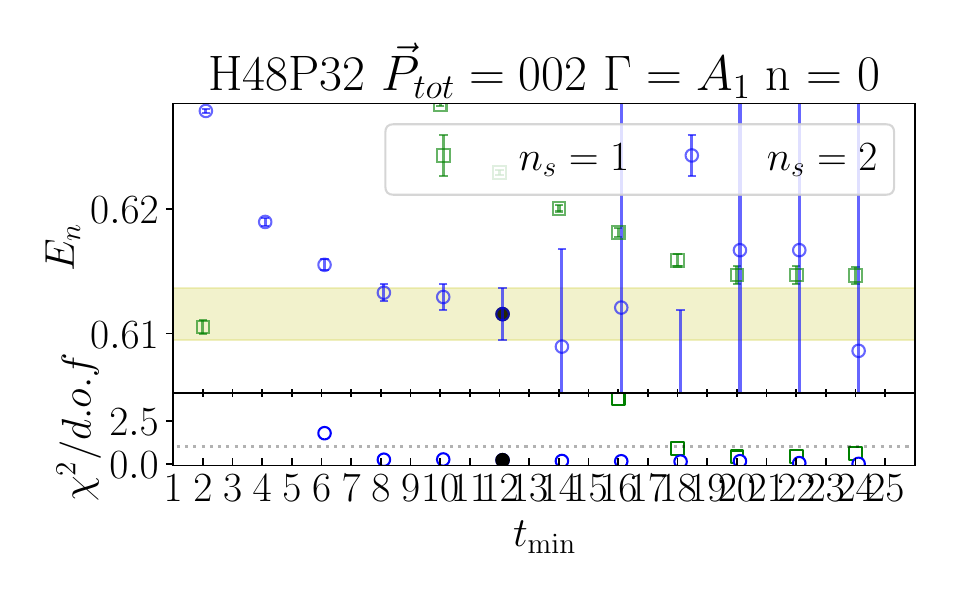}
\includegraphics[width=0.245\columnwidth]{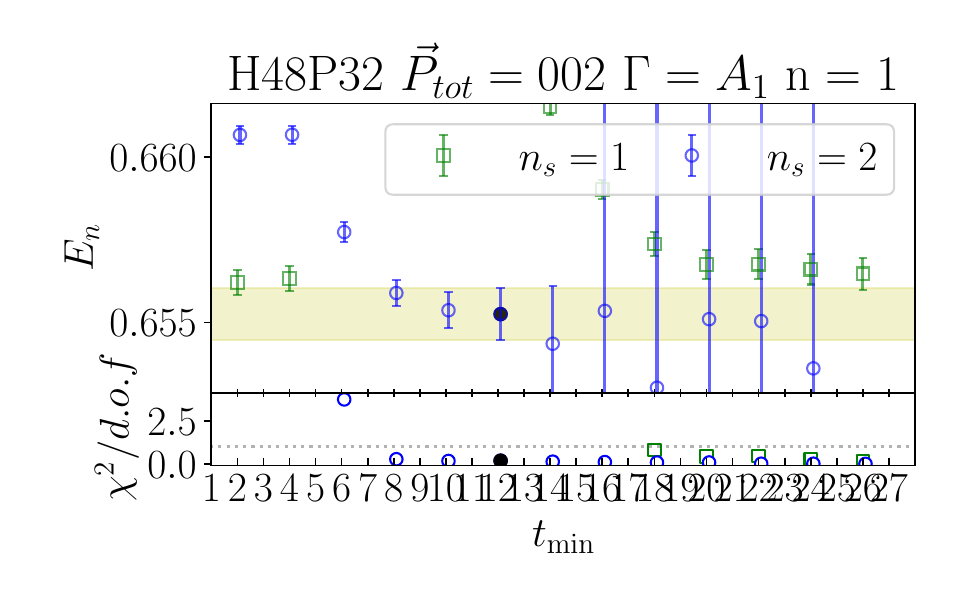}
\\
\includegraphics[width=0.245\columnwidth]{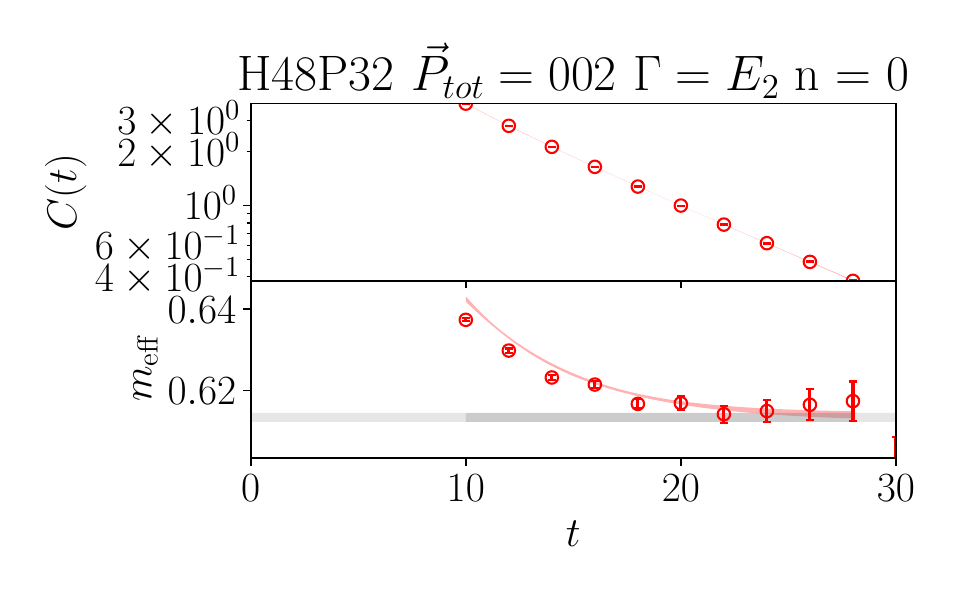}
\includegraphics[width=0.245\columnwidth]{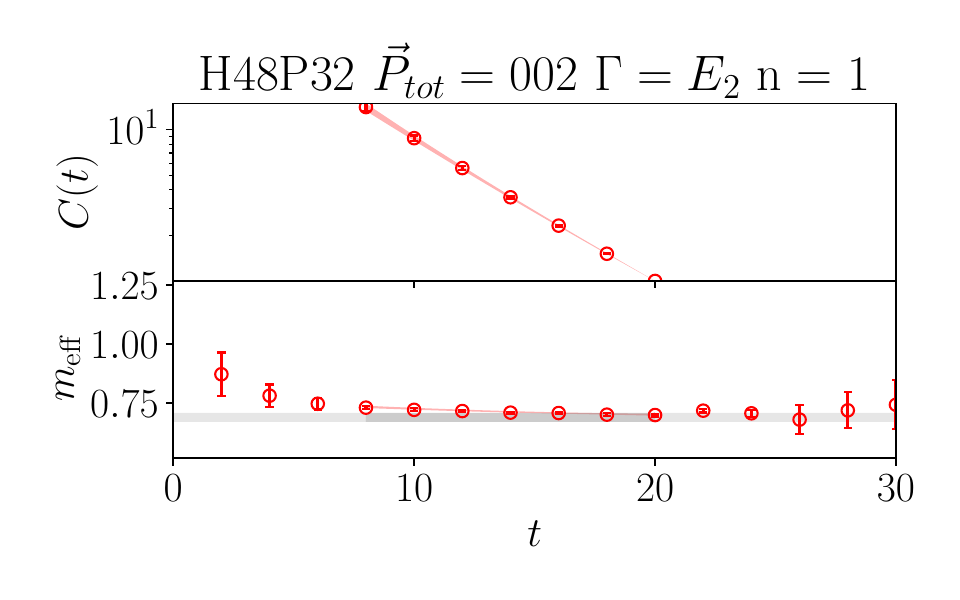}
\\
\includegraphics[width=0.245\columnwidth]{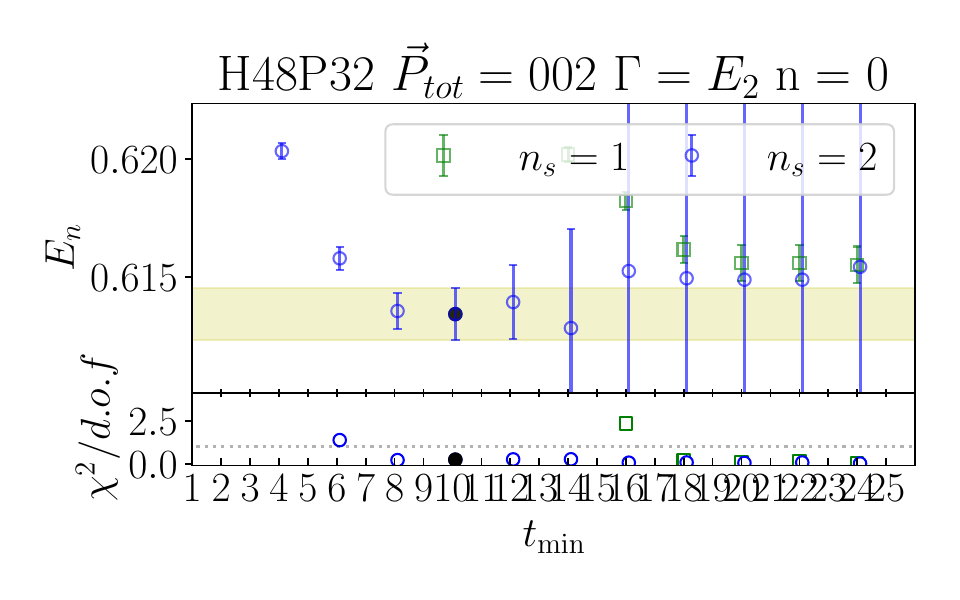}
\includegraphics[width=0.245\columnwidth]{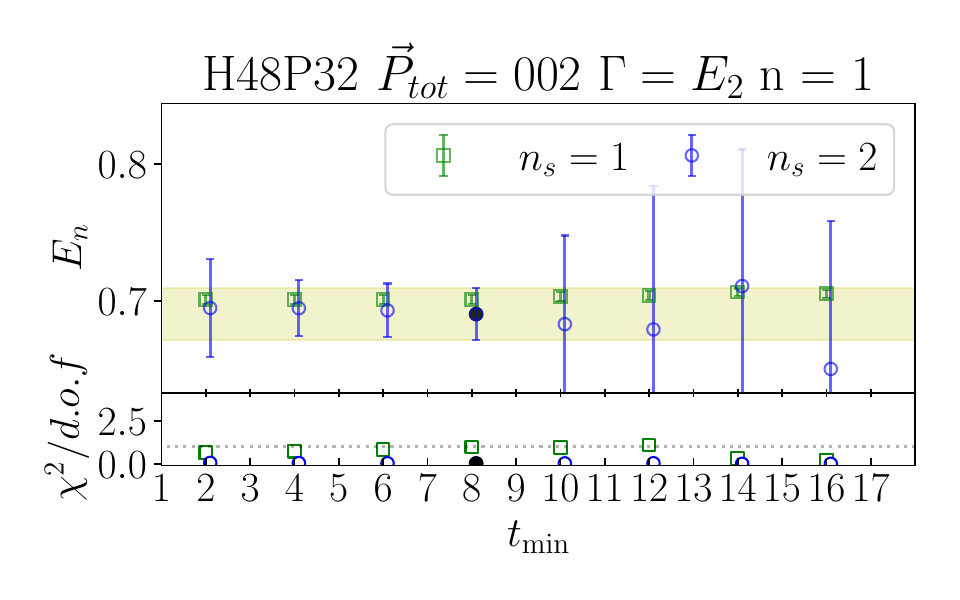}
\caption{Continued from Figure~\ref{fig:Dpi-fit-H48P32}. Energy-level fit results for the $I=\frac{1}{2}$ $D\pi$ channel on the H48P32 ensemble.}
\label{fig:Dpi-fit-H48P322}
\end{figure}

\begin{figure}[htbp]
\centering
\includegraphics[width=0.245\columnwidth]{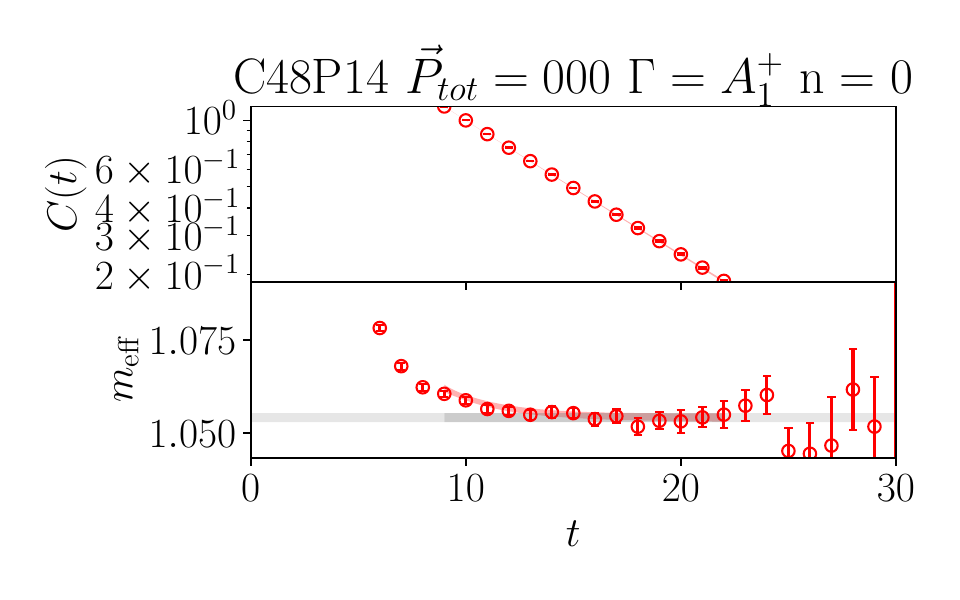}
\includegraphics[width=0.245\columnwidth]{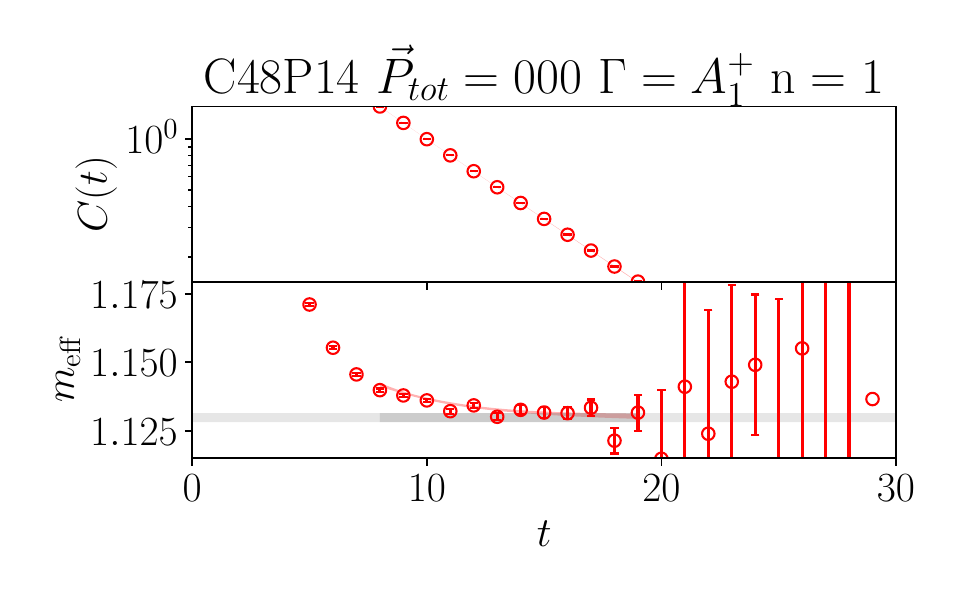}
\includegraphics[width=0.245\columnwidth]{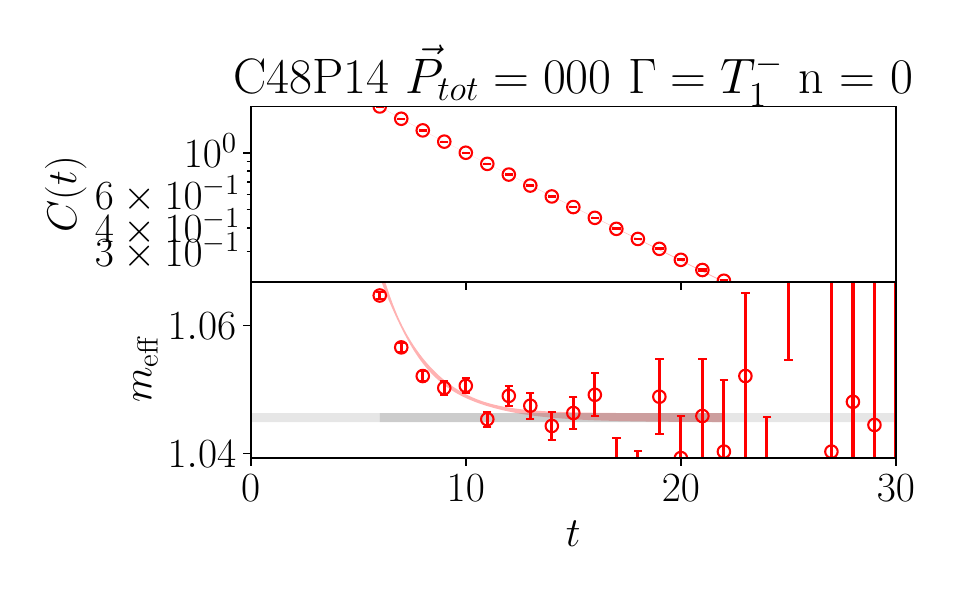}
\includegraphics[width=0.245\columnwidth]{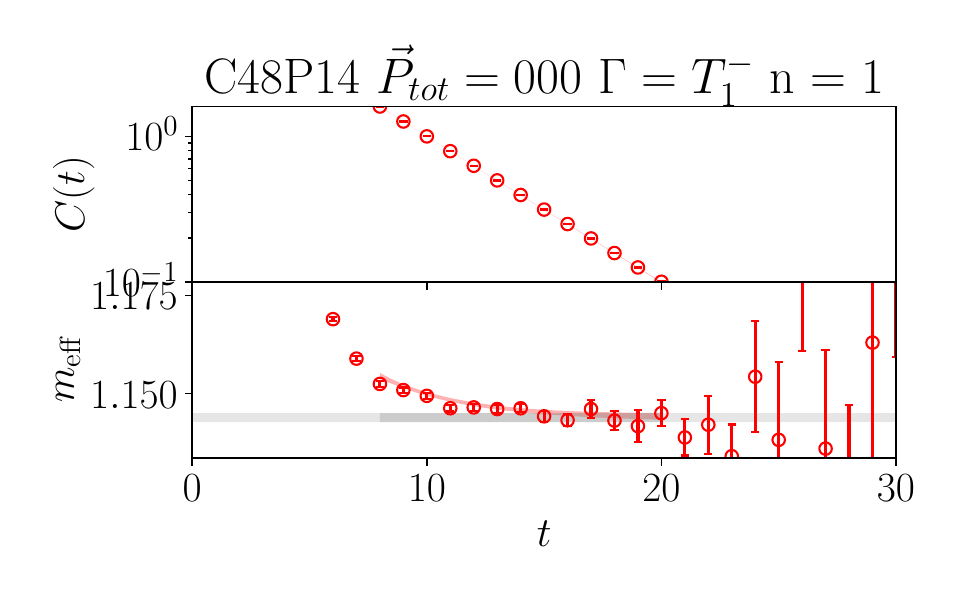}
\\
\includegraphics[width=0.245\columnwidth]{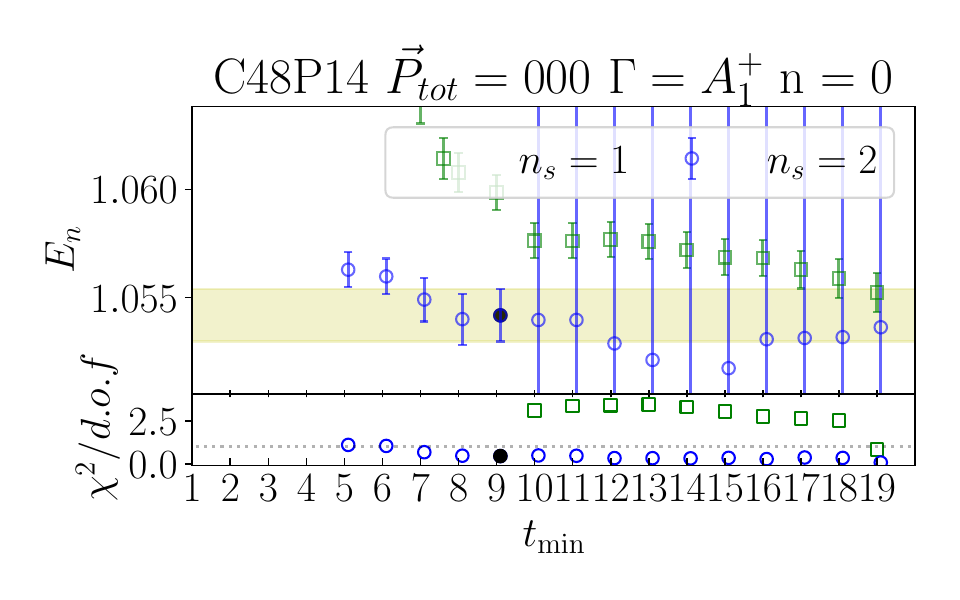}
\includegraphics[width=0.245\columnwidth]{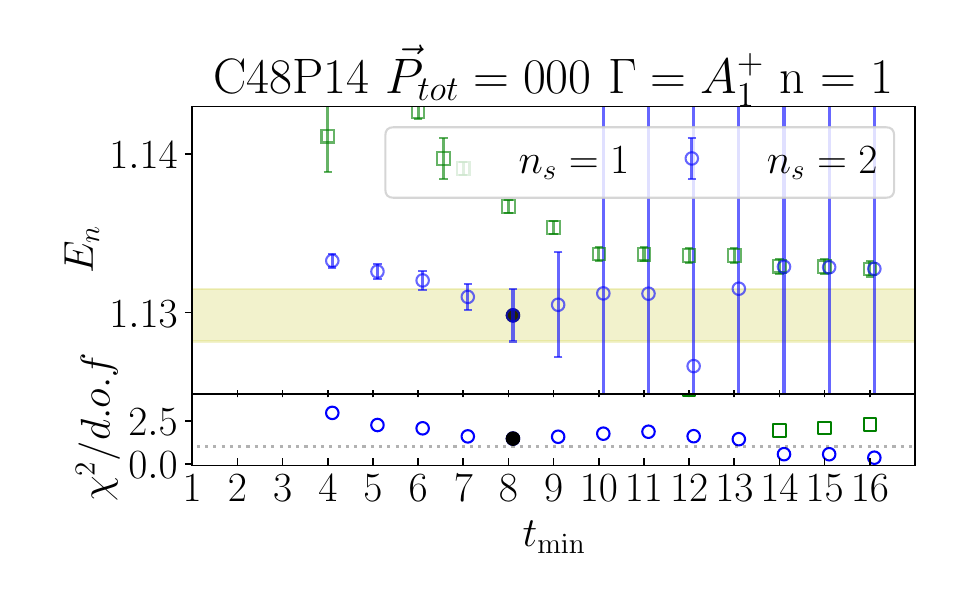}
\includegraphics[width=0.245\columnwidth]{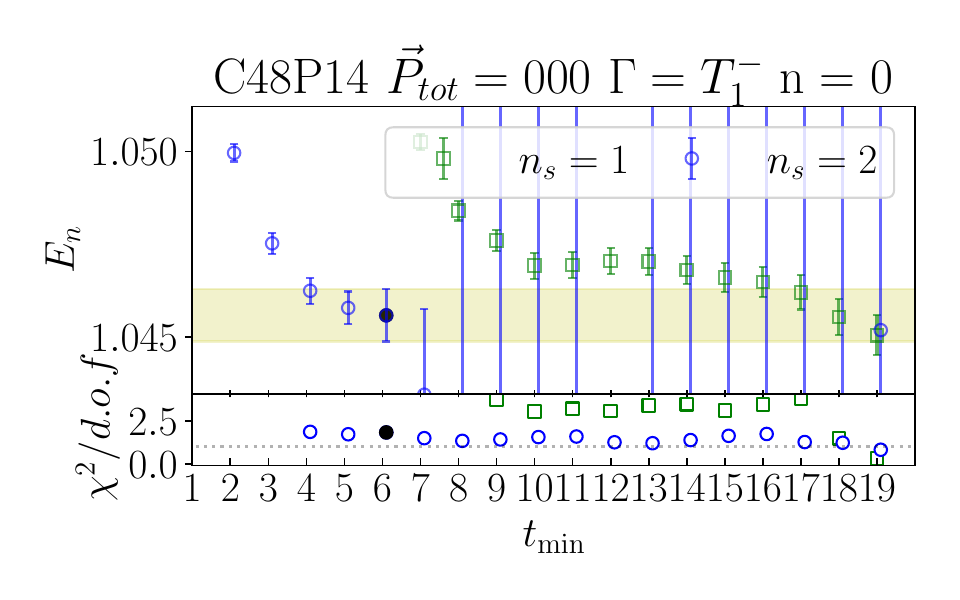}
\includegraphics[width=0.245\columnwidth]{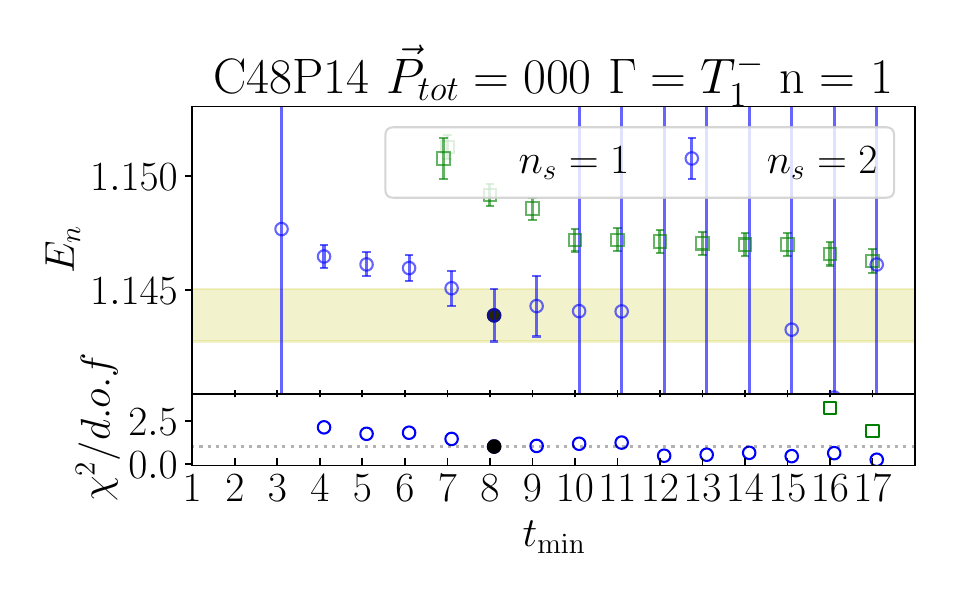}
\\
\includegraphics[width=0.245\columnwidth]{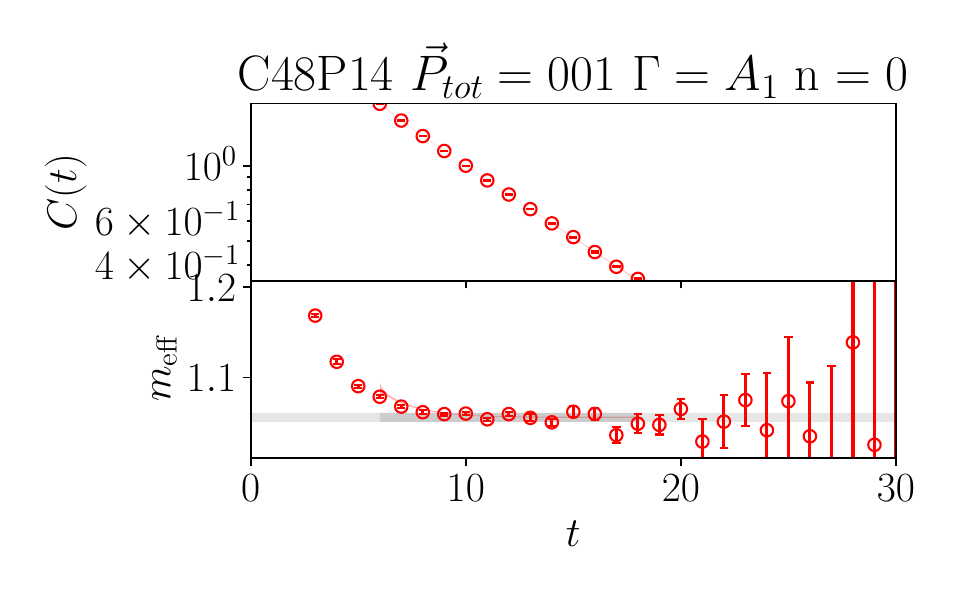}
\includegraphics[width=0.245\columnwidth]{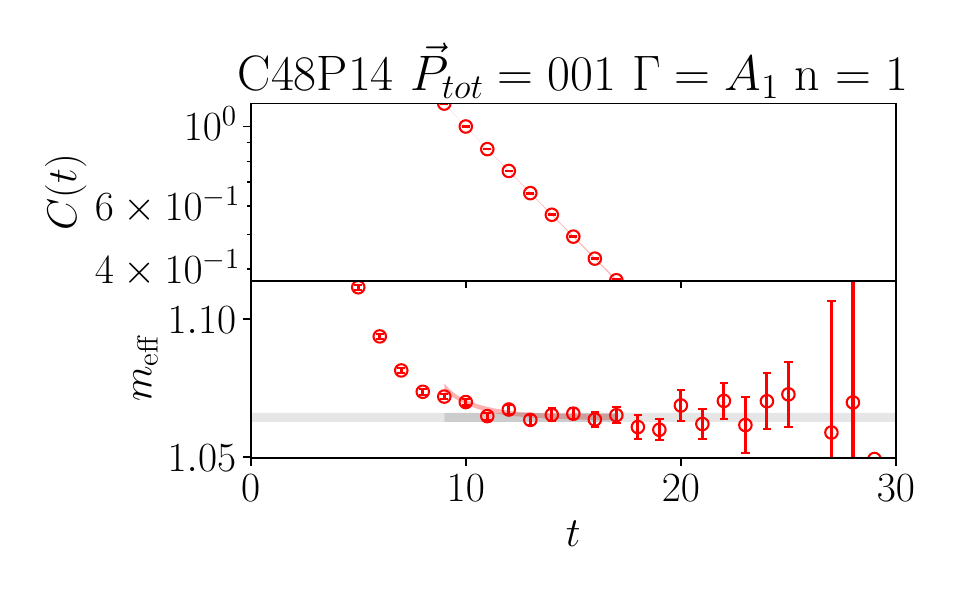}
\includegraphics[width=0.245\columnwidth]{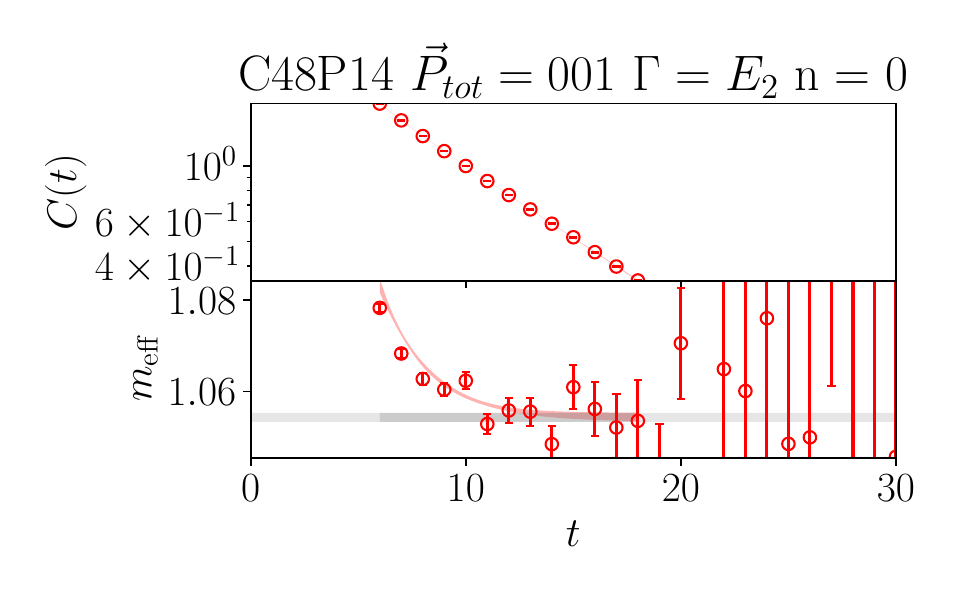}
\includegraphics[width=0.245\columnwidth]{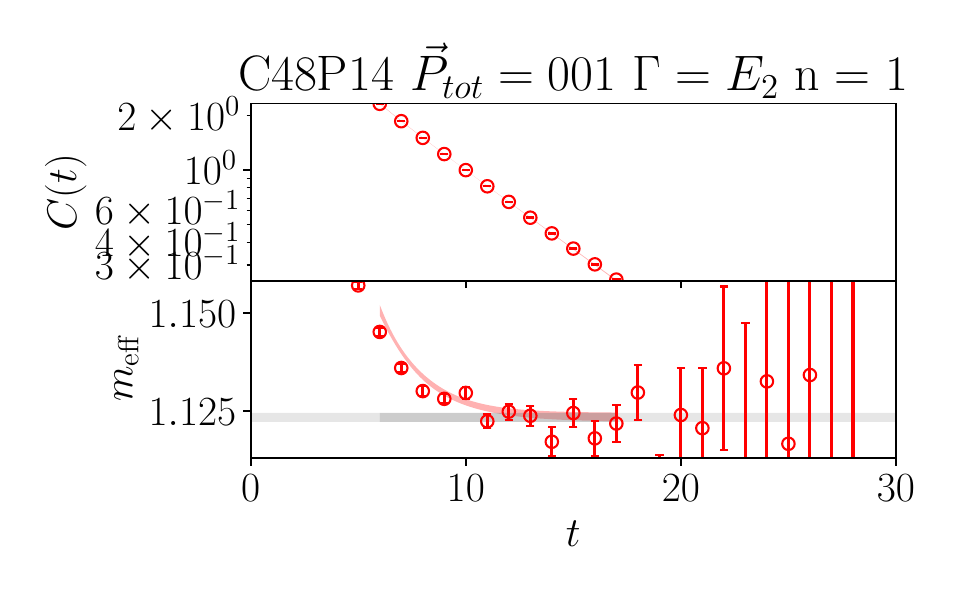}
\\
\includegraphics[width=0.245\columnwidth]{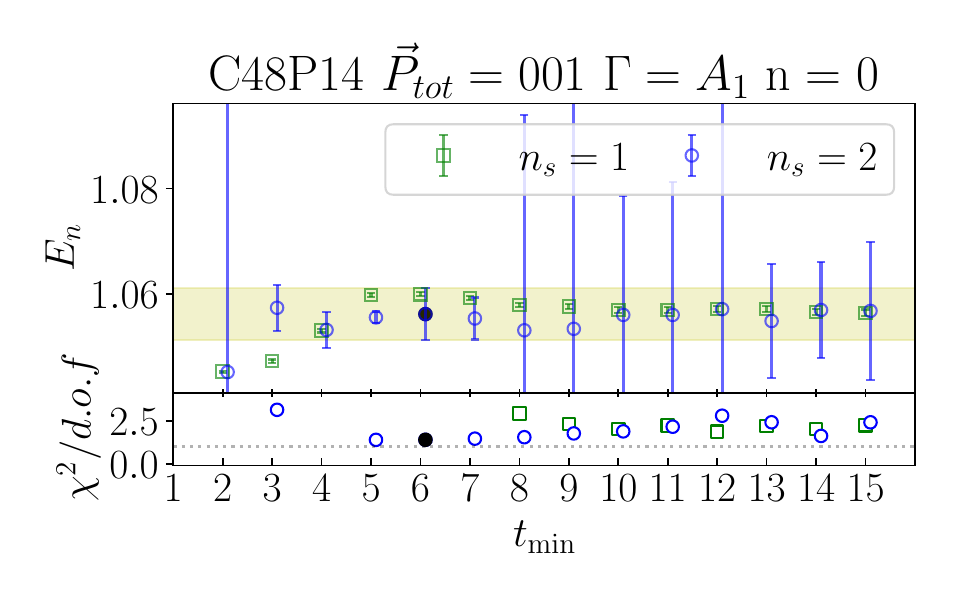}
\includegraphics[width=0.245\columnwidth]{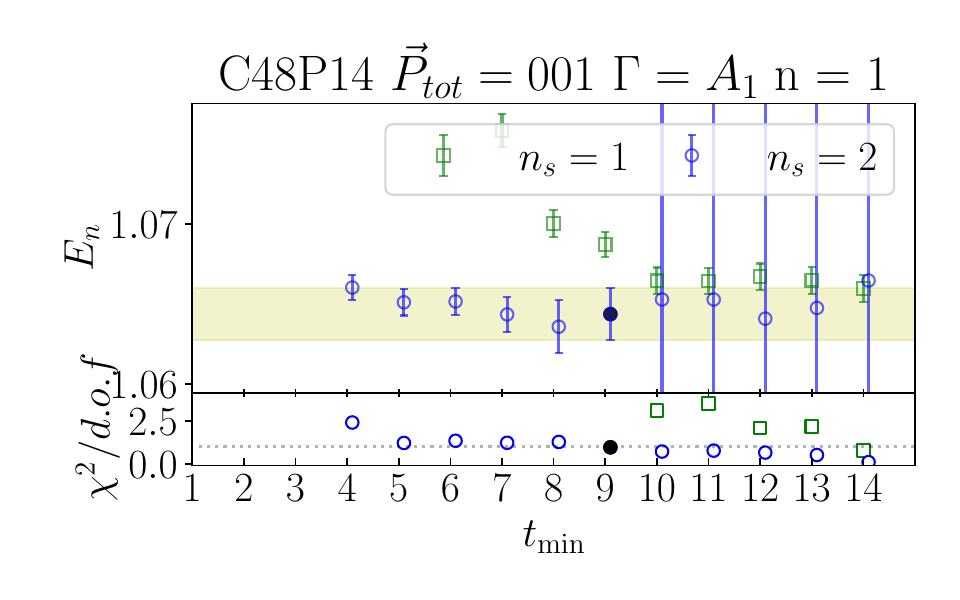}
\includegraphics[width=0.245\columnwidth]{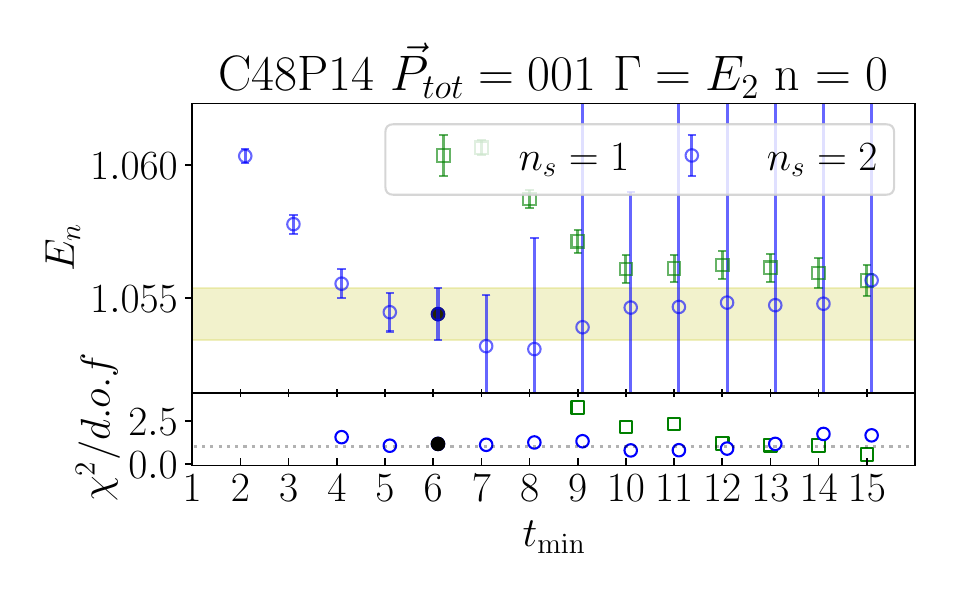}
\includegraphics[width=0.245\columnwidth]{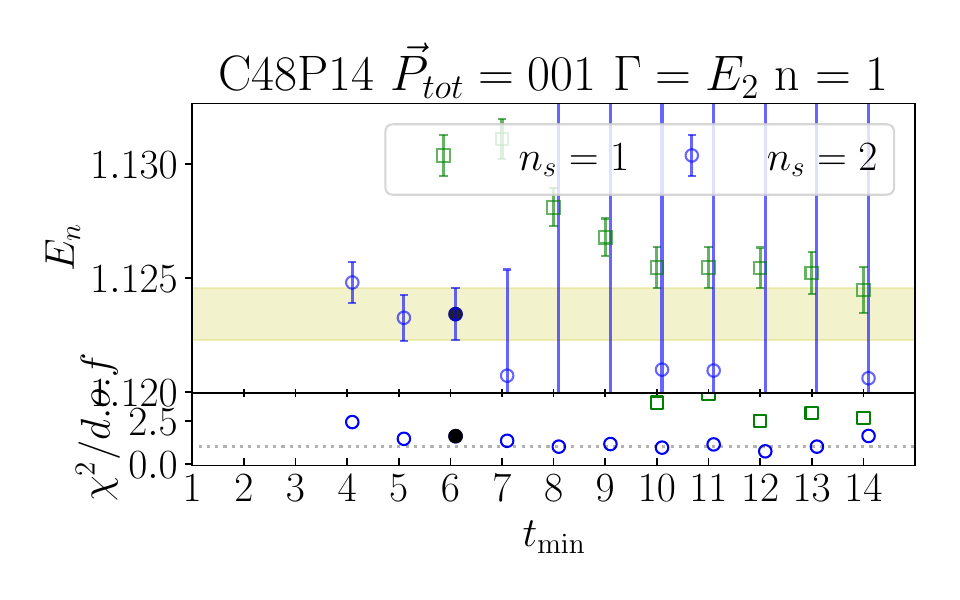}
\\
\includegraphics[width=0.245\columnwidth]{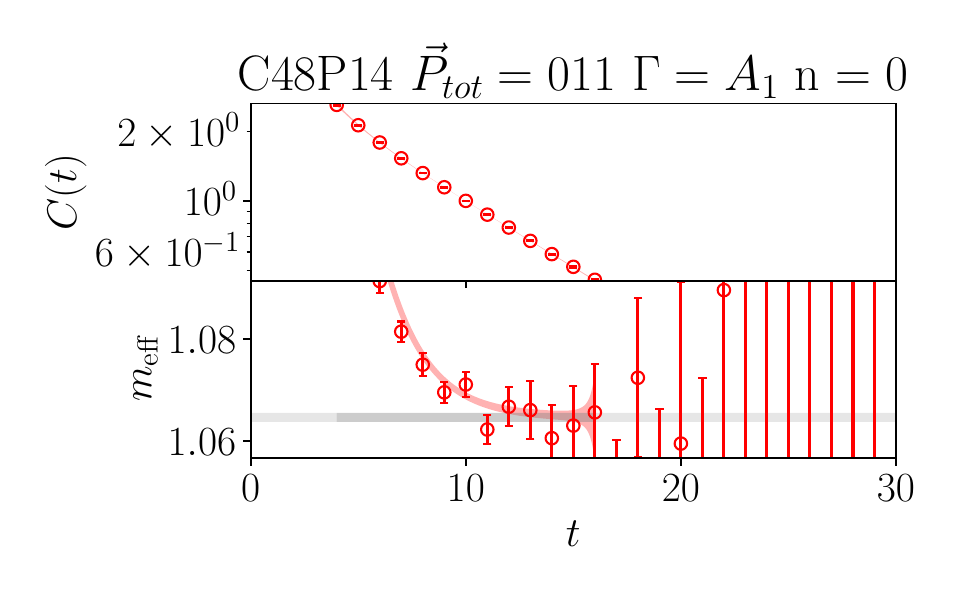}
\includegraphics[width=0.245\columnwidth]{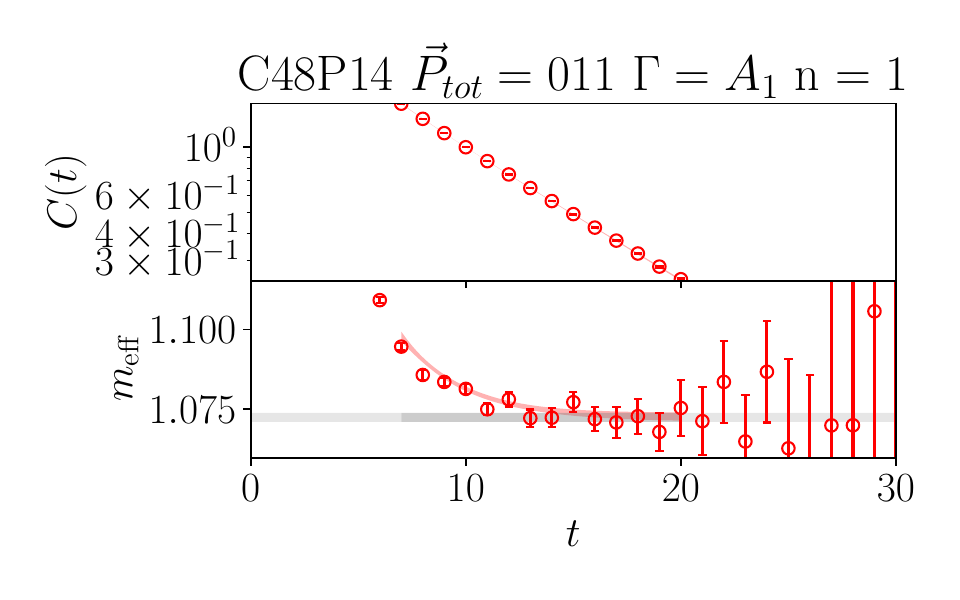}
\includegraphics[width=0.245\columnwidth]{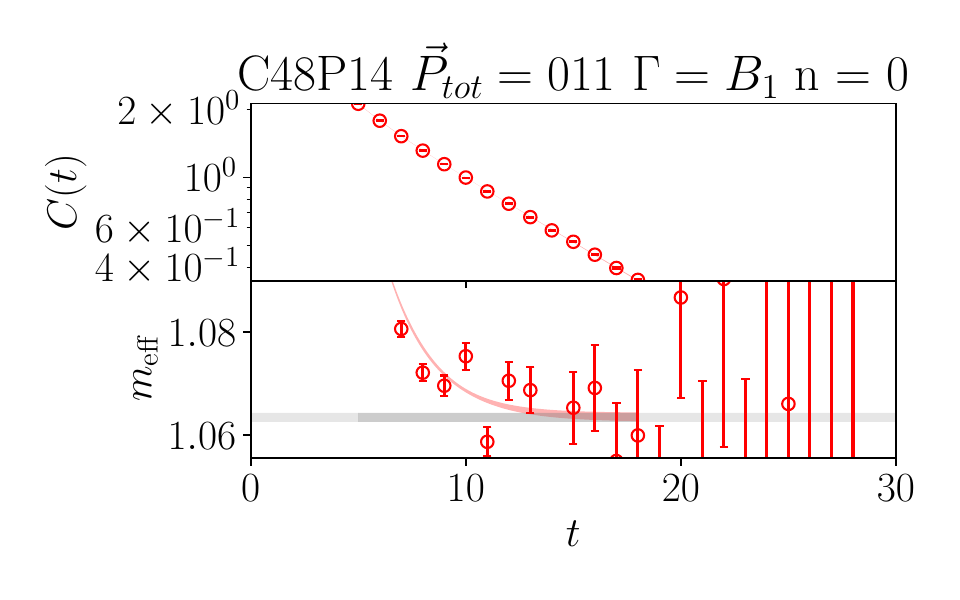}
\includegraphics[width=0.245\columnwidth]{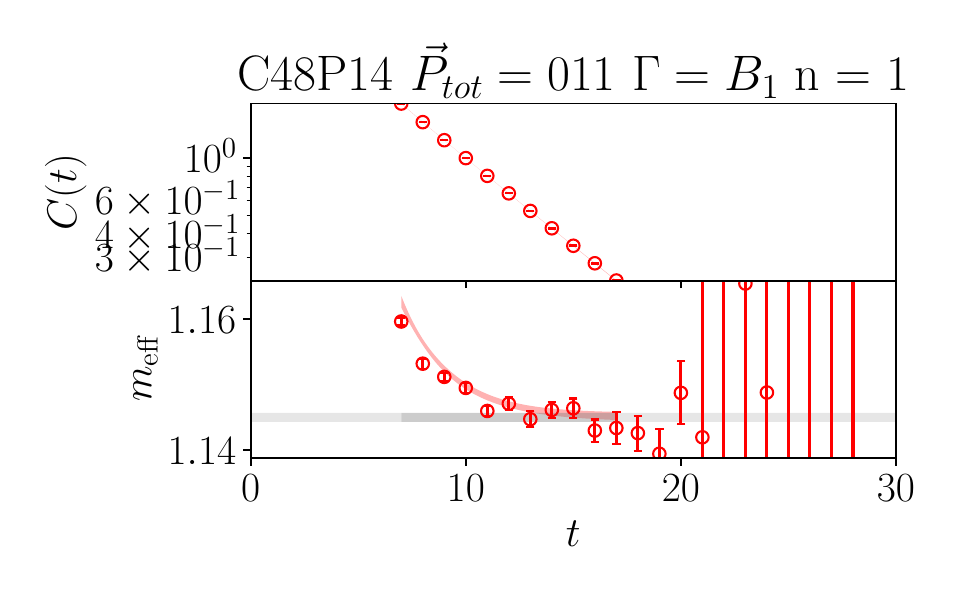}
\\
\includegraphics[width=0.245\columnwidth]{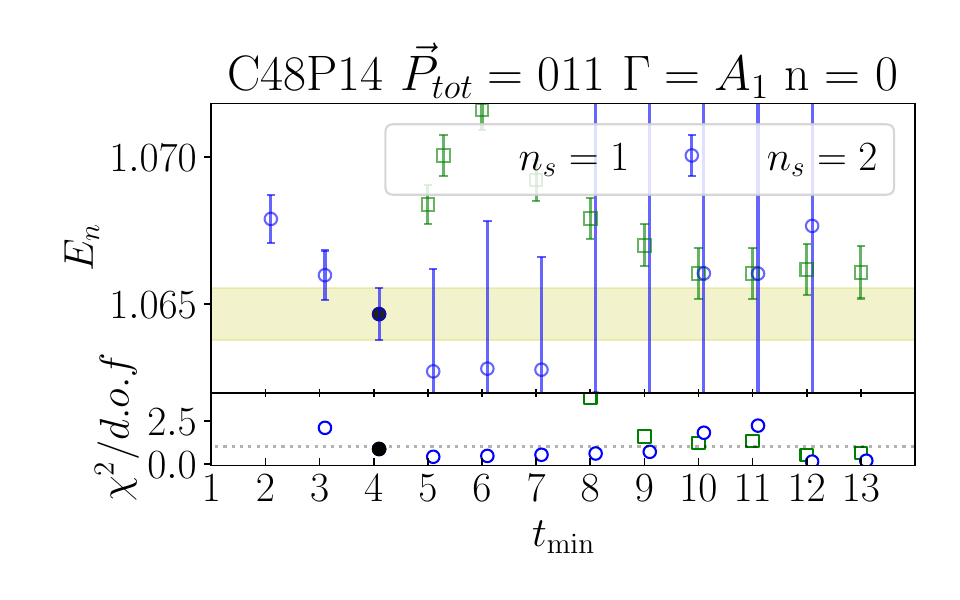}
\includegraphics[width=0.245\columnwidth]{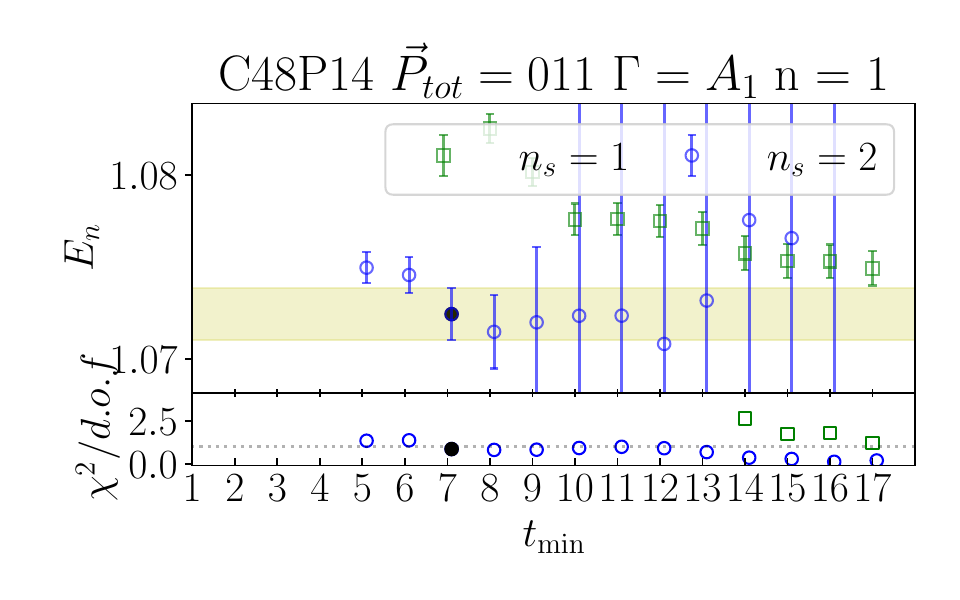}
\includegraphics[width=0.245\columnwidth]{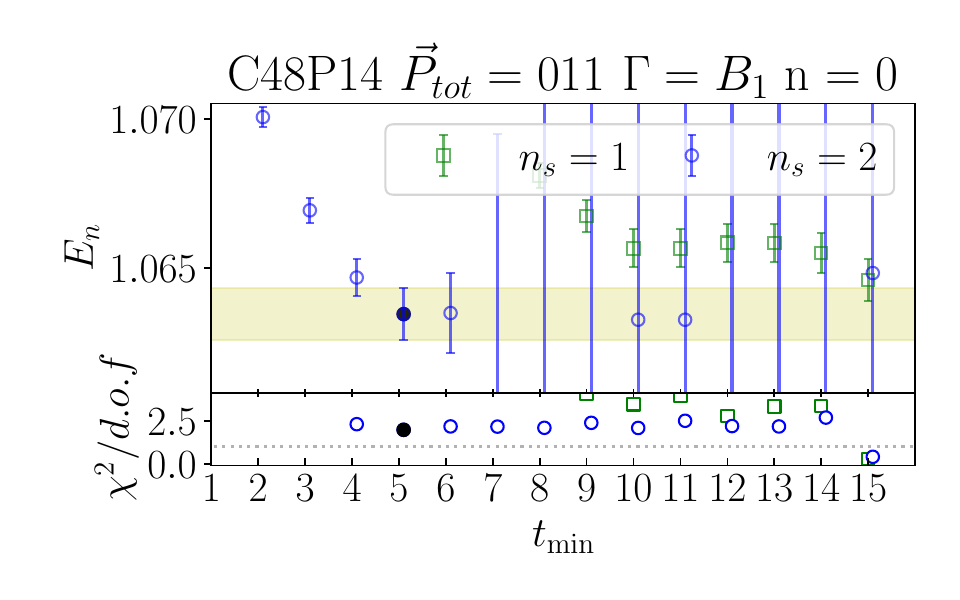}
\includegraphics[width=0.245\columnwidth]{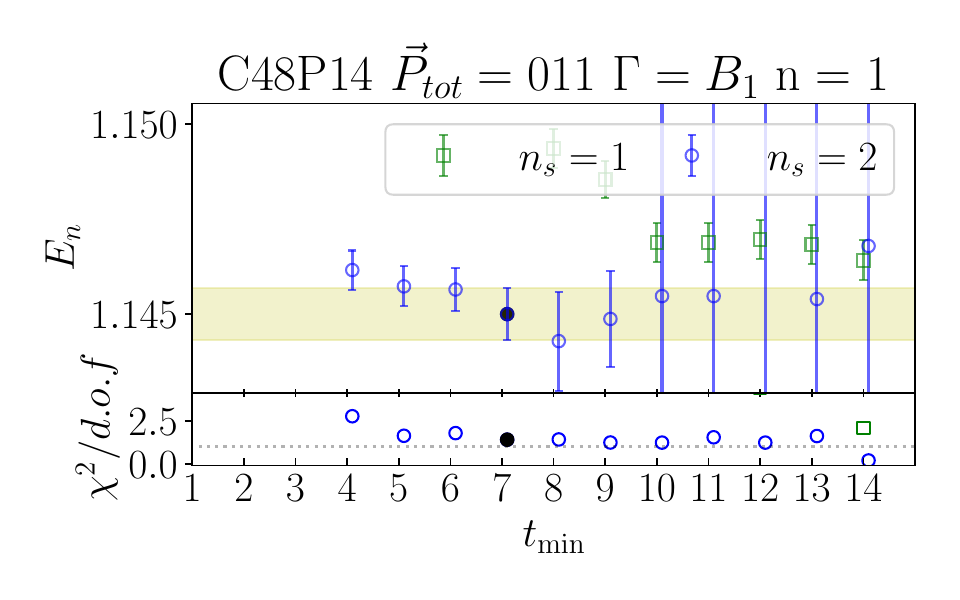}
\\
\includegraphics[width=0.245\columnwidth]{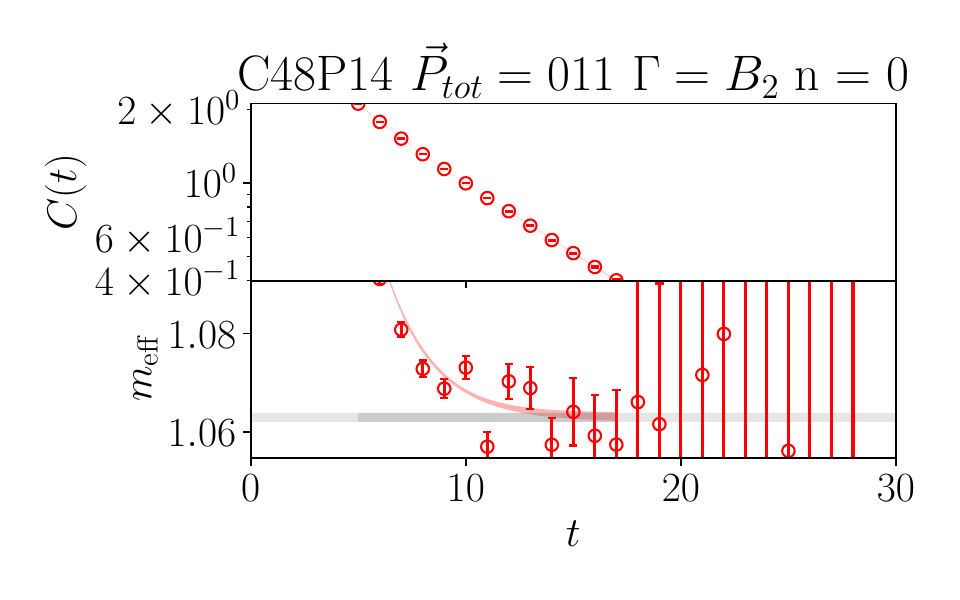}
\includegraphics[width=0.245\columnwidth]{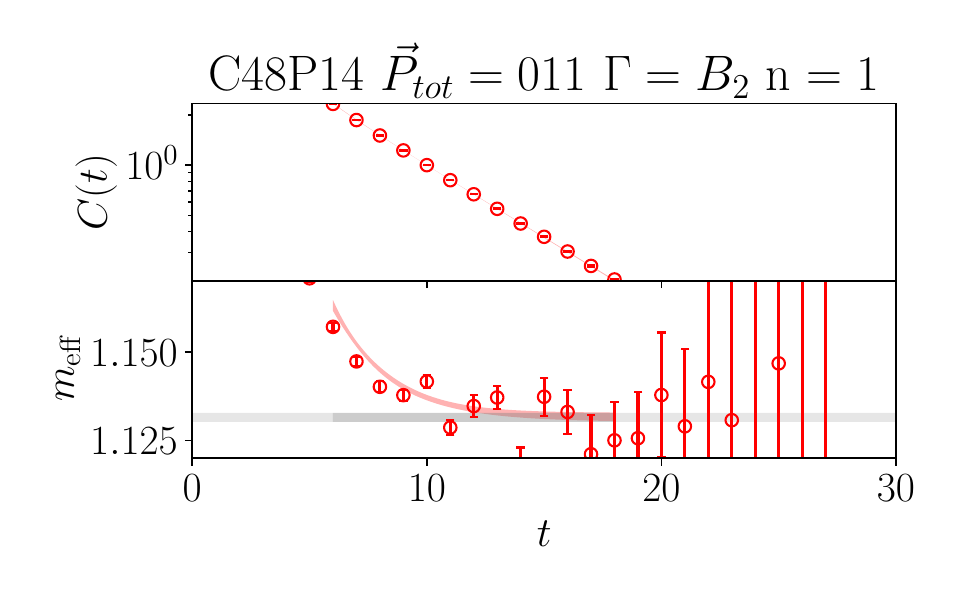}
\includegraphics[width=0.245\columnwidth]{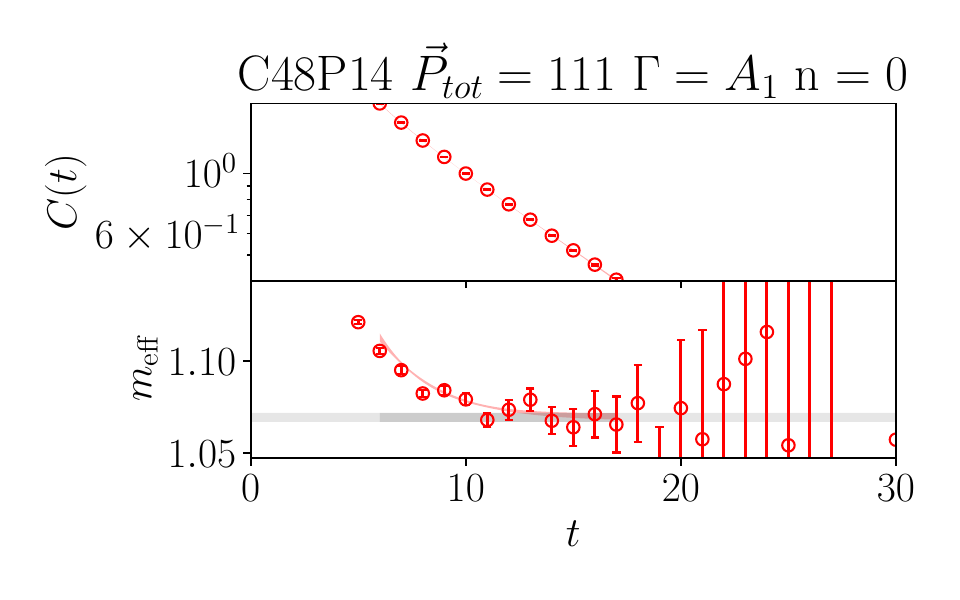}
\includegraphics[width=0.245\columnwidth]{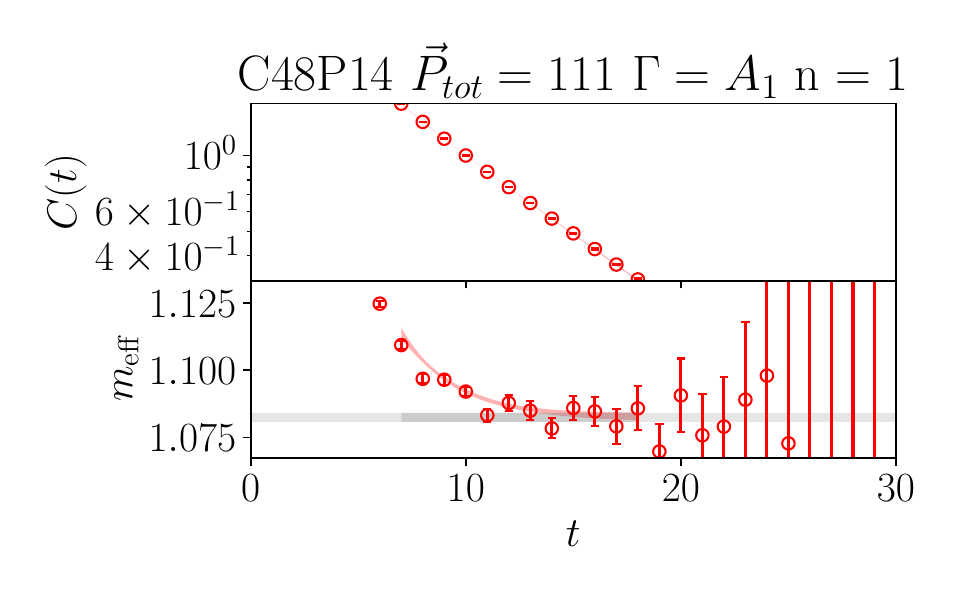}
\\
\includegraphics[width=0.245\columnwidth]{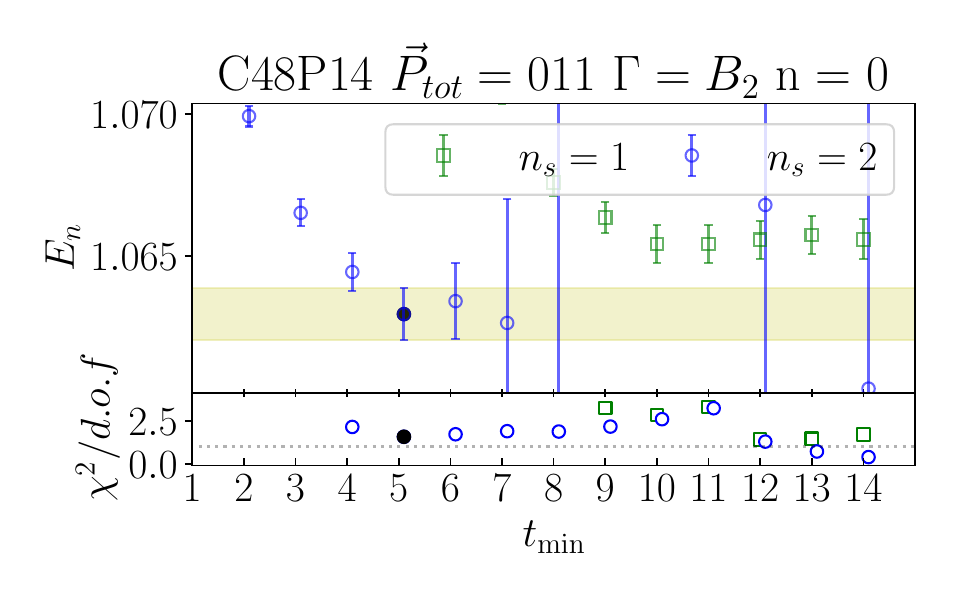}
\includegraphics[width=0.245\columnwidth]{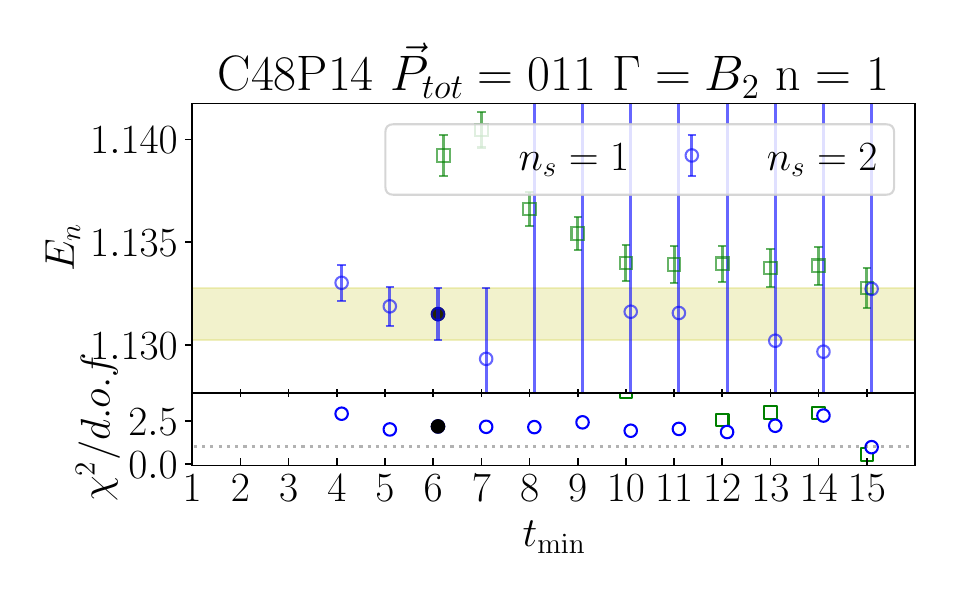}
\includegraphics[width=0.245\columnwidth]{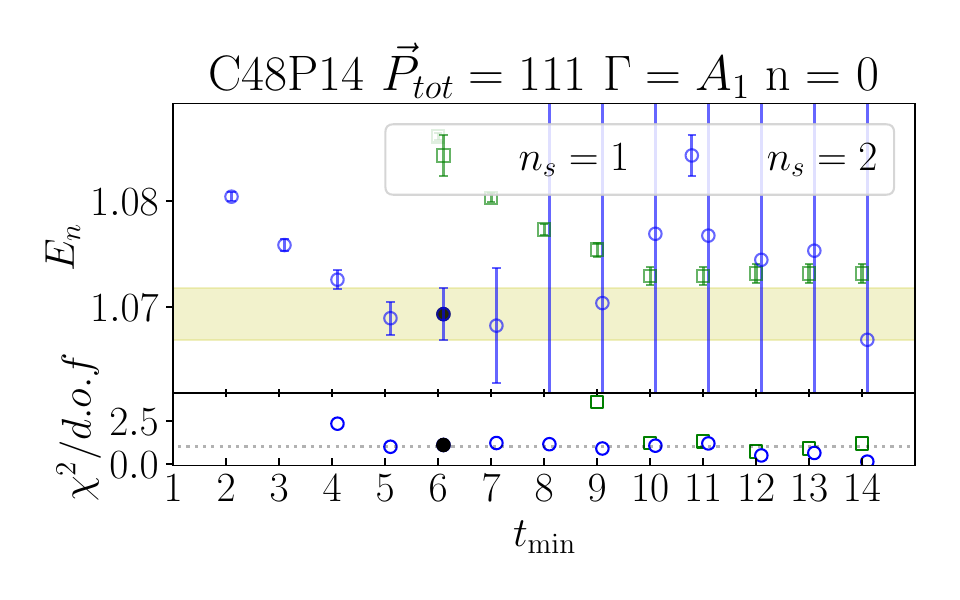}
\includegraphics[width=0.245\columnwidth]{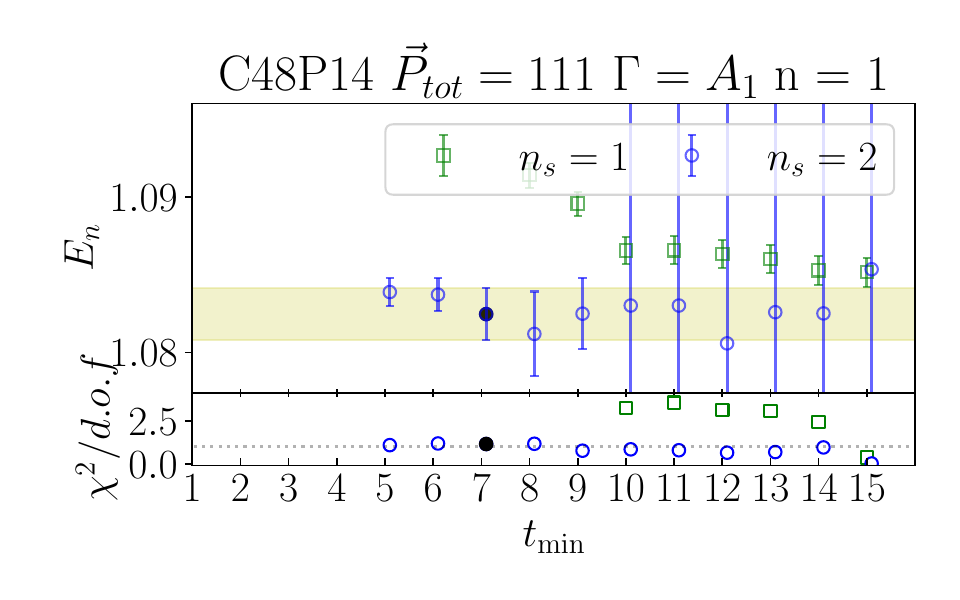}
\caption{Energy-level fit results for the $I=\frac{1}{2}$ $D\pi$ channel on the C48P14 ensemble. The description follows Figure~\ref{fig:Dpi-fit-F32P30}.}
\label{fig:Dpi-fit-C48P14}
\end{figure}

\begin{figure}[htbp]
\centering
\includegraphics[width=0.245\columnwidth]{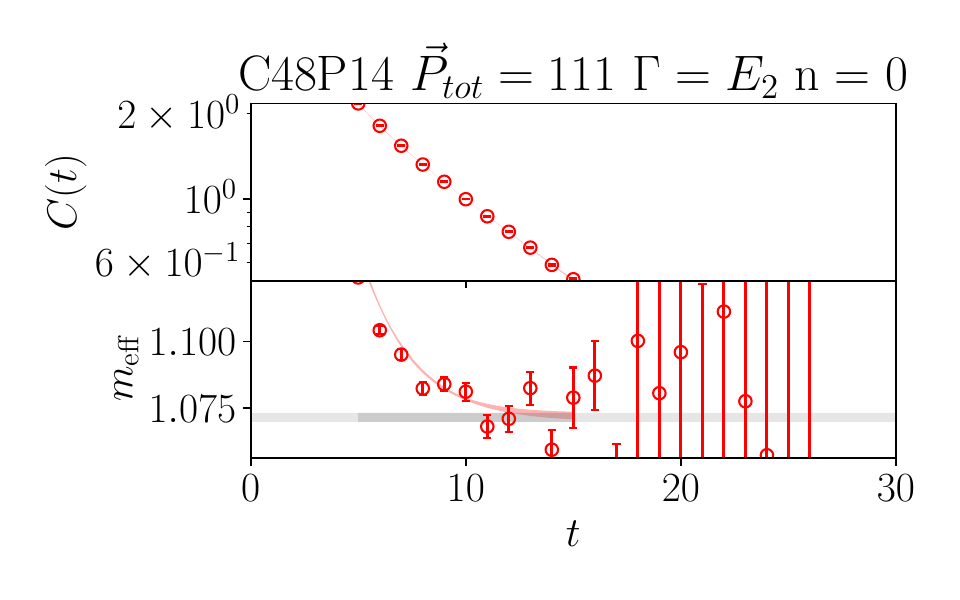}
\includegraphics[width=0.245\columnwidth]{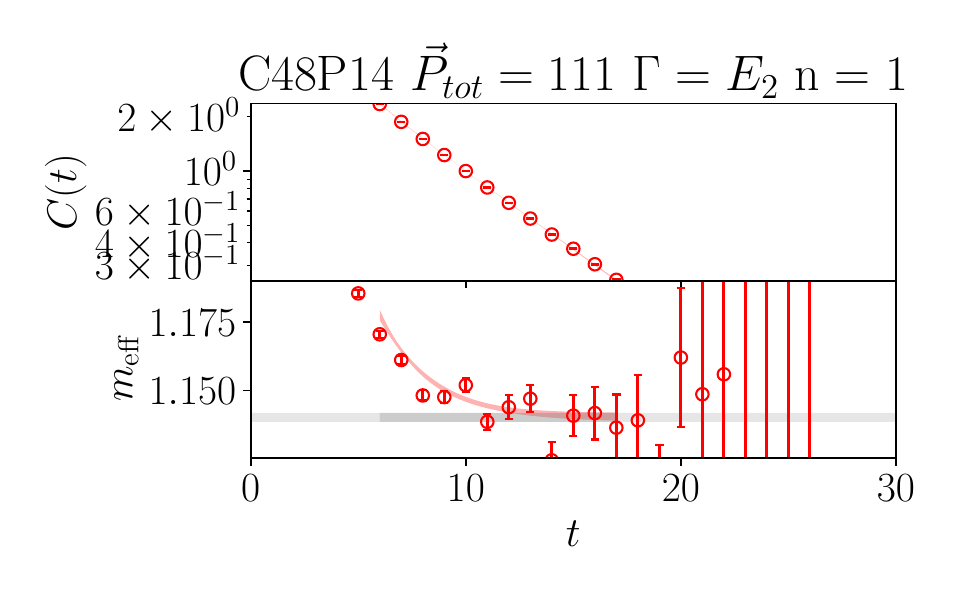}
\includegraphics[width=0.245\columnwidth]{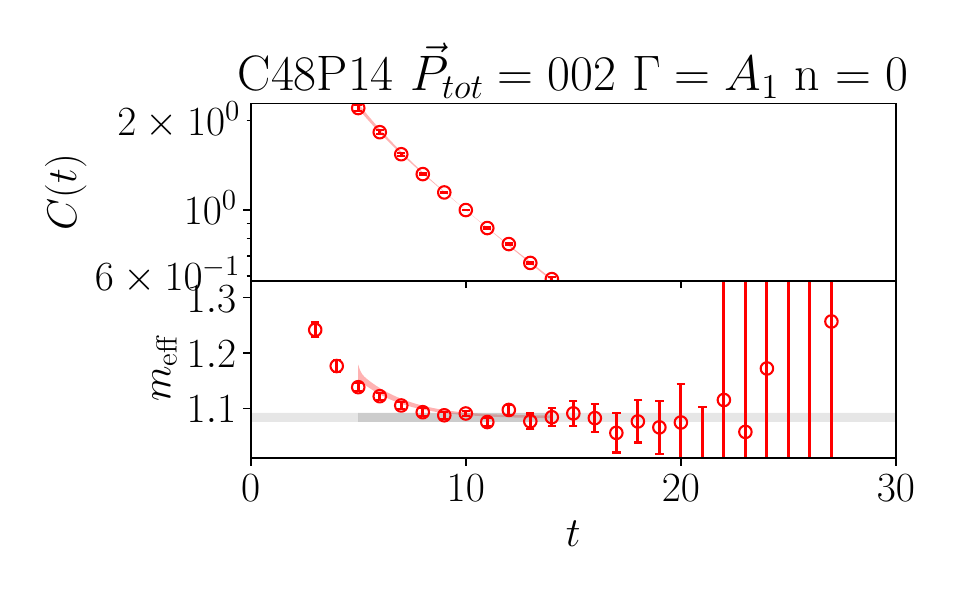}
\includegraphics[width=0.245\columnwidth]{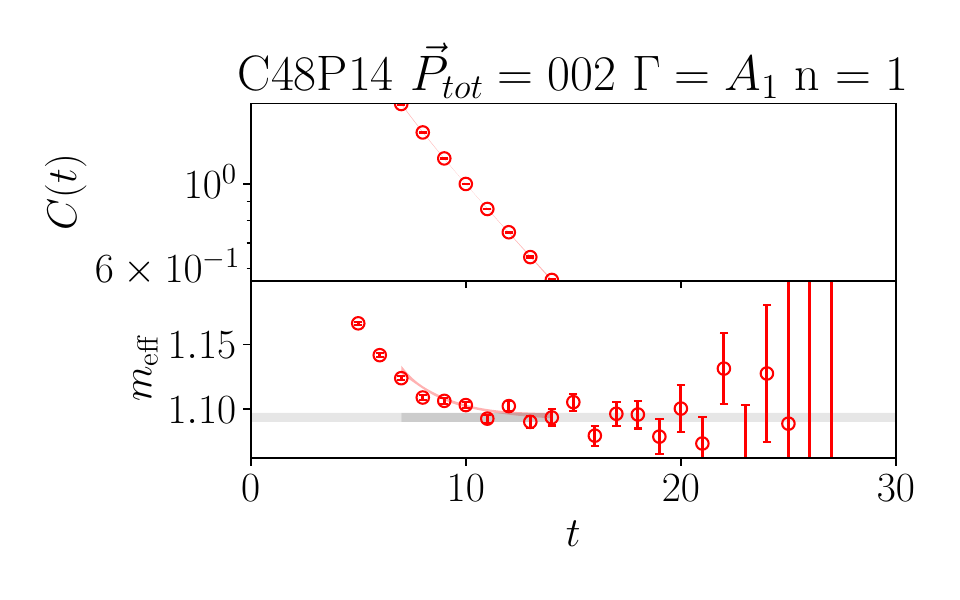}
\\
\includegraphics[width=0.245\columnwidth]{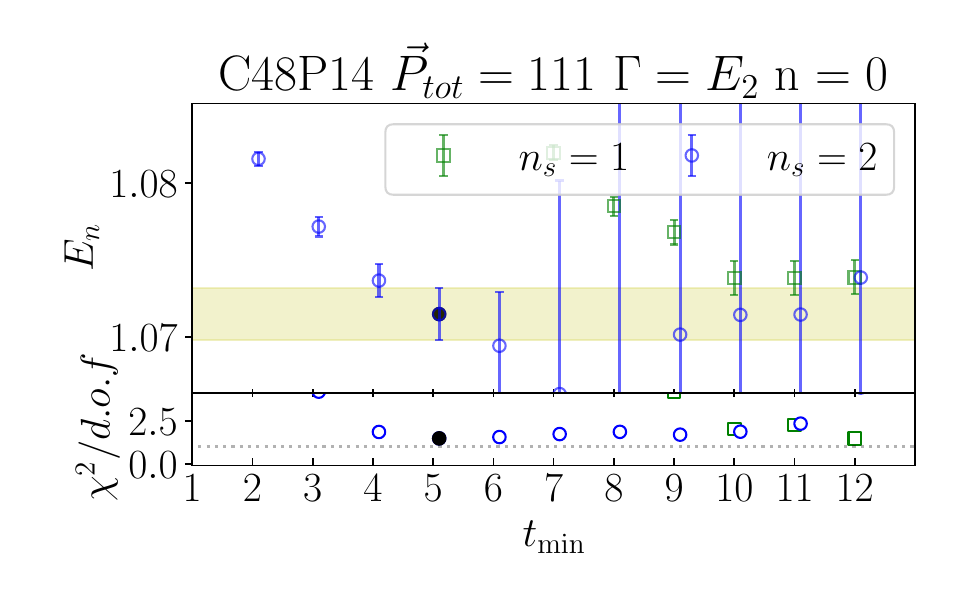}
\includegraphics[width=0.245\columnwidth]{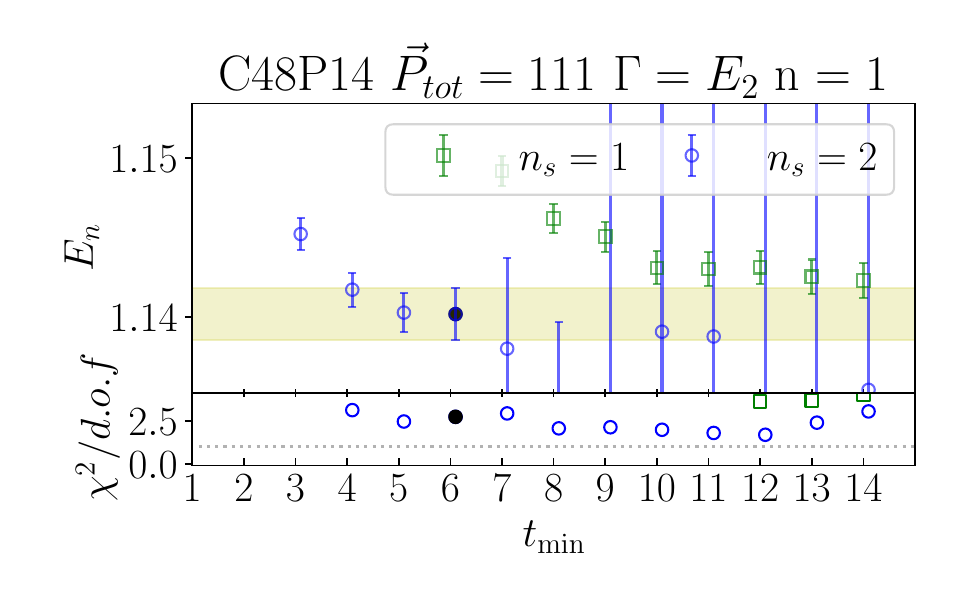}
\includegraphics[width=0.245\columnwidth]{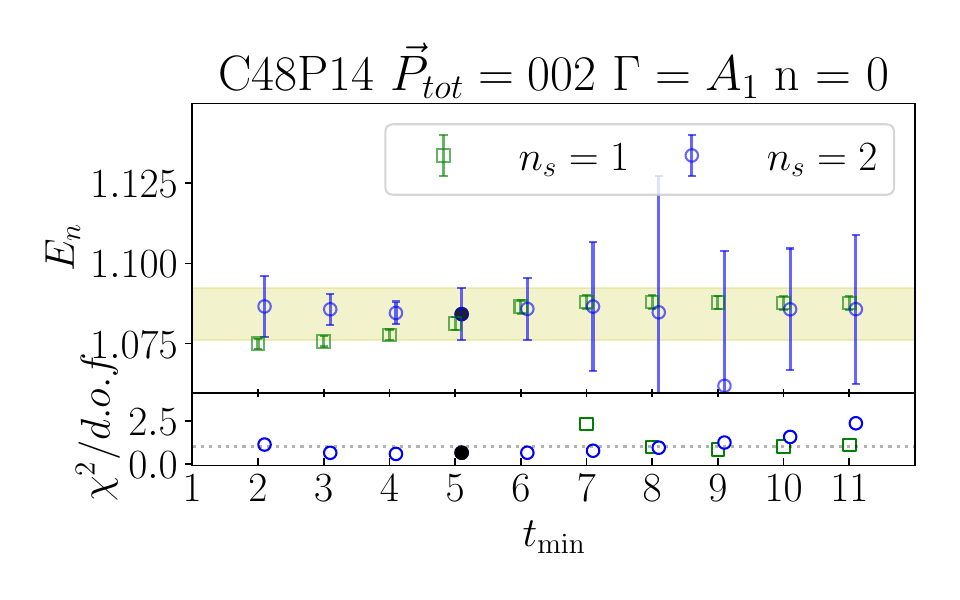}
\includegraphics[width=0.245\columnwidth]{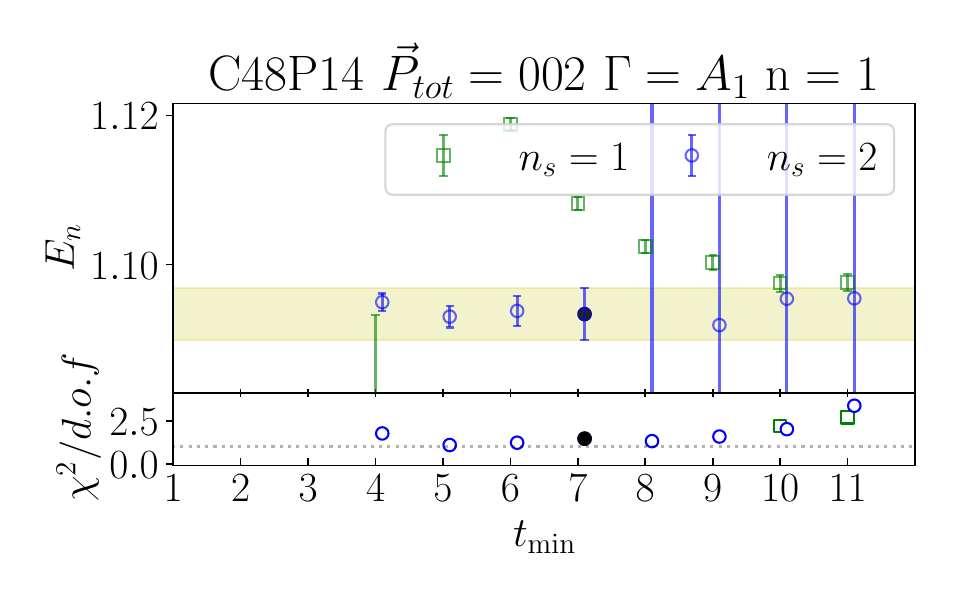}
\\
\includegraphics[width=0.245\columnwidth]{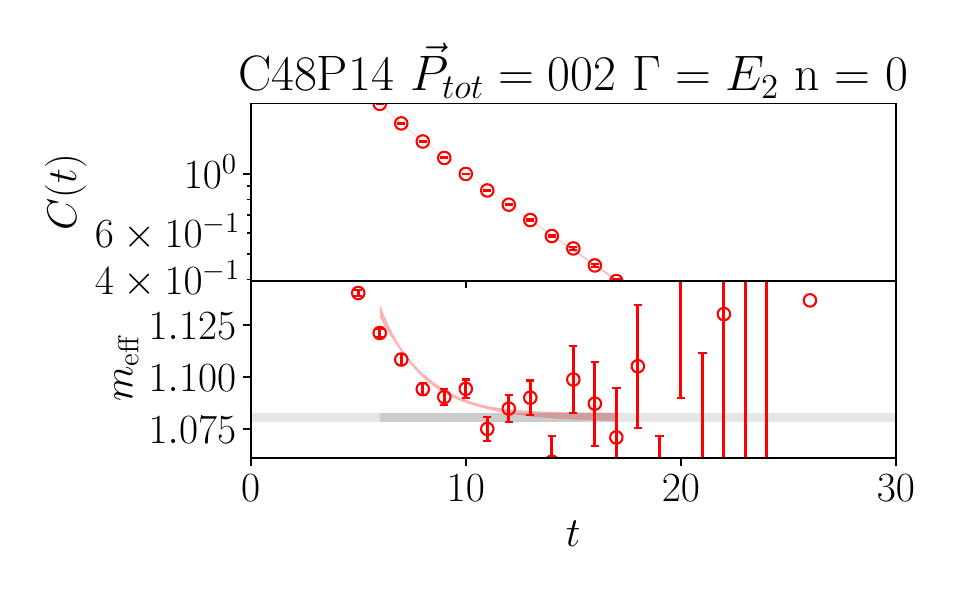}
\includegraphics[width=0.245\columnwidth]{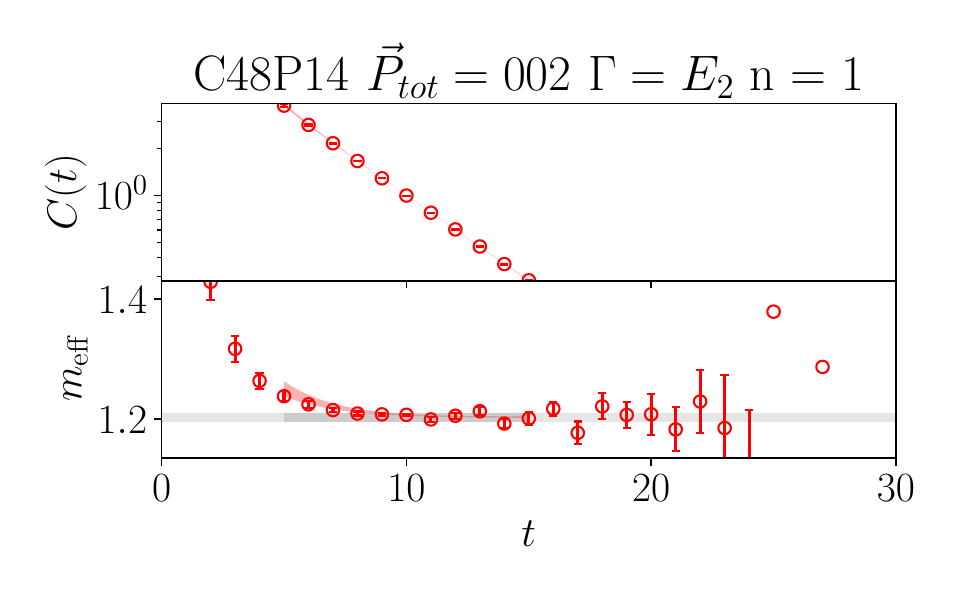}
\\
\includegraphics[width=0.245\columnwidth]{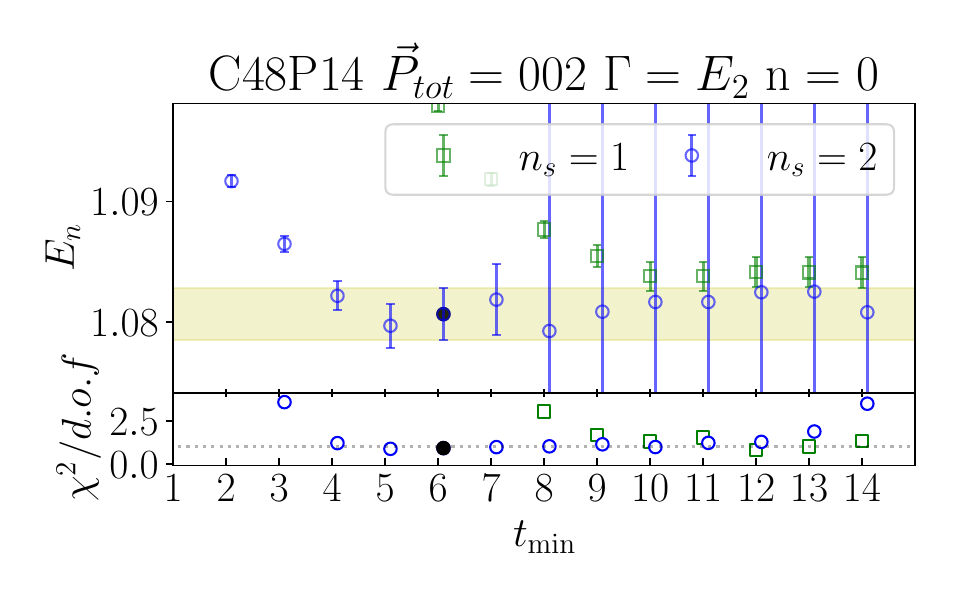}
\includegraphics[width=0.245\columnwidth]{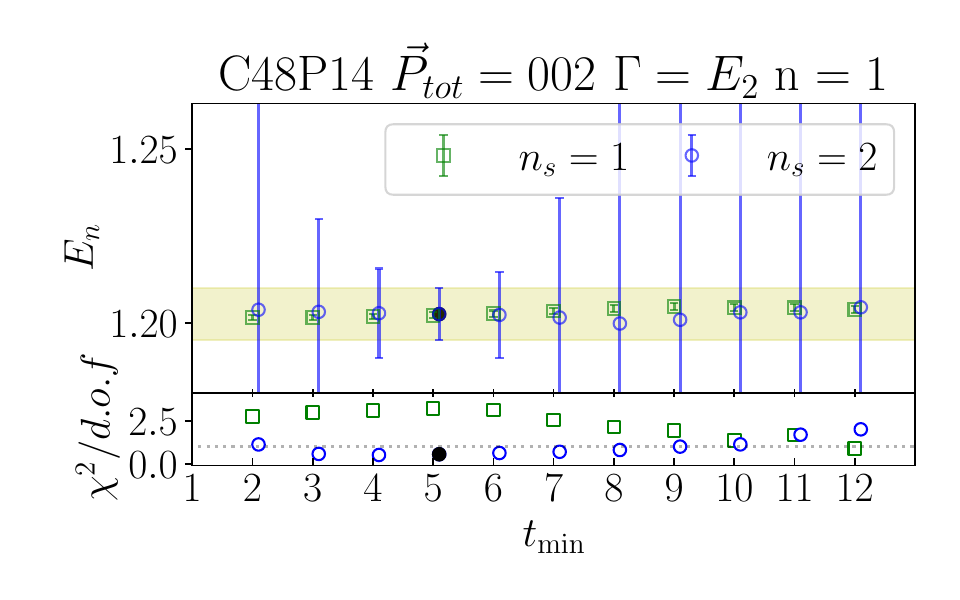}
\caption{Continued from Figure~\ref{fig:Dpi-fit-C48P14}. Energy-level fit results for the $I=\frac{1}{2}$ $D\pi$ channel on the C48P14 ensemble.}
\label{fig:Dpi-fit-C48P142}
\end{figure}

\cleardoublepage
\chapter{Finite-Volume Energy-Level Fits for $\pi\pi\pi \to \omega(782)$ Scattering}
\label{appendix:three_body_problems}

\begin{figure}[htbp]
\centering
\includegraphics[width=0.49\columnwidth]{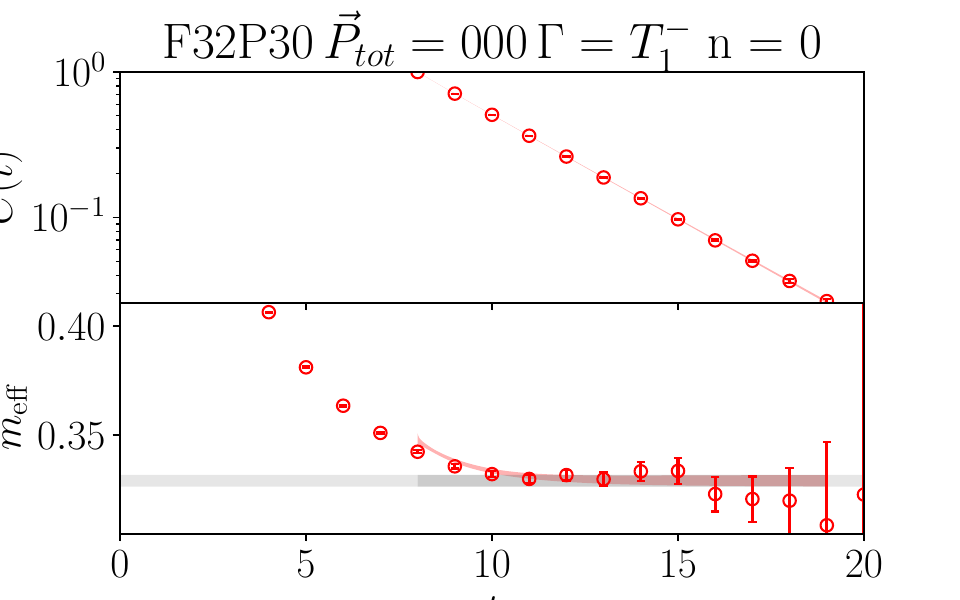}
\includegraphics[width=0.49\columnwidth]{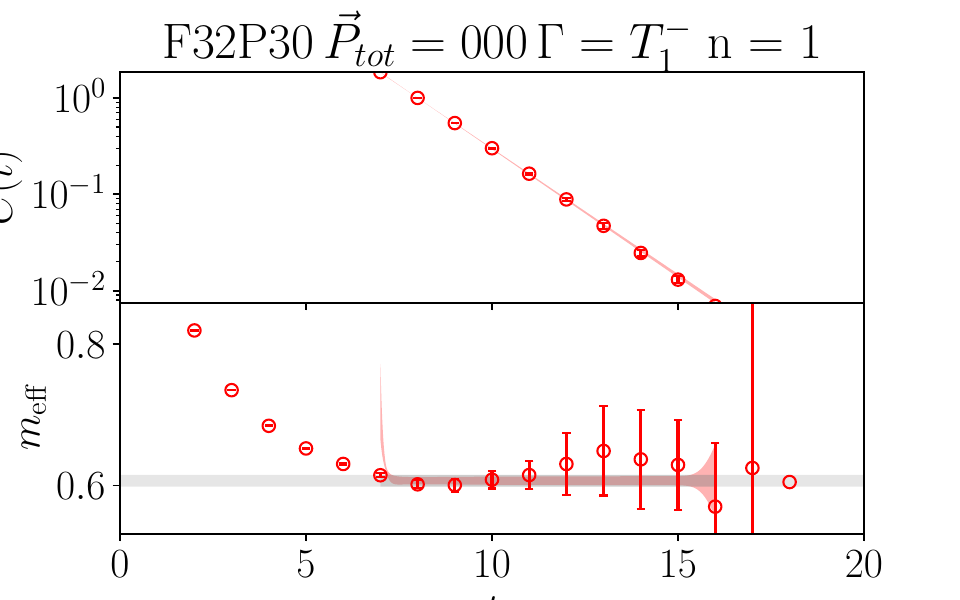}
\\
\includegraphics[width=0.49\columnwidth]{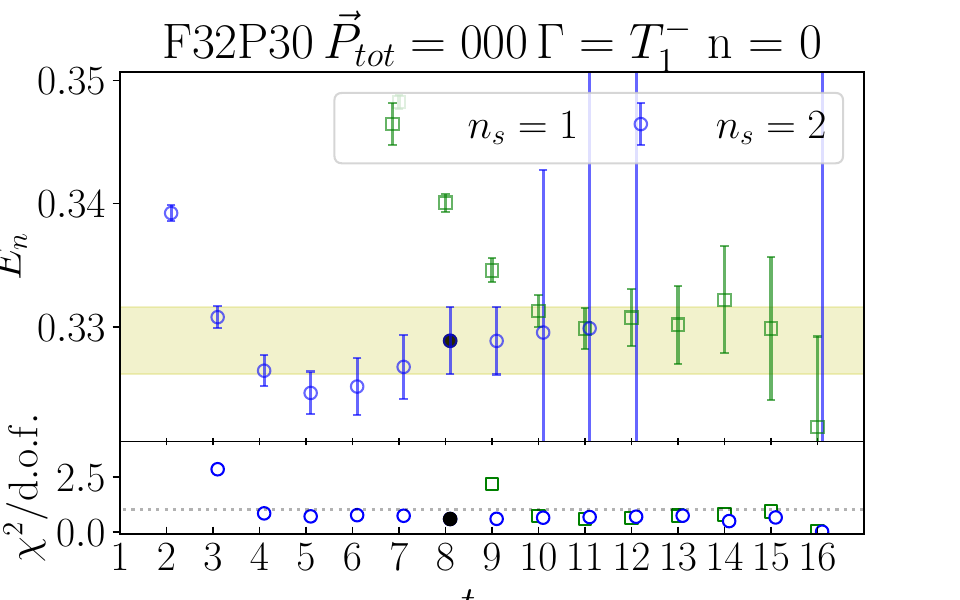}
\includegraphics[width=0.49\columnwidth]{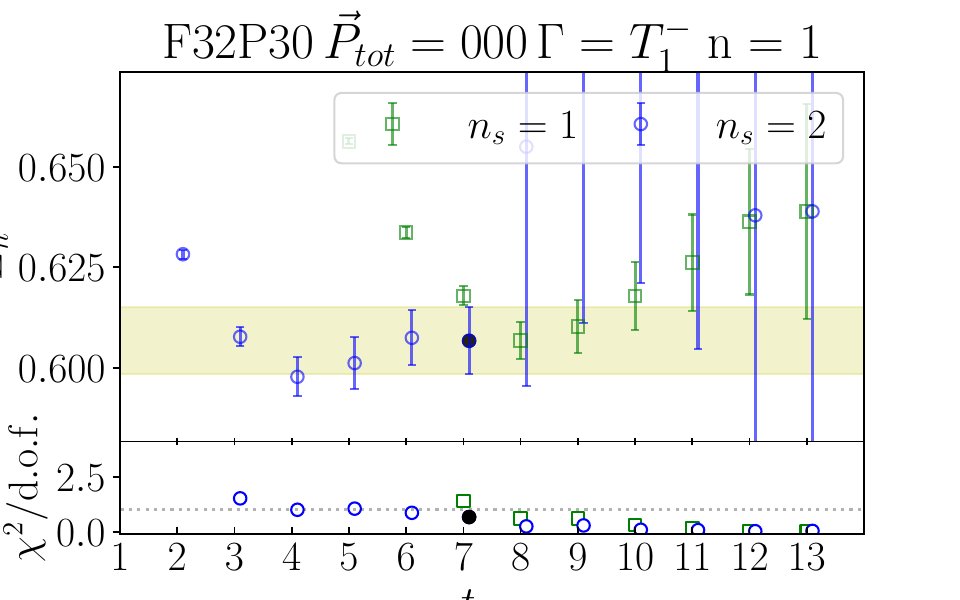}
\\
\includegraphics[width=0.49\columnwidth]{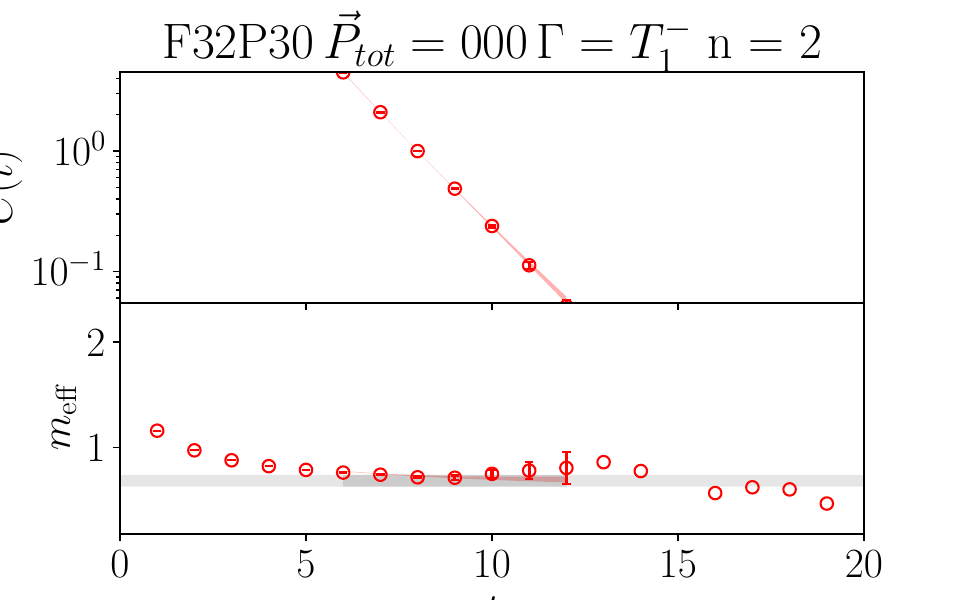}
\includegraphics[width=0.49\columnwidth]{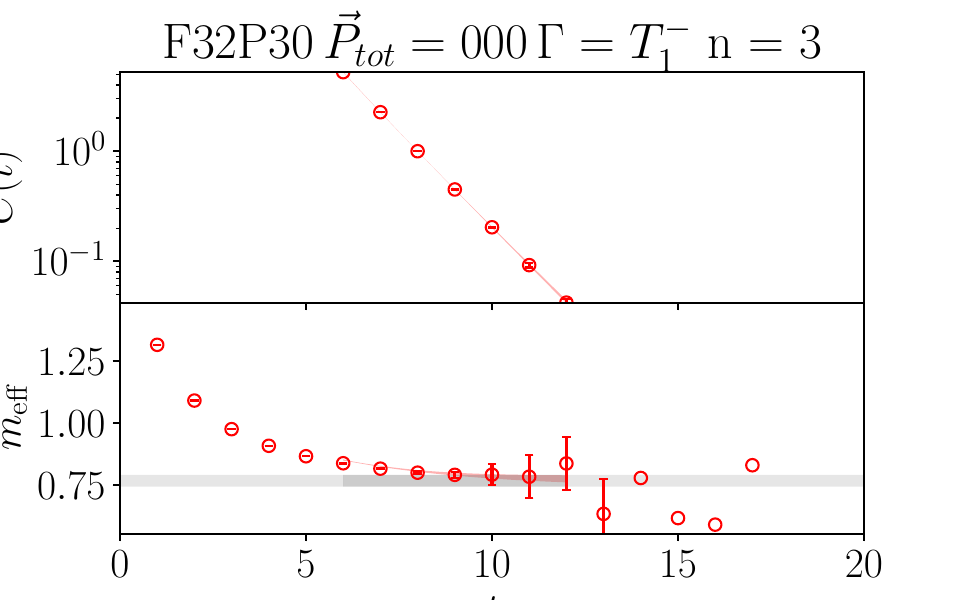}
\\
\includegraphics[width=0.49\columnwidth]{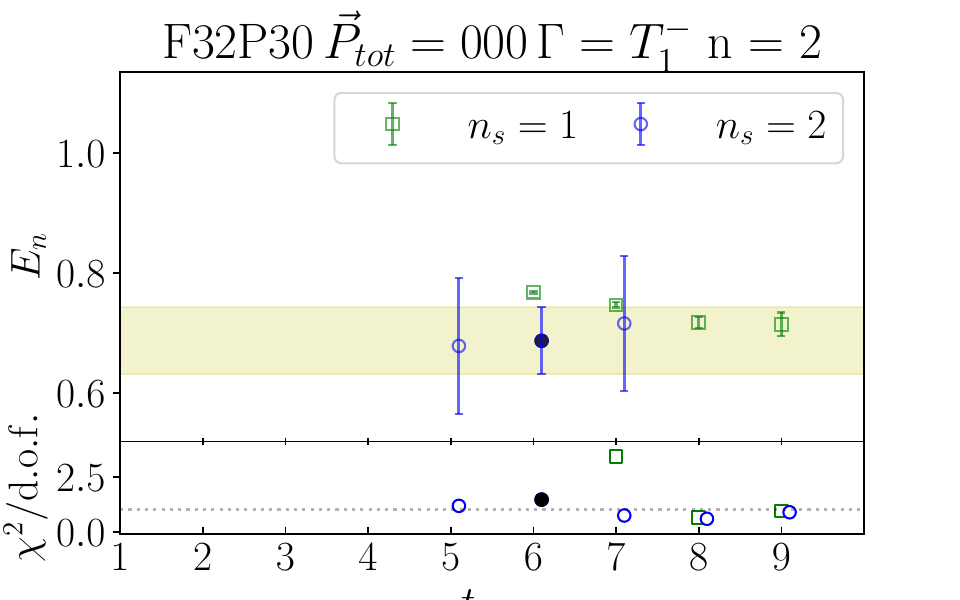}
\includegraphics[width=0.49\columnwidth]{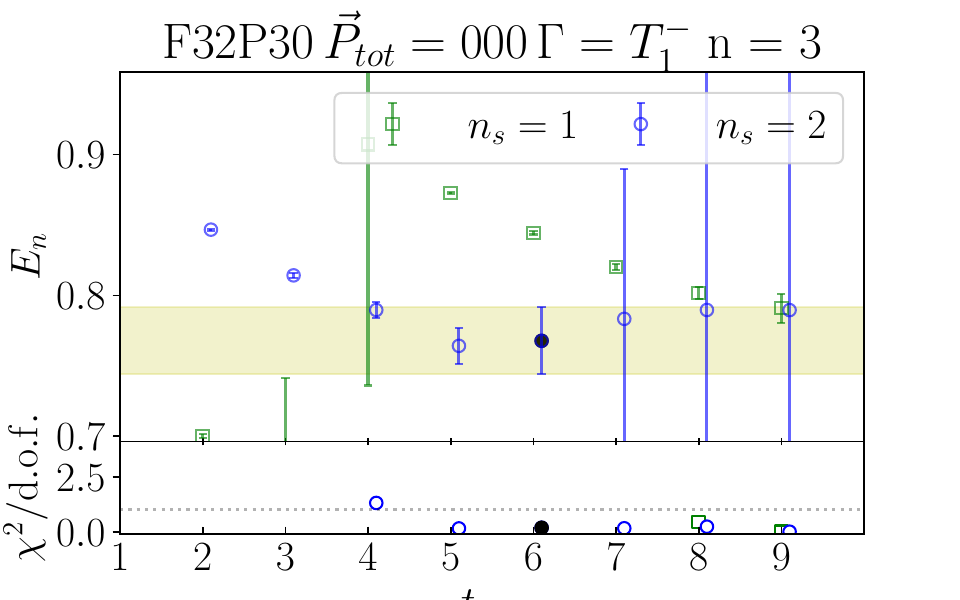}
\caption{Fit results for the $I=0$ $\pi\pi\pi$ channel on the F32P30 ensemble.}
\label{fig:pipipi-I=0-fit-F32P30}
\end{figure}

\begin{figure}[htbp]
\centering
\includegraphics[width=0.49\columnwidth]{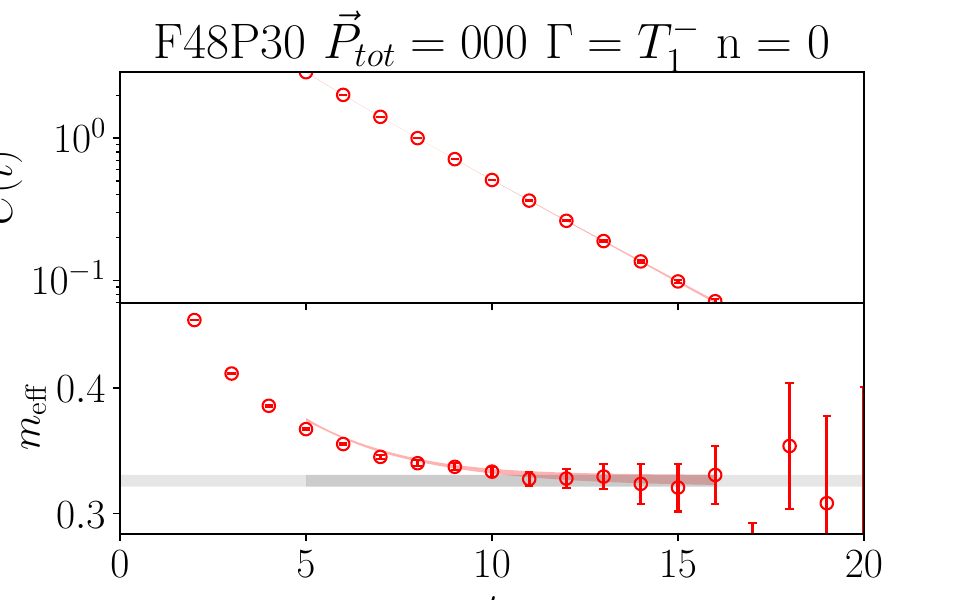}
\includegraphics[width=0.49\columnwidth]{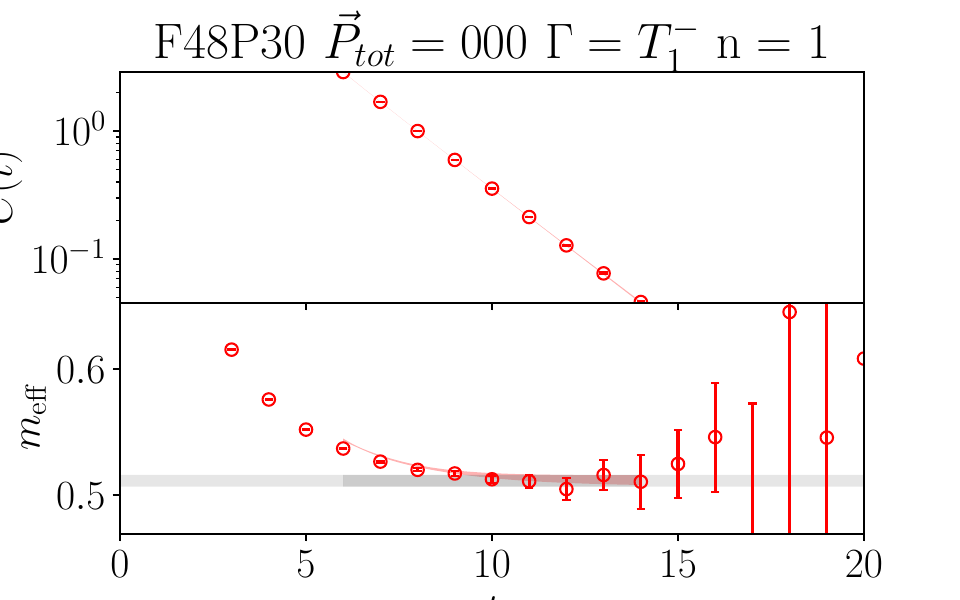}
\\
\includegraphics[width=0.49\columnwidth]{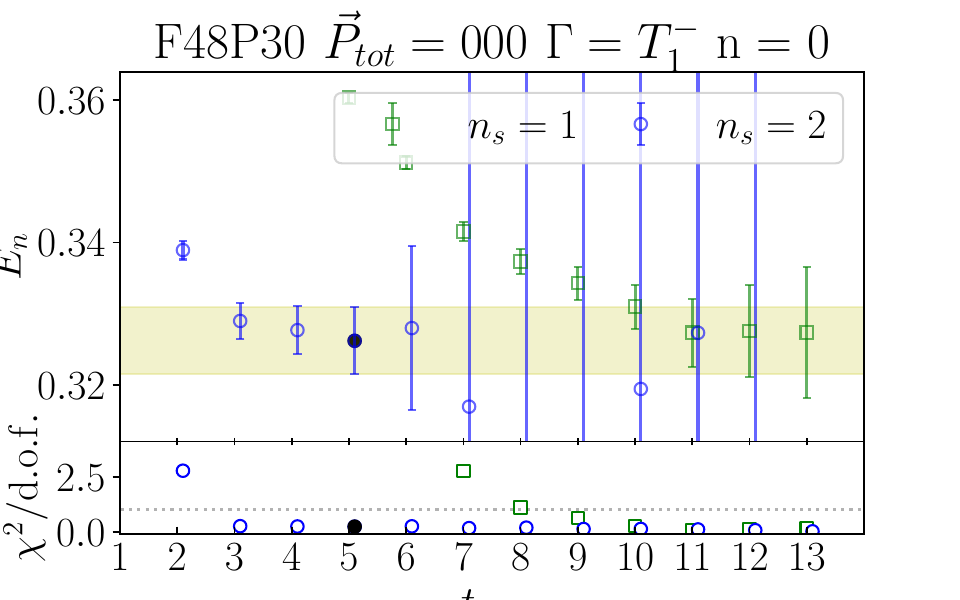}
\includegraphics[width=0.49\columnwidth]{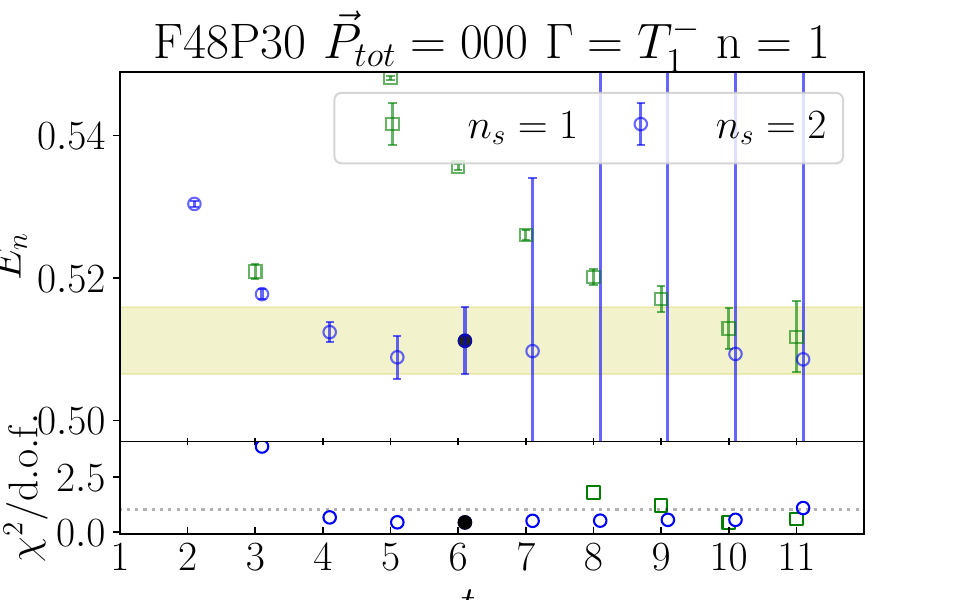}
\\
\includegraphics[width=0.49\columnwidth]{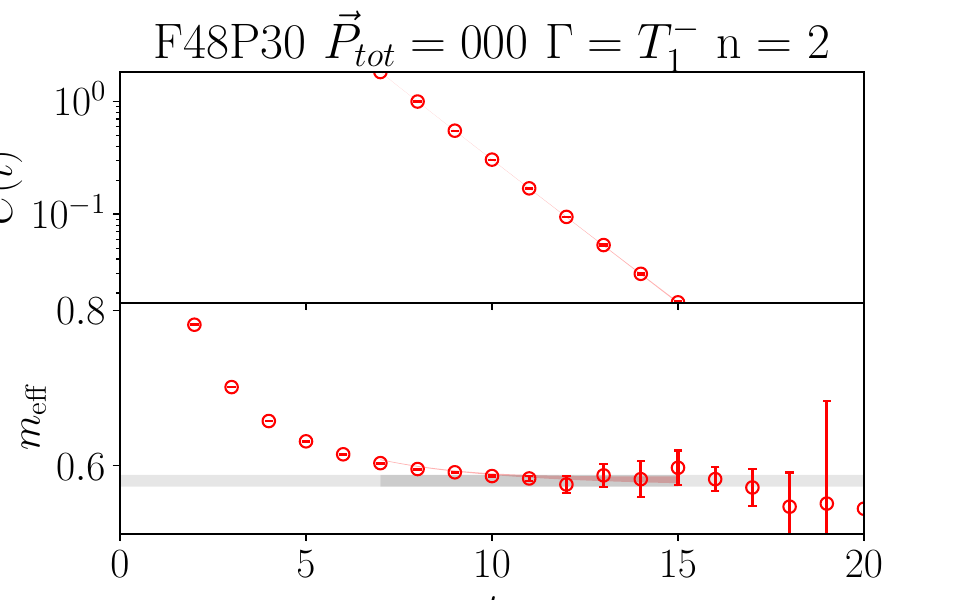}
\includegraphics[width=0.49\columnwidth]{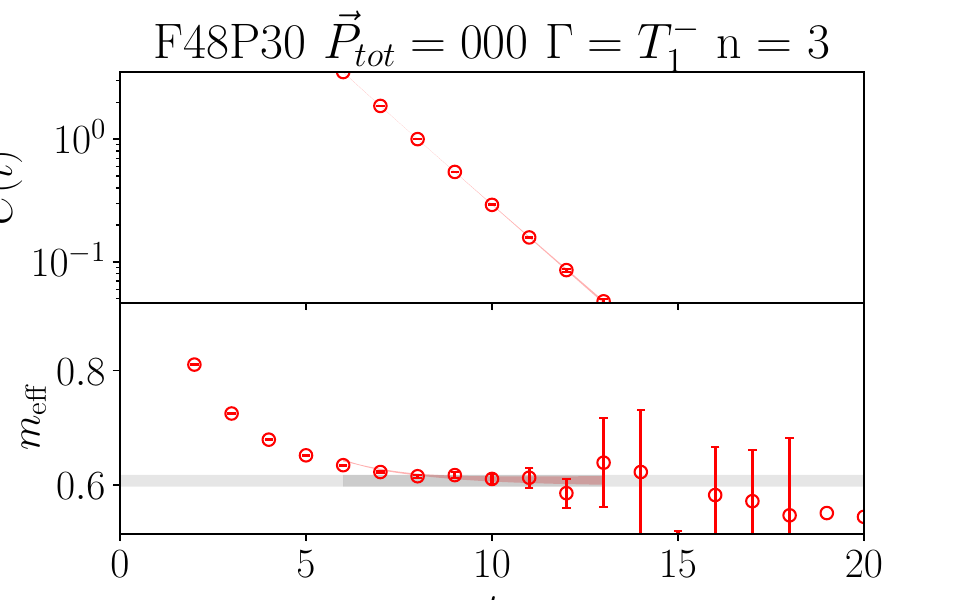}
\\
\includegraphics[width=0.49\columnwidth]{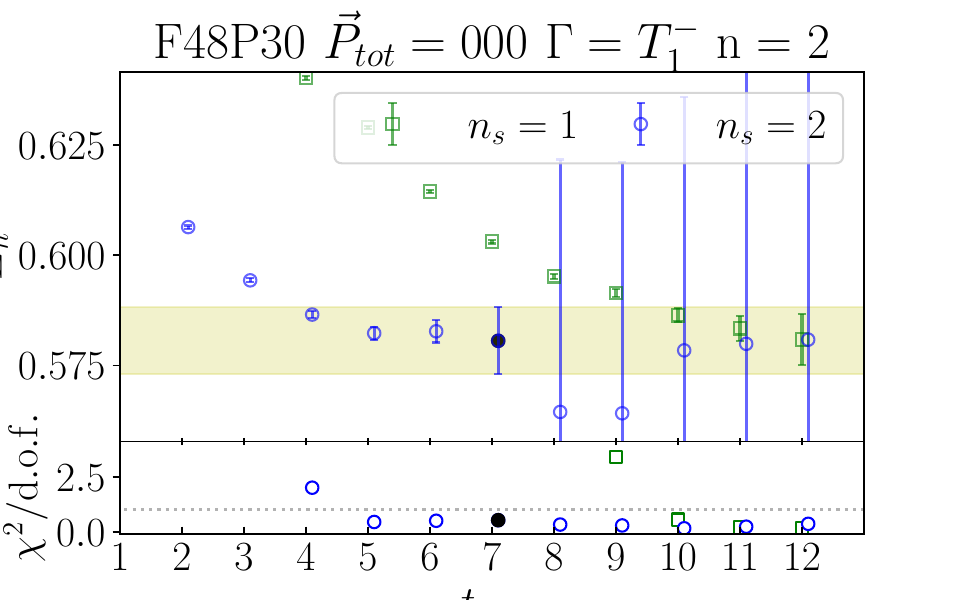}
\includegraphics[width=0.49\columnwidth]{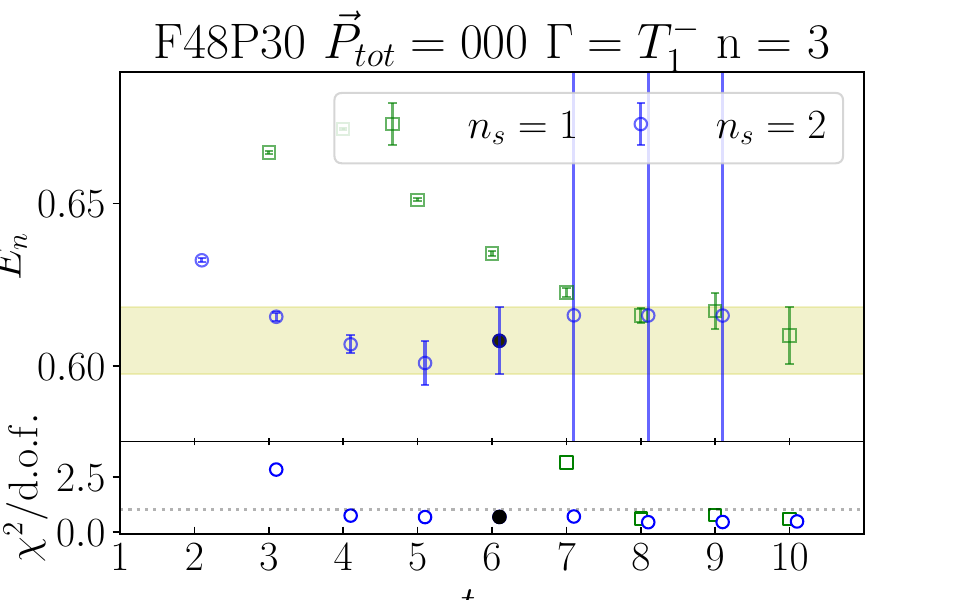}
\caption{Fit results for the $I=0$ $\pi\pi\pi$ channel on the F48P30 ensemble.}
\label{fig:pipipi-I=0-fit-F48P30}
\end{figure}

\begin{figure}[htbp]
\centering
\includegraphics[width=0.49\columnwidth]{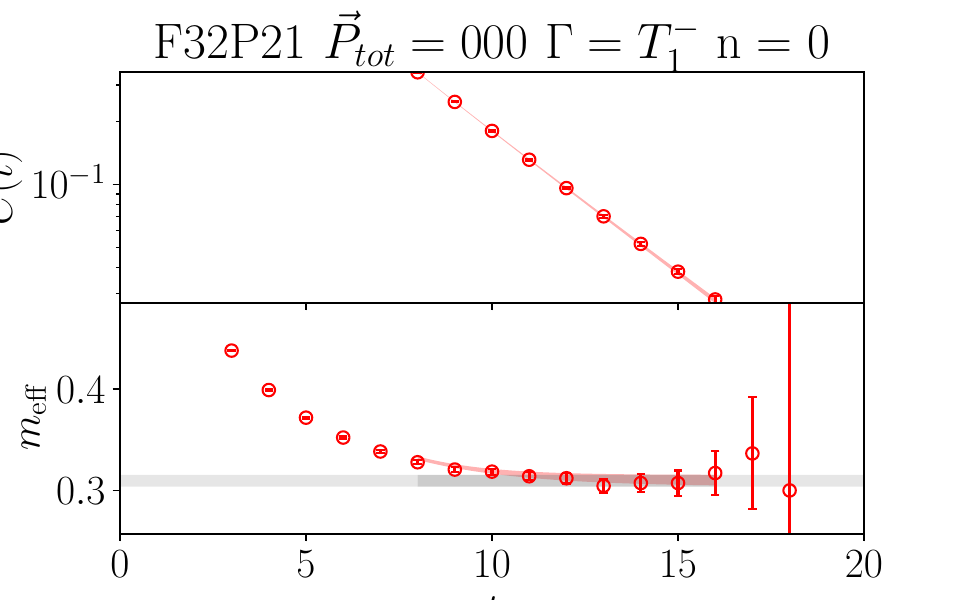}
\includegraphics[width=0.49\columnwidth]{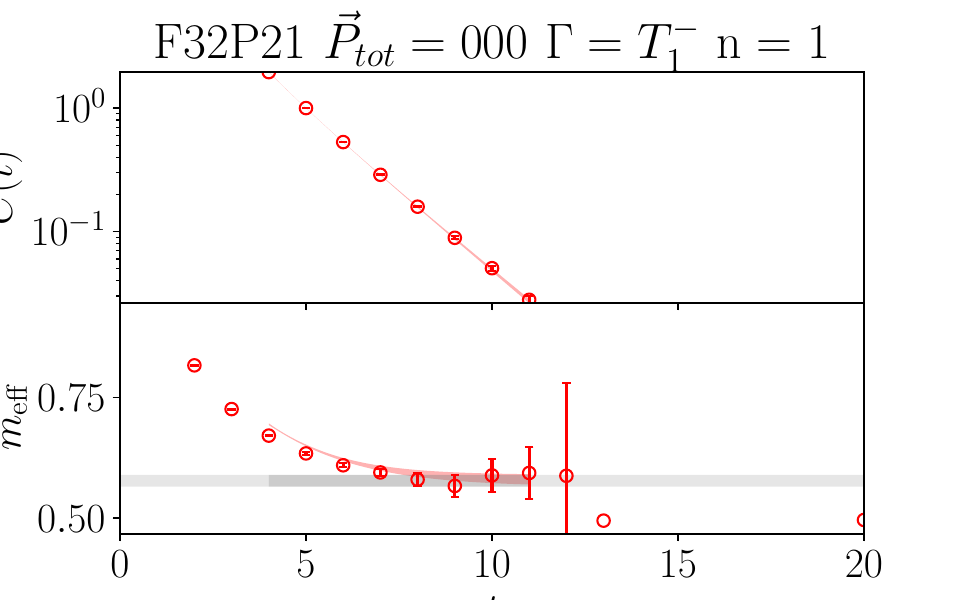}
\\
\includegraphics[width=0.49\columnwidth]{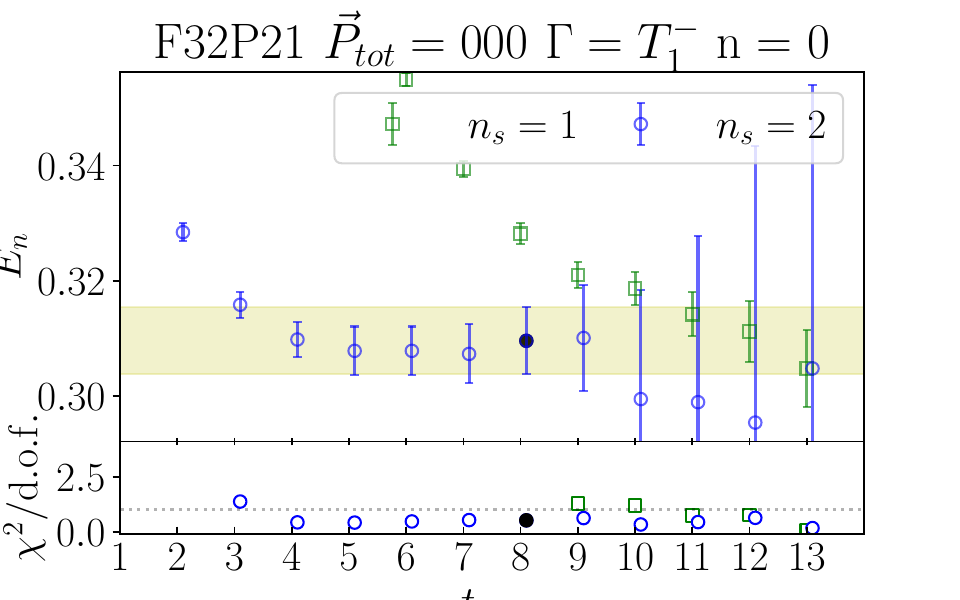}
\includegraphics[width=0.49\columnwidth]{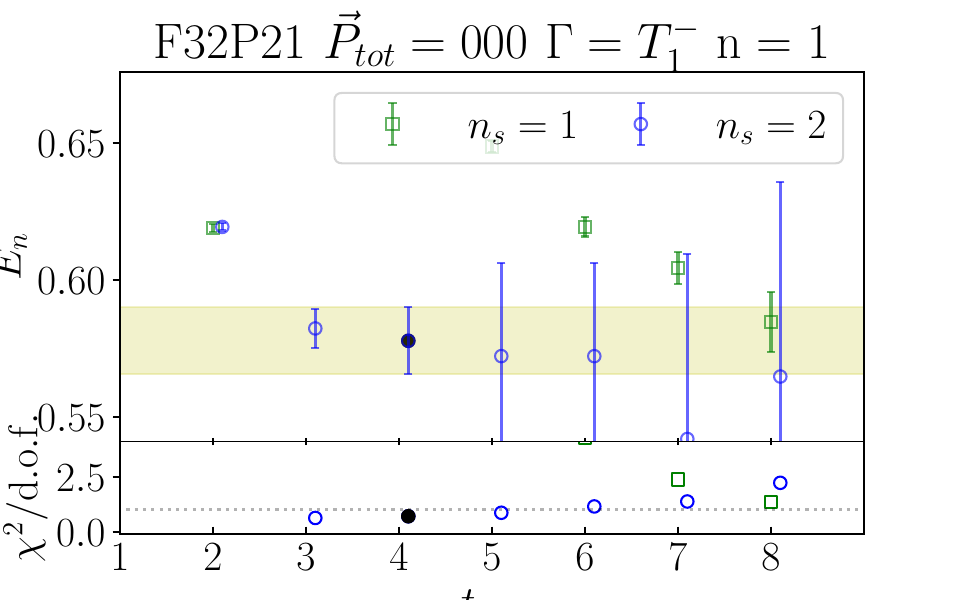}
\\
\includegraphics[width=0.49\columnwidth]{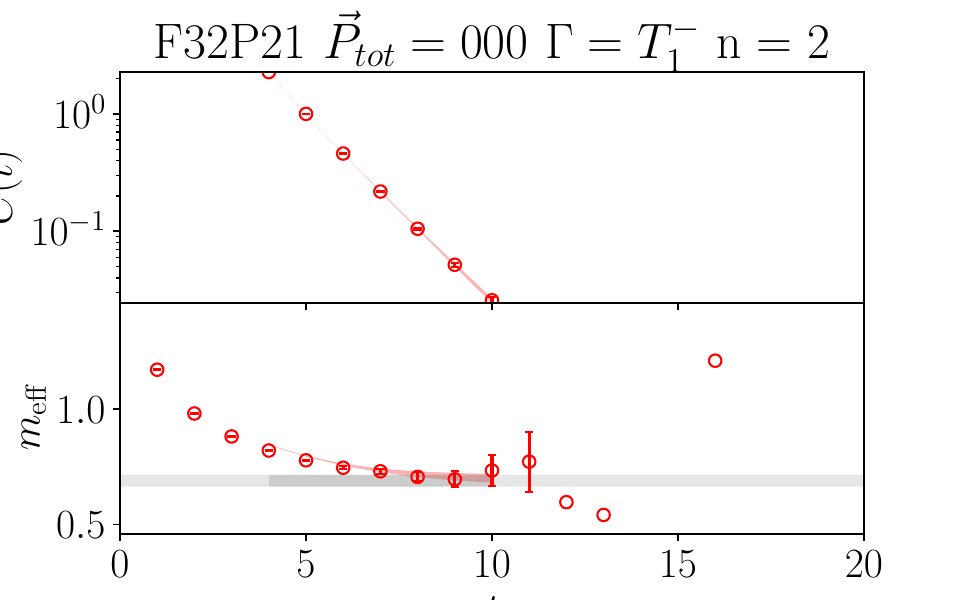}
\includegraphics[width=0.49\columnwidth]{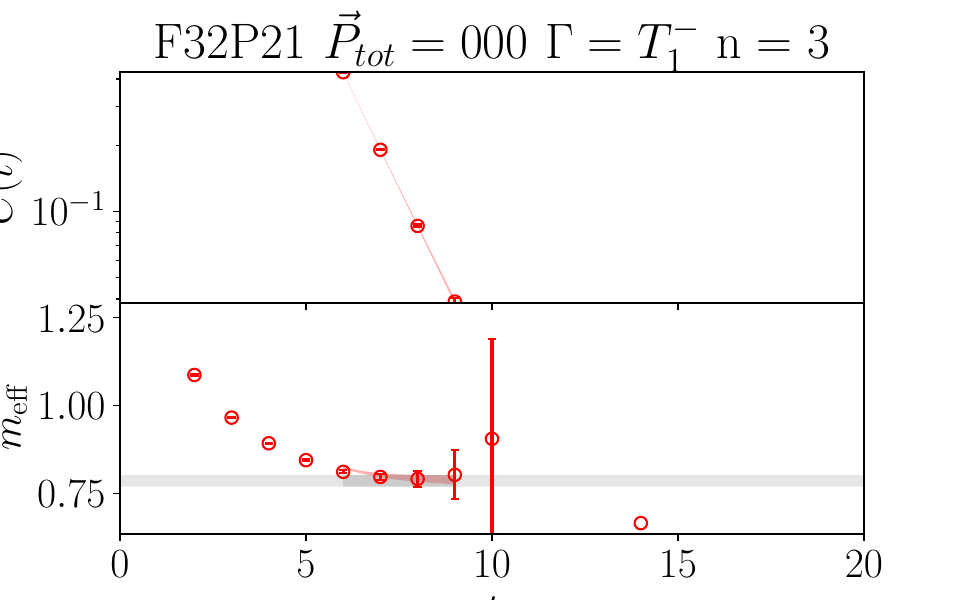}
\\
\includegraphics[width=0.49\columnwidth]{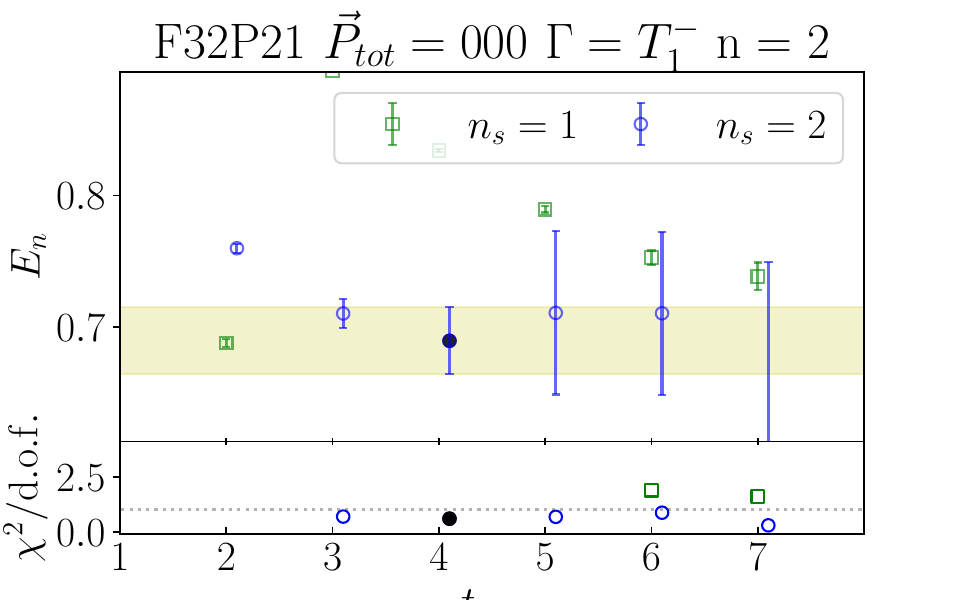}
\includegraphics[width=0.49\columnwidth]{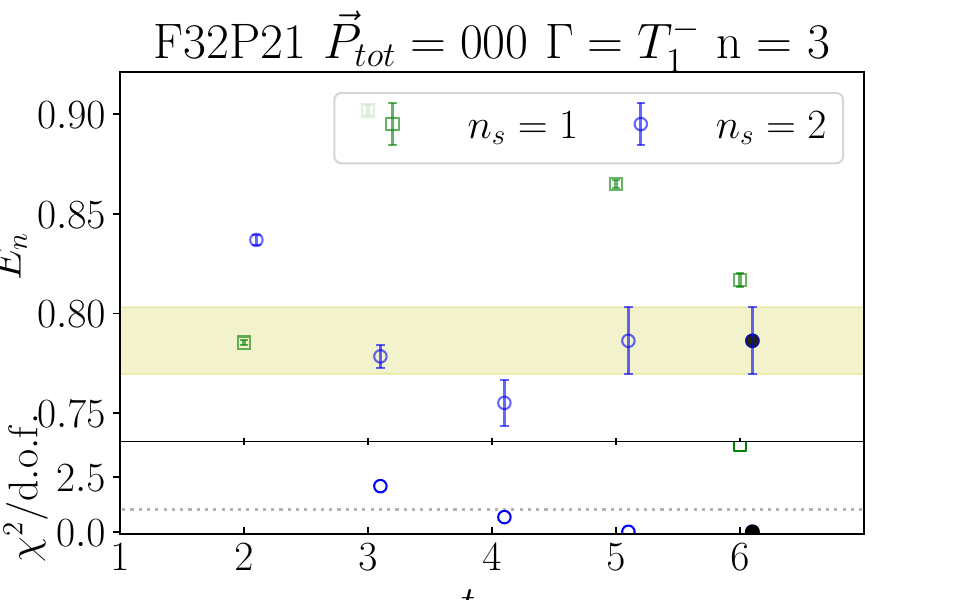}
\caption{Fit results for the $I=0$ $\pi\pi\pi$ channel on the F32P21 ensemble.}
\label{fig:pipipi-I=0-fit-F32P21}
\end{figure}

\begin{figure}[htbp]
\centering
\includegraphics[width=0.49\columnwidth]{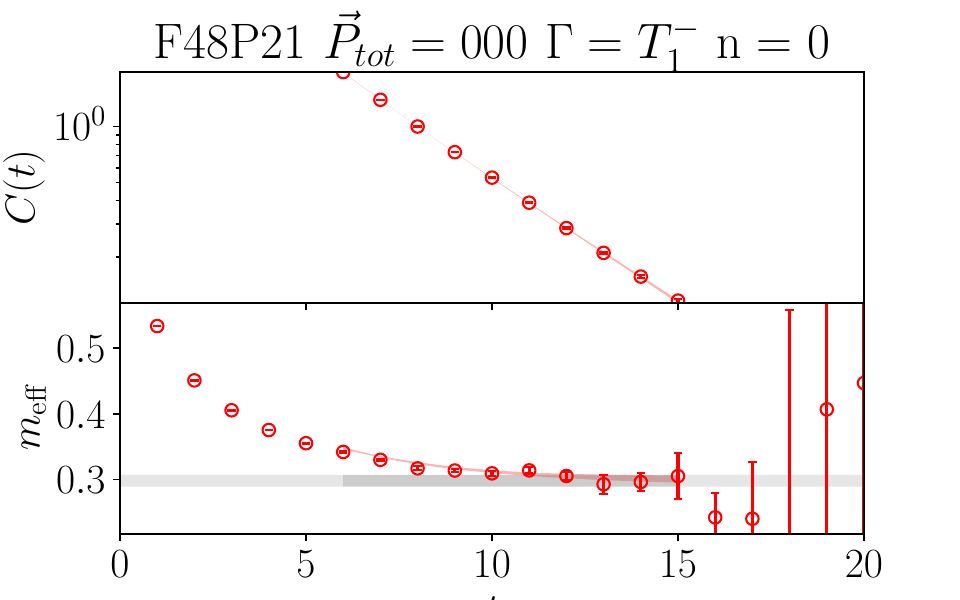}
\includegraphics[width=0.49\columnwidth]{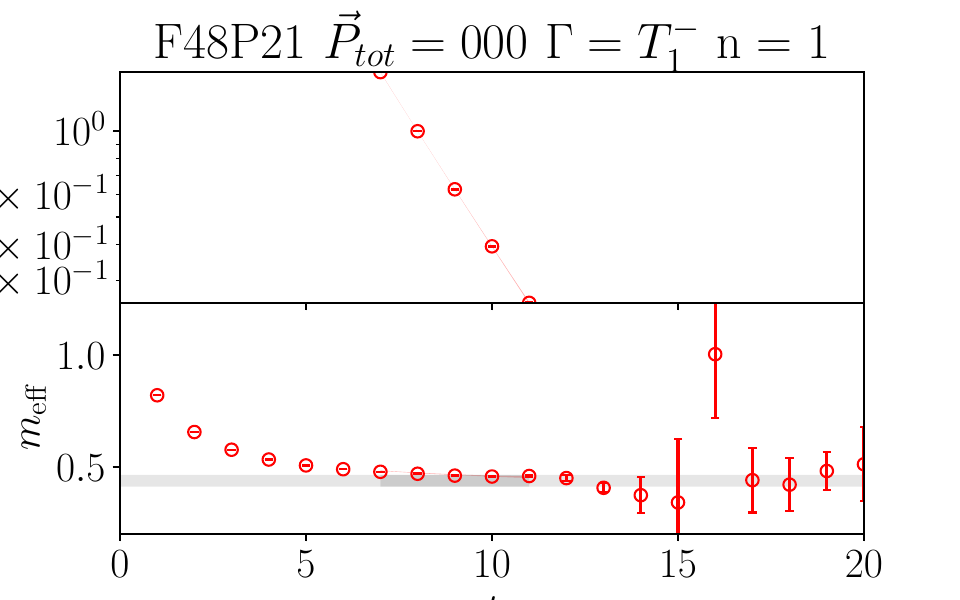}
\\
\includegraphics[width=0.49\columnwidth]{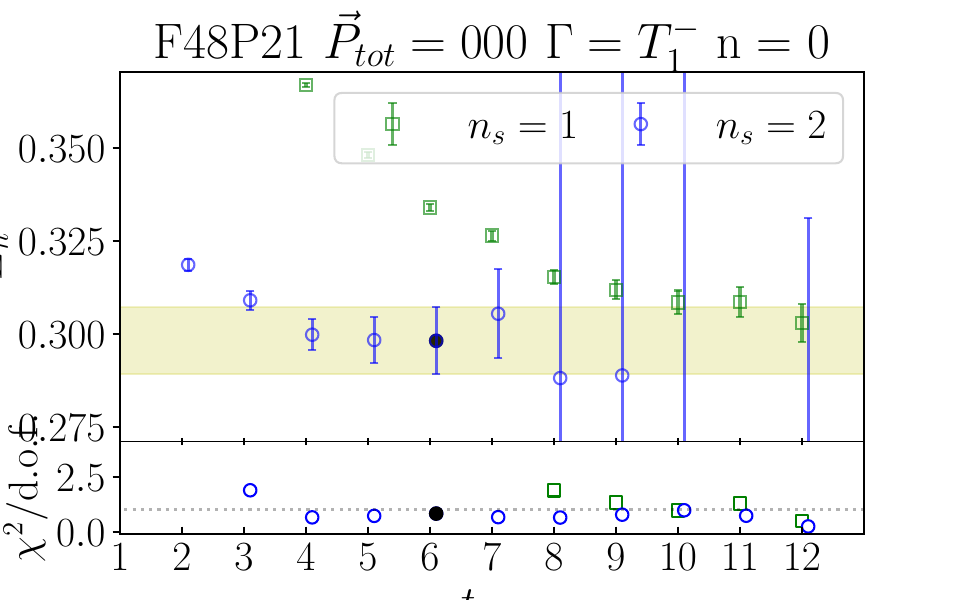}
\includegraphics[width=0.49\columnwidth]{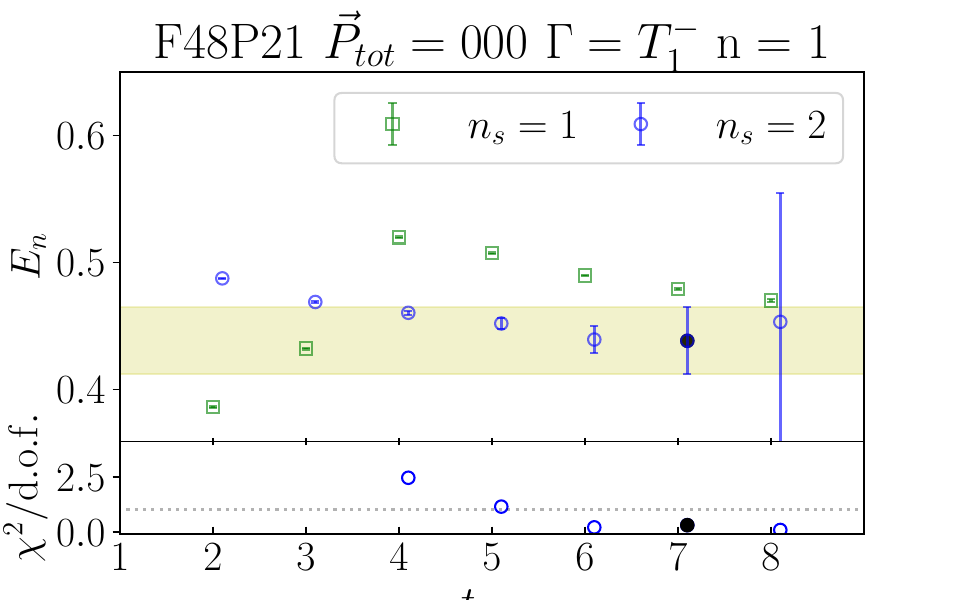}
\\
\includegraphics[width=0.49\columnwidth]{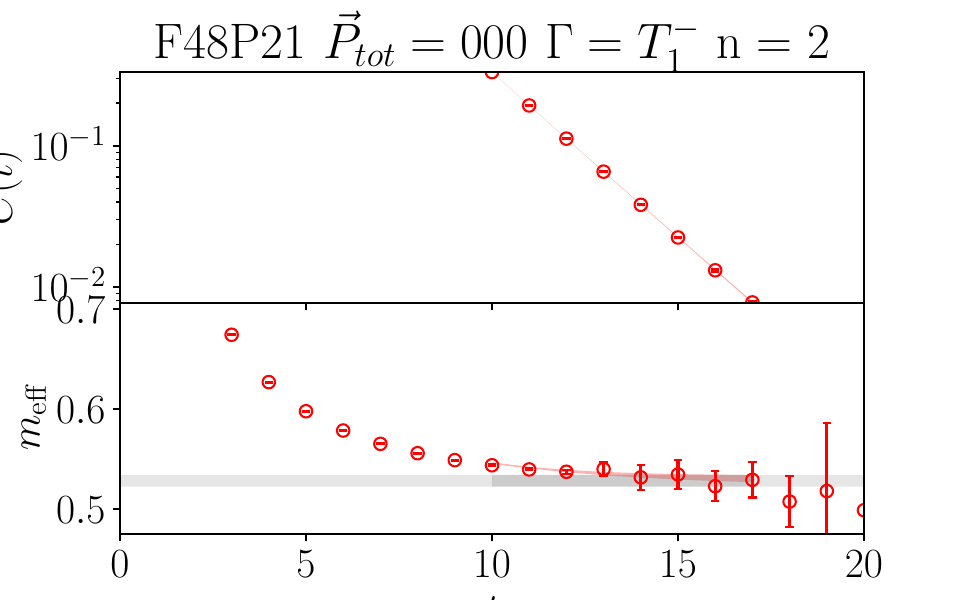}
\includegraphics[width=0.49\columnwidth]{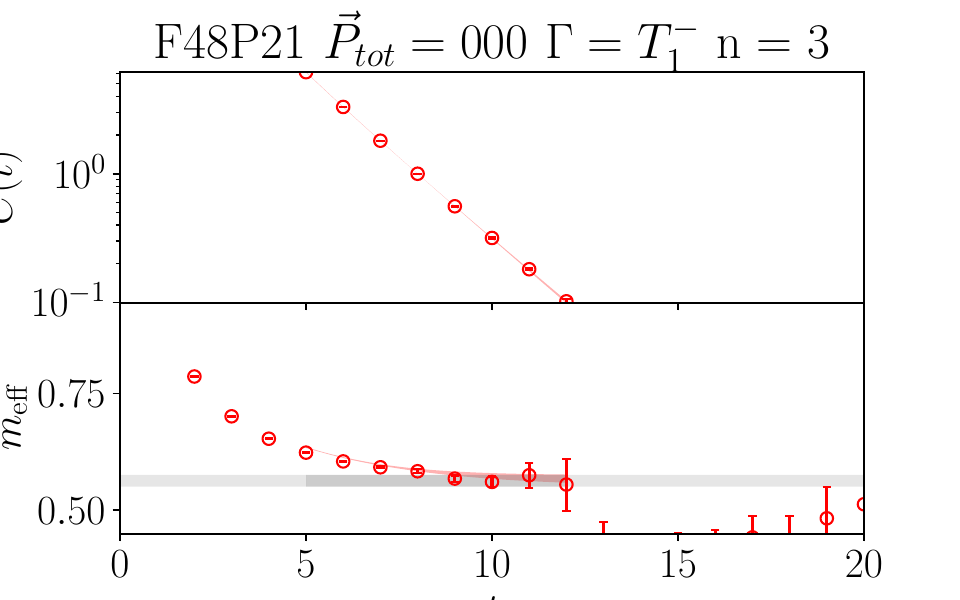}
\\
\includegraphics[width=0.49\columnwidth]{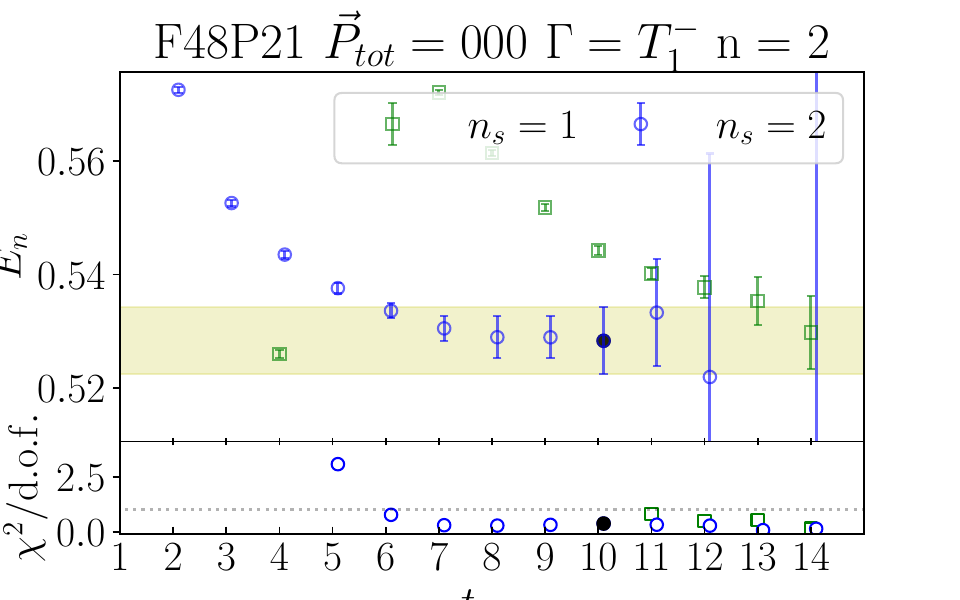}
\includegraphics[width=0.49\columnwidth]{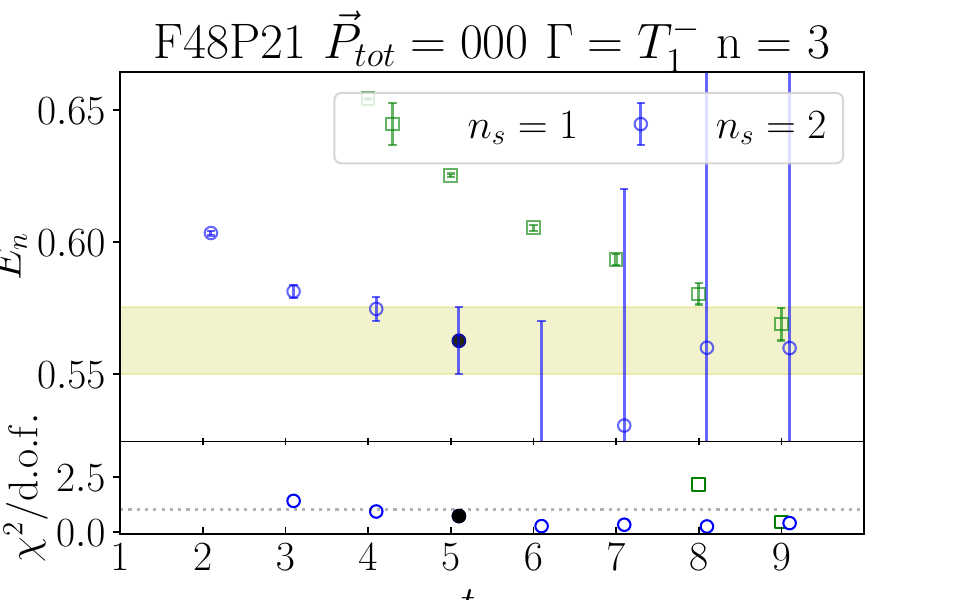}
\caption{Fit results for the $I=0$ $\pi\pi\pi$ channel on the F48P21 ensemble.}
\label{fig:pipipi-I=0-fit-F48P21}
\end{figure}

Figures~\ref{fig:pipipi-I=0-fit-F32P30}, \ref{fig:pipipi-I=0-fit-F48P30}, \ref{fig:pipipi-I=0-fit-F32P21}, and \ref{fig:pipipi-I=0-fit-F48P21} show the energy-level fits for the $I=0$ $\pi\pi\pi$ system on the F32P30, F48P30, F32P21, and F48P21 ensembles used in the $\omega(782)$ study. Statistical uncertainties are estimated from $2000$ bootstrap resamples.

\cleardoublepage
\chapter{Finite-Volume Energy-Level Fits for $\pi\pi\pi \to \pi(1300)$ Scattering}
\label{appendix:three_body_problems2}

\begin{figure}[htbp]
\centering
\includegraphics[width=0.32\columnwidth]{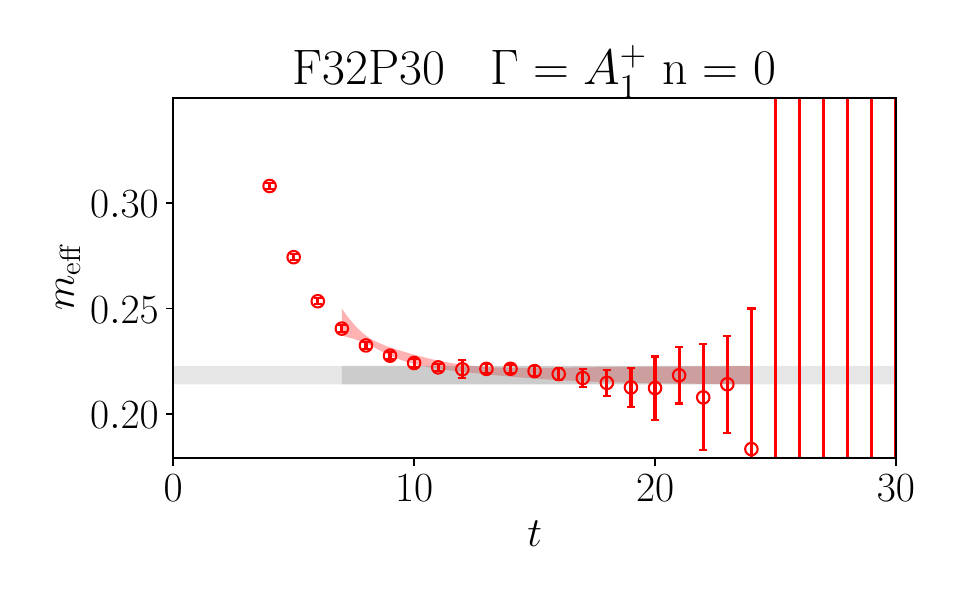}
\includegraphics[width=0.32\columnwidth]{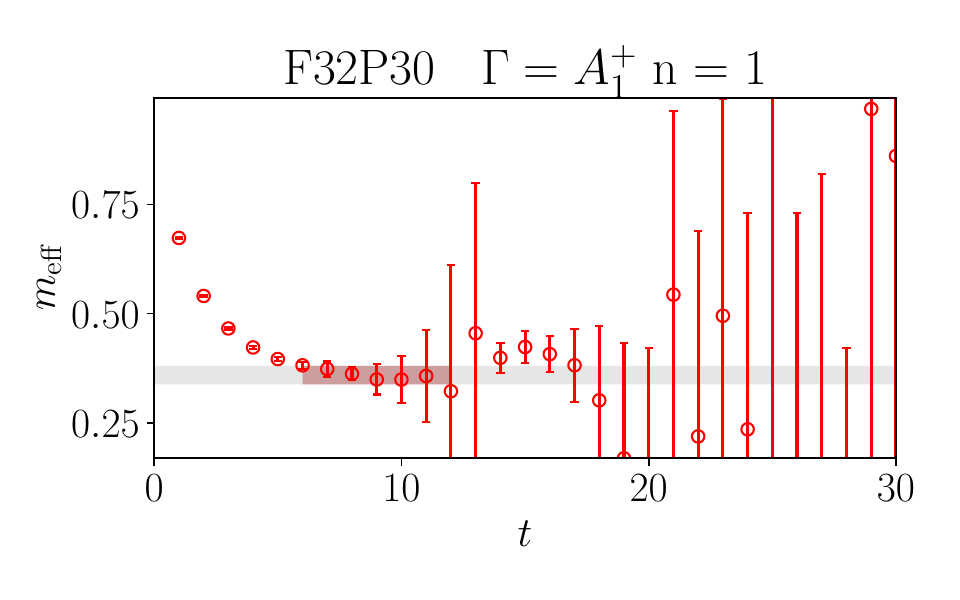}
\\
\includegraphics[width=0.32\columnwidth]{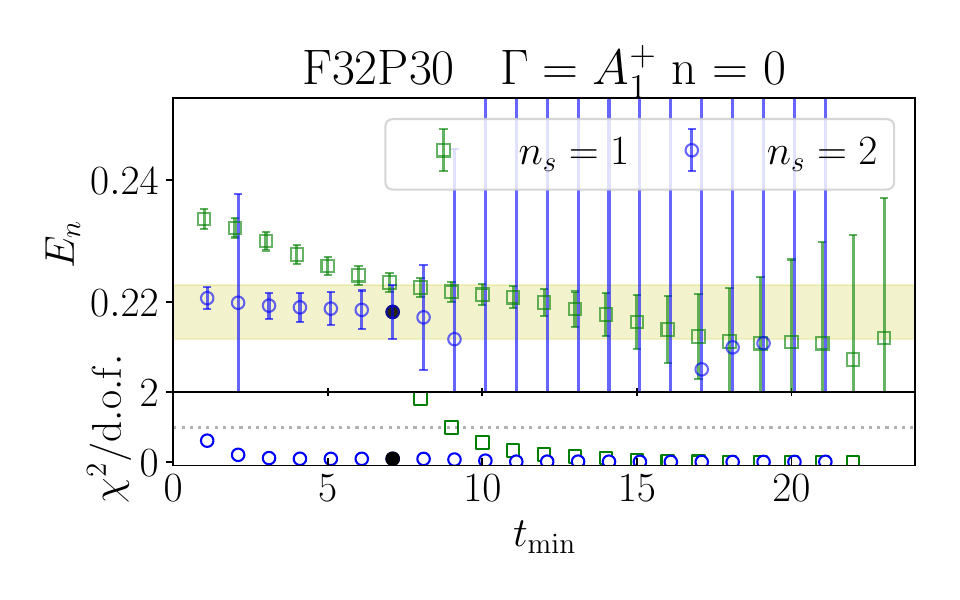}
\includegraphics[width=0.32\columnwidth]{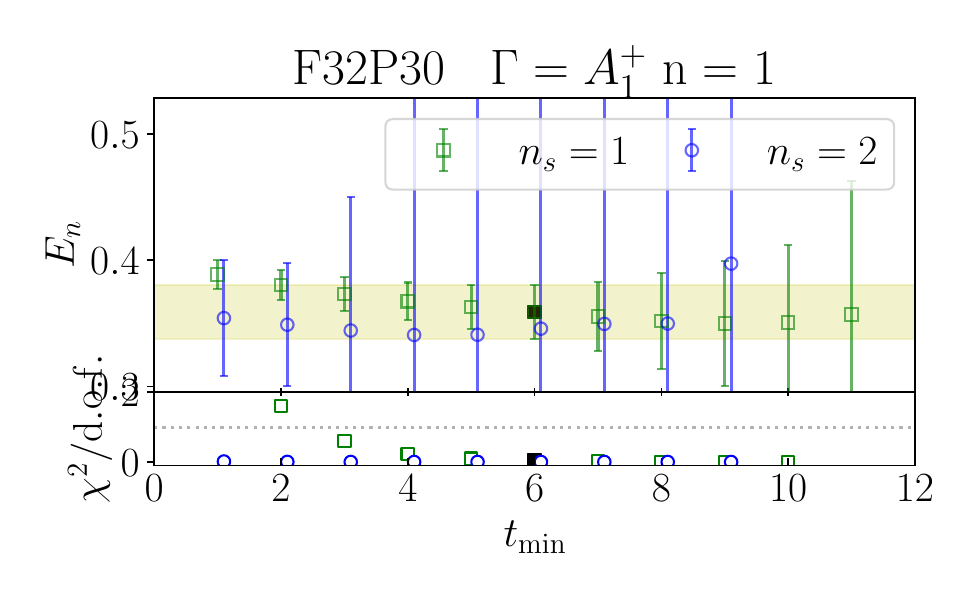}
\\
\includegraphics[width=0.32\columnwidth]{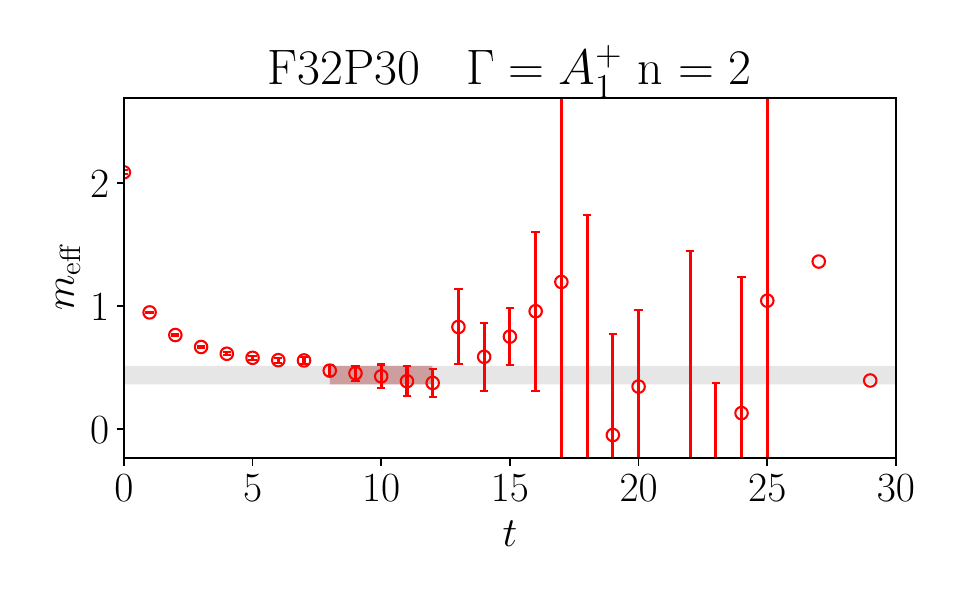}
\includegraphics[width=0.32\columnwidth]{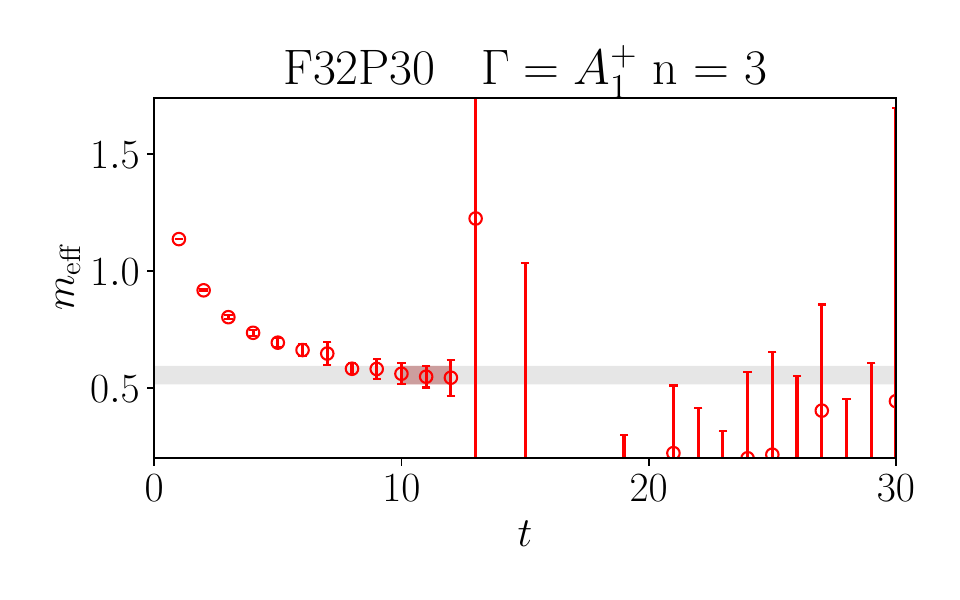}
\\
\includegraphics[width=0.32\columnwidth]{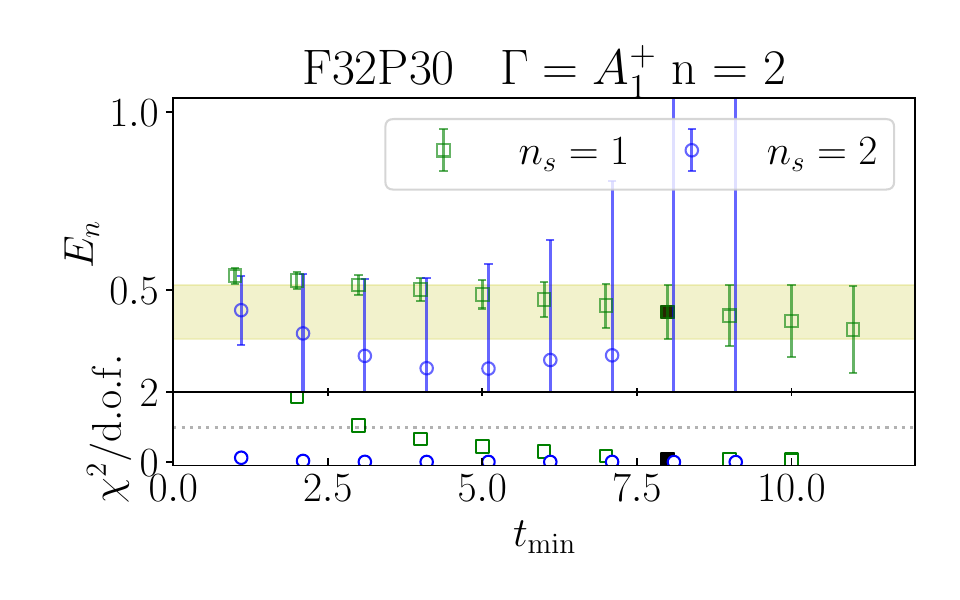}
\includegraphics[width=0.32\columnwidth]{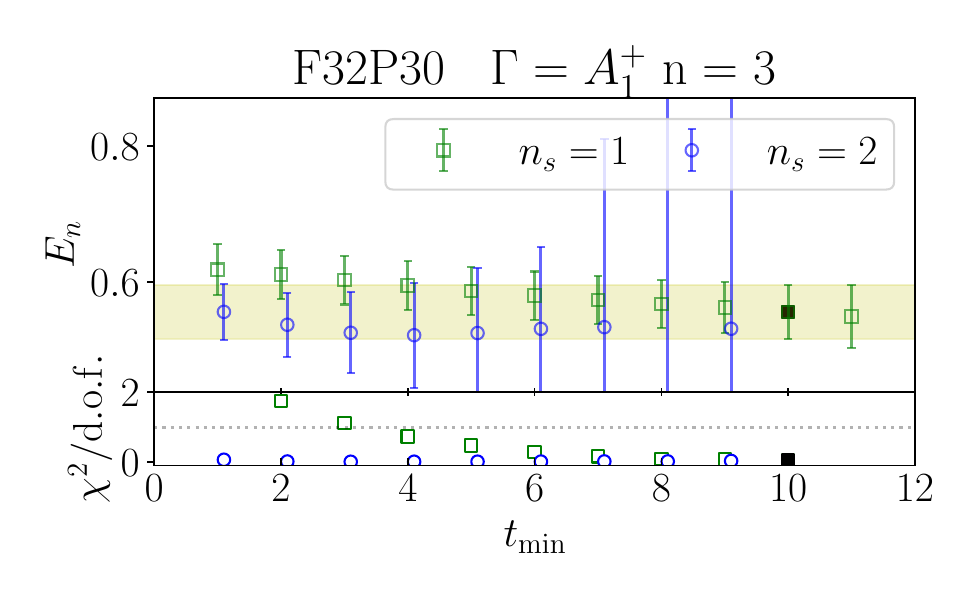}
\caption{Fit results for the $I=0$ $\pi\pi$ channel on the F32P30 ensemble. For each energy level, the fit is shown in two panels: in the upper panel, the red band denotes the reconstructed effective mass, and the gray band denotes the extracted energy; the lower panel shows the stability of the fit with respect to the starting time, where the green and blue points denote one-state and two-state fits, respectively, and the black error bar marks the chosen starting time. The lower panel also gives the $\chi^2/\mathrm{d.o.f.}$ value.}
\label{fig:pipi-I=0-fit-F32P30}
\end{figure}

\begin{figure}[htbp]
\centering
\includegraphics[width=0.32\columnwidth]{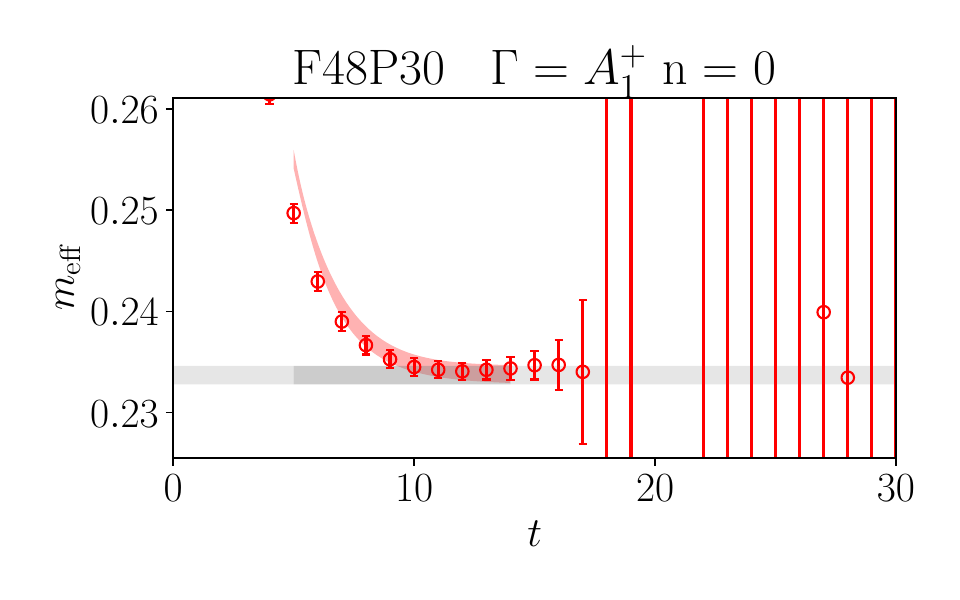}
\includegraphics[width=0.32\columnwidth]{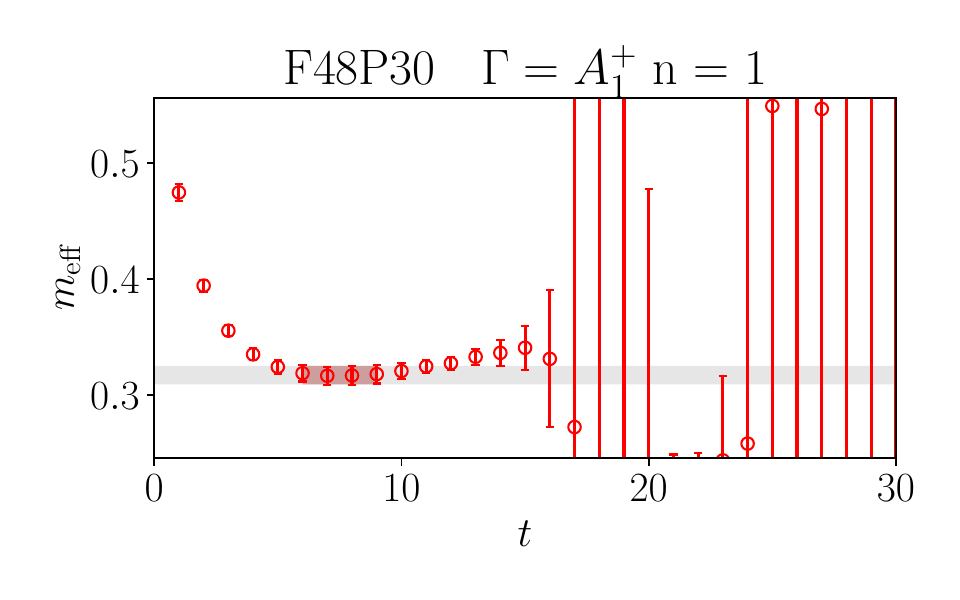}
\\
\includegraphics[width=0.32\columnwidth]{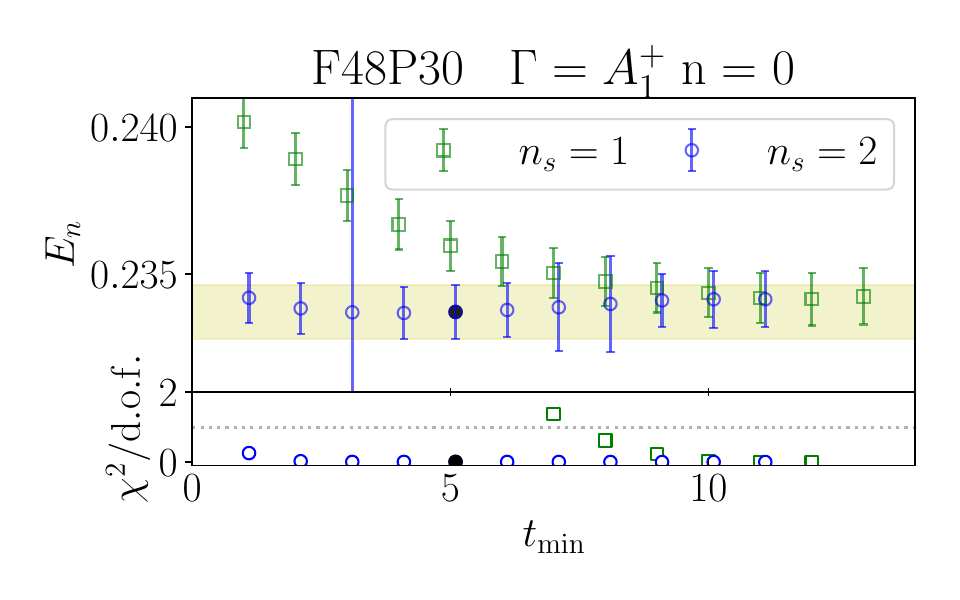}
\includegraphics[width=0.32\columnwidth]{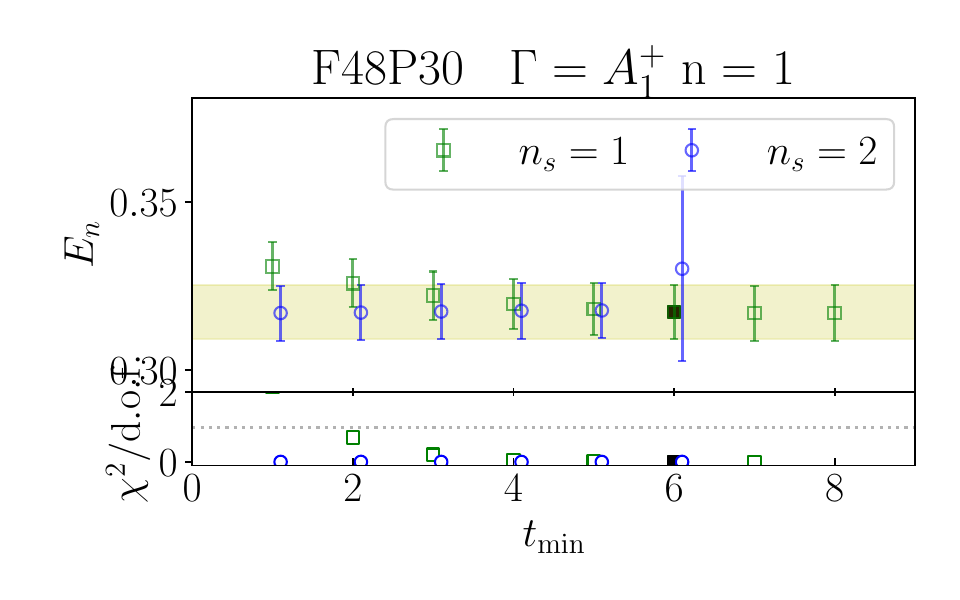}
\\
\includegraphics[width=0.32\columnwidth]{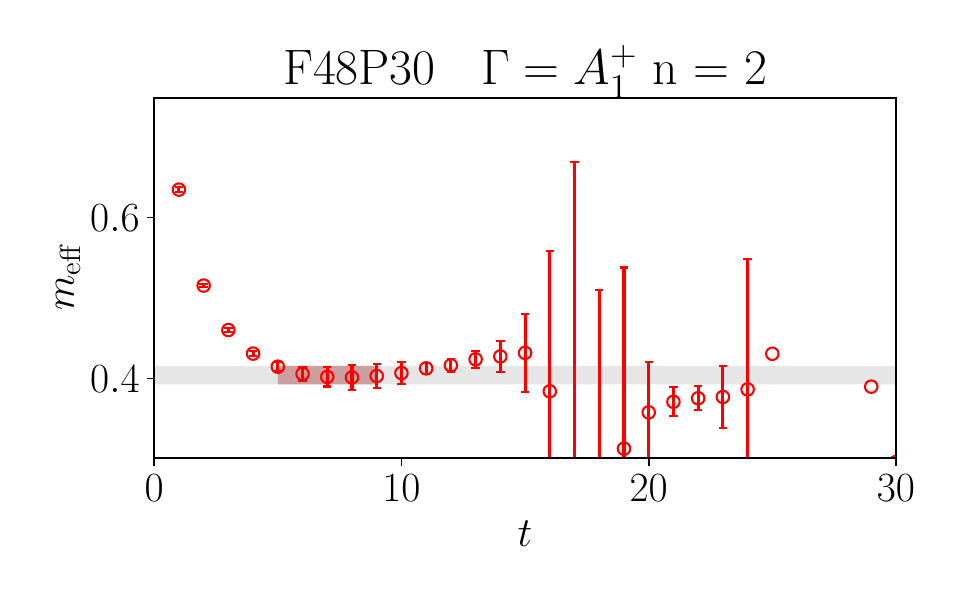}
\includegraphics[width=0.32\columnwidth]{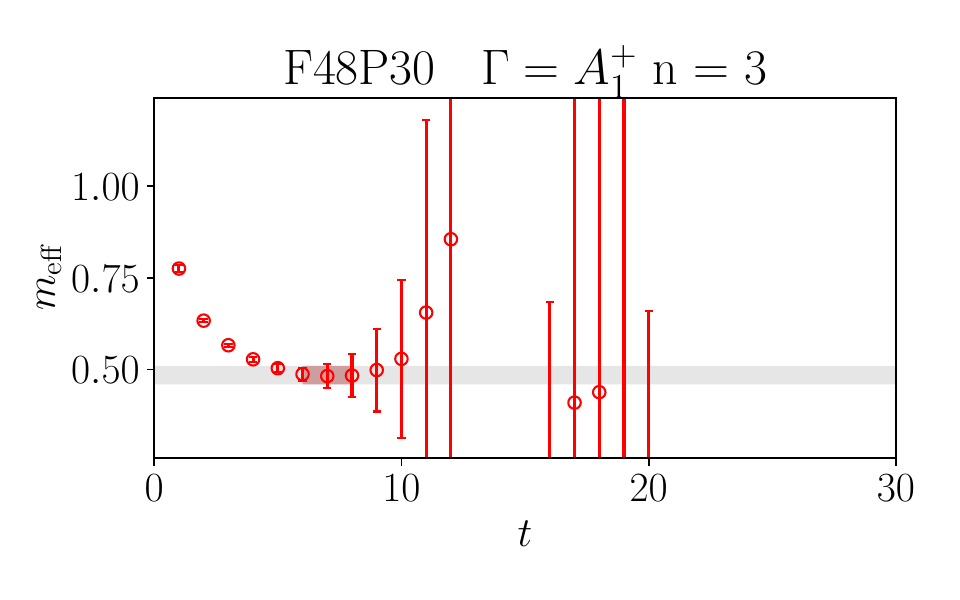}
\\
\includegraphics[width=0.32\columnwidth]{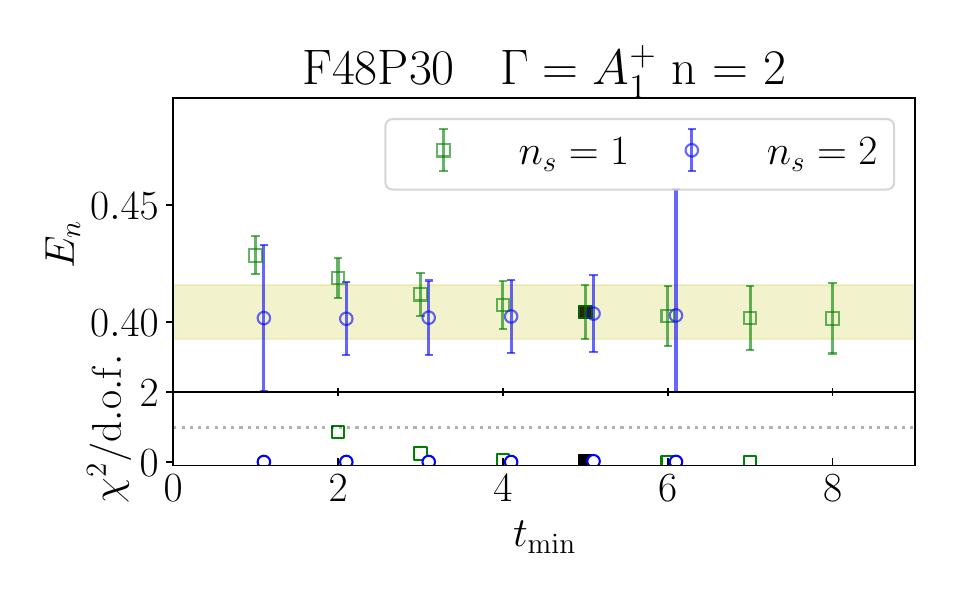}
\includegraphics[width=0.32\columnwidth]{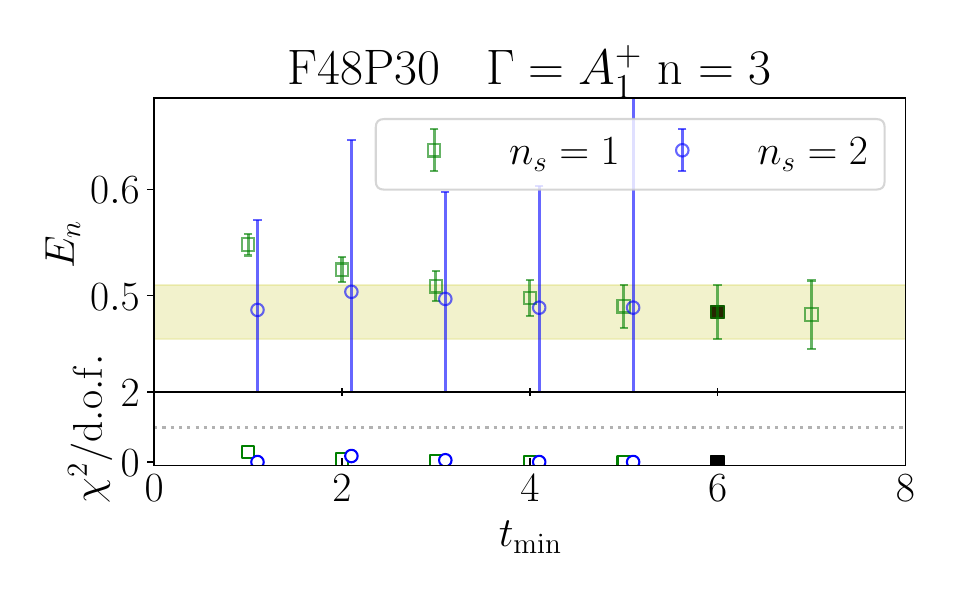}
\caption{Fit results for the $I=0$ $\pi\pi$ channel on the F48P30 ensemble.}
\label{fig:pipi-I=0-fit-F48P30}
\end{figure}

\begin{figure}[htbp]
\centering
\includegraphics[width=0.32\columnwidth]{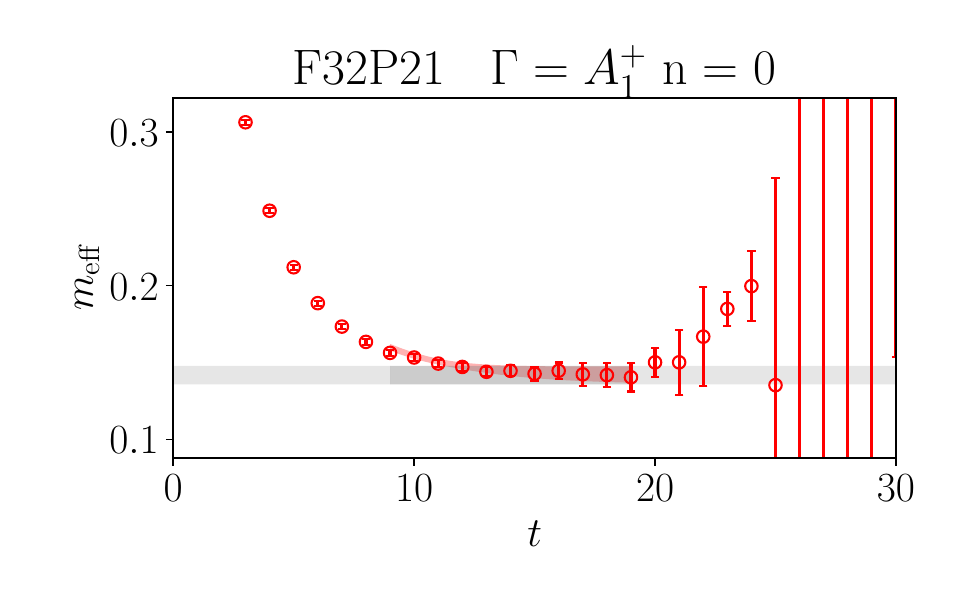}
\includegraphics[width=0.32\columnwidth]{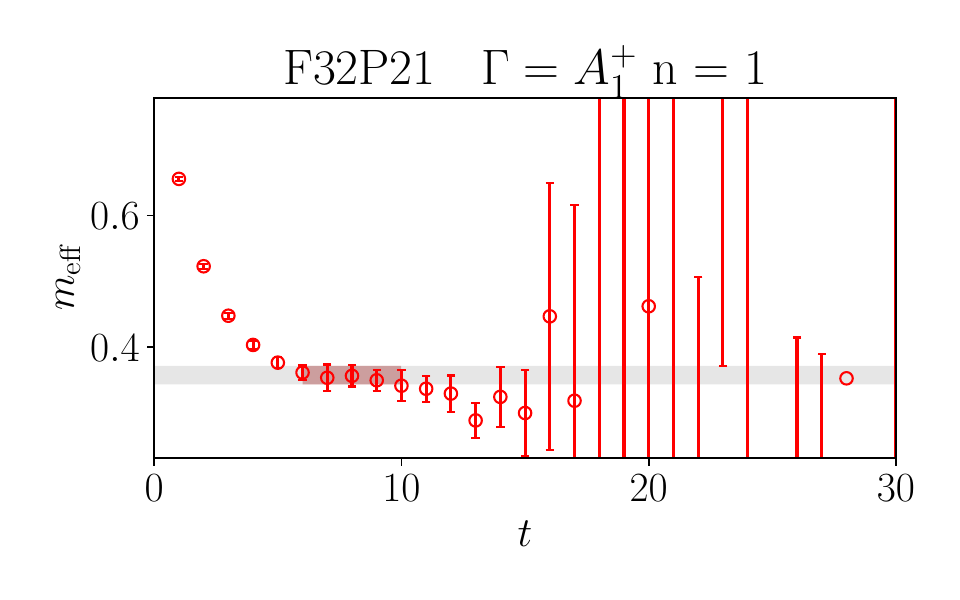}
\\
\includegraphics[width=0.32\columnwidth]{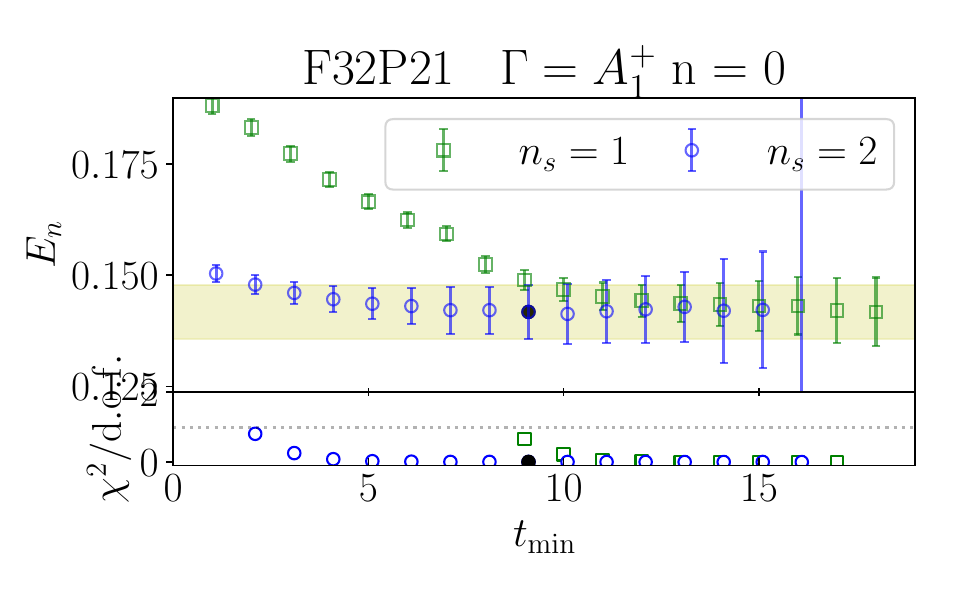}
\includegraphics[width=0.32\columnwidth]{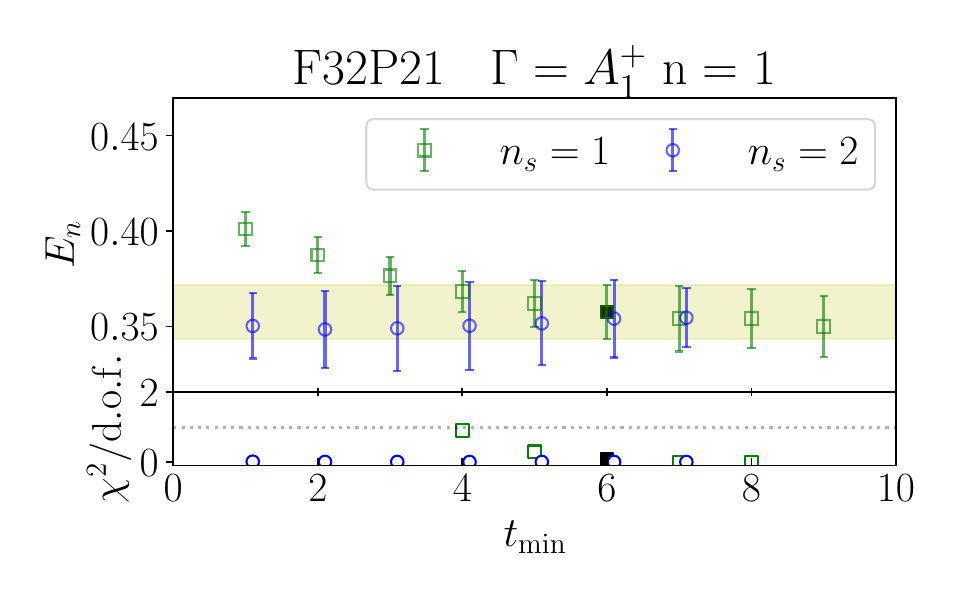}
\\
\includegraphics[width=0.32\columnwidth]{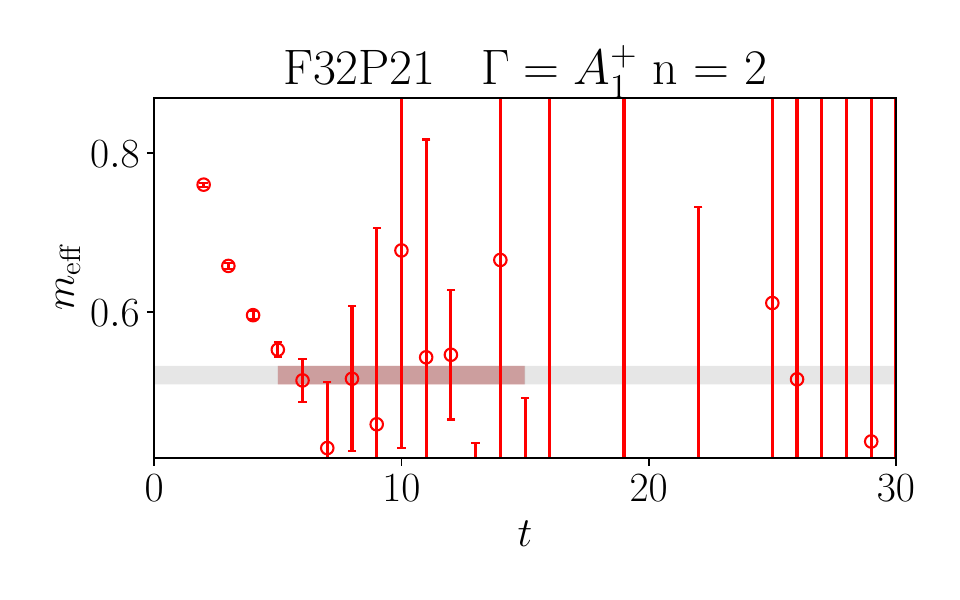}
\includegraphics[width=0.32\columnwidth]{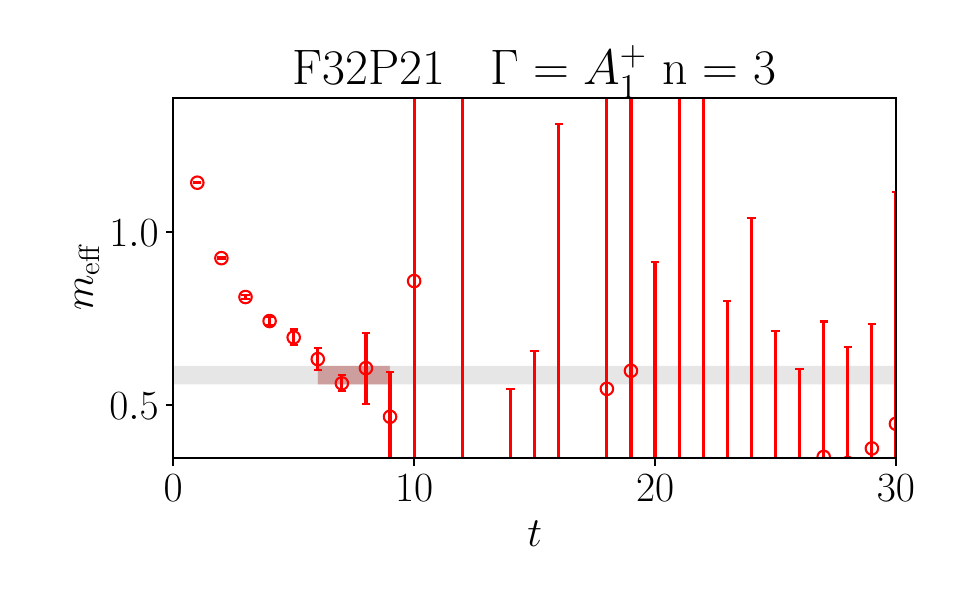}
\\
\includegraphics[width=0.32\columnwidth]{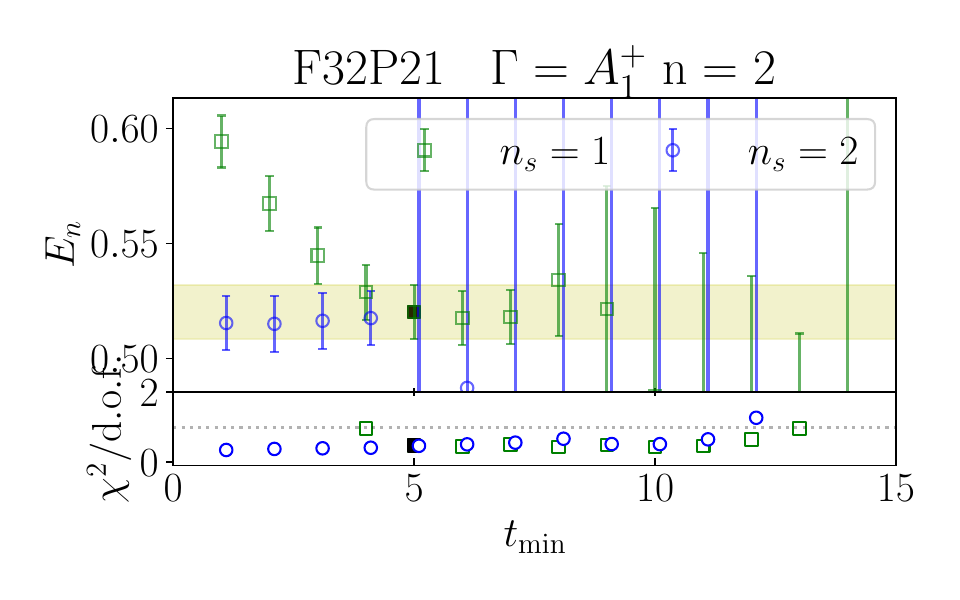}
\includegraphics[width=0.32\columnwidth]{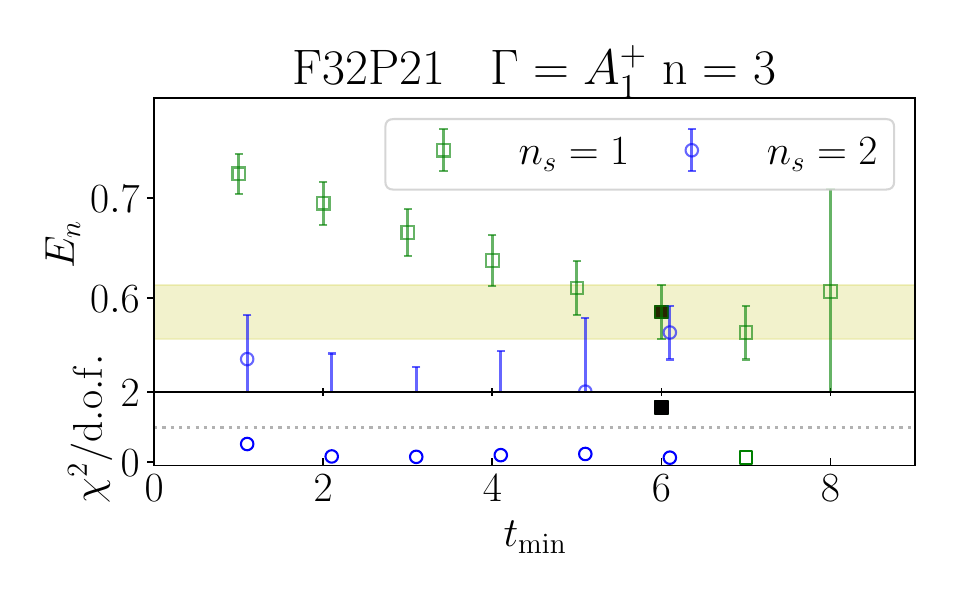}
\caption{Fit results for the $I=0$ $\pi\pi$ channel on the F32P21 ensemble.}
\label{fig:pipi-I=0-fit-F32P21}
\end{figure}

\begin{figure}[htbp]
\centering
\includegraphics[width=0.32\columnwidth]{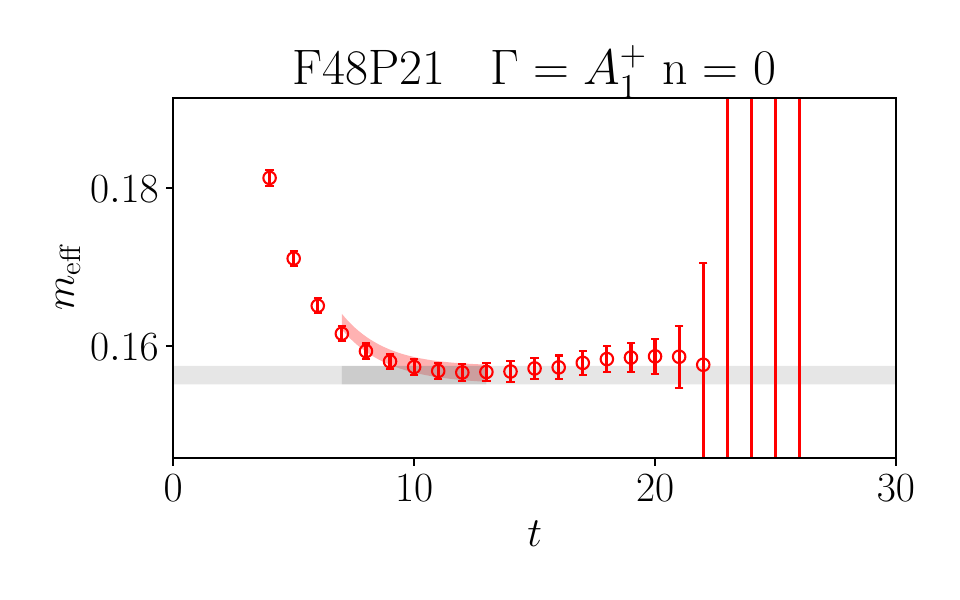}
\includegraphics[width=0.32\columnwidth]{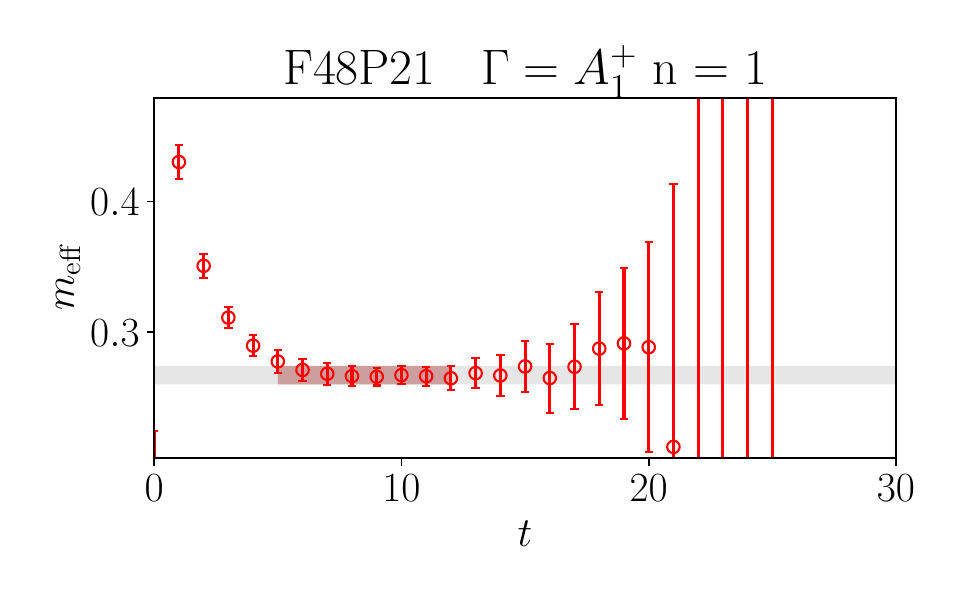}
\\
\includegraphics[width=0.32\columnwidth]{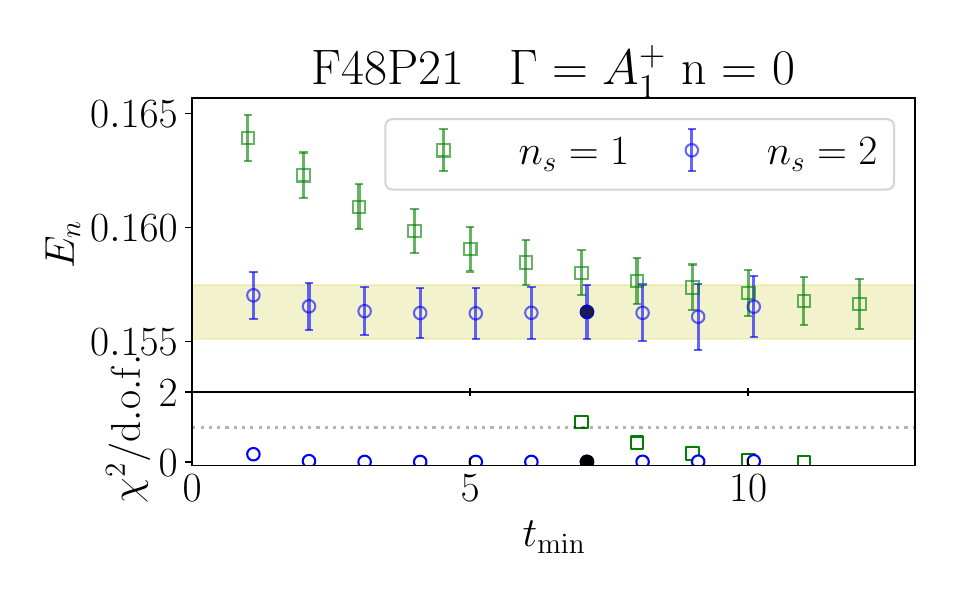}
\includegraphics[width=0.32\columnwidth]{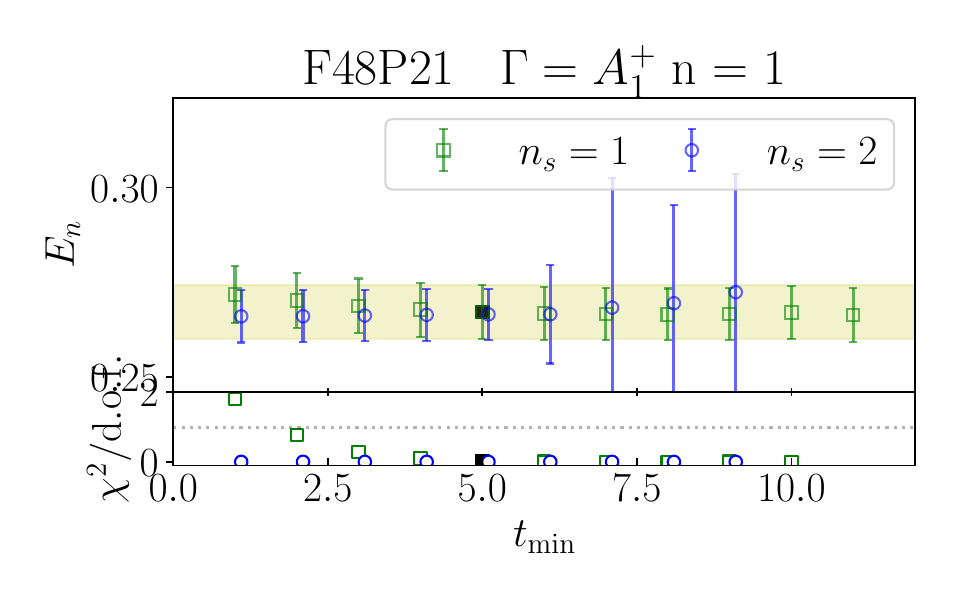}
\\
\includegraphics[width=0.32\columnwidth]{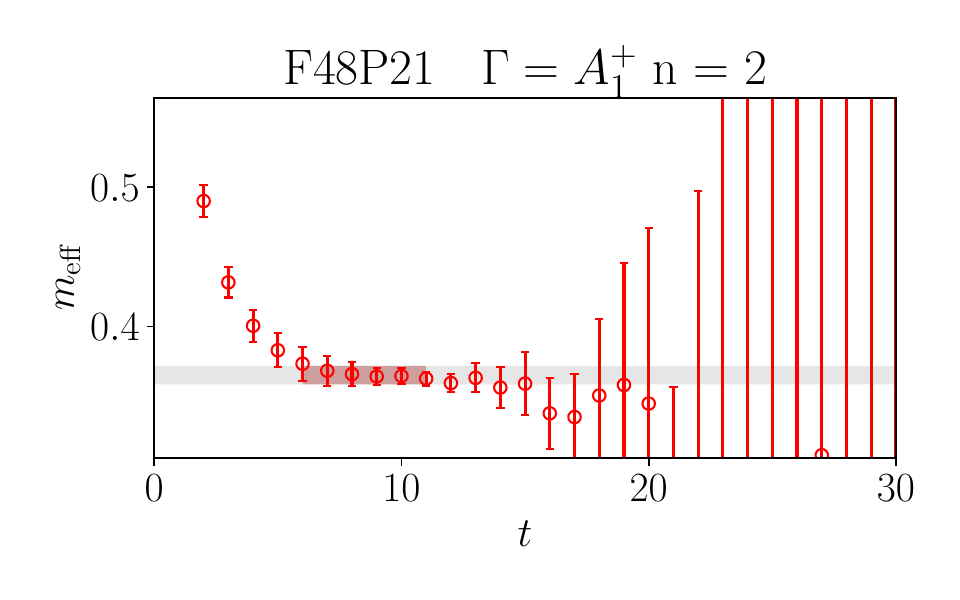}
\includegraphics[width=0.32\columnwidth]{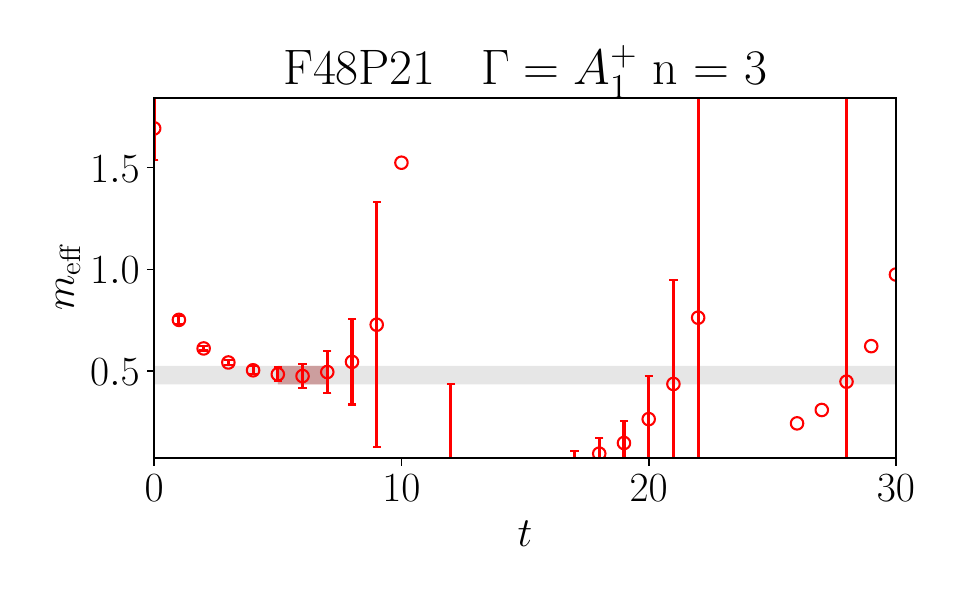}
\\
\includegraphics[width=0.32\columnwidth]{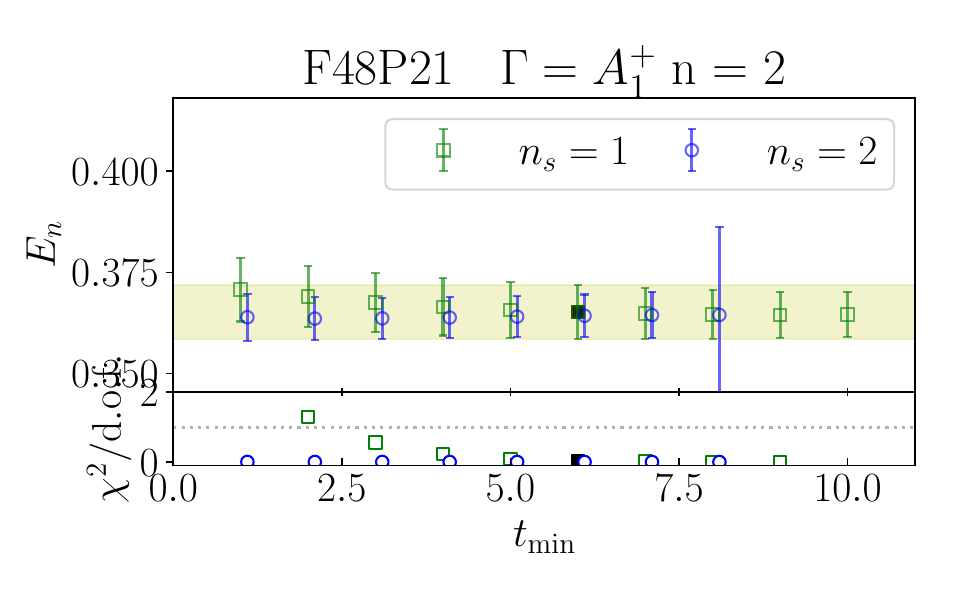}
\includegraphics[width=0.32\columnwidth]{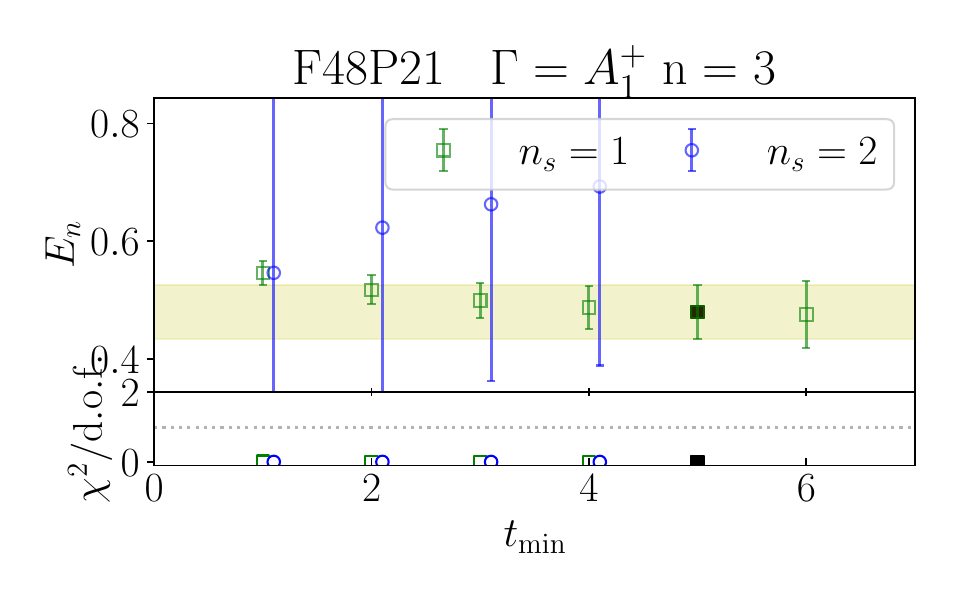}
\caption{Fit results for the $I=0$ $\pi\pi$ channel on the F48P21 ensemble.}
\label{fig:pipi-I=0-fit-F48P21}
\end{figure}

\begin{figure}[htbp]
\centering
\includegraphics[width=0.32\columnwidth]{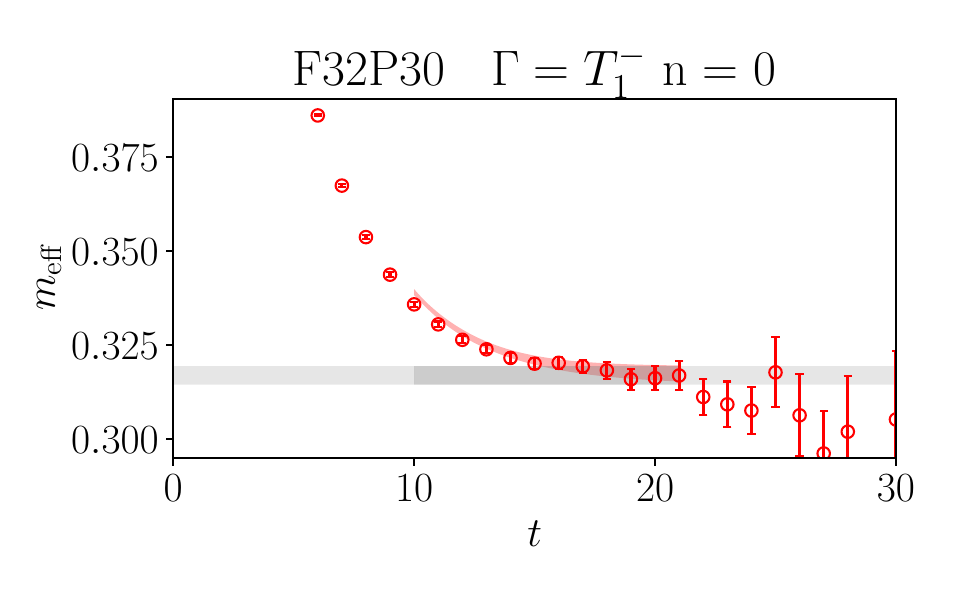}
\includegraphics[width=0.32\columnwidth]{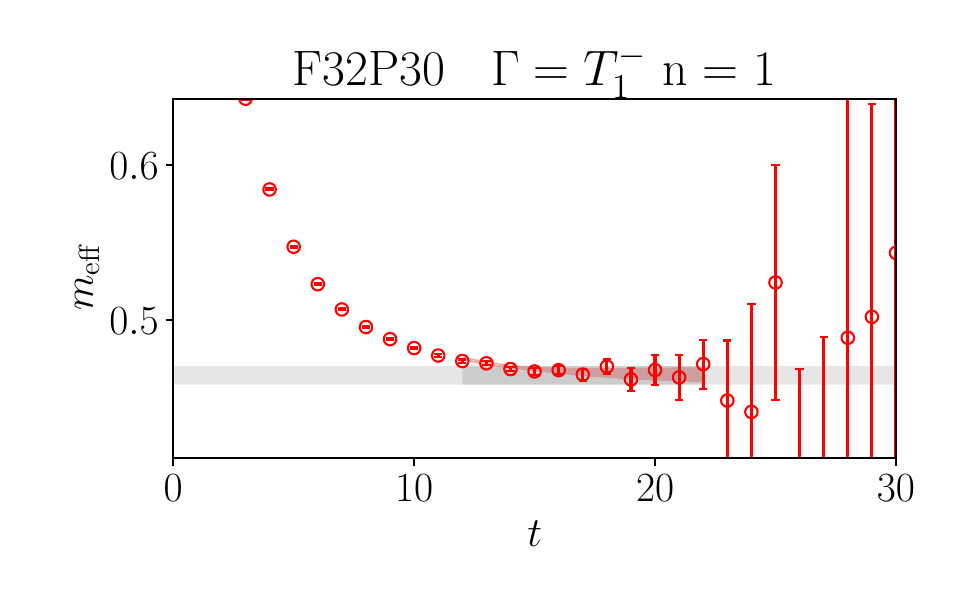}
\includegraphics[width=0.32\columnwidth]{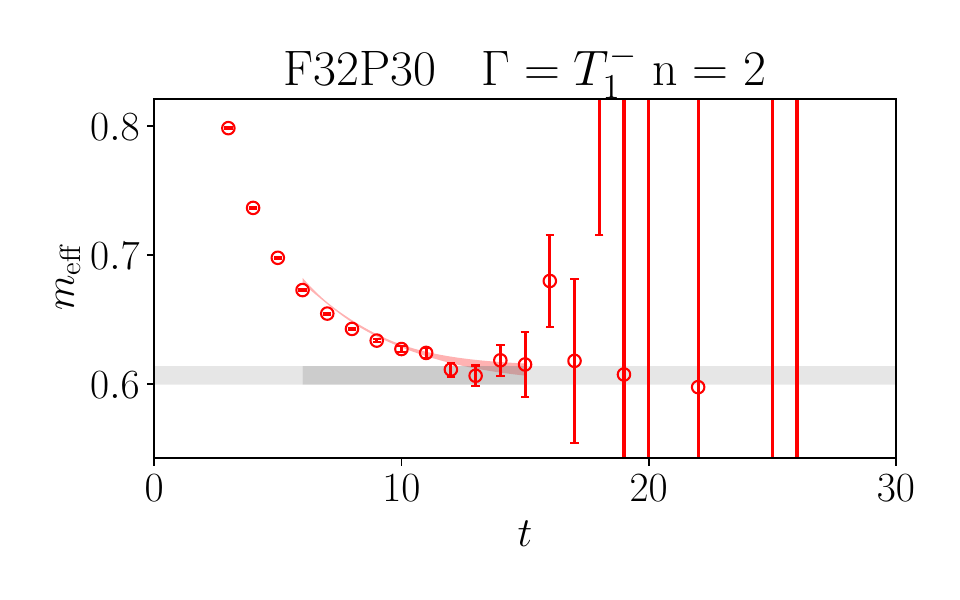}
\\
\includegraphics[width=0.32\columnwidth]{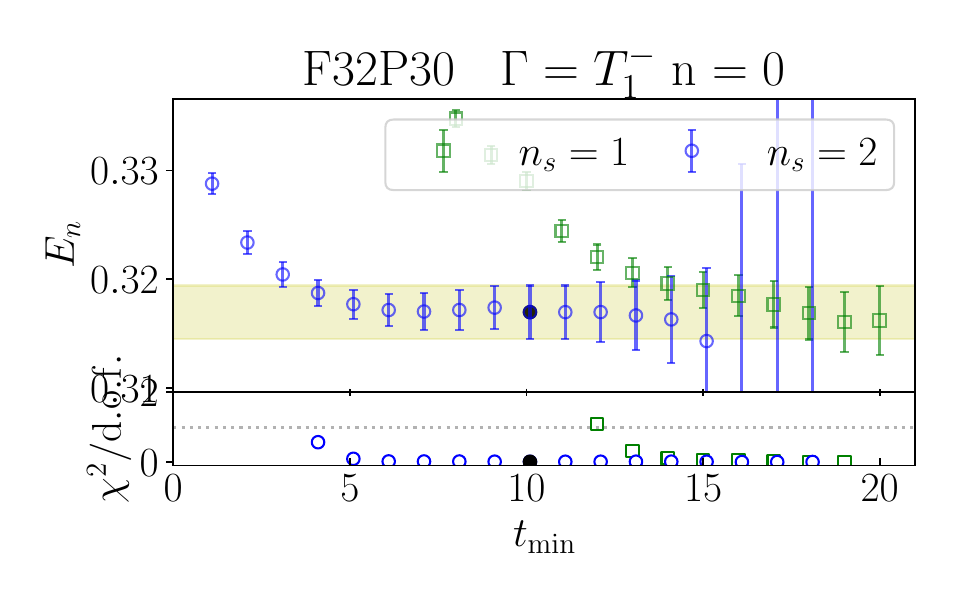}
\includegraphics[width=0.32\columnwidth]{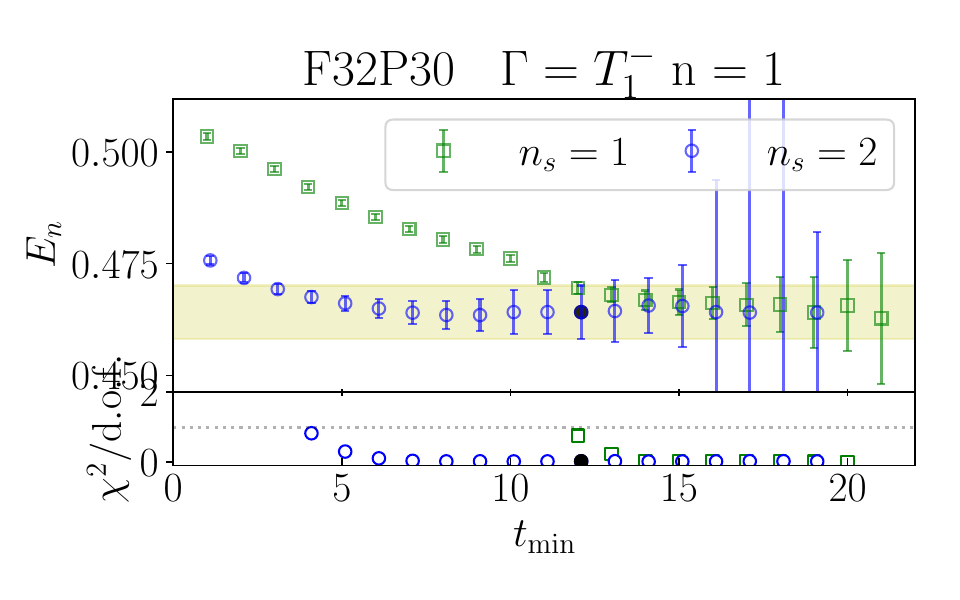}
\includegraphics[width=0.32\columnwidth]{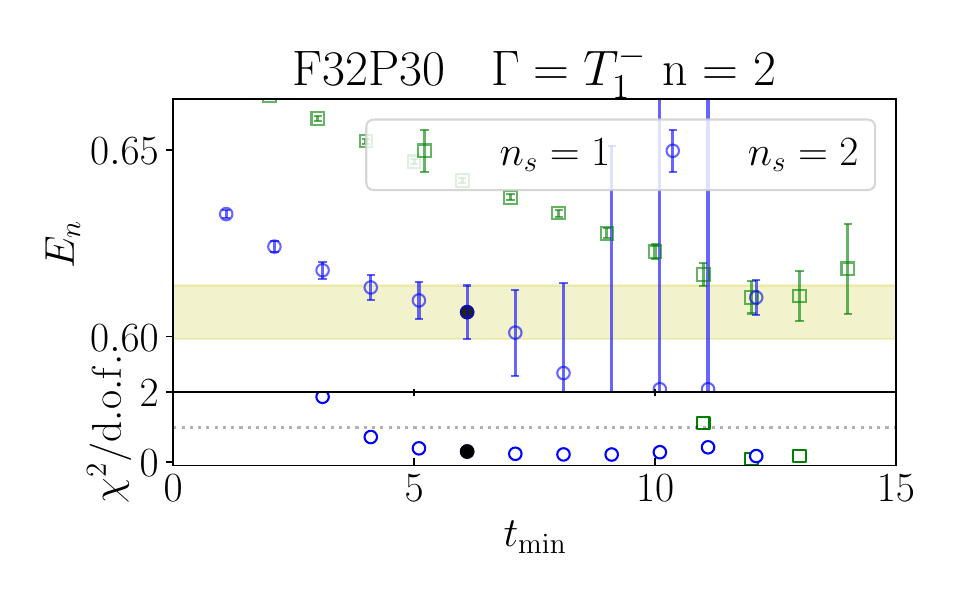}
\caption{Fit results for the $I=1$ $\pi\pi$ channel on the F32P30 ensemble.}
\label{fig:pipi-I=1-fit-F32P30}
\end{figure}

\begin{figure}[htbp]
\centering
\includegraphics[width=0.32\columnwidth]{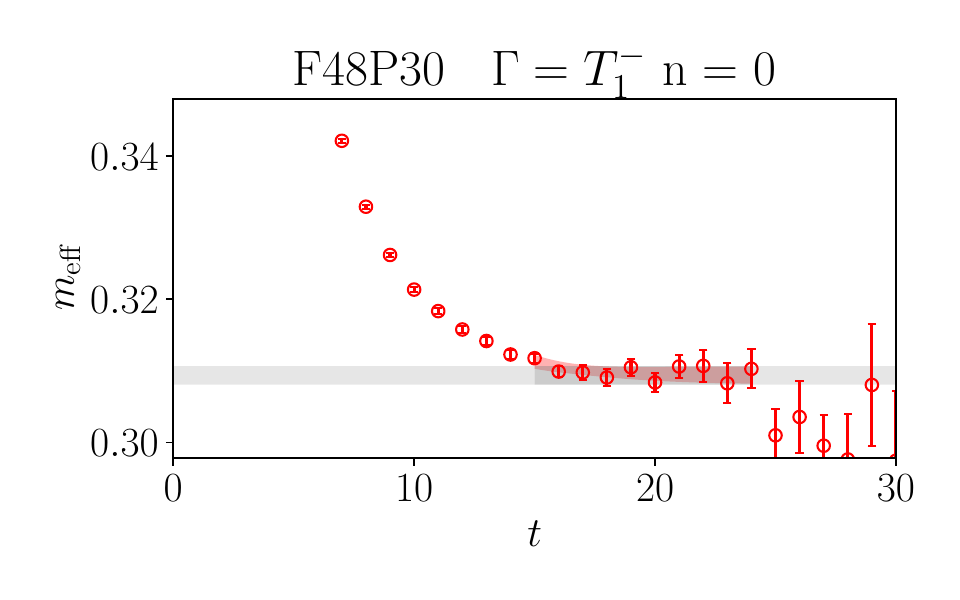}
\includegraphics[width=0.32\columnwidth]{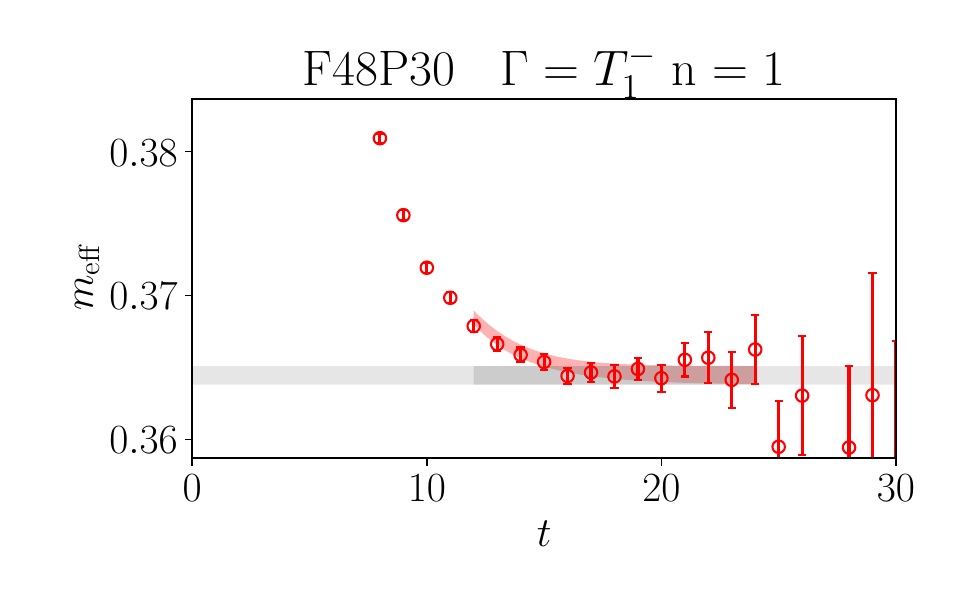}
\includegraphics[width=0.32\columnwidth]{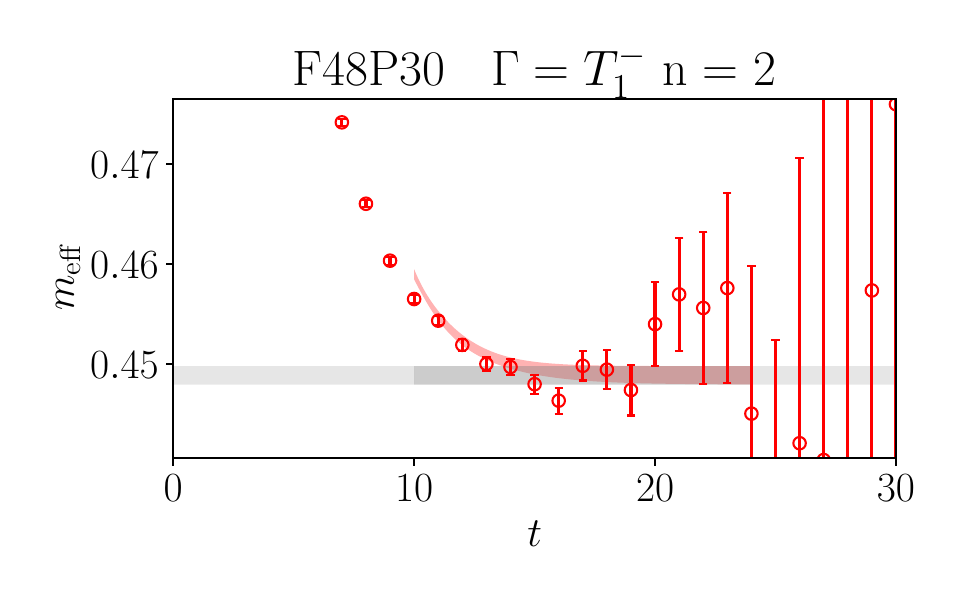}
\\
\includegraphics[width=0.32\columnwidth]{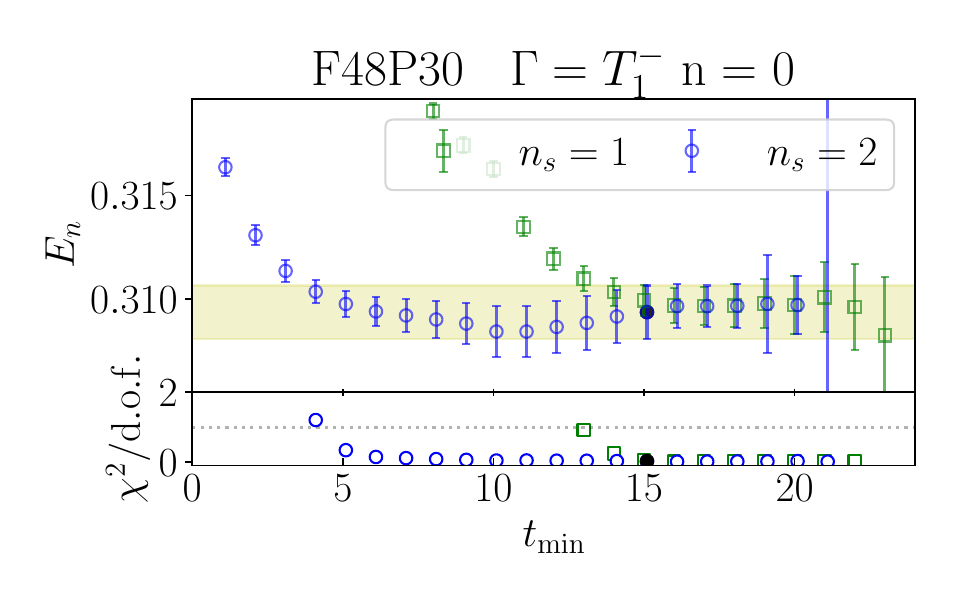}
\includegraphics[width=0.32\columnwidth]{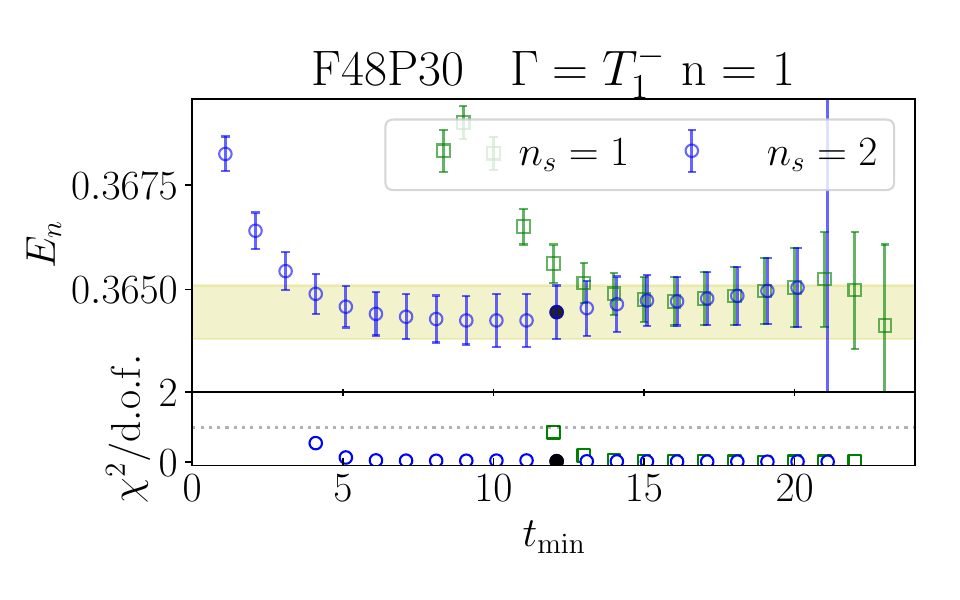}
\includegraphics[width=0.32\columnwidth]{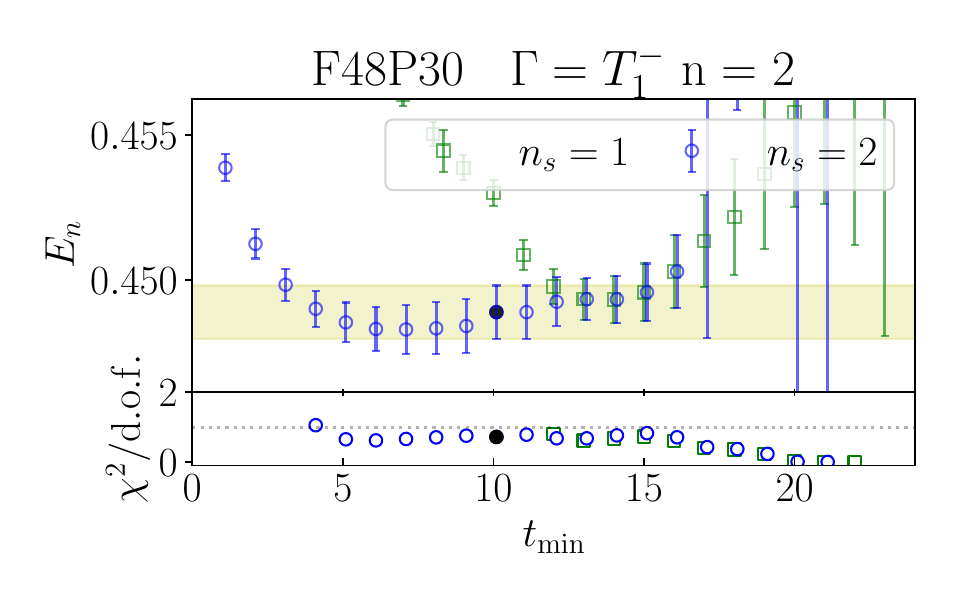}
\caption{Fit results for the $I=1$ $\pi\pi$ channel on the F48P30 ensemble.}
\label{fig:pipi-I=1-fit-F48P30}
\end{figure}

\begin{figure}[htbp]
\centering
\includegraphics[width=0.32\columnwidth]{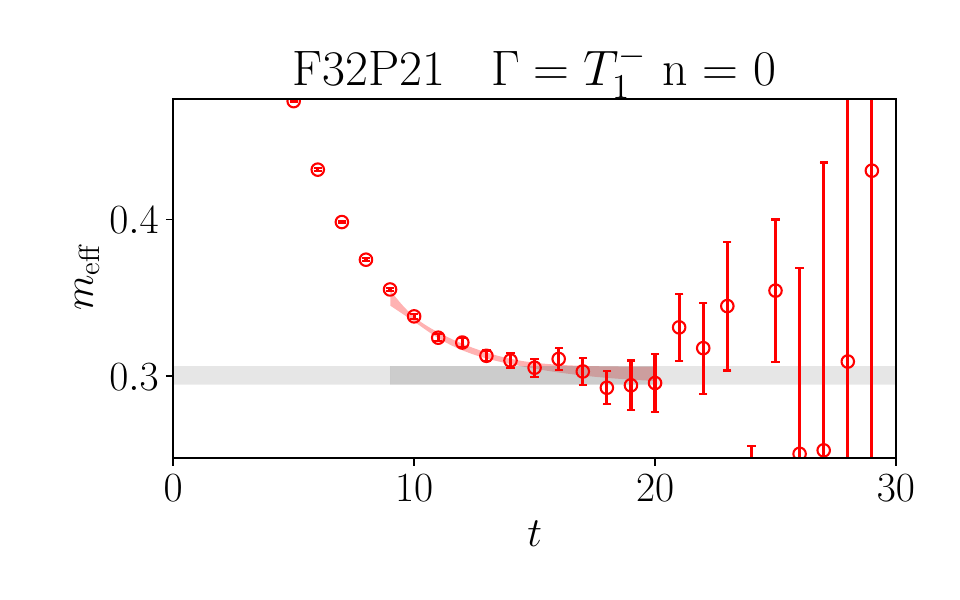}
\includegraphics[width=0.32\columnwidth]{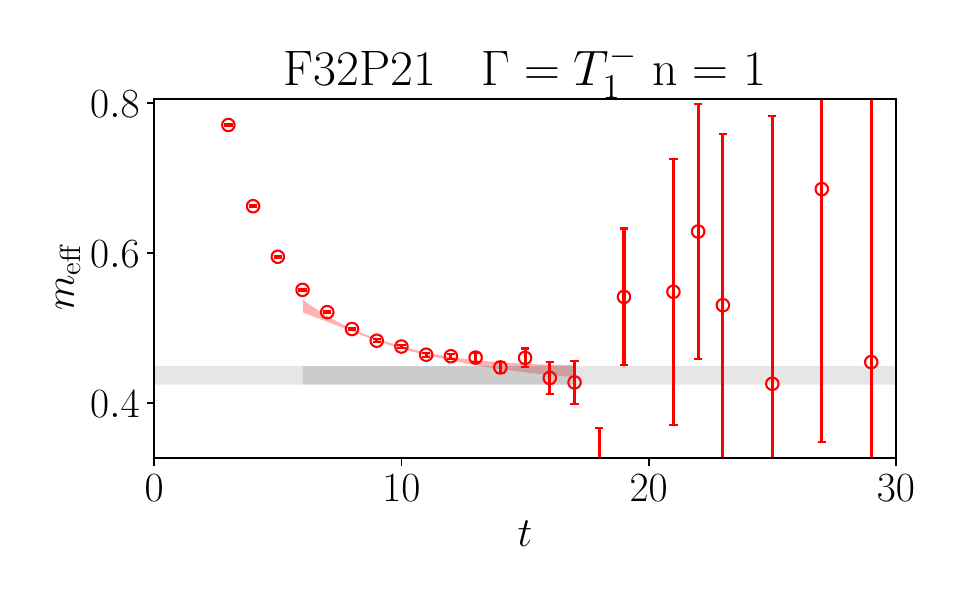}
\includegraphics[width=0.32\columnwidth]{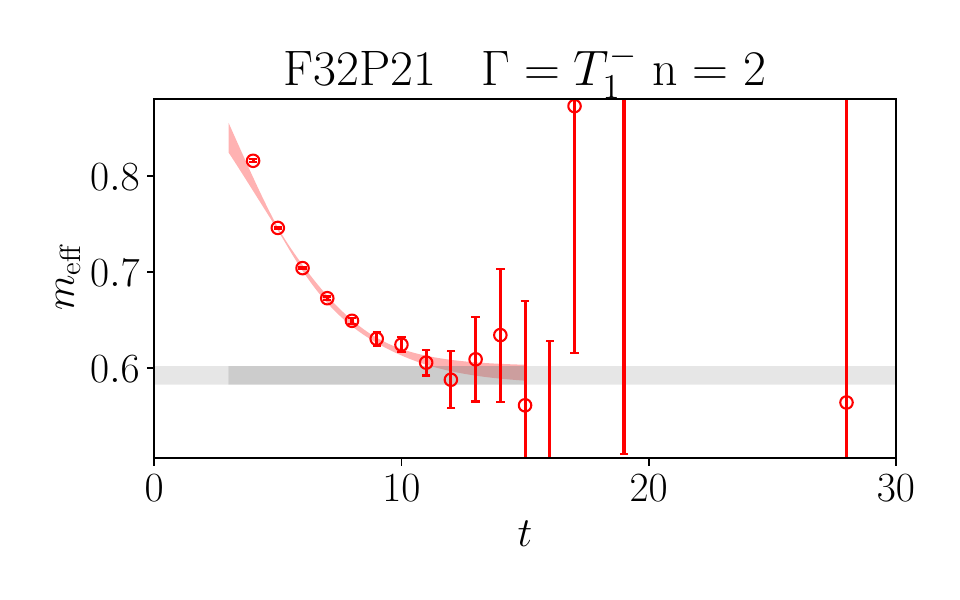}
\\
\includegraphics[width=0.32\columnwidth]{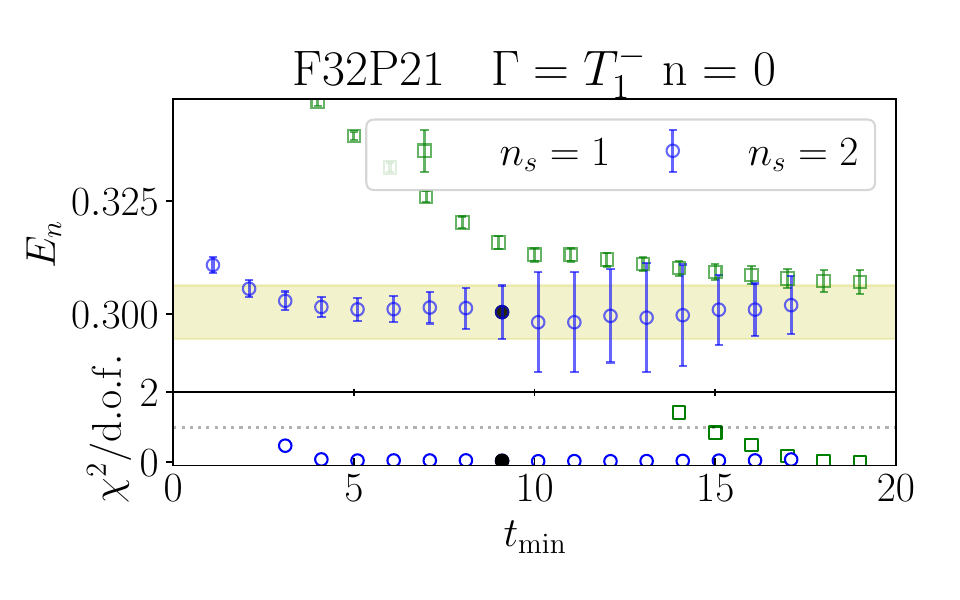}
\includegraphics[width=0.32\columnwidth]{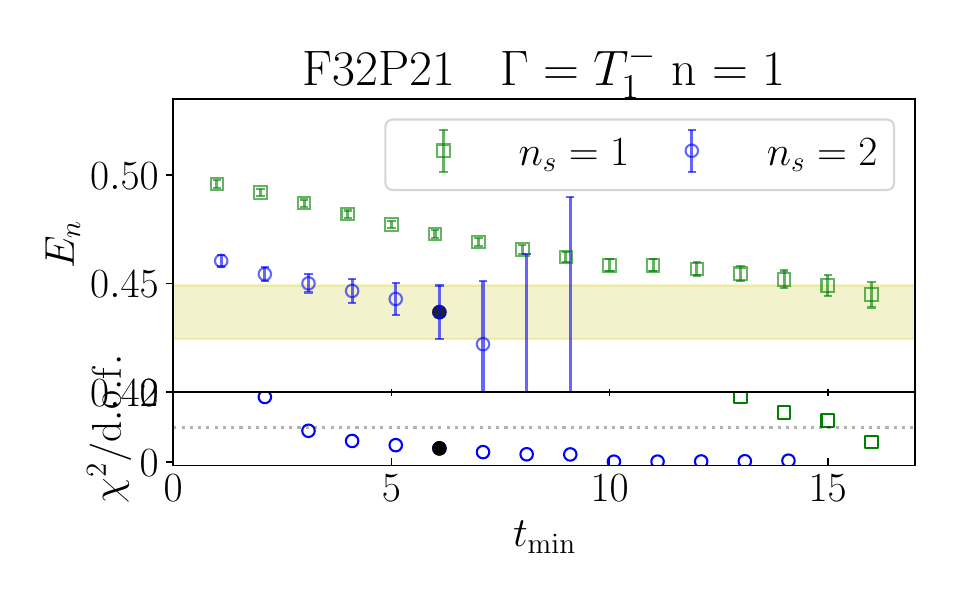}
\includegraphics[width=0.32\columnwidth]{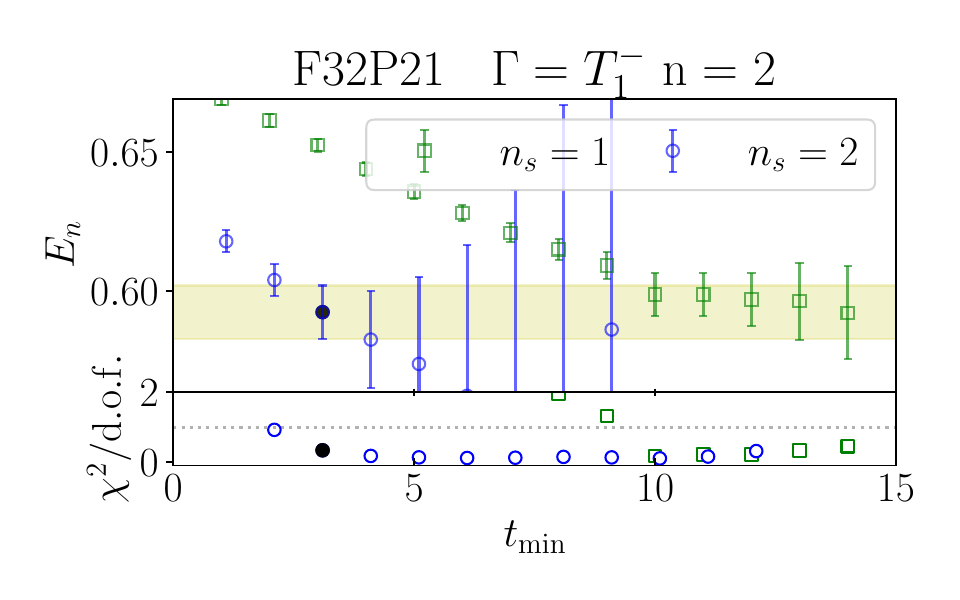}
\caption{Fit results for the $I=1$ $\pi\pi$ channel on the F32P21 ensemble.}
\label{fig:pipi-I=1-fit-F32P21}
\end{figure}

\begin{figure}[htbp]
\centering
\includegraphics[width=0.32\columnwidth]{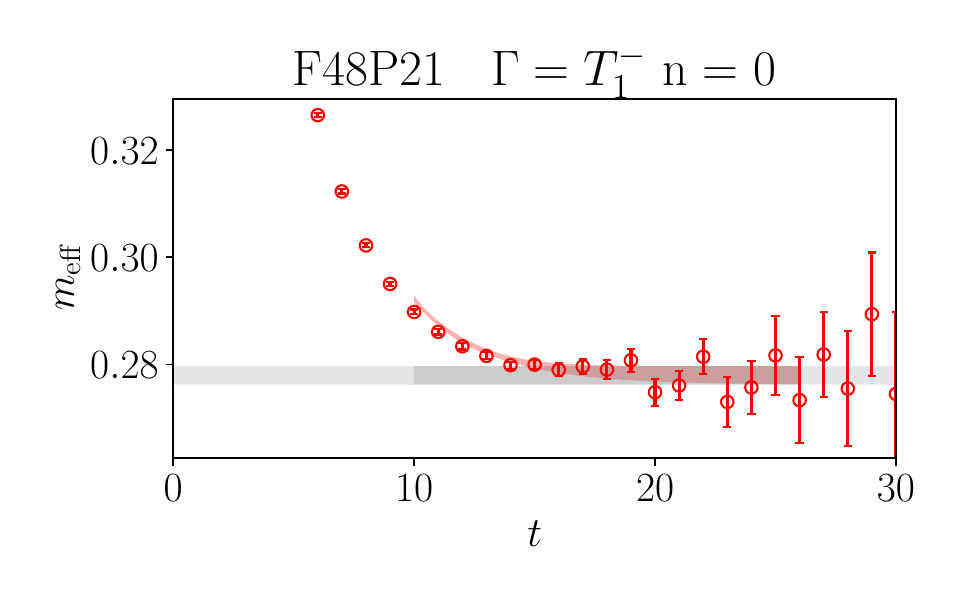}
\includegraphics[width=0.32\columnwidth]{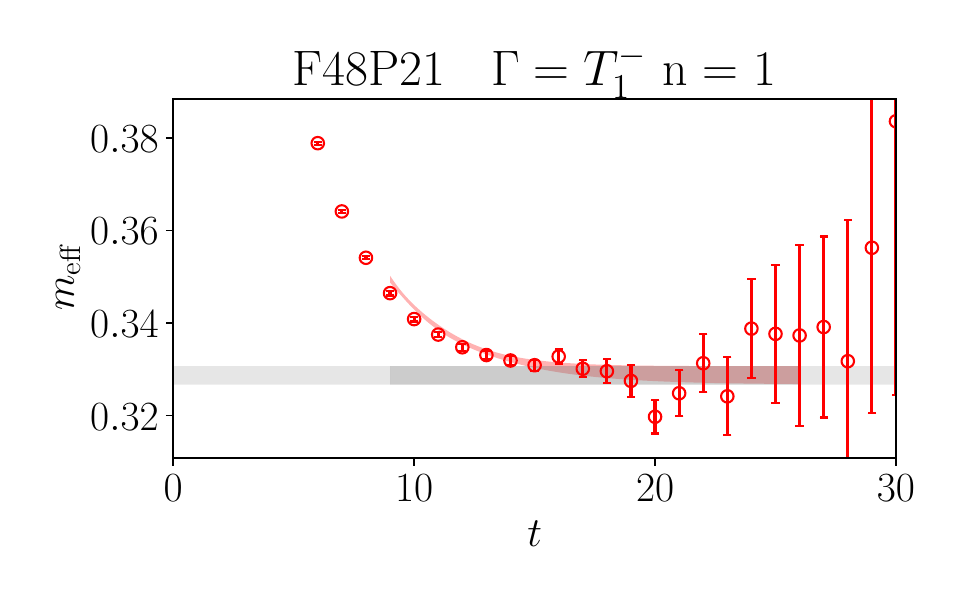}
\includegraphics[width=0.32\columnwidth]{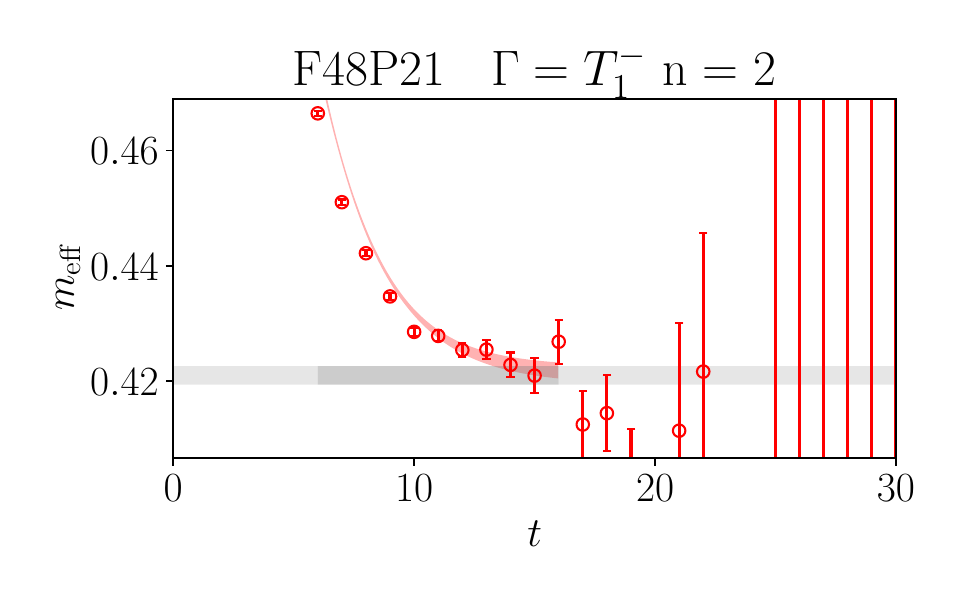}
\\
\includegraphics[width=0.32\columnwidth]{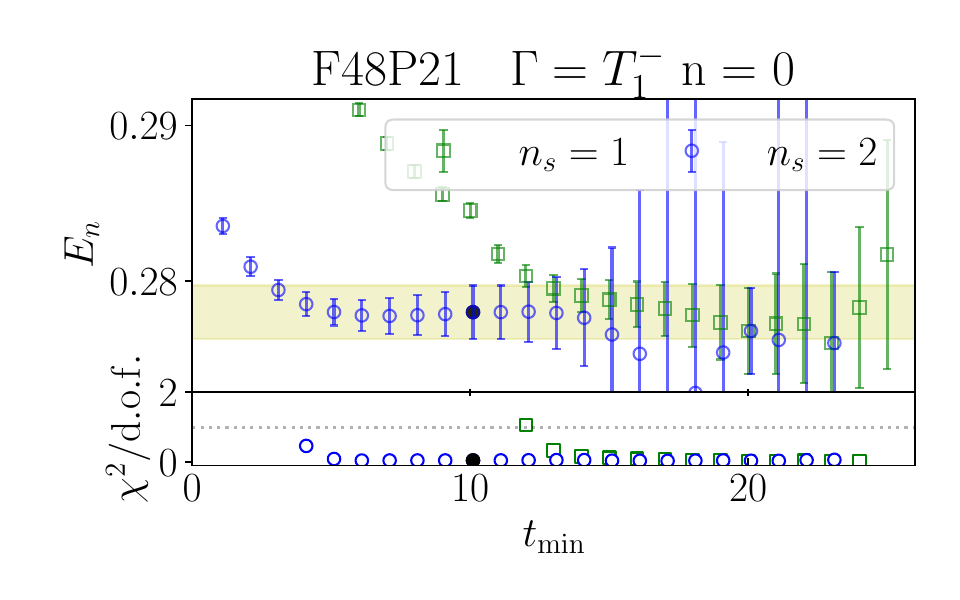}
\includegraphics[width=0.32\columnwidth]{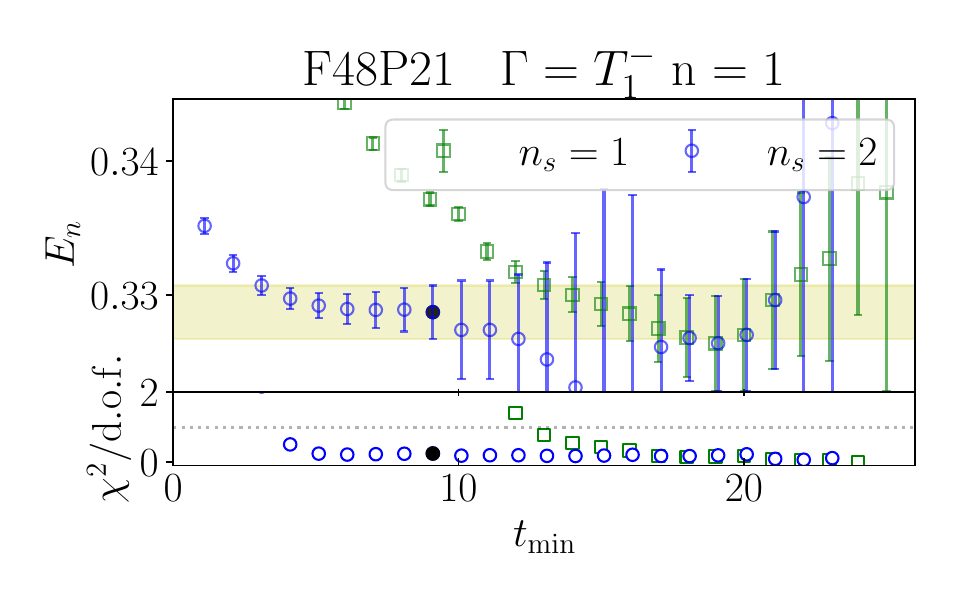}
\includegraphics[width=0.32\columnwidth]{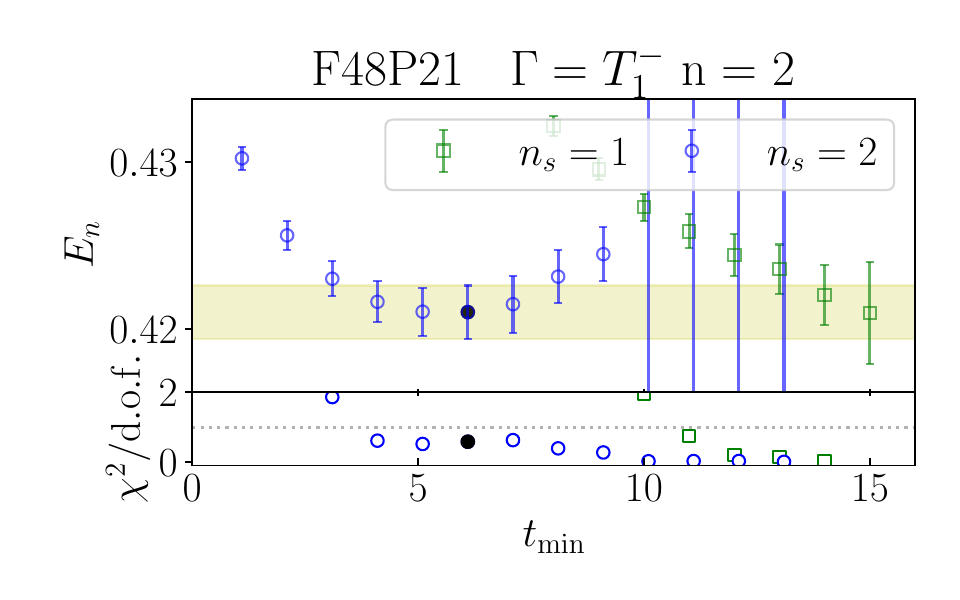}
\caption{Fit results for the $I=1$ $\pi\pi$ channel on the F48P21 ensemble.}
\label{fig:pipi-I=1-fit-F48P21}
\end{figure}

\begin{figure}[htbp]
\centering
\includegraphics[width=0.32\columnwidth]{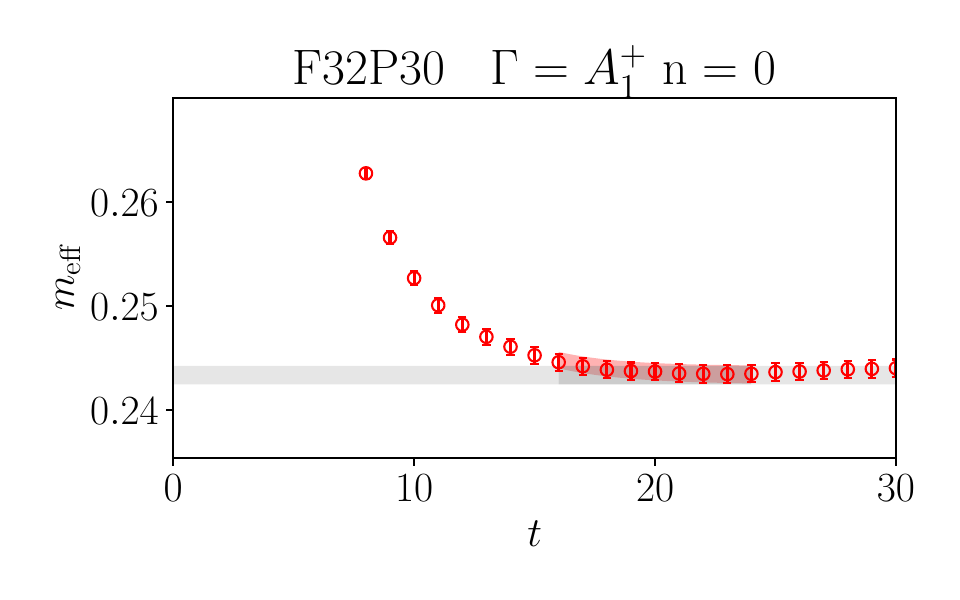}
\includegraphics[width=0.32\columnwidth]{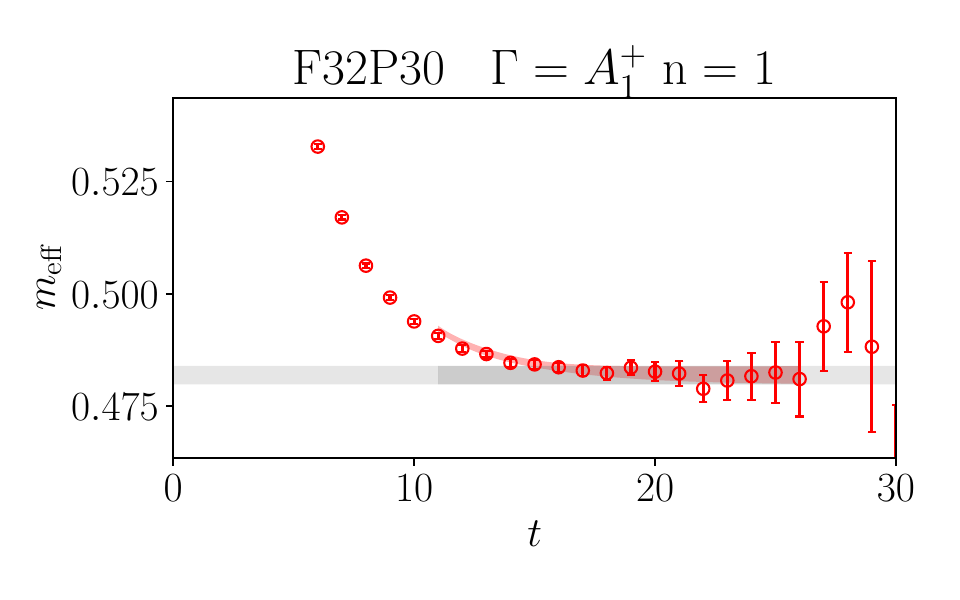}
\includegraphics[width=0.32\columnwidth]{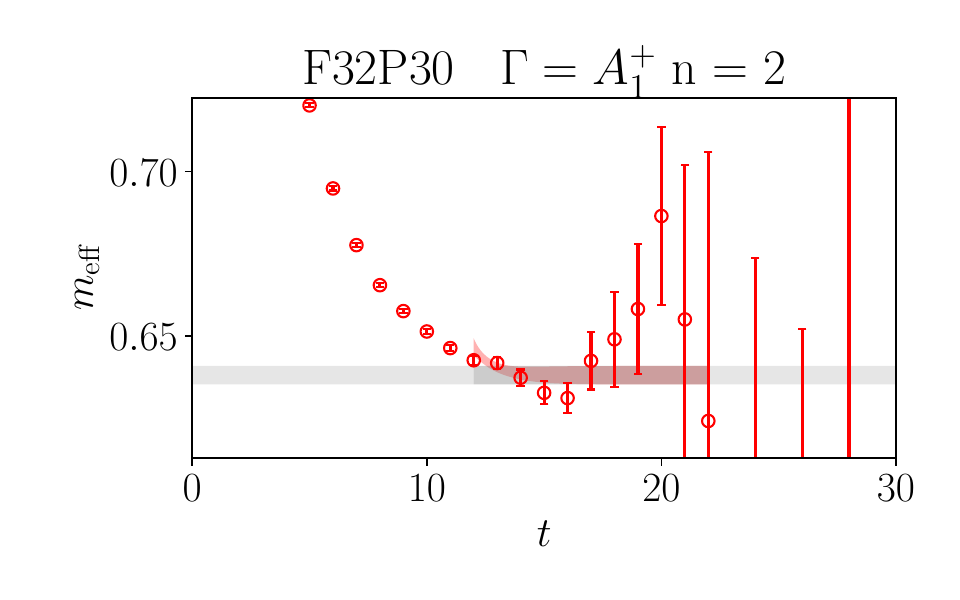}
\\
\includegraphics[width=0.32\columnwidth]{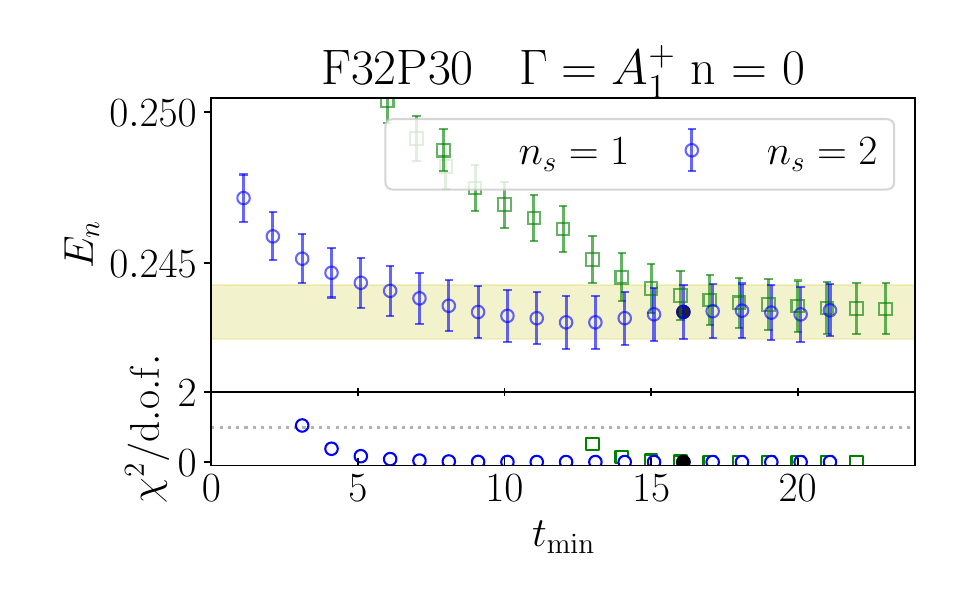}
\includegraphics[width=0.32\columnwidth]{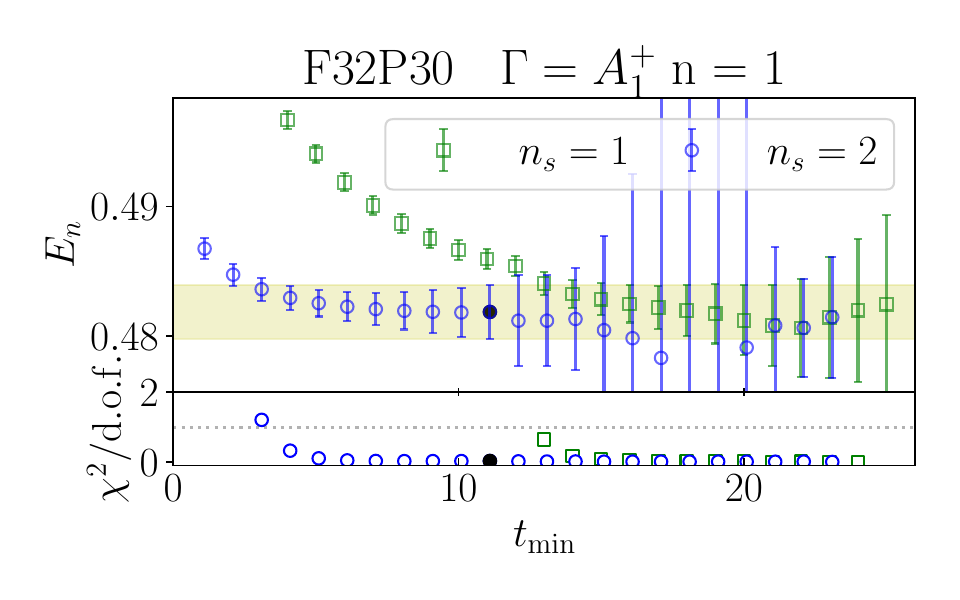}
\includegraphics[width=0.32\columnwidth]{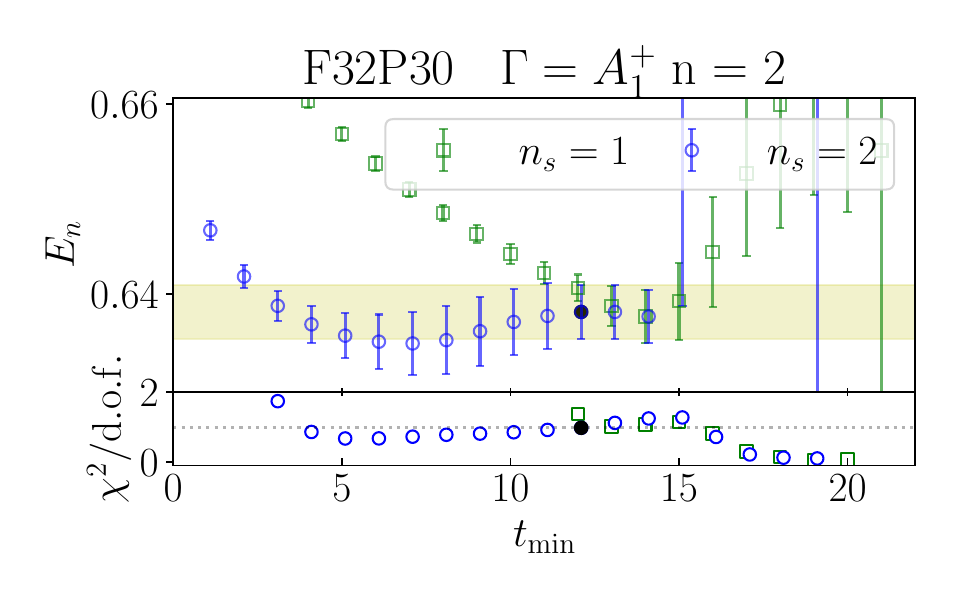}
\caption{Fit results for the $I=2$ $\pi\pi$ channel on the F32P30 ensemble.}
\label{fig:pipi-I=2-fit-F32P30}
\end{figure}

\begin{figure}[htbp]
\centering
\includegraphics[width=0.32\columnwidth]{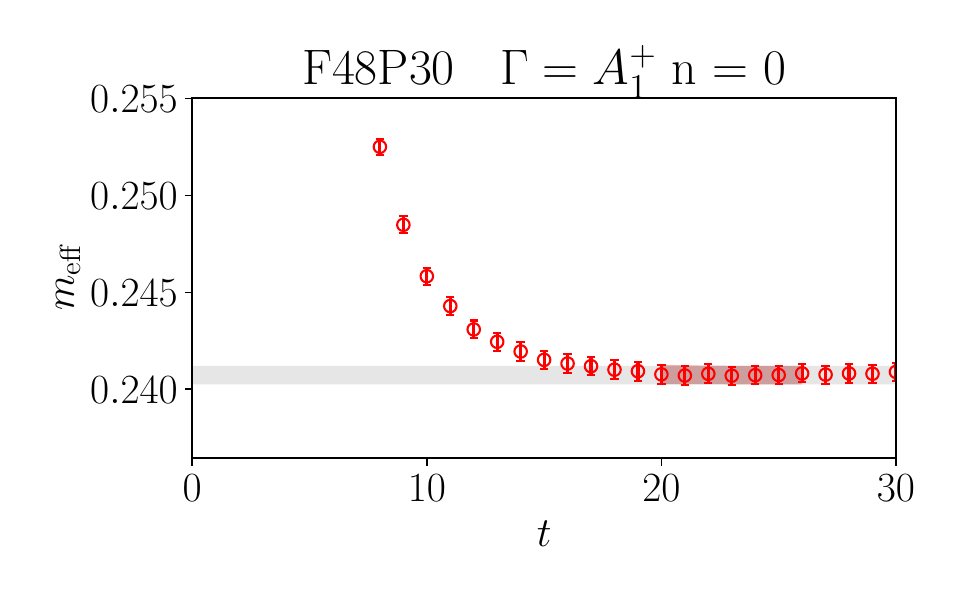}
\includegraphics[width=0.32\columnwidth]{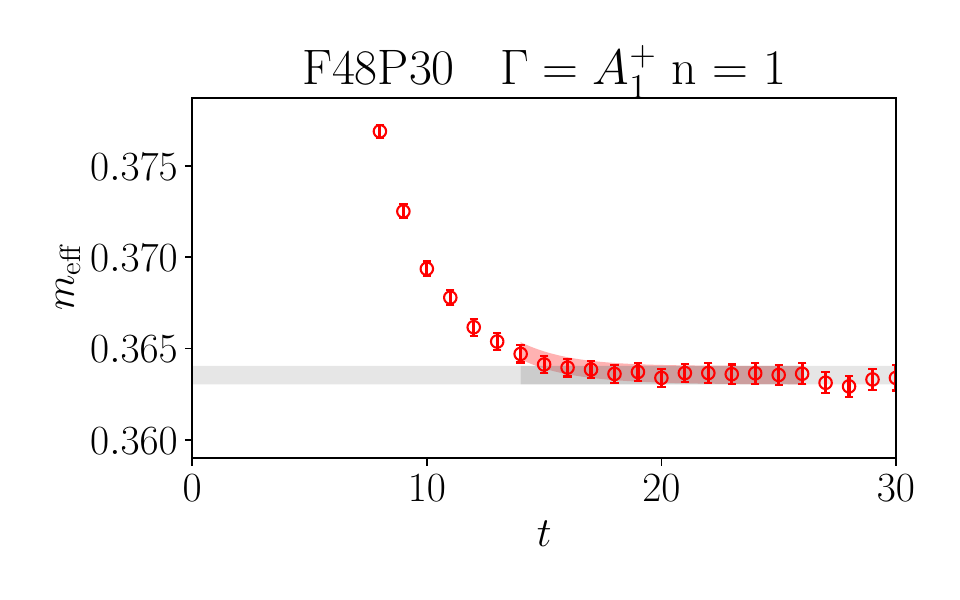}
\includegraphics[width=0.32\columnwidth]{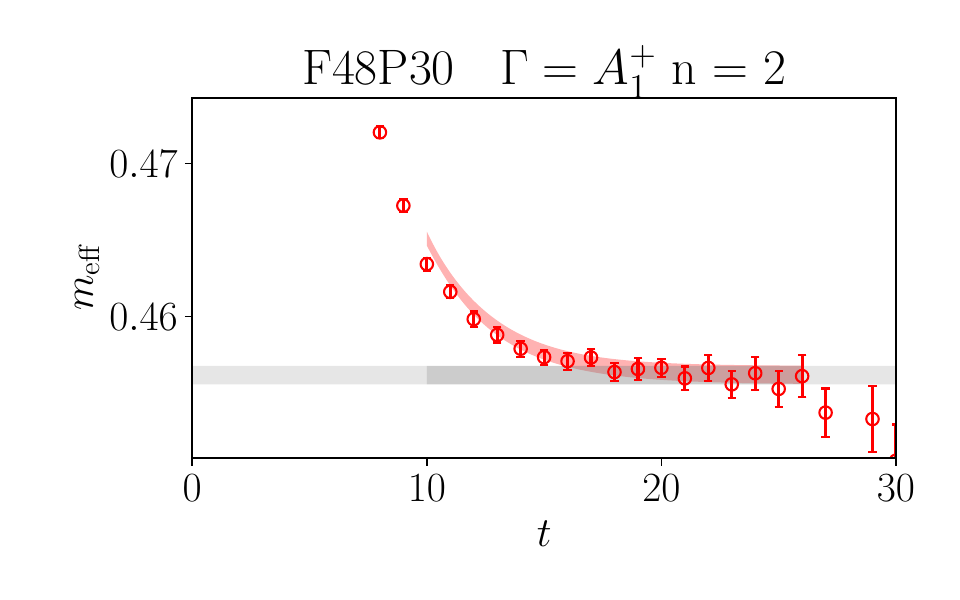}
\\
\includegraphics[width=0.32\columnwidth]{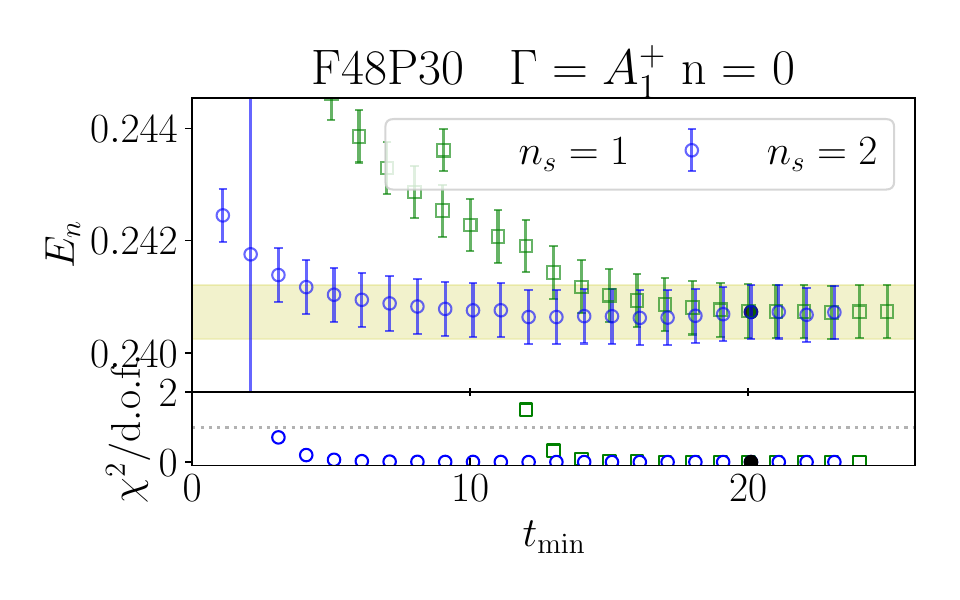}
\includegraphics[width=0.32\columnwidth]{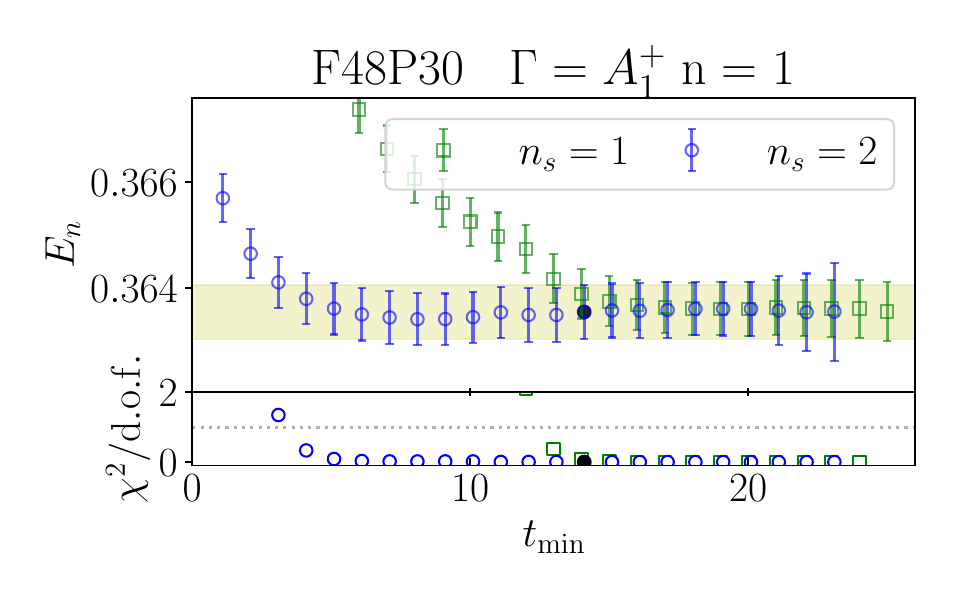}
\includegraphics[width=0.32\columnwidth]{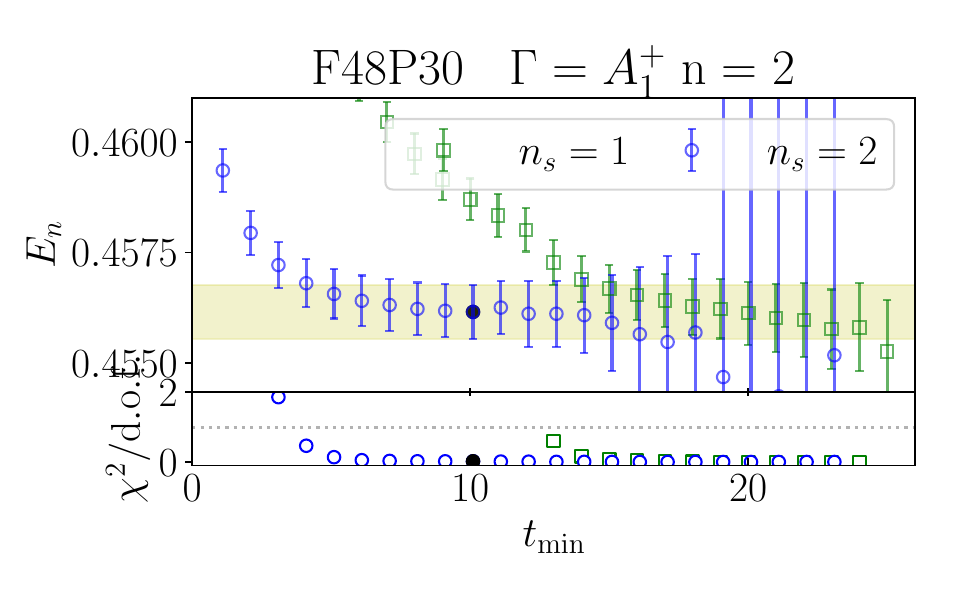}
\caption{Fit results for the $I=2$ $\pi\pi$ channel on the F48P30 ensemble.}
\label{fig:pipi-I=2-fit-F48P30}
\end{figure}

\begin{figure}[htbp]
\centering
\includegraphics[width=0.32\columnwidth]{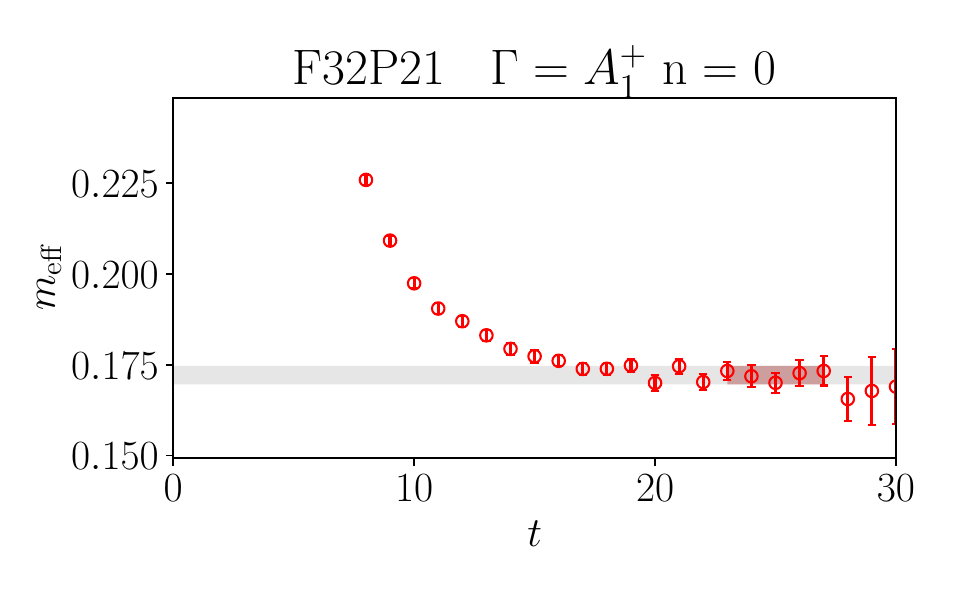}
\includegraphics[width=0.32\columnwidth]{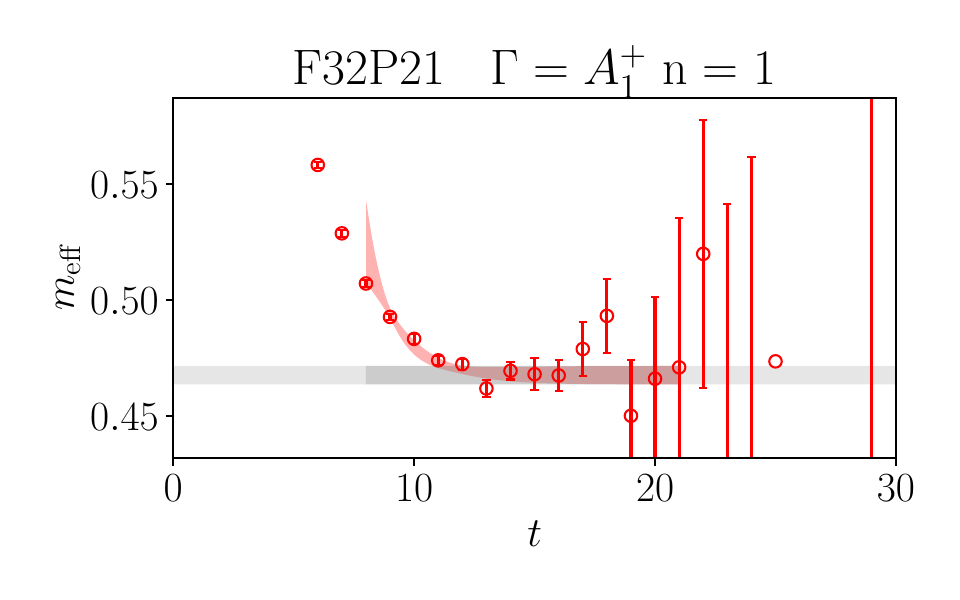}
\includegraphics[width=0.32\columnwidth]{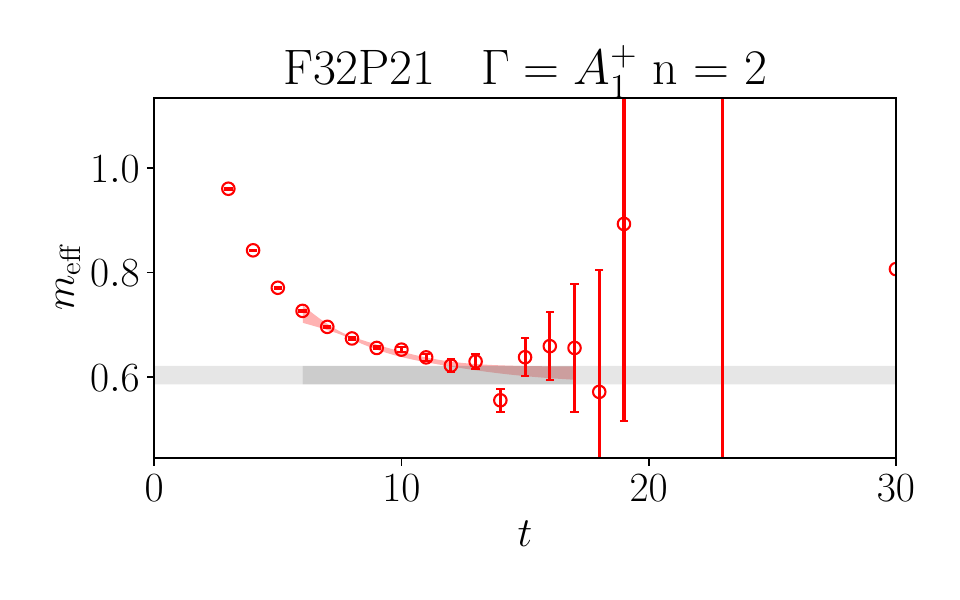}
\\
\includegraphics[width=0.32\columnwidth]{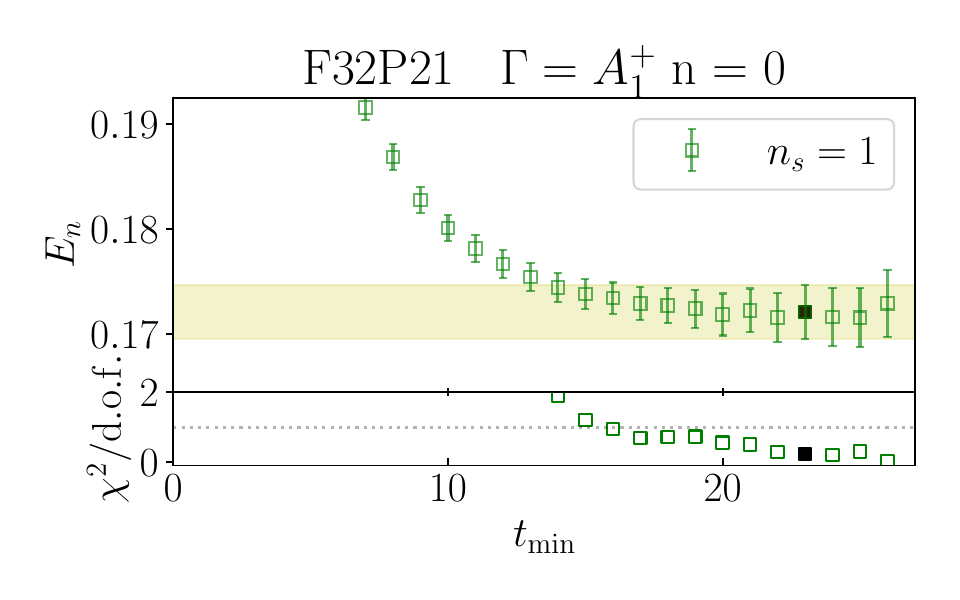}
\includegraphics[width=0.32\columnwidth]{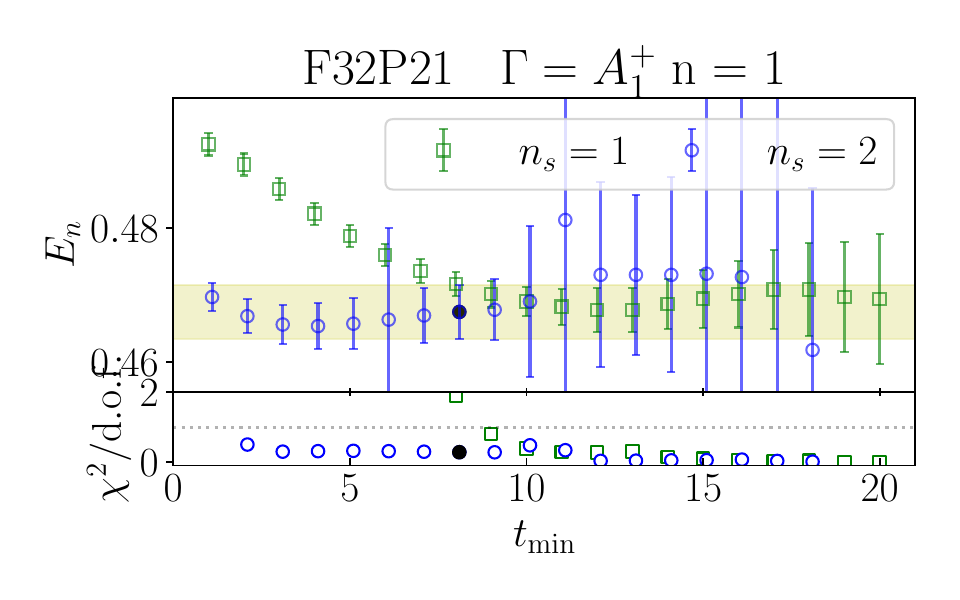}
\includegraphics[width=0.32\columnwidth]{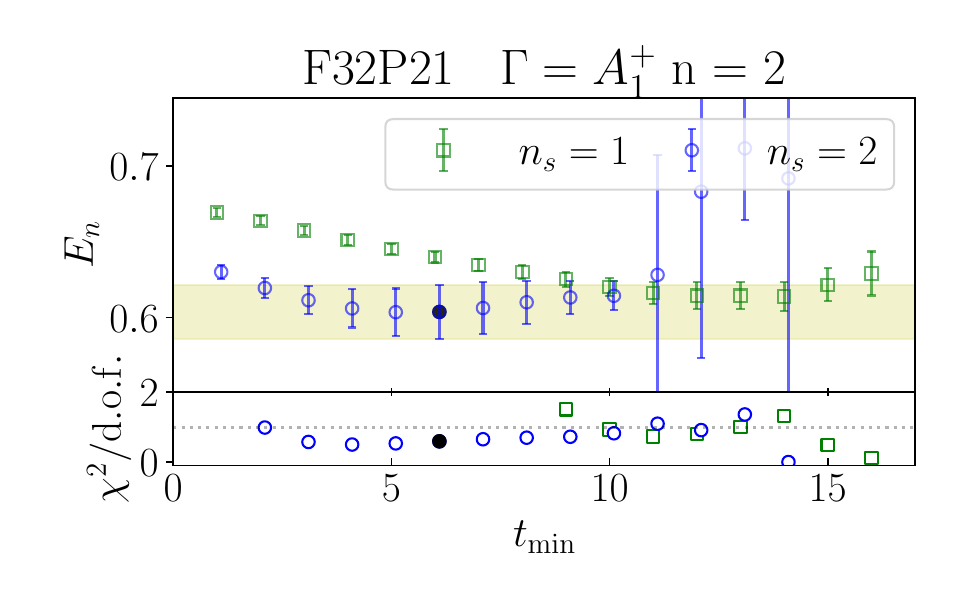}
\caption{Fit results for the $I=2$ $\pi\pi$ channel on the F32P21 ensemble.}
\label{fig:pipi-I=2-fit-F32P21}
\end{figure}

\begin{figure}[htbp]
\centering
\includegraphics[width=0.32\columnwidth]{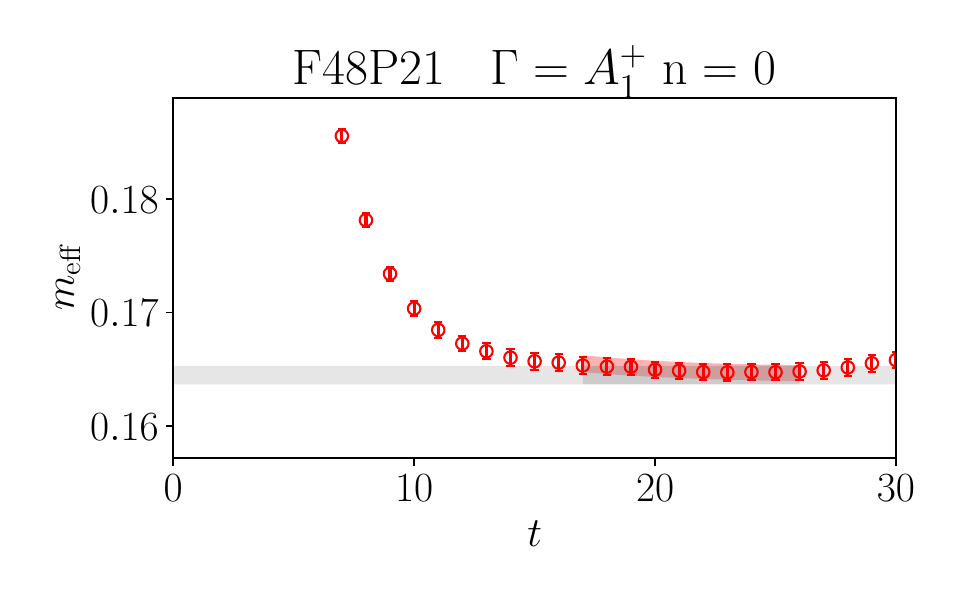}
\includegraphics[width=0.32\columnwidth]{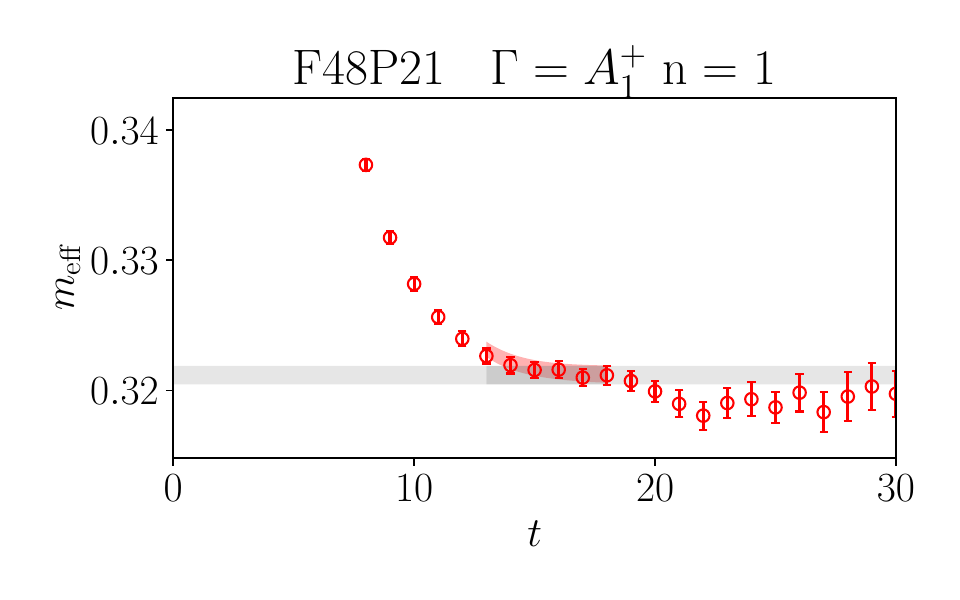}
\includegraphics[width=0.32\columnwidth]{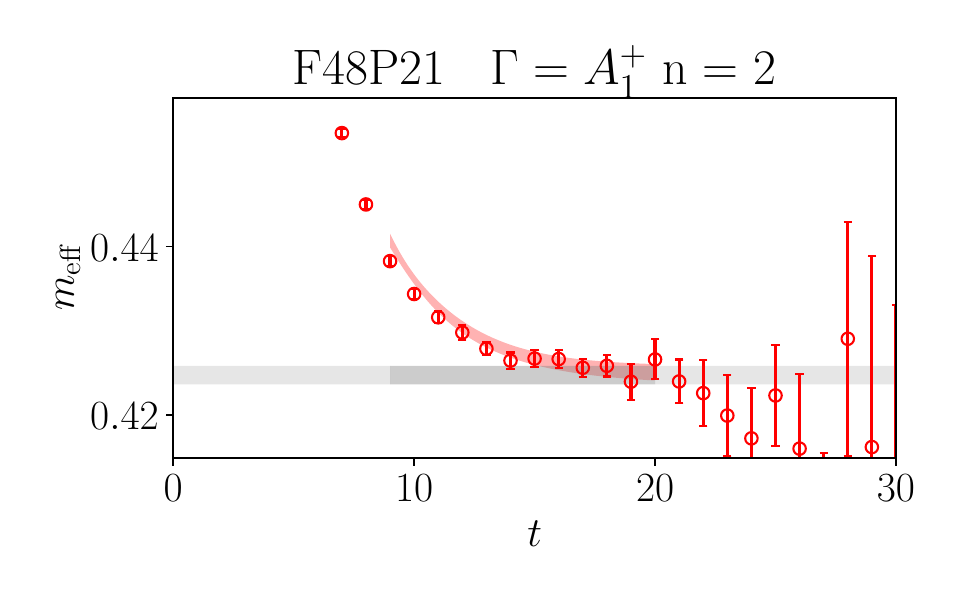}
\\
\includegraphics[width=0.32\columnwidth]{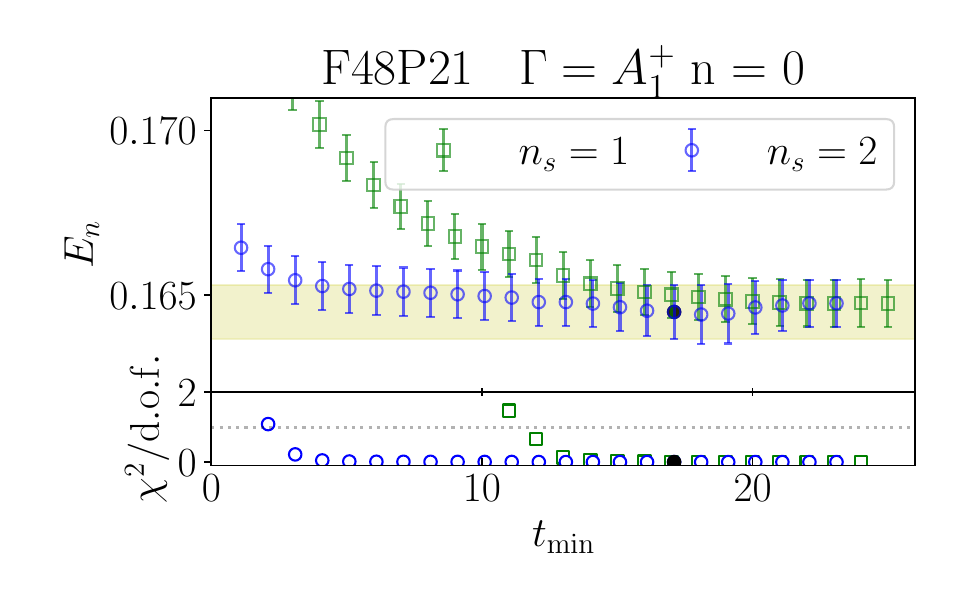}
\includegraphics[width=0.32\columnwidth]{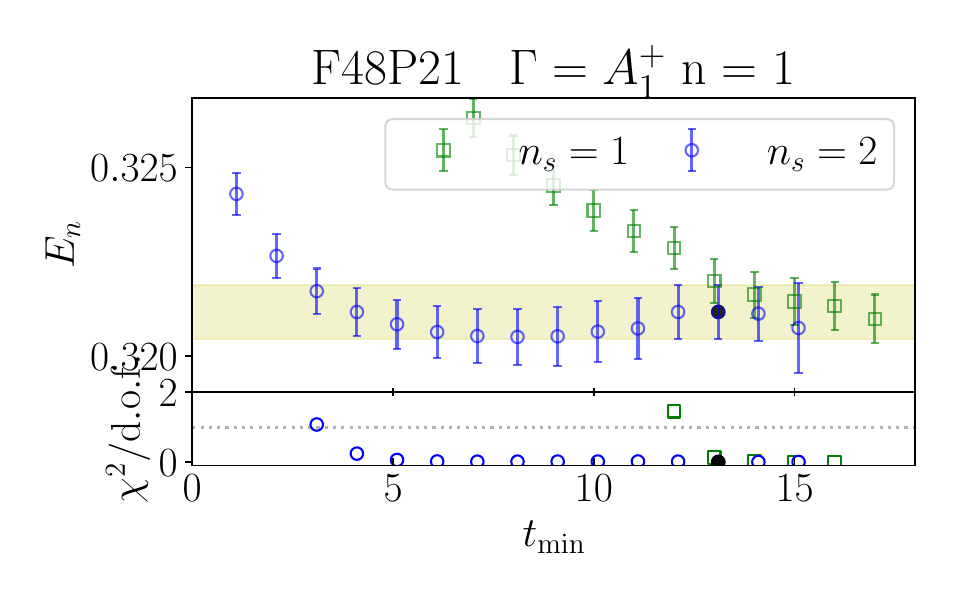}
\includegraphics[width=0.32\columnwidth]{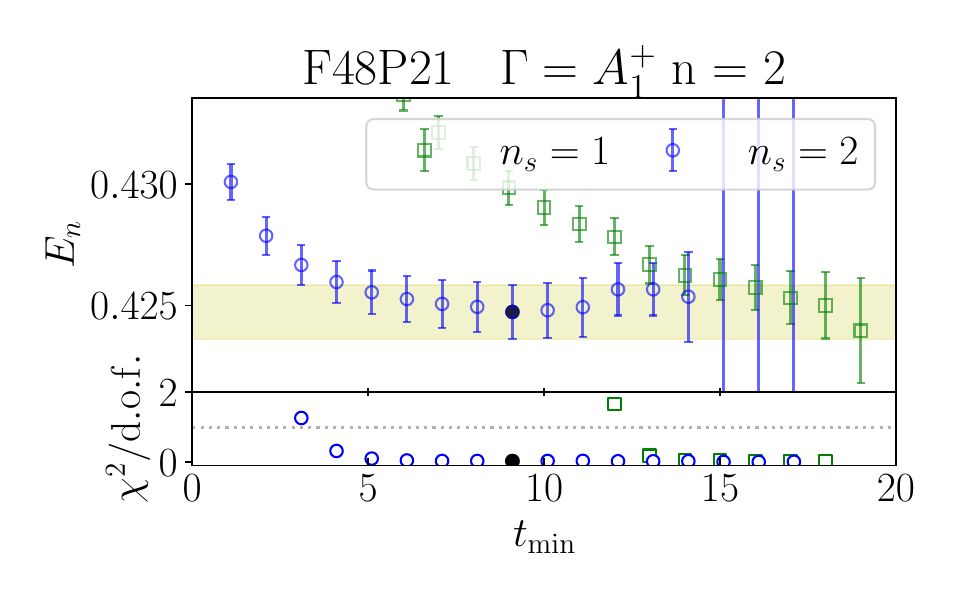}
\caption{Fit results for the $I=2$ $\pi\pi$ channel on the F48P21 ensemble.}
\label{fig:pipi-I=2-fit-F48P21}
\end{figure}

\begin{figure}[htbp]
\centering
\includegraphics[width=0.32\columnwidth]{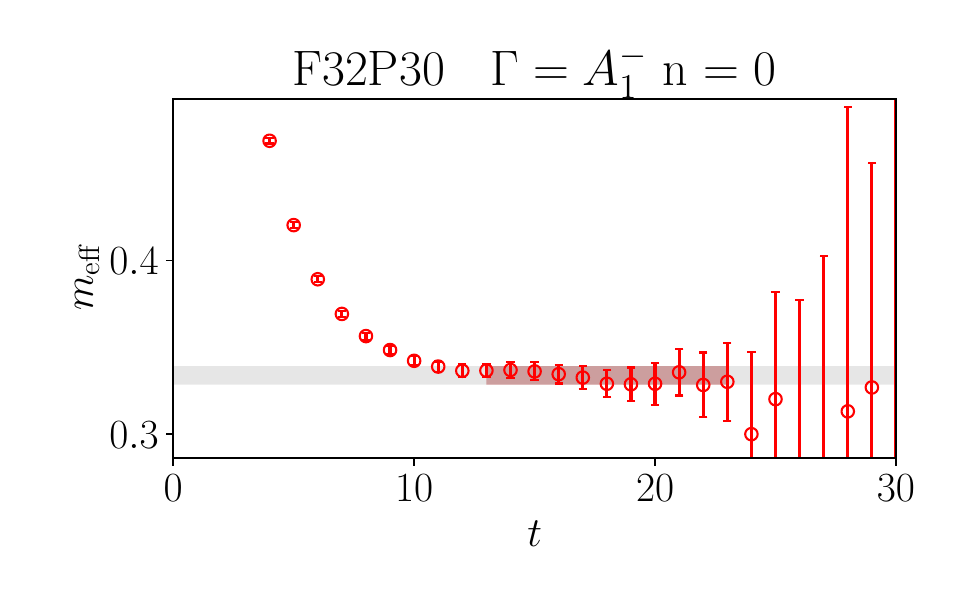}
\includegraphics[width=0.32\columnwidth]{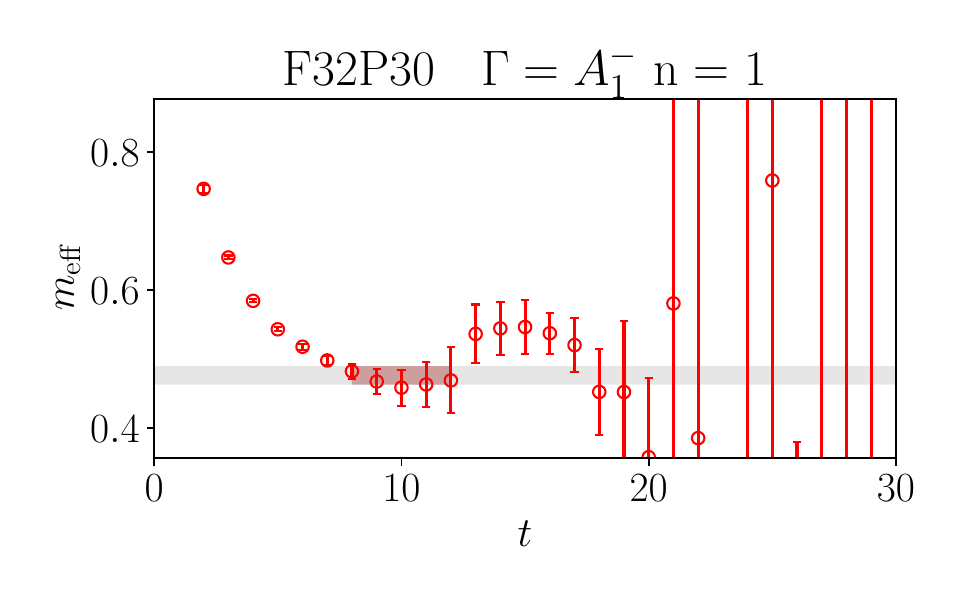}
\includegraphics[width=0.32\columnwidth]{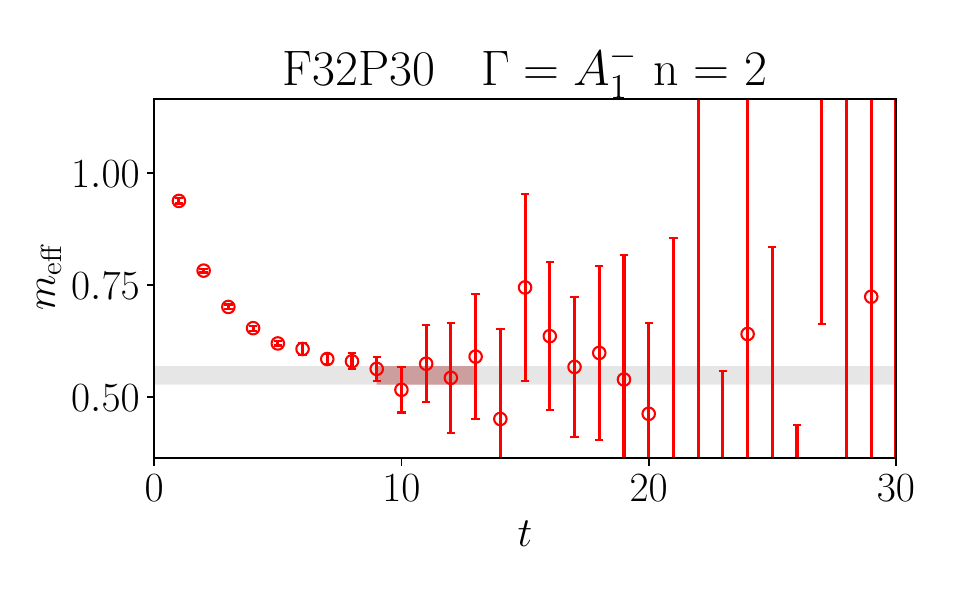}
\\
\includegraphics[width=0.32\columnwidth]{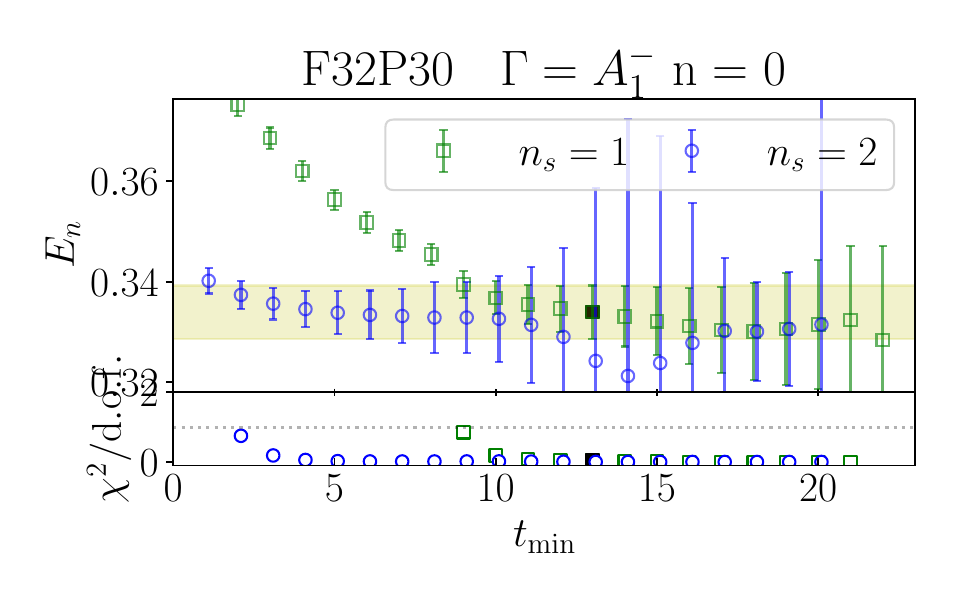}
\includegraphics[width=0.32\columnwidth]{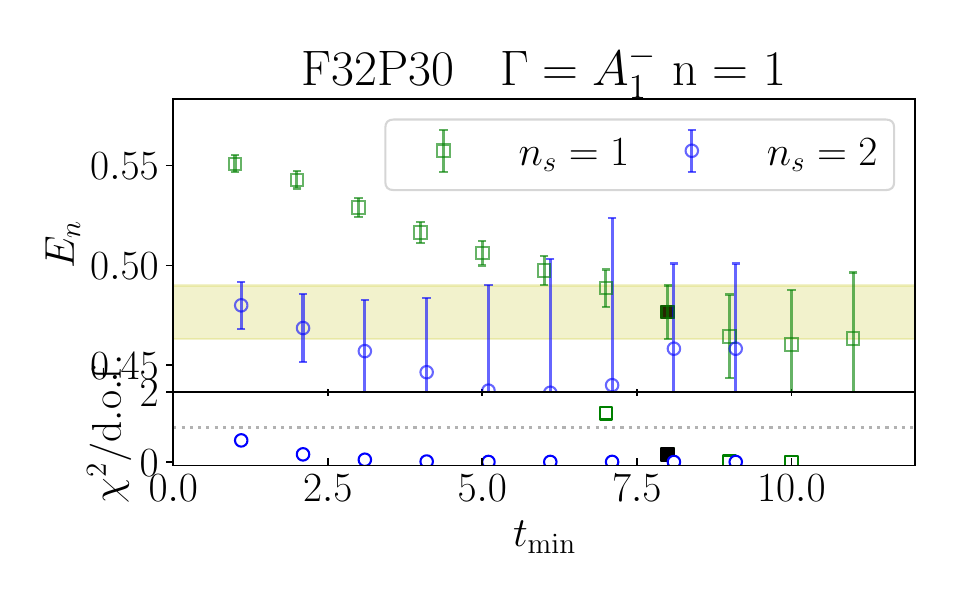}
\includegraphics[width=0.32\columnwidth]{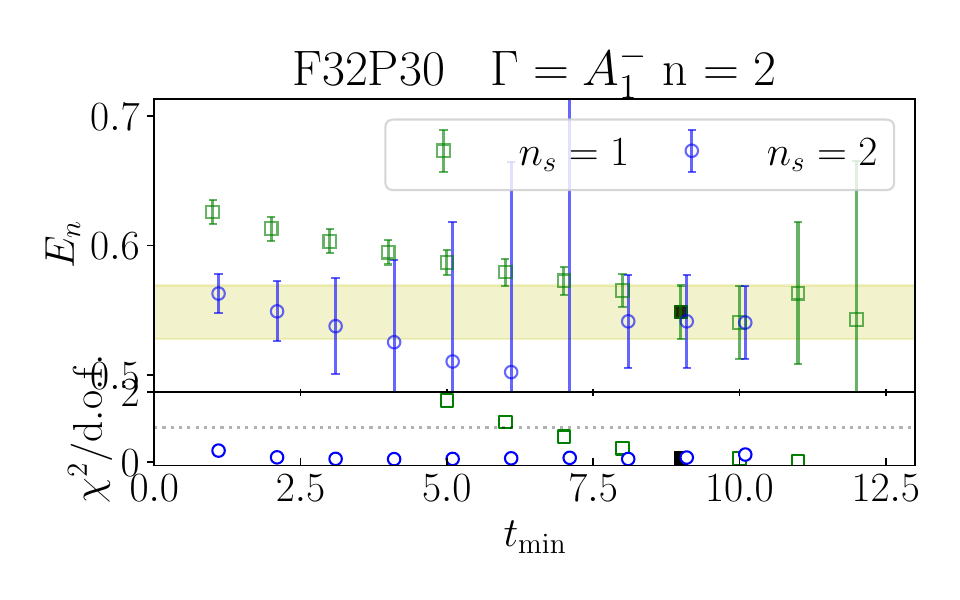}
\\
\includegraphics[width=0.32\columnwidth]{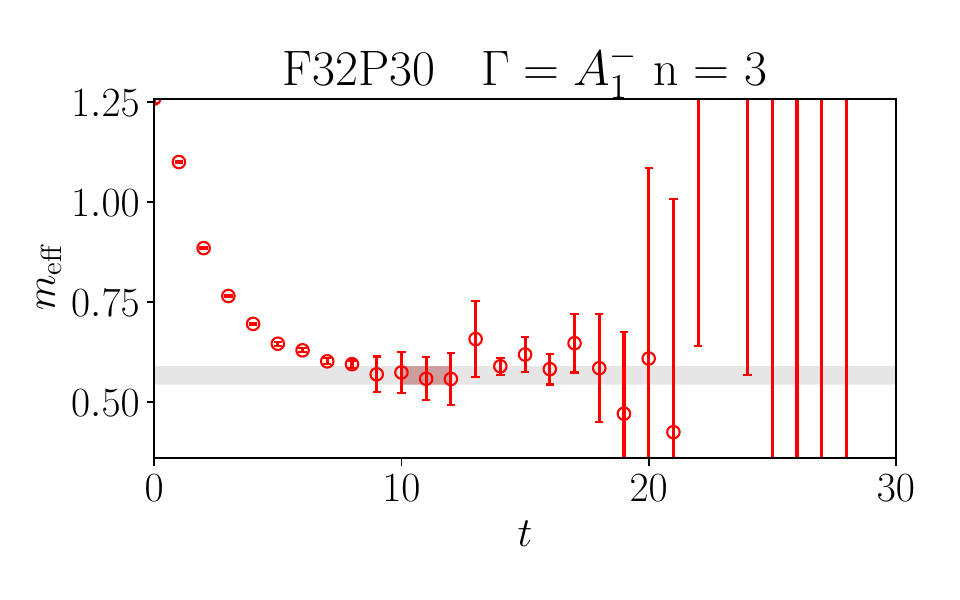}
\includegraphics[width=0.32\columnwidth]{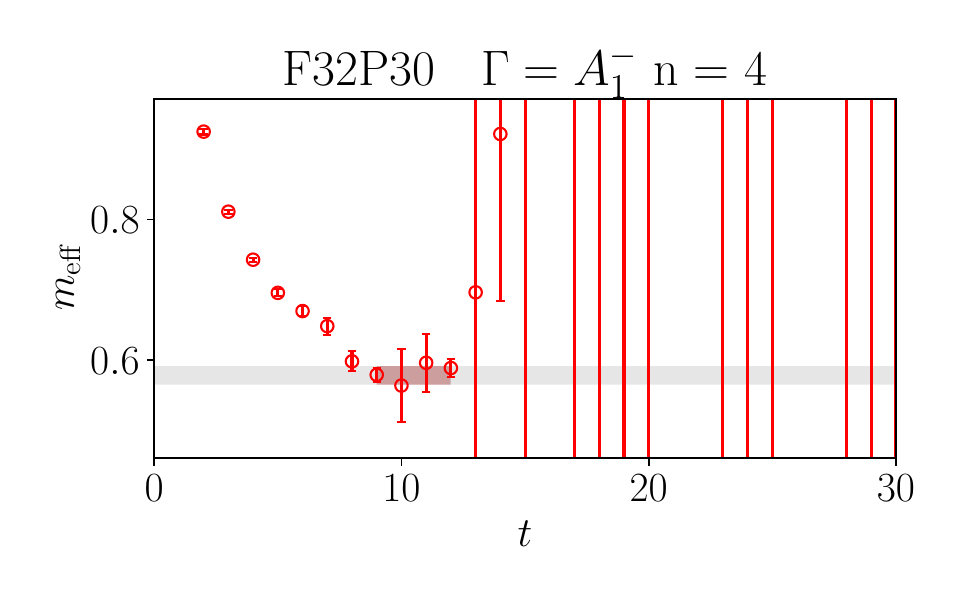}
\includegraphics[width=0.32\columnwidth]{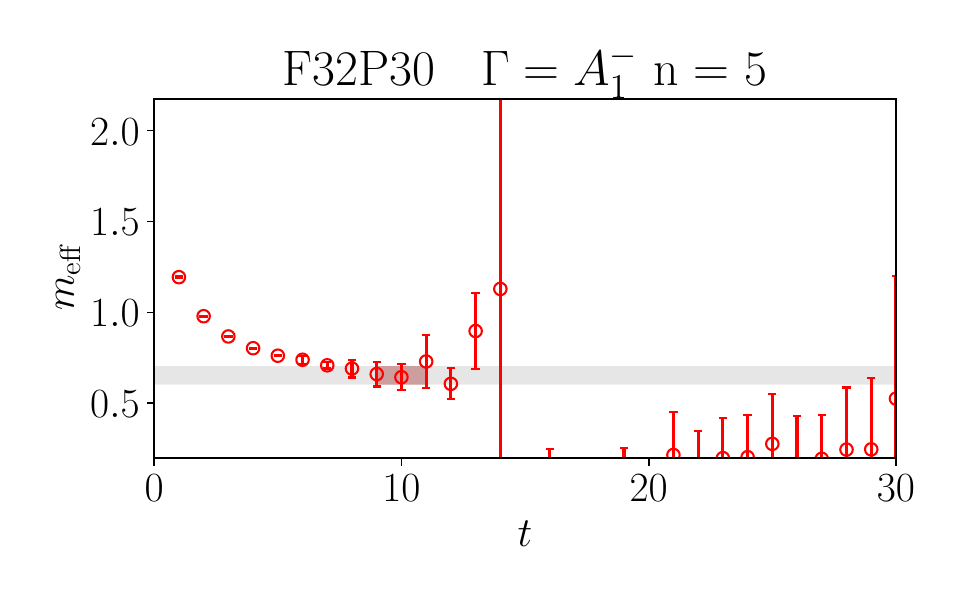}
\\
\includegraphics[width=0.32\columnwidth]{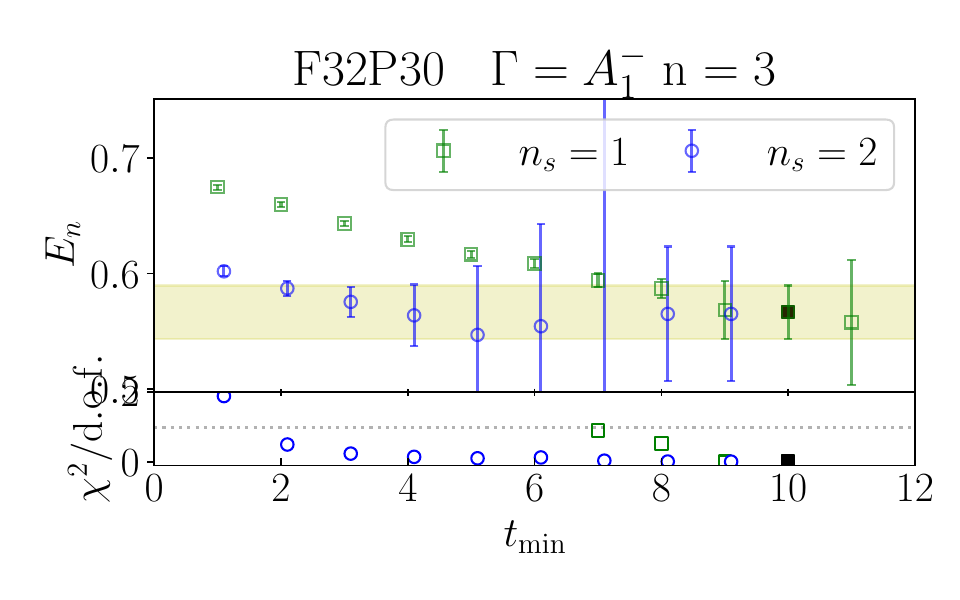}
\includegraphics[width=0.32\columnwidth]{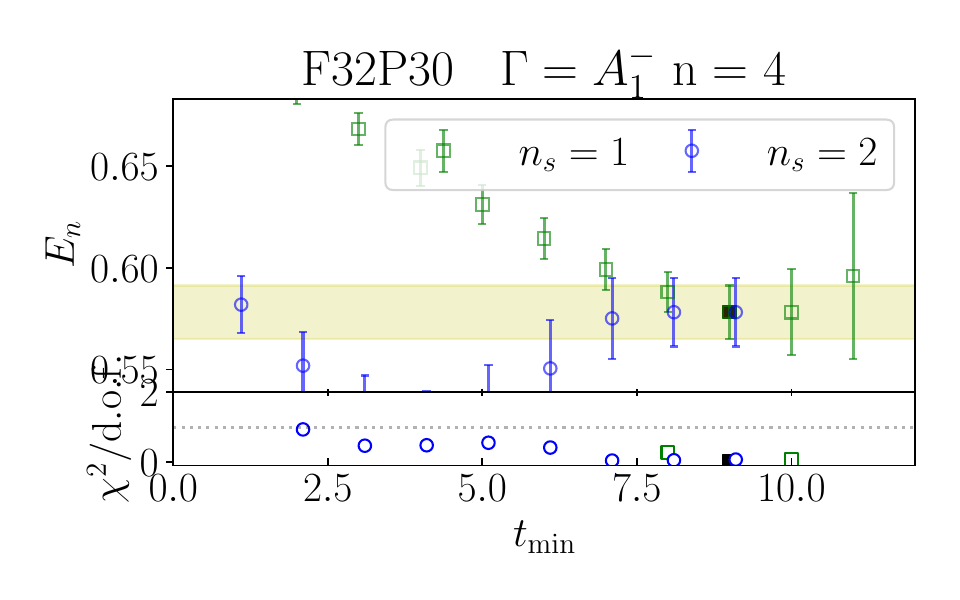}
\includegraphics[width=0.32\columnwidth]{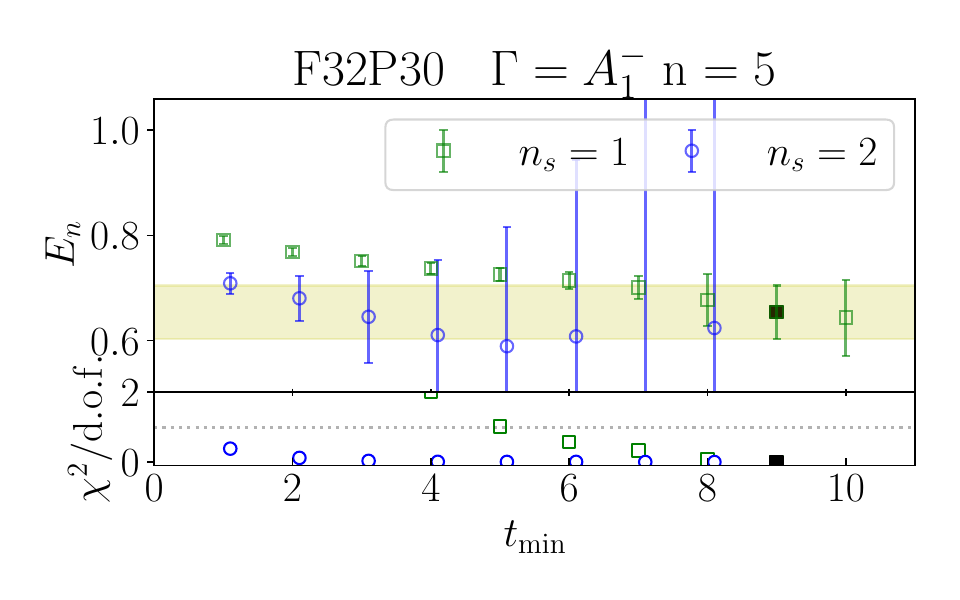}
\\
\includegraphics[width=0.32\columnwidth]{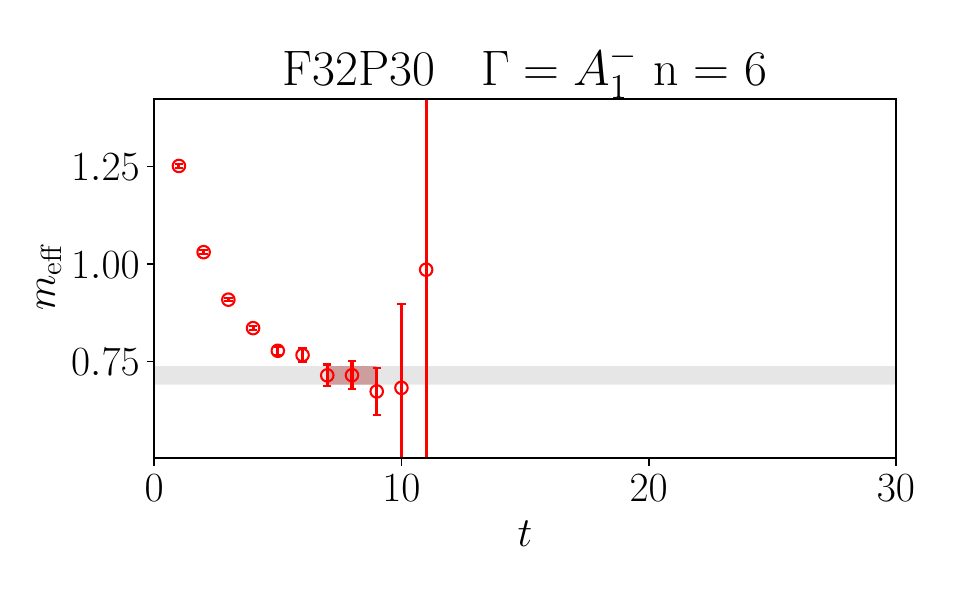}
\\
\includegraphics[width=0.32\columnwidth]{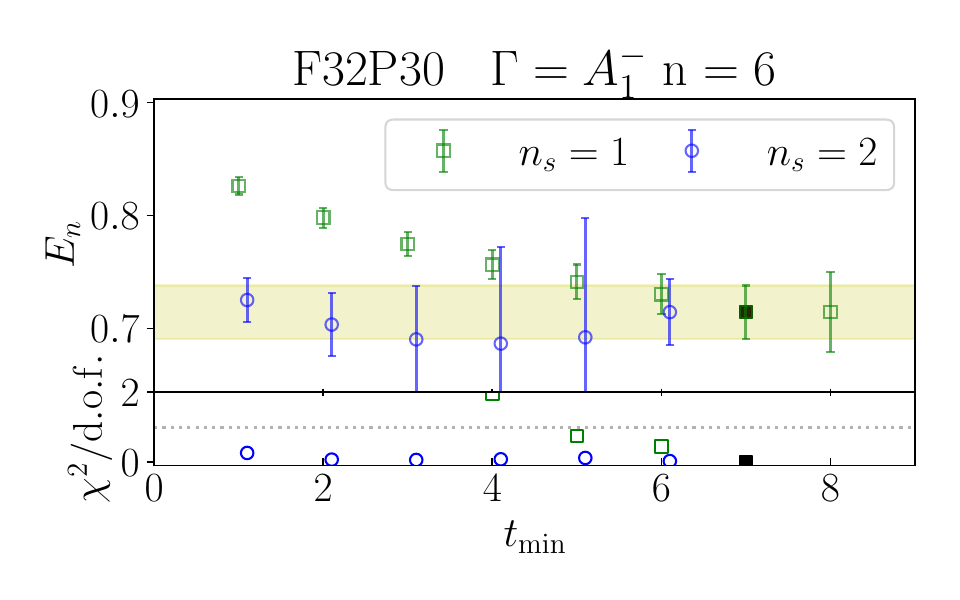}
\caption{Fit results for the $I=1$ $\pi\pi\pi$ channel on the F32P30 ensemble.}
\label{fig:pipipi-I=1-fit-F32P30}
\end{figure}

\begin{figure}[htbp]
\centering
\includegraphics[width=0.32\columnwidth]{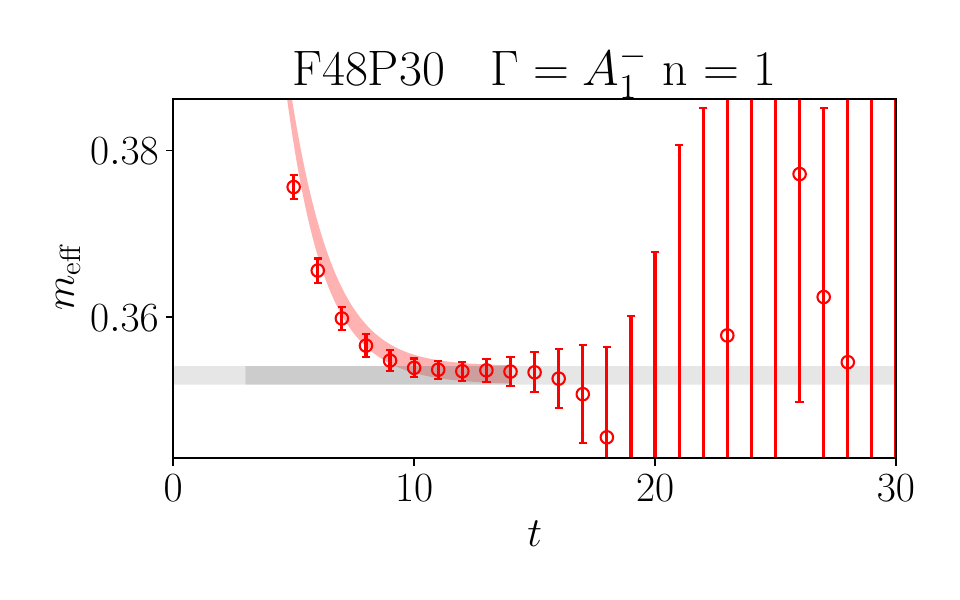}
\includegraphics[width=0.32\columnwidth]{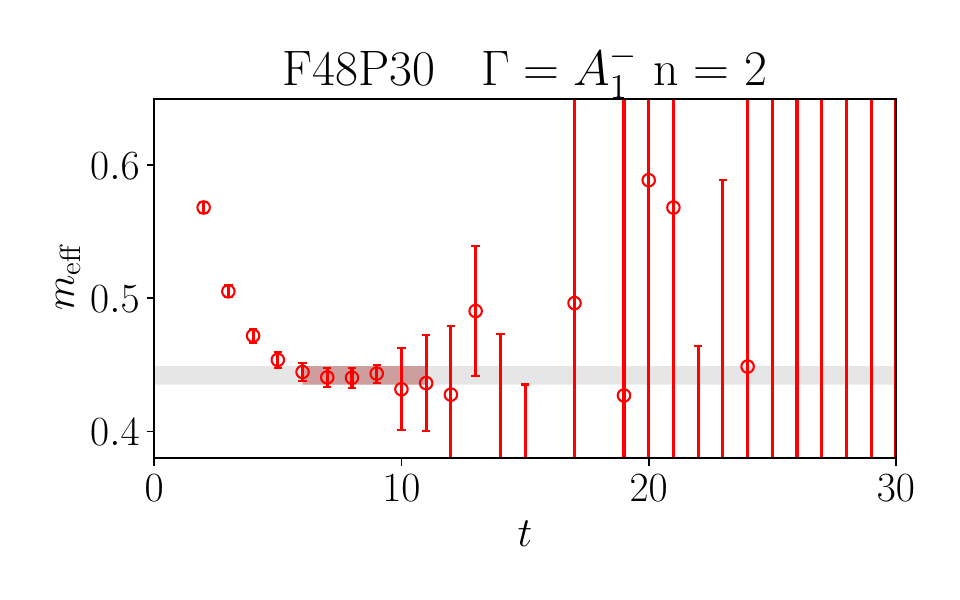}
\includegraphics[width=0.32\columnwidth]{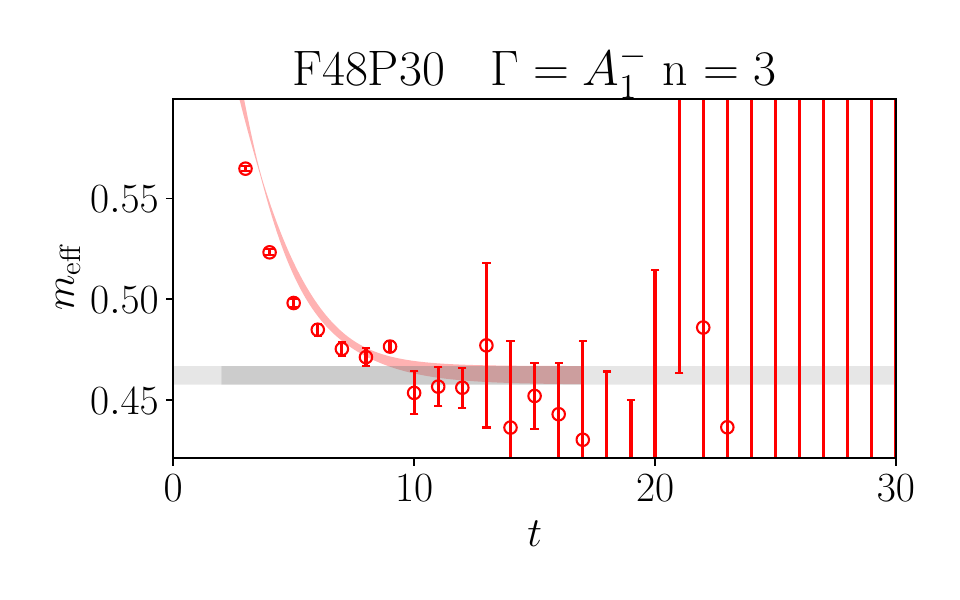}
\\
\includegraphics[width=0.32\columnwidth]{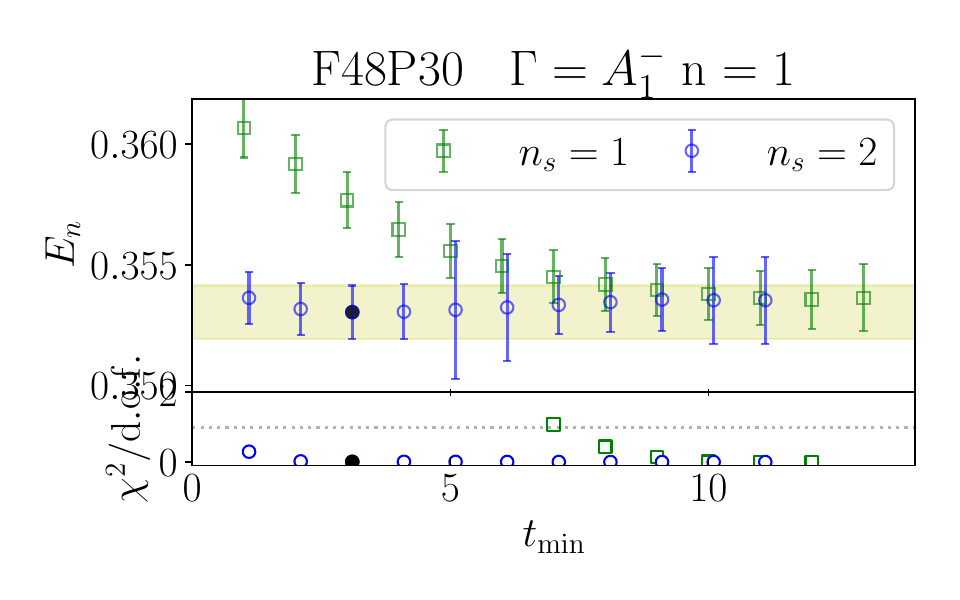}
\includegraphics[width=0.32\columnwidth]{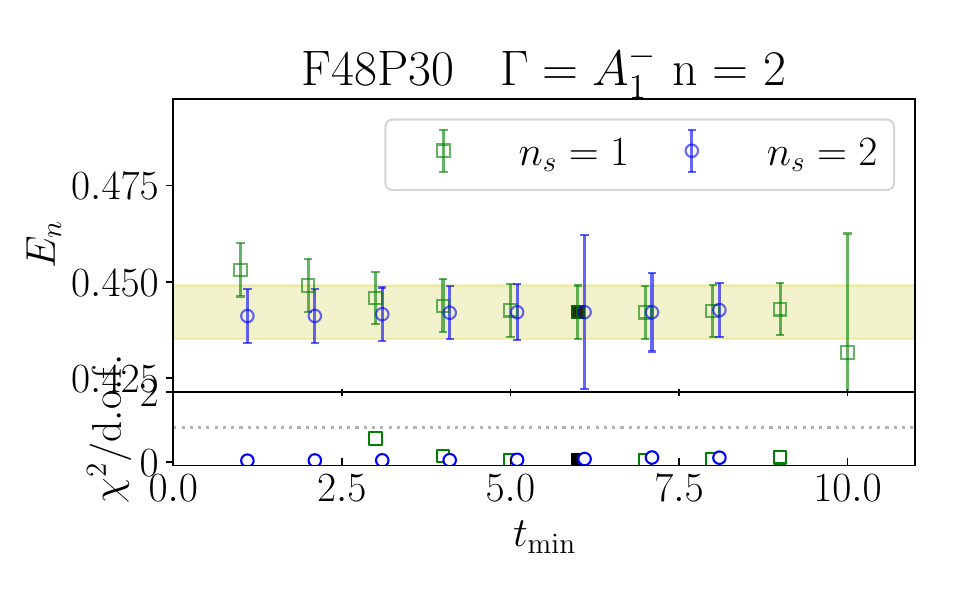}
\includegraphics[width=0.32\columnwidth]{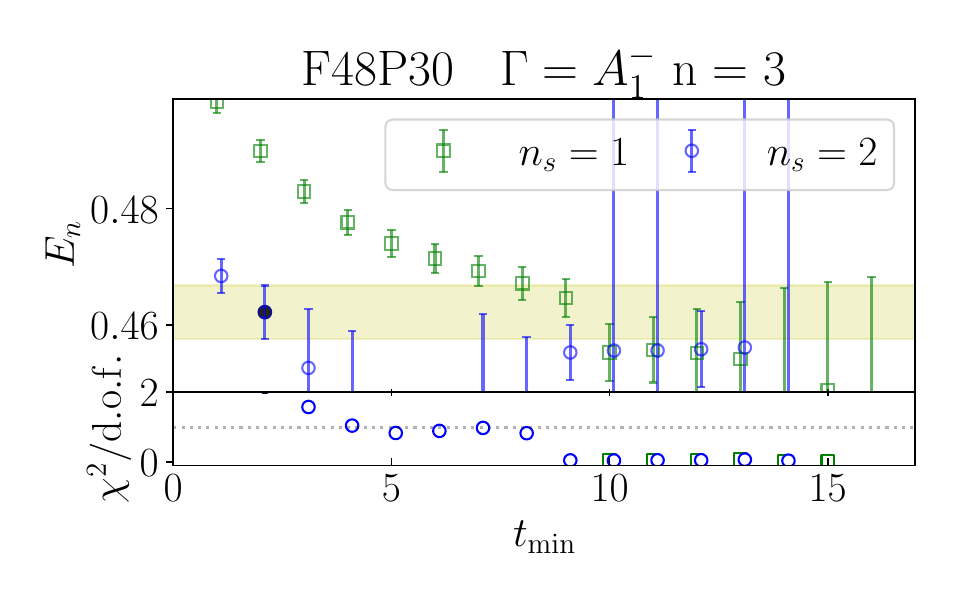}
\\
\includegraphics[width=0.32\columnwidth]{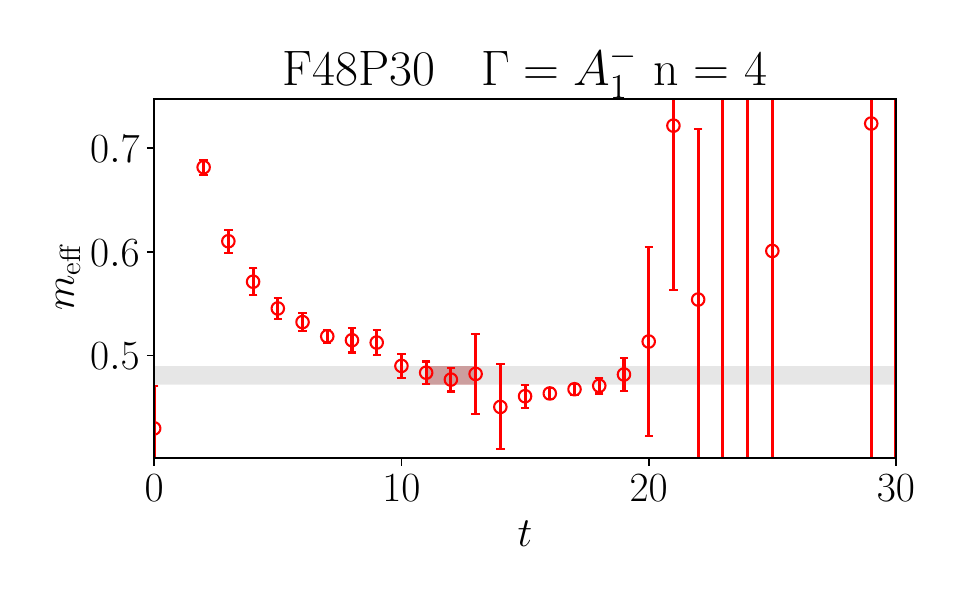}
\includegraphics[width=0.32\columnwidth]{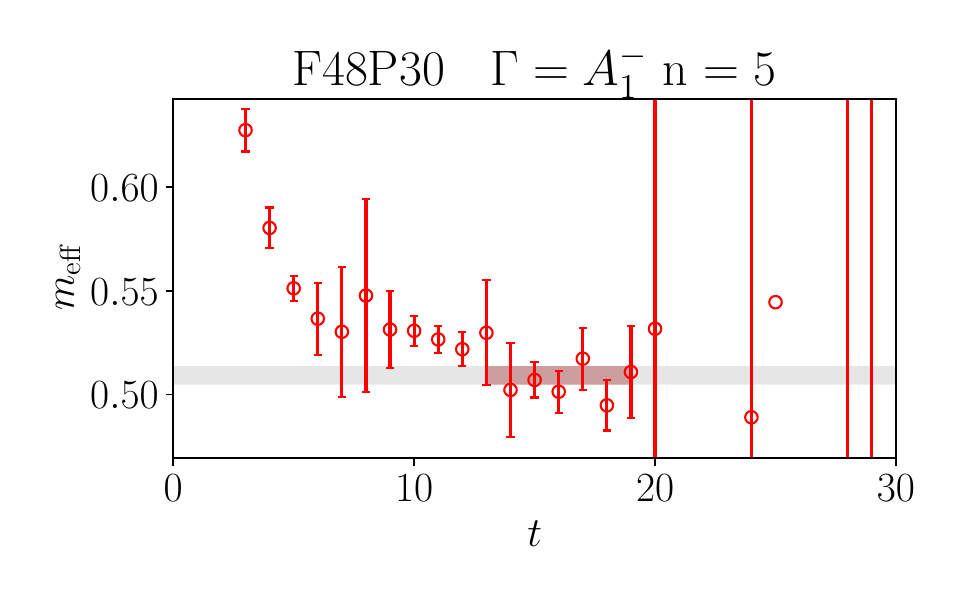}
\includegraphics[width=0.32\columnwidth]{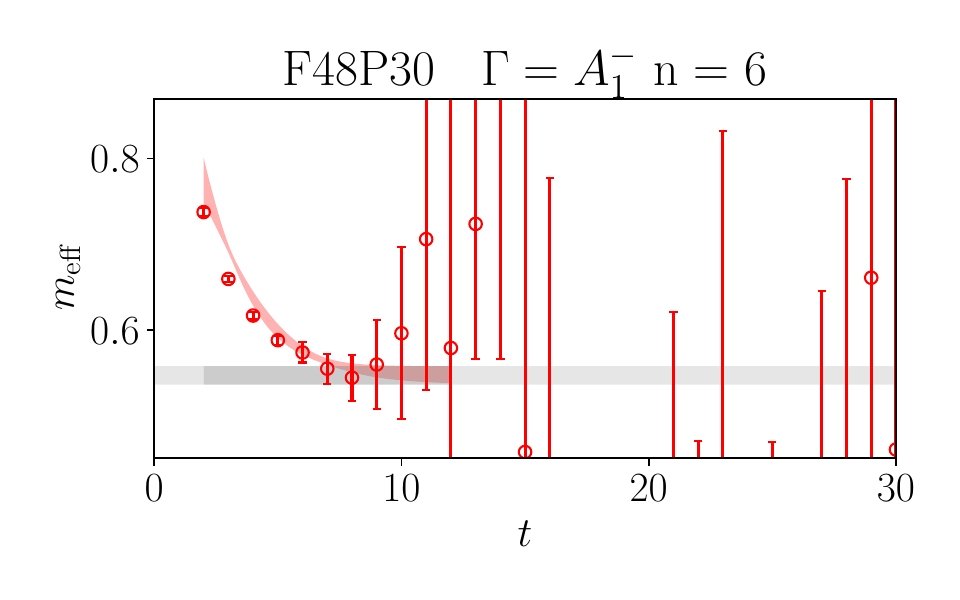}
\\
\includegraphics[width=0.32\columnwidth]{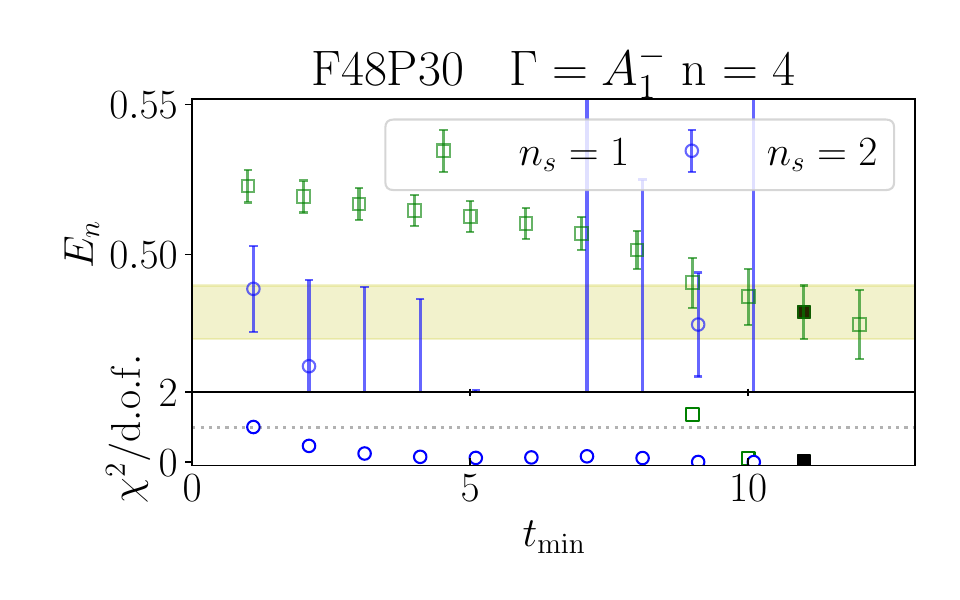}
\includegraphics[width=0.32\columnwidth]{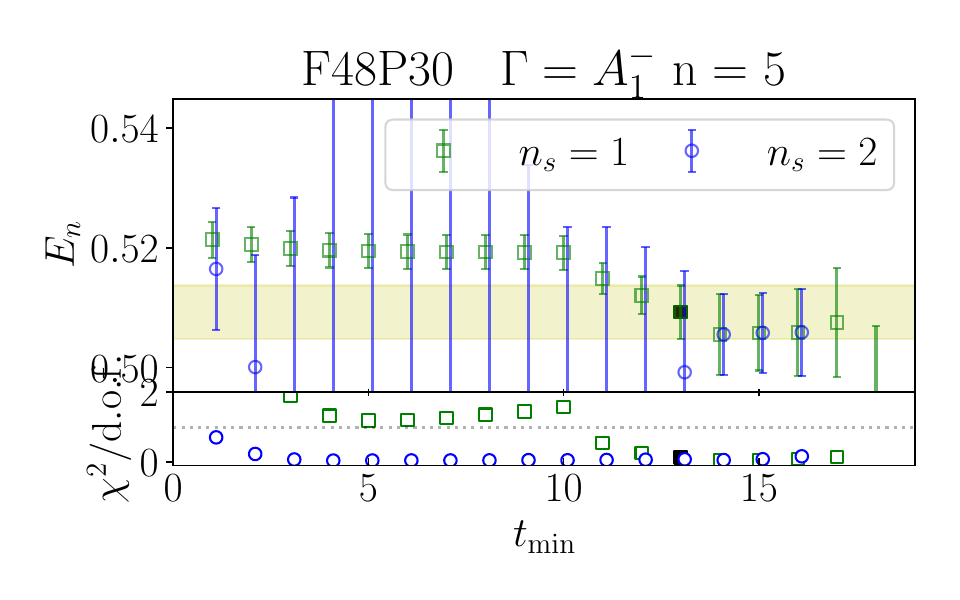}
\includegraphics[width=0.32\columnwidth]{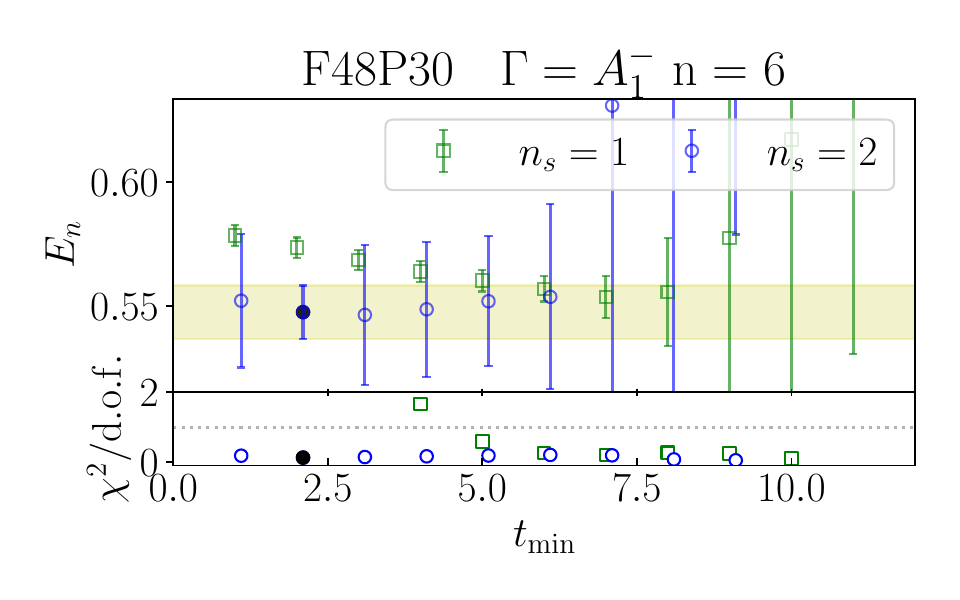}
\\
\includegraphics[width=0.32\columnwidth]{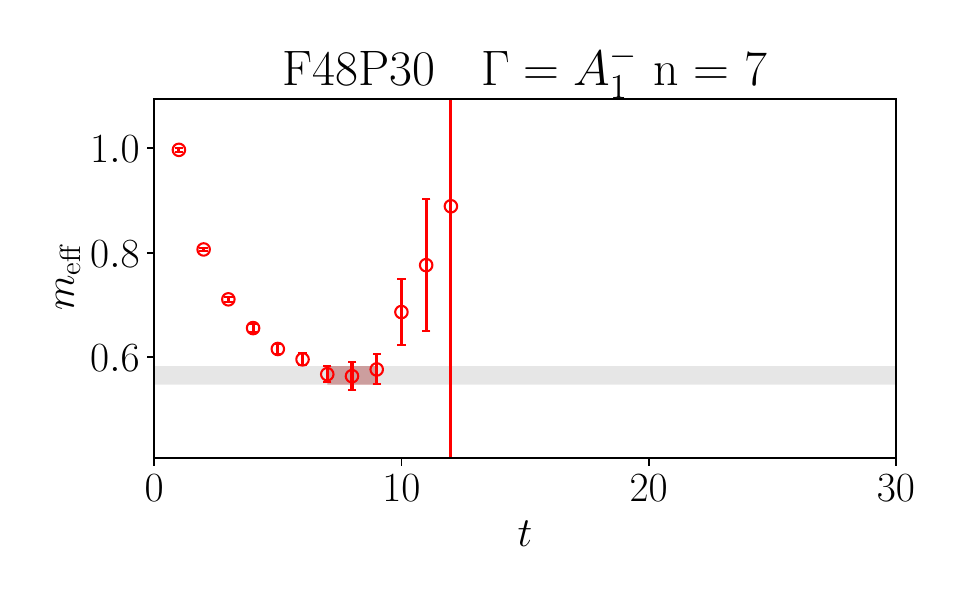}
\\
\includegraphics[width=0.32\columnwidth]{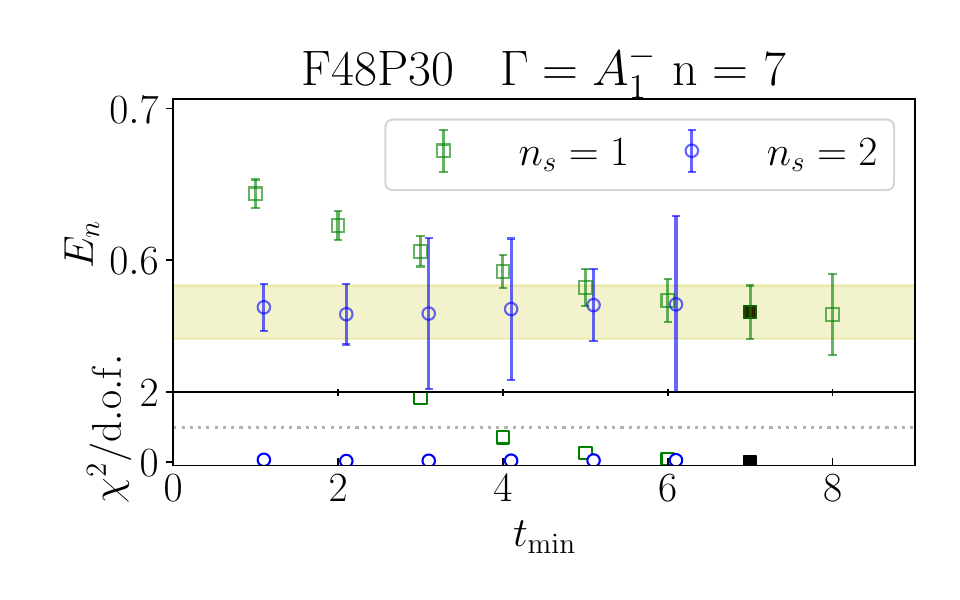}
\caption{Fit results for the $I=1$ $\pi\pi\pi$ channel on the F48P30 ensemble.}
\label{fig:pipipi-I=1-fit-F48P30}
\end{figure}

\begin{figure}[htbp]
\centering
\includegraphics[width=0.32\columnwidth]{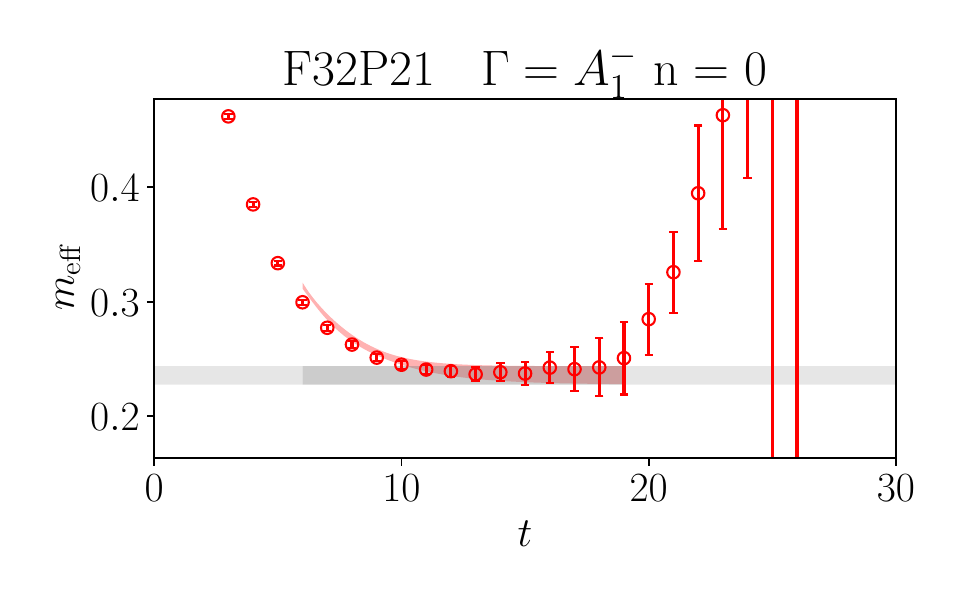}
\includegraphics[width=0.32\columnwidth]{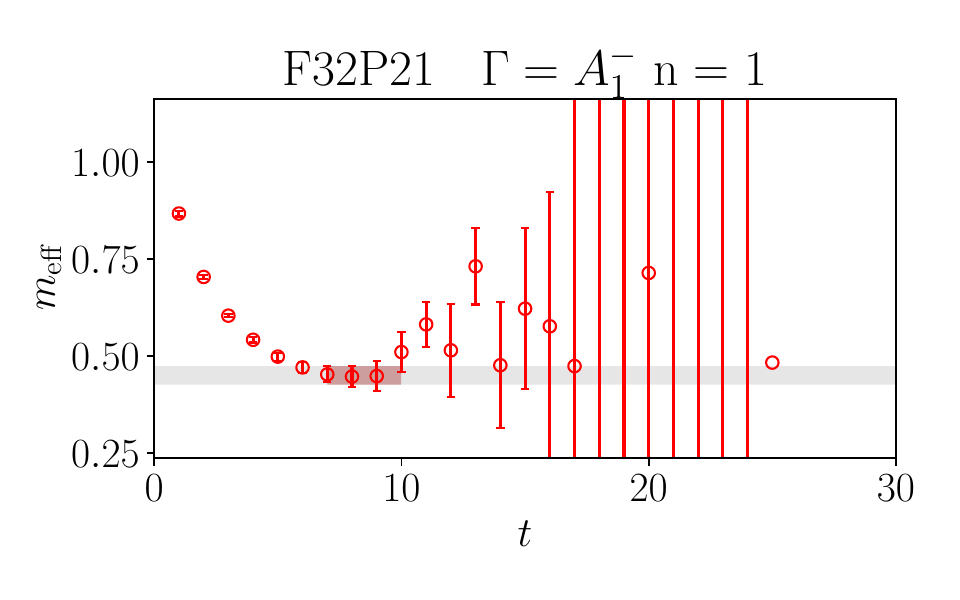}
\includegraphics[width=0.32\columnwidth]{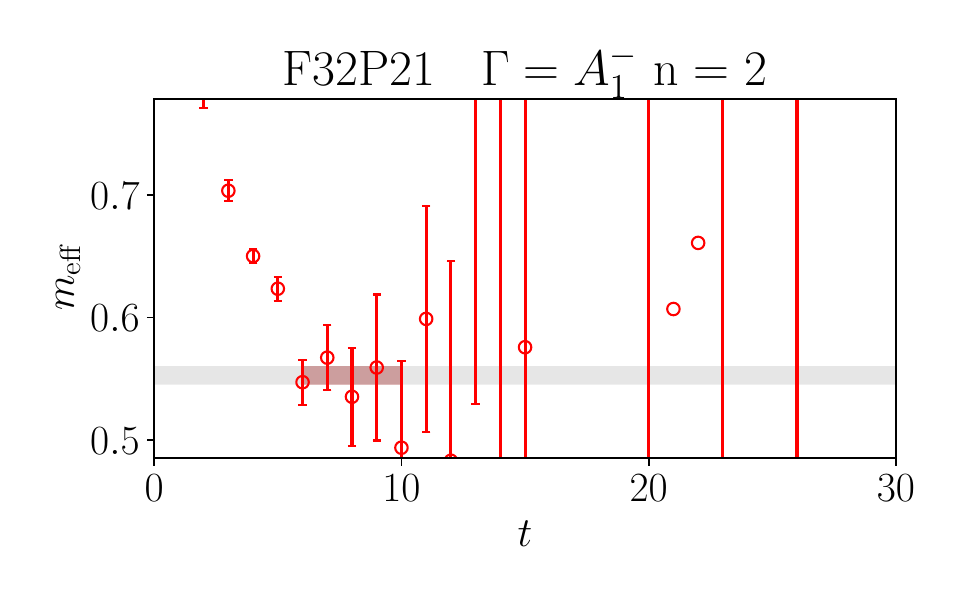}
\\
\includegraphics[width=0.32\columnwidth]{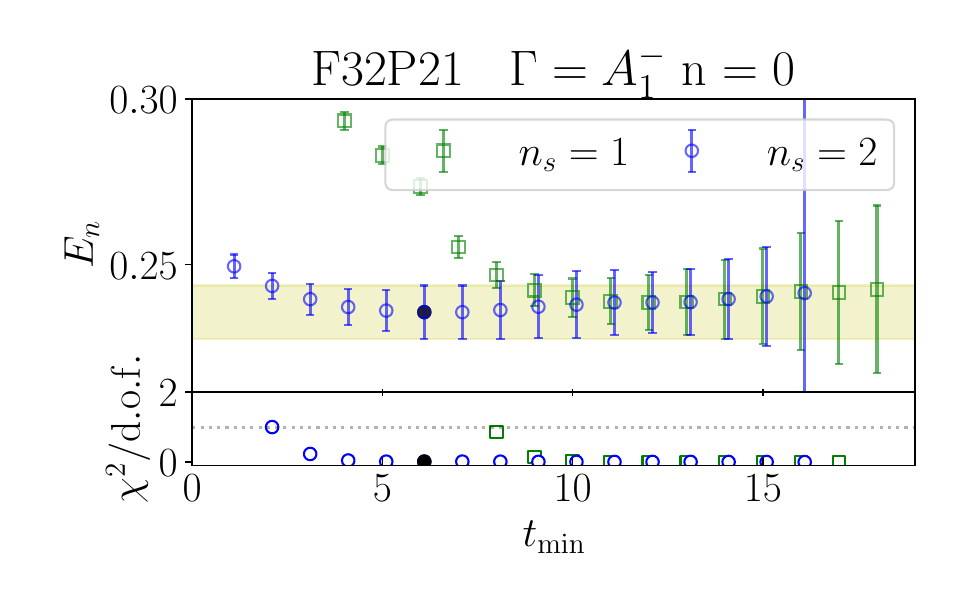}
\includegraphics[width=0.32\columnwidth]{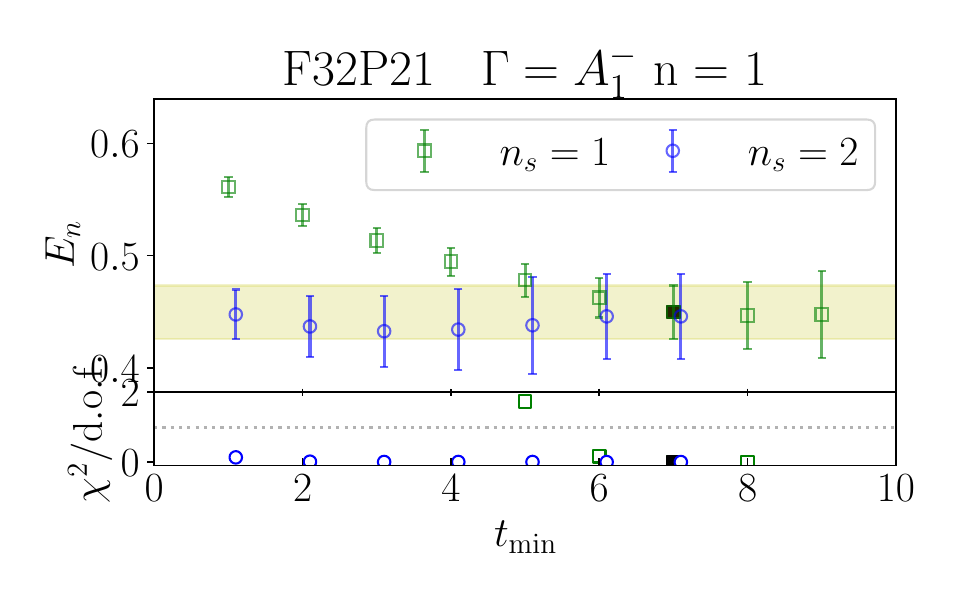}
\includegraphics[width=0.32\columnwidth]{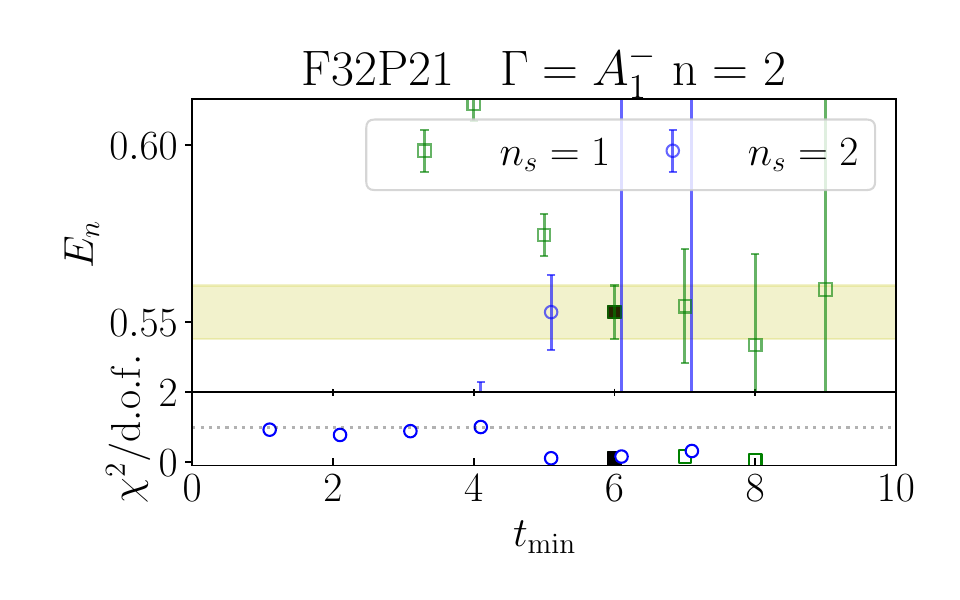}
\\
\includegraphics[width=0.32\columnwidth]{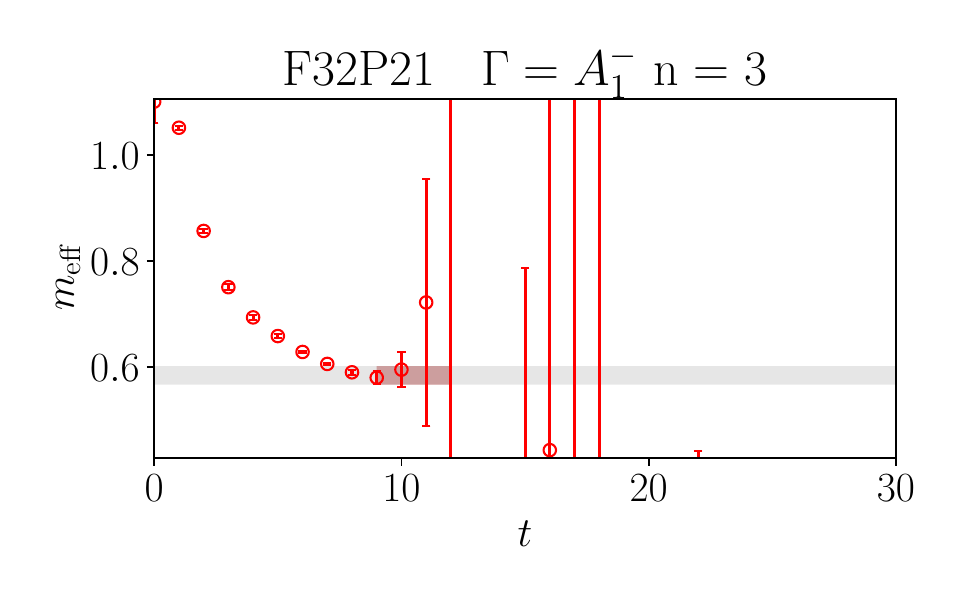}
\includegraphics[width=0.32\columnwidth]{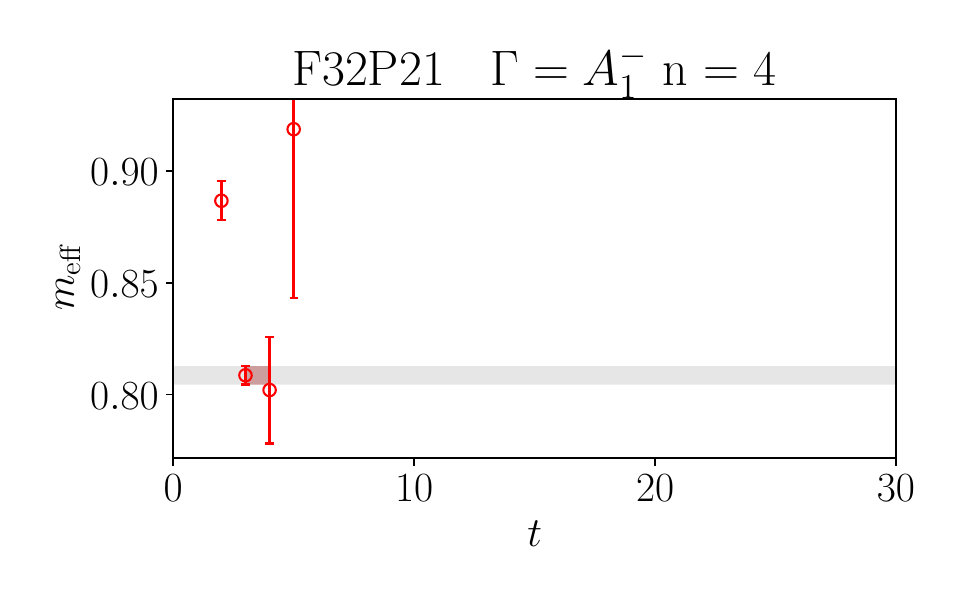}
\\
\includegraphics[width=0.32\columnwidth]{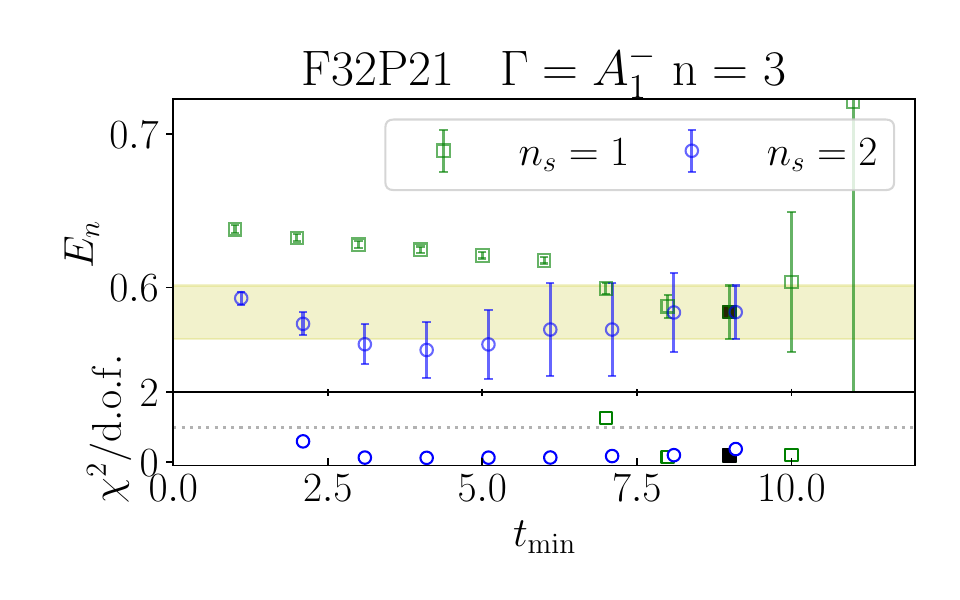}
\includegraphics[width=0.32\columnwidth]{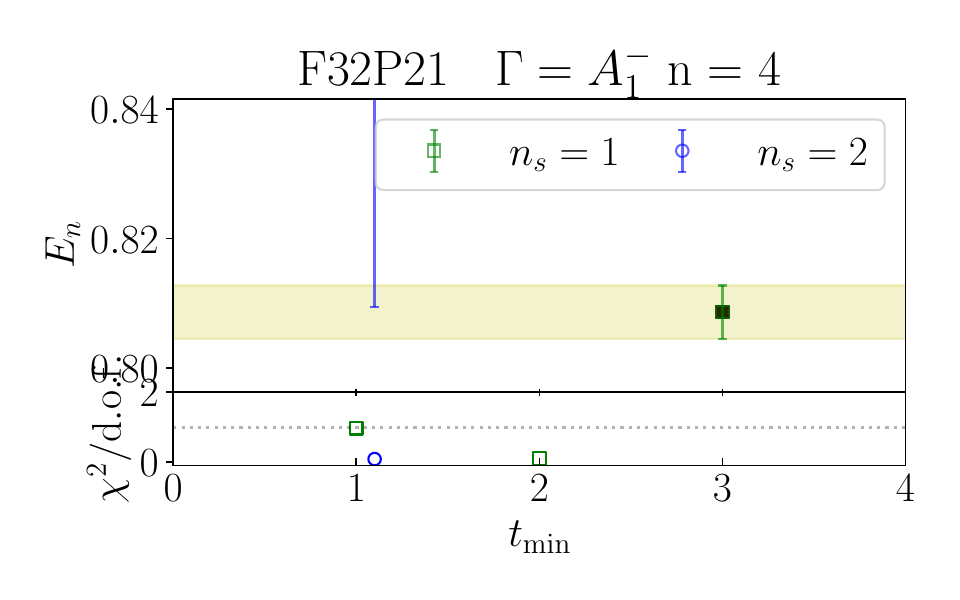}
\caption{Fit results for the $I=1$ $\pi\pi\pi$ channel on the F32P21 ensemble.}
\label{fig:pipipi-I=1-fit-F32P21}
\end{figure}

\begin{figure}[htbp]
\centering
\includegraphics[width=0.32\columnwidth]{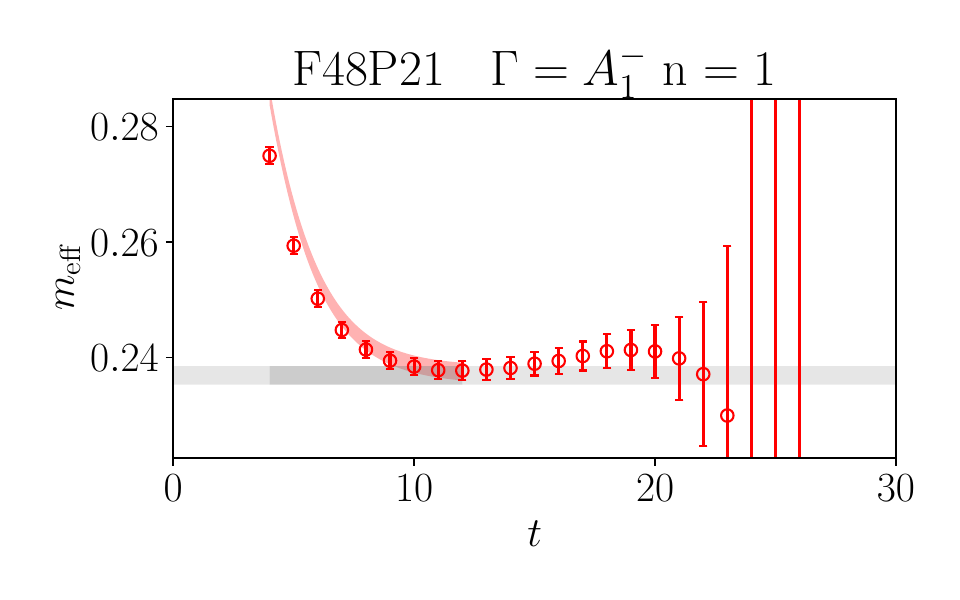}
\includegraphics[width=0.32\columnwidth]{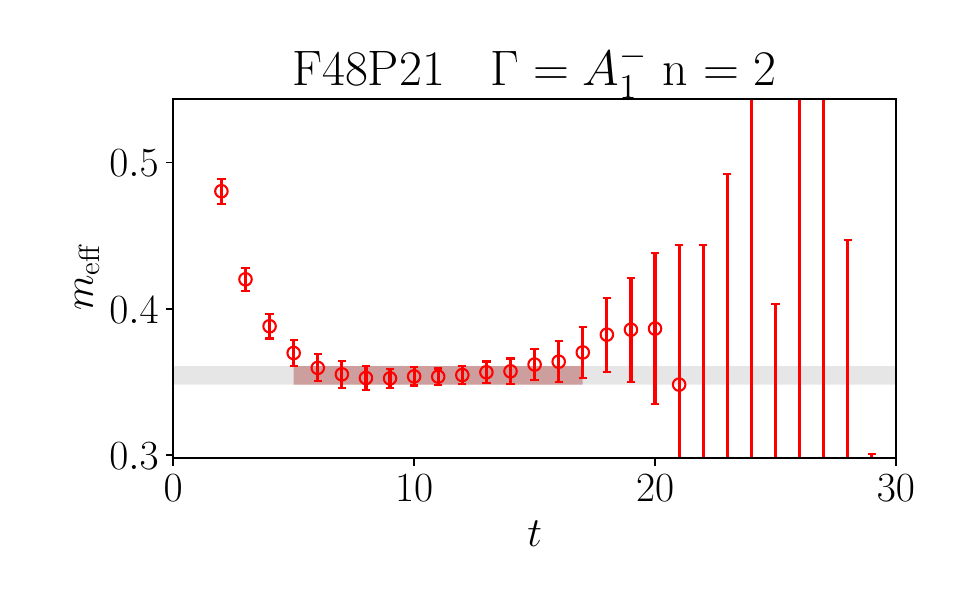}
\includegraphics[width=0.32\columnwidth]{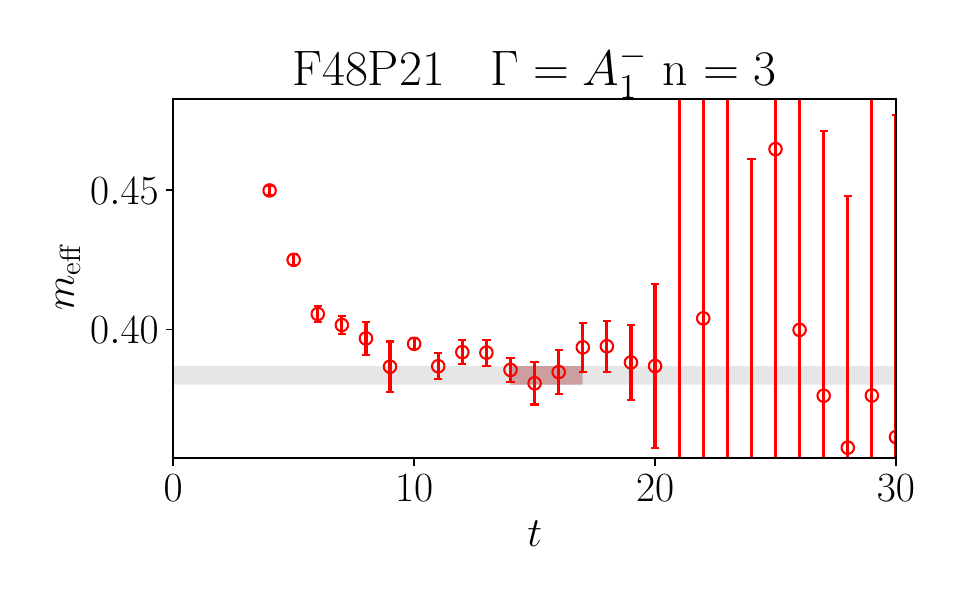}
\\
\includegraphics[width=0.32\columnwidth]{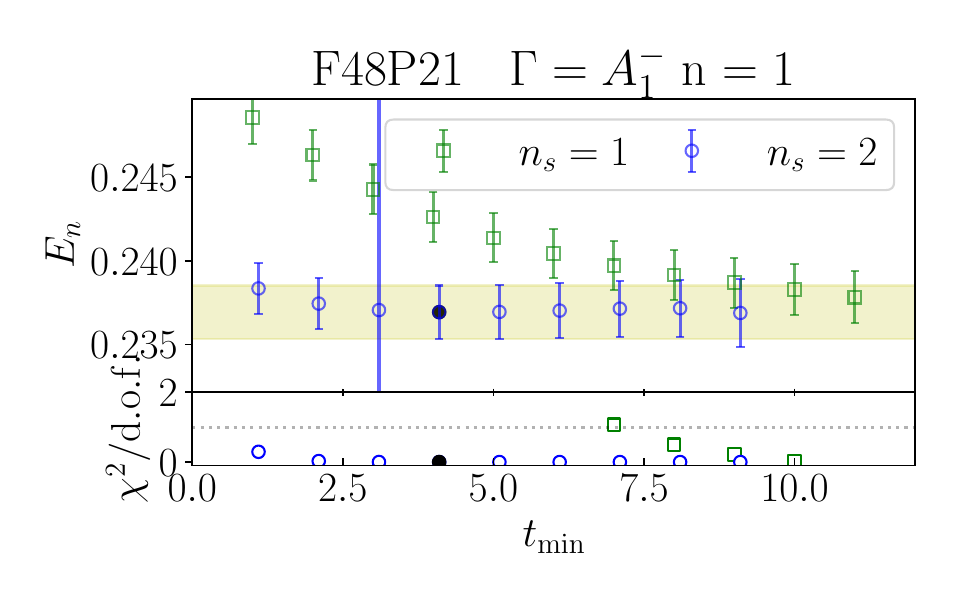}
\includegraphics[width=0.32\columnwidth]{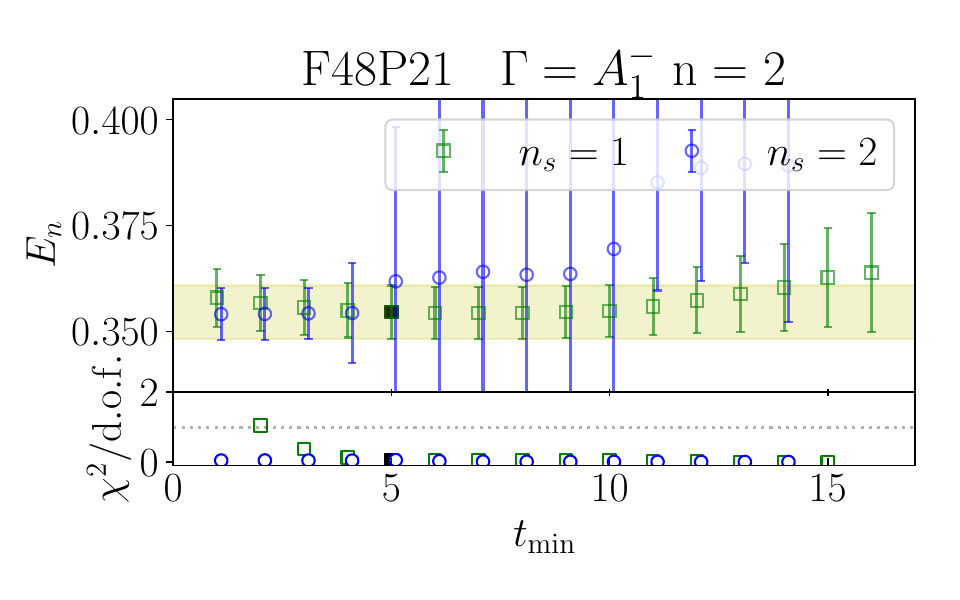}
\includegraphics[width=0.32\columnwidth]{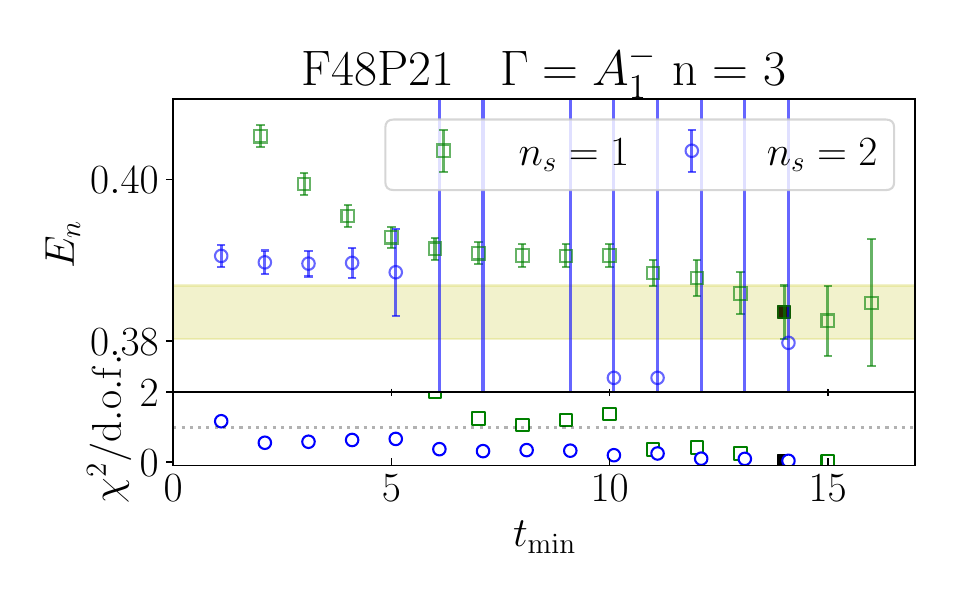}
\\
\includegraphics[width=0.32\columnwidth]{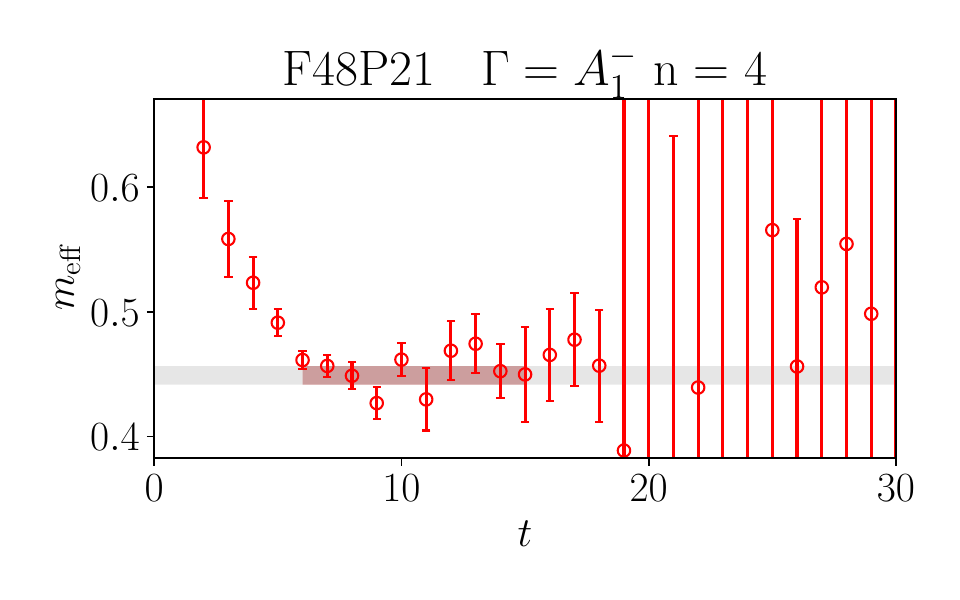}
\includegraphics[width=0.32\columnwidth]{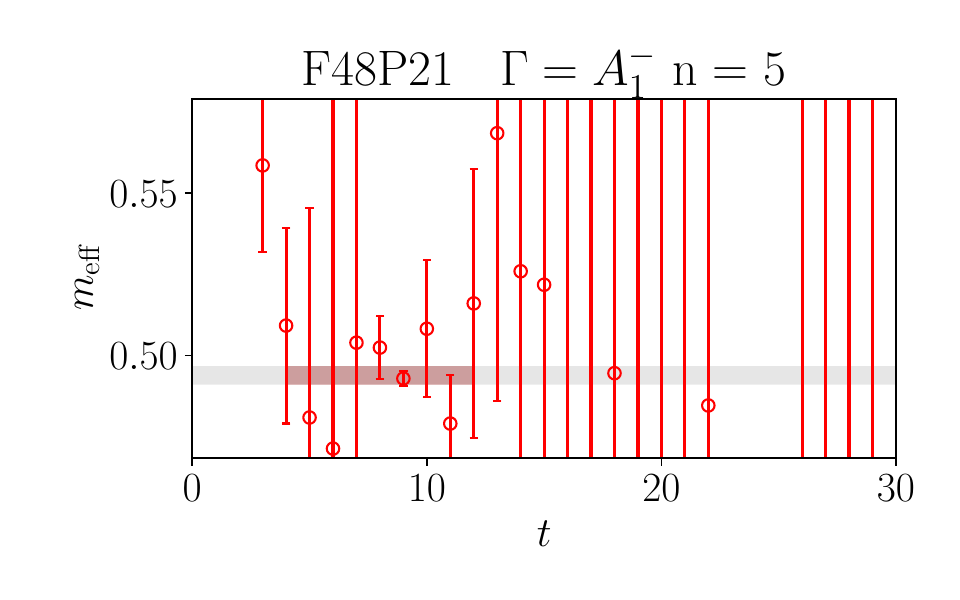}
\includegraphics[width=0.32\columnwidth]{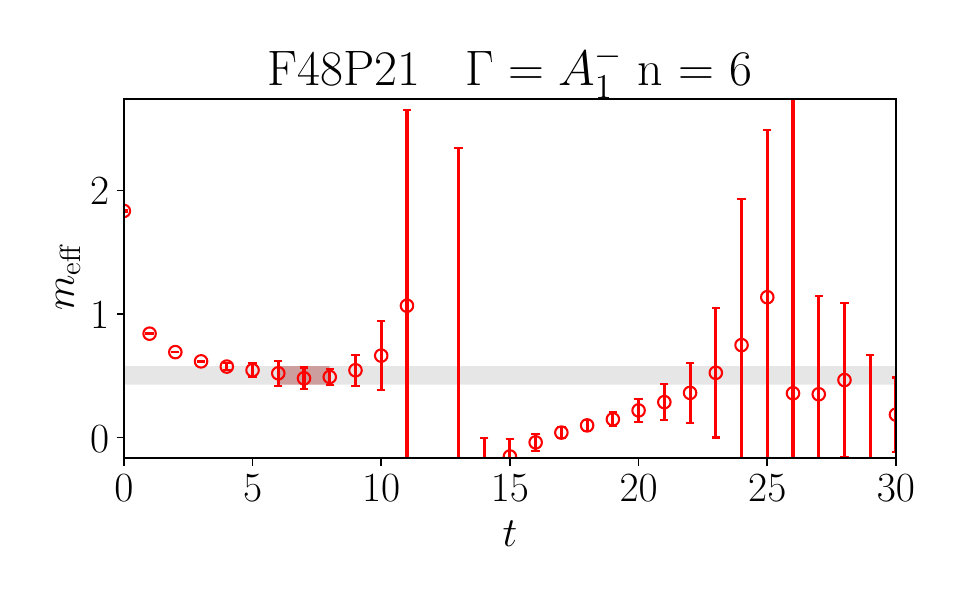}
\\
\includegraphics[width=0.32\columnwidth]{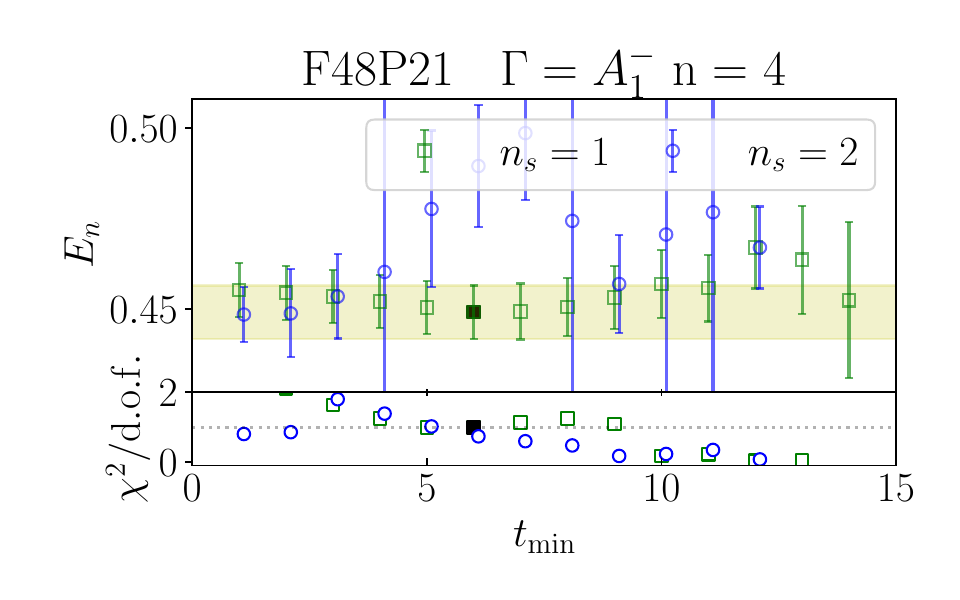}
\includegraphics[width=0.32\columnwidth]{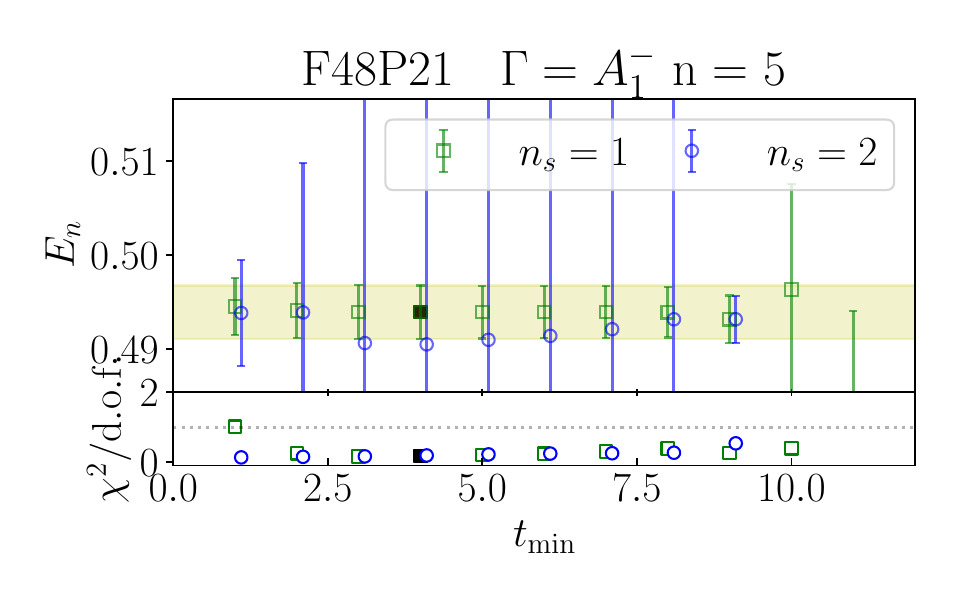}
\includegraphics[width=0.32\columnwidth]{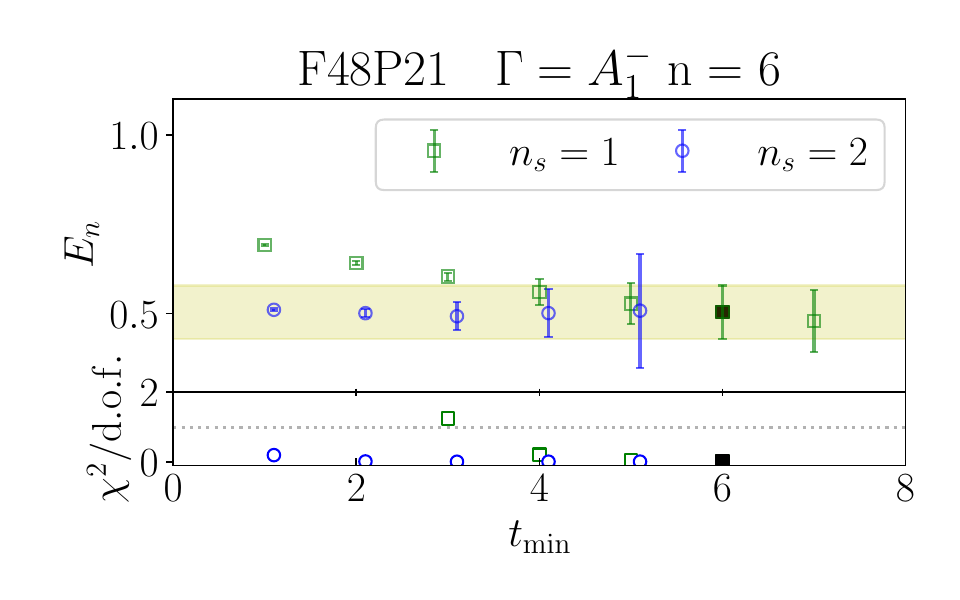}
\caption{Fit results for the $I=1$ $\pi\pi\pi$ channel on the F48P21 ensemble.}
\label{fig:pipipi-I=1-fit-F48P21}
\end{figure}

Figures~\ref{fig:pipi-I=0-fit-F32P30}--\ref{fig:pipipi-I=1-fit-F48P21} show the energy-level fits for the two- and three-body channels on the F32P30, F48P30, F32P21, and F48P21 ensembles used in the $\pi(1300)$ study. Statistical uncertainties are estimated using the jackknife method.

\cleardoublepage

\end{appendix}
\cleardoublepage

\renewcommand{\baselinestretch}{1.5}
\fontsize{12pt}{13pt}\selectfont

\chapter*{Acknowledgements}
\markboth{Acknowledgements}{Acknowledgements}
\phantomsection
\addcontentsline{toc}{chapter}{\hei Acknowledgements}

{
\kaishu
\begin{center}
    红楼飞雪，一时英杰，先哲曾书写，爱国进步民主科学。\\
    忆昔长别，阳关千叠，狂歌曾竟夜，收拾山河待百年约。
\end{center}
\hfill ——《燕园情》
}

Dear reader, if you have come this far, you may already have guessed that whatever I have been fortunate enough to accomplish, I owe to these courageous, upright, warm-hearted, kind, learned, and delightful people.

With my parents' support, I have been free to pursue my ideals; because of them, I have feared nothing.

My doctoral advisor is Professor Chuan Liu. A man of unblemished integrity, exemplary in word and deed, Professor Liu possesses a breadth of learning that commands my deepest admiration. He has always given me unstinting support, a democratic academic environment, freedom of thought\footnotecircle{This phrase is taken from a senior-high-school physics textbook published by the People's Education Press. The original reads: ``Without academic democracy and freedom of thought, science cannot flourish.'' I sincerely hope that the scientific environment in China will continue to improve.}, and unfailing optimism. Over these five years, both my way of conducting myself and my approach to research have increasingly come to resemble his; in this I count myself most fortunate. I love Yanyuan deeply: it has shown me boundless forbearance and brought me endless wonder. Now, on the eve of my departure, every tree and every blade of grass at Peking University moves me beyond words. Yet places of beauty do not last, and splendid gatherings are hard to renew. The great feast of the last century has drawn to a close, but countless feasts of the twenty-first have already begun, as brilliant scientists continue to break through the frontiers of knowledge.

André Walker-Loud and Ben Hörz were my mentors during the time I spent at Berkeley in my junior year. They are among the foremost scientists in the field. It was their guidance and encouragement that first enabled me to pursue lattice QCD---the foundation on which I have built my work in the past and on which I shall continue to build my future.

Professors Xu Feng, Liuming Liu, Ying Chen, Peng Sun, Yi-Bo Yang, Jiajun Wu, Feng-Kun Guo, Liupan An, Yanxi Zhang, and many others patiently answered most of my naive questions about lattice field theory, phenomenology, and experiment in high-energy physics. Yuan Zhong and Jun Feng, who mentored me as an undergraduate, offered me invaluable advice on both research and life.

In the third year of my doctorate, Professor Liu recommended that I undertake a research visit to the University of Bonn in Germany. There, I worked under the guidance of Carsten Urbach and Ulf Meißner. As eminent scientists in lattice QCD and phenomenology, they afforded me extraordinary freedom and ample space to explore. Maxim Mai and Marco Garofalo have been my closest collaborators, as much friends as mentors. I have discussed many of life's most consequential decisions with them, and my career has been far smoother for their presence. Within and beyond the Bonn group, I have had the good fortune to befriend Aniket Sen, Javi Suárez-Sucunza, Christiane Groß, Johann Ostmeyer, Marcus Petschlies, Ajay Sakthivasan, Akaki Rusetsky, Juncui Tao, Hailong Fu, Liping He, Bin Guo, Yonghui Lin, Hui Tong, Jing Gao, Xiangkun Dong, Teng Ji, Sonja Schäfer, Amina Rammeh, Narinder Kaur, Nicholas Kluge, and Sophia Falk. Through them, I have become a kinder person.

At Peking University, my path often crossed those of Zhanlin Wang, Ning Li, Hang Liu, Ting Chen, Yahui Chai, Yuan Li, Kelong Zhang, Chao Xiong, Yu Meng, Zuoheng Zou, and Jinlong Dang in our group; Yang Fu, Xinyu Tuo, Xuanhe Wang, Ziyu Wang, Xiaoyang Wang, Pengxiang Ma, Teng Wang, Yusheng Gao, Chenfei Lu, Tian Lin, Congling Fan, Zhaolong Zhang, Qiyuan Luo, Guanqi Wang, and Bohao Jian in Professor Xu Feng's group; Yan Lyu in Professor Jie Meng's group; Hanyang Xing and Zhicheng Hu at the Institute of Modern Physics; and Xueying Han at the Institute of High Energy Physics. Our conversations about science, ideals, and society form many of the most vivid memories of my doctoral years. I am especially grateful to Yu Meng for his generous guidance in research, and to Jinlong Dang, Pengxiang Ma, Chenfei Lu, Tian Lin, Congling Fan, and Zhaolong Zhang for the countless long conversations in which we spoke our minds about the troubles of our times, for the journeys we took together, and for all that we shared without reserve over the years.

The dear friends I have met at each stage of my life have gradually become one of its most essential parts. They include Hao Wen, Wenhua Shang, and Renquan Hu from Wanbailin No. 8 Middle School; Siyao Zhou, Yang Wang, Yuchen Zhu, Jingyu Lu, Yanfei Huang, Junjie Liu, Yuchen Zhao, Ruocheng Jia, Zelin Zhang, and Shengchen Liu from Taiyuan No. 5 Middle School; Zhen Chen, Hengjie Shi, Minghui Sun, Yingzhuo Han, and Yajing Wei from Xi'an Jiaotong University; and Hanlin Song, Junchen Wang, Teng Wang, Binghui Ge, Jingjun Zhang, Haohao Zhang, Ruiqi Wang, Liangzhen Wen, Xiaoxuan Ding, Kexin Liu, Chenjia Zhang, Guanyue Wan, and Xinran Wang from Peking University. For most of my life, they have shared in both my joys and my sorrows. The force of their characters has shaped me and made me a more complete person.

The stories I share with many of them would each merit a book of their own. But space is finite, and I must rein in my thoughts before they wander too far.

I was twenty-six when I wrote this dissertation, and fortunate to have been born into such a time. These people have tenderly guarded both my youthful audacity and my ideals. With them beside me, I look forward to the challenges ahead with all my heart.



\end{document}